# The Schulze Method of Voting

Markus Schulze
Markus.Schulze@Alumni.TU-Berlin.de

**Summary.** In recent years, the Pirate Party of Sweden, the Wikimedia Foundation, the Debian project, the Gentoo project, and many other private organizations adopted a new single-winner election method for internal elections and referendums. In this paper, we will introduce this method, demonstrate that it satisfies e.g. resolvability, Condorcet, Schwartz, Smith-IIA, Pareto, reversal symmetry, monotonicity, prudence, and independence of clones and present an O(*C*^3) algorithm to calculate the winner, where *C* is the number of alternatives.

## Summary

In this paper, we will propose a new single-winner election method (section 2). This method will be illustrated on a large number of examples (section 3). We will show that this method satisfies a large number of desirable criteria (section 4), e.g. *resolvability* (section 4.3), *reversal symmetry* (section 4.5), *monotonicity* (section 4.6), *independence of clones* (section 4.7), *prudence* (section 4.10), *k-consistency* (section 4.14), and *composition consistency* (section 5). In those cases, where the proposed method violates a criterion, this will be shown by using concrete counterexamples; see e.g. sections 3.7, 3.8, and 3.9.

In section 4, we use a deterministic model for election methods. In section 5, we will show what we have to take into consideration when we use a probabilistic model instead.

In sections 8 and 9, the proposed method will be generalized to "Proportional Representation by the Single Transferable Vote" (STV) and to methods to calculate a proportional ranking. Sections 8 and 9 differ significantly from the other sections. This is caused by the fact that, while we have the McGarvey method to easily create instances for single-winner elections (McGarvey, 1953), we don't have a similar method to create instances for multi-winner elections; therefore, large real-life instances are usually used to illustrate STV methods (Tideman, 2000); see section 8.2 of this dissertation. Furthermore, while there is a giant literature on mathematical aspects of single-winner election methods, there are no established criteria for STV methods; however in section 8.3, we will propose generalizations of the Condorcet criterion and the Smith criterion to STV methods.

## Contents

## Symbols

| | |
|---|---|
| $\wedge$ | ... and ... |
| $\vee$ | ... or ... |
| $\forall$ | ... for all ... |
| $\exists$ | ... there is at least one ... |
| $\in$ | ... element of ... |
| $\notin$ | ... not element of ... |
| $\setminus$ | "$a \in B_1 \setminus B_2$" means "$a \in B_1$ and $a \notin B_2$". |
| $\cap$ | "$a \in B_1 \cap B_2$" means "$a \in B_1$ and $a \in B_2$". |
| $\cup$ | "$a \in B_1 \cup B_2$" means "$a \in B_1$ or $a \in B_2$". |
| $\subseteq$ | "$B_1 \subseteq B_2$" means<br>"$B_1$ is a subset of $B_2$ or even identical to $B_2$". |
| $\subsetneq$ , $\subset$ | "$B_1 \subsetneq B_2$" means<br>"$B_1$ is a subset of $B_2$, but not identical to $B_2$". |
| $\Rightarrow$ | ... then ... |
| $\Leftrightarrow$ | ... then and only then ... |
| $:\Leftrightarrow$ | ... by definition then and only then ... |
| $\succ_v$ | "better than" or "greater than" according to $v$ |
| $\succsim_v$ | "better than or equal to" or "greater than or equal to" according to $v$ |
| $\prec_v$ | "worse than" or "smaller than" according to $v$ |
| $\precsim_v$ | "worse than or equal to" or "smaller than or equal to" according to $v$ |
| $\approx_v$ | "equal to" according to $v$ |
| $:=$ | ... equal by definition ... |
| $\times$ | a tuple<br>For example: "$a \in \mathbb{N}_0 \times \mathbb{N}_0$" means "$a$ is an ordered pair of two (not necessarily different) elements of $\mathbb{N}_0$". |
| $A$ | set of alternatives |
| $A_M$ | $A_M$ is the set of the $(C!)/((M!)\cdot((C-M)!))$ possible ways to choose $M$ different alternatives from the set $A$ of $C$ alternatives. |
| $C$ | $C \in \mathbb{N}$ with $0 < M < C < \infty$ is the number of alternatives. |
| $M$ | $M \in \mathbb{N}$ with $0 < M < C < \infty$ is the number of seats. |
| $\max_D B$ | maximum element in set $B$ according to measure $D$ |
| $\min_D B$ | minimum element in set $B$ according to measure $D$ |
| $N$ | $N \in \mathbb{N}$ with $0 < N < \infty$ is the number of voters. |
| $N[a,b]$ | $N[a,b] \in \mathbb{N}$ with $0 \leq N[a,b] \leq N$ is the number of voters who prefer alternative $a \in A$ to alternative $b \in A \setminus \{a\}$. |
| $N[\{a_1,...,a_M\};b]$ | $N[\{a_1,...,a_M\};b] \in \mathbb{R}_{\geq 0}$ with $0 \leq N[\{a_1,...,a_M\};b] \leq N / M$ is the largest value such that each alternative in $\{a_1,...,a_M\} \subset A$ has a separate quota of this value against alternative $b \in A \setminus \{a_1,...,a_M\}$. |
| $\mathbb{N}$ | natural numbers without zero, $\mathbb{N} = \{1, 2, 3, ...\}$ |
| $\mathbb{N}_0$ | natural numbers with zero, $\mathbb{N}_0 = \{0, 1, 2, 3, ...\}$ |
| $O$ | collective ranking |
| $\emptyset$ | the empty set |
| $P_D[a,b]$ | $P_D[a,b] \in \mathbb{N}_0 \times \mathbb{N}_0$ is the strength of the strongest path from alternative $a \in A$ to alternative $b \in A \setminus \{a\}$ when the strength is measured by $D$ ( where $D$ can e.g. be *margin*, *ratio*, *winning votes* or *losing votes* ). |
| $p[O]$ | $p[O] \in \mathbb{R}_{\geq 0}$ with $0 \leq p[O] \leq 1$ is the probability that the binary relation $O$ wins. |
| $q[a,b]$ | $q[a,b] \in \mathbb{R}_{\geq 0}$ with $0 \leq q[a,b] \leq 1$ is the probability that alternative $a \in A$ is ranked above alternative $b \in A \setminus \{a\}$ in the collective ranking. |
| $r[a]$ | $r[a] \in \mathbb{R}_{\geq 0}$ with $0 \leq r[a] \leq 1$ is the probability that alternative $a \in A$ wins. |
| $\mathbb{R}$ | real numbers |
| $\mathbb{R}_{\geq 0}$ | real numbers larger than or equal to zero |
| $\mathcal{S}$ | set of (potential) winners |
| $V$ | set of voters |
| $O\|_B$, $\mathcal{S}\|_B$, $q[a,b]\|_B$, $r[a]\|_B$ | ... when applied only to the alternatives in set $B$ ... |
| $\\| B \\|$ | number of elements in set $B$ |
| $\{ a \in B \mid property \}$ | set of elements of set $B$ with *property* |
| $\\| \{ a \in B \mid property \} \\|$ | number of elements of set $B$ with *property* |

## Abbreviations

| | |
|---|---|
| DDP | Dummett-Droop proportionality |
| DOI | digital object identifier (with open access) |
| DOI | digital object identifier (without open access) |
| IIA | independence of irrelevant alternatives |
| MMP | mixed member proportional representation |
| PRPL | proportional representation by party lists |
| STV | proportional representation by the single transferable vote |
| TBRC | tie-breaking ranking of the candidates |
| TBRL | tie-breaking ranking of the links |
| TBRV | tie-breaking ranking of the voters |

# 1. Introduction

One important property of a good single-winner election method is that it minimizes the number of “overruled” voters (according to some heuristic). Because of this reason, the Simpson-Kramer MinMax method, that always chooses that alternative whose worst pairwise defeat is the weakest, was very popular over a long time (Simpson, 1969; Kramer, 1977). However, in recent years, the Simpson-Kramer MinMax method has been criticized by many social choice theorists. J.H. Smith (1973) criticizes that this method doesn’t choose from the top-set of alternatives. Tideman (1987) complains that this method is vulnerable to the strategic nomination of a large number of similar alternatives, so-called *clones*. And Saari (1994) rejects this method for violating *reversal symmetry*. A violation of reversal symmetry can lead to strange situations where still the same alternative is chosen when all ballots are reversed, meaning that the same alternative is identified as best one and simultaneously as worst one.

In this paper, we will show that only a slight modification (section 4.9) of the Simpson-Kramer MinMax method is needed so that the resulting method satisfies the criteria proposed by J.H. Smith (section 4.8), Tideman (section 4.7), and Saari (section 4.5). The resulting method will be called *Schulze method*. Random simulations by Wright (2009) confirmed that, in almost 99% of all instances, the Schulze method conforms with the Simpson-Kramer MinMax method (table 11.1). In this paper, we will prove that, nevertheless, the Schulze method still satisfies all important criteria that are also satisfied by the Simpson-Kramer MinMax method, like resolvability (section 4.3), Pareto (section 4.4), monotonicity (section 4.6), and prudence (section 4.10). Because of these reasons, already several private organizations have adopted the Schulze method.

**1997 – 2006:** In 1997, I proposed the Schulze method to a large number of people, who are interested in mathematical aspects of election methods. This method was discussed for the first time in a public mailing list in 1997/1998 (e.g. Schulze, 1997, 1998a, 1998b, 1998c; Ossipoff, 1998; Petry, 1998), when it was discussed at the *Election-Methods mailing list*. In June 2003, Debian, a software project with about 1,000 members, adopted this method in a referendum with 144 against 16 votes (www001); Debian GNU/Linux is the largest and most popular non-commercial Linux distribution. In May 2005, the Gentoo Foundation, a software project with about 200 members, adopted this method (www002); Gentoo Linux is another wide-spread Linux distribution.

**2007 – 2012:** In 2008, 2009, and 2011, the Wikimedia Foundation, a non-profit charitable organization with about 43,000 members (in 2011), used the proposed method for the election of its Board of Trustees (www003); the Wikimedia Foundation is the umbrella organization e.g. for Wikipedia, Wiktionary, Wikiquote, Wikidata, Wikibooks, Wikisource, Wikinews, Wikivoyage, Wikiversity, and Wikispecies; in 2011, Wikimedia was the fourth most important Internet corporation (after Alphabet/Google/YouTube, Facebook/WhatsApp/Instagram, and Yahoo!). In July 2008, Ubuntu, a software project with about 700 members, adopted this method (www004). In August 2008, “K Desktop Environment” (KDE), a software project with about 150 members, adopted this method (www005). In October 2009, the “Pirate Party of Sweden” (about 1,800 members) adopted this method (www006). In November 2009, the “European Democratic Education Community” (EUDEC; about 8,000 members) adopted this method (www007). In May 2010, the “Pirate Party of Germany” (about 6,000 members) adopted this method. In November 2010, OpenStack, a software project with about 3,500 members, adopted this method (www008). Since February 2011, the “Pirate Party of Austria” (about 200 members) uses this method (www009); in 2013, the “Pirate Party of Austria”, the “Communist Party of Austria”, and “The Change” used the proposed method to create a joint list (“A Different Europe”) for the 2014 elections to the European Parliament (www010). Since November 2011, the “Pirate Party of Australia” (about 1,300 members) uses this method (www011). In December 2011, the “Pirate Party of Italy” adopted this method (www012). In July 2012, the “United States Pirate Party” (about 3,000 members) adopted this method (www013).

**2013 – 2019:** Since January 2013, the “Pirate Party of Iceland” (about 5,000 members) uses this method (www014). Since April 2013, the associated student government at Northwestern University (about 20,000 members; www015) uses this method. Since October 2013, the “Five Star Movement” (“Movimento 5 Stelle”, M5S), a political party in Italy with about 120,000 members, uses this method (www016). In December 2013, the “Club of the Alumni of the German Student Academies” (“Club der Ehemaligen der Deutschen SchülerAkademien”; CdE; about 4,500 members) adopted this method (www017). Since January 2014, the “Pirate Party of Belgium” (about 400 members) uses this method (www018); frequently, the “Pirate Party of Belgium” and the “Volt Party of Belgium” are running with a joint list (“Purple Party of Belgium”) which is then also generated with the proposed method (www019). Since May 2014, the associated student government at Albert Ludwig University of Freiburg (about 25,000 members) uses this method (www020). Since January 2015, the “Pirate Party of the Netherlands” (about 1,800 members) uses this method (www021). In February 2016, the city of Silla (about 20,000 inhabitants) in Spain adopted the Schulze method for referendums (www022; Gómez Álvarez, 2018). In July 2016, the “European Students’ Forum” (“Association des états généraux des étudiants de l’Europe”, AEGEE), a student organization with about 13,000 members, adopted this method (www023). Since January 2017, Podemos, a political party in Spain with about 520,000 members, uses this method (www024). In March 2017, the “Internet Corporation for Assigned Names and Numbers” (ICANN) adopted the Schulze method for the election of its board and the board of the “Address Supporting Organization” (ASO), a supporting organization affiliated with ICANN (www025). In December 2017, the “Associated Students of Minerva Schools at Keck Graduate Institute” (about 1,000 members) adopted the Schulze method (www026). In June 2019, “Volt Europa”, a pan-European political party with about 3,000 members, adopted the Schulze method for internal decisions (www027).

**Today (October 2026),** the proposed method is used by more than 200 organizations with more than 4,000,000 members in total. Therefore, the proposed method is more wide-spread than all other Condorcet-consistent single-winner election methods combined. This method is also used by Kleros (www028), Knative (www029), Kubernetes (www030), Kumoricon (www031), GraphQL (www032), Dapr (www033), Helm (www034), Hacksburg (www035), Homebrew (www036), StarlingX (www037), Airship (www038), Brigade (www039), NetBSD (www040), FOSDEM (www041), PETSc (www042), Ergodis (www043), Évriichi (www044), Squeak (www045), Tekton (www046), Sugar Labs (www047), RLLMUK (www048), FFmpeg (www049), foOpgp (www050), OpenSwitch (www051), OpenEmbedded (www052), OCaml (www053), DataSHIELD (www054), D-CENT (www055), TestPAC (www056), Ithaca Generator (www057), Fair Trade Northwest (www058), “Gothenburg Hackerspace” (www059), “Olten now!” (“Olten jetzt!”, www060), “Free Geek” (www061), “Board Game Geek” (www062), the “Vidya Gaem Awards” (www063), the “Free Software Foundation Latin America” (FSFLA, www064), the “Open Neural Network Exchange” (ONNX, www065), the “Python Software Foundation” (www066), the “Cloud Foundry Foundation” (www067), the “Cloud Native Computing Foundation” (www068), the “Computational Geometry Society” (www069), the “LIGO Scientific Collaboration” (www070), the “FinOps Foundation” (www071), the Boise, Idaho, section of the Democratic Socialists of America (DSA, www072), the Pirate Party of Portugal (www073), the “Australian Better Families Party” (www074), the Burgerlijst movement in Belgium (www075), the “Democratic Foundation of Austria” (“demokratisches bündnis österreich”; dbö; www076), the “Democracy Centre Vienna” (“Demokratiezentrum Wien”, www077), the Mathematics Department of Tufts University (Merow, 2016; www078), the “Yale Postdoctoral Association” (www079), the “International Union for the Study of Social Insects” (IUSSI; www080), the Mathematics Society of the University of Waterloo (www081), the Math Club at Western University (www082), the University of Michigan Fencing Club (www083), the Open Computing Facility of the University of California at Berkeley (www084), the associated student government at the Catholic University of Leuven (about 68,000 members; www085), the associated student government at the Computer Sciences Department of Rhineland-Palatinate University of Technology at Kaiserslautern-Landau (RPTU; www086), the Innocenzo Manzetti Institute of Technology at Aosta (www087), the associated student government of the High School of Paris (“délégation générale des élèves de l’école normale supérieure de Paris”; about 2,700 members; www088), the “Canadian Undergraduate Mathematics Conference” (CUMC, www089), the “International Olympiad in Informatics” (www090), and the “Groupe francophone des Utilisateurs de TEX” (GUT or GUTenberg; Chupin, 2022). It is used by the “Swedish National Union of Students” (“Sveriges förenade studentkårer”; SFS; about 380,000 members; www091), the “Union of Computer Science Students at the Swedish Royal Institute of Technology” (www092), the “Union of Media Technology Students at the Swedish Royal Institute of Technology” (www093), the “Union of Physics and Mathematics Students at the Swedish Royal Institute of Technology” (www094), the “Union of Students of Energy and Environmental Sciences at the Swedish Royal Institute of Technology” (www095), the “Union of Students of Aerospace Engineering at the Swedish Royal Institute of Technology” (www096), the Swedish Tourist Association (“Svenska Turistföreningen”; about 300,000 members; www097), the Swedish Kennel Club (“Svenska Kennelklubben”; SKK; about 300,000 members; www098), the Swedish Society for Nature Conservation (Naturskyddsföreningen; about 230,000 members; www099), the Swedish Trade Union of Public Employees (Vision; about 200,000 members; www100), the Swedish Trade Union for Academics (Akavia; about 135,000 members; www101), the Swedish Association for Nursery and Midwifery (Vårdförbundet; about 114,000 members; www102), the Swedish Union for Service and Communications Employees (“Service- och

Kommunikationsfacket”; SEKO; about 71,000 members; www103), the Swedish Scouts (Scouterna; about 86,000 members; www104), the Stockholm Disability Rights Federation (“Funktionsrätt Stockholms län”; about 70,000 members; www105), the Swedish Uniting Church (Equmeniakyrkan; about 65,000 members; www106), the Swedish Gaming Association (Sverok; about 55,000 members; www107), the Animal Rights Association of Sweden (“Djurens Rätt”; about 50,000 members; www108), the Swedish section of the “International Organisation of Good Templars” (IOGT-NTO; about 46,000 members; www109), the Stockholm County section of the Swedish Social Democratic Party (“Socialdemokraterna i Stockholm”; about 25,000 members; www110; www111), the Swedish Association for Culture and Communication (“dokumentation, information och kultur”; DIK; about 21,000 members; www112), the Swedish Adult Education Association (“Studieförbundet Vuxenskolan”; about 19,000 members; www113), the Swedish Sports Club (“Allmänna Idrottsklubben”; AIK; about 18,000 members; www114), the Church of Sweden Youth (“Svenska Kyrkans Unga”; about 13,000 members; www115), the Norden Association of Sweden (“Föreningen Norden”; about 13,000 members; www116), the Swedish Association of Young Physicians (“Sveriges Yngre Läkares Förbund”; SYLF; about 12,000 members; www117), the Green Party of Sweden (“Miljöpartiet de Gröna”; about 10,000 members; www110), the Swedish Youth Sobriety Association (“Ungdomens Nykterhetsförbunds”; UNF; about 7,000 members; www118), the “Swedish Federation for the Rights of Homosexuals, Bisexuals, Transgender, Queer and Intersex People” (“Riksförbundet för homosexuellas, bisexuellas, transpersoners, queeras och intersexpersoners rättigheter”; RFSL; about 7,000 members; www119), the Swedish Library Association (“Svensk Biblioteksförening”; about 3,000 members; www120), the Swedish Climate Trade Association (“Föreningen Klimatriksdagen”; www121), the Swedish Association for Sexuality Education (“Riksförbundet för sexuell upplysning”; RFSU; www122), 4H Sweden (www123), and the Skåne section of the Swedish Centre Party (“Centerpartiet i Skåne”; www124) through their use of the VoteIT decision tool. It is also used by the “Association for Computing Machinery” (ACM; about 110,000 members), by the “Institute of Electrical and Electronics Engineers” (IEEE; about 486,000 members), by the “Internet Society” (ISOC; about 70,000 members), and by USENIX to manage their conference review processes (Mao, 2018; Wenisch, 2018). The IEEE also uses the Schulze method to elect its conference chairs (www125). It is also used by many housing cooperatives, like “Kingman Hall” (Poundstone, 2008, page 224; www126), “Technology House” (www127), “Hillegass-Parker House” (www128), and “3HäuserProjekt” (www129). It is used by the Russian (www130) and the Persian (www131) Wikipedia sections for the elections of their Arbitration Committees. It is used by the English (Poundstone, 2008, pages 221–222; www132), the German (www133), the French, and the Hebrew Wikipedia sections for their internal decision-making processes. It is used by the Pirate Parties of Brazil, Mexico (“wikiPartido Pirata Mexicano”), and Switzerland, by “Slow Food Germany” (about 14,000 members), and by the Synaxon company (about 300 employees; Potschka, 2013; www134) through their use of the LiquidFeedback decision tool. It is used by the Pirate Party of France through their use of the Congressus decision tool (www135). It is used by the cities of Turin (about 900,000 inhabitants) and San Donà di Piave (about 40,000 inhabitants) and by the London Borough of Southwark (about 300,000 inhabitants) through their use of the WeGovNow platform, which in turn uses LiquidFeedback (www136). It is used in Paris (about 2,200,000 inhabitants), Athens (about 660,000 inhabitants), and (again) Turin through the CO3 project, which in turn (again) uses LiquidFeedback (www137). It is used by the city of Desio (about 42,000 inhabitants) in Italy for participatory budgeting (Comparetto, 2022; www138). Hardt (2015) writes that the proposed method is used among the Google staff (about 85,000 employees) for internal decisions. Chandler (2008) and B.M. Hill (2008) even write that MTV uses this method to decide which music videos go into rotation.

Furthermore, the proposed method is used by many Internet decision support systems, like the “Debian Vote Engine” (devotee), “Condorcet Internet Voting Service” (CIVS), “Modern Ballots”, GoogleVotes (Hardt, 2015; Paulin, 2019), Google DeepMind’s “Habermas Machine” (Tessler, 2024), Habermolt (Low, 2026; www139), “L’Expérience Démocratique” (www140), LiquidFeedback (Behrens, 2014a), Selectricity (B.M. Hill, 2008), Votator, schulzevote@DokuWiki (www141), ForceRank (www142), Zeus, CondorcetVote, JungleVote, LunchVote, VoteIT, Airesis, Agreeder, Belenios, Ordinos, Elekto, preftools, HotCRP (www143), Open Assembly, OpenAgora, and OpenSTV/OpaVote.

There has been some debate about an appropriate name for this method. Some people suggested names like “beatpath(s)”, “beatpath method”, “beatpath winner”, “beatpath matrix”, “beatpath tournament”, “path method”, “path voting”, “path winner”, “path matrix”, “widest path method”, “Schwartz sequential dropping” (SSD), and “cloneproof Schwartz sequential dropping” (CSSD). Brearley (1999) suggested names like “descending minimum gross score” (DminGS), “descending minimum augmented gross score” (DminAGS), and “descending minimum doubly augmented gross score” (DminDAGS), depending on how the strength of a pairwise link is measured. Heitzig (2002) suggested names like “strong immunity from binary arguments” ($sIm_A$) and “sequential dropping towards a spanning tree” (SDST). However, I prefer the name “Schulze method”, not because of academic arrogance, but because the other names do not refer to the method itself but to specific heuristics for implementing it, and so may mislead readers into believing that no other method for implementing it is possible. The calculation of the result of the Schulze method is called “all-pairs maximum capacity paths problem”, “all-pairs bottleneck paths problem” (Sornat, 2021) or “all-pairs directed widest paths problem” (Eppstein, 2021).

In section 2 of this paper, the Schulze method is defined. In section 3, this method is applied to concrete examples. In section 4, this method is analyzed. Detailed descriptions of this method can also be found in publications by Schulze (2003, 2011a, 2016a, 2016b), Tideman (2006, pages 228–232; 2019), Stahl (2006, 2017), McCaffrey (2008a, 2008b), Börgers (2009, pages 37–42), Camps (2008, 2012a, 2012b, 2013, 2014a, 2014b, 2015), Behrens (2014a), D. Müller (2014, 2015, 2019, 2022), and Okumura (2026).

Short descriptions and discussions of the Schulze method can also be found in papers by Green-Armytage (2004), P.A. Taylor (2004), Meskanen (2006a, 2006b, 2008), Wilke (2007), Yue (2007), Grabmeier (2009), Wright (2009), Jennings (2010), Rivest (2010), Švec (2011), Abisheva (2012), Bucovetsky (2012), Gaspers (2012), Grünheid (2012, 2014, 2015, 2016), Negriu (2012), Parkes (2012), Peyre (2012), A. Taylor (2012), Basilico (2013), Lawrence (2013), Menton (2013a, 2013b), J. Müller (2013, 2020), Parkes (2013), Felsenthal (2014), Li (2014), Mattei (2014), Reisch (2014), Gracia-Saz (2015), Schend (2015), V.A. Tushavin (2015, 2016, 2018, 2020),

Ádám (2016), Baumeister (2016), Caragiannis (2016), Contucci (2016, 2019), Darlington (2016, 2018, 2023; see also: Schulze, 2024a), Diethelm (2016), Fischer (2016), Hemaspaandra (2016), Pan (2016), Ruiz-Padillo (2016, 2018), Shah (2016), Sosnick (2016), Becirovic (2017), Golyshkov (2017), Hazra (2017), Hoang (2017), Izetta (2017), D.M. Khudoley (2017, 2021), Kostenko (2017), Louridas (2017), Mathur (2017), Ohrn (2017), Pérez-Fernández (2017a, 2017b, 2019), Salamon (2017), Sekar (2017), Skowron (2017), Tozer (2017), Bray (2018), Bubboloni (2018, 2026), Fernandez (2018), Hausberger (2018), Mayer (2018), Milosz (2018), Savvateev (2018), Tran (2018), Trapanese (2018), Wilkinson (2018), Abramov (2019), Alekseev (2019), Belayadi (2019, 2021), Budd (2019), Chechliński (2019), Devlin (2019), Krenn (2019), Kurz (2019), McInnes (2019), Aziz (2020), Cellier (2020), Costa (2020), Cummings (2020), Etesamipour (2020), Heilmaier (2020), Mori (2020), Serafini (2020), Sychev (2020), Yaremenko (2020, 2025), Agon (2021), Cachel (2021, 2022), Charignon (2021), Holliday (2021a–2021c, 2023a, 2023b, 2025), Moreno Calderón (2021), Parzen (2021), Podverbnyy (2021), Xia (2021a–2021c, 2022, 2023), Zovko (2021), Frolova (2022), Goaoc (2022), Jelić (2022), Paszkowska (2022), Pintar (2022), Yujie Wang (2022), Aazami (2023), Gori (2023, 2025, 2026), Kang (2023), Lanctot (2023), Siahaan (2023), Y.V. Tushavin (2023), Cindio (2024), Matone (2024), Ngoie (2024), Puhan (2024), Skowron (2024), Bochkov (2025), Merino (2025), Vansprengel (2025), and Tuynman (2026).

Applications of the Schulze method are described in papers by Narizzano (2006a–2006d, 2007), Ghersi (2007), Callison-Burch (2009), Lommatzsch (2009), Souza (2009), Prati (2010, 2012), Arguello (2011a–2011c, 2013, 2017), Audhkhasi (2011, 2013), Gelder (2011), Maheswari (2012, 2013), Muldoon (2012), Oryńczak (2012, 2014), Bohne (2013, 2015), Zhou (2013, 2014), Akbib (2014a, 2014b), Garg (2014), Lawonn (2014), S. Wang (2014), Baer (2015a, 2015b), Bountris (2015a, 2015b, 2017), Degeest (2015, 2021), Nguyen (2015), Plösch (2015), Proag (2015), Sompolinsky (2015), Zohar (2015), Aswatha (2016), Barradas-Bautista (2016, 2017), Cai (2016), Chen (2016), Mangeli (2016), Vargas (2016), Verdiesen (2016, 2018), Xexéo (2016), Avramenko (2017), Işıklı (2017, 2023), Moal (2017), Moroney (2018), Bagheri (2019), Laslier (2019), Lotfi (2019), T. Thomas (2019), Lihu (2020), Melekhov (2020), Mohan (2020), Alves da Silveira (2021), Chambers (2021), Lesaege (2021), Ghajar (2022), Majd (2022), Lambert (2023), Stodt (2023), Corina (2024), Staffieri (2024), Tosi (2025), Axenie (2026), and Venkatesan (2026).

Cases, where the Schulze method is used to evaluate empirical data, are mentioned by Katirai (2006), Morales (2008, 2010), Nanayakkara (2009), Wimmer (2009, 2010), Foale (2010), Gordevičius (2010a, 2010b), Hennecke (2013), Kowalski (2013), Casadebaig (2014), Pallett (2014), Chua (2015), Evita (2015), Rijnsburger (2015, 2017), Bilalov (2016), Dell (2016, 2017), Maio (2016), Vaughan (2016), Erhardt (2017), Gervits (2017), Strasser (2017), Wan (2017), Xue (2017), Chang (2018, 2021), Darras (2018a–2018d, 2020), Eng (2018), McKenna (2018), Muiños (2018), Al-Rousan (2018), Zarbafian (2018), Zheng (2018, 2020), the BBGLab (2019), Marian (2019, 2020, 2021), McGovern (2019), Sageder (2019), Forster (2020), Fortin (2020), Martínez-Jiménez (2020), Rajeh (2020a–2020d, 2021a, 2021b), Bener (2021), Bureš (2021, 2022), Fanourakis (2021), Freitas (2021, 2022), Mahmoudpour (2021), Pereira (2021), Safdar (2021), Shen (2021, 2023), Cawood (2022), Reynolds (2022), Santos (2022, 2023a, 2023b), Stanke (2022, 2023, 2024), Zabidov (2022), Zizien (2022), Barbour (2023, 2024), Biswas (2023), Han (2023), Jing (2023), Sheshko (2023), Sterzik

(2023), Yan (2023a, 2023b), Oliver Zendel (2023), Zyl (2023), Campello (2024), Cornish (2024), Culliford (2024), Hosseini (2024), Kinnersley (2024), Maheta (2024), Mills (2024), Rapp (2024), Tessler (2024), Oleg Zendel (2024), Gao (2025), Skinner (2025), Yun Wang (2025), Yun (2025), Cisse (2026), Kim (2026), Meyer (2026), Oei (2026), Revel (2026), and E. Thomas (2026).

Implementations of the Schulze method are decribed in papers by Csar (2018a, 2018b), Holtkamp (2018), Anand (2019), Hertel (2020, 2021), Yadav (2020), Sornat (2021), Cortier (2022), Harrison (2022, 2024a, 2024b), Liedtke (2023), Yang (2023), Anandita (2024), Vardanian (2024), Tokar (2025), Arora (2026), and Fujishige (2026). Aspects of computational social choice, with regard to the Schulze method, are discussed by Gaspers (2012), Parkes (2012), Menton (2013a, 2013b), J. Müller (2013, 2020), Mattei (2014), Reisch (2014), Durand (2015, 2026), Schend (2015), Baumeister (2016), Conitzer (2016), Faliszewski (2016), Hemaspaandra (2016), Maushagen (2020, 2021, 2024), Blokhina (2024a, 2024b), Chen (2025), Anurova (2026a, 2026b), Cechlárová (2026), Chavrimootoo (2026), and Woitaschik (2026). Metric distortion of the Schulze method is discussed by Goel (2017), Krishnaswamy (2019), Munagala (2019), Pierczyński (2019a, 2019b, 2024), and Kempe (2020). An interesting observation about the Schulze method is that it is Coq-verifiable (Moses, 2017; Pattinson, 2017; Ghale, 2019, 2021; Haines, 2019; Tiwari, 2021a, 2021b, 2022), which is a very strong form of verifiability.

# 2. Definition of the Schulze Method

## 2.1. Preliminaries

We presume that $A$ is a finite and non-empty set of alternatives. $C \in \mathbb{N}$ with $1 < C < \infty$ is the number of alternatives in $A$.

A binary relation $\succ$ on $A$ is *asymmetric* if it has the following property:

$\forall\ a,b \in A$, exactly one of the following three statements is valid:

1. $a \succ b$.
2. $b \succ a$.
3. $a \approx b$ (where “$a \approx b$” means “neither $a \succ b$ nor $b \succ a$”).

A binary relation $\succ$ on $A$ is *irreflexive* if it has the following property:

$\forall\ a \in A$: $a \approx a$.

A binary relation $\succ$ on $A$ is *transitive* if it has the following property:

$\forall\ a,b,c \in A$: ( ( $a \succ b$ and $b \succ c$ ) $\Rightarrow$ $a \succ c$ ).

A binary relation $\succ$ on $A$ is *negatively transitive* if it has the following property (where “$a \succsim b$” means “not $b \succ a$”):

$\forall\ a,b,c \in A$: ( ( $a \succsim b$ and $b \succsim c$ ) $\Rightarrow$ $a \succsim c$ ).

A binary relation $\succ$ on $A$ is *linear* (or *total* or *complete*) if it has the following property:

$\forall\ a,b \in A$: ( $b \in A \setminus \{a\}$ $\Rightarrow$ ( $a \succ b$ or $b \succ a$ ) ).

A *strict partial order* is an asymmetric, irreflexive, and transitive relation. A *strict weak order* is a strict partial order that is also negatively transitive. A *linear order* (or *total order* or *complete order*) is a strict weak order that is also linear. A *profile* is a finite and non-empty list of strict weak orders each on $A$.

Input of the proposed method is a profile $V$. $N \in \mathbb{N}$ with $0 < N < \infty$ is the number of strict weak orders in $V := \{ \succ_1, ..., \succ_N \}$. These strict weak orders will sometimes be called "voters" or "ballots".

Suppose $V_1 := \{ \succ_1, ..., \succ_{N_1} \}$ and $V_2 := \{ \succ_{1'}, ..., \succ_{N_2'} \}$ are two profiles each on the same set of alternatives $A$. Then the concatenation of these two profiles will be denoted $V_1 + V_2 := \{ \succ_1, ..., \succ_{N_1}, \succ_{1'}, ..., \succ_{N_2'} \}$.

"$a \succ_v b$" means "voter $v \in V$ strictly prefers alternative $a \in A$ to alternative $b$". "$a \approx_v b$" means "voter $v \in V$ is indifferent between alternative $a$ and alternative $b$". "$a \succsim_v b$" means "$a \succ_v b$ or $a \approx_v b$".

Output of the proposed method is (1) a strict partial order $O$ on $A$ and (2) a set $\emptyset \neq \mathcal{S} \subseteq A$ of potential winners.

A possible implementation of the Schulze method looks as follows:

> Each voter gets a complete list of all alternatives and ranks these alternatives in order of preference. The individual voter may give the same preference to more than one alternative and he may keep alternatives unranked. When a given voter does not rank all alternatives, then this means (1) that this voter strictly prefers all ranked alternatives to all not ranked alternatives and (2) that this voter is indifferent between all not ranked alternatives. The individual voter may also skip preferences; however, skipping preferences has no impact on the result of the elections since only the cast order of the preferences matters, not the absolute numbers.

Suppose $N[e,f] := \| \{ v \in V \mid e \succ_v f \} \|$ is the number of voters who strictly prefer alternative $e$ to alternative $f$. We presume that the strength of the link $ef$ depends only on $N[e,f]$ and $N[f,e]$. Therefore, the strength of the link $ef$ can be denoted $(N[e,f],N[f,e])$. We presume that a binary relation $\succ_D$ on $\mathbb{N}_0 \times \mathbb{N}_0$ is defined such that the link $ef$ is stronger than the link $gh$ if and only if $(N[e,f],N[f,e]) \succ_D (N[g,h],N[h,g])$. $N[e,f]$ is the *support* for the link $ef$; $N[f,e]$ is its *opposition*.

Example 1 (*margin*):

> When the strength of the link $ef$ is measured by *margin*, then its strength is the difference $N[e,f] - N[f,e]$ between its support $N[e,f]$ and its opposition $N[f,e]$.
>
> $(N[e,f],N[f,e]) \succ_{margin} (N[g,h],N[h,g])$ if and only if $N[e,f] - N[f,e] > N[g,h] - N[h,g]$.

Example 2 (*ratio*):

When the strength of the link $ef$ is measured by *ratio*, then its strength is the ratio $N[e,f] / N[f,e]$ between its support $N[e,f]$ and its opposition $N[f,e]$.

$(N[e,f],N[f,e]) \succ_{ratio} (N[g,h],N[h,g])$ if and only if at least one of the following conditions is satisfied:

1. $N[e,f] > N[f,e]$ and $N[g,h] \leq N[h,g]$.
2. $N[e,f] \geq N[f,e]$ and $N[g,h] < N[h,g]$.
3. $N[e,f] \cdot N[h,g] > N[f,e] \cdot N[g,h]$.
4. $N[e,f] > N[g,h]$ and $N[f,e] \leq N[h,g]$.
5. $N[e,f] \geq N[g,h]$ and $N[f,e] < N[h,g]$.

Comment: Condition 4 in the definition for $\succ_{ratio}$ is needed e.g. to say that $(N[e,f],N[f,e]) = (5,0)$ is stronger than $(N[g,h],N[h,g]) = (3,0)$; this doesn't follow from conditions 1, 2, 3, and 5. Condition 5 in the definition for $\succ_{ratio}$ is needed e.g. to say that $(N[e,f],N[f,e]) = (0,3)$ is stronger than $(N[g,h],N[h,g]) = (0,5)$; this doesn't follow from conditions 1 – 4.

Example 3 (*winning votes*):

When the strength of the link $ef$ is measured by *winning votes*, then its strength is measured primarily by its support $N[e,f]$.

$(N[e,f],N[f,e]) \succ_{win} (N[g,h],N[h,g])$ if and only if at least one of the following conditions is satisfied:

1. $N[e,f] > N[f,e]$ and $N[g,h] \leq N[h,g]$.
2. $N[e,f] \geq N[f,e]$ and $N[g,h] < N[h,g]$.
3. $N[e,f] > N[f,e]$ and $N[g,h] > N[h,g]$ and $N[e,f] > N[g,h]$.
4. $N[e,f] > N[f,e]$ and $N[g,h] > N[h,g]$ and $N[e,f] = N[g,h]$ and $N[f,e] < N[h,g]$.
5. $N[e,f] < N[f,e]$ and $N[g,h] < N[h,g]$ and $N[f,e] < N[h,g]$.
6. $N[e,f] < N[f,e]$ and $N[g,h] < N[h,g]$ and $N[f,e] = N[h,g]$ and $N[e,f] > N[g,h]$.

Example 4 (*losing votes*):

When the strength of the link $ef$ is measured by *losing votes*, then its strength is measured primarily by its opposition $N[f,e]$.

$(N[e,f],N[f,e]) \succ_{los} (N[g,h],N[h,g])$ if and only if at least one of the following conditions is satisfied:

1. $N[e,f] > N[f,e]$ and $N[g,h] \leq N[h,g]$.
2. $N[e,f] \geq N[f,e]$ and $N[g,h] < N[h,g]$.
3. $N[e,f] > N[f,e]$ and $N[g,h] > N[h,g]$ and $N[f,e] < N[h,g]$.
4. $N[e,f] > N[f,e]$ and $N[g,h] > N[h,g]$ and $N[f,e] = N[h,g]$ and $N[e,f] > N[g,h]$.
5. $N[e,f] < N[f,e]$ and $N[g,h] < N[h,g]$ and $N[e,f] > N[g,h]$.
6. $N[e,f] < N[f,e]$ and $N[g,h] < N[h,g]$ and $N[e,f] = N[g,h]$ and $N[f,e] < N[h,g]$.

The most intuitive definitions for the strength of a link are its *margin* and its *ratio*. However, we only presume that $\succ_D$ is a strict weak order on $\mathbb{N}_0 \times \mathbb{N}_0$.

For some proofs, we have to make additional presumptions for $\succ_D$. We will state explicitly when and where we take use of additional presumptions. Typical additional presumptions for $\succ_D$ are:

(2.1.1) (*positive responsiveness*)

$\forall\ (x_1,x_2),(y_1,y_2) \in \mathbb{N}_0 \times \mathbb{N}_0$:
$(\ (\ x_1 > y_1 \wedge x_2 \leq y_2\ ) \vee (\ x_1 \geq y_1 \wedge x_2 < y_2\ )\ ) \Rightarrow (x_1,x_2) \succ_D (y_1,y_2)$.

(2.1.2) (*reversal symmetry*)

$\forall\ (x_1,x_2),(y_1,y_2) \in \mathbb{N}_0 \times \mathbb{N}_0$:
$(x_1,x_2) \succ_D (y_1,y_2) \Rightarrow (y_2,y_1) \succ_D (x_2,x_1)$.

(2.1.3) (*homogeneity*)

$\forall\ (x_1,x_2),(y_1,y_2) \in \mathbb{N}_0 \times \mathbb{N}_0\ \forall\ c_1,c_2 \in \mathbb{N}$:
$(c_1 \cdot x_1,c_1 \cdot x_2) \succ_D (c_1 \cdot y_1,c_1 \cdot y_2) \Rightarrow (c_2 \cdot x_1,c_2 \cdot x_2) \succ_D (c_2 \cdot y_1,c_2 \cdot y_2)$.

The presumption, that the strength of the link *ef* depends only on $N[e,f]$ and $N[f,e]$, guarantees (1) that the proposed method satisfies anonymity and neutrality, (2) that adding a ballot, on which all alternatives are ranked equally, cannot change the result of the elections, and (3) that the proposed method is a C2 *Condorcet social choice function* (CSCF) according to Fishburn's (1977) terminology.

Presumption (2.1.1) says that, when the support of a link increases and its opposition doesn't increase or when its opposition decreases and its support doesn't decrease, then the strength of this link increases. So presumption (2.1.1) says that the strength of a link responds to a change of its support or its opposition in the correct manner. Presumption (2.1.1) guarantees that the proposed method satisfies decisiveness (section 4.2), resolvability (section 4.3), Pareto (section 4.4), and monotonicity (section 4.6). When each voter $v \in V$ casts a linear order $\succ_v$ on $A$, then all definitions for $\succ_D$, that satisfy presumption (2.1.1), are identical.

Presumption (2.1.2) says that, the stronger the link $(x_1,x_2)$ gets, the weaker the opposite link $(x_2,x_1)$ gets. Presumption (2.1.2) basically says that, when the individual ballots $\succ_v$ are reversed for all voters $v \in V$, then also the order of the links $(x_1,x_2) \succ_D (y_1,y_2)$ is reversed.

*Homogeneity* means that the result depends only on the proportion of ballots of each type, not on their absolute numbers. Presumption (2.1.3) guarantees that the proposed method satisfies homogeneity.

Presumptions (2.1.1) and (2.1.2) together guarantee that presumption (2.1.5) is satisfied. Presumption (2.1.5) guarantees that the proposed method satisfies majority criteria.

$\succ_{margin}$, $\succ_{ratio}$, $\succ_{win}$, and $\succ_{los}$ each satisfy (2.1.1) – (2.1.3).

**Definitions:**

Suppose $(x_1,x_2) \in \mathbb{N}_0 \times \mathbb{N}_0$.

If $x_1 > x_2$, then $(x_1,x_2)$ is a *victory*.

If $x_1 = x_2$, then $(x_1,x_2)$ is a *tie*.

If $x_1 < x_2$, then $(x_1,x_2)$ is a *defeat*.

**Corollary (2.1.4):**

If $\succ_D$ satisfies presumption (2.1.2), then all ties have equivalent strengths. In short:

(2.1.4) (*equivalence of ties*)

$$\forall\ x,y \in \mathbb{N}_0\text{: } (x,x) \approx_D (y,y).$$

**Proof of corollary (2.1.4):**

Suppose $(x,x) \succ_D (y,y)$ for some $x,y \in \mathbb{N}_0$. Then with (2.1.2), we get $(y,y) \succ_D (x,x)$. But this is a contradiction to the presumption $(x,x) \succ_D (y,y)$ and to the presumption that $\succ_D$ is a strict weak order. □

**Corollary (2.1.5):**

If $\succ_D$ satisfies presumptions (2.1.1) and (2.1.2), then (i) every victory is stronger than every tie and (ii) every tie is stronger than every pairwise defeat. In short:

(2.1.5) (*majority*)

$$\forall\ (x_1,x_2),(y_1,y_2) \in \mathbb{N}_0 \times \mathbb{N}_0\text{:}$$
$$(\ (\ x_1 > x_2 \wedge y_1 \leq y_2\ ) \vee (\ x_1 \geq x_2 \wedge y_1 < y_2\ )\ ) \Rightarrow (x_1,x_2) \succ_D (y_1,y_2).$$

**Proof of corollary (2.1.5):**

Suppose $(x_1,x_2) \in \mathbb{N}_0 \times \mathbb{N}_0$ with $x_1 > x_2$ is a victory.

Suppose $(y_1,y_2) \in \mathbb{N}_0 \times \mathbb{N}_0$ with $y_1 = y_2$ is a tie.

Suppose $(z_1,z_2) \in \mathbb{N}_0 \times \mathbb{N}_0$ with $z_1 < z_2$ is a defeat.

With (2.1.1), we get: $(x_1,x_2) \succ_D (x_2,x_2)$.

With (2.1.4), we get: $(x_2,x_2) \approx_D (y_1,y_2)$.

With (2.1.4), we get: $(y_1,y_2) \approx_D (z_1,z_1)$.

With (2.1.1), we get: $(z_1,z_1) \succ_D (z_1,z_2)$.

Therefore, we get: $(x_1,x_2) \succ_D (x_2,x_2) \approx_D (y_1,y_2) \approx_D (z_1,z_1) \succ_D (z_1,z_2)$.

Thus, we get (2.1.5). □

Suppose $\emptyset \neq \mathcal{M} \subset \mathbb{N}_0 \times \mathbb{N}_0$ is finite and non-empty. Then “$\max_D \mathcal{M}$”, the *set of maximum elements* of $\mathcal{M}$, and “$\min_D \mathcal{M}$”, the *set of minimum elements* of $\mathcal{M}$, are defined as follows: $(\beta_1,\beta_2) \in \max_D \mathcal{M}$ if and only if (1) $(\beta_1,\beta_2) \in \mathcal{M}$ and (2) $(\beta_1,\beta_2) \succsim_D (\delta_1,\delta_2)$ $\forall$ $(\delta_1,\delta_2) \in \mathcal{M}$. $(\gamma_1,\gamma_2) \in \min_D \mathcal{M}$ if and only if (1) $(\gamma_1,\gamma_2) \in \mathcal{M}$ and (2) $(\gamma_1,\gamma_2) \precsim_D (\delta_1,\delta_2)$ $\forall$ $(\delta_1,\delta_2) \in \mathcal{M}$.

We write “$(\beta_1,\beta_2) := \max_D \mathcal{M}$” and “$(\gamma_1,\gamma_2) := \min_D \mathcal{M}$” for “$(\beta_1,\beta_2)$ is an arbitrarily chosen element of $\max_D \mathcal{M}$” and “$(\gamma_1,\gamma_2)$ is an arbitrarily chosen element of $\min_D \mathcal{M}$”.

## 2.2. Basic Definitions

In this section, the Schulze method is defined. Concrete examples can be found in section 3.

Basic idea of the Schulze method is that the *strength* of the indirect comparison “alternative $a$ vs. alternative $b$” is the *strength* of the *strongest path* $a \equiv c(1),\ldots,c(n) \equiv b$ from alternative $a \in A$ to alternative $b \in A \setminus \{a\}$ and that the *strength* of a path is the *strength* $(N[c(i),c(i+1)],N[c(i+1),c(i)])$ of its *weakest link* $c(i),c(i+1)$.

The Schulze method is defined as follows:

A *path* from alternative $x \in A$ to alternative $y \in A \setminus \{x\}$ is a sequence of alternatives $c(1),\ldots,c(n) \in A$ with the following properties:

1. $x \equiv c(1)$.
2. $y \equiv c(n)$.
3. $n \in \mathbb{N}$ with $2 \leq n \leq C$.
4. For all $i,j \in \{1,\ldots,n\}$: $i \neq j \Rightarrow c(i) \in A \setminus \{c(j)\}$.

The *strength* of the path $c(1),\ldots,c(n)$ is

$$\min_D \{ (N[c(i),c(i+1)],N[c(i+1),c(i)]) \mid i = 1,\ldots,(n-1) \}.$$

In other words: The strength of a path is the strength of its weakest link.

When a path $c(1),\ldots,c(n)$ has the strength $(z_1,z_2) \in \mathbb{N}_0 \times \mathbb{N}_0$, then the *critical links* of this path are the links with $(N[c(i),c(i+1)], N[c(i+1),c(i)]) \approx_D (z_1,z_2)$.

$$P_D[a,b] := \max_D \{ \min_D \{ (N[c(i),c(i+1)],N[c(i+1),c(i)]) \mid i = 1,\ldots,(n-1) \} \mid c(1),\ldots,c(n) \text{ is a path from alternative } a \text{ to alternative } b \}.$$

In other words: $P_D[a,b] \in \mathbb{N}_0 \times \mathbb{N}_0$ is the strength of the strongest path from alternative $a \in A$ to alternative $b \in A \setminus \{a\}$.

(2.2.1) The binary relation $\mathcal{O}$ on $A$ is defined as follows:

$$ab \in \mathcal{O} :\Leftrightarrow P_D[a,b] >_D P_D[b,a].$$

(2.2.2) $\mathcal{S} := \{ a \in A \mid \forall\, b \in A \setminus \{a\}: ba \notin \mathcal{O} \}$ is the *set of potential winners*.

When there is only one potential winner $\mathcal{S} = \{a\}$, then this alternative is a *unique winner*.

When $P_D[a,b] \succ_D P_D[b,a]$, then we say "alternative $a$ disqualifies alternative $b$" or "alternative $a$ dominates alternative $b$".

As the link $ab$ is already a path from alternative $a$ to alternative $b$ of strength $(N[a,b],N[b,a])$, we get

(2.2.3) $\forall\ a,b \in A$: $P_D[a,b] \succsim_D (N[a,b],N[b,a])$.

With (2.2.1) and (2.2.3), we get

(2.2.4) $(N[a,b],N[b,a]) \succ_D P_D[b,a] \Rightarrow ab \in \mathcal{O}$.

A *route* from alternative $x \in A$ to alternative $y \in A \setminus \{x\}$ is a sequence of alternatives $c(1),\ldots,c(n) \in A$ with the following properties:

1. $x \equiv c(1)$.
2. $y \equiv c(n)$.
3. $n \in \mathbb{N}$ with $2 \leq n < \infty$.
4. For all $i = 1,\ldots,(n-1)$: $c(i+1) \in A \setminus \{c(i)\}$.

We have

(2.2.5) $\forall\ a,b,c \in A$: $\min_D \{ P_D[a,b], P_D[b,c] \} \precsim_D P_D[a,c]$.

Otherwise, if $\min_D \{ P_D[a,b], P_D[b,c] \}$ was strictly larger than $P_D[a,c]$, then this would be a contradiction to the definition of $P_D[a,c]$ since there would be a route from alternative $a$ to alternative $c$ via alternative $b$ with a strength of more than $P_D[a,c]$; and if this route was not itself a path (because it passed through some alternatives more than once) then some subset of its links would form a path from alternative $a$ to altenative $c$ with a strength of more than $P_D[a,c]$.

The asymmetry of $\mathcal{O}$ follows directly from the asymmetry of $\succ_D$. The irreflexivity of $\mathcal{O}$ follows directly from the irreflexivity of $\succ_D$. Furthermore, in section 4.1, we will see that the binary relation $\mathcal{O}$ is transitive. This guarantees that there is always at least one potential winner.

Suppose $\mathcal{LO}_A$ is the set of linear orders on $A$. Suppose $\mathcal{O}_0$ is the binary relation defined in (2.2.1). Then we define *potential winning rankings* as follows:

(2.2.6) $\mathcal{O}_1 \in \mathcal{LO}_A$ is a *potential winning ranking* of the Schulze method: $\Leftrightarrow \mathcal{O}_0 \subseteq \mathcal{O}_1$.

In section 4.1, we will prove that the binary relation $\mathcal{O}_0$, as defined in (2.2.1), is a strict partial order. Therefore, it is guaranteed that there is always at least one potential winning ranking.

## 2.3. Implementation

### 2.3.1. Calculation of the Schulze Winners Using the Floyd-Warshall Algorithm

In section 2.3.1, we explain how to calculate (1) the strict partial order $O$ on $A$ and (2) the set $\emptyset \neq \mathcal{S} \subseteq A$ of potential winners, as defined in section 2.2.

The strength $P_D[i,j]$ of the strongest path from alternative $i \in A$ to alternative $j \in A \setminus \{i\}$ can be calculated with the Floyd-Warshall (Floyd, 1962; Warshall, 1962) algorithm. The runtime to calculate the strengths of all strongest paths is O($C$^3), where $C$ is the number of alternatives in $A$.

Input: $N[i,j] \in \mathbb{N}_0$ is the number of voters who strictly prefer alternative $i \in A$ to alternative $j \in A \setminus \{i\}$.

Output: $P_D[i,j] \in \mathbb{N}_0 \times \mathbb{N}_0$ is the strength of the strongest path from alternative $i \in A$ to alternative $j \in A \setminus \{i\}$.

$pred[i,j] \in A \setminus \{j\}$ is the predecessor of alternative $j$ in the strongest path from alternative $i \in A$ to alternative $j \in A \setminus \{i\}$.

$O$ is the binary relation as defined in (2.2.1).

“$winner[i] = true$” if and only if $i \in \mathcal{S}$.

Stage 1 (initialization):

```
 1 | for i : = 1 to C
 2 | begin
 3 |    for j : = 1 to C
 4 |    begin
 5 |       if ( i ≠ j ) then
 6 |       begin
 7 |          P_D[i,j] : = (N[i,j],N[j,i])
 8 |          pred[i,j] : = i
 9 |       end
10 |    end
11 | end
```

Stage 2 (calculation of the strengths of the strongest paths):

```
for i : = 1 to C
begin
    for j : = 1 to C
    begin
        if ( i ≠ j ) then
        begin
            for k : = 1 to C
            begin
                if ( i ≠ k ) then
                begin
                    if ( j ≠ k ) then
                    begin
                        if ( P_D[j,k] ≺_D min_D { P_D[j,i], P_D[i,k] } ) then
                        begin
                            P_D[j,k] : = min_D { P_D[j,i], P_D[i,k] }
                            pred[j,k] : = pred[i,k]
                        end
                    end
                end
            end
        end
    end
end
```

Stage 3 (calculation of the binary relation $\mathcal{O}$ and the set of potential winners):

```
for i : = 1 to C
begin
    winner[i] : = true
    for j : = 1 to C
    begin
        if ( i ≠ j ) then
        begin
            if ( P_D[j,i] ≻_D P_D[i,j] ) then
            begin
                ji ∈ O
                winner[i] : = false
            end
            else
            begin
                ji ∉ O
            end
        end
    end
end
```

(α) It cannot be stressed frequently enough that the order of the indices in the triple-loop of the Floyd-Warshall algorithm is ***not*** irrelevant. When $i$ is the index of the outer loop of the triple-loop of the Floyd-Warshall algorithm, then the clause (line 24) must be " if ( $P_D[j,k] \prec_D \min_D \{ P_D[j,i], P_D[i,k] \}$ ) ". Otherwise, it is not guaranteed that a single pass through the triple-loop of the Floyd-Warshall algorithm is sufficient to find the strongest paths.

(β) With the predecessor matrix $pred[i,j]$, we can recursively determine the strongest paths. Suppose we want to determine the strongest path $c(1),...,c(n)$ from alternative $a \in A$ to alternative $b \in A \setminus \{a\}$. Then we start with

$$n := 1$$

$$d(1) := b$$

We repeat

$$n := n + 1$$

$$d(n) := pred[a,d(n-1)]$$

until we get $d(n) = a$ for some $n \in \mathbb{N}$. The strongest path $c(1),...,c(n)$ from alternative $a$ to alternative $b$ is then given by $d(n),...,d(1)$.

(γ) The runtime to calculate the pairwise matrix is O($N$·($C$^2)). The runtime of the Floyd-Warshall algorithm, as defined in this section, is O($C$^3). Therefore, the total runtime to calculate the binary relation $O$, as defined in (2.2.1), and the set $\mathcal{S}$, as defined in (2.2.2), is O($N$·($C$^2) + $C$^3).

### 2.3.2. Checking Whether Alternative *m* is a Schulze Winner Using the Dijkstra Algorithm

In section 2.3.2, we explain how to check whether a concrete alternative $m \in A$ is a potential winner.

Sometimes, we don't want to calculate all potential winners. We only want to check for a concrete alternative *m* whether it is a potential winner. In this case, we don't have to calculate the strengths $P_D[i,j]$ of the strongest paths from every alternative $i \in A$ to every other alternative $j \in A \setminus \{i\}$. It is sufficient to calculate the strengths of the strongest paths from alternative *m* to every other alternative $i \in A \setminus \{m\}$ and the strengths of the strongest paths from every other alternative $i \in A \setminus \{m\}$ to alternative *m*. This can be done with the Dijkstra (1959) algorithm in a runtime O(*C*^2).

Input: $N[i,j] \in \mathbb{N}_0$ is the number of voters who strictly prefer alternative $i \in A$ to alternative $j \in A \setminus \{i\}$.

$m \in A$ is that alternative for which we want to check whether it is a potential winner.

Output: $P_D[m,i] \in \mathbb{N}_0 \times \mathbb{N}_0$ is the strength of the strongest path from alternative *m* to alternative $i \in A \setminus \{m\}$.

$P_D[i,m] \in \mathbb{N}_0 \times \mathbb{N}_0$ is the strength of the strongest path from alternative $i \in A \setminus \{m\}$ to alternative *m*.

"*winner* = *true*" if and only if *m* is a potential winner.

Stage 1 (calculation of the strengths of the strongest paths from alternative $m$ to every other alternative $i \in A \setminus \{m\}$):

```
for i : = 1 to C
begin
    if ( i ≠ m ) then
    begin
        P_D[m,i] : = (N[m,i],N[i,m])
    end
end

for i : = 1 to C
begin
    list[i] : = i
end

list[m] : = list[C]

for i : = 1 to ( C – 1 )
begin
    j : = 1
    for k : = 2 to ( C – i )
    begin
        if  ( P_D[m,list[j]] ≺_D P_D[m,list[k]] ) then
        begin
            j : = k
        end
    end
    for k : = 1 to ( C – i )
    begin
        if ( j ≠ k ) then
        begin
            if  ( P_D[m,list[k]] ≺_D min_D { P_D[m,list[j]], (N[list[j],list[k]],N[list[k],list[j]]) } ) then
            begin
                P_D[m,list[k]] : = min_D { P_D[m,list[j]], (N[list[j],list[k]],N[list[k],list[j]]) }
            end
        end
    end
    list[j] : = list[ C – i ]
end
```

Stage 2 (calculation of the strengths of the strongest paths from every other alternative $i \in A \setminus \{m\}$ to alternative $m$):

```
for i : = 1 to C
begin
    if ( i ≠ m ) then
    begin
        P_D[i,m] : = (N[i,m],N[m,i])
    end
end

for i : = 1 to C
begin
    list[i] : = i
end

list[m] : = list[C]

for i : = 1 to ( C – 1 )
begin
    j : = 1
    for k : = 2 to ( C – i )
    begin
        if  ( P_D[list[j],m] ≺_D P_D[list[k],m] ) then
        begin
            j : = k
        end
    end
    for k : = 1 to ( C – i )
    begin
        if ( j ≠ k ) then
        begin
            if  ( P_D[list[k],m] ≺_D min_D { P_D[list[j],m], (N[list[k],list[j]],N[list[j],list[k]]) } ) then
            begin
                P_D[list[k],m] : = min_D { P_D[list[j],m], (N[list[k],list[j]],N[list[j],list[k]]) }
            end
        end
    end
    list[j] : = list[ C – i ]
end
```

Stage 3 (checking whether alternative $m$ is a potential winner):

```
winner : = true
for i : = 1 to C
begin
    if ( i ≠ m ) then
    begin
        if ( P_D[i,m] ≻_D P_D[m,i] ) then
        begin
            winner : = false
        end
    end
end
```

### 2.3.3. Checking Whether Alternative *a* Disqualifies Alternative *b*

Suppose that we have already guessed or determined that the statement " $ab \in O$ " is true. In section 2.3.3, we will show how we can then demonstrate the correctness of this statement.

To demonstrate that the statement " $ab \in O$ " is true, we have to present a $(x_1,x_2) \in \mathbb{N}_0 \times \mathbb{N}_0$ such that (1) there is a path from alternative $a$ to alternative $b$ with a strength of at least $(x_1,x_2)$ and (2) there is no path from alternative $b$ to alternative $a$ with a strength of at least $(x_1,x_2)$.

To demonstrate that there is a path from alternative $a$ to alternative $b$ with a strength of at least $(x_1,x_2)$, we can simply use the sequence $c(1),...,c(n)$ as calculated in remark β of section 2.3.1 or the path as determined in section 2.3.2 or a path found by guesswork. The runtime to verify that a given sequence is really a path from alternative $a$ to alternative $b$ with a strength of at least $(x_1,x_2)$ is O($C$).

When there is no path from alternative $b$ to alternative $a$ with a strength of at least $(x_1,x_2)$, we can demonstrate this by presenting two sets $B_1$ and $B_2$ such that

(2.3.3.1) $b \in B_1$.

(2.3.3.2) $a \in B_2$.

(2.3.3.3) $B_1 \cup B_2 = A$.

(2.3.3.4) $B_1 \cap B_2 = \varnothing$.

(2.3.3.5) $\forall\ i \in B_1\ \forall\ j \in B_2$: $(N[i,j],N[j,i]) \prec_D (x_1,x_2)$.

When $B_1$ and $B_2$ are given, then the runtime to verify that (2.3.3.1) – (2.3.3.5) are satisfied is O($C$^2).

(α) Suppose that we have *not* calculated the strengths of the strongest paths from every alternative $i \in A$ to every other alternative $j \in A \setminus \{i\}$, but that we have found a path from alternative $a$ to alternative $b$ of strength $(x_1,x_2) \in \mathbb{N}_0 \times \mathbb{N}_0$ and want to check whether this path is sufficient so that alternative $a$ disqualifies alternative $b$ ( i.e. $ab \in \mathcal{O}$ ).

Then we can calculate the sets $B_1$ and $B_2$, for example, with the "breadth-first search" (BFS) algorithm as follows. The runtime to calculate the sets $B_1$ and $B_2$ is O($C$^2).

Input: $N[i,j] \in \mathbb{N}_0$ is the number of voters who strictly prefer alternative $i \in A$ to alternative $j \in A \setminus \{i\}$.

$(x_1,x_2) \in \mathbb{N}_0 \times \mathbb{N}_0$.

$a,b \in A$ are those alternatives for which we want to show that there is no path from alternative $b$ to alternative $a$ with a strength of at least $(x_1,x_2)$.

Output: the sets $B_1$ and $B_2$ as described above

```
 1 | B1 : = {b}
 2 | m : = 1
 3 | array1 [1] : = b
 4 | while ( m > 0 ) do
 5 | begin
 6 |    n : = m
 7 |    for k : = 1 to m
 8 |    begin
 9 |       array2 [k] : = array1 [k]
10 |    end
11 |    m : = 0
12 |    for i : = 1 to n
13 |    begin
14 |       j : = array2 [i]
15 |       for k : = 1 to C
16 |       begin
17 |          if ( k ∉ B1 ) then
18 |          begin
19 |             if ( (N[j,k],N[k,j]) ≿D (x1,x2) ) then
20 |             begin
21 |                B1 : = B1 ∪ {k}
22 |                m : = m + 1
23 |                array1 [m] : = k
24 |             end
25 |          end
26 |       end
27 |    end
28 | end
29 | B2 : = A \ B1
```

When, at some point, alternative $a$ is added to the set $B_1$, then this means that a path from alternative $a$ to alternative $b$ of strength $(x_1,x_2)$ is *not* sufficient so that alternative $a$ disqualifies alternative $b$.

(β) Suppose (1) that we have calculated the strengths of the strongest paths from every alternative $i \in A$ to every other alternative $j \in A \setminus \{i\}$, as described in section 2.3.1, and (2) that the statement " $ab \in O$ " is true. Then $B_1$ and $B_2$ are given as follows:

$B_1 := ( \{b\} \cup \{ c \in A \mid P_D[b,c] \gtrsim_D (x_1,x_2) \} )$.

$B_2 := A \setminus B_1$.

### 2.3.4. Fast Calculation of the Schulze Winners

The MinMax score $\Gamma_D(\{a\})$ of alternative $a \in A$ is its worst pairwise defeat.

$$\Gamma_D(\{a\}) := \max_D \{ (N[b,a],N[a,b]) \mid b \in A \setminus \{a\} \}.$$

The Simpson-Kramer MinMax method always chooses an alternative $a$ with minimum $\Gamma_D(\{a\})$. In section 11, we will see that, in random simulations, the Schulze winner is almost always identical to the MinMax winner. In section 4.9, we will see an explanation for this observation. In section 2.3.2, we have seen that, with the Dijkstra algorithm, we can check whether a given alternative $a \in A$ is a Schulze winner in a runtime O($C$^2).

These considerations suggest that, in random simulations, we can calculate the Schulze winners in an average runtime O($C$^2) as follows:

Step 1: Calculate the MinMax scores $\Gamma_D(\{a\})$ for all alternatives $a \in A$.

Step 2: Sort the alternatives according to their MinMax scores:

$$\eta(i) < \eta(j) \Rightarrow \Gamma_D(\{a_{\eta(i)}\}) \precsim_D \Gamma_D(\{a_{\eta(j)}\}).$$

Step 3: For $a_{\eta(1)},\ldots,a_{\eta(C)}$, we set $winner[a_{\eta(i)}] := true$.

Step 4: From the alternative $a_{\eta(1)}$ with minimum MinMax score $\Gamma_D(\{a_{\eta(1)}\})$ to the alternative $a_{\eta(C)}$ with maximum MinMax score $\Gamma_D(\{a_{\eta(C)}\})$ proceed as follows:

Suppose $a_{\eta(i)} \in A$ is the currently considered alternative.

If $winner[a_{\eta(i)}] = true$ then

(α) Calculate the strengths $P_D[a_{\eta(i)},a_{\eta(j)}]$ of the strongest paths from alternative $a_{\eta(i)}$ to every other alternative $a_{\eta(j)} \in A \setminus \{a_{\eta(i)}\}$ and the strengths $P_D[a_{\eta(j)},a_{\eta(i)}]$ of the strongest paths from every other alternative $a_{\eta(j)}$ to alternative $a_{\eta(i)}$ using the Dijkstra algorithm, as described in section 2.3.2.

(β) Then, for $k := i + 1$ to $C$, we proceed:
If $P_D[a_{\eta(i)},a_{\eta(k)}] \succ_D P_D[a_{\eta(k)},a_{\eta(i)}]$, then we set $winner[a_{\eta(k)}] := false$.
If $P_D[a_{\eta(i)},a_{\eta(k)}] \prec_D P_D[a_{\eta(k)},a_{\eta(i)}]$, then we set $winner[a_{\eta(i)}] := false$.

This algorithm also has the advantage (compared to the Floyd-Warshall algorithm, as defined in section 2.3.1) that it can easily be parallelized because ...

... the strengths of the strongest paths from alternative $a \in A$ to every other alternative $b \in A \setminus \{a\}$ and the strengths of the strongest paths from alternative $c \in A \setminus \{a\}$ to every other alternative $d \in A \setminus \{c\}$ can be calculated independently of each other.

... the strengths of the strongest paths from alternative $a \in A$ to every other alternative $b \in A \setminus \{a\}$ and the strengths of the strongest paths from every other alternative $b$ to alternative $a$ can be calculated independently of each other.

## 2.4. Arborescences and Anti-Arborescences

An *arborescence* $\mathcal{A} \subset A \times A$ is a set of $C$–1 links such that there is a vertex $x \in A$, the so-called *root*, such that, for every other vertex $y \in A \setminus \{x\}$, there is a directed path in $\mathcal{A}$ from vertex $x$ to vertex $y$.

An *anti-arborescence* $\mathcal{D} \subset A \times A$ is a set of $C$–1 links such that there is a vertex $x \in A$, the so-called *anti-root*, such that, for every other vertex $y \in A \setminus \{x\}$, there is a directed path in $\mathcal{D}$ from vertex $y$ to vertex $x$.

When the Floyd-Warshall algorithm is used, as defined in section 2.3.1, then the links of the strongest paths from alternative $x$ to every other alternative form an arborescence with alternative $x$ as its root. This fact will be used in sections 4.3, 4.13, 4.14.3, 4.14.4, and 4.14.5. Throughout section 3, this arborescence will be shown in the column "... to every other alternative" in the tables with the strongest paths.

When the Floyd-Warshall algorithm is used, as defined in section 2.3.1, then the links of the strongest paths from every other alternative to alternative $x$ form an anti-arborescence with alternative $x$ as its anti-root. Throughout section 3, this anti-arborescence will be shown in the row "from every other alternative ..." in the tables with the strongest paths.

## 3. Examples

Throughout section 3, we presume that $\succ_D$ satisfies (2.1.1) so that, when each voter $v \in V$ casts a linear order $\succ_v$ on $A$, all definitions for $\succ_D$ are identical.

### 3.1. Example 1

Example 1:

| | |
|---|---|
| 8 voters | $a \succ_v c \succ_v d \succ_v b$ |
| 2 voters | $b \succ_v a \succ_v d \succ_v c$ |
| 4 voters | $c \succ_v d \succ_v b \succ_v a$ |
| 4 voters | $d \succ_v b \succ_v a \succ_v c$ |
| 3 voters | $d \succ_v c \succ_v b \succ_v a$ |

$N[i,j] \in \mathbb{N}_0$ is the number of voters who strictly prefer alternative $i \in A$ to alternative $j \in A \setminus \{i\}$. In example 1, the pairwise matrix $N$ looks as follows:

| | $N[*,a]$ | $N[*,b]$ | $N[*,c]$ | $N[*,d]$ |
|---|---|---|---|---|
| $N[a,*]$ | --- | 8 | 14 | 10 |
| $N[b,*]$ | 13 | --- | 6 | 2 |
| $N[c,*]$ | 7 | 15 | --- | 12 |
| $N[d,*]$ | 11 | 19 | 9 | --- |

The following digraph illustrates the graph theoretic interpretation of pairwise elections. If $N[i,j] > N[j,i]$, then there is a link from vertex $i$ to vertex $j$ of strength $(N[i,j],N[j,i])$:

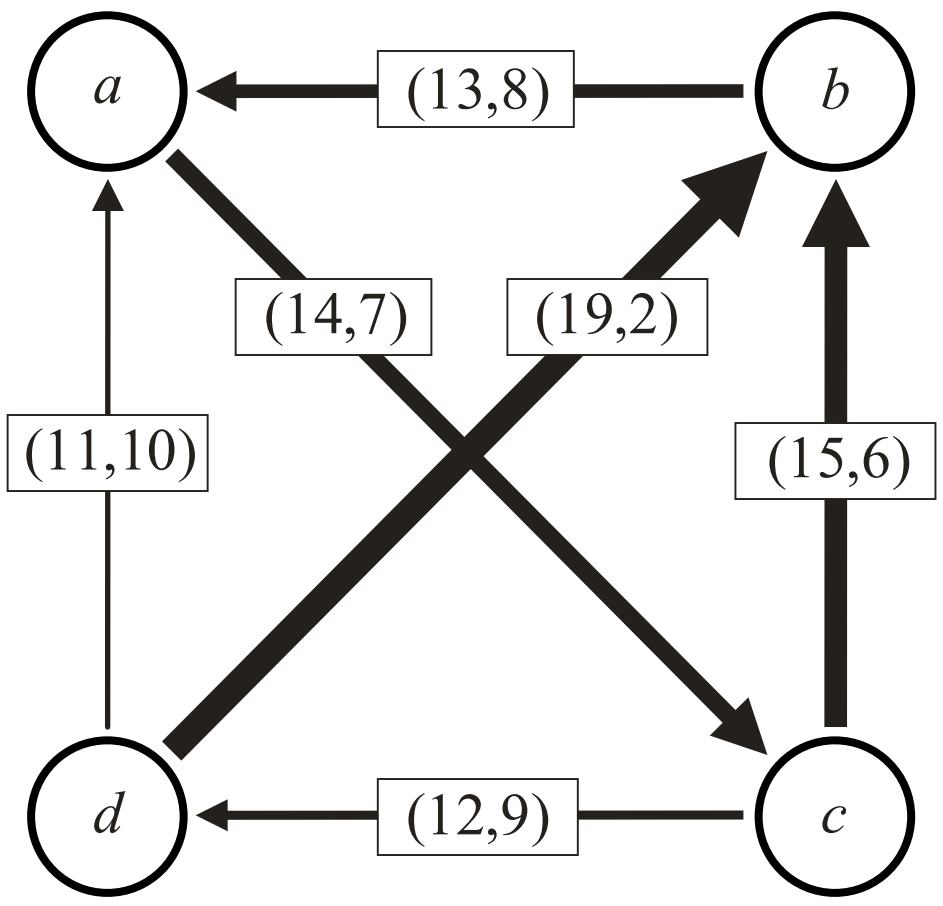

The above digraph can be used to determine the strengths of the strongest paths. In the following, "$x$, $(Z_1,Z_2)$, $y$" means "$(N[x,y],N[y,x]) = (Z_1,Z_2)$".

$a \rightarrow b$: There are 2 paths from alternative $a$ to alternative $b$.

Path 1: $a$, (14,7), $c$, (15,6), $b$
with a strength of $\min_D$ { (14,7), (15,6) } $\approx_D$ (14,7).

Path 2: $a$, (14,7), $c$, (12,9), $d$, (19,2), $b$
with a strength of $\min_D$ { (14,7), (12,9), (19,2) } $\approx_D$ (12,9).

So the strength of the strongest path from alternative $a$ to alternative $b$ is $\max_D$ { (14,7), (12,9) } $\approx_D$ (14,7).

$a \rightarrow c$: There is only one path from alternative $a$ to alternative $c$.

Path 1: $a$, (14,7), $c$
with a strength of (14,7).

So the strength of the strongest path from alternative $a$ to alternative $c$ is (14,7).

$a \rightarrow d$: There is only one path from alternative $a$ to alternative $d$.

Path 1: $a$, (14,7), $c$, (12,9), $d$
with a strength of $\min_D$ { (14,7), (12,9) } $\approx_D$ (12,9).

So the strength of the strongest path from alternative $a$ to alternative $d$ is (12,9).

$b \rightarrow a$: There is only one path from alternative $b$ to alternative $a$.

Path 1: $b$, (13,8), $a$
with a strength of (13,8).

So the strength of the strongest path from alternative $b$ to alternative $a$ is (13,8).

$b \rightarrow c$: There is only one path from alternative *b* to alternative *c*.

Path 1: *b*, (13,8), *a*, (14,7), *c*
with a strength of $\min_D$ { (13,8), (14,7) } $\approx_D$ (13,8).

So the strength of the strongest path from alternative *b* to alternative *c* is (13,8).

$b \rightarrow d$: There is only one path from alternative *b* to alternative *d*.

Path 1: *b*, (13,8), *a*, (14,7), *c*, (12,9), *d*
with a strength of $\min_D$ { (13,8), (14,7), (12,9) } $\approx_D$ (12,9).

So the strength of the strongest path from alternative *b* to alternative *d* is (12,9).

$c \rightarrow a$: There are 3 paths from alternative *c* to alternative *a*.

Path 1: *c*, (15,6), *b*, (13,8), *a*
with a strength of $\min_D$ { (15,6), (13,8) } $\approx_D$ (13,8).

Path 2: *c*, (12,9), *d*, (11,10), *a*
with a strength of $\min_D$ { (12,9), (11,10) } $\approx_D$ (11,10).

Path 3: *c*, (12,9), *d*, (19,2), *b*, (13,8), *a*
with a strength of $\min_D$ { (12,9), (19,2), (13,8) } $\approx_D$ (12,9).

So the strength of the strongest path from alternative *c* to alternative *a* is $\max_D$ { (13,8), (11,10), (12,9) } $\approx_D$ (13,8).

$c \rightarrow b$: There are 2 paths from alternative *c* to alternative *b*.

Path 1: *c*, (15,6), *b*
with a strength of (15,6).

Path 2: *c*, (12,9), *d*, (19,2), *b*
with a strength of $\min_D$ { (12,9), (19,2) } $\approx_D$ (12,9).

So the strength of the strongest path from alternative *c* to alternative *b* is $\max_D$ { (15,6), (12,9) } $\approx_D$ (15,6).

$c \rightarrow d$: There is only one path from alternative *c* to alternative *d*.

Path 1: *c*, (12,9), *d*
with a strength of (12,9).

So the strength of the strongest path from alternative *c* to alternative *d* is (12,9).

$d \rightarrow a$: There are 2 paths from alternative *d* to alternative *a*.

Path 1: *d*, (11,10), *a*
with a strength of (11,10).

Path 2: *d*, (19,2), *b*, (13,8), *a*
with a strength of $\min_D$ { (19,2), (13,8) } $\approx_D$ (13,8).

So the strength of the strongest path from alternative *d* to alternative *a* is $\max_D$ { (11,10), (13,8) } $\approx_D$ (13,8).

$d \rightarrow b$: There are 2 paths from alternative *d* to alternative *b*.

Path 1: *d*, (11,10), *a*, (14,7), *c*, (15,6), *b*
with a strength of $\min_D$ { (11,10), (14,7), (15,6) } $\approx_D$ (11,10).

Path 2: *d*, (19,2), *b*
with a strength of (19,2).

So the strength of the strongest path from alternative *d* to alternative *b* is $\max_D$ { (11,10), (19,2) } $\approx_D$ (19,2).

$d \rightarrow c$: There are 2 paths from alternative *d* to alternative *c*.

Path 1: *d*, (11,10), *a*, (14,7), *c*
with a strength of $\min_D$ { (11,10), (14,7) } $\approx_D$ (11,10).

Path 2: *d*, (19,2), *b*, (13,8), *a*, (14,7), *c*
with a strength of $\min_D$ { (19,2), (13,8), (14,7) } $\approx_D$ (13,8).

So the strength of the strongest path from alternative *d* to alternative *c* is $\max_D$ { (11,10), (13,8) } $\approx_D$ (13,8).

The following table lists the strongest paths, as determined by the Floyd-Warshall algorithm, as defined in section 2.3.1. The critical links of the strongest paths are <u>underlined</u>:

<table>
<tr><th></th><th>... to a</th><th>... to b</th><th>... to c</th><th>... to d</th><th>... to every other alternative</th></tr>
<tr><td>from a ...</td><td>---</td><td>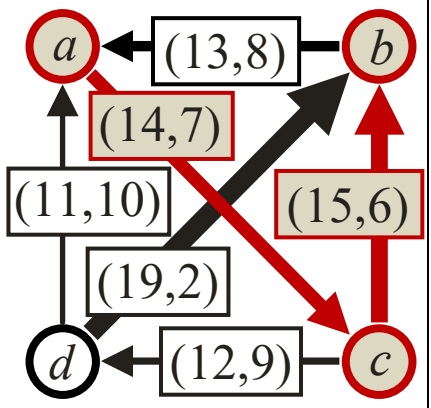

a, <u>(14,7)</u>, c,<br>(15,6), b</td><td>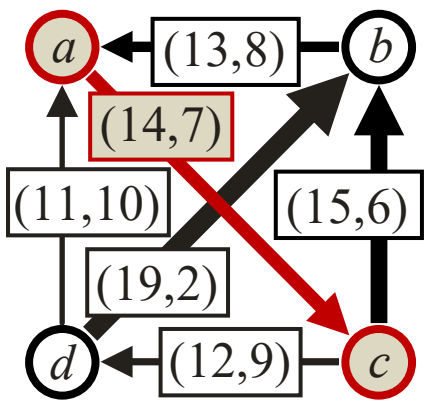

a, <u>(14,7)</u>, c</td><td>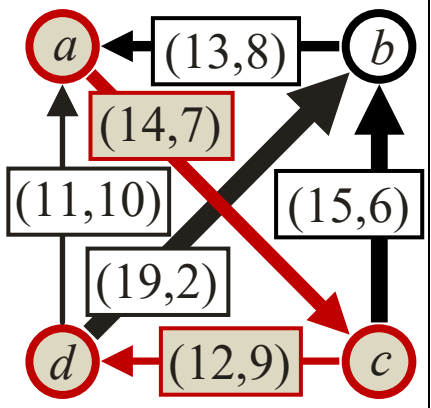

a, (14,7), c,<br><u>(12,9)</u>, d</td><td>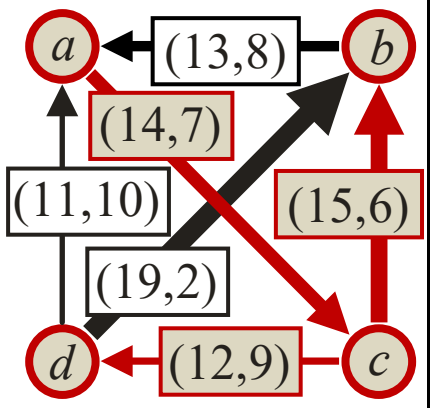
</td></tr>
<tr><td>from b ...</td><td>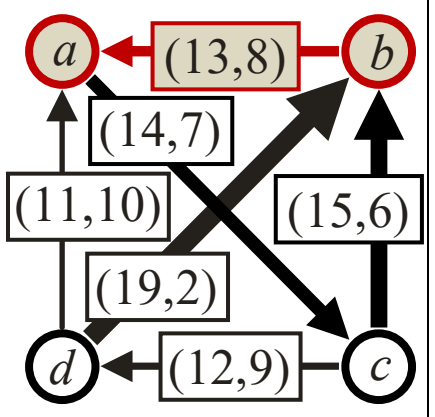

b, <u>(13,8)</u>, a</td><td>---</td><td>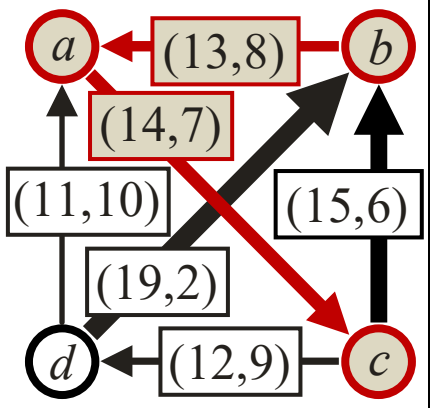

b, <u>(13,8)</u>, a,<br>(14,7), c</td><td>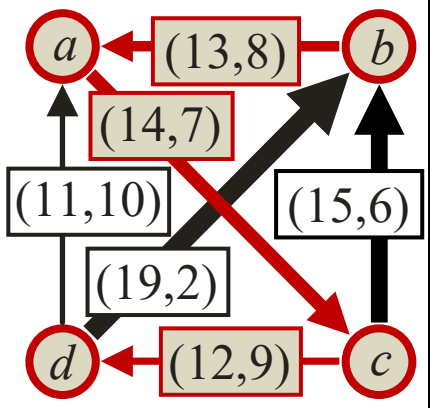

b, (13,8), a,<br>(14,7), c,<br><u>(12,9)</u>, d</td><td>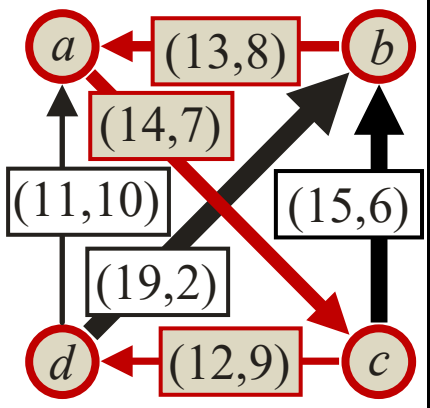
</td></tr>
<tr><td>from c ...</td><td>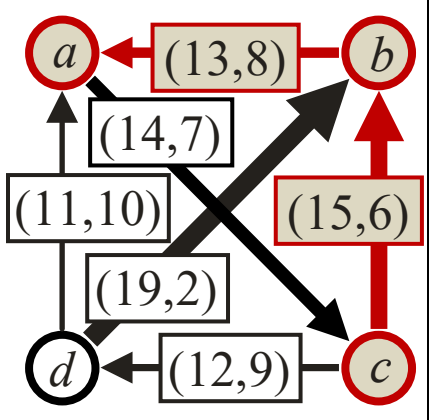

c, (15,6), b,<br><u>(13,8)</u>, a</td><td>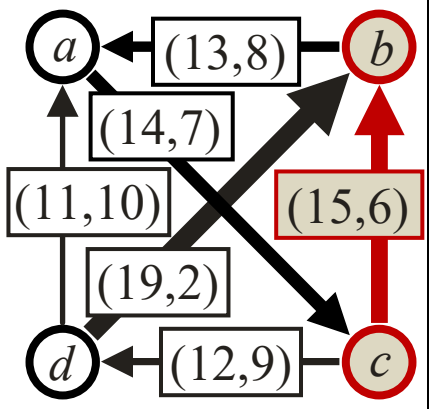

c, <u>(15,6)</u>, b</td><td>---</td><td>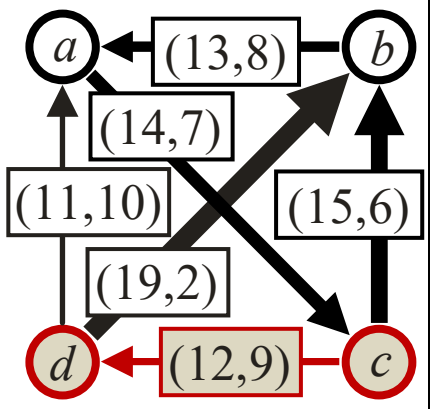

c, <u>(12,9)</u>, d</td><td>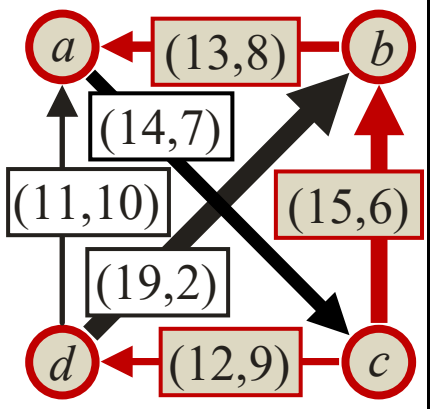
</td></tr>
<tr><td>from d ...</td><td>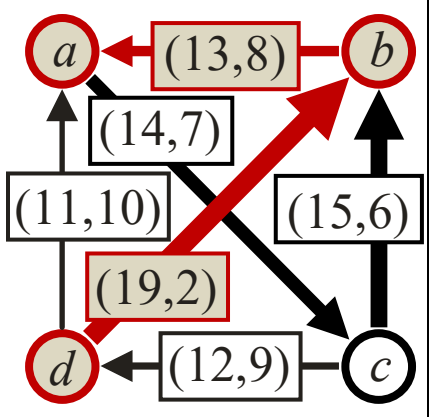

d, (19,2), b,<br><u>(13,8)</u>, a</td><td>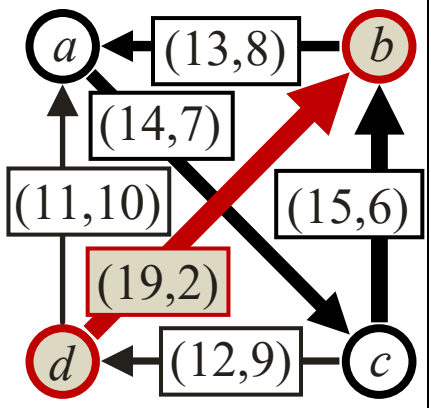

d, <u>(19,2)</u>, b</td><td>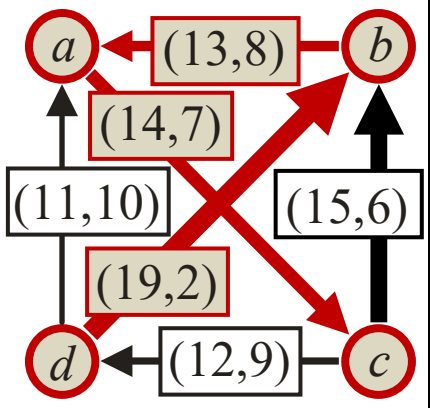

d, (19,2), b,<br><u>(13,8)</u>, a,<br>(14,7), c</td><td>---</td><td>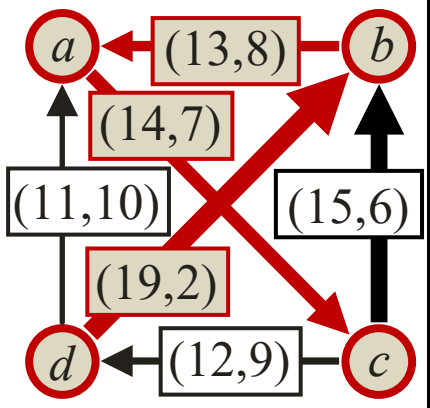
</td></tr>
<tr><td>from every other alternative ...</td><td>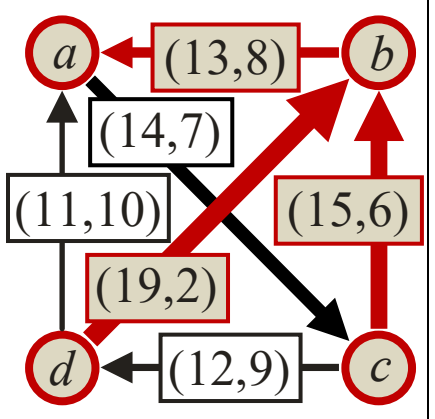
</td><td>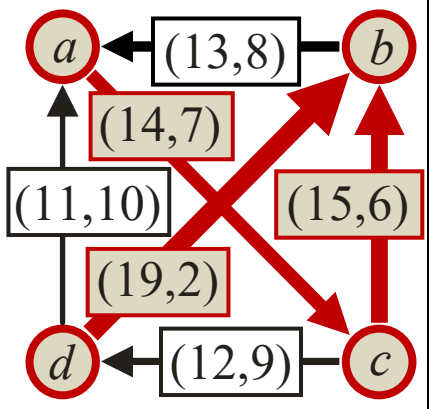
</td><td>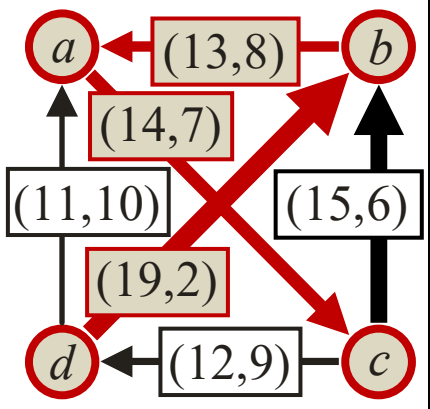
</td><td>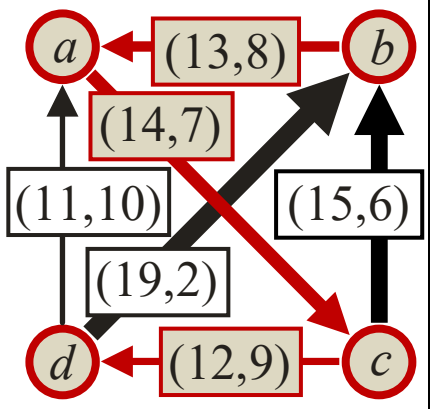
</td><td>---</td></tr>
</table>

The strengths of the strongest paths are:

| | $P_D[*,a]$ | $P_D[*,b]$ | $P_D[*,c]$ | $P_D[*,d]$ |
|---|---|---|---|---|
| $P_D[a,*]$ | --- | (14,7) | (14,7) | (12,9) |
| $P_D[b,*]$ | (13,8) | --- | (13,8) | (12,9) |
| $P_D[c,*]$ | (13,8) | (15,6) | --- | (12,9) |
| $P_D[d,*]$ | (13,8) | (19,2) | (13,8) | --- |

$xy \in \mathcal{O}$ if and only if $P_D[x,y] \succ_D P_D[y,x]$. So in example 1, we get $\mathcal{O} = \{ab, ac, cb, da, db, dc\}$.

$x \in \mathcal{S}$ if and only if $yx \notin \mathcal{O}$ for all $y \in A \setminus \{x\}$. So in example 1, we get $\mathcal{S} = \{d\}$.

Suppose, the strongest paths are calculated with the Floyd-Warshall algorithm, as defined in section 2.3.1. Then the following table documents the $C \cdot (C–1) \cdot (C–2) = 24$ steps of the Floyd-Warshall algorithm.

We start with

- $P_D[i,j] := (N[i,j],N[j,i])$ for all $i \in A$ and $j \in A \setminus \{i\}$.
- $pred[i,j] := i$ for all $i \in A$ and $j \in A \setminus \{i\}$.

| | $i$ | $j$ | $k$ | $P_D[j,k]$ | $P_D[j,i]$ | $P_D[i,k]$ | *pred*[$j$,$k$] | *pred*[$i$,$k$] | result |
|---|---|---|---|---|---|---|---|---|---|
| 1 | *a* | *b* | *c* | (6,15) | (13,8) | (14,7) | *b* | *a* | $P_D[b,c]$ is updated from (6,15) to (13,8); *pred*[*b*,*c*] is updated from *b* to *a*. |
| 2 | *a* | *b* | *d* | (2,19) | (13,8) | (10,11) | *b* | *a* | $P_D[b,d]$ is updated from (2,19) to (10,11); *pred*[*b*,*d*] is updated from *b* to *a*. |
| 3 | *a* | *c* | *b* | (15,6) | (7,14) | (8,13) | *c* | *a* | |
| 4 | *a* | *c* | *d* | (12,9) | (7,14) | (10,11) | *c* | *a* | |
| 5 | *a* | *d* | *b* | (19,2) | (11,10) | (8,13) | *d* | *a* | |
| 6 | *a* | *d* | *c* | (9,12) | (11,10) | (14,7) | *d* | *a* | $P_D[d,c]$ is updated from (9,12) to (11,10); *pred*[*d*,*c*] is updated from *d* to *a*. |
| 7 | *b* | *a* | *c* | (14,7) | (8,13) | (13,8) | *a* | *a* | |
| 8 | *b* | *a* | *d* | (10,11) | (8,13) | (10,11) | *a* | *a* | |
| 9 | *b* | *c* | *a* | (7,14) | (15,6) | (13,8) | *c* | *b* | $P_D[c,a]$ is updated from (7,14) to (13,8); *pred*[*c*,*a*] is updated from *c* to *b*. |
| 10 | *b* | *c* | *d* | (12,9) | (15,6) | (10,11) | *c* | *a* | |
| 11 | *b* | *d* | *a* | (11,10) | (19,2) | (13,8) | *d* | *b* | $P_D[d,a]$ is updated from (11,10) to (13,8); *pred*[*d*,*a*] is updated from *d* to *b*. |
| 12 | *b* | *d* | *c* | (11,10) | (19,2) | (13,8) | *a* | *a* | $P_D[d,c]$ is updated from (11,10) to (13,8). |
| 13 | *c* | *a* | *b* | (8,13) | (14,7) | (15,6) | *a* | *c* | $P_D[a,b]$ is updated from (8,13) to (14,7); *pred*[*a*,*b*] is updated from *a* to *c*. |
| 14 | *c* | *a* | *d* | (10,11) | (14,7) | (12,9) | *a* | *c* | $P_D[a,d]$ is updated from (10,11) to (12,9); *pred*[*a*,*d*] is updated from *a* to *c*. |
| 15 | *c* | *b* | *a* | (13,8) | (13,8) | (13,8) | *b* | *b* | |
| 16 | *c* | *b* | *d* | (10,11) | (13,8) | (12,9) | *a* | *c* | $P_D[b,d]$ is updated from (10,11) to (12,9); *pred*[*b*,*d*] is updated from *a* to *c*. |
| 17 | *c* | *d* | *a* | (13,8) | (13,8) | (13,8) | *b* | *b* | |
| 18 | *c* | *d* | *b* | (19,2) | (13,8) | (15,6) | *d* | *c* | |
| 19 | *d* | *a* | *b* | (14,7) | (12,9) | (19,2) | *c* | *d* | |
| 20 | *d* | *a* | *c* | (14,7) | (12,9) | (13,8) | *a* | *a* | |
| 21 | *d* | *b* | *a* | (13,8) | (12,9) | (13,8) | *b* | *b* | |
| 22 | *d* | *b* | *c* | (13,8) | (12,9) | (13,8) | *a* | *a* | |
| 23 | *d* | *c* | *a* | (13,8) | (12,9) | (13,8) | *b* | *b* | |
| 24 | *d* | *c* | *b* | (15,6) | (12,9) | (19,2) | *c* | *d* | |

### 3.2. Example 2

Example 2:

| | |
|---|---|
| 3 voters | $a \succ_v c \succ_v d \succ_v b$ |
| 9 voters | $b \succ_v a \succ_v c \succ_v d$ |
| 8 voters | $c \succ_v d \succ_v a \succ_v b$ |
| 5 voters | $d \succ_v a \succ_v b \succ_v c$ |
| 5 voters | $d \succ_v b \succ_v c \succ_v a$ |

$N[i,j] \in \mathbb{N}_0$ is the number of voters who strictly prefer alternative $i \in A$ to alternative $j \in A \setminus \{i\}$. In example 2, the pairwise matrix $N$ looks as follows:

| | $N[*,a]$ | $N[*,b]$ | $N[*,c]$ | $N[*,d]$ |
|---|---|---|---|---|
| $N[a,*]$ | --- | 16 | 17 | 12 |
| $N[b,*]$ | 14 | --- | 19 | 9 |
| $N[c,*]$ | 13 | 11 | --- | 20 |
| $N[d,*]$ | 18 | 21 | 10 | --- |

The following digraph illustrates the graph theoretic interpretation of pairwise elections. If $N[i,j] > N[j,i]$, then there is a link from vertex $i$ to vertex $j$ of strength $(N[i,j],N[j,i])$:

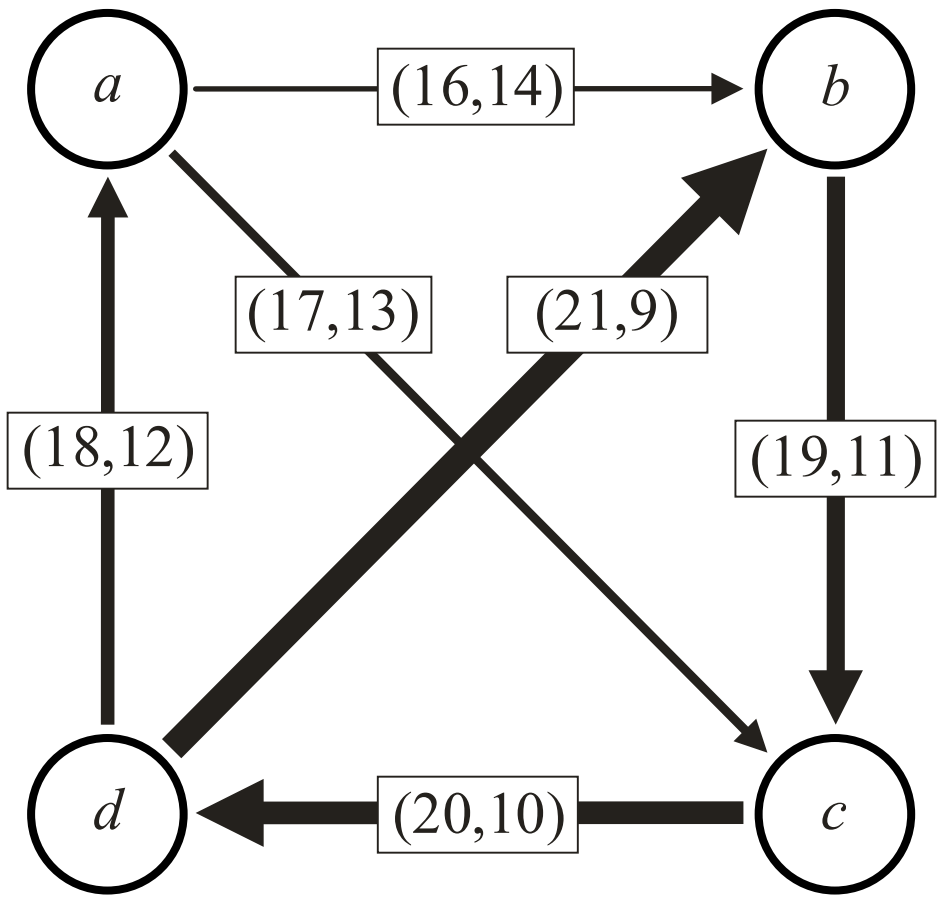

The above digraph can be used to determine the strengths of the strongest paths. In the following, "$x$, $(Z_1,Z_2)$, $y$" means "$(N[x,y],N[y,x]) = (Z_1,Z_2)$".

$a \rightarrow b$: There are 2 paths from alternative $a$ to alternative $b$.

Path 1: $a$, (16,14), $b$
with a strength of (16,14).

Path 2: $a$, (17,13), $c$, (20,10), $d$, (21,9), $b$
with a strength of $\min_D$ { (17,13), (20,10), (21,9) } $\approx_D$ (17,13).

So the strength of the strongest path from alternative $a$ to alternative $b$ is $\max_D$ { (16,14), (17,13) } $\approx_D$ (17,13).

$a \rightarrow c$: There are 2 paths from alternative $a$ to alternative $c$.

Path 1: $a$, (16,14), $b$, (19,11), $c$
with a strength of $\min_D$ { (16,14), (19,11) } $\approx_D$ (16,14).

Path 2: $a$, (17,13), $c$
with a strength of (17,13).

So the strength of the strongest path from alternative $a$ to alternative $c$ is $\max_D$ { (16,14), (17,13) } $\approx_D$ (17,13).

$a \rightarrow d$: There are 2 paths from alternative $a$ to alternative $d$.

Path 1: $a$, (16,14), $b$, (19,11), $c$, (20,10), $d$
with a strength of $\min_D$ { (16,14), (19,11), (20,10) } $\approx_D$ (16,14).

Path 2: $a$, (17,13), $c$, (20,10), $d$
with a strength of $\min_D$ { (17,13), (20,10) } $\approx_D$ (17,13).

So the strength of the strongest path from alternative $a$ to alternative $d$ is $\max_D$ { (16,14), (17,13) } $\approx_D$ (17,13).

$b \rightarrow a$: There is only one path from alternative $b$ to alternative $a$.

Path 1: $b$, (19,11), $c$, (20,10), $d$, (18,12), $a$
with a strength of $\min_D$ { (19,11), (20,10), (18,12) } $\approx_D$ (18,12).

So the strength of the strongest path from alternative $b$ to alternative $a$ is (18,12).

$b \to c$: There is only one path from alternative $b$ to alternative $c$.

Path 1: $b$, (19,11), $c$
with a strength of (19,11).

So the strength of the strongest path from alternative $b$ to alternative $c$ is (19,11).

$b \to d$: There is only one path from alternative $b$ to alternative $d$.

Path 1: $b$, (19,11), $c$, (20,10), $d$
with a strength of $\min_D$ { (19,11), (20,10) } $\approx_D$ (19,11).

So the strength of the strongest path from alternative $b$ to alternative $d$ is (19,11).

$c \to a$: There is only one path from alternative $c$ to alternative $a$.

Path 1: $c$, (20,10), $d$, (18,12), $a$
with a strength of $\min_D$ { (20,10), (18,12) } $\approx_D$ (18,12).

So the strength of the strongest path from alternative $c$ to alternative $a$ is (18,12).

$c \to b$: There are 2 paths from alternative $c$ to alternative $b$.

Path 1: $c$, (20,10), $d$, (21,9), $b$
with a strength of $\min_D$ { (20,10), (21,9) } $\approx_D$ (20,10).

Path 2: $c$, (20,10), $d$, (18,12), $a$, (16,14), $b$
with a strength of $\min_D$ { (20,10), (18,12), (16,14) } $\approx_D$ (16,14).

So the strength of the strongest path from alternative $c$ to alternative $b$ is $\max_D$ { (20,10), (16,14) } $\approx_D$ (20,10).

$c \to d$: There is only one path from alternative $c$ to alternative $d$.

Path 1: $c$, (20,10), $d$
with a strength of (20,10).

So the strength of the strongest path from alternative $c$ to alternative $d$ is (20,10).

$d \rightarrow a$: There is only one path from alternative *d* to alternative *a*.

Path 1: *d*, (18,12), *a*
with a strength of (18,12).

So the strength of the strongest path from alternative *d* to alternative *a* is (18,12).

$d \rightarrow b$: There are 2 paths from alternative *d* to alternative *b*.

Path 1: *d*, (18,12), *a*, (16,14), *b*
with a strength of $\min_D$ { (18,12), (16,14) } $\approx_D$ (16,14).

Path 2: *d*, (21,9), *b*
with a strength of (21,9).

So the strength of the strongest path from alternative *d* to alternative *b* is $\max_D$ { (16,14), (21,9) } $\approx_D$ (21,9).

$d \rightarrow c$: There are 3 paths from alternative *d* to alternative *c*.

Path 1: *d*, (18,12), *a*, (16,14), *b*, (19,11), *c*
with a strength of $\min_D$ { (18,12), (16,14), (19,11) } $\approx_D$ (16,14).

Path 2: *d*, (18,12), *a*, (17,13), *c*
with a strength of $\min_D$ { (18,12), (17,13) } $\approx_D$ (17,13).

Path 3: *d*, (21,9), *b*, (19,11), *c*
with a strength of $\min_D$ { (21,9), (19,11) } $\approx_D$ (19,11).

So the strength of the strongest path from alternative *d* to alternative *c* is $\max_D$ { (16,14), (17,13), (19,11) } $\approx_D$ (19,11).

The following table lists the strongest paths, as determined by the Floyd-Warshall algorithm, as defined in section 2.3.1. The critical links of the strongest paths are <u>underlined</u>:

| | ... to *a* | ... to *b* | ... to *c* | ... to *d* | ... to every other alternative |
|---|---|---|---|---|---|
| from *a* ... | --- | 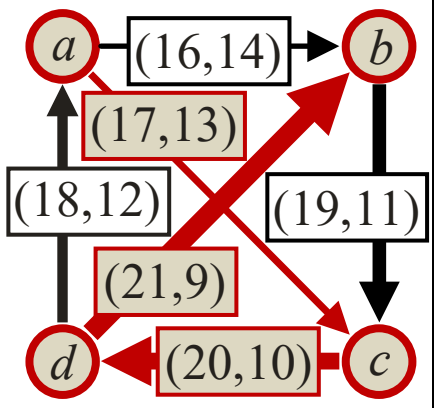<br>*a*, <u>(17,13)</u>, *c*, (20,10), *d*, (21,9), *b* | 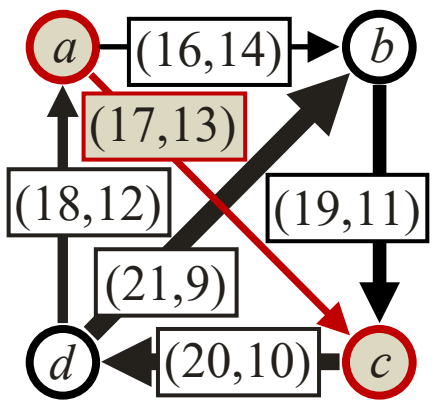<br>*a*, <u>(17,13)</u>, *c* | 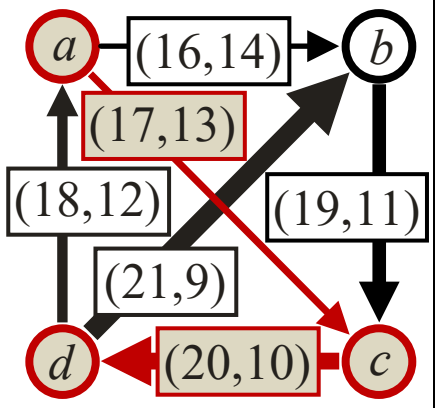<br>*a*, <u>(17,13)</u>, *c*, (20,10), *d* | 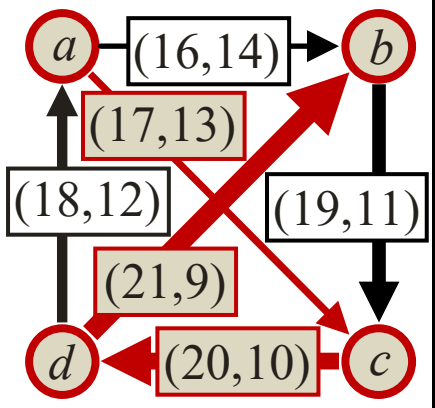 |
| from *b* ... | 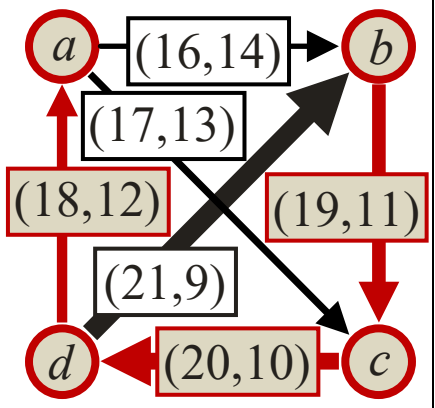<br>*b*, (19,11), *c*, (20,10), *d*, <u>(18,12)</u>, *a* | --- | 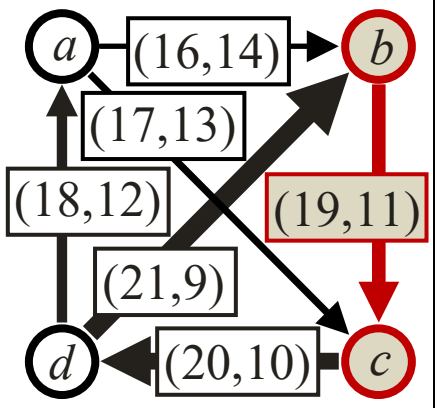<br>*b*, <u>(19,11)</u>, *c* | 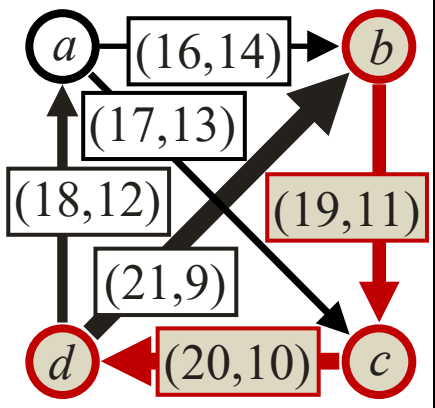<br>*b*, <u>(19,11)</u>, *c*, (20,10), *d* | 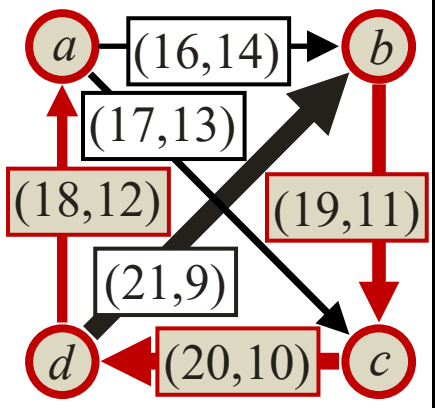 |
| from *c* ... | 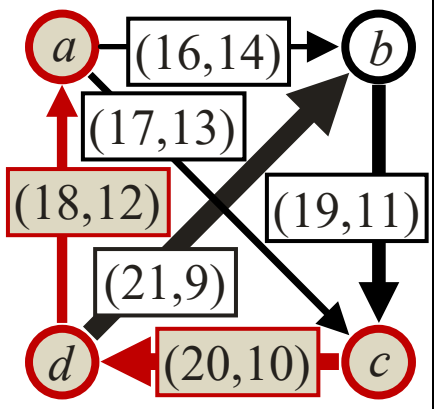<br>*c*, (20,10), *d*, <u>(18,12)</u>, *a* | 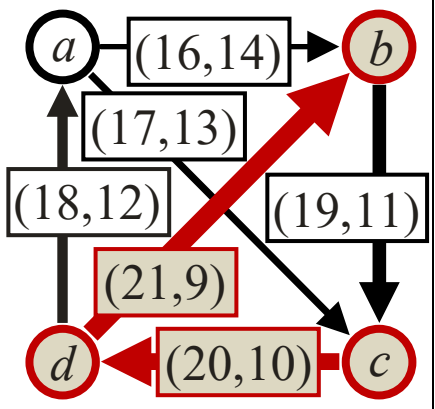<br>*c*, <u>(20,10)</u>, *d*, (21,9), *b* | --- | 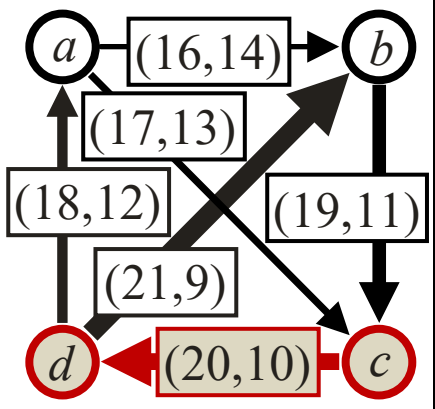<br>*c*, <u>(20,10)</u>, *d* | 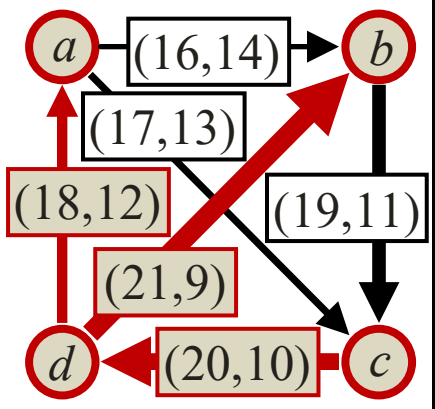 |
| from *d* ... | 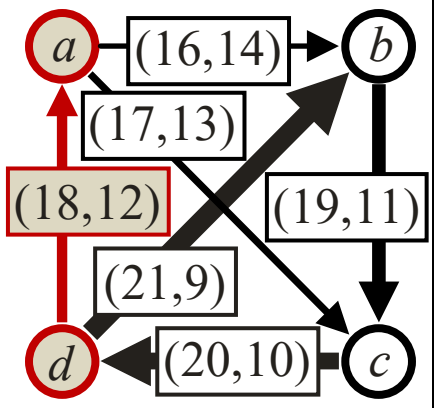<br>*d*, <u>(18,12)</u>, *a* | 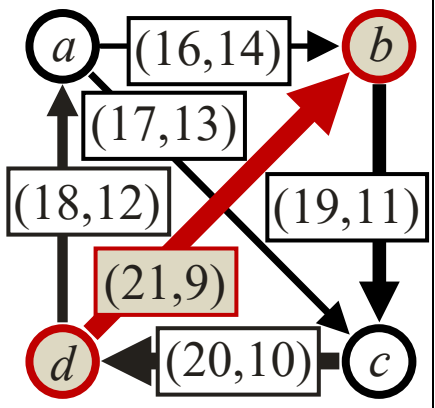<br>*d*, <u>(21,9)</u>, *b* | 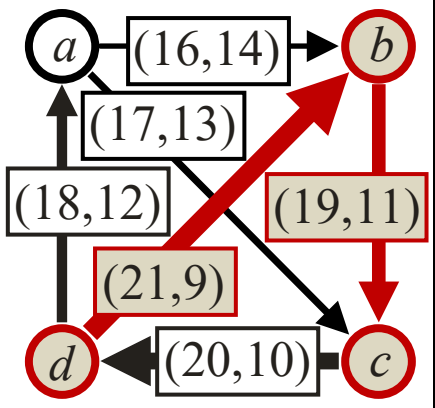<br>*d*, (21,9), *b*, <u>(19,11)</u>, *c* | --- | 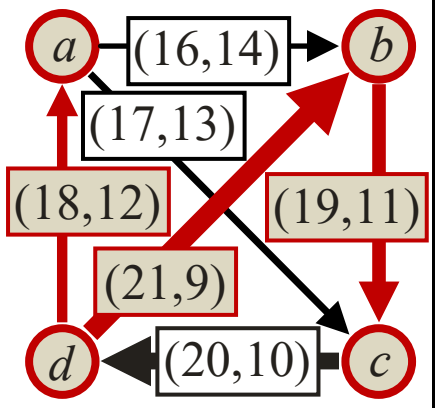 |
| from every other alternative ... | 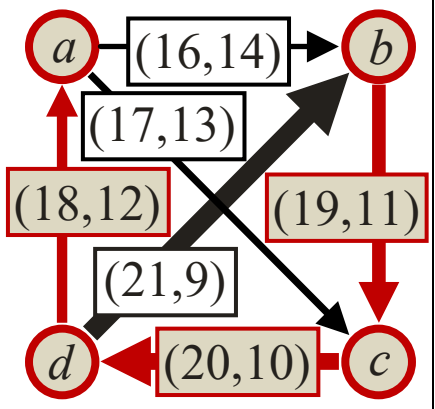 | 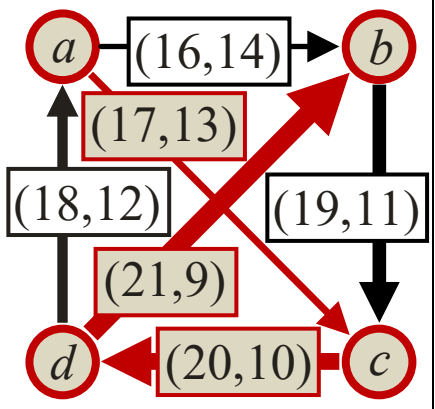 | 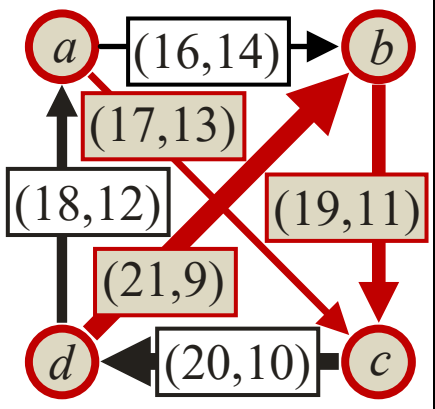 | 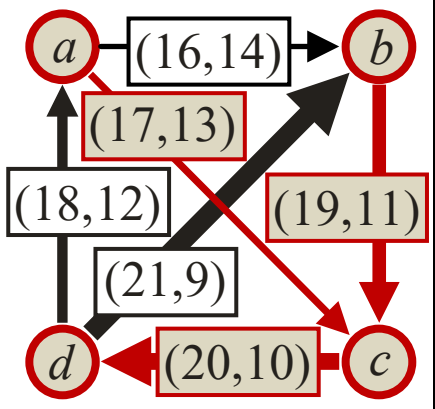 | --- |

The strengths of the strongest paths are:

| | $P_D[*,a]$ | $P_D[*,b]$ | $P_D[*,c]$ | $P_D[*,d]$ |
|---|---|---|---|---|
| $P_D[a,*]$ | --- | (17,13) | (17,13) | (17,13) |
| $P_D[b,*]$ | (18,12) | --- | (19,11) | (19,11) |
| $P_D[c,*]$ | (18,12) | (20,10) | --- | (20,10) |
| $P_D[d,*]$ | (18,12) | (21,9) | (19,11) | --- |

We get $\mathcal{O} = \{ba, ca, cb, cd, da, db\}$ and $\mathcal{S} = \{c\}$.

Suppose, the strongest paths are calculated with the Floyd-Warshall algorithm, as defined in section 2.3.1. Then the following table documents the $C \cdot (C–1) \cdot (C–2) = 24$ steps of the Floyd-Warshall algorithm.

We start with

- $P_D[i,j] := (N[i,j],N[j,i])$ for all $i \in A$ and $j \in A \setminus \{i\}$.
- $pred[i,j] := i$ for all $i \in A$ and $j \in A \setminus \{i\}$.

| | *i* | *j* | *k* | $P_D[j,k]$ | $P_D[j,i]$ | $P_D[i,k]$ | *pred*[*j*,*k*] | *pred*[*i*,*k*] | result |
|---|---|---|---|---|---|---|---|---|---|
| 1 | *a* | *b* | *c* | (19,11) | (14,16) | (17,13) | *b* | *a* | |
| 2 | *a* | *b* | *d* | (9,21) | (14,16) | (12,18) | *b* | *a* | $P_D[b,d]$ is updated from (9,21) to (12,18); *pred*[*b*,*d*] is updated from *b* to *a*. |
| 3 | *a* | *c* | *b* | (11,19) | (13,17) | (16,14) | *c* | *a* | $P_D[c,b]$ is updated from (11,19) to (13,17); *pred*[*c*,*b*] is updated from *c* to *a*. |
| 4 | *a* | *c* | *d* | (20,10) | (13,17) | (12,18) | *c* | *a* | |
| 5 | *a* | *d* | *b* | (21,9) | (18,12) | (16,14) | *d* | *a* | |
| 6 | *a* | *d* | *c* | (10,20) | (18,12) | (17,13) | *d* | *a* | $P_D[d,c]$ is updated from (10,20) to (17,13); *pred*[*d*,*c*] is updated from *d* to *a*. |
| 7 | *b* | *a* | *c* | (17,13) | (16,14) | (19,11) | *a* | *b* | |
| 8 | *b* | *a* | *d* | (12,18) | (16,14) | (12,18) | *a* | *a* | |
| 9 | *b* | *c* | *a* | (13,17) | (13,17) | (14,16) | *c* | *b* | |
| 10 | *b* | *c* | *d* | (20,10) | (13,17) | (12,18) | *c* | *a* | |
| 11 | *b* | *d* | *a* | (18,12) | (21,9) | (14,16) | *d* | *b* | |
| 12 | *b* | *d* | *c* | (17,13) | (21,9) | (19,11) | *a* | *b* | $P_D[d,c]$ is updated from (17,13) to (19,11); *pred*[*d*,*c*] is updated from *a* to *b*. |
| 13 | *c* | *a* | *b* | (16,14) | (17,13) | (13,17) | *a* | *a* | |
| 14 | *c* | *a* | *d* | (12,18) | (17,13) | (20,10) | *a* | *c* | $P_D[a,d]$ is updated from (12,18) to (17,13); *pred*[*a*,*d*] is updated from *a* to *c*. |
| 15 | *c* | *b* | *a* | (14,16) | (19,11) | (13,17) | *b* | *c* | |
| 16 | *c* | *b* | *d* | (12,18) | (19,11) | (20,10) | *a* | *c* | $P_D[b,d]$ is updated from (12,18) to (19,11); *pred*[*b*,*d*] is updated from *a* to *c*. |
| 17 | *c* | *d* | *a* | (18,12) | (19,11) | (13,17) | *d* | *c* | |
| 18 | *c* | *d* | *b* | (21,9) | (19,11) | (13,17) | *d* | *a* | |
| 19 | *d* | *a* | *b* | (16,14) | (17,13) | (21,9) | *a* | *d* | $P_D[a,b]$ is updated from (16,14) to (17,13); *pred*[*a*,*b*] is updated from *a* to *d*. |
| 20 | *d* | *a* | *c* | (17,13) | (17,13) | (19,11) | *a* | *b* | |
| 21 | *d* | *b* | *a* | (14,16) | (19,11) | (18,12) | *b* | *d* | $P_D[b,a]$ is updated from (14,16) to (18,12); *pred*[*b*,*a*] is updated from *b* to *d*. |
| 22 | *d* | *b* | *c* | (19,11) | (19,11) | (19,11) | *b* | *b* | |
| 23 | *d* | *c* | *a* | (13,17) | (20,10) | (18,12) | *c* | *d* | $P_D[c,a]$ is updated from (13,17) to (18,12); *pred*[*c*,*a*] is updated from *c* to *d*. |
| 24 | *d* | *c* | *b* | (13,17) | (20,10) | (21,9) | *a* | *d* | $P_D[c,b]$ is updated from (13,17) to (20,10); *pred*[*c*,*b*] is updated from *a* to *d*. |

## 3.3. Example 3

Example 3:

| | | |
|---|---|---|
| 5 | voters | $a \succ_v c \succ_v b \succ_v e \succ_v d$ |
| 5 | voters | $a \succ_v d \succ_v e \succ_v c \succ_v b$ |
| 8 | voters | $b \succ_v e \succ_v d \succ_v a \succ_v c$ |
| 3 | voters | $c \succ_v a \succ_v b \succ_v e \succ_v d$ |
| 7 | voters | $c \succ_v a \succ_v e \succ_v b \succ_v d$ |
| 2 | voters | $c \succ_v b \succ_v a \succ_v d \succ_v e$ |
| 7 | voters | $d \succ_v c \succ_v e \succ_v b \succ_v a$ |
| 8 | voters | $e \succ_v b \succ_v a \succ_v d \succ_v c$ |

The pairwise matrix *N* looks as follows:

| | *N*[*,*a*] | *N*[*,*b*] | *N*[*,*c*] | *N*[*,*d*] | *N*[*,*e*] |
|---|---|---|---|---|---|
| *N*[*a*,*] | --- | 20 | 26 | 30 | 22 |
| *N*[*b*,*] | 25 | --- | 16 | 33 | 18 |
| *N*[*c*,*] | 19 | 29 | --- | 17 | 24 |
| *N*[*d*,*] | 15 | 12 | 28 | --- | 14 |
| *N*[*e*,*] | 23 | 27 | 21 | 31 | --- |

The corresponding digraph looks as follows:

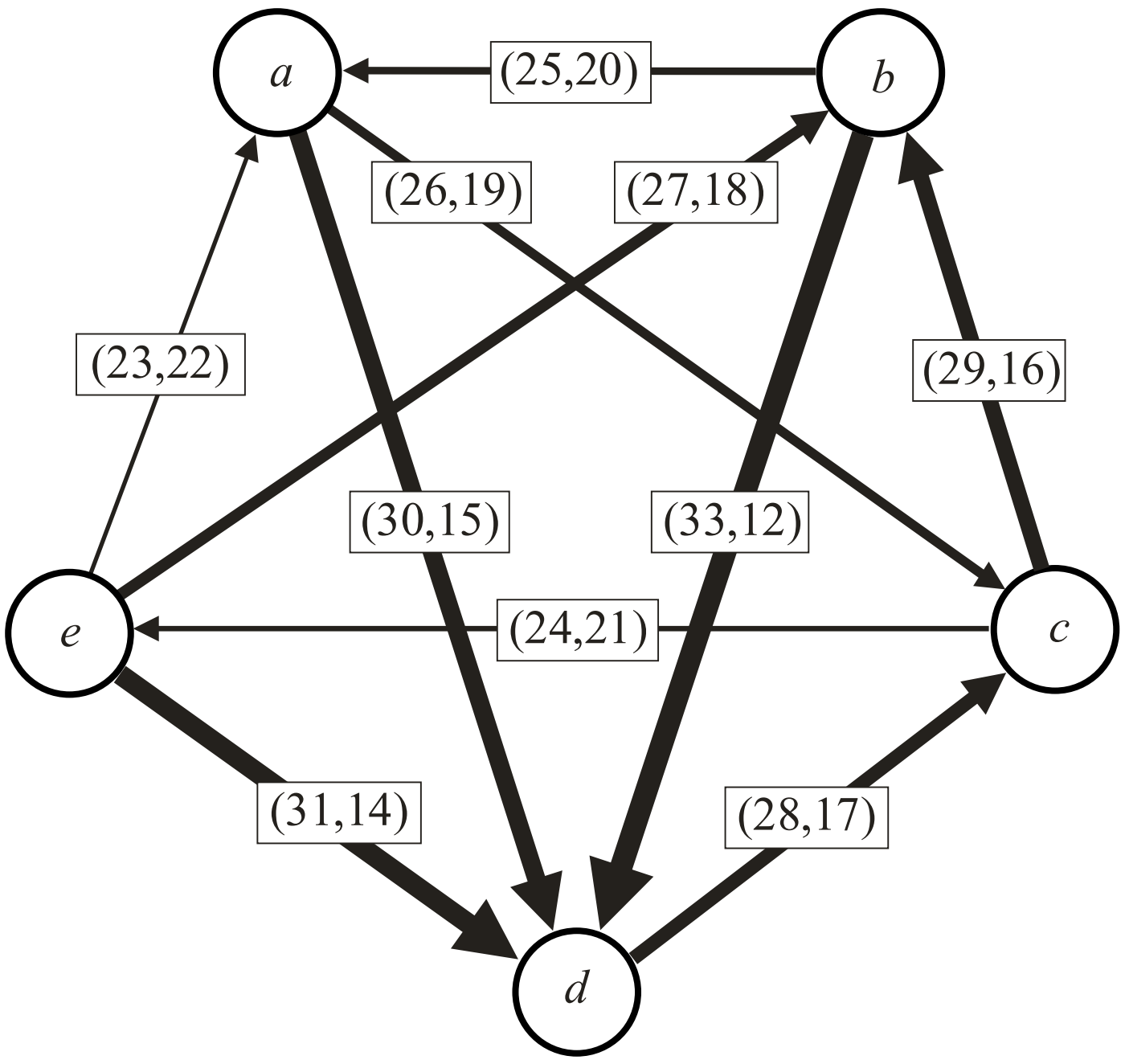

The above digraph can be used to determine the strengths of the strongest paths. In the following, “$x$, $(Z_1,Z_2)$, $y$” means “$(N[x,y],N[y,x]) = (Z_1,Z_2)$”.

$a \rightarrow b$: There are 4 paths from alternative $a$ to alternative $b$.

Path 1: $a$, (26,19), $c$, (29,16), $b$
with a strength of $\min_D$ { (26,19), (29,16) } $\approx_D$ (26,19).

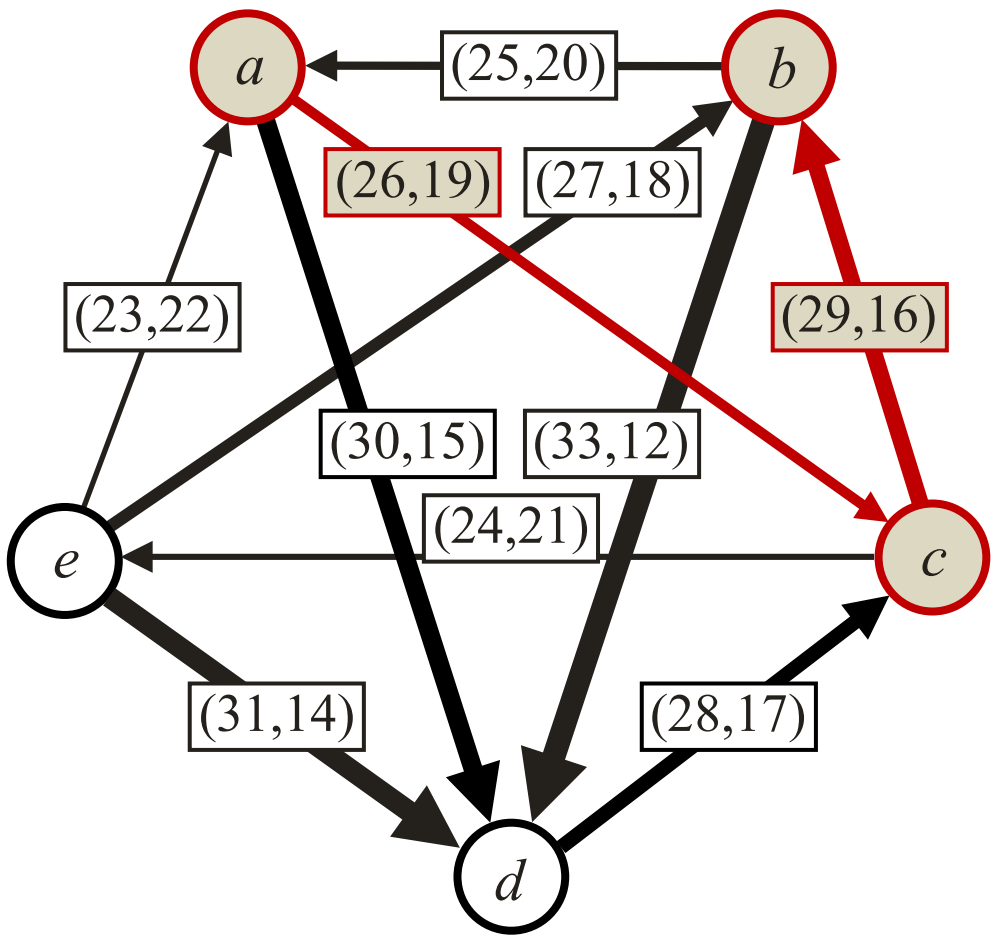

Path 2: $a$, (26,19), $c$, (24,21), $e$, (27,18), $b$
with a strength of $\min_D$ { (26,19), (24,21), (27,18) } $\approx_D$ (24,21).

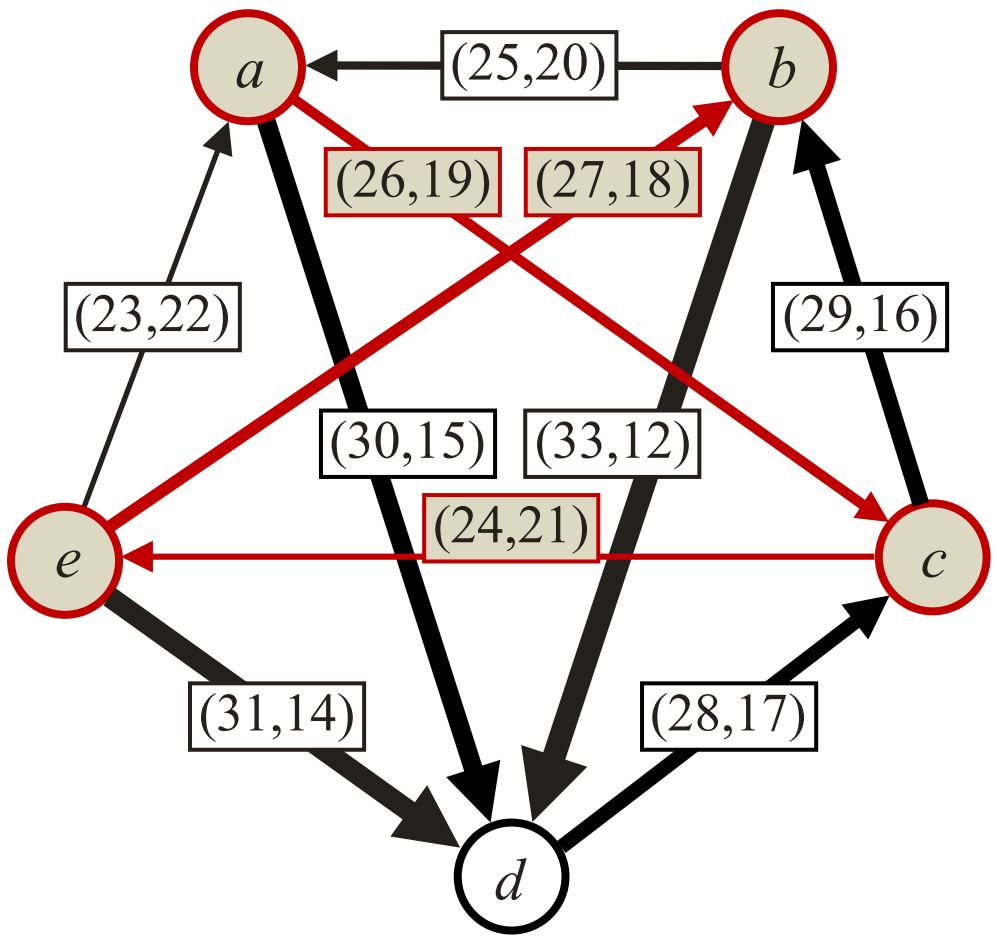

Path 3: $a$, (30,15), $d$, (28,17), $c$, (29,16), $b$
with a strength of $\min_D$ { (30,15), (28,17), (29,16) } $\approx_D$ (28,17).

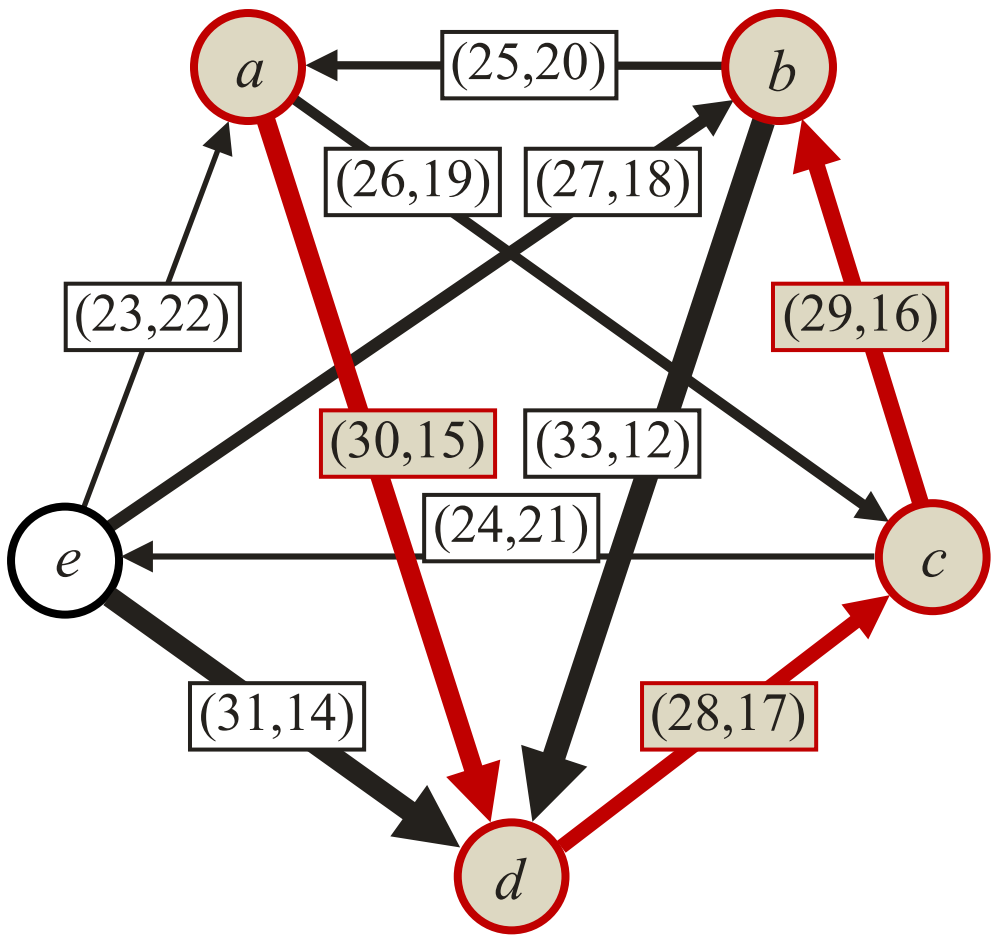

Path 4: $a$, (30,15), $d$, (28,17), $c$, (24,21), $e$, (27,18), $b$
with a strength of $\min_D$ { (30,15), (28,17), (24,21), (27,18) } $\approx_D$ (24,21).

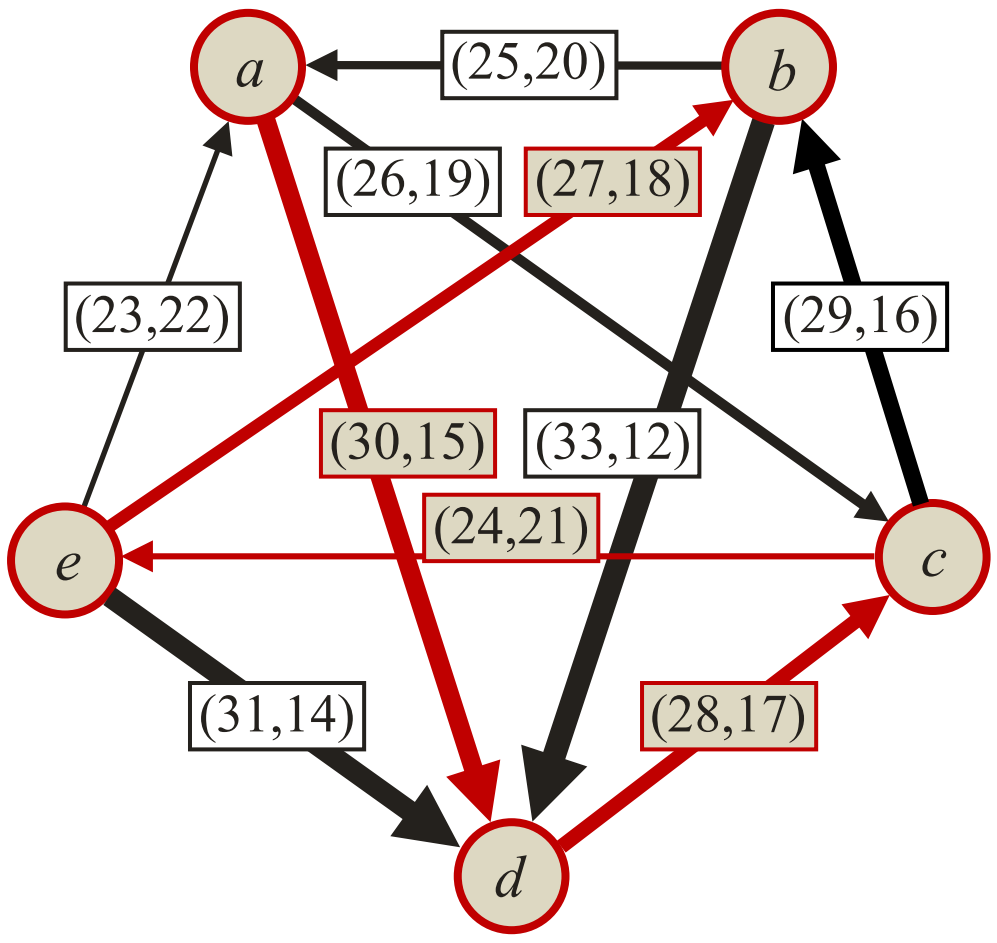

So the strength of the strongest path from alternative $a$ to alternative $b$
is $\max_D$ { (26,19), (24,21), (28,17), (24,21) } $\approx_D$ (28,17).

$a \rightarrow c$: There are 2 paths from alternative $a$ to alternative $c$.

Path 1: $a$, (26,19), $c$
with a strength of (26,19).

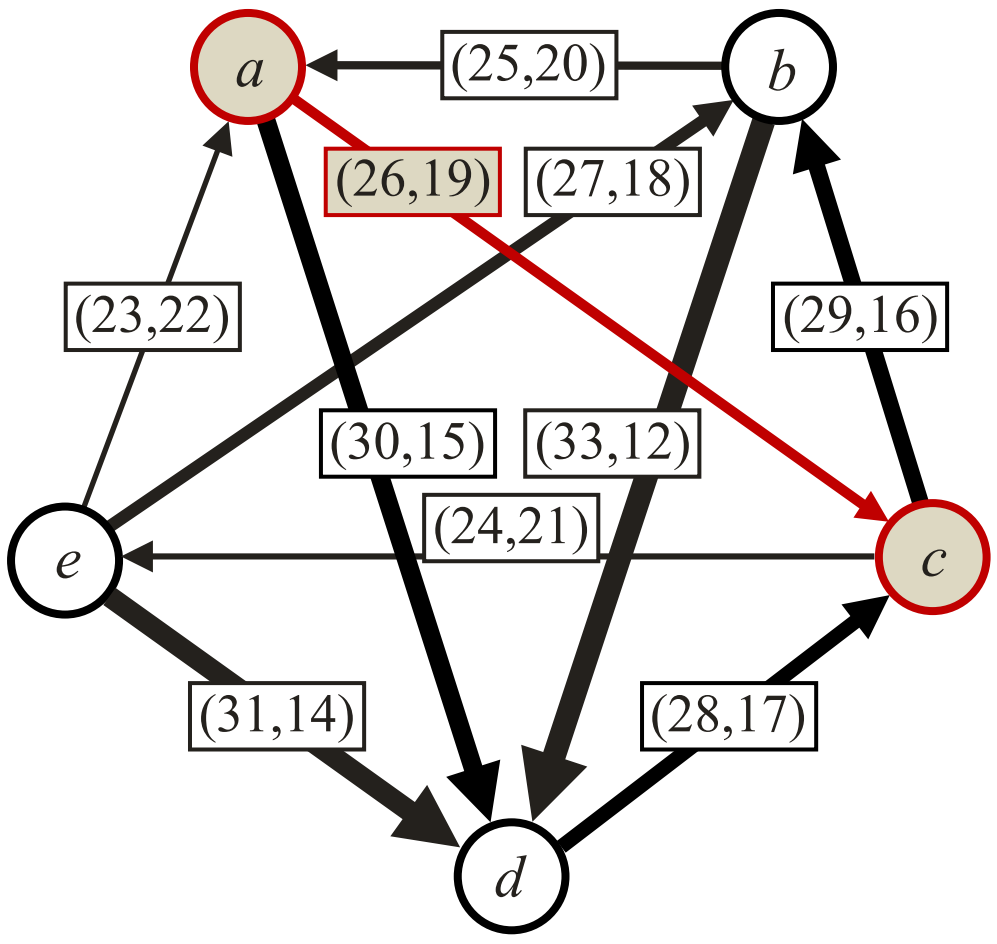

Path 2: $a$, (30,15), $d$, (28,17), $c$
with a strength of $\min_D$ { (30,15), (28,17) } $\approx_D$ (28,17).

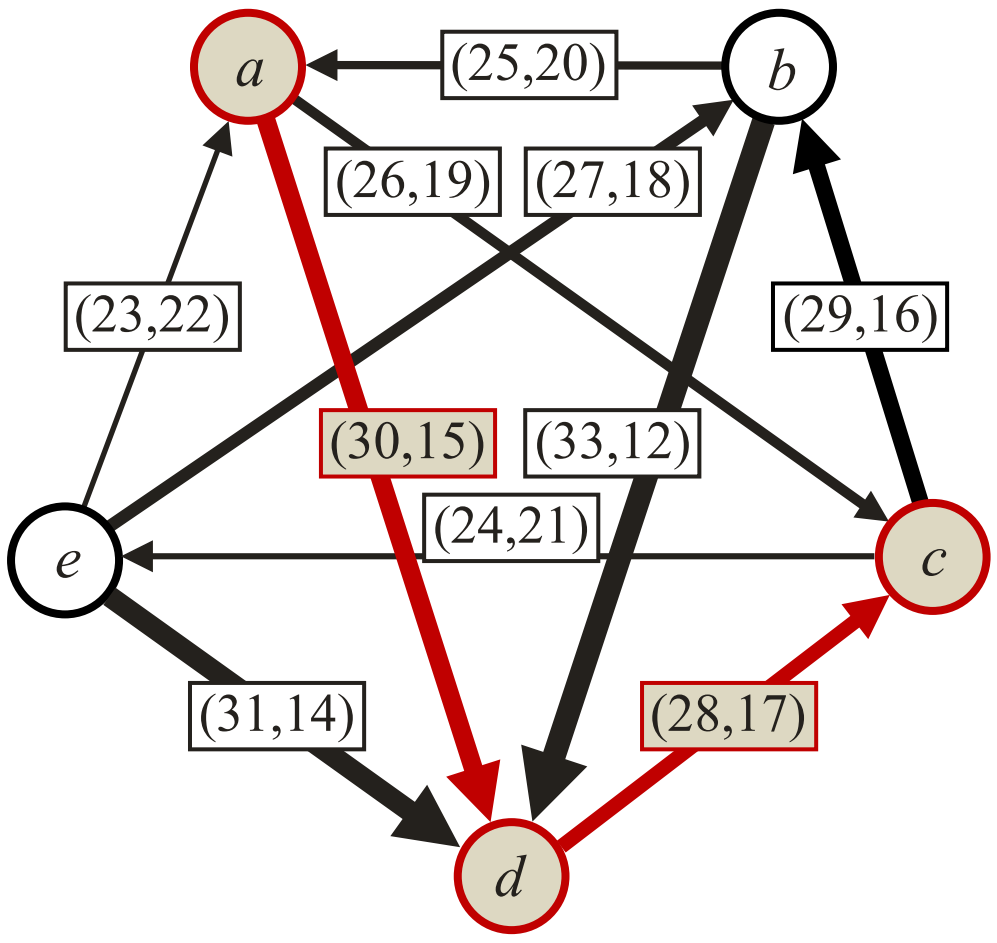

So the strength of the strongest path from alternative $a$ to alternative $c$ is $\max_D$ { (26,19), (28,17) } $\approx_D$ (28,17).

$a \rightarrow d$: There are 4 paths from alternative *a* to alternative *d*.

Path 1: *a*, (26,19), *c*, (29,16), *b*, (33,12), *d*
with a strength of $\min_D$ { (26,19), (29,16), (33,12) } $\approx_D$ (26,19).

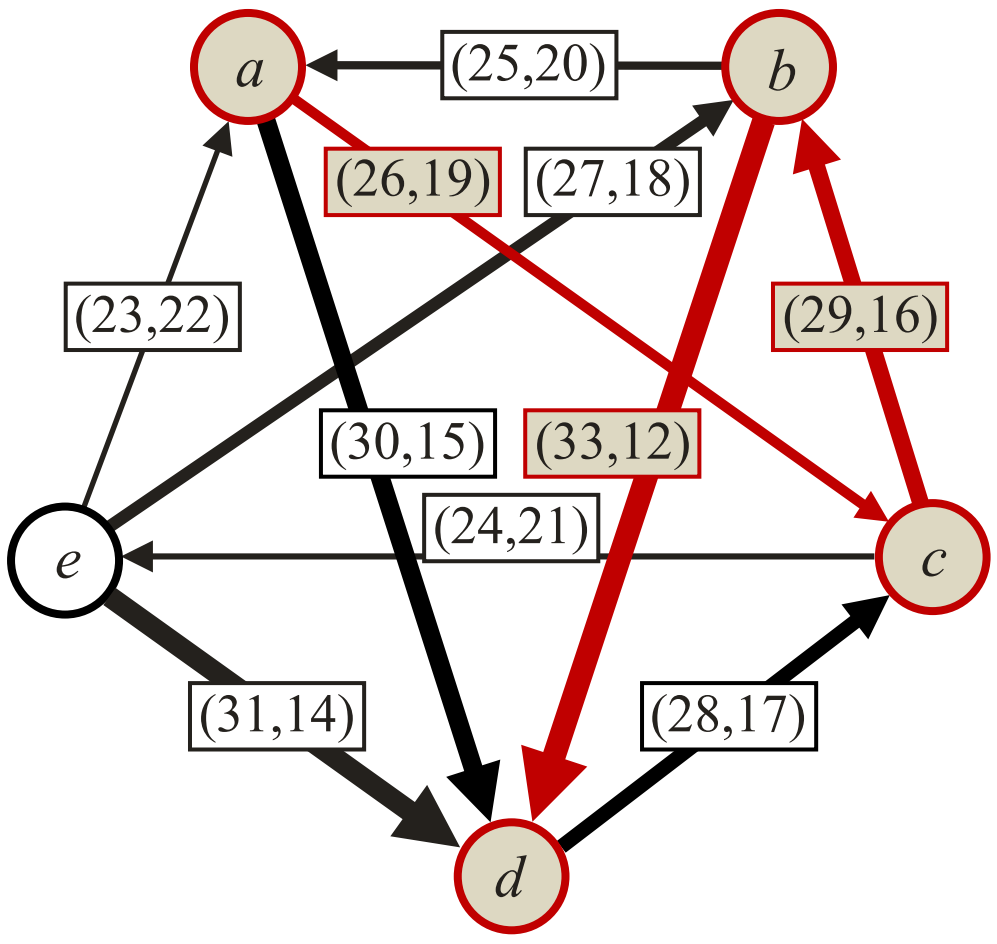

Path 2: *a*, (26,19), *c*, (24,21), *e*, (27,18), *b*, (33,12), *d*
with a strength of $\min_D$ { (26,19), (24,21), (27,18), (33,12) } $\approx_D$ (24,21).

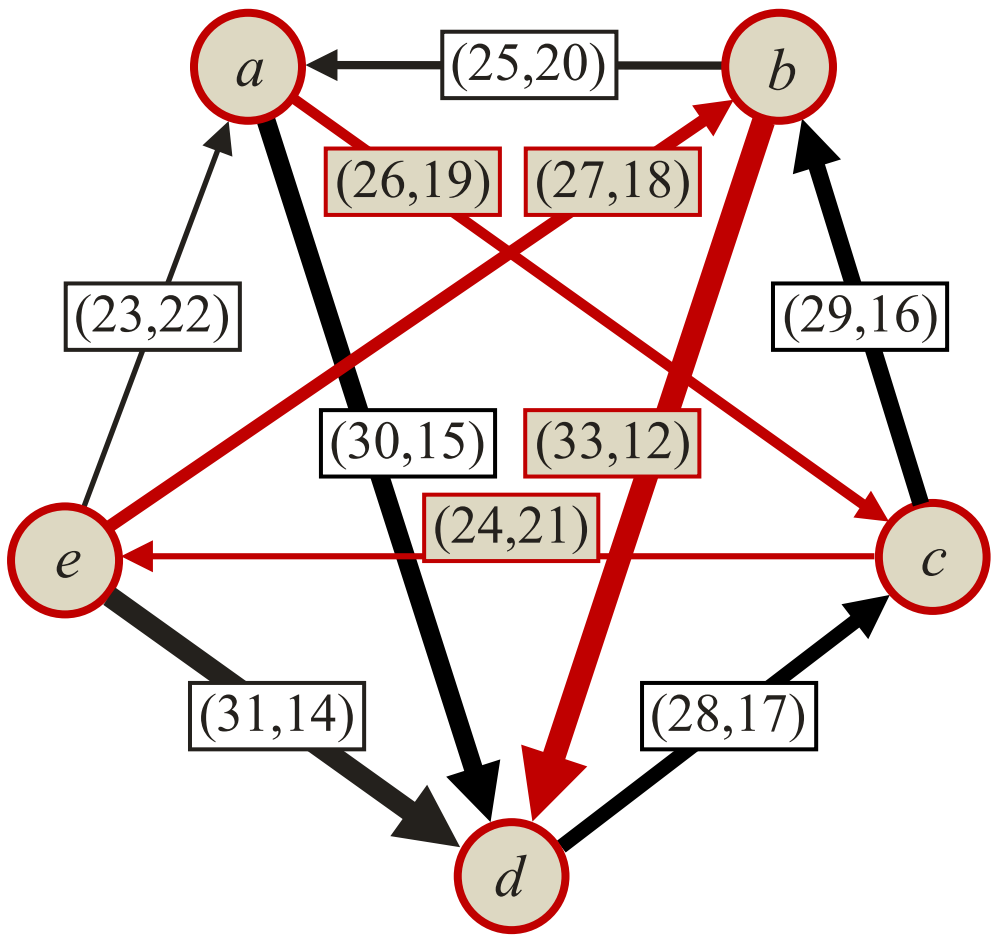

Path 3: *a*, (26,19), *c*, (24,21), *e*, (31,14), *d*
with a strength of $\min_D$ { (26,19), (24,21), (31,14) } $\approx_D$ (24,21).

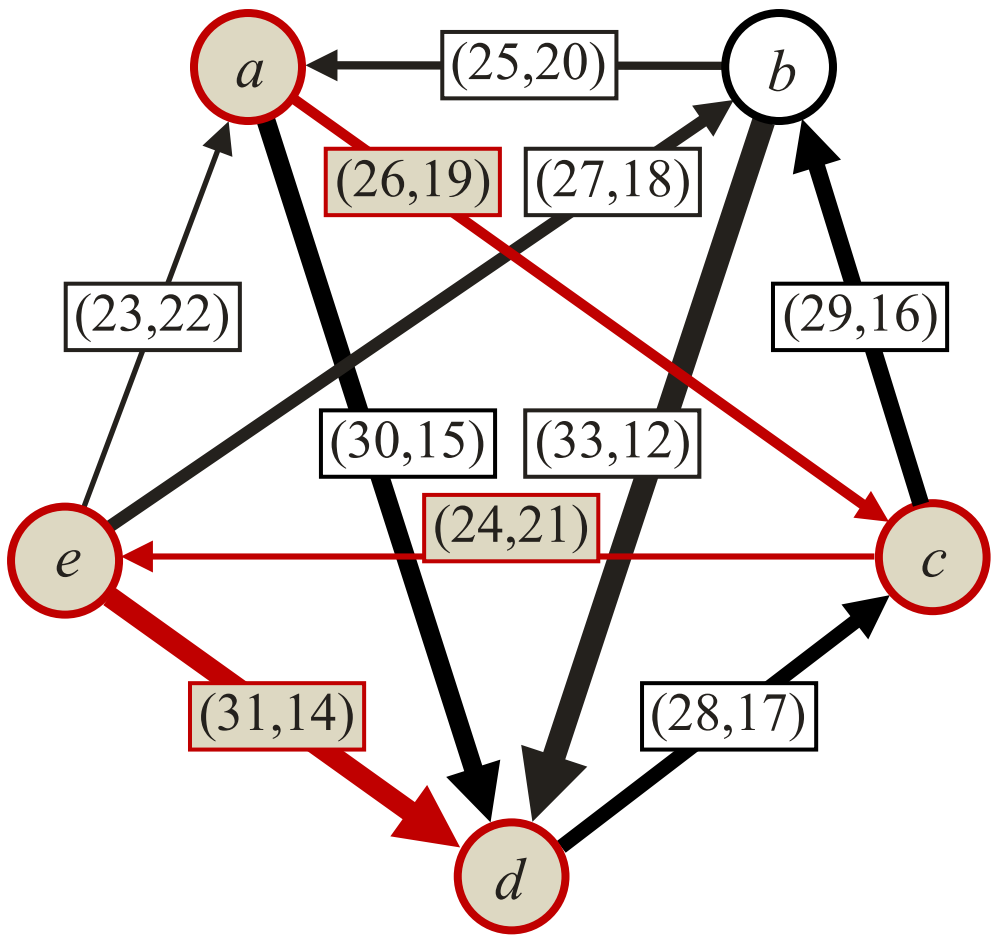

Path 4: *a*, (30,15), *d*
with a strength of (30,15).

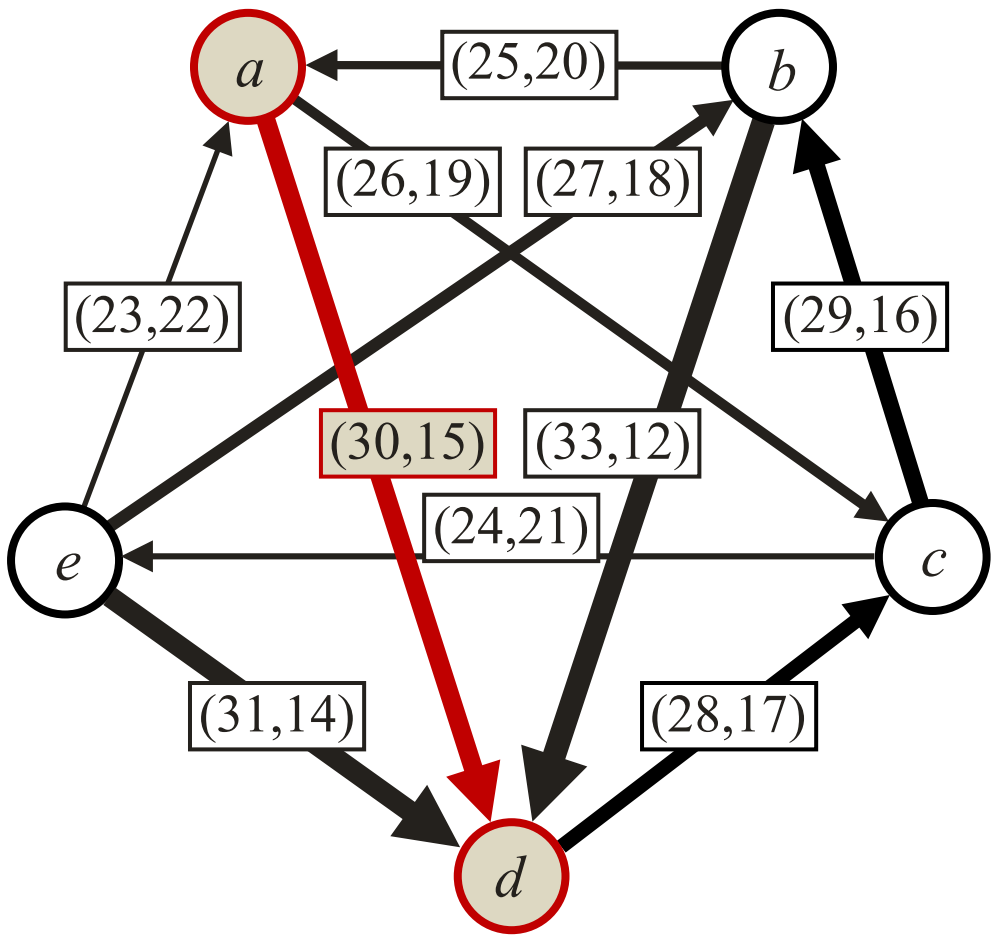

So the strength of the strongest path from alternative *a* to alternative *d* is $\max_D$ { (26,19), (24,21), (24,21), (30,15) } $\approx_D$ (30,15).

$a \rightarrow e$: There are 2 paths from alternative $a$ to alternative $e$.

Path 1: $a$, (26,19), $c$, (24,21), $e$
with a strength of $\min_D$ { (26,19), (24,21) } $\approx_D$ (24,21).

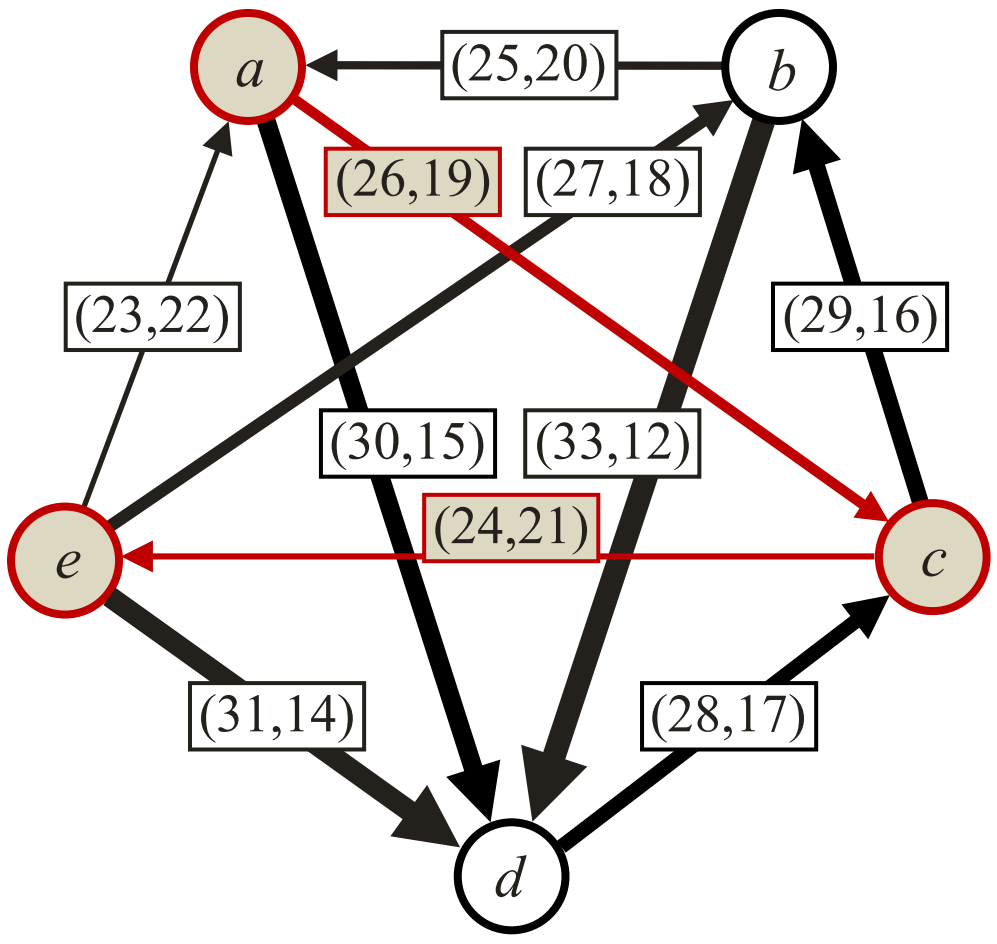

Path 2: $a$, (30,15), $d$, (28,17), $c$, (24,21), $e$
with a strength of $\min_D$ { (30,15), (28,17), (24,21) } $\approx_D$ (24,21).

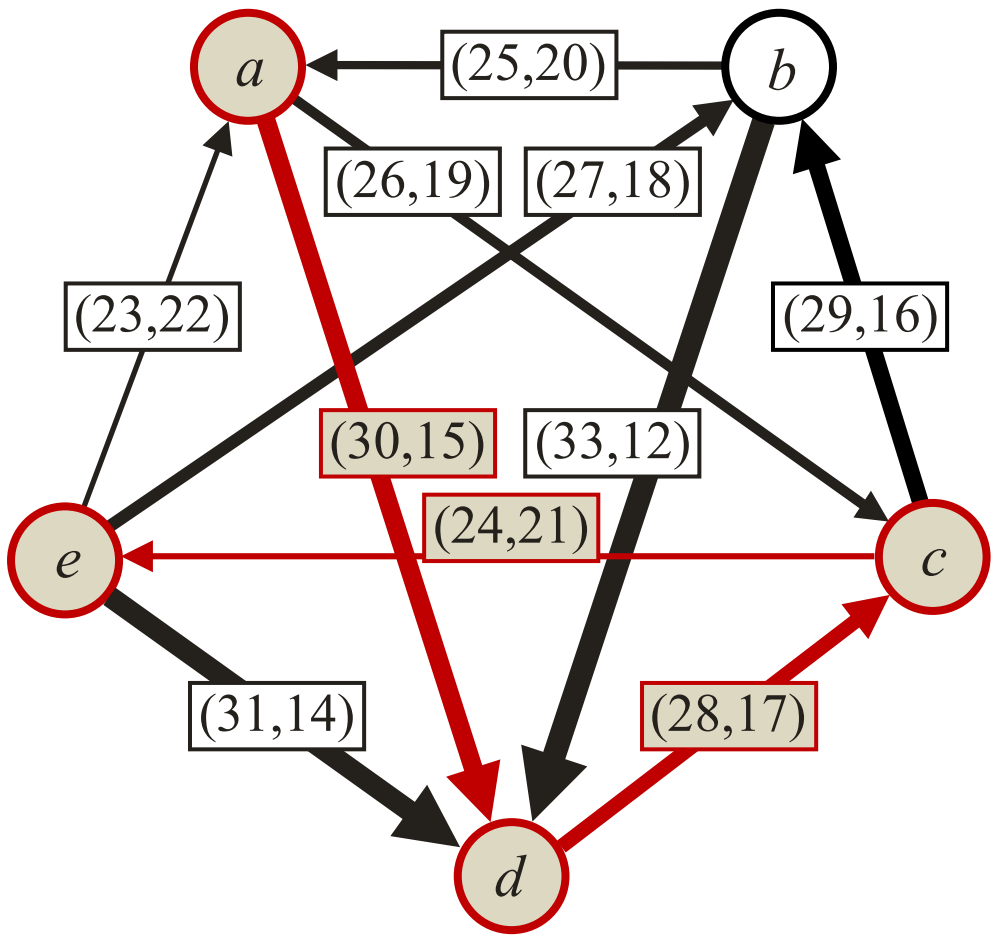

So the strength of the strongest path from alternative $a$ to alternative $e$
is $\max_D$ { (24,21), (24,21) } $\approx_D$ (24,21).

$b \to a$: There are 2 paths from alternative *b* to alternative *a*.

Path 1: *b*, (25,20), *a*
with a strength of (25,20).

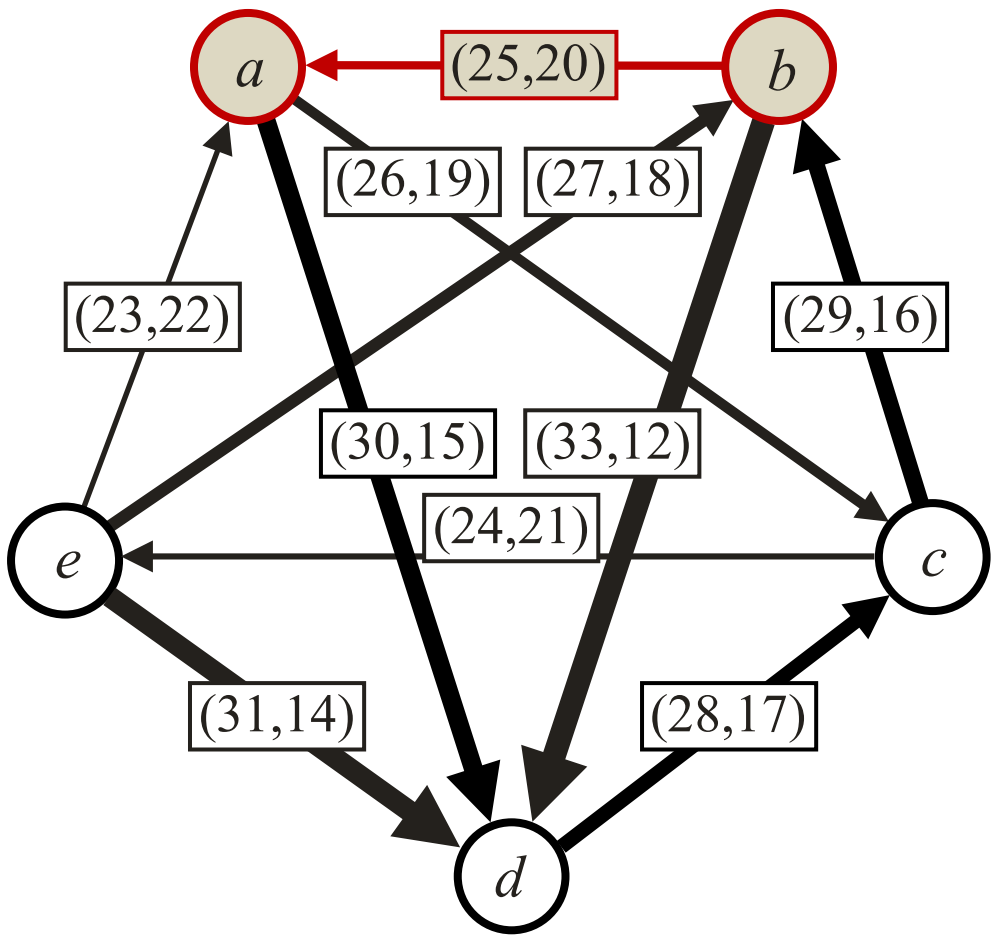

Path 2: *b*, (33,12), *d*, (28,17), *c*, (24,21), *e*, (23,22), *a*
with a strength of $\min_D$ { (33,12), (28,17), (24,21), (23,22) } $\approx_D$ (23,22).

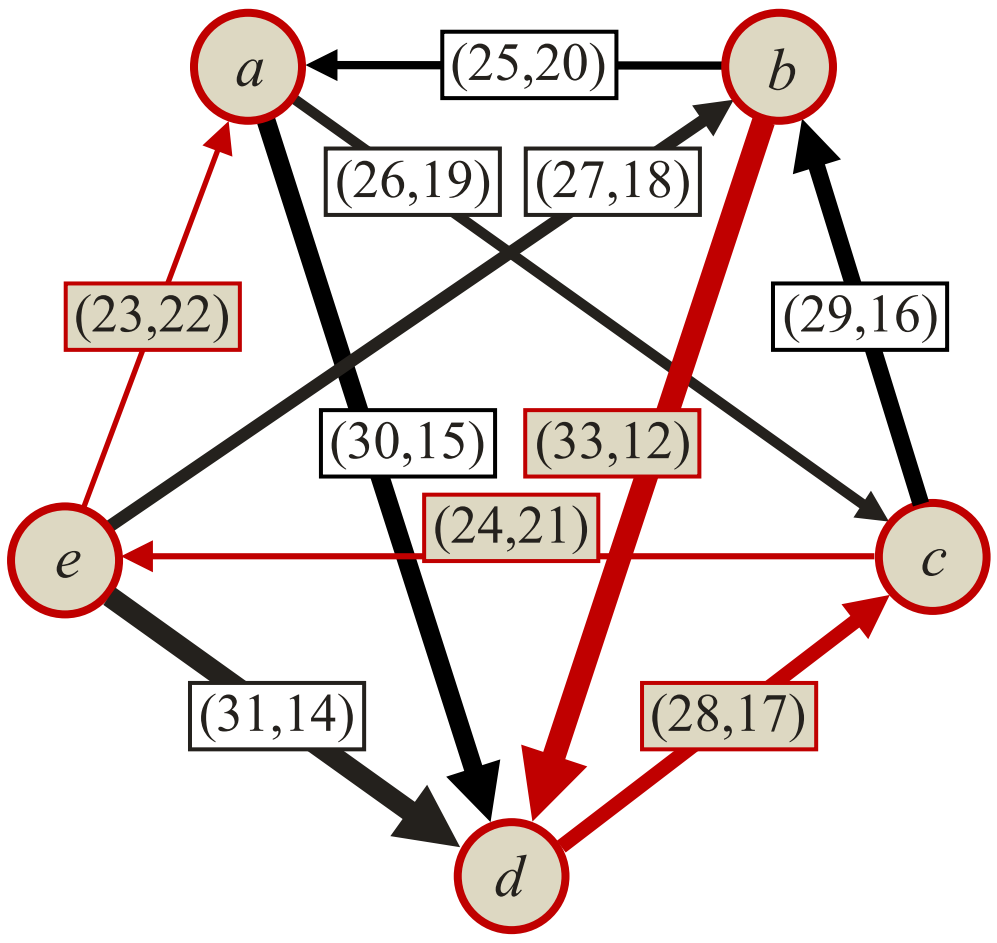

So the strength of the strongest path from alternative *b* to alternative *a* is $\max_D$ { (25,20), (23,22) } $\approx_D$ (25,20).

$b \rightarrow c$: There are 3 paths from alternative $b$ to alternative $c$.

Path 1: $b$, (25,20), $a$, (26,19), $c$
with a strength of $\min_D$ { (25,20), (26,19) } $\approx_D$ (25,20).

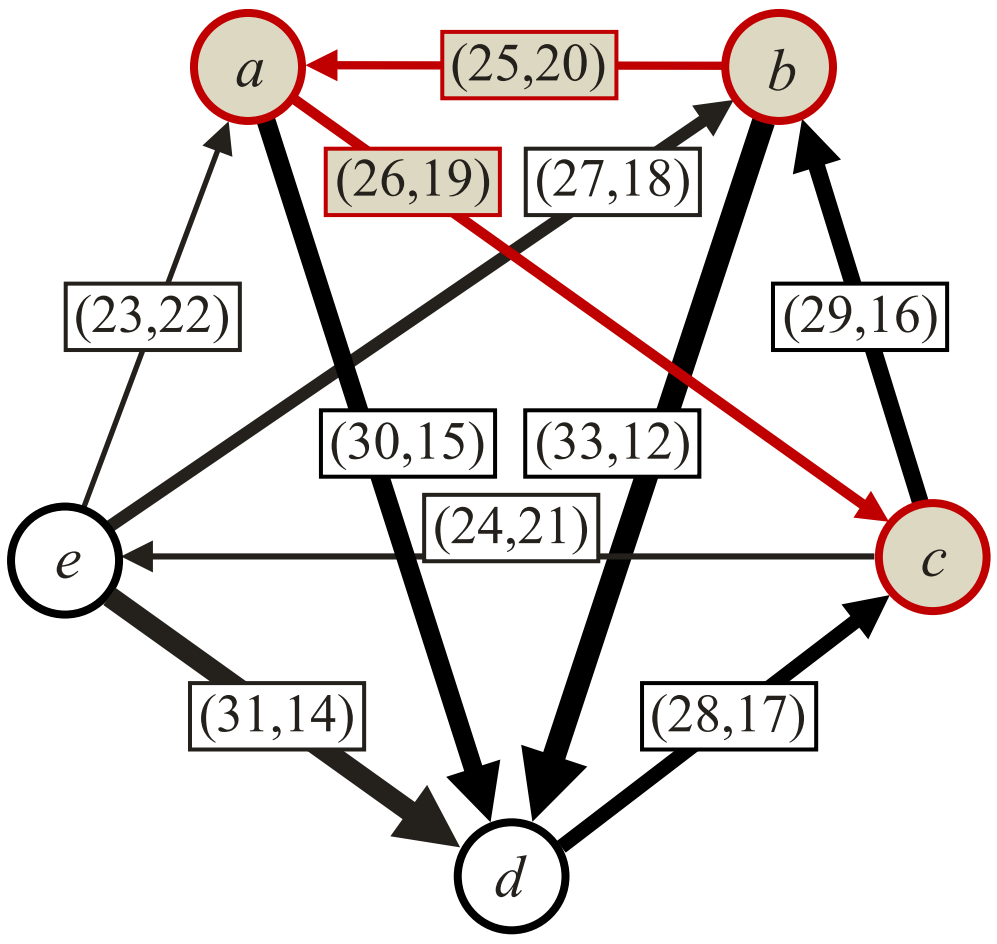

Path 2: $b$, (25,20), $a$, (30,15), $d$, (28,17), $c$
with a strength of $\min_D$ { (25,20), (30,15), (28,17) } $\approx_D$ (25,20).

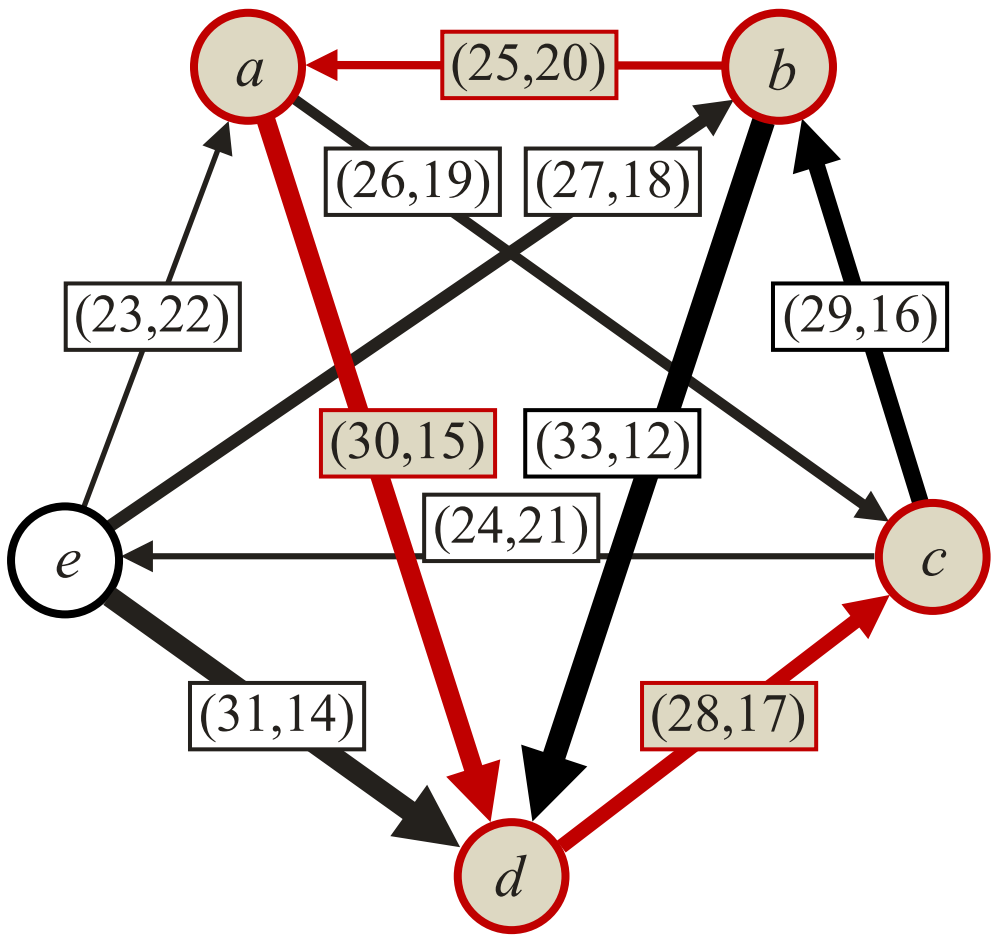

Path 3: *b*, (33,12), *d*, (28,17), *c*
with a strength of $\min_D$ { (33,12), (28,17) } $\approx_D$ (28,17).

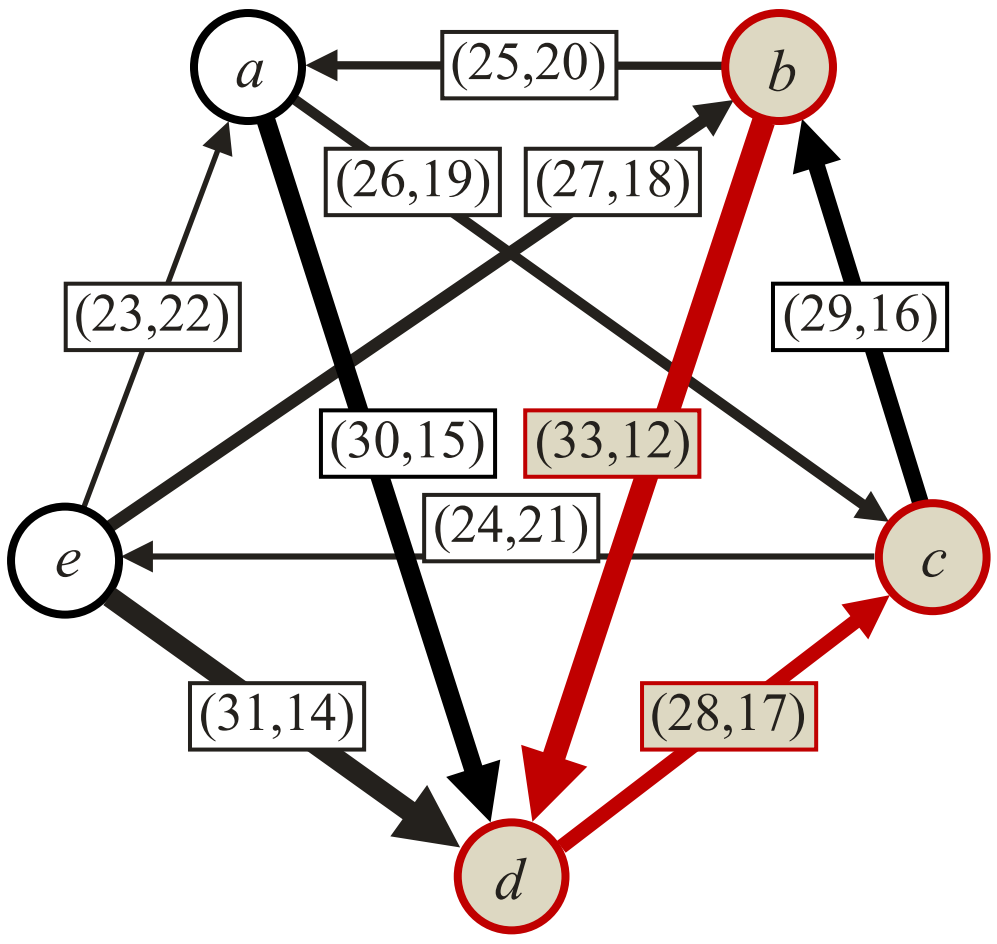

So the strength of the strongest path from alternative *b* to alternative *c* is $\max_D$ { (25,20), (25,20), (28,17) } $\approx_D$ (28,17).

*b* → *d*: There are 3 paths from alternative *b* to alternative *d*.

Path 1: *b*, (25,20), *a*, (26,19), *c*, (24,21), *e*, (31,14), *d*
with a strength of $\min_D$ { (25,20), (26,19), (24,21), (31,14) } $\approx_D$ (24,21).

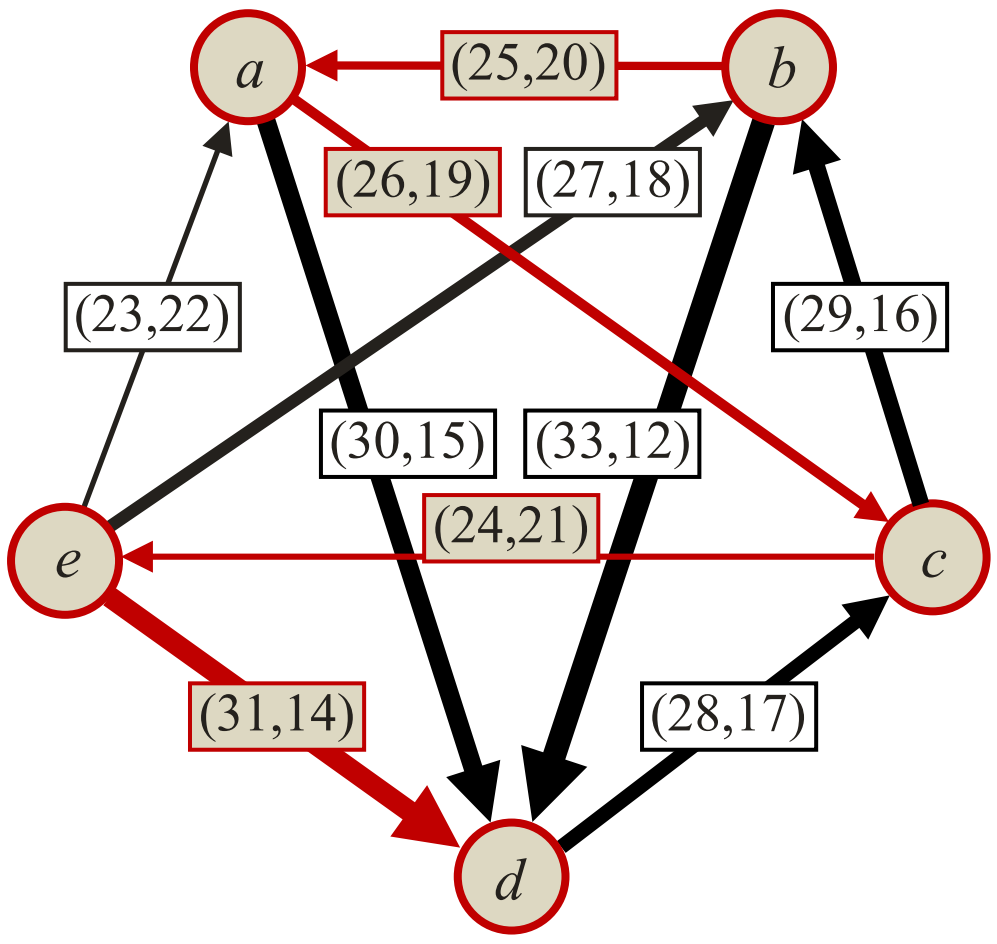

Path 2: *b*, (25,20), *a*, (30,15), *d*
with a strength of $\min_D$ { (25,20), (30,15) } $\approx_D$ (25,20).

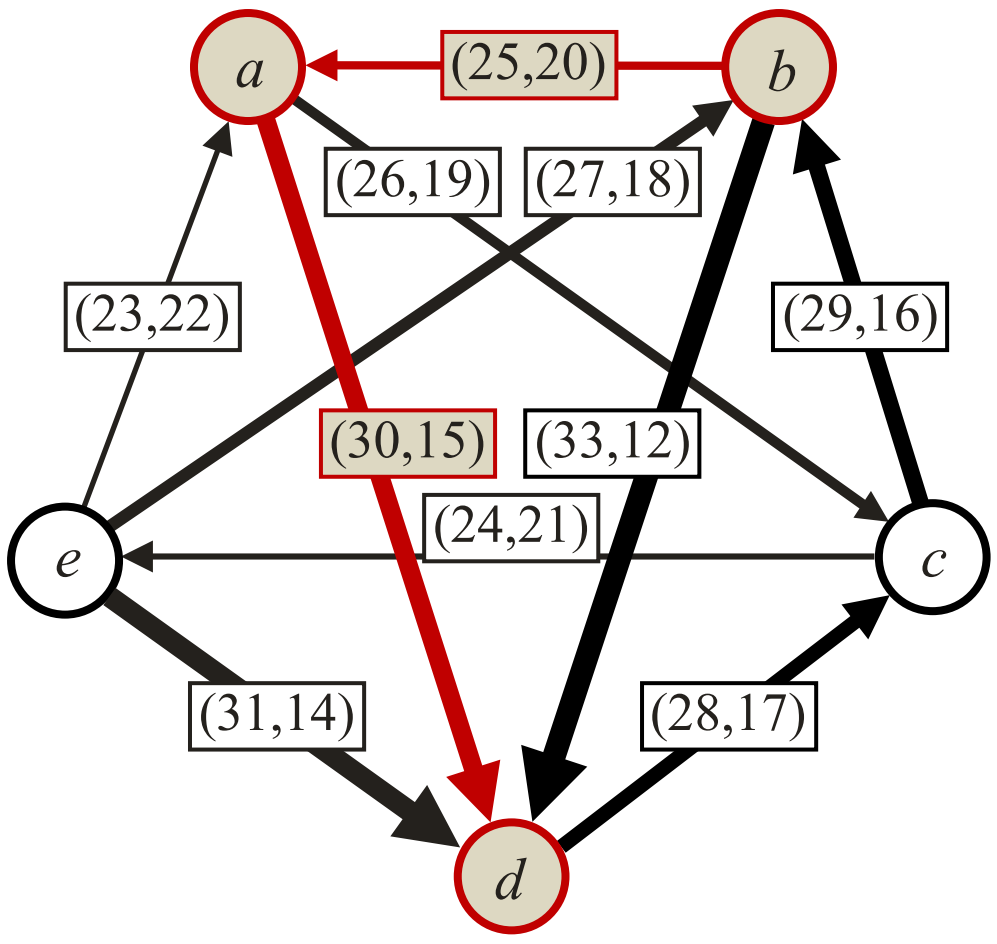

Path 3: *b*, (33,12), *d*
with a strength of (33,12).

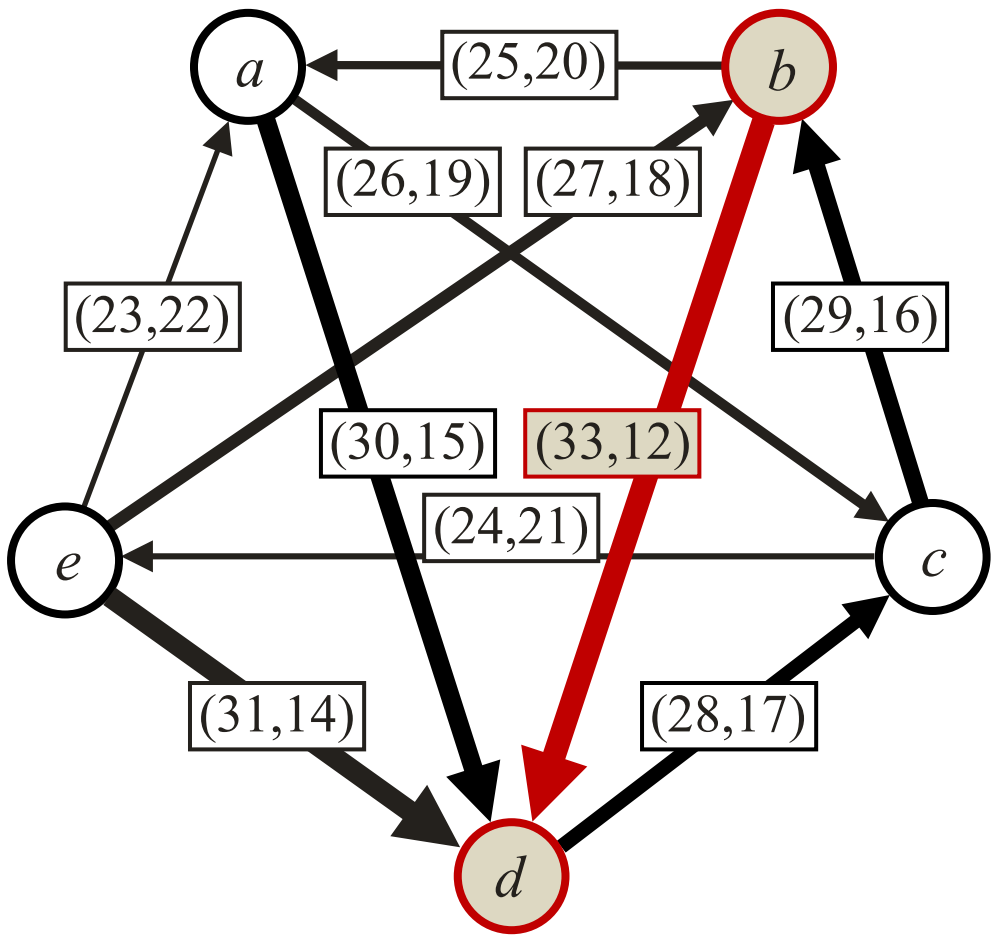

So the strength of the strongest path from alternative *b* to alternative *d* is $\max_D$ { (24,21), (25,20), (33,12) } $\approx_D$ (33,12).

*b* → *e*: There are 3 paths from alternative *b* to alternative *e*.

Path 1: *b*, (25,20), *a*, (26,19), *c*, (24,21), *e*
with a strength of $\min_D$ { (25,20), (26,19), (24,21) } $\approx_D$ (24,21).

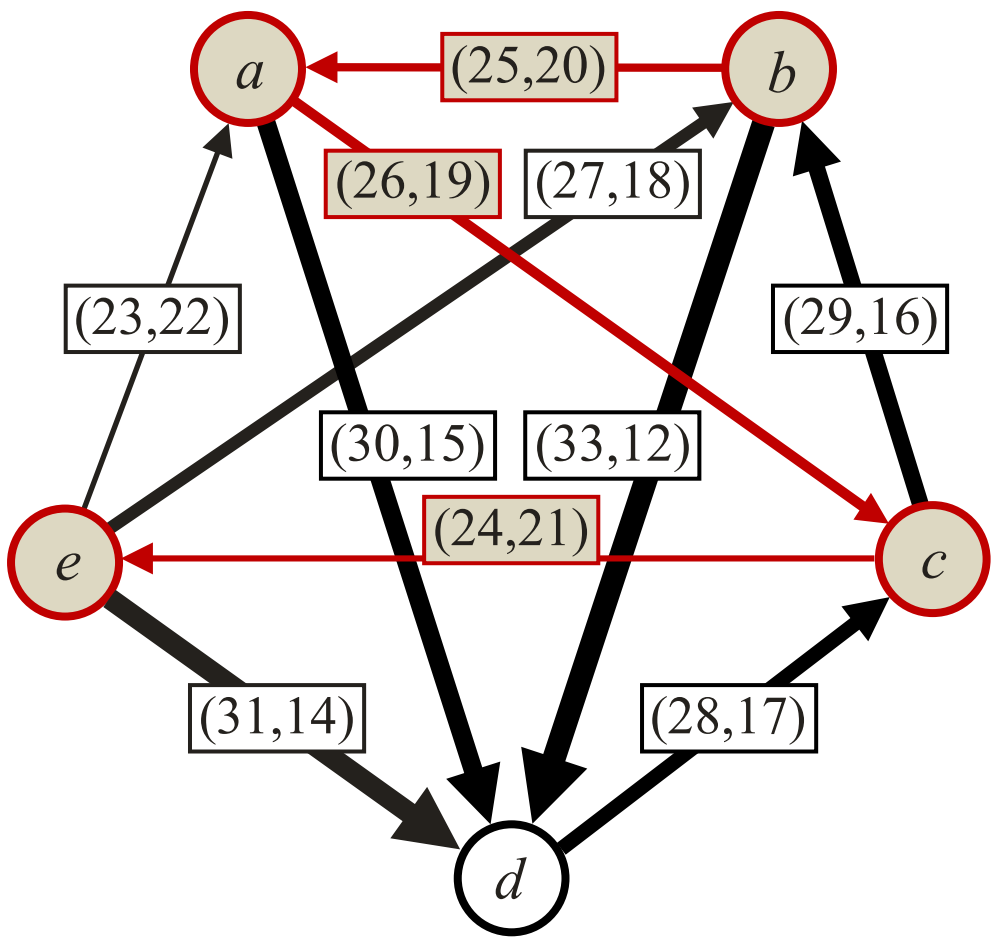

Path 2: *b*, (25,20), *a*, (30,15), *d*, (28,17), *c*, (24,21), *e*
with a strength of $\min_D$ { (25,20), (30,15), (28,17), (24,21) } $\approx_D$ (24,21).

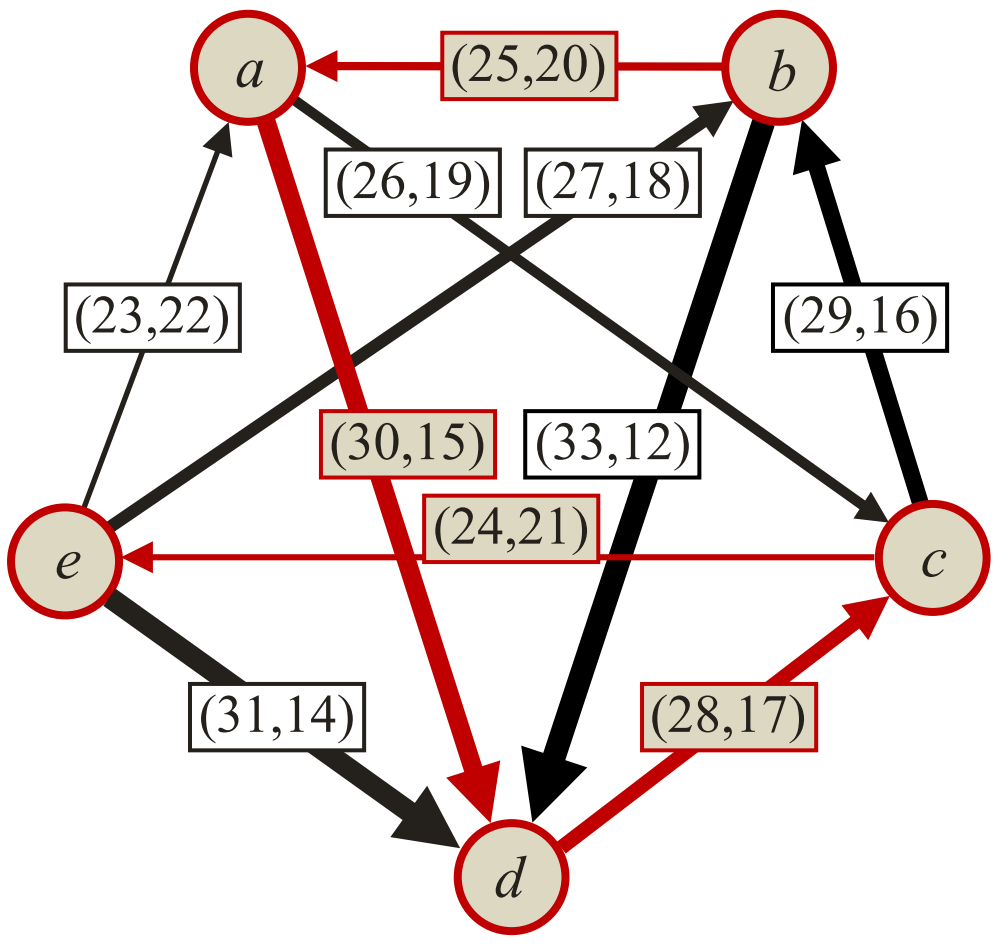

Path 3: *b*, (33,12), *d*, (28,17), *c*, (24,21), *e*
with a strength of $\min_D$ { (33,12), (28,17), (24,21) } $\approx_D$ (24,21).

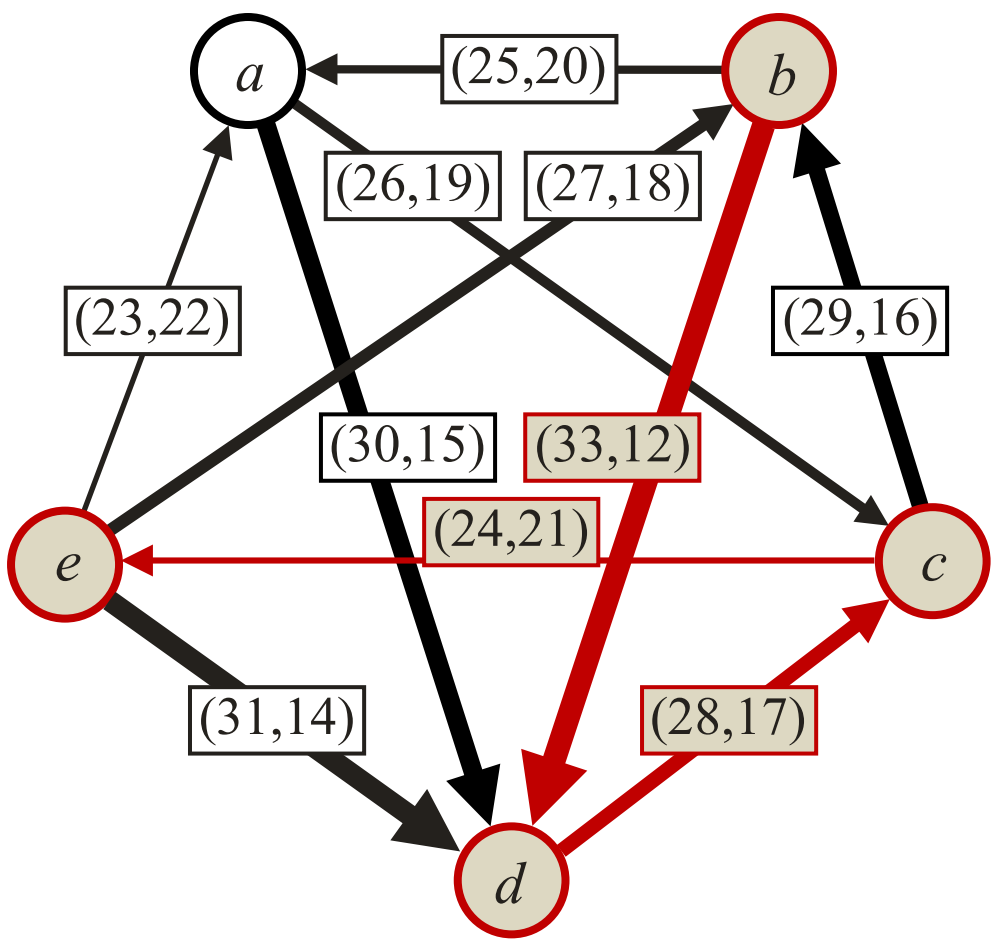

So the strength of the strongest path from alternative *b* to alternative *e* is $\max_D$ { (24,21), (24,21), (24,21) } $\approx_D$ (24,21).

*c* → *a*: There are 3 paths from alternative *c* to alternative *a*.

Path 1: *c*, (29,16), *b*, (25,20), *a*
with a strength of $\min_D$ { (29,16), (25,20) } $\approx_D$ (25,20).

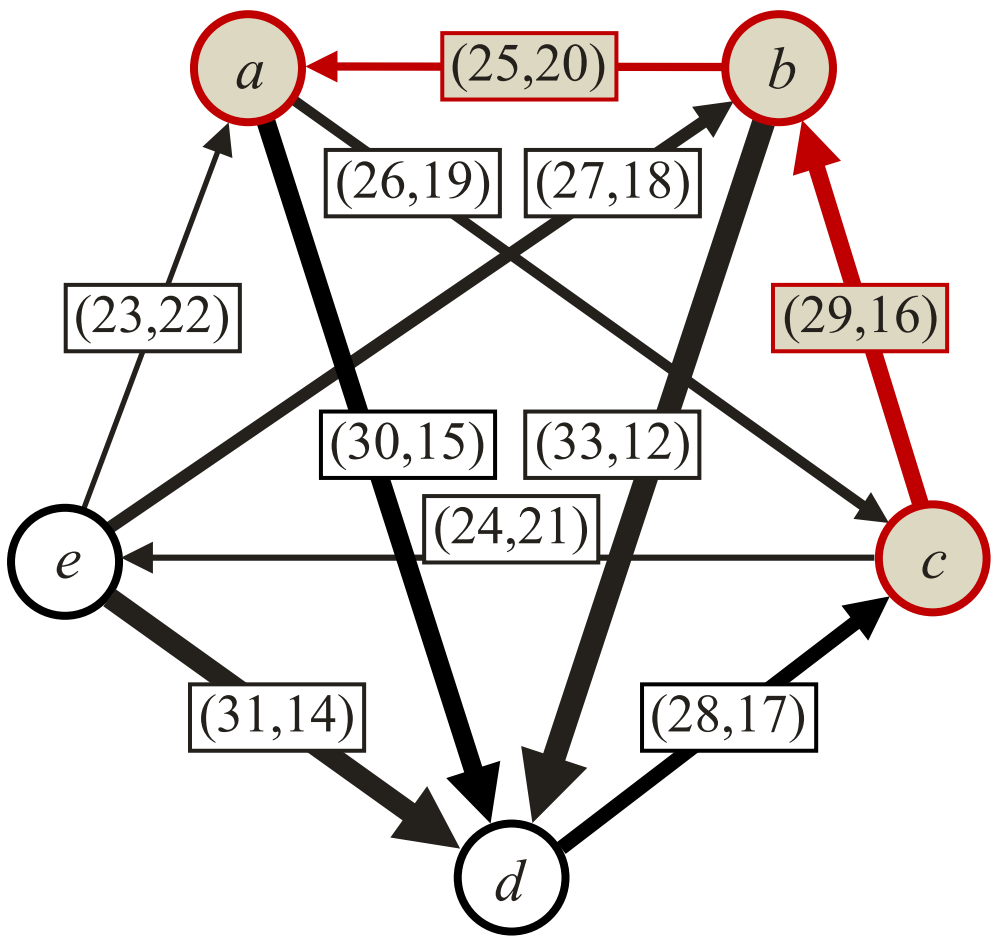

Path 2: *c*, (24,21), *e*, (23,22), *a*
with a strength of $\min_D$ { (24,21), (23,22) } $\approx_D$ (23,22).

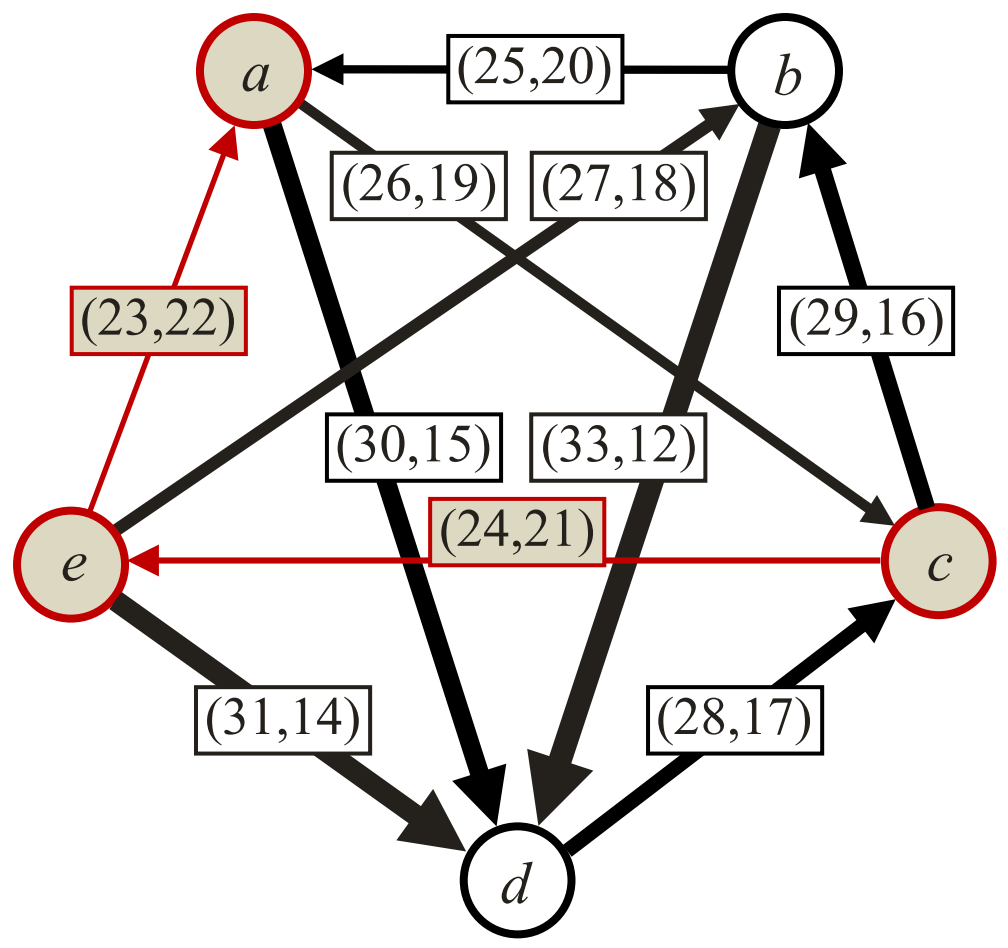

Path 3: *c*, (24,21), *e*, (27,18), *b*, (25,20), *a*
with a strength of $\min_D$ { (24,21), (27,18), (25,20) } $\approx_D$ (24,21).

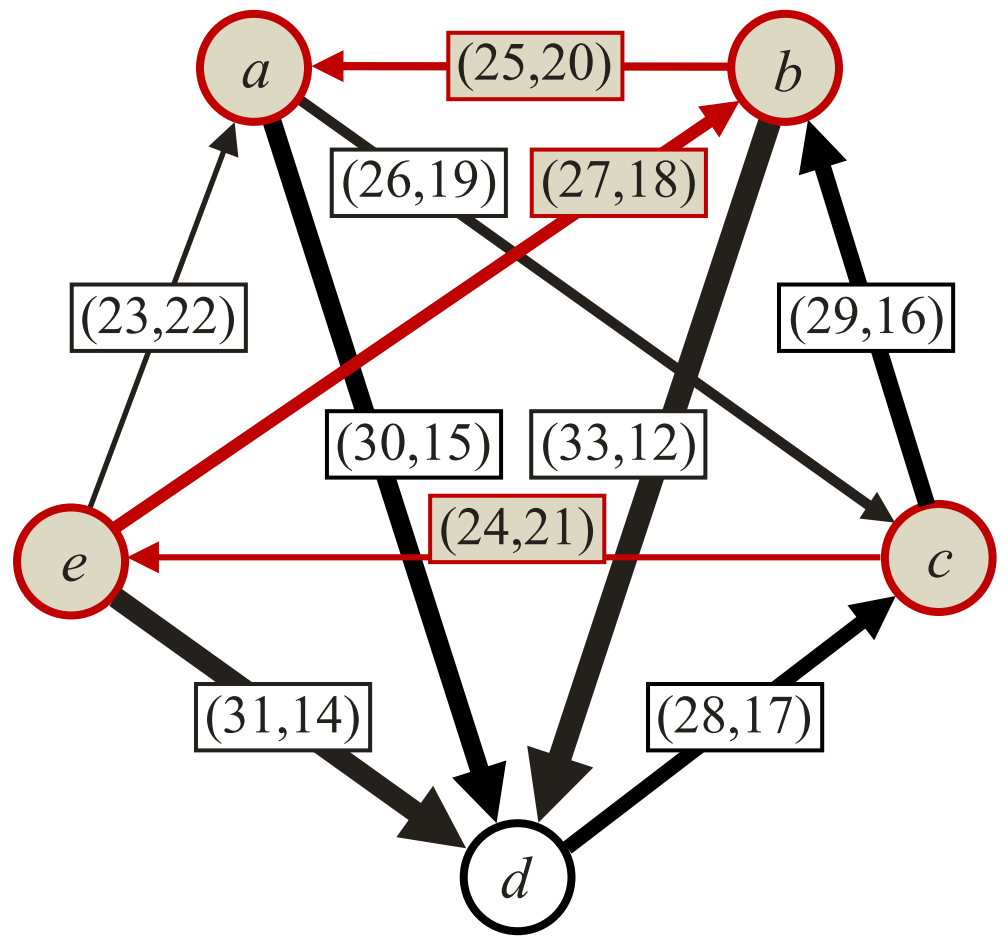

So the strength of the strongest path from alternative *c* to alternative *a* is $\max_D$ { (25,20), (23,22), (24,21) } $\approx_D$ (25,20).

$c \to b$: There are 2 paths from alternative $c$ to alternative $b$.

Path 1: $c$, (29,16), $b$
with a strength of (29,16).

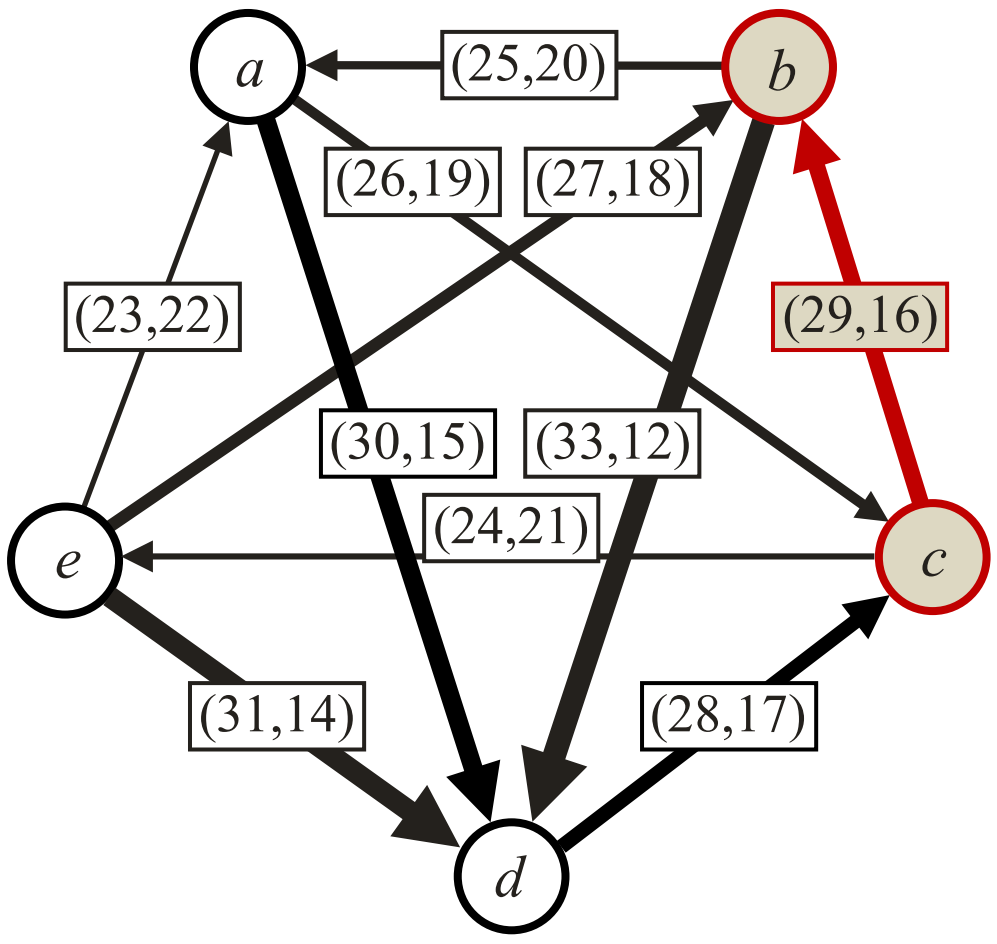

Path 2: $c$, (24,21), $e$, (27,18), $b$
with a strength of $\min_D$ { (24,21), (27,18) } $\approx_D$ (24,21).

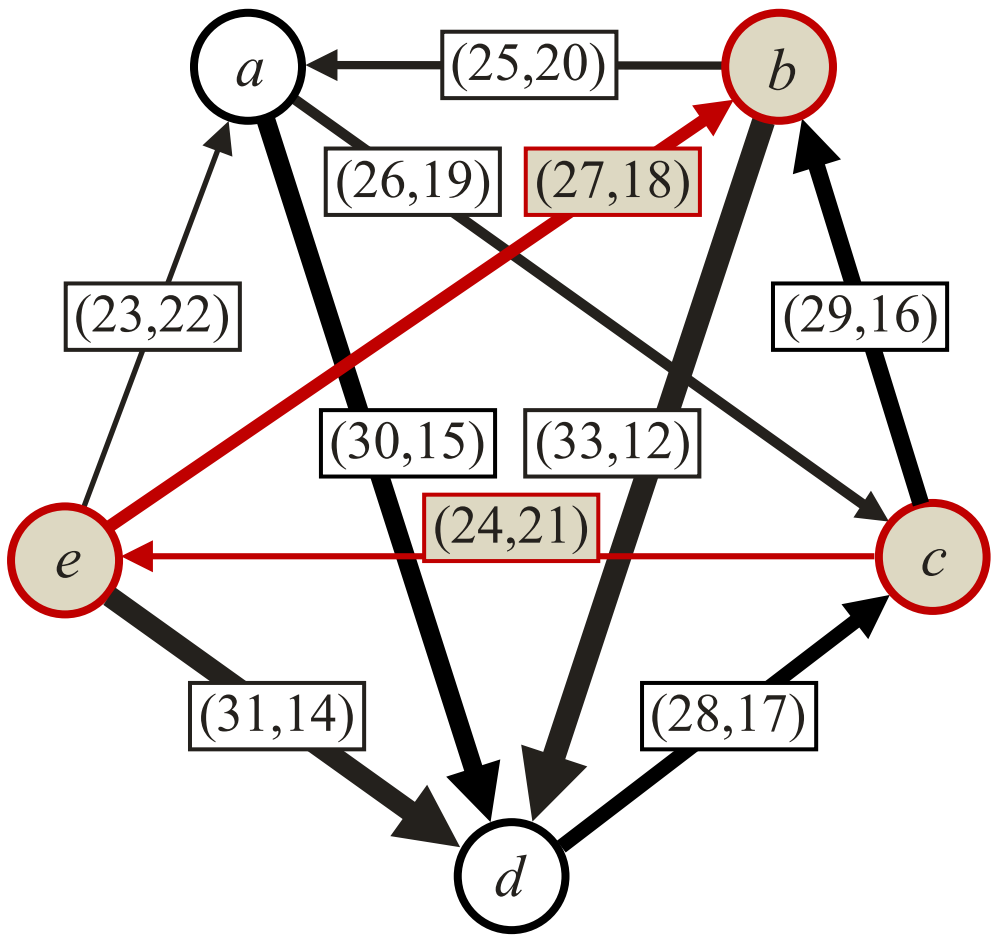

So the strength of the strongest path from alternative $c$ to alternative $b$ is $\max_D$ { (29,16), (24,21) } $\approx_D$ (29,16).

$c \rightarrow d$: There are 6 paths from alternative $c$ to alternative $d$.

Path 1: $c$, (29,16), $b$, (25,20), $a$, (30,15), $d$
with a strength of $\min_D$ { (29,16), (25,20), (30,15) } $\approx_D$ (25,20).

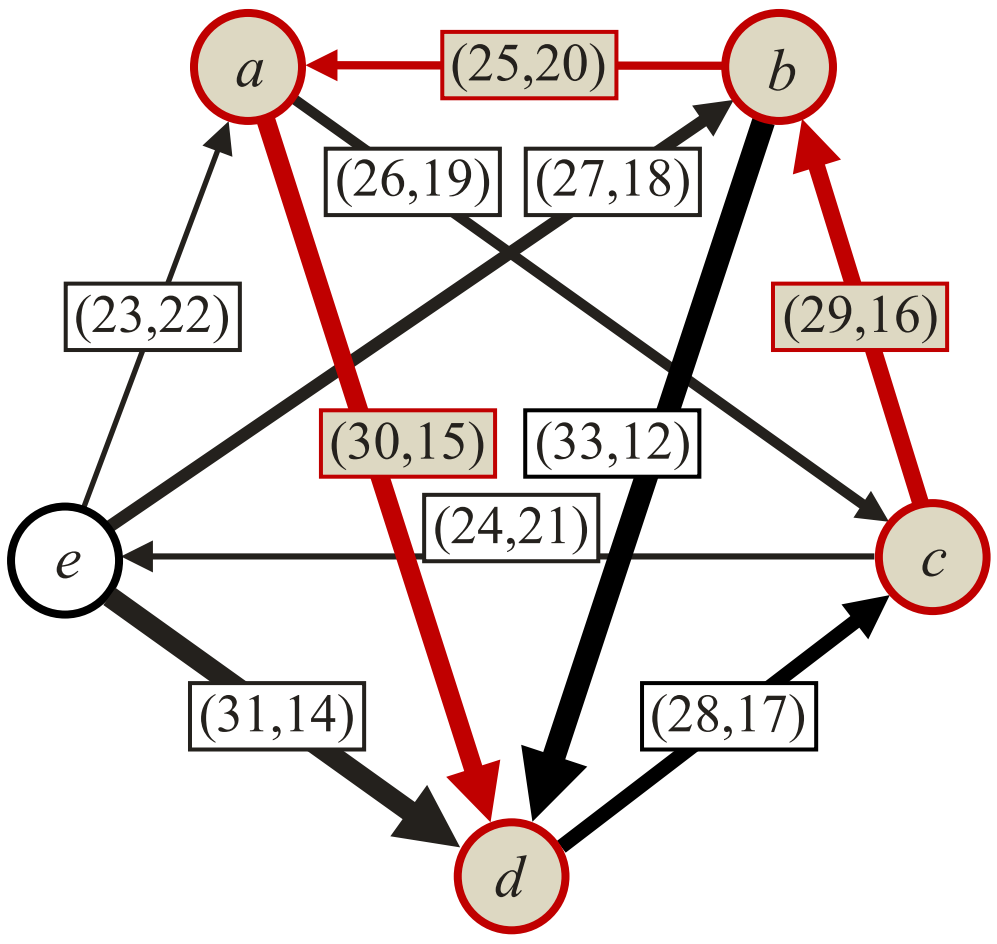

Path 2: $c$, (29,16), $b$, (33,12), $d$
with a strength of $\min_D$ { (29,16), (33,12) } $\approx_D$ (29,16).

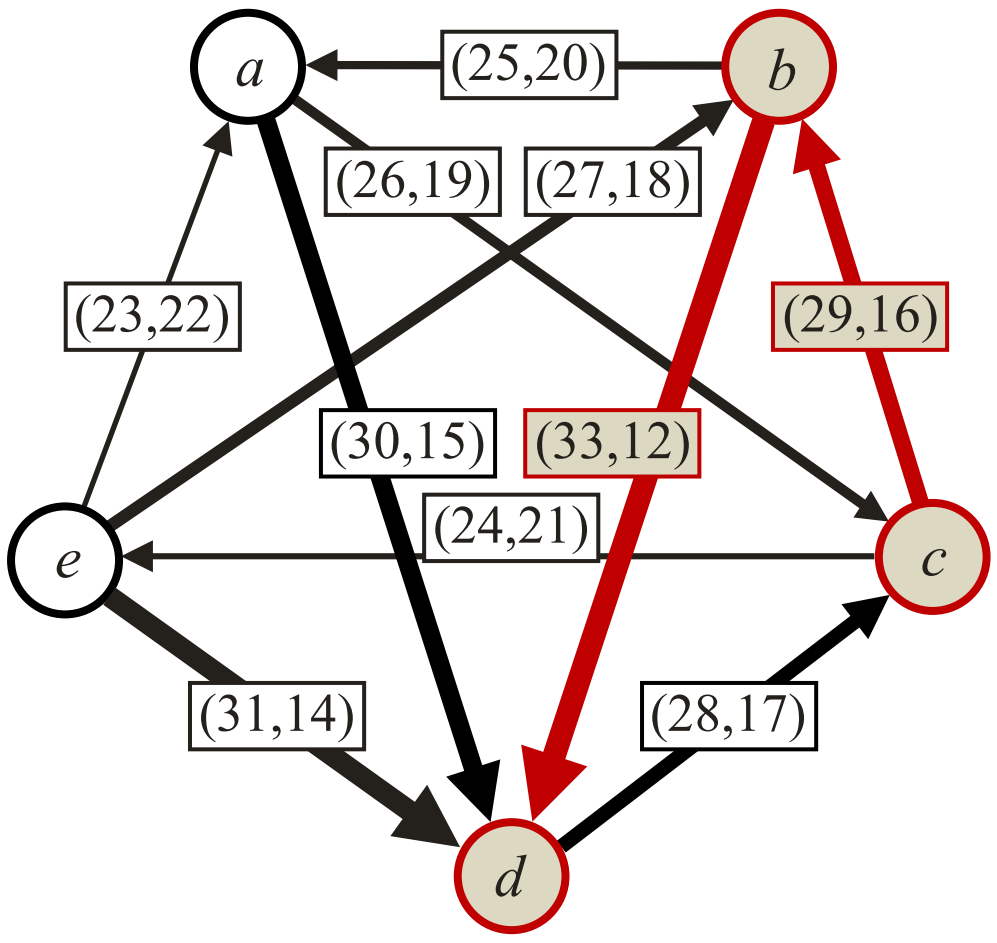

Path 3: *c*, (24,21), *e*, (23,22), *a*, (30,15), *d*
with a strength of $\min_D$ { (24,21), (23,22), (30,15) } $\approx_D$ (23,22).

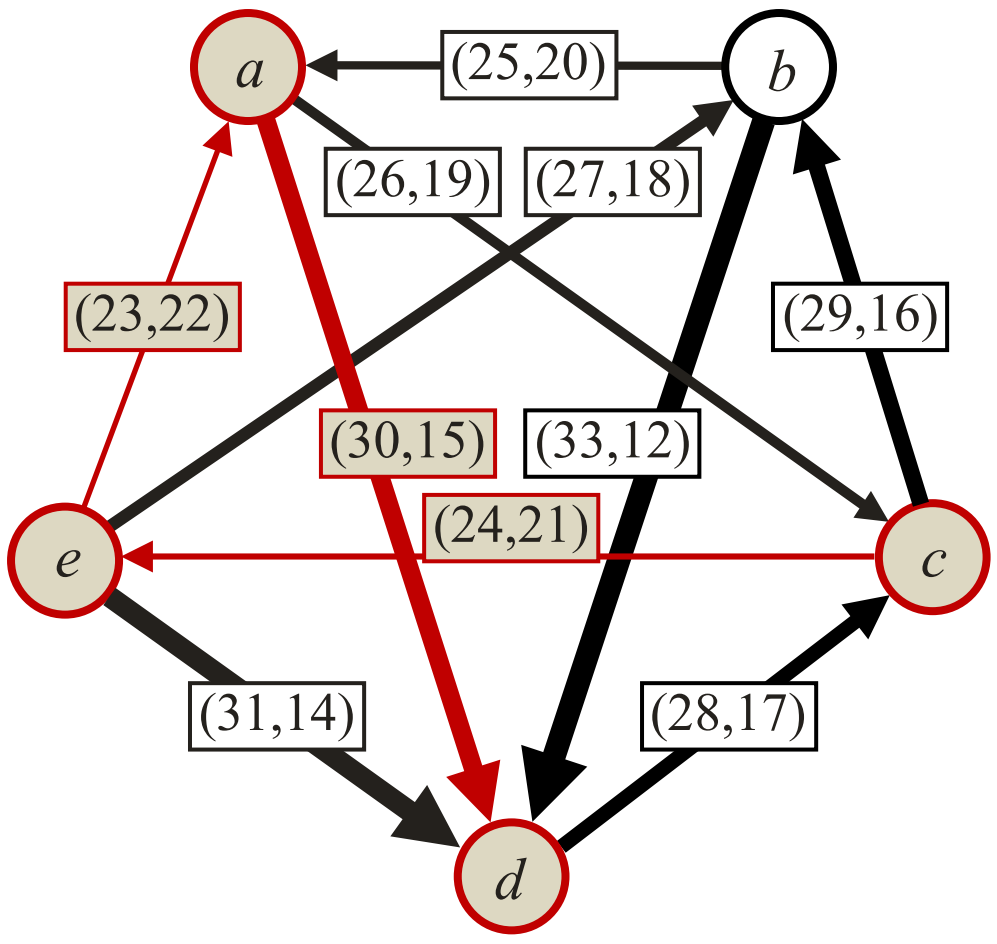

Path 4: *c*, (24,21), *e*, (27,18), *b*, (25,20), *a*, (30,15), *d*
with a strength of $\min_D$ { (24,21), (27,18), (25,20), (30,15) } $\approx_D$ (24,21).

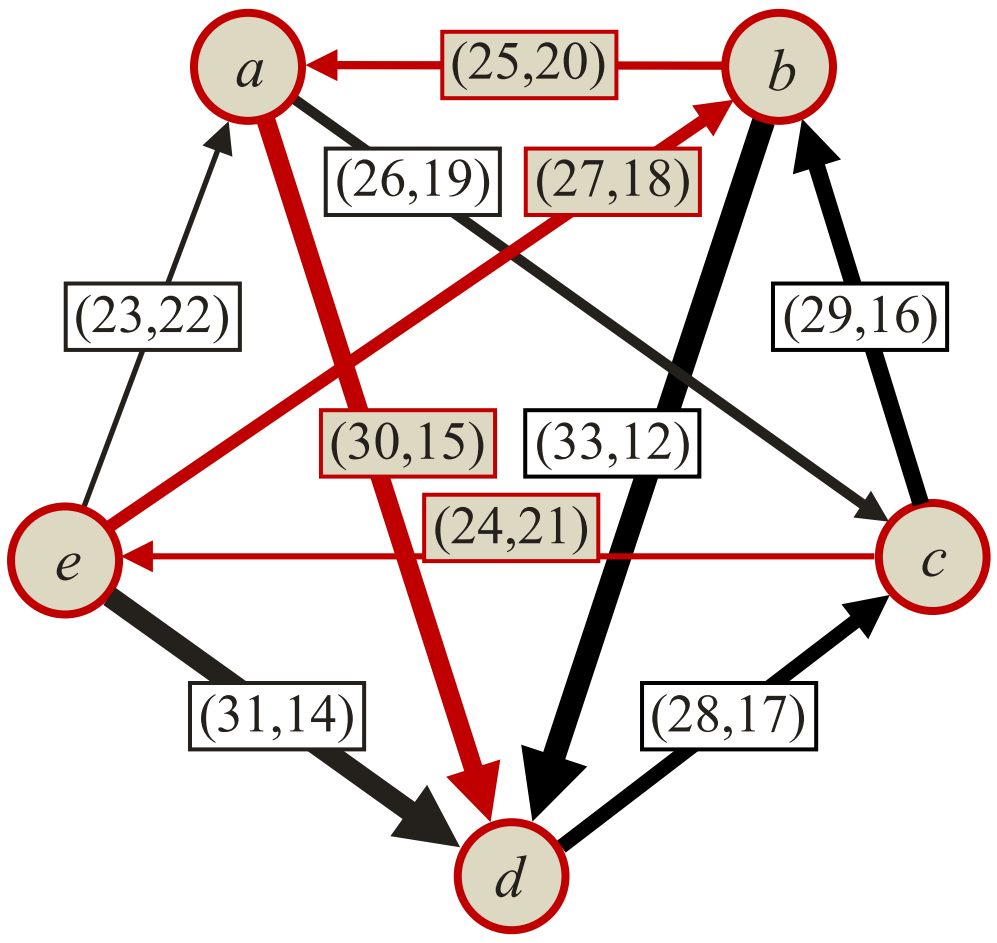

Path 5: *c*, (24,21), *e*, (27,18), *b*, (33,12), *d*
with a strength of $\min_D$ { (24,21), (27,18), (33,12) } $\approx_D$ (24,21).

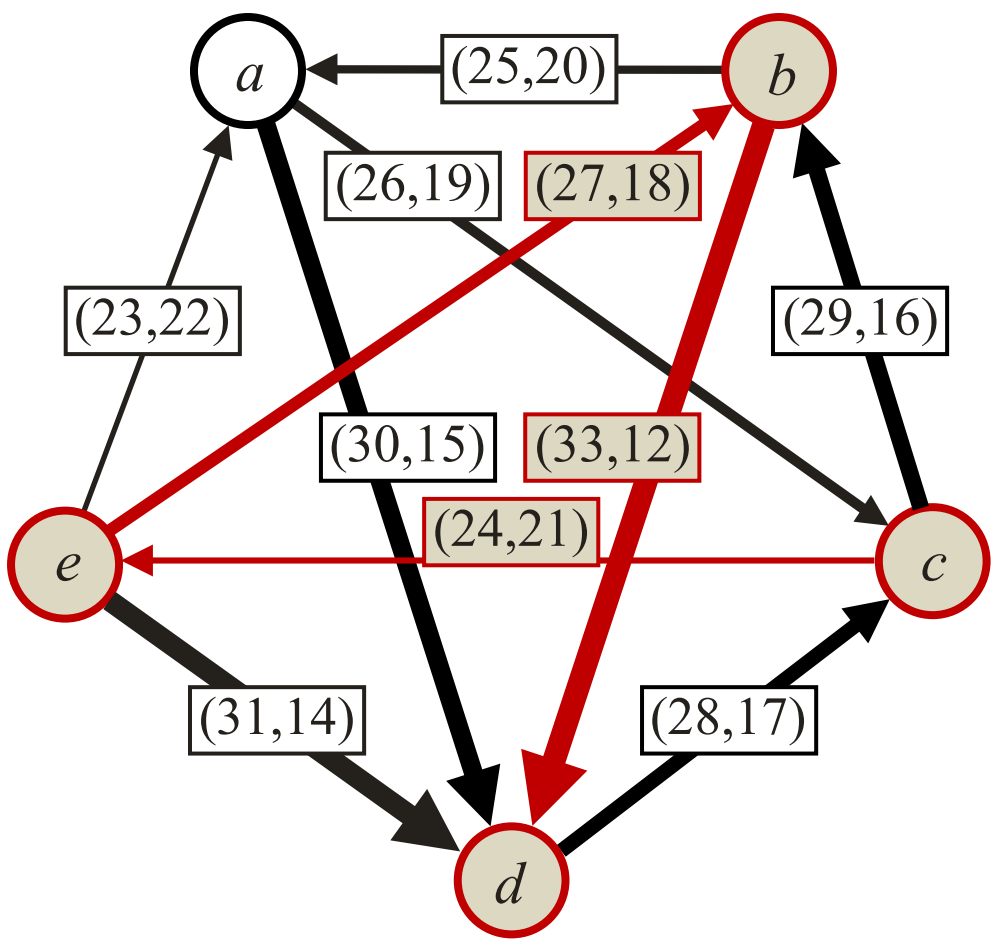

Path 6: *c*, (24,21), *e*, (31,14), *d*
with a strength of $\min_D$ { (24,21), (31,14) } $\approx_D$ (24,21).

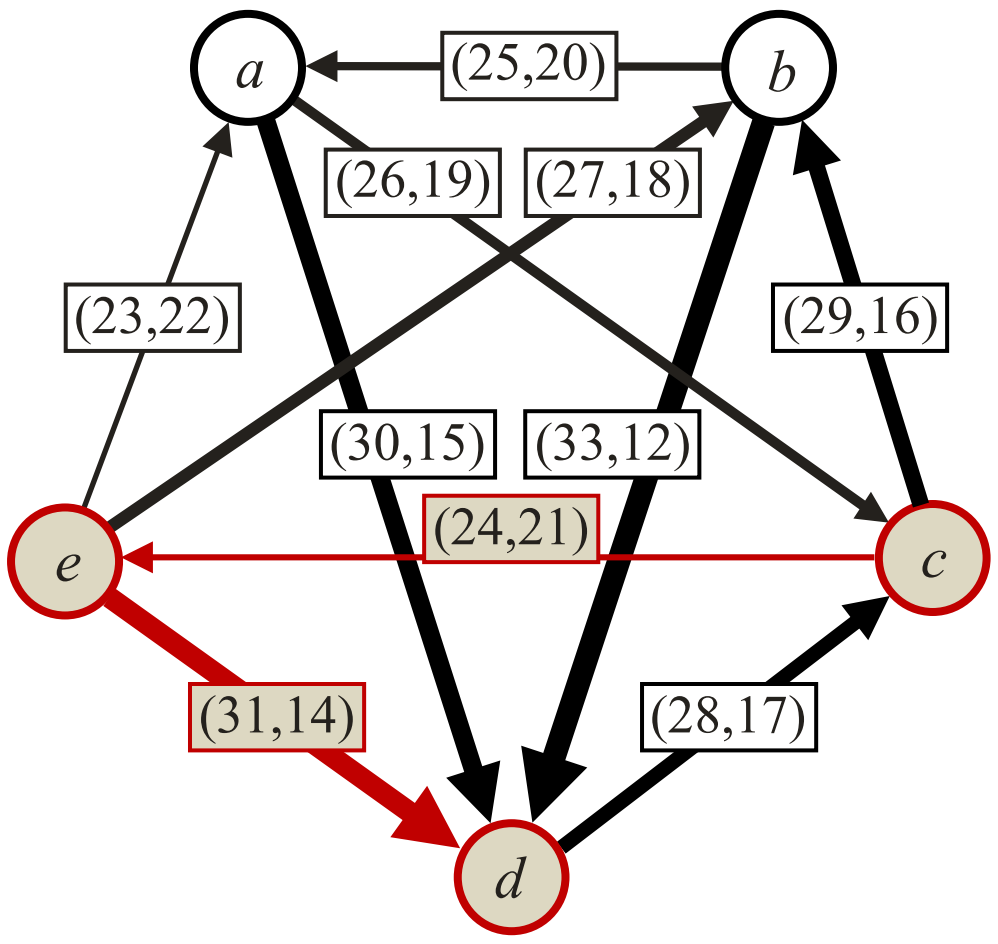

So the strength of the strongest path from alternative *c* to alternative *d* is $\max_D$ { (25,20), (29,16), (23,22), (24,21), (24,21), (24,21) } $\approx_D$ (29,16).

*c* → *e*: There is only one path from alternative *c* to alternative *e*.

Path 1: *c*, (24,21), *e*
with a strength of (24,21).

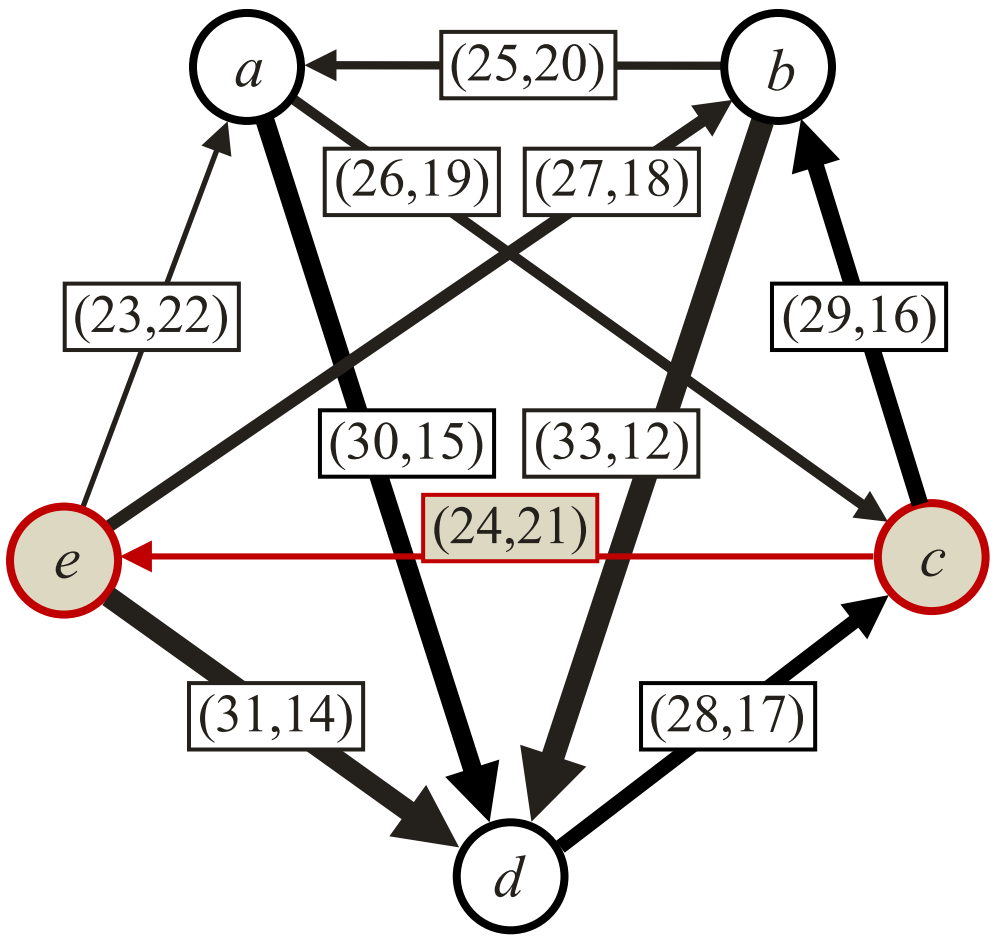

So the strength of the strongest path from alternative *c* to alternative *e* is (24,21).

*d* → *a*: There are 3 paths from alternative *d* to alternative *a*.

Path 1: *d*, (28,17), *c*, (29,16), *b*, (25,20), *a*
with a strength of $\min_D$ { (28,17), (29,16), (25,20) } $\approx_D$ (25,20).

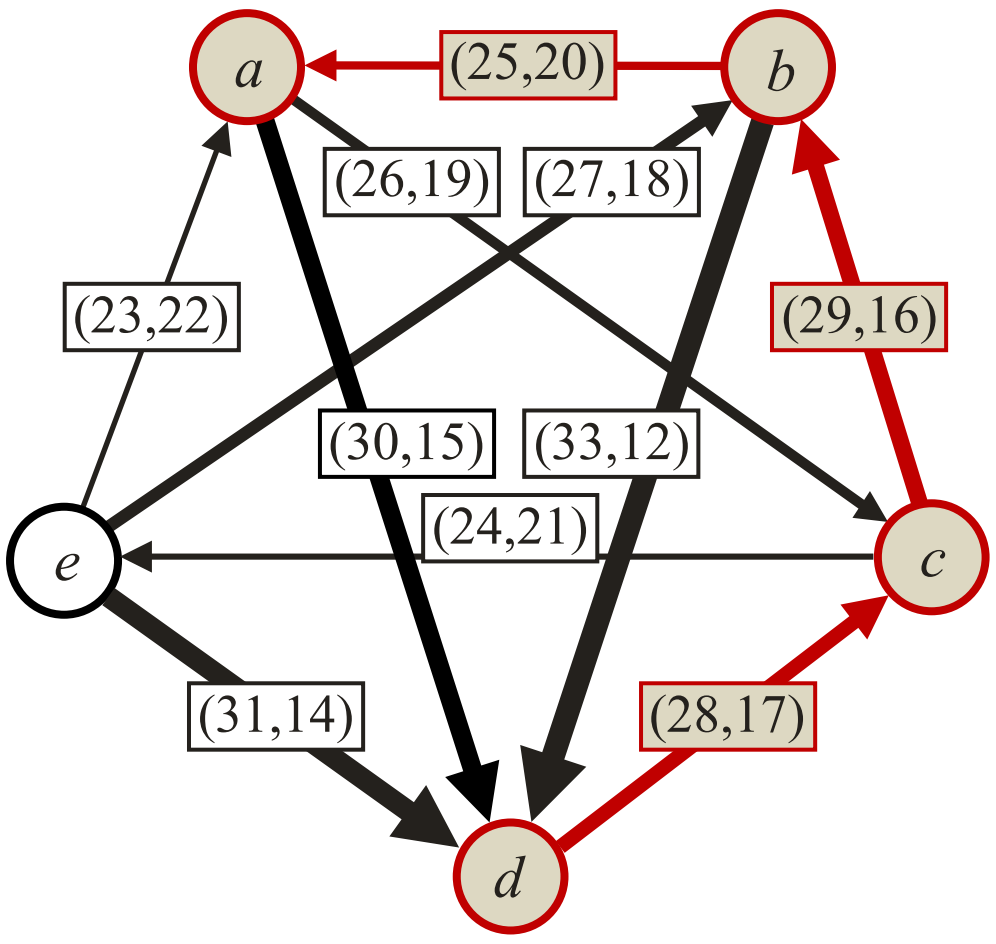

Path 2: *d*, (28,17), *c*, (24,21), *e*, (23,22), *a*
with a strength of $\min_D$ { (28,17), (24,21), (23,22) } $\approx_D$ (23,22).

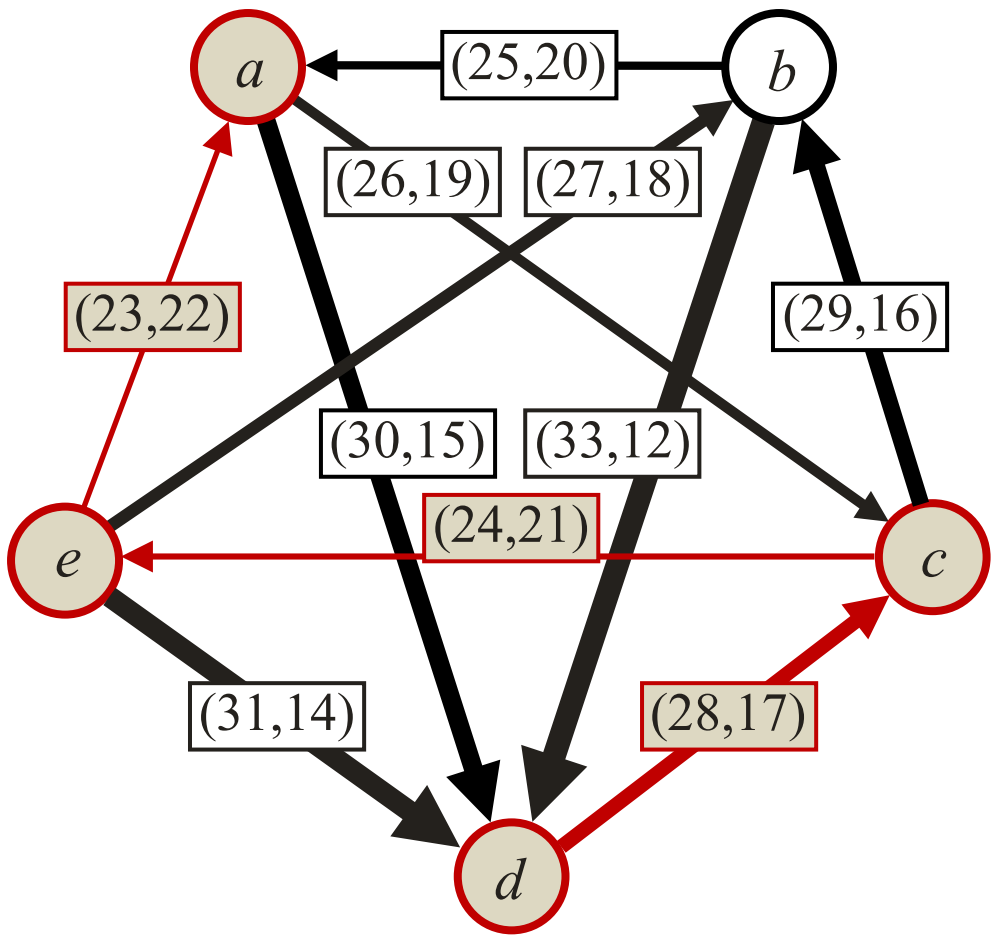

Path 3: *d*, (28,17), *c*, (24,21), *e*, (27,18), *b*, (25,20), *a*
with a strength of $\min_D$ { (28,17), (24,21), (27,18), (25,20) } $\approx_D$ (24,21).

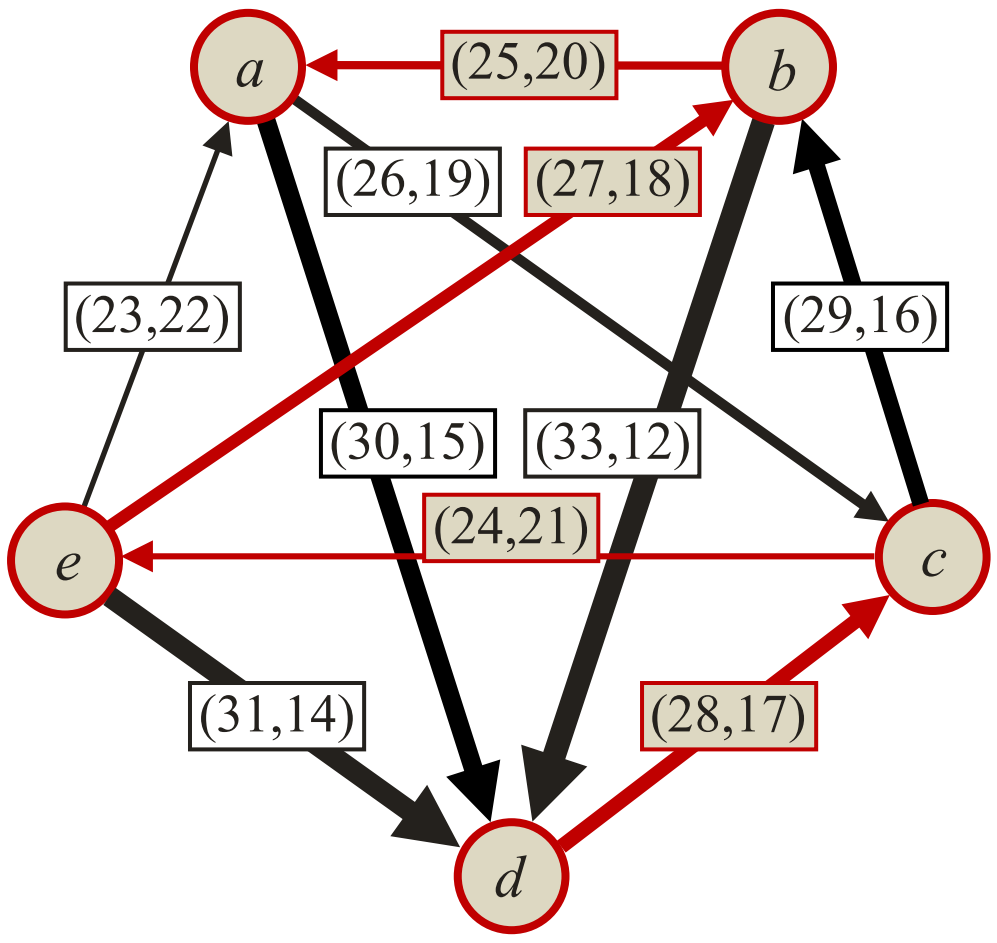

So the strength of the strongest path from alternative *d* to alternative *a* is $\max_D$ { (25,20), (23,22), (24,21) } $\approx_D$ (25,20).

$d \rightarrow b$: There are 2 paths from alternative $d$ to alternative $b$.

Path 1: $d$, (28,17), $c$, (29,16), $b$
with a strength of $\min_D$ { (28,17), (29,16) } $\approx_D$ (28,17).

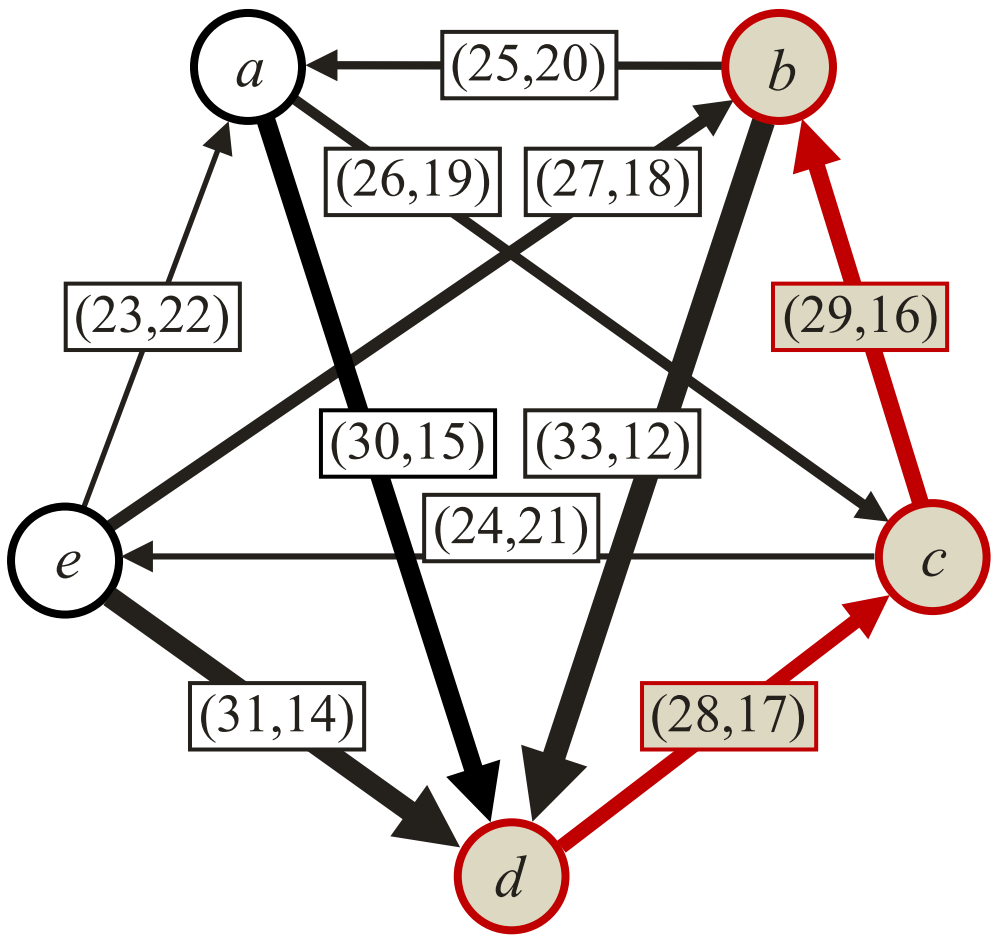

Path 2: $d$, (28,17), $c$, (24,21), $e$, (27,18), $b$
with a strength of $\min_D$ { (28,17), (24,21), (27,18) } $\approx_D$ (24,21).

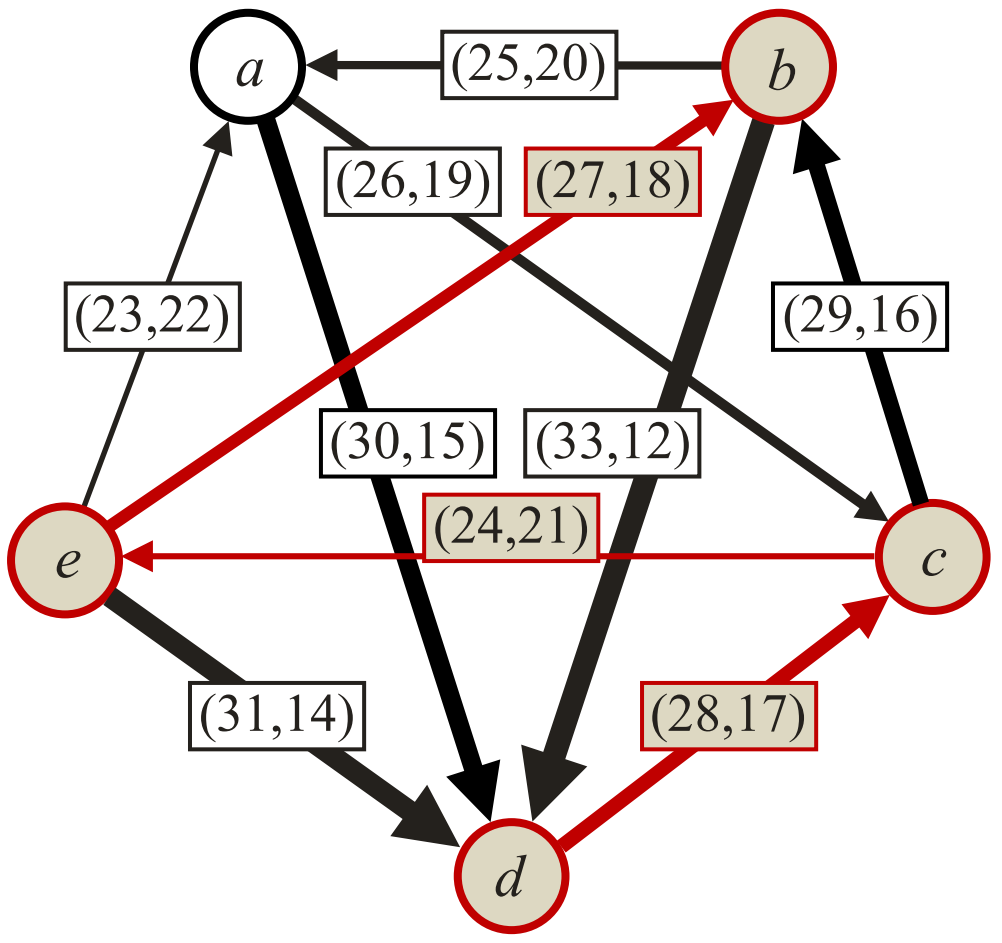

So the strength of the strongest path from alternative $d$ to alternative $b$ is $\max_D$ { (28,17), (24,21) } $\approx_D$ (28,17).

*d* → *c*: There is only one path from alternative *d* to alternative *c*.

Path 1: *d*, (28,17), *c*
with a strength of (28,17).

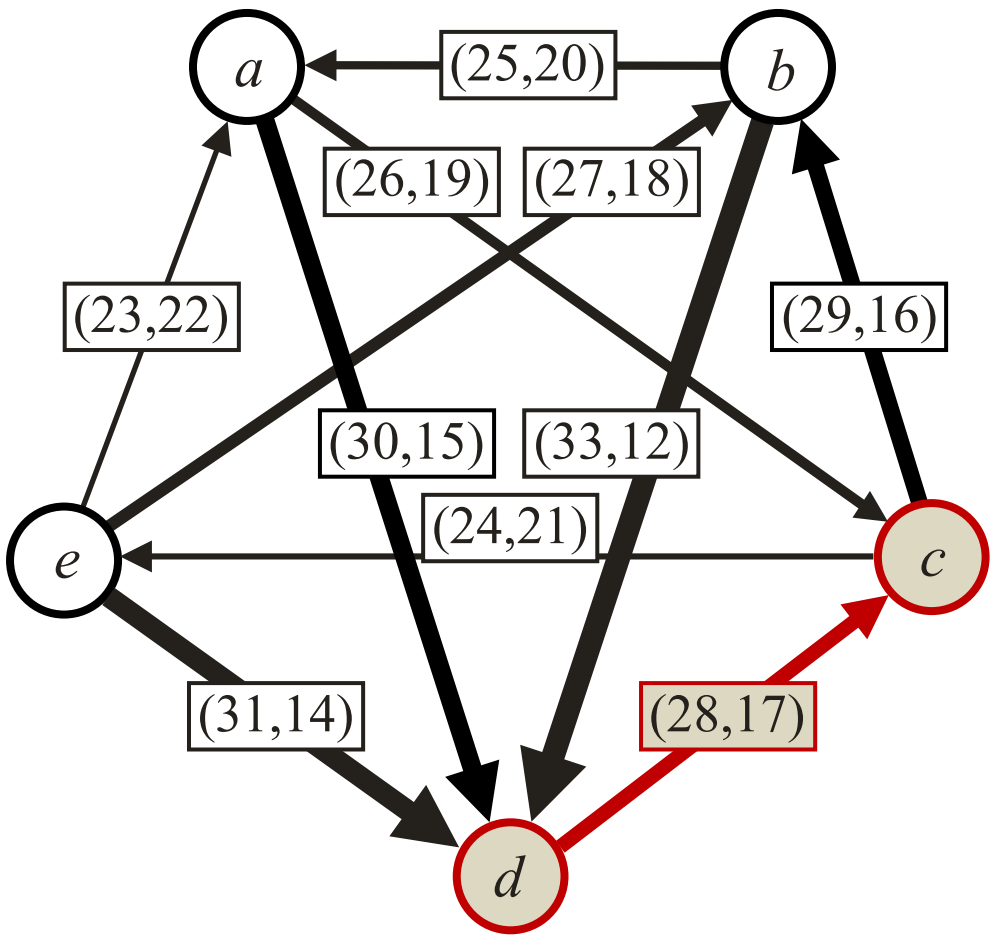

So the strength of the strongest path from alternative *d* to alternative *c* is (28,17).

*d* → *e*: There is only one path from alternative *d* to alternative *e*.

Path 1: *d*, (28,17), *c*, (24,21), *e*
with a strength of $\min_D$ { (28,17), (24,21) } $\approx_D$ (24,21).

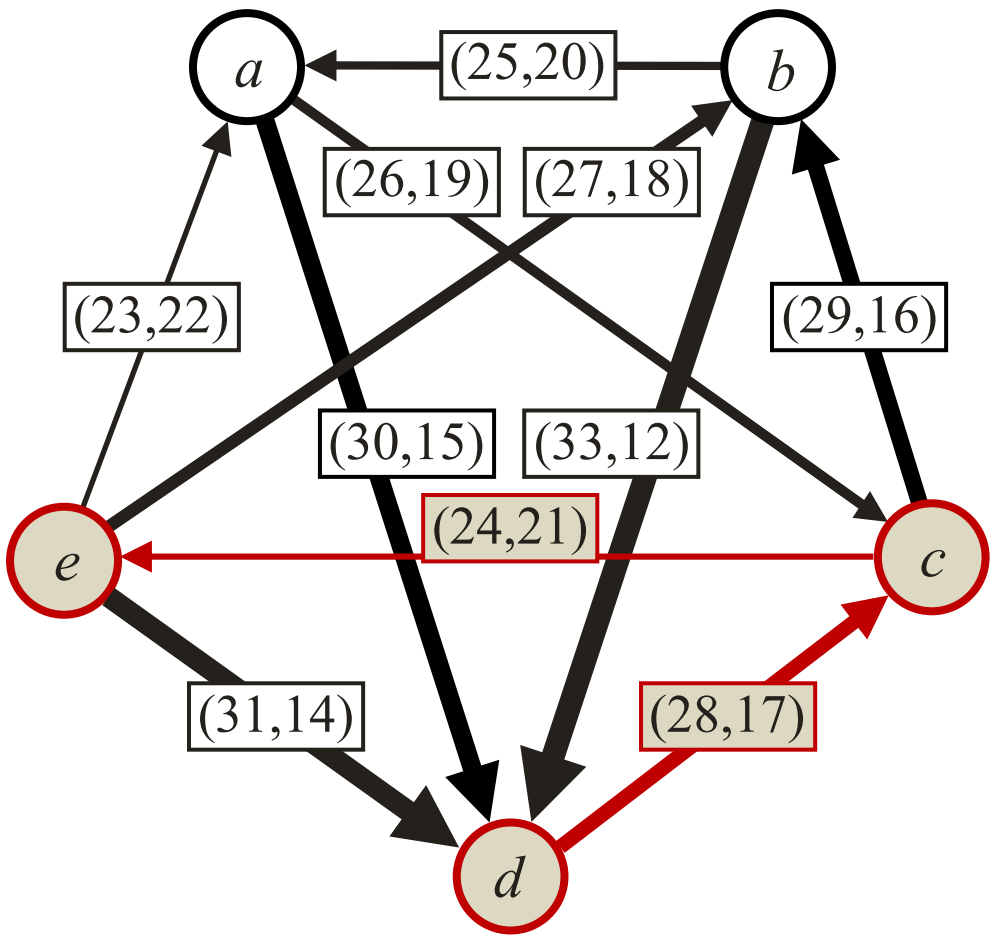

So the strength of the strongest path from alternative *d* to alternative *e* is (24,21).

$e \rightarrow a$: There are 3 paths from alternative $e$ to alternative $a$.

Path 1: *e*, (23,22), *a*
with a strength of (23,22).

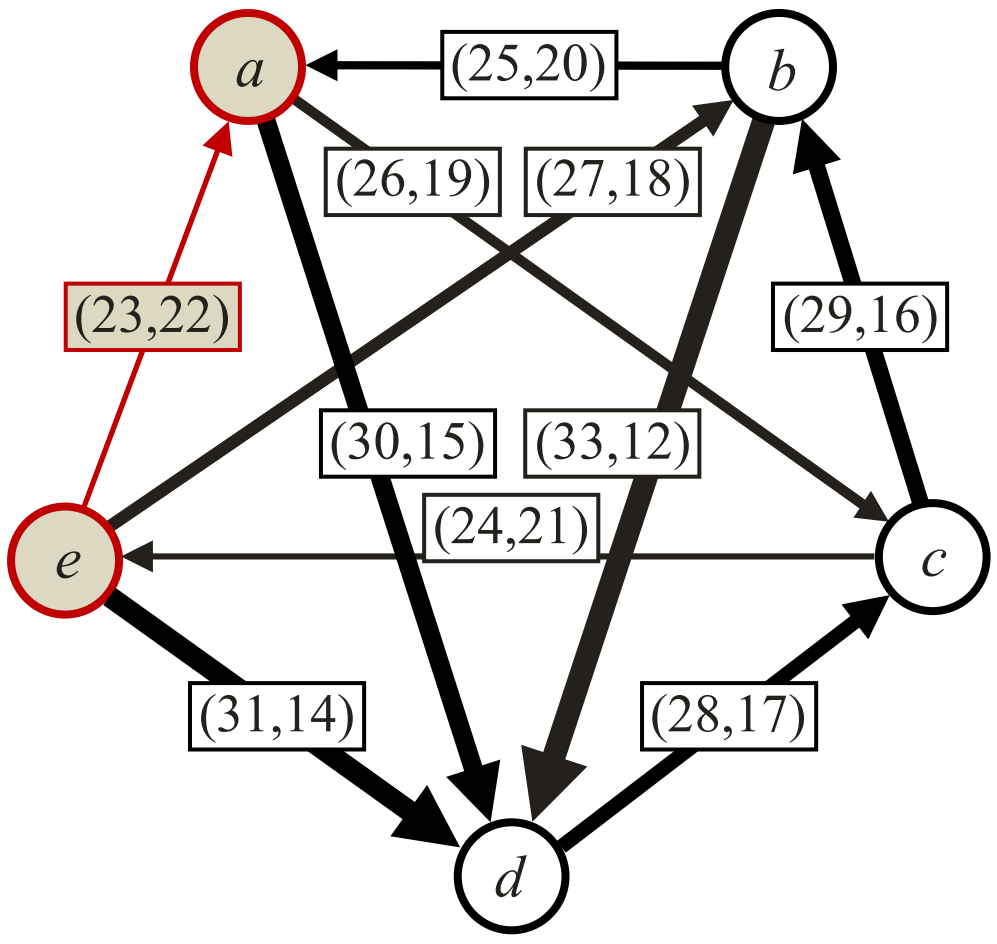

Path 2: *e*, (27,18), *b*, (25,20), *a*
with a strength of $\min_D$ { (27,18), (25,20) } $\approx_D$ (25,20).

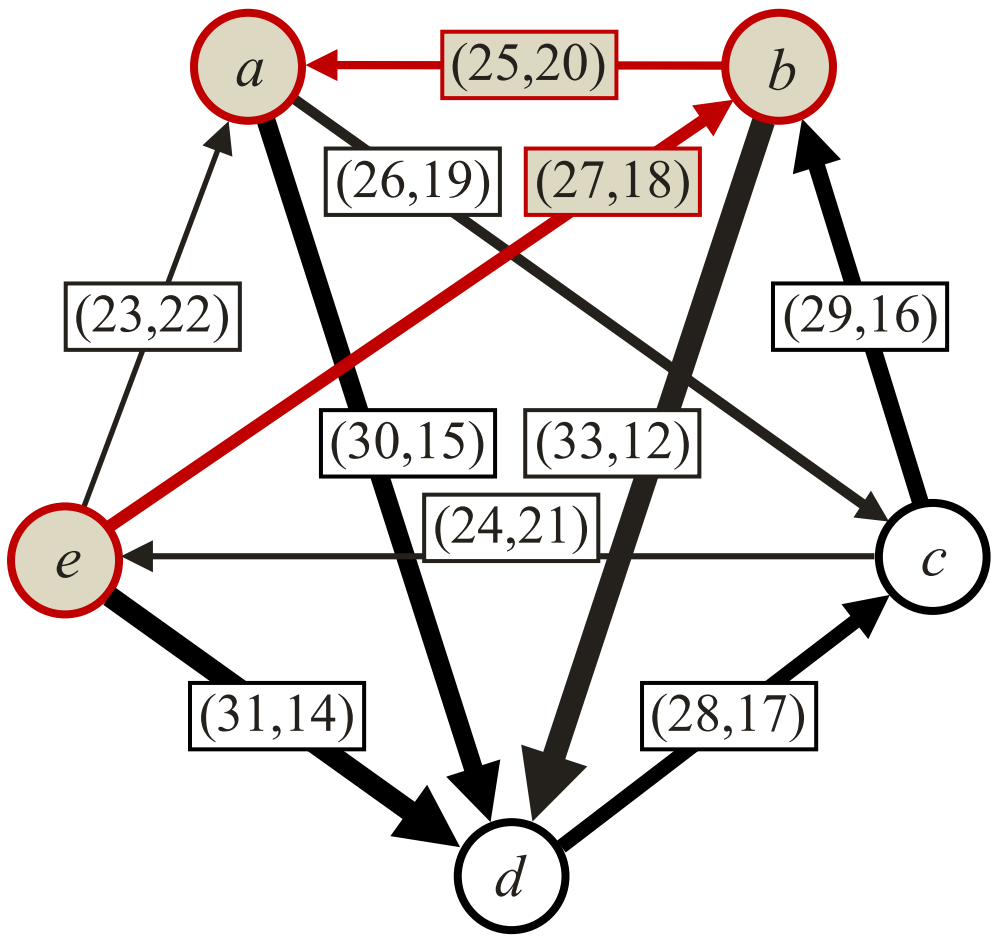

Path 3: *e*, (31,14), *d*, (28,17), *c*, (29,16), *b*, (25,20), *a*
with a strength of $\min_D$ { (31,14), (28,17), (29,16), (25,20) } $\approx_D$ (25,20).

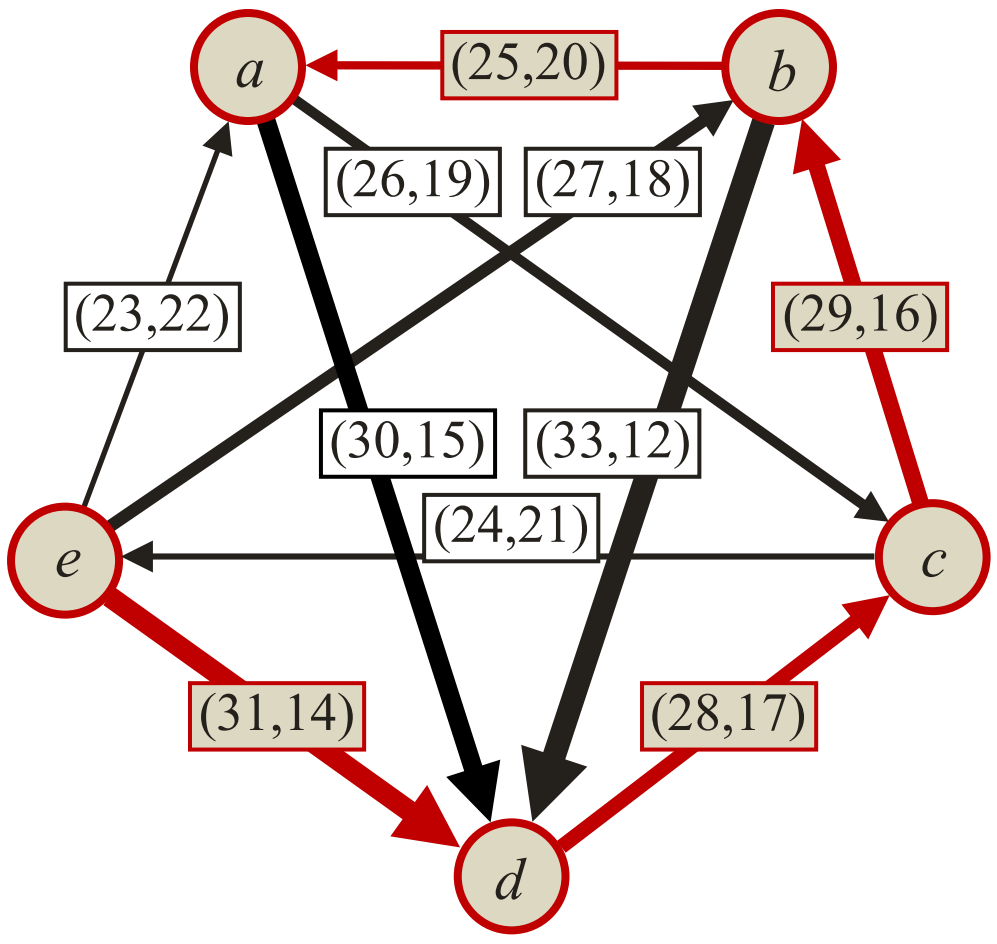

So the strength of the strongest path from alternative *e* to alternative *a* is $\max_D$ { (23,22), (25,20), (25,20) } $\approx_D$ (25,20).

*e* → *b*: There are 4 paths from alternative *e* to alternative *b*.

Path 1: *e*, (23,22), *a*, (26,19), *c*, (29,16), *b*
with a strength of $\min_D$ { (23,22), (26,19), (29,16) } $\approx_D$ (23,22).

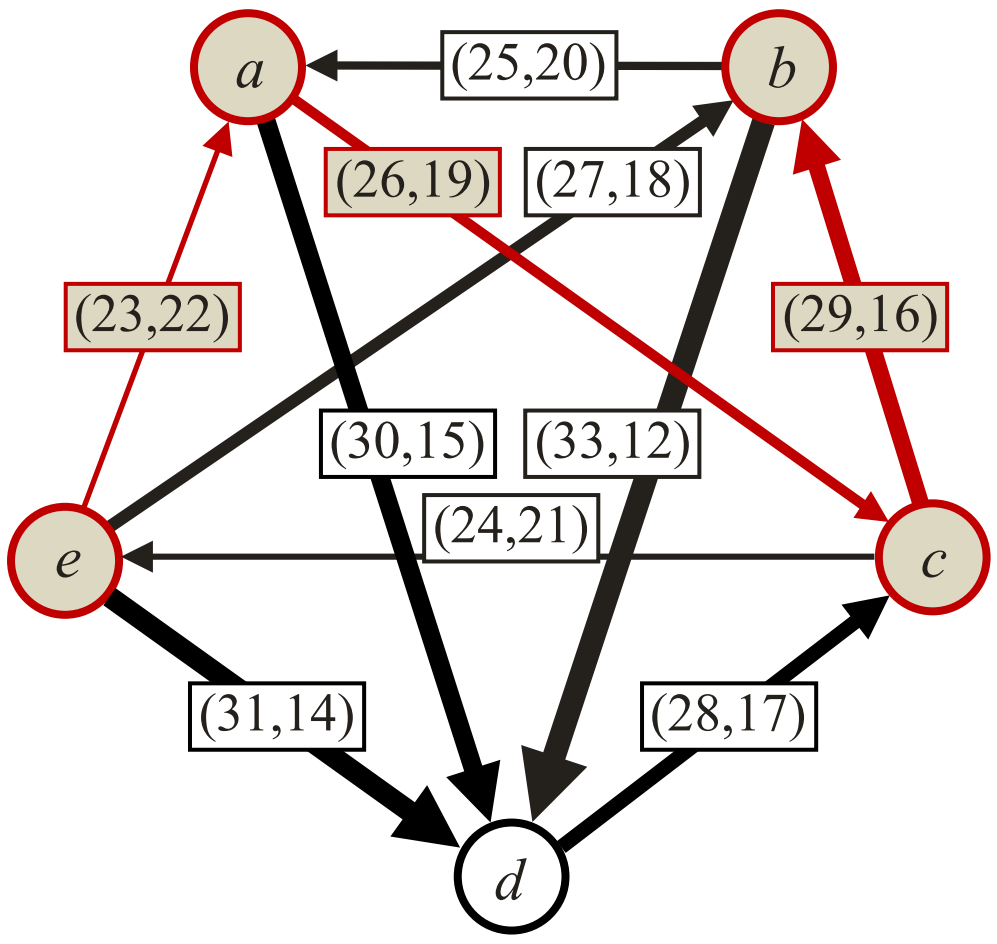

Path 2: *e*, (23,22), *a*, (30,15), *d*, (28,17), *c*, (29,16), *b*
with a strength of $\min_D$ { (23,22), (30,15), (28,17), (29,16) } $\approx_D$ (23,22).

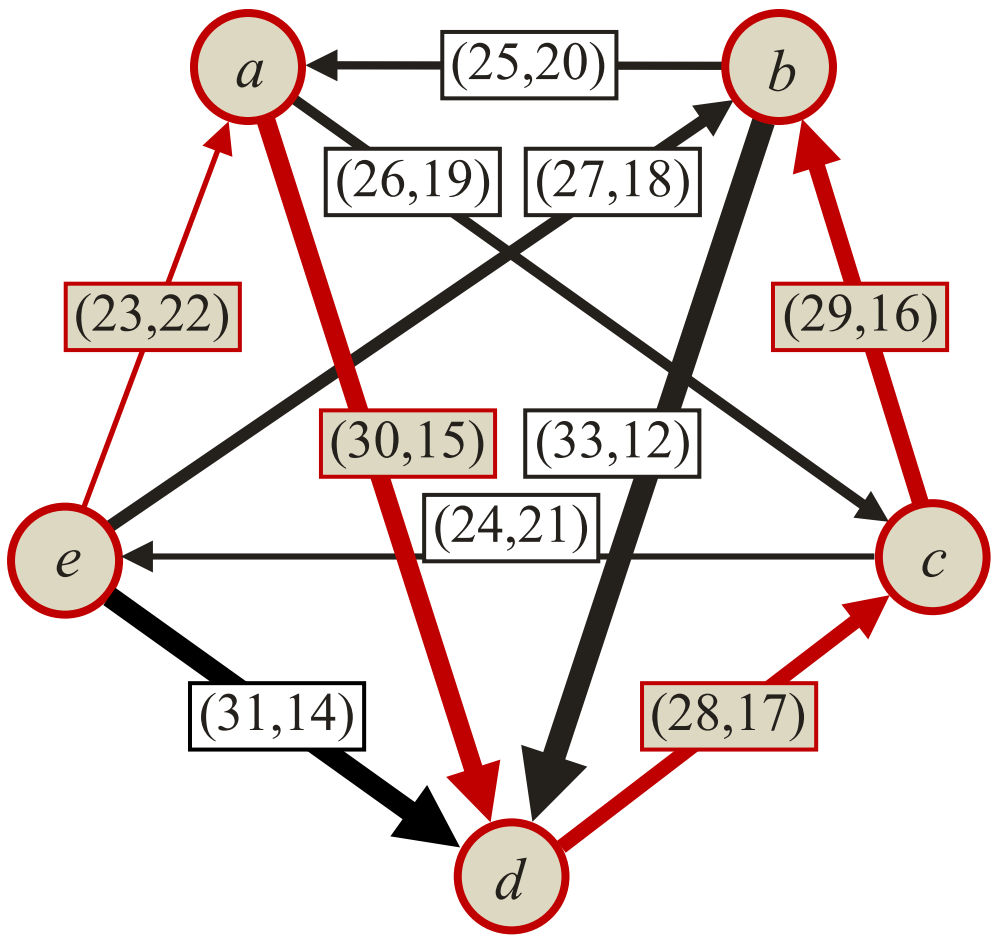

Path 3: *e*, (27,18), *b*
with a strength of (27,18).

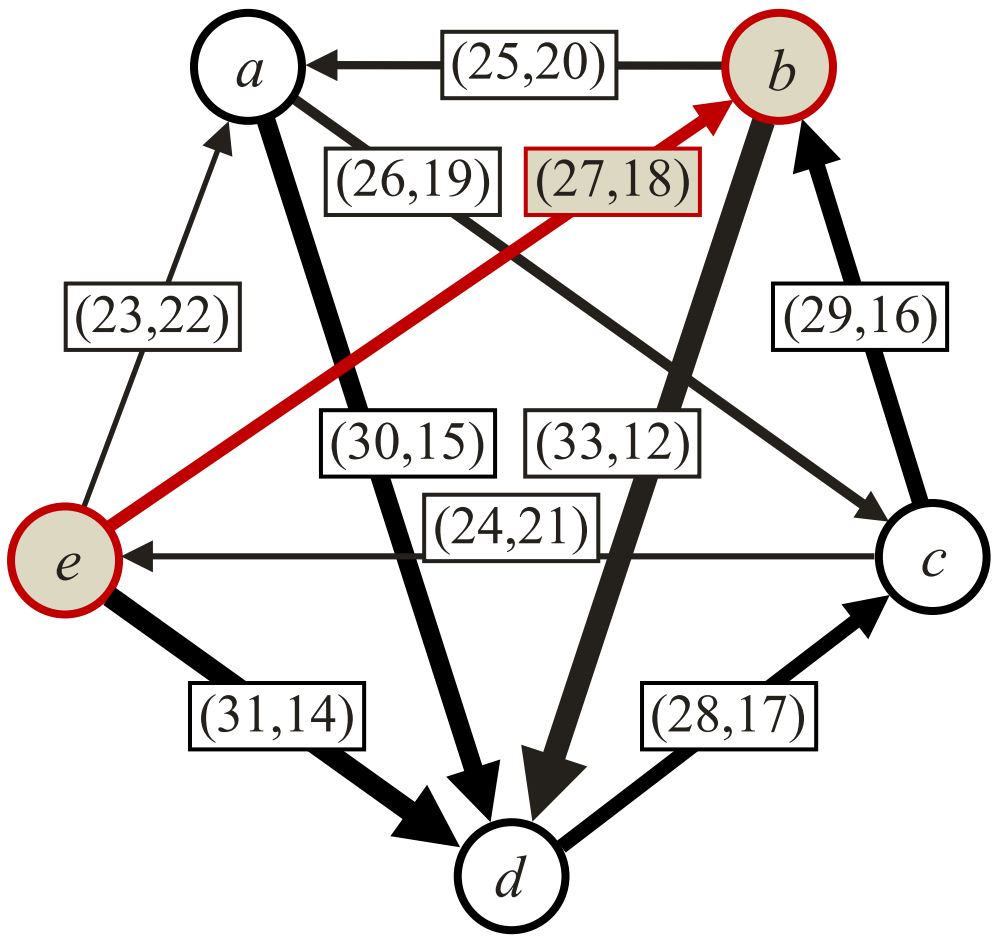

Path 4: *e*, (31,14), *d*, (28,17), *c*, (29,16), *b*
with a strength of $\min_D$ { (31,14), (28,17), (29,16) } $\approx_D$ (28,17).

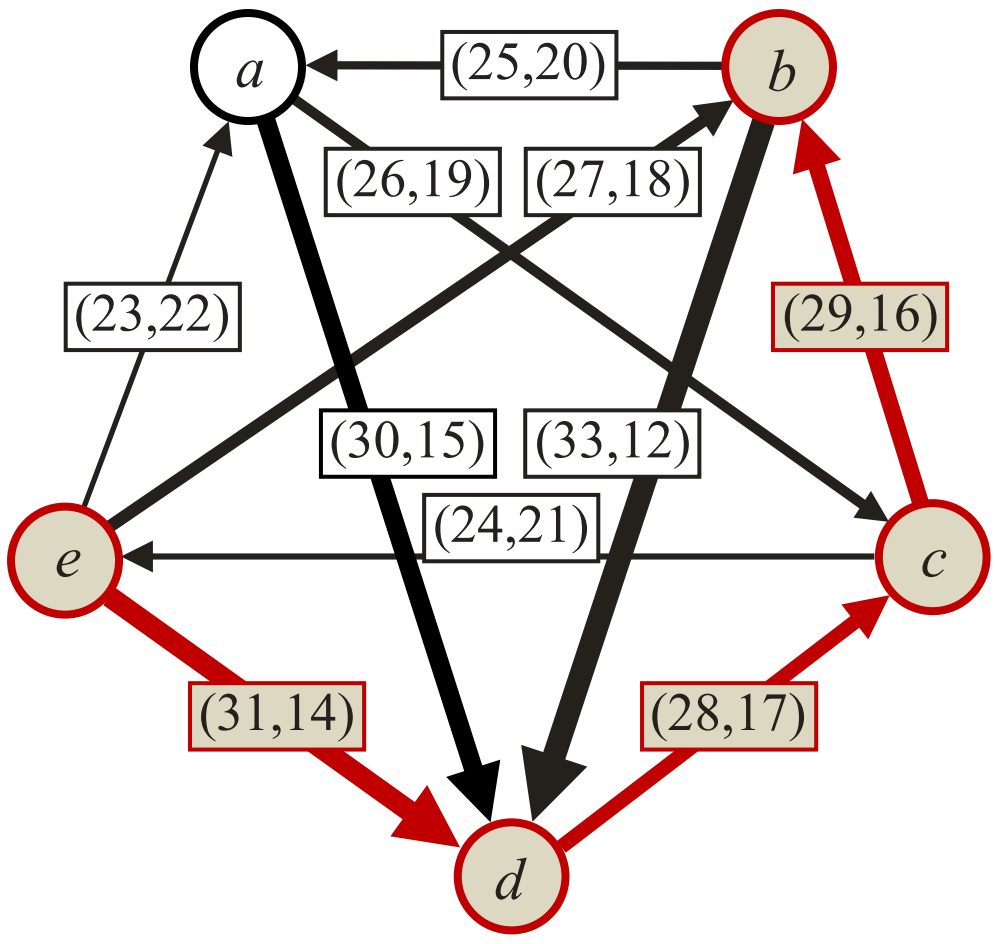

So the strength of the strongest path from alternative *e* to alternative *b* is $\max_D$ { (23,22), (23,22), (27,18), (28,17) } $\approx_D$ (28,17).

*e* → *c*: There are 6 paths from alternative *e* to alternative *c*.

Path 1: *e*, (23,22), *a*, (26,19), *c*
with a strength of $\min_D$ { (23,22), (26,19) } $\approx_D$ (23,22).

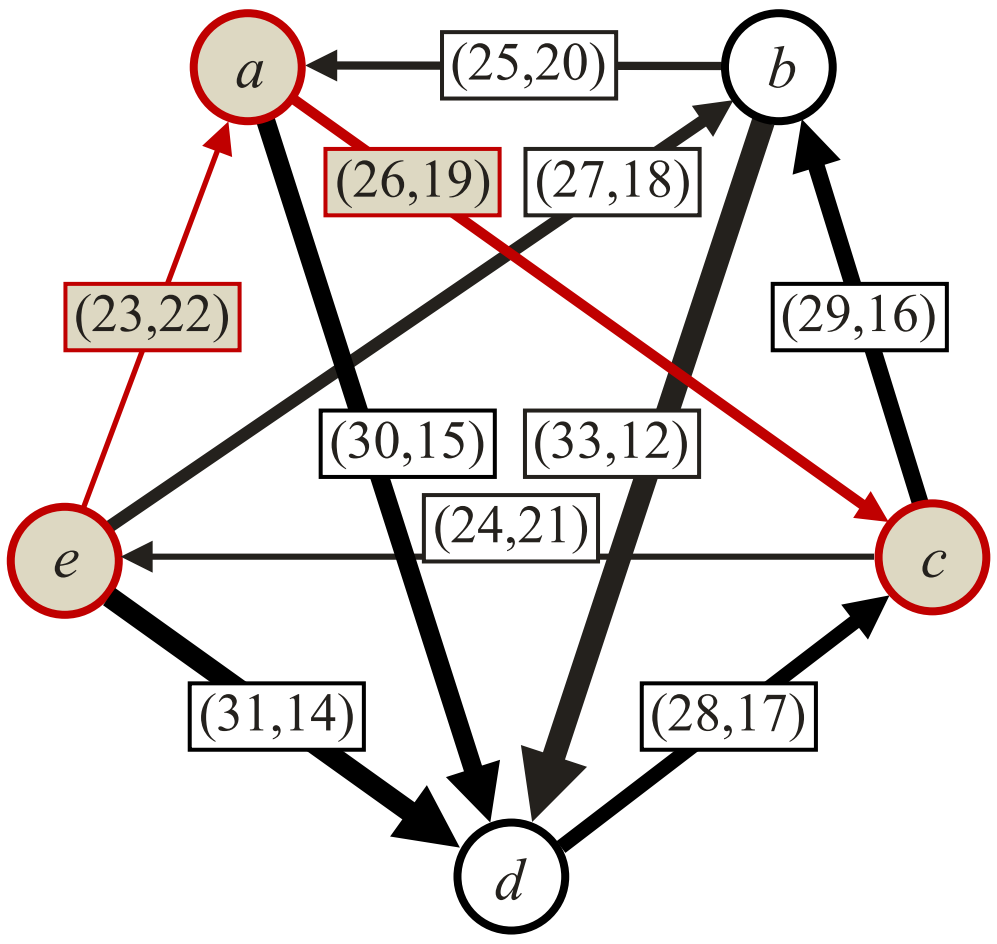

Path 2: *e*, (23,22), *a*, (30,15), *d*, (28,17), *c*
with a strength of $\min_D$ { (23,22), (30,15), (28,17) } $\approx_D$ (23,22).

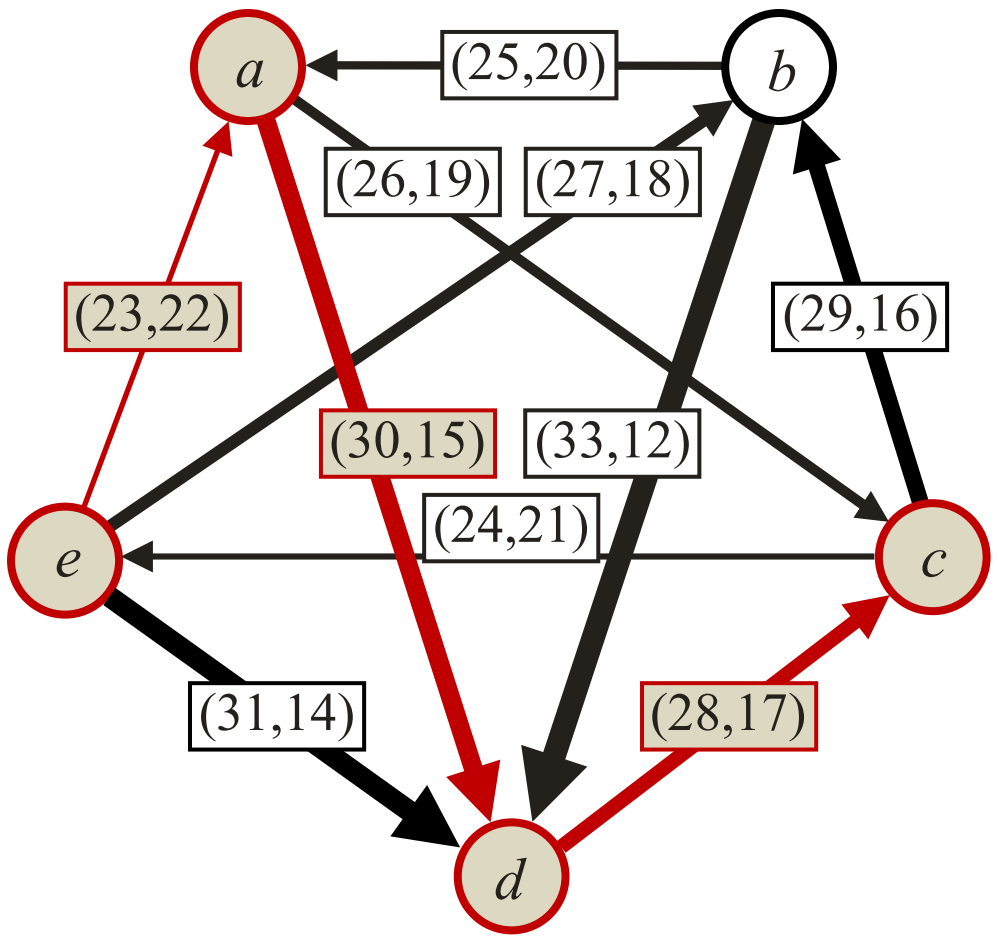

Path 3: *e*, (27,18), *b*, (25,20), *a*, (26,19), *c*
with a strength of $\min_D$ { (27,18), (25,20), (26,19) } $\approx_D$ (25,20).

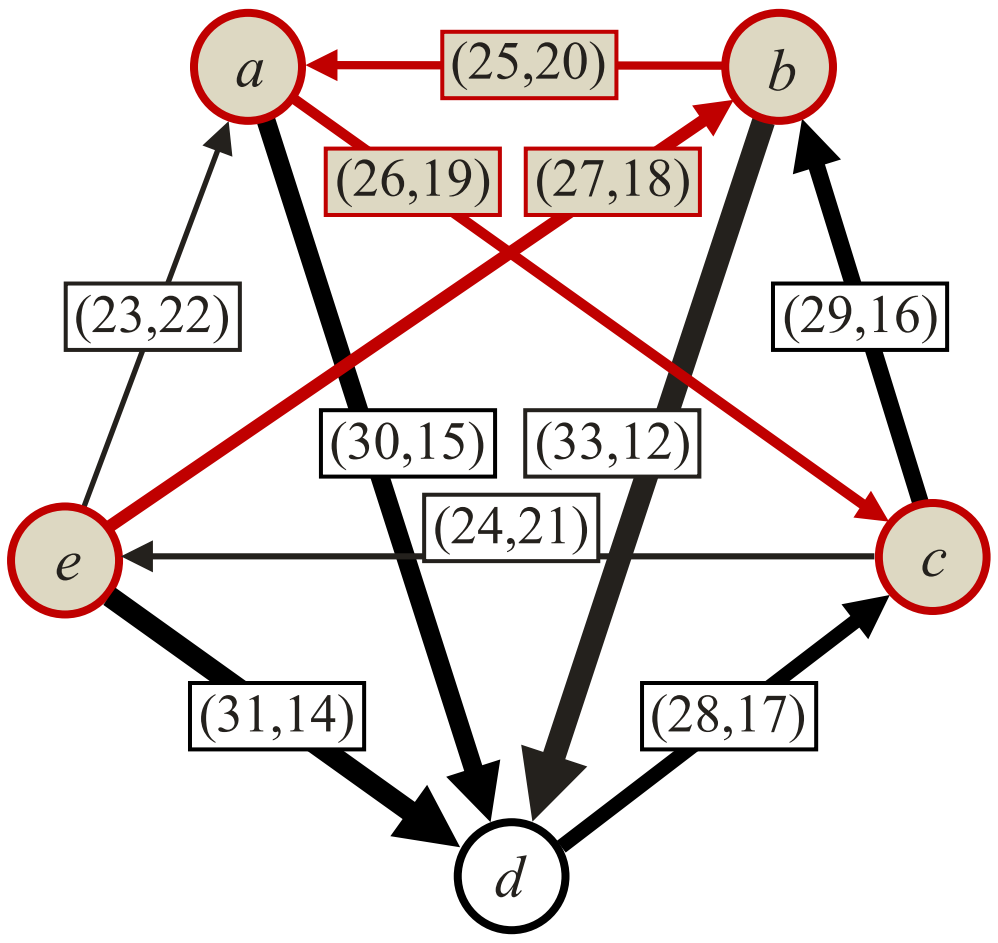

Path 4: *e*, (27,18), *b*, (25,20), *a*, (30,15), *d*, (28,17), *c*
with a strength of $\min_D$ { (27,18), (25,20), (30,15), (28,17) } $\approx_D$ (25,20).

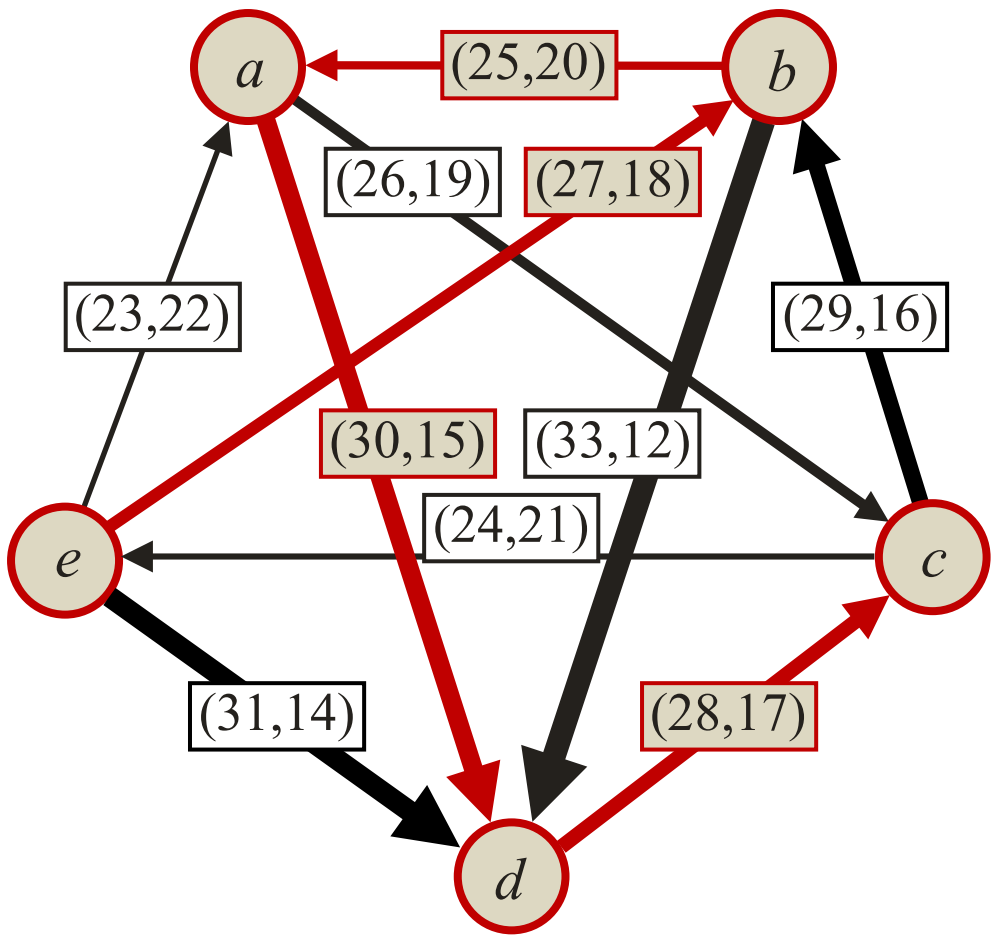

Path 5: *e*, (27,18), *b*, (33,12), *d*, (28,17), *c*
with a strength of $\min_D$ { (27,18), (33,12), (28,17) } $\approx_D$ (27,18).

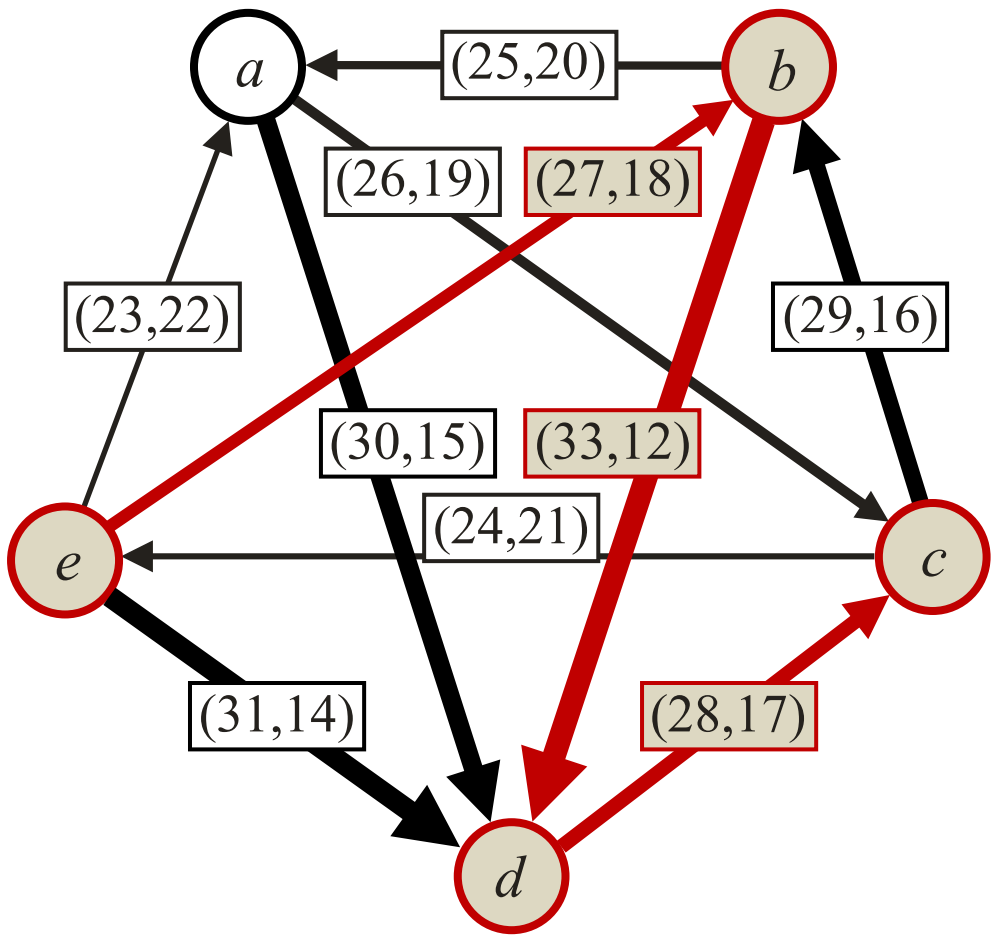

Path 6: *e*, (31,14), *d*, (28,17), *c*
with a strength of $\min_D$ { (31,14), (28,17) } $\approx_D$ (28,17).

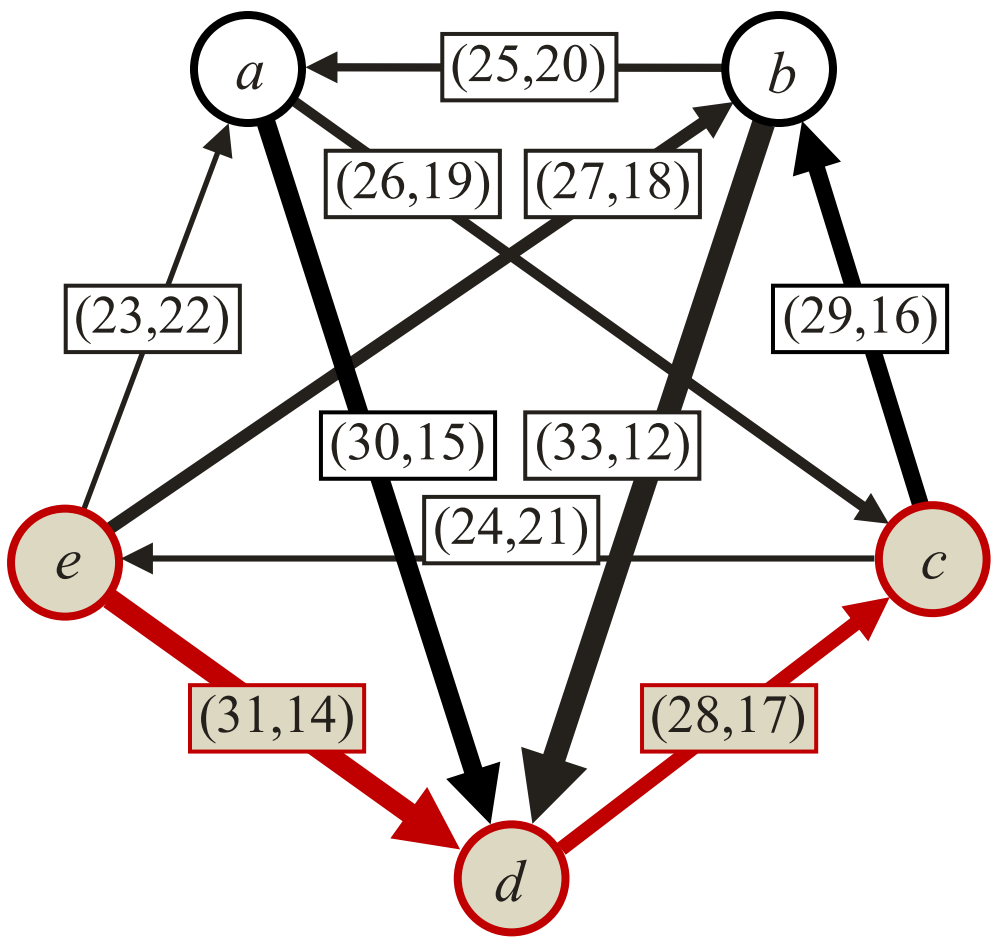

So the strength of the strongest path from alternative *e* to alternative *c* is $\max_D$ { (23,22), (23,22), (25,20), (25,20), (27,18), (28,17) } $\approx_D$ (28,17).

*e* → *d*: There are 5 paths from alternative *e* to alternative *d*.

Path 1: *e*, (23,22), *a*, (26,19), *c*, (29,16), *b*, (33,12), *d*
with a strength of $\min_D$ { (23,22), (26,19), (29,16), (33,12) } $\approx_D$ (23,22).

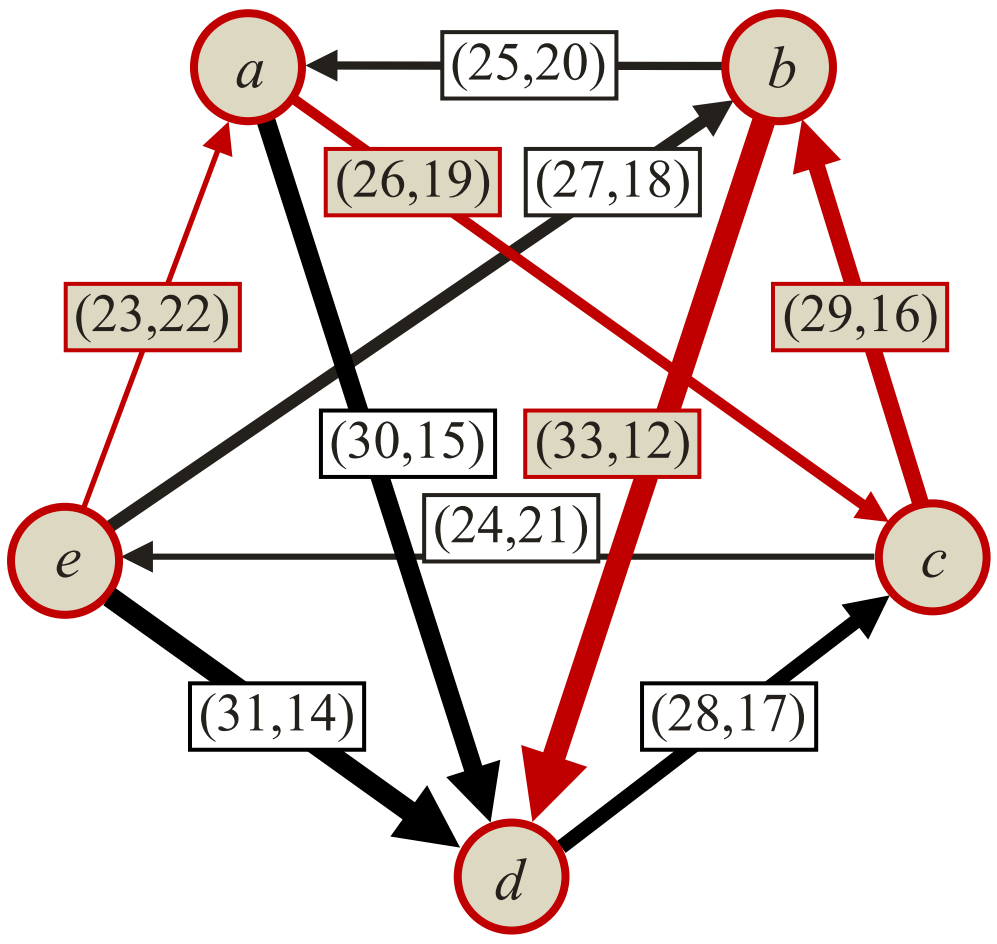

Path 2: *e*, (23,22), *a*, (30,15), *d*
with a strength of $\min_D$ { (23,22), (30,15) } $\approx_D$ (23,22).

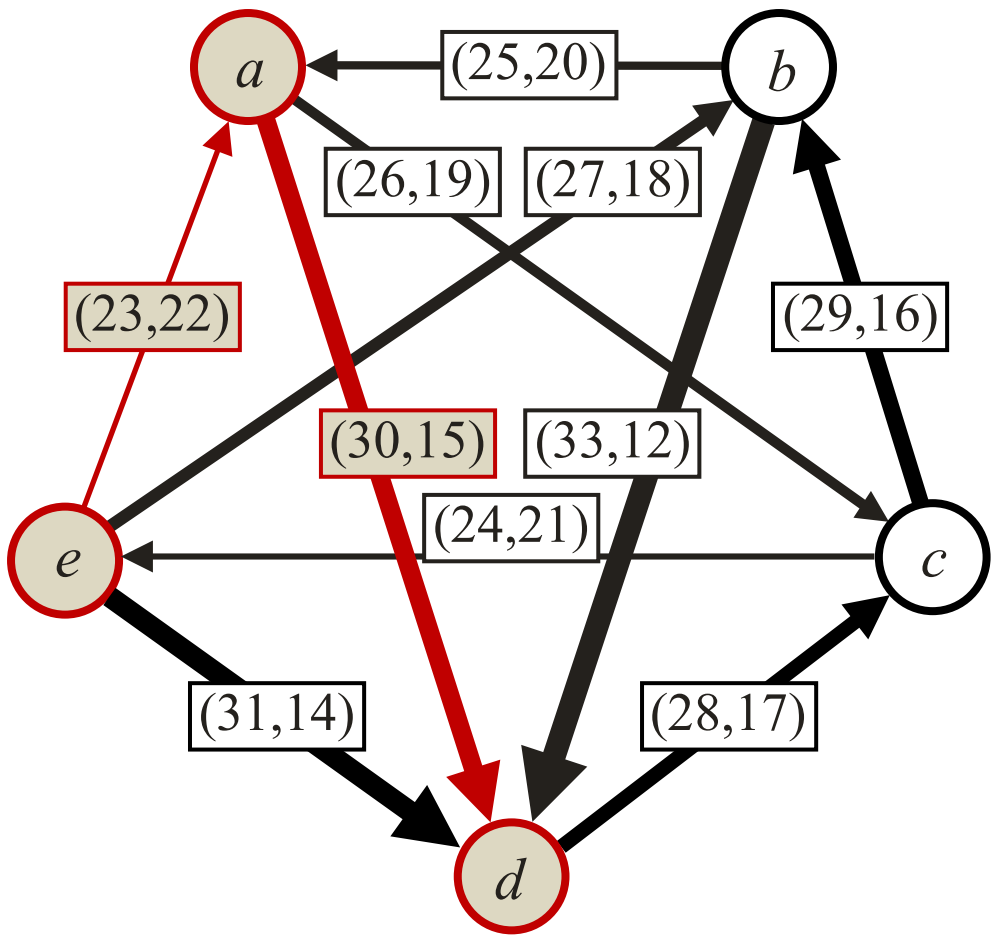

Path 3: *e*, (27,18), *b*, (25,20), *a*, (30,15), *d*
with a strength of $\min_D$ { (27,18), (25,20), (30,15) } $\approx_D$ (25,20).

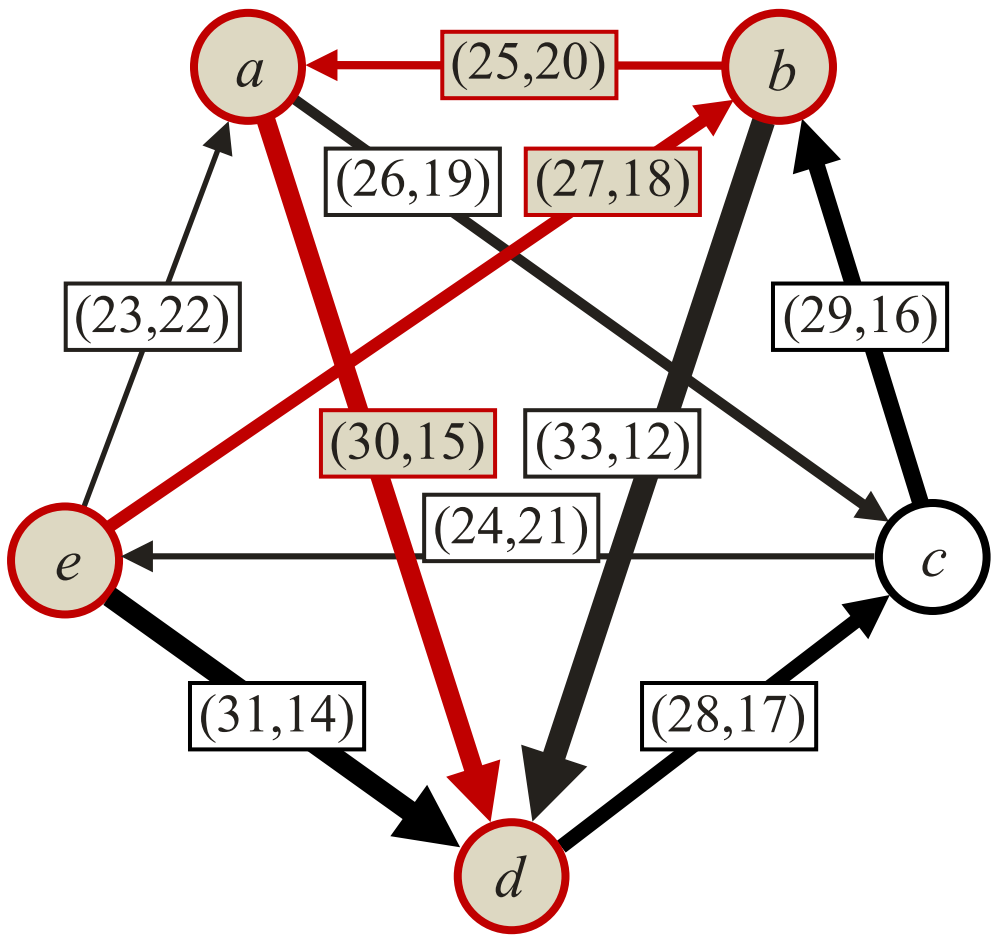

Path 4: *e*, (27,18), *b*, (33,12), *d*
with a strength of $\min_D$ { (27,18), (33,12) } $\approx_D$ (27,18).

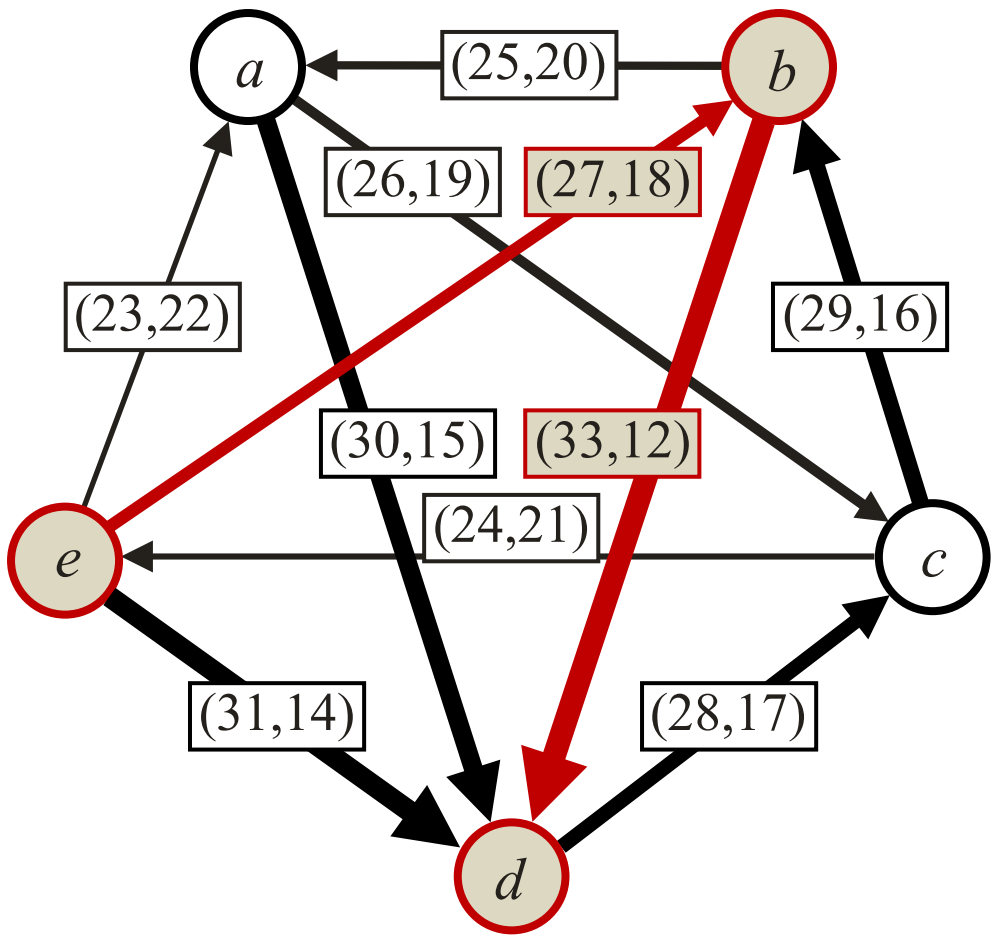

Path 5: *e*, (31,14), *d*
with a strength of (31,14).

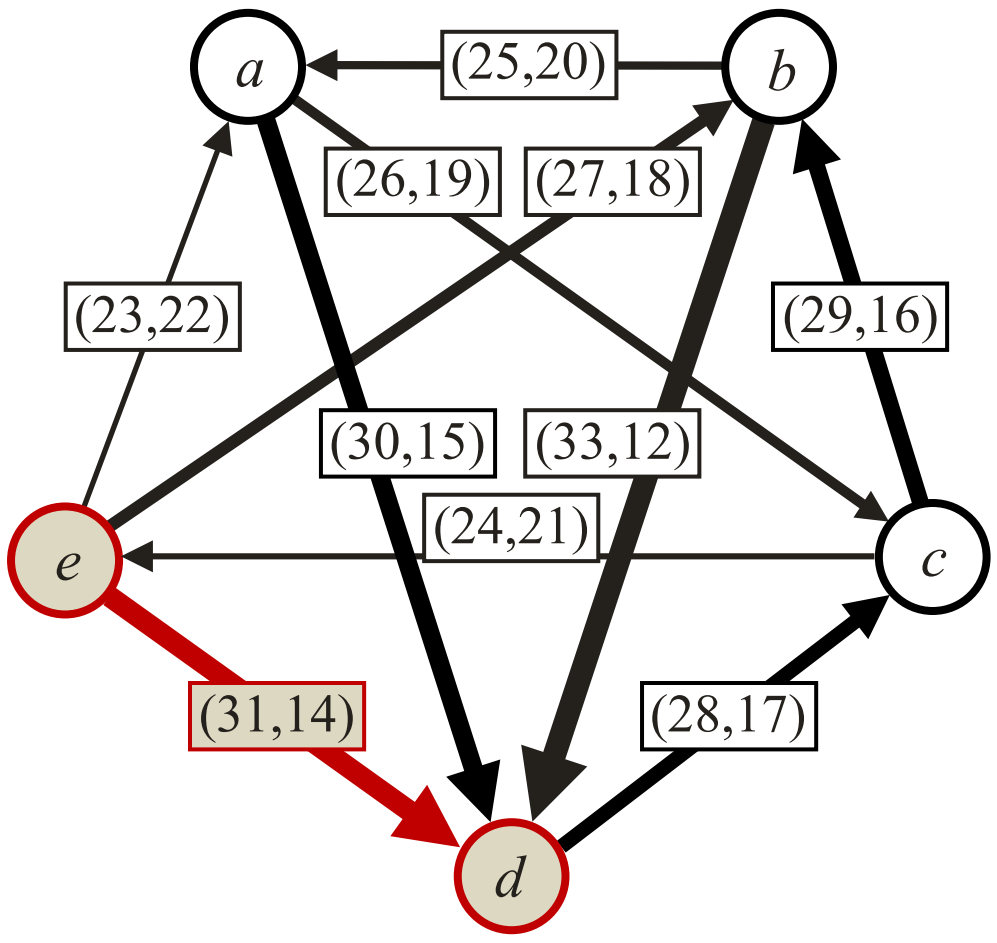

So the strength of the strongest path from alternative *e* to alternative *d* is $\max_D$ { (23,22), (23,22), (25,20), (27,18), (31,14) } $\approx_D$ (31,14).

The following table lists the strongest paths, as determined by the Floyd-Warshall algorithm, as defined in section 2.3.1. The critical links of the strongest paths are <u>underlined</u>:

| | ... to *a* | ... to *b* | ... to *c* | ... to *d* | ... to *e* |
|---|---|---|---|---|---|
| from *a* ... | --- | *a*, (30,15), *d*, <u>(28,17)</u>, *c*, (29,16), *b* | *a*, (30,15), *d*, <u>(28,17)</u>, *c* | *a*, <u>(30,15)</u>, *d* | *a*, (30,15), *d*, (28,17), *c*, <u>(24,21)</u>, *e* |
| from *b* ... | *b*, <u>(25,20)</u>, *a* | --- | *b*, (33,12), *d*, <u>(28,17)</u>, *c* | *b*, <u>(33,12)</u>, *d* | *b*, (33,12), *d*, (28,17), *c*, <u>(24,21)</u>, *e* |
| from *c* ... | *c*, (29,16), *b*, <u>(25,20)</u>, *a* | *c*, <u>(29,16)</u>, *b* | --- | *c*, <u>(29,16)</u>, *b*, (33,12), *d* | *c*, <u>(24,21)</u>, *e* |
| from *d* ... | *d*, (28,17), *c*, (29,16), *b*, <u>(25,20)</u>, *a* | *d*, <u>(28,17)</u>, *c*, (29,16), *b* | *d*, <u>(28,17)</u>, *c* | --- | *d*, (28,17), *c*, <u>(24,21)</u>, *e* |
| from *e* ... | *e*, (31,14), *d*, (28,17), *c*, (29,16), *b*, <u>(25,20)</u>, *a* | *e*, (31,14), *d*, <u>(28,17)</u>, *c*, (29,16), *b* | *e*, (31,14), *d*, <u>(28,17)</u>, *c* | *e*, <u>(31,14)</u>, *d* | --- |

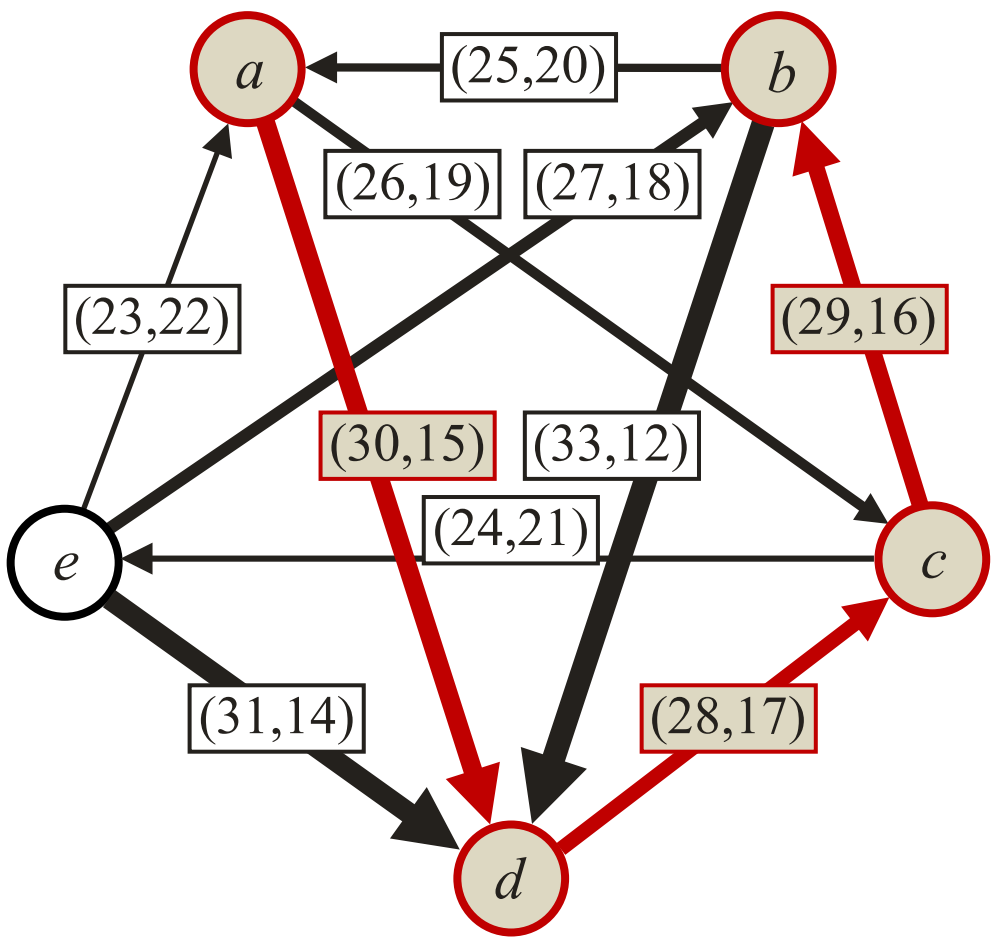

The strongest path from *a* to *b* is:
*a*, (30,15), *d*, (28,17), *c*, (29,16), *b*

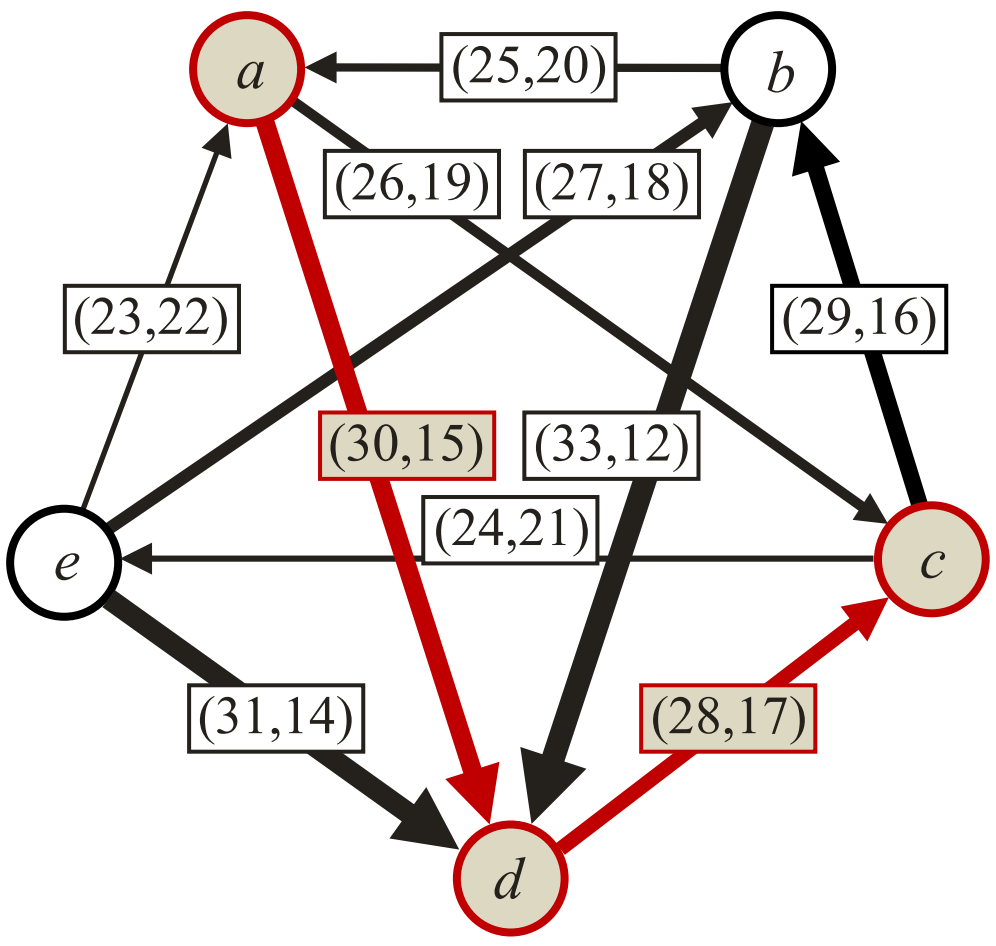

The strongest path from *a* to *c* is:
*a*, (30,15), *d*, (28,17), *c*

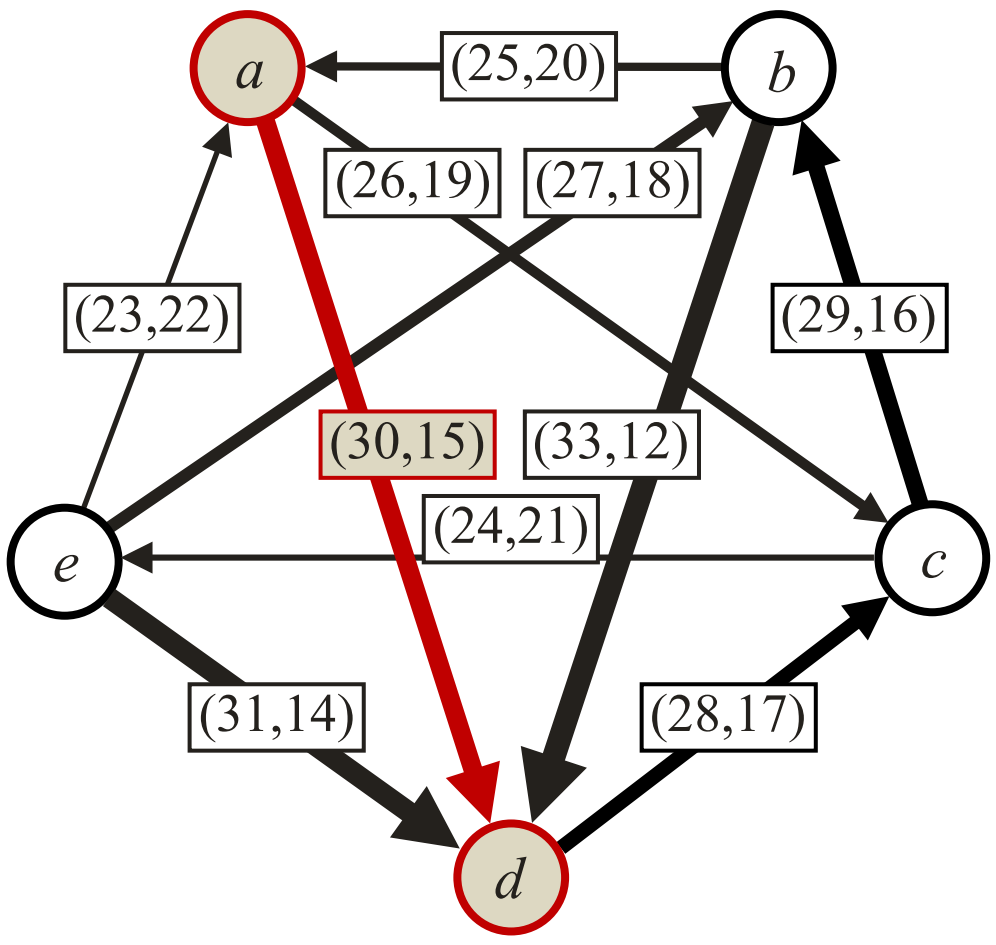

The strongest path from *a* to *d* is:
*a*, (30,15), *d*

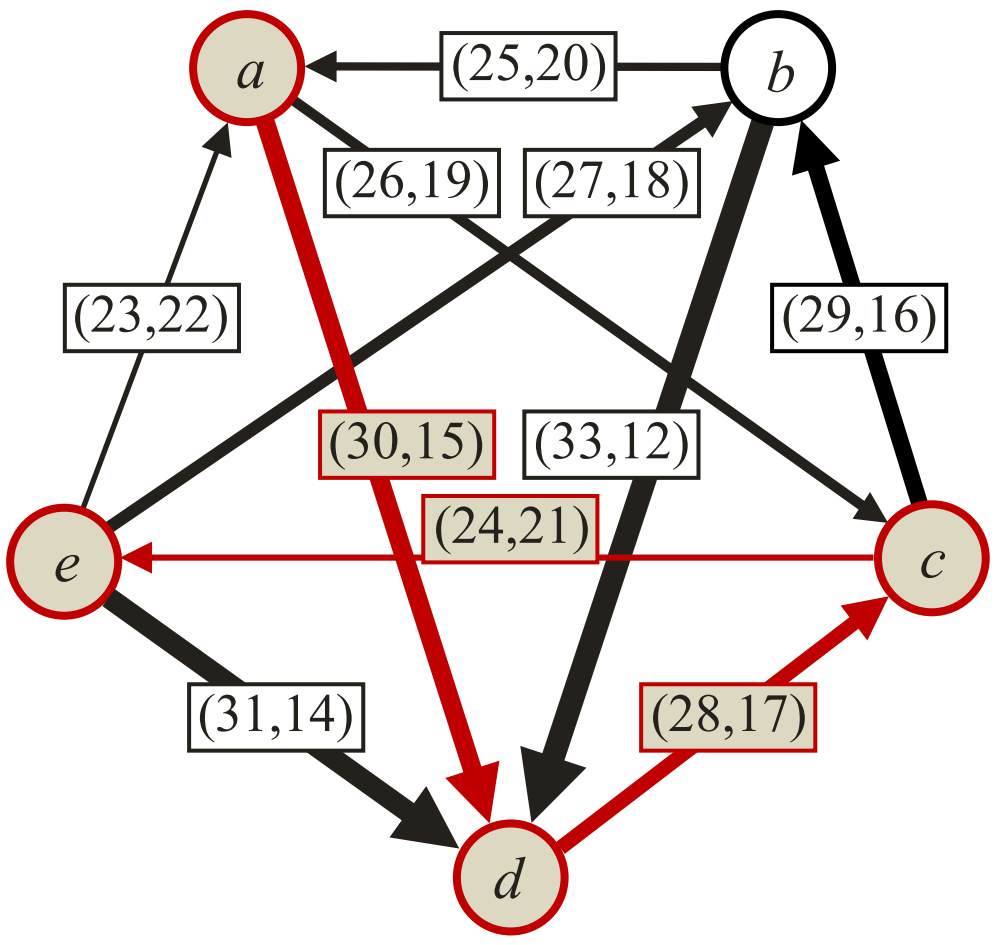

The strongest path from *a* to *e* is:
*a*, (30,15), *d*, (28,17), *c*, (24,21), *e*

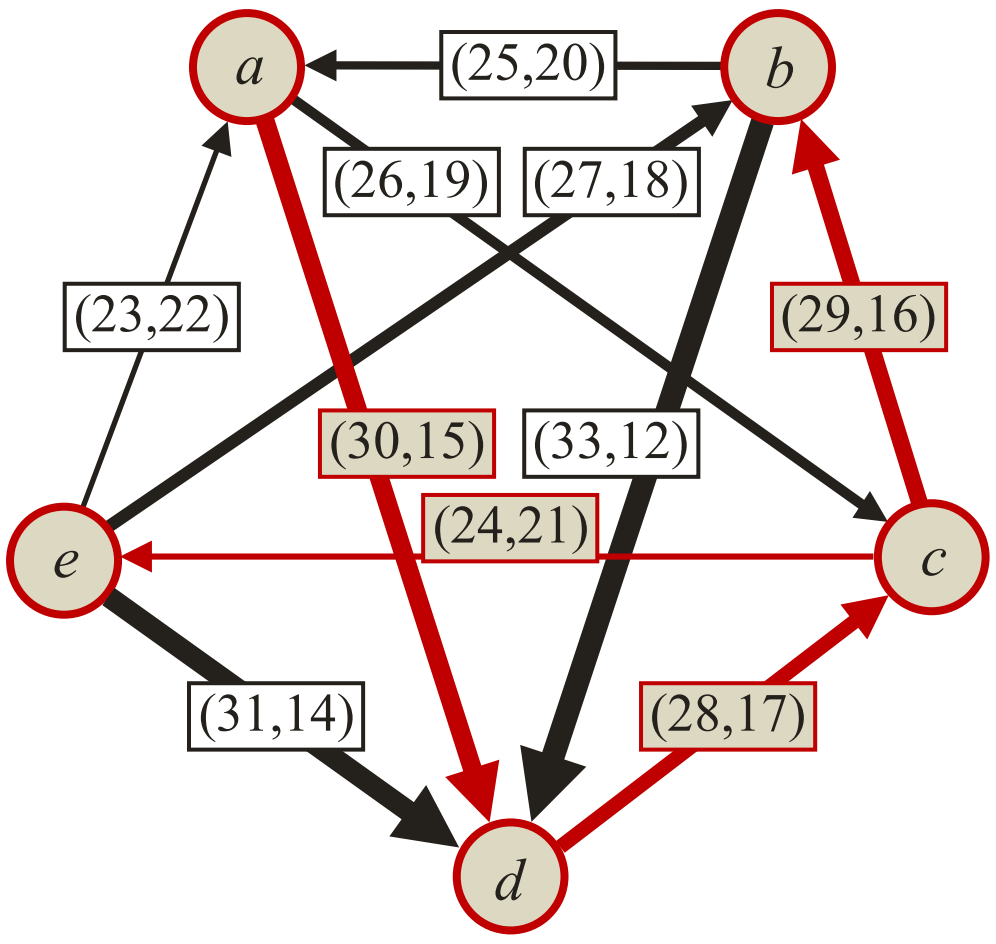

These are the strongest paths
from *a* to every other alternative.

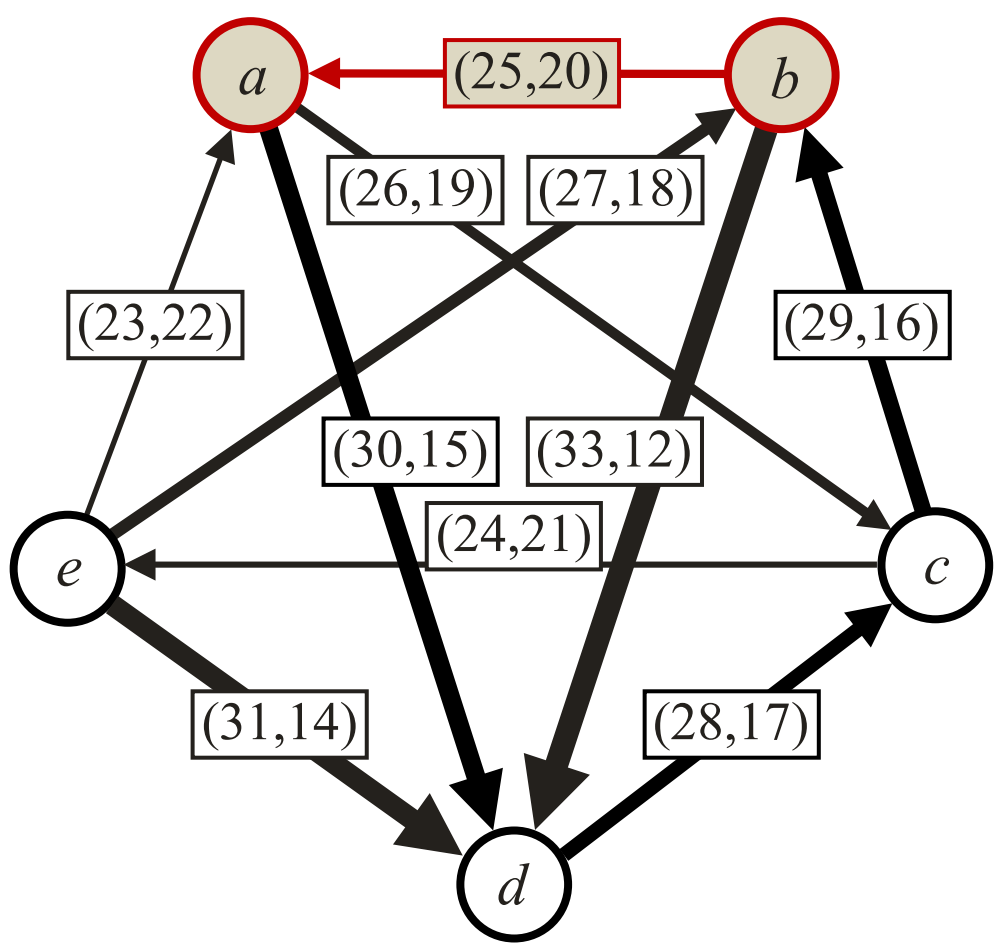

The strongest path from *b* to *a* is:
*b*, (25,20), *a*

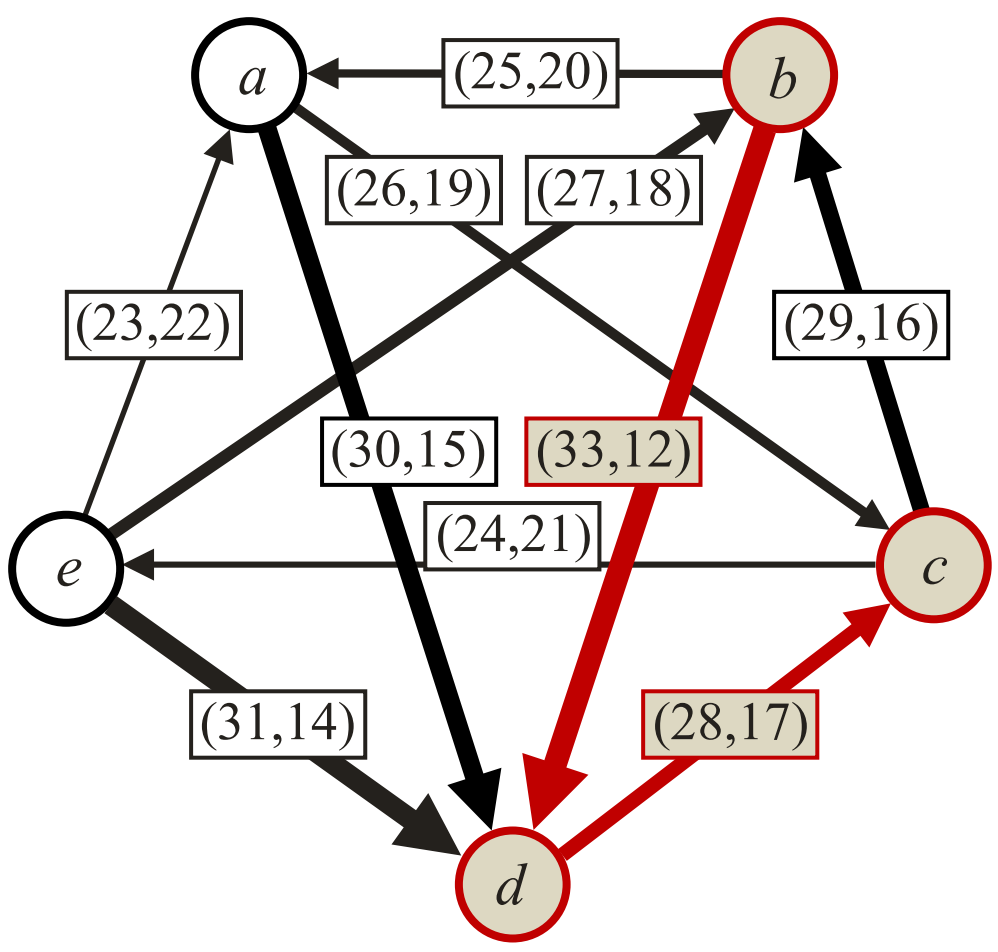

The strongest path from *b* to *c* is:
*b*, (33,12), *d*, (28,17), *c*

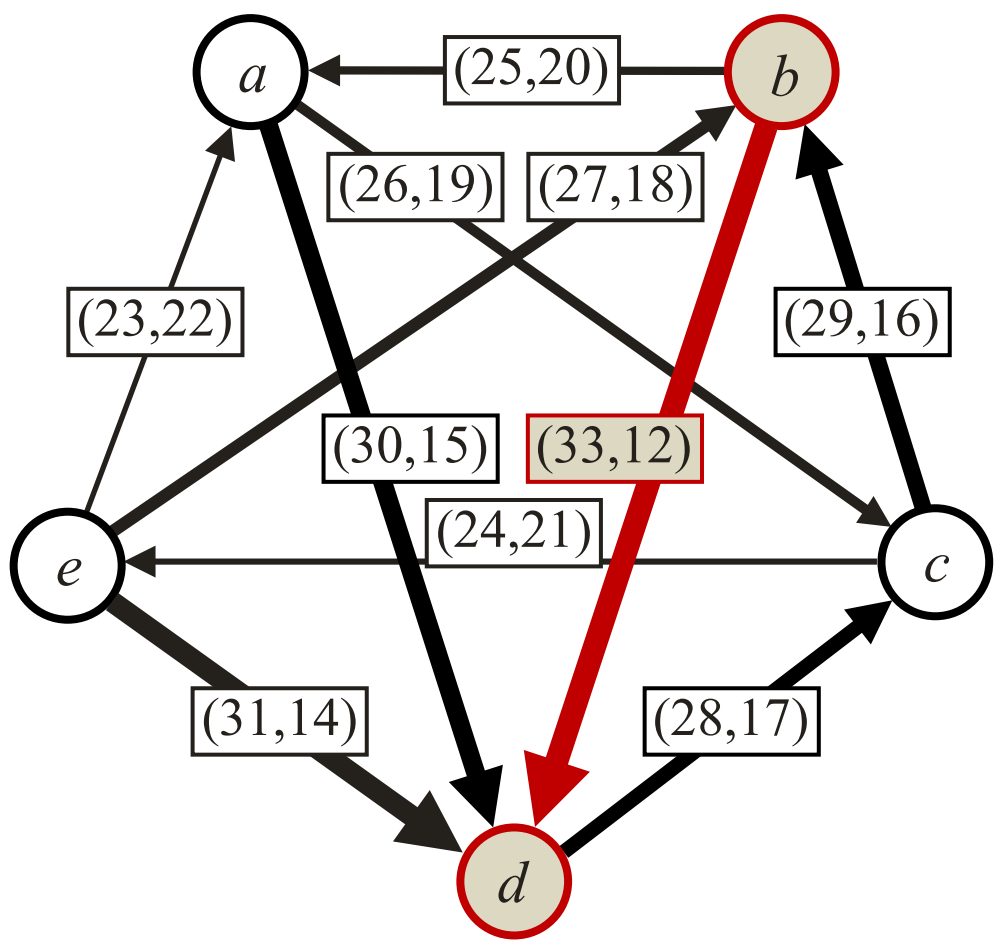

The strongest path from *b* to *d* is:
*b*, (33,12), *d*

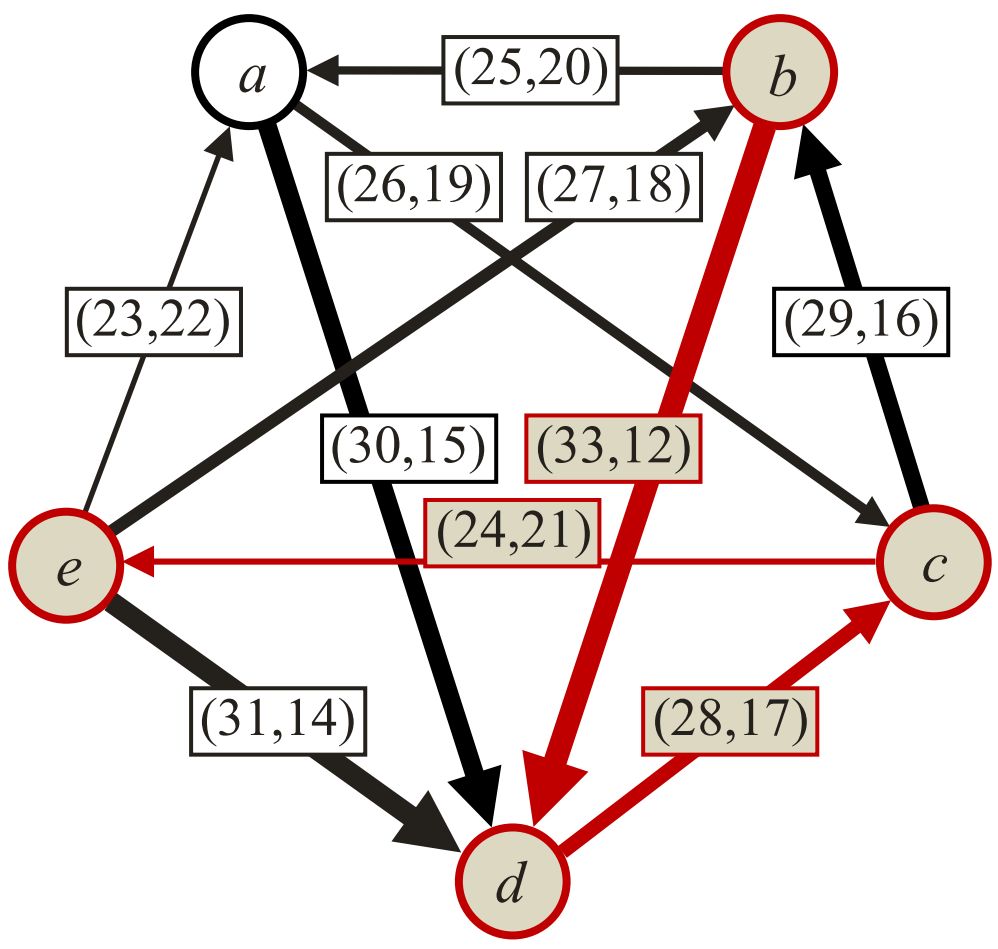

The strongest path from *b* to *e* is:
*b*, (33,12), *d*, (28,17), *c*, (24,21), *e*

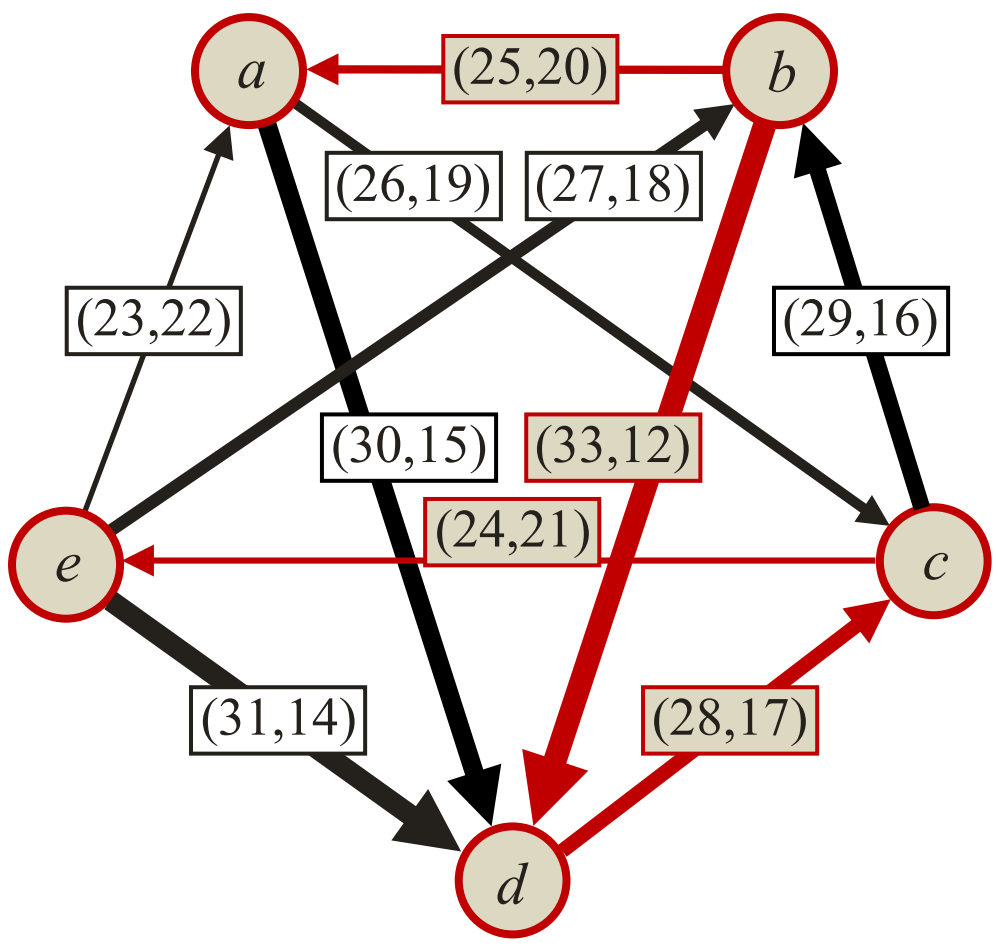

These are the strongest paths
from *b* to every other alternative.

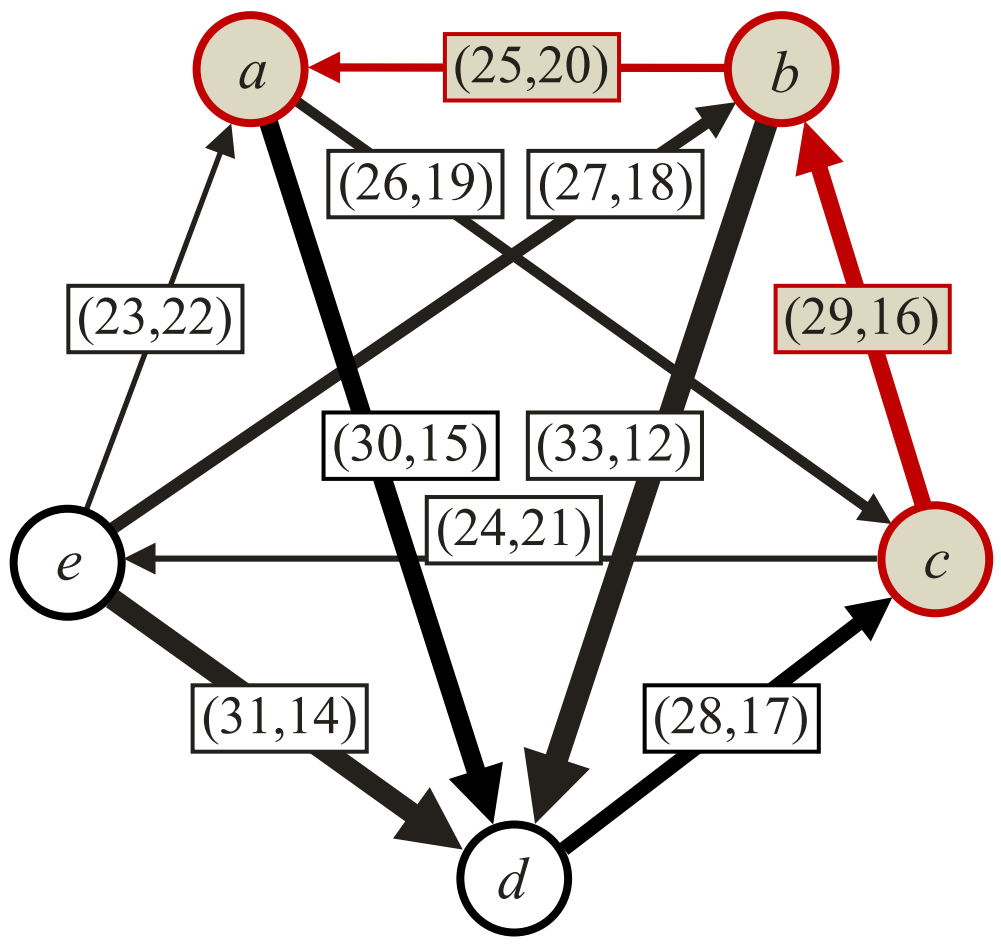

The strongest path from *c* to *a* is:
*c*, (29,16), *b*, (25,20), *a*

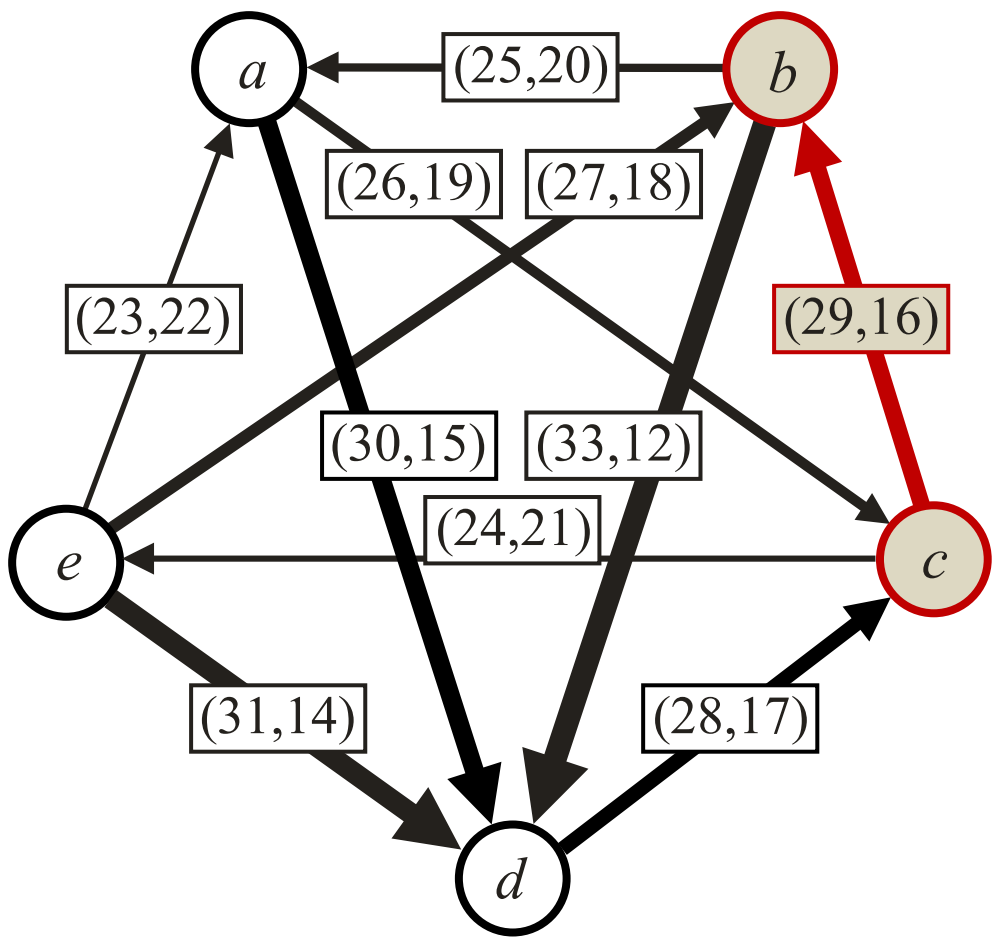

The strongest path from *c* to *b* is:
*c*, (29,16), *b*

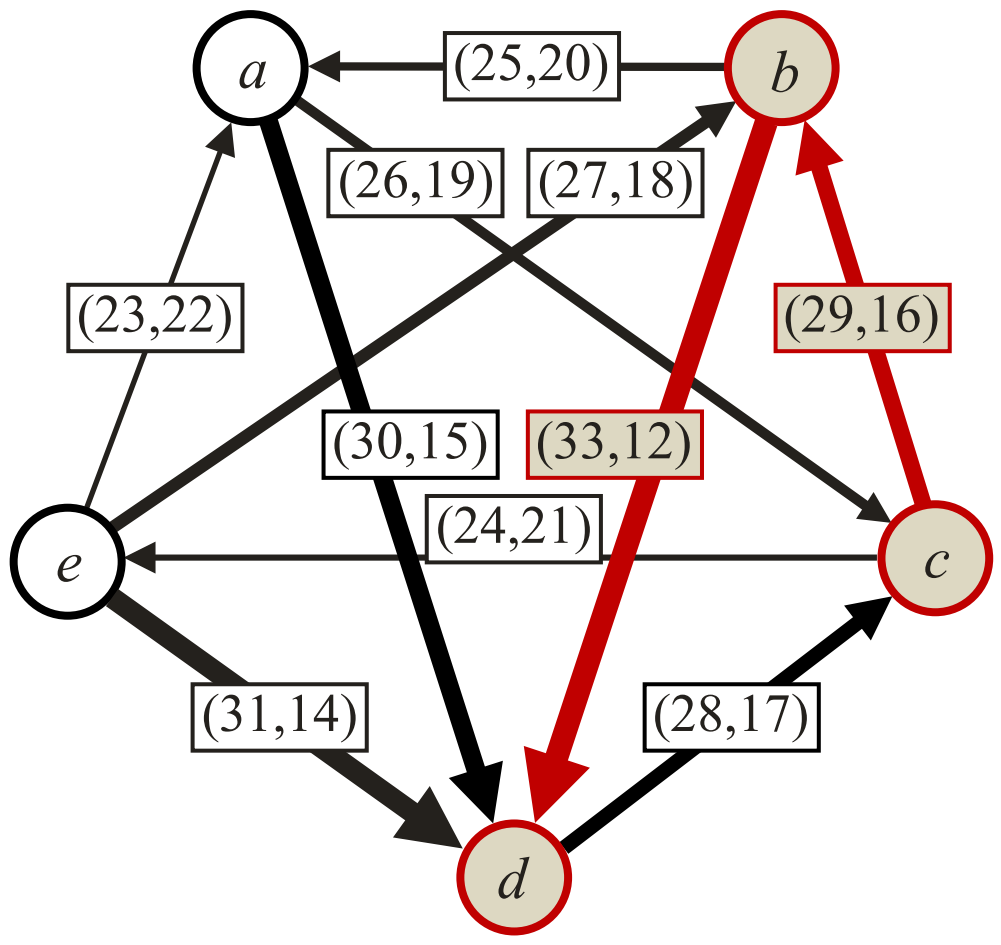

The strongest path from *c* to *d* is:
*c*, (29,16), *b*, (33,12), *d*

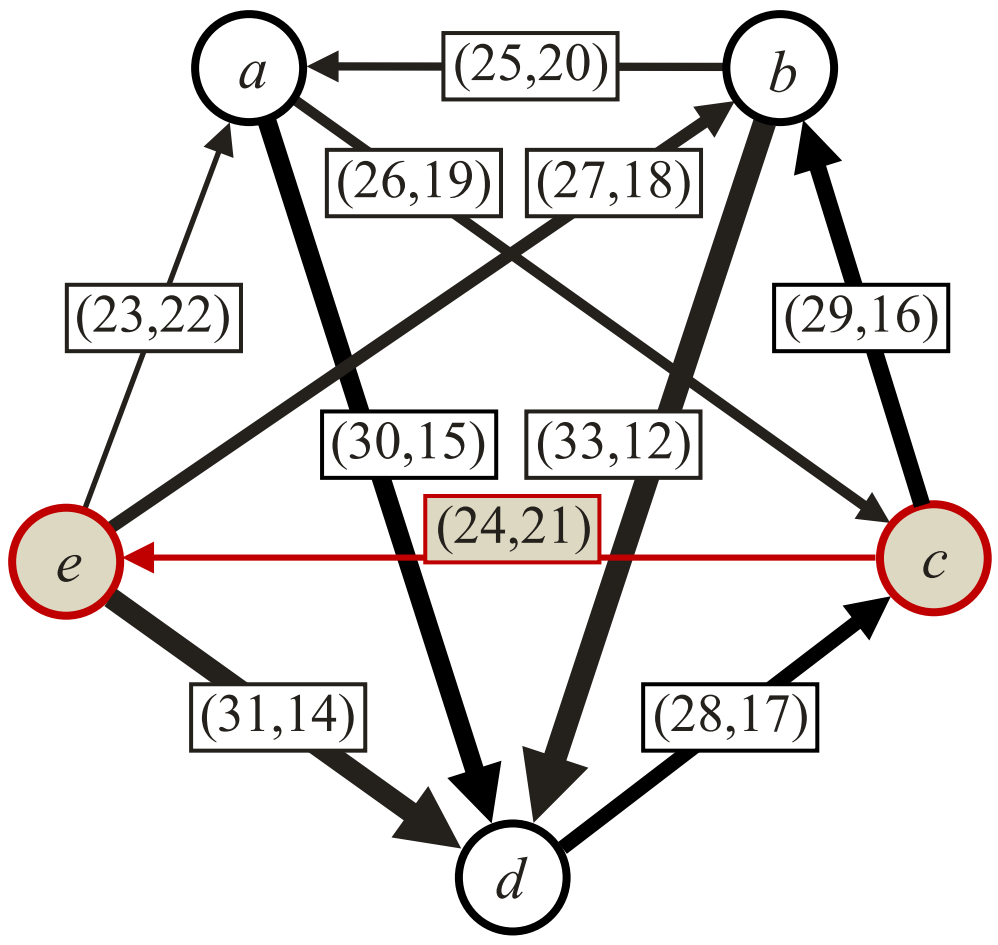

The strongest path from *c* to *e* is:
*c*, (24,21), *e*

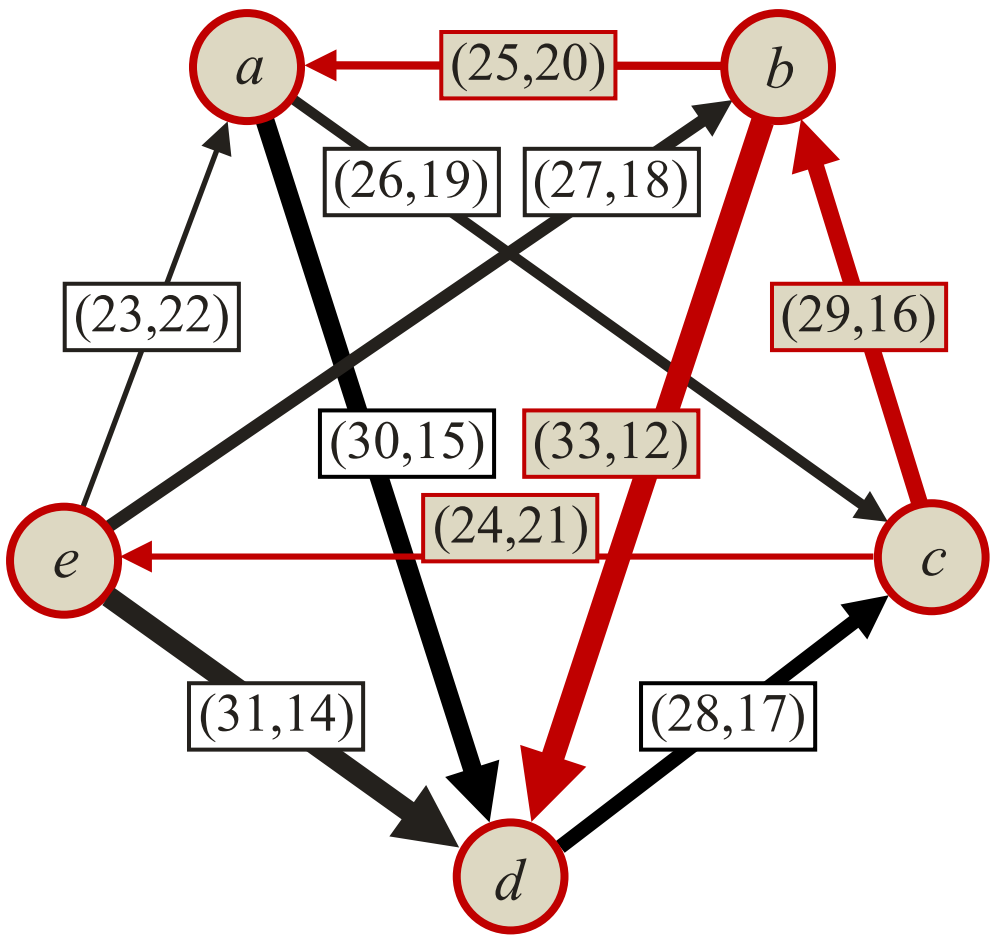

These are the strongest paths
from *c* to every other alternative.

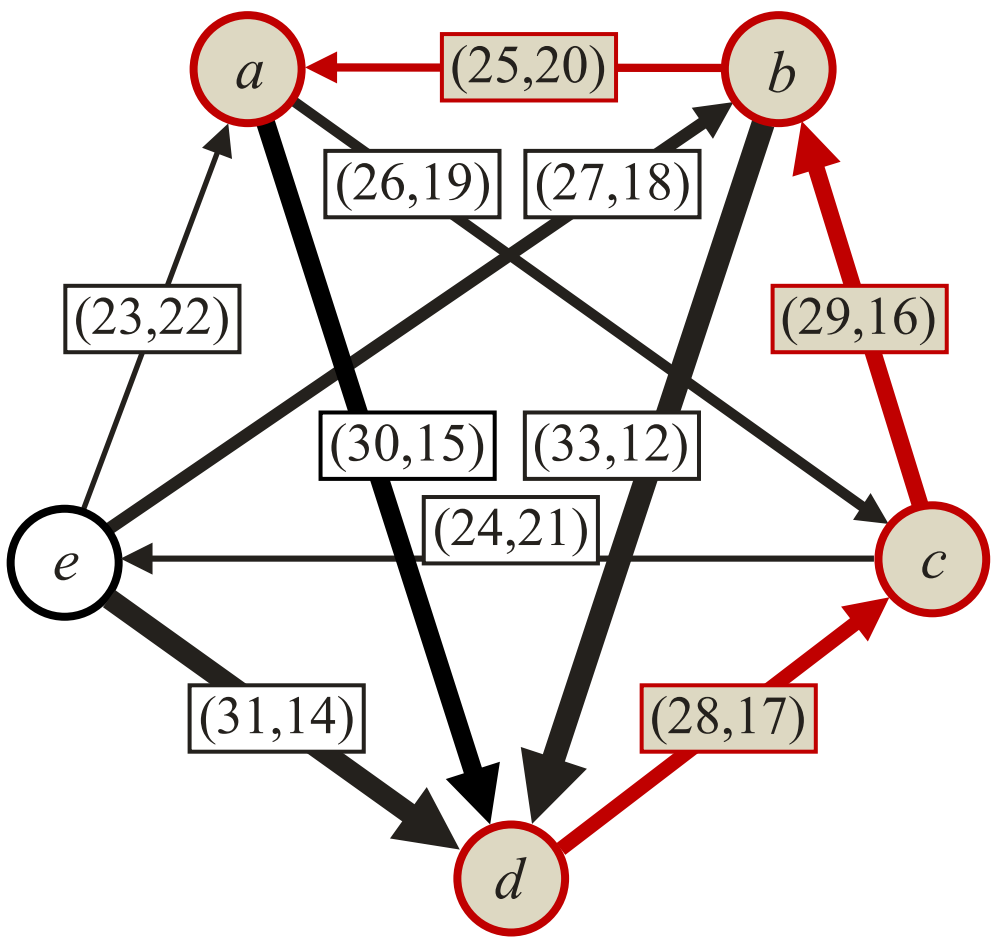

The strongest path from *d* to *a* is:
*d*, (28,17), *c*, (29,16), *b*, (25,20), *a*

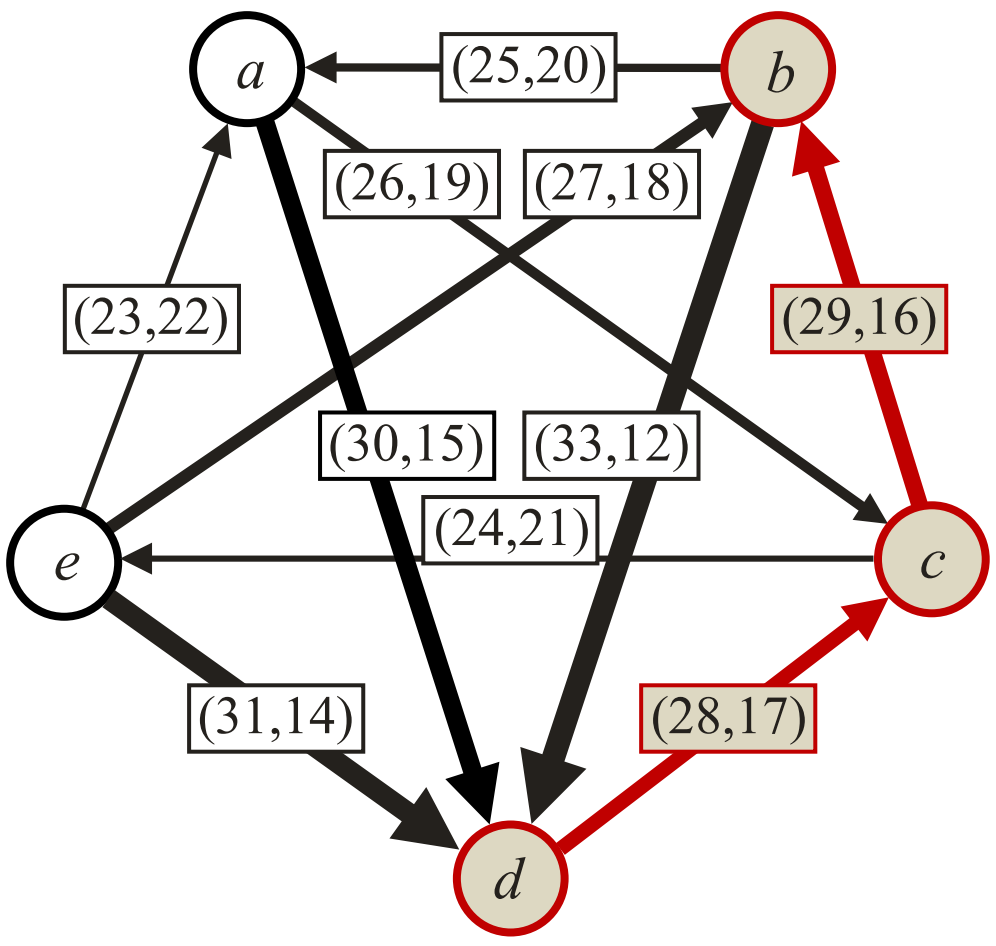

The strongest path from *d* to *b* is:
*d*, (28,17), *c*, (29,16), *b*

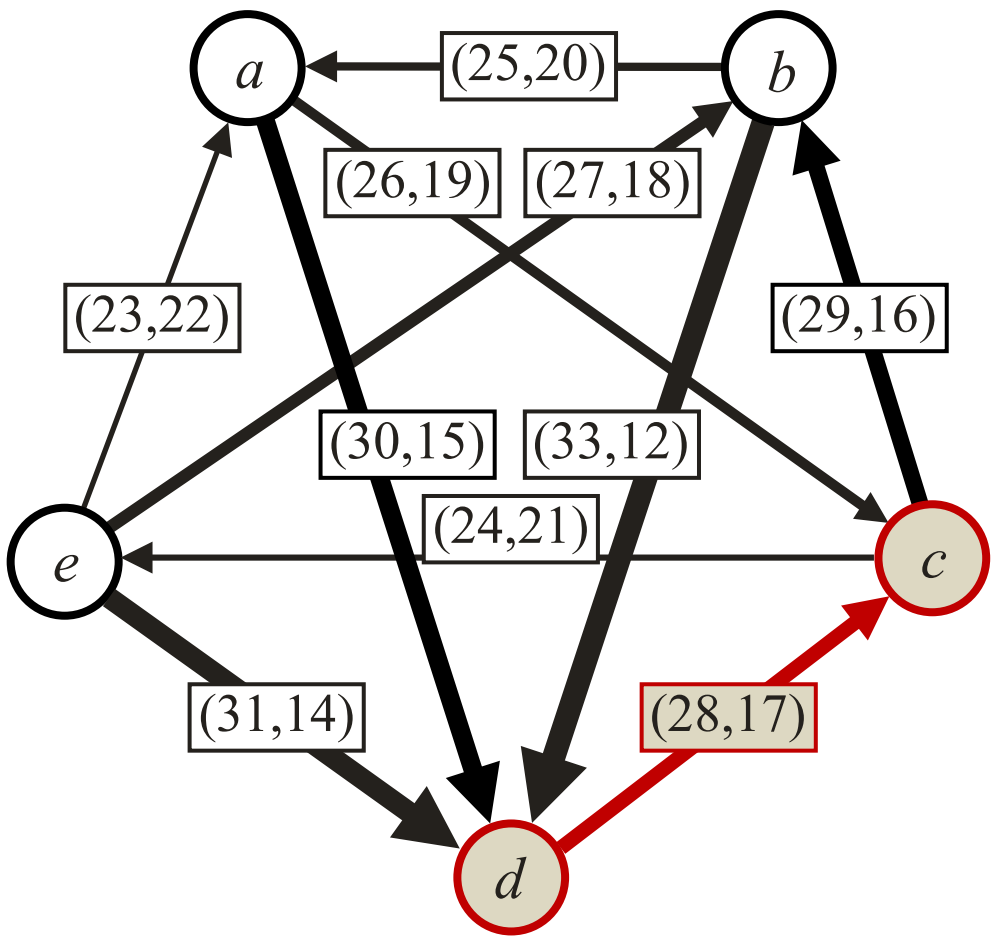

The strongest path from *d* to *c* is:
*d*, (28,17), *c*

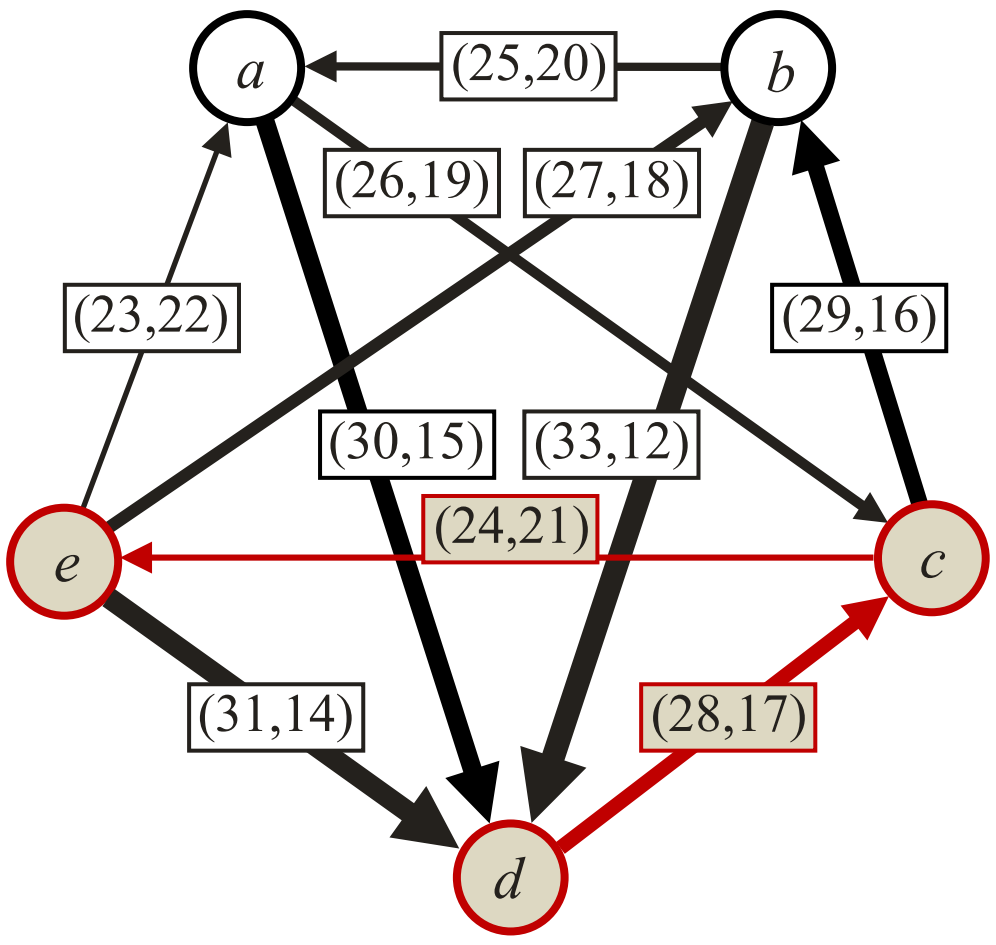

The strongest path from *d* to *e* is:
*d*, (28,17), *c*, (24,21), *e*

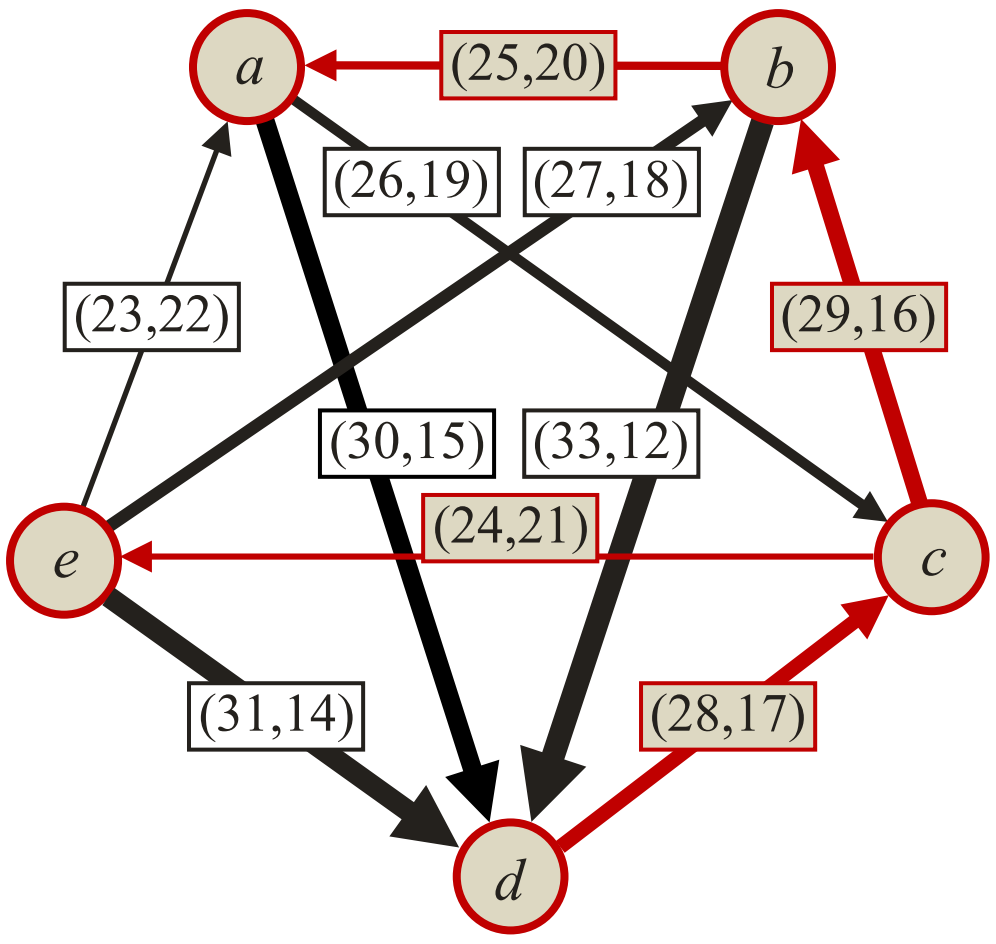

These are the strongest paths
from *d* to every other alternative.

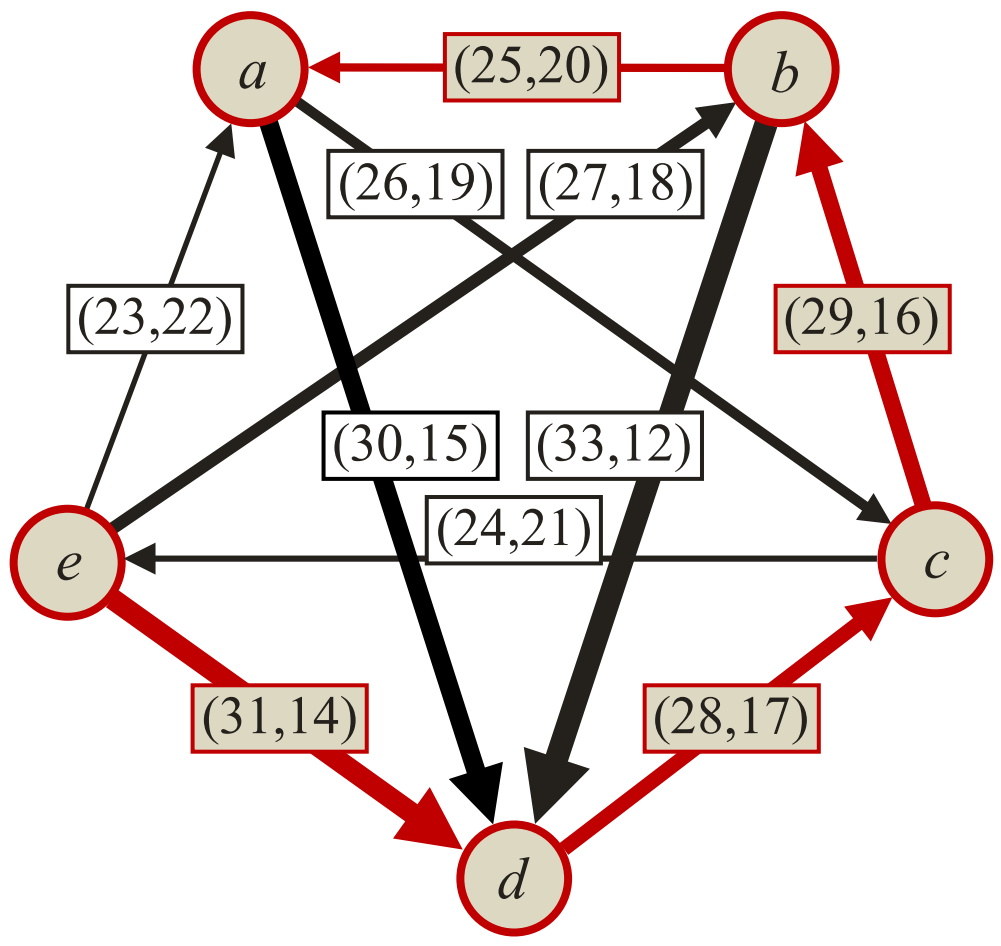

The strongest path from *e* to *a* is:
*e*, (31,14), *d*, (28,17), *c*,
(29,16), *b*, (25,20), *a*

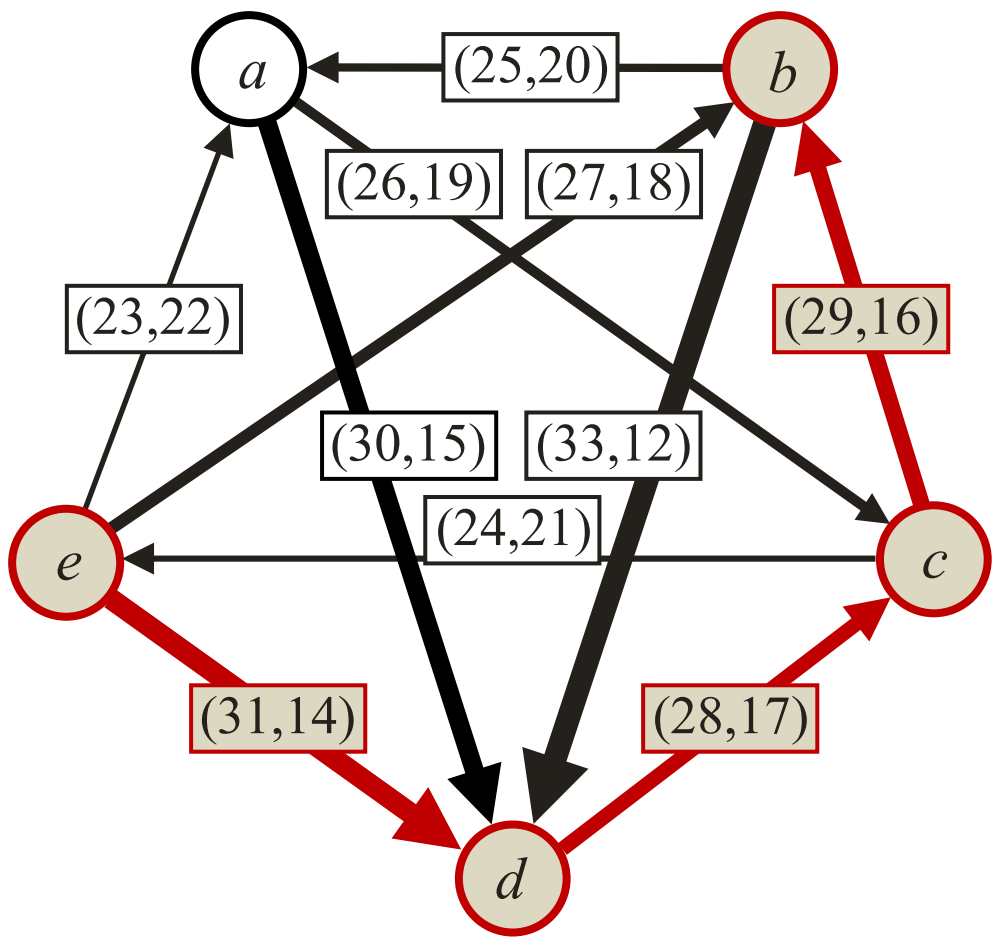

The strongest path from *e* to *b* is:
*e*, (31,14), *d*, (28,17), *c*, (29,16), *b*

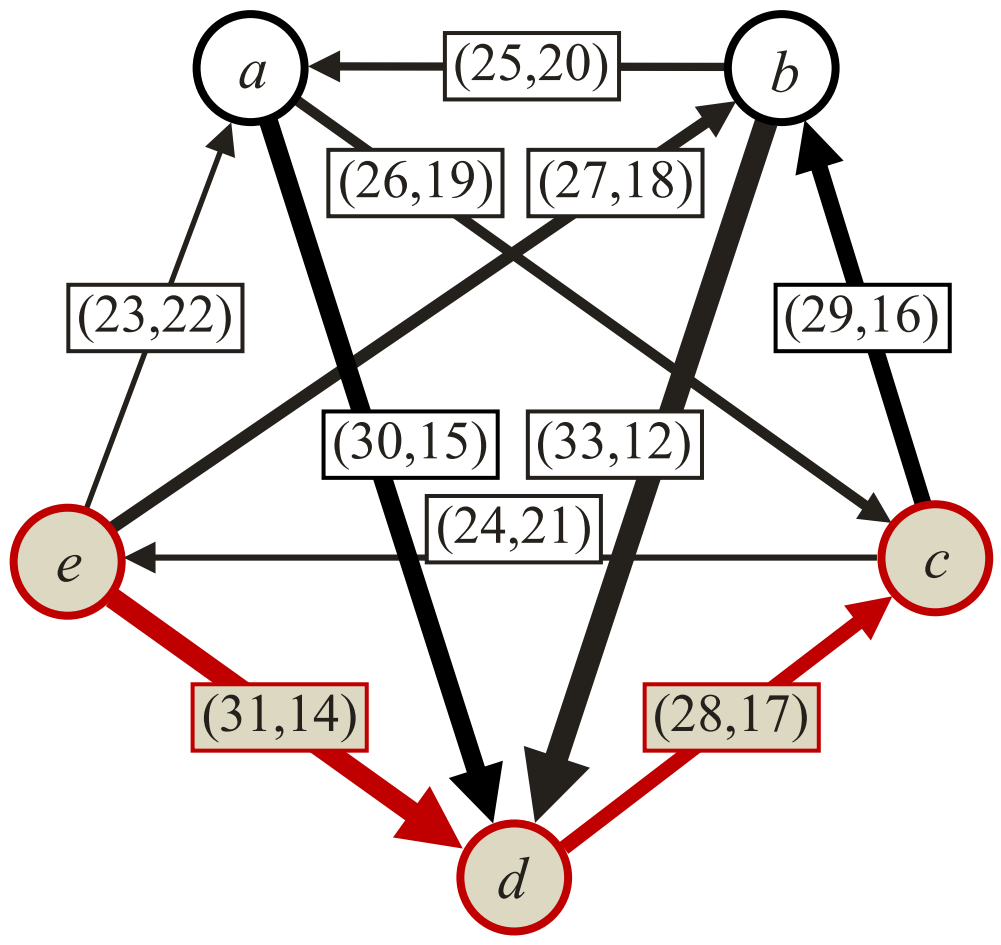

The strongest path from *e* to *c* is:
*e*, (31,14), *d*, (28,17), *c*

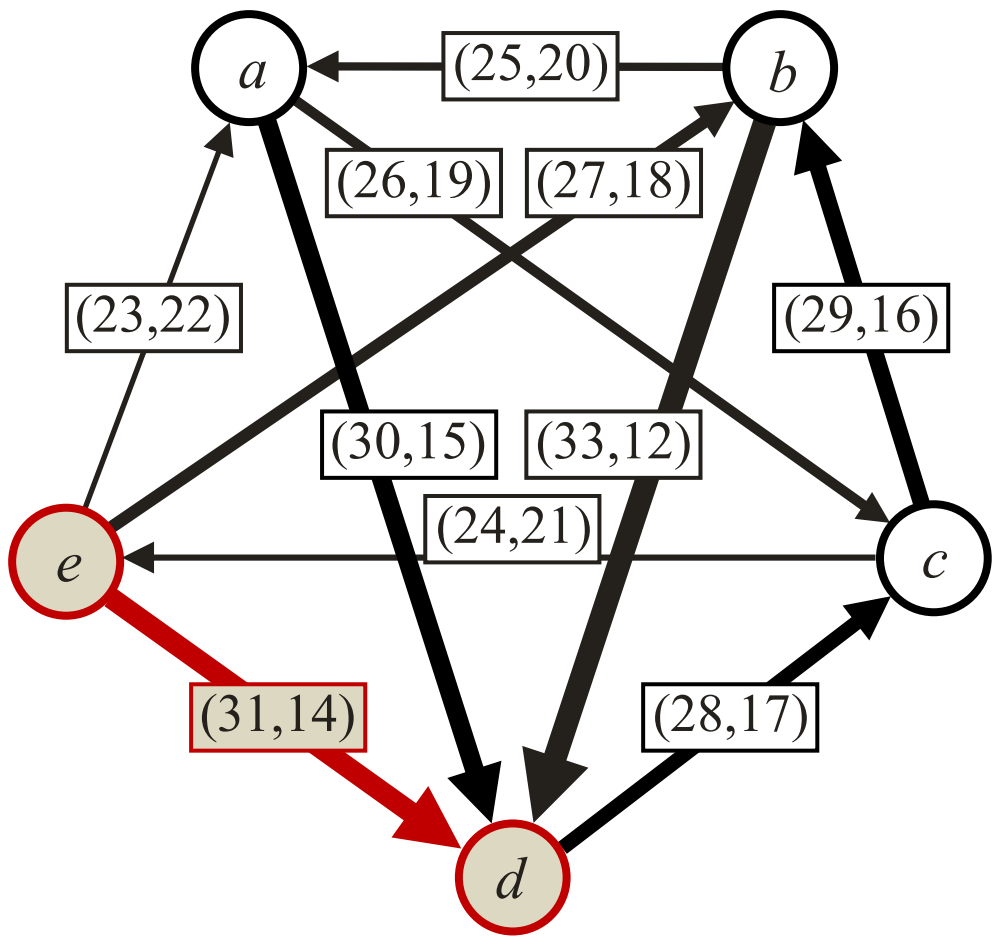

The strongest path from *e* to *d* is:
*e*, (31,14), *d*

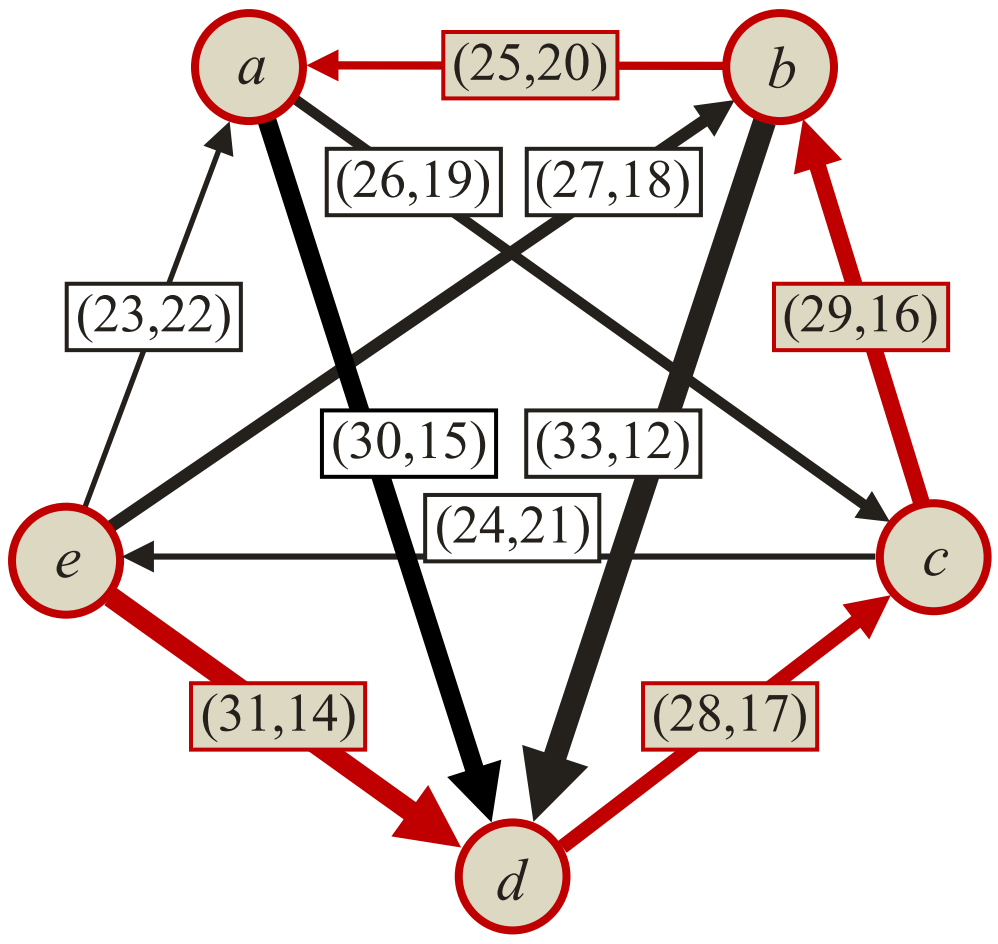

These are the strongest paths
from *e* to every other alternative.

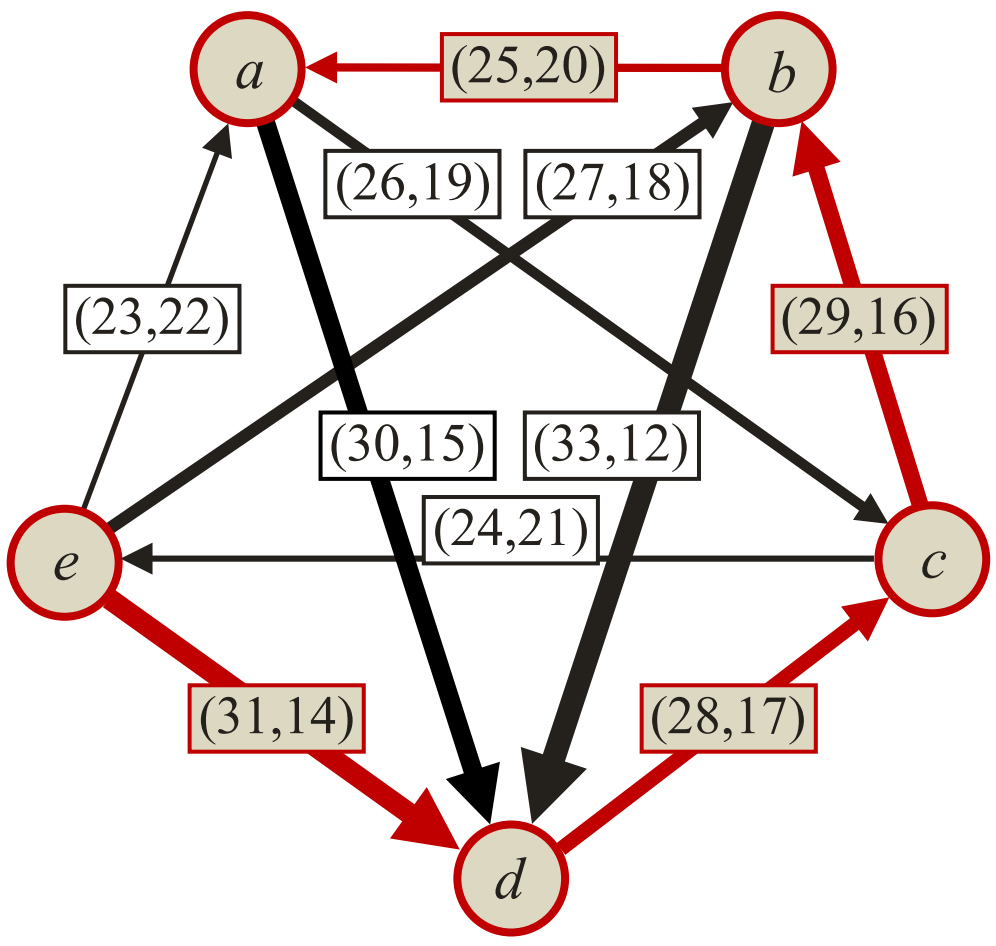

These are the strongest paths
from every other alternative to *a*.

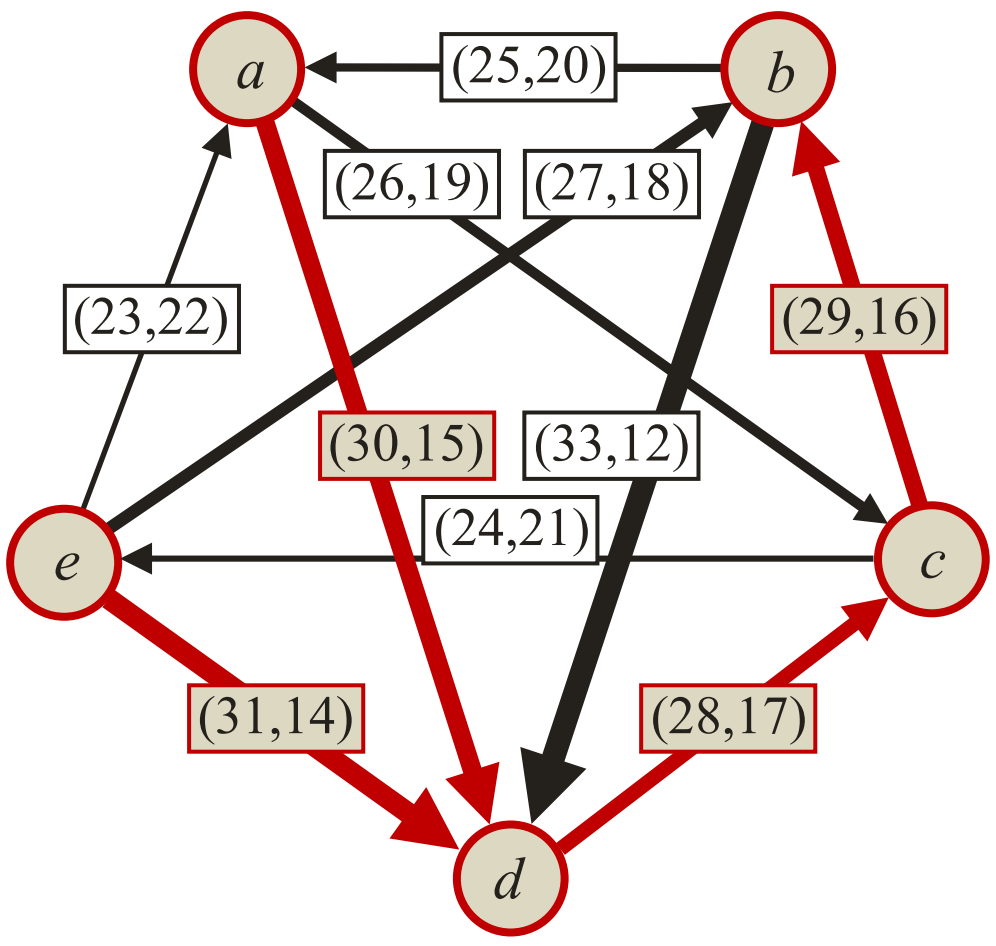

These are the strongest paths
from every other alternative to *b*.

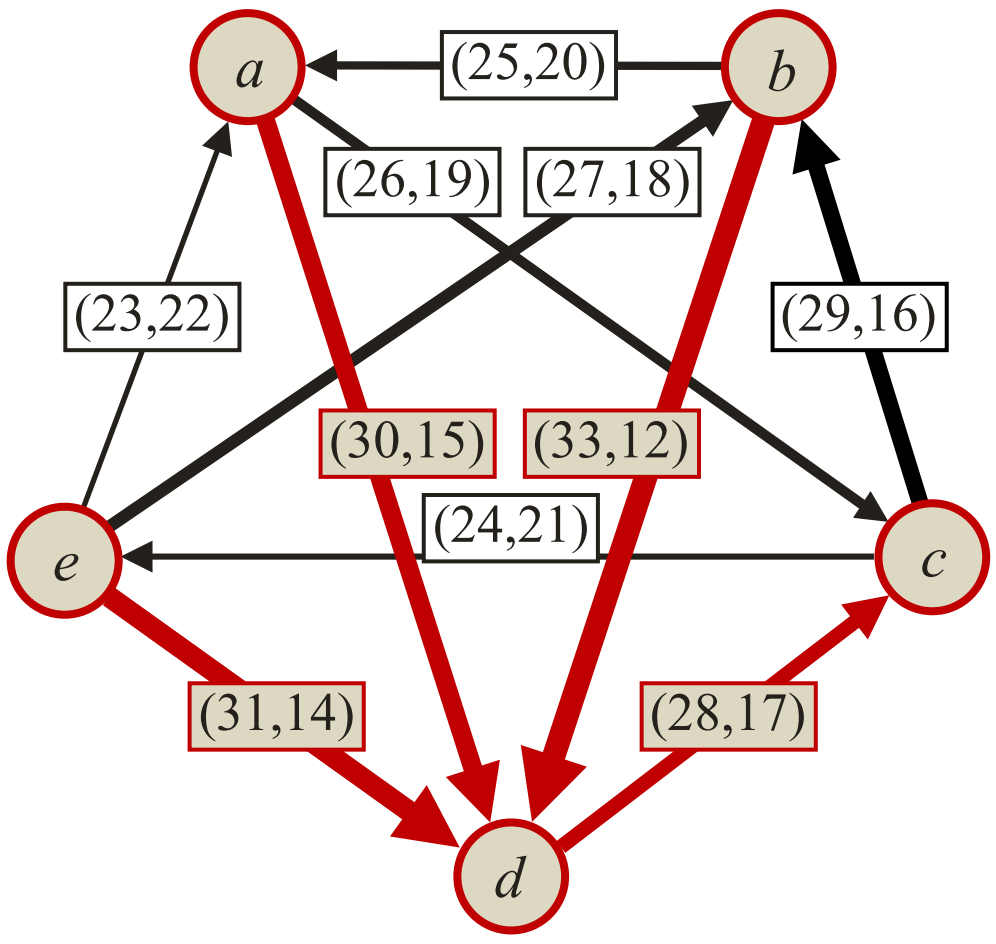

These are the strongest paths
from every other alternative to *c*.

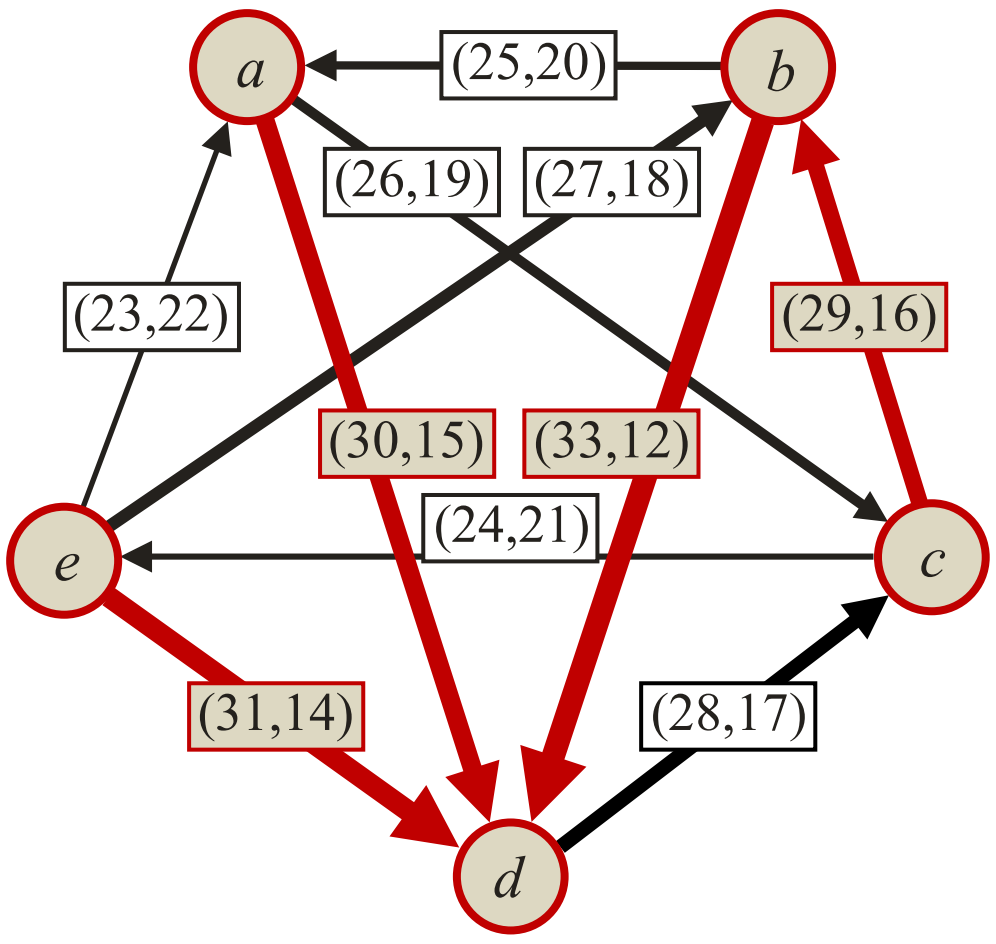

These are the strongest paths
from every other alternative to *d*.

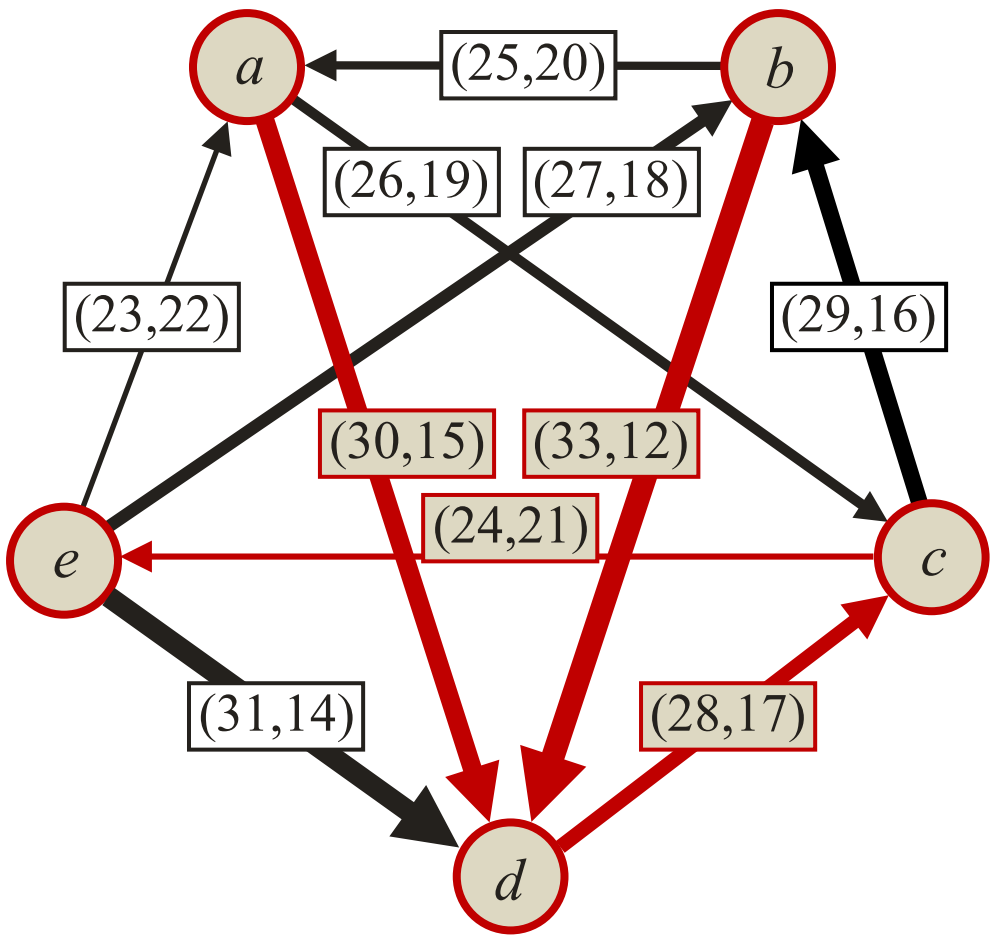

These are the strongest paths
from every other alternative to *e*.

Therefore, the strengths of the strongest paths are:

| | $P_D[*,a]$ | $P_D[*,b]$ | $P_D[*,c]$ | $P_D[*,d]$ | $P_D[*,e]$ |
|---|---|---|---|---|---|
| $P_D[a,*]$ | --- | (28,17) | (28,17) | (30,15) | (24,21) |
| $P_D[b,*]$ | (25,20) | --- | (28,17) | (33,12) | (24,21) |
| $P_D[c,*]$ | (25,20) | (29,16) | --- | (29,16) | (24,21) |
| $P_D[d,*]$ | (25,20) | (28,17) | (28,17) | --- | (24,21) |
| $P_D[e,*]$ | (25,20) | (28,17) | (28,17) | (31,14) | --- |

We get $\mathcal{O} = \{ab, ac, ad, bd, cb, cd, ea, eb, ec, ed\}$ and $\mathcal{S} = \{e\}$.

Suppose, the strongest paths are calculated with the Floyd-Warshall algorithm, as defined in section 2.3.1. Then the following table documents the $C \cdot (C–1) \cdot (C–2) = 60$ steps of the Floyd-Warshall algorithm.

We start with

- $P_D[i,j] := (N[i,j],N[j,i])$ for all $i \in A$ and $j \in A \setminus \{i\}$.
- $pred[i,j] := i$ for all $i \in A$ and $j \in A \setminus \{i\}$.

| | *i* | *j* | *k* | $P_D[j,k]$ | $P_D[j,i]$ | $P_D[i,k]$ | *pred*[*j*,*k*] | *pred*[*i*,*k*] | result |
|---|---|---|---|---|---|---|---|---|---|
| 1 | *a* | *b* | *c* | (16,29) | (25,20) | (26,19) | *b* | *a* | $P_D[b,c]$ is updated from (16,29) to (25,20); *pred*[*b*,*c*] is updated from *b* to *a*. |
| 2 | *a* | *b* | *d* | (33,12) | (25,20) | (30,15) | *b* | *a* | |
| 3 | *a* | *b* | *e* | (18,27) | (25,20) | (22,23) | *b* | *a* | $P_D[b,e]$ is updated from (18,27) to (22,23); *pred*[*b*,*e*] is updated from *b* to *a*. |
| 4 | *a* | *c* | *b* | (29,16) | (19,26) | (20,25) | *c* | *a* | |
| 5 | *a* | *c* | *d* | (17,28) | (19,26) | (30,15) | *c* | *a* | $P_D[c,d]$ is updated from (17,28) to (19,26); *pred*[*c*,*d*] is updated from *c* to *a*. |
| 6 | *a* | *c* | *e* | (24,21) | (19,26) | (22,23) | *c* | *a* | |
| 7 | *a* | *d* | *b* | (12,33) | (15,30) | (20,25) | *d* | *a* | $P_D[d,b]$ is updated from (12,33) to (15,30); *pred*[*d*,*b*] is updated from *d* to *a*. |
| 8 | *a* | *d* | *c* | (28,17) | (15,30) | (26,19) | *d* | *a* | |
| 9 | *a* | *d* | *e* | (14,31) | (15,30) | (22,23) | *d* | *a* | $P_D[d,e]$ is updated from (14,31) to (15,30); *pred*[*d*,*e*] is updated from *d* to *a*. |
| 10 | *a* | *e* | *b* | (27,18) | (23,22) | (20,25) | *e* | *a* | |
| 11 | *a* | *e* | *c* | (21,24) | (23,22) | (26,19) | *e* | *a* | $P_D[e,c]$ is updated from (21,24) to (23,22); *pred*[*e*,*c*] is updated from *e* to *a*. |
| 12 | *a* | *e* | *d* | (31,14) | (23,22) | (30,15) | *e* | *a* | |
| 13 | *b* | *a* | *c* | (26,19) | (20,25) | (25,20) | *a* | *a* | |
| 14 | *b* | *a* | *d* | (30,15) | (20,25) | (33,12) | *a* | *b* | |
| 15 | *b* | *a* | *e* | (22,23) | (20,25) | (22,23) | *a* | *a* | |
| 16 | *b* | *c* | *a* | (19,26) | (29,16) | (25,20) | *c* | *b* | $P_D[c,a]$ is updated from (19,26) to (25,20); *pred*[*c*,*a*] is updated from *c* to *b*. |
| 17 | *b* | *c* | *d* | (19,26) | (29,16) | (33,12) | *a* | *b* | $P_D[c,d]$ is updated from (19,26) to (29,16); *pred*[*c*,*d*] is updated from *a* to *b*. |
| 18 | *b* | *c* | *e* | (24,21) | (29,16) | (22,23) | *c* | *a* | |
| 19 | *b* | *d* | *a* | (15,30) | (15,30) | (25,20) | *d* | *b* | |
| 20 | *b* | *d* | *c* | (28,17) | (15,30) | (25,20) | *d* | *a* | |
| 21 | *b* | *d* | *e* | (15,30) | (15,30) | (22,23) | *a* | *a* | |
| 22 | *b* | *e* | *a* | (23,22) | (27,18) | (25,20) | *e* | *b* | $P_D[e,a]$ is updated from (23,22) to (25,20); *pred*[*e*,*a*] is updated from *e* to *b*. |
| 23 | *b* | *e* | *c* | (23,22) | (27,18) | (25,20) | *a* | *a* | $P_D[e,c]$ is updated from (23,22) to (25,20) |
| 24 | *b* | *e* | *d* | (31,14) | (27,18) | (33,12) | *e* | *b* | |
| 25 | *c* | *a* | *b* | (20,25) | (26,19) | (29,16) | *a* | *c* | $P_D[a,b]$ is updated from (20,25) to (26,19); *pred*[*a*,*b*] is updated from *a* to *c*. |
| 26 | *c* | *a* | *d* | (30,15) | (26,19) | (29,16) | *a* | *b* | |
| 27 | *c* | *a* | *e* | (22,23) | (26,19) | (24,21) | *a* | *c* | $P_D[a,e]$ is updated from (22,23) to (24,21); *pred*[*a*,*e*] is updated from *a* to *c*. |
| 28 | *c* | *b* | *a* | (25,20) | (25,20) | (25,20) | *b* | *b* | |
| 29 | *c* | *b* | *d* | (33,12) | (25,20) | (29,16) | *b* | *b* | |
| 30 | *c* | *b* | *e* | (22,23) | (25,20) | (24,21) | *a* | *c* | $P_D[b,e]$ is updated from (22,23) to (24,21); *pred*[*b*,*e*] is updated from *a* to *c*. |

| | $i$ | $j$ | $k$ | $P_D[j,k]$ | $P_D[j,i]$ | $P_D[i,k]$ | $pred[j,k]$ | $pred[i,k]$ | result |
|---|---|---|---|---|---|---|---|---|---|
| 31 | $c$ | $d$ | $a$ | (15,30) | (28,17) | (25,20) | $d$ | $b$ | $P_D[d,a]$ is updated from (15,30) to (25,20); $pred[d,a]$ is updated from $d$ to $b$. |
| 32 | $c$ | $d$ | $b$ | (15,30) | (28,17) | (29,16) | $a$ | $c$ | $P_D[d,b]$ is updated from (15,30) to (28,17); $pred[d,b]$ is updated from $a$ to $c$. |
| 33 | $c$ | $d$ | $e$ | (15,30) | (28,17) | (24,21) | $a$ | $c$ | $P_D[d,e]$ is updated from (15,30) to (24,21); $pred[d,e]$ is updated from $a$ to $c$. |
| 34 | $c$ | $e$ | $a$ | (25,20) | (25,20) | (25,20) | $b$ | $b$ | |
| 35 | $c$ | $e$ | $b$ | (27,18) | (25,20) | (29,16) | $e$ | $c$ | |
| 36 | $c$ | $e$ | $d$ | (31,14) | (25,20) | (29,16) | $e$ | $b$ | |
| 37 | $d$ | $a$ | $b$ | (26,19) | (30,15) | (28,17) | $c$ | $c$ | $P_D[a,b]$ is updated from (26,19) to (28,17). |
| 38 | $d$ | $a$ | $c$ | (26,19) | (30,15) | (28,17) | $a$ | $d$ | $P_D[a,c]$ is updated from (26,19) to (28,17); $pred[a,c]$ is updated from $a$ to $d$. |
| 39 | $d$ | $a$ | $e$ | (24,21) | (30,15) | (24,21) | $c$ | $c$ | |
| 40 | $d$ | $b$ | $a$ | (25,20) | (33,12) | (25,20) | $b$ | $b$ | |
| 41 | $d$ | $b$ | $c$ | (25,20) | (33,12) | (28,17) | $a$ | $d$ | $P_D[b,c]$ is updated from (25,20) to (28,17); $pred[b,c]$ is updated from $a$ to $d$. |
| 42 | $d$ | $b$ | $e$ | (24,21) | (33,12) | (24,21) | $c$ | $c$ | |
| 43 | $d$ | $c$ | $a$ | (25,20) | (29,16) | (25,20) | $b$ | $b$ | |
| 44 | $d$ | $c$ | $b$ | (29,16) | (29,16) | (28,17) | $c$ | $c$ | |
| 45 | $d$ | $c$ | $e$ | (24,21) | (29,16) | (24,21) | $c$ | $c$ | |
| 46 | $d$ | $e$ | $a$ | (25,20) | (31,14) | (25,20) | $b$ | $b$ | |
| 47 | $d$ | $e$ | $b$ | (27,18) | (31,14) | (28,17) | $e$ | $c$ | $P_D[e,b]$ is updated from (27,18) to (28,17); $pred[e,b]$ is updated from $e$ to $c$. |
| 48 | $d$ | $e$ | $c$ | (25,20) | (31,14) | (28,17) | $a$ | $d$ | $P_D[e,c]$ is updated from (25,20) to (28,17); $pred[e,c]$ is updated from $a$ to $d$. |
| 49 | $e$ | $a$ | $b$ | (28,17) | (24,21) | (28,17) | $c$ | $c$ | |
| 50 | $e$ | $a$ | $c$ | (28,17) | (24,21) | (28,17) | $d$ | $d$ | |
| 51 | $e$ | $a$ | $d$ | (30,15) | (24,21) | (31,14) | $a$ | $e$ | |
| 52 | $e$ | $b$ | $a$ | (25,20) | (24,21) | (25,20) | $b$ | $b$ | |
| 53 | $e$ | $b$ | $c$ | (28,17) | (24,21) | (28,17) | $d$ | $d$ | |
| 54 | $e$ | $b$ | $d$ | (33,12) | (24,21) | (31,14) | $b$ | $e$ | |
| 55 | $e$ | $c$ | $a$ | (25,20) | (24,21) | (25,20) | $b$ | $b$ | |
| 56 | $e$ | $c$ | $b$ | (29,16) | (24,21) | (28,17) | $c$ | $c$ | |
| 57 | $e$ | $c$ | $d$ | (29,16) | (24,21) | (31,14) | $b$ | $e$ | |
| 58 | $e$ | $d$ | $a$ | (25,20) | (24,21) | (25,20) | $b$ | $b$ | |
| 59 | $e$ | $d$ | $b$ | (28,17) | (24,21) | (28,17) | $c$ | $c$ | |
| 60 | $e$ | $d$ | $c$ | (28,17) | (24,21) | (28,17) | $d$ | $d$ | |

## 3.4. Example 4

The following example is by Hoag and Hallett (1926, page 502), where the authors use this example to illustrate their proposal (*Hallett count*).

Example 4:

| | |
|---|---|
| 3 voters | $a \succ_v b \succ_v c \succ_v d$ |
| 2 voters | $c \succ_v b \succ_v d \succ_v a$ |
| 2 voters | $d \succ_v a \succ_v b \succ_v c$ |
| 2 voters | $d \succ_v b \succ_v c \succ_v a$ |

The pairwise matrix *N* looks as follows:

| | *N*[*,*a*] | *N*[*,*b*] | *N*[*,*c*] | *N*[*,*d*] |
|---|---|---|---|---|
| *N*[*a*,*] | --- | 5 | 5 | 3 |
| *N*[*b*,*] | 4 | --- | 7 | 5 |
| *N*[*c*,*] | 4 | 2 | --- | 5 |
| *N*[*d*,*] | 6 | 4 | 4 | --- |

The corresponding digraph looks as follows:

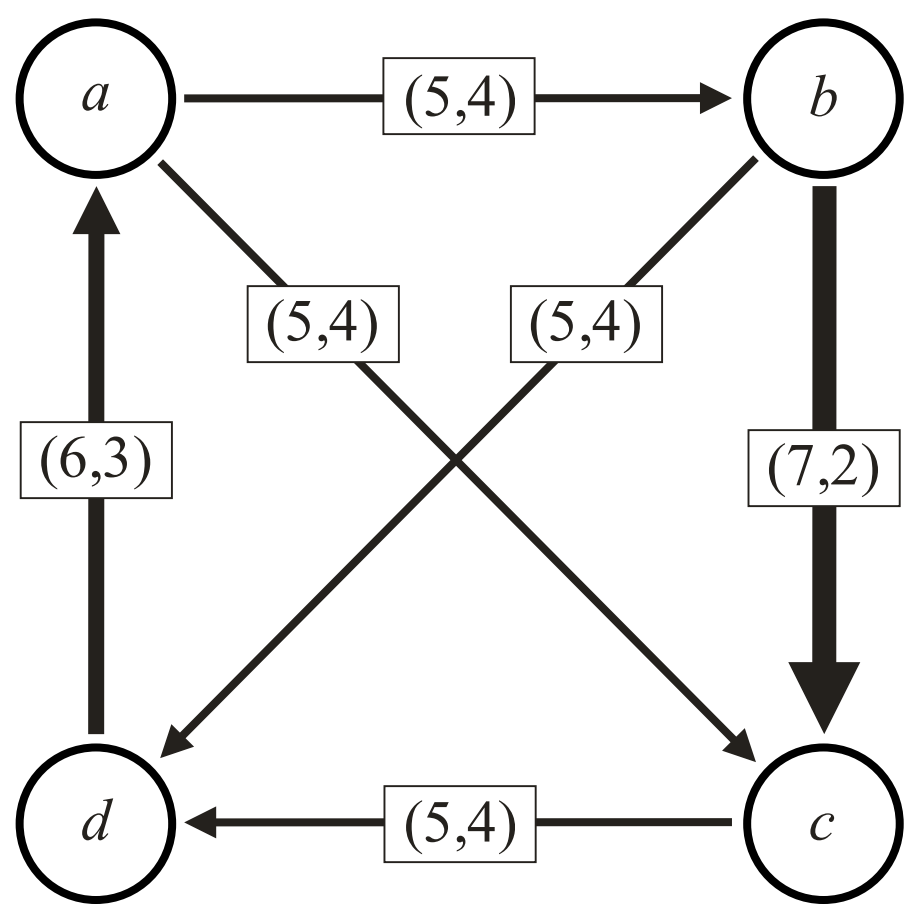

The following table lists the strongest paths, as determined by the Floyd-Warshall algorithm, as defined in section 2.3.1. The critical links of the strongest paths are underlined:

| | ... to *a* | ... to *b* | ... to *c* | ... to *d* | ... to every other alternative |
|---|---|---|---|---|---|
| from *a* ... | --- | 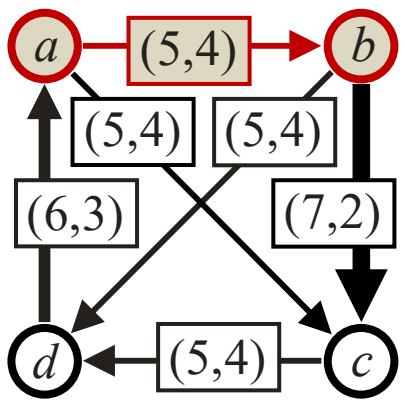<br>*a*, (5,4), *b* | 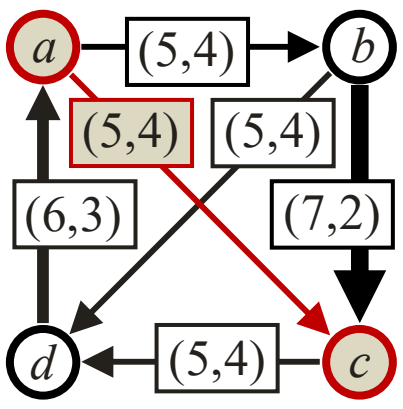<br>*a*, (5,4), *c* | 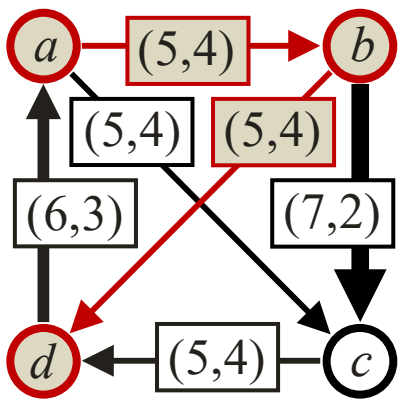<br>*a*, (5,4), *b*, (5,4), *d* | 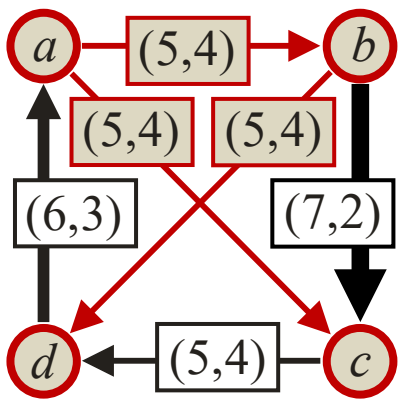 |
| from *b* ... | 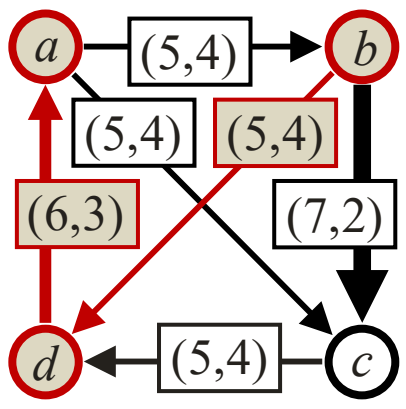<br>*b*, (5,4), *d*, (6,3), *a* | --- | 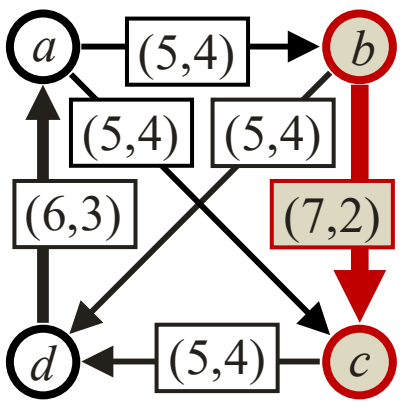<br>*b*, (7,2), *c* | 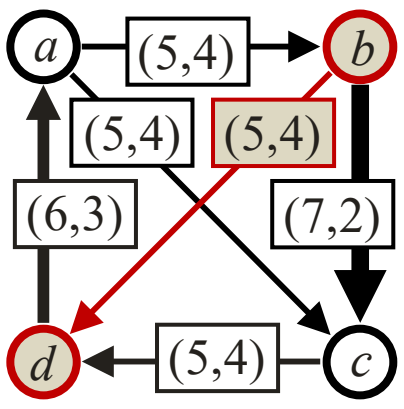<br>*b*, (5,4), *d* | 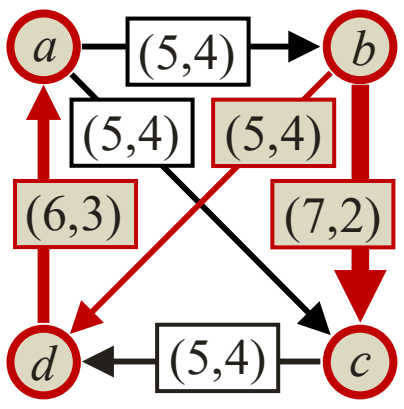 |
| from *c* ... | 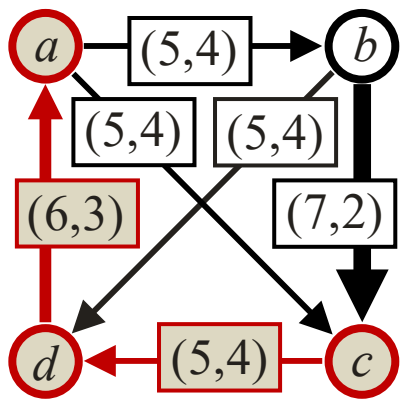<br>*c*, (5,4), *d*, (6,3), *a* | 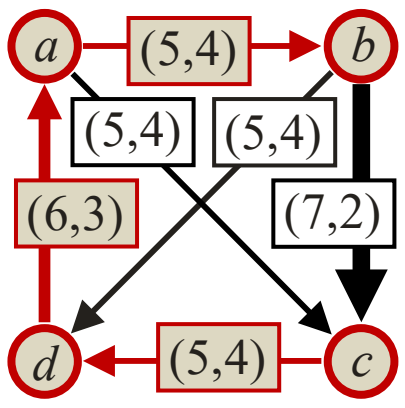<br>*c*, (5,4), *d*, (6,3), *a*, (5,4), *b* | --- | 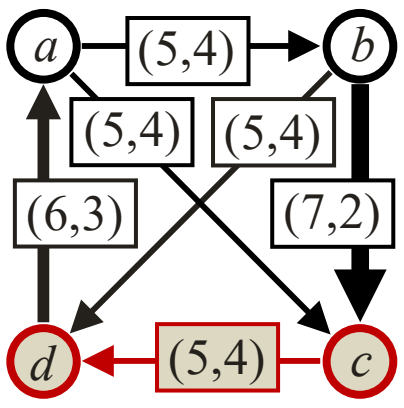<br>*c*, (5,4), *d* | 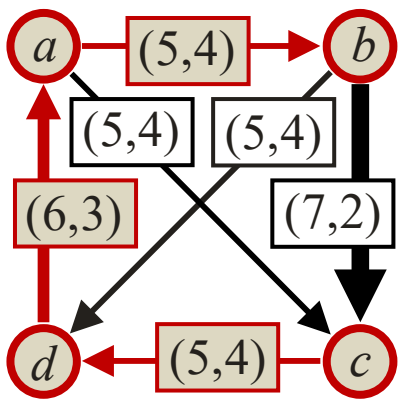 |
| from *d* ... | 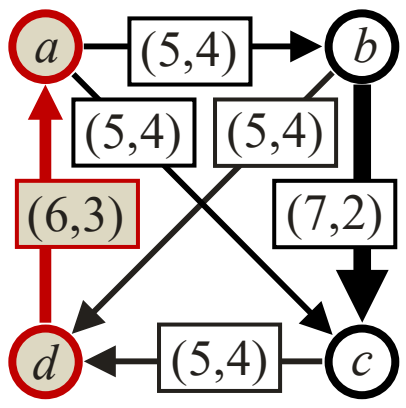<br>*d*, (6,3), *a* | 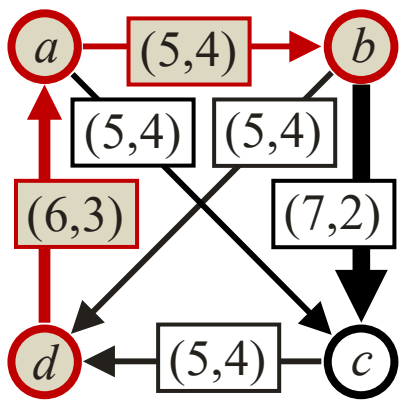<br>*d*, (6,3), *a*, (5,4), *b* | 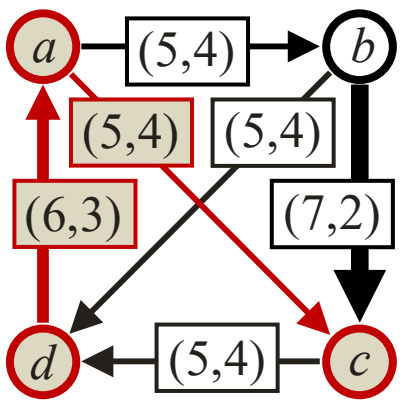<br>*d*, (6,3), *a*, (5,4), *c* | --- | 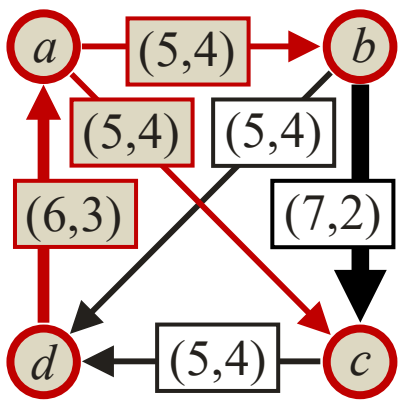 |
| from every other alternative ... | 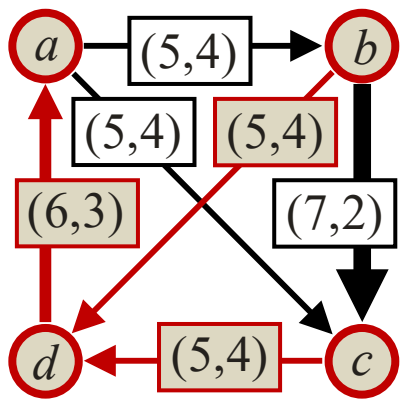 | 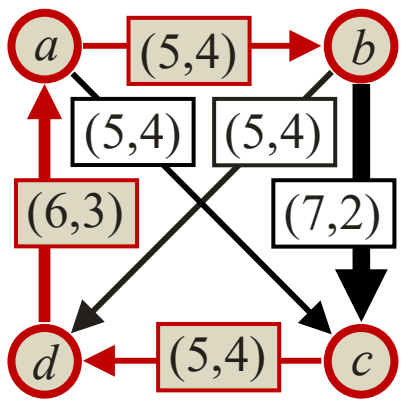 | 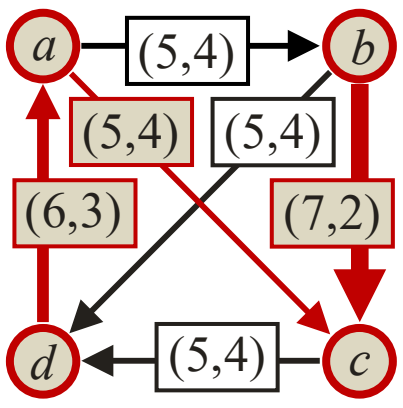 | 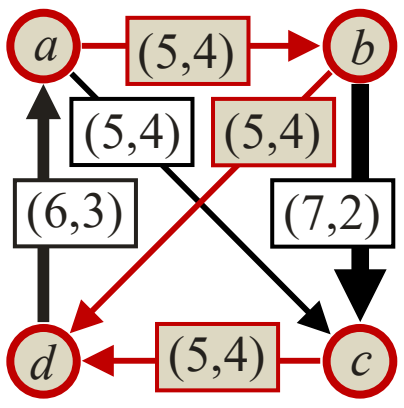 | --- |

Therefore, the strengths of the strongest paths are:

| | $P_D[*,a]$ | $P_D[*,b]$ | $P_D[*,c]$ | $P_D[*,d]$ |
|---|---|---|---|---|
| $P_D[a,*]$ | --- | (5,4) | (5,4) | (5,4) |
| $P_D[b,*]$ | (5,4) | --- | (7,2) | (5,4) |
| $P_D[c,*]$ | (5,4) | (5,4) | --- | (5,4) |
| $P_D[d,*]$ | (6,3) | (5,4) | (5,4) | --- |

We get $\mathcal{O} = \{bc, da\}$ and $\mathcal{S} = \{b, d\}$.

Suppose, the strongest paths are calculated with the Floyd-Warshall algorithm, as defined in section 2.3.1. Then the following table documents the $C \cdot (C–1) \cdot (C–2) = 24$ steps of the Floyd-Warshall algorithm.

We start with

- $P_D[i,j] := (N[i,j],N[j,i])$ for all $i \in A$ and $j \in A \setminus \{i\}$.
- $pred[i,j] := i$ for all $i \in A$ and $j \in A \setminus \{i\}$.

| | *i* | *j* | *k* | $P_D[j,k]$ | $P_D[j,i]$ | $P_D[i,k]$ | *pred*[*j*,*k*] | *pred*[*i*,*k*] | result |
|---|---|---|---|---|---|---|---|---|---|
| 1 | *a* | *b* | *c* | (7,2) | (4,5) | (5,4) | *b* | *a* | |
| 2 | *a* | *b* | *d* | (5,4) | (4,5) | (3,6) | *b* | *a* | |
| 3 | *a* | *c* | *b* | (2,7) | (4,5) | (5,4) | *c* | *a* | $P_D[c,b]$ is updated from (2,7) to (4,5); *pred*[*c*,*b*] is updated from *c* to *a*. |
| 4 | *a* | *c* | *d* | (5,4) | (4,5) | (3,6) | *c* | *a* | |
| 5 | *a* | *d* | *b* | (4,5) | (6,3) | (5,4) | *d* | *a* | $P_D[d,b]$ is updated from (4,5) to (5,4); *pred*[*d*,*b*] is updated from *d* to *a*. |
| 6 | *a* | *d* | *c* | (4,5) | (6,3) | (5,4) | *d* | *a* | $P_D[d,c]$ is updated from (4,5) to (5,4); *pred*[*d*,*c*] is updated from *d* to *a*. |
| 7 | *b* | *a* | *c* | (5,4) | (5,4) | (7,2) | *a* | *b* | |
| 8 | *b* | *a* | *d* | (3,6) | (5,4) | (5,4) | *a* | *b* | $P_D[a,d]$ is updated from (3,6) to (5,4); *pred*[*a*,*d*] is updated from *a* to *b*. |
| 9 | *b* | *c* | *a* | (4,5) | (4,5) | (4,5) | *c* | *b* | |
| 10 | *b* | *c* | *d* | (5,4) | (4,5) | (5,4) | *c* | *b* | |
| 11 | *b* | *d* | *a* | (6,3) | (5,4) | (4,5) | *d* | *b* | |
| 12 | *b* | *d* | *c* | (5,4) | (5,4) | (7,2) | *a* | *b* | |
| 13 | *c* | *a* | *b* | (5,4) | (5,4) | (4,5) | *a* | *a* | |
| 14 | *c* | *a* | *d* | (5,4) | (5,4) | (5,4) | *b* | *c* | |
| 15 | *c* | *b* | *a* | (4,5) | (7,2) | (4,5) | *b* | *c* | |
| 16 | *c* | *b* | *d* | (5,4) | (7,2) | (5,4) | *b* | *c* | |
| 17 | *c* | *d* | *a* | (6,3) | (5,4) | (4,5) | *d* | *c* | |
| 18 | *c* | *d* | *b* | (5,4) | (5,4) | (4,5) | *a* | *a* | |
| 19 | *d* | *a* | *b* | (5,4) | (5,4) | (5,4) | *a* | *a* | |
| 20 | *d* | *a* | *c* | (5,4) | (5,4) | (5,4) | *a* | *a* | |
| 21 | *d* | *b* | *a* | (4,5) | (5,4) | (6,3) | *b* | *d* | $P_D[b,a]$ is updated from (4,5) to (5,4); *pred*[*b*,*a*] is updated from *b* to *d*. |
| 22 | *d* | *b* | *c* | (7,2) | (5,4) | (5,4) | *b* | *a* | |
| 23 | *d* | *c* | *a* | (4,5) | (5,4) | (6,3) | *c* | *d* | $P_D[c,a]$ is updated from (4,5) to (5,4); *pred*[*c*,*a*] is updated from *c* to *d*. |
| 24 | *d* | *c* | *b* | (4,5) | (5,4) | (5,4) | *a* | *a* | $P_D[c,b]$ is updated from (4,5) to (5,4). |

## 3.5. Example 5

Example 5:

| | |
|---|---|
| 12 voters | $a \succ_v b \succ_v c \succ_v d$ |
| 6 voters | $a \succ_v d \succ_v b \succ_v c$ |
| 9 voters | $b \succ_v c \succ_v d \succ_v a$ |
| 15 voters | $c \succ_v d \succ_v a \succ_v b$ |
| 21 voters | $d \succ_v b \succ_v a \succ_v c$ |

The pairwise matrix *N* looks as follows:

| | *N*[*,*a*] | *N*[*,*b*] | *N*[*,*c*] | *N*[*,*d*] |
|---|---|---|---|---|
| *N*[*a*,*] | --- | 33 | 39 | 18 |
| *N*[*b*,*] | 30 | --- | 48 | 21 |
| *N*[*c*,*] | 24 | 15 | --- | 36 |
| *N*[*d*,*] | 45 | 42 | 27 | --- |

The corresponding digraph looks as follows:

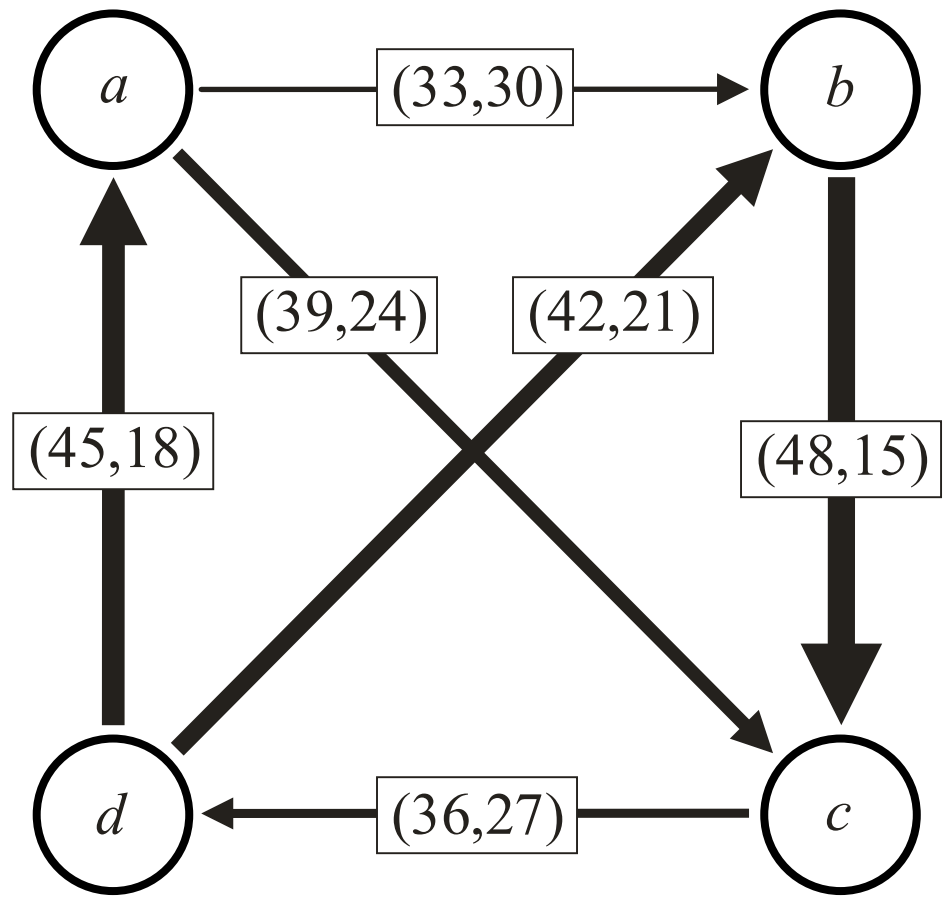

The following table lists the strongest paths, as determined by the Floyd-Warshall algorithm, as defined in section 2.3.1. The critical links of the strongest paths are underlined:

| | ... to *a* | ... to *b* | ... to *c* | ... to *d* | ... to every other alternative |
|---|---|---|---|---|---|
| from *a* ... | --- | 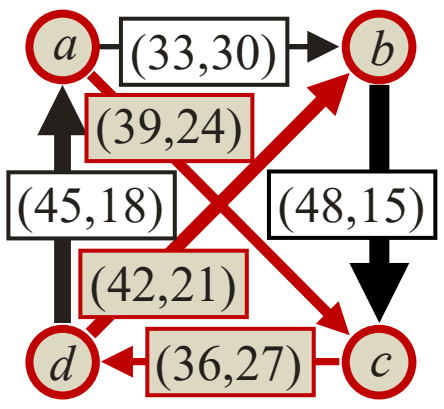<br>*a*, (39,24), *c*, (36,27), *d*, (42,21), *b* | 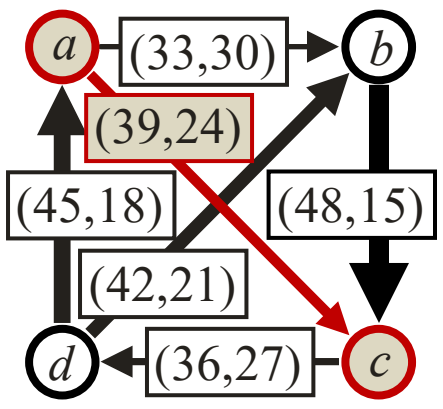<br>*a*, (39,24), *c* | 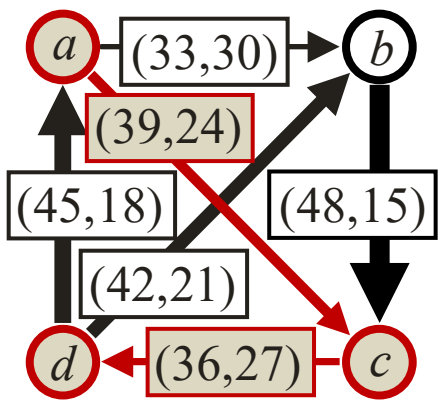<br>*a*, (39,24), *c*, (36,27), *d* | 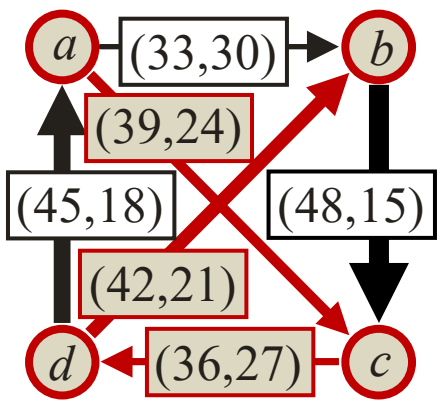 |
| from *b* ... | 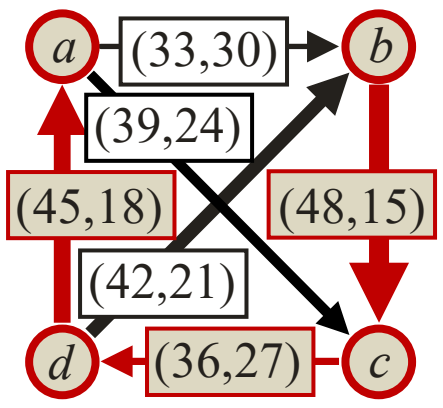<br>*b*, (48,15), *c*, (36,27), *d*, (45,18), *a* | --- | 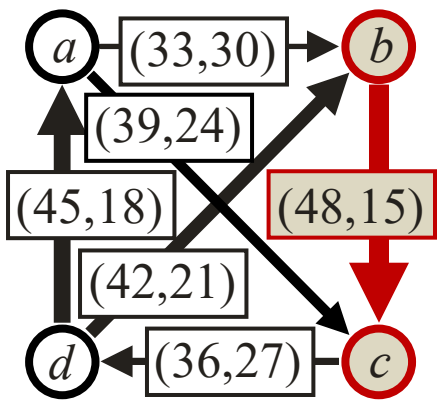<br>*b*, (48,15), *c* | 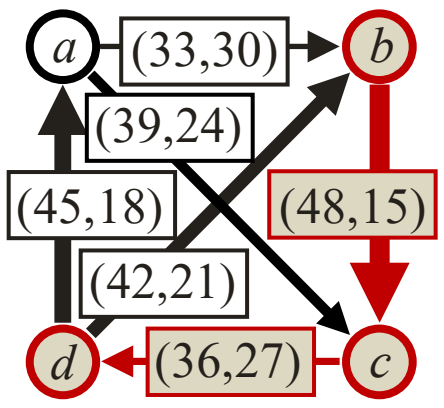<br>*b*, (48,15), *c*, (36,27), *d* | 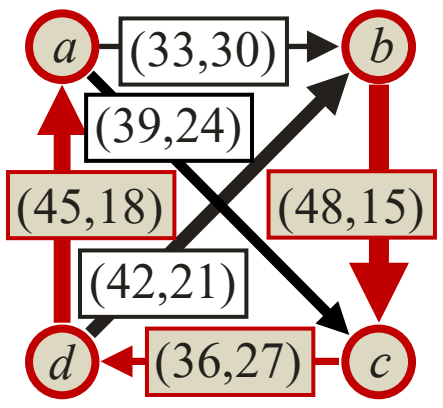 |
| from *c* ... | 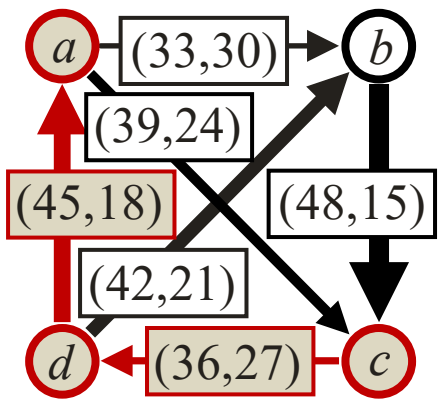<br>*c*, (36,27), *d*, (45,18), *a* | 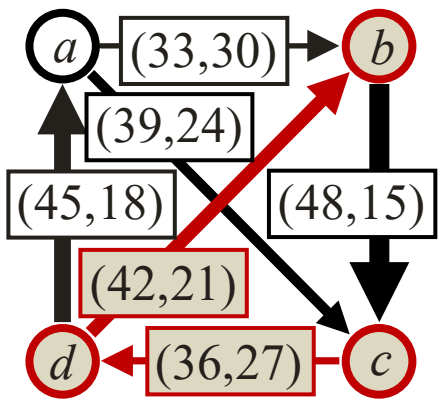<br>*c*, (36,27), *d*, (42,21), *b* | --- | 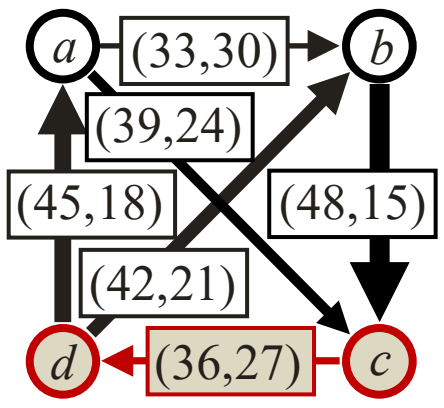<br>*c*, (36,27), *d* | 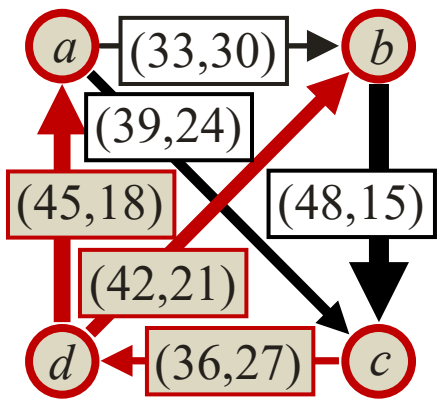 |
| from *d* ... | 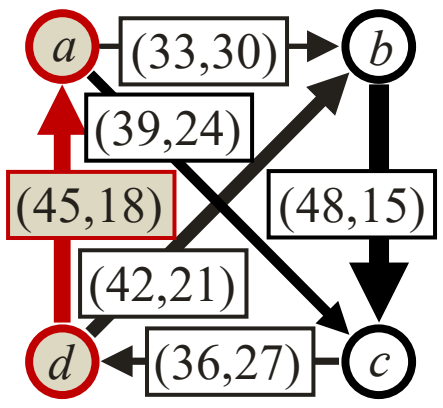<br>*d*, (45,18), *a* | 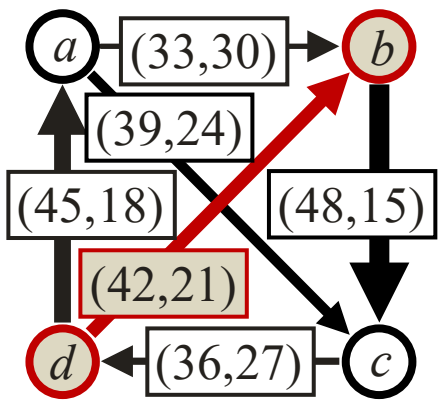<br>*d*, (42,21), *b* | 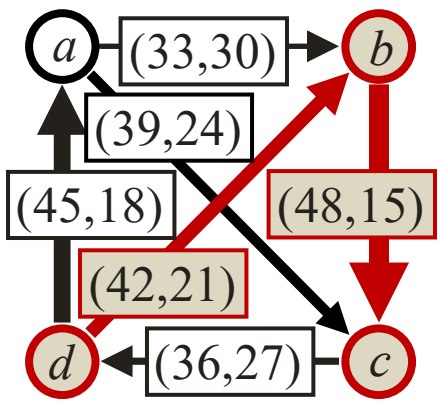<br>*d*, (42,21), *b*, (48,15), *c* | --- | 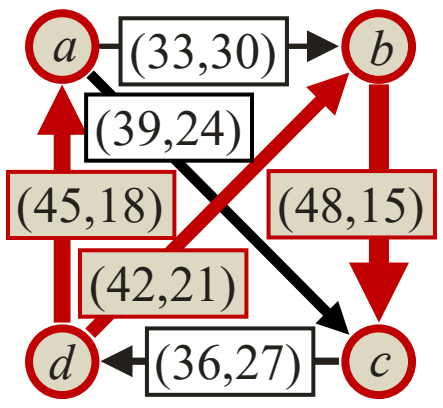 |
| from every other alternative ... | 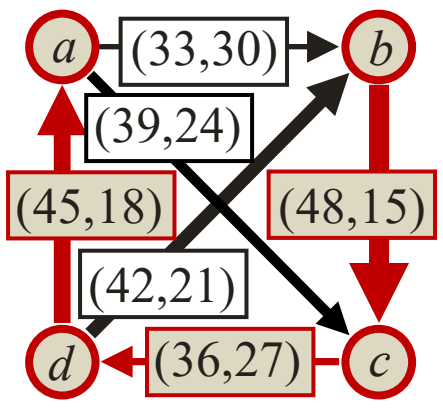 | 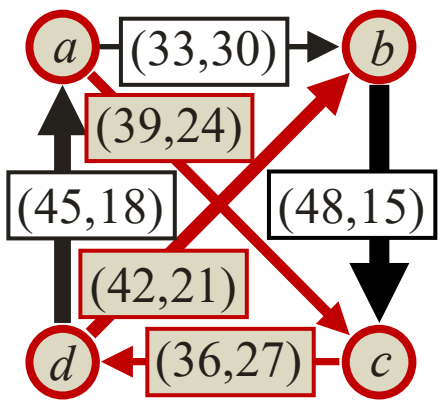 | 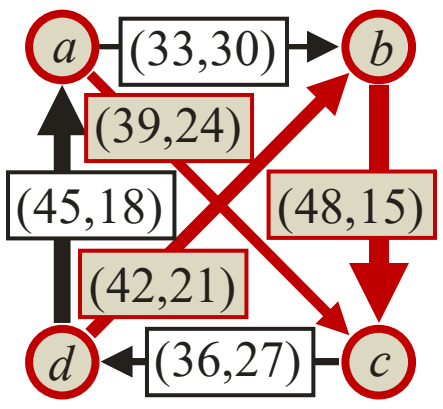 | 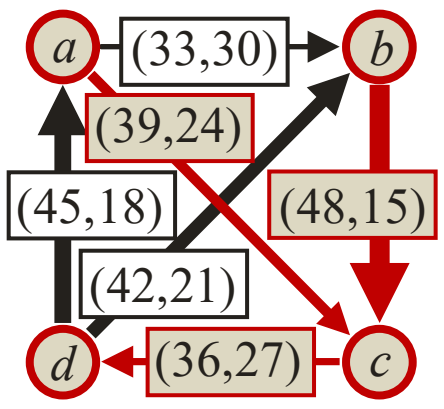 | --- |

Therefore, the strengths of the strongest paths are:

| | $P_D[*,a]$ | $P_D[*,b]$ | $P_D[*,c]$ | $P_D[*,d]$ |
|---|---|---|---|---|
| $P_D[a,*]$ | --- | (36,27) | (39,24) | (36,27) |
| $P_D[b,*]$ | (36,27) | --- | (48,15) | (36,27) |
| $P_D[c,*]$ | (36,27) | (36,27) | --- | (36,27) |
| $P_D[d,*]$ | (45,18) | (42,21) | (42,21) | --- |

We get $\mathcal{O} = \{ac, bc, da, db, dc\}$ and $\mathcal{S} = \{d\}$.

Suppose, the strongest paths are calculated with the Floyd-Warshall algorithm, as defined in section 2.3.1. Then the following table documents the $C \cdot (C–1) \cdot (C–2) = 24$ steps of the Floyd-Warshall algorithm.

We start with

- $P_D[i,j] := (N[i,j],N[j,i])$ for all $i \in A$ and $j \in A \setminus \{i\}$.
- $pred[i,j] := i$ for all $i \in A$ and $j \in A \setminus \{i\}$.

| | *i* | *j* | *k* | $P_D[j,k]$ | $P_D[j,i]$ | $P_D[i,k]$ | *pred*[*j*,*k*] | *pred*[*i*,*k*] | result |
|---|---|---|---|---|---|---|---|---|---|
| 1 | *a* | *b* | *c* | (48,15) | (30,33) | (39,24) | *b* | *a* | |
| 2 | *a* | *b* | *d* | (21,42) | (30,33) | (18,45) | *b* | *a* | |
| 3 | *a* | *c* | *b* | (15,48) | (24,39) | (33,30) | *c* | *a* | $P_D[c,b]$ is updated from (15,48) to (24,39); *pred*[*c*,*b*] is updated from *c* to *a*. |
| 4 | *a* | *c* | *d* | (36,27) | (24,39) | (18,45) | *c* | *a* | |
| 5 | *a* | *d* | *b* | (42,21) | (45,18) | (33,30) | *d* | *a* | |
| 6 | *a* | *d* | *c* | (27,36) | (45,18) | (39,24) | *d* | *a* | $P_D[d,c]$ is updated from (27,36) to (39,24); *pred*[*d*,*c*] is updated from *d* to *a*. |
| 7 | *b* | *a* | *c* | (39,24) | (33,30) | (48,15) | *a* | *b* | |
| 8 | *b* | *a* | *d* | (18,45) | (33,30) | (21,42) | *a* | *b* | $P_D[a,d]$ is updated from (18,45) to (21,42); *pred*[*a*,*d*] is updated from *a* to *b*. |
| 9 | *b* | *c* | *a* | (24,39) | (24,39) | (30,33) | *c* | *b* | |
| 10 | *b* | *c* | *d* | (36,27) | (24,39) | (21,42) | *c* | *b* | |
| 11 | *b* | *d* | *a* | (45,18) | (42,21) | (30,33) | *d* | *b* | |
| 12 | *b* | *d* | *c* | (39,24) | (42,21) | (48,15) | *a* | *b* | $P_D[d,c]$ is updated from (39,24) to (42,21); *pred*[*d*,*c*] is updated from *a* to *b*. |
| 13 | *c* | *a* | *b* | (33,30) | (39,24) | (24,39) | *a* | *a* | |
| 14 | *c* | *a* | *d* | (21,42) | (39,24) | (36,27) | *b* | *c* | $P_D[a,d]$ is updated from (21,42) to (36,27); *pred*[*a*,*d*] is updated from *b* to *c*. |
| 15 | *c* | *b* | *a* | (30,33) | (48,15) | (24,39) | *b* | *c* | |
| 16 | *c* | *b* | *d* | (21,42) | (48,15) | (36,27) | *b* | *c* | $P_D[b,d]$ is updated from (21,42) to (36,27); *pred*[*b*,*d*] is updated from *b* to *c*. |
| 17 | *c* | *d* | *a* | (45,18) | (42,21) | (24,39) | *d* | *c* | |
| 18 | *c* | *d* | *b* | (42,21) | (42,21) | (24,39) | *d* | *a* | |
| 19 | *d* | *a* | *b* | (33,30) | (36,27) | (42,21) | *a* | *d* | $P_D[a,b]$ is updated from (33,30) to (36,27); *pred*[*a*,*b*] is updated from *a* to *d*. |
| 20 | *d* | *a* | *c* | (39,24) | (36,27) | (42,21) | *a* | *b* | |
| 21 | *d* | *b* | *a* | (30,33) | (36,27) | (45,18) | *b* | *d* | $P_D[b,a]$ is updated from (30,33) to (36,27); *pred*[*b*,*a*] is updated from *b* to *d*. |
| 22 | *d* | *b* | *c* | (48,15) | (36,27) | (42,21) | *b* | *b* | |
| 23 | *d* | *c* | *a* | (24,39) | (36,27) | (45,18) | *c* | *d* | $P_D[c,a]$ is updated from (24,39) to (36,27); *pred*[*c*,*a*] is updated from *c* to *d*. |
| 24 | *d* | *c* | *b* | (24,39) | (36,27) | (42,21) | *a* | *d* | $P_D[c,b]$ is updated from (24,39) to (36,27); *pred*[*c*,*b*] is updated from *a* to *d*. |

## 3.6. Example 6

Example 6:

| | | |
|---|---|---|
| 8 | voters | $a \succ_v c \succ_v d \succ_v b$ |
| 2 | voter | $b \succ_v c \succ_v d \succ_v a$ |
| 3 | voters | $b \succ_v d \succ_v a \succ_v c$ |
| 5 | voters | $c \succ_v b \succ_v d \succ_v a$ |
| 1 | voters | $c \succ_v d \succ_v b \succ_v a$ |
| 3 | voters | $d \succ_v a \succ_v b \succ_v c$ |
| 1 | voters | $d \succ_v b \succ_v a \succ_v c$ |

The pairwise matrix *N* looks as follows:

| | *N*[*,*a*] | *N*[*,*b*] | *N*[*,*c*] | *N*[*,*d*] |
|---|---|---|---|---|
| *N*[*a*,*] | --- | 11 | 15 | 8 |
| *N*[*b*,*] | 12 | --- | 9 | 10 |
| *N*[*c*,*] | 8 | 14 | --- | 16 |
| *N*[*d*,*] | 15 | 13 | 7 | --- |

The corresponding digraph looks as follows:

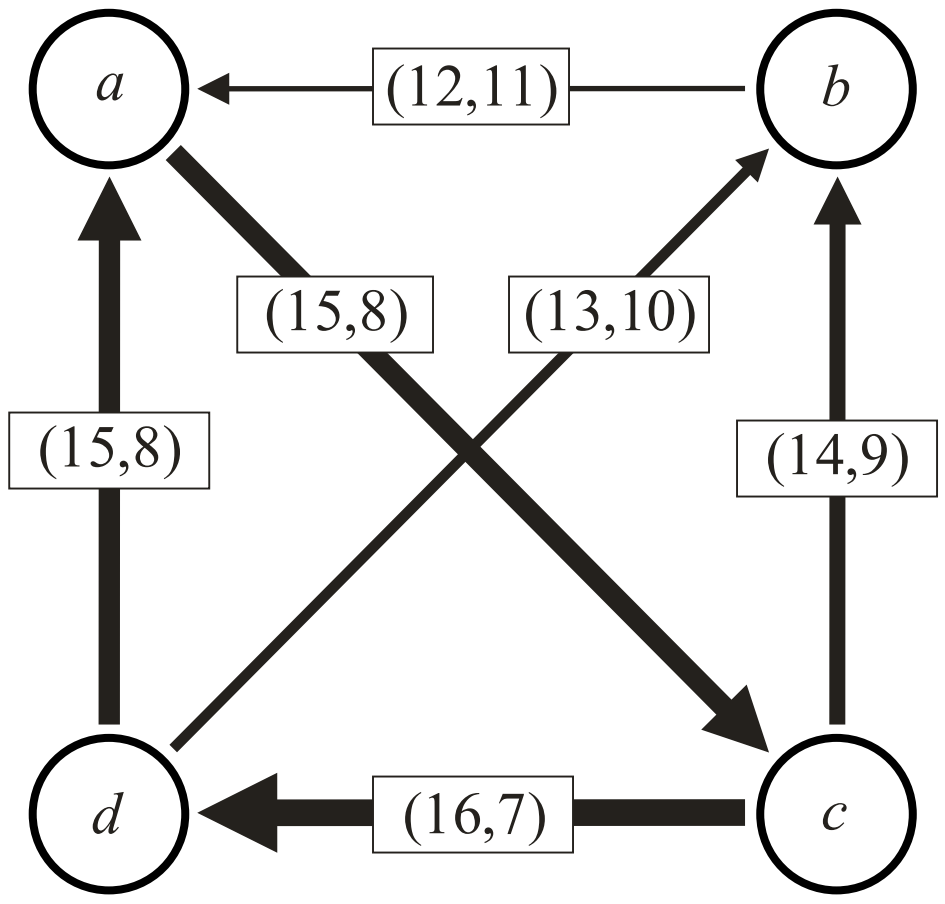

The following table lists the strongest paths, as determined by the Floyd-Warshall algorithm, as defined in section 2.3.1. The critical links of the strongest paths are underlined:

| | ... to *a* | ... to *b* | ... to *c* | ... to *d* | ... to every other alternative |
|---|---|---|---|---|---|
| from *a* ... | --- | 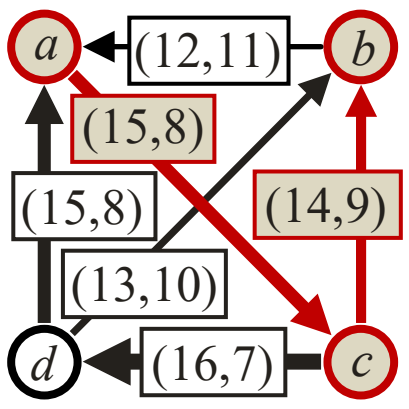 *a*, (15,8), *c*, (14,9), *b* | 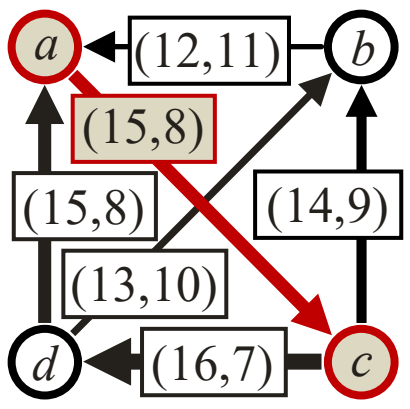 *a*, (15,8), *c* | 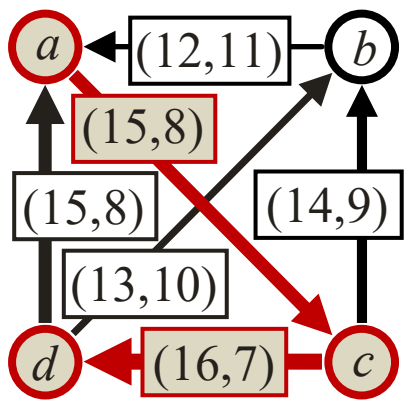 *a*, (15,8), *c*, (16,7), *d* | 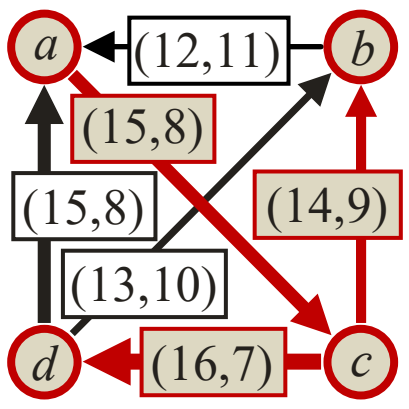 |
| from *b* ... | 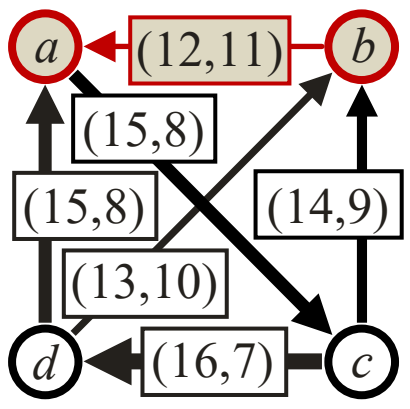 *b*, (12,11), *a* | --- | 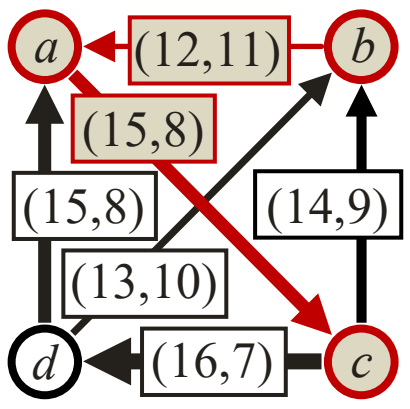 *b*, (12,11), *a*, (15,8), *c* | 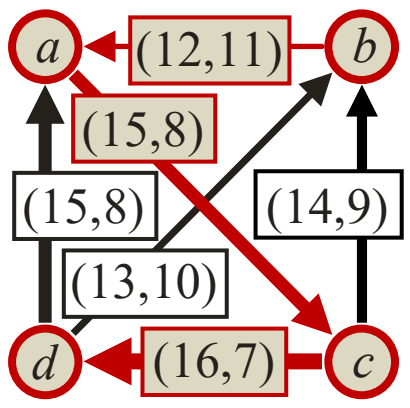 *b*, (12,11), *a*, (15,8), *c*, (16,7), *d* | 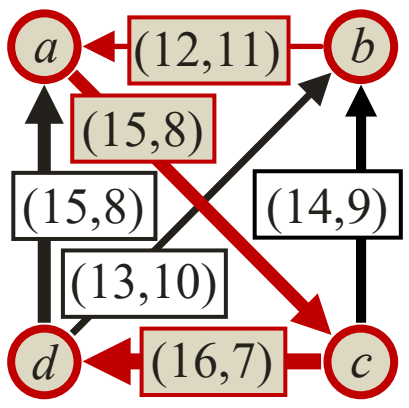 |
| from *c* ... | 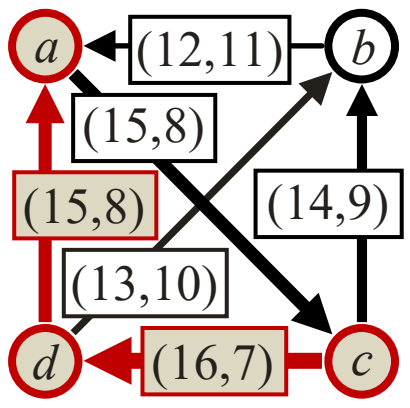 *c*, (16,7), *d*, (15,8), *a* | 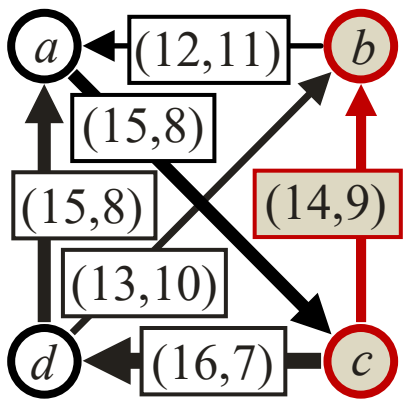 *c*, (14,9), *b* | --- | 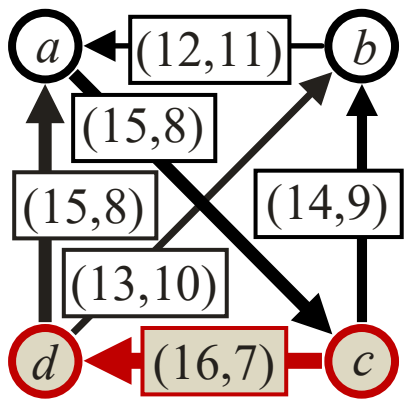 *c*, (16,7), *d* | 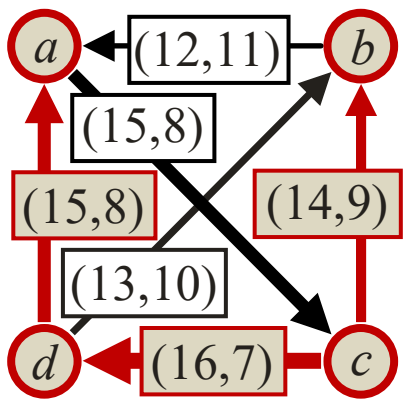 |
| from *d* ... | 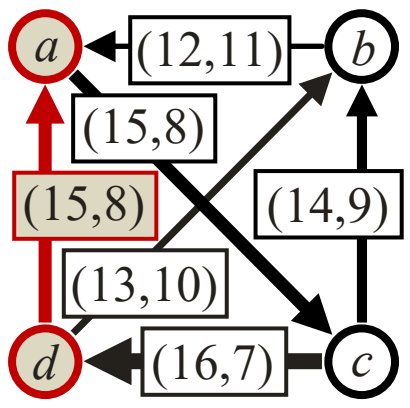 *d*, (15,8), *a* | 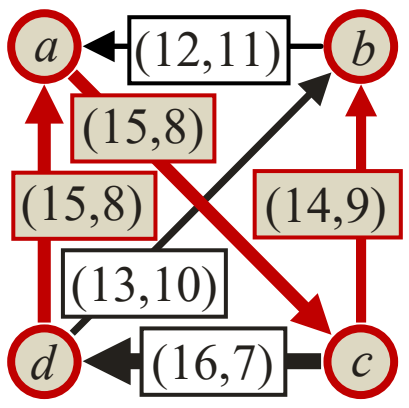 *d*, (15,8), *a*, (15,8), *c*, (14,9), *b* | 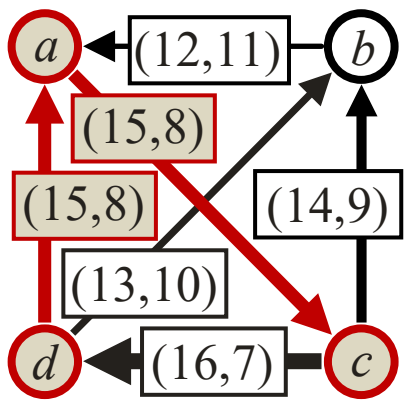 *d*, (15,8), *a*, (15,8), *c* | --- | 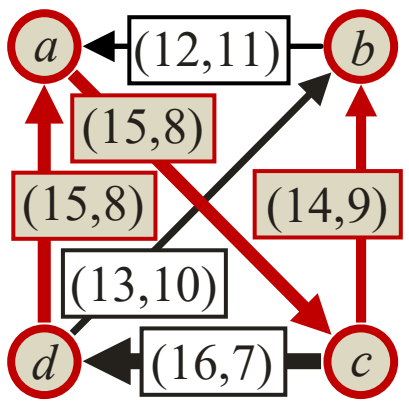 |
| from every other alternative ... | 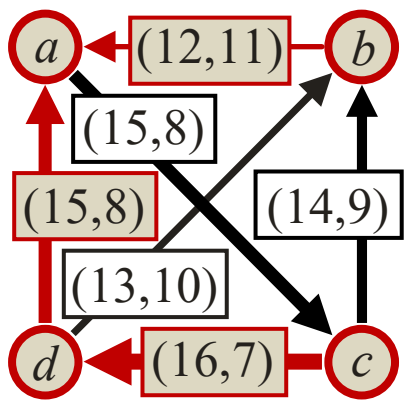 | 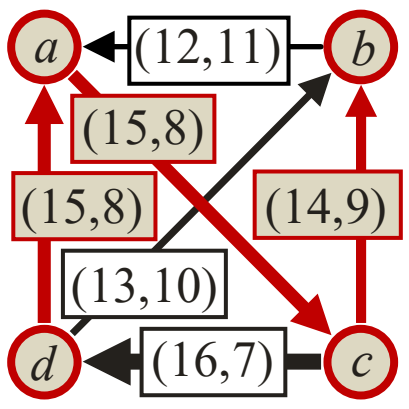 | 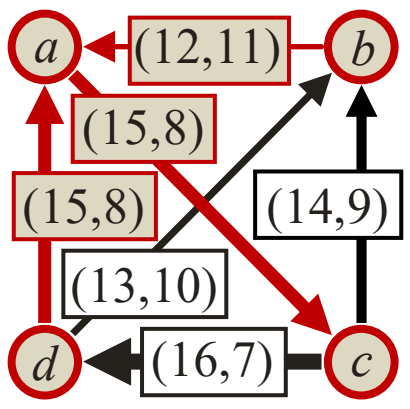 | 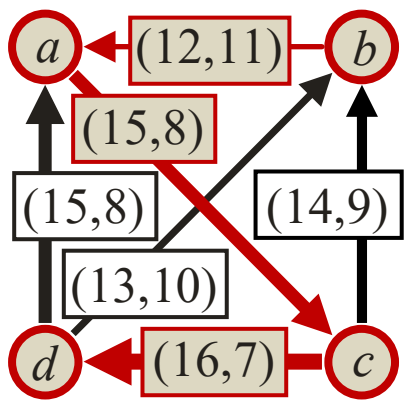 | --- |

Therefore, the strengths of the strongest paths are:

| | $P_D[*,a]$ | $P_D[*,b]$ | $P_D[*,c]$ | $P_D[*,d]$ |
|---|---|---|---|---|
| $P_D[a,*]$ | --- | (14,9) | (15,8) | (15,8) |
| $P_D[b,*]$ | (12,11) | --- | (12,11) | (12,11) |
| $P_D[c,*]$ | (15,8) | (14,9) | --- | (16,7) |
| $P_D[d,*]$ | (15,8) | (14,9) | (15,8) | --- |

We get $\mathcal{O} = \{ab, cb, cd, db\}$ and $\mathcal{S} = \{a, c\}$.

Suppose, the strongest paths are calculated with the Floyd-Warshall algorithm, as defined in section 2.3.1. Then the following table documents the $C \cdot (C–1) \cdot (C–2) = 24$ steps of the Floyd-Warshall algorithm.

We start with

- $P_D[i,j] := (N[i,j],N[j,i])$ for all $i \in A$ and $j \in A \setminus \{i\}$.
- $pred[i,j] := i$ for all $i \in A$ and $j \in A \setminus \{i\}$.

| | $i$ | $j$ | $k$ | $P_D[j,k]$ | $P_D[j,i]$ | $P_D[i,k]$ | $pred[j,k]$ | $pred[i,k]$ | result |
|---|---|---|---|---|---|---|---|---|---|
| 1 | $a$ | $b$ | $c$ | (9,14) | (12,11) | (15,8) | $b$ | $a$ | $P_D[b,c]$ is updated from (9,14) to (12,11); $pred[b,c]$ is updated from $b$ to $a$. |
| 2 | $a$ | $b$ | $d$ | (10,13) | (12,11) | (8,15) | $b$ | $a$ | |
| 3 | $a$ | $c$ | $b$ | (14,9) | (8,15) | (11,12) | $c$ | $a$ | |
| 4 | $a$ | $c$ | $d$ | (16,7) | (8,15) | (8,15) | $c$ | $a$ | |
| 5 | $a$ | $d$ | $b$ | (13,10) | (15,8) | (11,12) | $d$ | $a$ | |
| 6 | $a$ | $d$ | $c$ | (7,16) | (15,8) | (15,8) | $d$ | $a$ | $P_D[d,c]$ is updated from (7,16) to (15,8); $pred[d,c]$ is updated from $d$ to $a$. |
| 7 | $b$ | $a$ | $c$ | (15,8) | (11,12) | (12,11) | $a$ | $a$ | |
| 8 | $b$ | $a$ | $d$ | (8,15) | (11,12) | (10,13) | $a$ | $b$ | $P_D[a,d]$ is updated from (8,15) to (10,13); $pred[a,d]$ is updated from $a$ to $b$. |
| 9 | $b$ | $c$ | $a$ | (8,15) | (14,9) | (12,11) | $c$ | $b$ | $P_D[c,a]$ is updated from (8,15) to (12,11); $pred[c,a]$ is updated from $c$ to $b$. |
| 10 | $b$ | $c$ | $d$ | (16,7) | (14,9) | (10,13) | $c$ | $b$ | |
| 11 | $b$ | $d$ | $a$ | (15,8) | (13,10) | (12,11) | $d$ | $b$ | |
| 12 | $b$ | $d$ | $c$ | (15,8) | (13,10) | (12,11) | $a$ | $a$ | |
| 13 | $c$ | $a$ | $b$ | (11,12) | (15,8) | (14,9) | $a$ | $c$ | $P_D[a,b]$ is updated from (11,12) to (14,9); $pred[a,b]$ is updated from $a$ to $c$. |
| 14 | $c$ | $a$ | $d$ | (10,13) | (15,8) | (16,7) | $b$ | $c$ | $P_D[a,d]$ is updated from (10,13) to (15,8); $pred[a,d]$ is updated from $b$ to $c$. |
| 15 | $c$ | $b$ | $a$ | (12,11) | (12,11) | (12,11) | $b$ | $b$ | |
| 16 | $c$ | $b$ | $d$ | (10,13) | (12,11) | (16,7) | $b$ | $c$ | $P_D[b,d]$ is updated from (10,13) to (12,11); $pred[b,d]$ is updated from $b$ to $c$. |
| 17 | $c$ | $d$ | $a$ | (15,8) | (15,8) | (12,11) | $d$ | $b$ | |
| 18 | $c$ | $d$ | $b$ | (13,10) | (15,8) | (14,9) | $d$ | $c$ | $P_D[d,b]$ is updated from (13,10) to (14,9); $pred[d,b]$ is updated from $d$ to $c$. |
| 19 | $d$ | $a$ | $b$ | (14,9) | (15,8) | (14,9) | $c$ | $c$ | |
| 20 | $d$ | $a$ | $c$ | (15,8) | (15,8) | (15,8) | $a$ | $a$ | |
| 21 | $d$ | $b$ | $a$ | (12,11) | (12,11) | (15,8) | $b$ | $d$ | |
| 22 | $d$ | $b$ | $c$ | (12,11) | (12,11) | (15,8) | $a$ | $a$ | |
| 23 | $d$ | $c$ | $a$ | (12,11) | (16,7) | (15,8) | $b$ | $d$ | $P_D[c,a]$ is updated from (12,11) to (15,8); $pred[c,a]$ is updated from $b$ to $d$. |
| 24 | $d$ | $c$ | $b$ | (14,9) | (16,7) | (14,9) | $c$ | $c$ | |

## 3.7. Example 7

The basic idea for the following example has been proposed by Cretney (1998).

### 3.7.1. Situation #1

Example 7 (old):

| | |
|---|---|
| 3 voters | $a \succ_v d \succ_v e \succ_v b \succ_v c \succ_v f$ |
| 3 voters | $b \succ_v f \succ_v e \succ_v c \succ_v d \succ_v a$ |
| 4 voters | $c \succ_v a \succ_v b \succ_v f \succ_v d \succ_v e$ |
| 1 voter | $d \succ_v b \succ_v c \succ_v e \succ_v f \succ_v a$ |
| 4 voters | $d \succ_v e \succ_v f \succ_v a \succ_v b \succ_v c$ |
| 2 voters | $e \succ_v c \succ_v b \succ_v d \succ_v f \succ_v a$ |
| 2 voters | $f \succ_v a \succ_v c \succ_v d \succ_v b \succ_v e$ |

The pairwise matrix $N^{old}$ looks as follows:

| | $N^{old}[*,a]$ | $N^{old}[*,b]$ | $N^{old}[*,c]$ | $N^{old}[*,d]$ | $N^{old}[*,e]$ | $N^{old}[*,f]$ |
|---|---|---|---|---|---|---|
| $N^{old}[a,*]$ | --- | 13 | 9 | 9 | 9 | 7 |
| $N^{old}[b,*]$ | 6 | --- | 11 | 9 | 10 | 13 |
| $N^{old}[c,*]$ | 10 | 8 | --- | 11 | 7 | 10 |
| $N^{old}[d,*]$ | 10 | 10 | 8 | --- | 14 | 10 |
| $N^{old}[e,*]$ | 10 | 9 | 12 | 5 | --- | 10 |
| $N^{old}[f,*]$ | 12 | 6 | 9 | 9 | 9 | --- |

The corresponding digraph looks as follows:

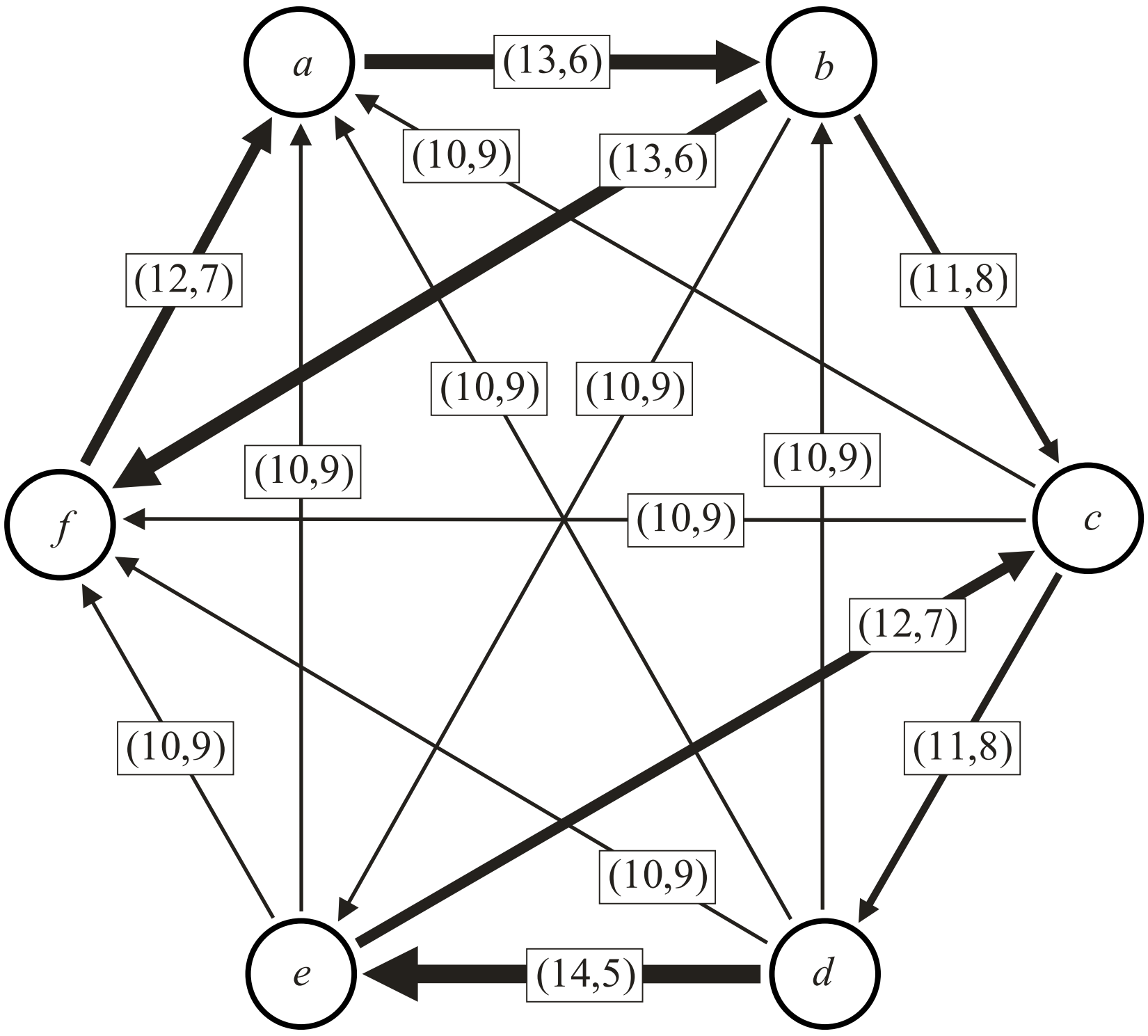

The following table lists the strongest paths, as determined by the Floyd-Warshall algorithm, as defined in section 2.3.1. The critical links of the strongest paths are underlined:

| | ... to *a* | ... to *b* | ... to *c* | ... to *d* | ... to *e* | ... to *f* |
|---|---|---|---|---|---|---|
| from *a* ... | --- | *a*, (13,6), *b* | *a*, (13,6), *b*, (11,8), *c* | *a*, (13,6), *b*, (11,8), *c*, (11,8), *d* | *a*, (13,6), *b*, (11,8), *c*, (11,8), *d*, (14,5), *e* | *a*, (13,6), *b*, (13,6), *f* |
| from *b* ... | *b*, (13,6), *f*, (12,7), *a* | --- | *b*, (11,8), *c* | *b*, (11,8), *c*, (11,8), *d* | *b*, (11,8), *c*, (11,8), *d*, (14,5), *e* | *b*, (13,6), *f* |
| from *c* ... | *c*, (10,9), *a* | *c*, (10,9), *a*, (13,6), *b* | --- | *c*, (11,8), *d* | *c*, (11,8), *d*, (14,5), *e* | *c*, (10,9), *f* |
| from *d* ... | *d*, (10,9), *a* | *d*, (10,9), *b* | *d*, (14,5), *e*, (12,7), *c* | --- | *d*, (14,5), *e* | *d*, (10,9), *f* |
| from *e* ... | *e*, (10,9), *a* | *e*, (10,9), *a*, (13,6), *b* | *e*, (12,7), *c* | *e*, (12,7), *c*, (11,8), *d* | --- | *e*, (10,9), *f* |
| from *f* ... | *f*, (12,7), *a* | *f*, (12,7), *a*, (13,6), *b* | *f*, (12,7), *a*, (13,6), *b*, (11,8), *c* | *f*, (12,7), *a*, (13,6), *b*, (11,8), *c*, (11,8), *d* | *f*, (12,7), *a*, (13,6), *b*, (11,8), *c*, (11,8), *d*, (14,5), *e* | --- |

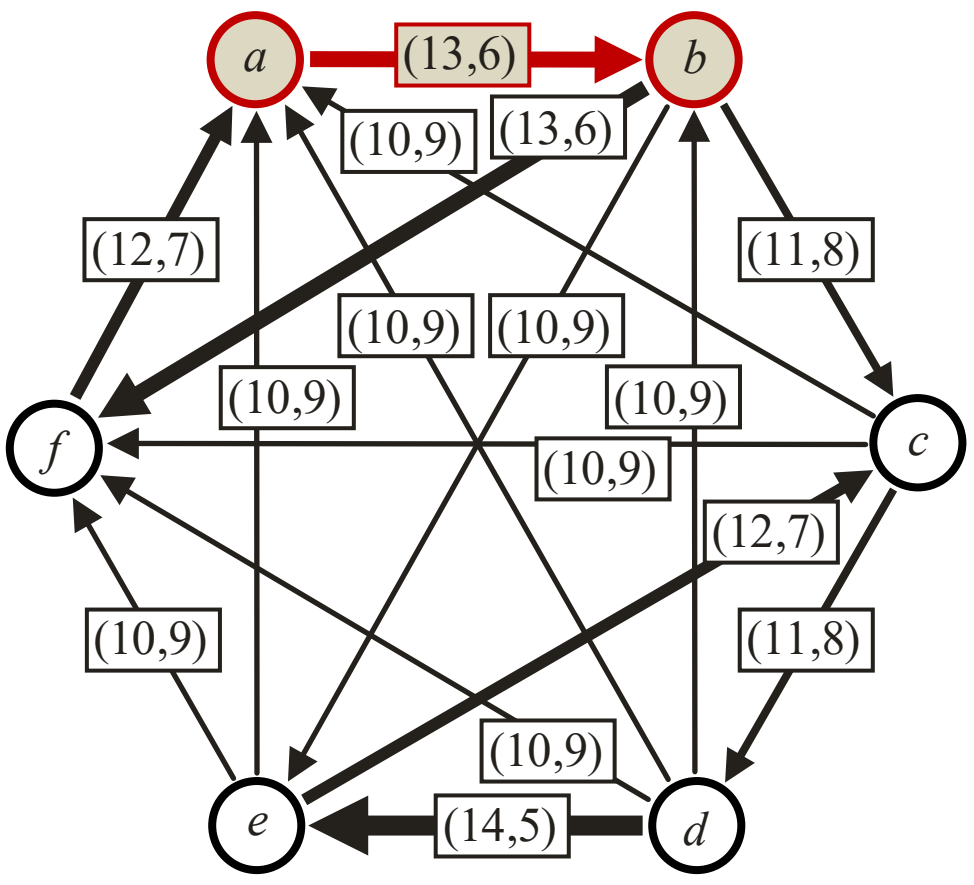

The strongest path from *a* to *b* is:
*a*, (13,6), *b*

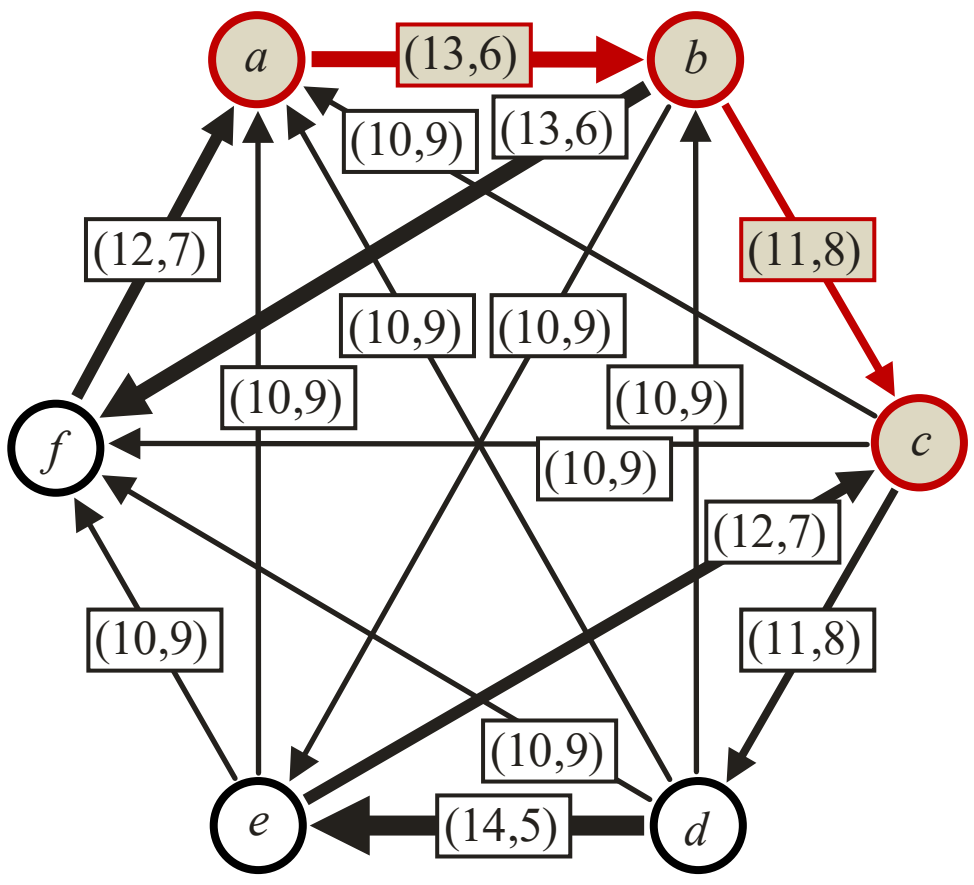

The strongest path from *a* to *c* is:
*a*, (13,6), *b*, (11,8), *c*

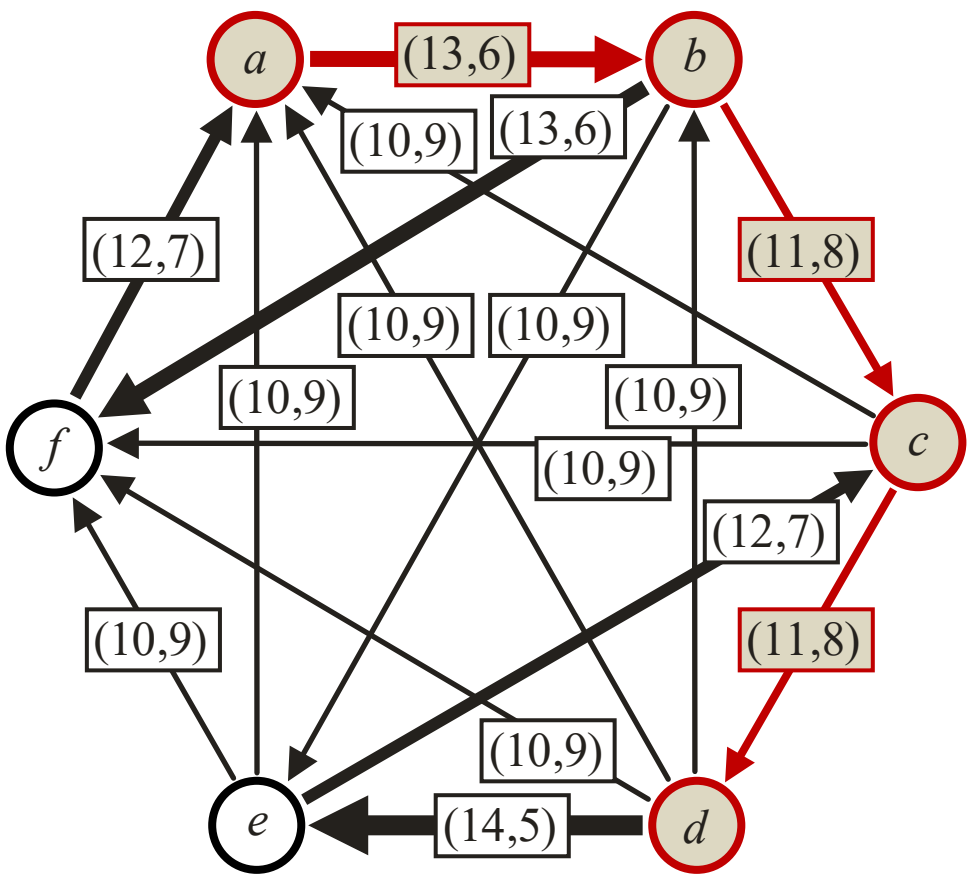

The strongest path from *a* to *d* is:
*a*, (13,6), *b*, (11,8), *c*, (11,8), *d*

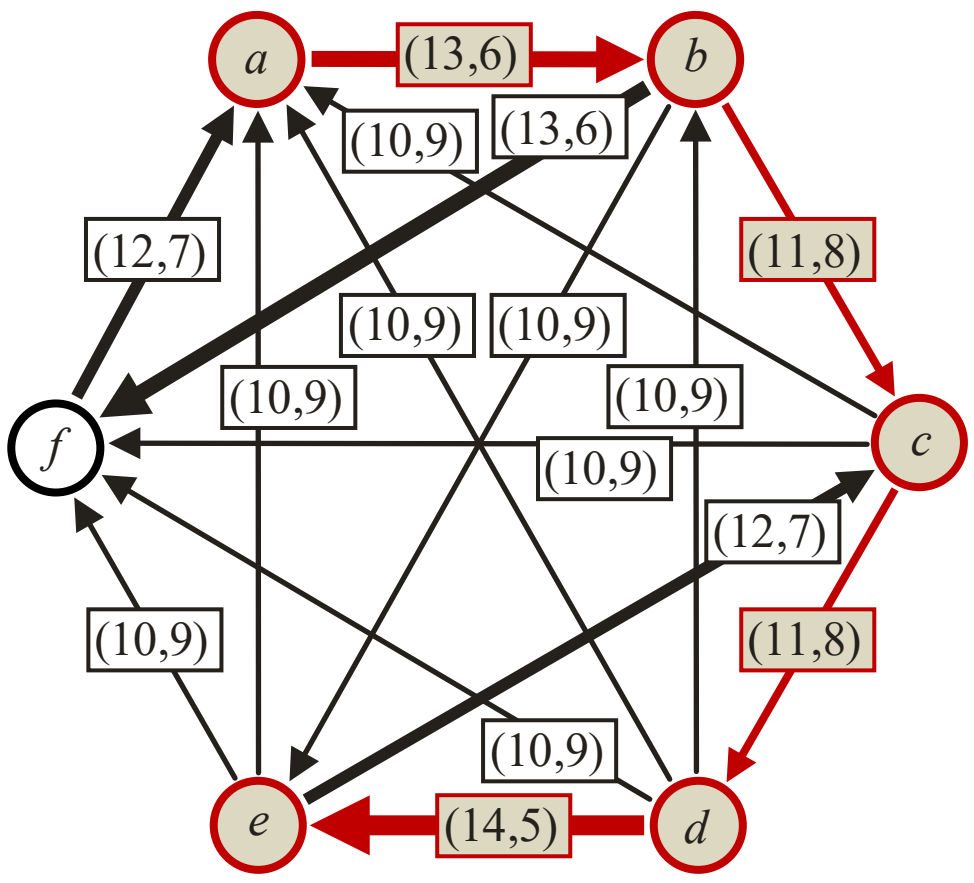

The strongest path from *a* to *e* is:
*a*, (13,6), *b*, (11,8), *c*, (11,8), *d*, (14,5), *e*

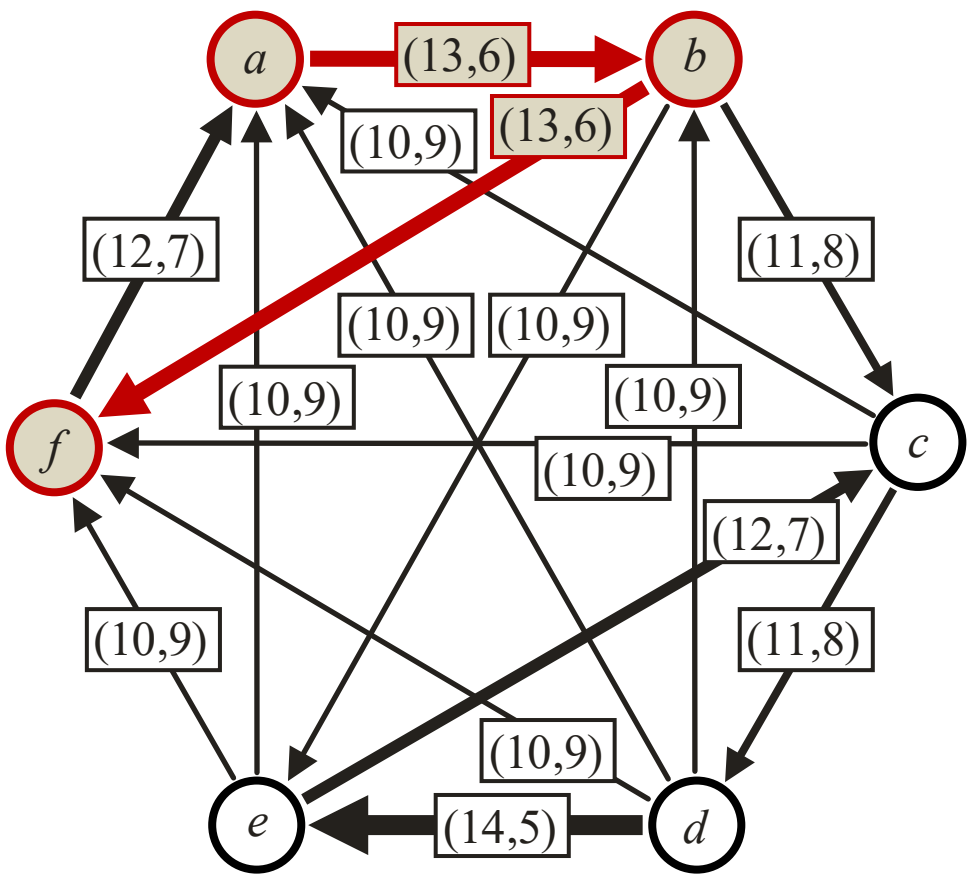

The strongest path from *a* to *f* is:
*a*, (13,6), *b*, (13,6), *f*

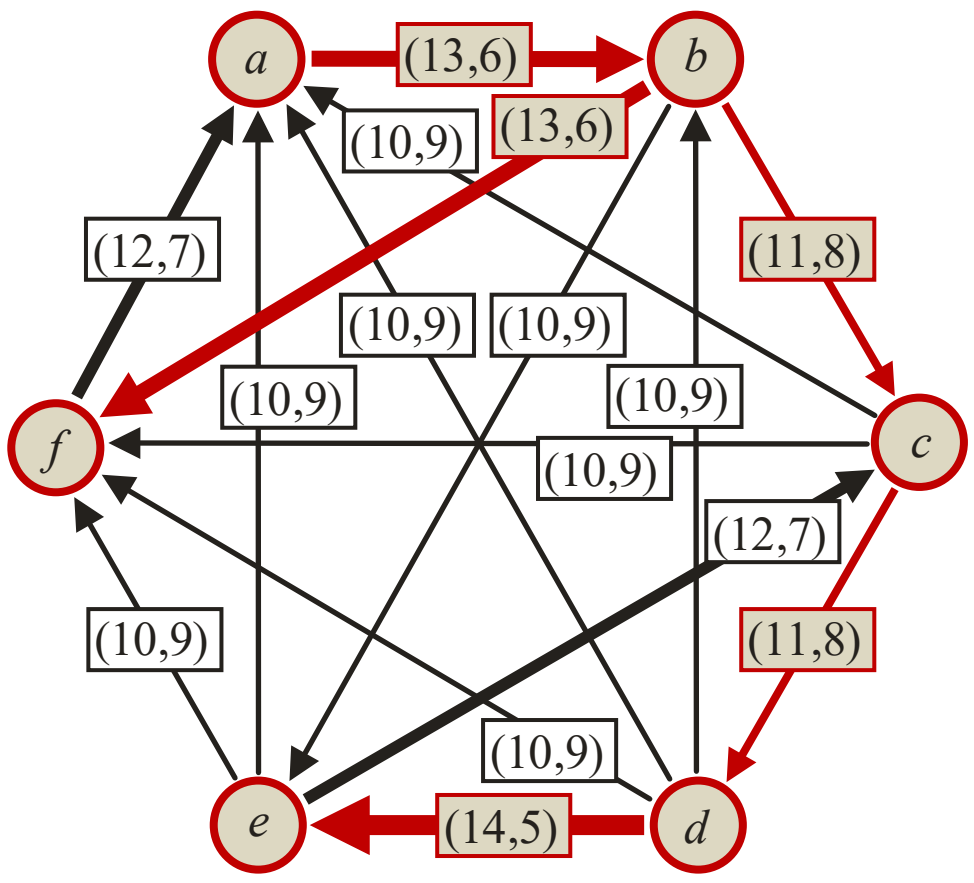

These are the strongest paths
from *a* to every other alternative.

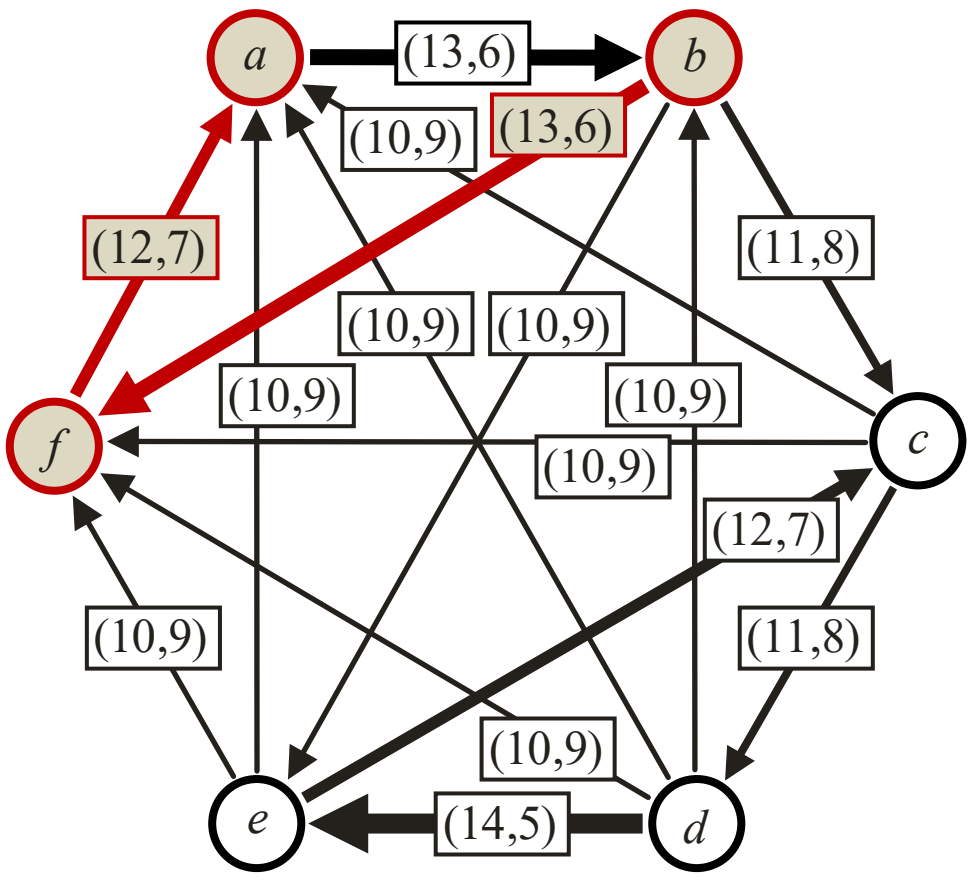

The strongest path from *b* to *a* is:
*b*, (13,6), *f*, (12,7), *a*

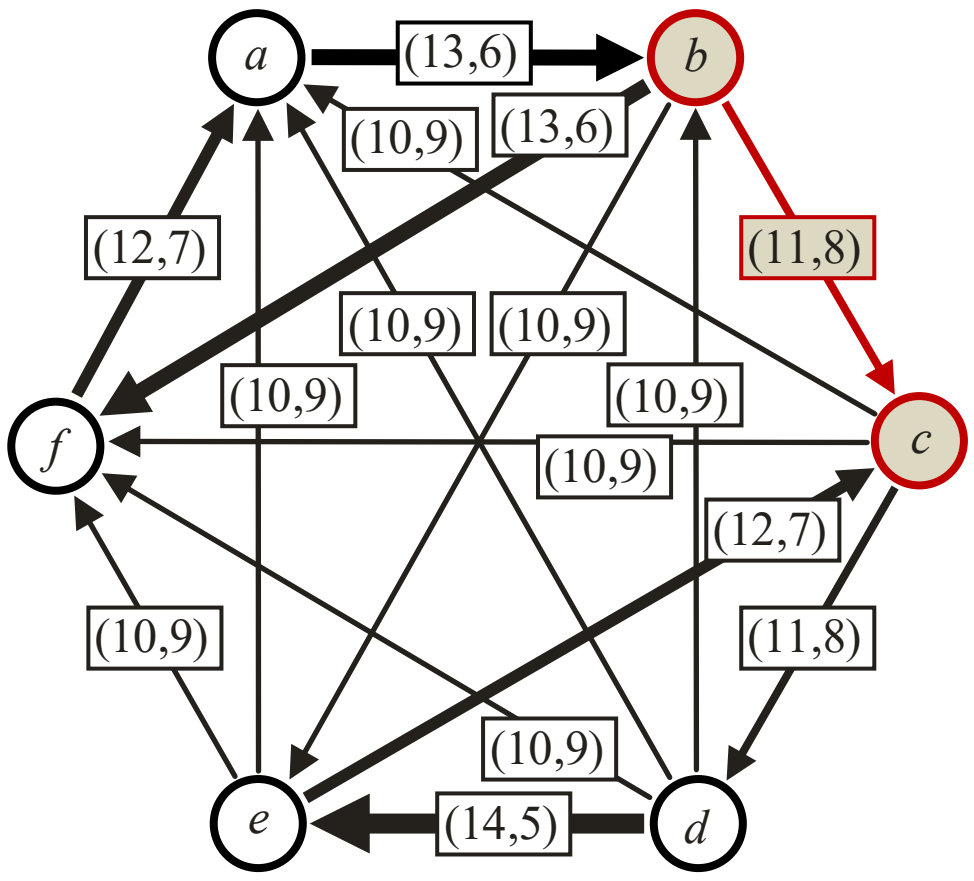

The strongest path from *b* to *c* is:
*b*, (11,8), *c*

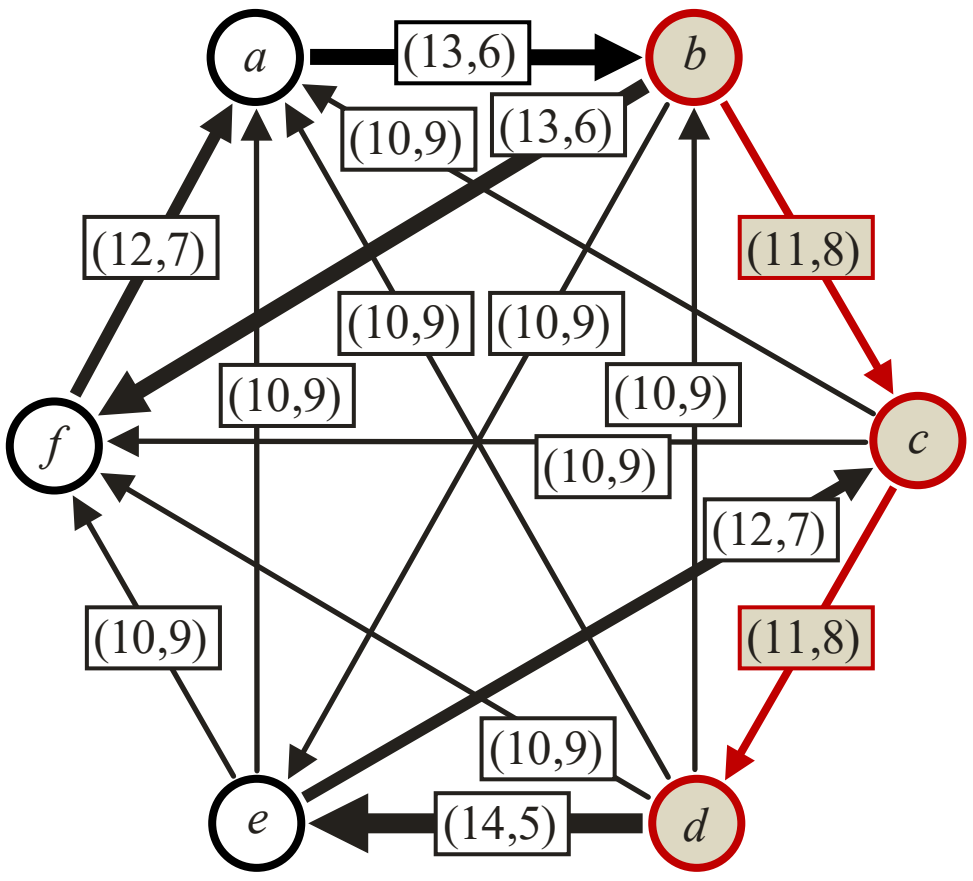

The strongest path from *b* to *d* is:
*b*, (11,8), *c*, (11,8), *d*

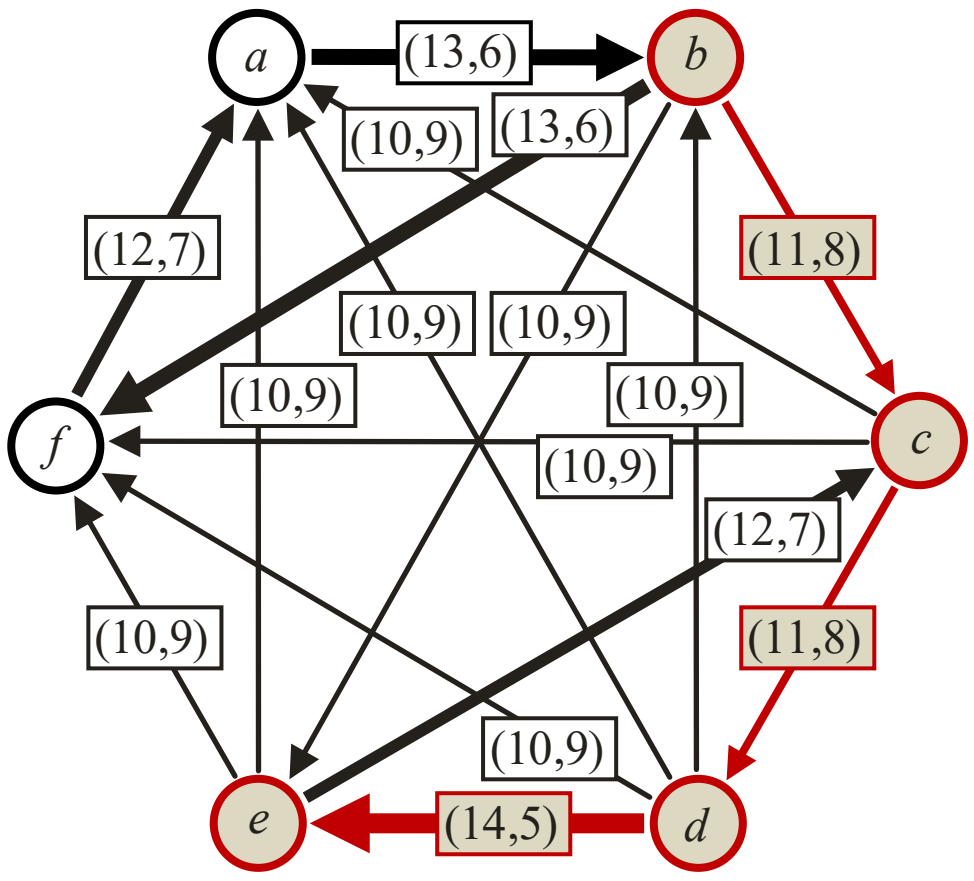

The strongest path from *b* to *e* is:
*b*, (11,8), *c*, (11,8), *d*, (14,5), *e*

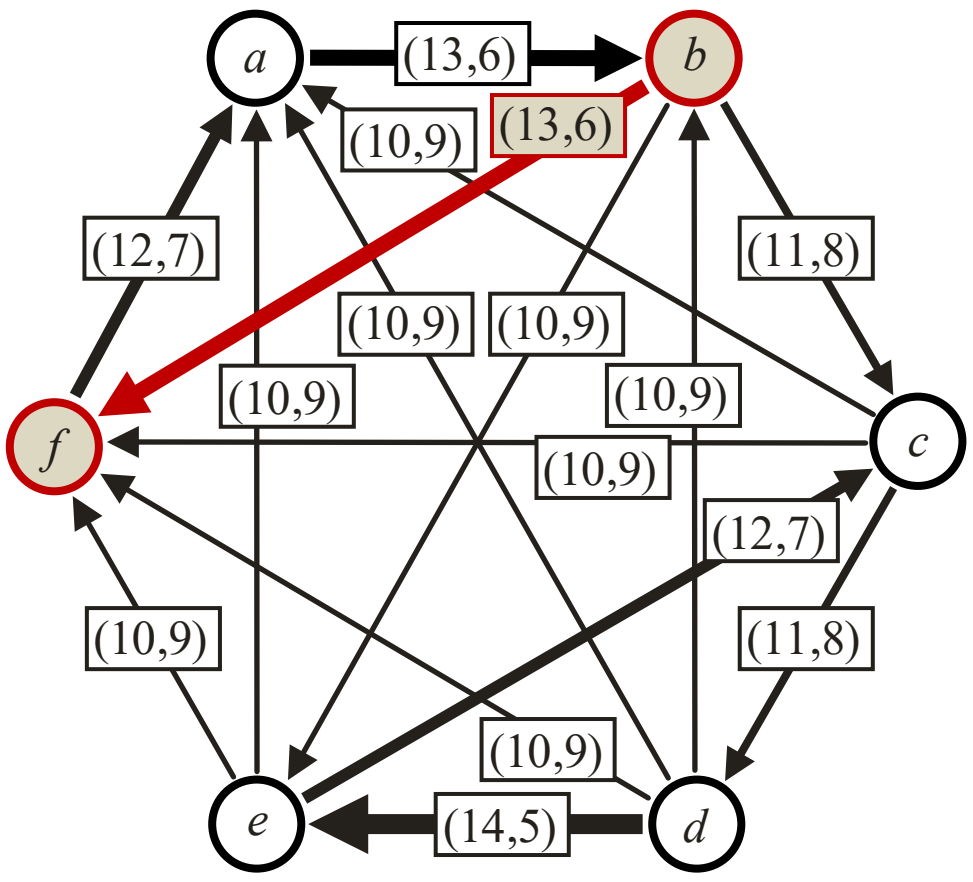

The strongest path from *b* to *f* is:
*b*, (13,6), *f*

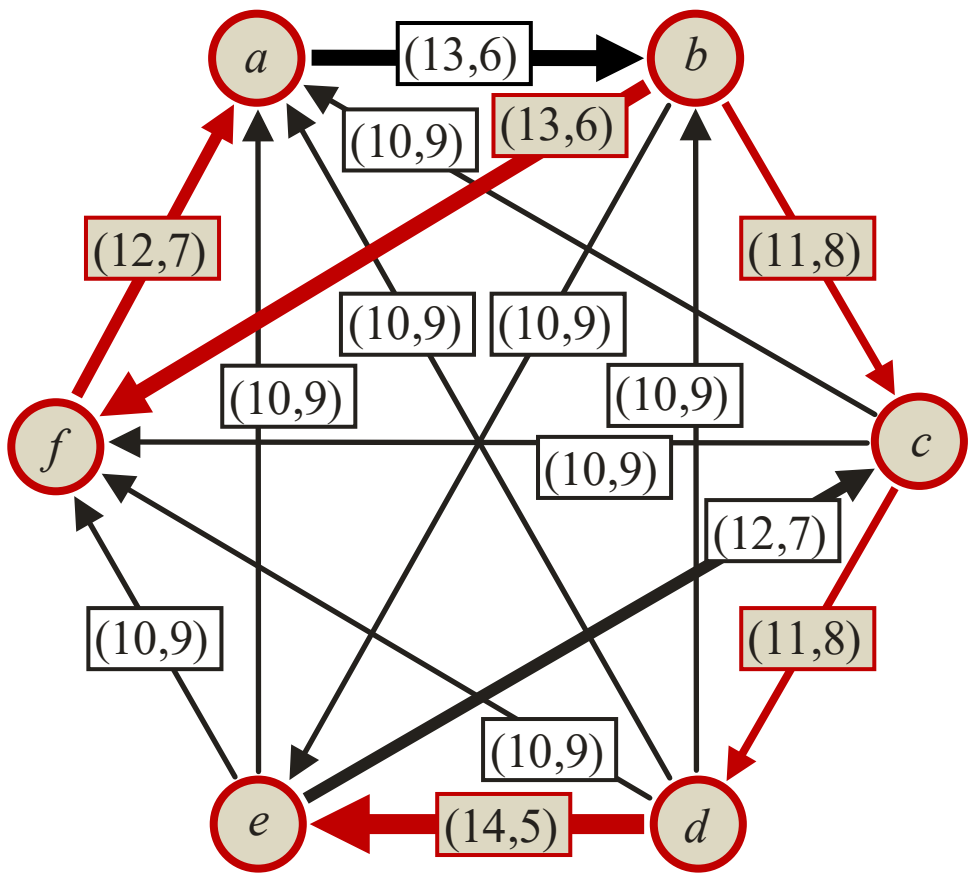

These are the strongest paths
from *b* to every other alternative.

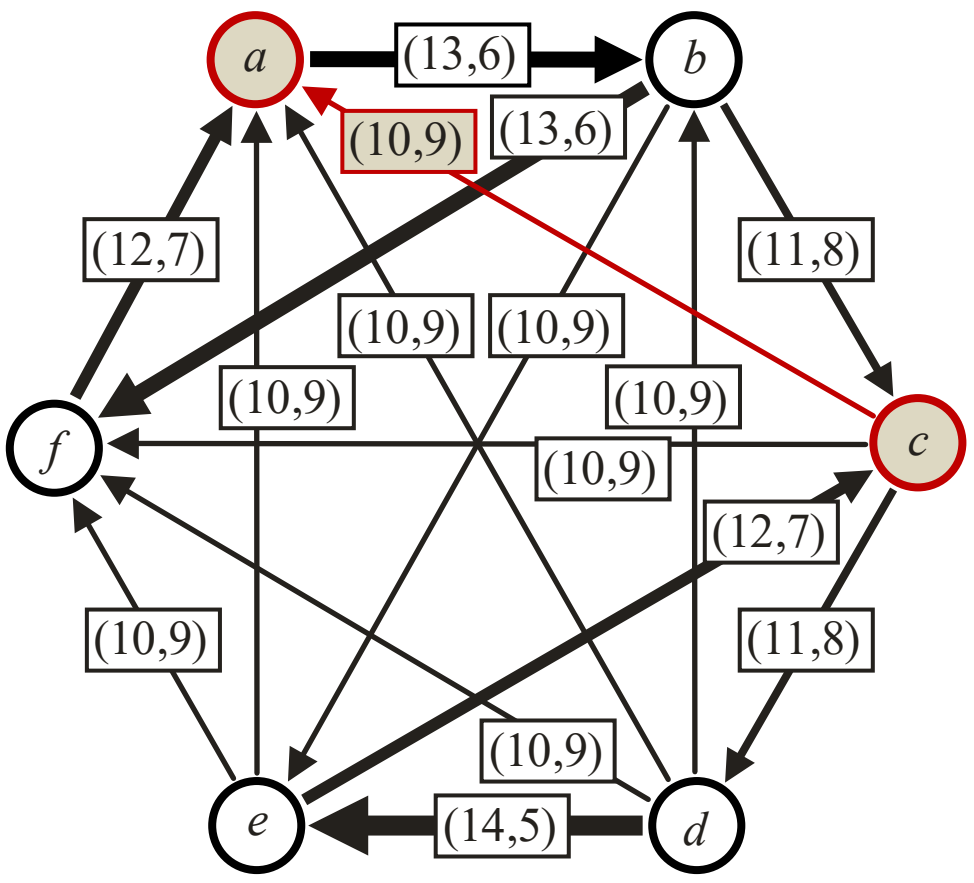

The strongest path from *c* to *a* is:
*c*, (10,9), *a*

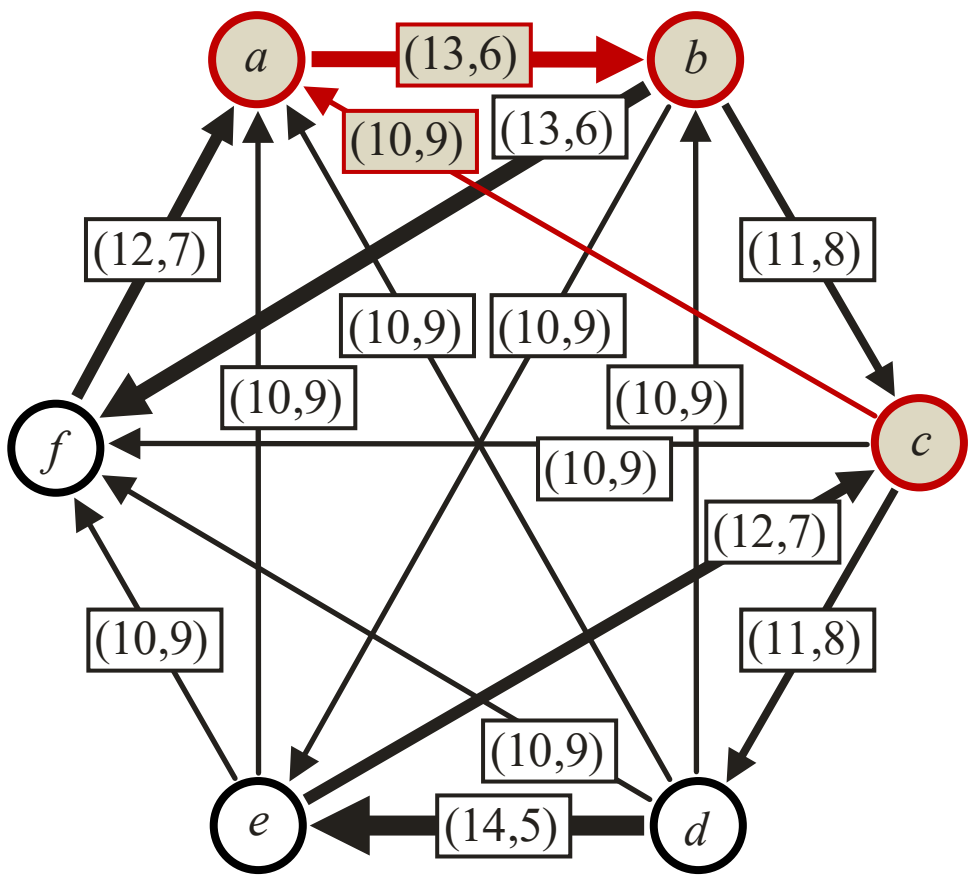

The strongest path from *c* to *b* is:
*c*, (10,9), *a*, (13,6), *b*

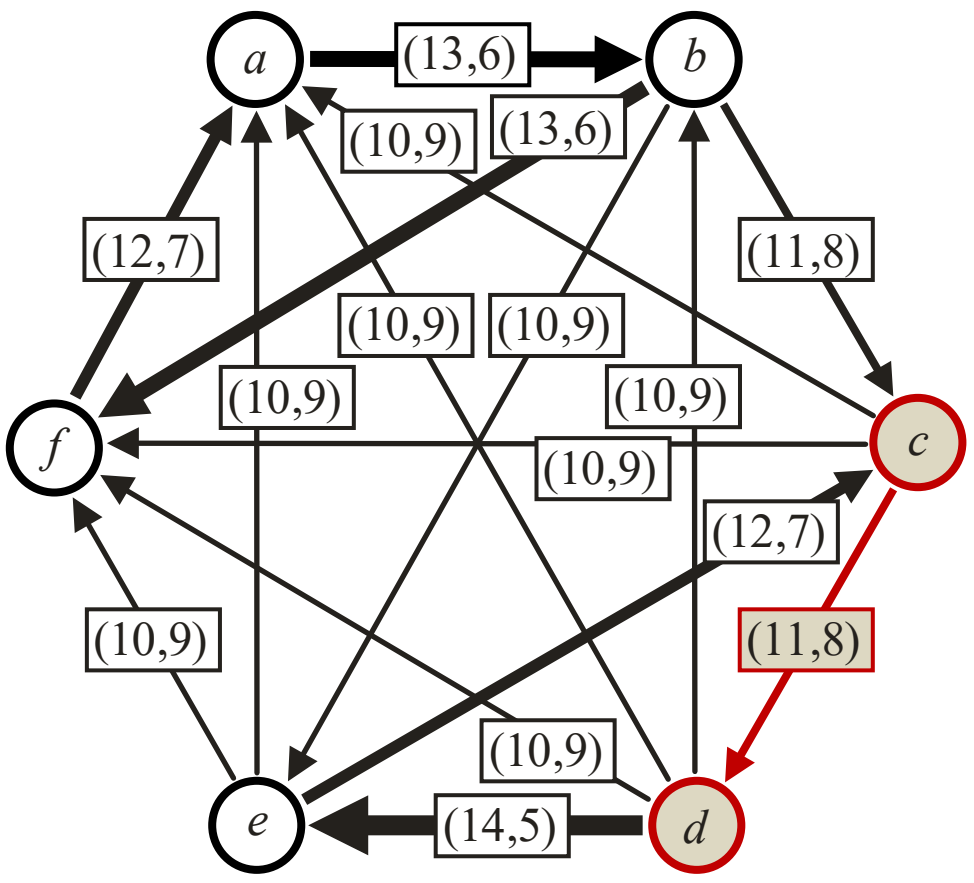

The strongest path from *c* to *d* is:
*c*, (11,8), *d*

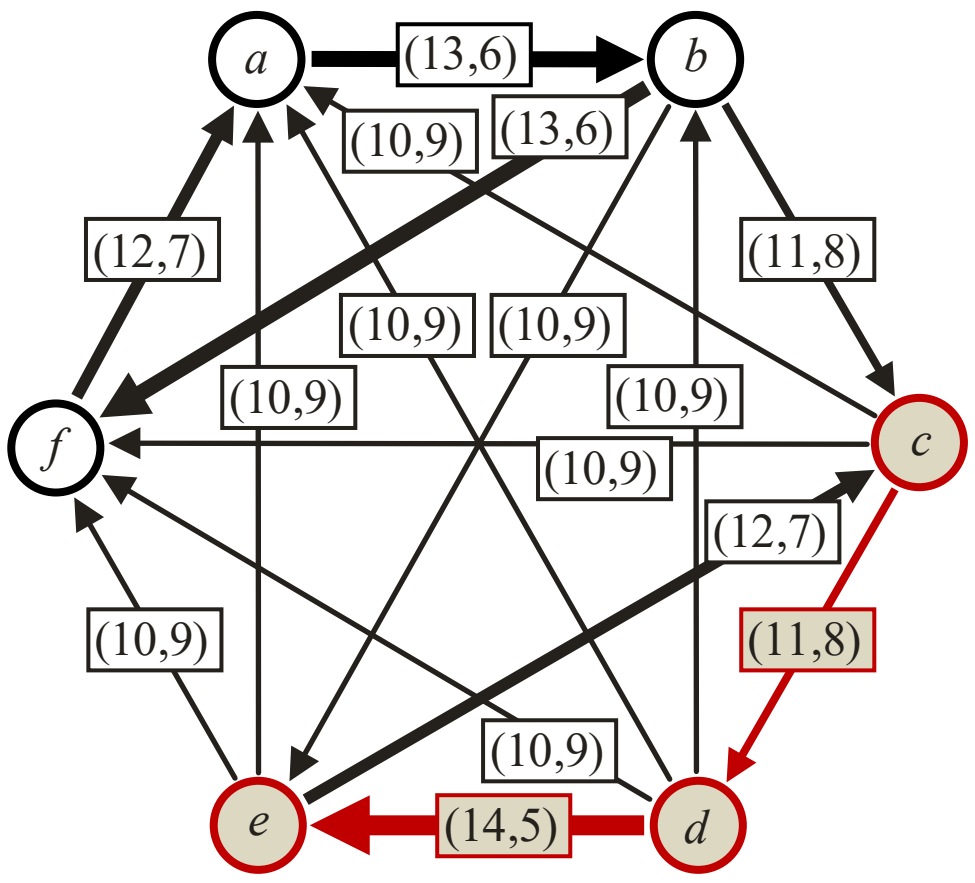

The strongest path from *c* to *e* is:
*c*, (11,8), *d*, (14,5), *e*

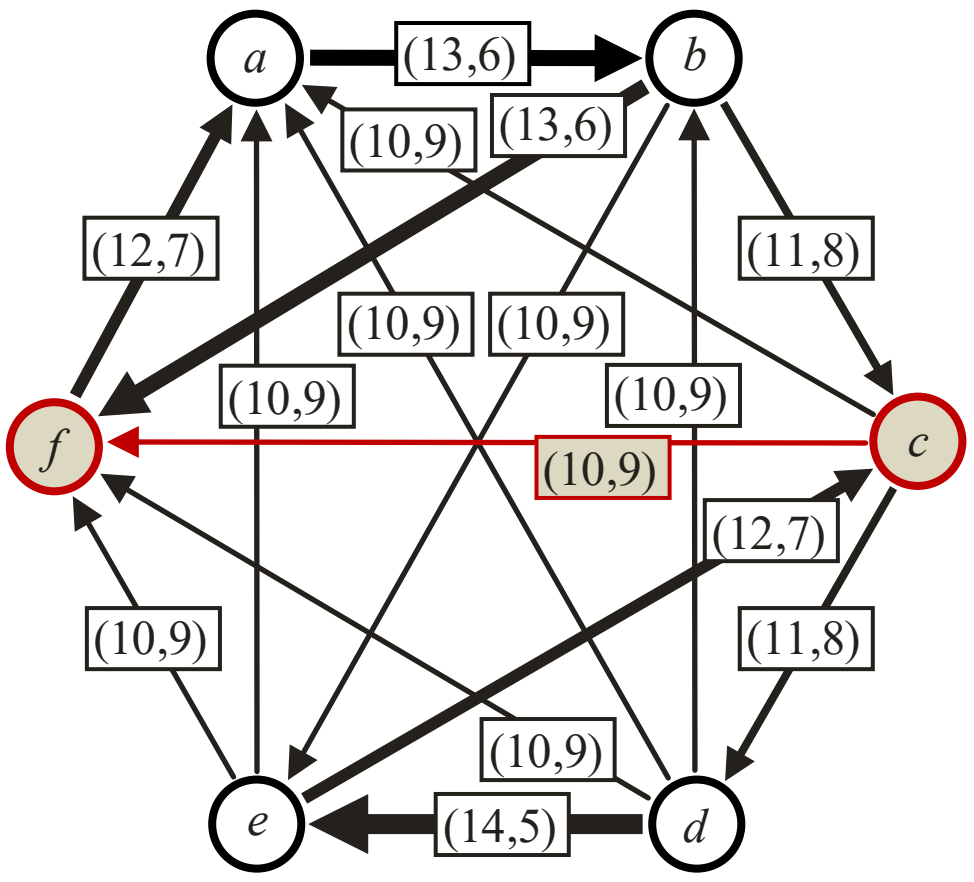

The strongest path from *c* to *f* is:
*c*, (10,9), *f*

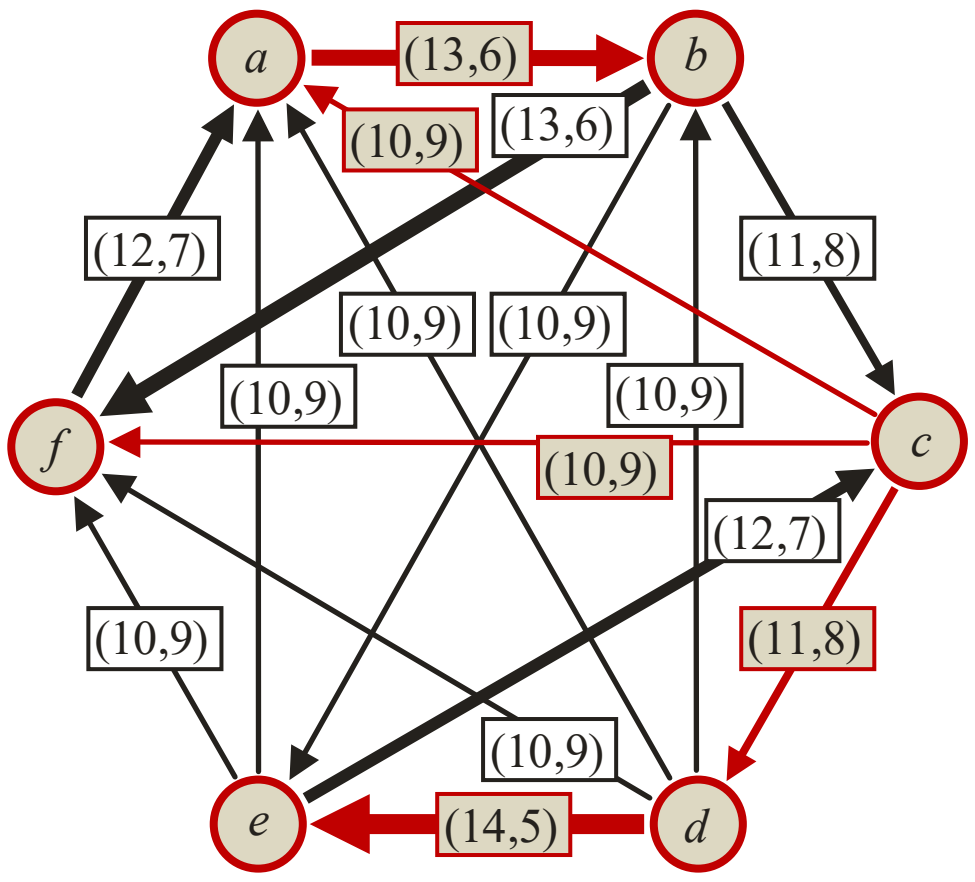

These are the strongest paths
from *c* to every other alternative.

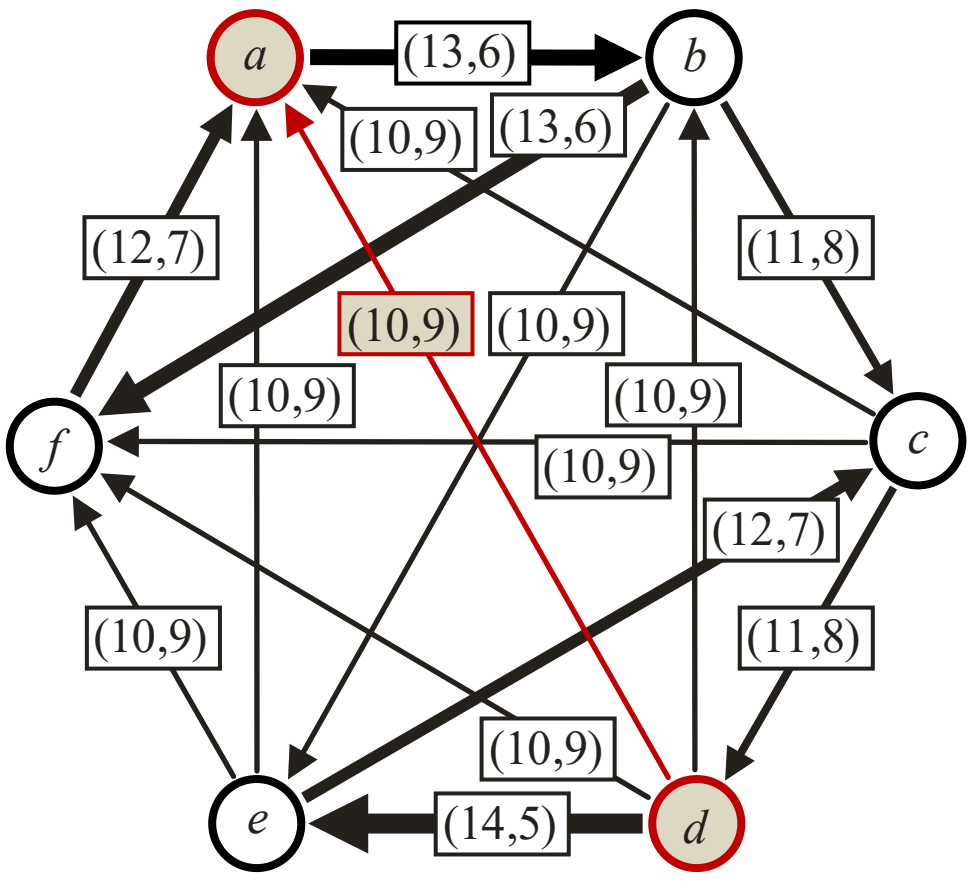

The strongest path from *d* to *a* is:
*d*, (10,9), *a*

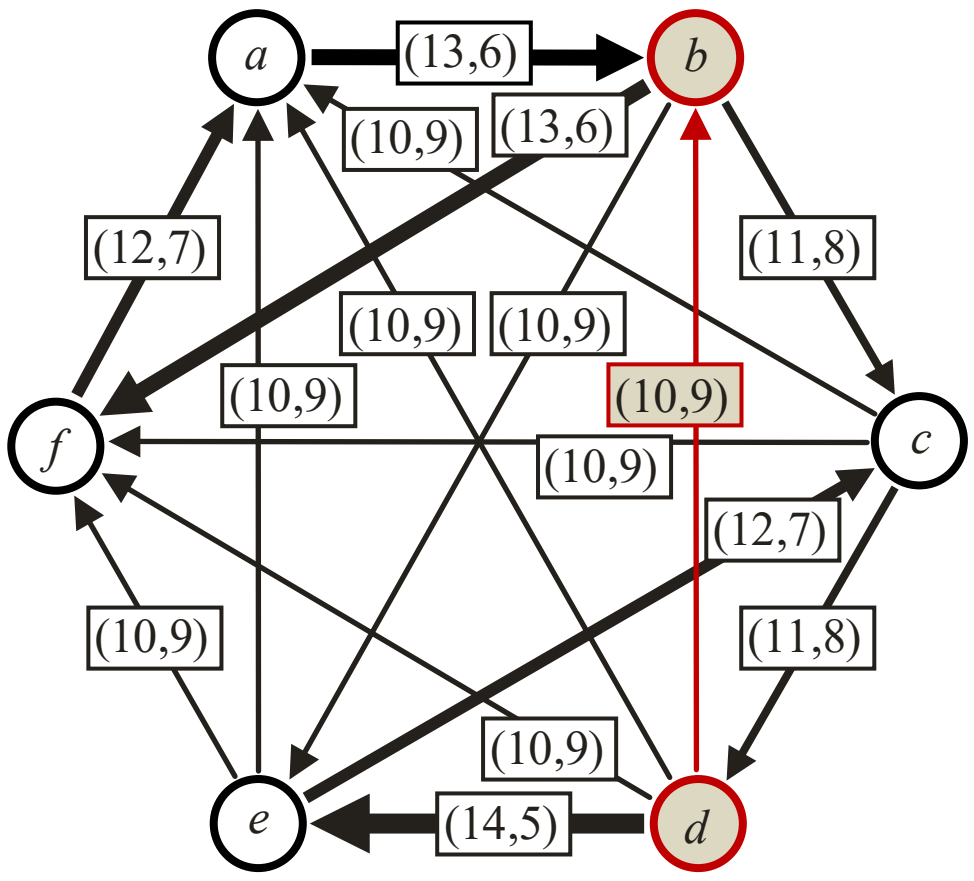

The strongest path from *d* to *b* is:
*d*, (10,9), *b*

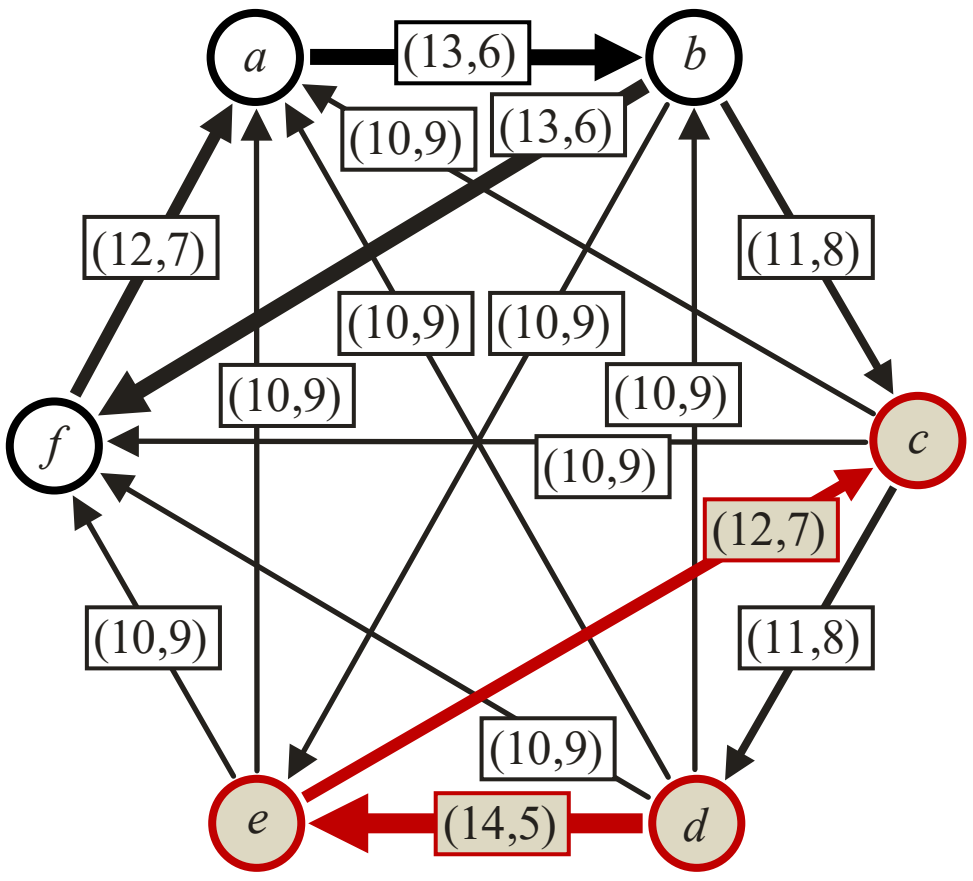

The strongest path from *d* to *c* is:
*d*, (14,5), *e*, (12,7), *c*

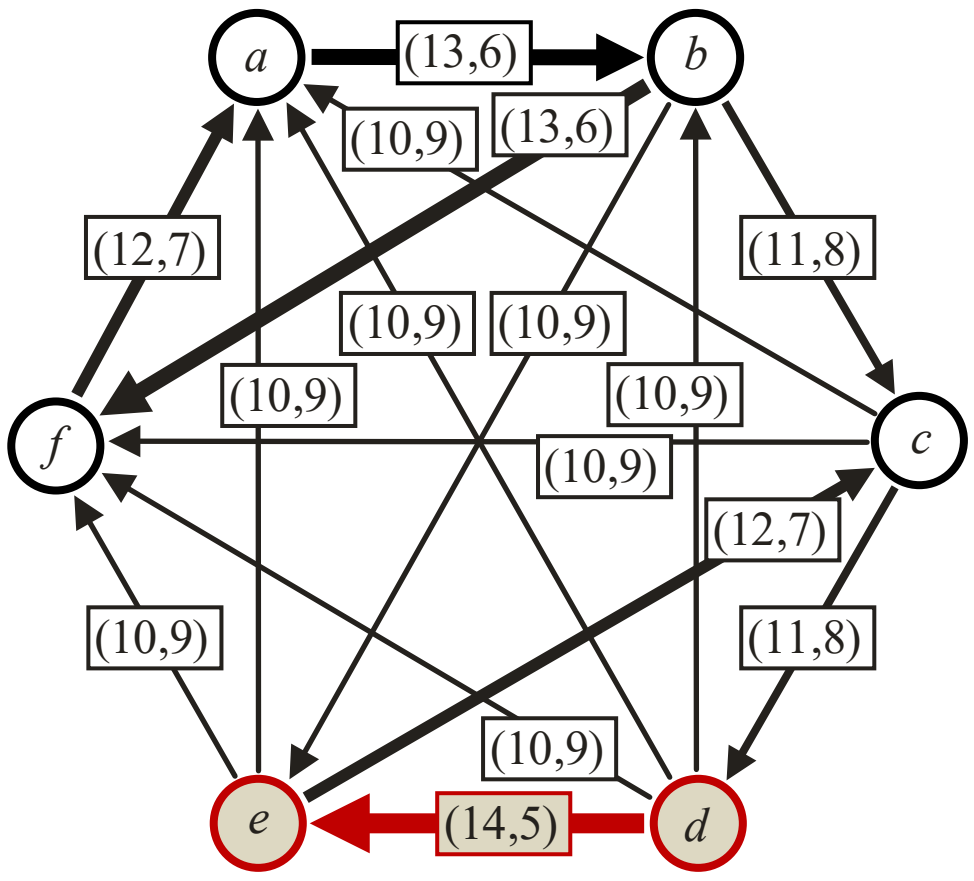

The strongest path from *d* to *e* is:
*d*, (14,5), *e*

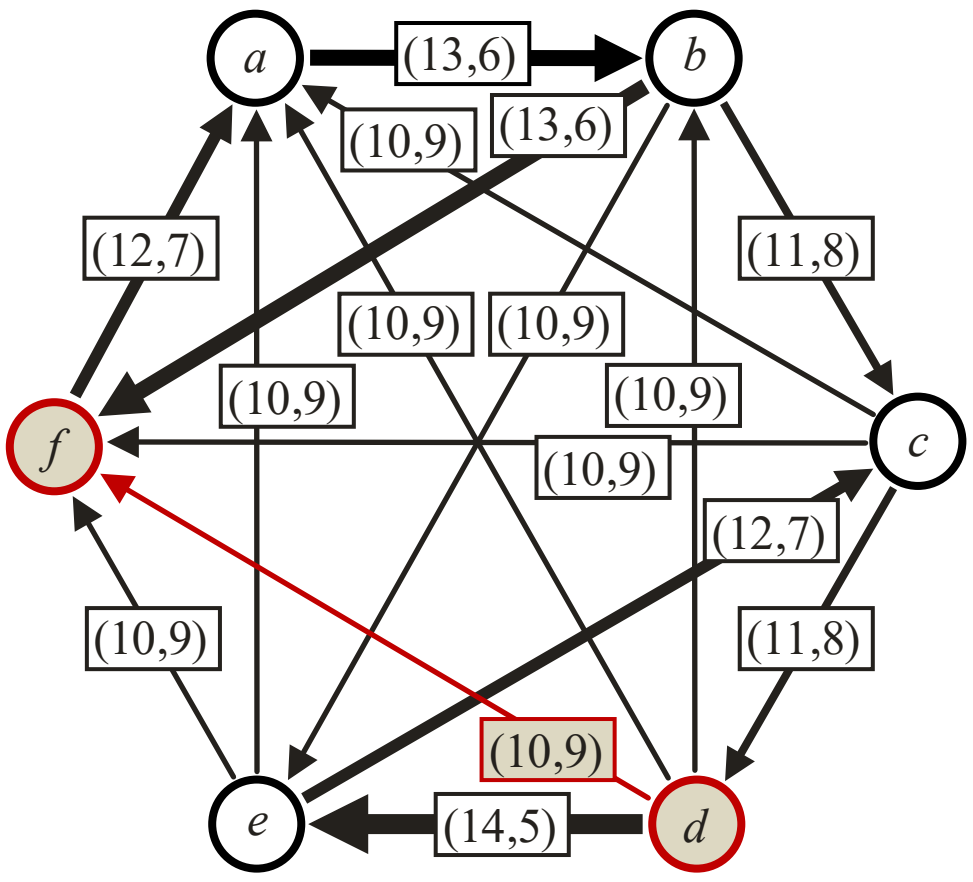

The strongest path from *d* to *f* is:
*d*, (10,9), *f*

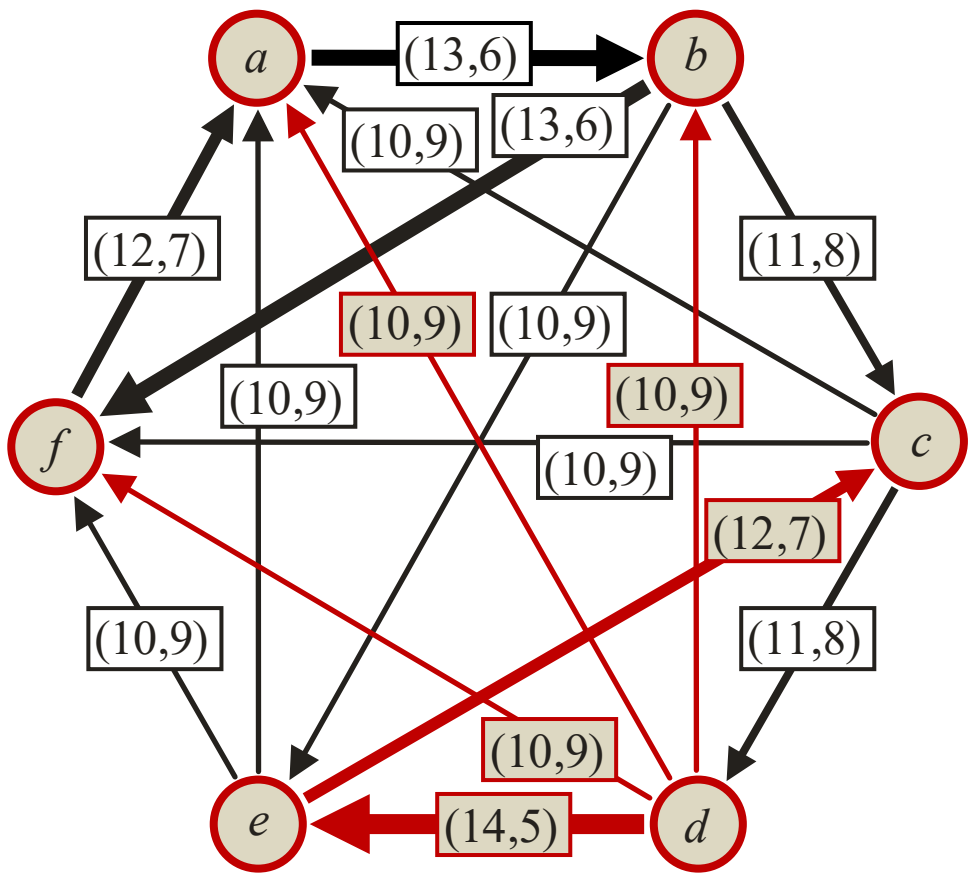

These are the strongest paths
from *d* to every other alternative.

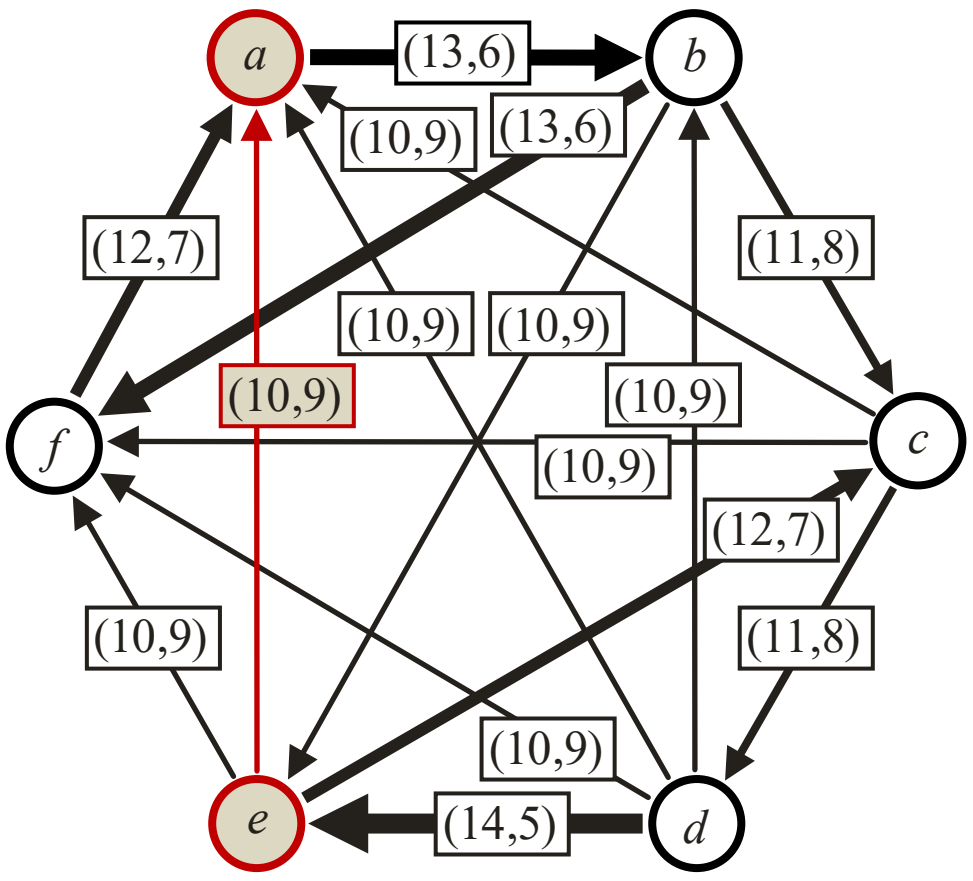

The strongest path from *e* to *a* is:
*e*, (10,9), *a*

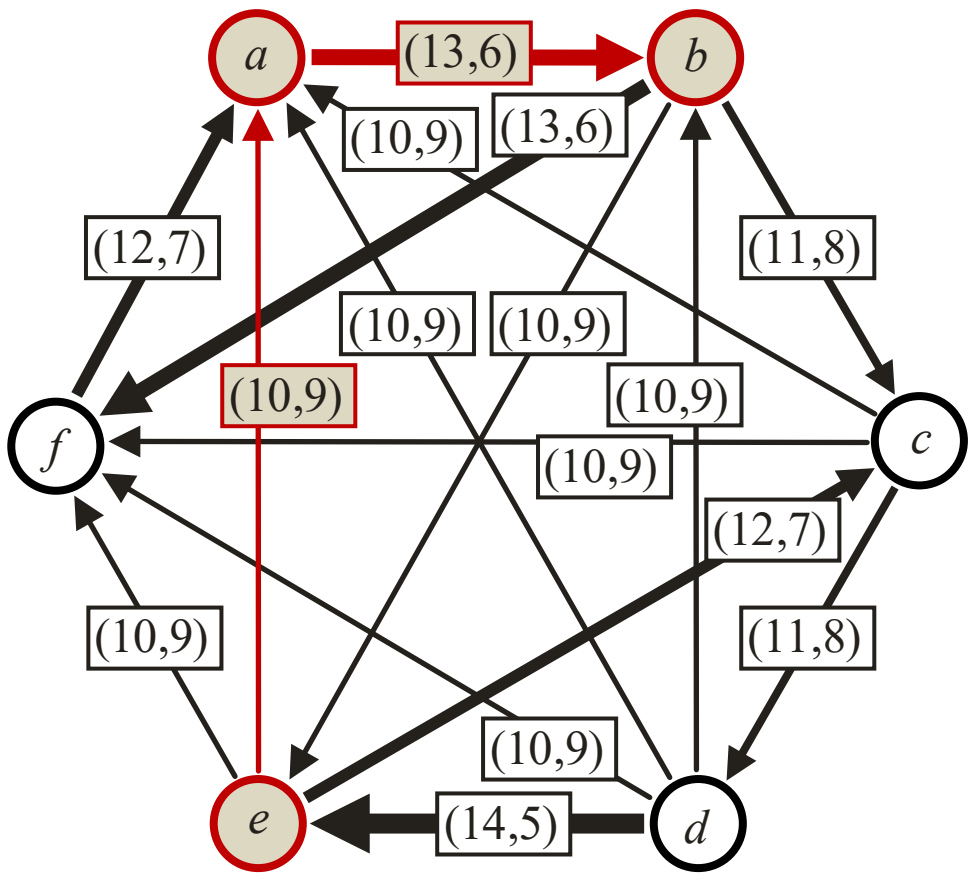

The strongest path from *e* to *b* is:
*e*, (10,9), *a*, (13,6), *b*

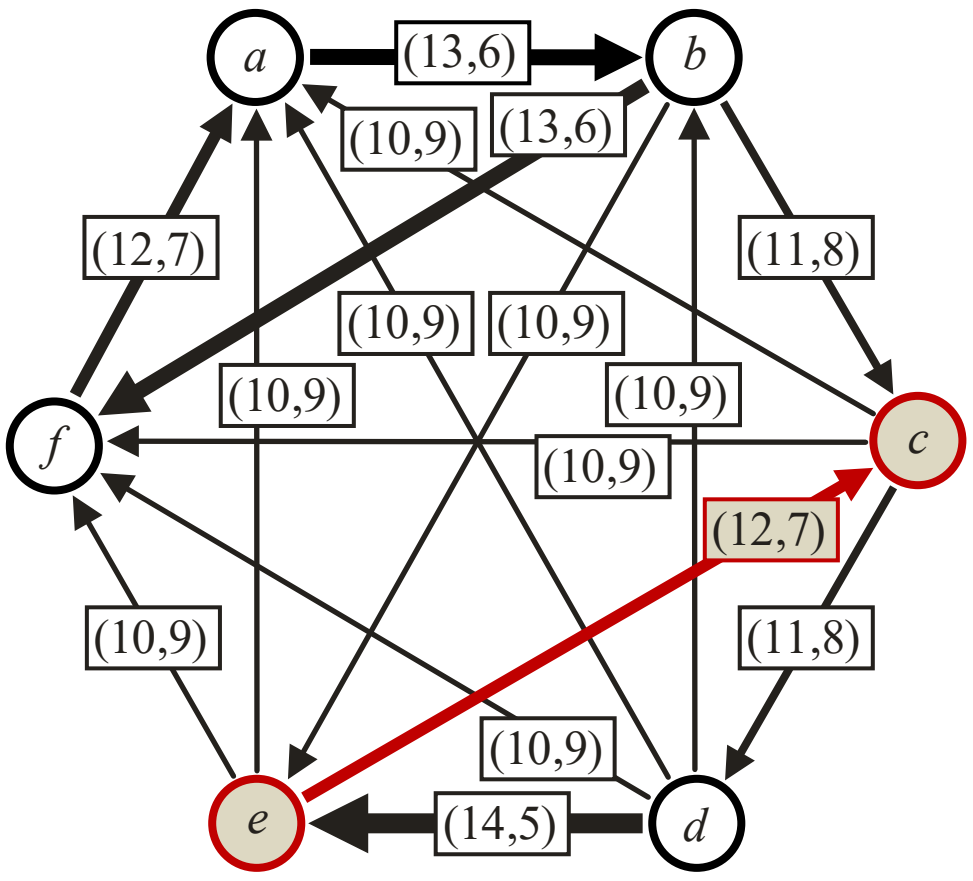

The strongest path from *e* to *c* is:
*e*, (12,7), *c*

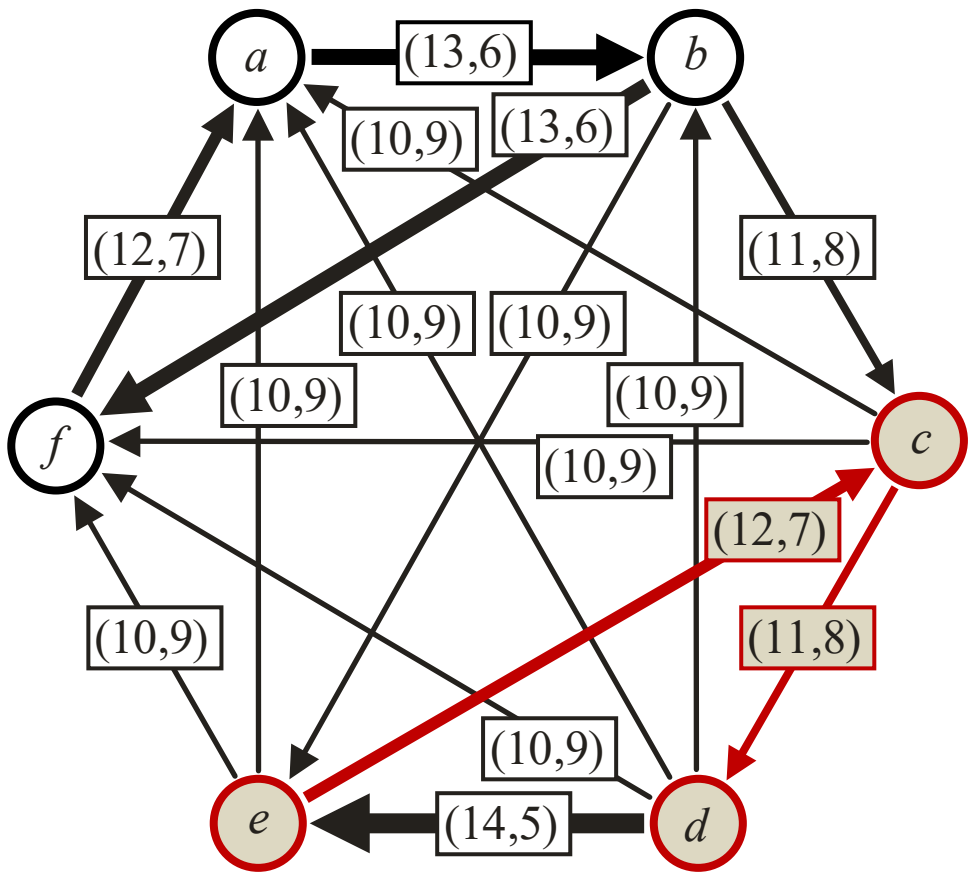

The strongest path from *e* to *d* is:
*e*, (12,7), *c*, (11,8), *d*

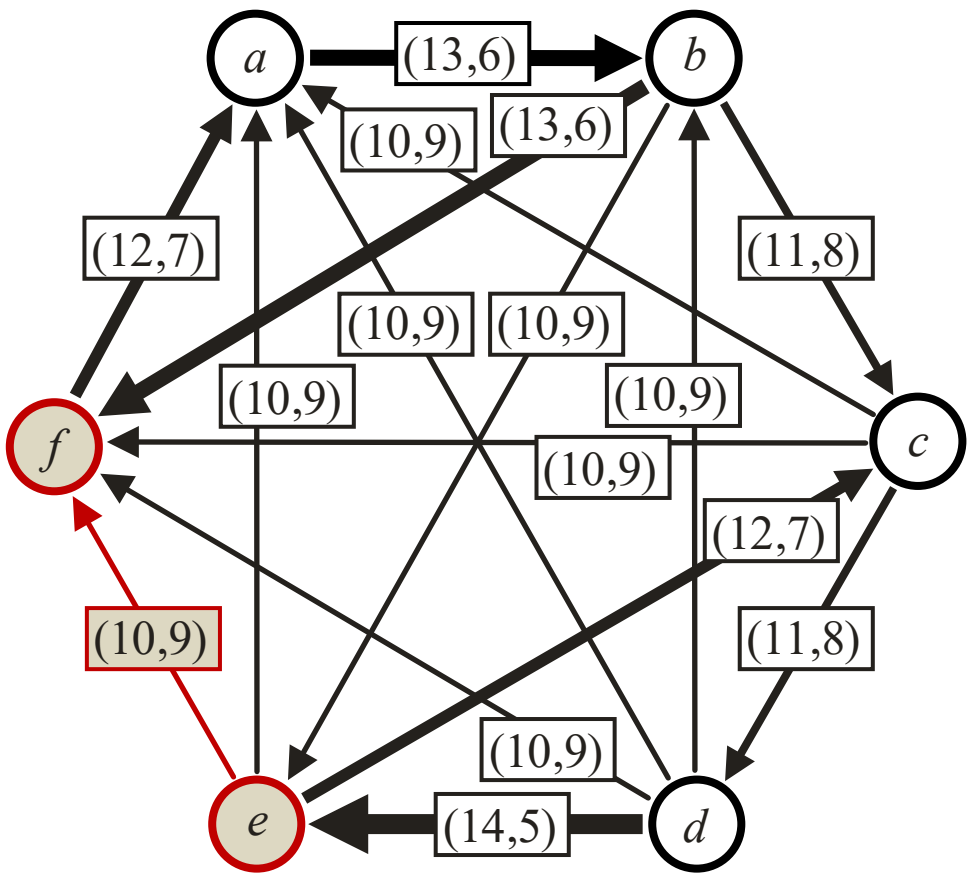

The strongest path from *e* to *f* is:
*e*, (10,9), *f*

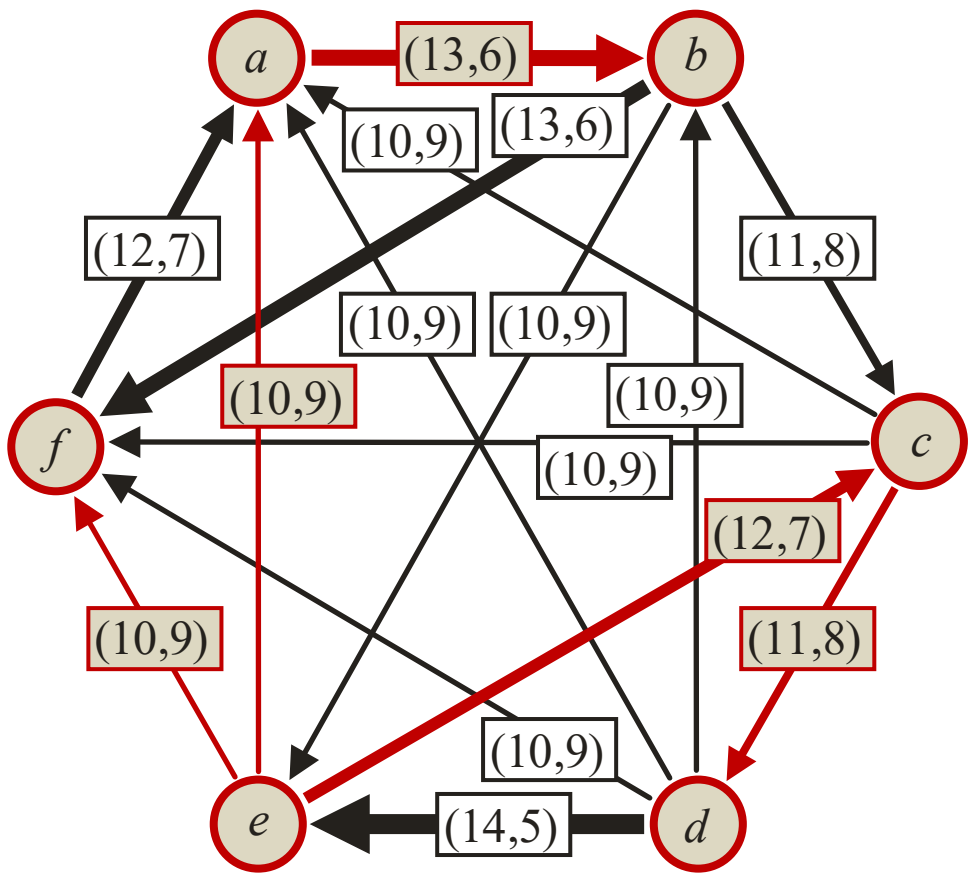

These are the strongest paths
from *e* to every other alternative.

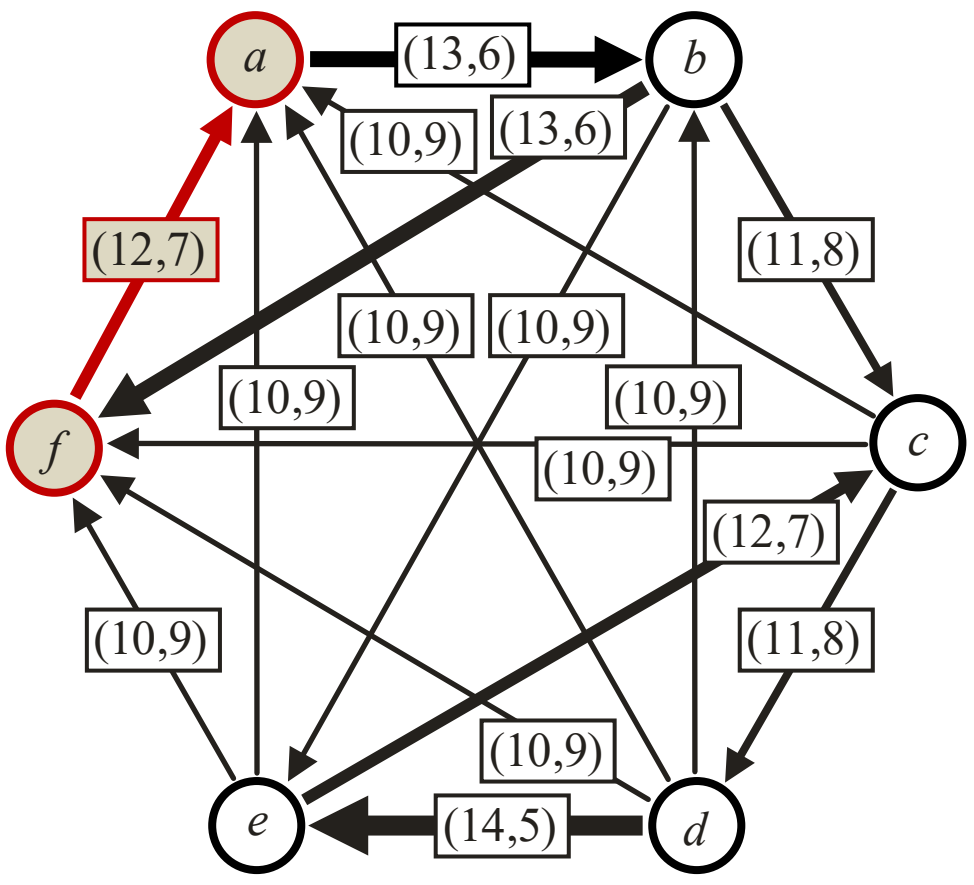

The strongest path from *f* to *a* is:
*f*, (12,7), *a*

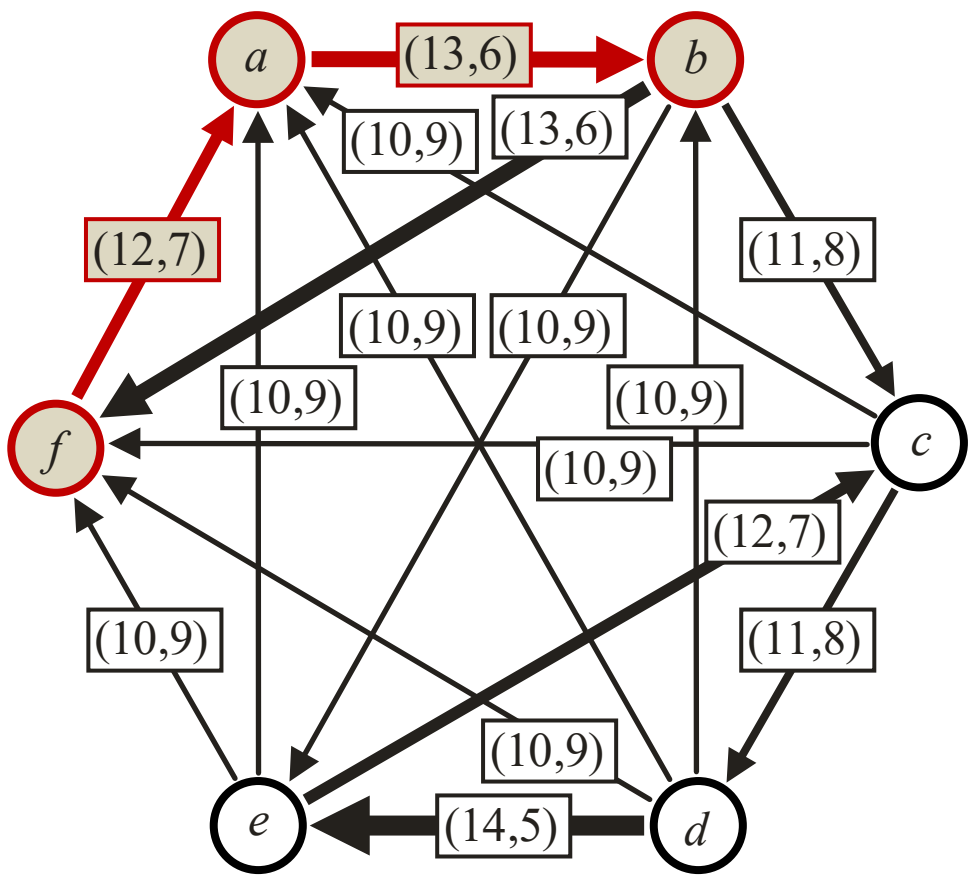

The strongest path from *f* to *b* is:
*f*, (12,7), *a*, (13,6), *b*

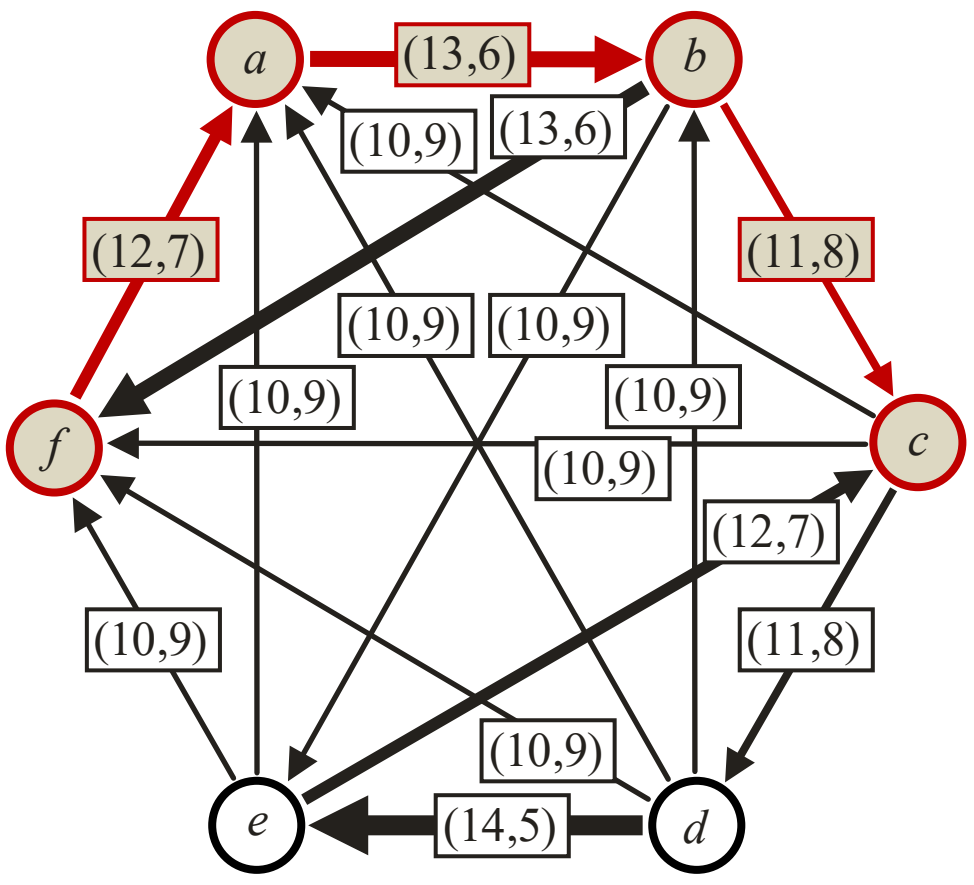

The strongest path from *f* to *c* is:
*f*, (12,7), *a*, (13,6), *b*, (11,8), *c*

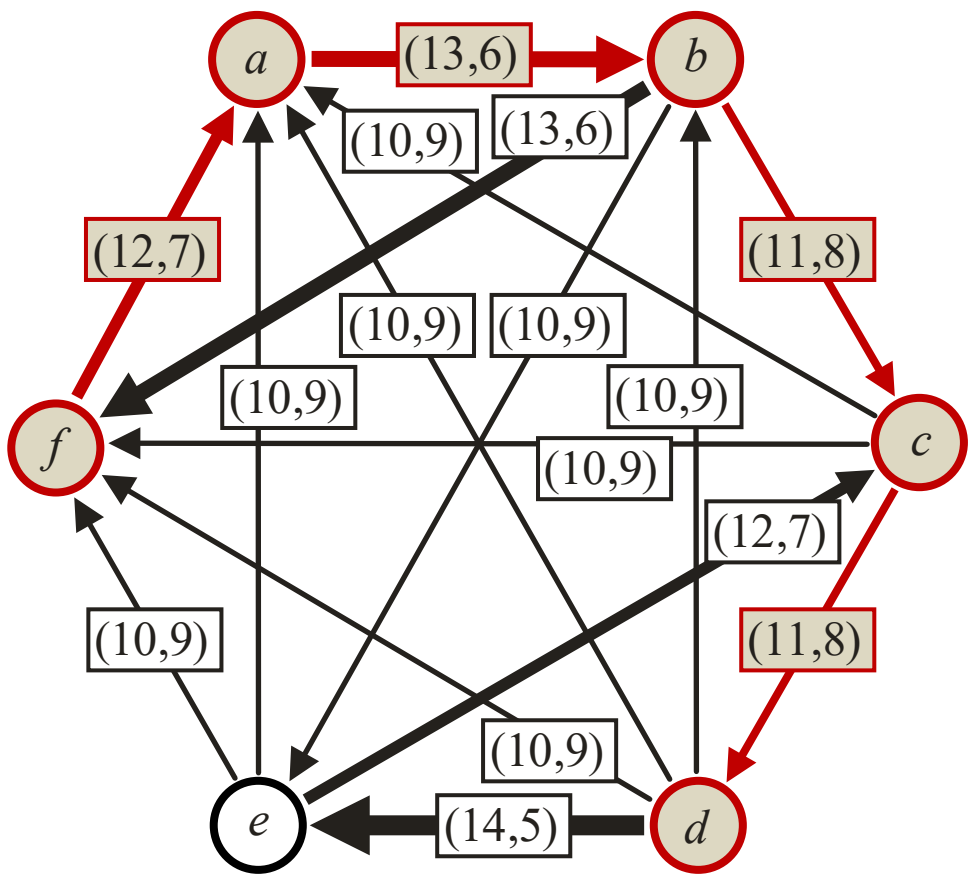

The strongest path from *f* to *d* is:
*f*, (12,7), *a*, (13,6), *b*, (11,8), *c*, (11,8), *d*

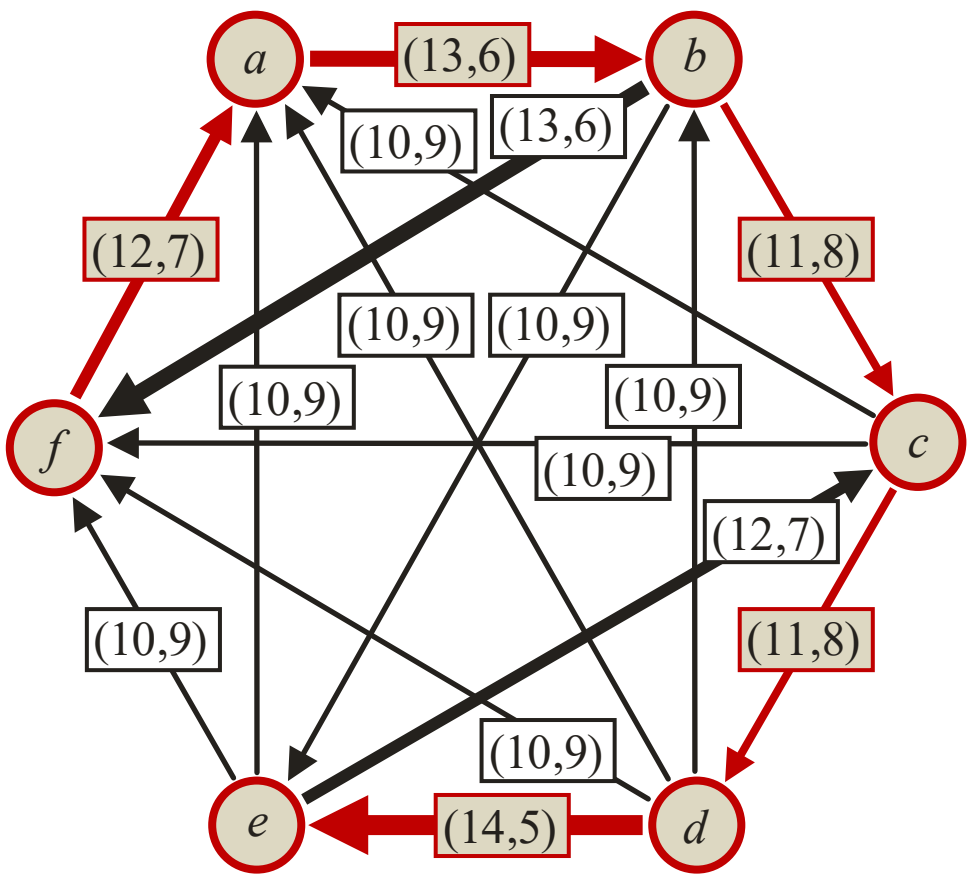

The strongest path from *f* to *e* is:
*f*, (12,7), *a*, (13,6), *b*, (11,8), *c*,
(11,8), *d*, (14,5), *e*

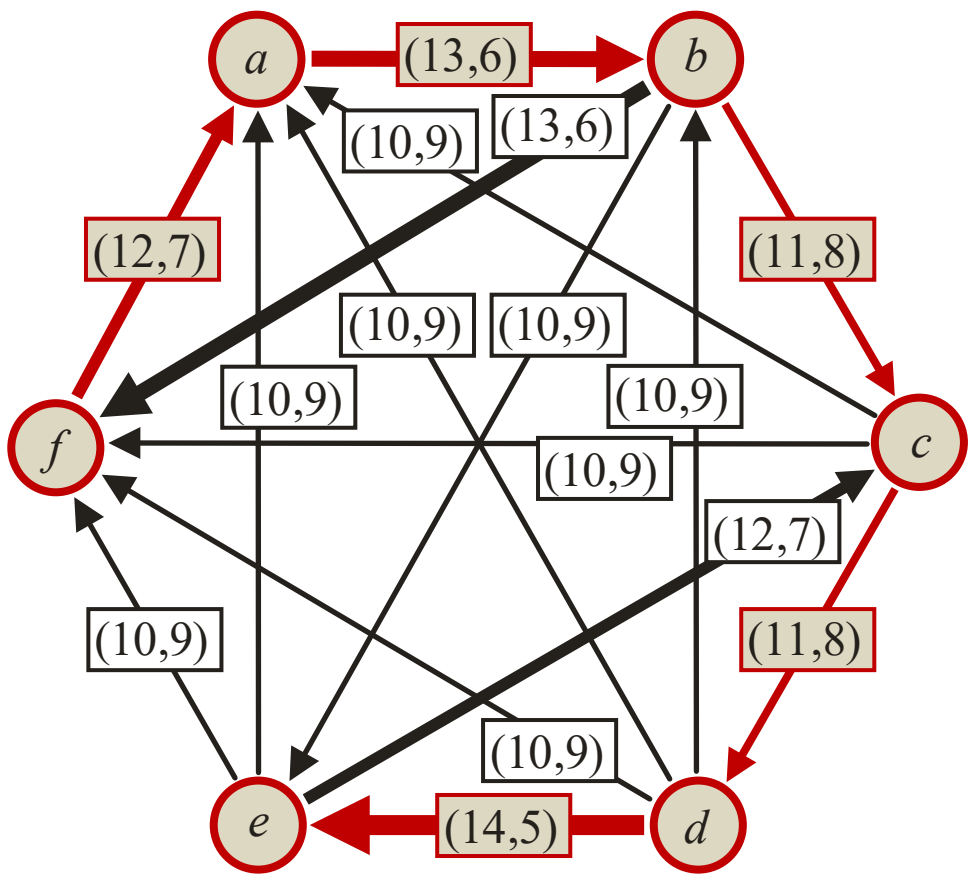

These are the strongest paths
from *f* to every other alternative.

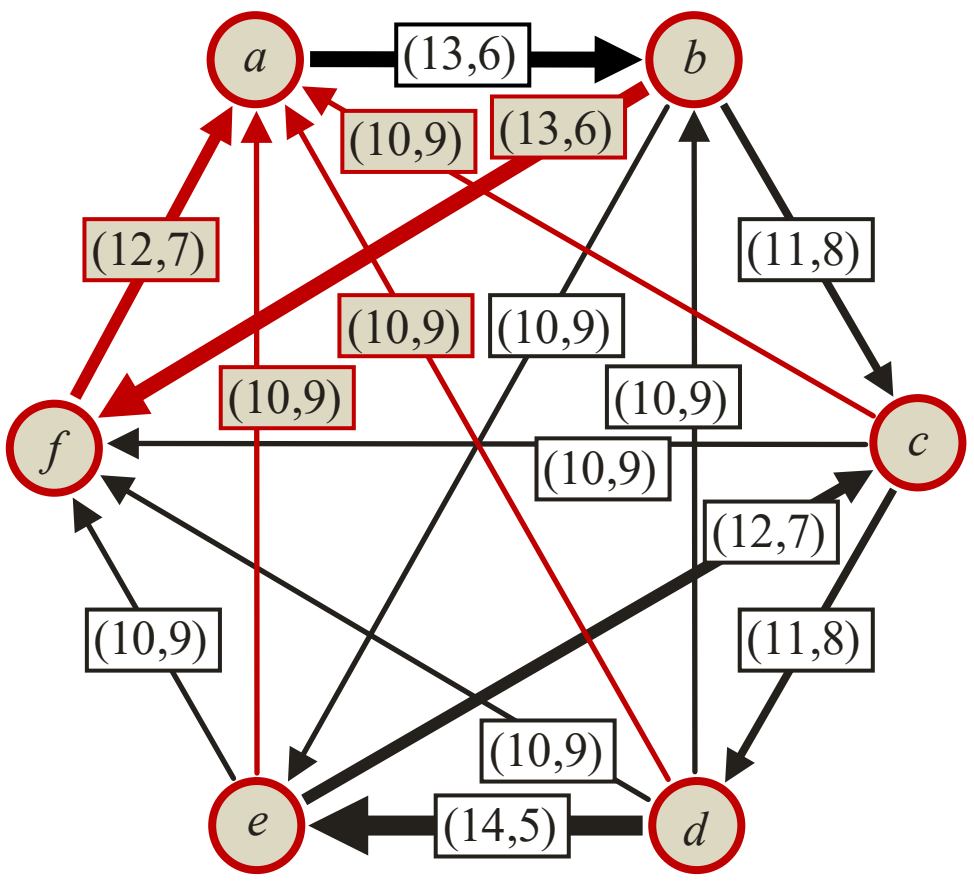

These are the strongest paths
from every other alternative to *a*.

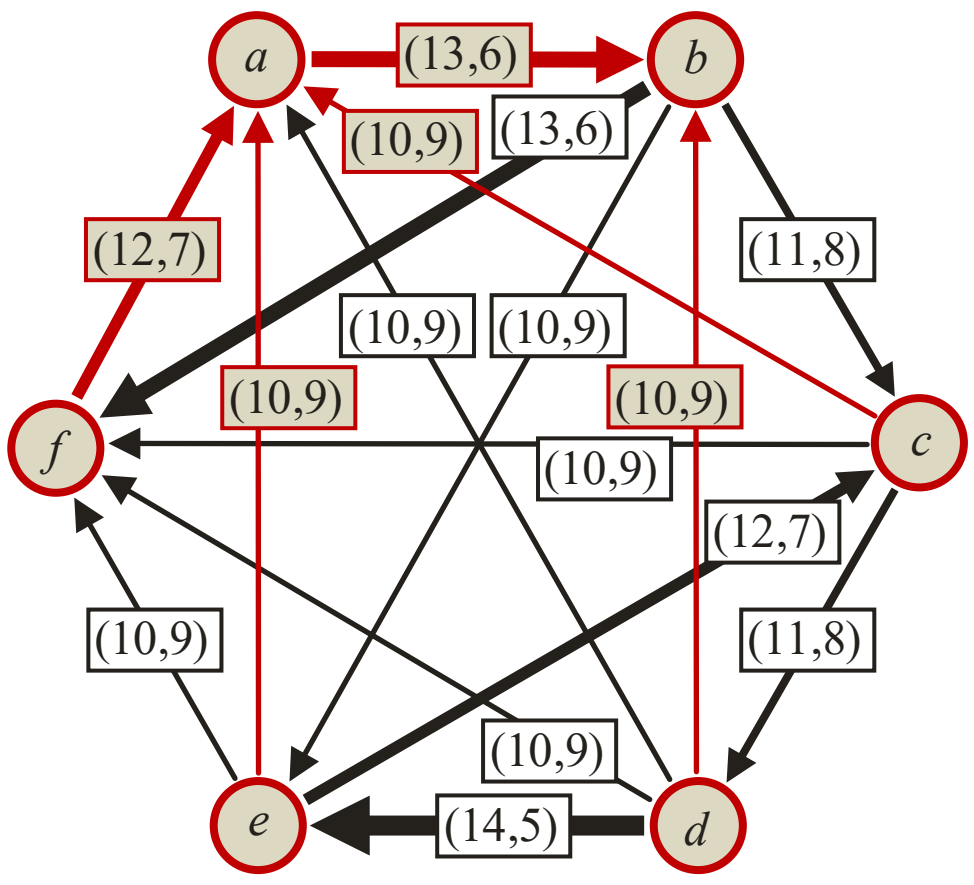

These are the strongest paths
from every other alternative to *b*.

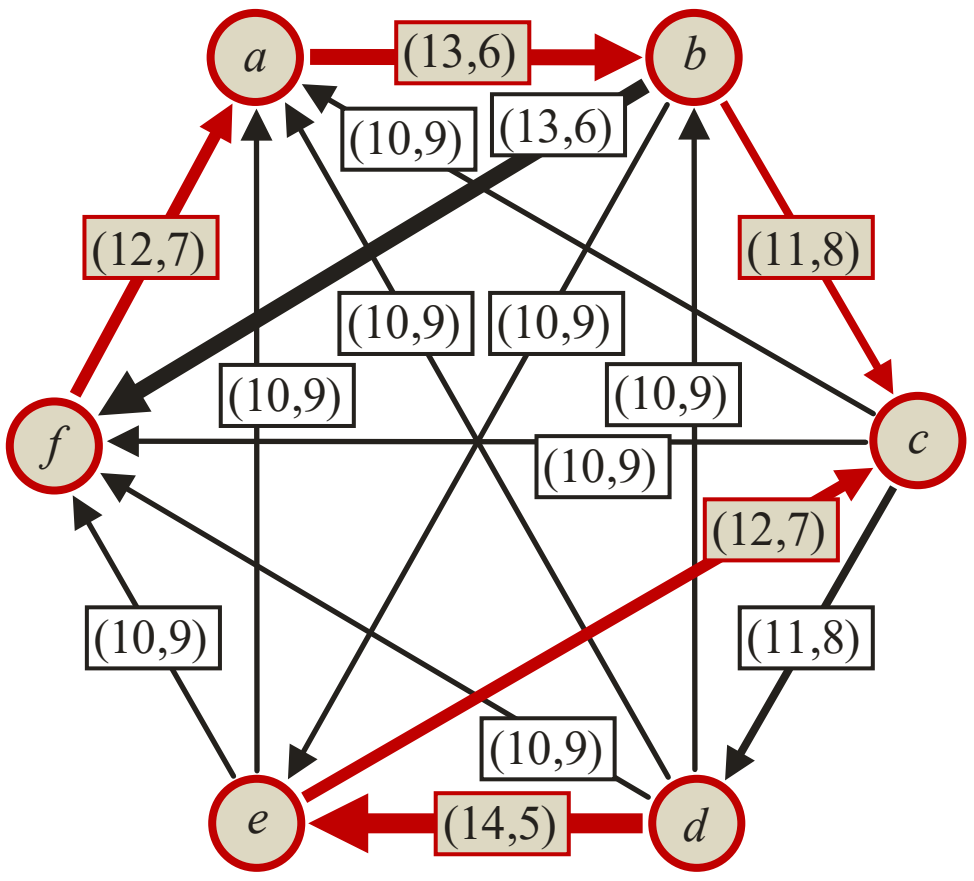

These are the strongest paths
from every other alternative to *c*.

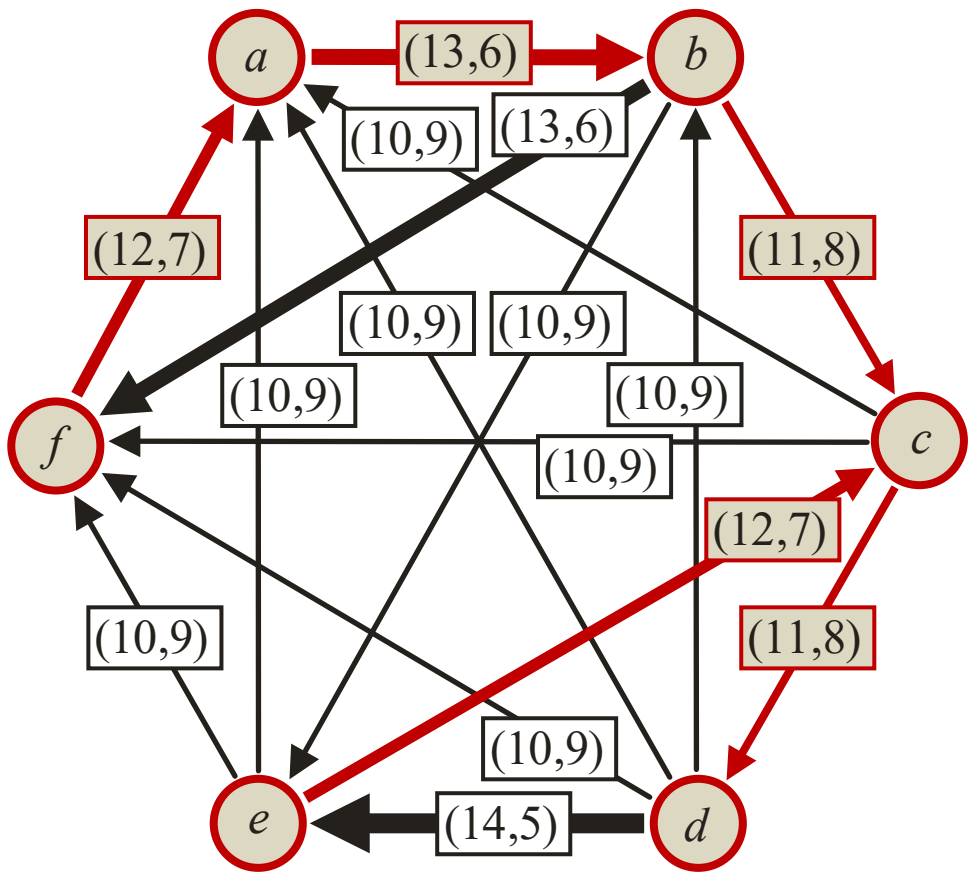

These are the strongest paths
from every other alternative to *d*.

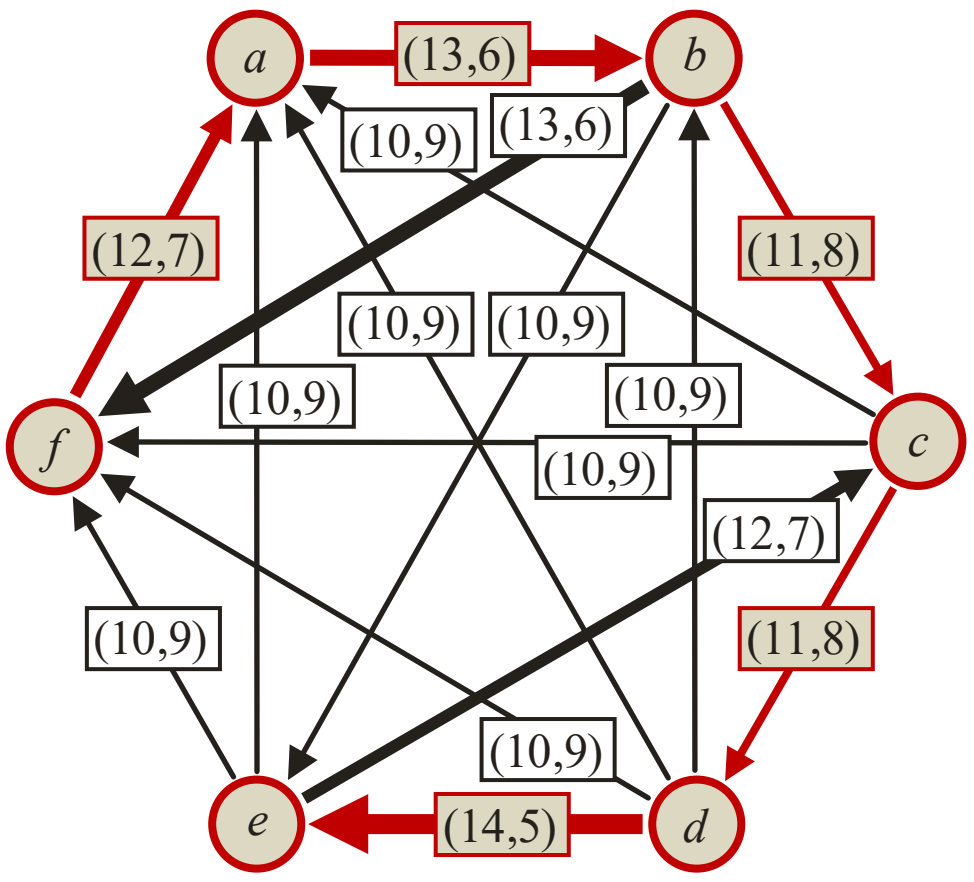

These are the strongest paths
from every other alternative to *e*.

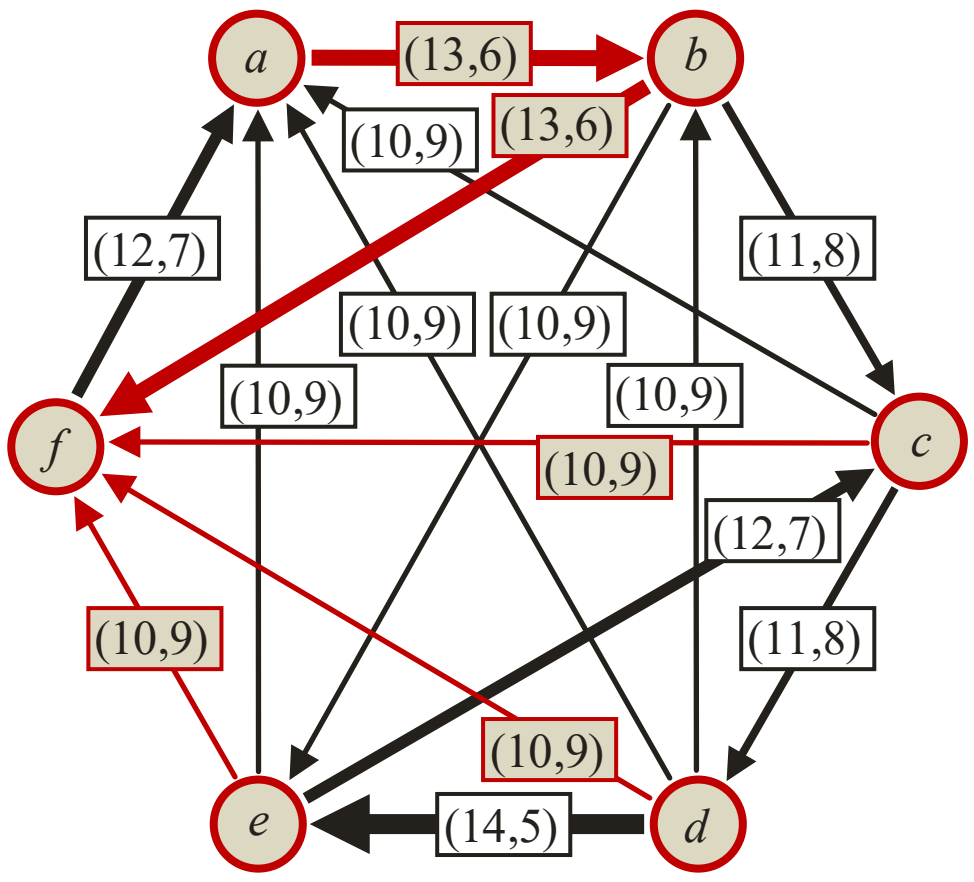

These are the strongest paths
from every other alternative to *f*.

Therefore, the strengths of the strongest paths are:

| | $P_D[*,a]$ | $P_D[*,b]$ | $P_D[*,c]$ | $P_D[*,d]$ | $P_D[*,e]$ | $P_D[*,f]$ |
|---|---|---|---|---|---|---|
| $P_D[a,*]$ | --- | (13,6) | (11,8) | (11,8) | (11,8) | (13,6) |
| $P_D[b,*]$ | (12,7) | --- | (11,8) | (11,8) | (11,8) | (13,6) |
| $P_D[c,*]$ | (10,9) | (10,9) | --- | (11,8) | (11,8) | (10,9) |
| $P_D[d,*]$ | (10,9) | (10,9) | (12,7) | --- | (14,5) | (10,9) |
| $P_D[e,*]$ | (10,9) | (10,9) | (12,7) | (11,8) | --- | (10,9) |
| $P_D[f,*]$ | (12,7) | (12,7) | (11,8) | (11,8) | (11,8) | --- |

We get $\mathcal{O}^{\text{old}} = \{ab, ac, ad, ae, af, bc, bd, be, bf, dc, de, ec, fc, fd, fe\}$ and $\mathcal{S}^{\text{old}} = \{a\}$.

Suppose, the strongest paths are calculated with the Floyd-Warshall algorithm, as defined in section 2.3.1. Then the following table documents the $C \cdot (C{-}1) \cdot (C{-}2) = 120$ steps of the Floyd-Warshall algorithm.

We start with

- $P_D[i,j] := (N^{\text{old}}[i,j], N^{\text{old}}[j,i])$ for all $i \in A$ and $j \in A \setminus \{i\}$.
- $pred[i,j] := i$ for all $i \in A$ and $j \in A \setminus \{i\}$.

| | *i* | *j* | *k* | $P_D[j,k]$ | $P_D[j,i]$ | $P_D[i,k]$ | *pred*[*j*,*k*] | *pred*[*i*,*k*] | result |
|---|---|---|---|---|---|---|---|---|---|
| 1 | *a* | *b* | *c* | (11,8) | (6,13) | (9,10) | *b* | *a* | |
| 2 | *a* | *b* | *d* | (9,10) | (6,13) | (9,10) | *b* | *a* | |
| 3 | *a* | *b* | *e* | (10,9) | (6,13) | (9,10) | *b* | *a* | |
| 4 | *a* | *b* | *f* | (13,6) | (6,13) | (7,12) | *b* | *a* | |
| 5 | *a* | *c* | *b* | (8,11) | (10,9) | (13,6) | *c* | *a* | $P_D[c,b]$ is updated from (8,11) to (10,9); *pred*[*c*,*b*] is updated from *c* to *a*. |
| 6 | *a* | *c* | *d* | (11,8) | (10,9) | (9,10) | *c* | *a* | |
| 7 | *a* | *c* | *e* | (7,12) | (10,9) | (9,10) | *c* | *a* | $P_D[c,e]$ is updated from (7,12) to (9,10); *pred*[*c*,*e*] is updated from *c* to *a*. |
| 8 | *a* | *c* | *f* | (10,9) | (10,9) | (7,12) | *c* | *a* | |
| 9 | *a* | *d* | *b* | (10,9) | (10,9) | (13,6) | *d* | *a* | |
| 10 | *a* | *d* | *c* | (8,11) | (10,9) | (9,10) | *d* | *a* | $P_D[d,c]$ is updated from (8,11) to (9,10); *pred*[*d*,*c*] is updated from *d* to *a*. |
| 11 | *a* | *d* | *e* | (14,5) | (10,9) | (9,10) | *d* | *a* | |
| 12 | *a* | *d* | *f* | (10,9) | (10,9) | (7,12) | *d* | *a* | |
| 13 | *a* | *e* | *b* | (9,10) | (10,9) | (13,6) | *e* | *a* | $P_D[e,b]$ is updated from (9,10) to (10,9); *pred*[*e*,*b*] is updated from *e* to *a*. |
| 14 | *a* | *e* | *c* | (12,7) | (10,9) | (9,10) | *e* | *a* | |
| 15 | *a* | *e* | *d* | (5,14) | (10,9) | (9,10) | *e* | *a* | $P_D[e,d]$ is updated from (5,14) to (9,10); *pred*[*e*,*d*] is updated from *e* to *a*. |
| 16 | *a* | *e* | *f* | (10,9) | (10,9) | (7,12) | *e* | *a* | |
| 17 | *a* | *f* | *b* | (6,13) | (12,7) | (13,6) | *f* | *a* | $P_D[f,b]$ is updated from (6,13) to (12,7); *pred*[*f*,*b*] is updated from *f* to *a*. |
| 18 | *a* | *f* | *c* | (9,10) | (12,7) | (9,10) | *f* | *a* | |
| 19 | *a* | *f* | *d* | (9,10) | (12,7) | (9,10) | *f* | *a* | |
| 20 | *a* | *f* | *e* | (9,10) | (12,7) | (9,10) | *f* | *a* | |
| 21 | *b* | *a* | *c* | (9,10) | (13,6) | (11,8) | *a* | *b* | $P_D[a,c]$ is updated from (9,10) to (11,8); *pred*[*a*,*c*] is updated from *a* to *b*. |
| 22 | *b* | *a* | *d* | (9,10) | (13,6) | (9,10) | *a* | *b* | |
| 23 | *b* | *a* | *e* | (9,10) | (13,6) | (10,9) | *a* | *b* | $P_D[a,e]$ is updated from (9,10) to (10,9); *pred*[*a*,*e*] is updated from *a* to *b*. |
| 24 | *b* | *a* | *f* | (7,12) | (13,6) | (13,6) | *a* | *b* | $P_D[a,f]$ is updated from (7,12) to (13,6); *pred*[*a*,*f*] is updated from *a* to *b*. |
| 25 | *b* | *c* | *a* | (10,9) | (10,9) | (6,13) | *c* | *b* | |
| 26 | *b* | *c* | *d* | (11,8) | (10,9) | (9,10) | *c* | *b* | |
| 27 | *b* | *c* | *e* | (9,10) | (10,9) | (10,9) | *a* | *b* | $P_D[c,e]$ is updated from (9,10) to (10,9); *pred*[*c*,*e*] is updated from *a* to *b*. |
| 28 | *b* | *c* | *f* | (10,9) | (10,9) | (13,6) | *c* | *b* | |
| 29 | *b* | *d* | *a* | (10,9) | (10,9) | (6,13) | *d* | *b* | |
| 30 | *b* | *d* | *c* | (9,10) | (10,9) | (11,8) | *a* | *b* | $P_D[d,c]$ is updated from (9,10) to (10,9); *pred*[*d*,*c*] is updated from *a* to *b*. |

| | *i* | *j* | *k* | $P_D[j,k]$ | $P_D[j,i]$ | $P_D[i,k]$ | *pred*[*j*,*k*] | *pred*[*i*,*k*] | result |
|---|---|---|---|---|---|---|---|---|---|
| 31 | *b* | *d* | *e* | (14,5) | (10,9) | (10,9) | *d* | *b* | |
| 32 | *b* | *d* | *f* | (10,9) | (10,9) | (13,6) | *d* | *b* | |
| 33 | *b* | *e* | *a* | (10,9) | (10,9) | (6,13) | *e* | *b* | |
| 34 | *b* | *e* | *c* | (12,7) | (10,9) | (11,8) | *e* | *b* | |
| 35 | *b* | *e* | *d* | (9,10) | (10,9) | (9,10) | *a* | *b* | |
| 36 | *b* | *e* | *f* | (10,9) | (10,9) | (13,6) | *e* | *b* | |
| 37 | *b* | *f* | *a* | (12,7) | (12,7) | (6,13) | *f* | *b* | |
| 38 | *b* | *f* | *c* | (9,10) | (12,7) | (11,8) | *f* | *b* | $P_D[f,c]$ is updated from (9,10) to (11,8); *pred*[*f*,*c*] is updated from *f* to *b*. |
| 39 | *b* | *f* | *d* | (9,10) | (12,7) | (9,10) | *f* | *b* | |
| 40 | *b* | *f* | *e* | (9,10) | (12,7) | (10,9) | *f* | *b* | $P_D[f,e]$ is updated from (9,10) to (10,9); *pred*[*f*,*e*] is updated from *f* to *b*. |
| 41 | *c* | *a* | *b* | (13,6) | (11,8) | (10,9) | *a* | *a* | |
| 42 | *c* | *a* | *d* | (9,10) | (11,8) | (11,8) | *a* | *c* | $P_D[a,d]$ is updated from (9,10) to (11,8); *pred*[*a*,*d*] is updated from *a* to *c*. |
| 43 | *c* | *a* | *e* | (10,9) | (11,8) | (10,9) | *b* | *b* | |
| 44 | *c* | *a* | *f* | (13,6) | (11,8) | (10,9) | *b* | *c* | |
| 45 | *c* | *b* | *a* | (6,13) | (11,8) | (10,9) | *b* | *c* | $P_D[b,a]$ is updated from (6,13) to (10,9); *pred*[*b*,*a*] is updated from *b* to *c*. |
| 46 | *c* | *b* | *d* | (9,10) | (11,8) | (11,8) | *b* | *c* | $P_D[b,d]$ is updated from (9,10) to (11,8); *pred*[*b*,*d*] is updated from *b* to *c*. |
| 47 | *c* | *b* | *e* | (10,9) | (11,8) | (10,9) | *b* | *b* | |
| 48 | *c* | *b* | *f* | (13,6) | (11,8) | (10,9) | *b* | *c* | |
| 49 | *c* | *d* | *a* | (10,9) | (10,9) | (10,9) | *d* | *c* | |
| 50 | *c* | *d* | *b* | (10,9) | (10,9) | (10,9) | *d* | *a* | |
| 51 | *c* | *d* | *e* | (14,5) | (10,9) | (10,9) | *d* | *b* | |
| 52 | *c* | *d* | *f* | (10,9) | (10,9) | (10,9) | *d* | *c* | |
| 53 | *c* | *e* | *a* | (10,9) | (12,7) | (10,9) | *e* | *c* | |
| 54 | *c* | *e* | *b* | (10,9) | (12,7) | (10,9) | *a* | *a* | |
| 55 | *c* | *e* | *d* | (9,10) | (12,7) | (11,8) | *a* | *c* | $P_D[e,d]$ is updated from (9,10) to (11,8); *pred*[*e*,*d*] is updated from *a* to *c*. |
| 56 | *c* | *e* | *f* | (10,9) | (12,7) | (10,9) | *e* | *c* | |
| 57 | *c* | *f* | *a* | (12,7) | (11,8) | (10,9) | *f* | *c* | |
| 58 | *c* | *f* | *b* | (12,7) | (11,8) | (10,9) | *a* | *a* | |
| 59 | *c* | *f* | *d* | (9,10) | (11,8) | (11,8) | *f* | *c* | $P_D[f,d]$ is updated from (9,10) to (11,8); *pred*[*f*,*d*] is updated from *f* to *c*. |
| 60 | *c* | *f* | *e* | (10,9) | (11,8) | (10,9) | *b* | *b* | |

| | *i* | *j* | *k* | $P_D[j,k]$ | $P_D[j,i]$ | $P_D[i,k]$ | *pred*[*j*,*k*] | *pred*[*i*,*k*] | result |
|---|---|---|---|---|---|---|---|---|---|
| 61 | *d* | *a* | *b* | (13,6) | (11,8) | (10,9) | *a* | *d* | |
| 62 | *d* | *a* | *c* | (11,8) | (11,8) | (10,9) | *b* | *b* | |
| 63 | *d* | *a* | *e* | (10,9) | (11,8) | (14,5) | *b* | *d* | $P_D[a,e]$ is updated from (10,9) to (11,8); *pred*[*a*,*e*] is updated from *b* to *d*. |
| 64 | *d* | *a* | *f* | (13,6) | (11,8) | (10,9) | *b* | *d* | |
| 65 | *d* | *b* | *a* | (10,9) | (11,8) | (10,9) | *c* | *d* | |
| 66 | *d* | *b* | *c* | (11,8) | (11,8) | (10,9) | *b* | *b* | |
| 67 | *d* | *b* | *e* | (10,9) | (11,8) | (14,5) | *b* | *d* | $P_D[b,e]$ is updated from (10,9) to (11,8); *pred*[*b*,*e*] is updated from *b* to *d*. |
| 68 | *d* | *b* | *f* | (13,6) | (11,8) | (10,9) | *b* | *d* | |
| 69 | *d* | *c* | *a* | (10,9) | (11,8) | (10,9) | *c* | *d* | |
| 70 | *d* | *c* | *b* | (10,9) | (11,8) | (10,9) | *a* | *d* | |
| 71 | *d* | *c* | *e* | (10,9) | (11,8) | (14,5) | *b* | *d* | $P_D[c,e]$ is updated from (10,9) to (11,8); *pred*[*c*,*e*] is updated from *b* to *d*. |
| 72 | *d* | *c* | *f* | (10,9) | (11,8) | (10,9) | *c* | *d* | |
| 73 | *d* | *e* | *a* | (10,9) | (11,8) | (10,9) | *e* | *d* | |
| 74 | *d* | *e* | *b* | (10,9) | (11,8) | (10,9) | *a* | *d* | |
| 75 | *d* | *e* | *c* | (12,7) | (11,8) | (10,9) | *e* | *b* | |
| 76 | *d* | *e* | *f* | (10,9) | (11,8) | (10,9) | *e* | *d* | |
| 77 | *d* | *f* | *a* | (12,7) | (11,8) | (10,9) | *f* | *d* | |
| 78 | *d* | *f* | *b* | (12,7) | (11,8) | (10,9) | *a* | *d* | |
| 79 | *d* | *f* | *c* | (11,8) | (11,8) | (10,9) | *b* | *b* | |
| 80 | *d* | *f* | *e* | (10,9) | (11,8) | (14,5) | *b* | *d* | $P_D[f,e]$ is updated from (10,9) to (11,8); *pred*[*f*,*e*] is updated from *b* to *d*. |
| 81 | *e* | *a* | *b* | (13,6) | (11,8) | (10,9) | *a* | *a* | |
| 82 | *e* | *a* | *c* | (11,8) | (11,8) | (12,7) | *b* | *e* | |
| 83 | *e* | *a* | *d* | (11,8) | (11,8) | (11,8) | *c* | *c* | |
| 84 | *e* | *a* | *f* | (13,6) | (11,8) | (10,9) | *b* | *e* | |
| 85 | *e* | *b* | *a* | (10,9) | (11,8) | (10,9) | *c* | *e* | |
| 86 | *e* | *b* | *c* | (11,8) | (11,8) | (12,7) | *b* | *e* | |
| 87 | *e* | *b* | *d* | (11,8) | (11,8) | (11,8) | *c* | *c* | |
| 88 | *e* | *b* | *f* | (13,6) | (11,8) | (10,9) | *b* | *e* | |
| 89 | *e* | *c* | *a* | (10,9) | (11,8) | (10,9) | *c* | *e* | |
| 90 | *e* | *c* | *b* | (10,9) | (11,8) | (10,9) | *a* | *a* | |

| | *i* | *j* | *k* | $P_D[j,k]$ | $P_D[j,i]$ | $P_D[i,k]$ | *pred*[*j*,*k*] | *pred*[*i*,*k*] | result |
|---|---|---|---|---|---|---|---|---|---|
| 91 | *e* | *c* | *d* | (11,8) | (11,8) | (11,8) | *c* | *c* | |
| 92 | *e* | *c* | *f* | (10,9) | (11,8) | (10,9) | *c* | *e* | |
| 93 | *e* | *d* | *a* | (10,9) | (14,5) | (10,9) | *d* | *e* | |
| 94 | *e* | *d* | *b* | (10,9) | (14,5) | (10,9) | *d* | *a* | |
| 95 | *e* | *d* | *c* | (10,9) | (14,5) | (12,7) | *b* | *e* | $P_D[d,c]$ is updated from (10,9) to (12,7); *pred*[*d*,*c*] is updated from *b* to *e*. |
| 96 | *e* | *d* | *f* | (10,9) | (14,5) | (10,9) | *d* | *e* | |
| 97 | *e* | *f* | *a* | (12,7) | (11,8) | (10,9) | *f* | *e* | |
| 98 | *e* | *f* | *b* | (12,7) | (11,8) | (10,9) | *a* | *a* | |
| 99 | *e* | *f* | *c* | (11,8) | (11,8) | (12,7) | *b* | *e* | |
| 100 | *e* | *f* | *d* | (11,8) | (11,8) | (11,8) | *c* | *c* | |
| 101 | *f* | *a* | *b* | (13,6) | (13,6) | (12,7) | *a* | *a* | |
| 102 | *f* | *a* | *c* | (11,8) | (13,6) | (11,8) | *b* | *b* | |
| 103 | *f* | *a* | *d* | (11,8) | (13,6) | (11,8) | *c* | *c* | |
| 104 | *f* | *a* | *e* | (11,8) | (13,6) | (11,8) | *d* | *d* | |
| 105 | *f* | *b* | *a* | (10,9) | (13,6) | (12,7) | *c* | *f* | $P_D[b,a]$ is updated from (10,9) to (12,7); *pred*[*b*,*a*] is updated from *c* to *f*. |
| 106 | *f* | *b* | *c* | (11,8) | (13,6) | (11,8) | *b* | *b* | |
| 107 | *f* | *b* | *d* | (11,8) | (13,6) | (11,8) | *c* | *c* | |
| 108 | *f* | *b* | *e* | (11,8) | (13,6) | (11,8) | *d* | *d* | |
| 109 | *f* | *c* | *a* | (10,9) | (10,9) | (12,7) | *c* | *f* | |
| 110 | *f* | *c* | *b* | (10,9) | (10,9) | (12,7) | *a* | *a* | |
| 111 | *f* | *c* | *d* | (11,8) | (10,9) | (11,8) | *c* | *c* | |
| 112 | *f* | *c* | *e* | (11,8) | (10,9) | (11,8) | *d* | *d* | |
| 113 | *f* | *d* | *a* | (10,9) | (10,9) | (12,7) | *d* | *f* | |
| 114 | *f* | *d* | *b* | (10,9) | (10,9) | (12,7) | *d* | *a* | |
| 115 | *f* | *d* | *c* | (12,7) | (10,9) | (11,8) | *e* | *b* | |
| 116 | *f* | *d* | *e* | (14,5) | (10,9) | (11,8) | *d* | *d* | |
| 117 | *f* | *e* | *a* | (10,9) | (10,9) | (12,7) | *e* | *f* | |
| 118 | *f* | *e* | *b* | (10,9) | (10,9) | (12,7) | *a* | *a* | |
| 119 | *f* | *e* | *c* | (12,7) | (10,9) | (11,8) | *e* | *b* | |
| 120 | *f* | *e* | *d* | (11,8) | (10,9) | (11,8) | *c* | *c* | |

## 3.7.2. Situation #2

When 2 $a \succ_v e \succ_v f \succ_v c \succ_v b \succ_v d$ ballots are added, then the pairwise matrix $N^{new}$ looks as follows:

| | $N^{new}[*,a]$ | $N^{new}[*,b]$ | $N^{new}[*,c]$ | $N^{new}[*,d]$ | $N^{new}[*,e]$ | $N^{new}[*,f]$ |
|---|---|---|---|---|---|---|
| $N^{new}[a,*]$ | --- | 15 | 11 | 11 | 11 | 9 |
| $N^{new}[b,*]$ | 6 | --- | 11 | 11 | 10 | 13 |
| $N^{new}[c,*]$ | 10 | 10 | --- | 13 | 7 | 10 |
| $N^{new}[d,*]$ | 10 | 10 | 8 | --- | 14 | 10 |
| $N^{new}[e,*]$ | 10 | 11 | 14 | 7 | --- | 12 |
| $N^{new}[f,*]$ | 12 | 8 | 11 | 11 | 9 | --- |

The corresponding digraph looks as follows:

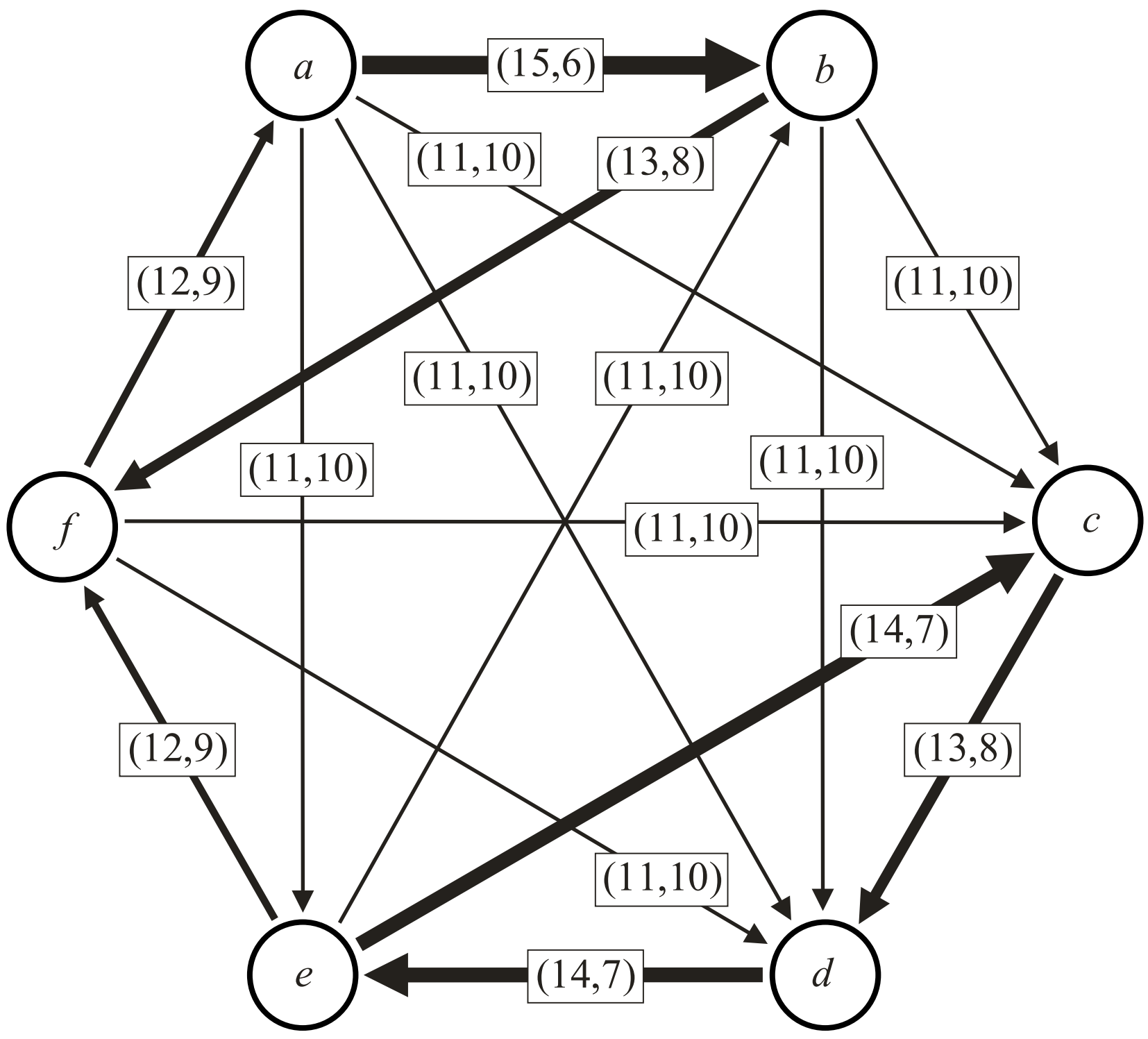

The following table lists the strongest paths, as determined by the Floyd-Warshall algorithm, as defined in section 2.3.1. The critical links of the strongest paths are <u>underlined</u>:

| | ... to *a* | ... to *b* | ... to *c* | ... to *d* | ... to *e* | ... to *f* |
|---|---|---|---|---|---|---|
| from *a* ... | --- | *a*, <u>(15,6)</u>, *b* | *a*, <u>(11,10)</u>, *c* | *a*, <u>(11,10)</u>, *d* | *a*, <u>(11,10)</u>, *e* | *a*, (15,6), *b*, <u>(13,8)</u>, *f* |
| from *b* ... | *b*, (13,8), *f*, <u>(12,9)</u>, *a* | --- | *b*, <u>(11,10)</u>, *c* | *b*, <u>(11,10)</u>, *d* | *b*, <u>(11,10)</u>, *d*, (14,7), *e* | *b*, <u>(13,8)</u>, *f* |
| from *c* ... | *c*, (13,8), *d*, (14,7), *e*, <u>(12,9)</u>, *f*, <u>(12,9)</u>, *a* | *c*, (13,8), *d*, (14,7), *e*, <u>(12,9)</u>, *f*, <u>(12,9)</u>, *a*, (15,6), *b* | --- | *c*, <u>(13,8)</u>, *d* | *c*, <u>(13,8)</u>, *d*, (14,7), *e* | *c*, (13,8), *d*, (14,7), *e*, <u>(12,9)</u>, *f* |
| from *d* ... | *d*, (14,7), *e*, <u>(12,9)</u>, *f*, <u>(12,9)</u>, *a* | *d*, (14,7), *e*, <u>(12,9)</u>, *f*, <u>(12,9)</u>, *a*, (15,6), *b* | *d*, <u>(14,7)</u>, *e*, <u>(14,7)</u>, *c* | --- | *d*, <u>(14,7)</u>, *e* | *d*, (14,7), *e*, <u>(12,9)</u>, *f* |
| from *e* ... | *e*, <u>(12,9)</u>, *f*, <u>(12,9)</u>, *a* | *e*, <u>(12,9)</u>, *f*, <u>(12,9)</u>, *a*, (15,6), *b* | *e*, <u>(14,7)</u>, *c* | *e*, (14,7), *c*, <u>(13,8)</u>, *d* | --- | *e*, <u>(12,9)</u>, *f* |
| from *f* ... | *f*, <u>(12,9)</u>, *a* | *f*, <u>(12,9)</u>, *a*, (15,6), *b* | *f*, <u>(11,10)</u>, *c* | *f*, <u>(11,10)</u>, *d* | *f*, (12,9), *a*, <u>(11,10)</u>, *e* | --- |

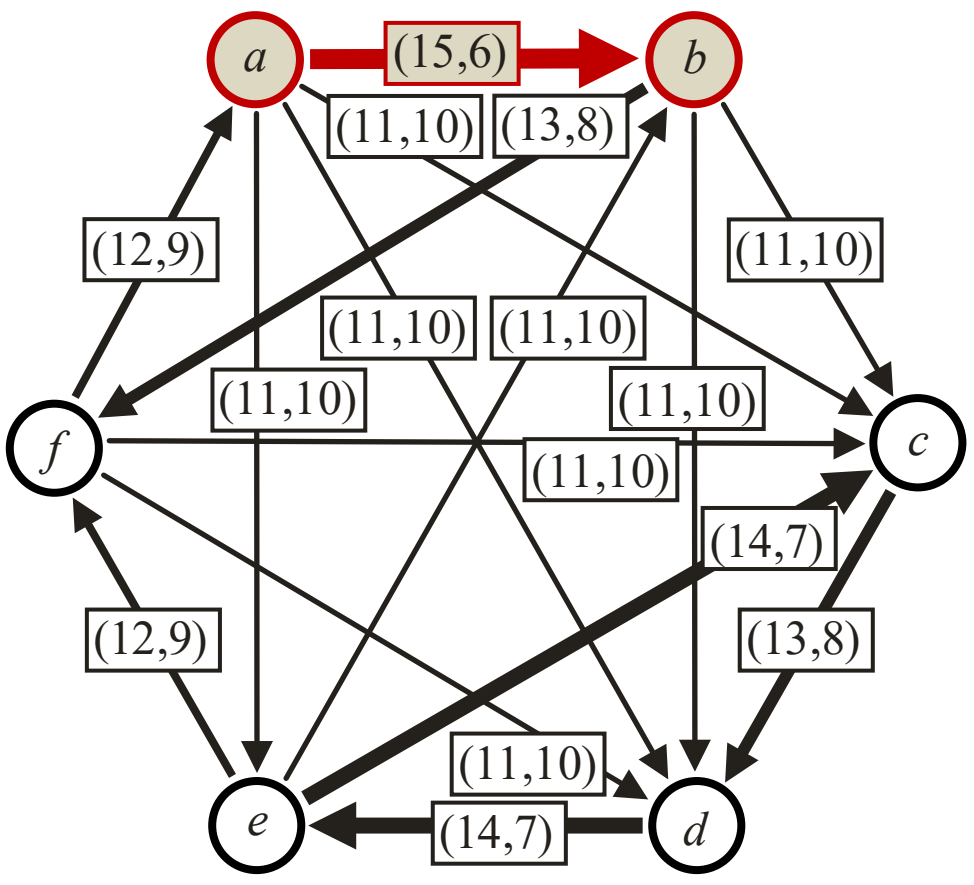

The strongest path from *a* to *b* is:
*a*, (15,6), *b*

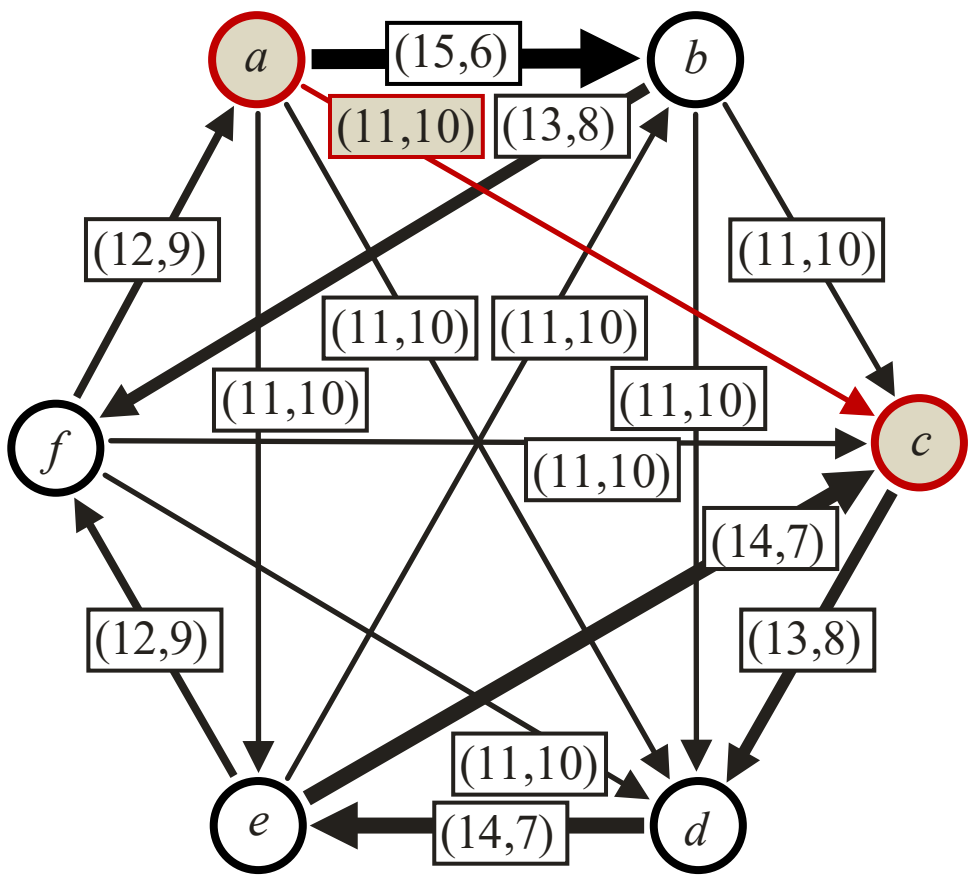

The strongest path from *a* to *c* is:
*a*, (11,10), *c*

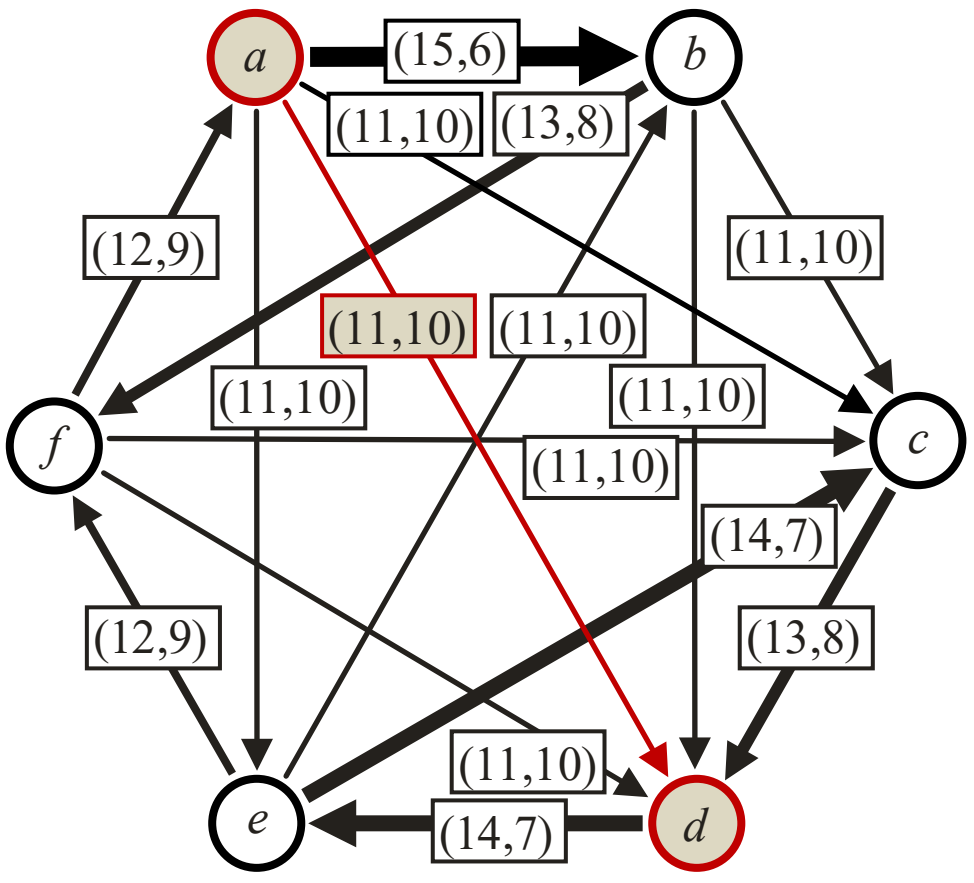

The strongest path from *a* to *d* is:
*a*, (11,10), *d*

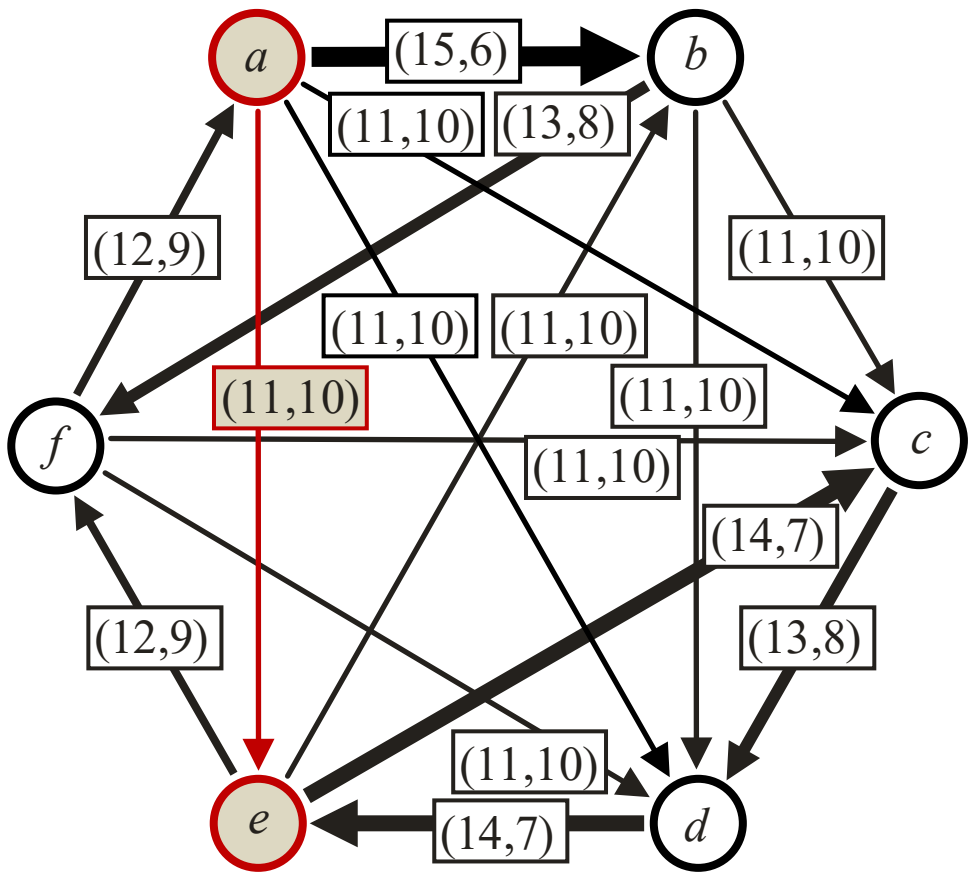

The strongest path from *a* to *e* is:
*a*, (11,10), *e*

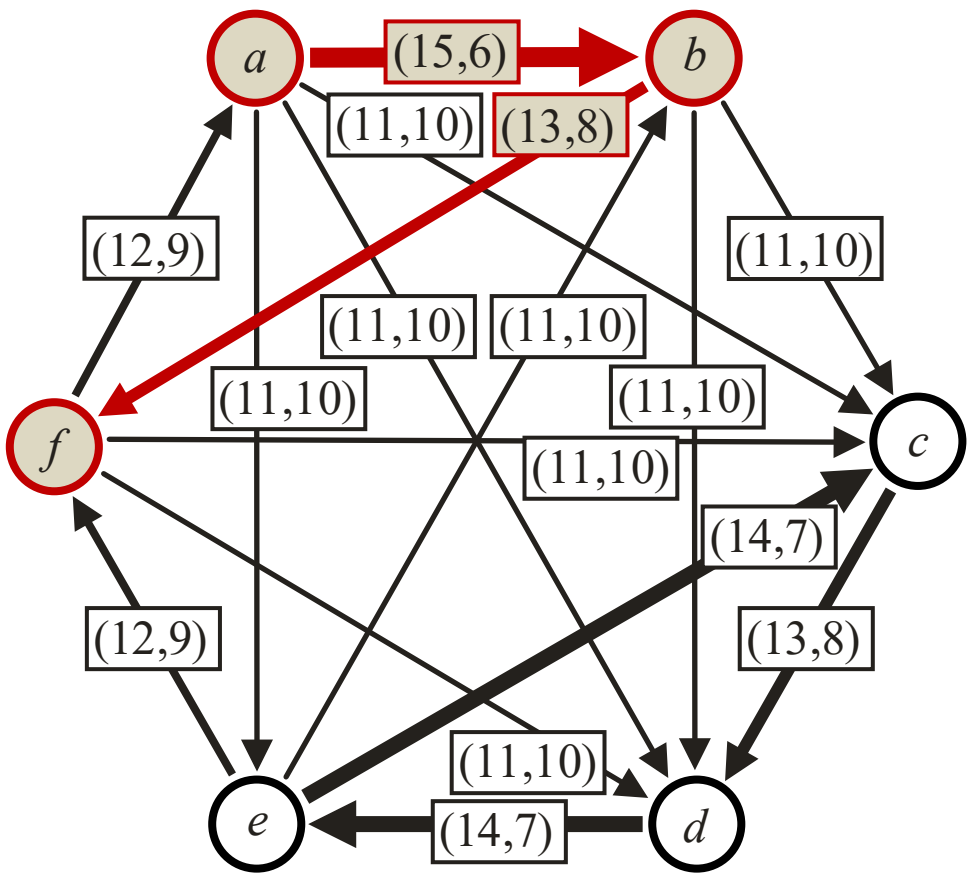

The strongest path from *a* to *f* is:
*a*, (15,6), *b*, (13,8), *f*

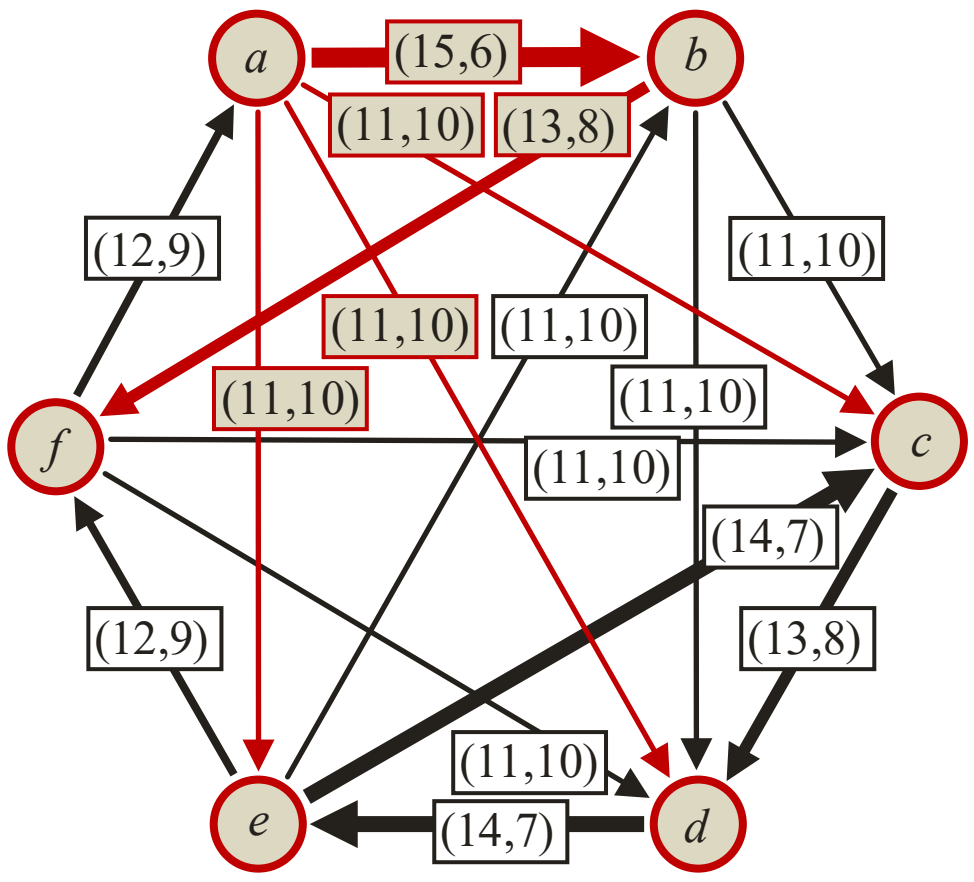

These are the strongest paths
from *a* to every other alternative.

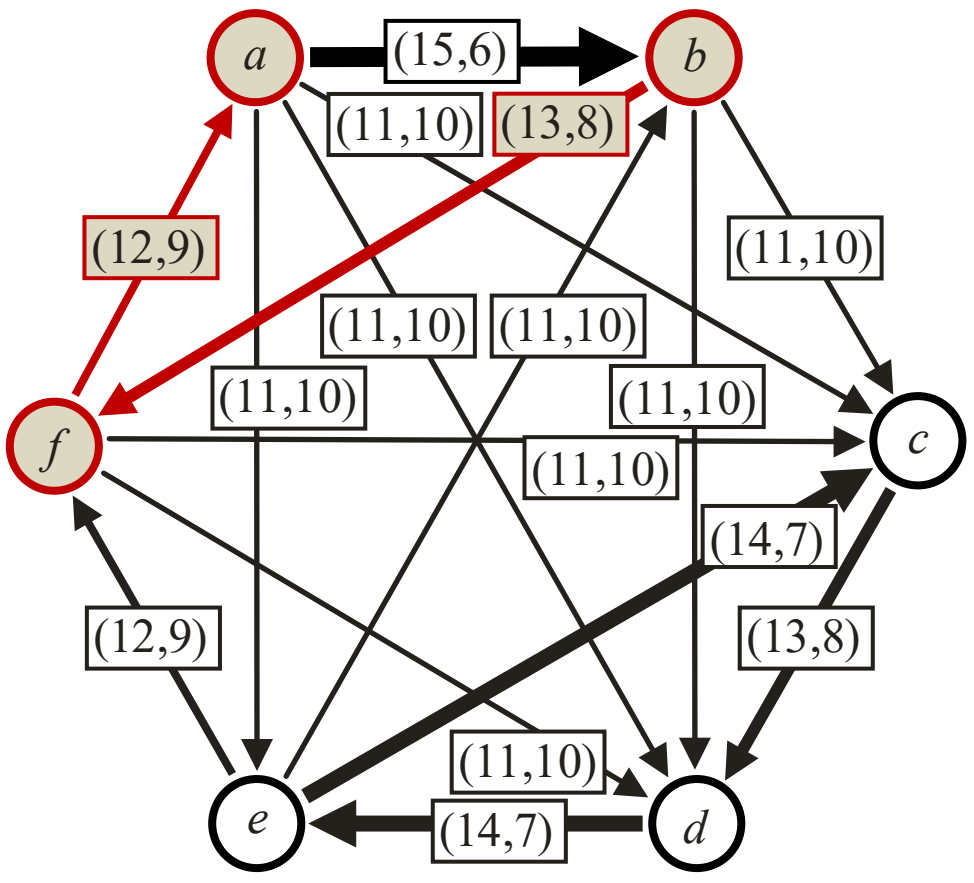

The strongest path from *b* to *a* is:
*b*, (13,8), *f*, (12,9), *a*

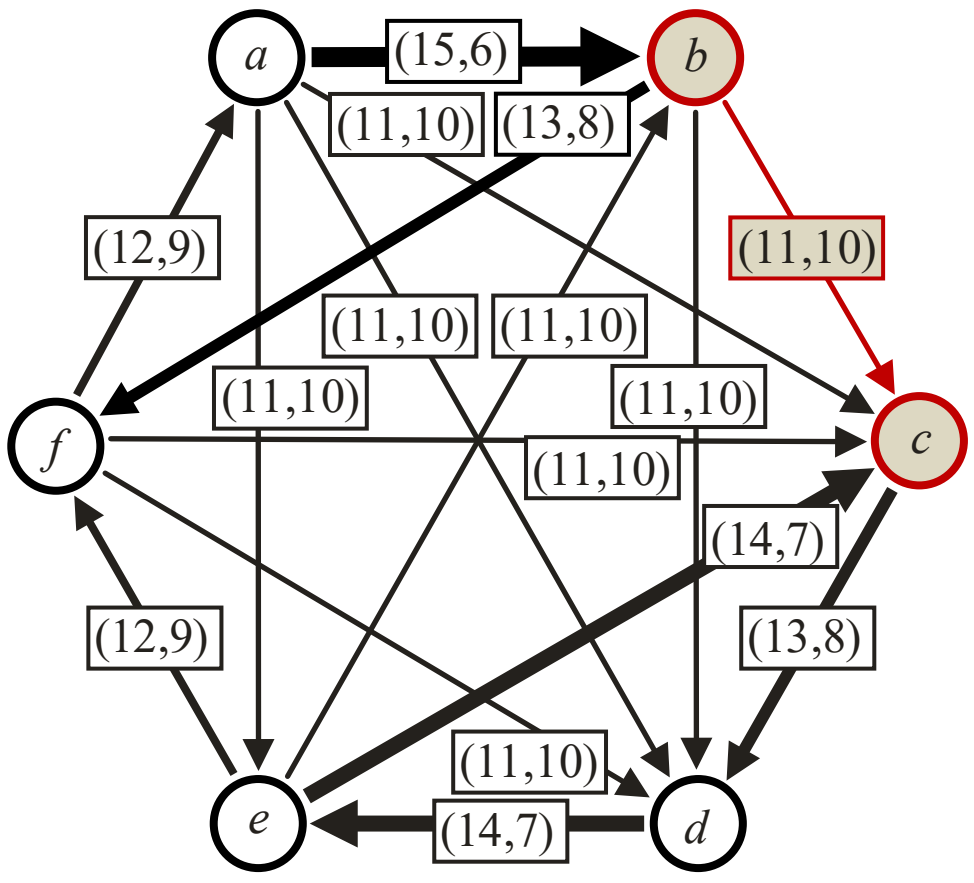

The strongest path from *b* to *c* is:
*b*, (11,10), *c*

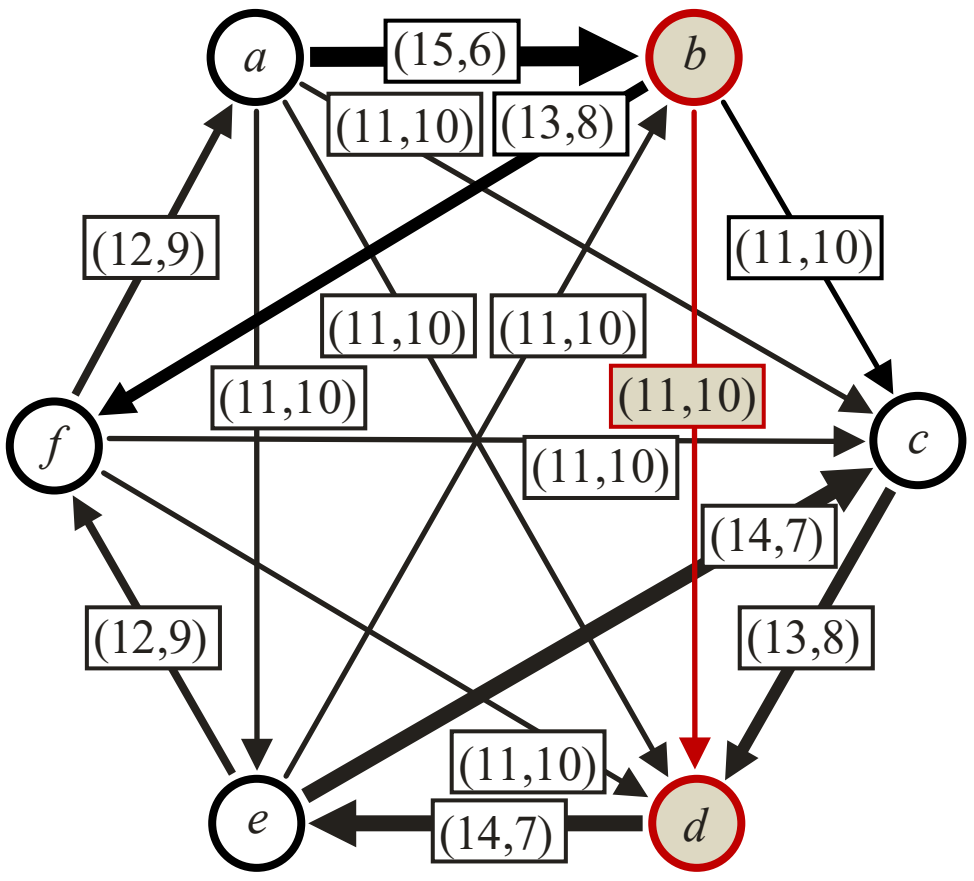

The strongest path from *b* to *d* is:
*b*, (11,10), *d*

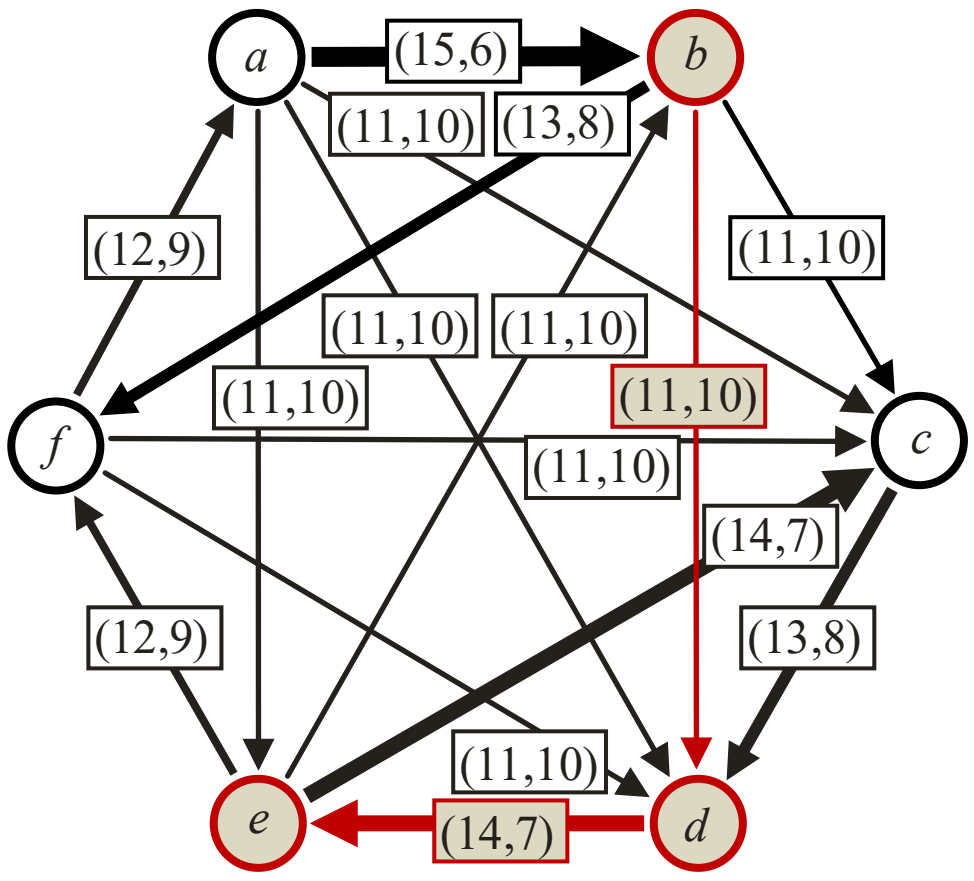

The strongest path from *b* to *e* is:
*b*, (11,10), *d*, (14,7), *e*

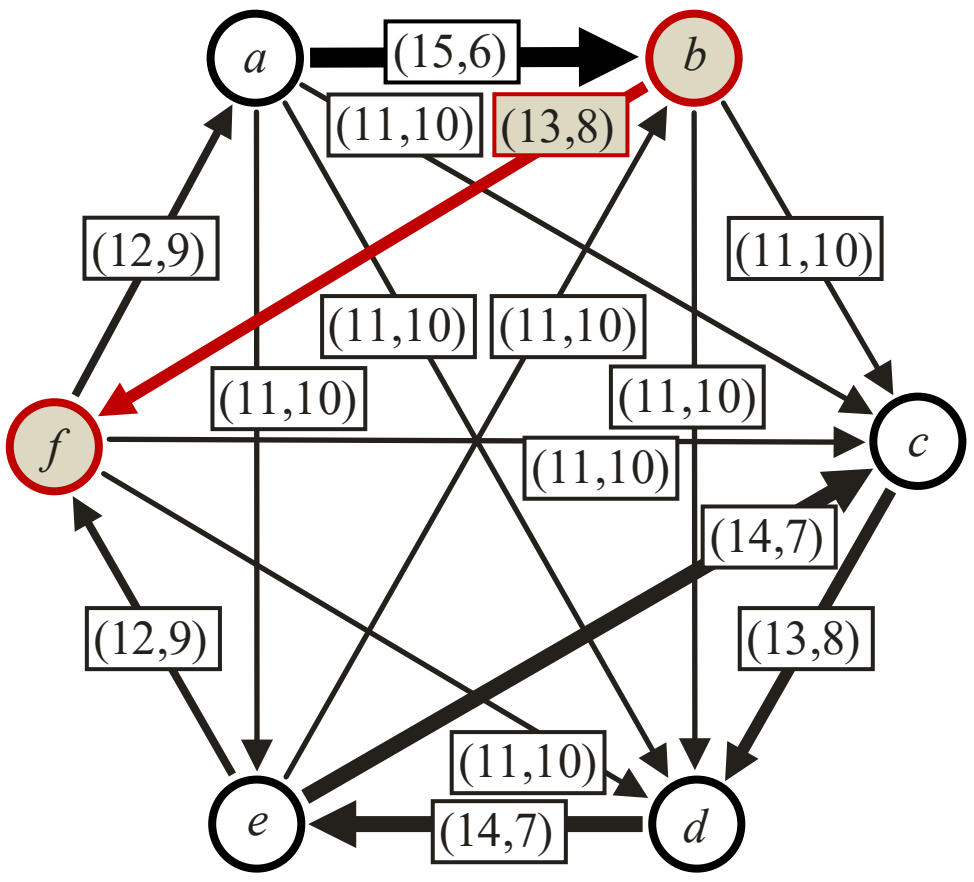

The strongest path from *b* to *f* is:
*b*, (13,8), *f*

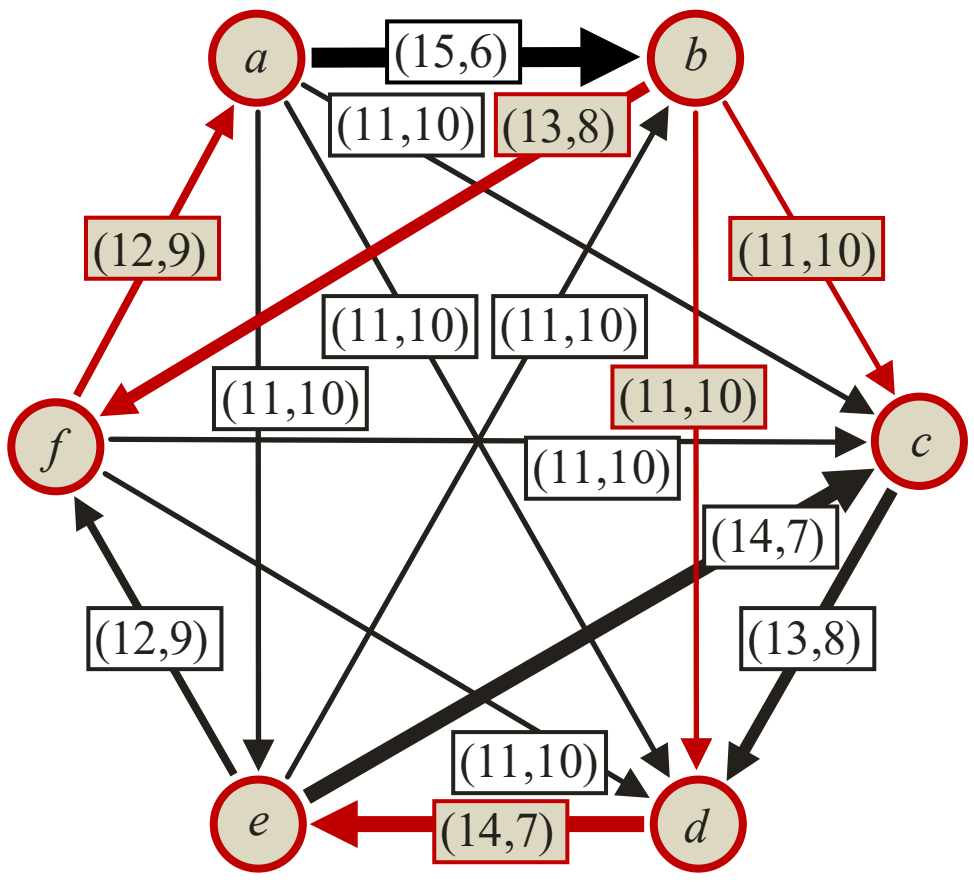

These are the strongest paths
from *b* to every other alternative.

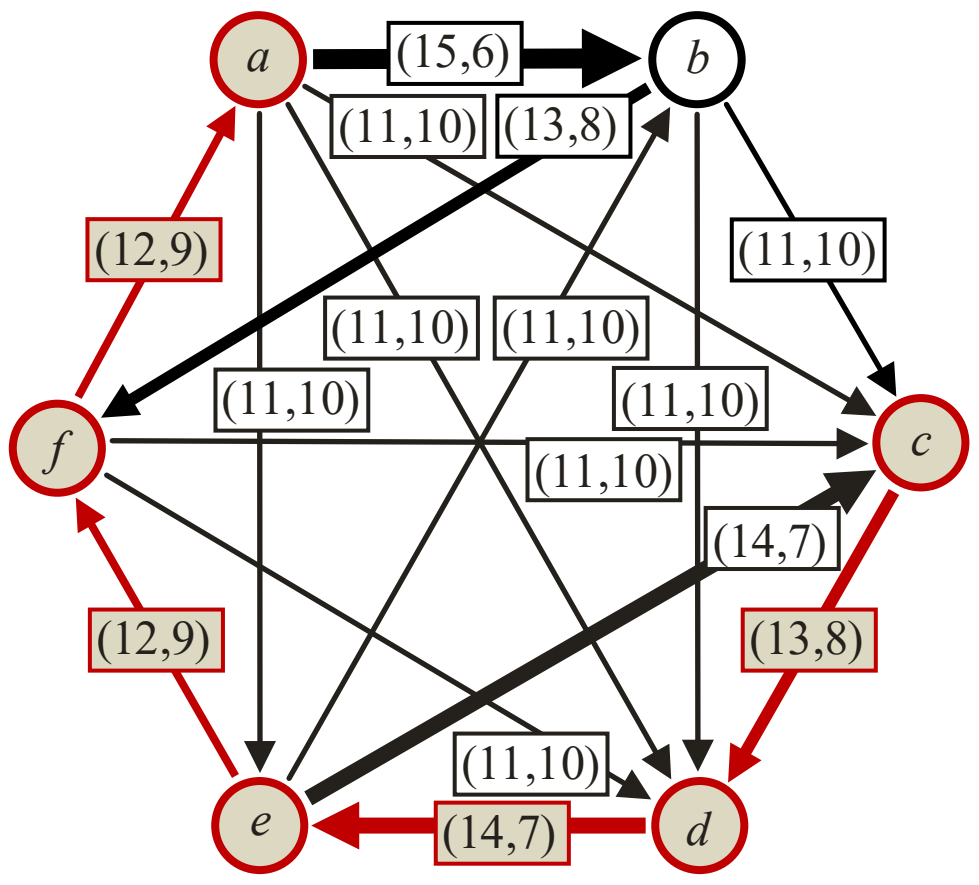

The strongest path from *c* to *a* is:
*c*, (13,8), *d*, (14,7), *e*, (12,9), *f*, (12,9), *a*

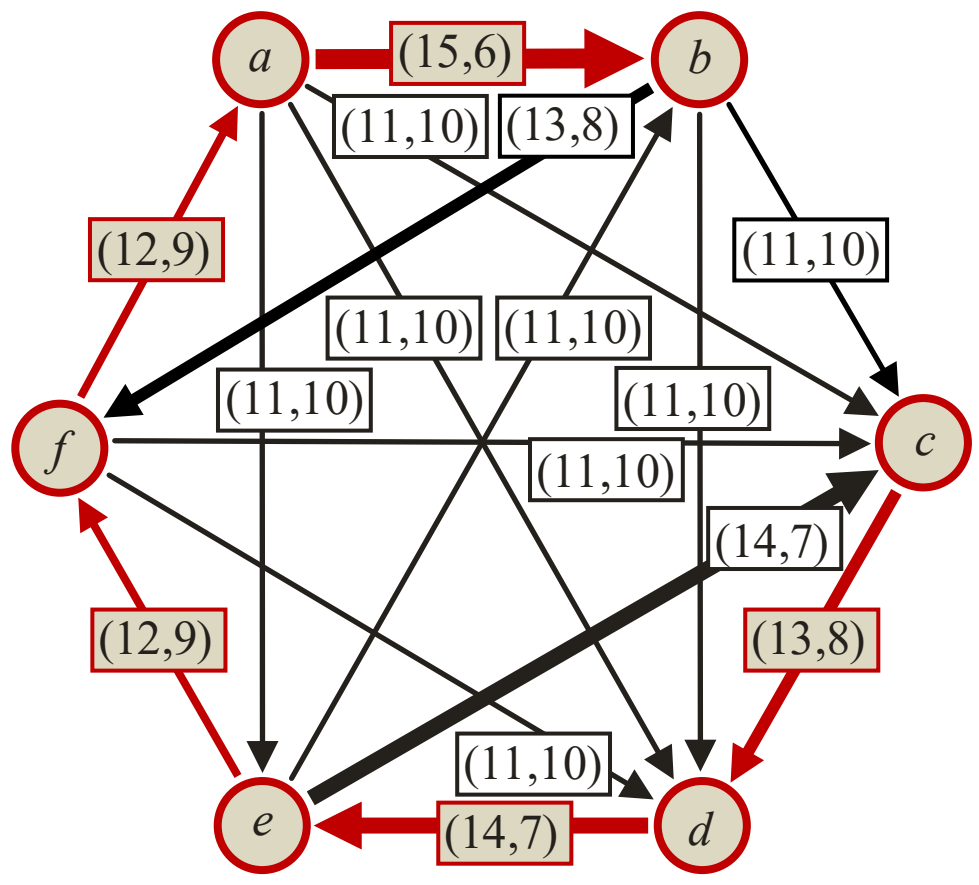

The strongest path from *c* to *b* is:
*c*, (13,8), *d*, (14,7), *e*, (12,9), *f*, (12,9), *a*, (15,6), *b*

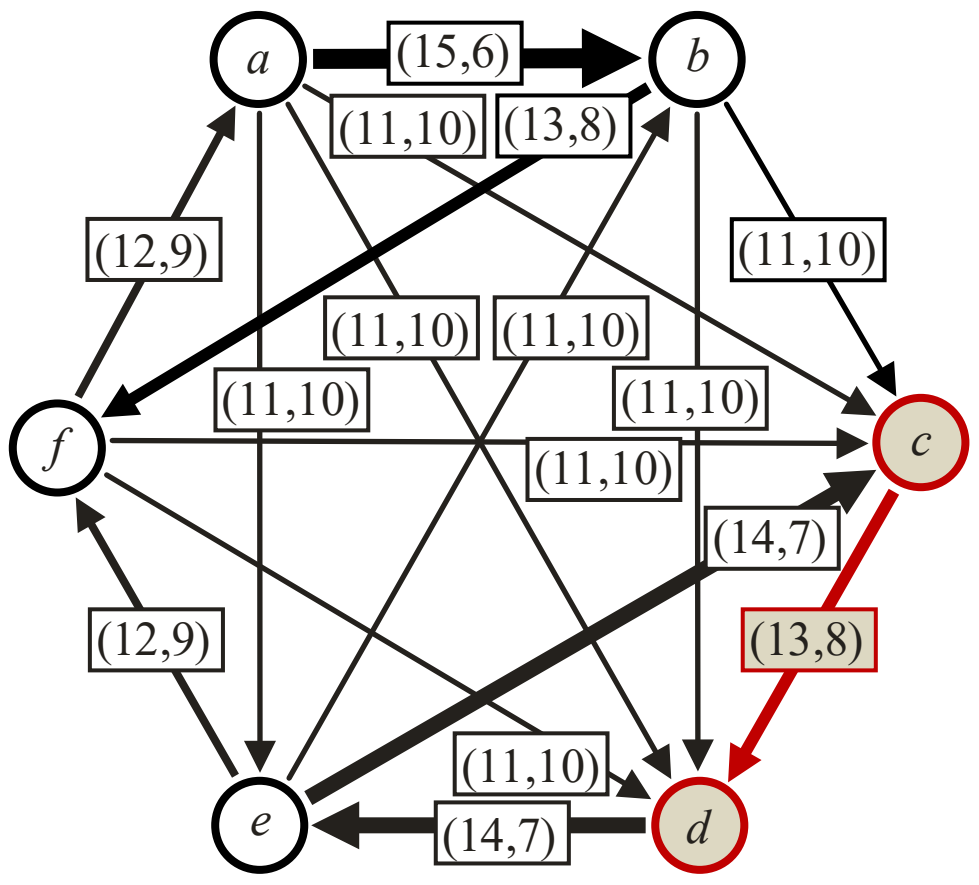

The strongest path from *c* to *d* is:
*c*, (13,8), *d*

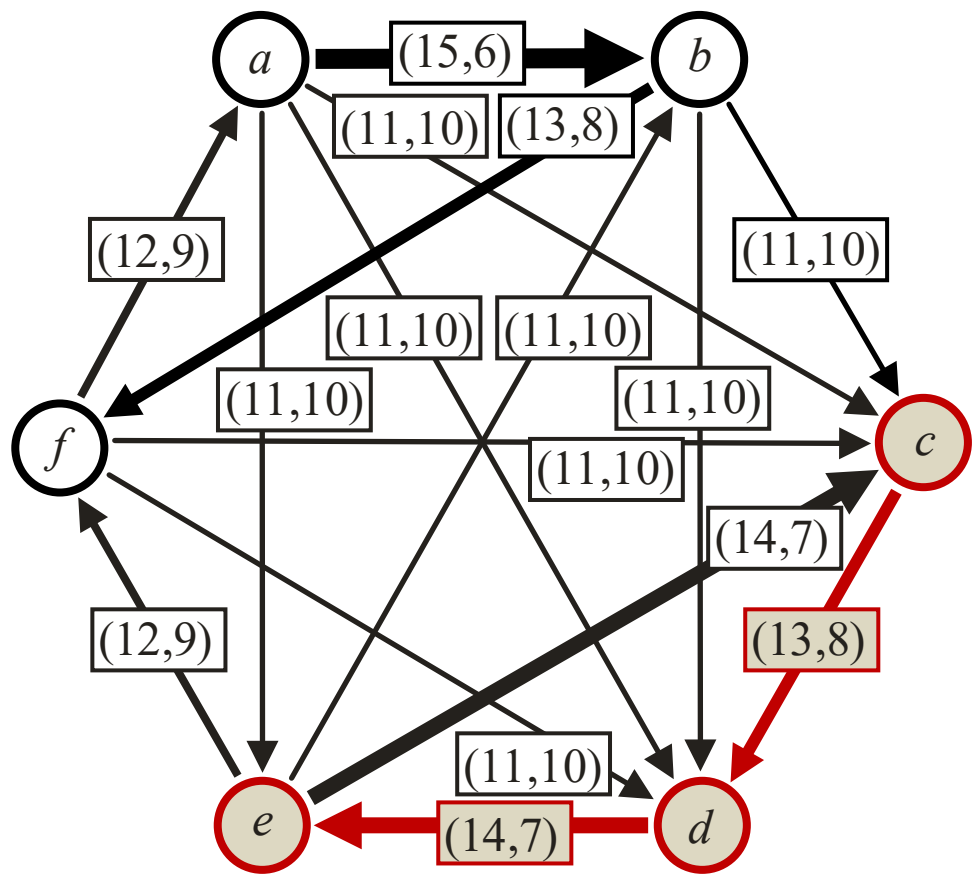

The strongest path from *c* to *e* is:
*c*, (13,8), *d*, (14,7), *e*

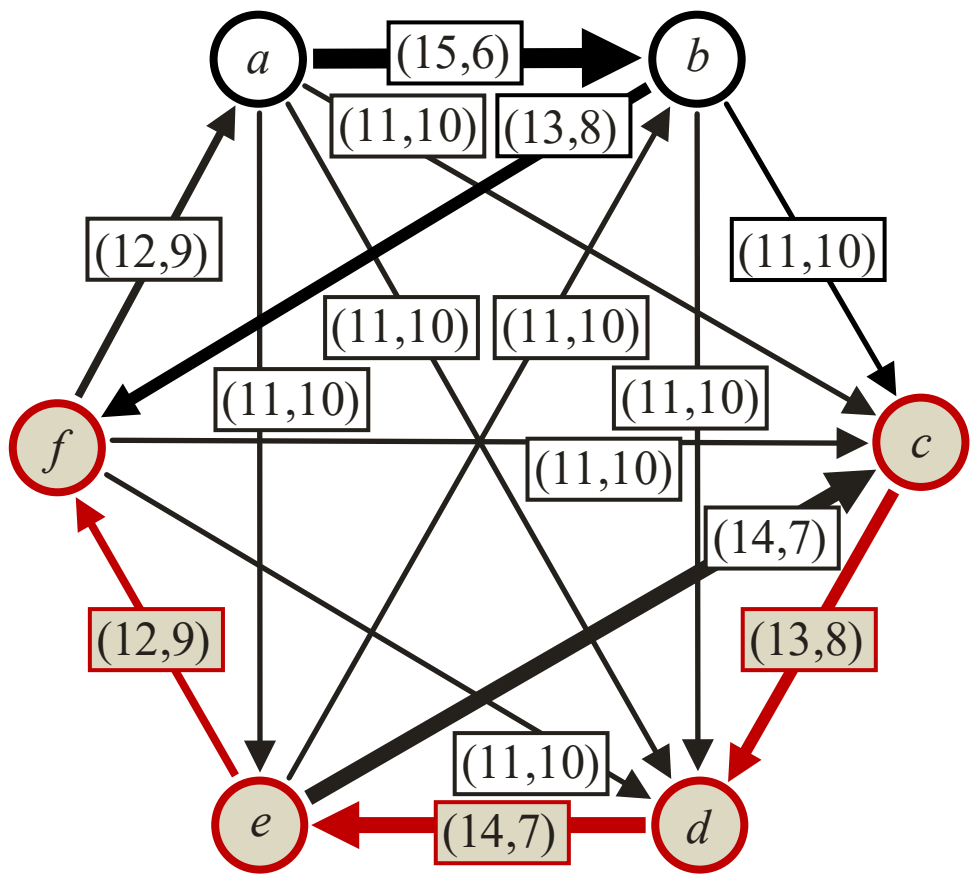

The strongest path from *c* to *f* is:
*c*, (13,8), *d*, (14,7), *e*, (12,9), *f*

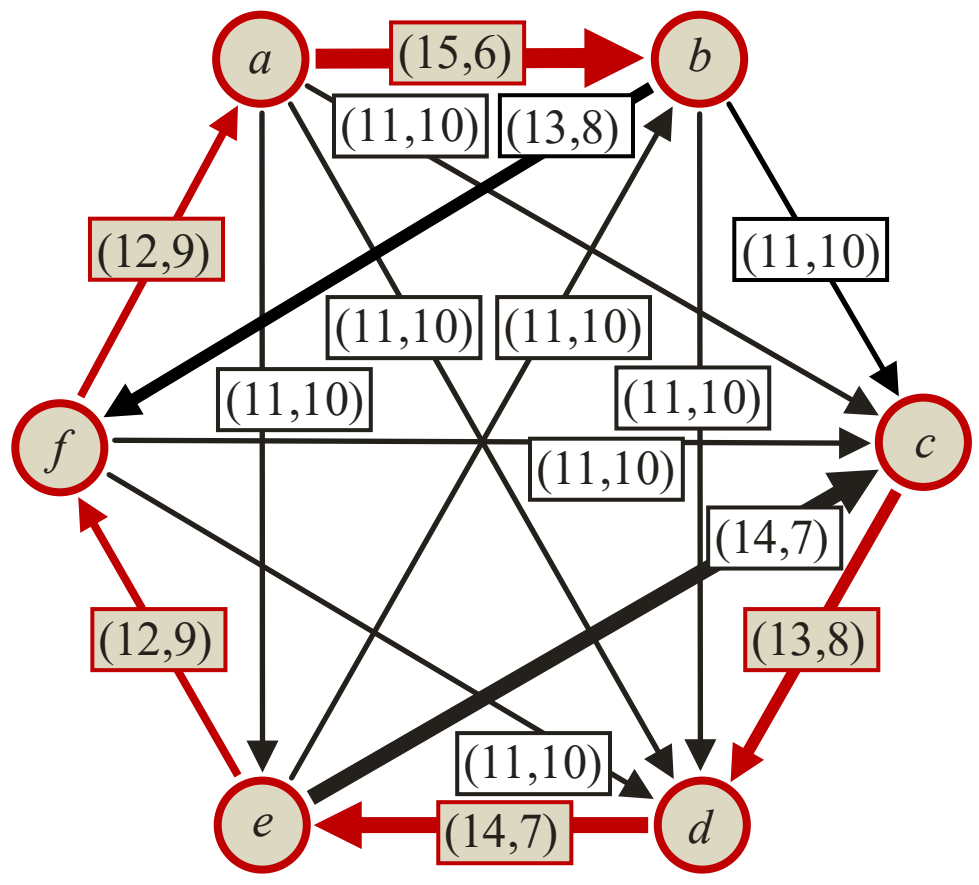

These are the strongest paths
from *c* to every other alternative.

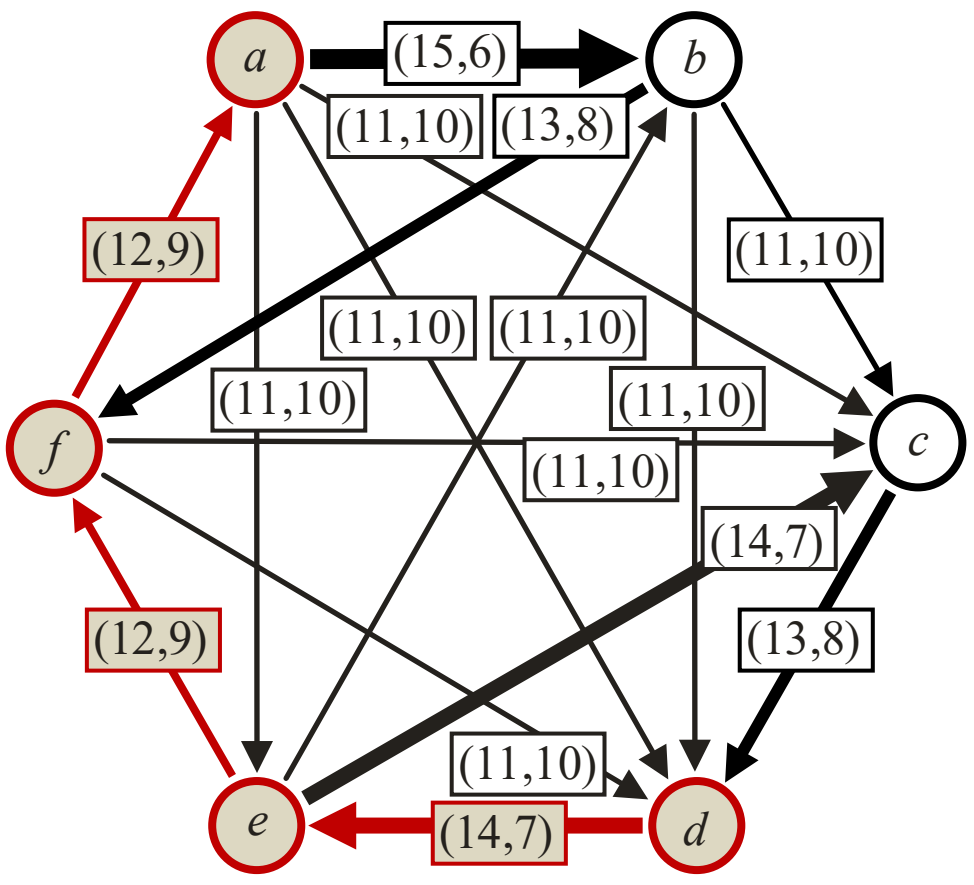

The strongest path from *d* to *a* is:
*d*, (14,7), *e*, (12,9), *f*, (12,9), *a*

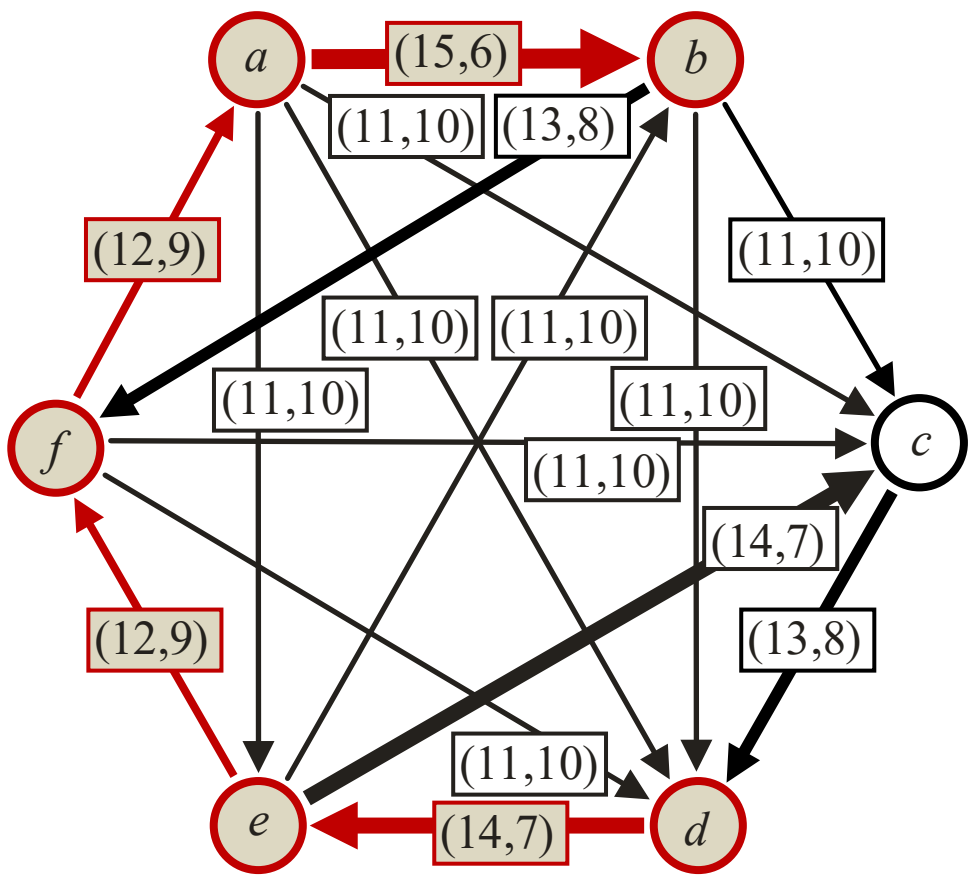

The strongest path from *d* to *b* is:
*d*, (14,7), *e*, (12,9), *f*, (12,9), *a*, (15,6), *b*

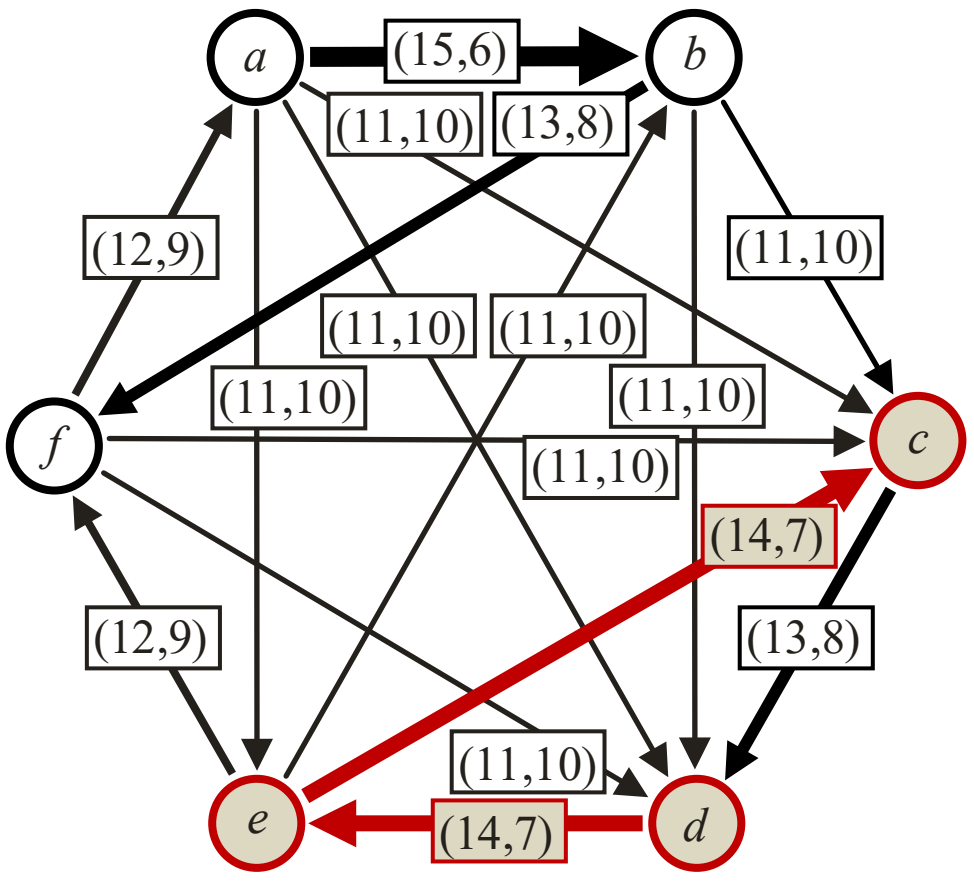

The strongest path from *d* to *c* is:
*d*, (14,7), *e*, (14,7), *c*

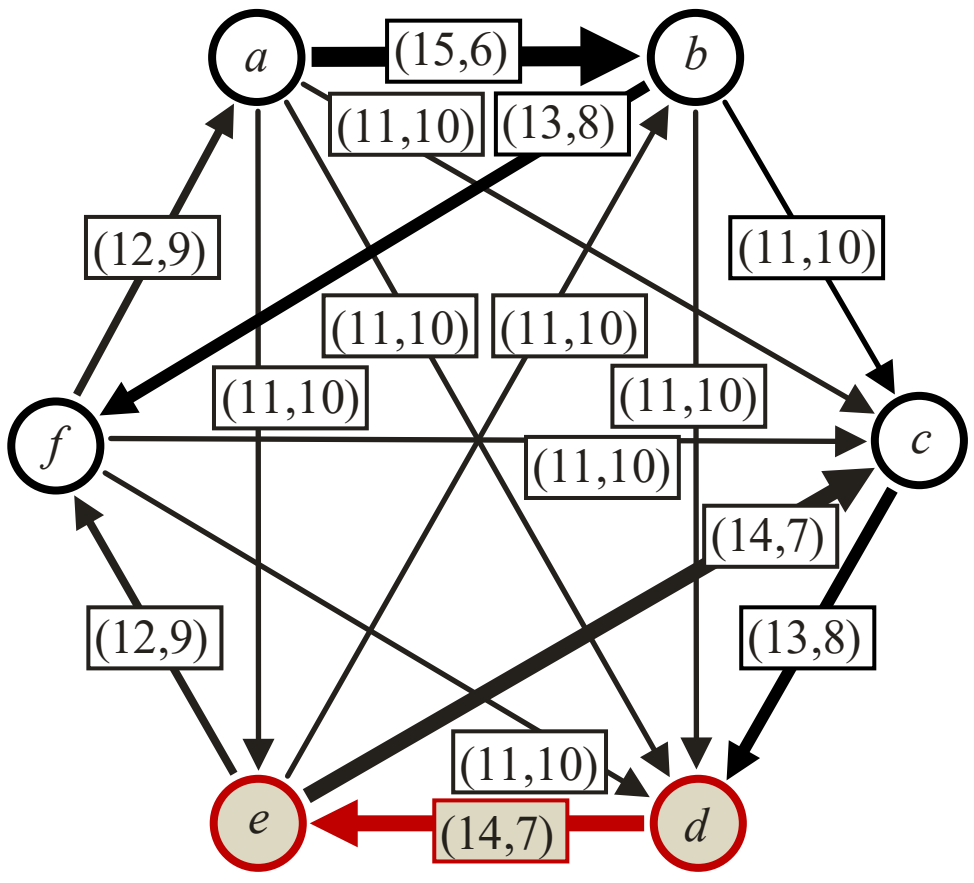

The strongest path from *d* to *e* is:
*d*, (14,7), *e*

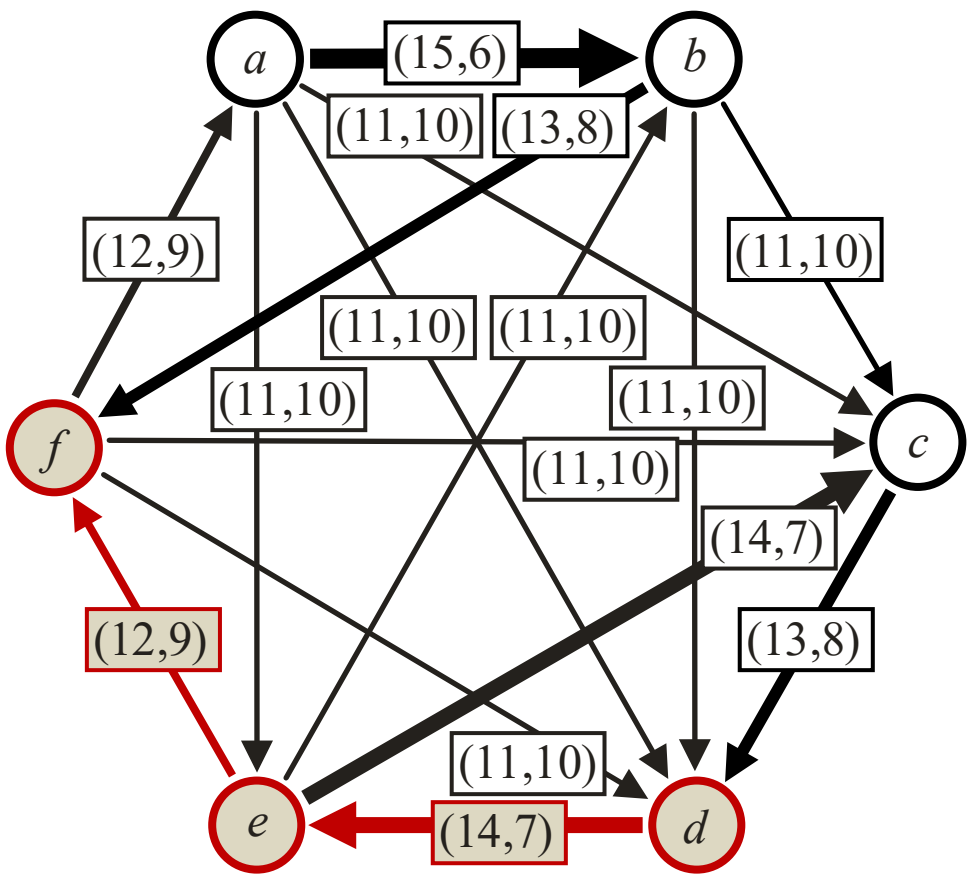

The strongest path from *d* to *f* is:
*d*, (14,7), *e*, (12,9), *f*

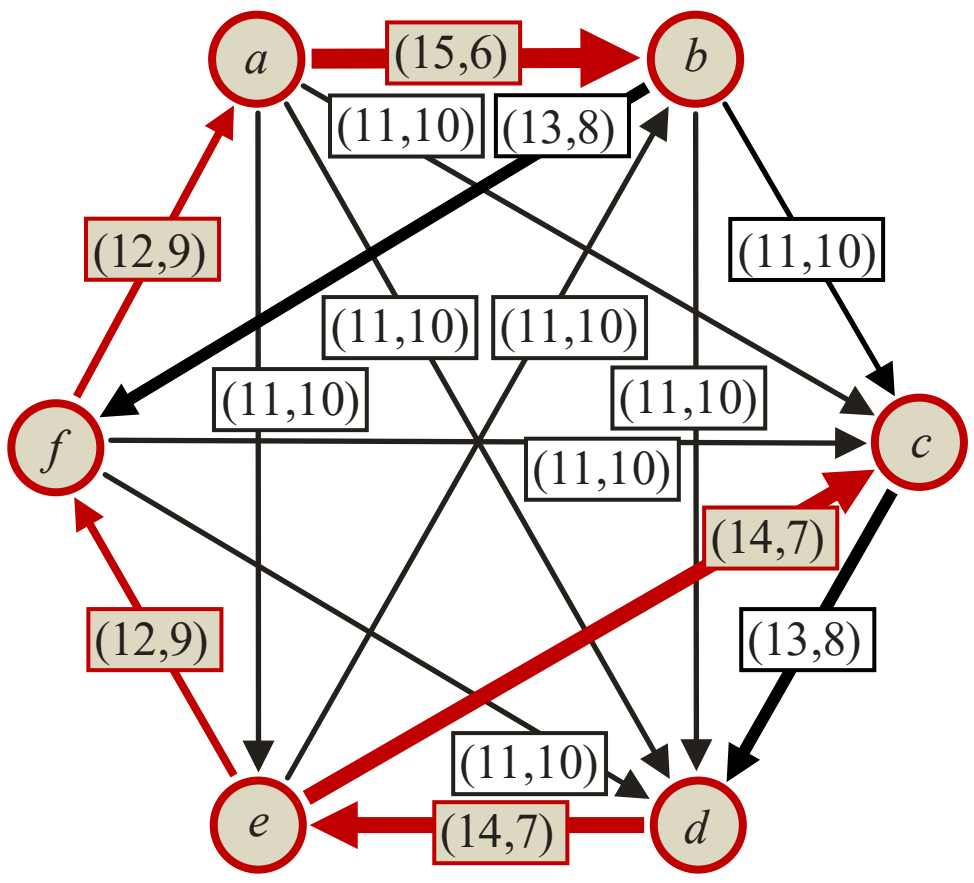

These are the strongest paths
from *d* to every other alternative.

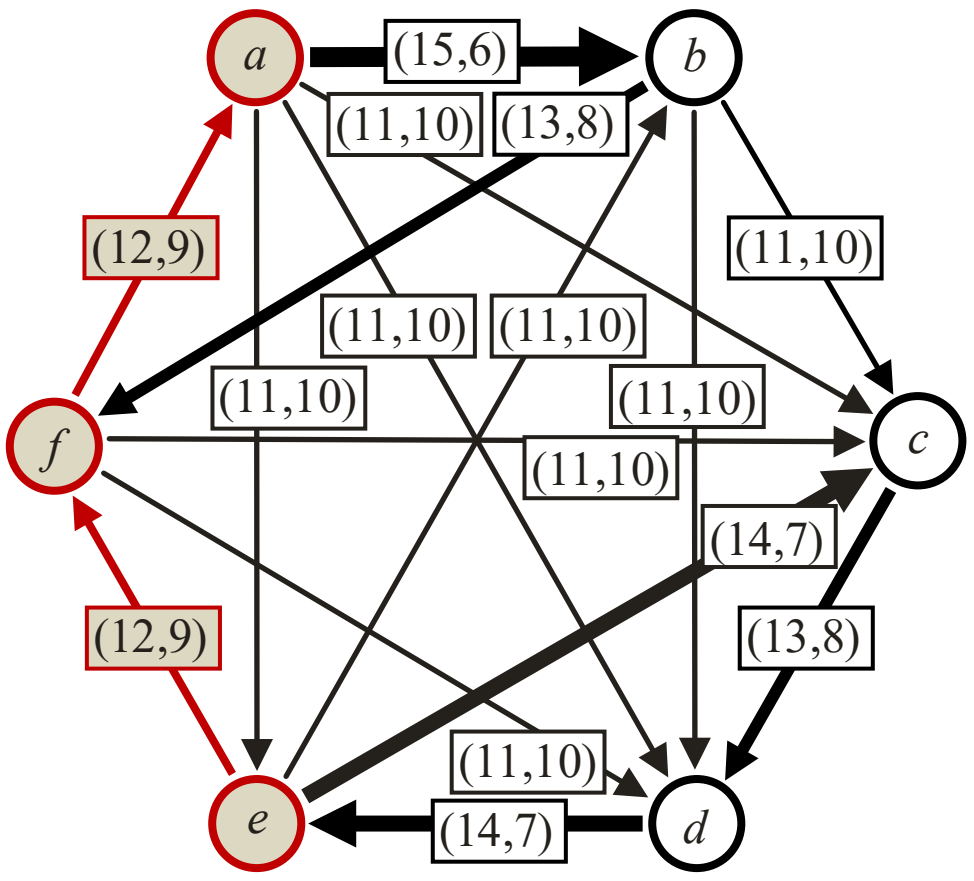

The strongest path from *e* to *a* is:
*e*, (12,9), *f*, (12,9), *a*

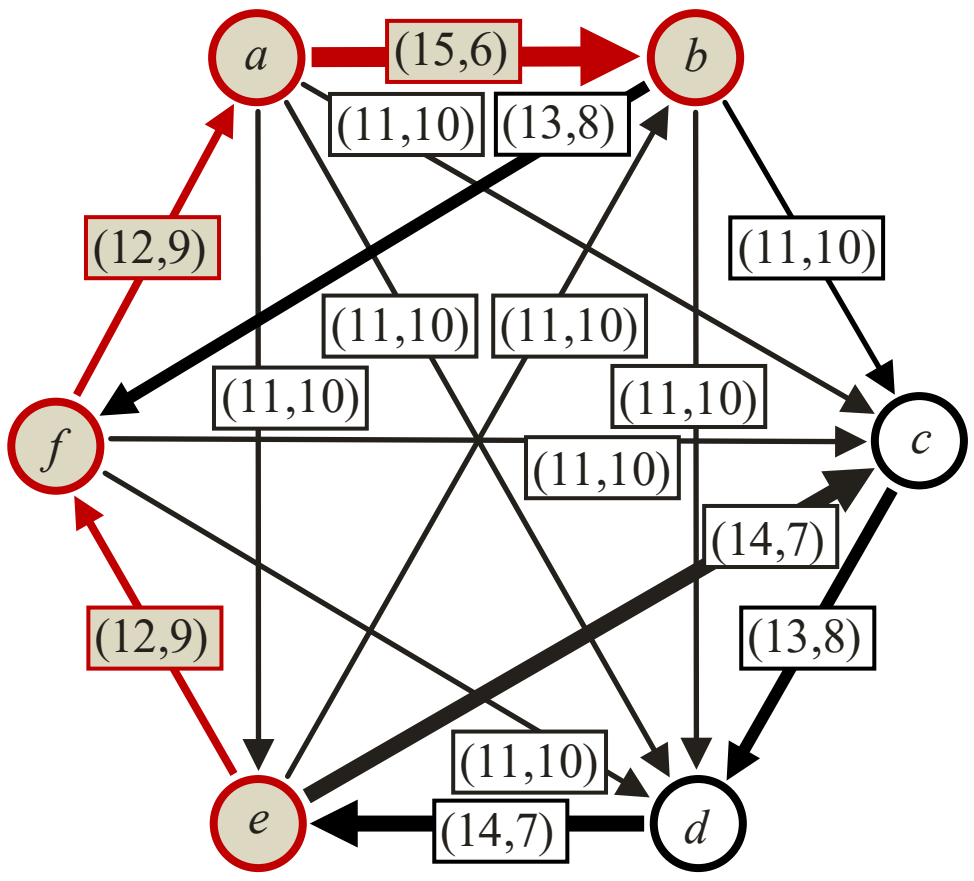

The strongest path from *e* to *b* is:
*e*, (12,9), *f*, (12,9), *a*, (15,6), *b*

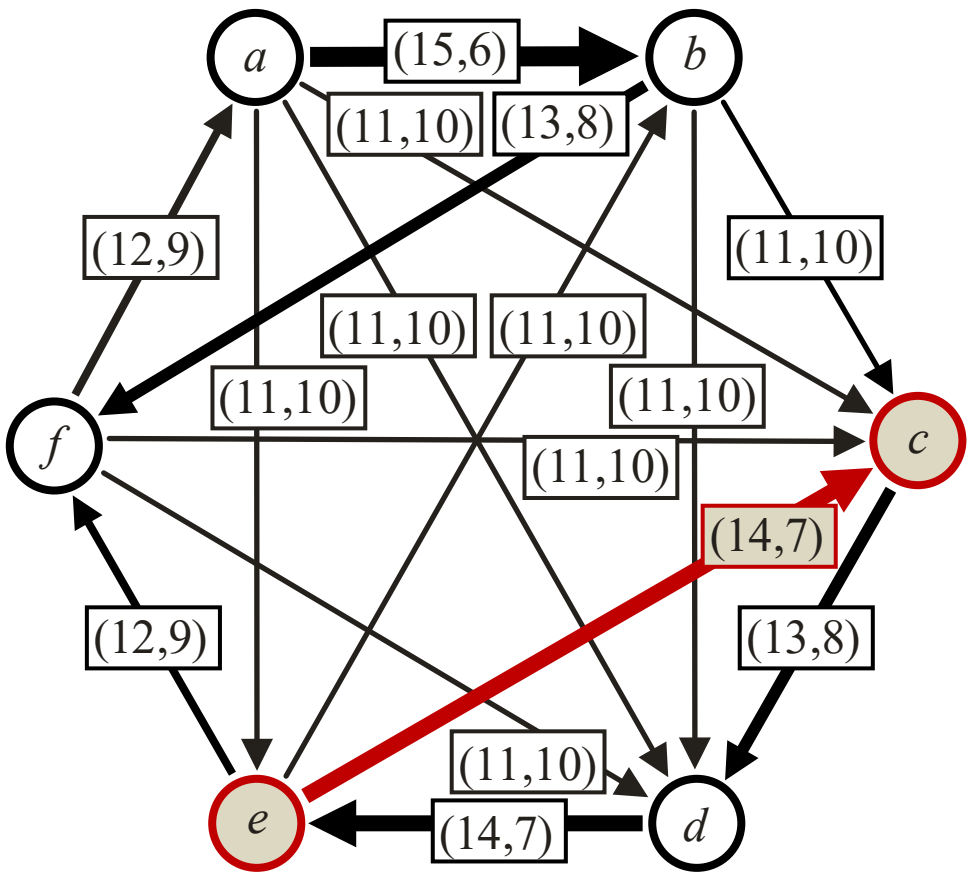

The strongest path from *e* to *c* is:
*e*, (14,7), *c*

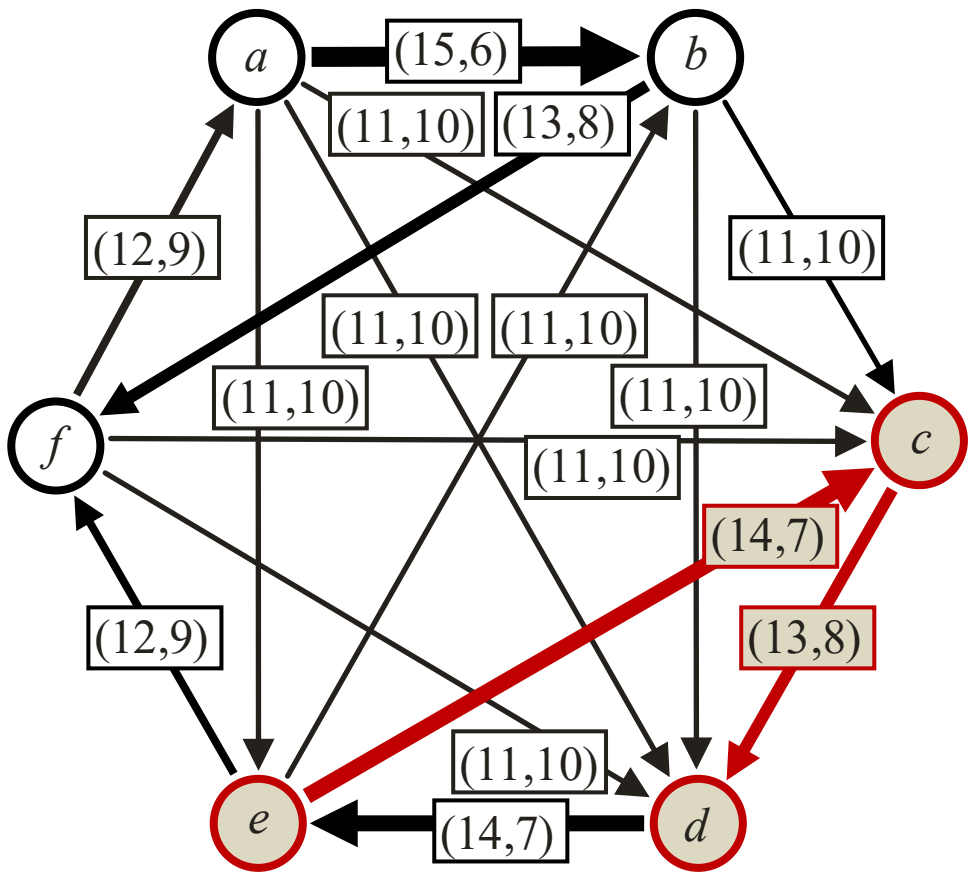

The strongest path from *e* to *d* is:
*e*, (14,7), *c*, (13,8), *d*

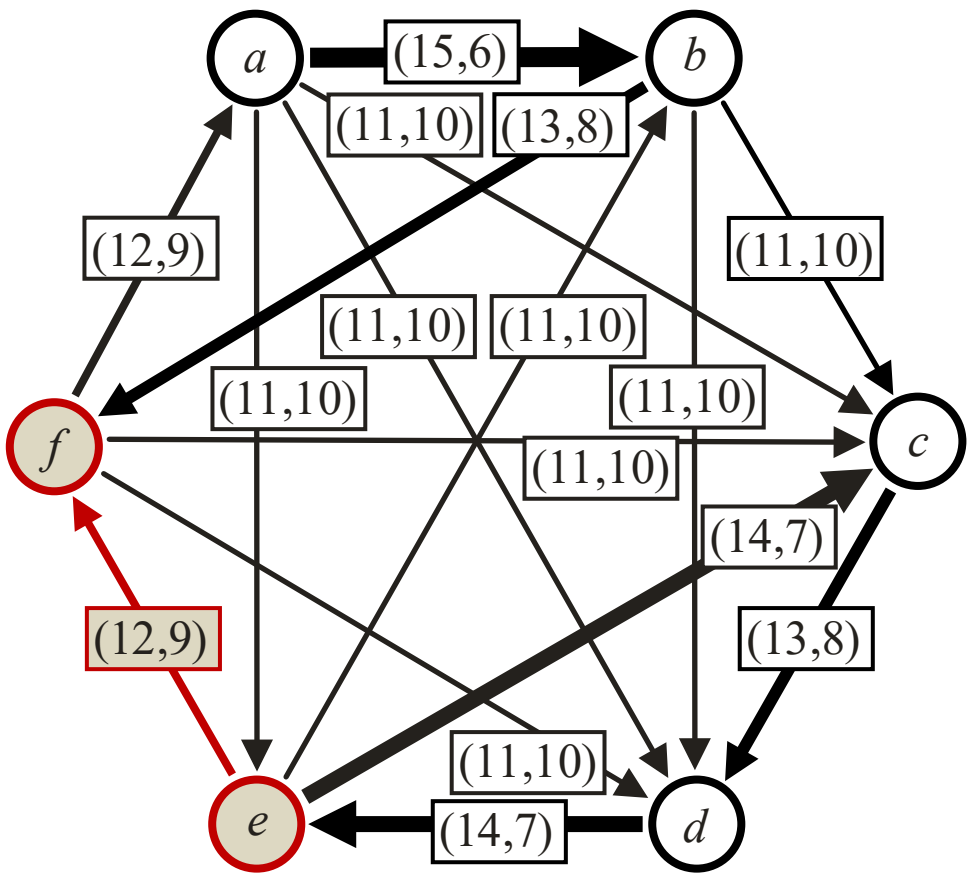

The strongest path from *e* to *f* is:
*e*, (12,9), *f*

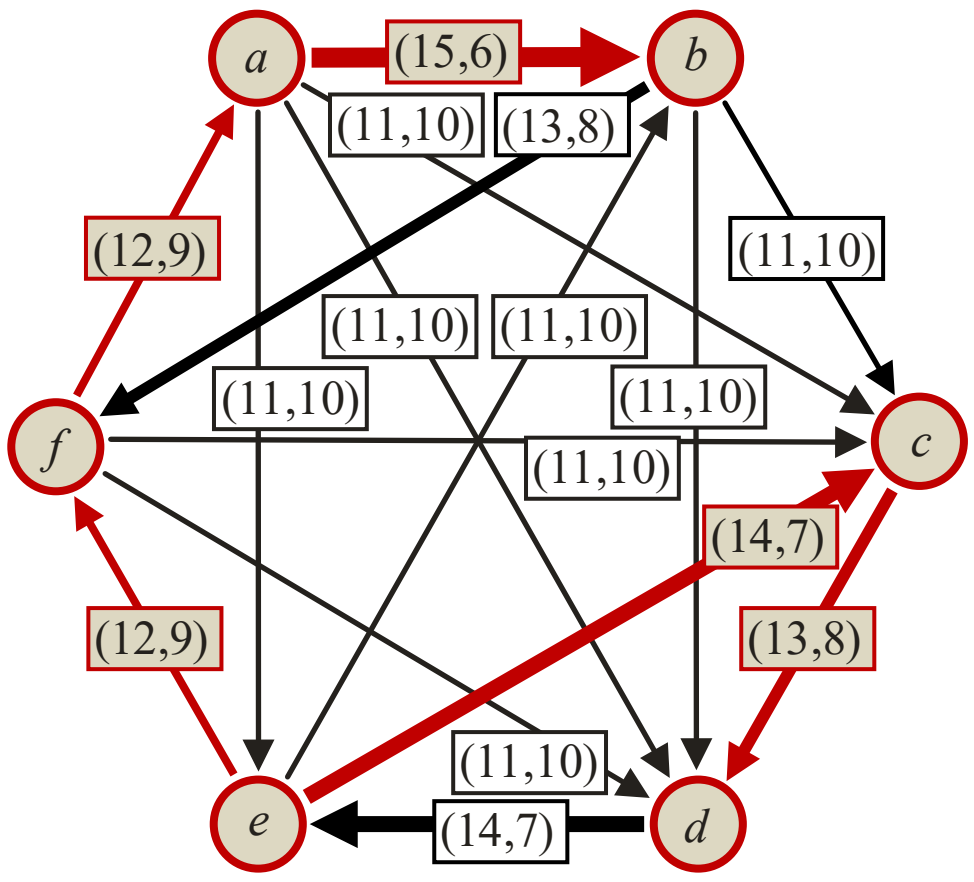

These are the strongest paths
from *e* to every other alternative.

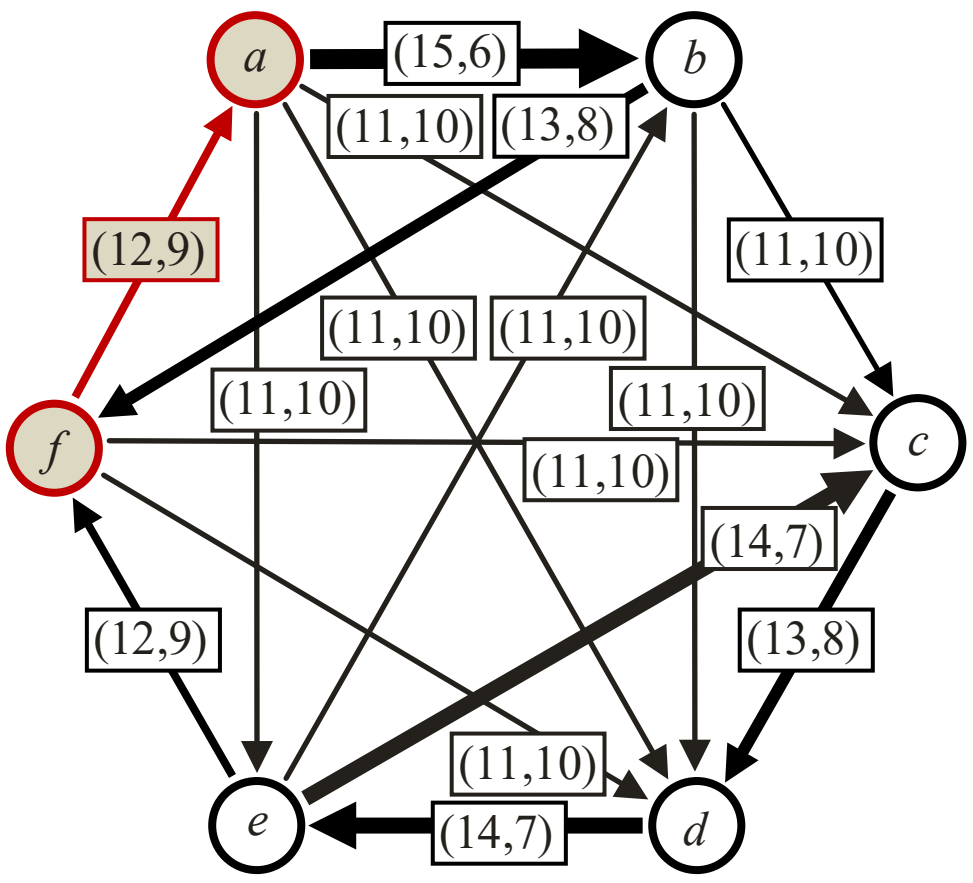

The strongest path from *f* to *a* is:
*f*, (12,9), *a*

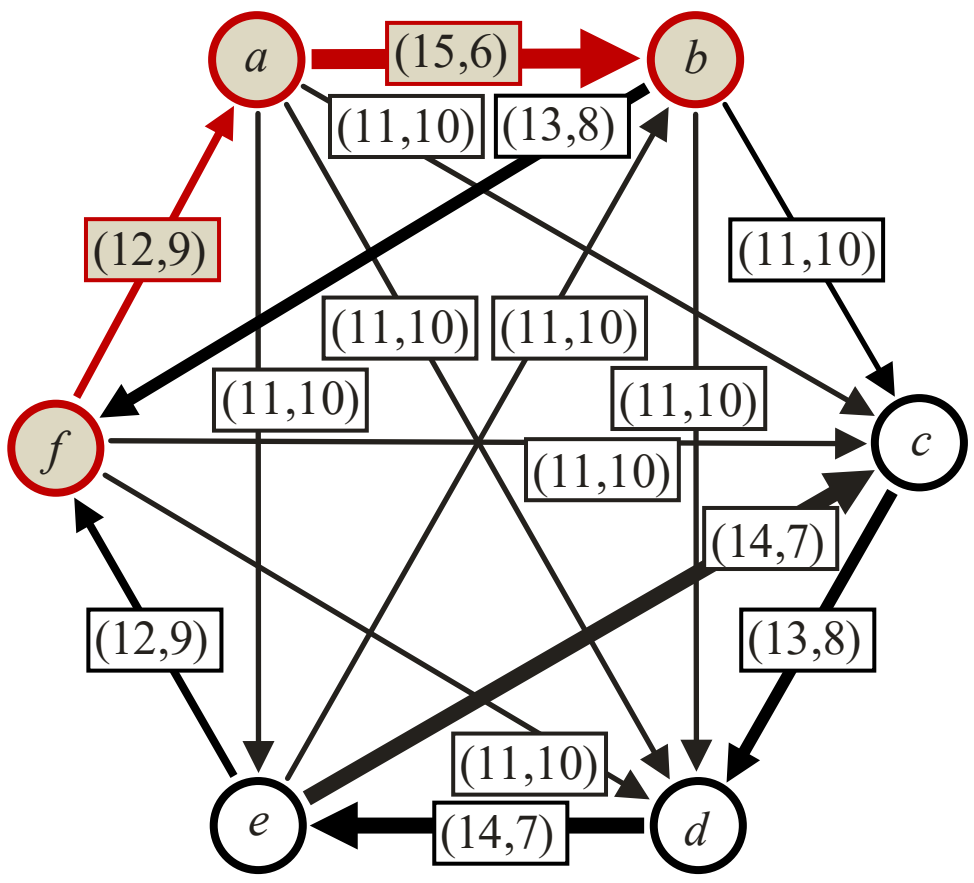

The strongest path from *f* to *b* is:
*f*, (12,9), *a*, (15,6), *b*

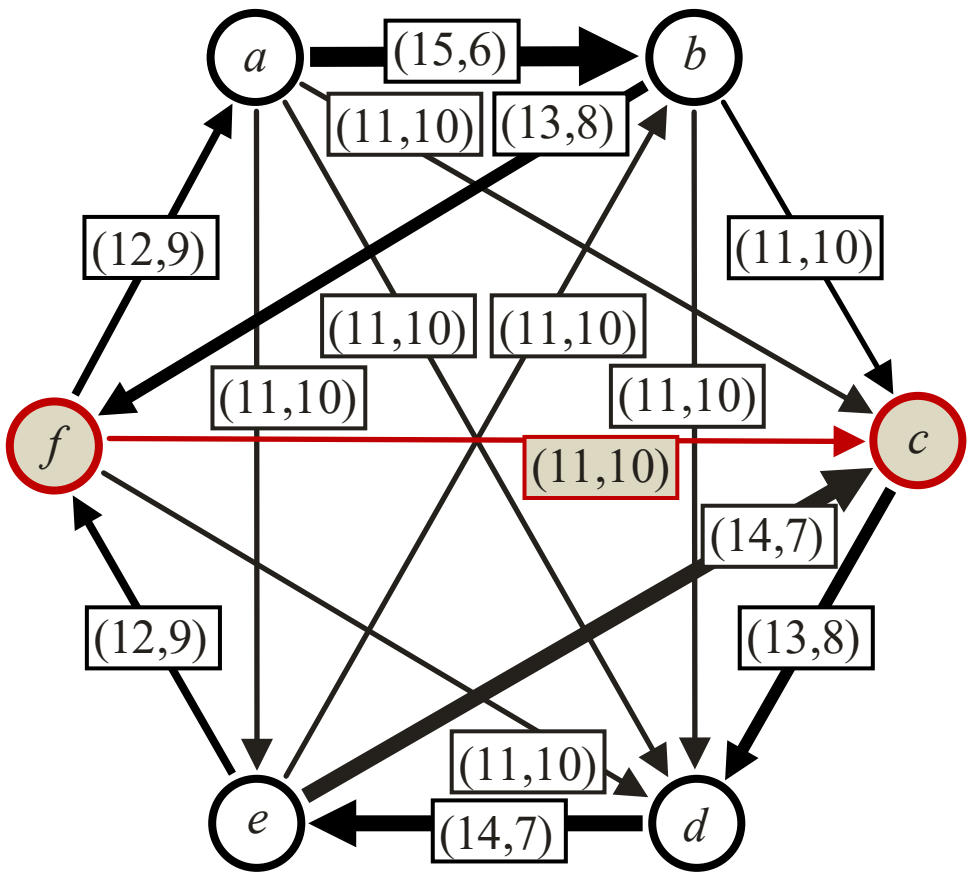

The strongest path from *f* to *c* is:
*f*, (11,10), *c*

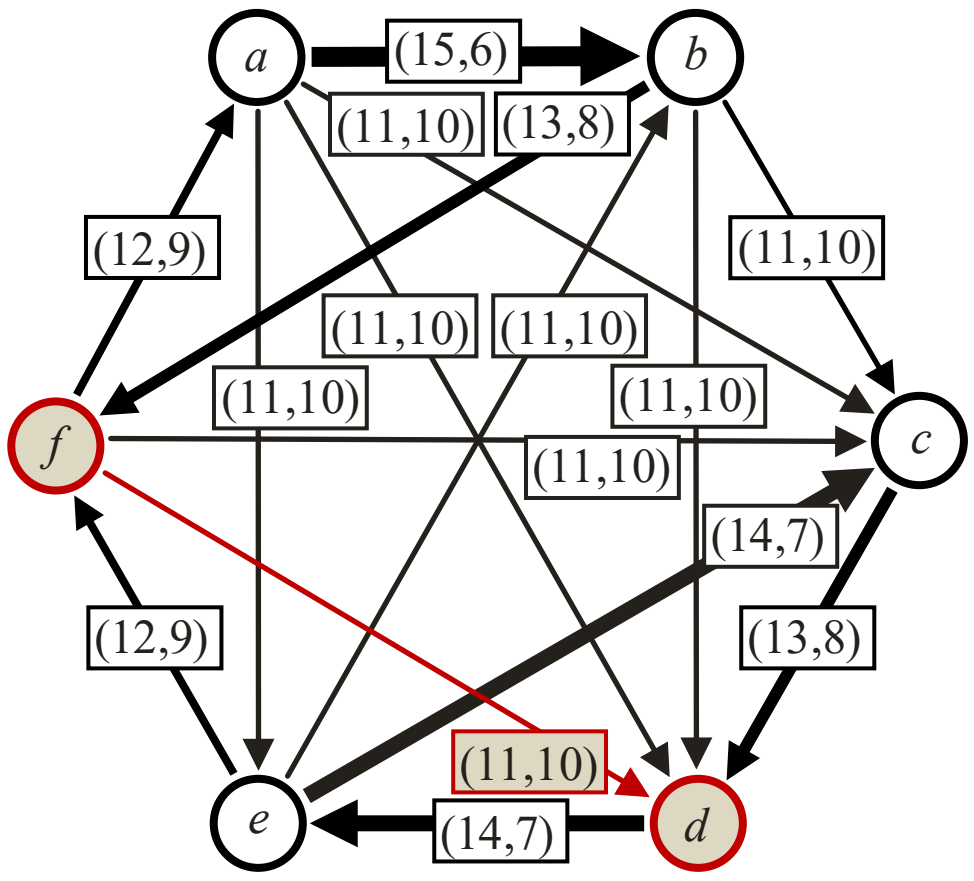

The strongest path from *f* to *d* is:
*f*, (11,10), *d*

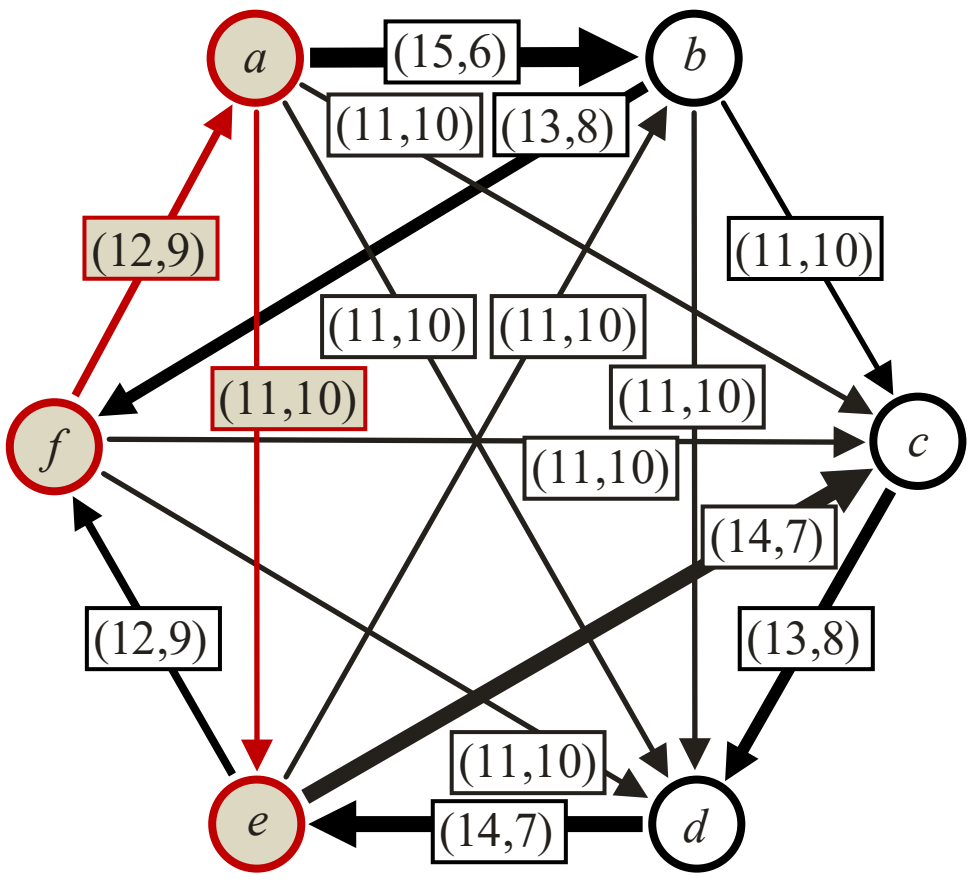

The strongest path from *f* to *e* is:
*f*, (12,9), *a*, (11,10), *e*

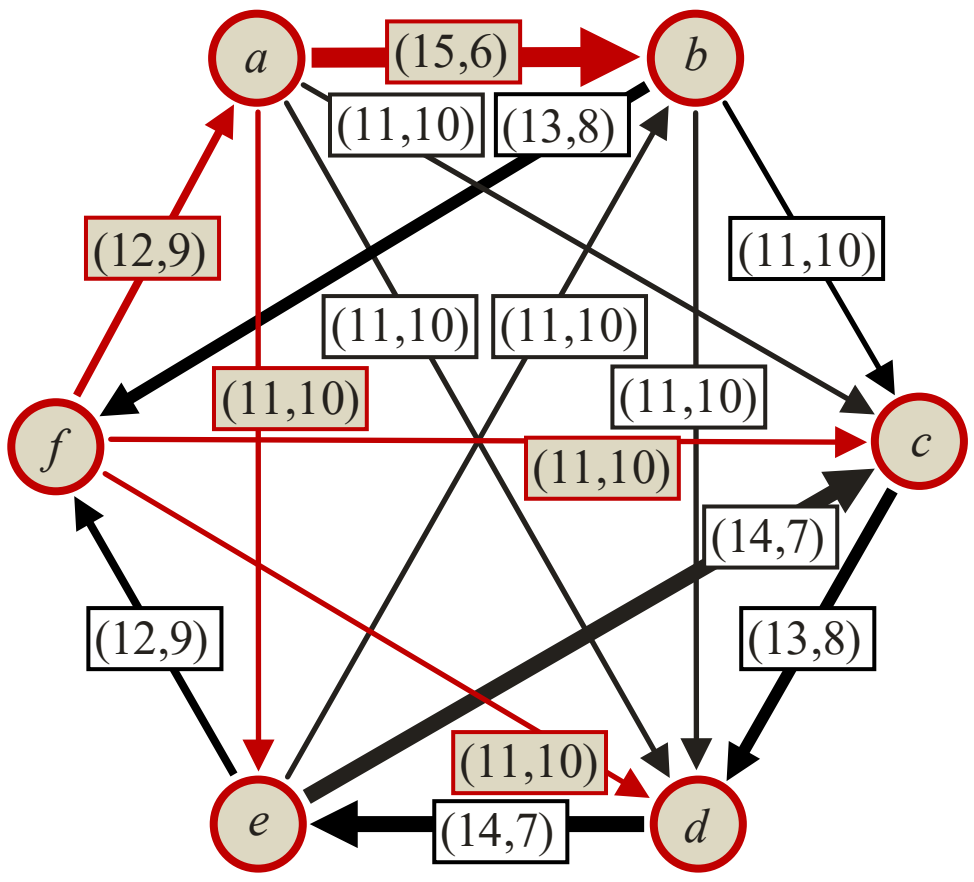

These are the strongest paths
from *f* to every other alternative.

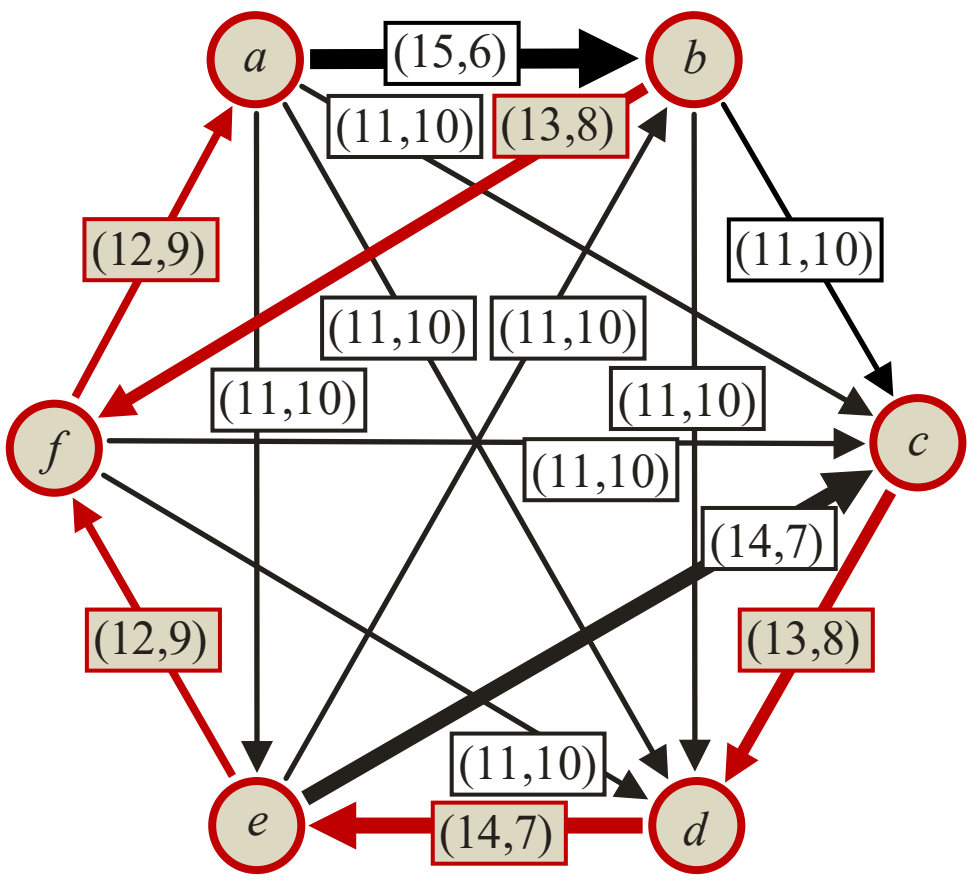

These are the strongest paths
from every other alternative to *a*.

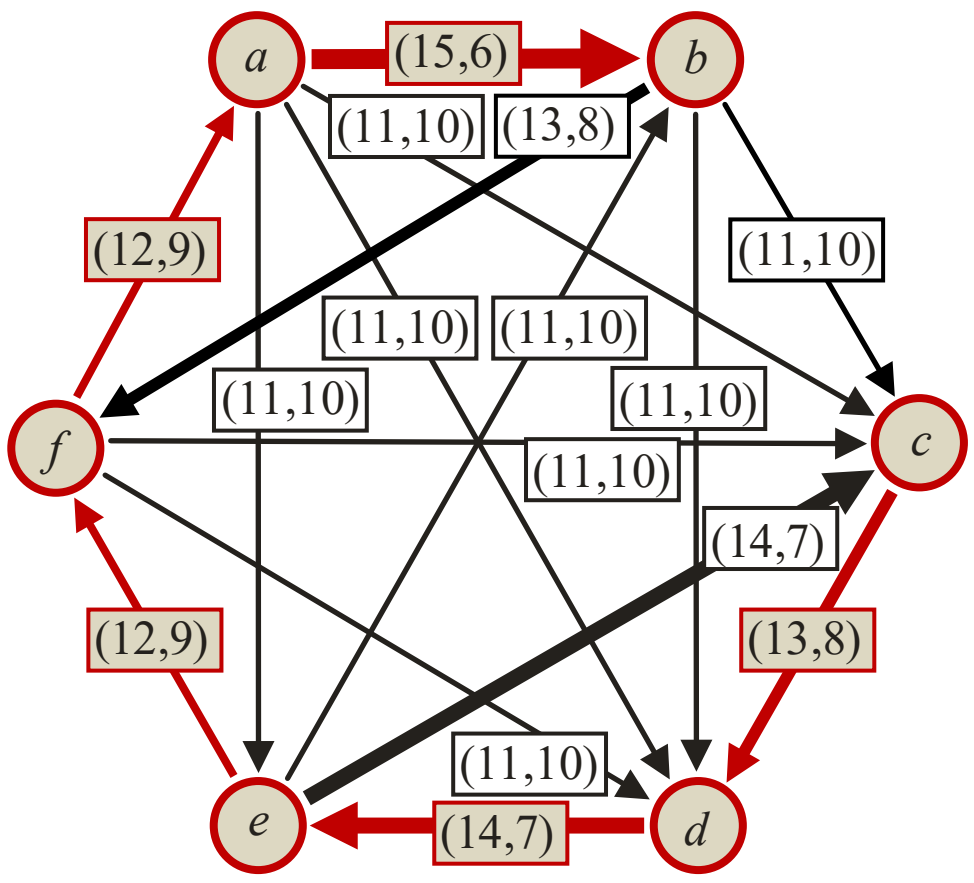

These are the strongest paths
from every other alternative to *b*.

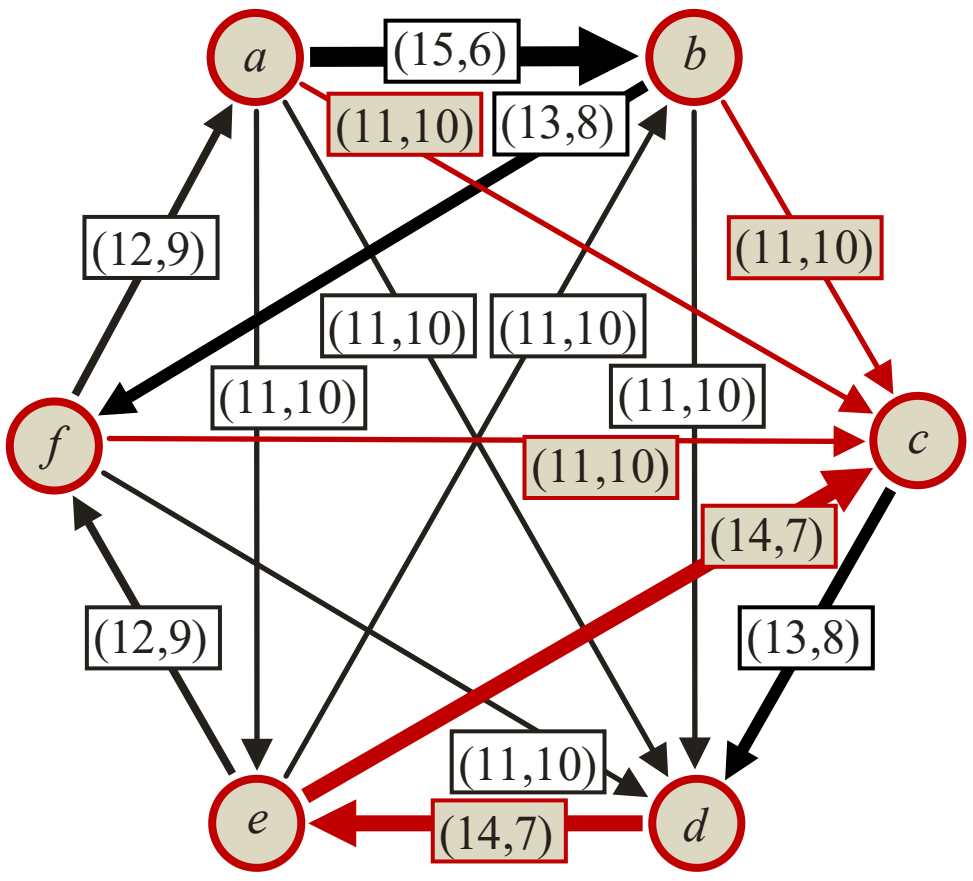

These are the strongest paths
from every other alternative to *c*.

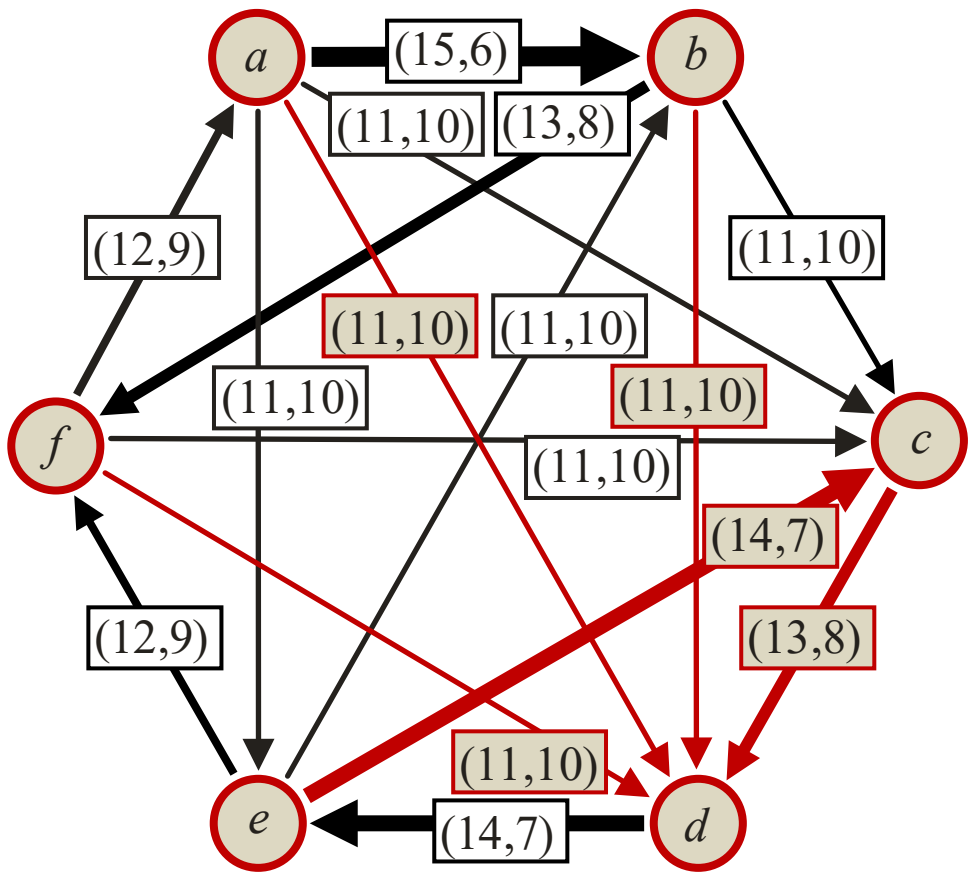

These are the strongest paths
from every other alternative to *d*.

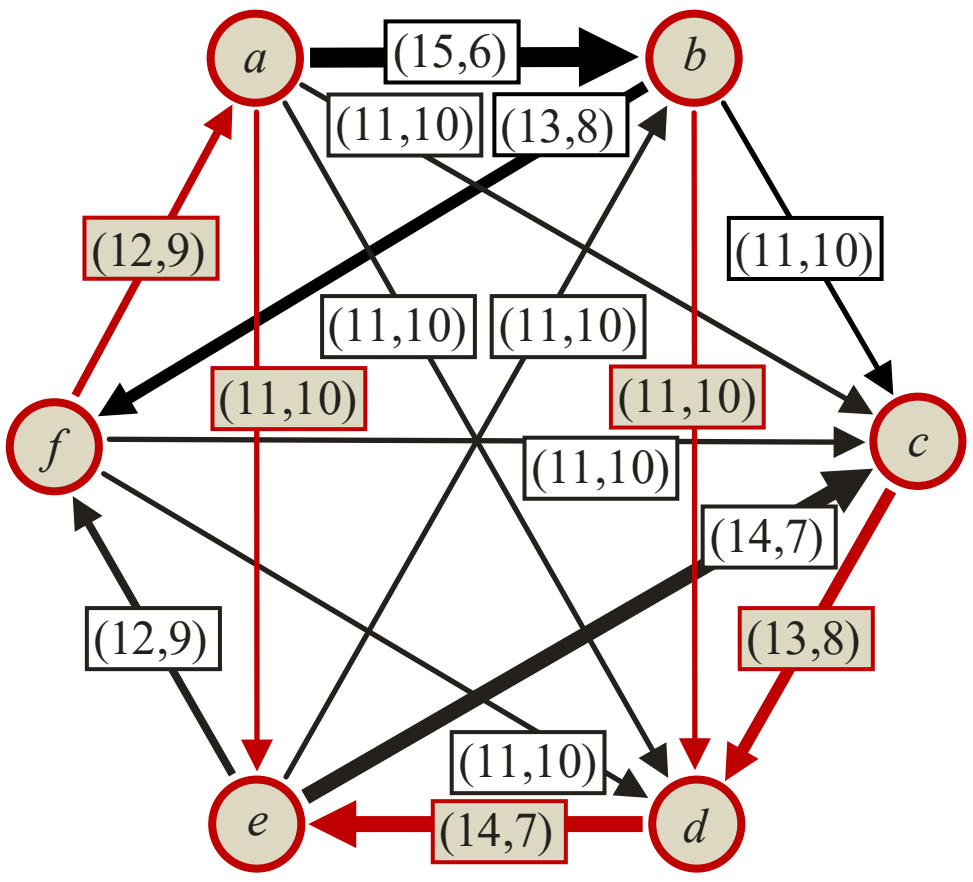

These are the strongest paths
from every other alternative to *e*.

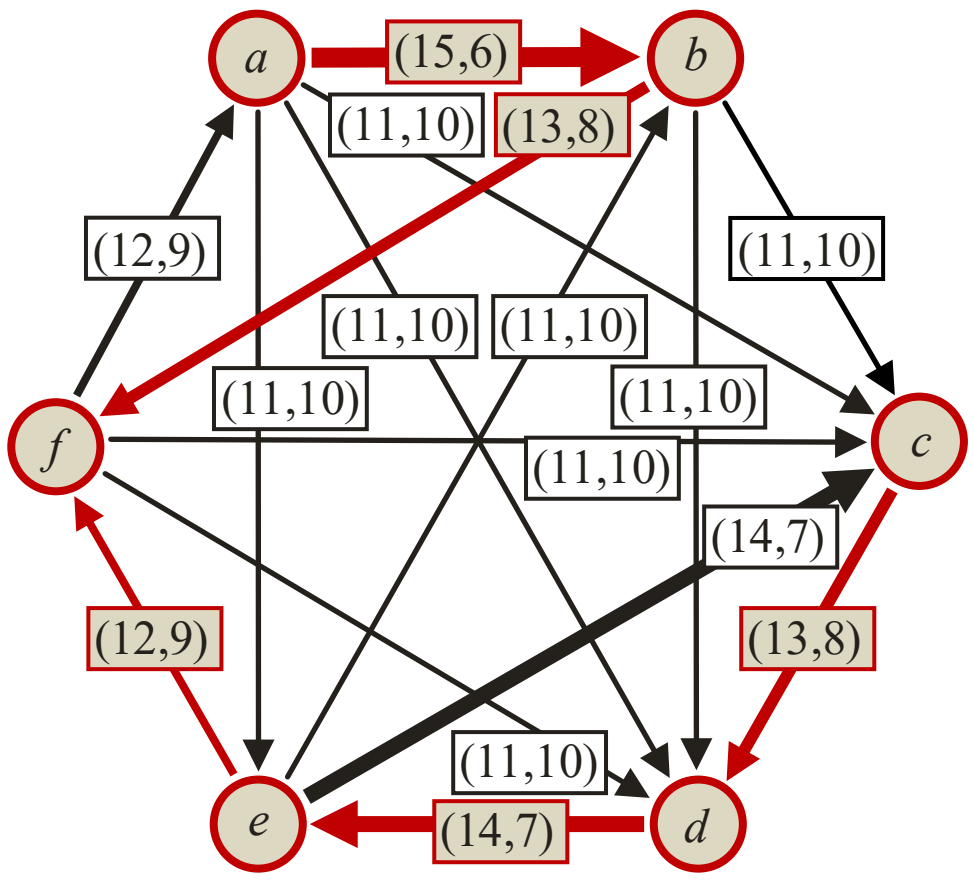

These are the strongest paths
from every other alternative to *f*.

Therefore, the strengths of the strongest paths are:

| | $P_D[*,a]$ | $P_D[*,b]$ | $P_D[*,c]$ | $P_D[*,d]$ | $P_D[*,e]$ | $P_D[*,f]$ |
|---|---|---|---|---|---|---|
| $P_D[a,*]$ | --- | (15,6) | (11,10) | (11,10) | (11,10) | (13,8) |
| $P_D[b,*]$ | (12,9) | --- | (11,10) | (11,10) | (11,10) | (13,8) |
| $P_D[c,*]$ | (12,9) | (12,9) | --- | (13,8) | (13,8) | (12,9) |
| $P_D[d,*]$ | (12,9) | (12,9) | (14,7) | --- | (14,7) | (12,9) |
| $P_D[e,*]$ | (12,9) | (12,9) | (14,7) | (13,8) | --- | (12,9) |
| $P_D[f,*]$ | (12,9) | (12,9) | (11,10) | (11,10) | (11,10) | --- |

We get $\mathcal{O}^{\text{new}}$ = {*ab*, *af*, *bf*, *ca*, *cb*, *cf*, *da*, *db*, *dc*, *de*, *df*, *ea*, *eb*, *ec*, *ef*} and $\mathcal{S}^{\text{new}}$ = {*d*}.

Thus the 2 $a \succ_v e \succ_v f \succ_v c \succ_v b \succ_v d$ voters change the unique winner from alternative *a* to alternative *d*.

Suppose, the strongest paths are calculated with the Floyd-Warshall algorithm, as defined in section 2.3.1. Then the following table documents the $C \cdot (C-1) \cdot (C-2) = 120$ steps of the Floyd-Warshall algorithm.

We start with

- $P_D[i,j] := (N^{\text{new}}[i,j], N^{\text{new}}[j,i])$ for all $i \in A$ and $j \in A \setminus \{i\}$.
- $pred[i,j] := i$ for all $i \in A$ and $j \in A \setminus \{i\}$.

| | *i* | *j* | *k* | $P_D[j,k]$ | $P_D[j,i]$ | $P_D[i,k]$ | *pred*[*j*,*k*] | *pred*[*i*,*k*] | result |
|---|---|---|---|---|---|---|---|---|---|
| 1 | *a* | *b* | *c* | (11,10) | (6,15) | (11,10) | *b* | *a* | |
| 2 | *a* | *b* | *d* | (11,10) | (6,15) | (11,10) | *b* | *a* | |
| 3 | *a* | *b* | *e* | (10,11) | (6,15) | (11,10) | *b* | *a* | |
| 4 | *a* | *b* | *f* | (13,8) | (6,15) | (9,12) | *b* | *a* | |
| 5 | *a* | *c* | *b* | (10,11) | (10,11) | (15,6) | *c* | *a* | |
| 6 | *a* | *c* | *d* | (13,8) | (10,11) | (11,10) | *c* | *a* | |
| 7 | *a* | *c* | *e* | (7,14) | (10,11) | (11,10) | *c* | *a* | $P_D[c,e]$ is updated from (7,14) to (10,11); *pred*[*c*,*e*] is updated from *c* to *a*. |
| 8 | *a* | *c* | *f* | (10,11) | (10,11) | (9,12) | *c* | *a* | |
| 9 | *a* | *d* | *b* | (10,11) | (10,11) | (15,6) | *d* | *a* | |
| 10 | *a* | *d* | *c* | (8,13) | (10,11) | (11,10) | *d* | *a* | $P_D[d,c]$ is updated from (8,13) to (10,11); *pred*[*d*,*c*] is updated from *d* to *a*. |
| 11 | *a* | *d* | *e* | (14,7) | (10,11) | (11,10) | *d* | *a* | |
| 12 | *a* | *d* | *f* | (10,11) | (10,11) | (9,12) | *d* | *a* | |
| 13 | *a* | *e* | *b* | (11,10) | (10,11) | (15,6) | *e* | *a* | |
| 14 | *a* | *e* | *c* | (14,7) | (10,11) | (11,10) | *e* | *a* | |
| 15 | *a* | *e* | *d* | (7,14) | (10,11) | (11,10) | *e* | *a* | $P_D[e,d]$ is updated from (7,14) to (10,11); *pred*[*e*,*d*] is updated from *e* to *a*. |
| 16 | *a* | *e* | *f* | (12,9) | (10,11) | (9,12) | *e* | *a* | |
| 17 | *a* | *f* | *b* | (8,13) | (12,9) | (15,6) | *f* | *a* | $P_D[f,b]$ is updated from (8,13) to (12,9); *pred*[*f*,*b*] is updated from *f* to *a*. |
| 18 | *a* | *f* | *c* | (11,10) | (12,9) | (11,10) | *f* | *a* | |
| 19 | *a* | *f* | *d* | (11,10) | (12,9) | (11,10) | *f* | *a* | |
| 20 | *a* | *f* | *e* | (9,12) | (12,9) | (11,10) | *f* | *a* | $P_D[f,e]$ is updated from (9,12) to (11,10); *pred*[*f*,*e*] is updated from *f* to *a*. |
| 21 | *b* | *a* | *c* | (11,10) | (15,6) | (11,10) | *a* | *b* | |
| 22 | *b* | *a* | *d* | (11,10) | (15,6) | (11,10) | *a* | *b* | |
| 23 | *b* | *a* | *e* | (11,10) | (15,6) | (10,11) | *a* | *b* | |
| 24 | *b* | *a* | *f* | (9,12) | (15,6) | (13,8) | *a* | *b* | $P_D[a,f]$ is updated from (9,12) to (13,8); *pred*[*a*,*f*] is updated from *a* to *b*. |
| 25 | *b* | *c* | *a* | (10,11) | (10,11) | (6,15) | *c* | *b* | |
| 26 | *b* | *c* | *d* | (13,8) | (10,11) | (11,10) | *c* | *b* | |
| 27 | *b* | *c* | *e* | (10,11) | (10,11) | (10,11) | *a* | *b* | |
| 28 | *b* | *c* | *f* | (10,11) | (10,11) | (13,8) | *c* | *b* | |
| 29 | *b* | *d* | *a* | (10,11) | (10,11) | (6,15) | *d* | *b* | |
| 30 | *b* | *d* | *c* | (10,11) | (10,11) | (11,10) | *a* | *b* | |

| | $i$ | $j$ | $k$ | $P_D[j,k]$ | $P_D[j,i]$ | $P_D[i,k]$ | $pred[j,k]$ | $pred[i,k]$ | result |
|---|---|---|---|---|---|---|---|---|---|
| 31 | *b* | *d* | *e* | (14,7) | (10,11) | (10,11) | *d* | *b* | |
| 32 | *b* | *d* | *f* | (10,11) | (10,11) | (13,8) | *d* | *b* | |
| 33 | *b* | *e* | *a* | (10,11) | (11,10) | (6,15) | *e* | *b* | |
| 34 | *b* | *e* | *c* | (14,7) | (11,10) | (11,10) | *e* | *b* | |
| 35 | *b* | *e* | *d* | (10,11) | (11,10) | (11,10) | *a* | *b* | $P_D[e,d]$ is updated from (10,11) to (11,10); *pred*[*e*,*d*] is updated from *a* to *b*. |
| 36 | *b* | *e* | *f* | (12,9) | (11,10) | (13,8) | *e* | *b* | |
| 37 | *b* | *f* | *a* | (12,9) | (12,9) | (6,15) | *f* | *b* | |
| 38 | *b* | *f* | *c* | (11,10) | (12,9) | (11,10) | *f* | *b* | |
| 39 | *b* | *f* | *d* | (11,10) | (12,9) | (11,10) | *f* | *b* | |
| 40 | *b* | *f* | *e* | (11,10) | (12,9) | (10,11) | *a* | *b* | |
| 41 | *c* | *a* | *b* | (15,6) | (11,10) | (10,11) | *a* | *c* | |
| 42 | *c* | *a* | *d* | (11,10) | (11,10) | (13,8) | *a* | *c* | |
| 43 | *c* | *a* | *e* | (11,10) | (11,10) | (10,11) | *a* | *a* | |
| 44 | *c* | *a* | *f* | (13,8) | (11,10) | (10,11) | *b* | *c* | |
| 45 | *c* | *b* | *a* | (6,15) | (11,10) | (10,11) | *b* | *c* | $P_D[b,a]$ is updated from (6,15) to (10,11); *pred*[*b*,*a*] is updated from *b* to *c*. |
| 46 | *c* | *b* | *d* | (11,10) | (11,10) | (13,8) | *b* | *c* | |
| 47 | *c* | *b* | *e* | (10,11) | (11,10) | (10,11) | *b* | *a* | |
| 48 | *c* | *b* | *f* | (13,8) | (11,10) | (10,11) | *b* | *c* | |
| 49 | *c* | *d* | *a* | (10,11) | (10,11) | (10,11) | *d* | *c* | |
| 50 | *c* | *d* | *b* | (10,11) | (10,11) | (10,11) | *d* | *c* | |
| 51 | *c* | *d* | *e* | (14,7) | (10,11) | (10,11) | *d* | *a* | |
| 52 | *c* | *d* | *f* | (10,11) | (10,11) | (10,11) | *d* | *c* | |
| 53 | *c* | *e* | *a* | (10,11) | (14,7) | (10,11) | *e* | *c* | |
| 54 | *c* | *e* | *b* | (11,10) | (14,7) | (10,11) | *e* | *c* | |
| 55 | *c* | *e* | *d* | (11,10) | (14,7) | (13,8) | *b* | *c* | $P_D[e,d]$ is updated from (11,10) to (13,8); *pred*[*e*,*d*] is updated from *b* to *c*. |
| 56 | *c* | *e* | *f* | (12,9) | (14,7) | (10,11) | *e* | *c* | |
| 57 | *c* | *f* | *a* | (12,9) | (11,10) | (10,11) | *f* | *c* | |
| 58 | *c* | *f* | *b* | (12,9) | (11,10) | (10,11) | *a* | *c* | |
| 59 | *c* | *f* | *d* | (11,10) | (11,10) | (13,8) | *f* | *c* | |
| 60 | *c* | *f* | *e* | (11,10) | (11,10) | (10,11) | *a* | *a* | |

| | *i* | *j* | *k* | $P_D[j,k]$ | $P_D[j,i]$ | $P_D[i,k]$ | *pred*[*j*,*k*] | *pred*[*i*,*k*] | result |
|---|---|---|---|---|---|---|---|---|---|
| 61 | *d* | *a* | *b* | (15,6) | (11,10) | (10,11) | *a* | *d* | |
| 62 | *d* | *a* | *c* | (11,10) | (11,10) | (10,11) | *a* | *a* | |
| 63 | *d* | *a* | *e* | (11,10) | (11,10) | (14,7) | *a* | *d* | |
| 64 | *d* | *a* | *f* | (13,8) | (11,10) | (10,11) | *b* | *d* | |
| 65 | *d* | *b* | *a* | (10,11) | (11,10) | (10,11) | *c* | *d* | |
| 66 | *d* | *b* | *c* | (11,10) | (11,10) | (10,11) | *b* | *a* | |
| 67 | *d* | *b* | *e* | (10,11) | (11,10) | (14,7) | *b* | *d* | $P_D[b,e]$ is updated from (10,11) to (11,10); *pred*[*b*,*e*] is updated from *b* to *d*. |
| 68 | *d* | *b* | *f* | (13,8) | (11,10) | (10,11) | *b* | *d* | |
| 69 | *d* | *c* | *a* | (10,11) | (13,8) | (10,11) | *c* | *d* | |
| 70 | *d* | *c* | *b* | (10,11) | (13,8) | (10,11) | *c* | *d* | |
| 71 | *d* | *c* | *e* | (10,11) | (13,8) | (14,7) | *a* | *d* | $P_D[c,e]$ is updated from (10,11) to (13,8); *pred*[*c*,*e*] is updated from *a* to *d*. |
| 72 | *d* | *c* | *f* | (10,11) | (13,8) | (10,11) | *c* | *d* | |
| 73 | *d* | *e* | *a* | (10,11) | (13,8) | (10,11) | *e* | *d* | |
| 74 | *d* | *e* | *b* | (11,10) | (13,8) | (10,11) | *e* | *d* | |
| 75 | *d* | *e* | *c* | (14,7) | (13,8) | (10,11) | *e* | *a* | |
| 76 | *d* | *e* | *f* | (12,9) | (13,8) | (10,11) | *e* | *d* | |
| 77 | *d* | *f* | *a* | (12,9) | (11,10) | (10,11) | *f* | *d* | |
| 78 | *d* | *f* | *b* | (12,9) | (11,10) | (10,11) | *a* | *d* | |
| 79 | *d* | *f* | *c* | (11,10) | (11,10) | (10,11) | *f* | *a* | |
| 80 | *d* | *f* | *e* | (11,10) | (11,10) | (14,7) | *a* | *d* | |
| 81 | *e* | *a* | *b* | (15,6) | (11,10) | (11,10) | *a* | *e* | |
| 82 | *e* | *a* | *c* | (11,10) | (11,10) | (14,7) | *a* | *e* | |
| 83 | *e* | *a* | *d* | (11,10) | (11,10) | (13,8) | *a* | *c* | |
| 84 | *e* | *a* | *f* | (13,8) | (11,10) | (12,9) | *b* | *e* | |
| 85 | *e* | *b* | *a* | (10,11) | (11,10) | (10,11) | *c* | *e* | |
| 86 | *e* | *b* | *c* | (11,10) | (11,10) | (14,7) | *b* | *e* | |
| 87 | *e* | *b* | *d* | (11,10) | (11,10) | (13,8) | *b* | *c* | |
| 88 | *e* | *b* | *f* | (13,8) | (11,10) | (12,9) | *b* | *e* | |
| 89 | *e* | *c* | *a* | (10,11) | (13,8) | (10,11) | *c* | *e* | |
| 90 | *e* | *c* | *b* | (10,11) | (13,8) | (11,10) | *c* | *e* | $P_D[c,b]$ is updated from (10,11) to (11,10); *pred*[*c*,*b*] is updated from *c* to *e*. |

| | $i$ | $j$ | $k$ | $P_D[j,k]$ | $P_D[j,i]$ | $P_D[i,k]$ | $pred[j,k]$ | $pred[i,k]$ | result |
|---|---|---|---|---|---|---|---|---|---|
| 91 | $e$ | $c$ | $d$ | (13,8) | (13,8) | (13,8) | $c$ | $c$ | |
| 92 | $e$ | $c$ | $f$ | (10,11) | (13,8) | (12,9) | $c$ | $e$ | $P_D[c,f]$ is updated from (10,11) to (12,9); $pred[c,f]$ is updated from $c$ to $e$. |
| 93 | $e$ | $d$ | $a$ | (10,11) | (14,7) | (10,11) | $d$ | $e$ | |
| 94 | $e$ | $d$ | $b$ | (10,11) | (14,7) | (11,10) | $d$ | $e$ | $P_D[d,b]$ is updated from (10,11) to (11,10); $pred[d,b]$ is updated from $d$ to $e$. |
| 95 | $e$ | $d$ | $c$ | (10,11) | (14,7) | (14,7) | $a$ | $e$ | $P_D[d,c]$ is updated from (10,11) to (14,7); $pred[d,c]$ is updated from $a$ to $e$. |
| 96 | $e$ | $d$ | $f$ | (10,11) | (14,7) | (12,9) | $d$ | $e$ | $P_D[d,f]$ is updated from (10,11) to (12,9); $pred[d,f]$ is updated from $d$ to $e$. |
| 97 | $e$ | $f$ | $a$ | (12,9) | (11,10) | (10,11) | $f$ | $e$ | |
| 98 | $e$ | $f$ | $b$ | (12,9) | (11,10) | (11,10) | $a$ | $e$ | |
| 99 | $e$ | $f$ | $c$ | (11,10) | (11,10) | (14,7) | $f$ | $e$ | |
| 100 | $e$ | $f$ | $d$ | (11,10) | (11,10) | (13,8) | $f$ | $c$ | |
| 101 | $f$ | $a$ | $b$ | (15,6) | (13,8) | (12,9) | $a$ | $a$ | |
| 102 | $f$ | $a$ | $c$ | (11,10) | (13,8) | (11,10) | $a$ | $f$ | |
| 103 | $f$ | $a$ | $d$ | (11,10) | (13,8) | (11,10) | $a$ | $f$ | |
| 104 | $f$ | $a$ | $e$ | (11,10) | (13,8) | (11,10) | $a$ | $a$ | |
| 105 | $f$ | $b$ | $a$ | (10,11) | (13,8) | (12,9) | $c$ | $f$ | $P_D[b,a]$ is updated from (10,11) to (12,9); $pred[b,a]$ is updated from $c$ to $f$. |
| 106 | $f$ | $b$ | $c$ | (11,10) | (13,8) | (11,10) | $b$ | $f$ | |
| 107 | $f$ | $b$ | $d$ | (11,10) | (13,8) | (11,10) | $b$ | $f$ | |
| 108 | $f$ | $b$ | $e$ | (11,10) | (13,8) | (11,10) | $d$ | $a$ | |
| 109 | $f$ | $c$ | $a$ | (10,11) | (12,9) | (12,9) | $c$ | $f$ | $P_D[c,a]$ is updated from (10,11) to (12,9); $pred[c,a]$ is updated from $c$ to $f$. |
| 110 | $f$ | $c$ | $b$ | (11,10) | (12,9) | (12,9) | $e$ | $a$ | $P_D[c,b]$ is updated from (11,10) to (12,9); $pred[c,b]$ is updated from $e$ to $a$. |
| 111 | $f$ | $c$ | $d$ | (13,8) | (12,9) | (11,10) | $c$ | $f$ | |
| 112 | $f$ | $c$ | $e$ | (13,8) | (12,9) | (11,10) | $d$ | $a$ | |
| 113 | $f$ | $d$ | $a$ | (10,11) | (12,9) | (12,9) | $d$ | $f$ | $P_D[d,a]$ is updated from (10,11) to (12,9); $pred[d,a]$ is updated from $d$ to $f$. |
| 114 | $f$ | $d$ | $b$ | (11,10) | (12,9) | (12,9) | $e$ | $a$ | $P_D[d,b]$ is updated from (11,10) to (12,9); $pred[d,b]$ is updated from $e$ to $a$. |
| 115 | $f$ | $d$ | $c$ | (14,7) | (12,9) | (11,10) | $e$ | $f$ | |
| 116 | $f$ | $d$ | $e$ | (14,7) | (12,9) | (11,10) | $d$ | $a$ | |
| 117 | $f$ | $e$ | $a$ | (10,11) | (12,9) | (12,9) | $e$ | $f$ | $P_D[e,a]$ is updated from (10,11) to (12,9); $pred[e,a]$ is updated from $e$ to $f$. |
| 118 | $f$ | $e$ | $b$ | (11,10) | (12,9) | (12,9) | $e$ | $a$ | $P_D[e,b]$ is updated from (11,10) to (12,9); $pred[e,b]$ is updated from $e$ to $a$. |
| 119 | $f$ | $e$ | $c$ | (14,7) | (12,9) | (11,10) | $e$ | $f$ | |
| 120 | $f$ | $e$ | $d$ | (13,8) | (12,9) | (11,10) | $c$ | $f$ | |

## 3.8. Example 8

*Independence from Pareto-dominated alternatives* (IPDA) as a criterion for single-winner election methods has been proposed by Fishburn (1973). This criterion is also called *reduction* (Fishburn, 1973; Richelson, 1978).

When $i \succsim_v j$ for every $v \in V$, then we say “ alternative $i$ Pareto-dominates alternative $j$ ”.

Suppose an alternative $j$ is added such that:

(3.8.1) $\exists\, i \in A^{\text{old}}\ \forall\, v \in V: i \succsim_v^{\text{new}} j.$

(3.8.2) $\forall\, g,h \in A^{\text{old}}\ \forall\, v \in V: g \succ_v^{\text{old}} h \Leftrightarrow g \succ_v^{\text{new}} h.$

Then *independence from Pareto-dominated alternatives* says that we must get:

(3.8.3) $\forall\, g,h \in A^{\text{old}}: gh \in \mathcal{O}^{\text{old}} \Leftrightarrow gh \in \mathcal{O}^{\text{new}}.$

(3.8.4) $\forall\, g \in A^{\text{old}}: g \in \mathcal{S}^{\text{old}} \Leftrightarrow g \in \mathcal{S}^{\text{new}}.$

The following example demonstrates that the Schulze method, as defined in section 2.2, does not satisfy IPDA. This example has been proposed by Eppley (2003).

### 3.8.1. Situation #1

Example 8 (old):

| | |
|---|---|
| 3 voters | $a \succ_v b \succ_v d \succ_v c$ |
| 5 voters | $a \succ_v d \succ_v b \succ_v c$ |
| 1 voter | $a \succ_v d \succ_v c \succ_v b$ |
| 2 voters | $b \succ_v a \succ_v d \succ_v c$ |
| 2 voters | $b \succ_v d \succ_v c \succ_v a$ |
| 4 voters | $c \succ_v a \succ_v b \succ_v d$ |
| 6 voters | $c \succ_v b \succ_v a \succ_v d$ |
| 2 voters | $d \succ_v b \succ_v c \succ_v a$ |
| 5 voters | $d \succ_v c \succ_v a \succ_v b$ |

The pairwise matrix $N^{old}$ looks as follows:

| | $N^{old}[*,a]$ | $N^{old}[*,b]$ | $N^{old}[*,c]$ | $N^{old}[*,d]$ |
|---|---|---|---|---|
| $N^{old}[a,*]$ | --- | 18 | 11 | 21 |
| $N^{old}[b,*]$ | 12 | --- | 14 | 17 |
| $N^{old}[c,*]$ | 19 | 16 | --- | 10 |
| $N^{old}[d,*]$ | 9 | 13 | 20 | --- |

The corresponding digraph looks as follows:

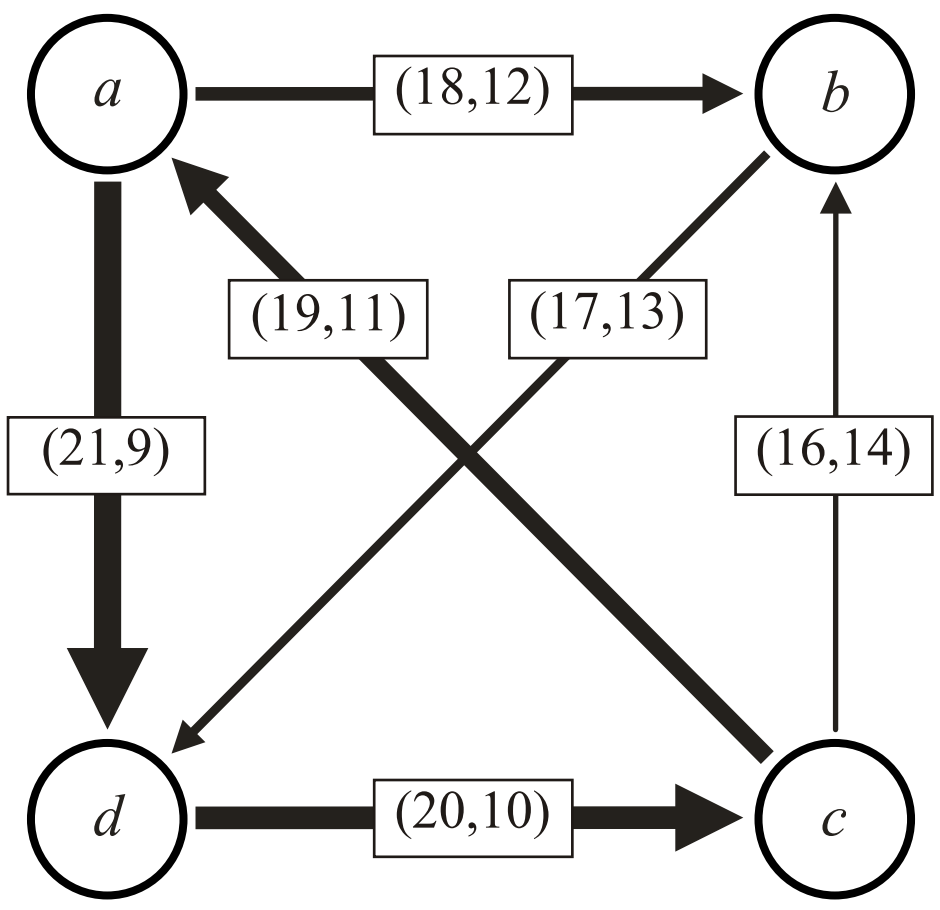

The following table lists the strongest paths, as determined by the Floyd-Warshall algorithm, as defined in section 2.3.1. The critical links of the strongest paths are underlined:

| | ... to *a* | ... to *b* | ... to *c* | ... to *d* | ... to every other alternative |
|---|---|---|---|---|---|
| from *a* ... | --- | 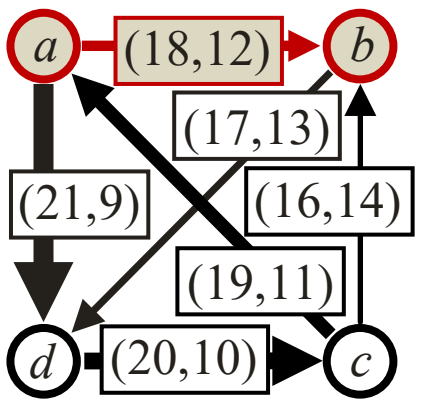 *a*, (18,12), *b* | 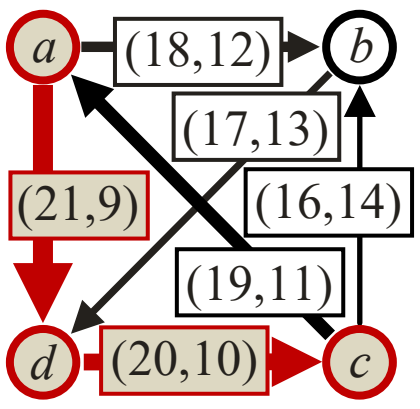 *a*, (21,9), *d*, (20,10), *c* | 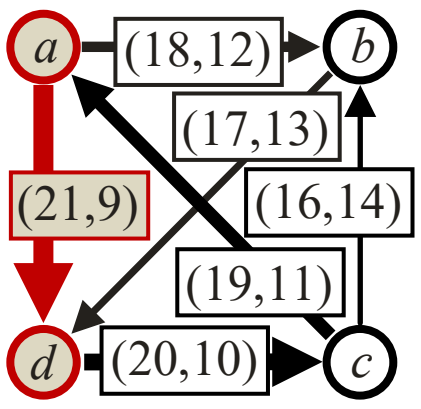 *a*, (21,9), *d* | 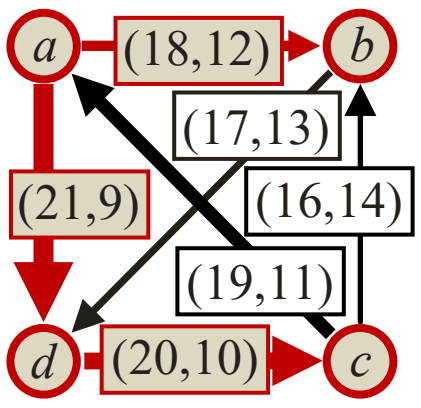 |
| from *b* ... | 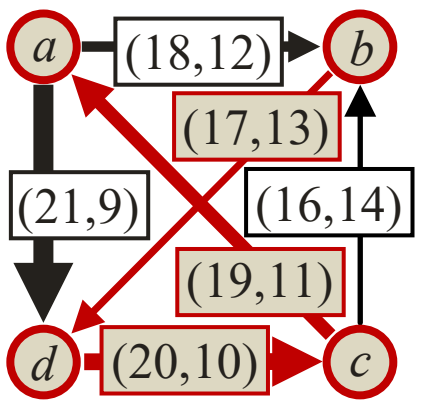 *b*, (17,13), *d*, (20,10), *c*, (19,11), *a* | --- | 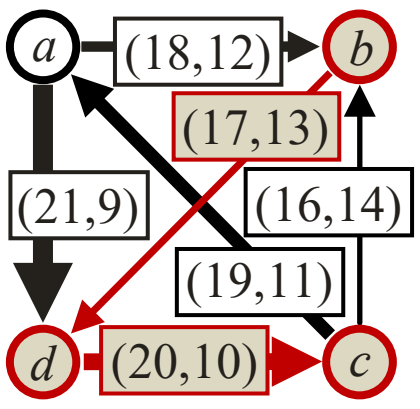 *b*, (17,13), *d*, (20,10), *c* | 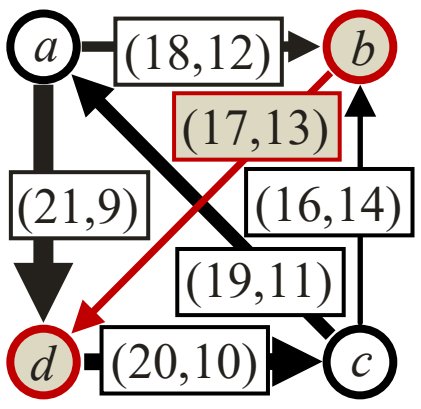 *b*, (17,13), *d* | 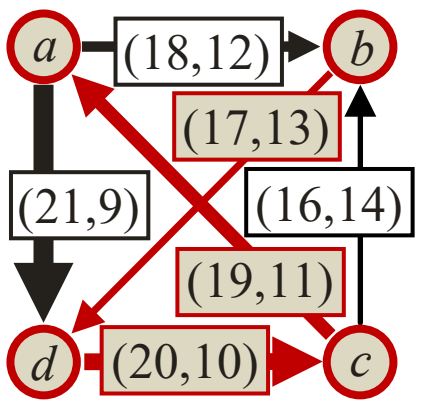 |
| from *c* ... | 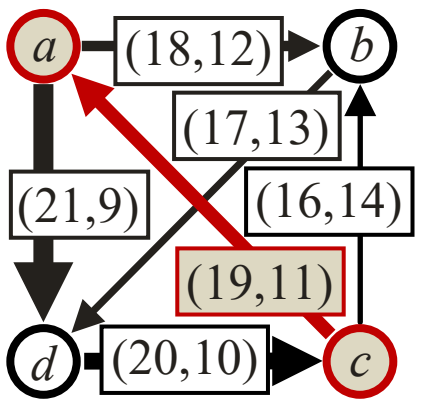 *c*, (19,11), *a* | 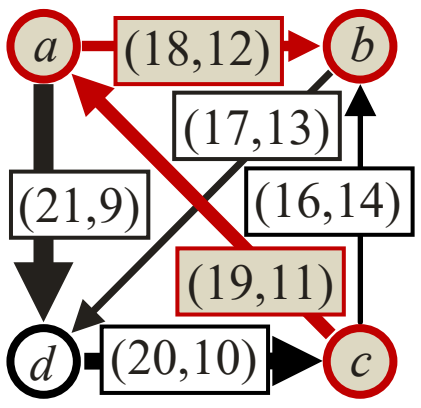 *c*, (19,11), *a*, (18,12), *b* | --- | 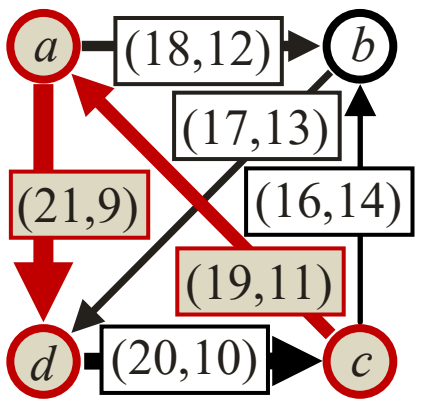 *c*, (19,11), *a*, (21,9), *d* | 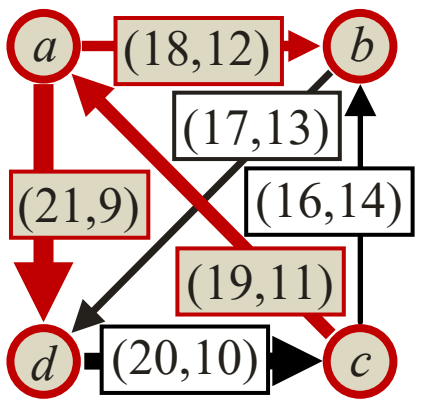 |
| from *d* ... | 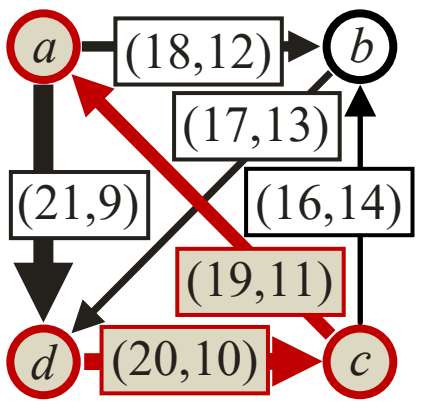 *d*, (20,10), *c*, (19,11), *a* | 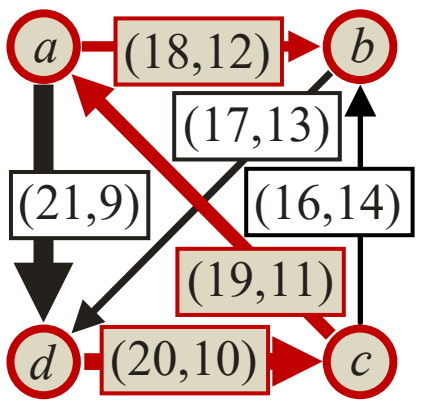 *d*, (20,10), *c*, (19,11), *a*, (18,12), *b* | 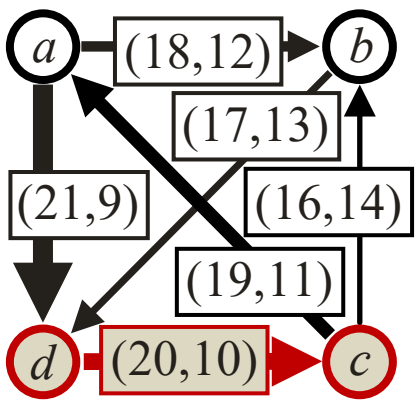 *d*, (20,10), *c* | --- | 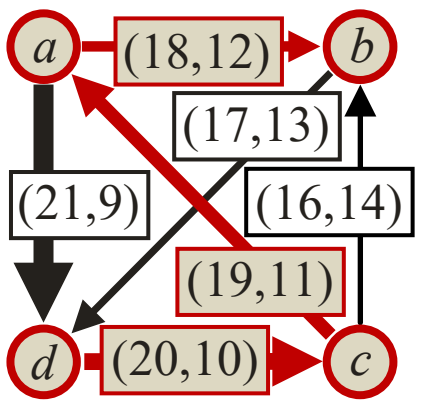 |
| from every other alternative ... | 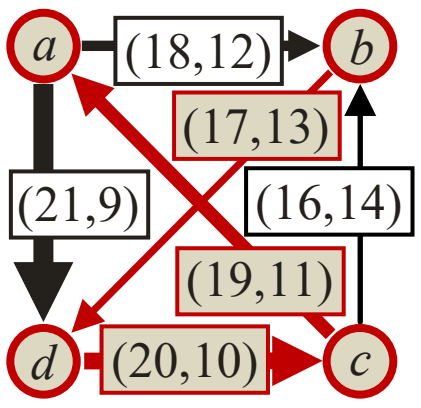 | 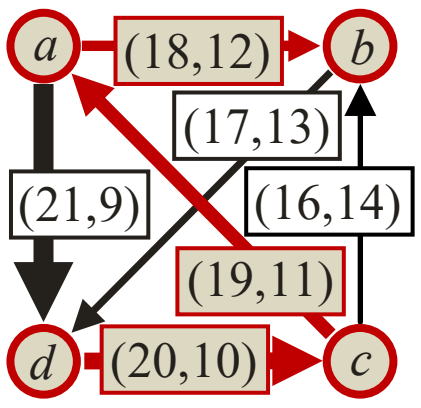 | 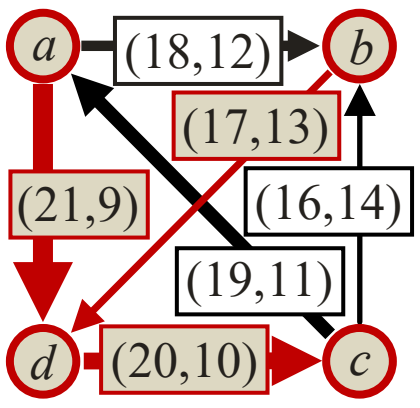 | 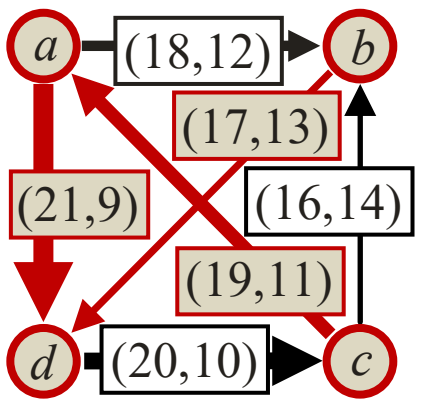 | --- |

The strengths of the strongest paths are:

| | $P_D[*,a]$ | $P_D[*,b]$ | $P_D[*,c]$ | $P_D[*,d]$ |
|---|---|---|---|---|
| $P_D[a,*]$ | --- | (18,12) | (20,10) | (21,9) |
| $P_D[b,*]$ | (17,13) | --- | (17,13) | (17,13) |
| $P_D[c,*]$ | (19,11) | (18,12) | --- | (19,11) |
| $P_D[d,*]$ | (19,11) | (18,12) | (20,10) | --- |

We get $\mathcal{O}^{\text{old}} = \{ab, ac, ad, cb, db, dc\}$ and $\mathcal{S}^{\text{old}} = \{a\}$.

Suppose, the strongest paths are calculated with the Floyd-Warshall algorithm, as defined in section 2.3.1. Then the following table documents the $C \cdot (C–1) \cdot (C–2) = 24$ steps of the Floyd-Warshall algorithm.

We start with

- $P_D[i,j] := (N^{\text{old}}[i,j],N^{\text{old}}[j,i])$ for all $i \in A$ and $j \in A \setminus \{i\}$.
- $pred[i,j] := i$ for all $i \in A$ and $j \in A \setminus \{i\}$.

| | *i* | *j* | *k* | $P_D[j,k]$ | $P_D[j,i]$ | $P_D[i,k]$ | *pred*[*j*,*k*] | *pred*[*i*,*k*] | result |
|---|---|---|---|---|---|---|---|---|---|
| 1 | *a* | *b* | *c* | (14,16) | (12,18) | (11,19) | *b* | *a* | |
| 2 | *a* | *b* | *d* | (17,13) | (12,18) | (21,9) | *b* | *a* | |
| 3 | *a* | *c* | *b* | (16,14) | (19,11) | (18,12) | *c* | *a* | $P_D[c,b]$ is updated from (16,14) to (18,12); *pred*[*c*,*b*] is updated from *c* to *a*. |
| 4 | *a* | *c* | *d* | (10,20) | (19,11) | (21,9) | *c* | *a* | $P_D[c,d]$ is updated from (10,20) to (19,11); *pred*[*c*,*d*] is updated from *c* to *a*. |
| 5 | *a* | *d* | *b* | (13,17) | (9,21) | (18,12) | *d* | *a* | |
| 6 | *a* | *d* | *c* | (20,10) | (9,21) | (11,19) | *d* | *a* | |
| 7 | *b* | *a* | *c* | (11,19) | (18,12) | (14,16) | *a* | *b* | $P_D[a,c]$ is updated from (11,19) to (14,16); *pred*[*a*,*c*] is updated from *a* to *b*. |
| 8 | *b* | *a* | *d* | (21,9) | (18,12) | (17,13) | *a* | *b* | |
| 9 | *b* | *c* | *a* | (19,11) | (18,12) | (12,18) | *c* | *b* | |
| 10 | *b* | *c* | *d* | (19,11) | (18,12) | (17,13) | *a* | *b* | |
| 11 | *b* | *d* | *a* | (9,21) | (13,17) | (12,18) | *d* | *b* | $P_D[d,a]$ is updated from (9,21) to (12,18); *pred*[*d*,*a*] is updated from *d* to *b*. |
| 12 | *b* | *d* | *c* | (20,10) | (13,17) | (14,16) | *d* | *b* | |
| 13 | *c* | *a* | *b* | (18,12) | (14,16) | (18,12) | *a* | *a* | |
| 14 | *c* | *a* | *d* | (21,9) | (14,16) | (19,11) | *a* | *a* | |
| 15 | *c* | *b* | *a* | (12,18) | (14,16) | (19,11) | *b* | *c* | $P_D[b,a]$ is updated from (12,18) to (14,16); *pred*[*b*,*a*] is updated from *b* to *c*. |
| 16 | *c* | *b* | *d* | (17,13) | (14,16) | (19,11) | *b* | *a* | |
| 17 | *c* | *d* | *a* | (12,18) | (20,10) | (19,11) | *b* | *c* | $P_D[d,a]$ is updated from (12,18) to (19,11); *pred*[*d*,*a*] is updated from *b* to *c*. |
| 18 | *c* | *d* | *b* | (13,17) | (20,10) | (18,12) | *d* | *a* | $P_D[d,b]$ is updated from (13,17) to (18,12); *pred*[*d*,*b*] is updated from *d* to *a*. |
| 19 | *d* | *a* | *b* | (18,12) | (21,9) | (18,12) | *a* | *a* | |
| 20 | *d* | *a* | *c* | (14,16) | (21,9) | (20,10) | *b* | *d* | $P_D[a,c]$ is updated from (14,16) to (20,10); *pred*[*a*,*c*] is updated from *b* to *d*. |
| 21 | *d* | *b* | *a* | (14,16) | (17,13) | (19,11) | *c* | *c* | $P_D[b,a]$ is updated from (14,16) to (17,13). |
| 22 | *d* | *b* | *c* | (14,16) | (17,13) | (20,10) | *b* | *d* | $P_D[b,c]$ is updated from (14,16) to (17,13); *pred*[*b*,*c*] is updated from *b* to *d*. |
| 23 | *d* | *c* | *a* | (19,11) | (19,11) | (19,11) | *c* | *c* | |
| 24 | *d* | *c* | *b* | (18,12) | (19,11) | (18,12) | *a* | *a* | |

### 3.8.2. Situation #2

Suppose alternative $e$ is added as follows:

Example 8 (new):

| | |
|---|---|
| 3 voters | $a \succ_v b \succ_v d \succ_v e \succ_v c$ |
| 5 voters | $a \succ_v d \succ_v e \succ_v b \succ_v c$ |
| 1 voter | $a \succ_v d \succ_v e \succ_v c \succ_v b$ |
| 2 voters | $b \succ_v a \succ_v d \succ_v e \succ_v c$ |
| 2 voters | $b \succ_v d \succ_v e \succ_v c \succ_v a$ |
| 4 voters | $c \succ_v a \succ_v b \succ_v d \succ_v e$ |
| 6 voters | $c \succ_v b \succ_v a \succ_v d \succ_v e$ |
| 2 voters | $d \succ_v b \succ_v e \succ_v c \succ_v a$ |
| 5 voters | $d \succ_v e \succ_v c \succ_v a \succ_v b$ |

The newly added alternative $e$ is Pareto-dominated by alternative $d$, because $d \succ_v e$ for every voter $v \in V$. Therefore, (3.8.1) – (3.8.4) say that the result should not change.

The pairwise matrix $N^{new}$ looks as follows:

| | $N^{new}[*,a]$ | $N^{new}[*,b]$ | $N^{new}[*,c]$ | $N^{new}[*,d]$ | $N^{new}[*,e]$ |
|---|---|---|---|---|---|
| $N^{new}[a,*]$ | --- | 18 | 11 | 21 | 21 |
| $N^{new}[b,*]$ | 12 | --- | 14 | 17 | 19 |
| $N^{new}[c,*]$ | 19 | 16 | --- | 10 | 10 |
| $N^{new}[d,*]$ | 9 | 13 | 20 | --- | 30 |
| $N^{new}[e,*]$ | 9 | 11 | 20 | 0 | --- |

The corresponding digraph looks as follows:

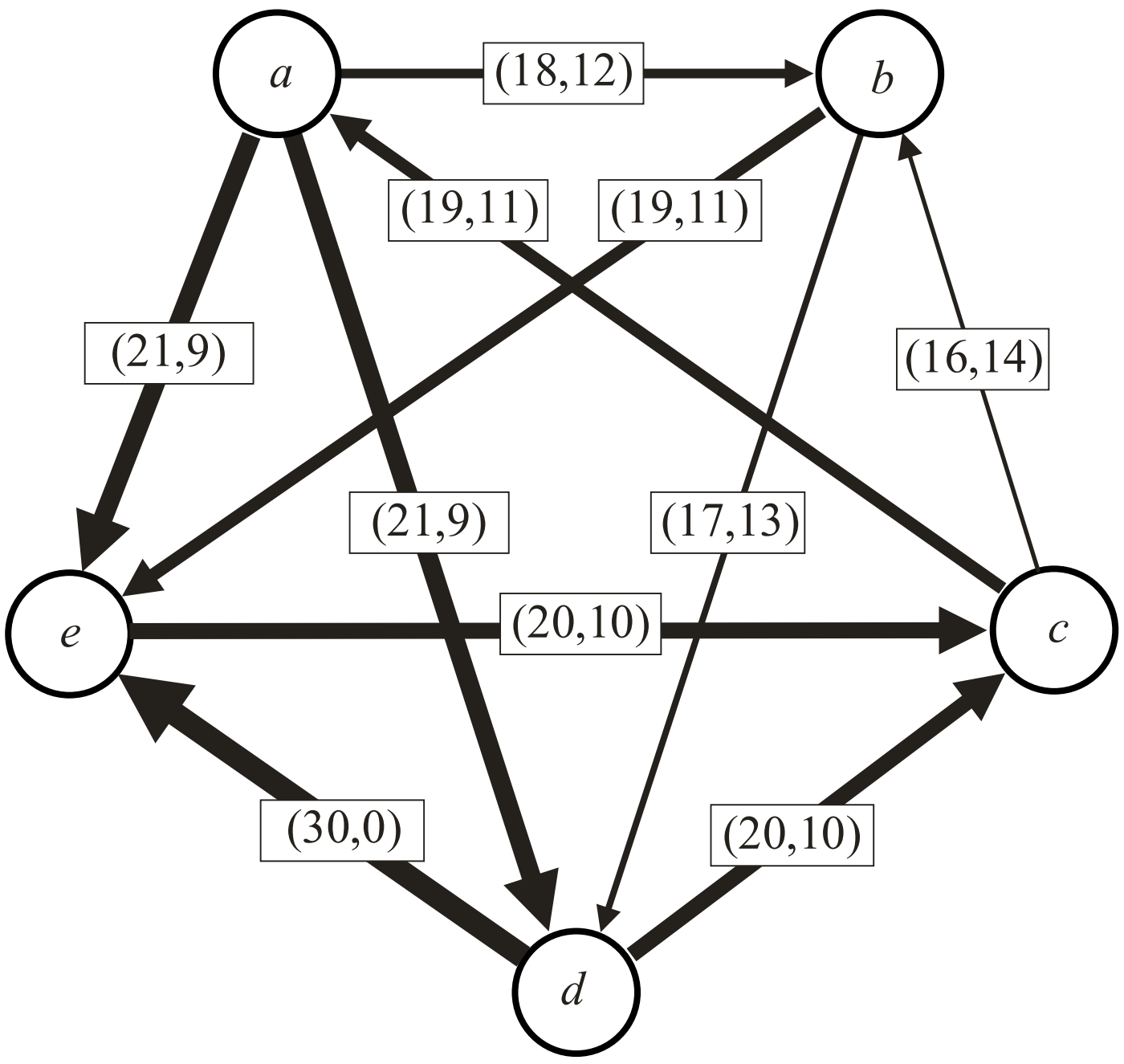

The following table lists the strongest paths, as determined by the Floyd-Warshall algorithm, as defined in section 2.3.1. The critical links of the strongest paths are underlined:

| | ... to *a* | ... to *b* | ... to *c* | ... to *d* | ... to *e* |
|---|---|---|---|---|---|
| from *a* ... | --- | *a*, (18,12), *b* | *a*, (21,9), *d*, (20,10), *c* | *a*, (21,9), *d* | *a*, (21,9), *e* |
| from *b* ... | *b*, (19,11), *e*, (20,10), *c*, (19,11), *a* | --- | *b*, (19,11), *e*, (20,10), *c* | *b*, (19,11), *e*, (20,10), *c*, (19,11), *a*, (21,9), *d* | *b*, (19,11), *e* |
| from *c* ... | *c*, (19,11), *a* | *c*, (19,11), *a*, (18,12), *b* | --- | *c*, (19,11), *a*, (21,9), *d* | *c*, (19,11), *a*, (21,9), *e* |
| from *d* ... | *d*, (20,10), *c*, (19,11), *a* | *d*, (20,10), *c*, (19,11), *a*, (18,12), *b* | *d*, (20,10), *c* | --- | *d*, (30,0), *e* |
| from *e* ... | *e*, (20,10), *c*, (19,11), *a* | *e*, (20,10), *c*, (19,11), *a*, (18,12), *b* | *e*, (20,10), *c* | *e*, (20,10), *c*, (19,11), *a*, (21,9), *d* | --- |

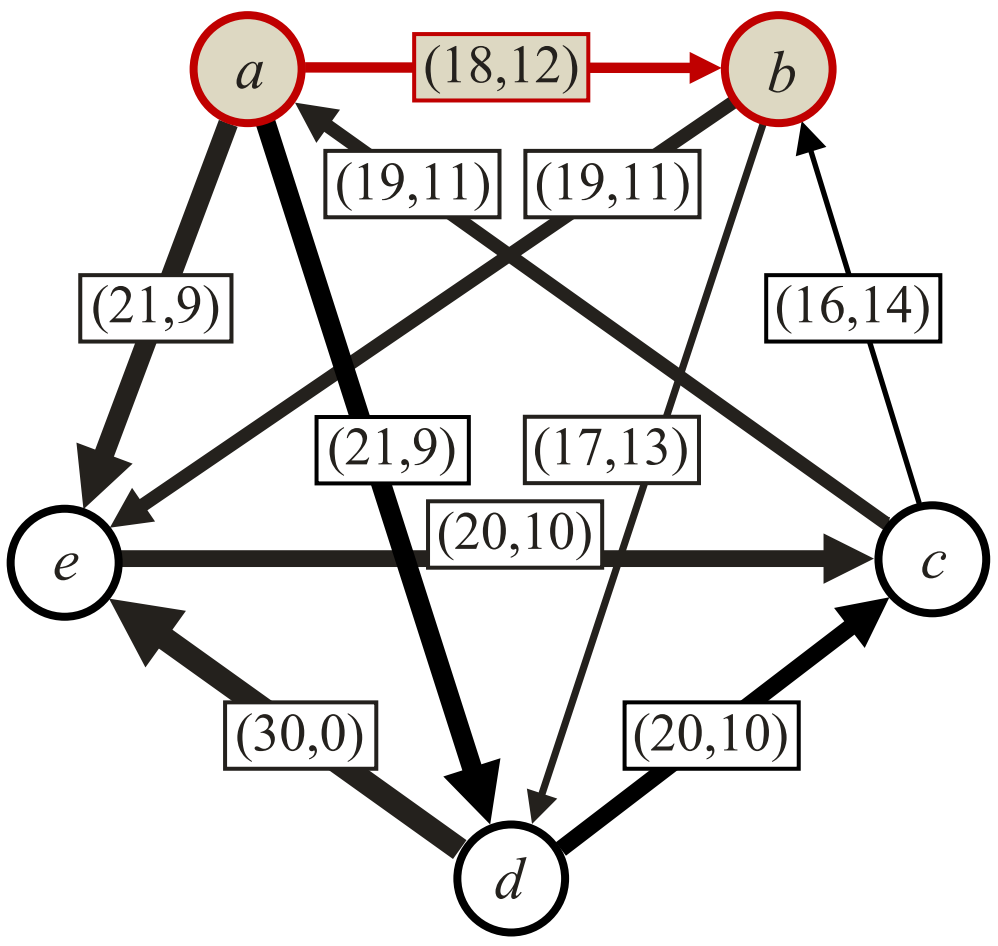

The strongest path from *a* to *b* is:
*a*, (18,12), *b*

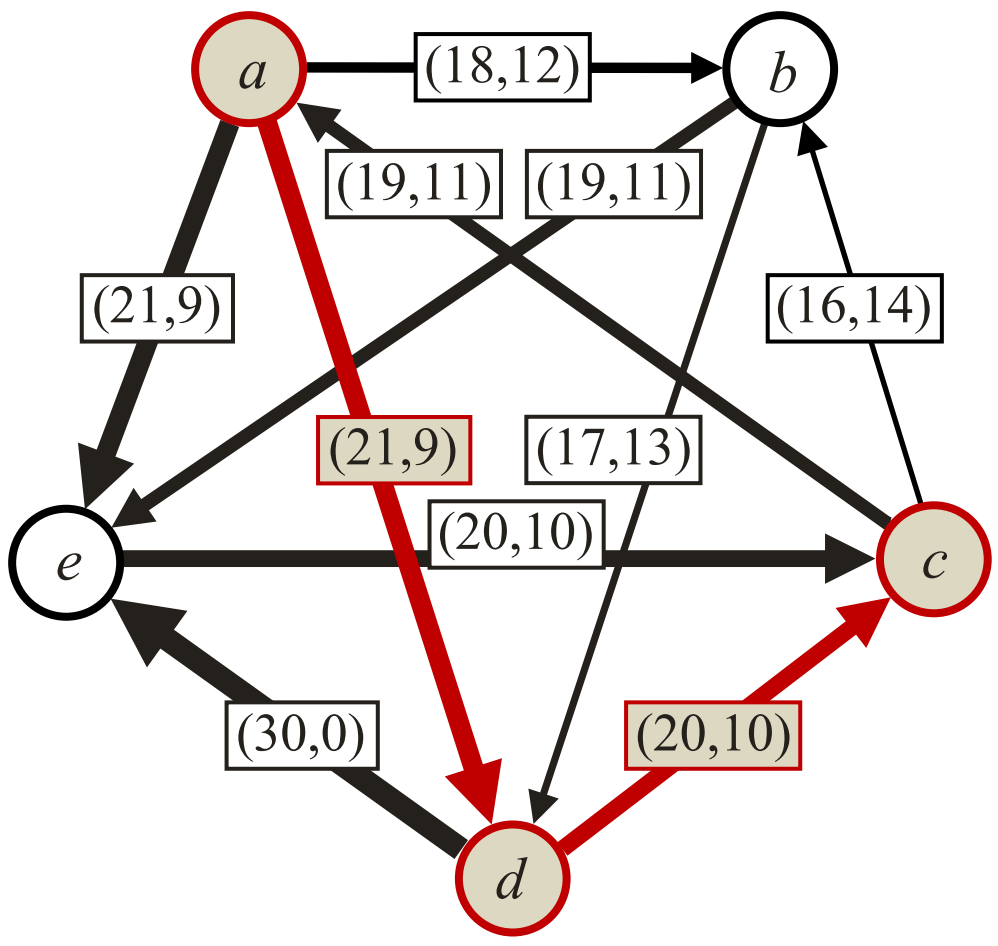

The strongest path from *a* to *c* is:
*a*, (21,9), *d*, (20,10), *c*

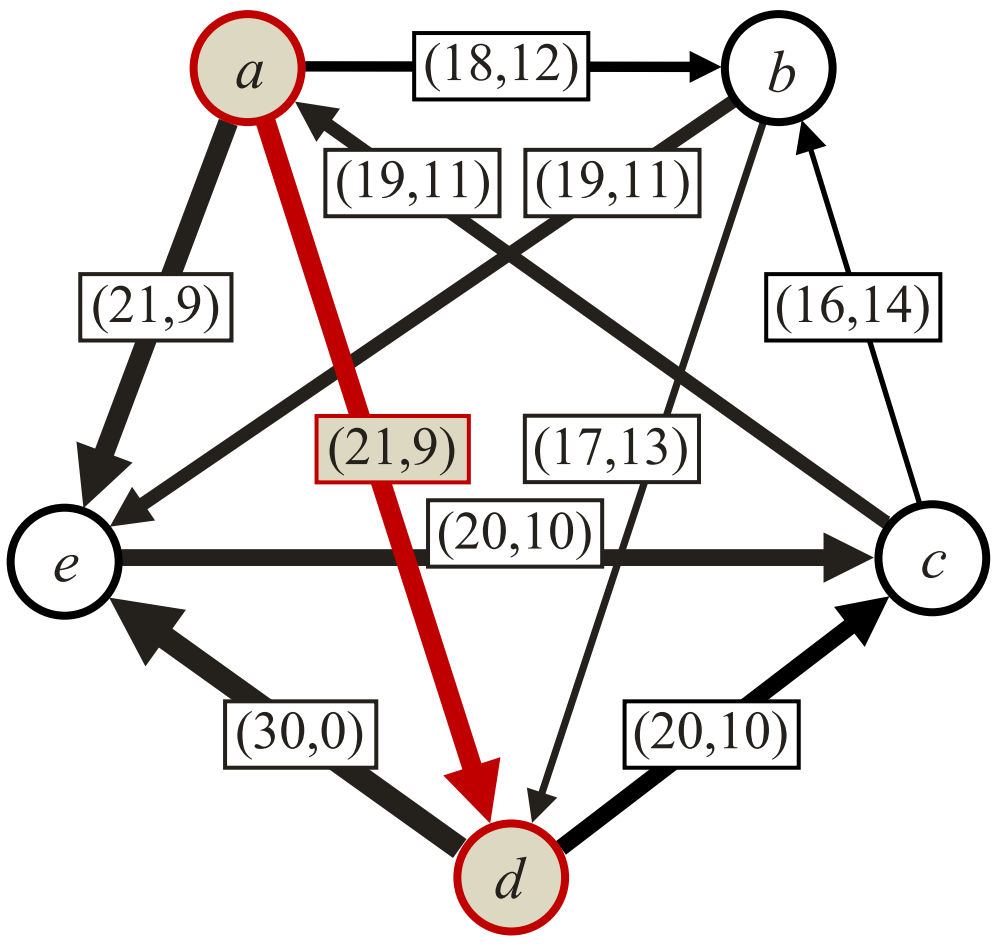

The strongest path from *a* to *d* is:
*a*, (21,9), *d*

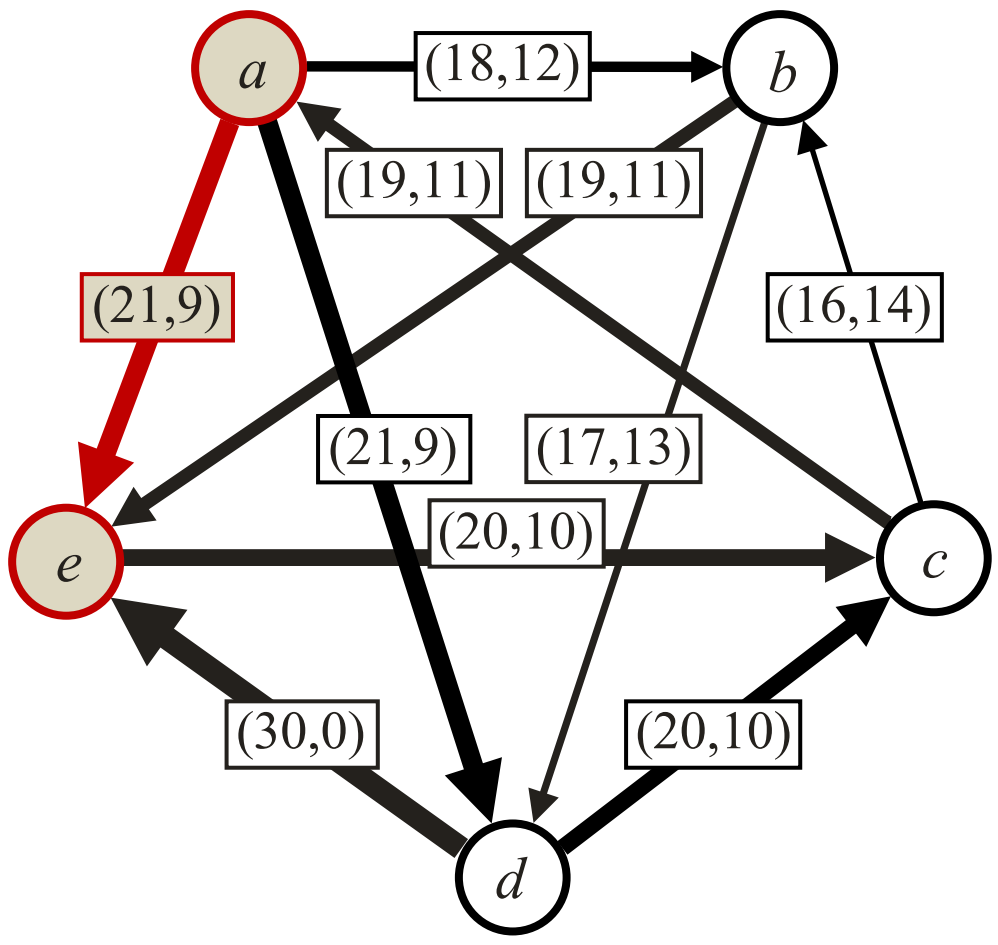

The strongest path from *a* to *e* is:
*a*, (21,9), *e*

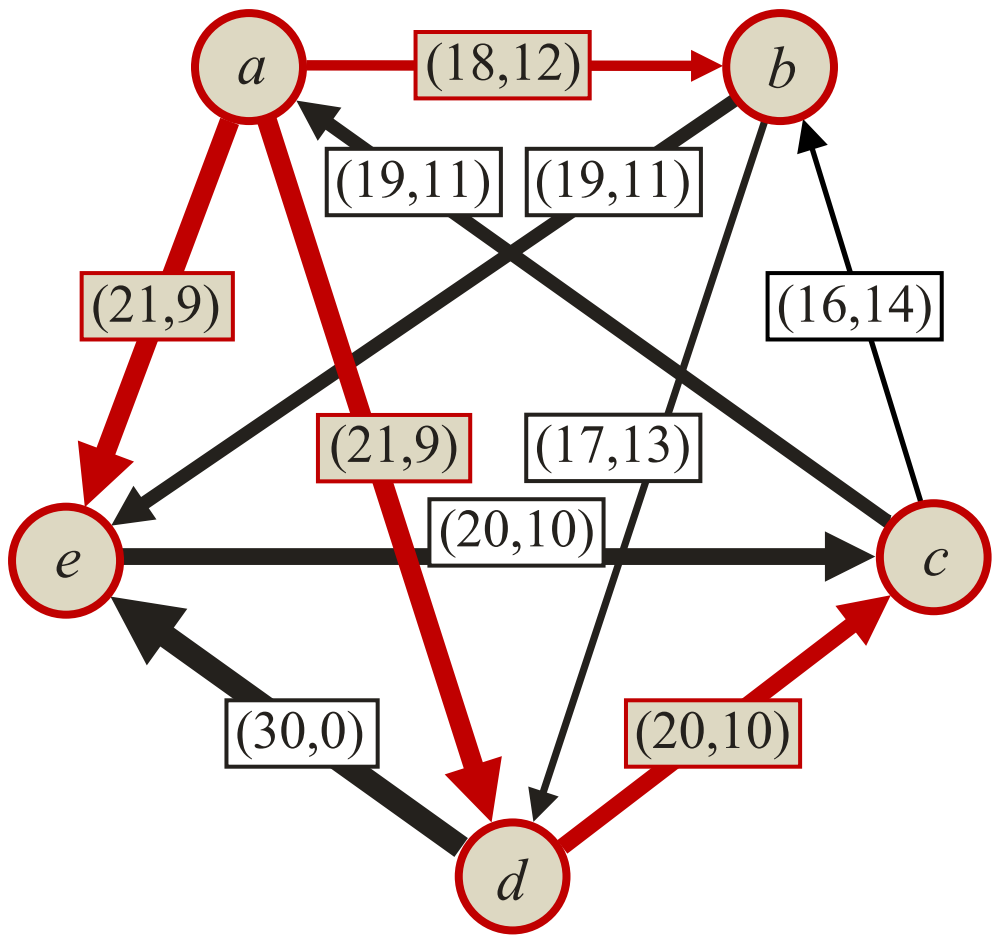

These are the strongest paths
from *a* to every other alternative.

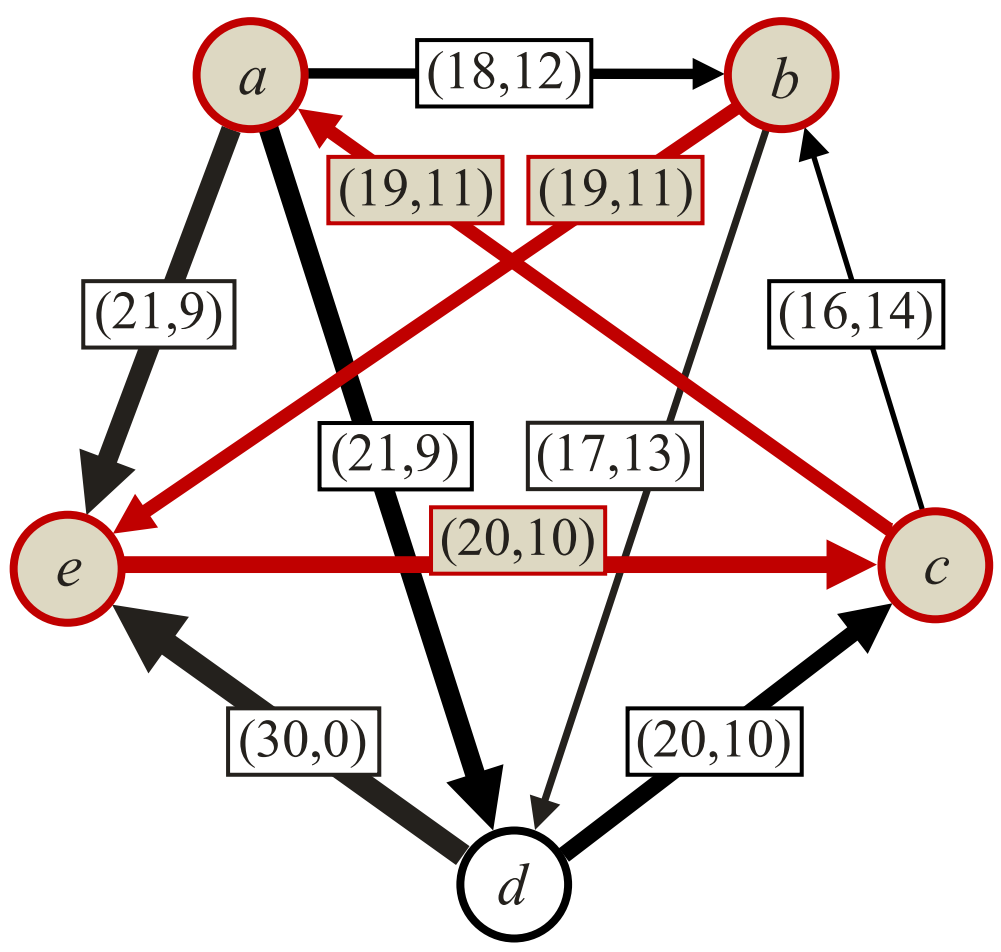

The strongest path from *b* to *a* is:
*b*, (19,11), *e*, (20,10), *c*, (19,11), *a*

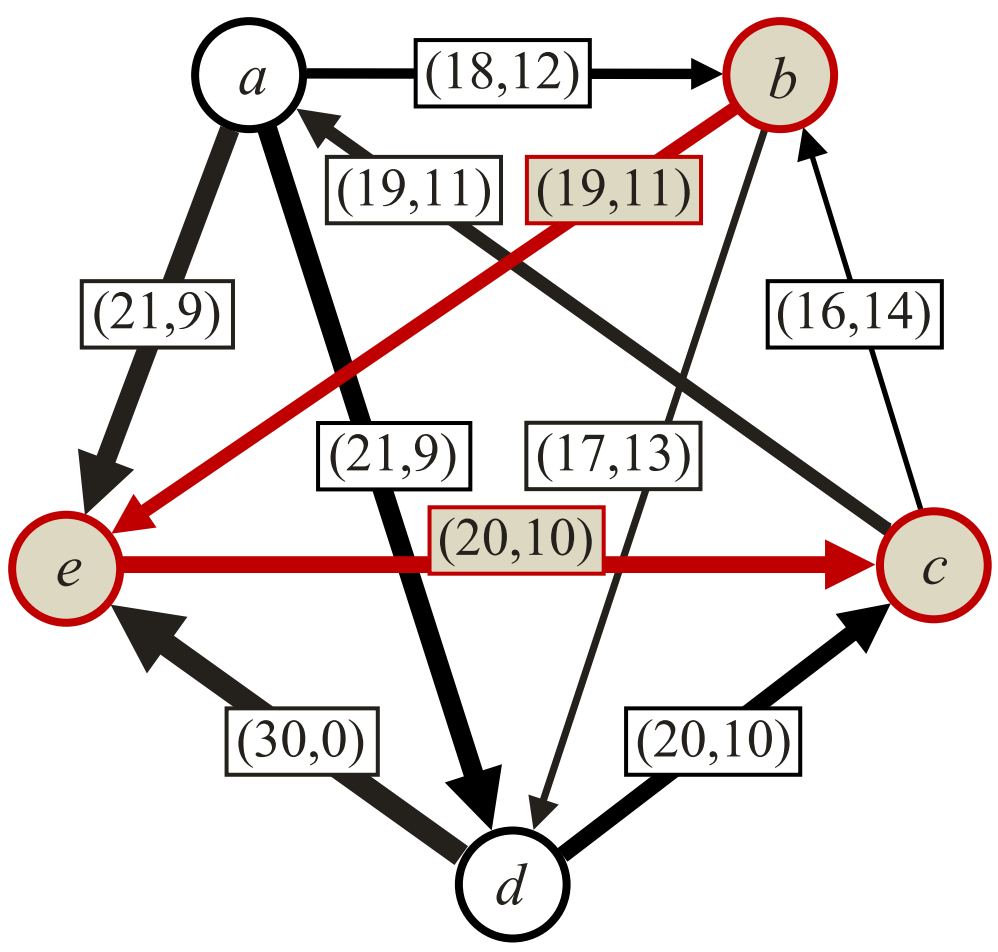

The strongest path from *b* to *c* is:
*b*, (19,11), *e*, (20,10), *c*

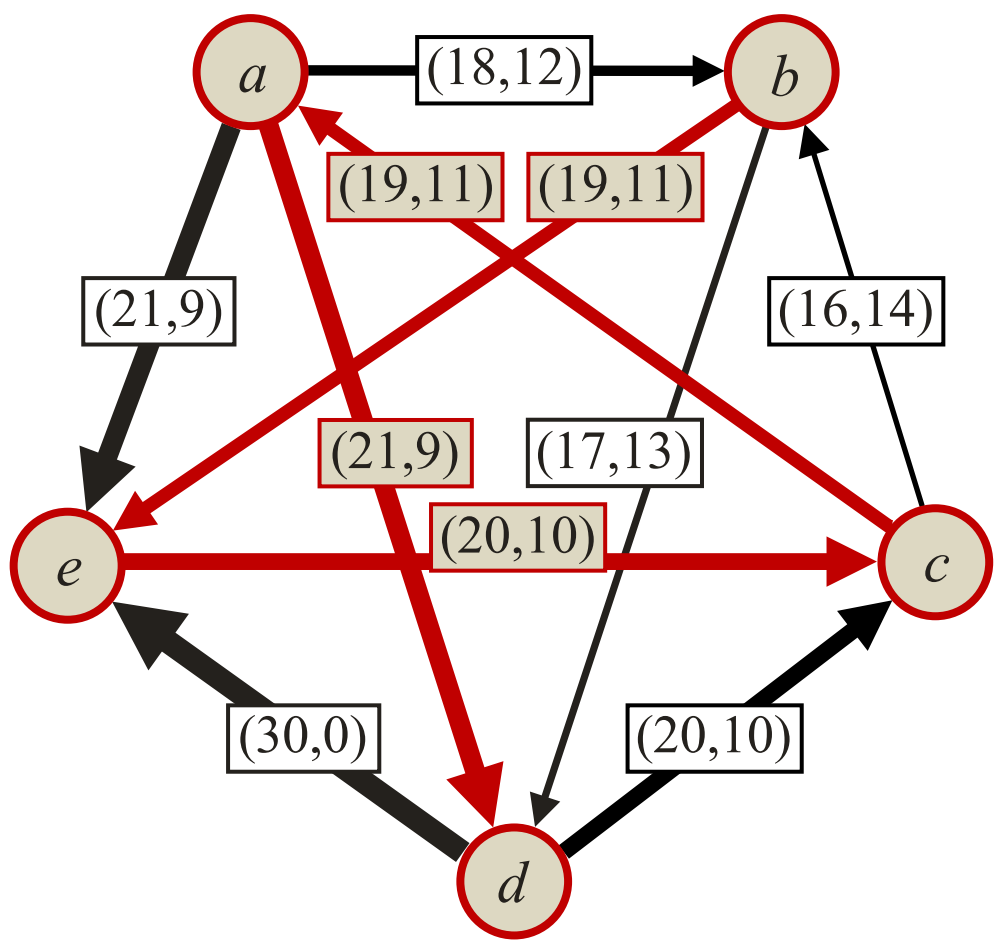

The strongest path from *b* to *d* is:
*b*, (19,11), *e*, (20,10), *c*,
(19,11), *a*, (21,9), *d*

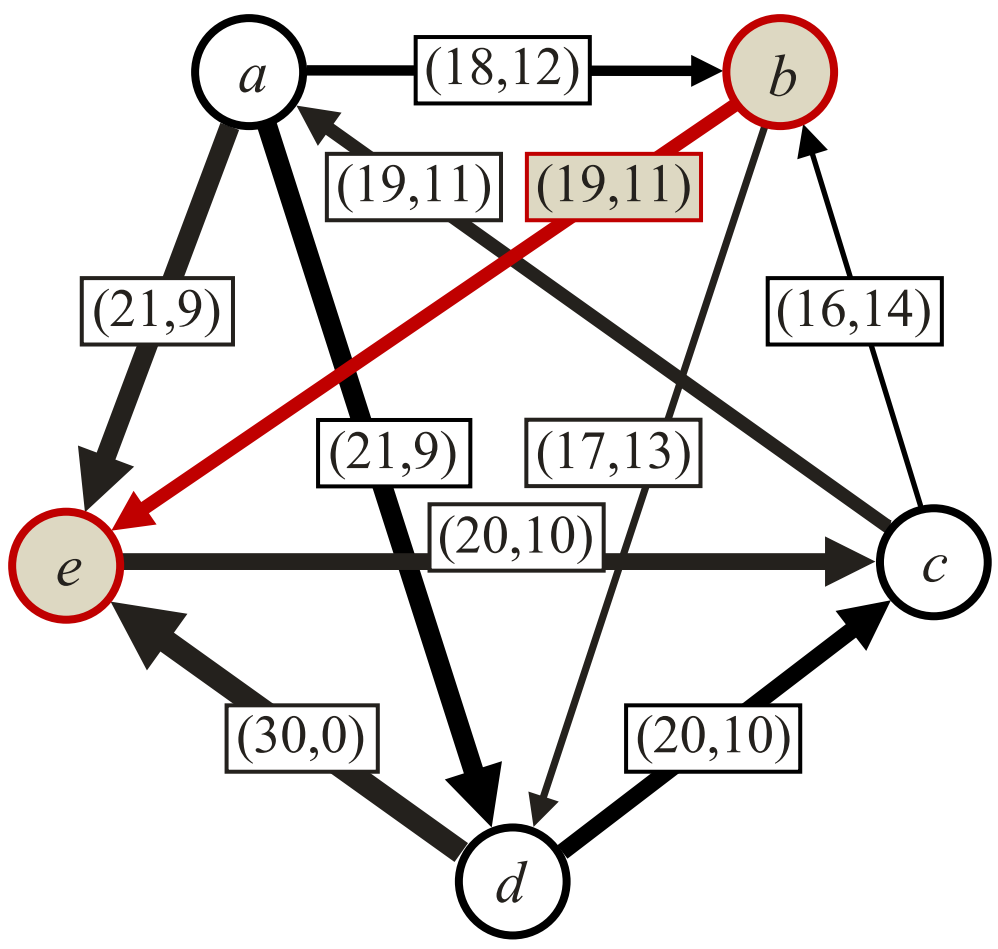

The strongest path from *b* to *e* is:
*b*, (19,11), *e*

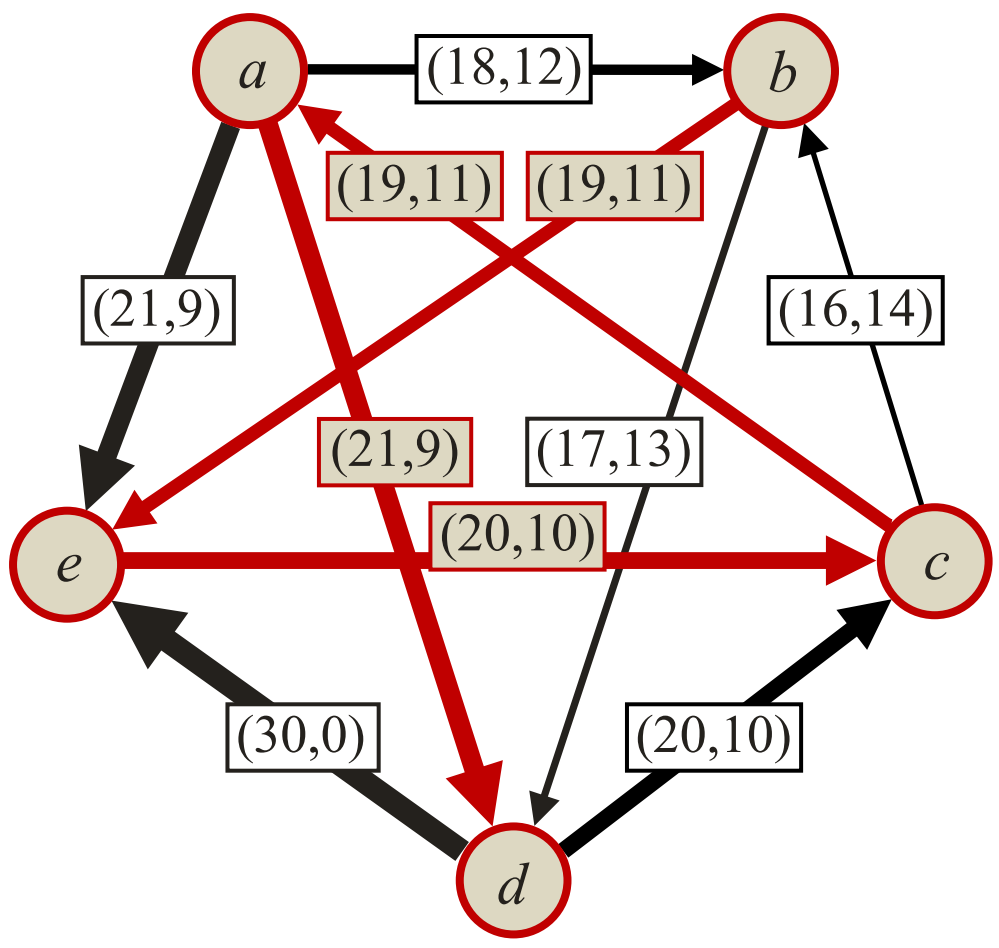

These are the strongest paths
from *b* to every other alternative.

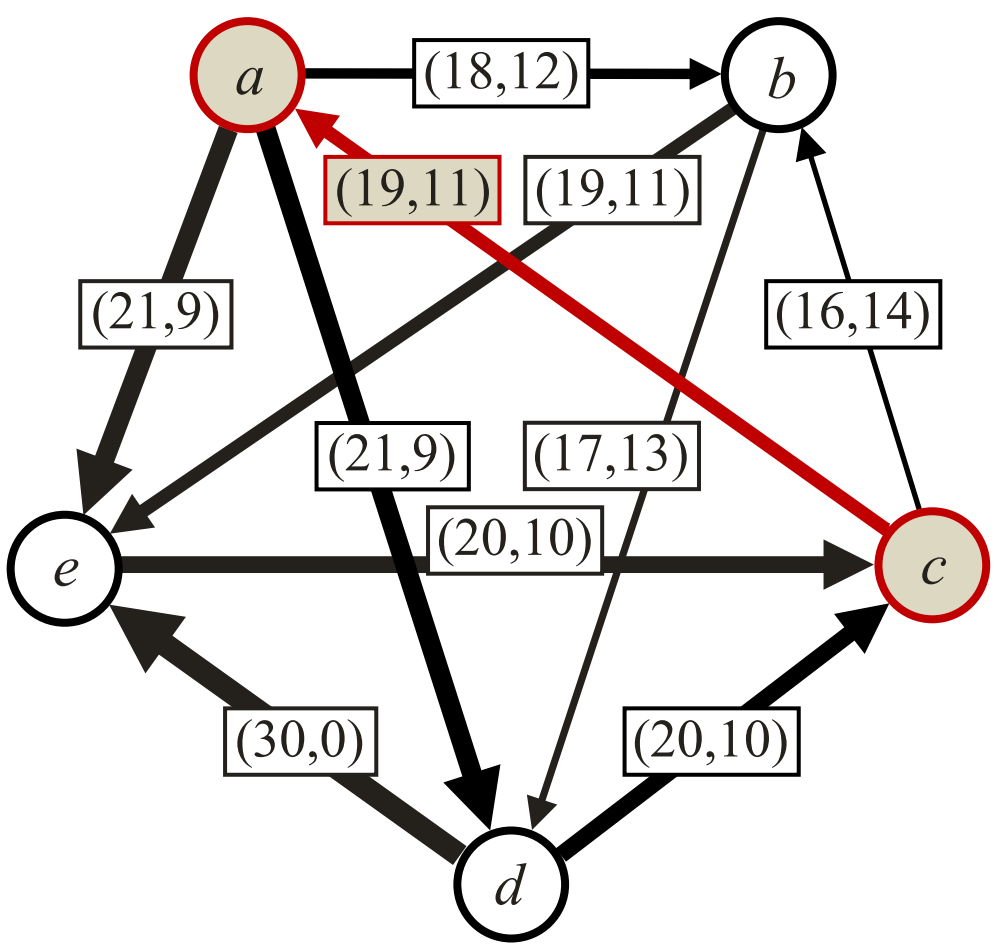

The strongest path from *c* to *a* is:
*c*, (19,11), *a*

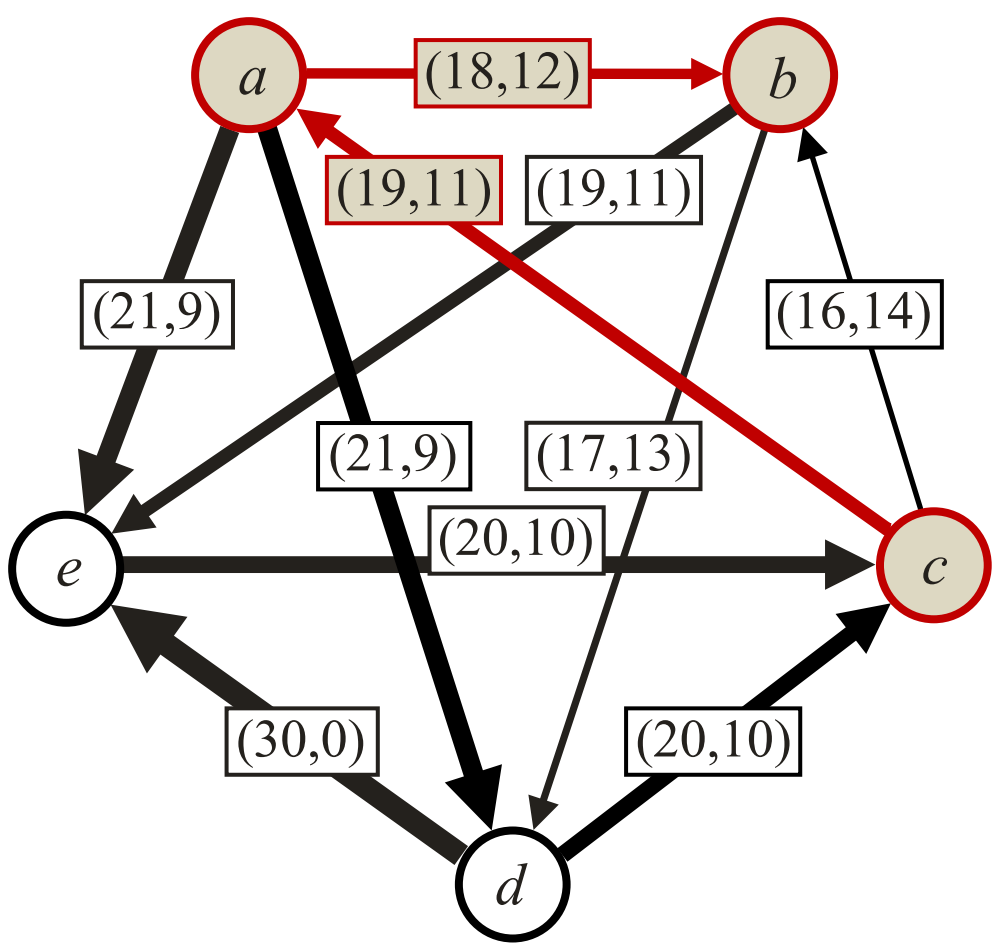

The strongest path from *c* to *b* is:
*c*, (19,11), *a*, (18,12), *b*

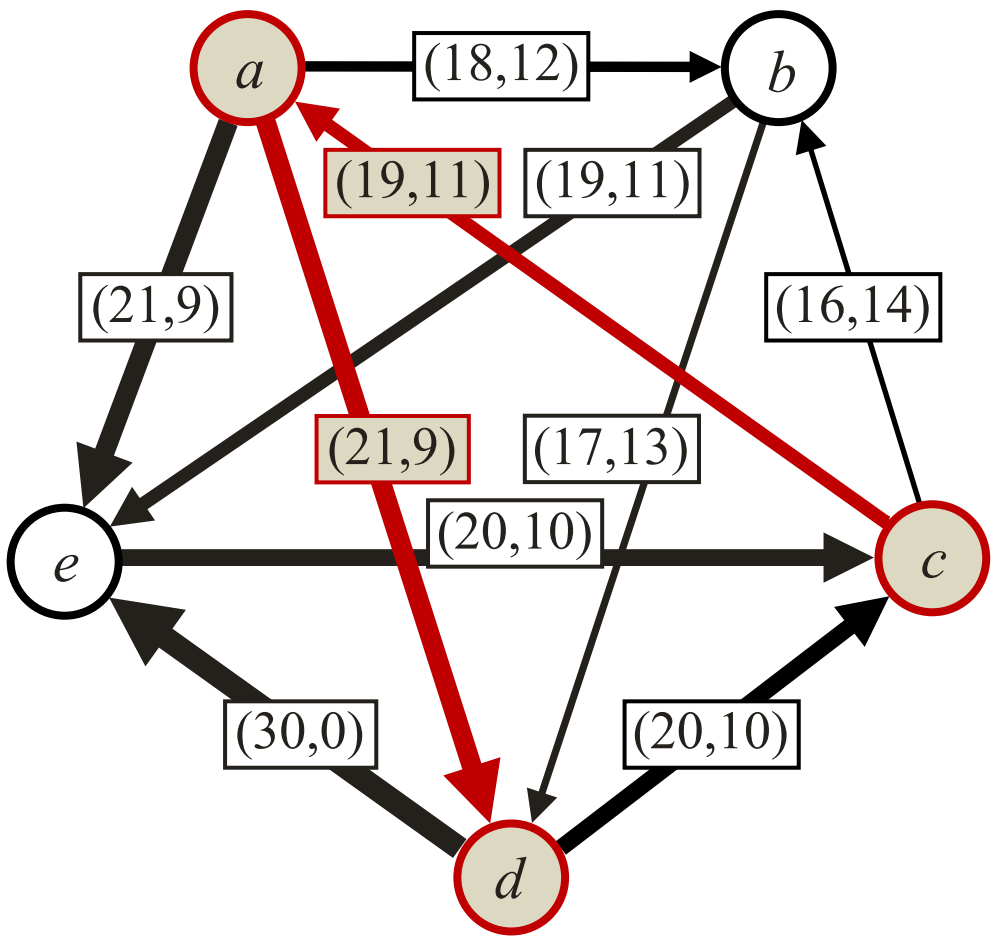

The strongest path from *c* to *d* is:
*c*, (19,11), *a*, (21,9), *d*

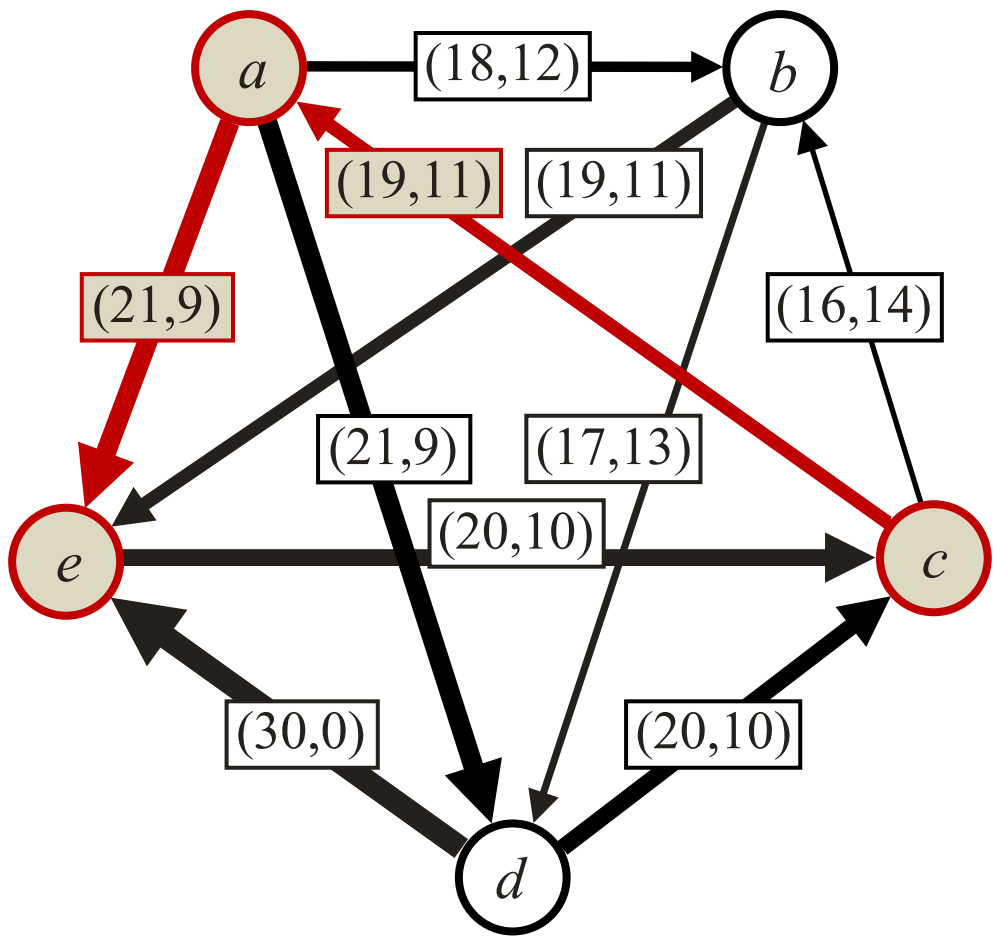

The strongest path from *c* to *e* is:
*c*, (19,11), *a*, (21,9), *e*

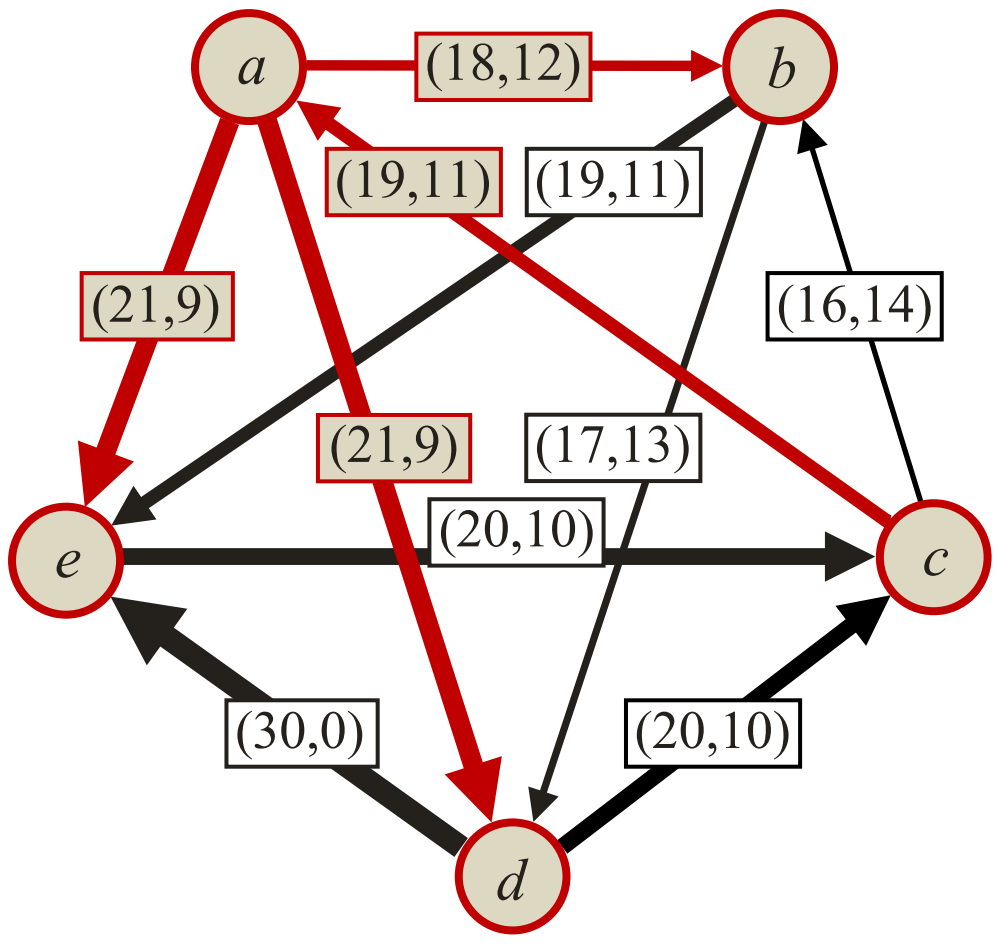

These are the strongest paths
from *c* to every other alternative.

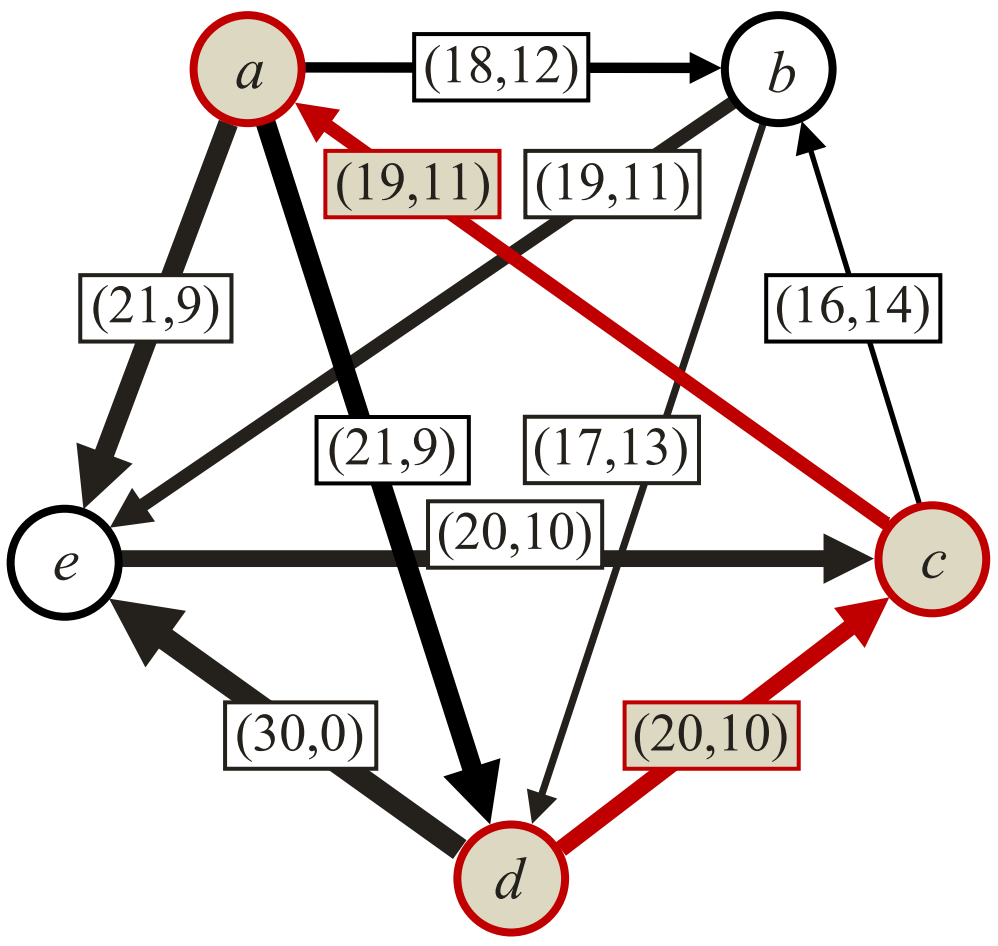

The strongest path from *d* to *a* is:
*d*, (20,10), *c*, (19,11), *a*

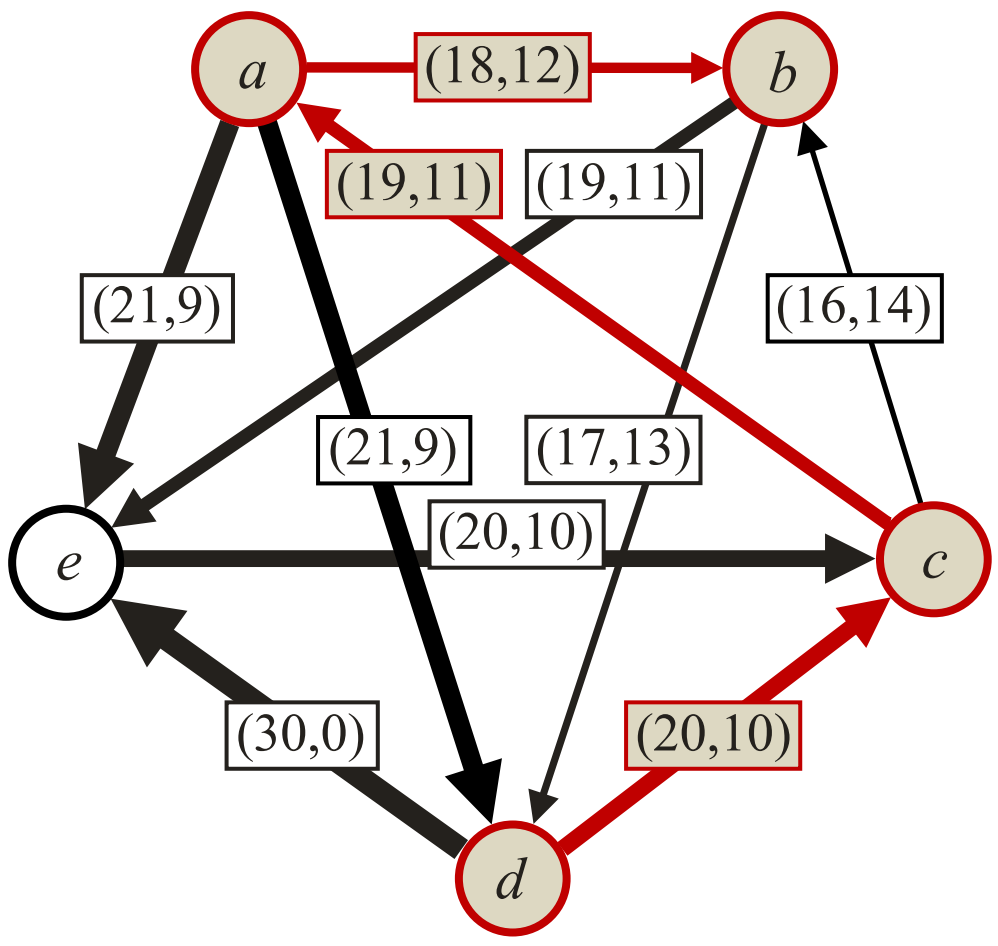

The strongest path from *d* to *b* is:
*d*, (20,10), *c*, (19,11), *a*, (18,12), *b*

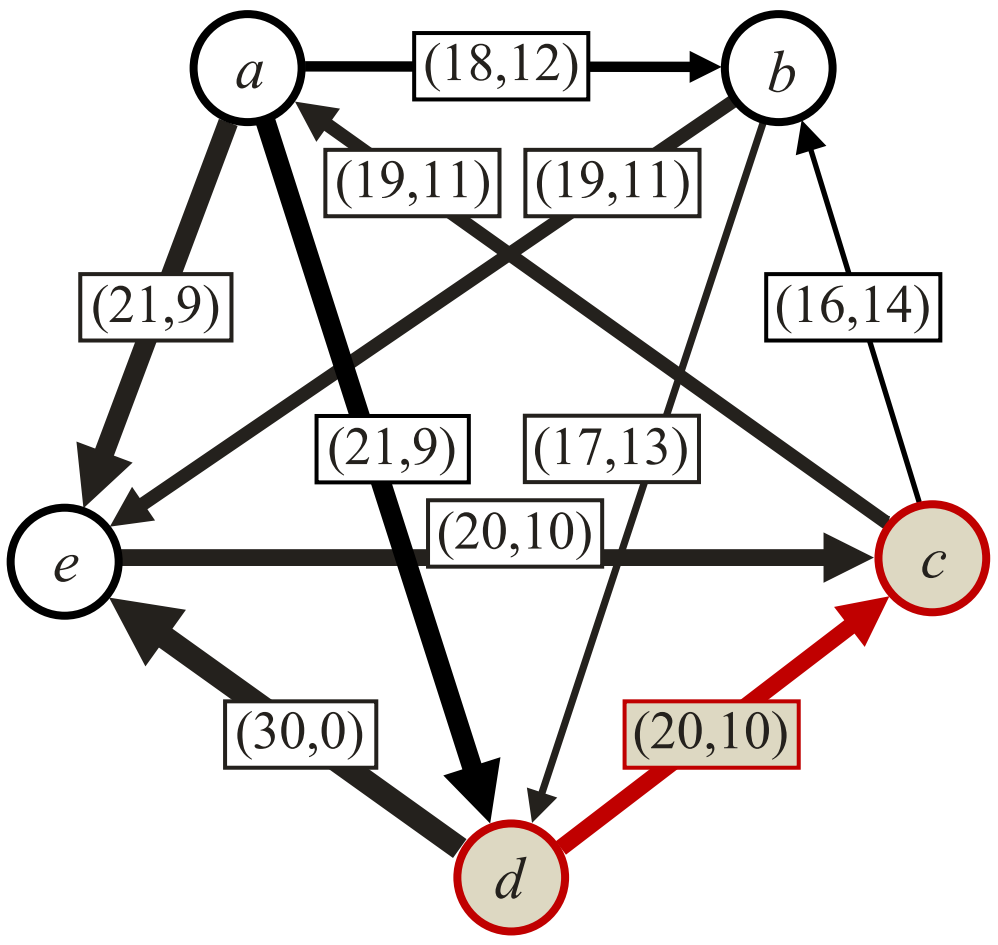

The strongest path from *d* to *c* is:
*d*, (20,10), *c*

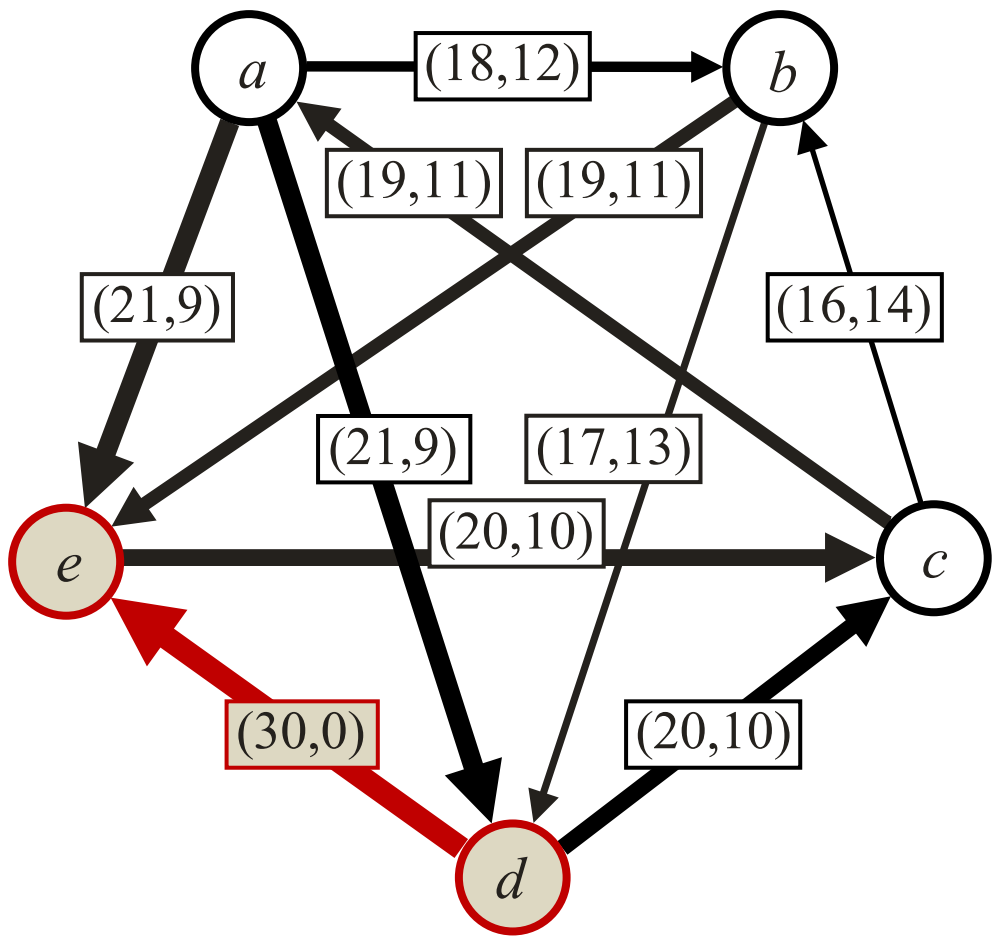

The strongest path from *d* to *e* is:
*d*, (30,0), *e*

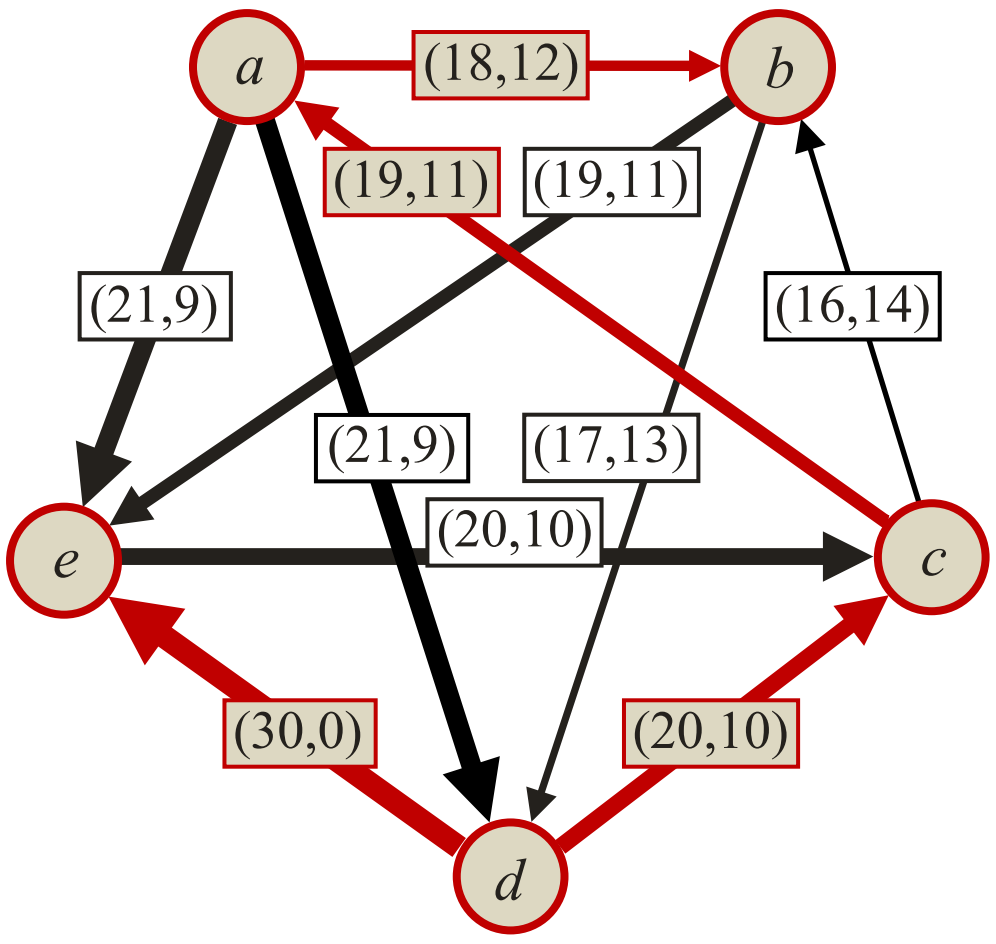

These are the strongest paths
from *d* to every other alternative.

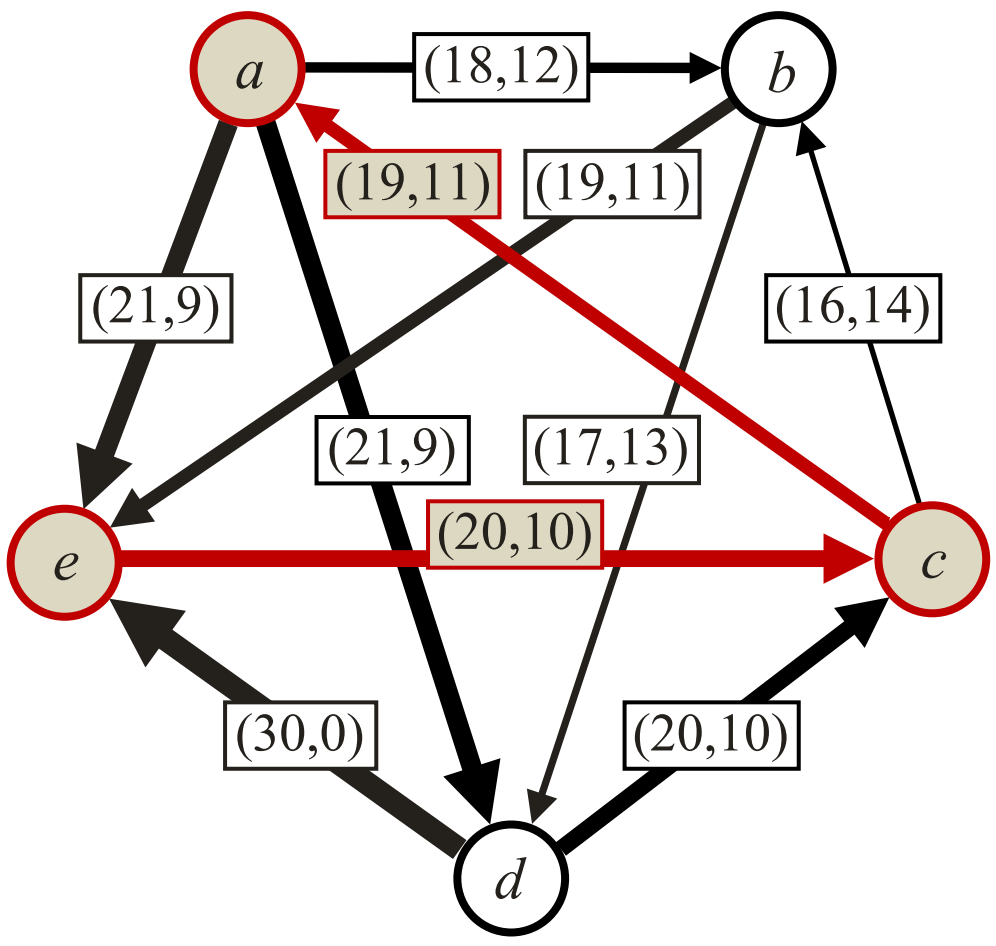

The strongest path from *e* to *a* is:
*e*, (20,10), *c*, (19,11), *a*

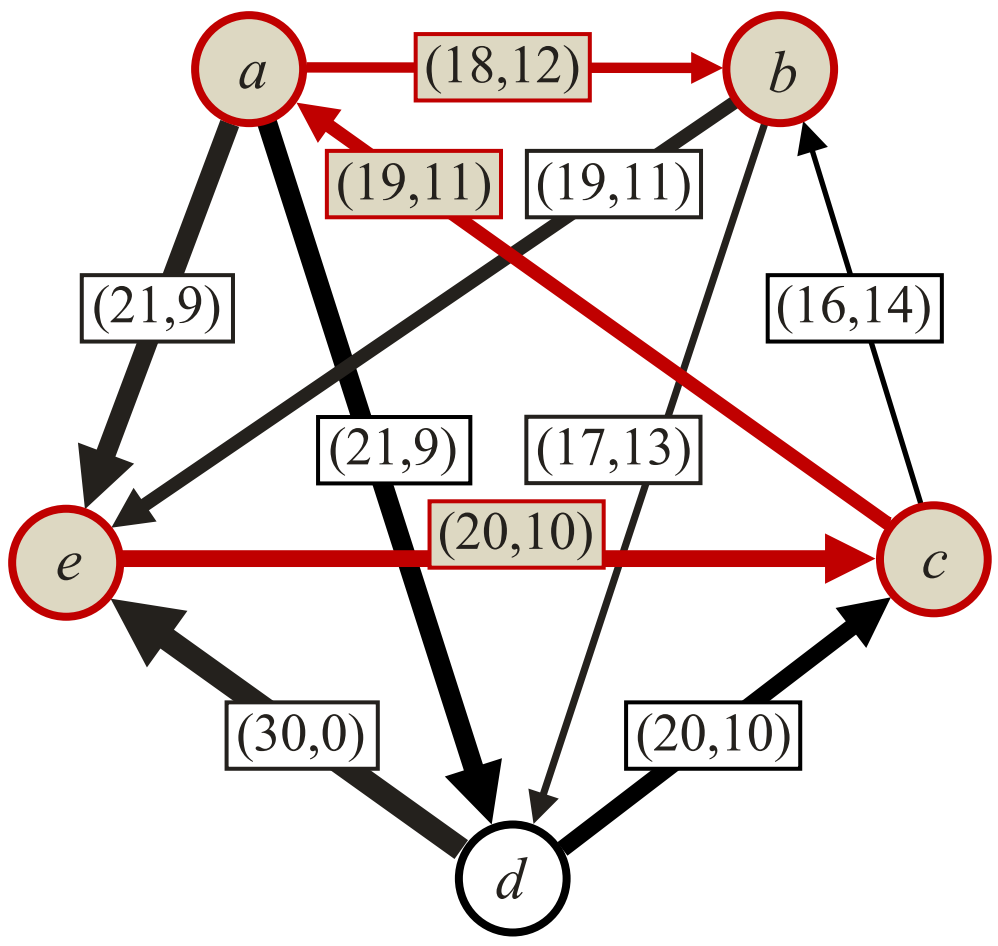

The strongest path from *e* to *b* is:
*e*, (20,10), *c*, (19,11), *a*, (18,12), *b*

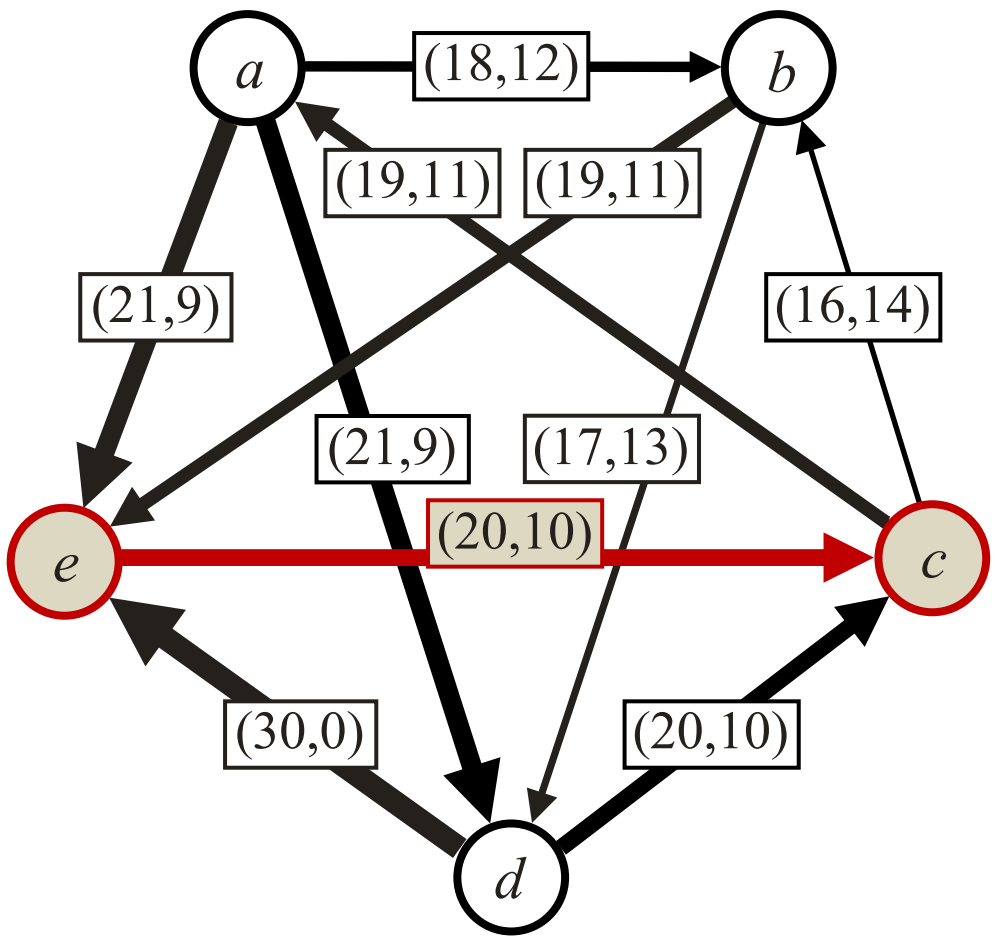

The strongest path from *e* to *c* is:
*e*, (20,10), *c*

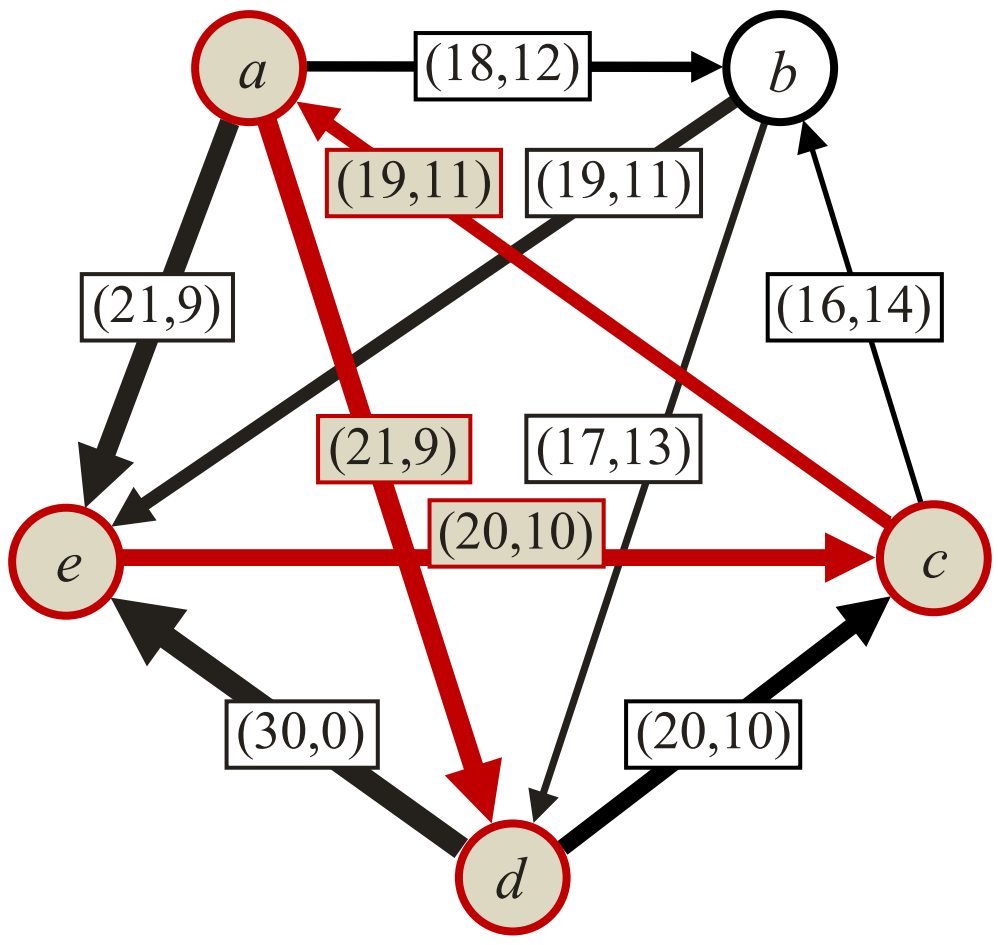

The strongest path from *e* to *d* is:
*e*, (20,10), *c*, (19,11), *a*, (21,9), *d*

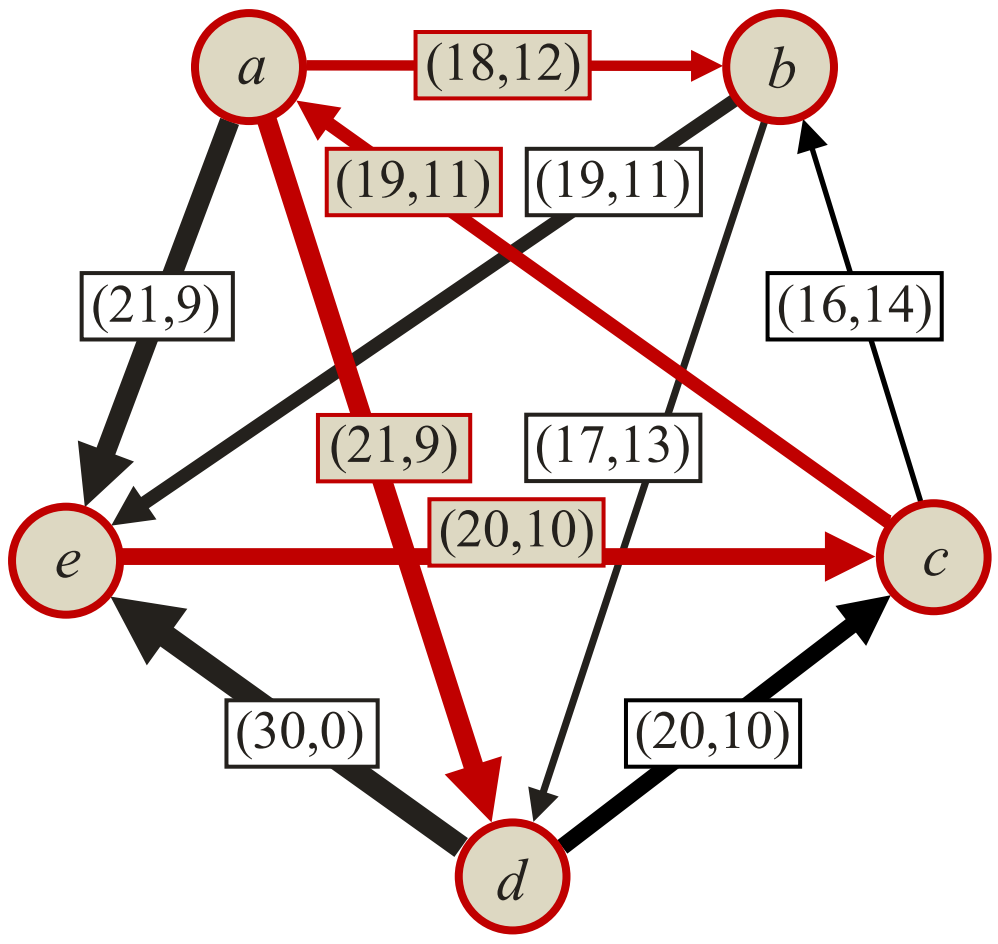

These are the strongest paths
from *e* to every other alternative.

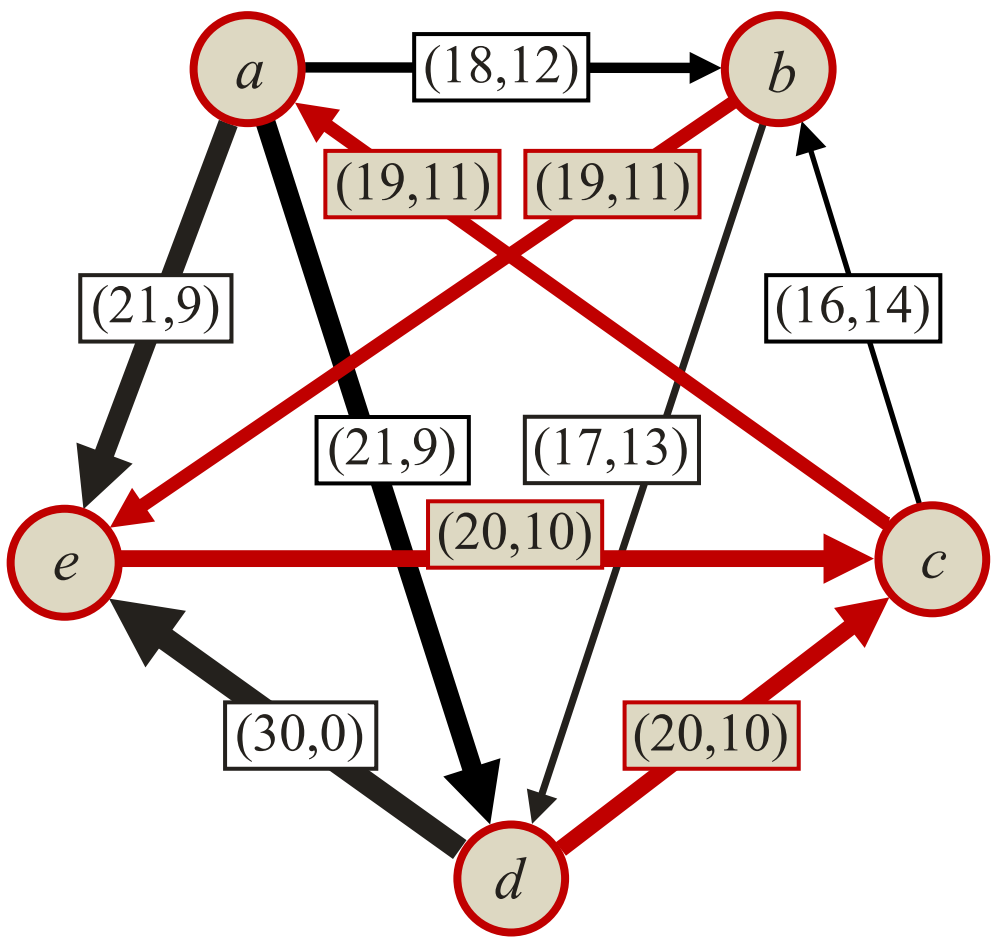

These are the strongest paths
from every other alternative to *a*.

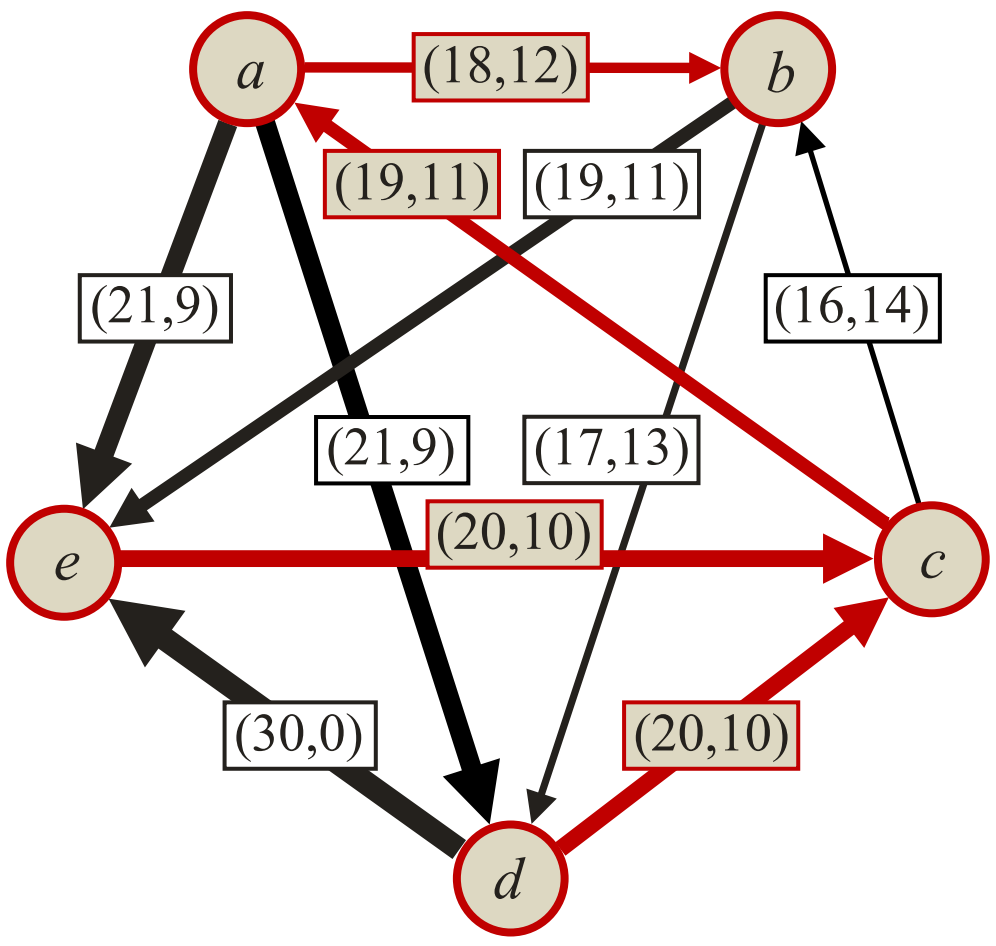

These are the strongest paths
from every other alternative to *b*.

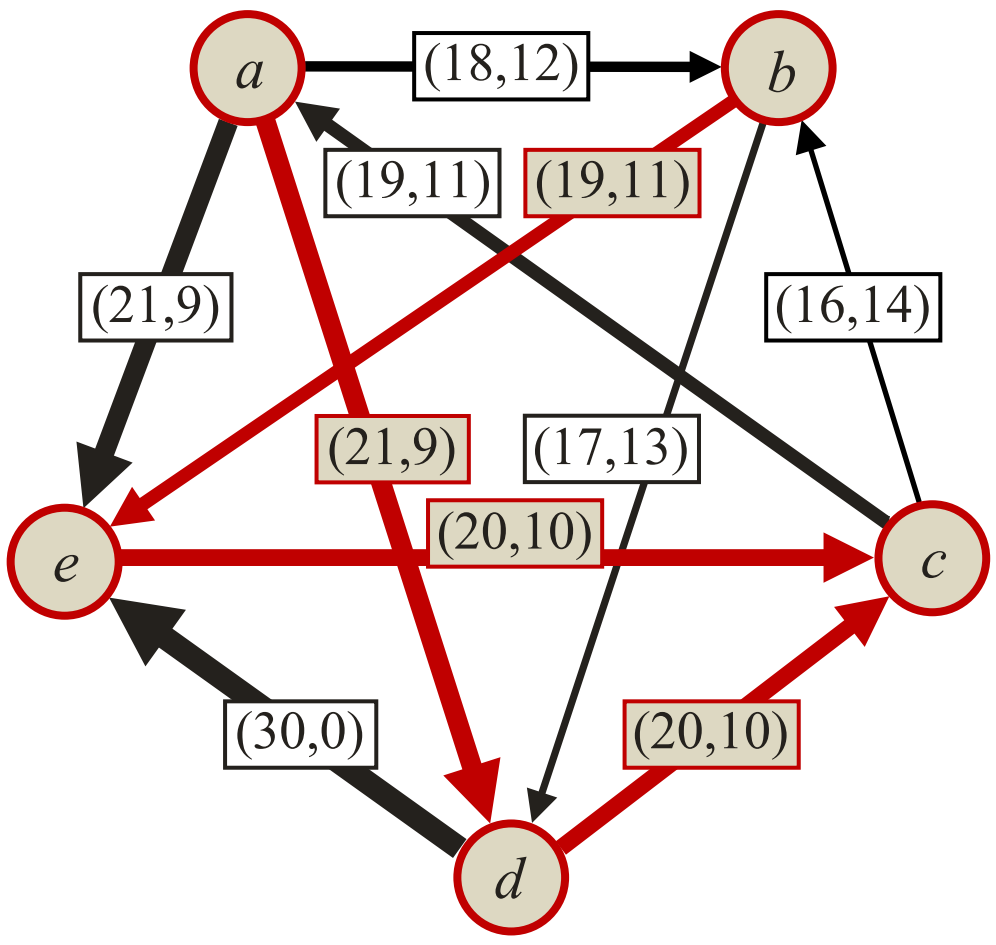

These are the strongest paths
from every other alternative to *c*.

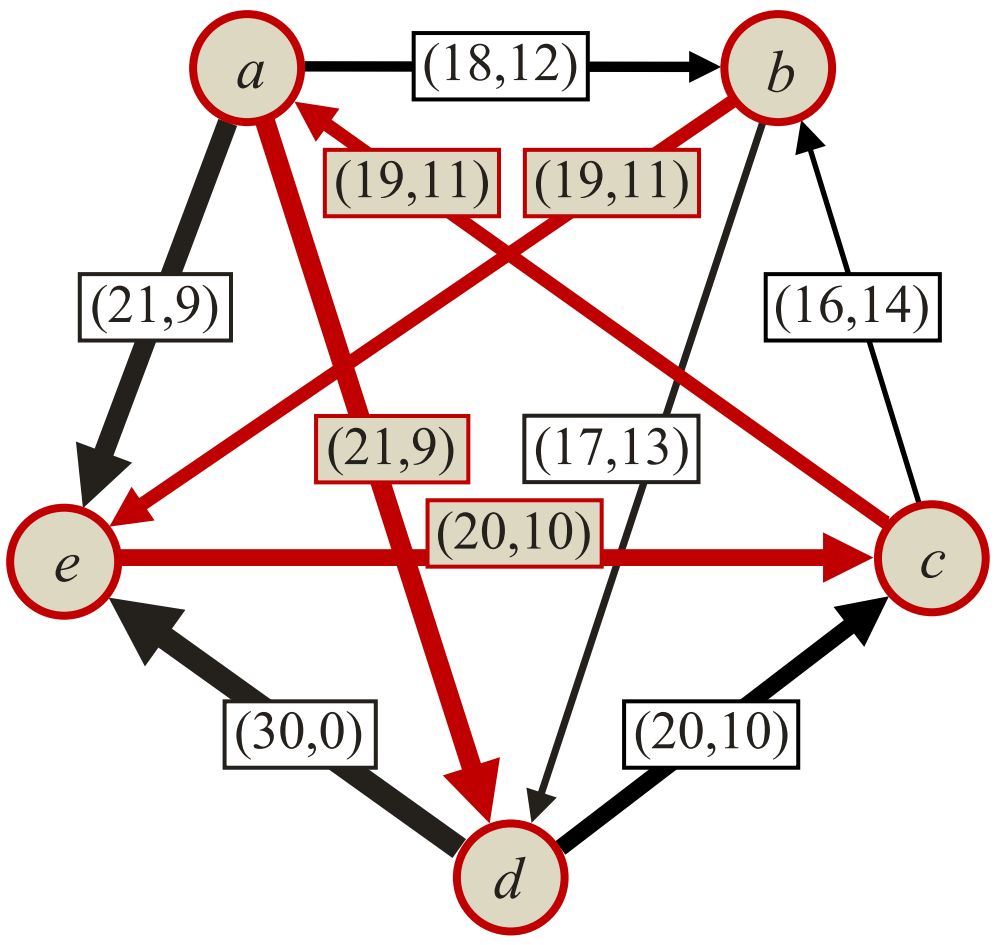

These are the strongest paths
from every other alternative to *d*.

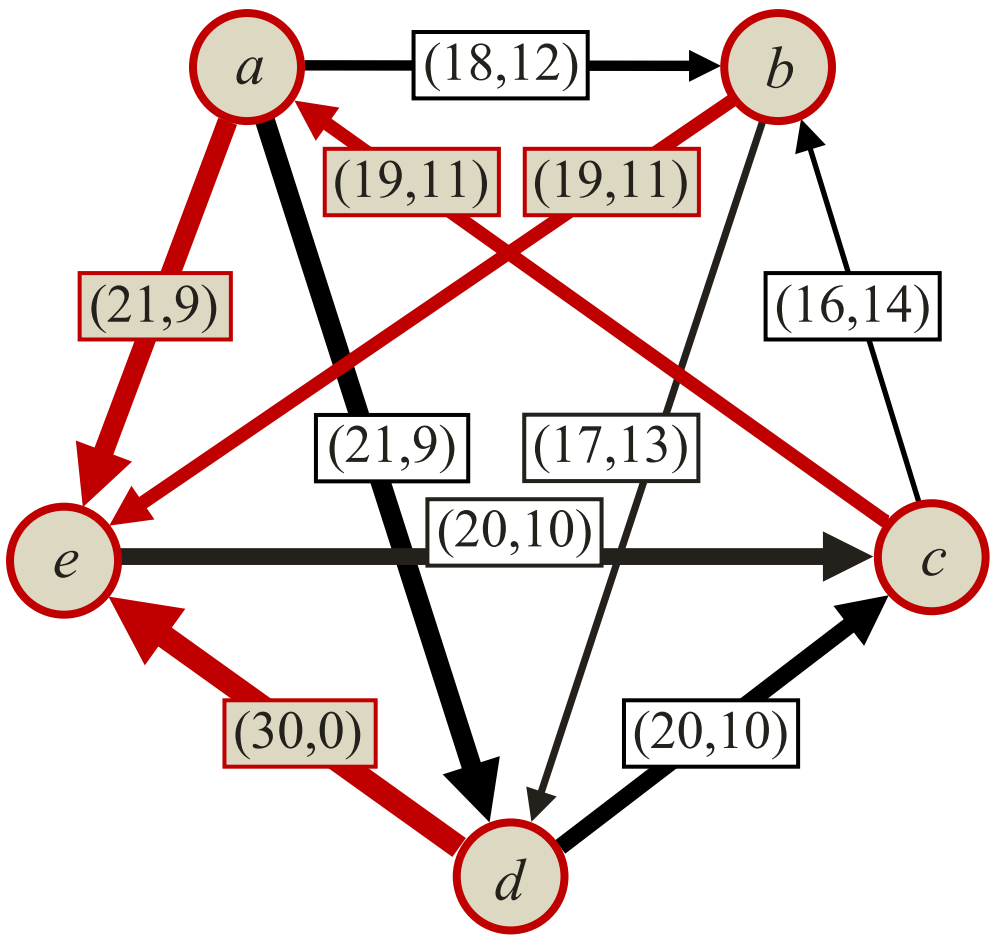

These are the strongest paths
from every other alternative to *e*.

Therefore, the strengths of the strongest paths are:

| | $P_D[*,a]$ | $P_D[*,b]$ | $P_D[*,c]$ | $P_D[*,d]$ | $P_D[*,e]$ |
|---|---|---|---|---|---|
| $P_D[a,*]$ | --- | (18,12) | (20,10) | (21,9) | (21,9) |
| $P_D[b,*]$ | (19,11) | --- | (19,11) | (19,11) | (19,11) |
| $P_D[c,*]$ | (19,11) | (18,12) | --- | (19,11) | (19,11) |
| $P_D[d,*]$ | (19,11) | (18,12) | (20,10) | --- | (30,0) |
| $P_D[e,*]$ | (19,11) | (18,12) | (20,10) | (19,11) | --- |

We get $\mathcal{O}^{\text{new}} = \{ac, ad, ae, ba, bc, bd, be, dc, de, ec\}$ and $\mathcal{S}^{\text{new}} = \{b\}$.

Example 8 shows that the Schulze method, as defined in section 2.2, violates IPDA, as defined in (3.8.1) – (3.8.4). For example, we have (1) $ab \in \mathcal{O}^{\text{old}}$ and $ba \in \mathcal{O}^{\text{new}}$, (2) $cb \in \mathcal{O}^{\text{old}}$ and $bc \in \mathcal{O}^{\text{new}}$, (3) $db \in \mathcal{O}^{\text{old}}$ and $bd \in \mathcal{O}^{\text{new}}$, (4) $a \in \mathcal{S}^{\text{old}}$ and $a \notin \mathcal{S}^{\text{new}}$, and (5) $b \notin \mathcal{S}^{\text{old}}$ and $b \in \mathcal{S}^{\text{new}}$.

Suppose, the strongest paths are calculated with the Floyd-Warshall algorithm, as defined in section 2.3.1. Then the following table documents the $C \cdot (C\text{–}1) \cdot (C\text{–}2) = 60$ steps of the Floyd-Warshall algorithm.

We start with

- $P_D[i,j] := (N^{\text{new}}[i,j],N^{\text{new}}[j,i])$ for all $i \in A$ and $j \in A \setminus \{i\}$.
- $pred[i,j] := i$ for all $i \in A$ and $j \in A \setminus \{i\}$.

| | *i* | *j* | *k* | $P_D[j,k]$ | $P_D[j,i]$ | $P_D[i,k]$ | *pred*[*j*,*k*] | *pred*[*i*,*k*] | result |
|---|---|---|---|---|---|---|---|---|---|
| 1 | *a* | *b* | *c* | (14,16) | (12,18) | (11,19) | *b* | *a* | |
| 2 | *a* | *b* | *d* | (17,13) | (12,18) | (21,9) | *b* | *a* | |
| 3 | *a* | *b* | *e* | (19,11) | (12,18) | (21,9) | *b* | *a* | |
| 4 | *a* | *c* | *b* | (16,14) | (19,11) | (18,12) | *c* | *a* | $P_D[c,b]$ is updated from (16,14) to (18,12); *pred*[*c*,*b*] is updated from *c* to *a*. |
| 5 | *a* | *c* | *d* | (10,20) | (19,11) | (21,9) | *c* | *a* | $P_D[c,d]$ is updated from (10,20) to (19,11); *pred*[*c*,*d*] is updated from *c* to *a*. |
| 6 | *a* | *c* | *e* | (10,20) | (19,11) | (21,9) | *c* | *a* | $P_D[c,e]$ is updated from (10,20) to (19,11); *pred*[*c*,*e*] is updated from *c* to *a*. |
| 7 | *a* | *d* | *b* | (13,17) | (9,21) | (18,12) | *d* | *a* | |
| 8 | *a* | *d* | *c* | (20,10) | (9,21) | (11,19) | *d* | *a* | |
| 9 | *a* | *d* | *e* | (30,0) | (9,21) | (21,9) | *d* | *a* | |
| 10 | *a* | *e* | *b* | (11,19) | (9,21) | (18,12) | *e* | *a* | |
| 11 | *a* | *e* | *c* | (20,10) | (9,21) | (11,19) | *e* | *a* | |
| 12 | *a* | *e* | *d* | (0,30) | (9,21) | (21,9) | *e* | *a* | $P_D[e,d]$ is updated from (0,30) to (9,21); *pred*[*e*,*d*] is updated from *e* to *a*. |
| 13 | *b* | *a* | *c* | (11,19) | (18,12) | (14,16) | *a* | *b* | $P_D[a,c]$ is updated from (11,19) to (14,16); *pred*[*a*,*c*] is updated from *a* to *b*. |
| 14 | *b* | *a* | *d* | (21,9) | (18,12) | (17,13) | *a* | *b* | |
| 15 | *b* | *a* | *e* | (21,9) | (18,12) | (19,11) | *a* | *b* | |
| 16 | *b* | *c* | *a* | (19,11) | (18,12) | (12,18) | *c* | *b* | |
| 17 | *b* | *c* | *d* | (19,11) | (18,12) | (17,13) | *a* | *b* | |
| 18 | *b* | *c* | *e* | (19,11) | (18,12) | (19,11) | *a* | *b* | |
| 19 | *b* | *d* | *a* | (9,21) | (13,17) | (12,18) | *d* | *b* | $P_D[d,a]$ is updated from (9,21) to (12,18); *pred*[*d*,*a*] is updated from *d* to *b*. |
| 20 | *b* | *d* | *c* | (20,10) | (13,17) | (14,16) | *d* | *b* | |
| 21 | *b* | *d* | *e* | (30,0) | (13,17) | (19,11) | *d* | *b* | |
| 22 | *b* | *e* | *a* | (9,21) | (11,19) | (12,18) | *e* | *b* | $P_D[e,a]$ is updated from (9,21) to (11,19); *pred*[*e*,*a*] is updated from *e* to *b*. |
| 23 | *b* | *e* | *c* | (20,10) | (11,19) | (14,16) | *e* | *b* | |
| 24 | *b* | *e* | *d* | (9,21) | (11,19) | (17,13) | *a* | *b* | $P_D[e,d]$ is updated from (9,21) to (11,19); *pred*[*e*,*d*] is updated from *a* to *b*. |
| 25 | *c* | *a* | *b* | (18,12) | (14,16) | (18,12) | *a* | *a* | |
| 26 | *c* | *a* | *d* | (21,9) | (14,16) | (19,11) | *a* | *a* | |
| 27 | *c* | *a* | *e* | (21,9) | (14,16) | (19,11) | *a* | *a* | |
| 28 | *c* | *b* | *a* | (12,18) | (14,16) | (19,11) | *b* | *c* | $P_D[b,a]$ is updated from (12,18) to (14,16); *pred*[*b*,*a*] is updated from *b* to *c*. |
| 29 | *c* | *b* | *d* | (17,13) | (14,16) | (19,11) | *b* | *a* | |
| 30 | *c* | *b* | *e* | (19,11) | (14,16) | (19,11) | *b* | *a* | |

| | *i* | *j* | *k* | $P_D[j,k]$ | $P_D[j,i]$ | $P_D[i,k]$ | *pred*[*j*,*k*] | *pred*[*i*,*k*] | result |
|---|---|---|---|---|---|---|---|---|---|
| 31 | *c* | *d* | *a* | (12,18) | (20,10) | (19,11) | *b* | *c* | $P_D[d,a]$ is updated from (12,18) to (19,11); *pred*[*d*,*a*] is updated from *b* to *c*. |
| 32 | *c* | *d* | *b* | (13,17) | (20,10) | (18,12) | *d* | *a* | $P_D[d,b]$ is updated from (13,17) to (18,12); *pred*[*d*,*b*] is updated from *d* to *a*. |
| 33 | *c* | *d* | *e* | (30,0) | (20,10) | (19,11) | *d* | *a* | |
| 34 | *c* | *e* | *a* | (11,19) | (20,10) | (19,11) | *b* | *c* | $P_D[e,a]$ is updated from (11,19) to (19,11); *pred*[*e*,*a*] is updated from *b* to *c*. |
| 35 | *c* | *e* | *b* | (11,19) | (20,10) | (18,12) | *e* | *a* | $P_D[e,b]$ is updated from (11,19) to (18,12); *pred*[*e*,*b*] is updated from *e* to *a*. |
| 36 | *c* | *e* | *d* | (11,19) | (20,10) | (19,11) | *b* | *a* | $P_D[e,d]$ is updated from (11,19) to (19,11); *pred*[*e*,*d*] is updated from *b* to *a*. |
| 37 | *d* | *a* | *b* | (18,12) | (21,9) | (18,12) | *a* | *a* | |
| 38 | *d* | *a* | *c* | (14,16) | (21,9) | (20,10) | *b* | *d* | $P_D[a,c]$ is updated from (14,16) to (20,10); *pred*[*a*,*c*] is updated from *b* to *d*. |
| 39 | *d* | *a* | *e* | (21,9) | (21,9) | (30,0) | *a* | *d* | |
| 40 | *d* | *b* | *a* | (14,16) | (17,13) | (19,11) | *c* | *c* | $P_D[b,a]$ is updated from (14,16) to (17,13). |
| 41 | *d* | *b* | *c* | (14,16) | (17,13) | (20,10) | *b* | *d* | $P_D[b,c]$ is updated from (14,16) to (17,13); *pred*[*b*,*c*] is updated from *b* to *d*. |
| 42 | *d* | *b* | *e* | (19,11) | (17,13) | (30,0) | *b* | *d* | |
| 43 | *d* | *c* | *a* | (19,11) | (19,11) | (19,11) | *c* | *c* | |
| 44 | *d* | *c* | *b* | (18,12) | (19,11) | (18,12) | *a* | *a* | |
| 45 | *d* | *c* | *e* | (19,11) | (19,11) | (30,0) | *a* | *d* | |
| 46 | *d* | *e* | *a* | (19,11) | (19,11) | (19,11) | *c* | *c* | |
| 47 | *d* | *e* | *b* | (18,12) | (19,11) | (18,12) | *a* | *a* | |
| 48 | *d* | *e* | *c* | (20,10) | (19,11) | (20,10) | *e* | *d* | |
| 49 | *e* | *a* | *b* | (18,12) | (21,9) | (18,12) | *a* | *a* | |
| 50 | *e* | *a* | *c* | (20,10) | (21,9) | (20,10) | *d* | *e* | |
| 51 | *e* | *a* | *d* | (21,9) | (21,9) | (19,11) | *a* | *a* | |
| 52 | *e* | *b* | *a* | (17,13) | (19,11) | (19,11) | *c* | *c* | $P_D[b,a]$ is updated from (17,13) to (19,11). |
| 53 | *e* | *b* | *c* | (17,13) | (19,11) | (20,10) | *d* | *e* | $P_D[b,c]$ is updated from (17,13) to (19,11); *pred*[*b*,*c*] is updated from *d* to *e*. |
| 54 | *e* | *b* | *d* | (17,13) | (19,11) | (19,11) | *b* | *a* | $P_D[b,d]$ is updated from (17,13) to (19,11); *pred*[*b*,*d*] is updated from *b* to *a*. |
| 55 | *e* | *c* | *a* | (19,11) | (19,11) | (19,11) | *c* | *c* | |
| 56 | *e* | *c* | *b* | (18,12) | (19,11) | (18,12) | *a* | *a* | |
| 57 | *e* | *c* | *d* | (19,11) | (19,11) | (19,11) | *a* | *a* | |
| 58 | *e* | *d* | *a* | (19,11) | (30,0) | (19,11) | *c* | *c* | |
| 59 | *e* | *d* | *b* | (18,12) | (30,0) | (18,12) | *a* | *a* | |
| 60 | *e* | *d* | *c* | (20,10) | (30,0) | (20,10) | *d* | *e* | |

### 3.9. Example 9

#### 3.9.1. Situation #1

Example 9 (old):

| | |
|---|---|
| 5 voters | $a \succ_v c \succ_v d \succ_v b$ |
| 2 voters | $b \succ_v c \succ_v d \succ_v a$ |
| 4 voters | $b \succ_v d \succ_v a \succ_v c$ |
| 2 voters | $c \succ_v d \succ_v a \succ_v b$ |

The pairwise matrix $N^{old}$ looks as follows:

| | $N^{old}[*,a]$ | $N^{old}[*,b]$ | $N^{old}[*,c]$ | $N^{old}[*,d]$ |
|---|---|---|---|---|
| $N^{old}[a,*]$ | --- | 7 | 9 | 5 |
| $N^{old}[b,*]$ | 6 | --- | 6 | 6 |
| $N^{old}[c,*]$ | 4 | 7 | --- | 9 |
| $N^{old}[d,*]$ | 8 | 7 | 4 | --- |

The corresponding digraph looks as follows:

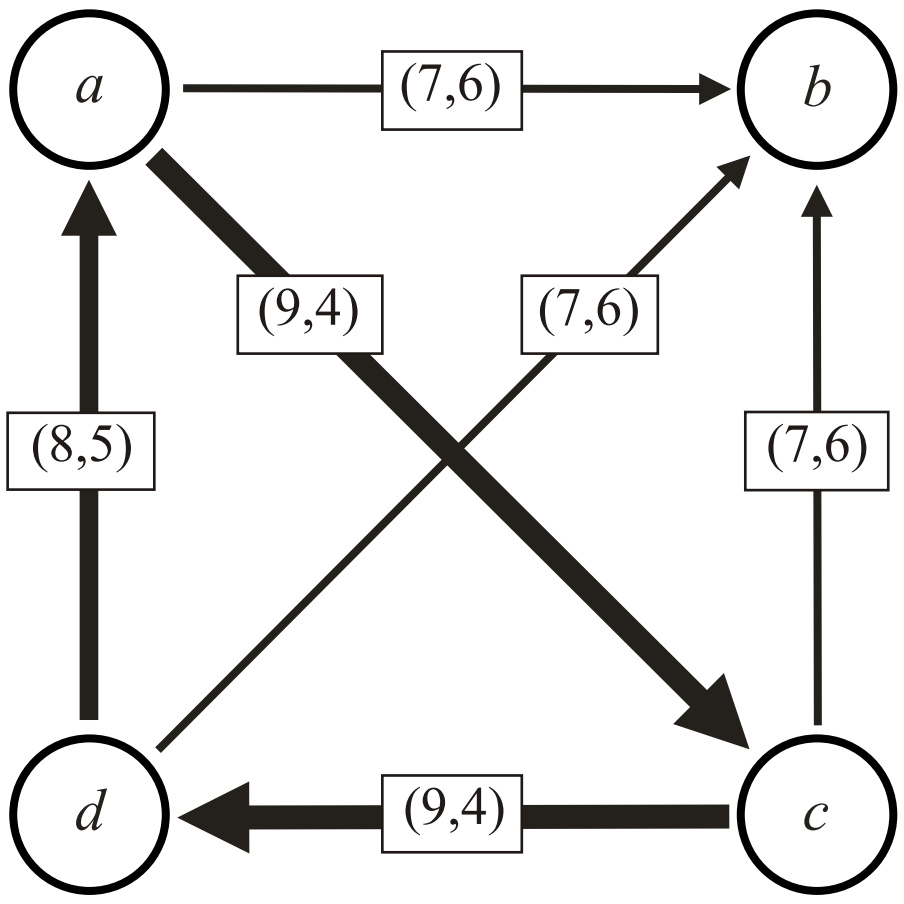

The following table lists the strongest paths, as determined by the Floyd-Warshall algorithm, as defined in section 2.3.1. The critical links of the strongest paths are underlined:

| | ... to *a* | ... to *b* | ... to *c* | ... to *d* | ... to every other alternative |
|---|---|---|---|---|---|
| from *a* ... | --- | 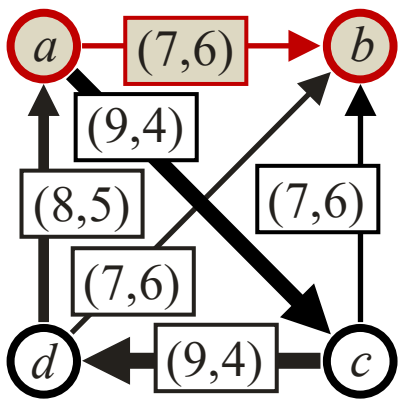 *a*, (7,6), *b* | 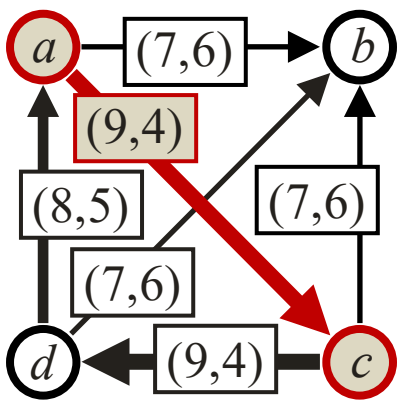 *a*, (9,4), *c* | 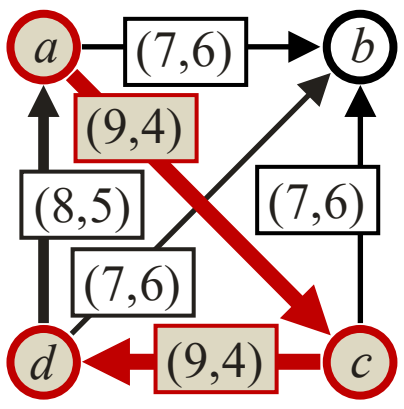 *a*, (9,4), *c*, (9,4), *d* | 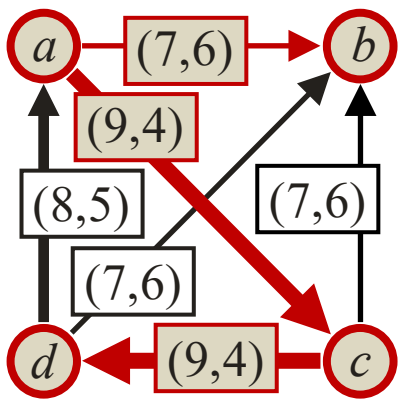 |
| from *b* ... | 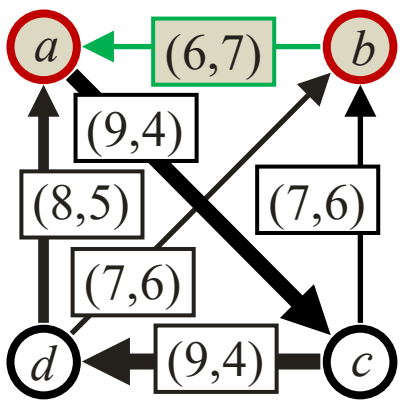 *b*, (6,7), *a* | --- | 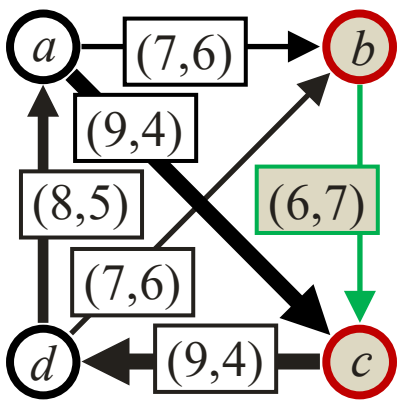 *b*, (6,7), *c* | 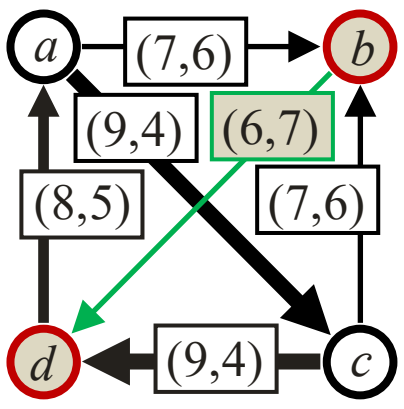 *b*, (6,7), *d* | 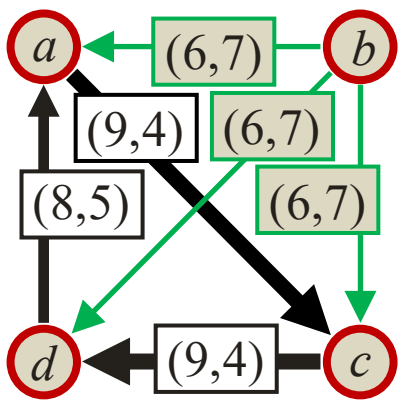 |
| from *c* ... | 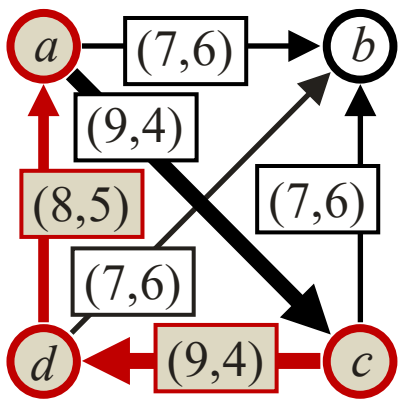 *c*, (9,4), *d*, (8,5), *a* | 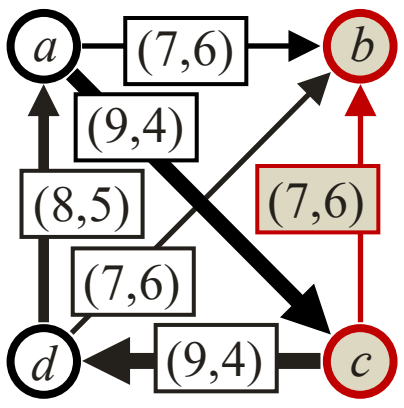 *c*, (7,6), *b* | --- | 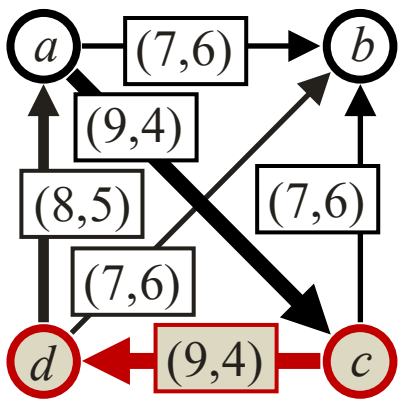 *c*, (9,4), *d* | 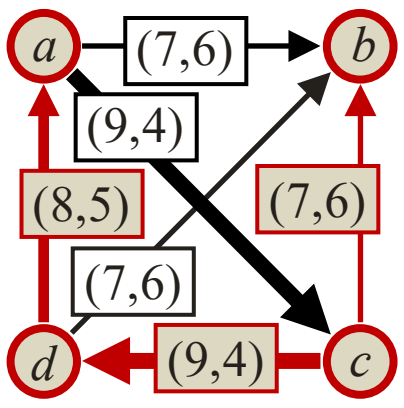 |
| from *d* ... | 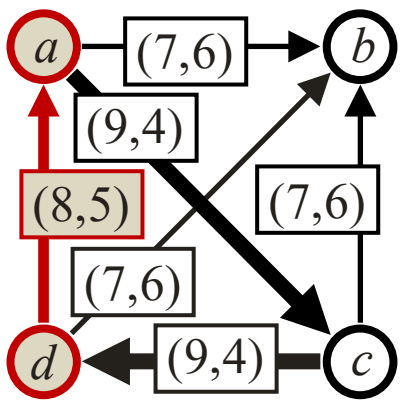 *d*, (8,5), *a* | 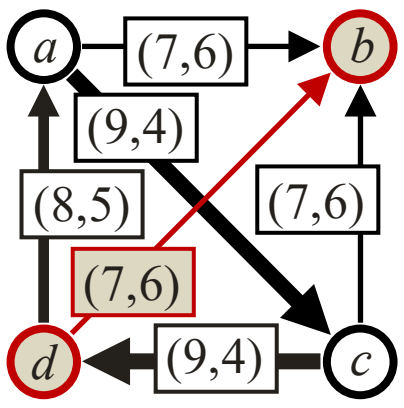 *d*, (7,6), *b* | 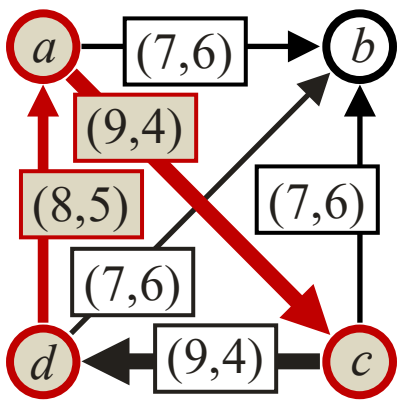 *d*, (8,5), *a*, (9,4), *c* | --- | 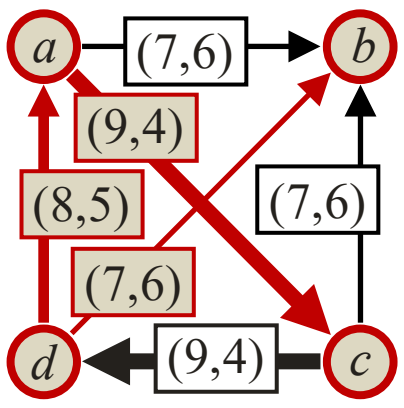 |
| from every other alternative ... | 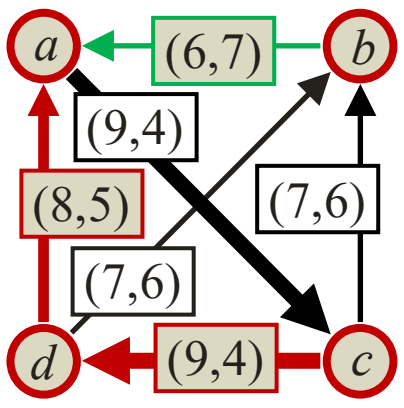 | 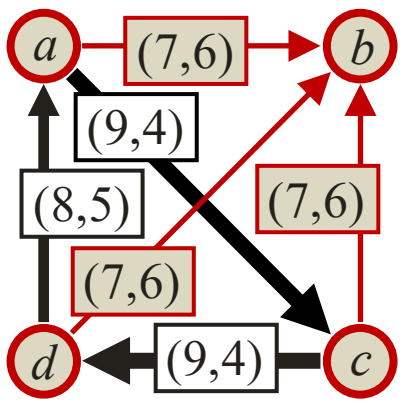 | 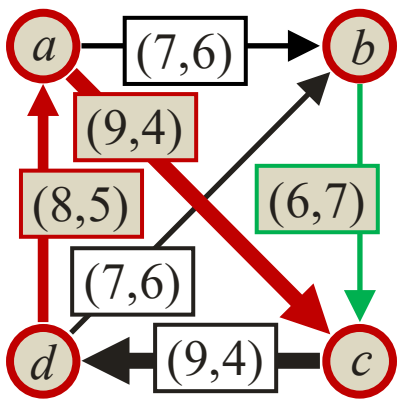 | 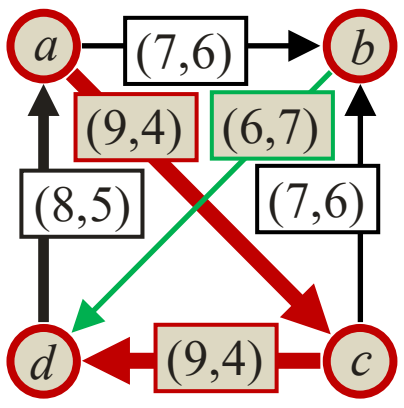 | --- |

The strengths of the strongest paths are:

| | $P_D[*,a]$ | $P_D[*,b]$ | $P_D[*,c]$ | $P_D[*,d]$ |
|---|---|---|---|---|
| $P_D[a,*]$ | --- | (7,6) | (9,4) | (9,4) |
| $P_D[b,*]$ | (6,7) | --- | (6,7) | (6,7) |
| $P_D[c,*]$ | (8,5) | (7,6) | --- | (9,4) |
| $P_D[d,*]$ | (8,5) | (7,6) | (8,5) | --- |

We get $\mathcal{O}^{\text{old}} = \{ab, ac, ad, cb, cd, db\}$ and $\mathcal{S}^{\text{old}} = \{a\}$.

As there are no paths from alternative *b* to the other alternatives that contain only wins or ties, these paths must necessarily contain losses. These losses are marked in green in the above table of the strongest paths.

Suppose, the strongest paths are calculated with the Floyd-Warshall algorithm, as defined in section 2.3.1. Then the following table documents the $C \cdot (C-1) \cdot (C-2) = 24$ steps of the Floyd-Warshall algorithm.

We start with

- $P_D[i,j] := (N^{\text{old}}[i,j],N^{\text{old}}[j,i])$ for all $i \in A$ and $j \in A \setminus \{i\}$.
- $pred[i,j] := i$ for all $i \in A$ and $j \in A \setminus \{i\}$.

| | *i* | *j* | *k* | $P_D[j,k]$ | $P_D[j,i]$ | $P_D[i,k]$ | *pred*[*j*,*k*] | *pred*[*i*,*k*] | result |
|---|---|---|---|---|---|---|---|---|---|
| 1 | *a* | *b* | *c* | (6,7) | (6,7) | (9,4) | *b* | *a* | |
| 2 | *a* | *b* | *d* | (6,7) | (6,7) | (5,8) | *b* | *a* | |
| 3 | *a* | *c* | *b* | (7,6) | (4,9) | (7,6) | *c* | *a* | |
| 4 | *a* | *c* | *d* | (9,4) | (4,9) | (5,8) | *c* | *a* | |
| 5 | *a* | *d* | *b* | (7,6) | (8,5) | (7,6) | *d* | *a* | |
| 6 | *a* | *d* | *c* | (4,9) | (8,5) | (9,4) | *d* | *a* | $P_D[d,c]$ is updated from (4,9) to (8,5); *pred*[*d*,*c*] is updated from *d* to *a*. |
| 7 | *b* | *a* | *c* | (9,4) | (7,6) | (6,7) | *a* | *b* | |
| 8 | *b* | *a* | *d* | (5,8) | (7,6) | (6,7) | *a* | *b* | $P_D[a,d]$ is updated from (5,8) to (6,7); *pred*[*a*,*d*] is updated from *a* to *b*. |
| 9 | *b* | *c* | *a* | (4,9) | (7,6) | (6,7) | *c* | *b* | $P_D[c,a]$ is updated from (4,9) to (6,7); *pred*[*c*,*a*] is updated from *c* to *b*. |
| 10 | *b* | *c* | *d* | (9,4) | (7,6) | (6,7) | *c* | *b* | |
| 11 | *b* | *d* | *a* | (8,5) | (7,6) | (6,7) | *d* | *b* | |
| 12 | *b* | *d* | *c* | (8,5) | (7,6) | (6,7) | *a* | *b* | |
| 13 | *c* | *a* | *b* | (7,6) | (9,4) | (7,6) | *a* | *c* | |
| 14 | *c* | *a* | *d* | (6,7) | (9,4) | (9,4) | *b* | *c* | $P_D[a,d]$ is updated from (6,7) to (9,4); *pred*[*a*,*d*] is updated from *b* to *c*. |
| 15 | *c* | *b* | *a* | (6,7) | (6,7) | (6,7) | *b* | *b* | |
| 16 | *c* | *b* | *d* | (6,7) | (6,7) | (9,4) | *b* | *c* | |
| 17 | *c* | *d* | *a* | (8,5) | (8,5) | (6,7) | *d* | *b* | |
| 18 | *c* | *d* | *b* | (7,6) | (8,5) | (7,6) | *d* | *c* | |
| 19 | *d* | *a* | *b* | (7,6) | (9,4) | (7,6) | *a* | *d* | |
| 20 | *d* | *a* | *c* | (9,4) | (9,4) | (8,5) | *a* | *a* | |
| 21 | *d* | *b* | *a* | (6,7) | (6,7) | (8,5) | *b* | *d* | |
| 22 | *d* | *b* | *c* | (6,7) | (6,7) | (8,5) | *b* | *a* | |
| 23 | *d* | *c* | *a* | (6,7) | (9,4) | (8,5) | *b* | *d* | $P_D[c,a]$ is updated from (6,7) to (8,5); *pred*[*c*,*a*] is updated from *b* to *d*. |
| 24 | *d* | *c* | *b* | (7,6) | (9,4) | (7,6) | *c* | *d* | |

### 3.9.2. Situation #2

Suppose alternative *e* is added as follows:

Example 9 (new):

| | |
|---|---|
| 5 voters | $a \succ_v e \succ_v c \succ_v d \succ_v b$ |
| 2 voters | $b \succ_v c \succ_v d \succ_v a \succ_v e$ |
| 4 voters | $b \succ_v d \succ_v a \succ_v e \succ_v c$ |
| 2 voters | $c \succ_v d \succ_v a \succ_v b \succ_v e$ |

The newly added alternative *e* is Pareto-dominated by alternative *a*, because $a \succ_v e$ for every voter $v \in V$. Therefore, (3.8.1) – (3.8.4) say that the result should not change.

The pairwise matrix $N^{new}$ looks as follows:

| | $N^{new}[*,a]$ | $N^{new}[*,b]$ | $N^{new}[*,c]$ | $N^{new}[*,d]$ | $N^{new}[*,e]$ |
|---|---|---|---|---|---|
| $N^{new}[a,*]$ | --- | 7 | 9 | 5 | 13 |
| $N^{new}[b,*]$ | 6 | --- | 6 | 6 | 8 |
| $N^{new}[c,*]$ | 4 | 7 | --- | 9 | 4 |
| $N^{new}[d,*]$ | 8 | 7 | 4 | --- | 8 |
| $N^{new}[e,*]$ | 0 | 5 | 9 | 5 | --- |

The corresponding digraph looks as follows:

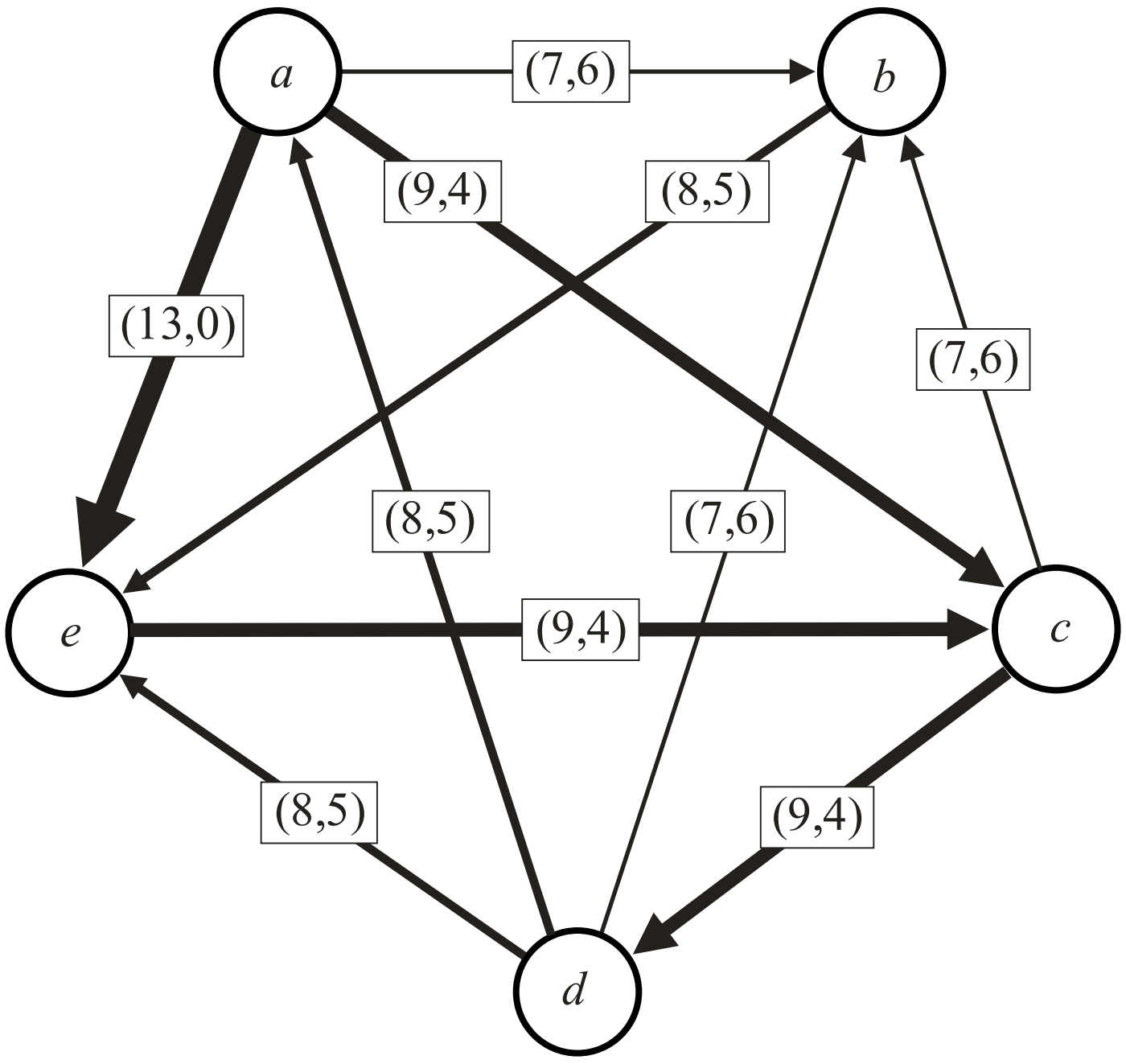

The following table lists the strongest paths, as determined by the Floyd-Warshall algorithm, as defined in section 2.3.1. The critical links of the strongest paths are underlined:

| | ... to *a* | ... to *b* | ... to *c* | ... to *d* | ... to *e* |
|---|---|---|---|---|---|
| from *a* ... | --- | *a*, (7,6), *b* | *a*, (9,4), *c* | *a*, (9,4), *c*, (9,4), *d* | *a*, (13,0), *e* |
| from *b* ... | *b*, (8,5), *e*, (9,4), *c*, (9,4), *d*, (8,5), *a* | --- | *b*, (8,5), *e*, (9,4), *c* | *b*, (8,5), *e*, (9,4), *c*, (9,4), *d* | *b*, (8,5), *e* |
| from *c* ... | *c*, (9,4), *d*, (8,5), *a* | *c*, (7,6), *b* | --- | *c*, (9,4), *d* | *c*, (9,4), *d*, (8,5), *e* |
| from *d* ... | *d*, (8,5), *a* | *d*, (7,6), *b* | *d*, (8,5), *a*, (9,4), *c* | --- | *d*, (8,5), *e* |
| from *e* ... | *e*, (9,4), *c*, (9,4), *d*, (8,5), *a* | *e*, (9,4), *c*, (7,6), *b* | *e*, (9,4), *c* | *e*, (9,4), *c*, (9,4), *d* | --- |

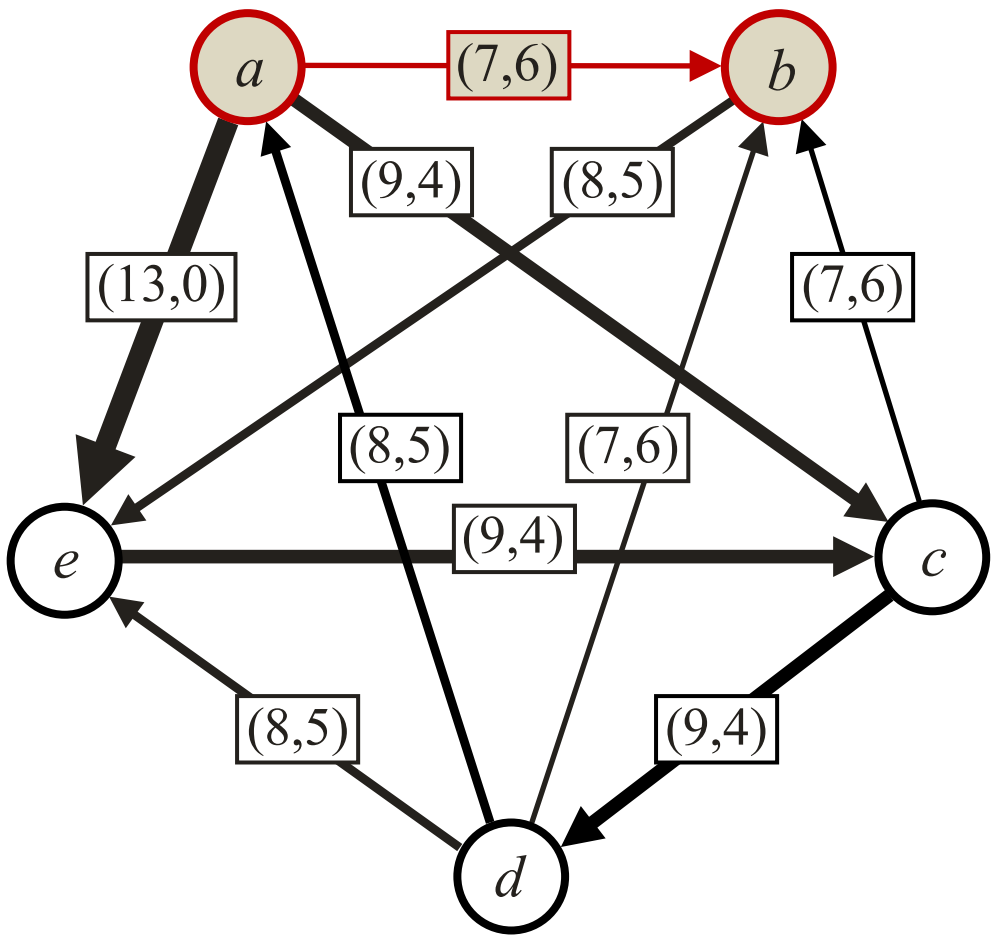

The strongest path from *a* to *b* is:
*a*, (7,6), *b*

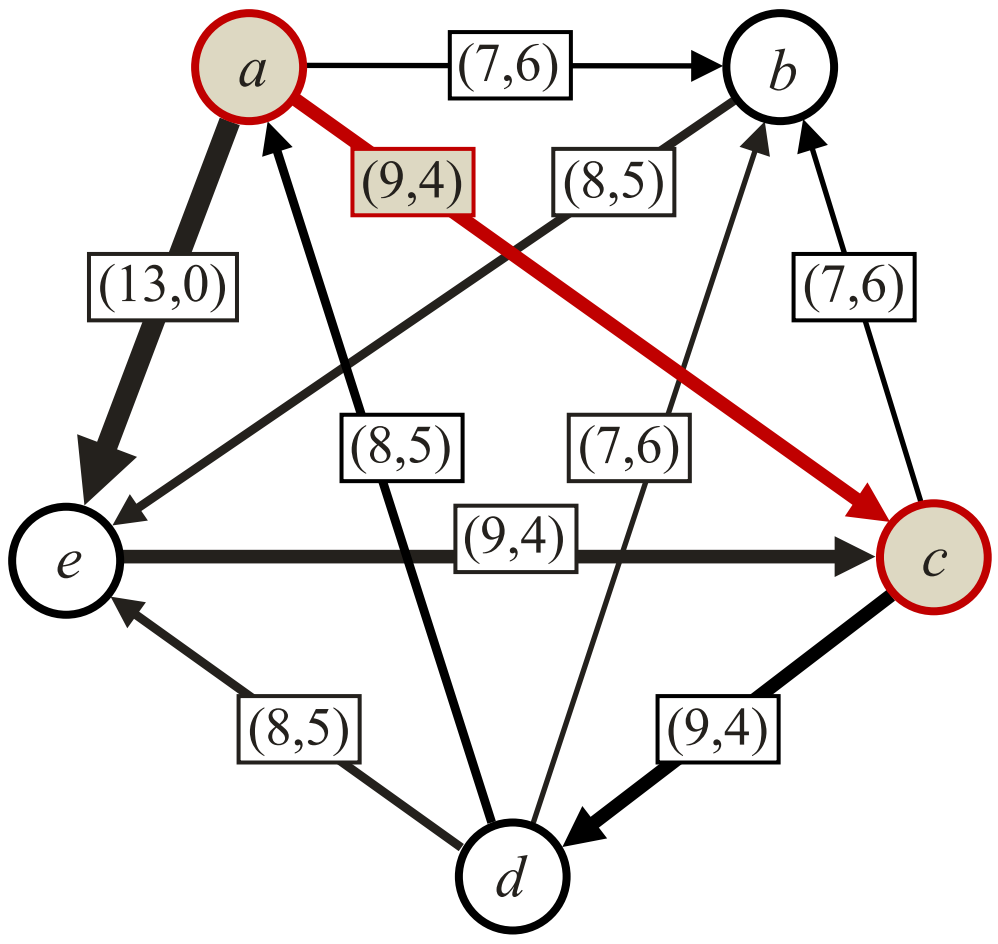

The strongest path from *a* to *c* is:
*a*, (9,4), *c*

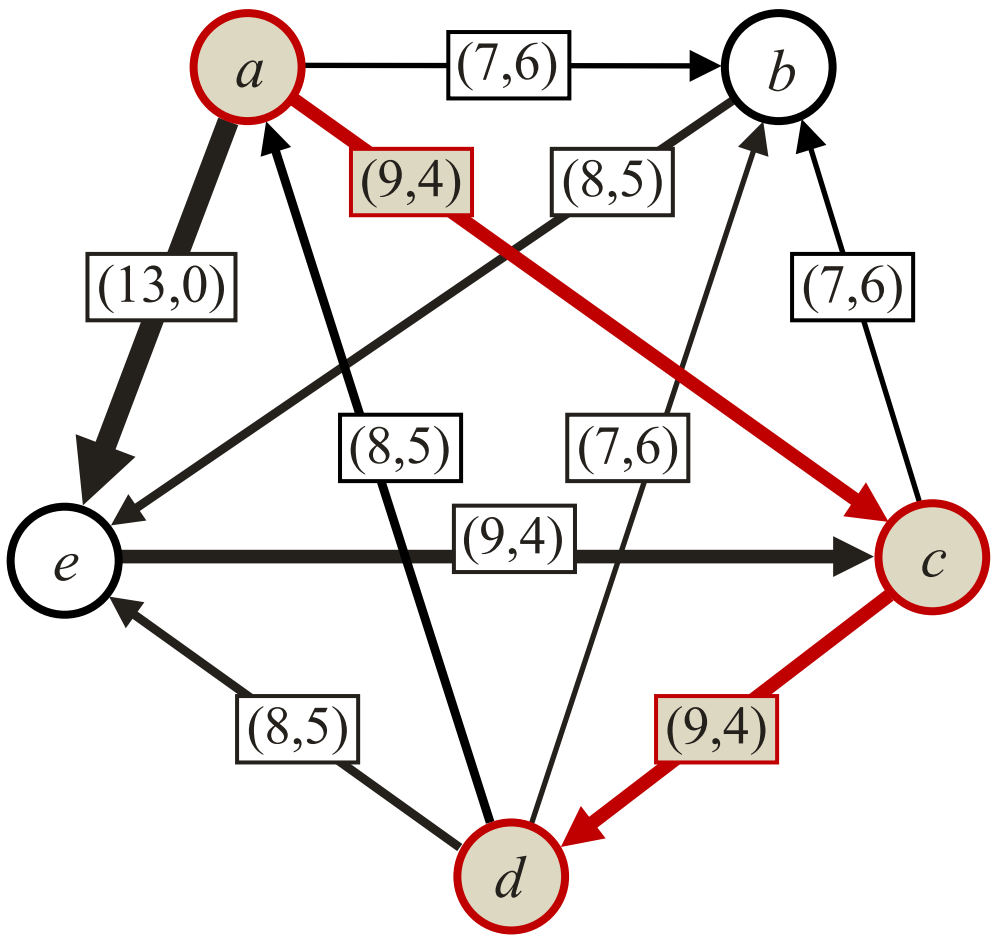

The strongest path from *a* to *d* is:
*a*, (9,4), *c*, (9,4), *d*

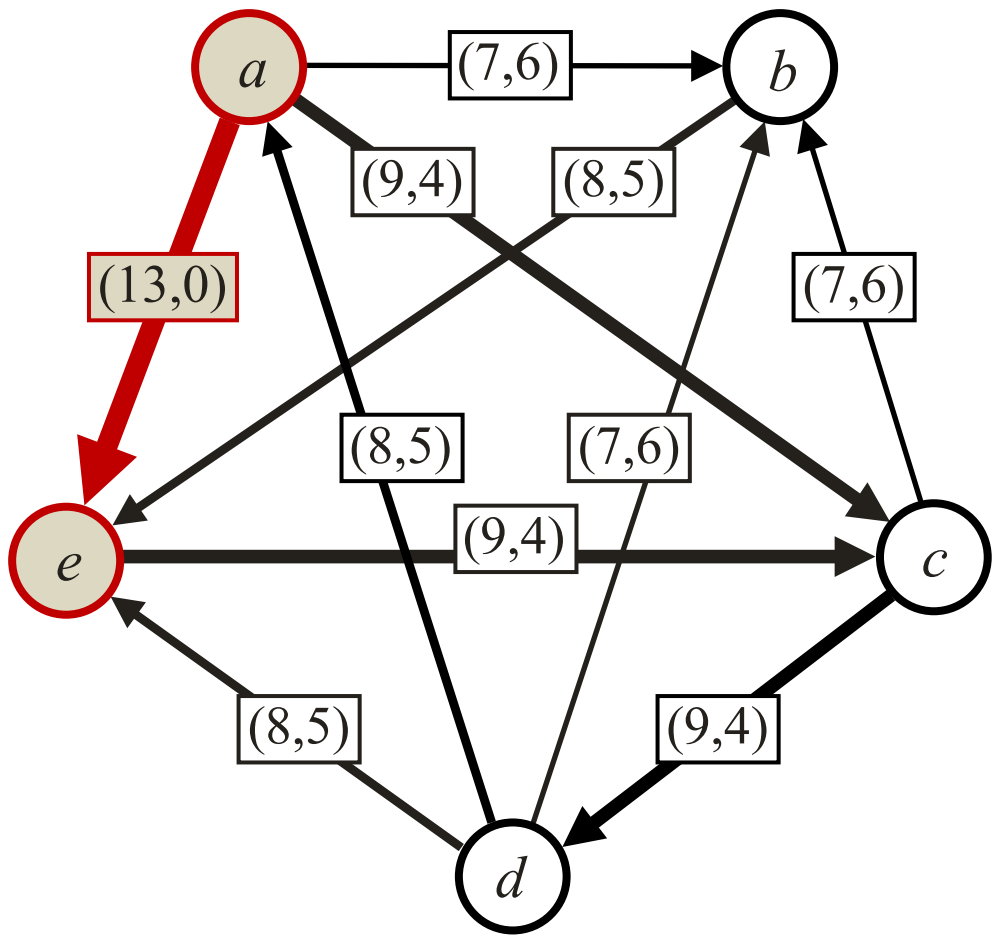

The strongest path from *a* to *e* is:
*a*, (13,0), *e*

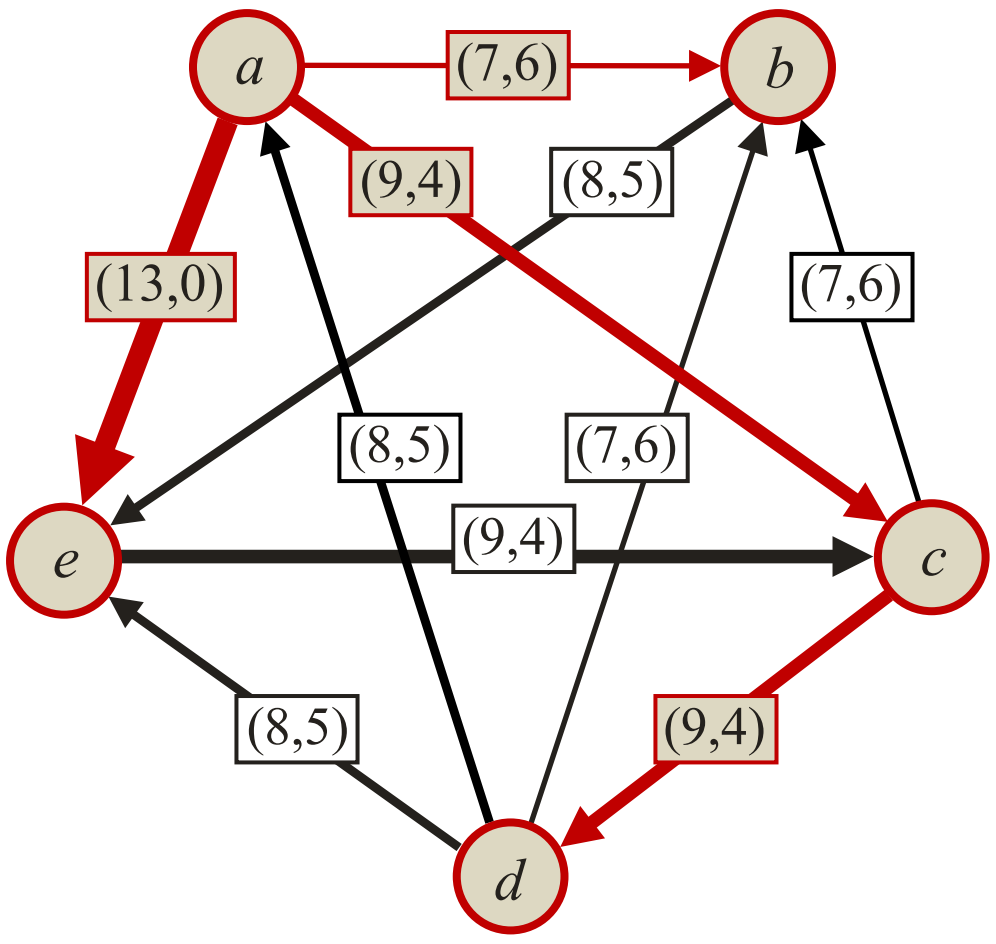

These are the strongest paths
from *a* to every other alternative.

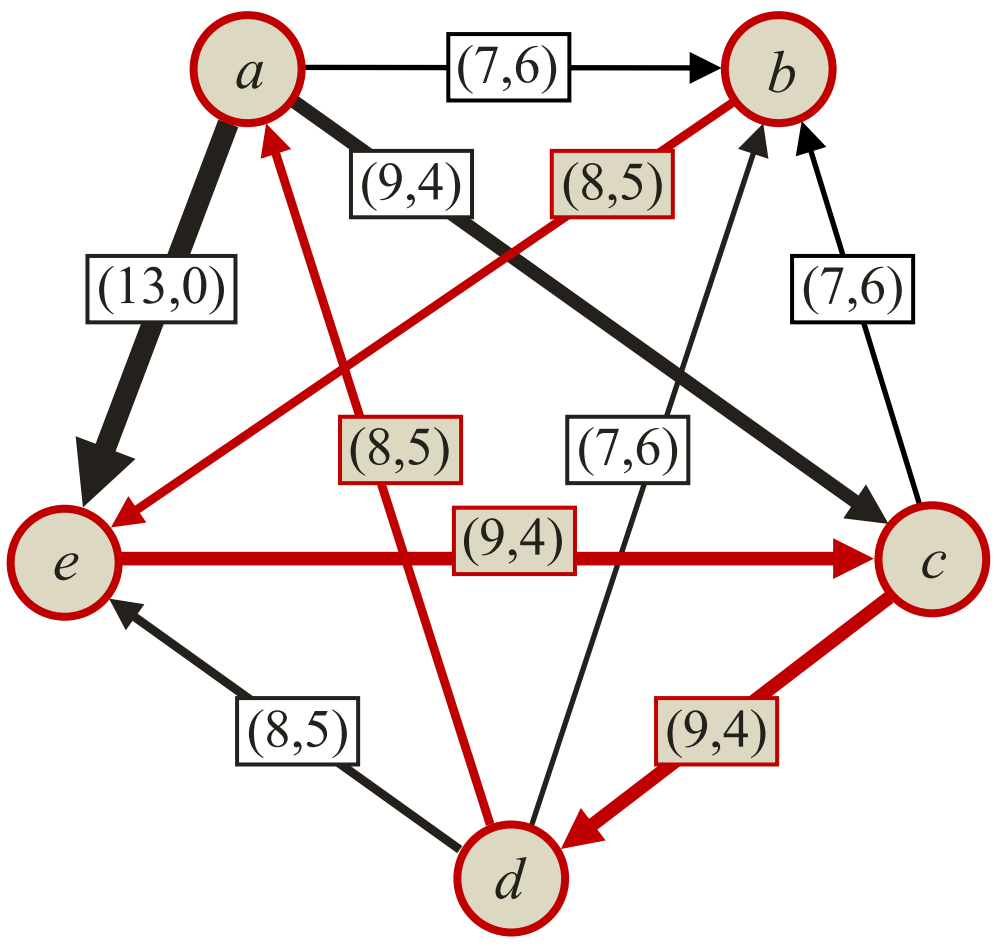

The strongest path from *b* to *a* is:
*b*, (8,5), *e*, (9,4), *c*, (9,4), *d*, (8,5), *a*

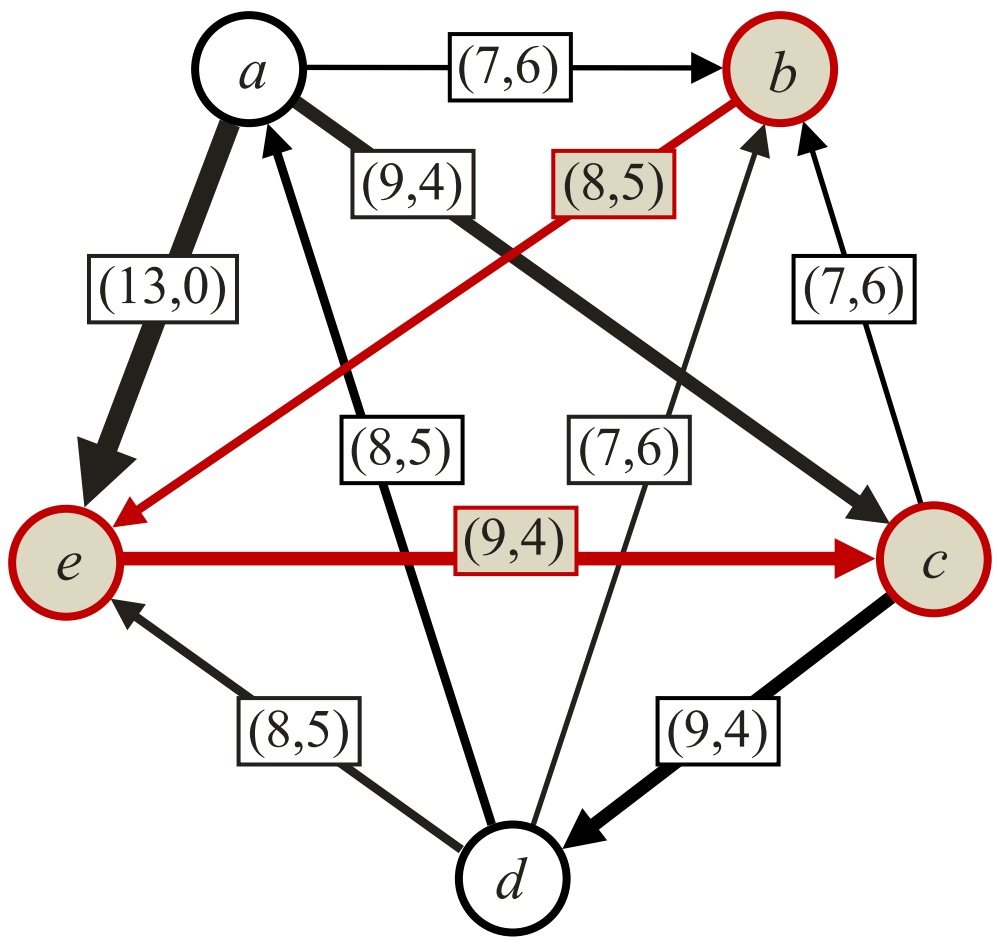

The strongest path from *b* to *c* is:
*b*, (8,5), *e*, (9,4), *c*

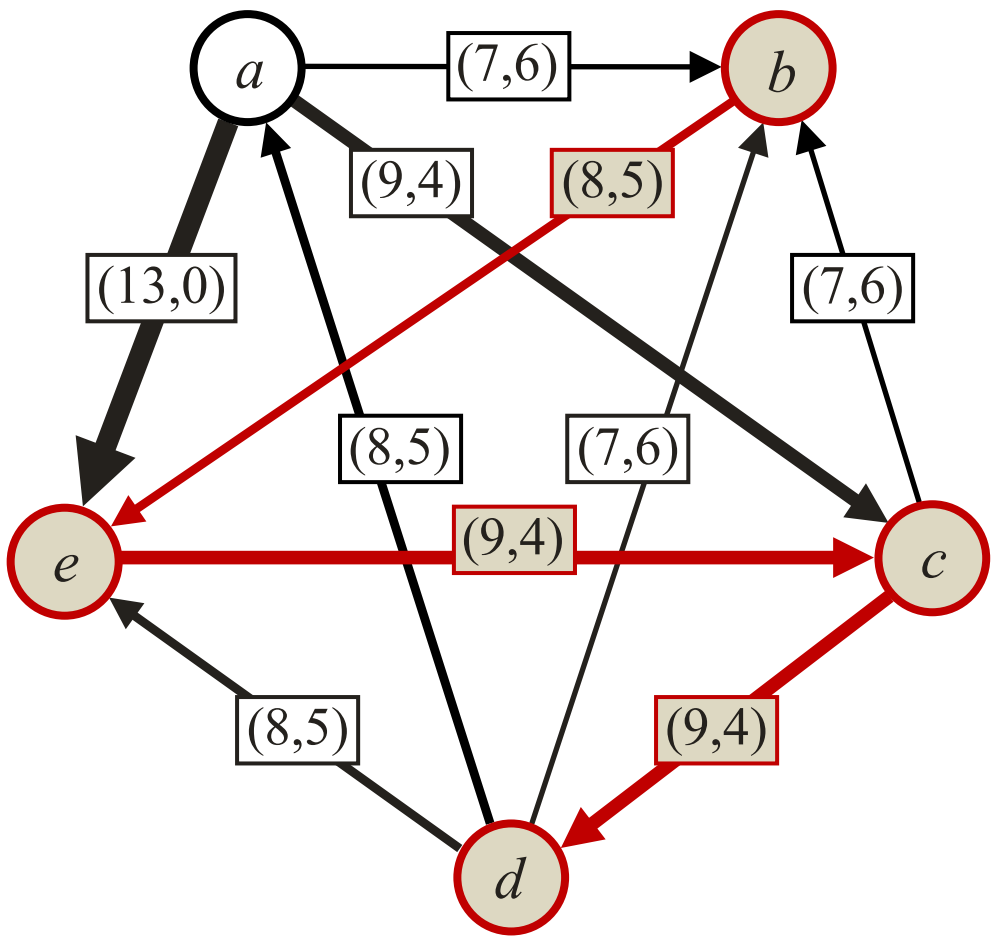

The strongest path from *b* to *d* is:
*b*, (8,5), *e*, (9,4), *c*, (9,4), *d*

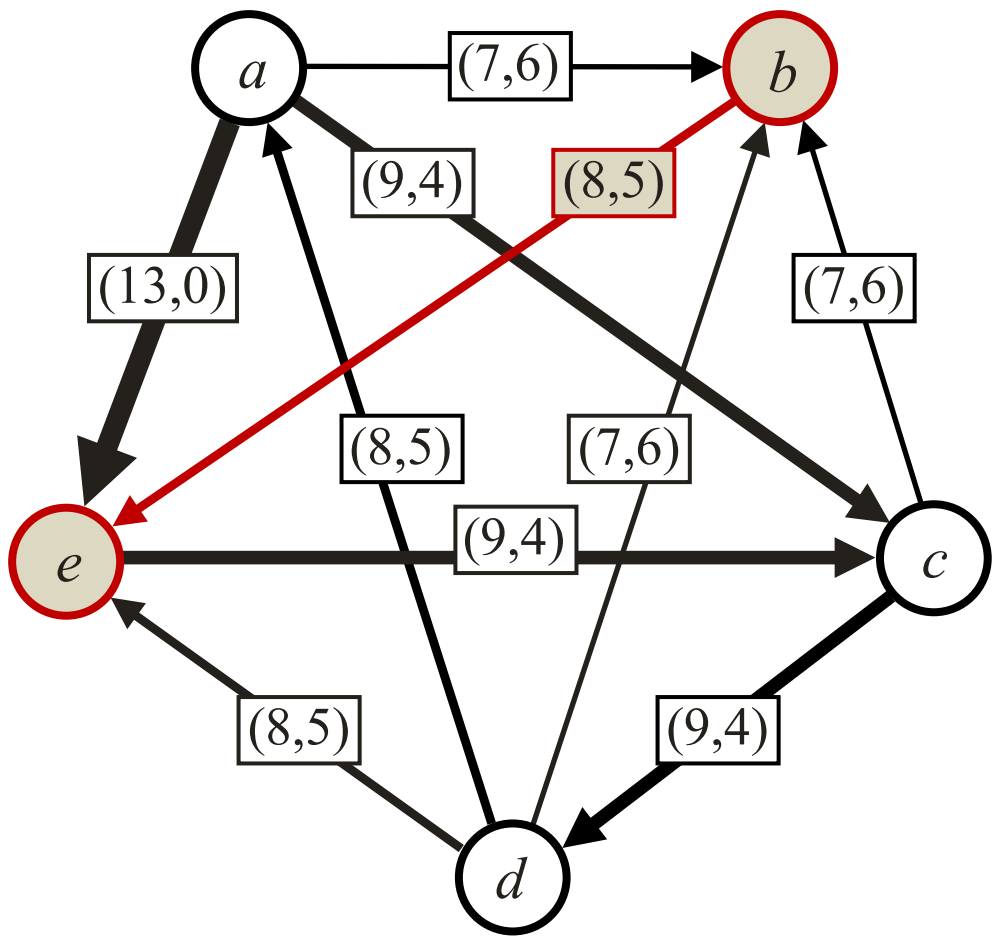

The strongest path from *b* to *e* is:
*b*, (8,5), *e*

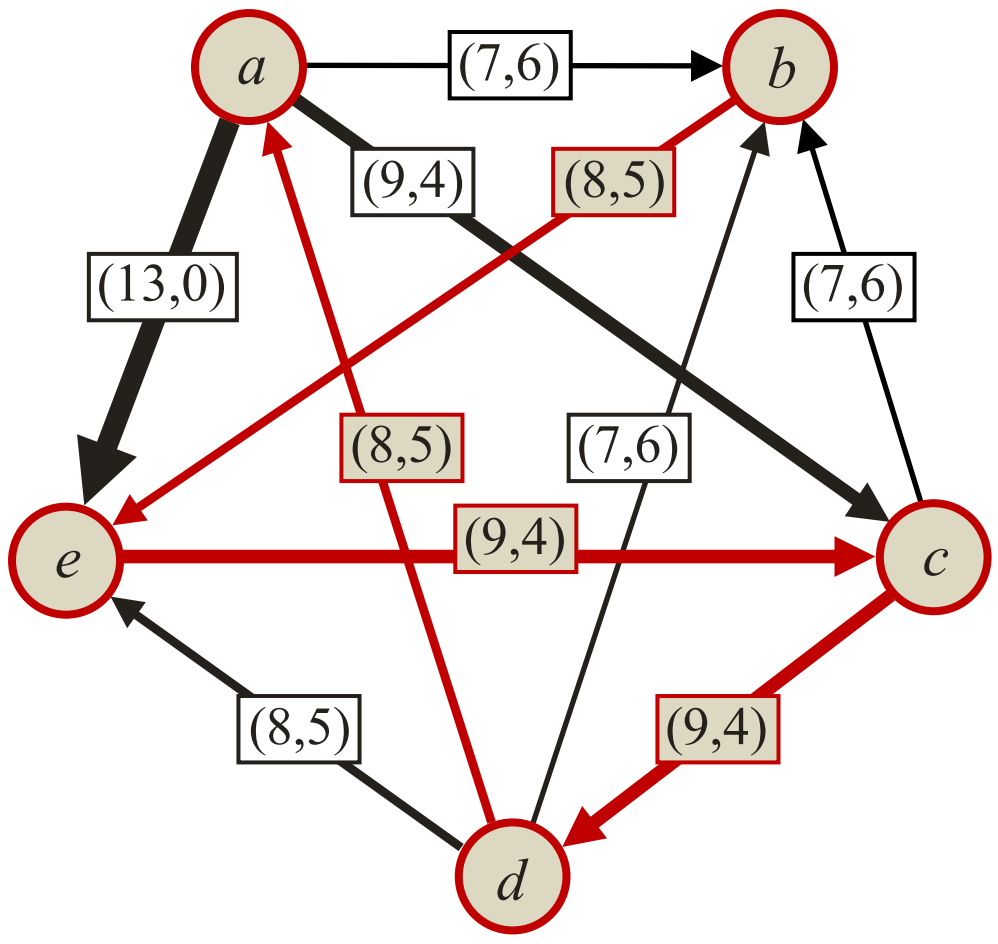

These are the strongest paths
from *b* to every other alternative.

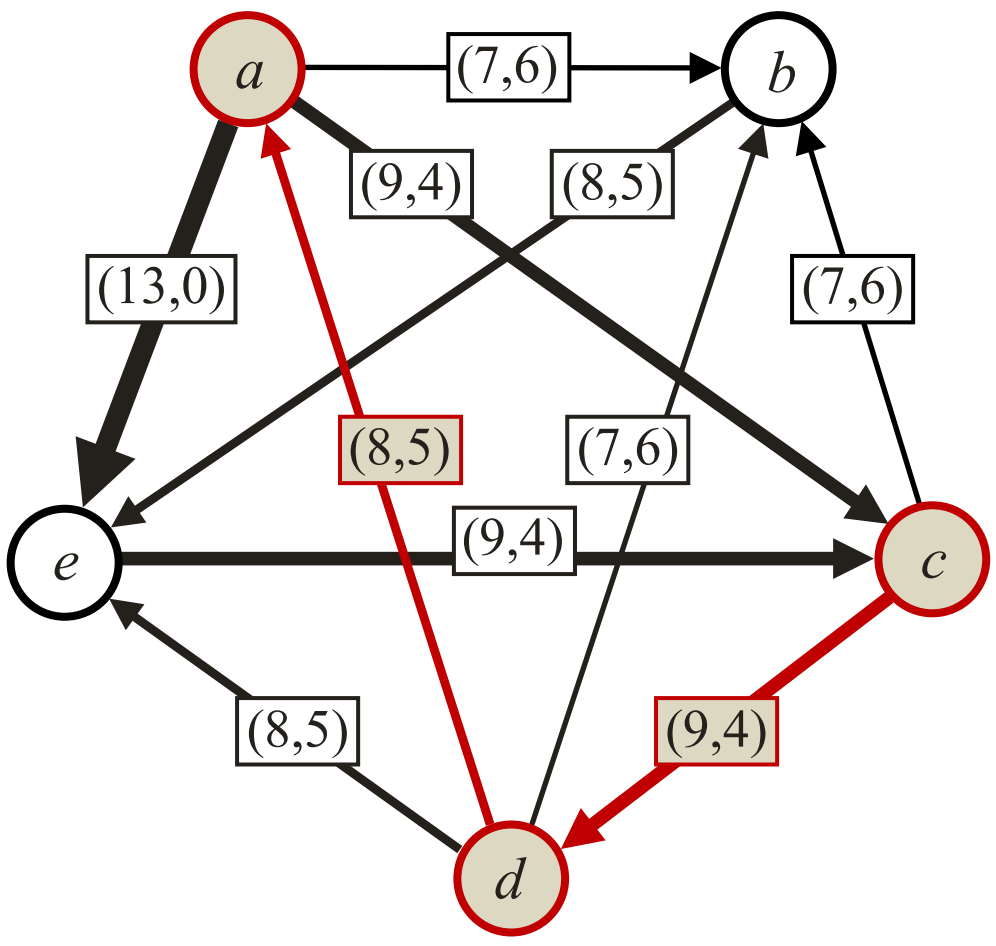

The strongest path from *c* to *a* is:
*c*, (9,4), *d*, (8,5), *a*

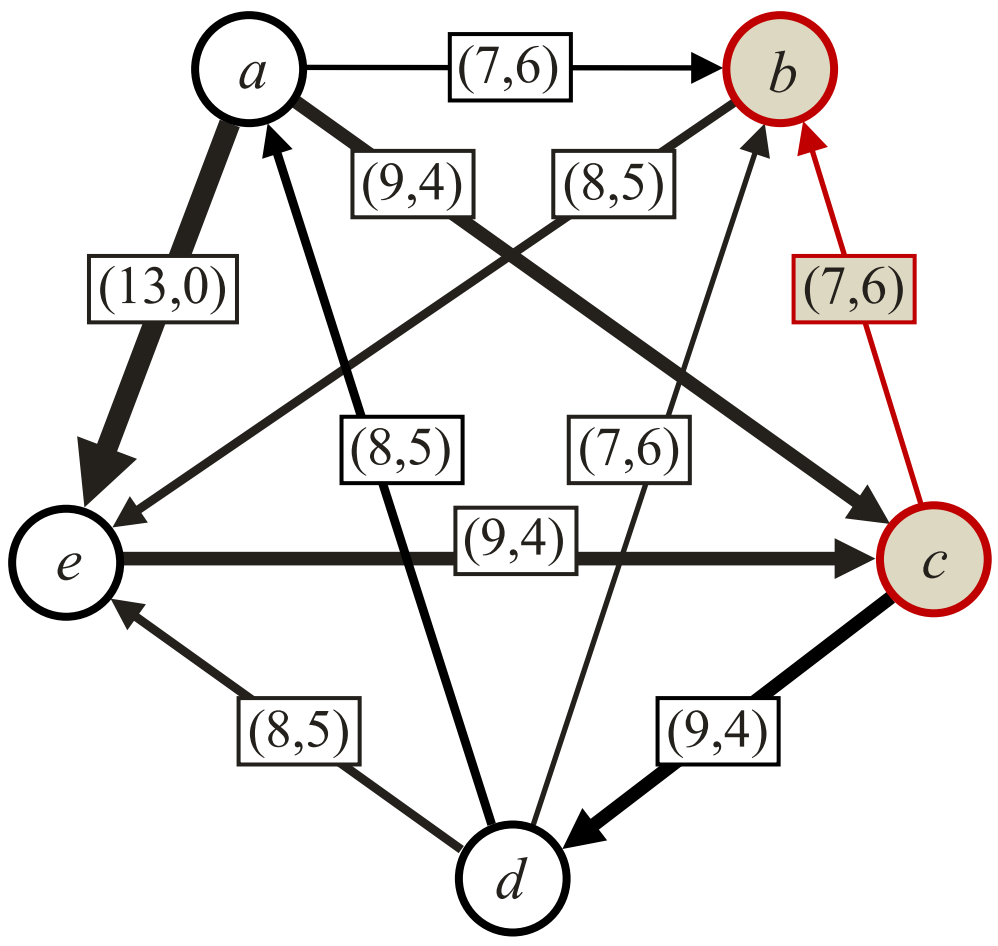

The strongest path from *c* to *b* is:
*c*, (7,6), *b*

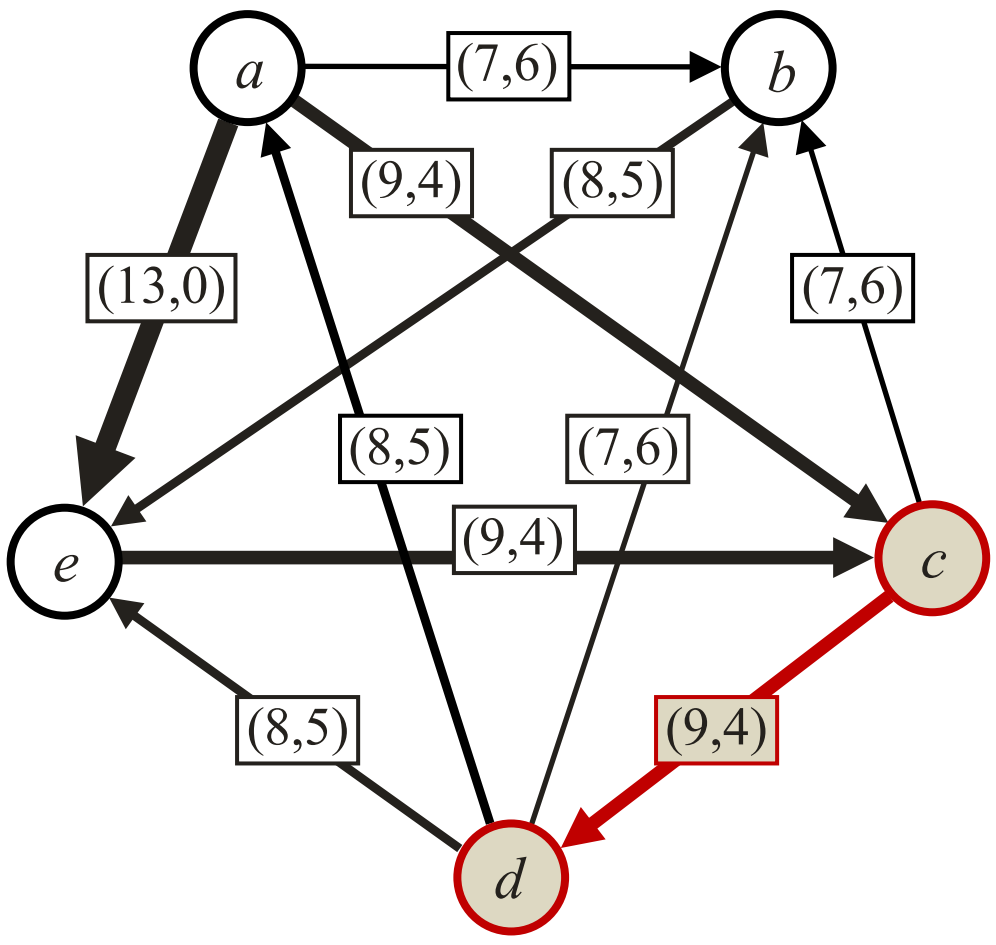

The strongest path from *c* to *d* is:
*c*, (9,4), *d*

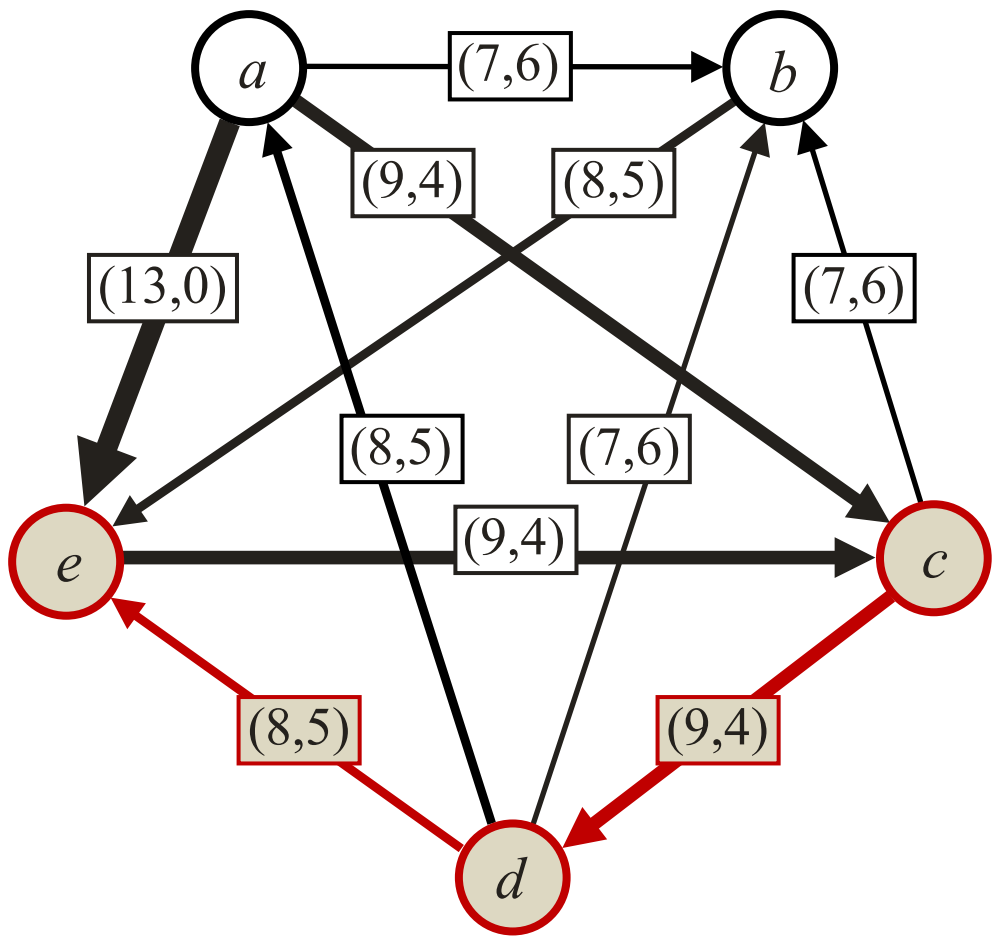

The strongest path from *c* to *e* is:
*c*, (9,4), *d*, (8,5), *e*

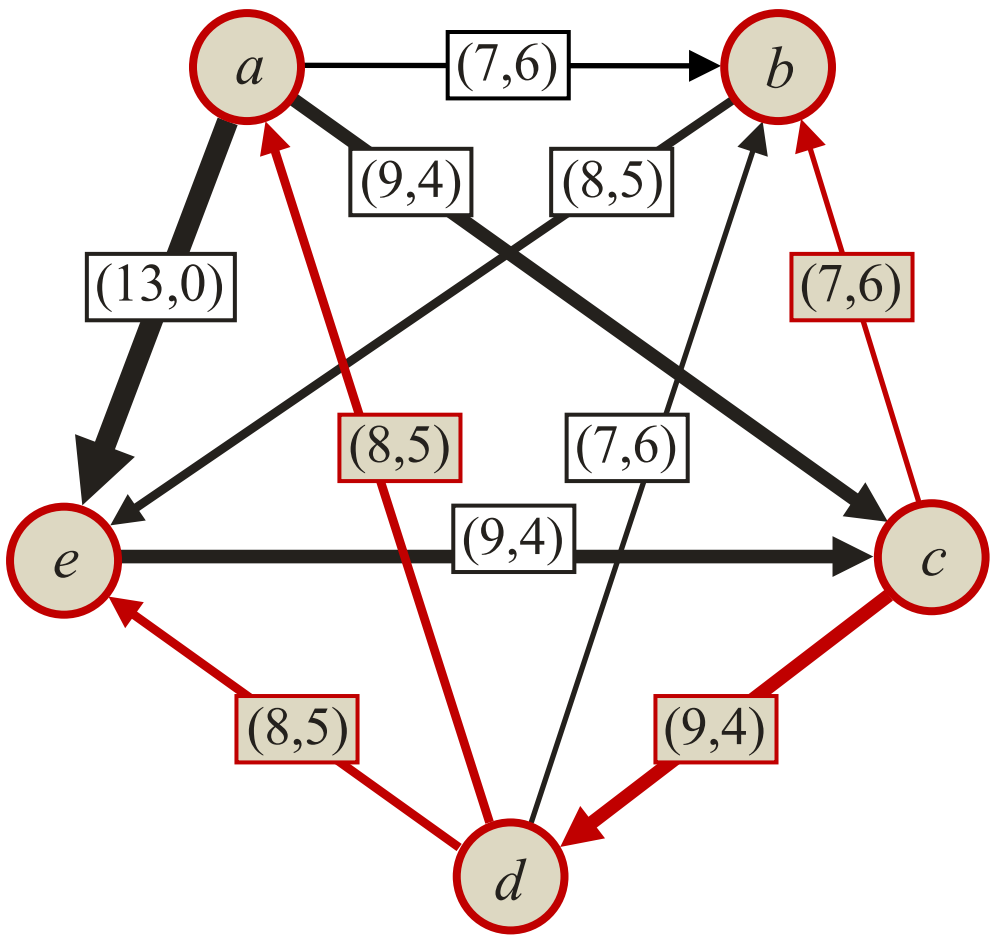

These are the strongest paths
from *c* to every other alternative.

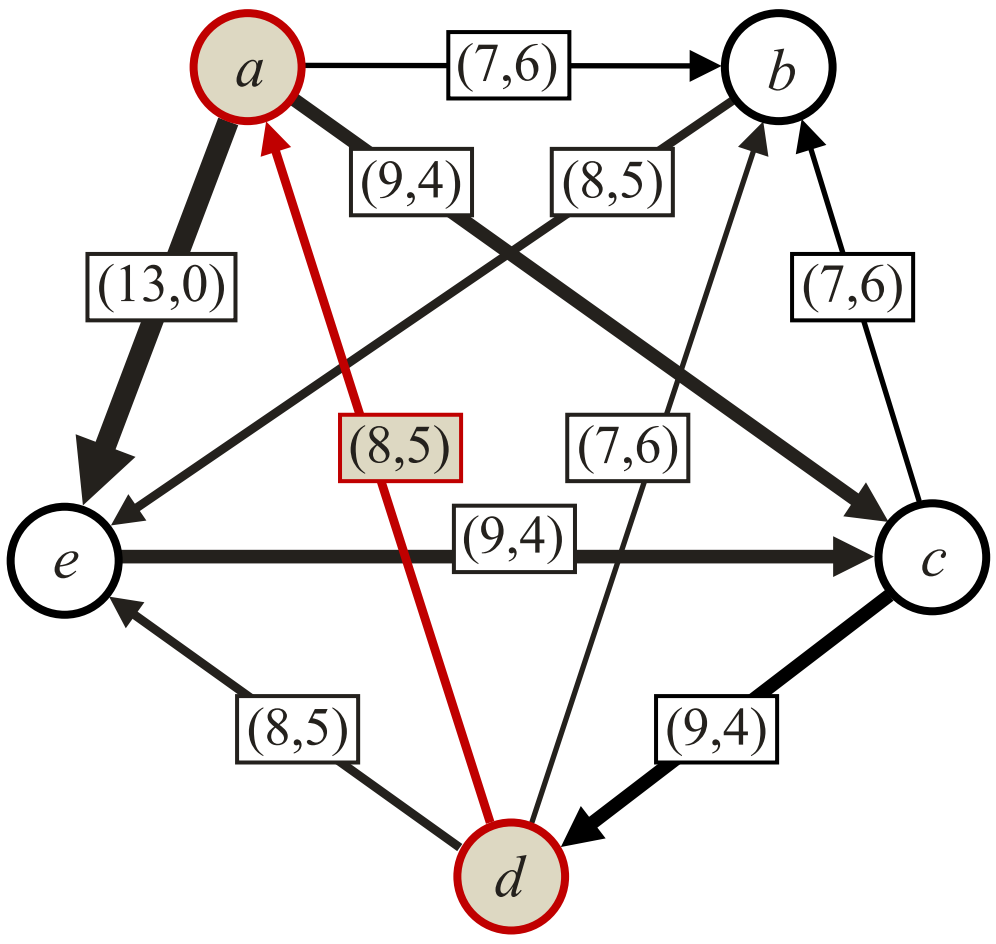

The strongest path from *d* to *a* is:
*d*, (8,5), *a*

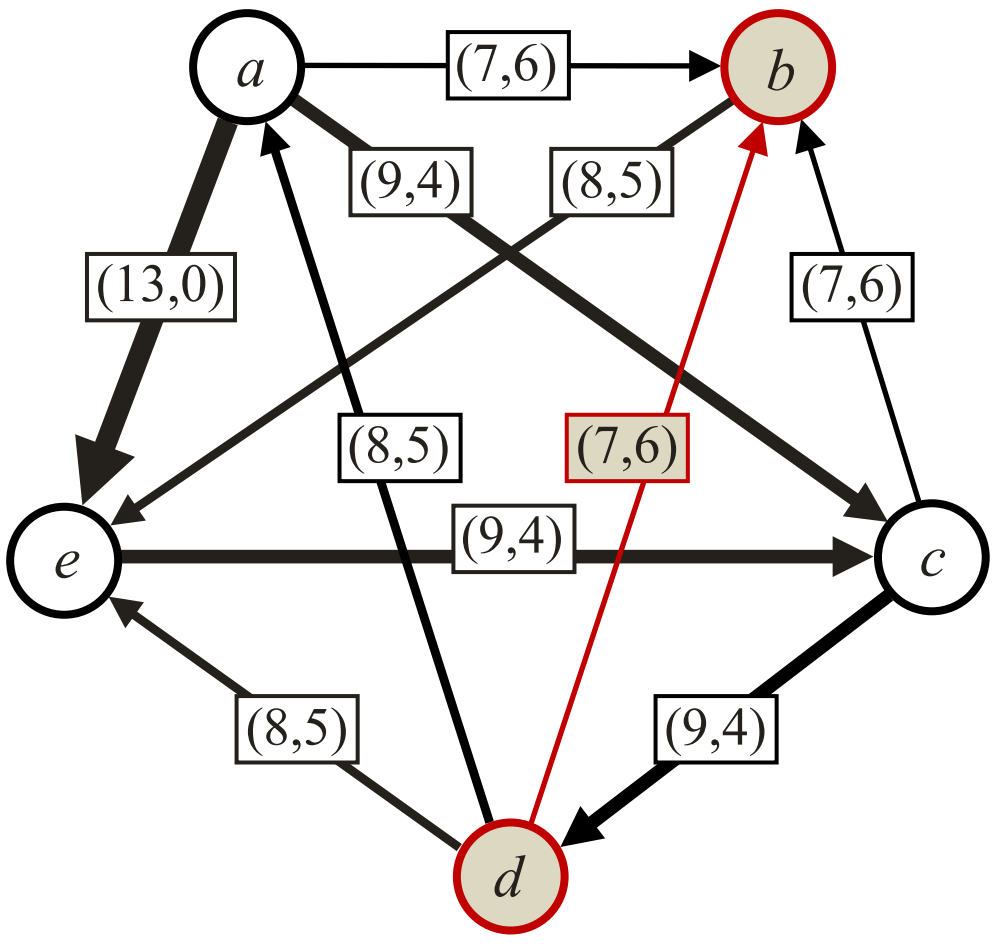

The strongest path from *d* to *b* is:
*d*, (7,6), *b*

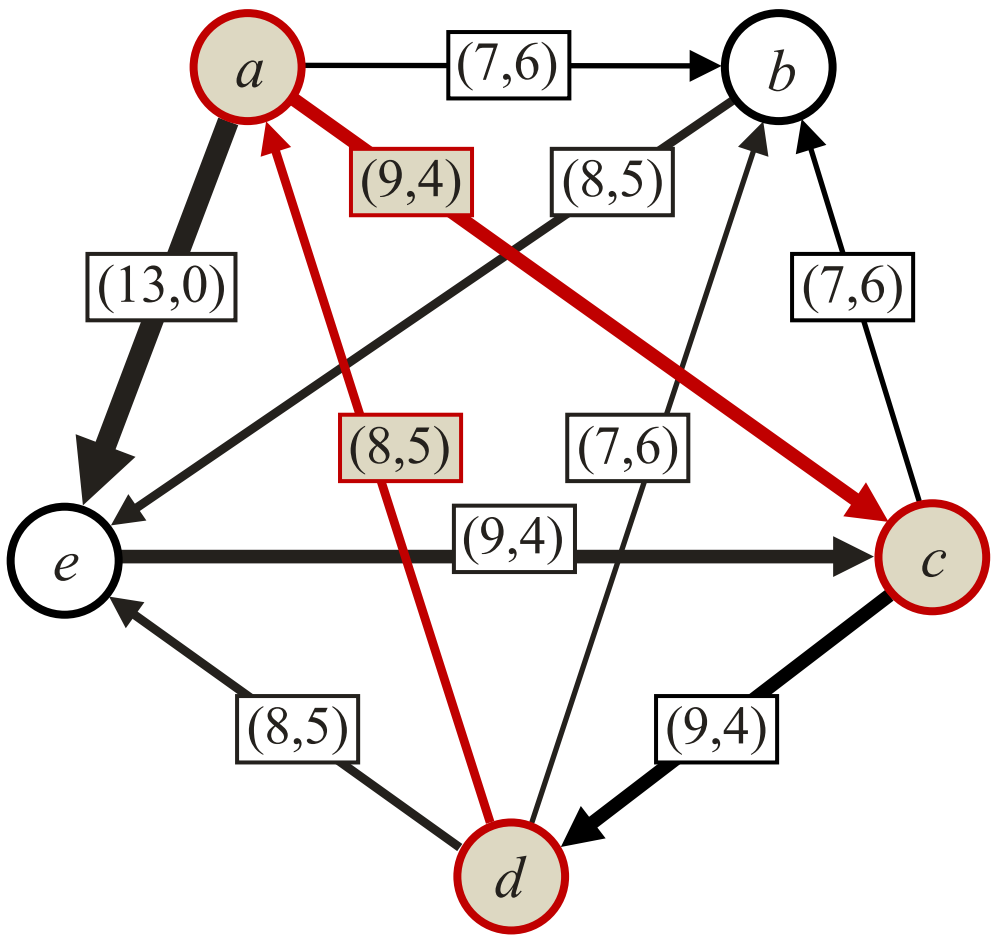

The strongest path from *d* to *c* is:
*d*, (8,5), *a*, (9,4), *c*

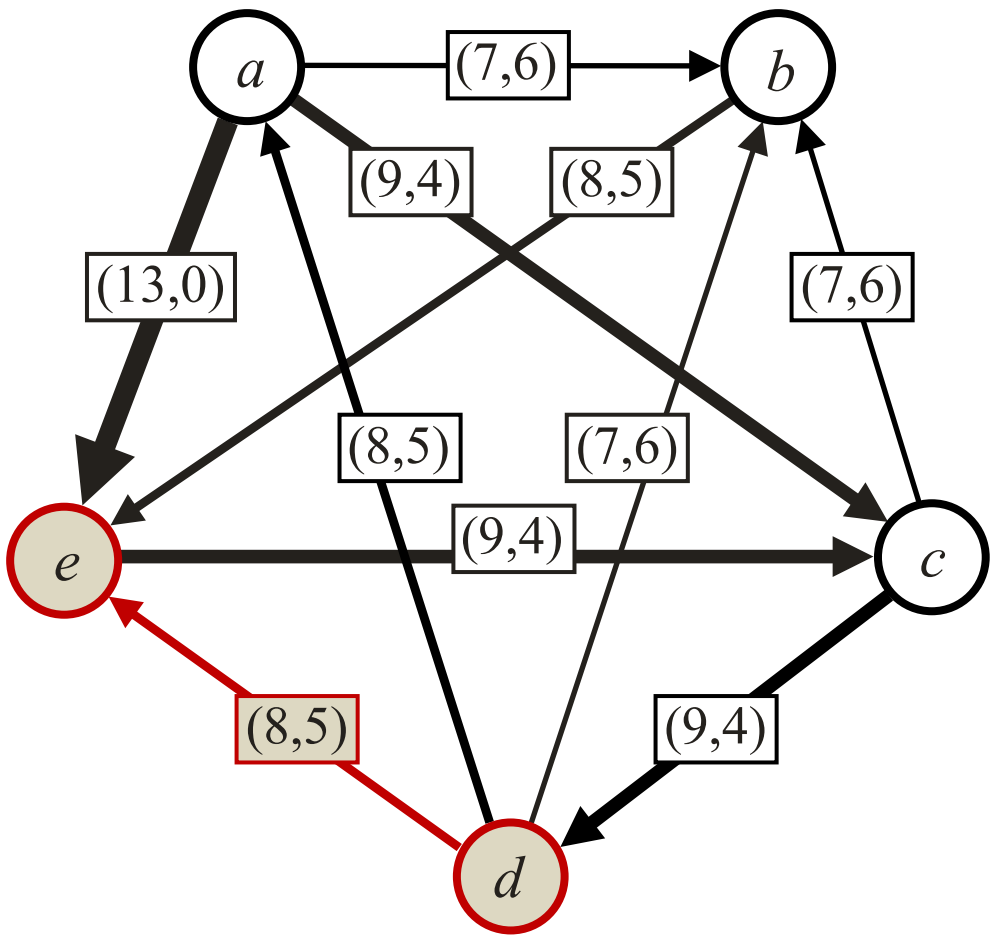

The strongest path from *d* to *e* is:
*d*, (8,5), *e*

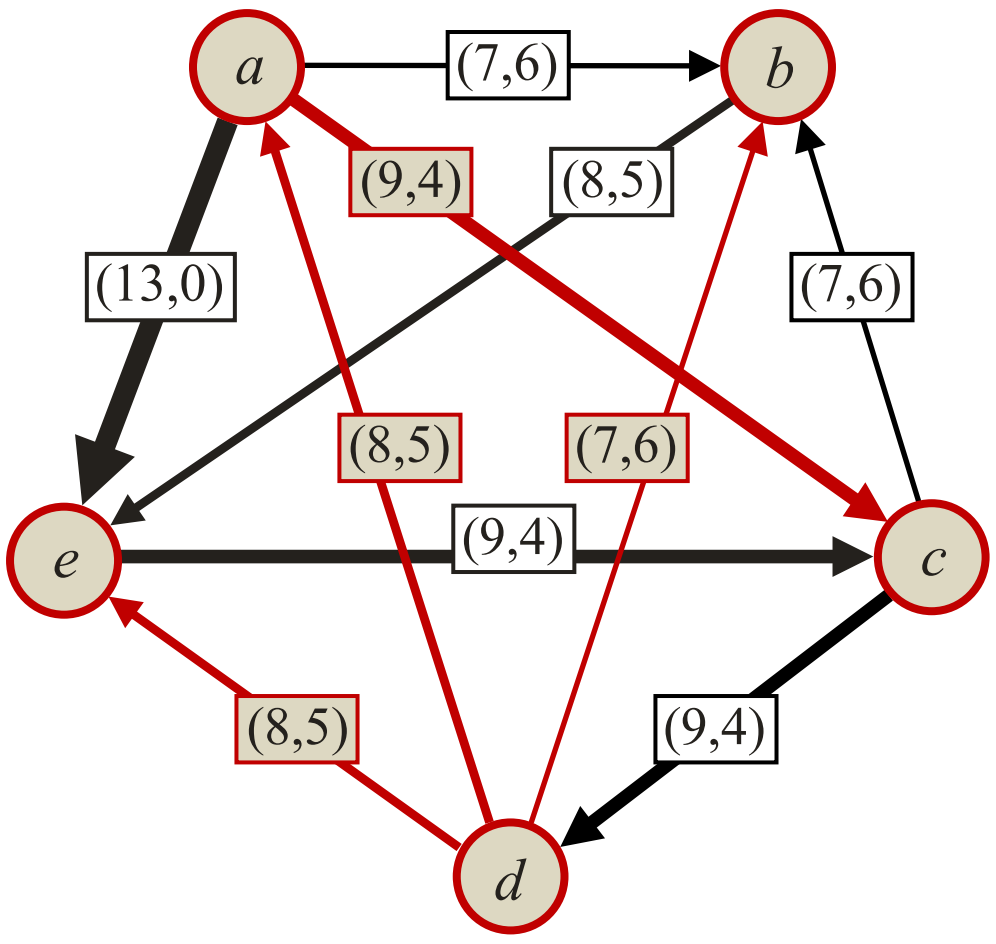

These are the strongest paths
from *d* to every other alternative.

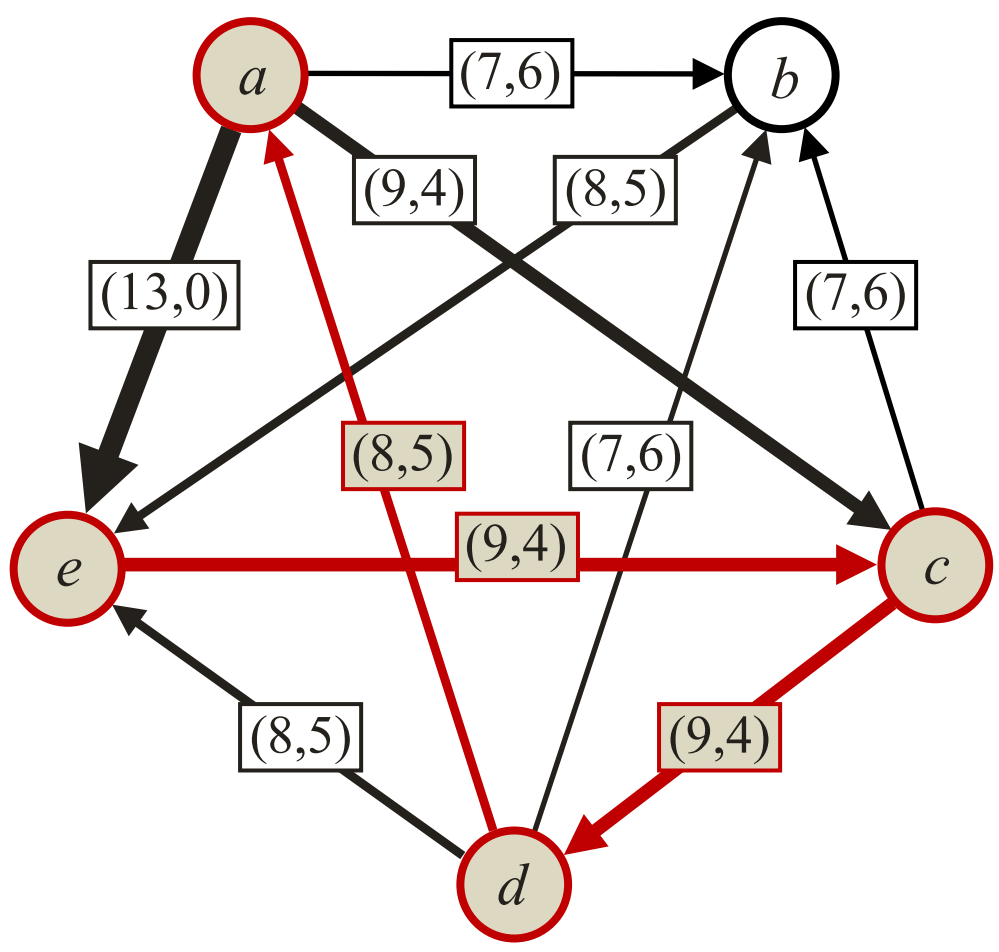

The strongest path from *e* to *a* is:
*e*, (9,4), *c*, (9,4), *d*, (8,5), *a*

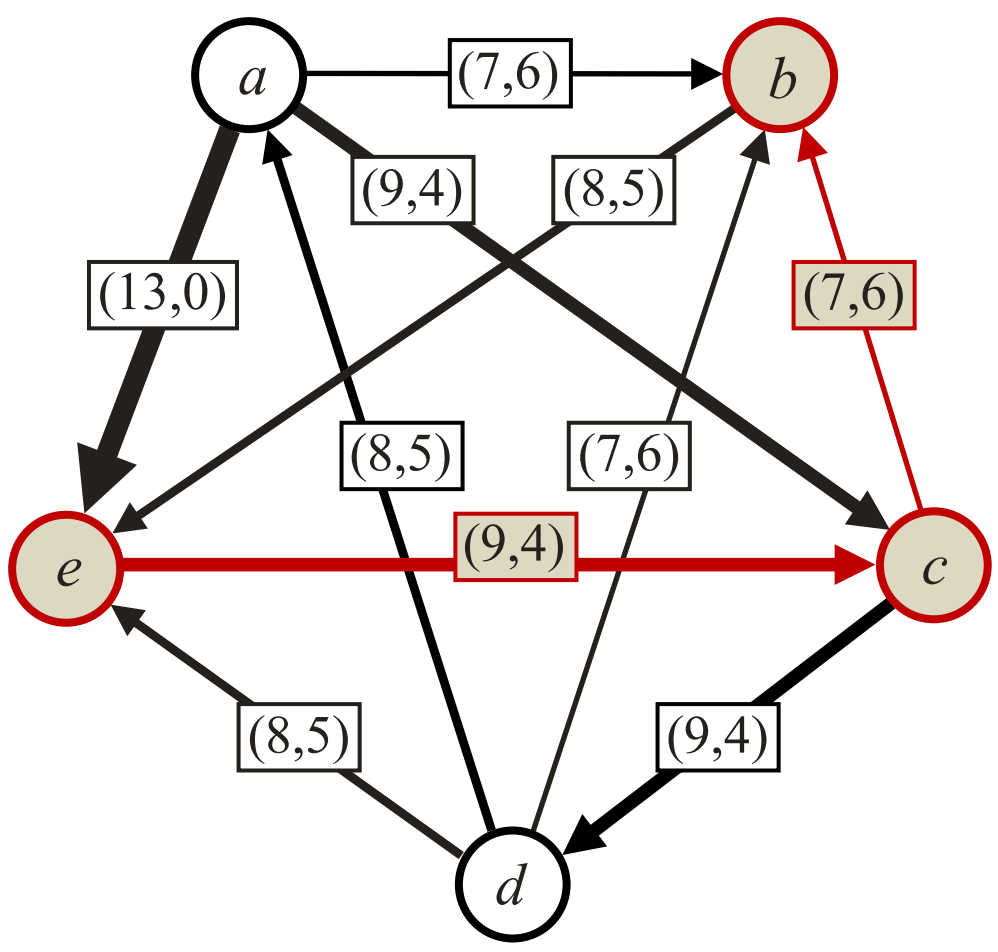

The strongest path from *e* to *b* is:
*e*, (9,4), *c*, (7,6), *b*

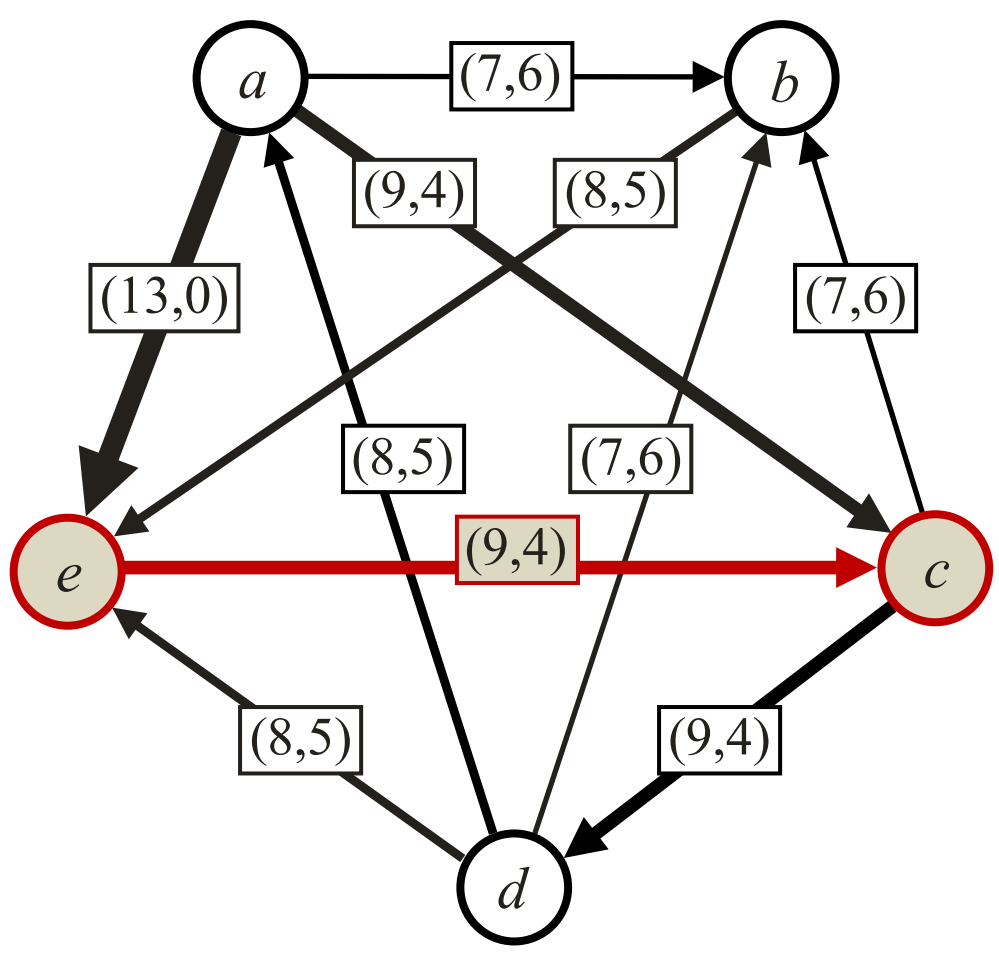

The strongest path from *e* to *c* is:
*e*, (9,4), *c*

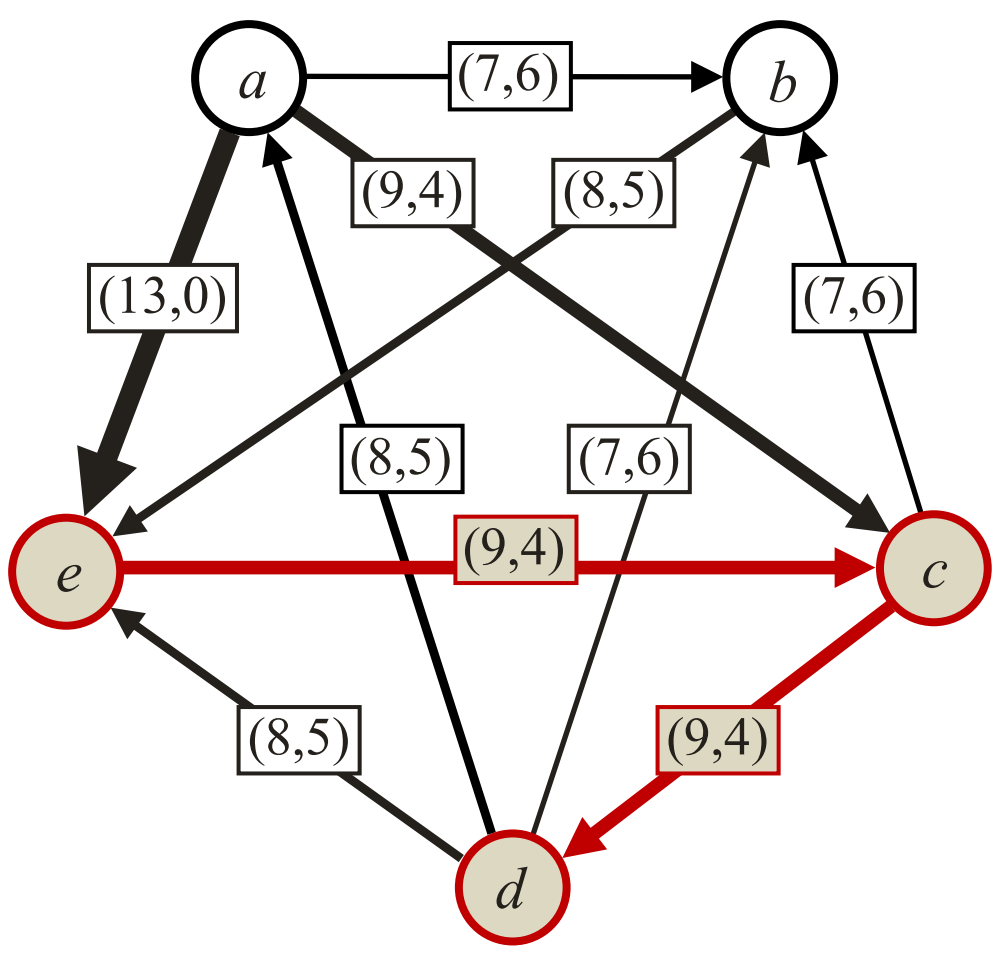

The strongest path from *e* to *d* is:
*e*, (9,4), *c*, (9,4), *d*

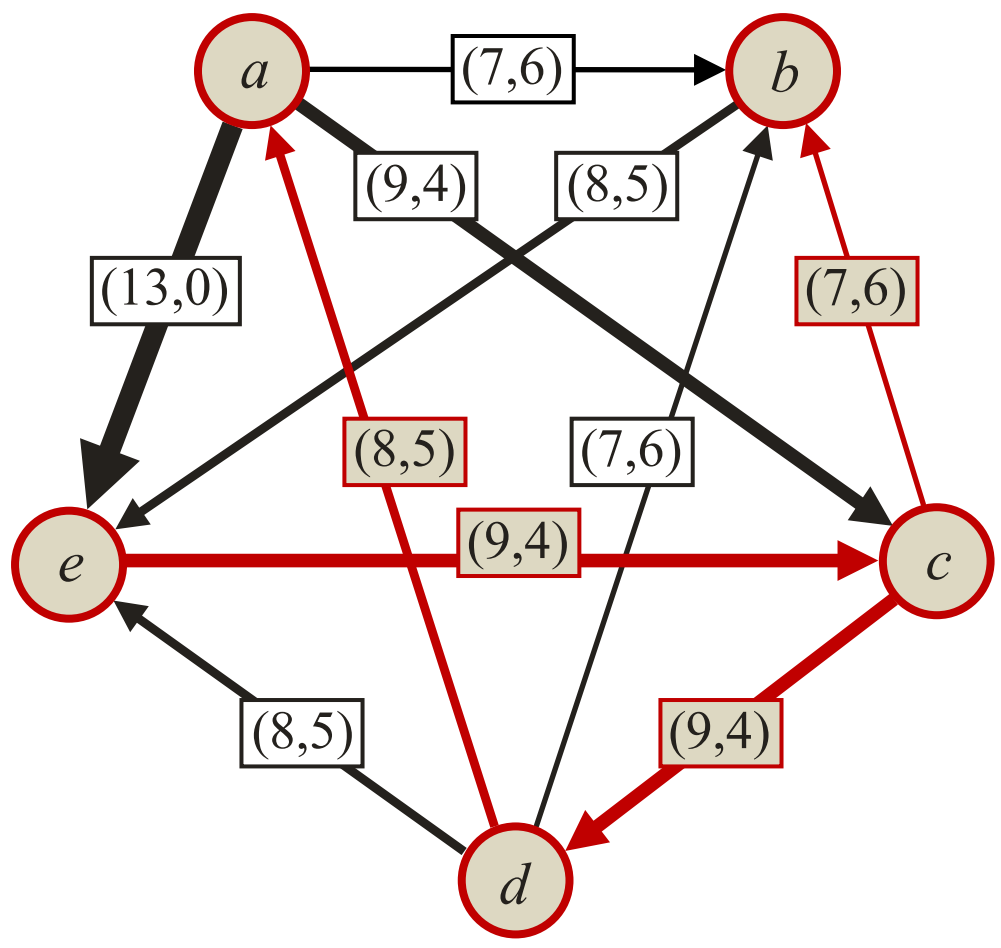

These are the strongest paths
from *e* to every other alternative.

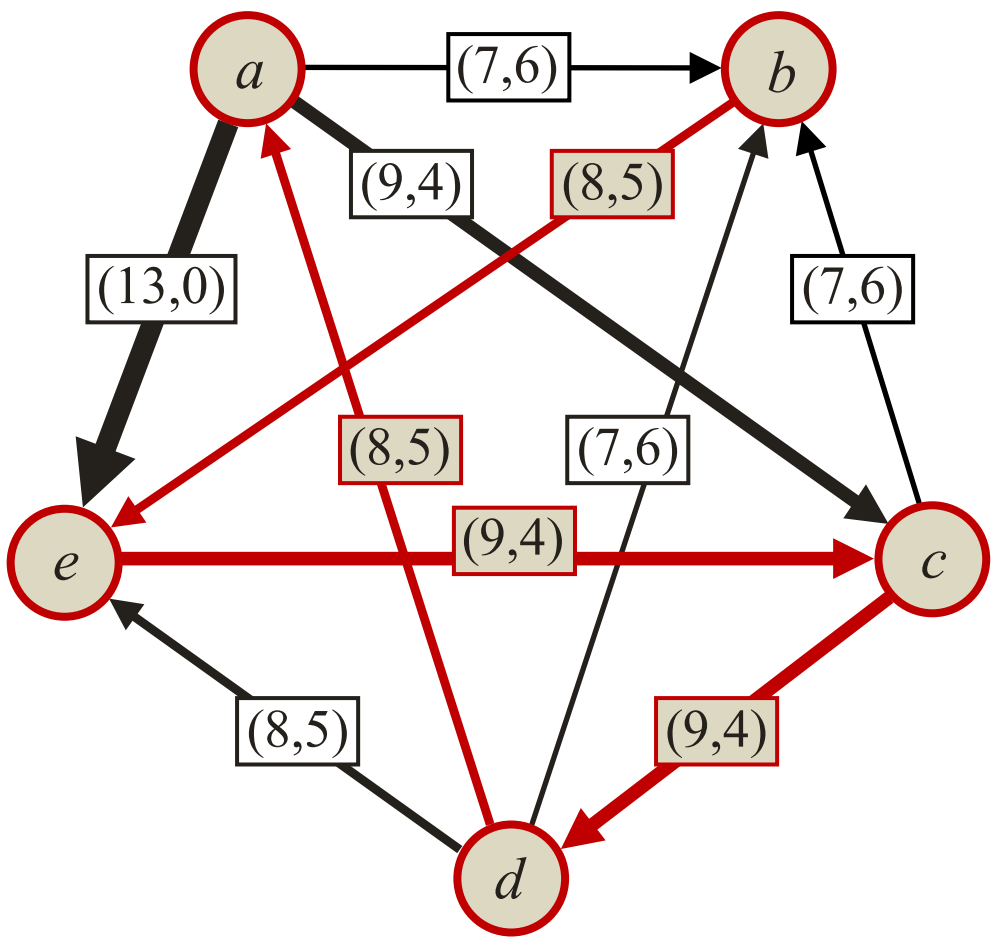

These are the strongest paths
from every other alternative to *a*.

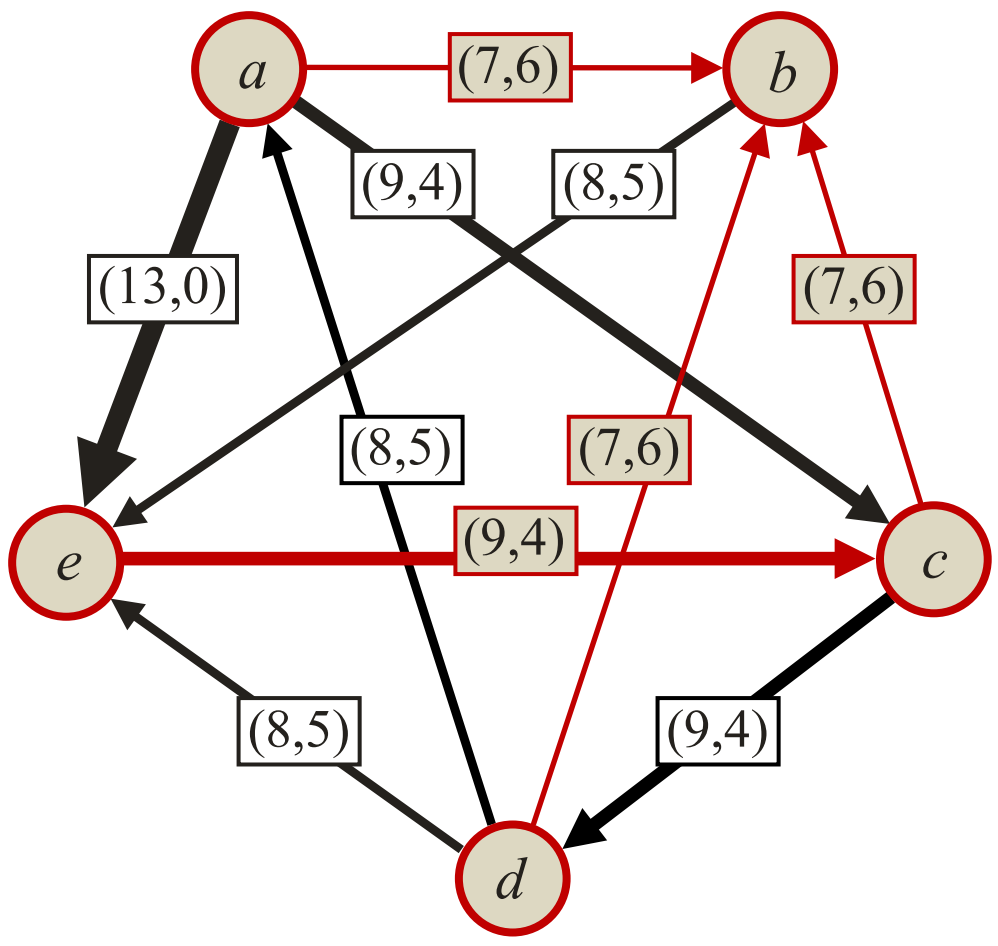

These are the strongest paths
from every other alternative to *b*.

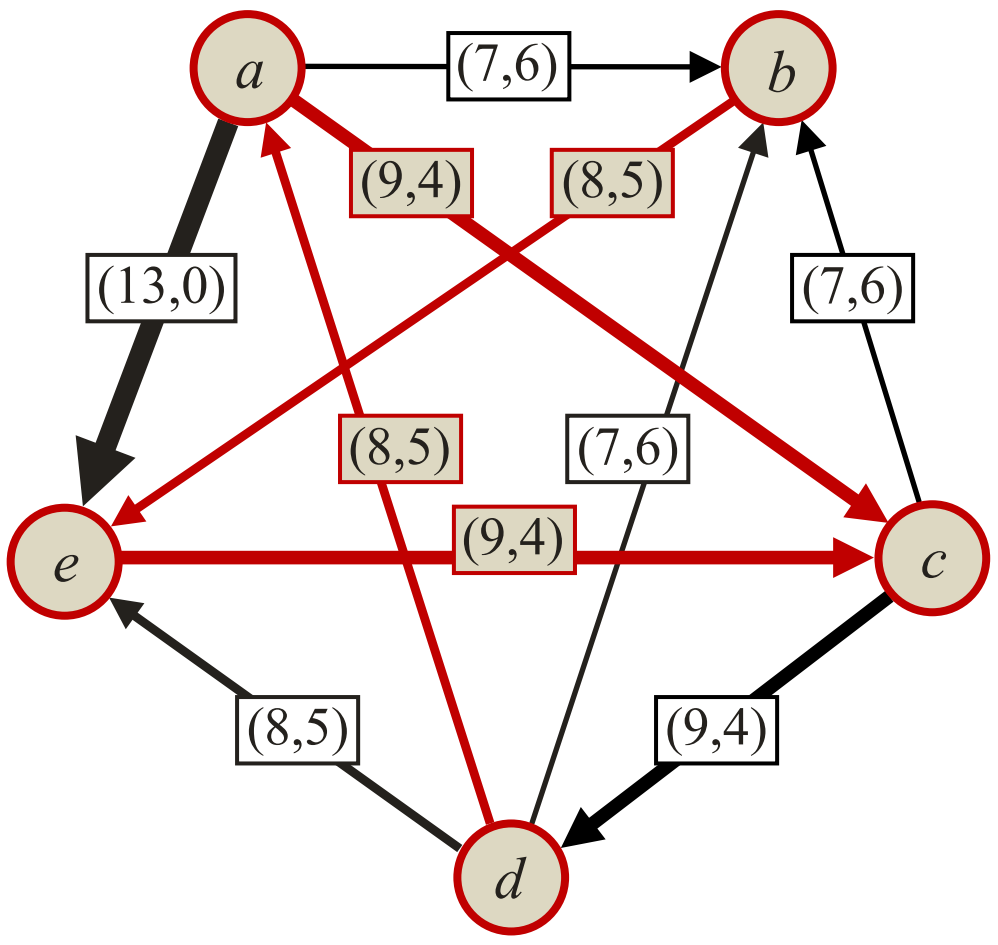

These are the strongest paths
from every other alternative to *c*.

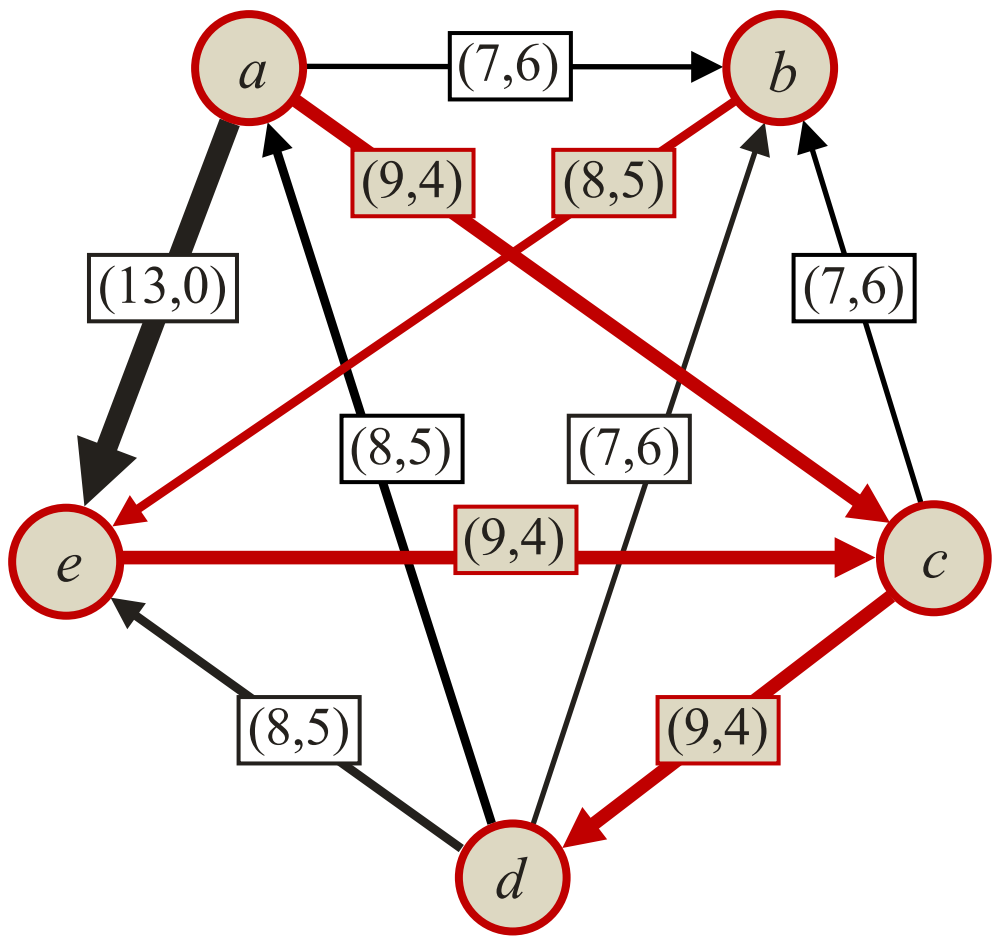

These are the strongest paths
from every other alternative to *d*.

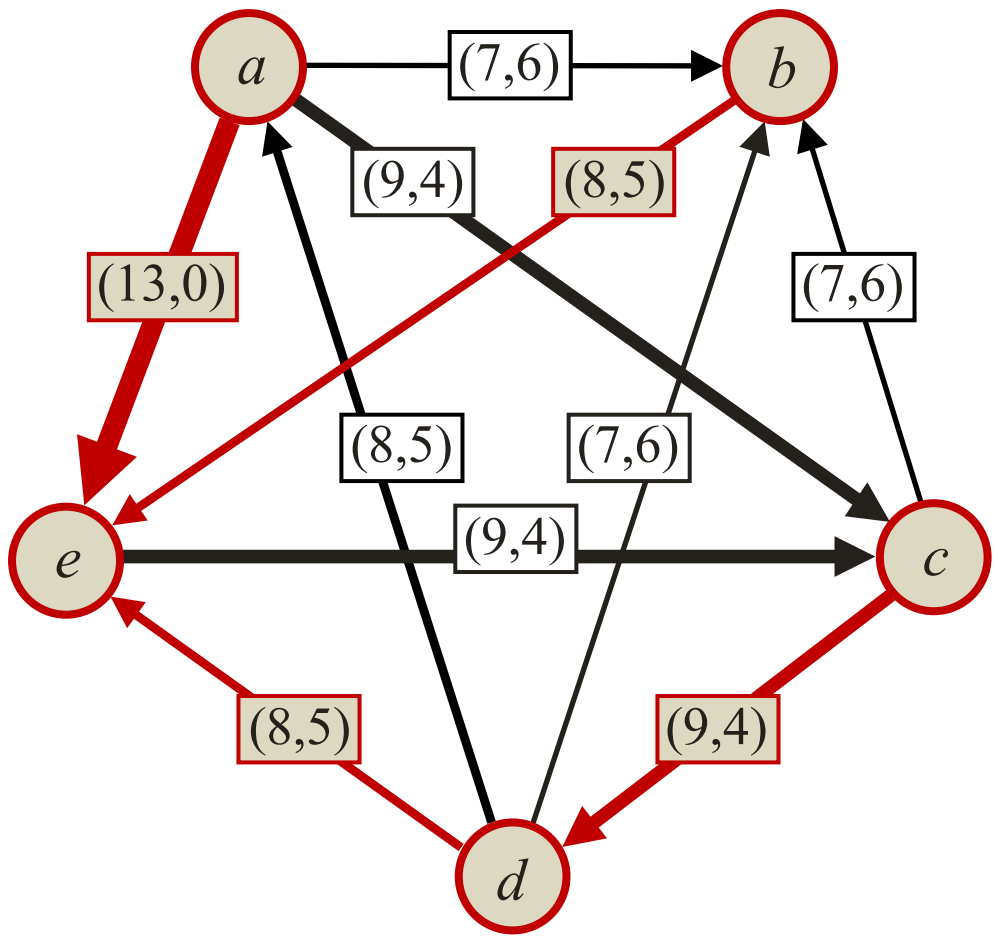

These are the strongest paths
from every other alternative to *e*.

Therefore, the strengths of the strongest paths are:

| | $P_D[*,a]$ | $P_D[*,b]$ | $P_D[*,c]$ | $P_D[*,d]$ | $P_D[*,e]$ |
|---|---|---|---|---|---|
| $P_D[a,*]$ | --- | (7,6) | (9,4) | (9,4) | (13,0) |
| $P_D[b,*]$ | (8,5) | --- | (8,5) | (8,5) | (8,5) |
| $P_D[c,*]$ | (8,5) | (7,6) | --- | (9,4) | (8,5) |
| $P_D[d,*]$ | (8,5) | (7,6) | (8,5) | --- | (8,5) |
| $P_D[e,*]$ | (8,5) | (7,6) | (9,4) | (9,4) | --- |

We get $\mathcal{O}^{new} = \{ac, ad, ae, ba, bc, bd, be, cd, ec, ed\}$ and $\mathcal{S}^{new} = \{b\}$.

Example 9 shows that the Schulze method, as defined in section 2.2, violates IPDA, as defined in (3.8.1) – (3.8.4). For example, we have (1) $ab \in \mathcal{O}^{old}$ and $ba \in \mathcal{O}^{new}$, (2) $cb \in \mathcal{O}^{old}$ and $bc \in \mathcal{O}^{new}$, (3) $db \in \mathcal{O}^{old}$ and $bd \in \mathcal{O}^{new}$, (4) $a \in \mathcal{S}^{old}$ and $a \notin \mathcal{S}^{new}$, and (5) $b \notin \mathcal{S}^{old}$ and $b \in \mathcal{S}^{new}$.

Suppose, the strongest paths are calculated with the Floyd-Warshall algorithm, as defined in section 2.3.1. Then the following table documents the $C \cdot (C\text{–}1) \cdot (C\text{–}2) = 60$ steps of the Floyd-Warshall algorithm.

We start with

- $P_D[i,j] := (N^{new}[i,j],N^{new}[j,i])$ for all $i \in A$ and $j \in A \setminus \{i\}$.
- $pred[i,j] := i$ for all $i \in A$ and $j \in A \setminus \{i\}$.

| | *i* | *j* | *k* | $P_D[j,k]$ | $P_D[j,i]$ | $P_D[i,k]$ | *pred*[*j*,*k*] | *pred*[*i*,*k*] | result |
|---|---|---|---|---|---|---|---|---|---|
| 1 | *a* | *b* | *c* | (6,7) | (6,7) | (9,4) | *b* | *a* | |
| 2 | *a* | *b* | *d* | (6,7) | (6,7) | (5,8) | *b* | *a* | |
| 3 | *a* | *b* | *e* | (8,5) | (6,7) | (13,0) | *b* | *a* | |
| 4 | *a* | *c* | *b* | (7,6) | (4,9) | (7,6) | *c* | *a* | |
| 5 | *a* | *c* | *d* | (9,4) | (4,9) | (5,8) | *c* | *a* | |
| 6 | *a* | *c* | *e* | (4,9) | (4,9) | (13,0) | *c* | *a* | |
| 7 | *a* | *d* | *b* | (7,6) | (8,5) | (7,6) | *d* | *a* | |
| 8 | *a* | *d* | *c* | (4,9) | (8,5) | (9,4) | *d* | *a* | $P_D[d,c]$ is updated from (4,9) to (8,5); *pred*[*d*,*c*] is updated from *d* to *a*. |
| 9 | *a* | *d* | *e* | (8,5) | (8,5) | (13,0) | *d* | *a* | |
| 10 | *a* | *e* | *b* | (5,8) | (0,13) | (7,6) | *e* | *a* | |
| 11 | *a* | *e* | *c* | (9,4) | (0,13) | (9,4) | *e* | *a* | |
| 12 | *a* | *e* | *d* | (5,8) | (0,13) | (5,8) | *e* | *a* | |
| 13 | *b* | *a* | *c* | (9,4) | (7,6) | (6,7) | *a* | *b* | |
| 14 | *b* | *a* | *d* | (5,8) | (7,6) | (6,7) | *a* | *b* | $P_D[a,d]$ is updated from (5,8) to (6,7); *pred*[*a*,*d*] is updated from *a* to *b*. |
| 15 | *b* | *a* | *e* | (13,0) | (7,6) | (8,5) | *a* | *b* | |
| 16 | *b* | *c* | *a* | (4,9) | (7,6) | (6,7) | *c* | *b* | $P_D[c,a]$ is updated from (4,9) to (6,7); *pred*[*c*,*a*] is updated from *c* to *b*. |
| 17 | *b* | *c* | *d* | (9,4) | (7,6) | (6,7) | *c* | *b* | |
| 18 | *b* | *c* | *e* | (4,9) | (7,6) | (8,5) | *c* | *b* | $P_D[c,e]$ is updated from (4,9) to (7,6); *pred*[*c*,*e*] is updated from *c* to *b*. |
| 19 | *b* | *d* | *a* | (8,5) | (7,6) | (6,7) | *d* | *b* | |
| 20 | *b* | *d* | *c* | (8,5) | (7,6) | (6,7) | *a* | *b* | |
| 21 | *b* | *d* | *e* | (8,5) | (7,6) | (8,5) | *d* | *b* | |
| 22 | *b* | *e* | *a* | (0,13) | (5,8) | (6,7) | *e* | *b* | $P_D[e,a]$ is updated from (0,13) to (5,8); *pred*[*e*,*a*] is updated from *e* to *b*. |
| 23 | *b* | *e* | *c* | (9,4) | (5,8) | (6,7) | *e* | *b* | |
| 24 | *b* | *e* | *d* | (5,8) | (5,8) | (6,7) | *e* | *b* | |
| 25 | *c* | *a* | *b* | (7,6) | (9,4) | (7,6) | *a* | *c* | |
| 26 | *c* | *a* | *d* | (6,7) | (9,4) | (9,4) | *b* | *c* | $P_D[a,d]$ is updated from (6,7) to (9,4); *pred*[*a*,*d*] is updated from *b* to *c*. |
| 27 | *c* | *a* | *e* | (13,0) | (9,4) | (7,6) | *a* | *b* | |
| 28 | *c* | *b* | *a* | (6,7) | (6,7) | (6,7) | *b* | *b* | |
| 29 | *c* | *b* | *d* | (6,7) | (6,7) | (9,4) | *b* | *c* | |
| 30 | *c* | *b* | *e* | (8,5) | (6,7) | (7,6) | *b* | *b* | |

| | *i* | *j* | *k* | $P_D[j,k]$ | $P_D[j,i]$ | $P_D[i,k]$ | *pred*[*j*,*k*] | *pred*[*i*,*k*] | result |
|---|---|---|---|---|---|---|---|---|---|
| 31 | *c* | *d* | *a* | (8,5) | (8,5) | (6,7) | *d* | *b* | |
| 32 | *c* | *d* | *b* | (7,6) | (8,5) | (7,6) | *d* | *c* | |
| 33 | *c* | *d* | *e* | (8,5) | (8,5) | (7,6) | *d* | *b* | |
| 34 | *c* | *e* | *a* | (5,8) | (9,4) | (6,7) | *b* | *b* | $P_D[e,a]$ is updated from (5,8) to (6,7). |
| 35 | *c* | *e* | *b* | (5,8) | (9,4) | (7,6) | *e* | *c* | $P_D[e,b]$ is updated from (5,8) to (7,6); *pred*[*e*,*b*] is updated from *e* to *c*. |
| 36 | *c* | *e* | *d* | (5,8) | (9,4) | (9,4) | *e* | *c* | $P_D[e,d]$ is updated from (5,8) to (9,4); *pred*[*e*,*d*] is updated from *e* to *c*. |
| 37 | *d* | *a* | *b* | (7,6) | (9,4) | (7,6) | *a* | *d* | |
| 38 | *d* | *a* | *c* | (9,4) | (9,4) | (8,5) | *a* | *a* | |
| 39 | *d* | *a* | *e* | (13,0) | (9,4) | (8,5) | *a* | *d* | |
| 40 | *d* | *b* | *a* | (6,7) | (6,7) | (8,5) | *b* | *d* | |
| 41 | *d* | *b* | *c* | (6,7) | (6,7) | (8,5) | *b* | *a* | |
| 42 | *d* | *b* | *e* | (8,5) | (6,7) | (8,5) | *b* | *d* | |
| 43 | *d* | *c* | *a* | (6,7) | (9,4) | (8,5) | *b* | *d* | $P_D[c,a]$ is updated from (6,7) to (8,5); *pred*[*c*,*a*] is updated from *b* to *d*. |
| 44 | *d* | *c* | *b* | (7,6) | (9,4) | (7,6) | *c* | *d* | |
| 45 | *d* | *c* | *e* | (7,6) | (9,4) | (8,5) | *b* | *d* | $P_D[c,e]$ is updated from (7,6) to (8,5); *pred*[*c*,*e*] is updated from *b* to *d*. |
| 46 | *d* | *e* | *a* | (6,7) | (9,4) | (8,5) | *b* | *d* | $P_D[e,a]$ is updated from (6,7) to (8,5); *pred*[*e*,*a*] is updated from *b* to *d*. |
| 47 | *d* | *e* | *b* | (7,6) | (9,4) | (7,6) | *c* | *d* | |
| 48 | *d* | *e* | *c* | (9,4) | (9,4) | (8,5) | *e* | *a* | |
| 49 | *e* | *a* | *b* | (7,6) | (13,0) | (7,6) | *a* | *c* | |
| 50 | *e* | *a* | *c* | (9,4) | (13,0) | (9,4) | *a* | *e* | |
| 51 | *e* | *a* | *d* | (9,4) | (13,0) | (9,4) | *c* | *c* | |
| 52 | *e* | *b* | *a* | (6,7) | (8,5) | (8,5) | *b* | *d* | $P_D[b,a]$ is updated from (6,7) to (8,5); *pred*[*b*,*a*] is updated from *b* to *d*. |
| 53 | *e* | *b* | *c* | (6,7) | (8,5) | (9,4) | *b* | *e* | $P_D[b,c]$ is updated from (6,7) to (8,5); *pred*[*b*,*c*] is updated from *b* to *e*. |
| 54 | *e* | *b* | *d* | (6,7) | (8,5) | (9,4) | *b* | *c* | $P_D[b,d]$ is updated from (6,7) to (8,5); *pred*[*b*,*d*] is updated from *b* to *c*. |
| 55 | *e* | *c* | *a* | (8,5) | (8,5) | (8,5) | *d* | *d* | |
| 56 | *e* | *c* | *b* | (7,6) | (8,5) | (7,6) | *c* | *c* | |
| 57 | *e* | *c* | *d* | (9,4) | (8,5) | (9,4) | *c* | *c* | |
| 58 | *e* | *d* | *a* | (8,5) | (8,5) | (8,5) | *d* | *d* | |
| 59 | *e* | *d* | *b* | (7,6) | (8,5) | (7,6) | *d* | *c* | |
| 60 | *e* | *d* | *c* | (8,5) | (8,5) | (9,4) | *a* | *e* | |

## 3.10. Example 10

When each voter $v \in V$ casts a linear order $\succ_v$ on $A$, then all definitions for $\succ_D$, that satisfy presumption (2.1.1), are equivalent. However, when some voters cast non-linear orders, then there are many possible definitions for the strength of a link. The following example illustrates how the different definitions for the strength of a link can lead to different winners.

Example 10:

| | | |
|---|---|---|
| 6 | voters | $a \succ_v b \succ_v c \succ_v d$ |
| 8 | voters | $a \approx_v b \succ_v c \approx_v d$ |
| 8 | voters | $a \approx_v c \succ_v b \approx_v d$ |
| 18 | voters | $a \approx_v c \succ_v d \succ_v b$ |
| 8 | voters | $a \approx_v c \approx_v d \succ_v b$ |
| 40 | voters | $b \succ_v a \approx_v c \approx_v d$ |
| 4 | voters | $c \succ_v b \succ_v d \succ_v a$ |
| 9 | voters | $c \succ_v d \succ_v a \succ_v b$ |
| 8 | voters | $c \approx_v d \succ_v a \approx_v b$ |
| 14 | voters | $d \succ_v a \succ_v b \succ_v c$ |
| 11 | voters | $d \succ_v b \succ_v c \succ_v a$ |
| 4 | voters | $d \succ_v c \succ_v a \succ_v b$ |

The pairwise matrix $N$ looks as follows:

| | $N$[*,$a$] | $N$[*,$b$] | $N$[*,$c$] | $N$[*,$d$] |
|---|---|---|---|---|
| $N$[$a$,*] | --- | 67 | 28 | 40 |
| $N$[$b$,*] | 55 | --- | 79 | 58 |
| $N$[$c$,*] | 36 | 59 | --- | 45 |
| $N$[$d$,*] | 50 | 72 | 29 | --- |

The corresponding digraph looks as follows:

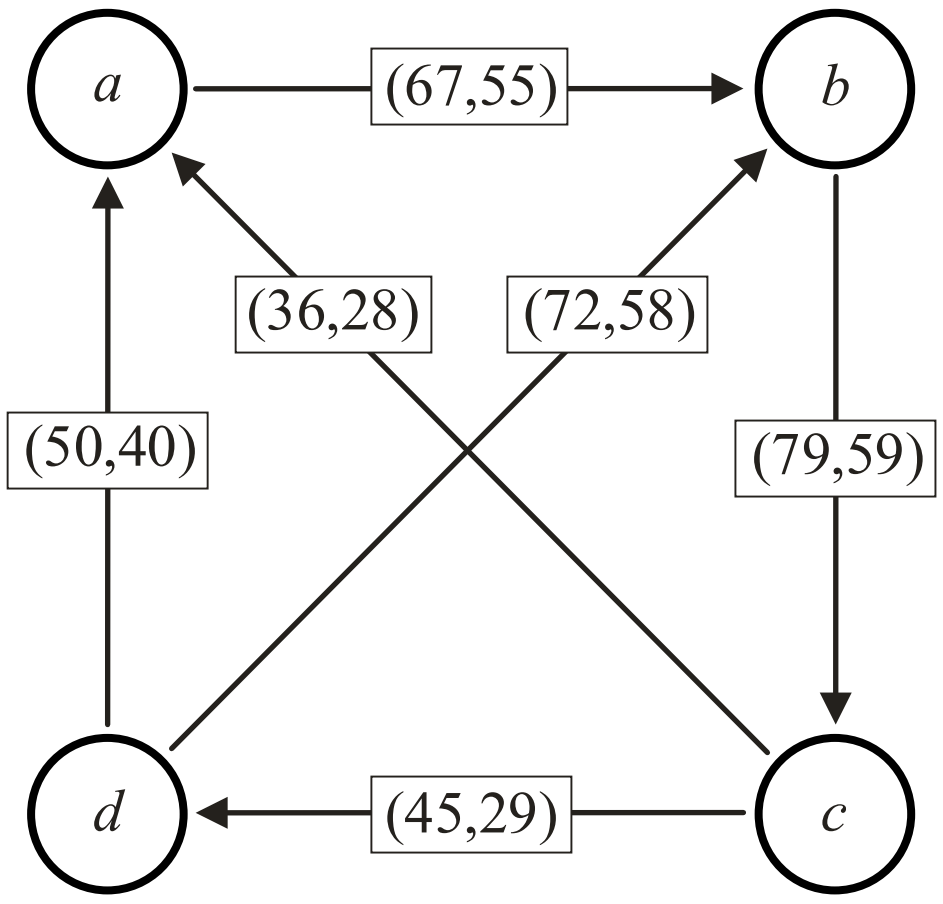

## a) margin

We get: $(N[b,c],N[c,b]) \succ_{margin} (N[c,d],N[d,c]) \succ_{margin} (N[d,b],N[b,d]) \succ_{margin} (N[a,b],N[b,a]) \succ_{margin} (N[d,a],N[a,d]) \succ_{margin} (N[c,a],N[a,c])$.

The pairwise victories are:

*bc* with a margin of $N[b,c] - N[c,b] = 20$
*cd* with a margin of $N[c,d] - N[d,c] = 16$
*db* with a margin of $N[d,b] - N[b,d] = 14$
*ab* with a margin of $N[a,b] - N[b,a] = 12$
*da* with a margin of $N[d,a] - N[a,d] = 10$
*ca* with a margin of $N[c,a] - N[a,c] = 8$

The following table lists the strongest paths, as determined by the Floyd-Warshall algorithm, as defined in section 2.3.1. The critical links of the strongest paths are underlined:

| | ... to *a* | ... to *b* | ... to *c* | ... to *d* | ... to every other alternative |
|---|---|---|---|---|---|
| from *a* ... | --- | 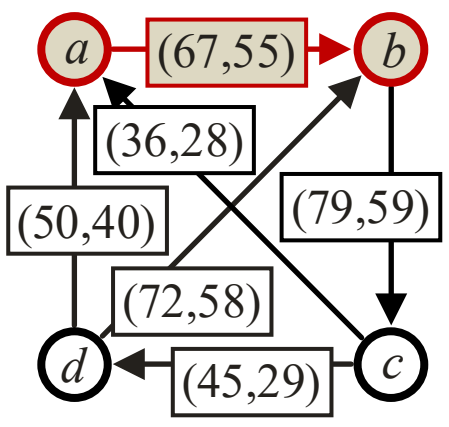 *a*, (67,55), *b* | 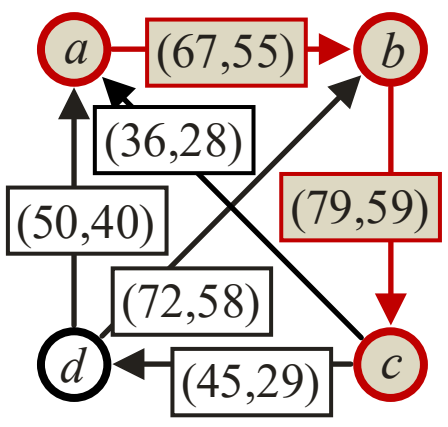 *a*, (67,55), *b*, (79,59), *c* | 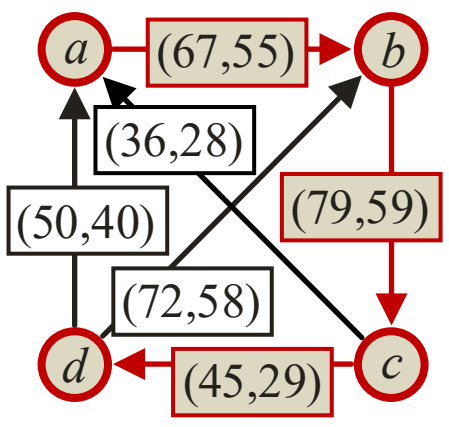 *a*, (67,55), *b*, (79,59), *c*, (45,29), *d* | 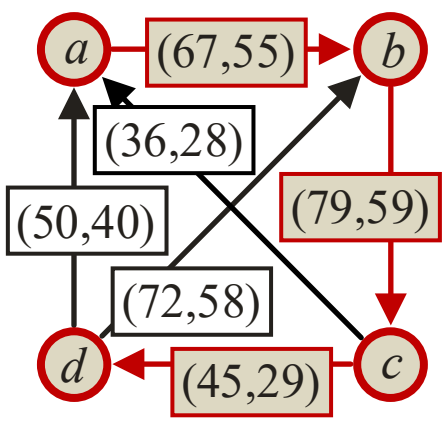 |
| from *b* ... | 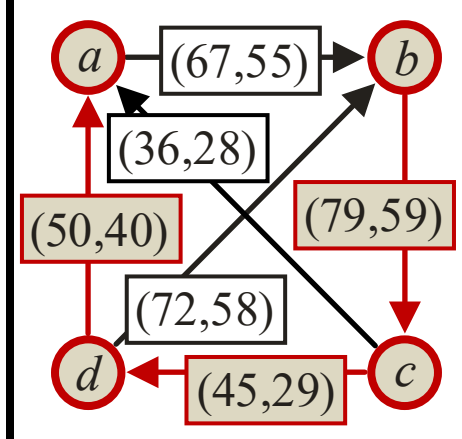 *b*, (79,59), *c*, (45,29), *d*, (50,40), *a* | --- | 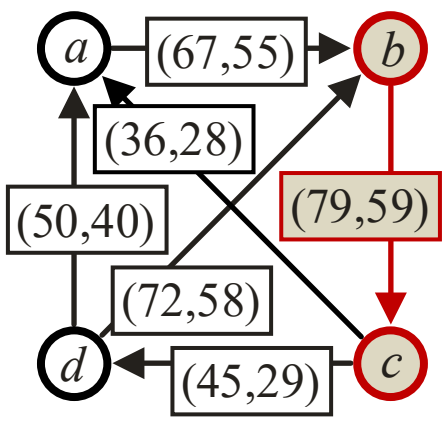 *b*, (79,59), *c* | 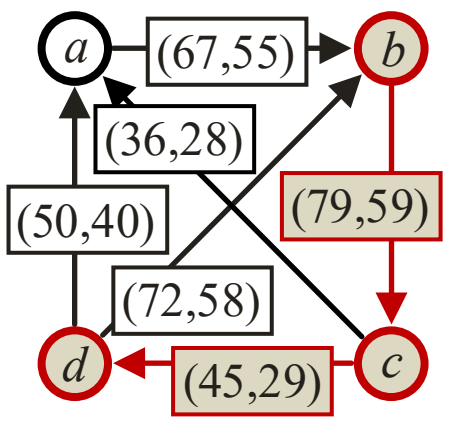 *b*, (79,59), *c*, (45,29), *d* | 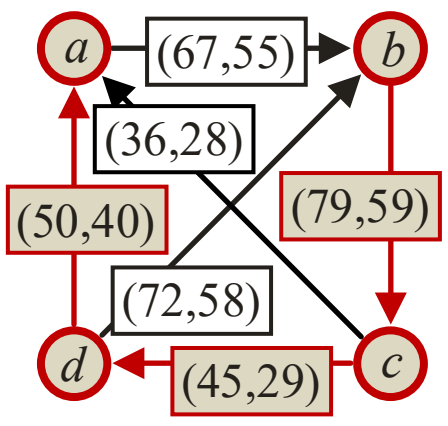 |
| from *c* ... | 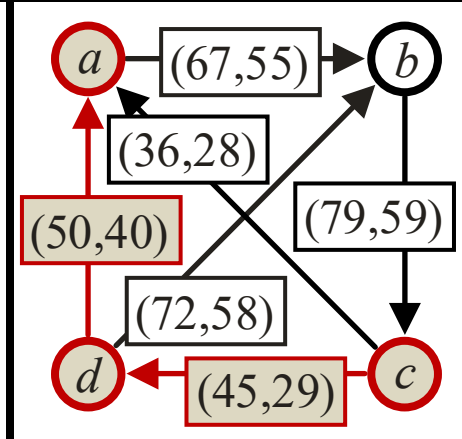 *c*, (45,29), *d*, (50,40), *a* | 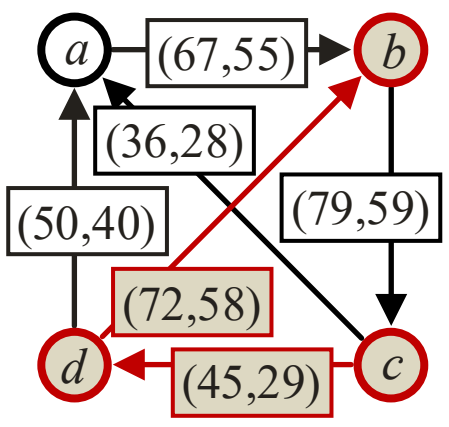 *c*, (45,29), *d*, (72,58), *b* | --- | 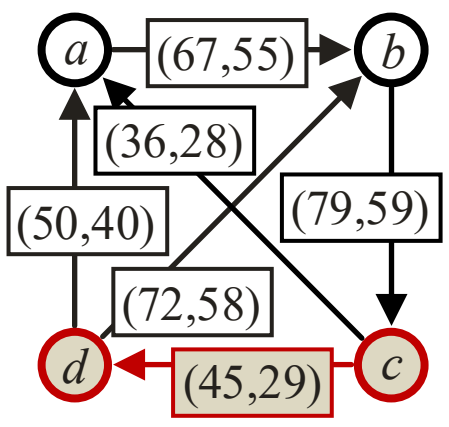 *c*, (45,29), *d* | 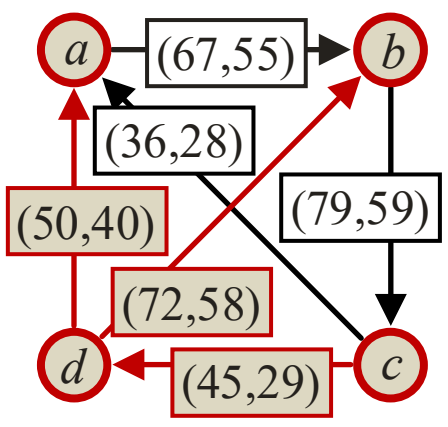 |
| from *d* ... | 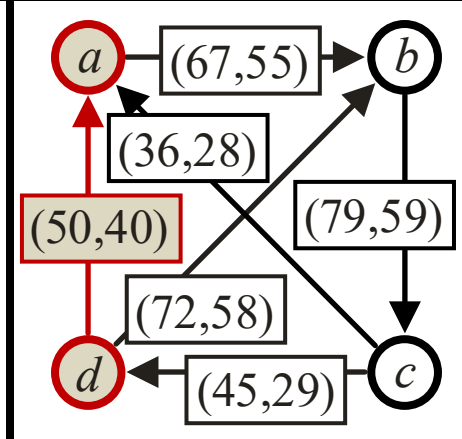 *d*, (50,40), *a* | 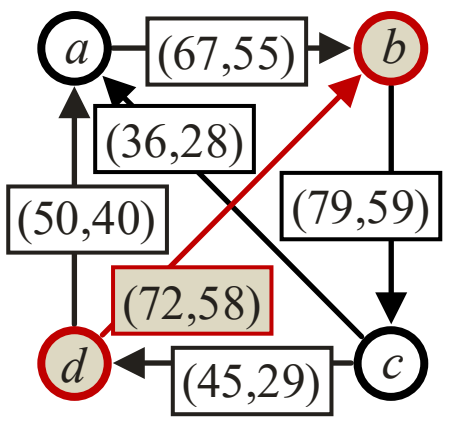 *d*, (72,58), *b* | 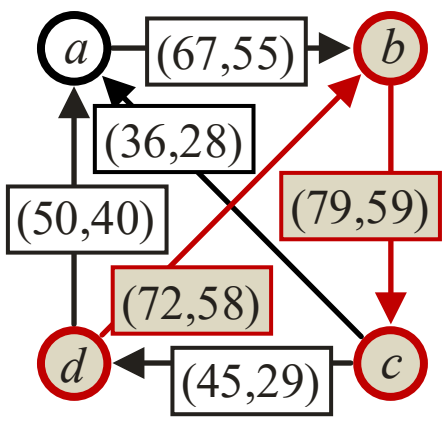 *d*, (72,58), *b*, (79,59), *c* | --- | 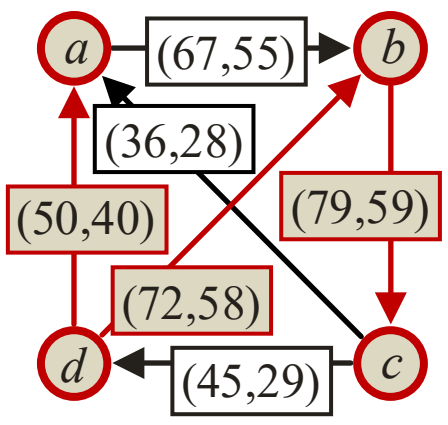 |
| from every other alternative ... | 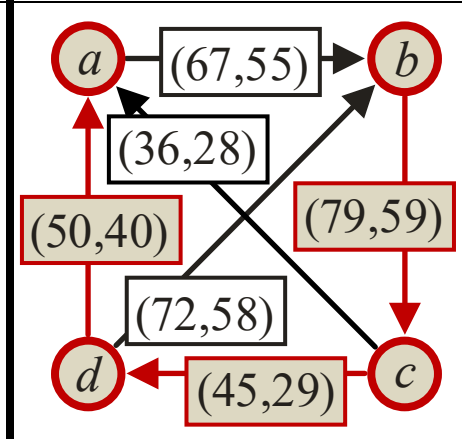 | 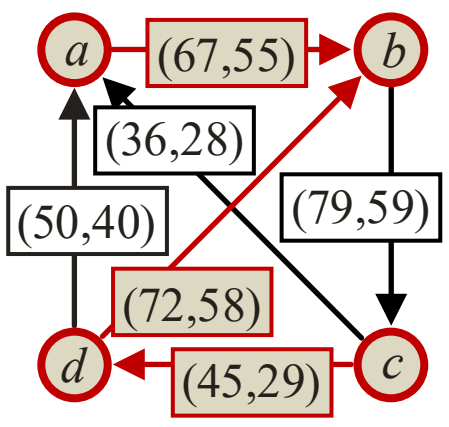 | 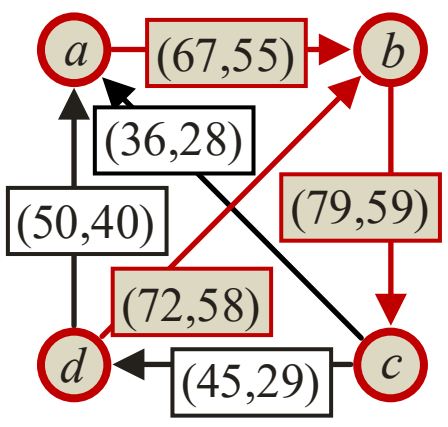 | 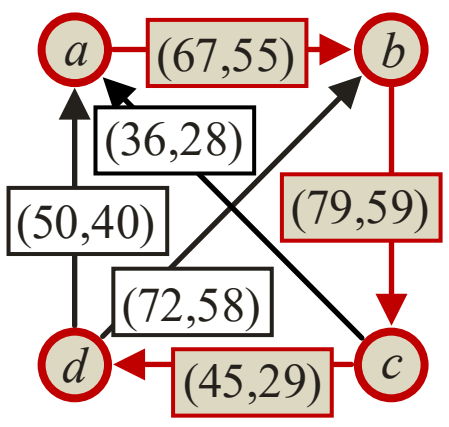 | --- |

The strengths of the strongest paths are:

| | $P_{margin}[*,a]$ | $P_{margin}[*,b]$ | $P_{margin}[*,c]$ | $P_{margin}[*,d]$ |
|---|---|---|---|---|
| $P_{margin}[a,*]$ | --- | (67,55) | (67,55) | (67,55) |
| $P_{margin}[b,*]$ | (50,40) | --- | (79,59) | (45,29) |
| $P_{margin}[c,*]$ | (50,40) | (72,58) | --- | (45,29) |
| $P_{margin}[d,*]$ | (50,40) | (72,58) | (72,58) | --- |

We get $\mathcal{O}_{margin} = \{ab, ac, ad, bc, bd, cd\}$ and $\mathcal{S}_{margin} = \{a\}$.

Suppose, the strongest paths are calculated with the Floyd-Warshall algorithm, as defined in section 2.3.1. Then the following table documents the $C \cdot (C–1) \cdot (C–2) = 24$ steps of the Floyd-Warshall algorithm.

We start with

- $P_{margin}[i,j] := (N[i,j],N[j,i])$ for all $i \in A$ and $j \in A \setminus \{i\}$.
- $pred[i,j] := i$ for all $i \in A$ and $j \in A \setminus \{i\}$.

| | $i$ | $j$ | $k$ | $P_{margin}[j,k]$ | $P_{margin}[j,i]$ | $P_{margin}[i,k]$ | $pred[j,k]$ | $pred[i,k]$ | result |
|---|---|---|---|---|---|---|---|---|---|
| 1 | *a* | *b* | *c* | (79,59) | (55,67) | (28,36) | *b* | *a* | |
| 2 | *a* | *b* | *d* | (58,72) | (55,67) | (40,50) | *b* | *a* | $P_{margin}[b,d]$ is updated from (58,72) to (55,67); *pred*[*b*,*d*] is updated from *b* to *a*. |
| 3 | *a* | *c* | *b* | (59,79) | (36,28) | (67,55) | *c* | *a* | $P_{margin}[c,b]$ is updated from (59,79) to (36,28); *pred*[*c*,*b*] is updated from *c* to *a*. |
| 4 | *a* | *c* | *d* | (45,29) | (36,28) | (40,50) | *c* | *a* | |
| 5 | *a* | *d* | *b* | (72,58) | (50,40) | (67,55) | *d* | *a* | |
| 6 | *a* | *d* | *c* | (29,45) | (50,40) | (28,36) | *d* | *a* | $P_{margin}[d,c]$ is updated from (29,45) to (28,36); *pred*[*d*,*c*] is updated from *d* to *a*. |
| 7 | *b* | *a* | *c* | (28,36) | (67,55) | (79,59) | *a* | *b* | $P_{margin}[a,c]$ is updated from (28,36) to (67,55); *pred*[*a*,*c*] is updated from *a* to *b*. |
| 8 | *b* | *a* | *d* | (40,50) | (67,55) | (55,67) | *a* | *a* | |
| 9 | *b* | *c* | *a* | (36,28) | (36,28) | (55,67) | *c* | *b* | |
| 10 | *b* | *c* | *d* | (45,29) | (36,28) | (55,67) | *c* | *a* | |
| 11 | *b* | *d* | *a* | (50,40) | (72,58) | (55,67) | *d* | *b* | |
| 12 | *b* | *d* | *c* | (28,36) | (72,58) | (79,59) | *a* | *b* | $P_{margin}[d,c]$ is updated from (28,36) to (72,58); *pred*[*d*,*c*] is updated from *a* to *b*. |
| 13 | *c* | *a* | *b* | (67,55) | (67,55) | (36,28) | *a* | *a* | |
| 14 | *c* | *a* | *d* | (40,50) | (67,55) | (45,29) | *a* | *c* | $P_{margin}[a,d]$ is updated from (40,50) to (67,55); *pred*[*a*,*d*] is updated from *a* to *c*. |
| 15 | *c* | *b* | *a* | (55,67) | (79,59) | (36,28) | *b* | *c* | $P_{margin}[b,a]$ is updated from (55,67) to (36,28); *pred*[*b*,*a*] is updated from *b* to *c*. |
| 16 | *c* | *b* | *d* | (55,67) | (79,59) | (45,29) | *a* | *c* | $P_{margin}[b,d]$ is updated from (55,67) to (45,29); *pred*[*b*,*d*] is updated from *a* to *c*. |
| 17 | *c* | *d* | *a* | (50,40) | (72,58) | (36,28) | *d* | *c* | |
| 18 | *c* | *d* | *b* | (72,58) | (72,58) | (36,28) | *d* | *a* | |
| 19 | *d* | *a* | *b* | (67,55) | (67,55) | (72,58) | *a* | *d* | |
| 20 | *d* | *a* | *c* | (67,55) | (67,55) | (72,58) | *b* | *b* | |
| 21 | *d* | *b* | *a* | (36,28) | (45,29) | (50,40) | *c* | *d* | $P_{margin}[b,a]$ is updated from (36,28) to (50,40); *pred*[*b*,*a*] is updated from *c* to *d*. |
| 22 | *d* | *b* | *c* | (79,59) | (45,29) | (72,58) | *b* | *b* | |
| 23 | *d* | *c* | *a* | (36,28) | (45,29) | (50,40) | *c* | *d* | $P_{margin}[c,a]$ is updated from (36,28) to (50,40); *pred*[*c*,*a*] is updated from *c* to *d*. |
| 24 | *d* | *c* | *b* | (36,28) | (45,29) | (72,58) | *a* | *d* | $P_{margin}[c,b]$ is updated from (36,28) to (72,58); *pred*[*c*,*b*] is updated from *a* to *d*. |

## b) ratio

We get: $(N[c,d],N[d,c]) \succ_{ratio} (N[b,c],N[c,b]) \succ_{ratio} (N[c,a],N[a,c]) \succ_{ratio} (N[d,a],N[a,d]) \succ_{ratio} (N[d,b],N[b,d]) \succ_{ratio} (N[a,b],N[b,a])$.

The pairwise victories are:

*cd* with a ratio of $N[c,d] / N[d,c] = 1.552$
*bc* with a ratio of $N[b,c] / N[c,b] = 1.339$
*ca* with a ratio of $N[c,a] / N[a,c] = 1.286$
*da* with a ratio of $N[d,a] / N[a,d] = 1.250$
*db* with a ratio of $N[d,b] / N[b,d] = 1.241$
*ab* with a ratio of $N[a,b] / N[b,a] = 1.218$

The following table lists the strongest paths, as determined by the Floyd-Warshall algorithm, as defined in section 2.3.1. The critical links of the strongest paths are underlined:

<table>
<tr><th></th><th>... to a</th><th>... to b</th><th>... to c</th><th>... to d</th><th>... to every other alternative</th></tr>
<tr><td>from a ...</td><td>---</td><td>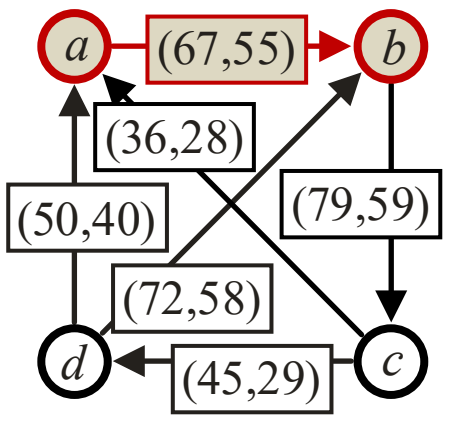<br>a, (67,55), b</td><td>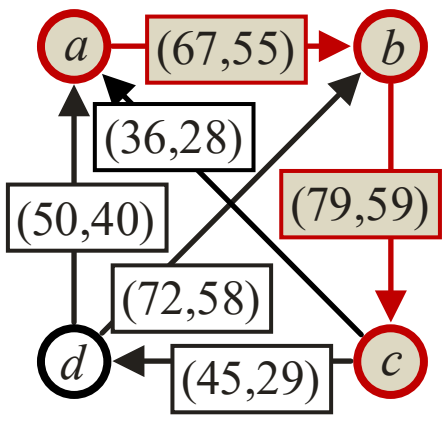<br>a, (67,55), b, (79,59), c</td><td>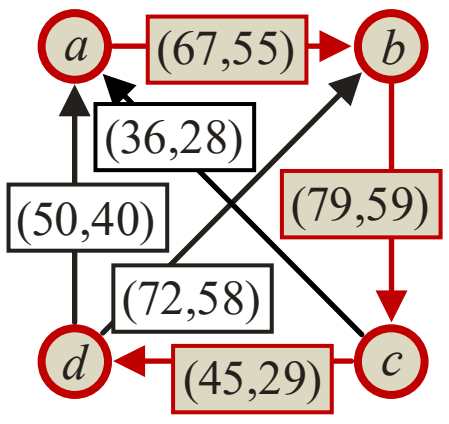<br>a, (67,55), b, (79,59), c, (45,29), d</td><td>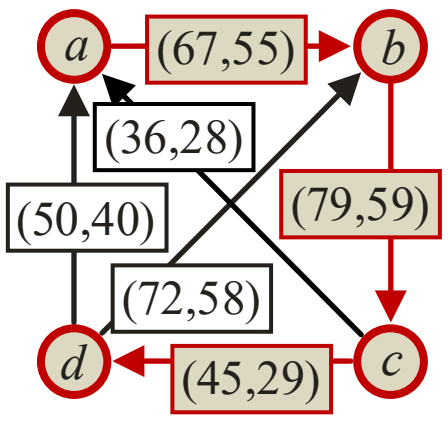</td></tr>
<tr><td>from b ...</td><td>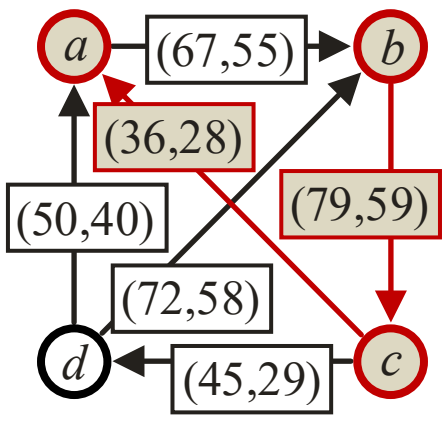<br>b, (79,59), c, (36,28), a</td><td>---</td><td>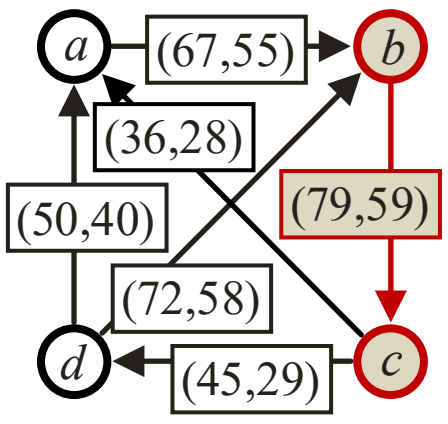<br>b, (79,59), c</td><td>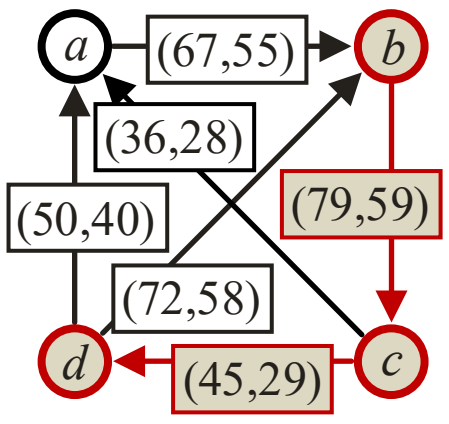<br>b, (79,59), c, (45,29), d</td><td>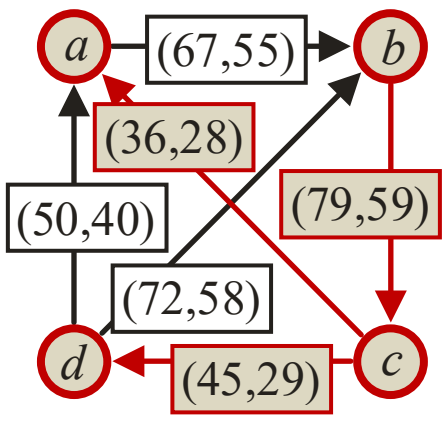</td></tr>
<tr><td>from c ...</td><td>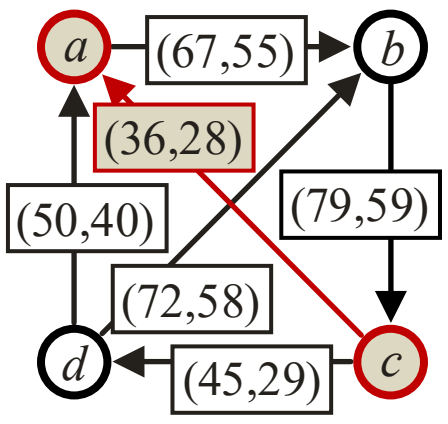<br>c, (36,28), a</td><td>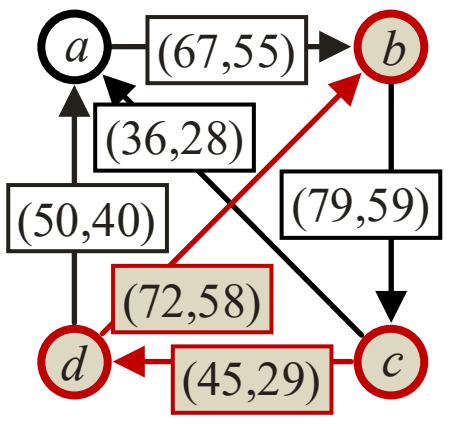<br>c, (45,29), d, (72,58), b</td><td>---</td><td>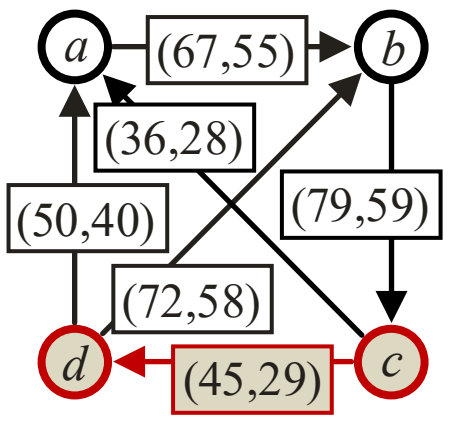<br>c, (45,29), d</td><td>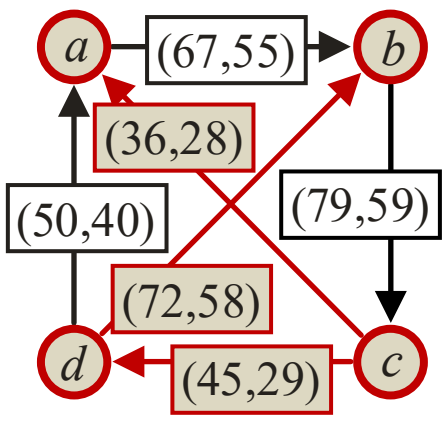</td></tr>
<tr><td>from d ...</td><td>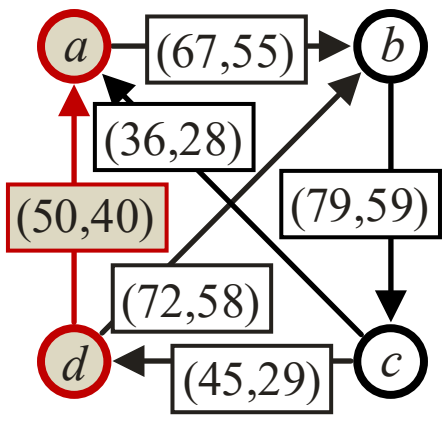<br>d, (50,40), a</td><td>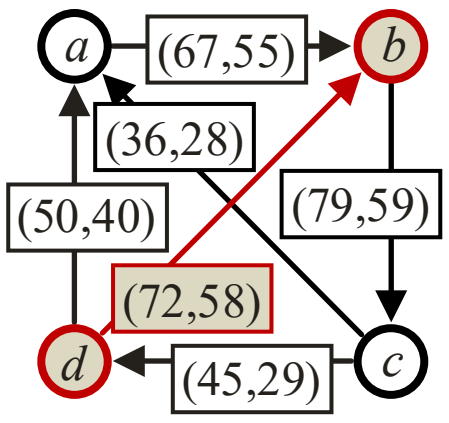<br>d, (72,58), b</td><td>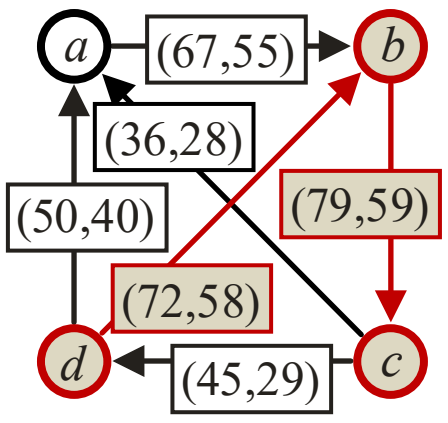<br>d, (72,58), b, (79,59), c</td><td>---</td><td>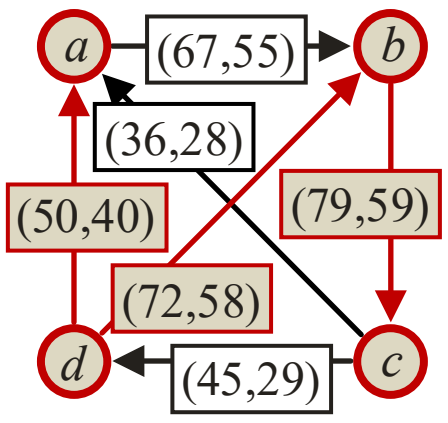</td></tr>
<tr><td>from every other alternative ...</td><td>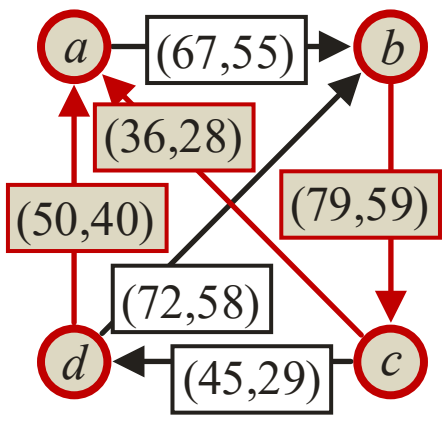</td><td>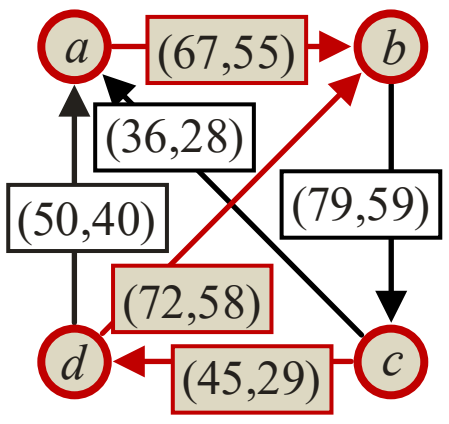</td><td>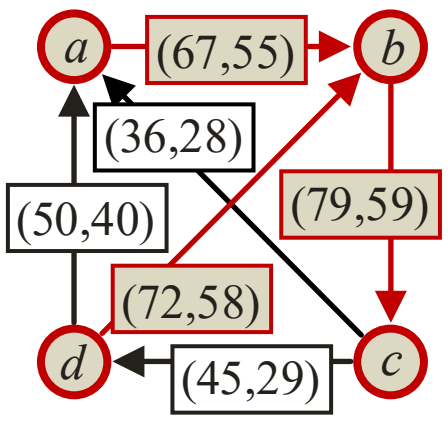</td><td>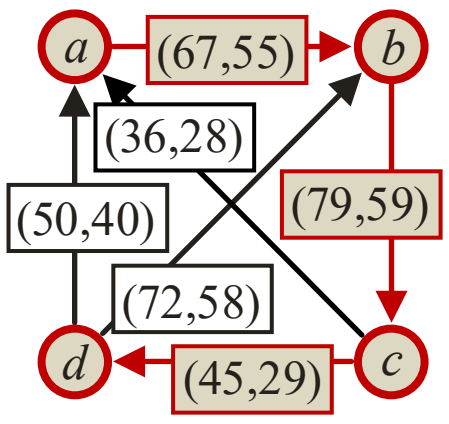</td><td>---</td></tr>
</table>

The strengths of the strongest paths are:

| | $P_{ratio}[*,a]$ | $P_{ratio}[*,b]$ | $P_{ratio}[*,c]$ | $P_{ratio}[*,d]$ |
|---|---|---|---|---|
| $P_{ratio}[a,*]$ | --- | (67,55) | (67,55) | (67,55) |
| $P_{ratio}[b,*]$ | (36,28) | --- | (79,59) | (79,59) |
| $P_{ratio}[c,*]$ | (36,28) | (72,58) | --- | (45,29) |
| $P_{ratio}[d,*]$ | (50,40) | (72,58) | (72,58) | --- |

We get $\mathcal{O}_{ratio} = \{ba, bc, bd, ca, cd, da\}$ and $\mathcal{S}_{ratio} = \{b\}$.

Suppose, the strongest paths are calculated with the Floyd-Warshall algorithm, as defined in section 2.3.1. Then the following table documents the $C \cdot (C–1) \cdot (C–2) = 24$ steps of the Floyd-Warshall algorithm.

We start with

- $P_{ratio}[i,j] := (N[i,j],N[j,i])$ for all $i \in A$ and $j \in A \setminus \{i\}$.
- $pred[i,j] := i$ for all $i \in A$ and $j \in A \setminus \{i\}$.

| | $i$ | $j$ | $k$ | $P_{ratio}[j,k]$ | $P_{ratio}[j,i]$ | $P_{ratio}[i,k]$ | $pred[j,k]$ | $pred[i,k]$ | result |
|---|---|---|---|---|---|---|---|---|---|
| 1 | $a$ | $b$ | $c$ | (79,59) | (55,67) | (28,36) | $b$ | $a$ | |
| 2 | $a$ | $b$ | $d$ | (58,72) | (55,67) | (40,50) | $b$ | $a$ | |
| 3 | $a$ | $c$ | $b$ | (59,79) | (36,28) | (67,55) | $c$ | $a$ | $P_{ratio}[c,b]$ is updated from (59,79) to (67,55); $pred[c,b]$ is updated from $c$ to $a$. |
| 4 | $a$ | $c$ | $d$ | (45,29) | (36,28) | (40,50) | $c$ | $a$ | |
| 5 | $a$ | $d$ | $b$ | (72,58) | (50,40) | (67,55) | $d$ | $a$ | |
| 6 | $a$ | $d$ | $c$ | (29,45) | (50,40) | (28,36) | $d$ | $a$ | $P_{ratio}[d,c]$ is updated from (29,45) to (28,36); $pred[d,c]$ is updated from $d$ to $a$. |
| 7 | $b$ | $a$ | $c$ | (28,36) | (67,55) | (79,59) | $a$ | $b$ | $P_{ratio}[a,c]$ is updated from (28,36) to (67,55); $pred[a,c]$ is updated from $a$ to $b$. |
| 8 | $b$ | $a$ | $d$ | (40,50) | (67,55) | (58,72) | $a$ | $b$ | $P_{ratio}[a,d]$ is updated from (40,50) to (58,72); $pred[a,d]$ is updated from $a$ to $b$. |
| 9 | $b$ | $c$ | $a$ | (36,28) | (67,55) | (55,67) | $c$ | $b$ | |
| 10 | $b$ | $c$ | $d$ | (45,29) | (67,55) | (58,72) | $c$ | $b$ | |
| 11 | $b$ | $d$ | $a$ | (50,40) | (72,58) | (55,67) | $d$ | $b$ | |
| 12 | $b$ | $d$ | $c$ | (28,36) | (72,58) | (79,59) | $a$ | $b$ | $P_{ratio}[d,c]$ is updated from (28,36) to (72,58); $pred[d,c]$ is updated from $a$ to $b$. |
| 13 | $c$ | $a$ | $b$ | (67,55) | (67,55) | (67,55) | $a$ | $a$ | |
| 14 | $c$ | $a$ | $d$ | (58,72) | (67,55) | (45,29) | $b$ | $c$ | $P_{ratio}[a,d]$ is updated from (58,72) to (67,55); $pred[a,d]$ is updated from $b$ to $c$. |
| 15 | $c$ | $b$ | $a$ | (55,67) | (79,59) | (36,28) | $b$ | $c$ | $P_{ratio}[b,a]$ is updated from (55,67) to (36,28); $pred[b,a]$ is updated from $b$ to $c$. |
| 16 | $c$ | $b$ | $d$ | (58,72) | (79,59) | (45,29) | $b$ | $c$ | $P_{ratio}[b,d]$ is updated from (58,72) to (79,59); $pred[b,d]$ is updated from $b$ to $c$. |
| 17 | $c$ | $d$ | $a$ | (50,40) | (72,58) | (36,28) | $d$ | $c$ | |
| 18 | $c$ | $d$ | $b$ | (72,58) | (72,58) | (67,55) | $d$ | $a$ | |
| 19 | $d$ | $a$ | $b$ | (67,55) | (67,55) | (72,58) | $a$ | $d$ | |
| 20 | $d$ | $a$ | $c$ | (67,55) | (67,55) | (72,58) | $b$ | $b$ | |
| 21 | $d$ | $b$ | $a$ | (36,28) | (79,59) | (50,40) | $c$ | $d$ | |
| 22 | $d$ | $b$ | $c$ | (79,59) | (79,59) | (72,58) | $b$ | $b$ | |
| 23 | $d$ | $c$ | $a$ | (36,28) | (45,29) | (50,40) | $c$ | $d$ | |
| 24 | $d$ | $c$ | $b$ | (67,55) | (45,29) | (72,58) | $a$ | $d$ | $P_{ratio}[c,b]$ is updated from (67,55) to (72,58); $pred[c,b]$ is updated from $a$ to $d$. |

## c) winning votes

We get: $(N[b,c],N[c,b]) \succ_{win} (N[d,b],N[b,d]) \succ_{win} (N[a,b],N[b,a]) \succ_{win} (N[d,a],N[a,d]) \succ_{win} (N[c,d],N[d,c]) \succ_{win} (N[c,a],N[a,c])$.

The pairwise victories are:

*bc* with a support of $N[b,c] = 79$
*db* with a support of $N[d,b] = 72$
*ab* with a support of $N[a,b] = 67$
*da* with a support of $N[d,a] = 50$
*cd* with a support of $N[c,d] = 45$
*ca* with a support of $N[c,a] = 36$

The following table lists the strongest paths, as determined by the Floyd-Warshall algorithm, as defined in section 2.3.1. The critical links of the strongest paths are <u>underlined</u>:

<table>
<tr><th></th><th>... to a</th><th>... to b</th><th>... to c</th><th>... to d</th><th>... to every other alternative</th></tr>
<tr><td>from a ...</td><td>---</td>
<td>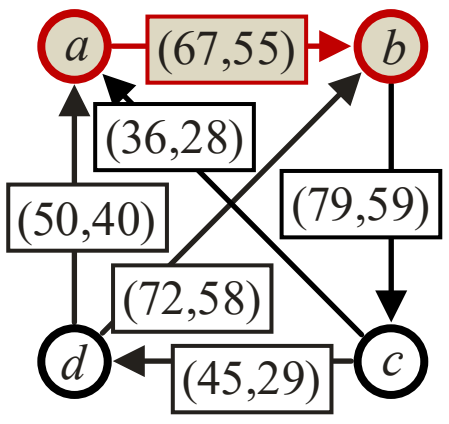

a, <u>(67,55)</u>, b</td>
<td>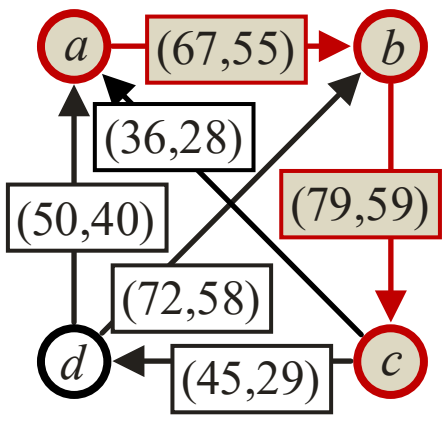

a, <u>(67,55)</u>, b, (79,59), c</td>
<td>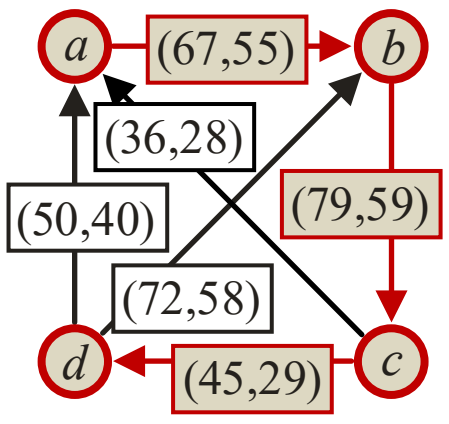

a, (67,55), b, (79,59), c, <u>(45,29)</u>, d</td>
<td>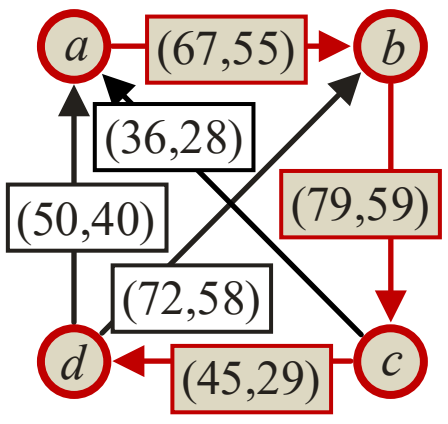
</td></tr>
<tr><td>from b ...</td>
<td>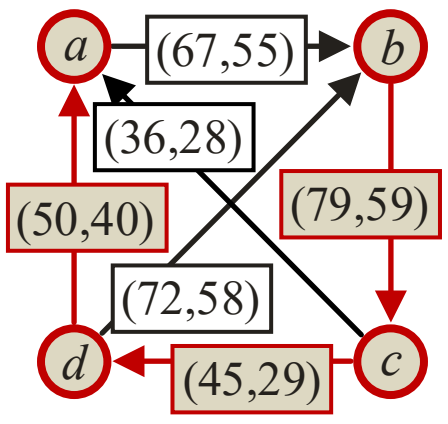

b, (79,59), c, <u>(45,29)</u>, d, (50,40), a</td>
<td>---</td>
<td>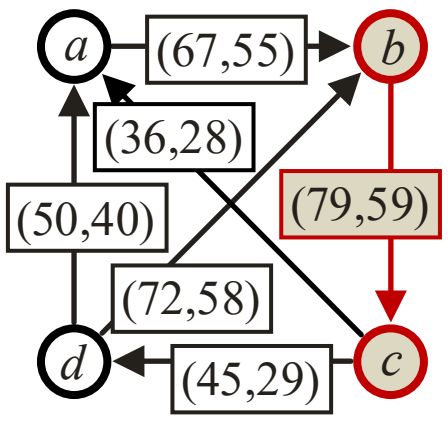

b, <u>(79,59)</u>, c</td>
<td>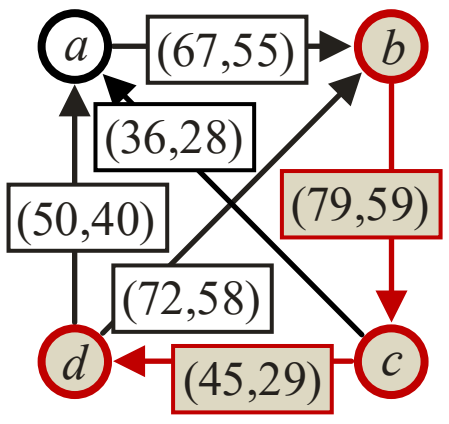

b, (79,59), c, <u>(45,29)</u>, d</td>
<td>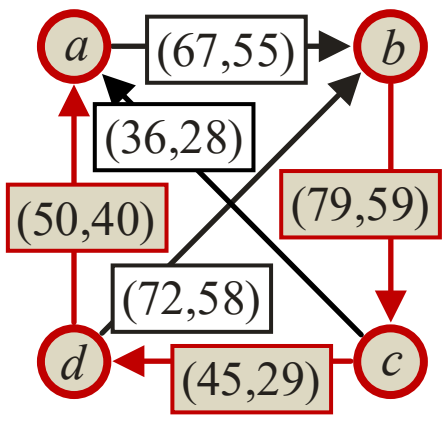
</td></tr>
<tr><td>from c ...</td>
<td>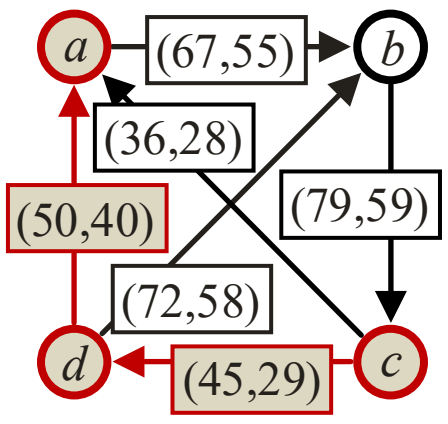

c, <u>(45,29)</u>, d, (50,40), a</td>
<td>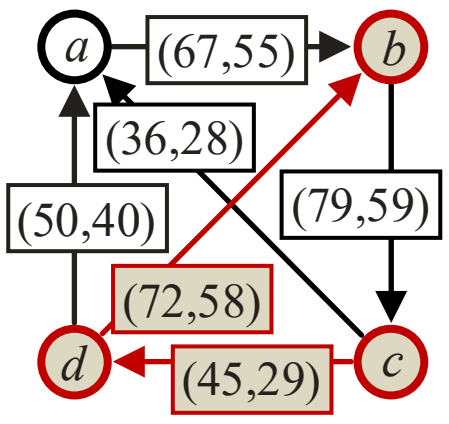

c, <u>(45,29)</u>, d, (72,58), b</td>
<td>---</td>
<td>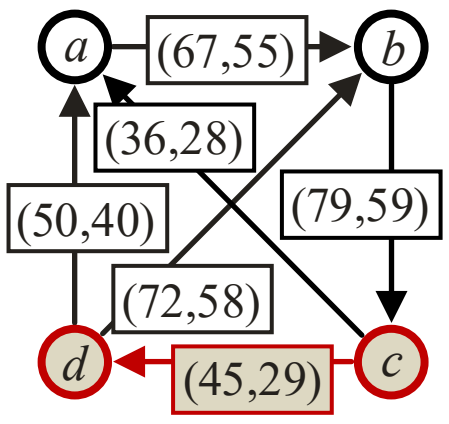

c, <u>(45,29)</u>, d</td>
<td>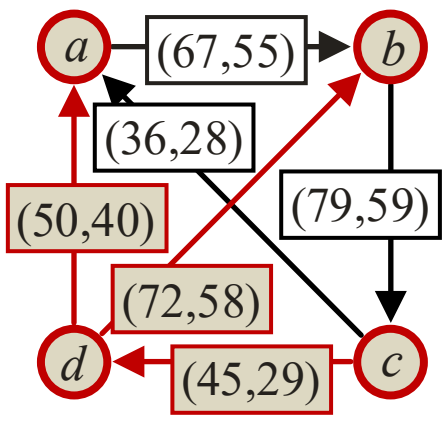
</td></tr>
<tr><td>from d ...</td>
<td>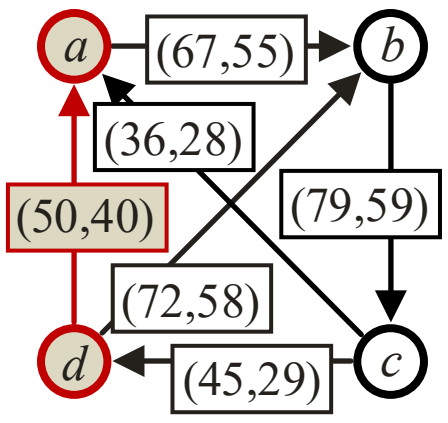

d, <u>(50,40)</u>, a</td>
<td>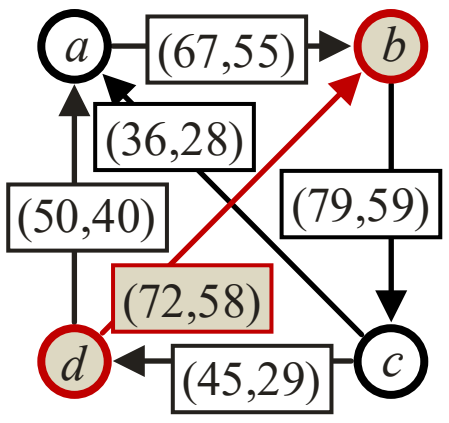

d, <u>(72,58)</u>, b</td>
<td>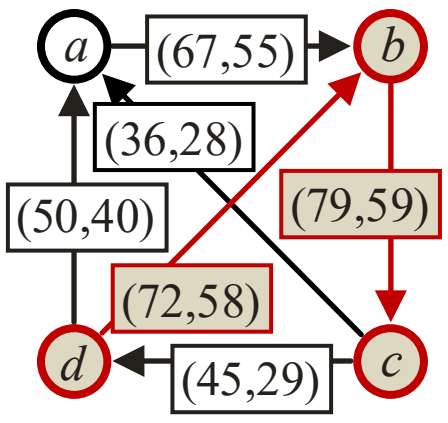

d, <u>(72,58)</u>, b, (79,59), c</td>
<td>---</td>
<td>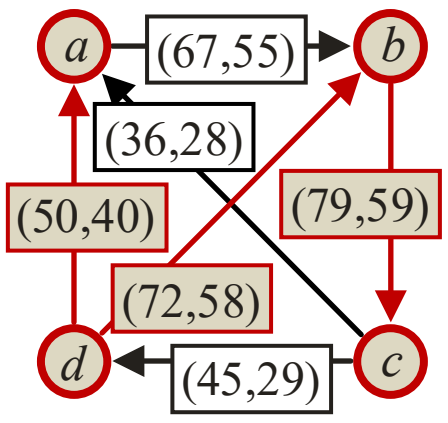
</td></tr>
<tr><td>from every other alternative ...</td>
<td>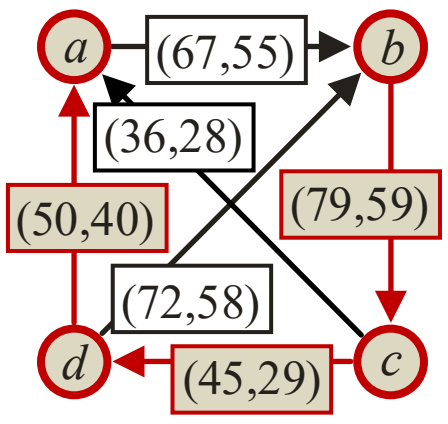
</td>
<td>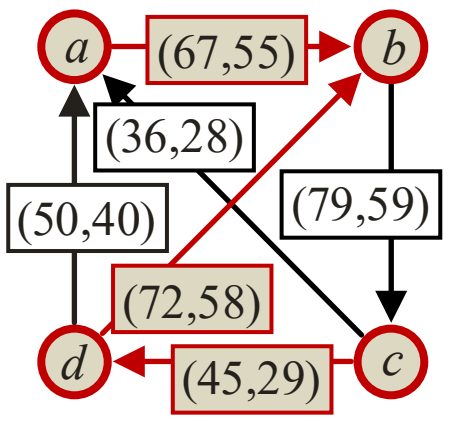
</td>
<td>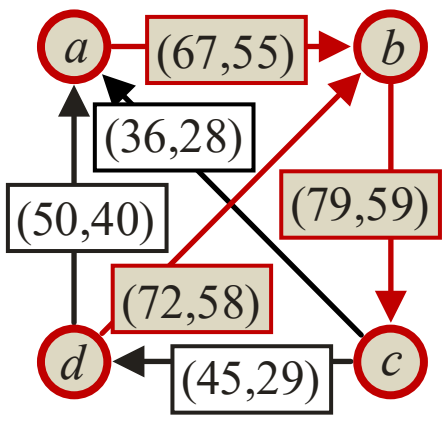
</td>
<td>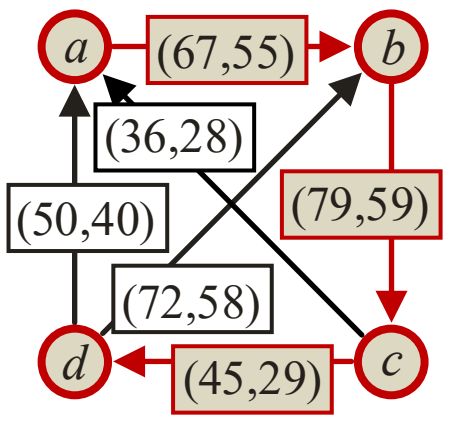
</td>
<td>---</td></tr>
</table>

The strengths of the strongest paths are:

| | $P_{win}[*,a]$ | $P_{win}[*,b]$ | $P_{win}[*,c]$ | $P_{win}[*,d]$ |
|---|---|---|---|---|
| $P_{win}[a,*]$ | --- | (67,55) | (67,55) | (45,29) |
| $P_{win}[b,*]$ | (45,29) | --- | (79,59) | (45,29) |
| $P_{win}[c,*]$ | (45,29) | (45,29) | --- | (45,29) |
| $P_{win}[d,*]$ | (50,40) | (72,58) | (72,58) | --- |

We get $\mathcal{O}_{win} = \{ab, ac, bc, da, db, dc\}$ and $\mathcal{S}_{win} = \{d\}$.

Suppose, the strongest paths are calculated with the Floyd-Warshall algorithm, as defined in section 2.3.1. Then the following table documents the $C \cdot (C–1) \cdot (C–2) = 24$ steps of the Floyd-Warshall algorithm.

We start with

- $P_{win}[i,j] := (N[i,j],N[j,i])$ for all $i \in A$ and $j \in A \setminus \{i\}$.
- $pred[i,j] := i$ for all $i \in A$ and $j \in A \setminus \{i\}$.

| | $i$ | $j$ | $k$ | $P_{win}[j,k]$ | $P_{win}[j,i]$ | $P_{win}[i,k]$ | $pred[j,k]$ | $pred[i,k]$ | result |
|---|---|---|---|---|---|---|---|---|---|
| 1 | *a* | *b* | *c* | (79,59) | (55,67) | (28,36) | *b* | *a* | |
| 2 | *a* | *b* | *d* | (58,72) | (55,67) | (40,50) | *b* | *a* | $P_{win}[b,d]$ is updated from (58,72) to (55,67); *pred*[*b*,*d*] is updated from *b* to *a*. |
| 3 | *a* | *c* | *b* | (59,79) | (36,28) | (67,55) | *c* | *a* | $P_{win}[c,b]$ is updated from (59,79) to (36,28); *pred*[*c*,*b*] is updated from *c* to *a*. |
| 4 | *a* | *c* | *d* | (45,29) | (36,28) | (40,50) | *c* | *a* | |
| 5 | *a* | *d* | *b* | (72,58) | (50,40) | (67,55) | *d* | *a* | |
| 6 | *a* | *d* | *c* | (29,45) | (50,40) | (28,36) | *d* | *a* | $P_{win}[d,c]$ is updated from (29,45) to (28,36); *pred*[*d*,*c*] is updated from *d* to *a*. |
| 7 | *b* | *a* | *c* | (28,36) | (67,55) | (79,59) | *a* | *b* | $P_{win}[a,c]$ is updated from (28,36) to (67,55); *pred*[*a*,*c*] is updated from *a* to *b*. |
| 8 | *b* | *a* | *d* | (40,50) | (67,55) | (55,67) | *a* | *a* | |
| 9 | *b* | *c* | *a* | (36,28) | (36,28) | (55,67) | *c* | *b* | |
| 10 | *b* | *c* | *d* | (45,29) | (36,28) | (55,67) | *c* | *a* | |
| 11 | *b* | *d* | *a* | (50,40) | (72,58) | (55,67) | *d* | *b* | |
| 12 | *b* | *d* | *c* | (28,36) | (72,58) | (79,59) | *a* | *b* | $P_{win}[d,c]$ is updated from (28,36) to (72,58); *pred*[*d*,*c*] is updated from *a* to *b*. |
| 13 | *c* | *a* | *b* | (67,55) | (67,55) | (36,28) | *a* | *a* | |
| 14 | *c* | *a* | *d* | (40,50) | (67,55) | (45,29) | *a* | *c* | $P_{win}[a,d]$ is updated from (40,50) to (45,29); *pred*[*a*,*d*] is updated from *a* to *c*. |
| 15 | *c* | *b* | *a* | (55,67) | (79,59) | (36,28) | *b* | *c* | $P_{win}[b,a]$ is updated from (55,67) to (36,28); *pred*[*b*,*a*] is updated from *b* to *c*. |
| 16 | *c* | *b* | *d* | (55,67) | (79,59) | (45,29) | *a* | *c* | $P_{win}[b,d]$ is updated from (55,67) to (45,29); *pred*[*b*,*d*] is updated from *a* to *c*. |
| 17 | *c* | *d* | *a* | (50,40) | (72,58) | (36,28) | *d* | *c* | |
| 18 | *c* | *d* | *b* | (72,58) | (72,58) | (36,28) | *d* | *a* | |
| 19 | *d* | *a* | *b* | (67,55) | (45,29) | (72,58) | *a* | *d* | |
| 20 | *d* | *a* | *c* | (67,55) | (45,29) | (72,58) | *b* | *b* | |
| 21 | *d* | *b* | *a* | (36,28) | (45,29) | (50,40) | *c* | *d* | $P_{win}[b,a]$ is updated from (36,28) to (45,29); *pred*[*b*,*a*] is updated from *c* to *d*. |
| 22 | *d* | *b* | *c* | (79,59) | (45,29) | (72,58) | *b* | *b* | |
| 23 | *d* | *c* | *a* | (36,28) | (45,29) | (50,40) | *c* | *d* | $P_{win}[c,a]$ is updated from (36,28) to (45,29); *pred*[*c*,*a*] is updated from *c* to *d*. |
| 24 | *d* | *c* | *b* | (36,28) | (45,29) | (72,58) | *a* | *d* | $P_{win}[c,b]$ is updated from (36,28) to (45,29); *pred*[*c*,*b*] is updated from *a* to *d*. |

## d) losing votes

We get: $(N[c,a],N[a,c]) \succ_{los} (N[c,d],N[d,c]) \succ_{los} (N[d,a],N[a,d]) \succ_{los} (N[a,b],N[b,a]) \succ_{los} (N[d,b],N[b,d]) \succ_{los} (N[b,c],N[c,b])$.

The pairwise victories are:

$ca$ with an opposition of $N[a,c] = 28$
$cd$ with an opposition of $N[d,c] = 29$
$da$ with an opposition of $N[a,d] = 40$
$ab$ with an opposition of $N[b,a] = 55$
$db$ with an opposition of $N[b,d] = 58$
$bc$ with an opposition of $N[c,b] = 59$

The following table lists the strongest paths, as determined by the Floyd-Warshall algorithm, as defined in section 2.3.1. The critical links of the strongest paths are underlined:

| | ... to *a* | ... to *b* | ... to *c* | ... to *d* | ... to every other alternative |
|---|---|---|---|---|---|
| from *a* ... | --- | 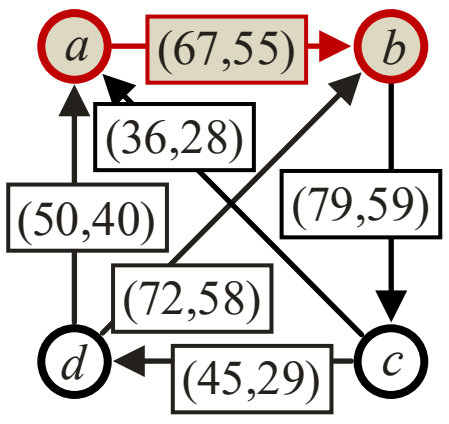<br>*a*, (67,55), *b* | 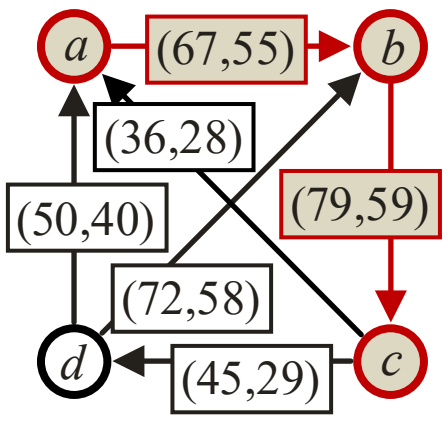<br>*a*, (67,55), *b*, (79,59), *c* | 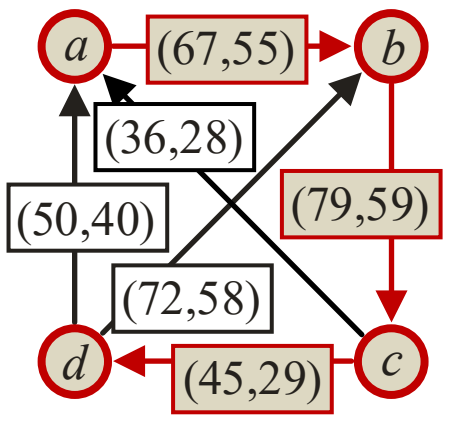<br>*a*, (67,55), *b*, (79,59), *c*, (45,29), *d* | 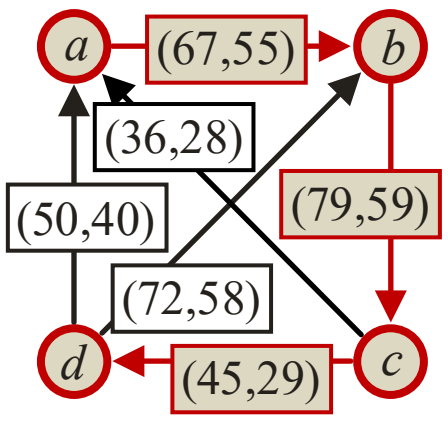 |
| from *b* ... | 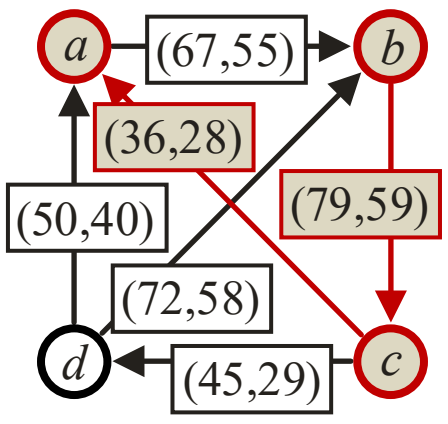<br>*b*, (79,59), *c*, (36,28), *a* | --- | 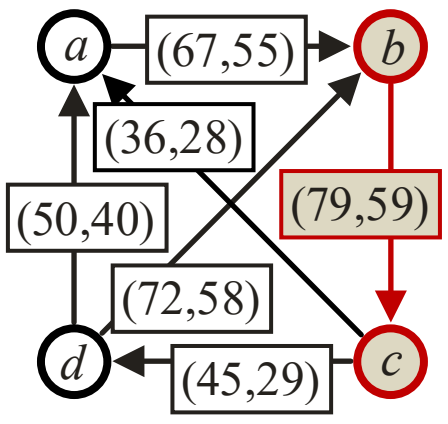<br>*b*, (79,59), *c* | 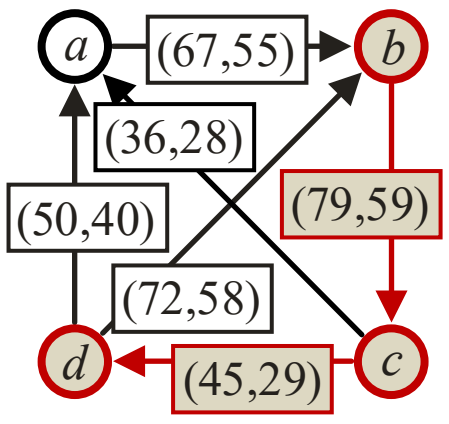<br>*b*, (79,59), *c*, (45,29), *d* | 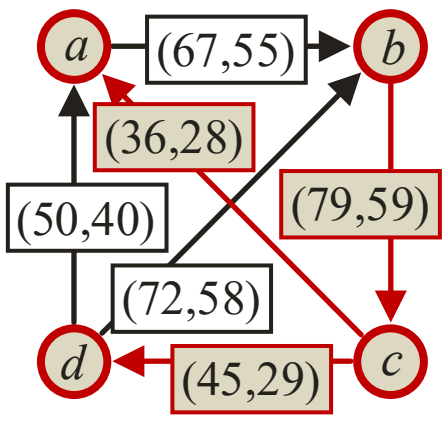 |
| from *c* ... | 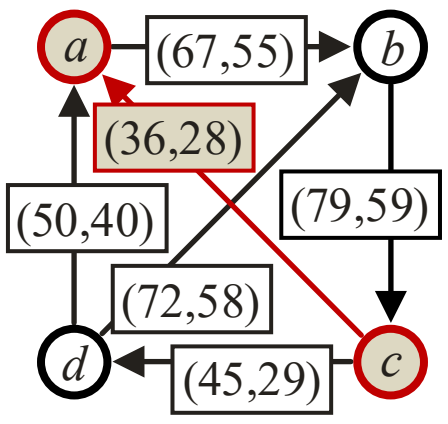<br>*c*, (36,28), *a* | 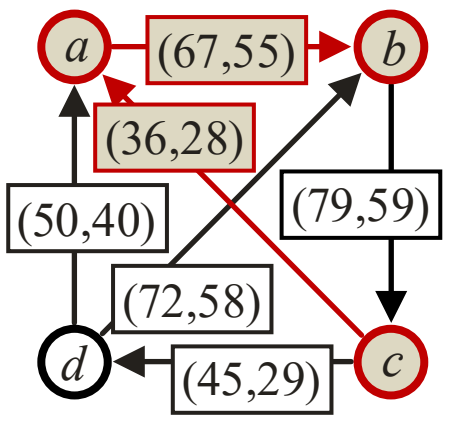<br>*c*, (36,28), *a*, (67,55), *b* | --- | 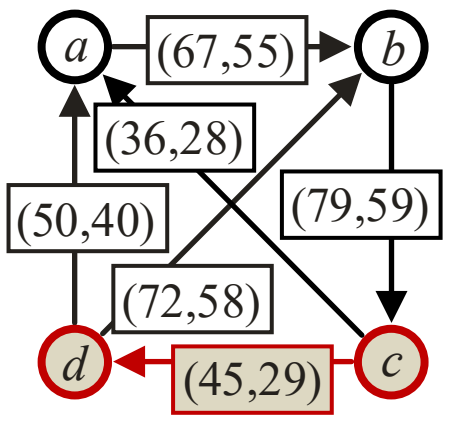<br>*c*, (45,29), *d* | 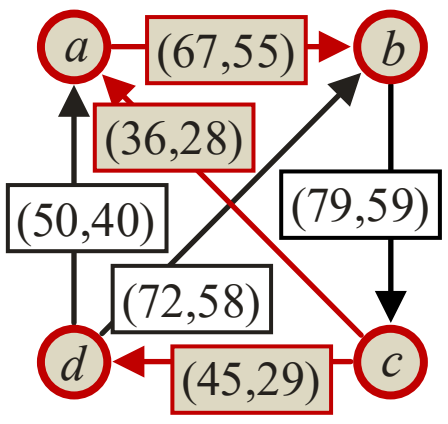 |
| from *d* ... | 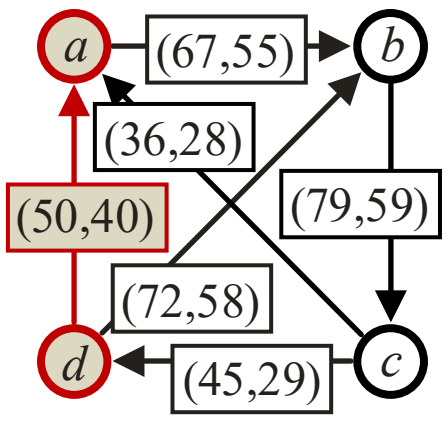<br>*d*, (50,40), *a* | 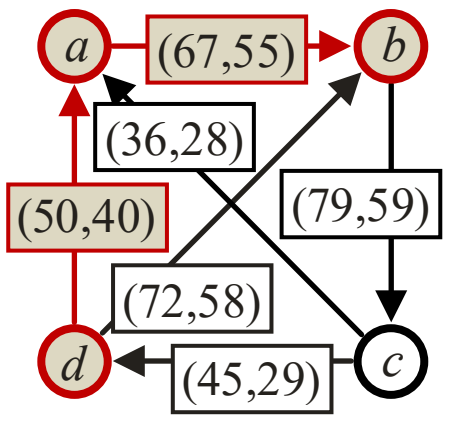<br>*d*, (50,40), *a* (67,55), *b* | 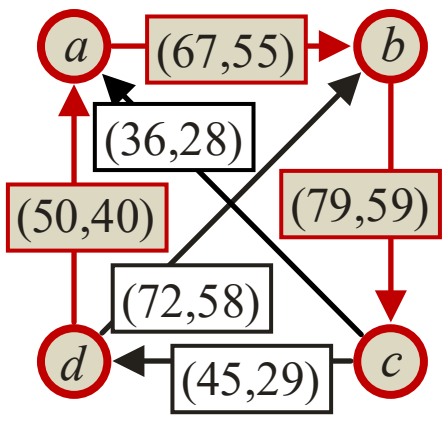<br>*d*, (50,40), *a* (67,55), *b*, (79,59), *c* | --- | 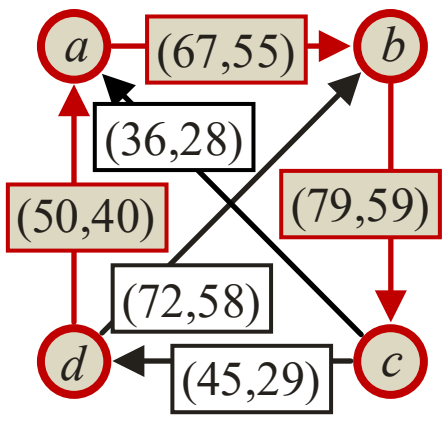 |
| from every other alternative ... | 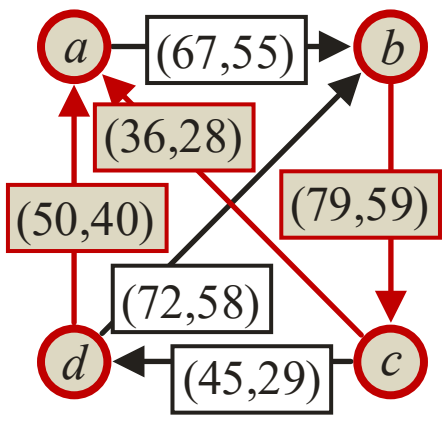 | 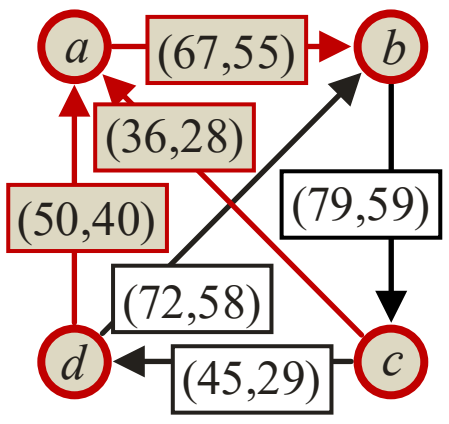 | 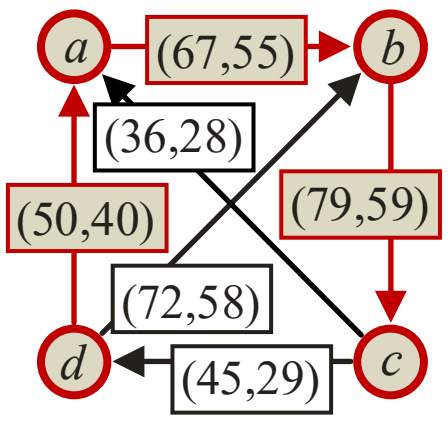 | 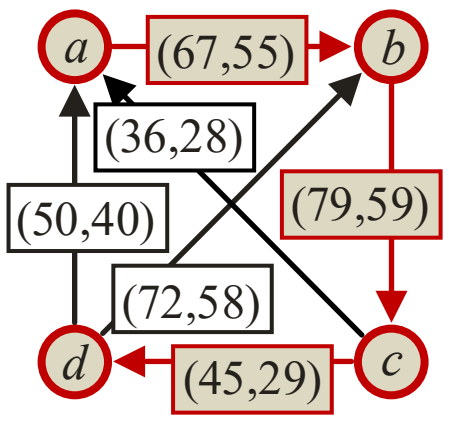 | --- |

The strengths of the strongest paths are:

| | $P_{los}[*,a]$ | $P_{los}[*,b]$ | $P_{los}[*,c]$ | $P_{los}[*,d]$ |
|---|---|---|---|---|
| $P_{los}[a,*]$ | --- | (67,55) | (79,59) | (79,59) |
| $P_{los}[b,*]$ | (79,59) | --- | (79,59) | (79,59) |
| $P_{los}[c,*]$ | (36,28) | (67,55) | --- | (45,29) |
| $P_{los}[d,*]$ | (50,40) | (67,55) | (79,59) | --- |

We get $\mathcal{O}_{los} = \{ab, ca, cb, cd, da, db\}$ and $\mathcal{S}_{los} = \{c\}$.

Suppose, the strongest paths are calculated with the Floyd-Warshall algorithm, as defined in section 2.3.1. Then the following table documents the $C \cdot (C–1) \cdot (C–2) = 24$ steps of the Floyd-Warshall algorithm.

We start with

- $P_{los}[i,j] := (N[i,j],N[j,i])$ for all $i \in A$ and $j \in A \setminus \{i\}$.
- $pred[i,j] := i$ for all $i \in A$ and $j \in A \setminus \{i\}$.

| | $i$ | $j$ | $k$ | $P_{los}[j,k]$ | $P_{los}[j,i]$ | $P_{los}[i,k]$ | $pred[j,k]$ | $pred[i,k]$ | result |
|---|---|---|---|---|---|---|---|---|---|
| 1 | *a* | *b* | *c* | (79,59) | (55,67) | (28,36) | *b* | *a* | |
| 2 | *a* | *b* | *d* | (58,72) | (55,67) | (40,50) | *b* | *a* | |
| 3 | *a* | *c* | *b* | (59,79) | (36,28) | (67,55) | *c* | *a* | $P_{los}[c,b]$ is updated from (59,79) to (67,55); *pred*[*c*,*b*] is updated from *c* to *a*. |
| 4 | *a* | *c* | *d* | (45,29) | (36,28) | (40,50) | *c* | *a* | |
| 5 | *a* | *d* | *b* | (72,58) | (50,40) | (67,55) | *d* | *a* | $P_{los}[d,b]$ is updated from (72,58) to (67,55); *pred*[*d*,*b*] is updated from *d* to *a*. |
| 6 | *a* | *d* | *c* | (29,45) | (50,40) | (28,36) | *d* | *a* | |
| 7 | *b* | *a* | *c* | (28,36) | (67,55) | (79,59) | *a* | *b* | $P_{los}[a,c]$ is updated from (28,36) to (79,59); *pred*[*a*,*c*] is updated from *a* to *b*. |
| 8 | *b* | *a* | *d* | (40,50) | (67,55) | (58,72) | *a* | *b* | $P_{los}[a,d]$ is updated from (40,50) to (58,72); *pred*[*a*,*d*] is updated from *a* to *b*. |
| 9 | *b* | *c* | *a* | (36,28) | (67,55) | (55,67) | *c* | *b* | |
| 10 | *b* | *c* | *d* | (45,29) | (67,55) | (58,72) | *c* | *b* | |
| 11 | *b* | *d* | *a* | (50,40) | (67,55) | (55,67) | *d* | *b* | |
| 12 | *b* | *d* | *c* | (29,45) | (67,55) | (79,59) | *d* | *b* | $P_{los}[d,c]$ is updated from (29,45) to (79,59); *pred*[*d*,*c*] is updated from *d* to *b*. |
| 13 | *c* | *a* | *b* | (67,55) | (79,59) | (67,55) | *a* | *a* | |
| 14 | *c* | *a* | *d* | (58,72) | (79,59) | (45,29) | *b* | *c* | $P_{los}[a,d]$ is updated from (58,72) to (79,59); *pred*[*a*,*d*] is updated from *b* to *c*. |
| 15 | *c* | *b* | *a* | (55,67) | (79,59) | (36,28) | *b* | *c* | $P_{los}[b,a]$ is updated from (55,67) to (79,59); *pred*[*b*,*a*] is updated from *b* to *c*. |
| 16 | *c* | *b* | *d* | (58,72) | (79,59) | (45,29) | *b* | *c* | $P_{los}[b,d]$ is updated from (58,72) to (79,59); *pred*[*b*,*d*] is updated from *b* to *c*. |
| 17 | *c* | *d* | *a* | (50,40) | (79,59) | (36,28) | *d* | *c* | |
| 18 | *c* | *d* | *b* | (67,55) | (79,59) | (67,55) | *a* | *a* | |
| 19 | *d* | *a* | *b* | (67,55) | (79,59) | (67,55) | *a* | *a* | |
| 20 | *d* | *a* | *c* | (79,59) | (79,59) | (79,59) | *b* | *b* | |
| 21 | *d* | *b* | *a* | (79,59) | (79,59) | (50,40) | *c* | *d* | |
| 22 | *d* | *b* | *c* | (79,59) | (79,59) | (79,59) | *b* | *b* | |
| 23 | *d* | *c* | *a* | (36,28) | (45,29) | (50,40) | *c* | *d* | |
| 24 | *d* | *c* | *b* | (67,55) | (45,29) | (67,55) | *a* | *a* | |

## 3.11. Example 11

Example 11:

| | | |
|---|---|---|
| 9 | voters | $a \succ_v d \succ_v b \succ_v e \succ_v c$ |
| 6 | voters | $b \succ_v c \succ_v a \succ_v d \succ_v e$ |
| 5 | voters | $b \succ_v c \succ_v d \succ_v e \succ_v a$ |
| 2 | voters | $c \succ_v d \succ_v b \succ_v e \succ_v a$ |
| 6 | voters | $d \succ_v e \succ_v c \succ_v b \succ_v a$ |
| 14 | voters | $e \succ_v a \succ_v c \succ_v b \succ_v d$ |
| 2 | voters | $e \succ_v c \succ_v a \succ_v b \succ_v d$ |
| 1 | voter | $e \succ_v d \succ_v a \succ_v c \succ_v b$ |

The pairwise matrix *N* looks as follows:

| | *N*[*,*a*] | *N*[*,*b*] | *N*[*,*c*] | *N*[*,*d*] | *N*[*,*e*] |
|---|---|---|---|---|---|
| *N*[*a*,*] | --- | 26 | 24 | 31 | 15 |
| *N*[*b*,*] | 19 | --- | 20 | 27 | 22 |
| *N*[*c*,*] | 21 | 25 | --- | 29 | 13 |
| *N*[*d*,*] | 14 | 18 | 16 | --- | 28 |
| *N*[*e*,*] | 30 | 23 | 32 | 17 | --- |

The corresponding digraph looks as follows:

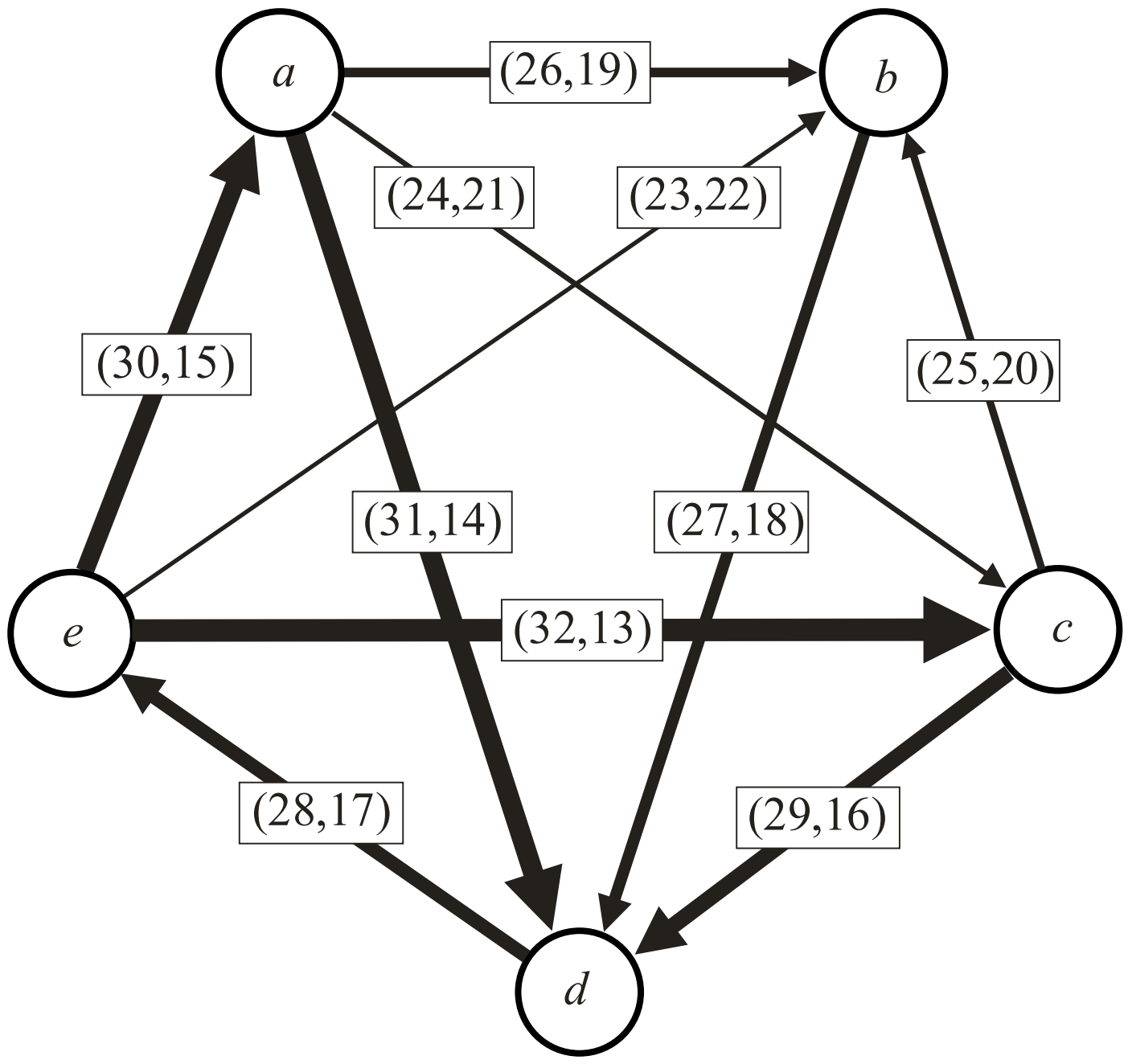

The following table lists the strongest paths, as determined by the Floyd-Warshall algorithm, as defined in section 2.3.1. The critical links of the strongest paths are underlined:

| | ... to *a* | ... to *b* | ... to *c* | ... to *d* | ... to *e* |
|---|---|---|---|---|---|
| from *a* ... | --- | *a*, (26,19), *b* | *a*, (31,14), *d*, (28,17), *e*, (32,13), *c* | *a*, (31,14), *d* | *a*, (31,14), *d*, (28,17), *e* |
| from *b* ... | *b*, (27,18), *d*, (28,17), *e*, (30,15), *a* | --- | *b*, (27,18), *d*, (28,17), *e*, (32,13), *c* | *b*, (27,18), *d* | *b*, (27,18), *d*, (28,17), *e* |
| from *c* ... | *c*, (29,16), *d*, (28,17), *e*, (30,15), *a* | *c*, (29,16), *d*, (28,17), *e*, (30,15), *a*, (26,19), *b* | --- | *c*, (29,16), *d* | *c*, (29,16), *d*, (28,17), *e* |
| from *d* ... | *d*, (28,17), *e*, (30,15), *a* | *d*, (28,17), *e*, (30,15), *a*, (26,19), *b* | *d*, (28,17), *e*, (32,13), *c* | --- | *d*, (28,17), *e* |
| from *e* ... | *e*, (30,15), *a* | *e*, (30,15), *a*, (26,19), *b* | *e*, (32,13), *c* | *e*, (30,15), *a*, (31,14), *d* | --- |

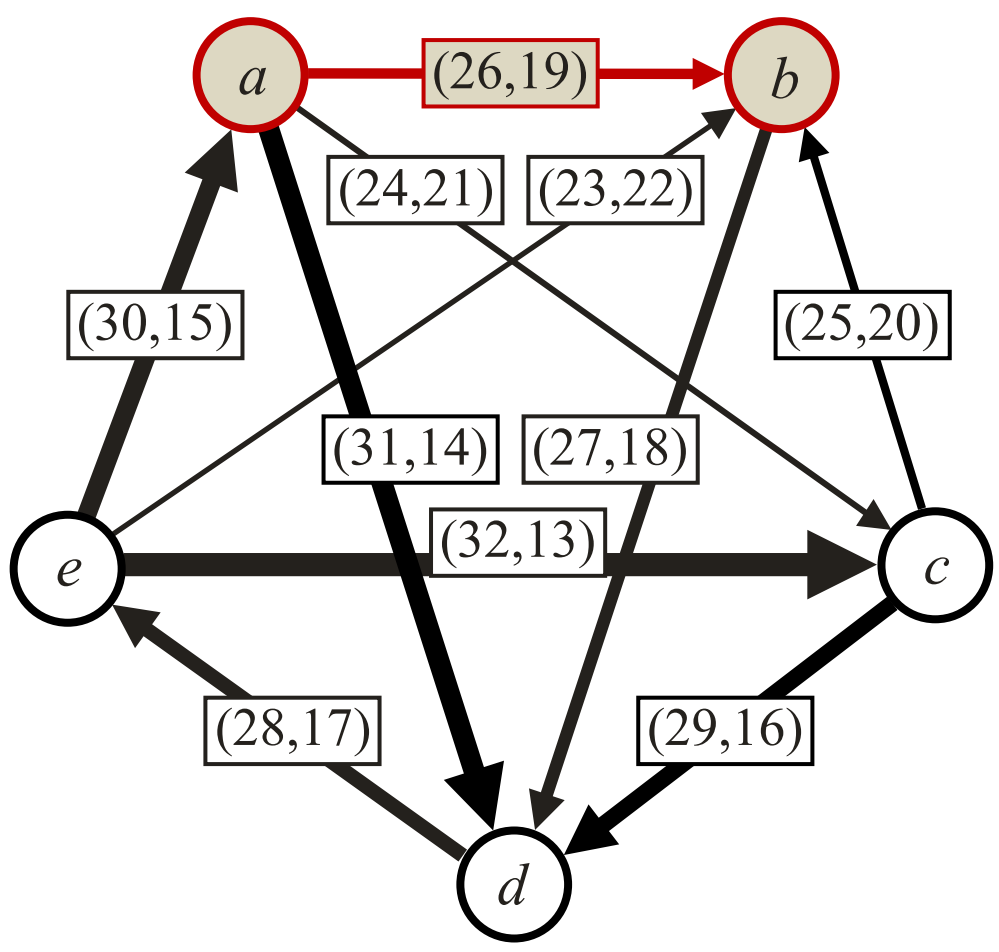

The strongest path from *a* to *b* is:
*a*, (26,19), *b*

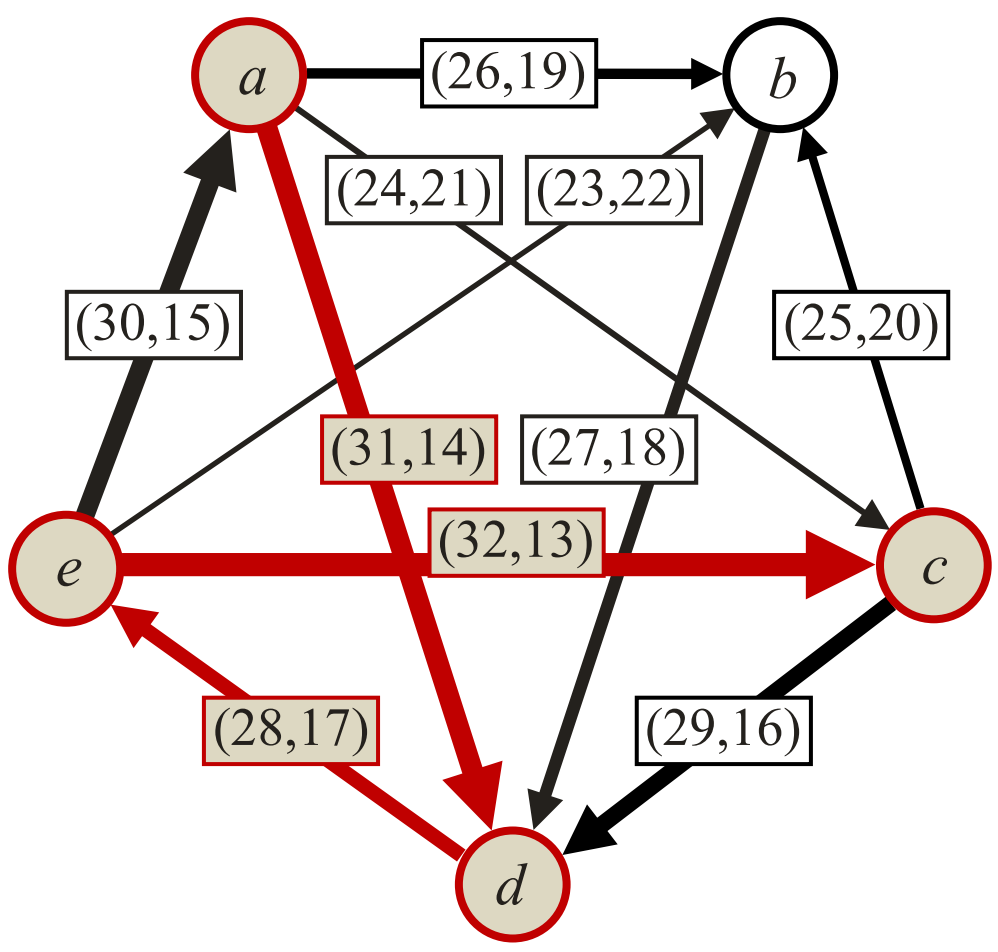

The strongest path from *a* to *c* is:
*a*, (31,14), *d*, (28,17), *e*, (32,13), *c*

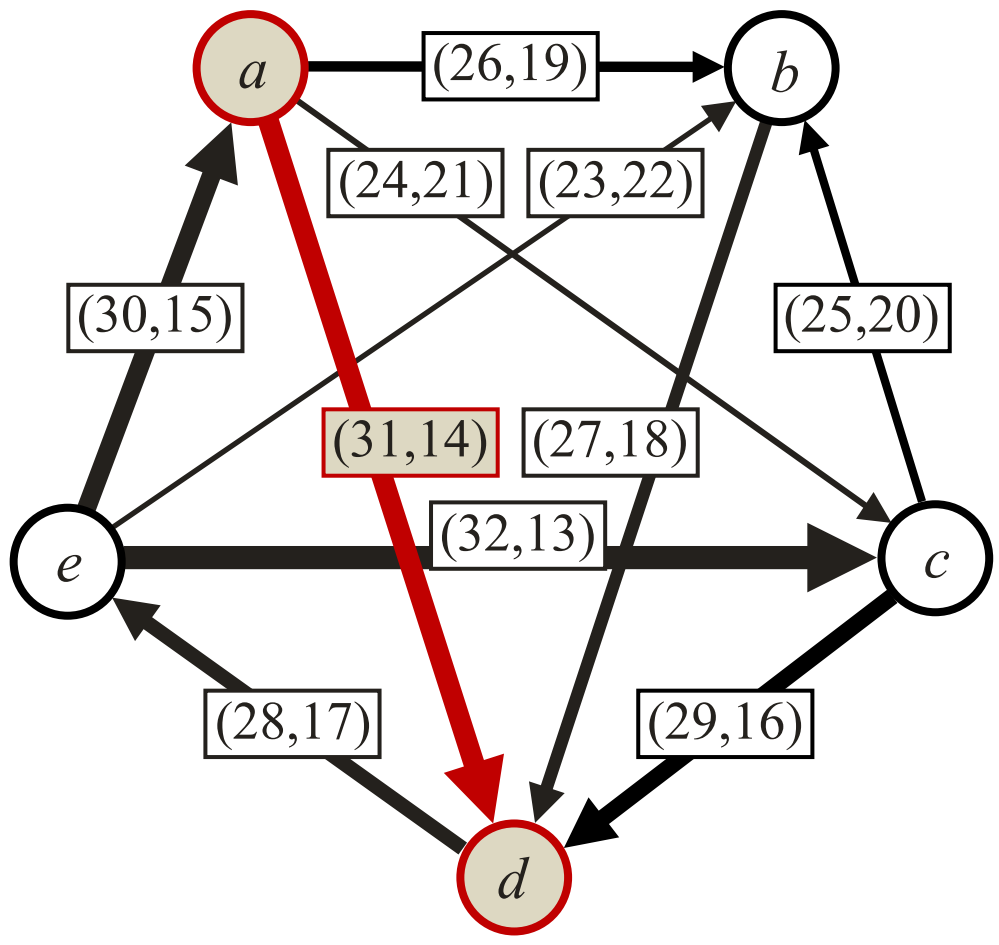

The strongest path from *a* to *d* is:
*a*, (31,14), *d*

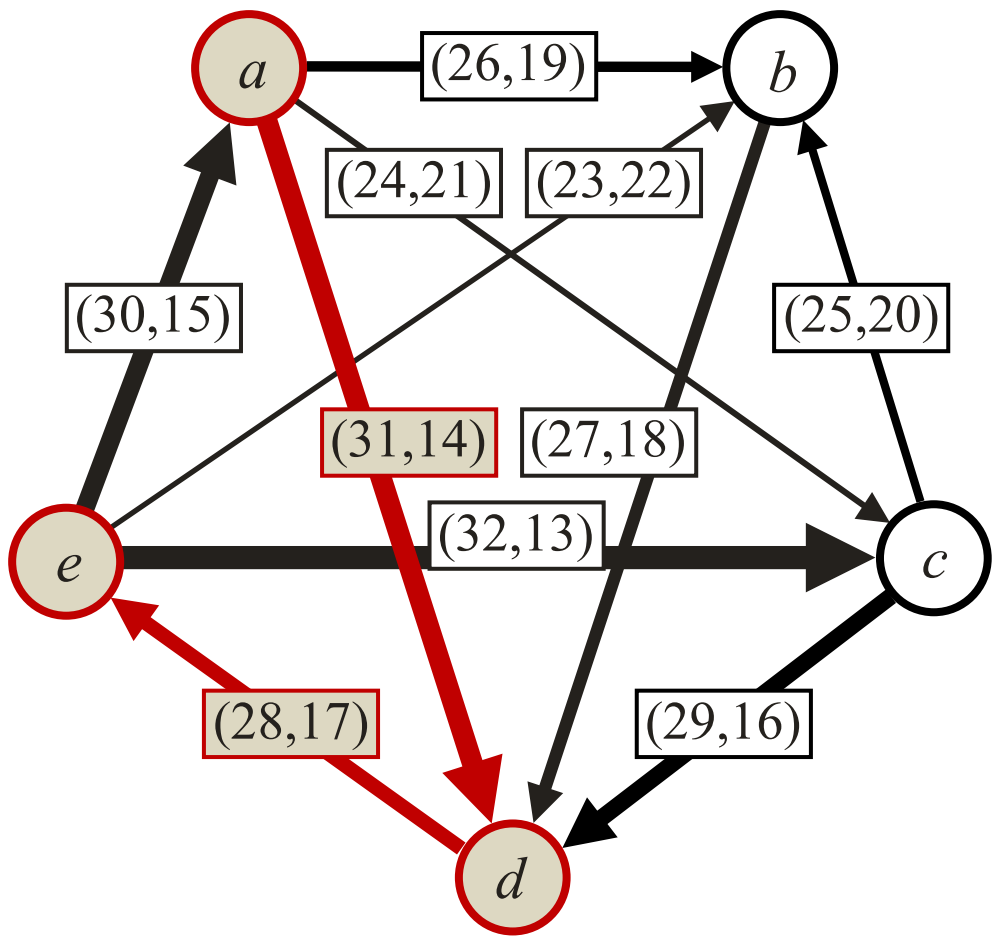

The strongest path from *a* to *e* is:
*a*, (31,14), *d*, (28,17), *e*

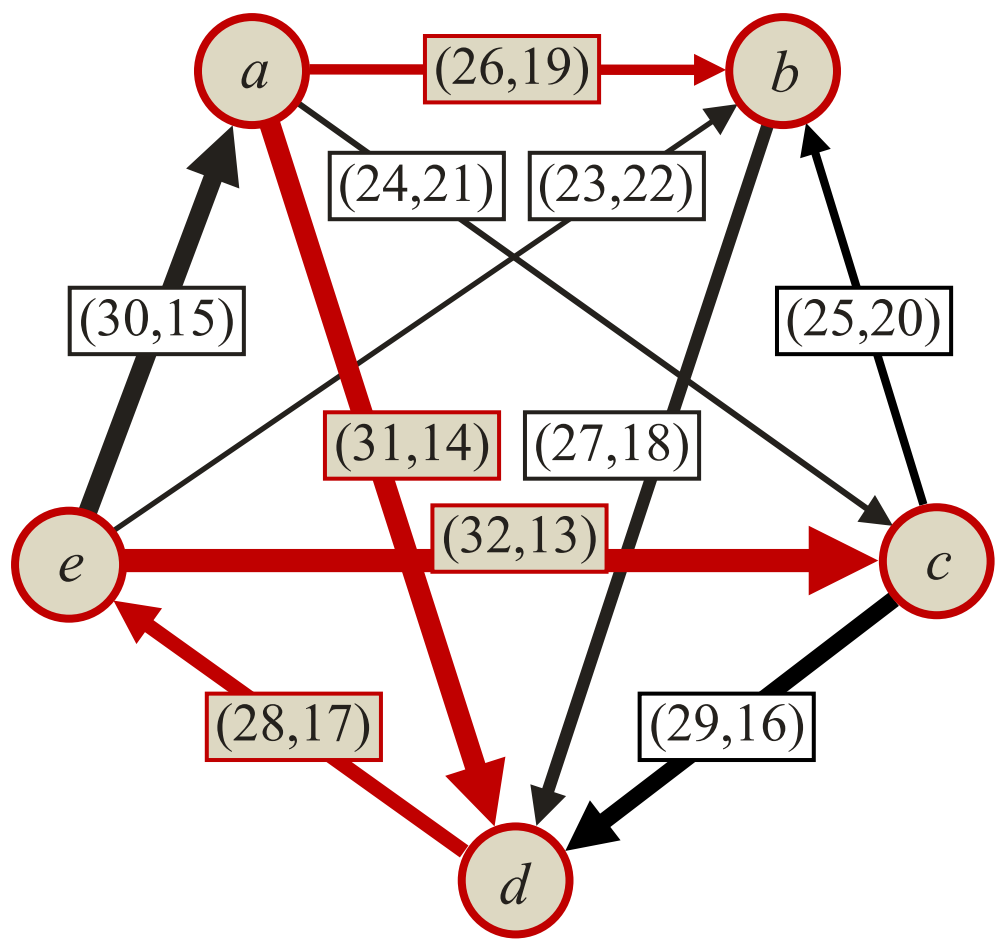

These are the strongest paths
from *a* to every other alternative.

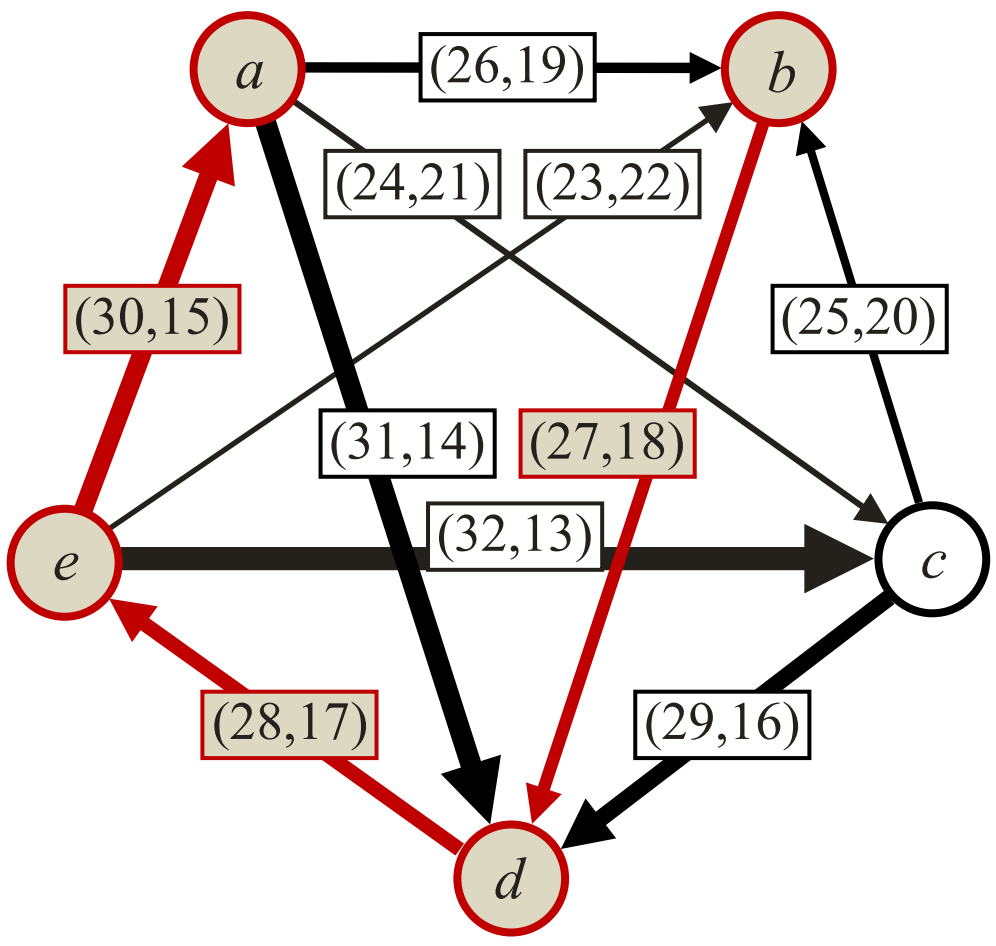

The strongest path from *b* to *a* is:
*b*, (27,18), *d*, (28,17), *e*, (30,15), *a*

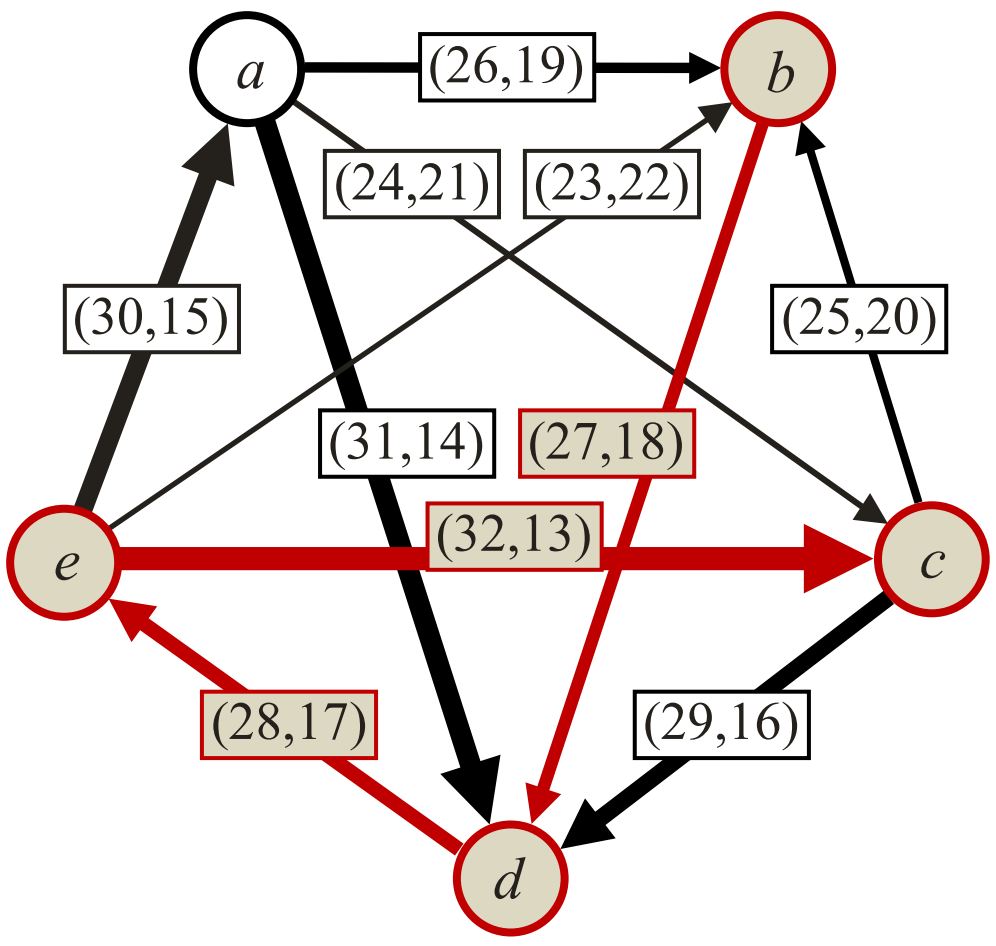

The strongest path from *b* to *c* is:
*b*, (27,18), *d*, (28,17), *e*, (32,13), *c*

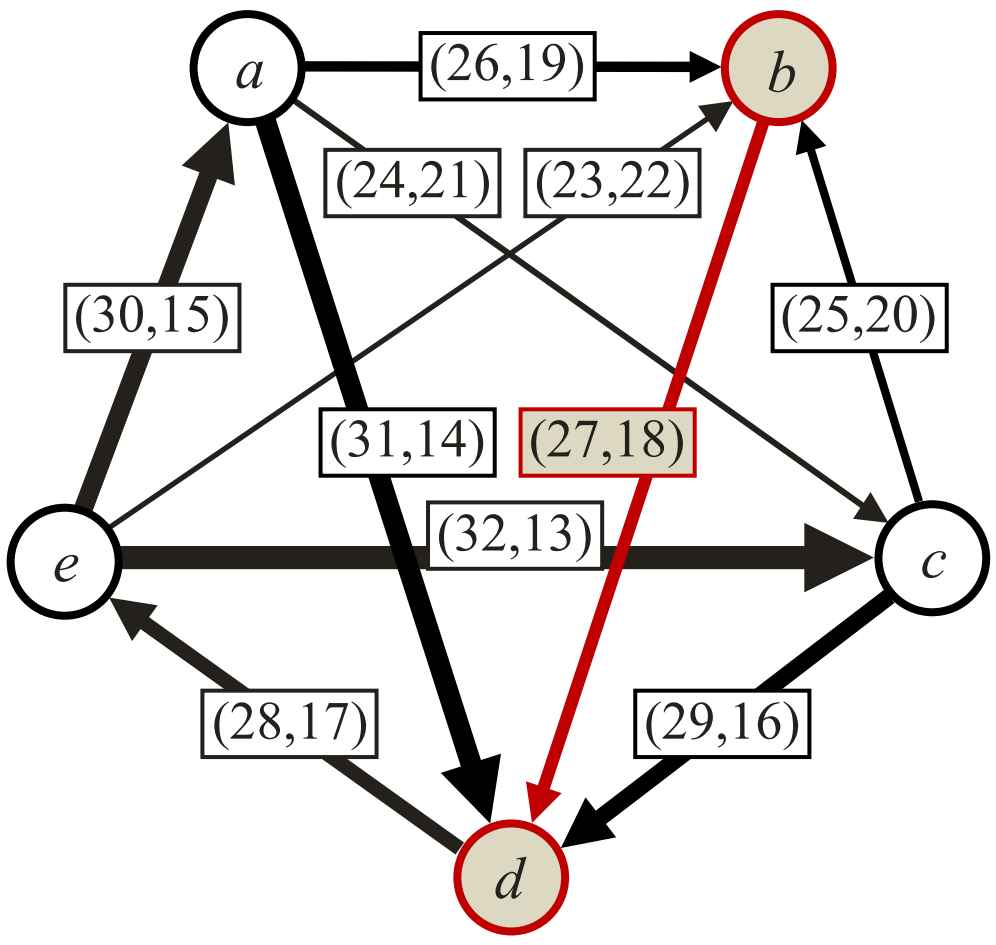

The strongest path from *b* to *d* is:
*b*, (27,18), *d*

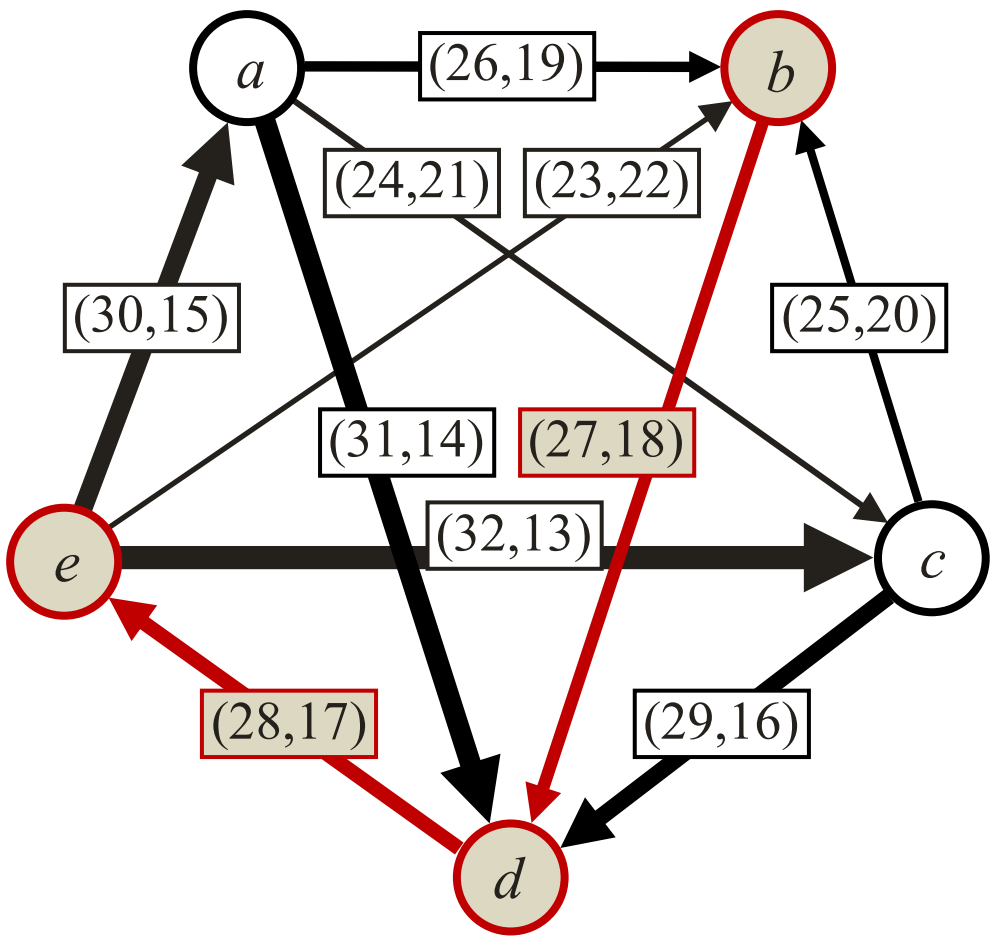

The strongest path from *b* to *e* is:
*b*, (27,18), *d*, (28,17), *e*

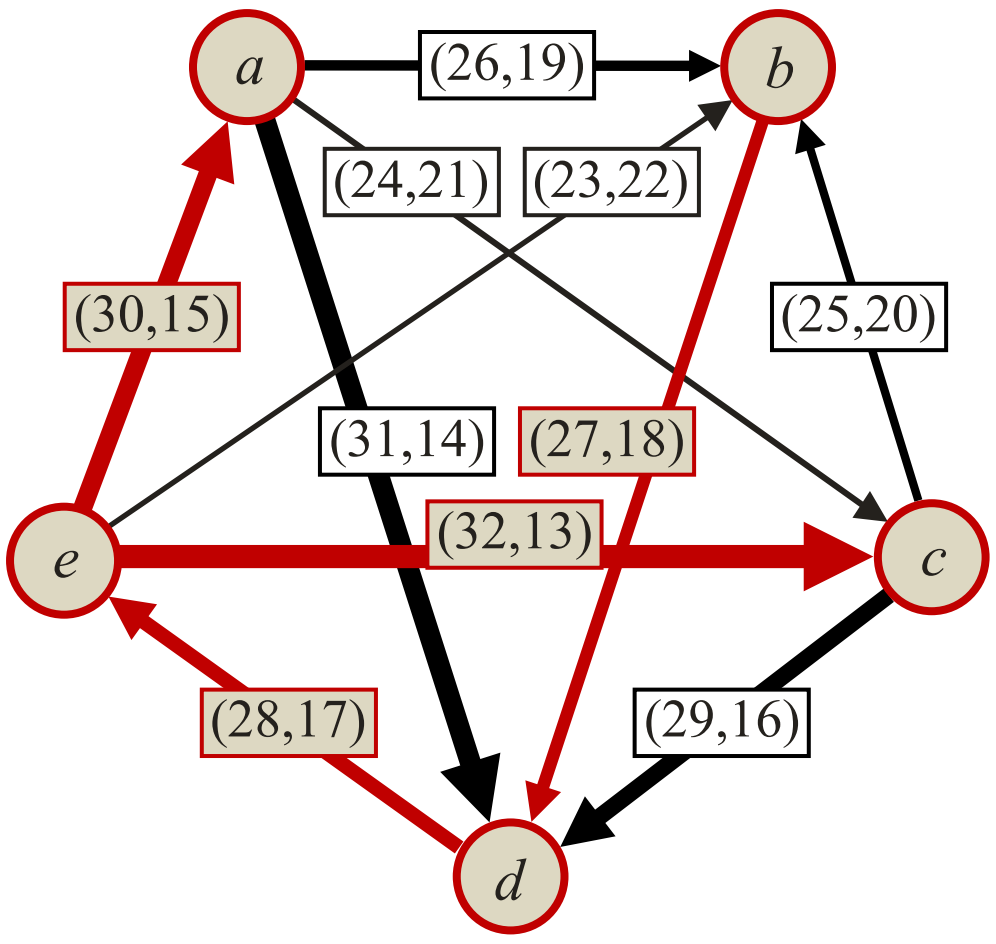

These are the strongest paths
from *b* to every other alternative.

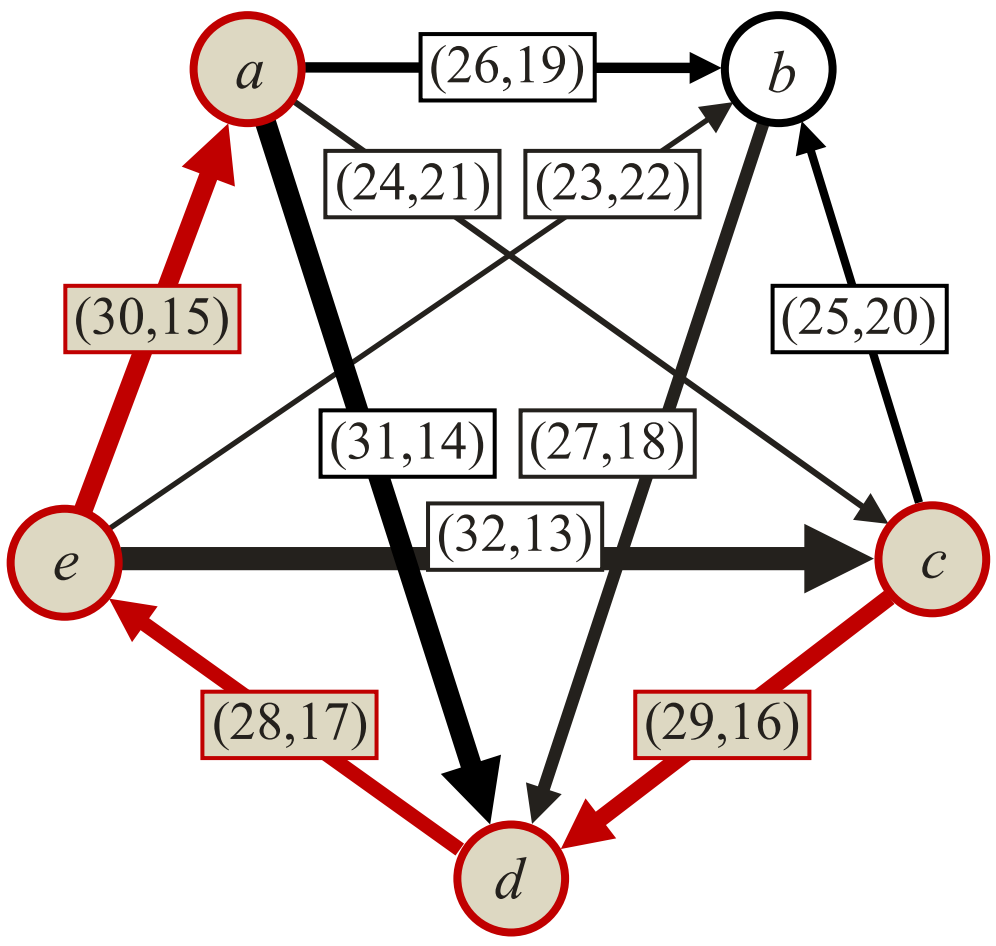

The strongest path from *c* to *a* is:
*c*, (29,16), *d*, (28,17), *e*, (30,15), *a*

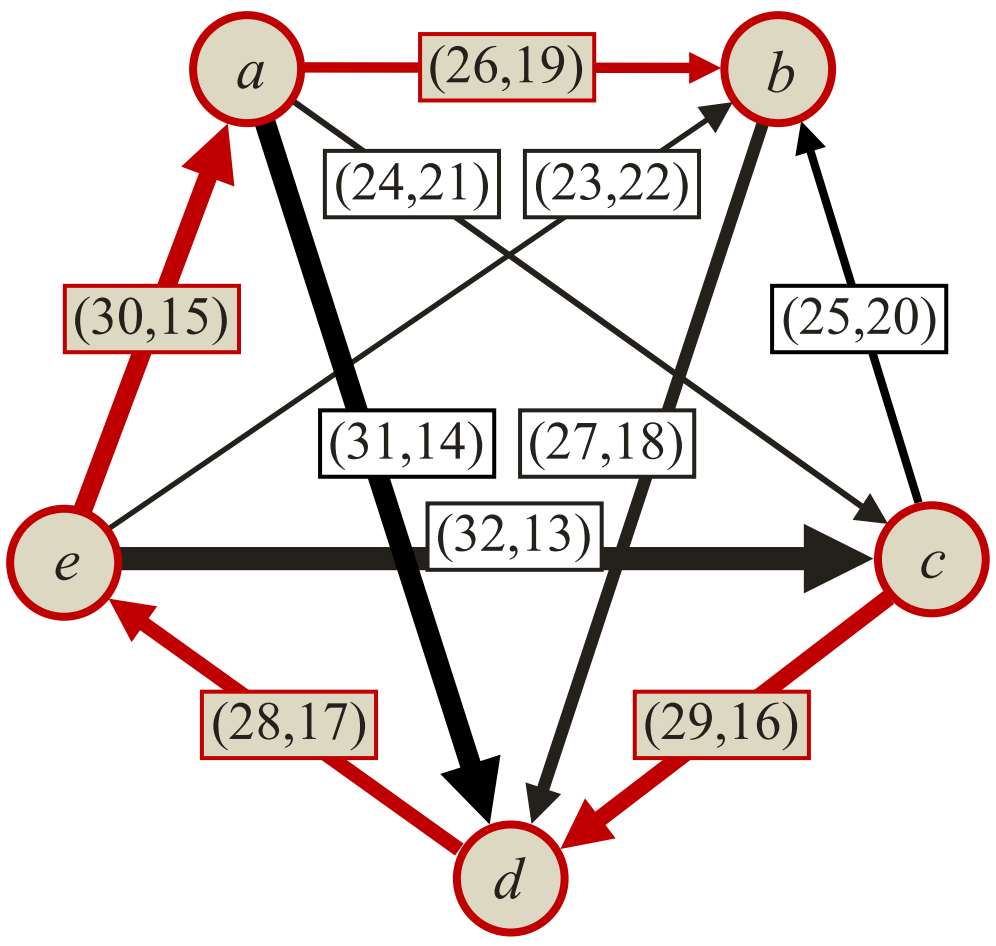

The strongest path from *c* to *b* is:
*c*, (29,16), *d*, (28,17), *e*,
(30,15), *a*, (26,19), *b*

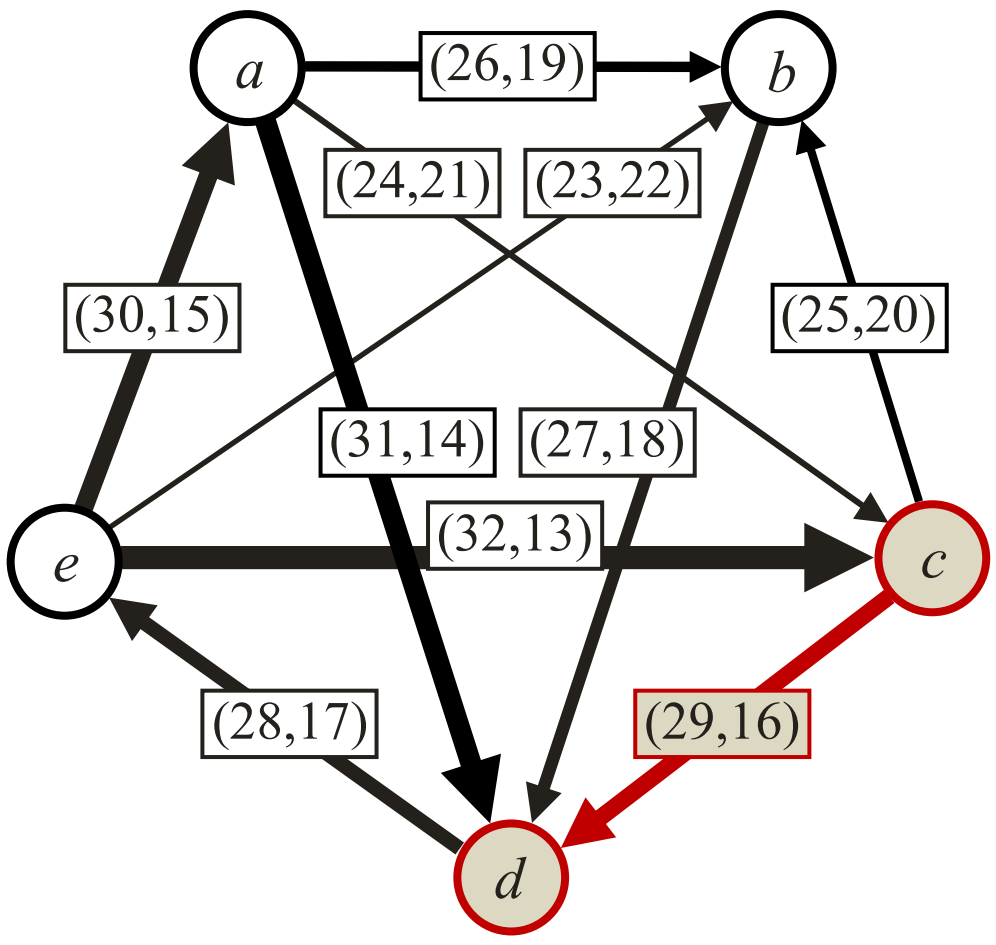

The strongest path from *c* to *d* is:
*c*, (29,16), *d*

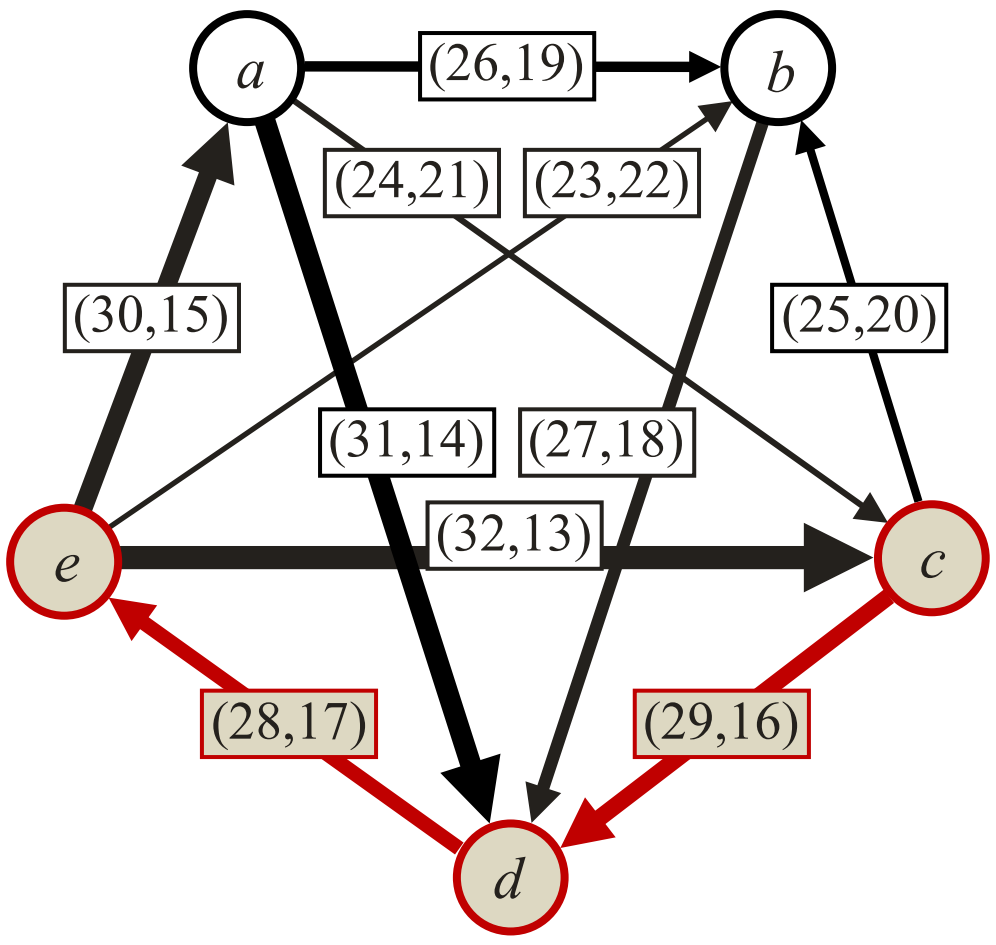

The strongest path from *c* to *e* is:
*c*, (29,16), *d*, (28,17), *e*

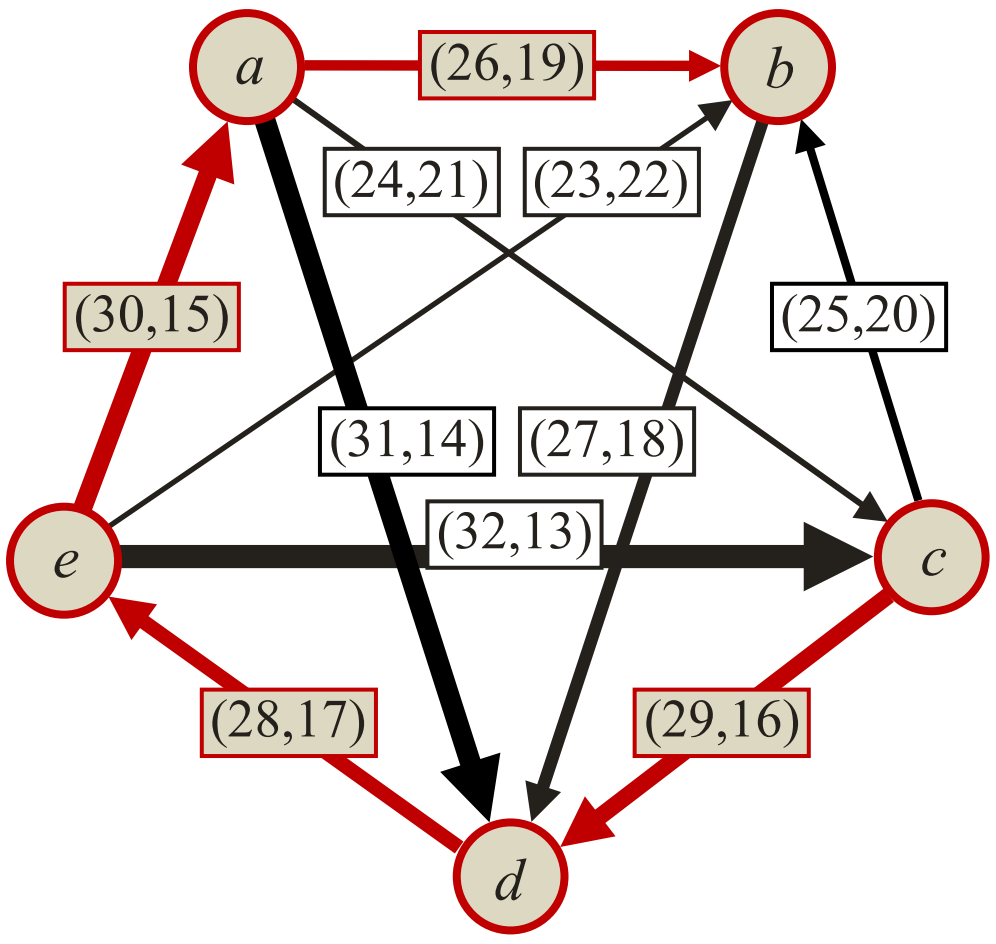

These are the strongest paths
from *c* to every other alternative.

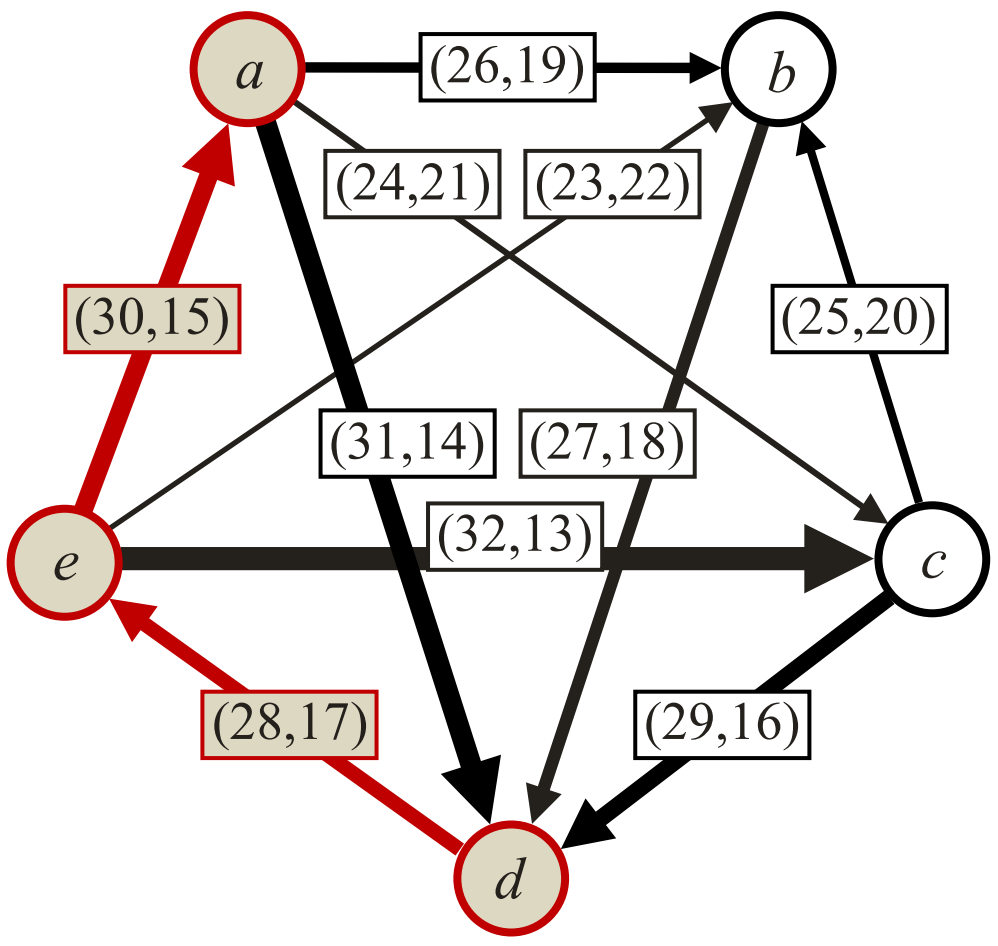

The strongest path from *d* to *a* is:
*d*, (28,17), *e*, (30,15), *a*

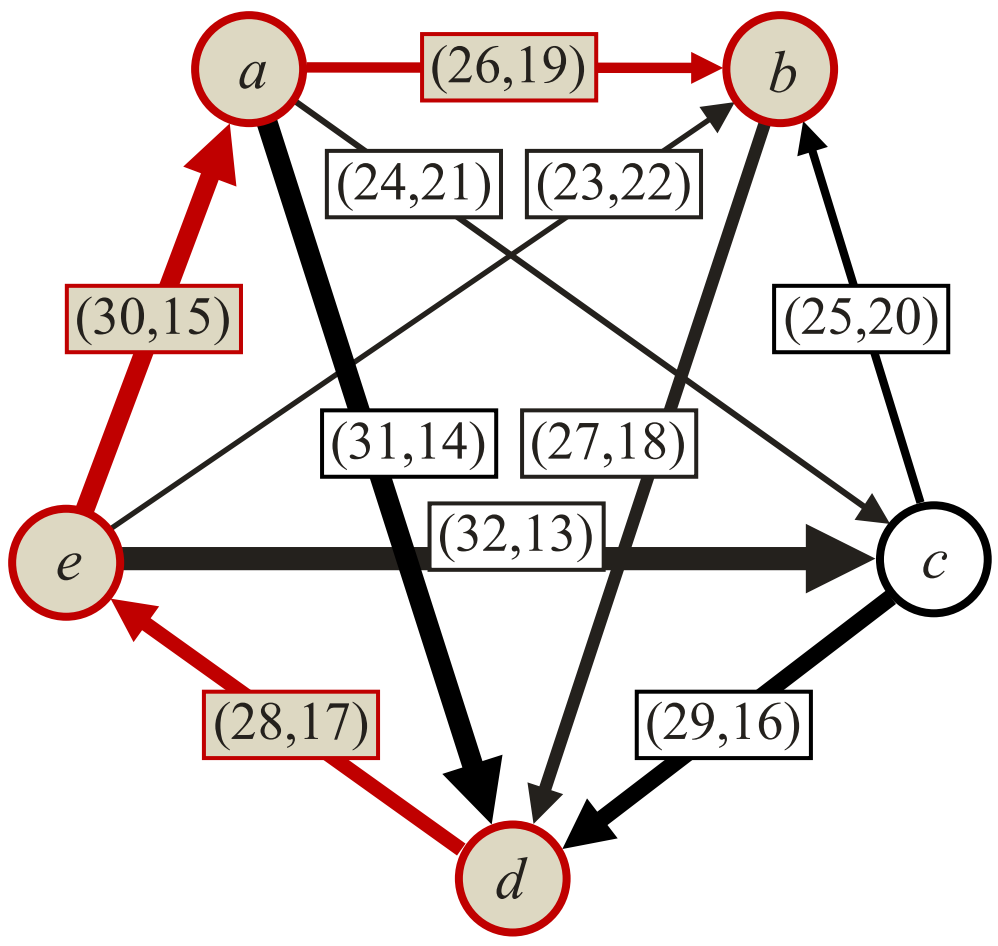

The strongest path from *d* to *b* is:
*d*, (28,17), *e*, (30,15), *a*, (26,19), *b*

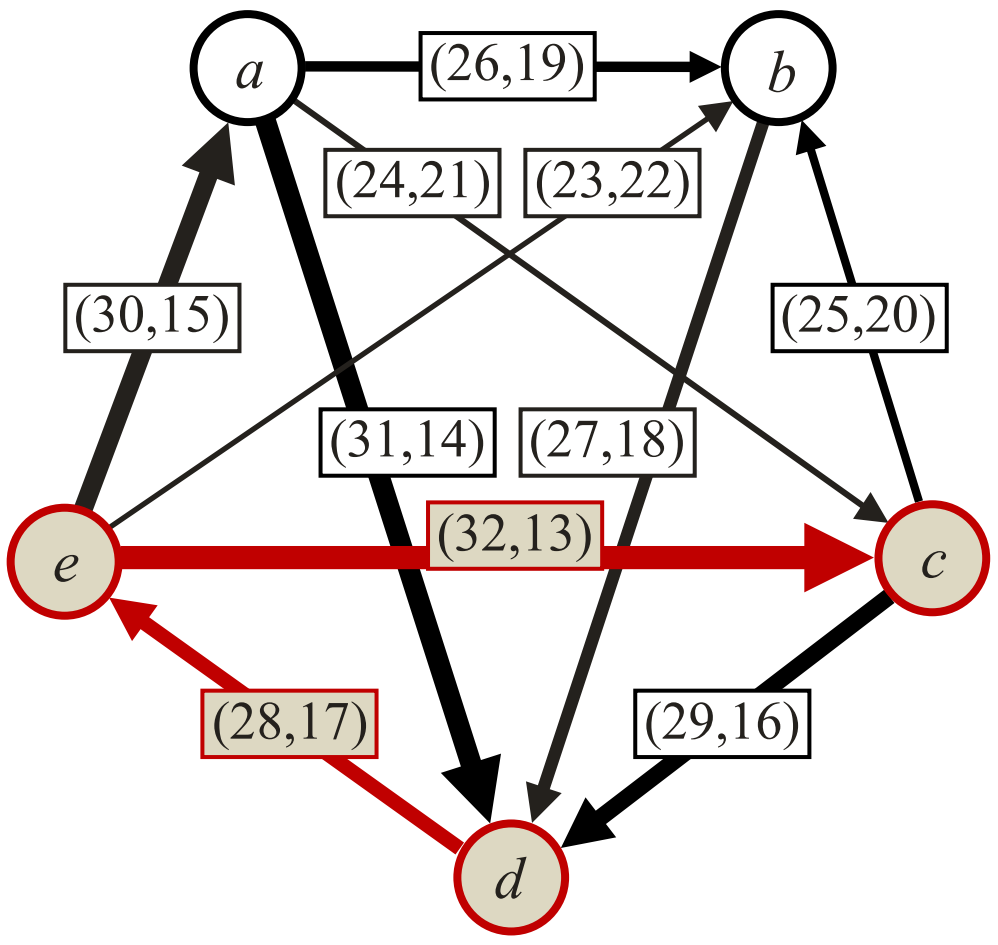

The strongest path from *d* to *c* is:
*d*, (28,17), *e*, (32,13), *c*

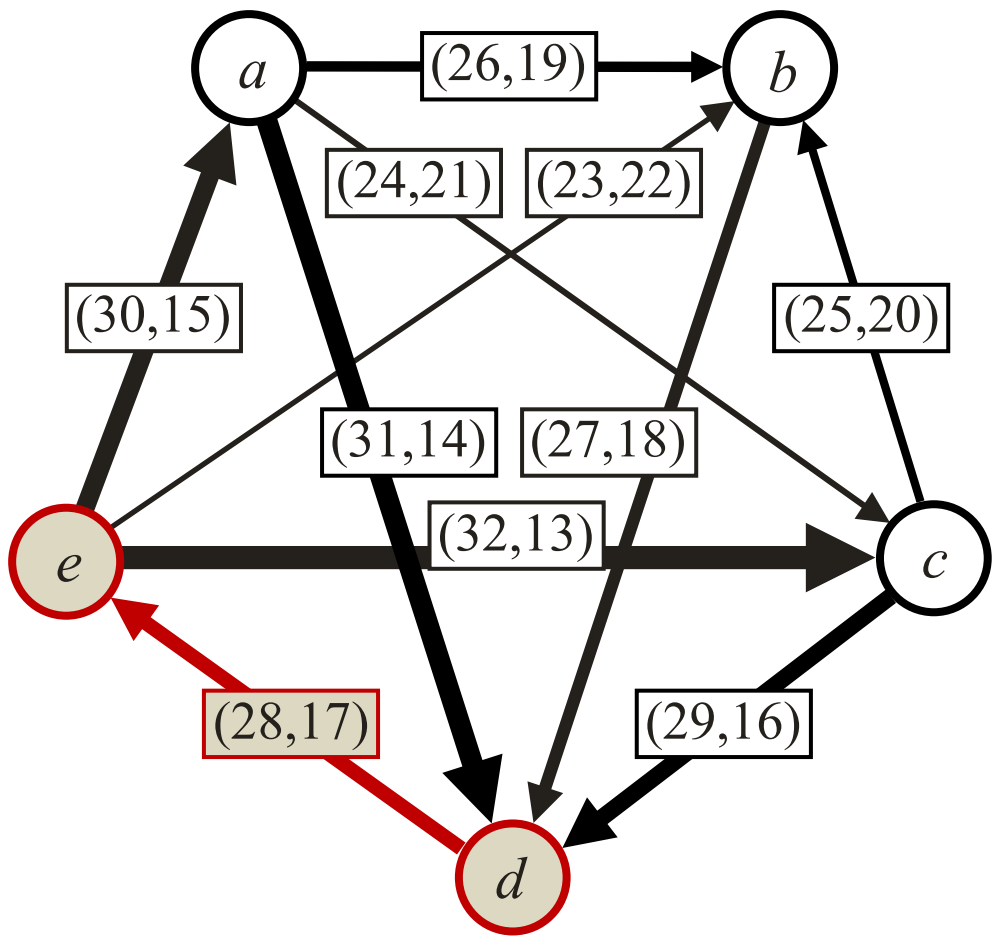

The strongest path from *d* to *e* is:
*d*, (28,17), *e*

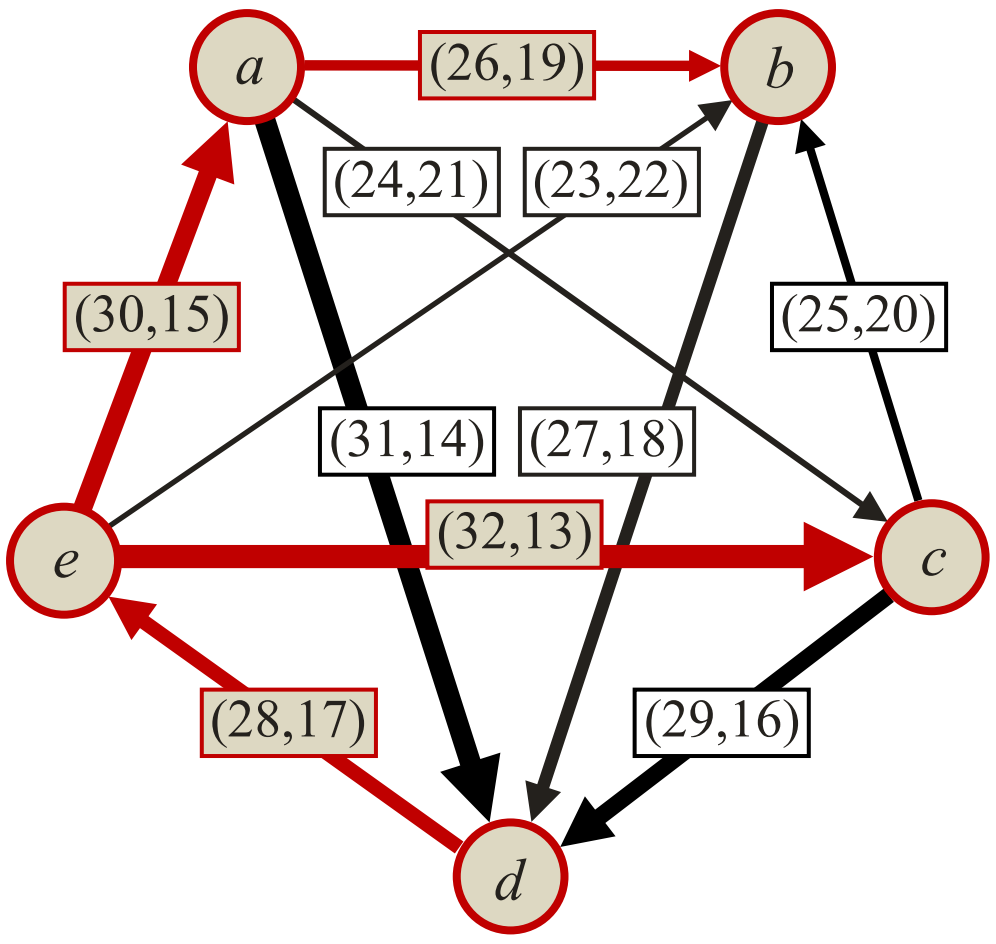

These are the strongest paths
from *d* to every other alternative.

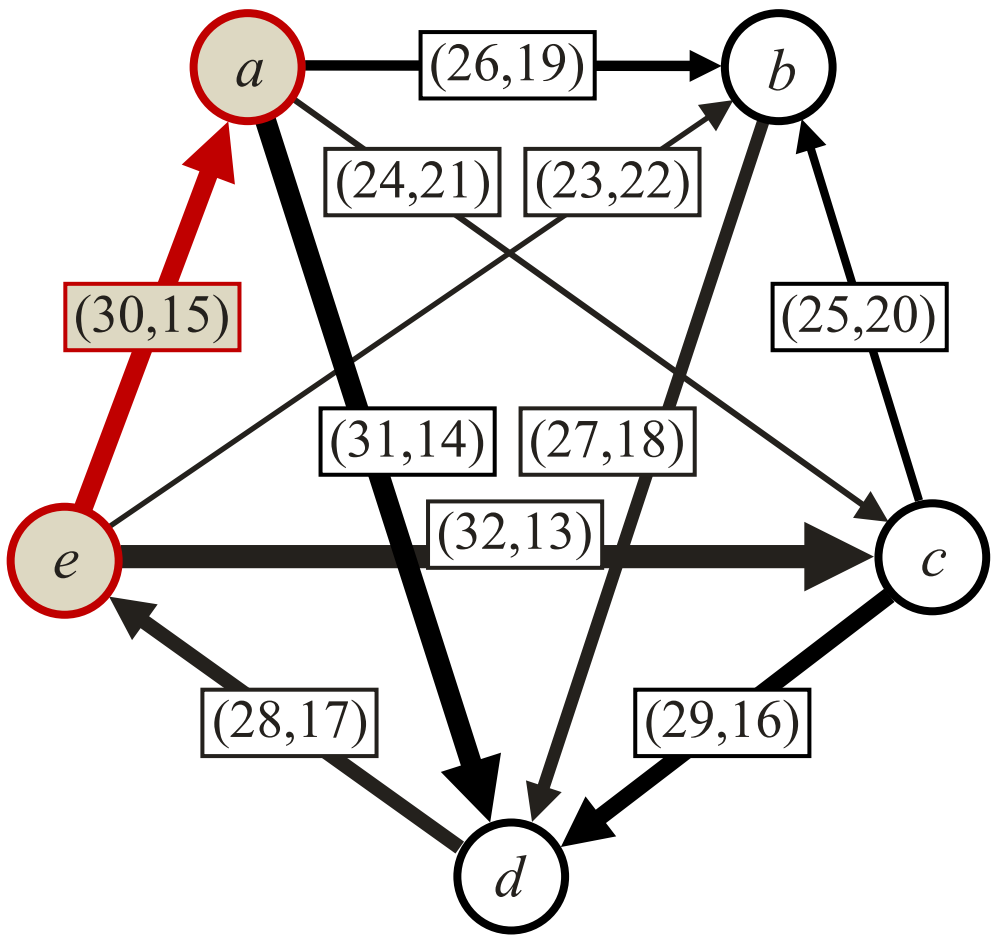

The strongest path from *e* to *a* is:
*e*, (30,15), *a*

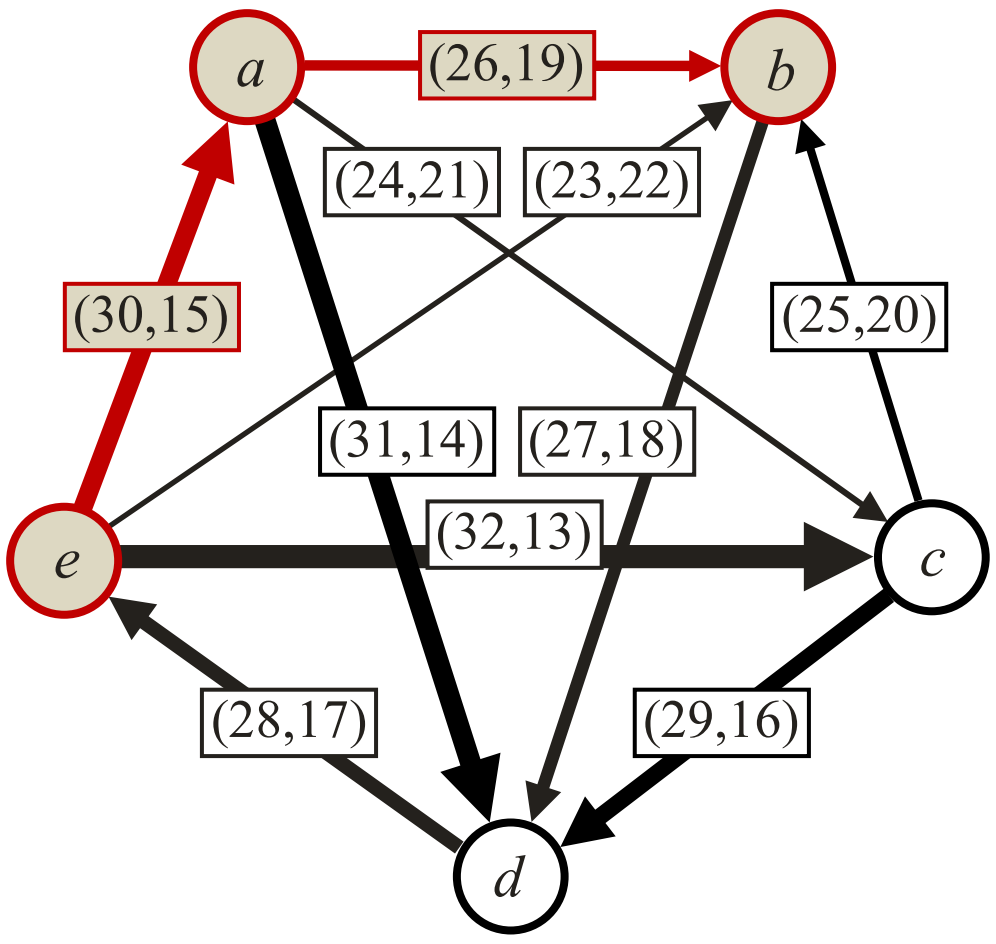

The strongest path from *e* to *b* is:
*e*, (30,15), *a*, (26,19), *b*

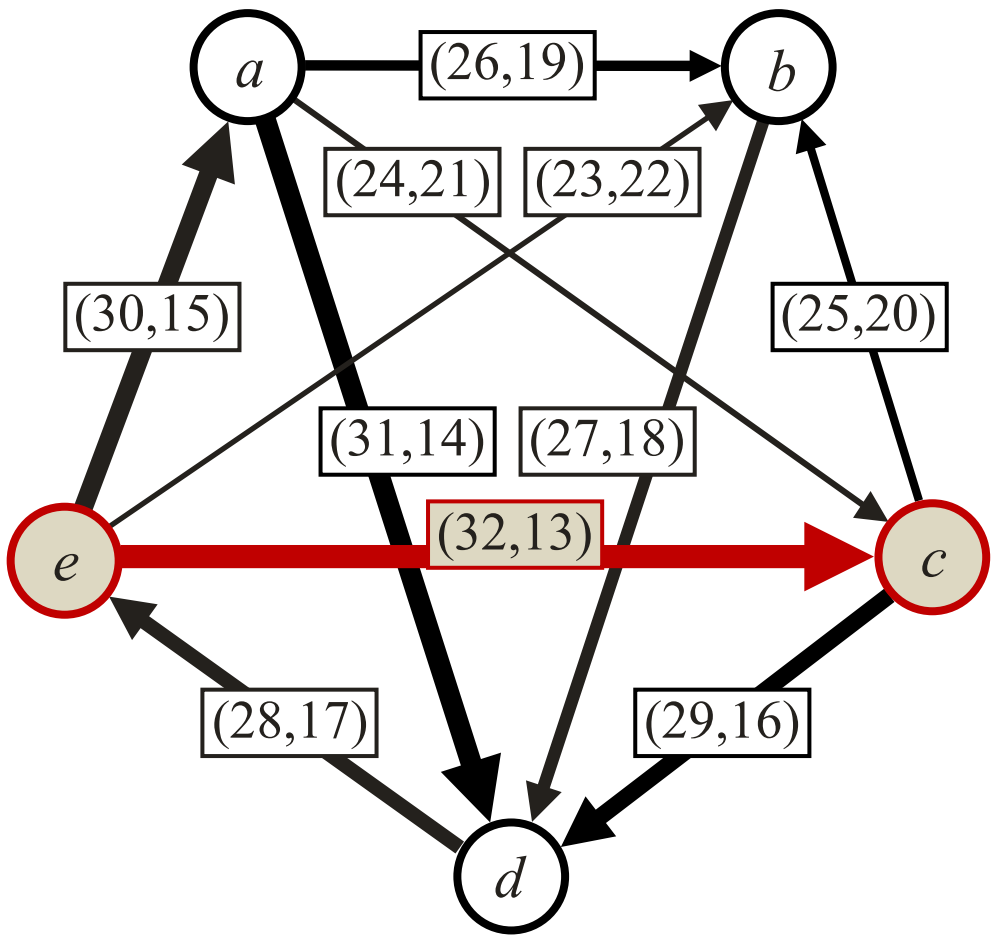

The strongest path from *e* to *c* is:
*e*, (32,13), *c*

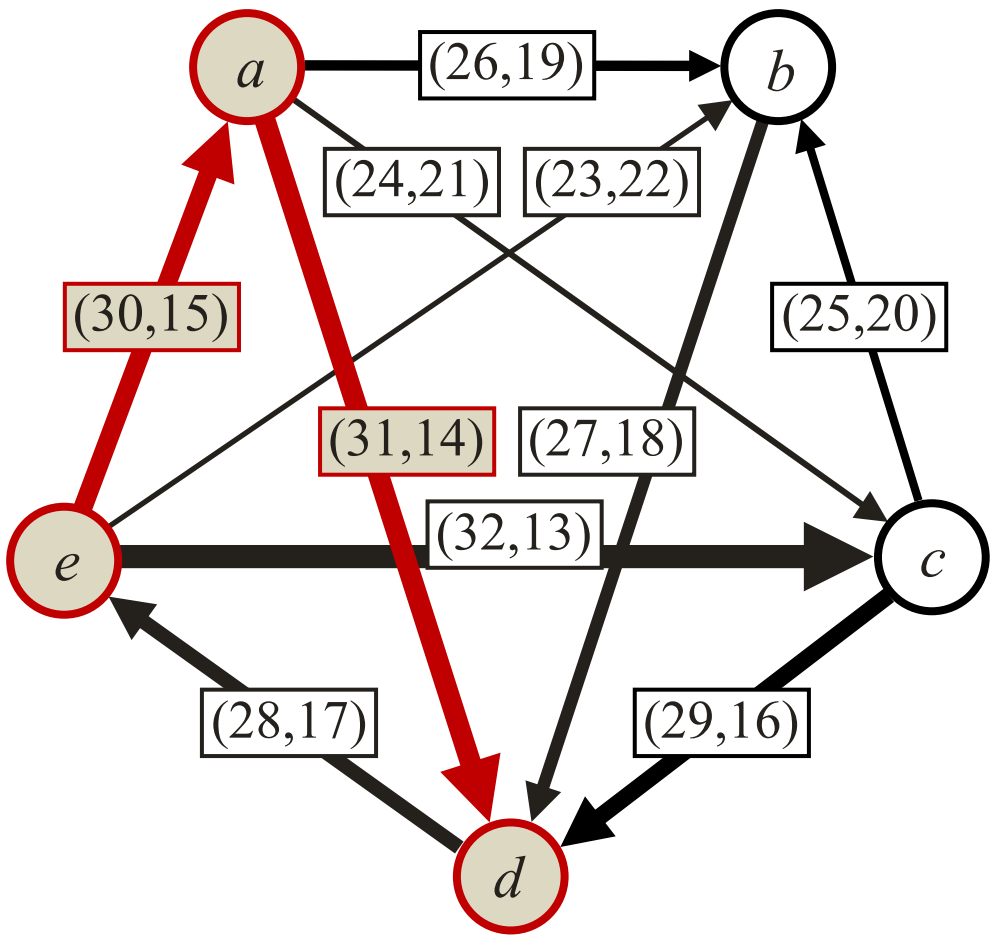

The strongest path from *e* to *d* is:
*e*, (30,15), *a*, (31,14), *d*

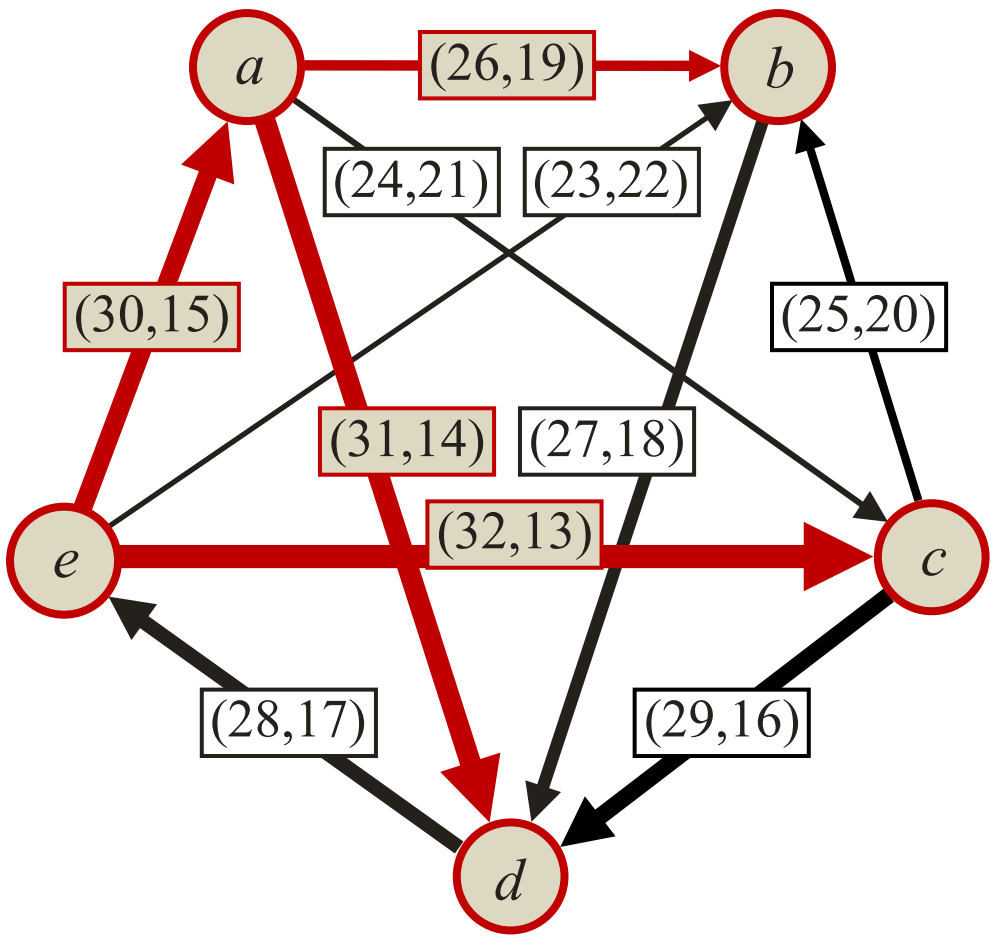

These are the strongest paths
from *e* to every other alternative.

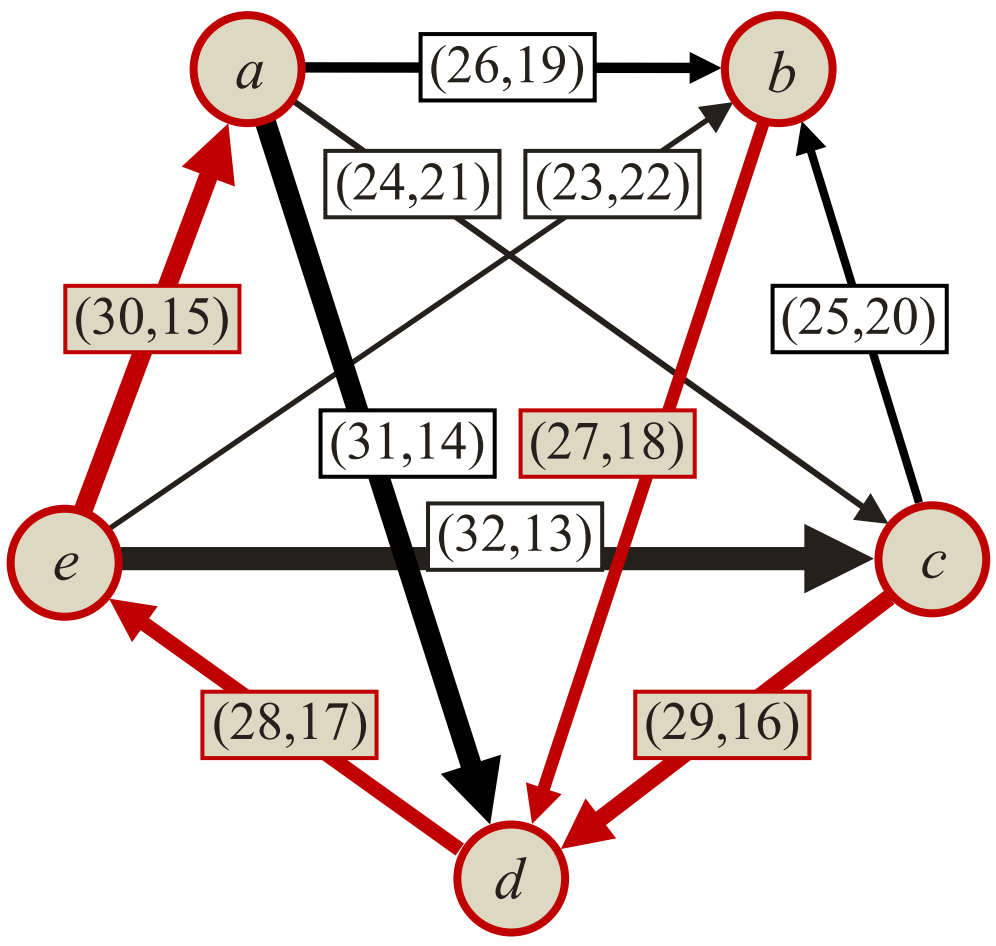

These are the strongest paths
from every other alternative to *a*.

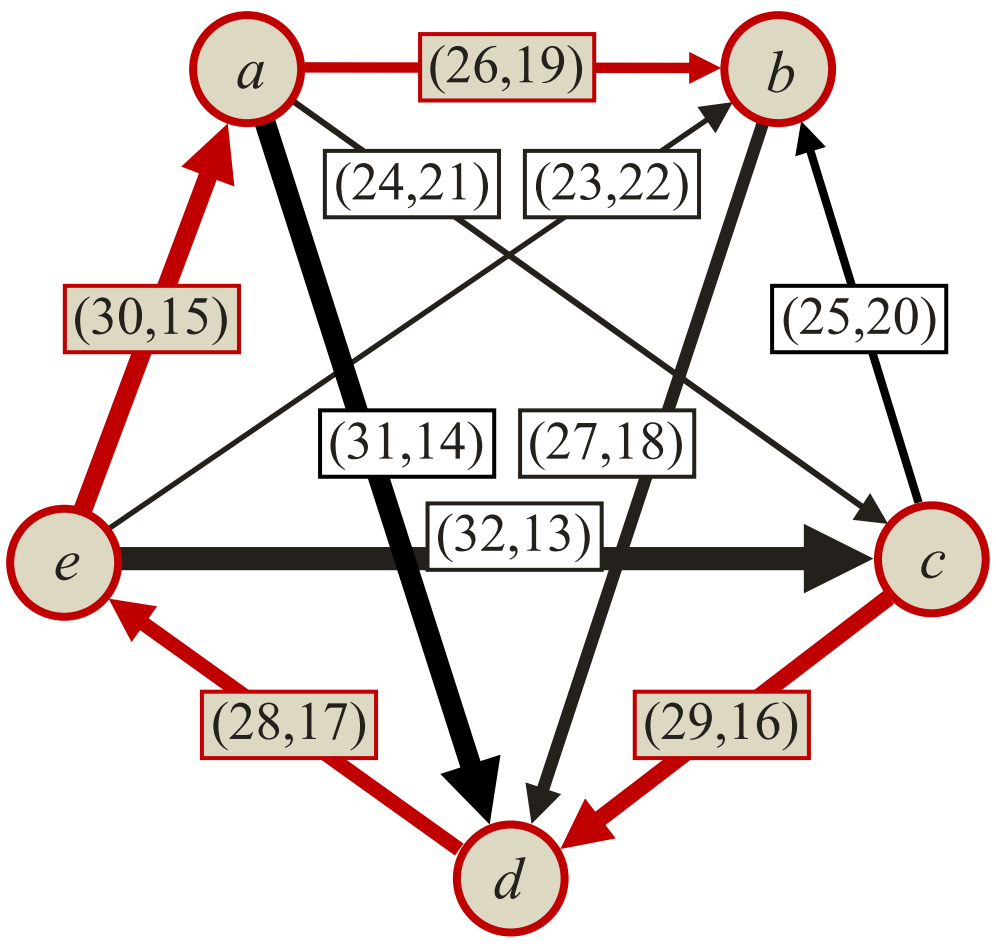

These are the strongest paths
from every other alternative to *b*.

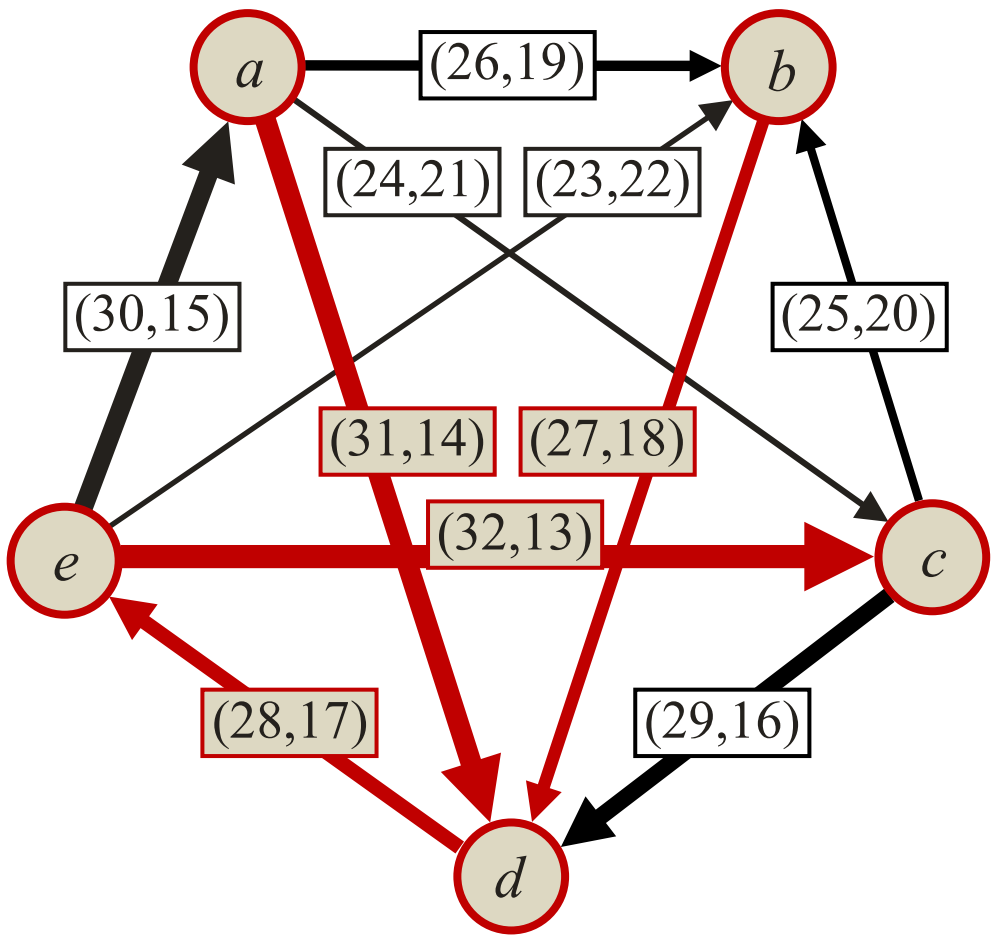

These are the strongest paths
from every other alternative to *c*.

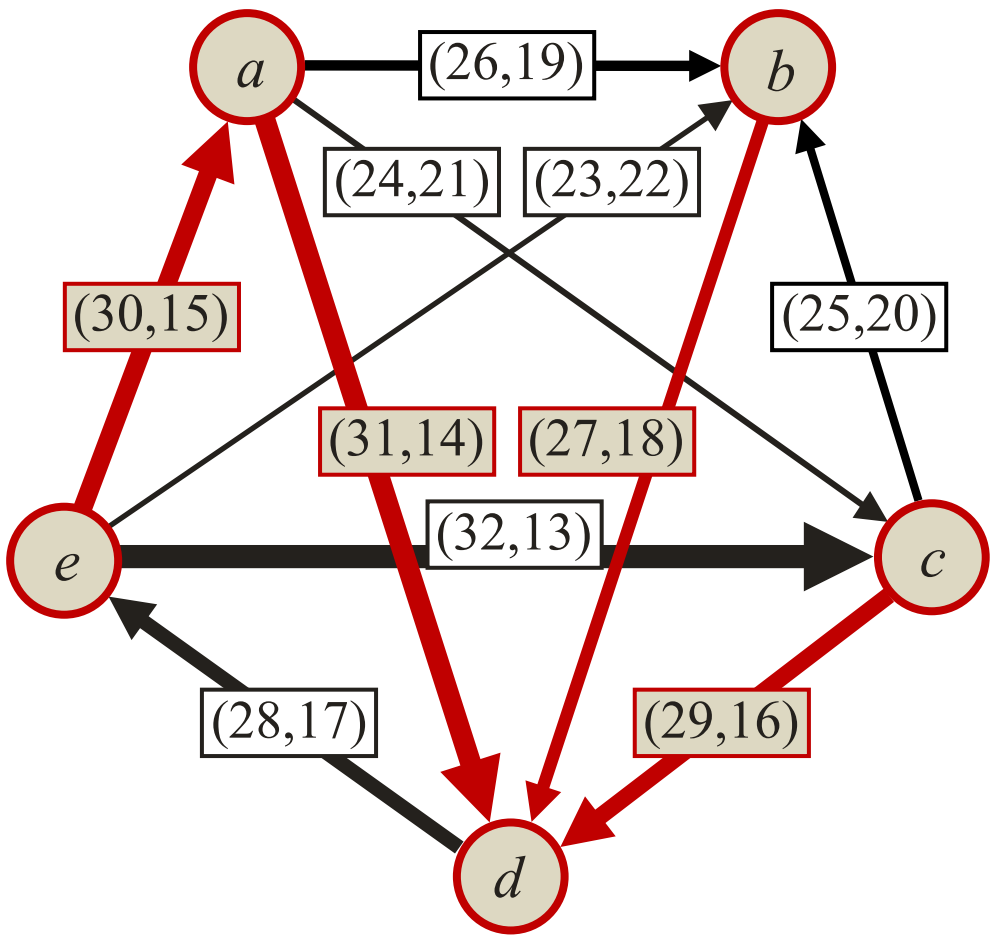

These are the strongest paths
from every other alternative to *d*.

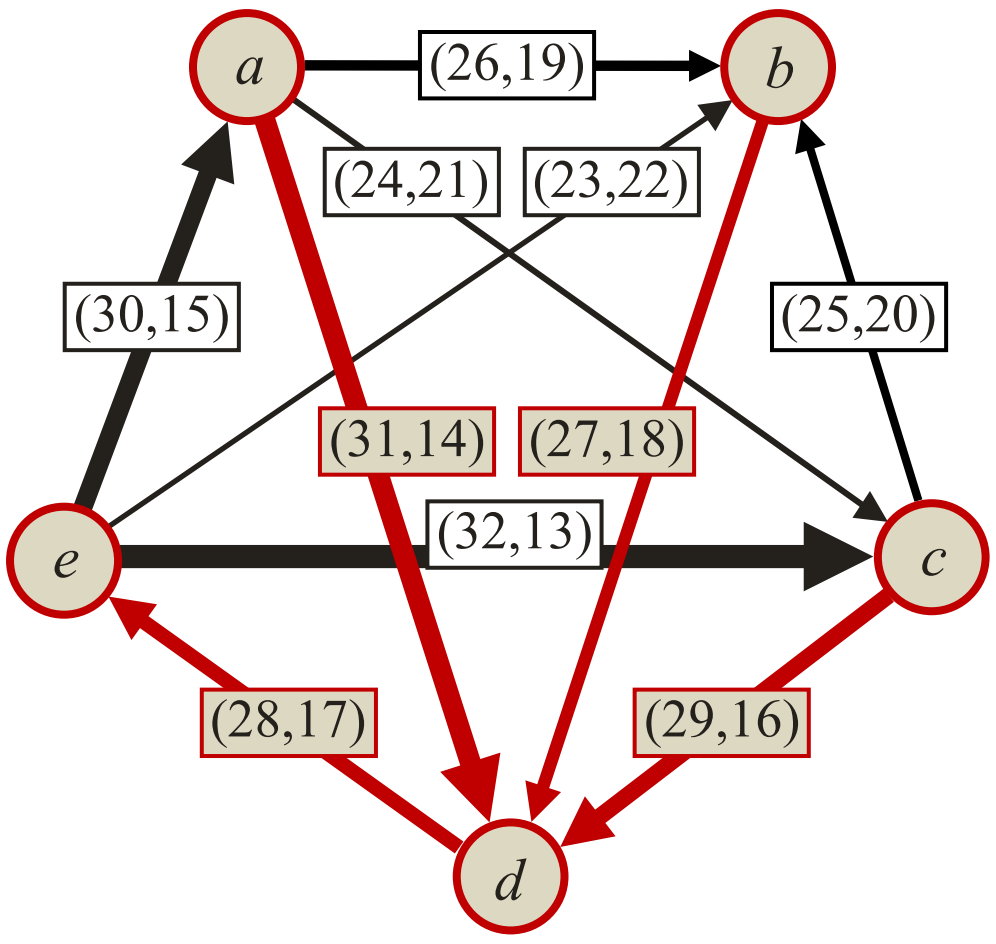

These are the strongest paths
from every other alternative to *e*.

Therefore, the strengths of the strongest paths are:

| | $P_D[*,a]$ | $P_D[*,b]$ | $P_D[*,c]$ | $P_D[*,d]$ | $P_D[*,e]$ |
|---|---|---|---|---|---|
| $P_D[a,*]$ | --- | (26,19) | (28,17) | (31,14) | (28,17) |
| $P_D[b,*]$ | (27,18) | --- | (27,18) | (27,18) | (27,18) |
| $P_D[c,*]$ | (28,17) | (26,19) | --- | (29,16) | (28,17) |
| $P_D[d,*]$ | (28,17) | (26,19) | (28,17) | --- | (28,17) |
| $P_D[e,*]$ | (30,15) | (26,19) | (32,13) | (30,15) | --- |

We get $\mathcal{O} = \{ad, ba, bc, bd, be, cd, ea, ec, ed\}$ and $\mathcal{S} = \{b\}$.

Suppose, the strongest paths are calculated with the Floyd-Warshall algorithm, as defined in section 2.3.1. Then the following table documents the $C \cdot (C-1) \cdot (C-2) = 60$ steps of the Floyd-Warshall algorithm.

We start with

- $P_D[i,j] := (N[i,j],N[j,i])$ for all $i \in A$ and $j \in A \setminus \{i\}$.
- $pred[i,j] := i$ for all $i \in A$ and $j \in A \setminus \{i\}$.

| | $i$ | $j$ | $k$ | $P_D[j,k]$ | $P_D[j,i]$ | $P_D[i,k]$ | $pred[j,k]$ | $pred[i,k]$ | result |
|---|---|---|---|---|---|---|---|---|---|
| 1 | $a$ | $b$ | $c$ | (20,25) | (19,26) | (24,21) | $b$ | $a$ | |
| 2 | $a$ | $b$ | $d$ | (27,18) | (19,26) | (31,14) | $b$ | $a$ | |
| 3 | $a$ | $b$ | $e$ | (22,23) | (19,26) | (15,30) | $b$ | $a$ | |
| 4 | $a$ | $c$ | $b$ | (25,20) | (21,24) | (26,19) | $c$ | $a$ | |
| 5 | $a$ | $c$ | $d$ | (29,16) | (21,24) | (31,14) | $c$ | $a$ | |
| 6 | $a$ | $c$ | $e$ | (13,32) | (21,24) | (15,30) | $c$ | $a$ | $P_D[c,e]$ is updated from (13,32) to (15,30); $pred[c,e]$ is updated from $c$ to $a$. |
| 7 | $a$ | $d$ | $b$ | (18,27) | (14,31) | (26,19) | $d$ | $a$ | |
| 8 | $a$ | $d$ | $c$ | (16,29) | (14,31) | (24,21) | $d$ | $a$ | |
| 9 | $a$ | $d$ | $e$ | (28,17) | (14,31) | (15,30) | $d$ | $a$ | |
| 10 | $a$ | $e$ | $b$ | (23,22) | (30,15) | (26,19) | $e$ | $a$ | $P_D[e,b]$ is updated from (23,22) to (26,19); $pred[e,b]$ is updated from $e$ to $a$. |
| 11 | $a$ | $e$ | $c$ | (32,13) | (30,15) | (24,21) | $e$ | $a$ | |
| 12 | $a$ | $e$ | $d$ | (17,28) | (30,15) | (31,14) | $e$ | $a$ | $P_D[e,d]$ is updated from (17,28) to (30,15); $pred[e,d]$ is updated from $e$ to $a$. |
| 13 | $b$ | $a$ | $c$ | (24,21) | (26,19) | (20,25) | $a$ | $b$ | |
| 14 | $b$ | $a$ | $d$ | (31,14) | (26,19) | (27,18) | $a$ | $b$ | |
| 15 | $b$ | $a$ | $e$ | (15,30) | (26,19) | (22,23) | $a$ | $b$ | $P_D[a,e]$ is updated from (15,30) to (22,23); $pred[a,e]$ is updated from $a$ to $b$. |
| 16 | $b$ | $c$ | $a$ | (21,24) | (25,20) | (19,26) | $c$ | $b$ | |
| 17 | $b$ | $c$ | $d$ | (29,16) | (25,20) | (27,18) | $c$ | $b$ | |
| 18 | $b$ | $c$ | $e$ | (15,30) | (25,20) | (22,23) | $a$ | $b$ | $P_D[c,e]$ is updated from (15,30) to (22,23); $pred[c,e]$ is updated from $a$ to $b$. |
| 19 | $b$ | $d$ | $a$ | (14,31) | (18,27) | (19,26) | $d$ | $b$ | $P_D[d,a]$ is updated from (14,31) to (18,27); $pred[d,a]$ is updated from $d$ to $b$. |
| 20 | $b$ | $d$ | $c$ | (16,29) | (18,27) | (20,25) | $d$ | $b$ | $P_D[d,c]$ is updated from (16,29) to (18,27); $pred[d,c]$ is updated from $d$ to $b$. |
| 21 | $b$ | $d$ | $e$ | (28,17) | (18,27) | (22,23) | $d$ | $b$ | |
| 22 | $b$ | $e$ | $a$ | (30,15) | (26,19) | (19,26) | $e$ | $b$ | |
| 23 | $b$ | $e$ | $c$ | (32,13) | (26,19) | (20,25) | $e$ | $b$ | |
| 24 | $b$ | $e$ | $d$ | (30,15) | (26,19) | (27,18) | $a$ | $b$ | |
| 25 | $c$ | $a$ | $b$ | (26,19) | (24,21) | (25,20) | $a$ | $c$ | |
| 26 | $c$ | $a$ | $d$ | (31,14) | (24,21) | (29,16) | $a$ | $c$ | |
| 27 | $c$ | $a$ | $e$ | (22,23) | (24,21) | (22,23) | $b$ | $b$ | |
| 28 | $c$ | $b$ | $a$ | (19,26) | (20,25) | (21,24) | $b$ | $c$ | $P_D[b,a]$ is updated from (19,26) to (20,25); $pred[b,a]$ is updated from $b$ to $c$. |
| 29 | $c$ | $b$ | $d$ | (27,18) | (20,25) | (29,16) | $b$ | $c$ | |
| 30 | $c$ | $b$ | $e$ | (22,23) | (20,25) | (22,23) | $b$ | $b$ | |

| | $i$ | $j$ | $k$ | $P_D[j,k]$ | $P_D[j,i]$ | $P_D[i,k]$ | $pred[j,k]$ | $pred[i,k]$ | result |
|---|---|---|---|---|---|---|---|---|---|
| 31 | *c* | *d* | *a* | (18,27) | (18,27) | (21,24) | *b* | *c* | |
| 32 | *c* | *d* | *b* | (18,27) | (18,27) | (25,20) | *d* | *c* | |
| 33 | *c* | *d* | *e* | (28,17) | (18,27) | (22,23) | *d* | *b* | |
| 34 | *c* | *e* | *a* | (30,15) | (32,13) | (21,24) | *e* | *c* | |
| 35 | *c* | *e* | *b* | (26,19) | (32,13) | (25,20) | *a* | *c* | |
| 36 | *c* | *e* | *d* | (30,15) | (32,13) | (29,16) | *a* | *c* | |
| 37 | *d* | *a* | *b* | (26,19) | (31,14) | (18,27) | *a* | *d* | |
| 38 | *d* | *a* | *c* | (24,21) | (31,14) | (18,27) | *a* | *b* | |
| 39 | *d* | *a* | *e* | (22,23) | (31,14) | (28,17) | *b* | *d* | $P_D[a,e]$ is updated from (22,23) to (28,17); *pred*[*a*,*e*] is updated from *b* to *d*. |
| 40 | *d* | *b* | *a* | (20,25) | (27,18) | (18,27) | *c* | *b* | |
| 41 | *d* | *b* | *c* | (20,25) | (27,18) | (18,27) | *b* | *b* | |
| 42 | *d* | *b* | *e* | (22,23) | (27,18) | (28,17) | *b* | *d* | $P_D[b,e]$ is updated from (22,23) to (27,18); *pred*[*b*,*e*] is updated from *b* to *d*. |
| 43 | *d* | *c* | *a* | (21,24) | (29,16) | (18,27) | *c* | *b* | |
| 44 | *d* | *c* | *b* | (25,20) | (29,16) | (18,27) | *c* | *d* | |
| 45 | *d* | *c* | *e* | (22,23) | (29,16) | (28,17) | *b* | *d* | $P_D[c,e]$ is updated from (22,23) to (28,17); *pred*[*c*,*e*] is updated from *b* to *d*. |
| 46 | *d* | *e* | *a* | (30,15) | (30,15) | (18,27) | *e* | *b* | |
| 47 | *d* | *e* | *b* | (26,19) | (30,15) | (18,27) | *a* | *d* | |
| 48 | *d* | *e* | *c* | (32,13) | (30,15) | (18,27) | *e* | *b* | |
| 49 | *e* | *a* | *b* | (26,19) | (28,17) | (26,19) | *a* | *a* | |
| 50 | *e* | *a* | *c* | (24,21) | (28,17) | (32,13) | *a* | *e* | $P_D[a,c]$ is updated from (24,21) to (28,17); *pred*[*a*,*c*] is updated from *a* to *e*. |
| 51 | *e* | *a* | *d* | (31,14) | (28,17) | (30,15) | *a* | *a* | |
| 52 | *e* | *b* | *a* | (20,25) | (27,18) | (30,15) | *c* | *e* | $P_D[b,a]$ is updated from (20,25) to (27,18); *pred*[*b*,*a*] is updated from *c* to *e*. |
| 53 | *e* | *b* | *c* | (20,25) | (27,18) | (32,13) | *b* | *e* | $P_D[b,c]$ is updated from (20,25) to (27,18); *pred*[*b*,*c*] is updated from *b* to *e*. |
| 54 | *e* | *b* | *d* | (27,18) | (27,18) | (30,15) | *b* | *a* | |
| 55 | *e* | *c* | *a* | (21,24) | (28,17) | (30,15) | *c* | *e* | $P_D[c,a]$ is updated from (21,24) to (28,17); *pred*[*c*,*a*] is updated from *c* to *e*. |
| 56 | *e* | *c* | *b* | (25,20) | (28,17) | (26,19) | *c* | *a* | $P_D[c,b]$ is updated from (25,20) to (26,19); *pred*[*c*,*b*] is updated from *c* to *a*. |
| 57 | *e* | *c* | *d* | (29,16) | (28,17) | (30,15) | *c* | *a* | |
| 58 | *e* | *d* | *a* | (18,27) | (28,17) | (30,15) | *b* | *e* | $P_D[d,a]$ is updated from (18,27) to (28,17); *pred*[*d*,*a*] is updated from *b* to *e*. |
| 59 | *e* | *d* | *b* | (18,27) | (28,17) | (26,19) | *d* | *a* | $P_D[d,b]$ is updated from (18,27) to (26,19); *pred*[*d*,*b*] is updated from *d* to *a*. |
| 60 | *e* | *d* | *c* | (18,27) | (28,17) | (32,13) | *b* | *e* | $P_D[d,c]$ is updated from (18,27) to (28,17); *pred*[*d*,*c*] is updated from *b* to *e*. |

## 3.12. Example 12

Example 12:

| | | |
|---|---|---|
| 9 | voters | $a \succ_v d \succ_v b \succ_v e \succ_v c$ |
| 1 | voter | $b \succ_v a \succ_v c \succ_v e \succ_v d$ |
| 6 | voters | $c \succ_v b \succ_v a \succ_v d \succ_v e$ |
| 2 | voters | $c \succ_v d \succ_v b \succ_v e \succ_v a$ |
| 5 | voters | $c \succ_v d \succ_v e \succ_v a \succ_v b$ |
| 6 | voters | $d \succ_v e \succ_v c \succ_v a \succ_v b$ |
| 14 | voters | $e \succ_v b \succ_v a \succ_v c \succ_v d$ |
| 2 | voters | $e \succ_v b \succ_v c \succ_v a \succ_v d$ |

The pairwise matrix *N* looks as follows:

| | *N*[*,*a*] | *N*[*,*b*] | *N*[*,*c*] | *N*[*,*d*] | *N*[*,*e*] |
|---|---|---|---|---|---|
| *N*[*a*,*] | --- | 20 | 24 | 32 | 16 |
| *N*[*b*,*] | 25 | --- | 26 | 23 | 18 |
| *N*[*c*,*] | 21 | 19 | --- | 30 | 14 |
| *N*[*d*,*] | 13 | 22 | 15 | --- | 28 |
| *N*[*e*,*] | 29 | 27 | 31 | 17 | --- |

The corresponding digraph looks as follows:

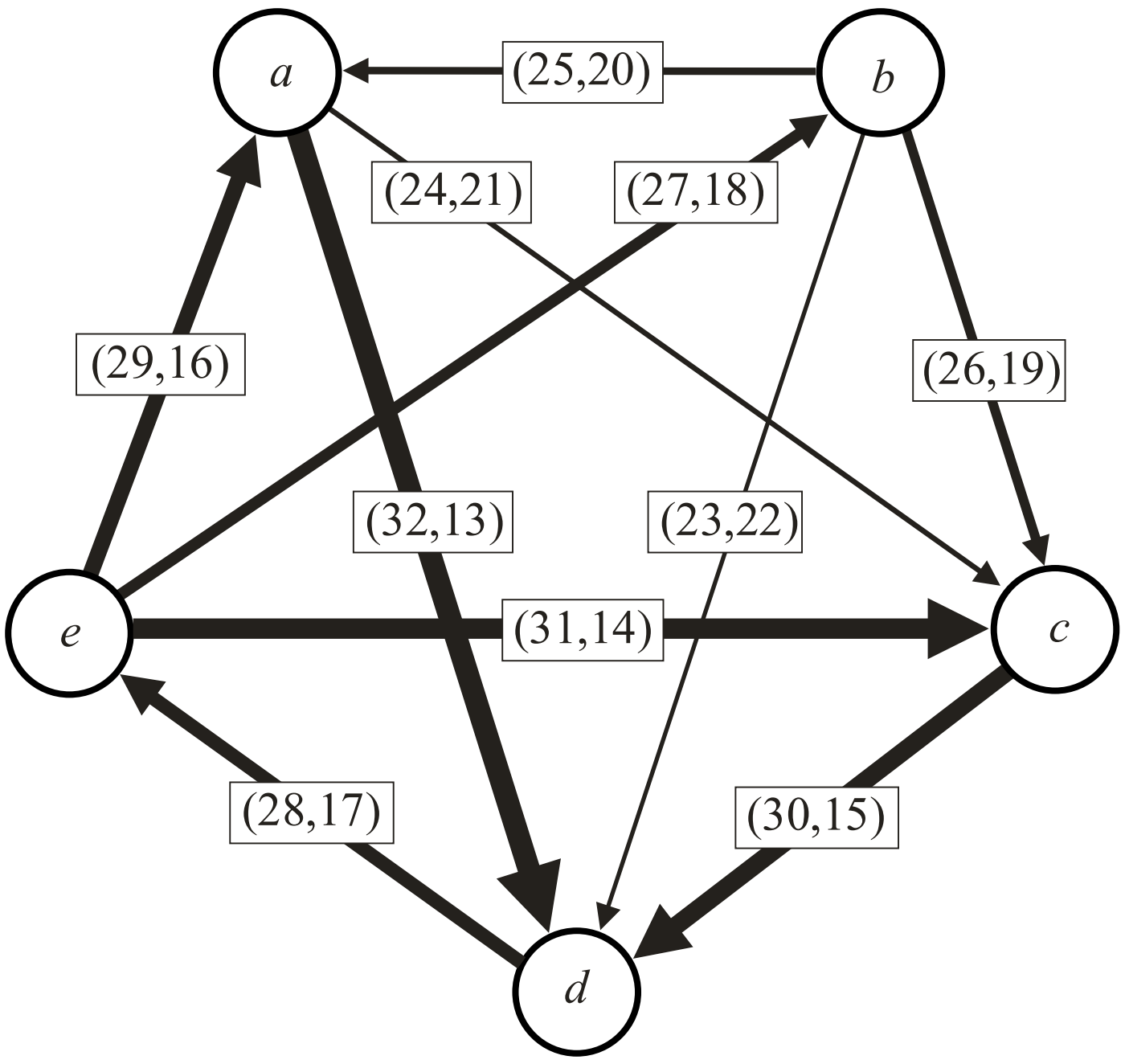

The following table lists the strongest paths, as determined by the Floyd-Warshall algorithm, as defined in section 2.3.1. The critical links of the strongest paths are underlined:

| | ... to *a* | ... to *b* | ... to *c* | ... to *d* | ... to *e* |
|---|---|---|---|---|---|
| from *a* ... | --- | *a*, (32,13), *d*, (28,17), *e*, <u>(27,18)</u>, *b* | *a*, (32,13), *d*, <u>(28,17)</u>, *e*, (31,14), *c* | *a*, <u>(32,13)</u>, *d* | *a*, (32,13), *d*, <u>(28,17)</u>, *e* |
| from *b* ... | *b*, <u>(26,19)</u>, *c*, (30,15), *d*, (28,17), *e*, (29,16), *a* | --- | *b*, <u>(26,19)</u>, *c* | *b*, <u>(26,19)</u>, *c*, (30,15), *d* | *b*, <u>(26,19)</u>, *c*, (30,15), *d*, (28,17), *e* |
| from *c* ... | *c*, (30,15), *d*, <u>(28,17)</u>, *e*, (29,16), *a* | *c*, (30,15), *d*, (28,17), *e*, <u>(27,18)</u>, *b* | --- | *c*, <u>(30,15)</u>, *d* | *c*, (30,15), *d*, <u>(28,17)</u>, *e* |
| from *d* ... | *d*, <u>(28,17)</u>, *e*, (29,16), *a* | *d*, (28,17), *e*, <u>(27,18)</u>, *b* | *d*, <u>(28,17)</u>, *e*, (31,14), *c* | --- | *d*, <u>(28,17)</u>, *e* |
| from *e* ... | *e*, <u>(29,16)</u>, *a* | *e*, <u>(27,18)</u>, *b* | *e*, <u>(31,14)</u>, *c* | *e*, (31,14), *c*, <u>(30,15)</u>, *d* | --- |

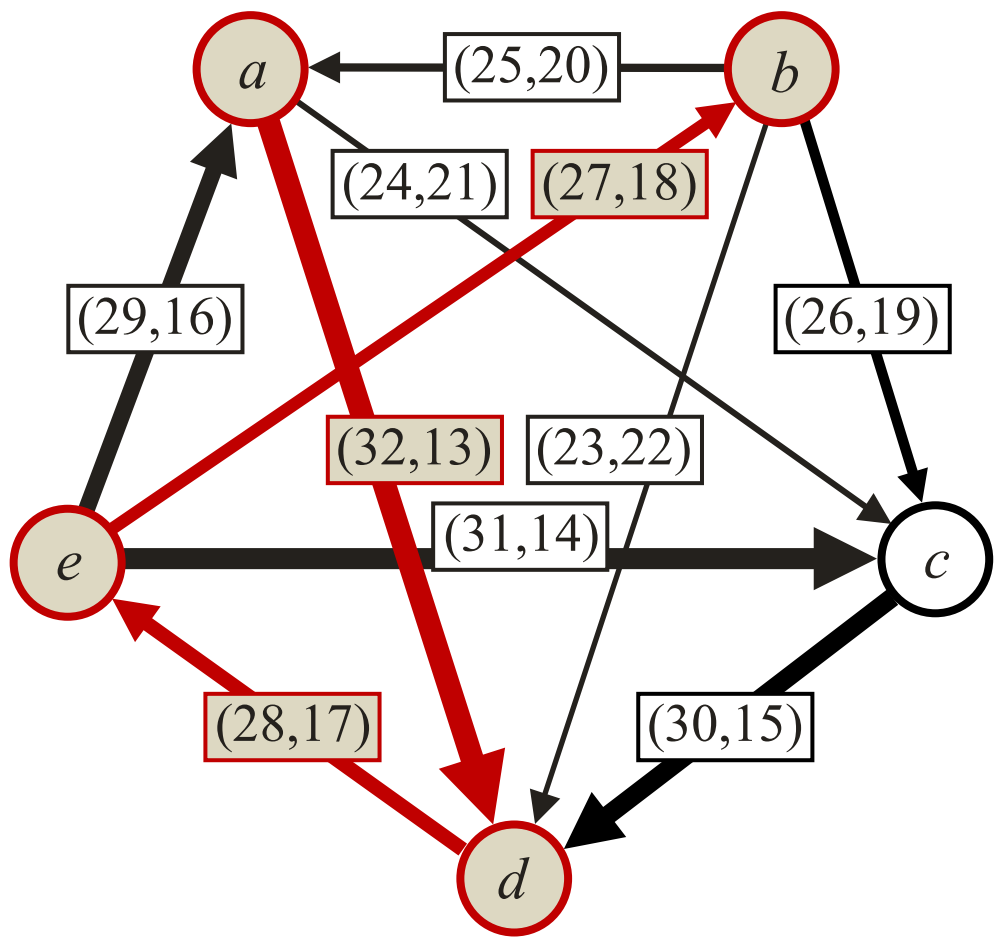

The strongest path from *a* to *b* is:
*a*, (32,13), *d*, (28,17), *e*, (27,18), *b*

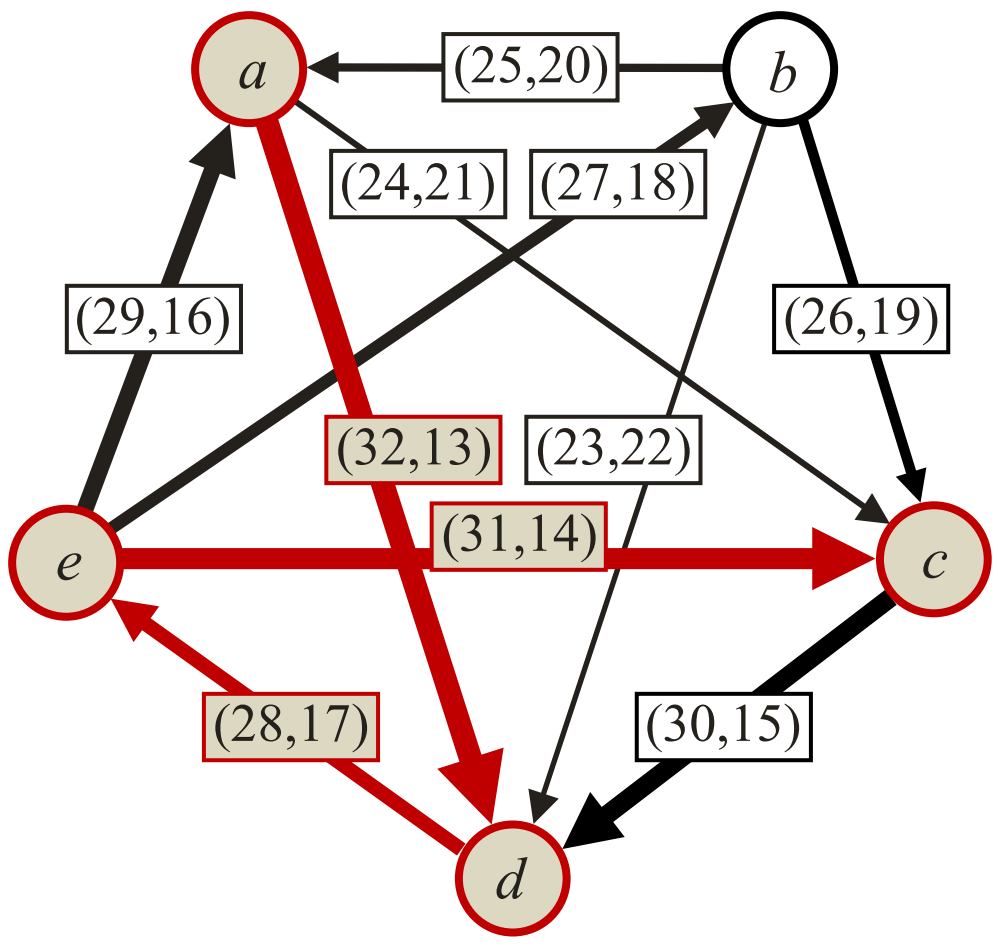

The strongest path from *a* to *c* is:
*a*, (32,13), *d*, (28,17), *e*, (31,14), *c*

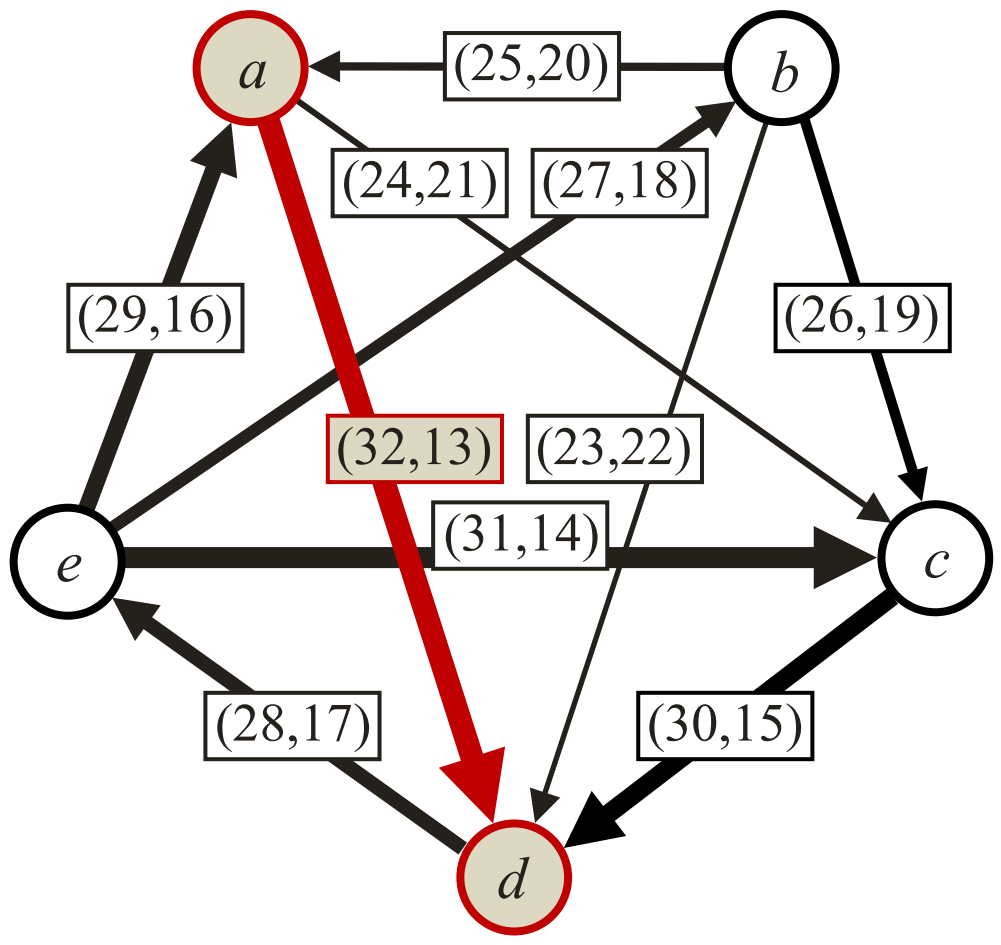

The strongest path from *a* to *d* is:
*a*, (32,13), *d*

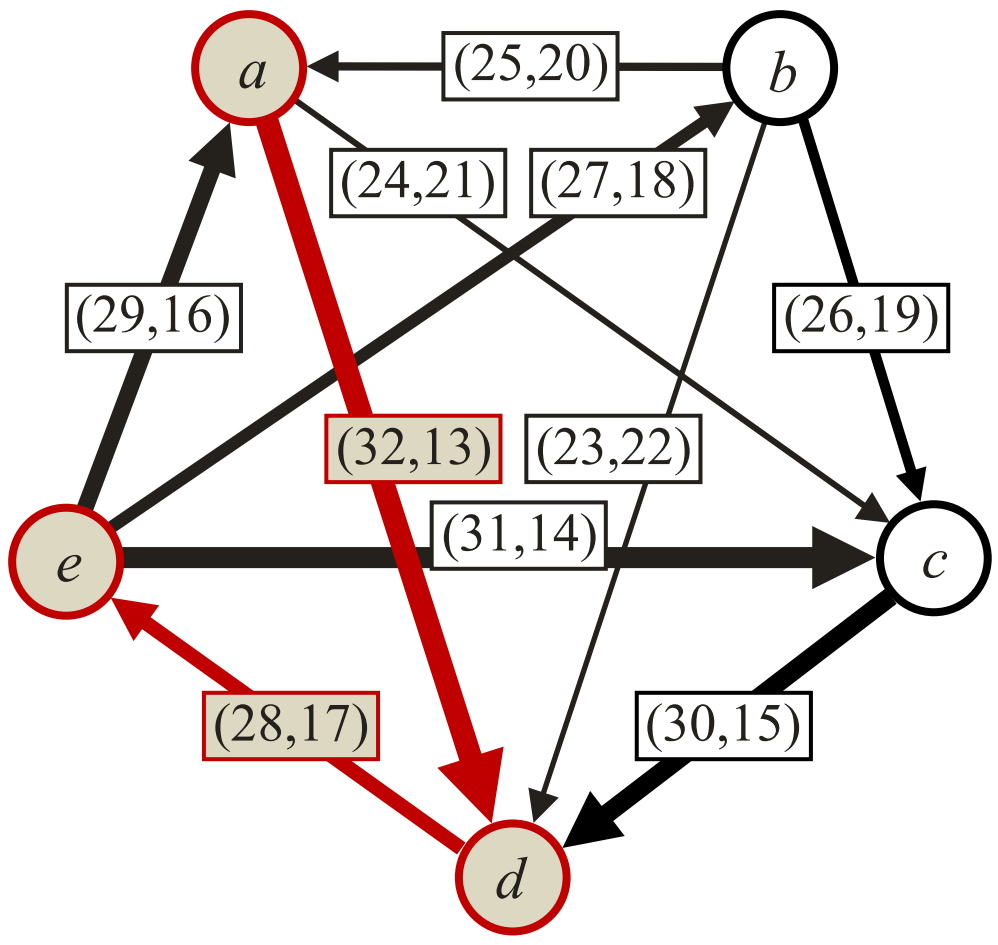

The strongest path from *a* to *e* is:
*a*, (32,13), *d*, (28,17), *e*

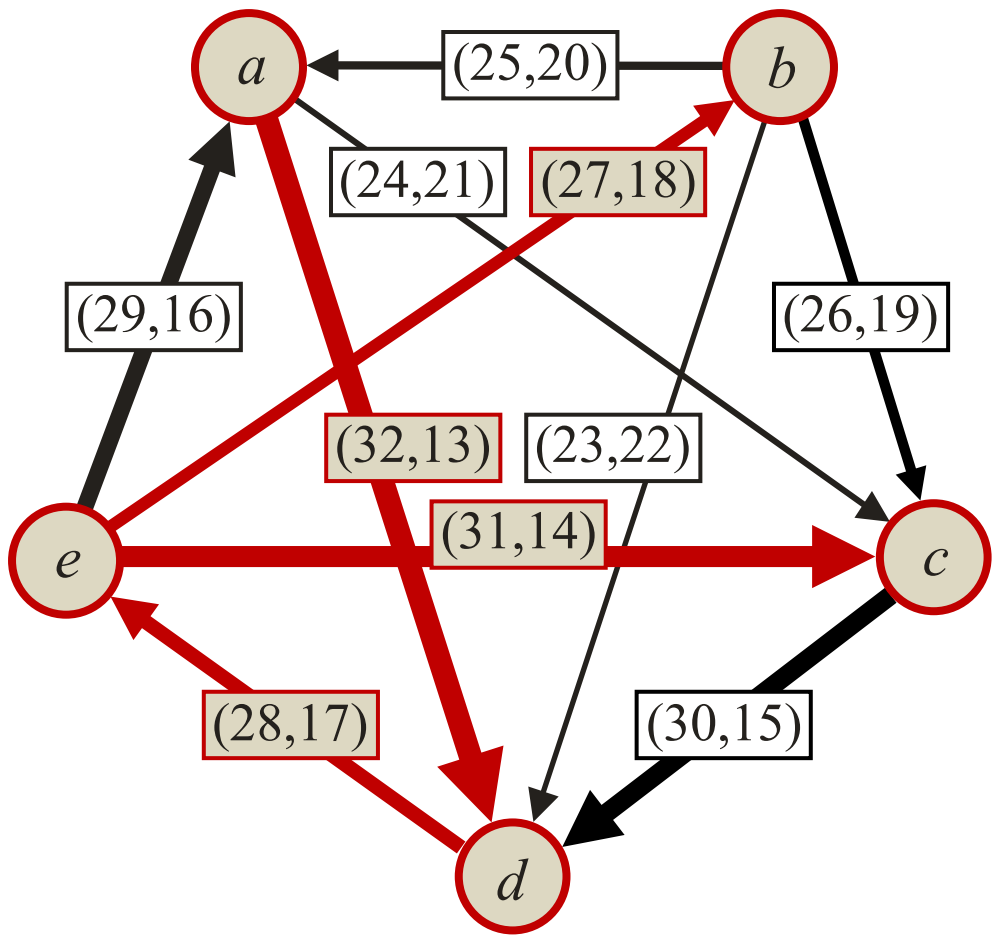

These are the strongest paths
from *a* to every other alternative.

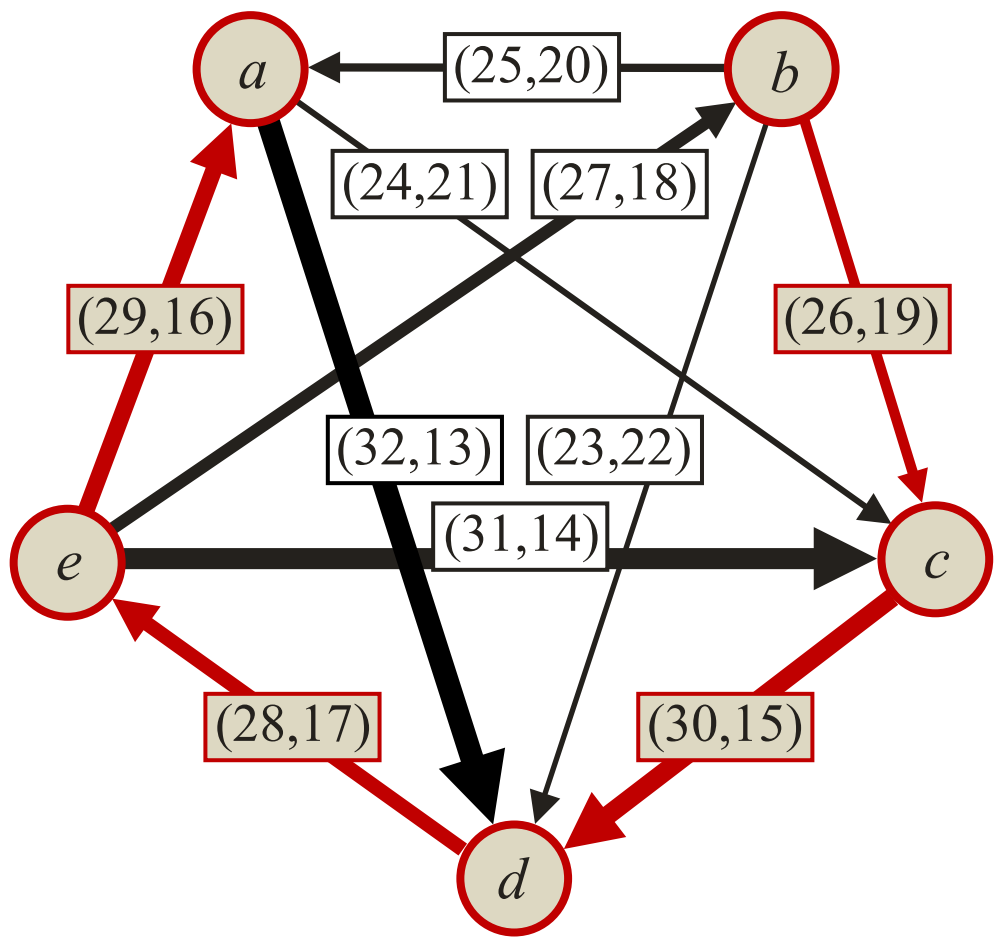

The strongest path from *b* to *a* is:
*b*, (26,19), *c*, (30,15), *d*,
(28,17), *e*, (29,16), *a*

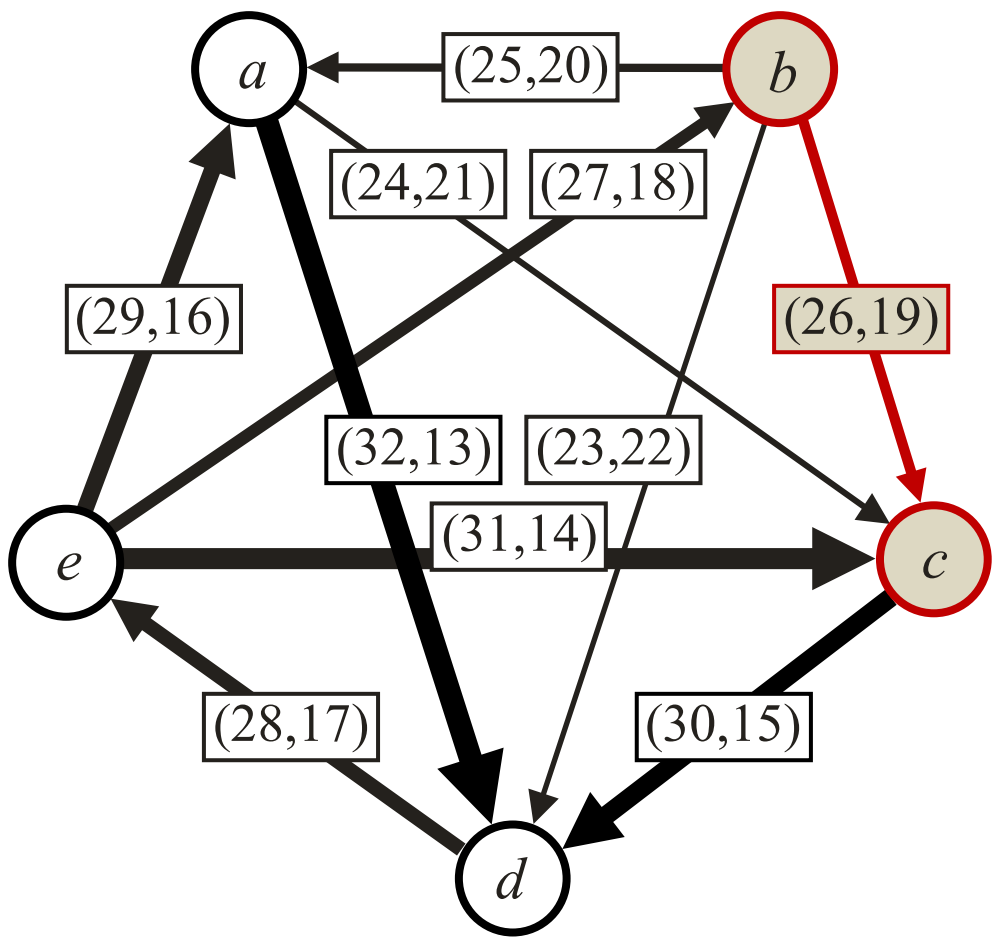

The strongest path from *b* to *c* is:
*b*, (26,19), *c*

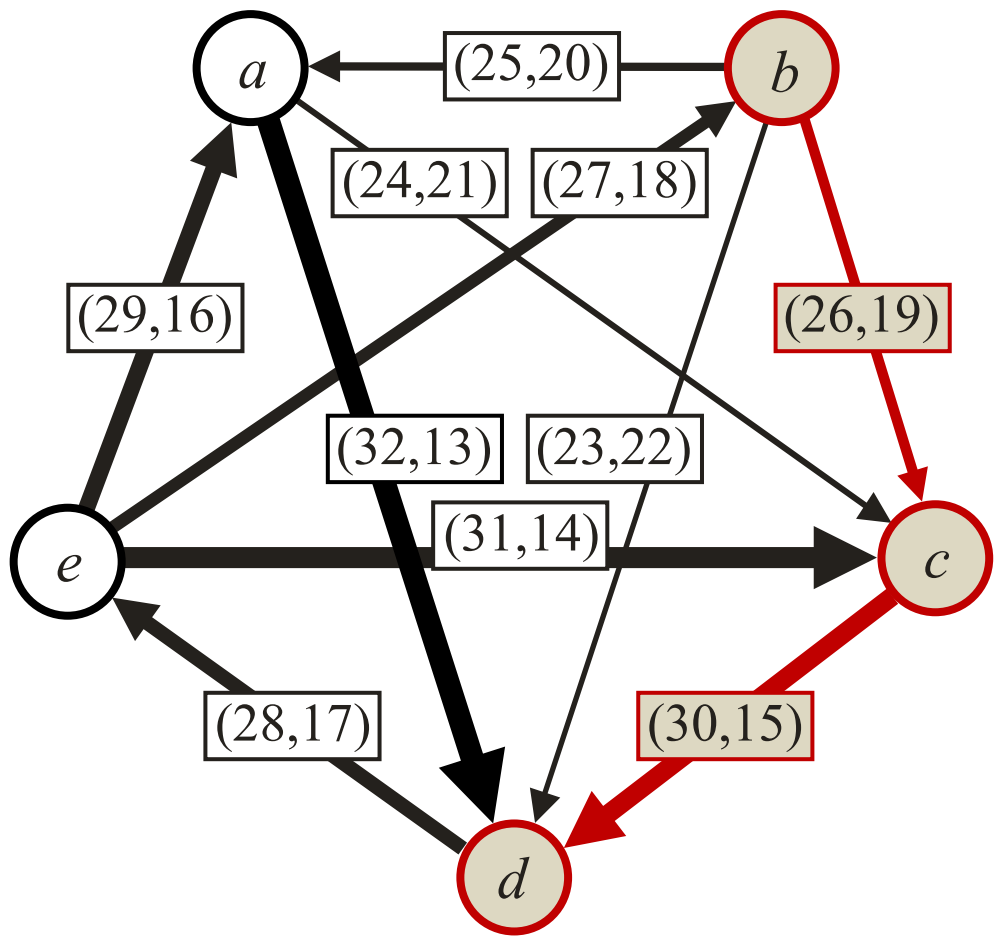

The strongest path from *b* to *d* is:
*b*, (26,19), *c*, (30,15), *d*

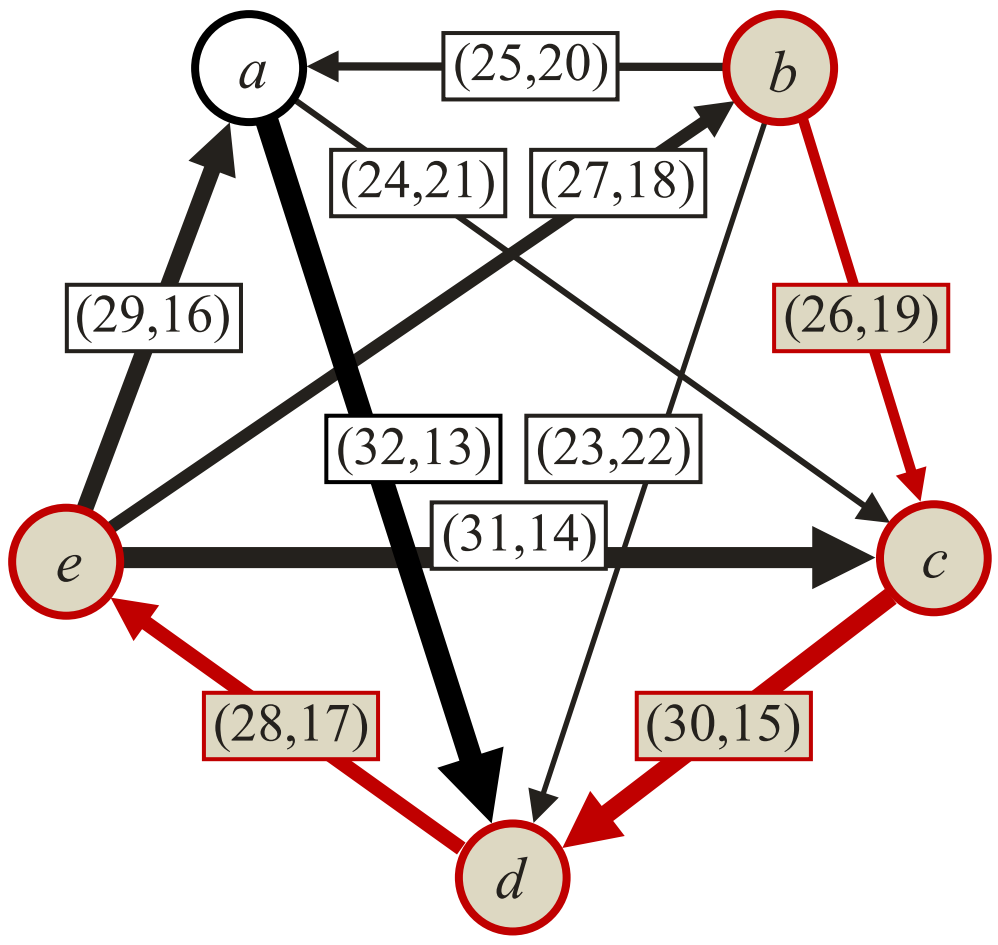

The strongest path from *b* to *e* is:
*b*, (26,19), *c*, (30,15), *d*, (28,17), *e*

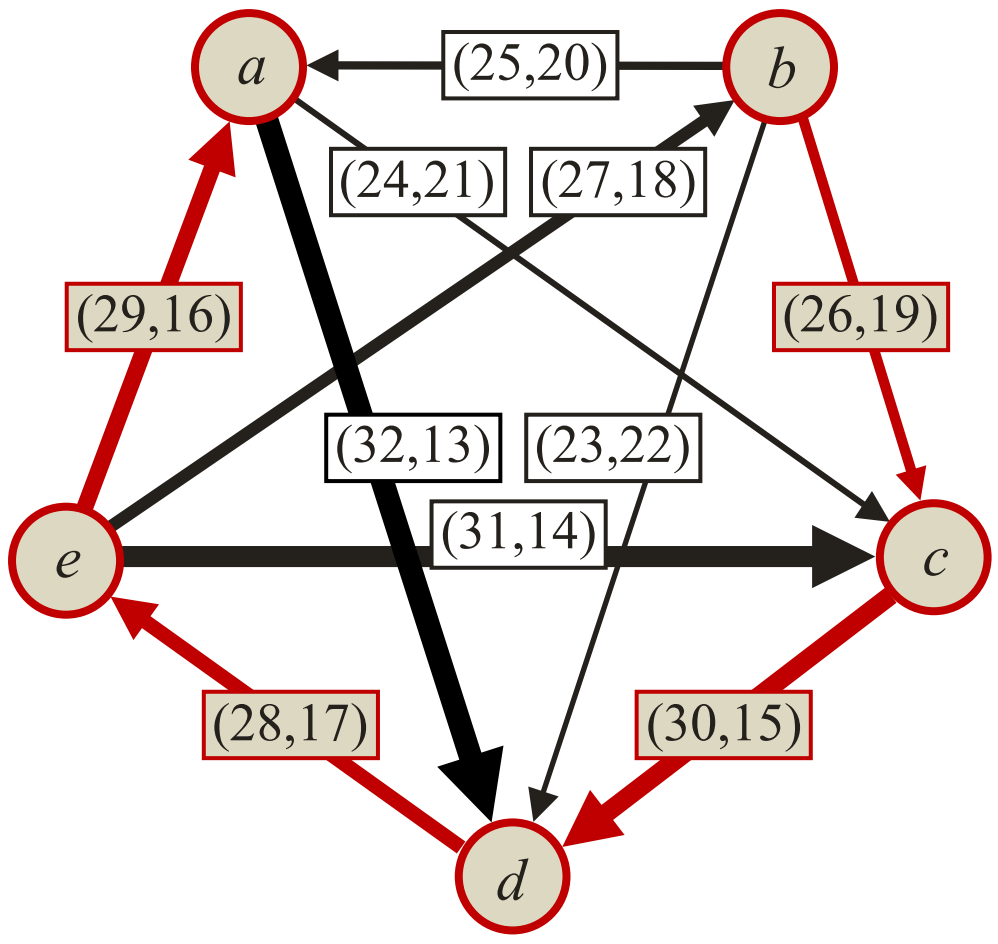

These are the strongest paths
from *b* to every other alternative.

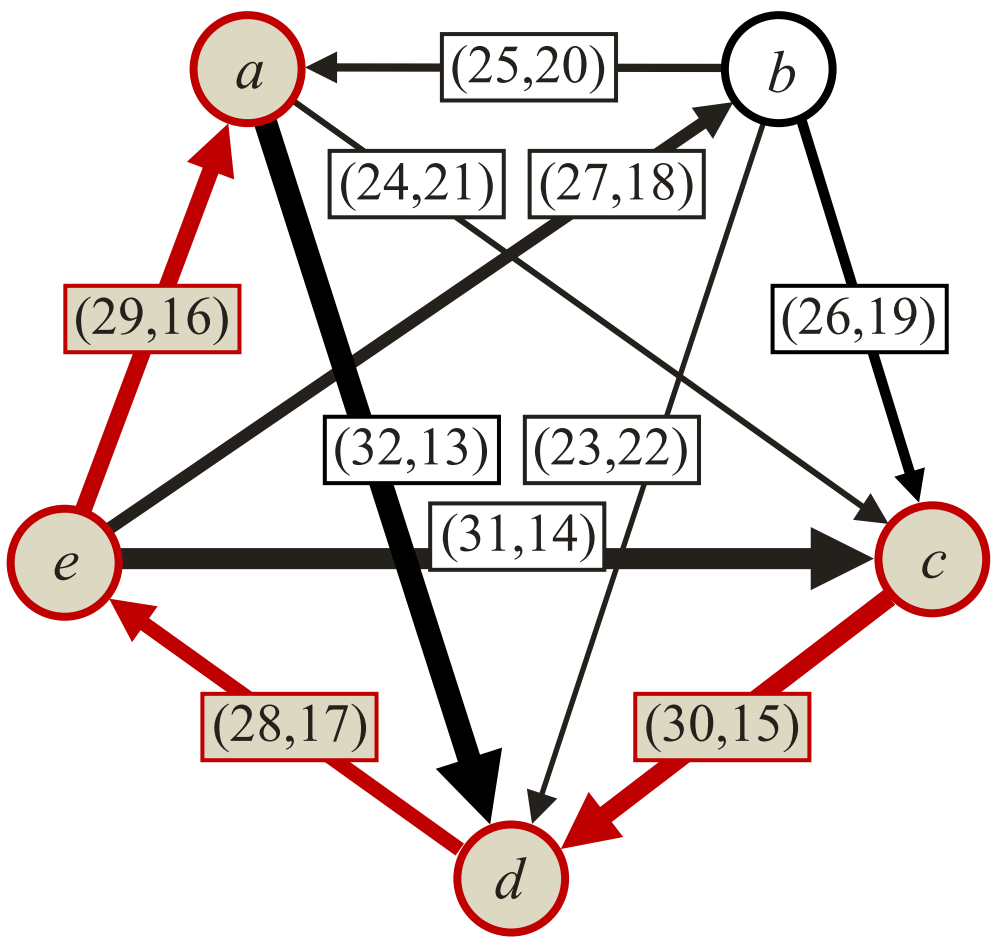

The strongest path from *c* to *a* is:
*c*, (30,15), *d*, (28,17), *e*, (29,16), *a*

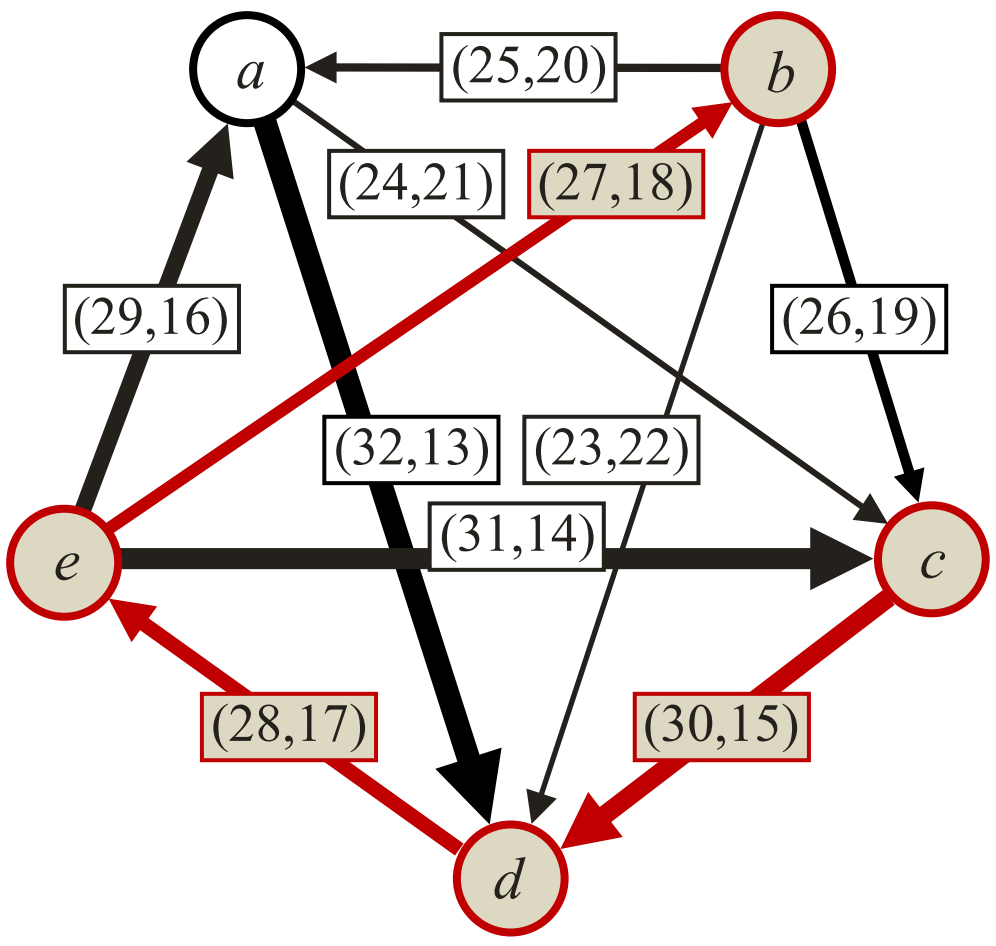

The strongest path from *c* to *b* is:
*c*, (30,15), *d*, (28,17), *e*, (27,18), *b*

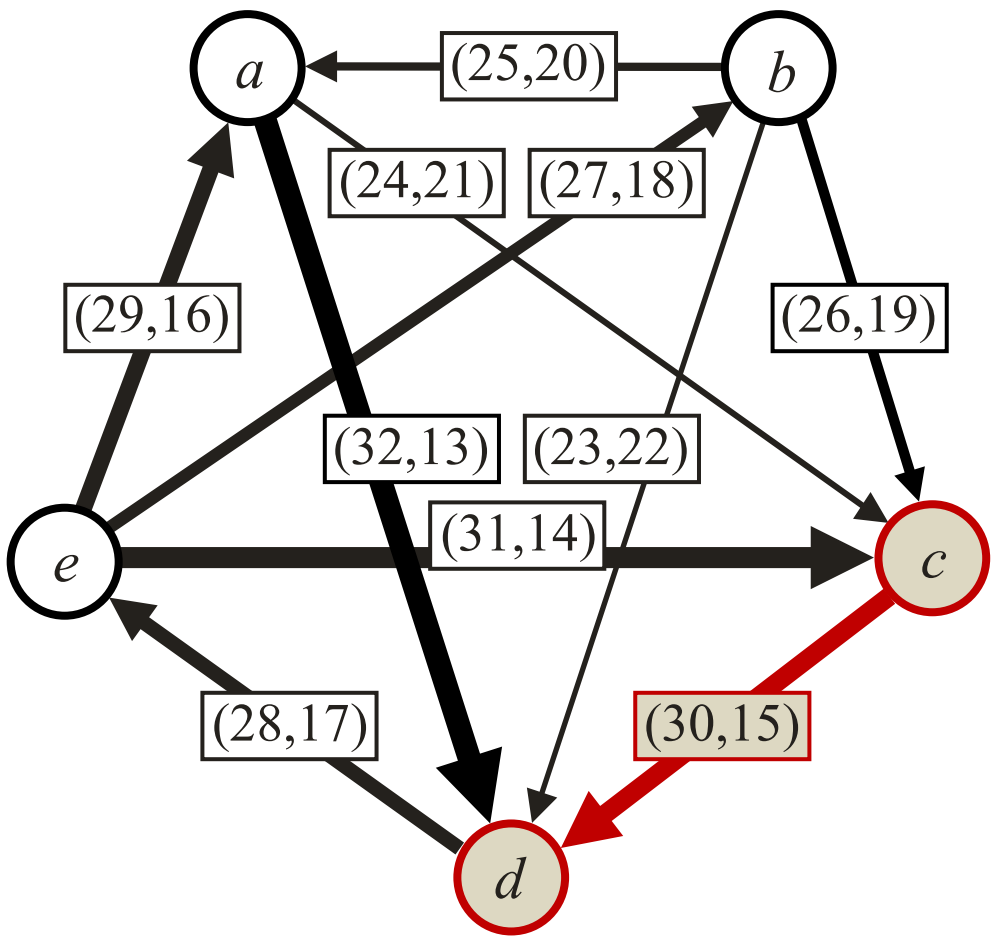

The strongest path from *c* to *d* is:
*c*, (30,15), *d*

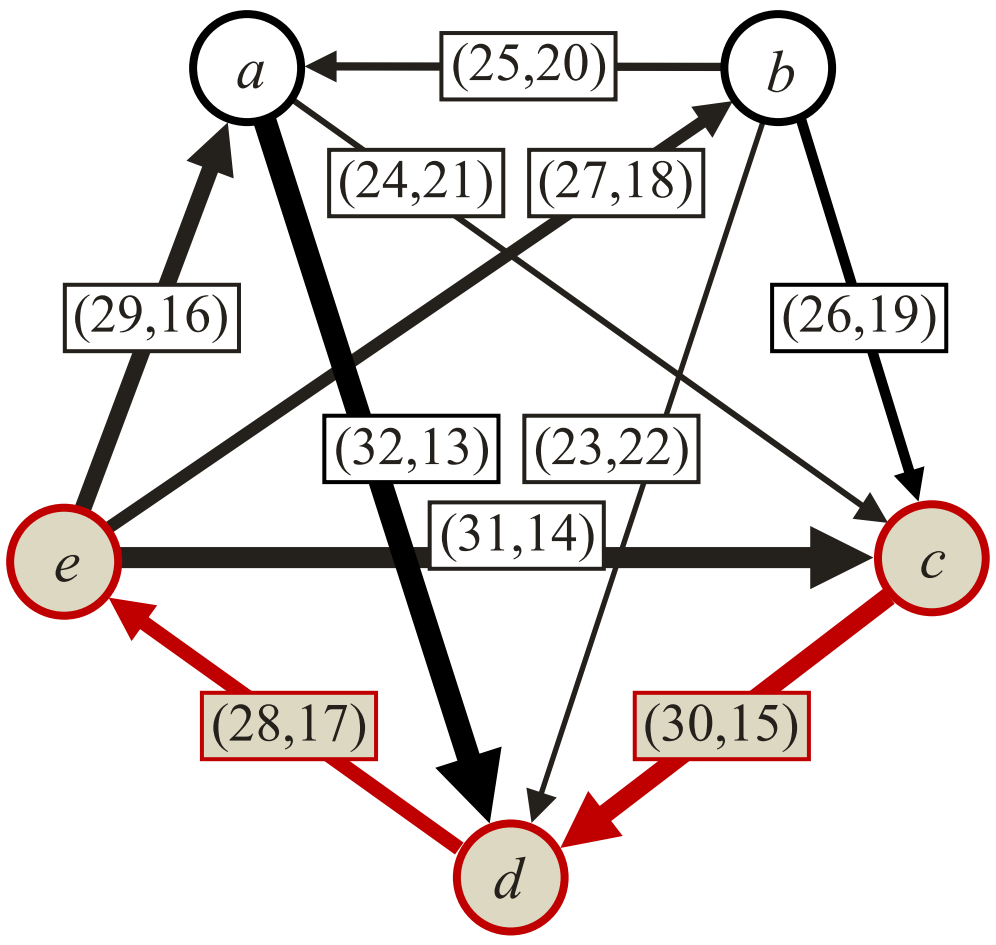

The strongest path from *c* to *e* is:
*c*, (30,15), *d*, (28,17), *e*

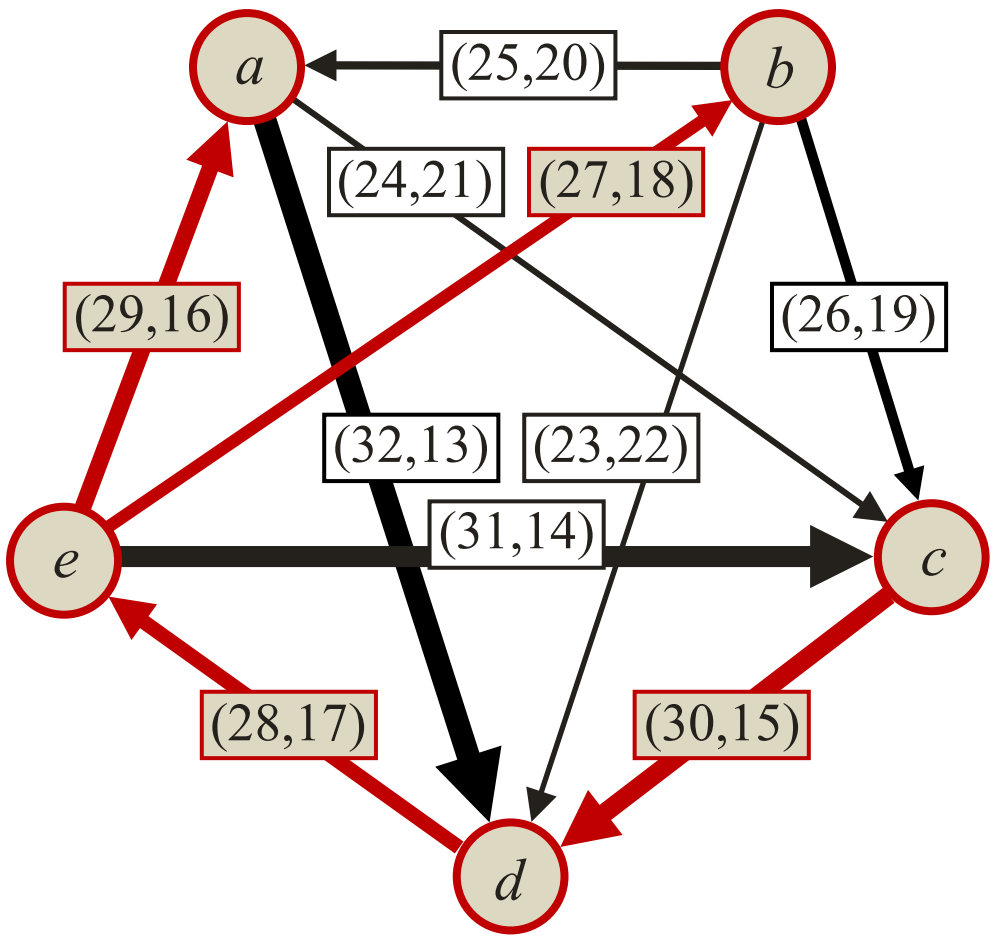

These are the strongest paths
from *c* to every other alternative.

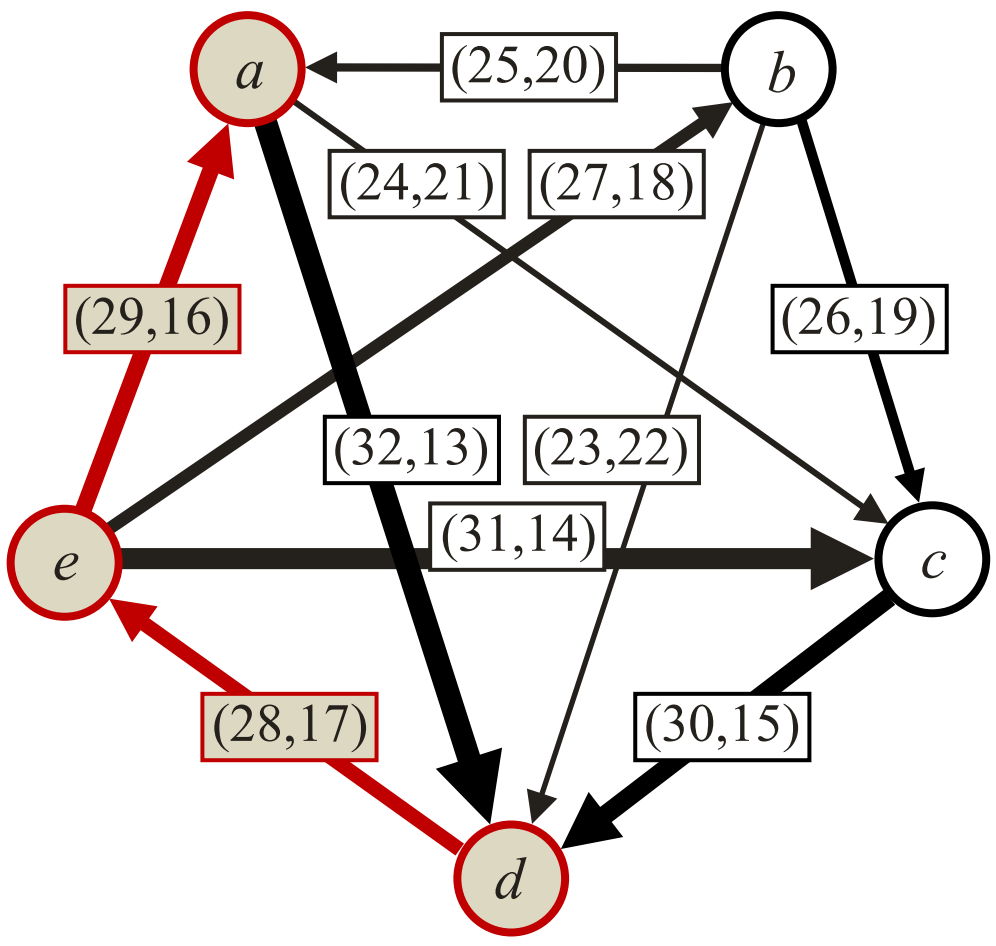

The strongest path from *d* to *a* is:
*d*, (28,17), *e*, (29,16), *a*

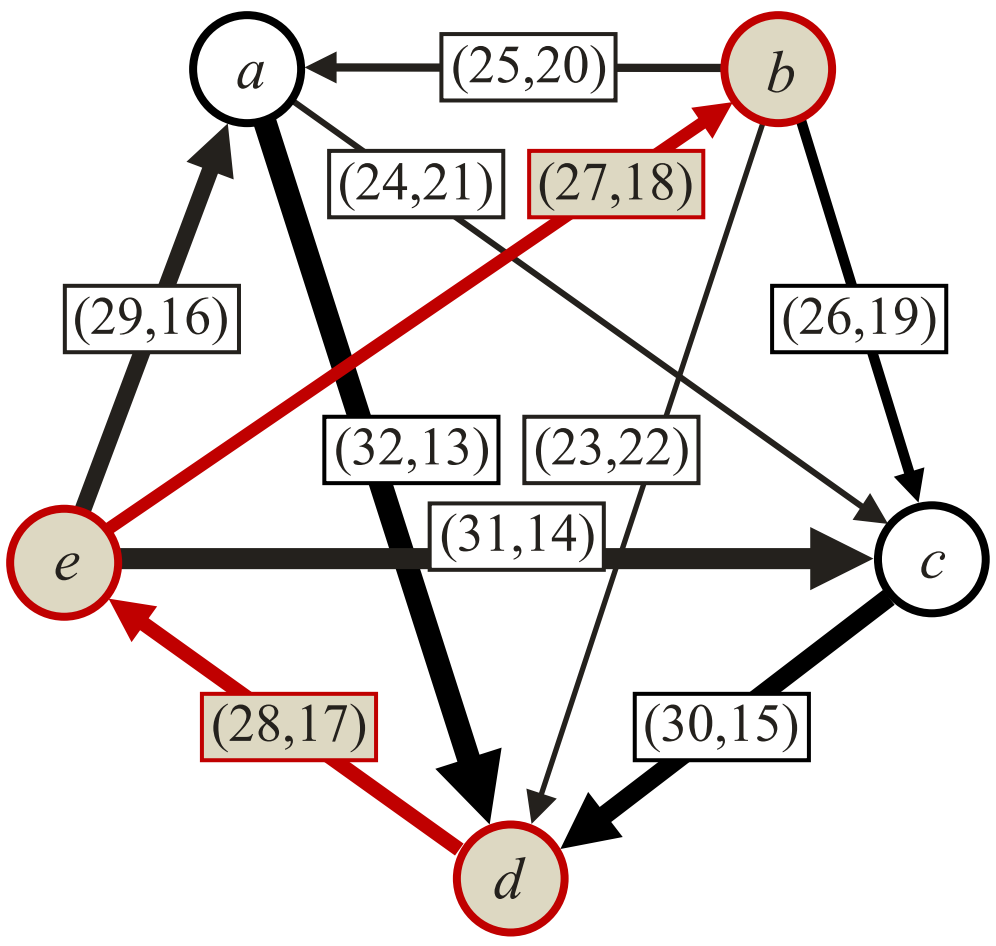

The strongest path from *d* to *b* is:
*d*, (28,17), *e*, (27,18), *b*

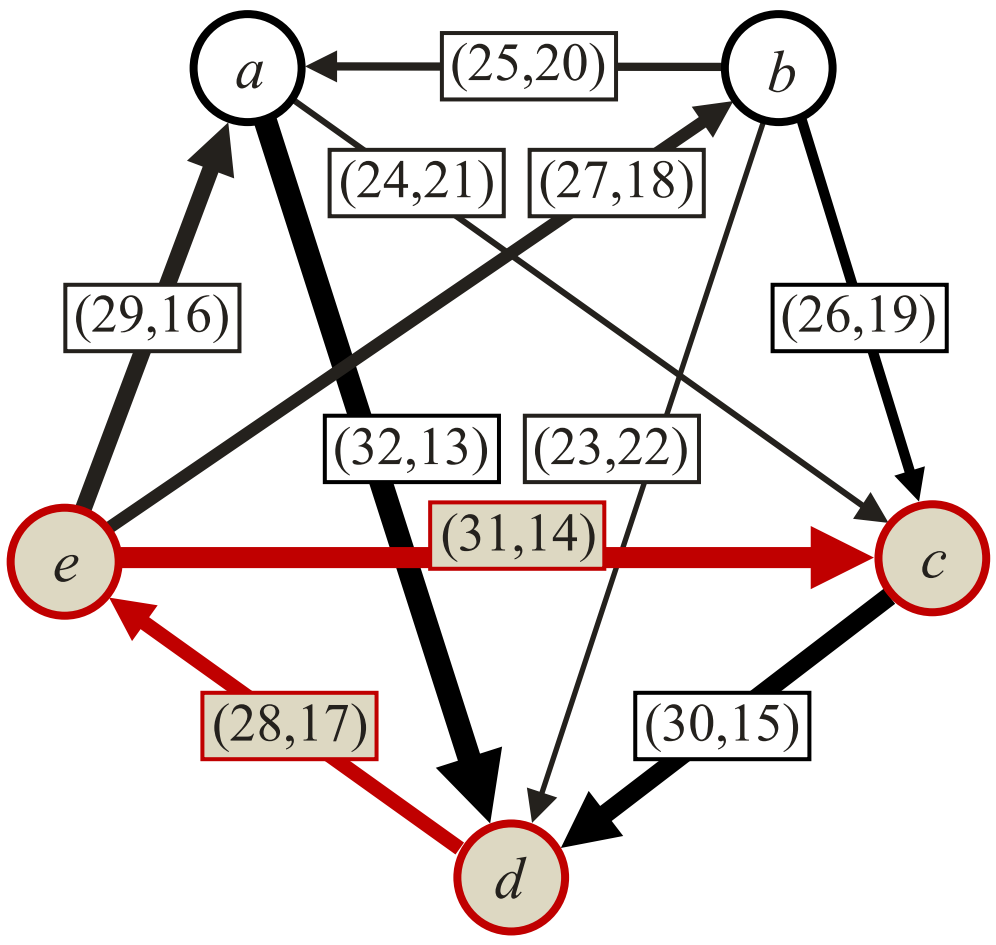

The strongest path from *d* to *c* is:
*d*, (28,17), *e*, (31,14), *c*

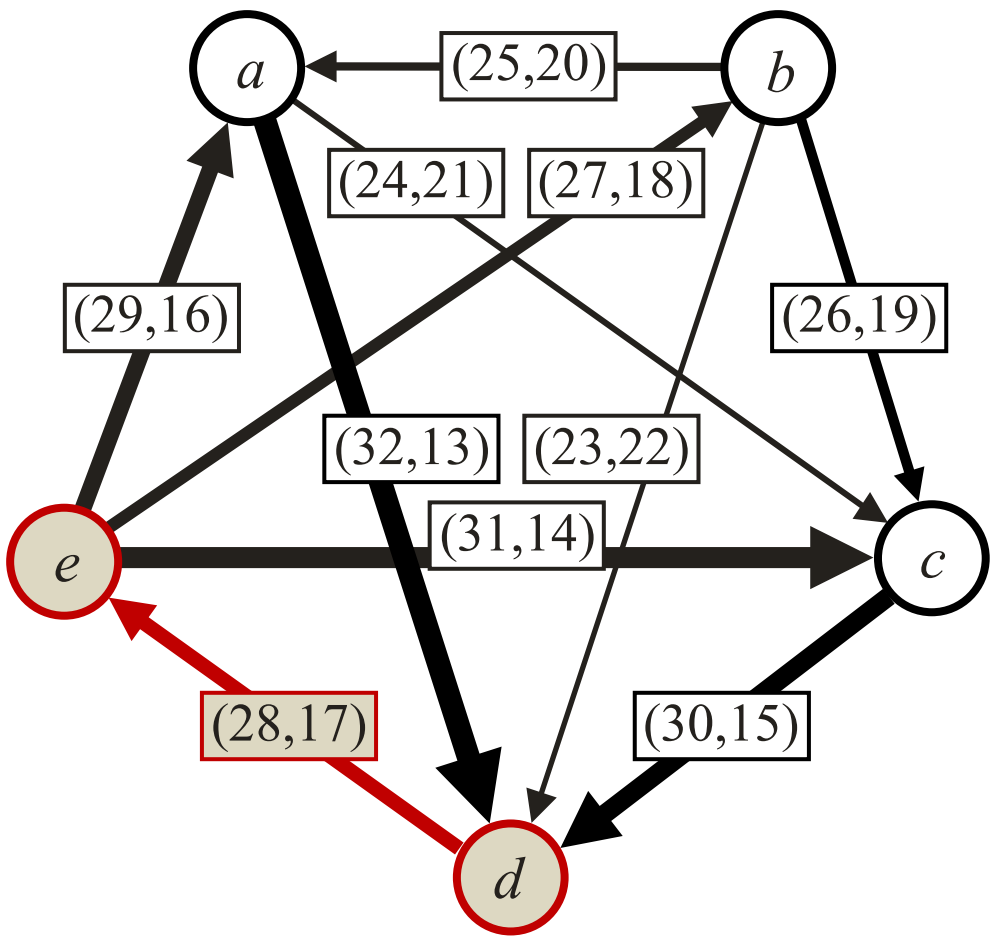

The strongest path from *d* to *e* is:
*d*, (28,17), *e*

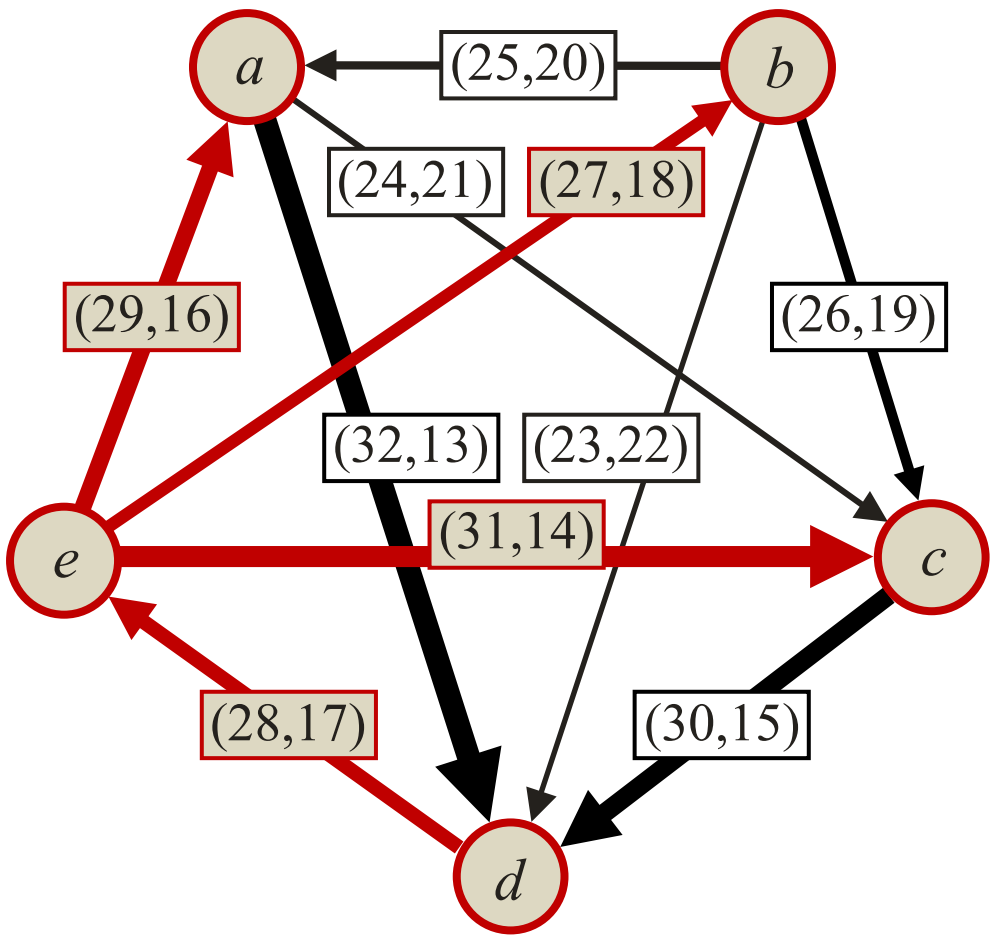

These are the strongest paths
from *d* to every other alternative.

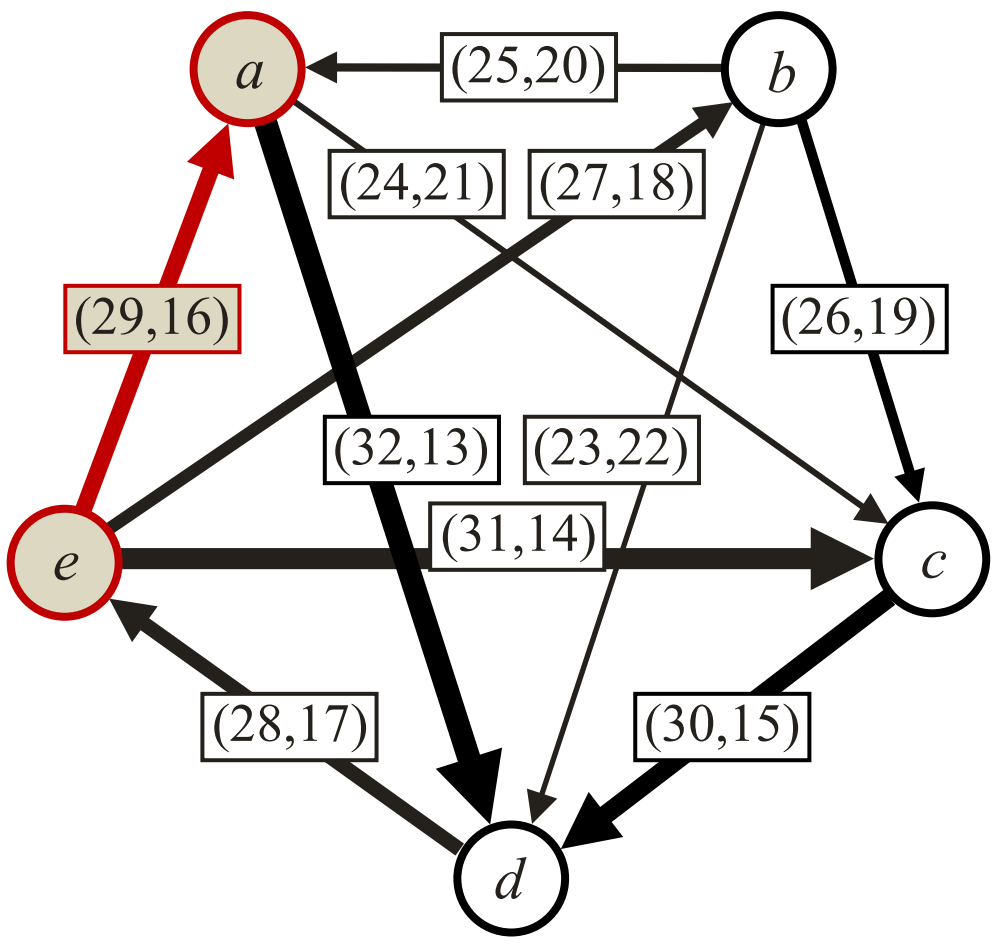

The strongest path from *e* to *a* is:
*e*, (29,16), *a*

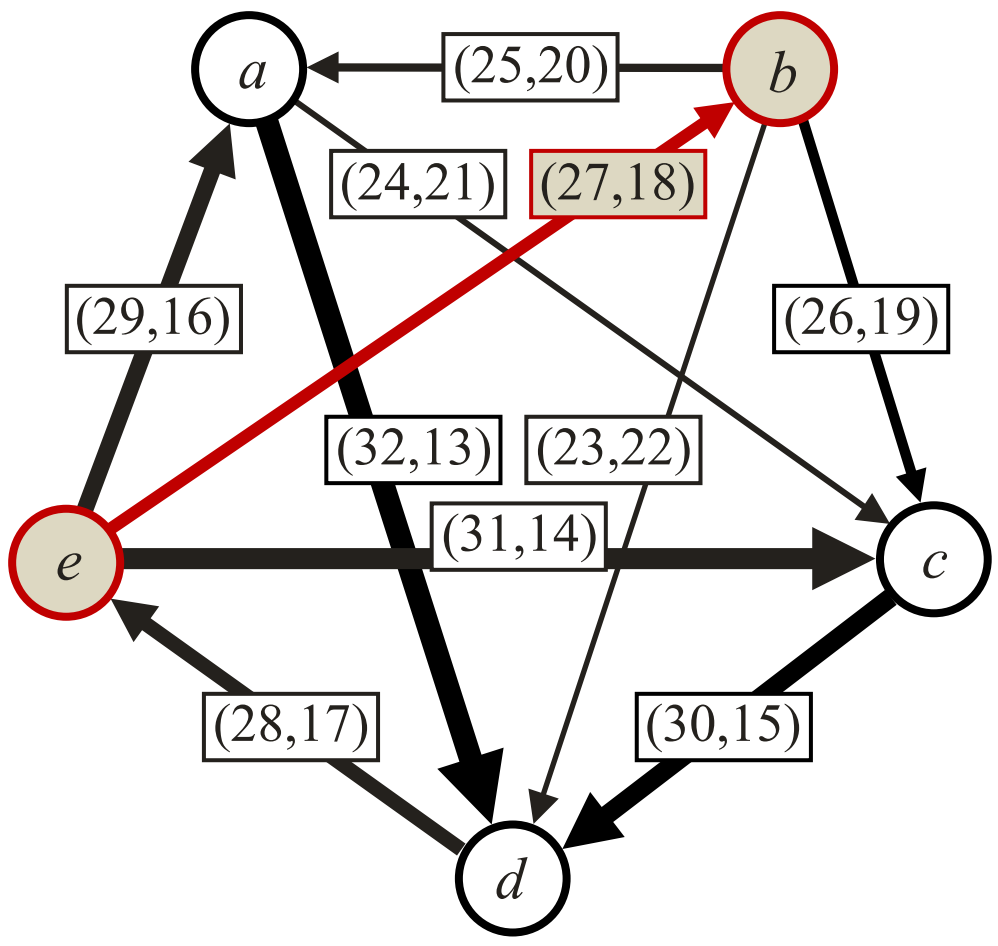

The strongest path from *e* to *b* is:
*e*, (27,18), *b*

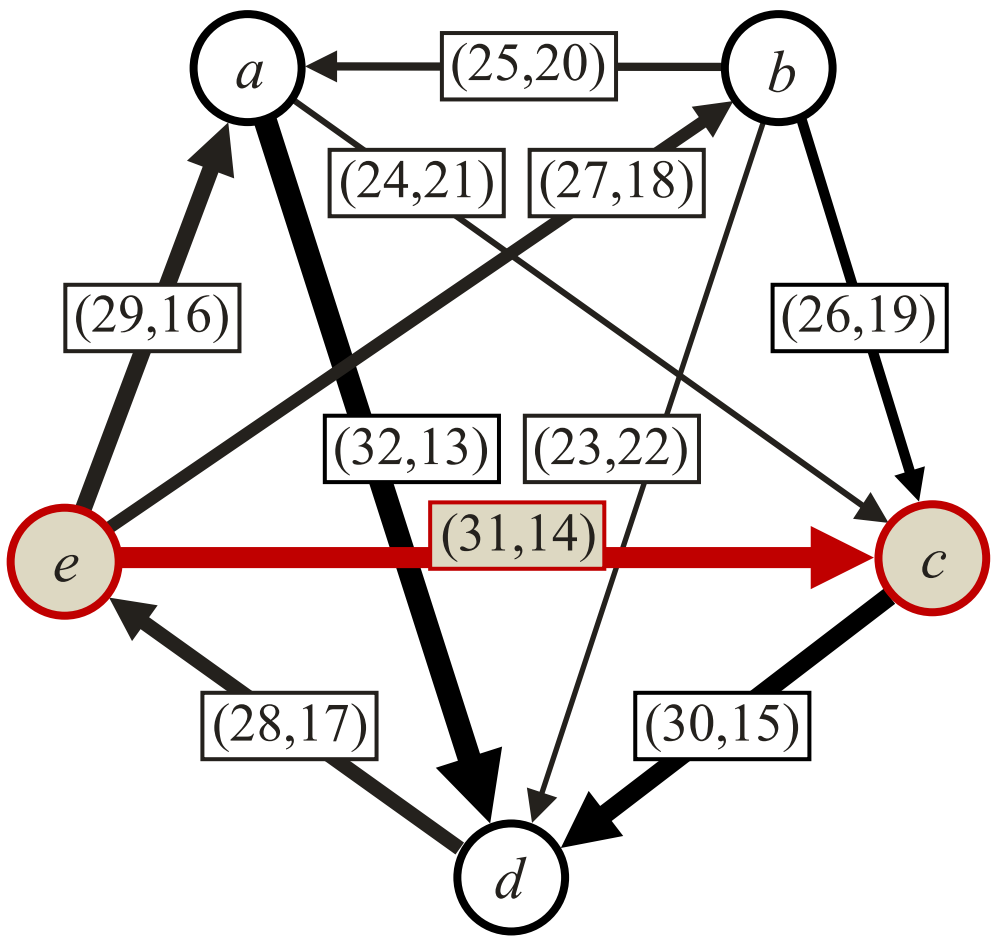

The strongest path from *e* to *c* is:
*e*, (31,14), *c*

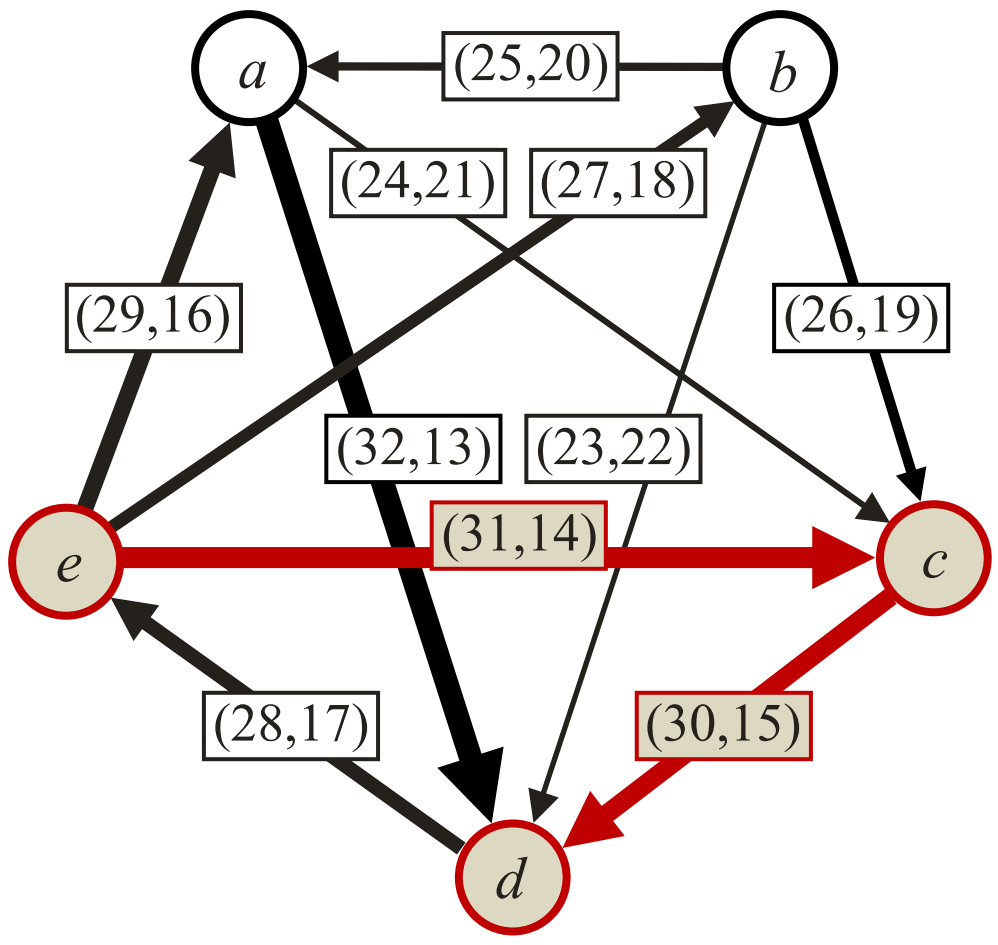

The strongest path from *e* to *d* is:
*e*, (31,14), *c*, (30,15), *d*

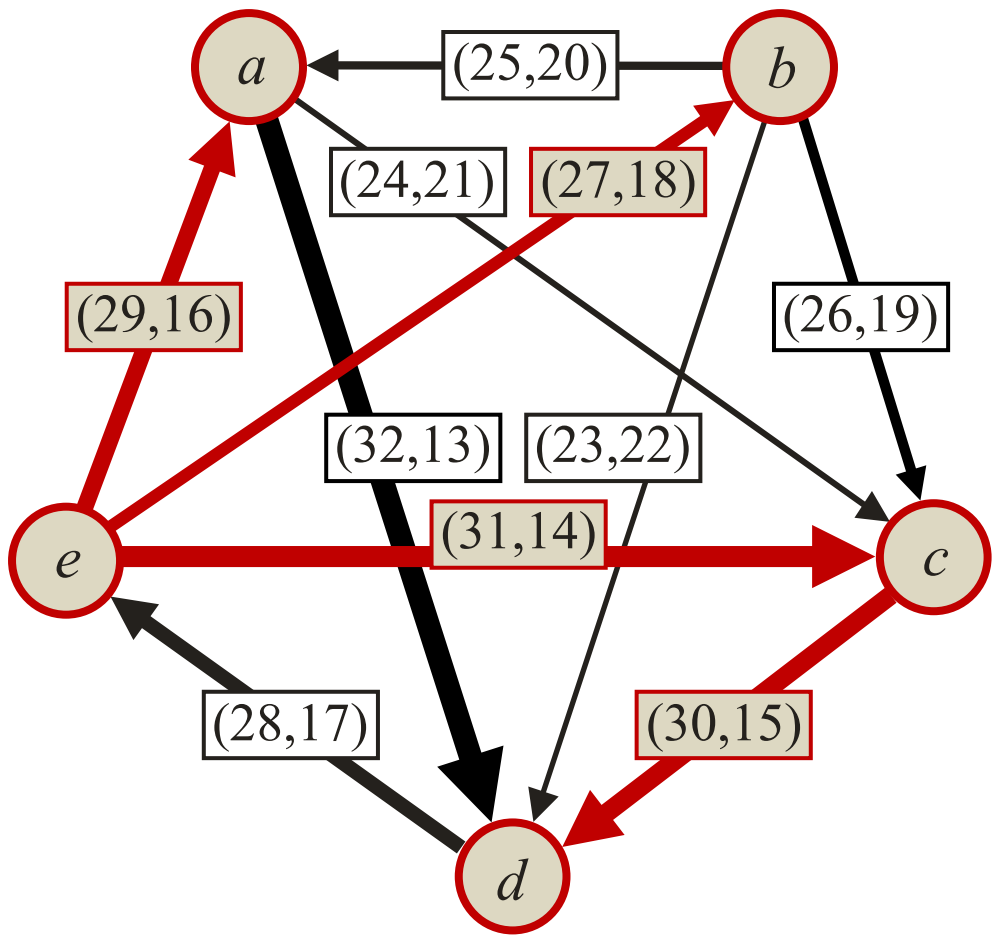

These are the strongest paths
from *e* to every other alternative.

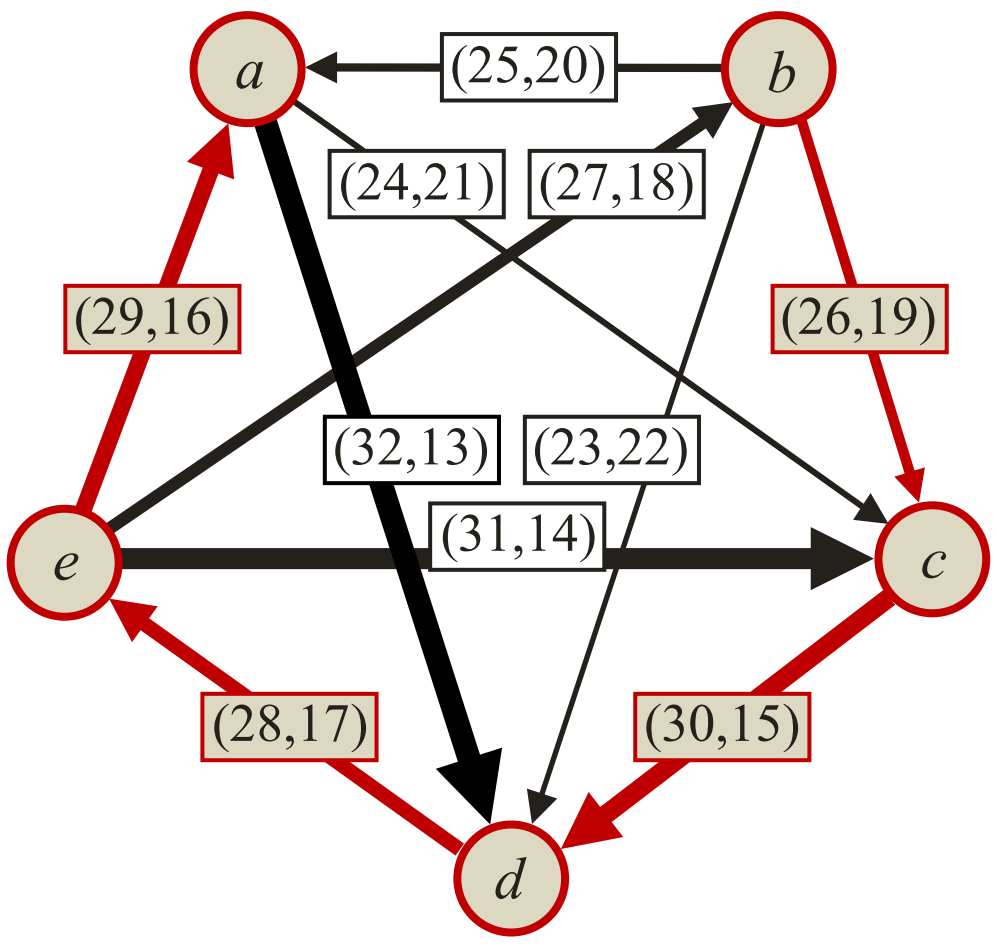

These are the strongest paths
from every other alternative to *a*.

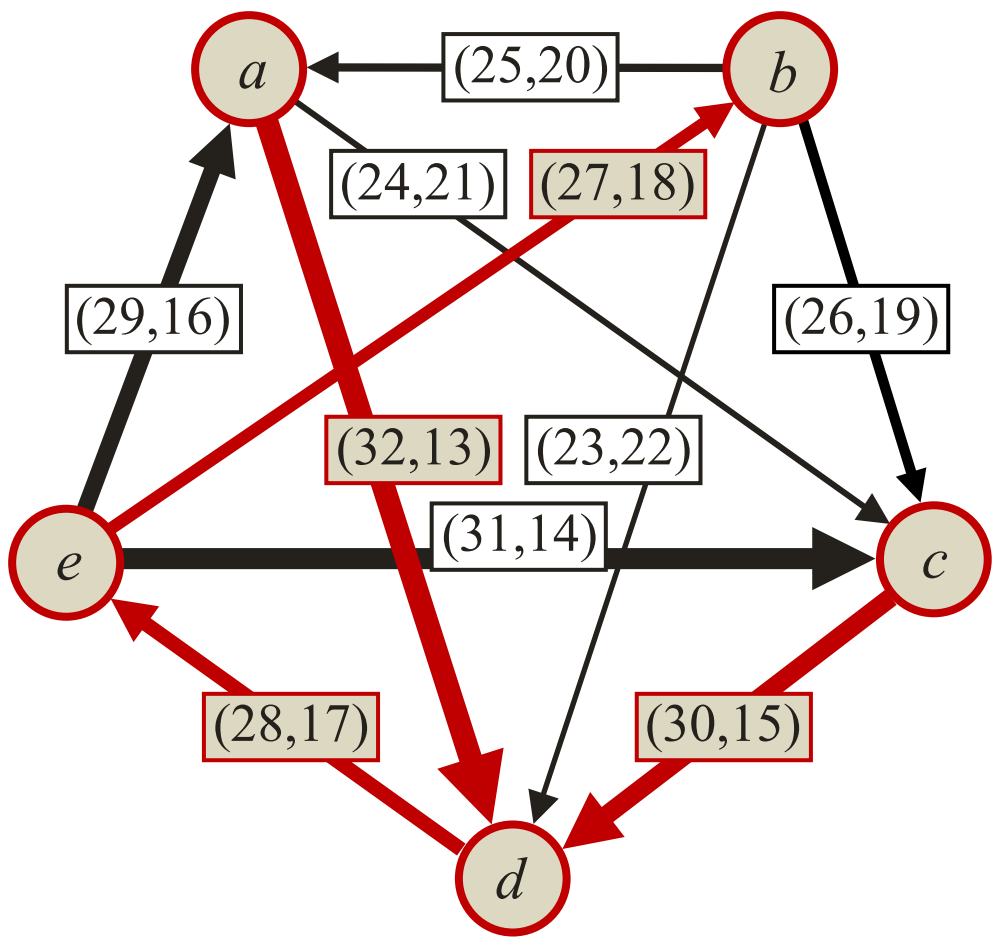

These are the strongest paths
from every other alternative to *b*.

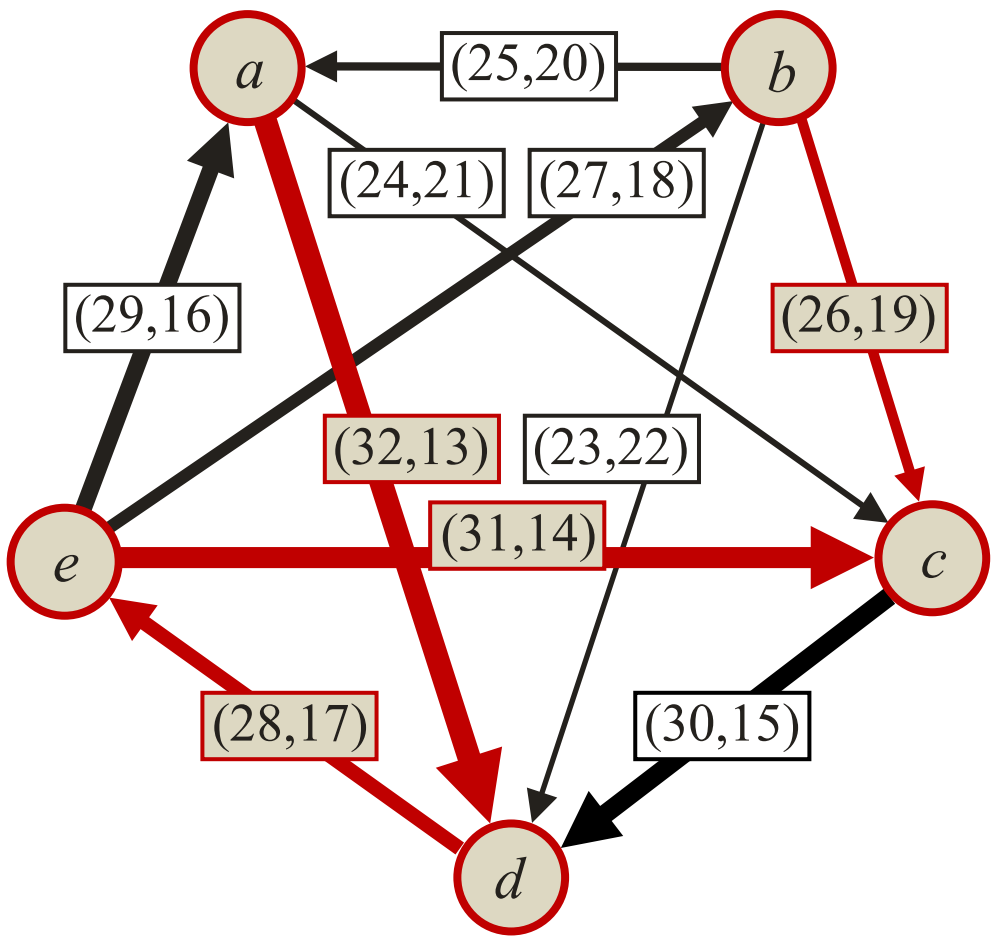

These are the strongest paths
from every other alternative to *c*.

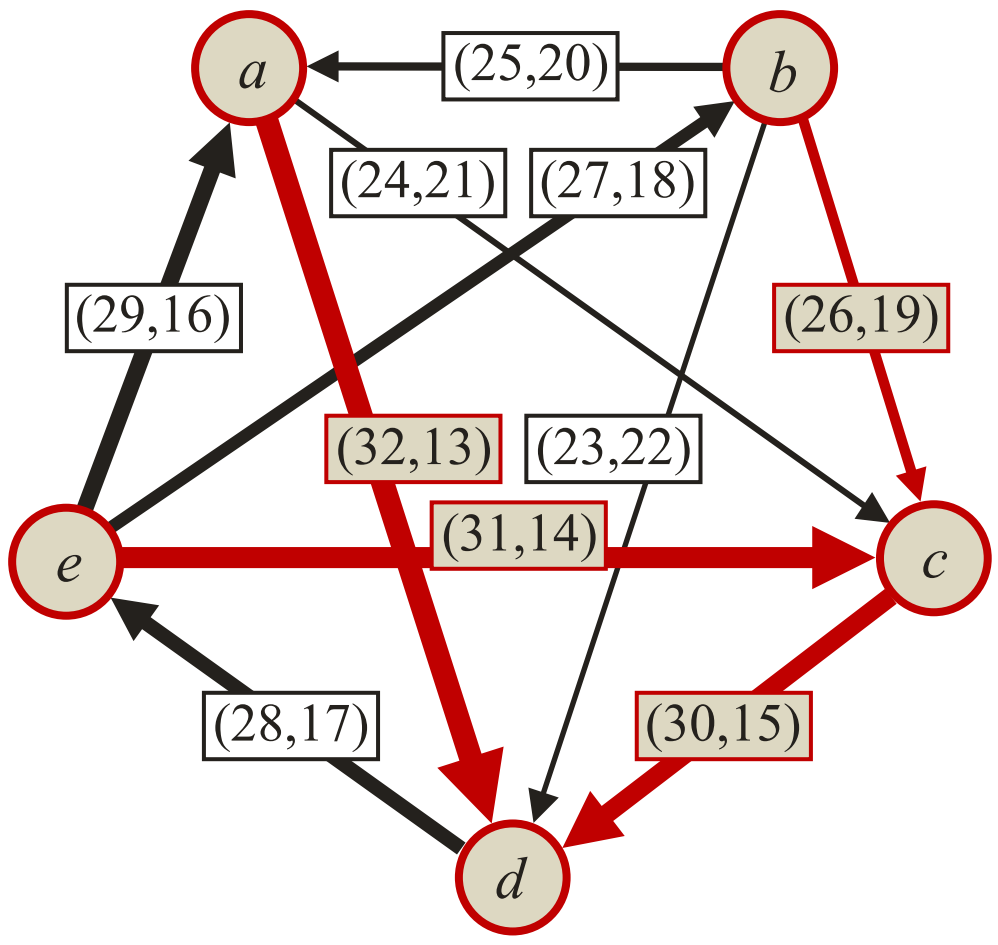

These are the strongest paths
from every other alternative to *d*.

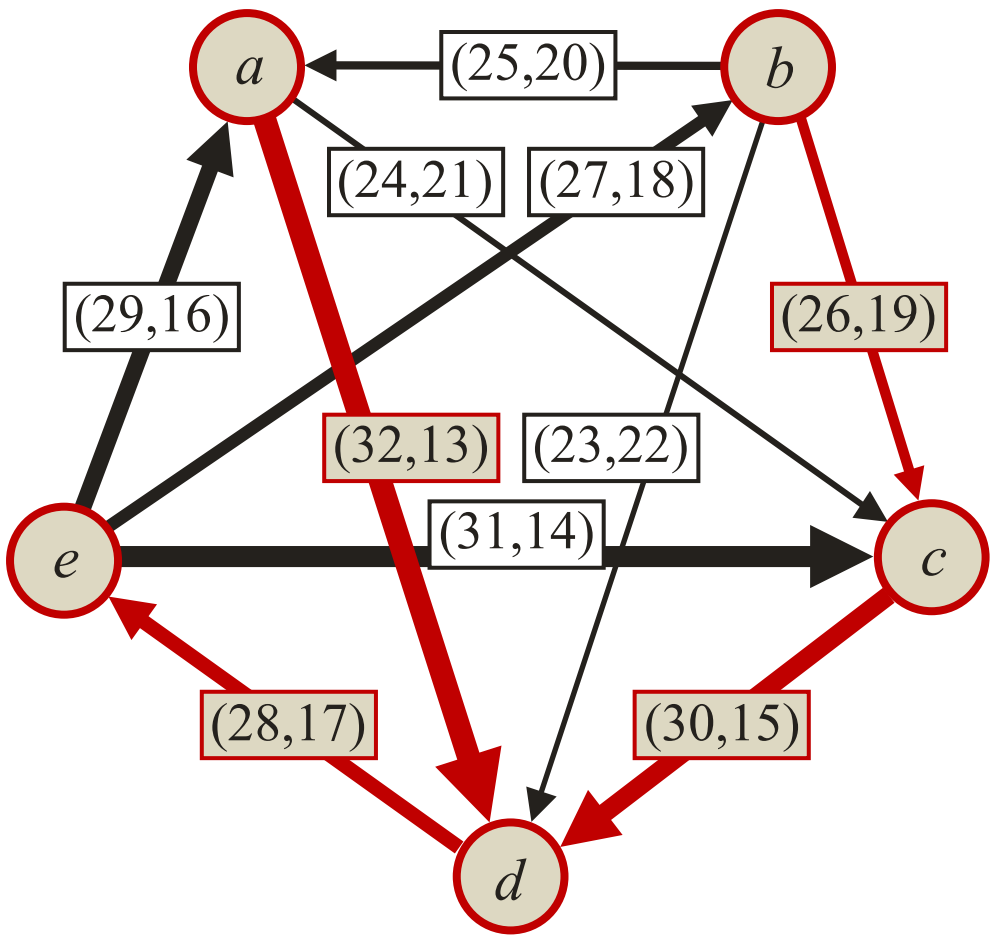

These are the strongest paths
from every other alternative to *e*.

Therefore, the strengths of the strongest paths are:

| | $P_D[*,a]$ | $P_D[*,b]$ | $P_D[*,c]$ | $P_D[*,d]$ | $P_D[*,e]$ |
|---|---|---|---|---|---|
| $P_D[a,*]$ | --- | (27,18) | (28,17) | (32,13) | (28,17) |
| $P_D[b,*]$ | (26,19) | --- | (26,19) | (26,19) | (26,19) |
| $P_D[c,*]$ | (28,17) | (27,18) | --- | (30,15) | (28,17) |
| $P_D[d,*]$ | (28,17) | (27,18) | (28,17) | --- | (28,17) |
| $P_D[e,*]$ | (29,16) | (27,18) | (31,14) | (30,15) | --- |

We get $\mathcal{O} = \{ab, ad, cb, cd, db, ea, eb, ec, ed\}$ and $\mathcal{S} = \{e\}$.

Suppose, the strongest paths are calculated with the Floyd-Warshall algorithm, as defined in section 2.3.1. Then the following table documents the $C \cdot (C-1) \cdot (C-2) = 60$ steps of the Floyd-Warshall algorithm.

We start with

- $P_D[i,j] := (N[i,j],N[j,i])$ for all $i \in A$ and $j \in A \setminus \{i\}$.
- $pred[i,j] := i$ for all $i \in A$ and $j \in A \setminus \{i\}$.

| | *i* | *j* | *k* | $P_D[j,k]$ | $P_D[j,i]$ | $P_D[i,k]$ | *pred*[*j*,*k*] | *pred*[*i*,*k*] | result |
|---|---|---|---|---|---|---|---|---|---|
| 1 | *a* | *b* | *c* | (26,19) | (25,20) | (24,21) | *b* | *a* | |
| 2 | *a* | *b* | *d* | (23,22) | (25,20) | (32,13) | *b* | *a* | $P_D[b,d]$ is updated from (23,22) to (25,20); *pred*[*b*,*d*] is updated from *b* to *a*. |
| 3 | *a* | *b* | *e* | (18,27) | (25,20) | (16,29) | *b* | *a* | |
| 4 | *a* | *c* | *b* | (19,26) | (21,24) | (20,25) | *c* | *a* | $P_D[c,b]$ is updated from (19,26) to (20,25); *pred*[*c*,*b*] is updated from *c* to *a*. |
| 5 | *a* | *c* | *d* | (30,15) | (21,24) | (32,13) | *c* | *a* | |
| 6 | *a* | *c* | *e* | (14,31) | (21,24) | (16,29) | *c* | *a* | $P_D[c,e]$ is updated from (14,31) to (16,29); *pred*[*c*,*e*] is updated from *c* to *a*. |
| 7 | *a* | *d* | *b* | (22,23) | (13,32) | (20,25) | *d* | *a* | |
| 8 | *a* | *d* | *c* | (15,30) | (13,32) | (24,21) | *d* | *a* | |
| 9 | *a* | *d* | *e* | (28,17) | (13,32) | (16,29) | *d* | *a* | |
| 10 | *a* | *e* | *b* | (27,18) | (29,16) | (20,25) | *e* | *a* | |
| 11 | *a* | *e* | *c* | (31,14) | (29,16) | (24,21) | *e* | *a* | |
| 12 | *a* | *e* | *d* | (17,28) | (29,16) | (32,13) | *e* | *a* | $P_D[e,d]$ is updated from (17,28) to (29,16); *pred*[*e*,*d*] is updated from *e* to *a*. |
| 13 | *b* | *a* | *c* | (24,21) | (20,25) | (26,19) | *a* | *b* | |
| 14 | *b* | *a* | *d* | (32,13) | (20,25) | (25,20) | *a* | *a* | |
| 15 | *b* | *a* | *e* | (16,29) | (20,25) | (18,27) | *a* | *b* | $P_D[a,e]$ is updated from (16,29) to (18,27); *pred*[*a*,*e*] is updated from *a* to *b*. |
| 16 | *b* | *c* | *a* | (21,24) | (20,25) | (25,20) | *c* | *b* | |
| 17 | *b* | *c* | *d* | (30,15) | (20,25) | (25,20) | *c* | *a* | |
| 18 | *b* | *c* | *e* | (16,29) | (20,25) | (18,27) | *a* | *b* | $P_D[c,e]$ is updated from (16,29) to (18,27); *pred*[*c*,*e*] is updated from *a* to *b*. |
| 19 | *b* | *d* | *a* | (13,32) | (22,23) | (25,20) | *d* | *b* | $P_D[d,a]$ is updated from (13,32) to (22,23); *pred*[*d*,*a*] is updated from *d* to *b*. |
| 20 | *b* | *d* | *c* | (15,30) | (22,23) | (26,19) | *d* | *b* | $P_D[d,c]$ is updated from (15,30) to (22,23); *pred*[*d*,*c*] is updated from *d* to *b*. |
| 21 | *b* | *d* | *e* | (28,17) | (22,23) | (18,27) | *d* | *b* | |
| 22 | *b* | *e* | *a* | (29,16) | (27,18) | (25,20) | *e* | *b* | |
| 23 | *b* | *e* | *c* | (31,14) | (27,18) | (26,19) | *e* | *b* | |
| 24 | *b* | *e* | *d* | (29,16) | (27,18) | (25,20) | *a* | *a* | |
| 25 | *c* | *a* | *b* | (20,25) | (24,21) | (20,25) | *a* | *a* | |
| 26 | *c* | *a* | *d* | (32,13) | (24,21) | (30,15) | *a* | *c* | |
| 27 | *c* | *a* | *e* | (18,27) | (24,21) | (18,27) | *b* | *b* | |
| 28 | *c* | *b* | *a* | (25,20) | (26,19) | (21,24) | *b* | *c* | |
| 29 | *c* | *b* | *d* | (25,20) | (26,19) | (30,15) | *a* | *c* | $P_D[b,d]$ is updated from (25,20) to (26,19); *pred*[*b*,*d*] is updated from *a* to *c*. |
| 30 | *c* | *b* | *e* | (18,27) | (26,19) | (18,27) | *b* | *b* | |

| | *i* | *j* | *k* | $P_D[j,k]$ | $P_D[j,i]$ | $P_D[i,k]$ | *pred*[*j*,*k*] | *pred*[*i*,*k*] | result |
|---|---|---|---|---|---|---|---|---|---|
| 31 | *c* | *d* | *a* | (22,23) | (22,23) | (21,24) | *b* | *c* | |
| 32 | *c* | *d* | *b* | (22,23) | (22,23) | (20,25) | *d* | *a* | |
| 33 | *c* | *d* | *e* | (28,17) | (22,23) | (18,27) | *d* | *b* | |
| 34 | *c* | *e* | *a* | (29,16) | (31,14) | (21,24) | *e* | *c* | |
| 35 | *c* | *e* | *b* | (27,18) | (31,14) | (20,25) | *e* | *a* | |
| 36 | *c* | *e* | *d* | (29,16) | (31,14) | (30,15) | *a* | *c* | $P_D[e,d]$ is updated from (29,16) to (30,15); *pred*[*e*,*d*] is updated from *a* to *c*. |
| 37 | *d* | *a* | *b* | (20,25) | (32,13) | (22,23) | *a* | *d* | $P_D[a,b]$ is updated from (20,25) to (22,23); *pred*[*a*,*b*] is updated from *a* to *d*. |
| 38 | *d* | *a* | *c* | (24,21) | (32,13) | (22,23) | *a* | *b* | |
| 39 | *d* | *a* | *e* | (18,27) | (32,13) | (28,17) | *b* | *d* | $P_D[a,e]$ is updated from (18,27) to (28,17); *pred*[*a*,*e*] is updated from *b* to *d*. |
| 40 | *d* | *b* | *a* | (25,20) | (26,19) | (22,23) | *b* | *b* | |
| 41 | *d* | *b* | *c* | (26,19) | (26,19) | (22,23) | *b* | *b* | |
| 42 | *d* | *b* | *e* | (18,27) | (26,19) | (28,17) | *b* | *d* | $P_D[b,e]$ is updated from (18,27) to (26,19); *pred*[*b*,*e*] is updated from *b* to *d*. |
| 43 | *d* | *c* | *a* | (21,24) | (30,15) | (22,23) | *c* | *b* | $P_D[c,a]$ is updated from (21,24) to (22,23); *pred*[*c*,*a*] is updated from *c* to *b*. |
| 44 | *d* | *c* | *b* | (20,25) | (30,15) | (22,23) | *a* | *d* | $P_D[c,b]$ is updated from (20,25) to (22,23); *pred*[*c*,*b*] is updated from *a* to *d*. |
| 45 | *d* | *c* | *e* | (18,27) | (30,15) | (28,17) | *b* | *d* | $P_D[c,e]$ is updated from (18,27) to (28,17); *pred*[*c*,*e*] is updated from *b* to *d*. |
| 46 | *d* | *e* | *a* | (29,16) | (30,15) | (22,23) | *e* | *b* | |
| 47 | *d* | *e* | *b* | (27,18) | (30,15) | (22,23) | *e* | *d* | |
| 48 | *d* | *e* | *c* | (31,14) | (30,15) | (22,23) | *e* | *b* | |
| 49 | *e* | *a* | *b* | (22,23) | (28,17) | (27,18) | *d* | *e* | $P_D[a,b]$ is updated from (22,23) to (27,18); *pred*[*a*,*b*] is updated from *d* to *e*. |
| 50 | *e* | *a* | *c* | (24,21) | (28,17) | (31,14) | *a* | *e* | $P_D[a,c]$ is updated from (24,21) to (28,17); *pred*[*a*,*c*] is updated from *a* to *e*. |
| 51 | *e* | *a* | *d* | (32,13) | (28,17) | (30,15) | *a* | *c* | |
| 52 | *e* | *b* | *a* | (25,20) | (26,19) | (29,16) | *b* | *e* | $P_D[b,a]$ is updated from (25,20) to (26,19); *pred*[*b*,*a*] is updated from *b* to *e*. |
| 53 | *e* | *b* | *c* | (26,19) | (26,19) | (31,14) | *b* | *e* | |
| 54 | *e* | *b* | *d* | (26,19) | (26,19) | (30,15) | *c* | *c* | |
| 55 | *e* | *c* | *a* | (22,23) | (28,17) | (29,16) | *b* | *e* | $P_D[c,a]$ is updated from (22,23) to (28,17); *pred*[*c*,*a*] is updated from *b* to *e*. |
| 56 | *e* | *c* | *b* | (22,23) | (28,17) | (27,18) | *d* | *e* | $P_D[c,b]$ is updated from (22,23) to (27,18); *pred*[*c*,*b*] is updated from *d* to *e*. |
| 57 | *e* | *c* | *d* | (30,15) | (28,17) | (30,15) | *c* | *c* | |
| 58 | *e* | *d* | *a* | (22,23) | (28,17) | (29,16) | *b* | *e* | $P_D[d,a]$ is updated from (22,23) to (28,17); *pred*[*d*,*a*] is updated from *b* to *e*. |
| 59 | *e* | *d* | *b* | (22,23) | (28,17) | (27,18) | *d* | *e* | $P_D[d,b]$ is updated from (22,23) to (27,18); *pred*[*d*,*b*] is updated from *d* to *e*. |
| 60 | *e* | *d* | *c* | (22,23) | (28,17) | (31,14) | *b* | *e* | $P_D[d,c]$ is updated from (22,23) to (28,17); *pred*[*d*,*c*] is updated from *b* to *e*. |

### 3.13. Example 13

Example 13:

| | | |
|---|---|---|
| 2 | voters | $a \succ_v b \succ_v c$ |
| 2 | voters | $b \succ_v c \succ_v a$ |
| 1 | voter | $c \succ_v a \succ_v b$ |

The pairwise matrix *N* looks as follows:

| | *N*[*,*a*] | *N*[*,*b*] | *N*[*,*c*] |
|---|---|---|---|
| *N*[*a*,*] | --- | 3 | 2 |
| *N*[*b*,*] | 2 | --- | 4 |
| *N*[*c*,*] | 3 | 1 | --- |

The corresponding digraph looks as follows:

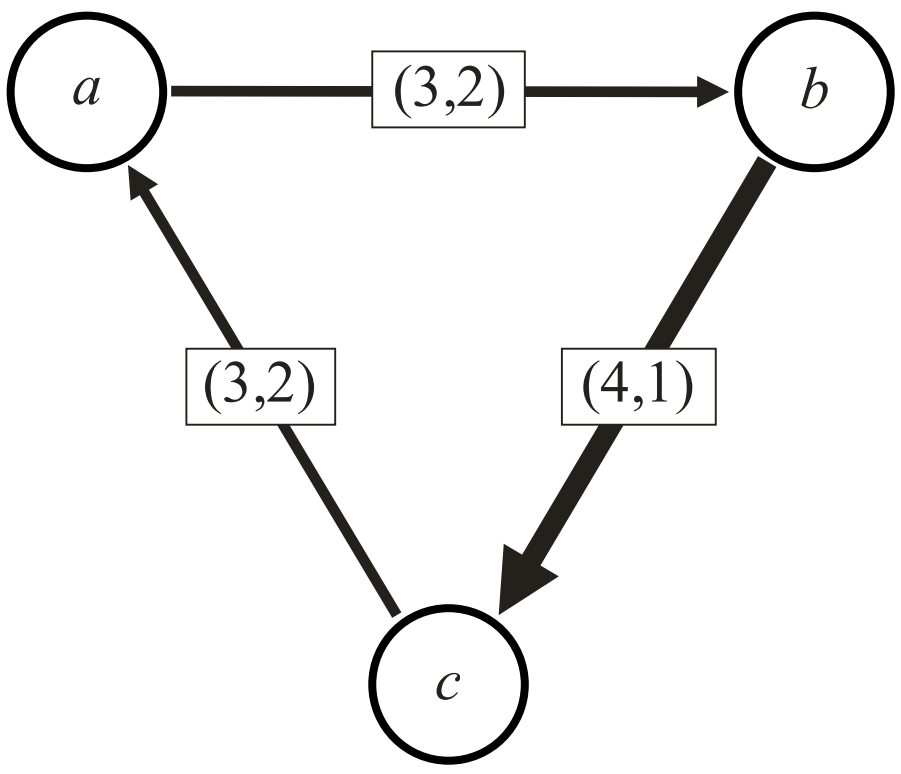

The following table lists the strongest paths, as determined by the Floyd-Warshall algorithm, as defined in section 2.3.1. The critical links of the strongest paths are underlined:

| | ... to *a* | ... to *b* | ... to *c* | ... to every other alternative |
|---|---|---|---|---|
| from *a* ... | --- | 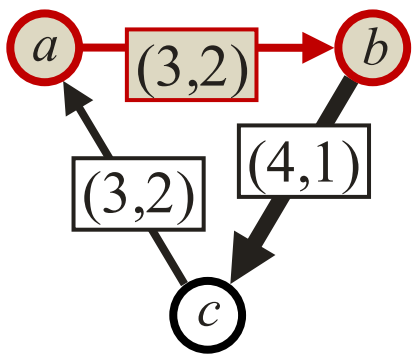 *a*, (3,2), *b* | 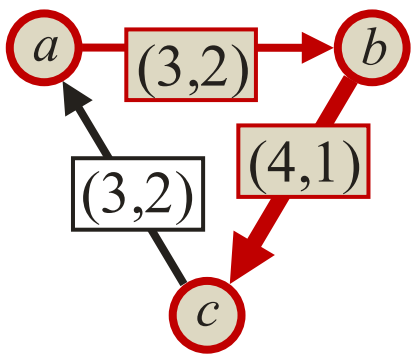 *a*, (3,2), *b*, (4,1), *c* | 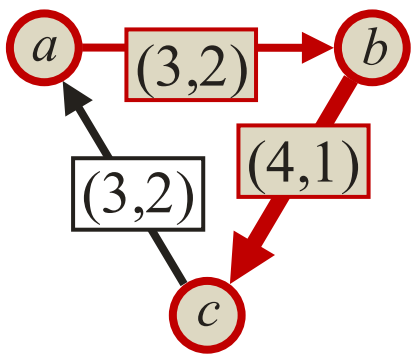 |
| from *b* ... | 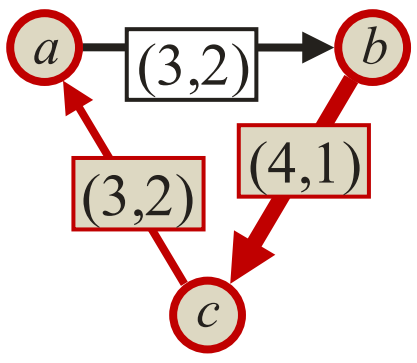 *b*, (4,1), *c*, (3,2), *a* | --- | 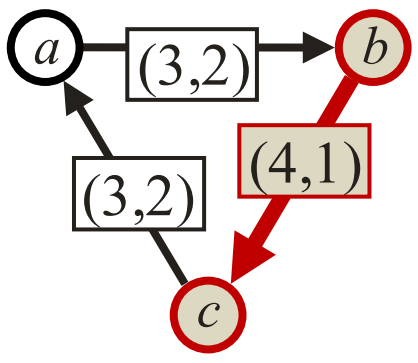 *b*, (4,1), *c* | 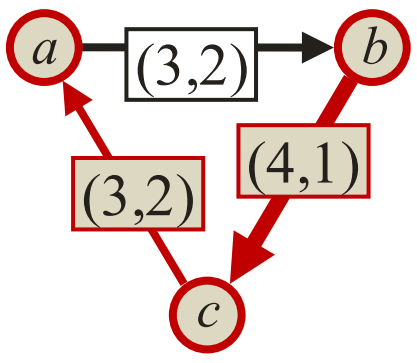 |
| from *c* ... | 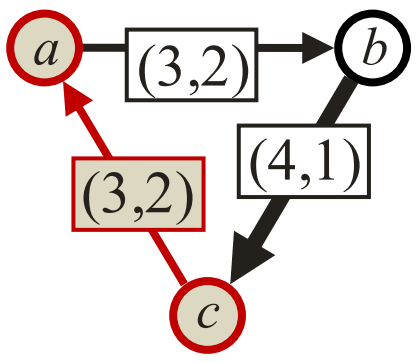 *c*, (3,2), *a* | 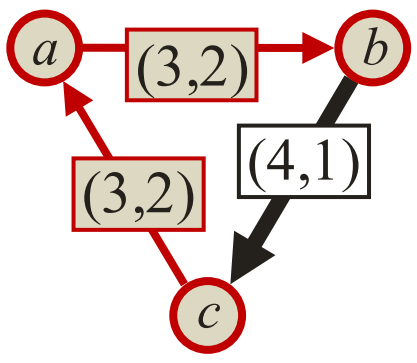 *c*, (3,2), *a*, (3,2), *b* | --- | 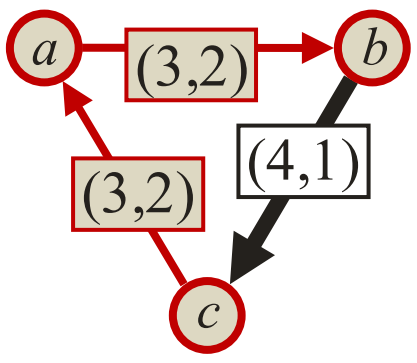 |
| from every other alternative ... | 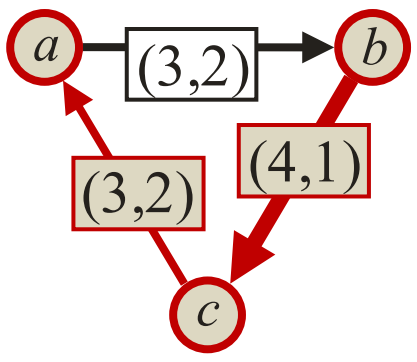 | 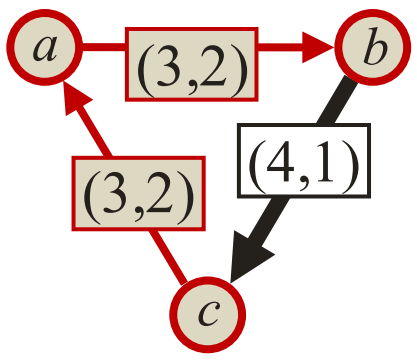 | 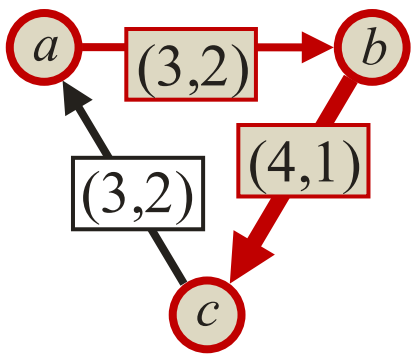 | --- |

The strengths of the strongest paths are:

| | $P_D[*,a]$ | $P_D[*,b]$ | $P_D[*,c]$ |
|---|---|---|---|
| $P_D[a,*]$ | --- | (3,2) | (3,2) |
| $P_D[b,*]$ | (3,2) | --- | (4,1) |
| $P_D[c,*]$ | (3,2) | (3,2) | --- |

We get $\mathcal{O} = \{bc\}$ and $\mathcal{S} = \{a, b\}$.

Suppose, the strongest paths are calculated with the Floyd-Warshall algorithm, as defined in section 2.3.1. Then the following table documents the $C \cdot (C-1) \cdot (C-2) = 6$ steps of the Floyd-Warshall algorithm.

We start with

- $P_D[i,j] := (N[i,j],N[j,i])$ for all $i \in A$ and $j \in A \setminus \{i\}$.
- $pred[i,j] := i$ for all $i \in A$ and $j \in A \setminus \{i\}$.

| | $i$ | $j$ | $k$ | $P_D[j,k]$ | $P_D[j,i]$ | $P_D[i,k]$ | $pred[j,k]$ | $pred[i,k]$ | result |
|---|---|---|---|---|---|---|---|---|---|
| 1 | $a$ | $b$ | $c$ | (4,1) | (2,3) | (2,3) | $b$ | $a$ | |
| 2 | $a$ | $c$ | $b$ | (1,4) | (3,2) | (3,2) | $c$ | $a$ | $P_D[c,b]$ is updated from (1,4) to (3,2); $pred[c,b]$ is updated from $c$ to $a$. |
| 3 | $b$ | $a$ | $c$ | (2,3) | (3,2) | (4,1) | $a$ | $b$ | $P_D[a,c]$ is updated from (2,3) to (3,2); $pred[a,c]$ is updated from $a$ to $b$. |
| 4 | $b$ | $c$ | $a$ | (3,2) | (3,2) | (2,3) | $c$ | $b$ | |
| 5 | $c$ | $a$ | $b$ | (3,2) | (3,2) | (3,2) | $a$ | $a$ | |
| 6 | $c$ | $b$ | $a$ | (2,3) | (4,1) | (3,2) | $b$ | $c$ | $P_D[b,a]$ is updated from (2,3) to (3,2); $pred[b,a]$ is updated from $b$ to $c$. |

## 3.14. Example 14

Example 14:

| | | |
|---|---|---|
| 1 | voter | $a >_v b >_v d >_v e >_v c >_v f$ |
| 11 | voters | $a >_v b >_v e >_v f >_v c >_v d$ |
| 5 | voters | $a >_v d >_v b >_v e >_v c >_v f$ |
| 12 | voters | $a >_v e >_v f >_v d >_v b >_v c$ |
| 7 | voters | $b >_v c >_v f >_v a >_v d >_v e$ |
| 1 | voter | $c >_v a >_v f >_v b >_v d >_v e$ |
| 15 | voters | $c >_v b >_v f >_v d >_v a >_v e$ |
| 3 | voters | $c >_v d >_v b >_v a >_v e >_v f$ |
| 3 | voters | $c >_v d >_v b >_v f >_v a >_v e$ |
| 2 | voters | $d >_v c >_v a >_v b >_v e >_v f$ |
| 11 | voters | $d >_v c >_v f >_v b >_v e >_v a$ |
| 14 | voters | $d >_v e >_v f >_v a >_v b >_v c$ |
| 1 | voter | $e >_v c >_v f >_v d >_v b >_v a$ |
| 5 | voters | $e >_v f >_v b >_v c >_v a >_v d$ |
| 5 | voters | $e >_v f >_v c >_v d >_v a >_v b$ |
| 1 | voter | $f >_v a >_v d >_v b >_v e >_v c$ |

The pairwise matrix $N$ looks as follows:

| | $N[*,a]$ | $N[*,b]$ | $N[*,c]$ | $N[*,d]$ | $N[*,e]$ | $N[*,f]$ |
|---|---|---|---|---|---|---|
| $N[a,*]$ | --- | 52 | 44 | 43 | 61 | 35 |
| $N[b,*]$ | 45 | --- | 56 | 40 | 60 | 47 |
| $N[c,*]$ | 53 | 41 | --- | 51 | 42 | 49 |
| $N[d,*]$ | 54 | 57 | 46 | --- | 63 | 39 |
| $N[e,*]$ | 36 | 37 | 55 | 34 | --- | 59 |
| $N[f,*]$ | 62 | 50 | 48 | 58 | 38 | --- |

The corresponding digraph looks as follows:

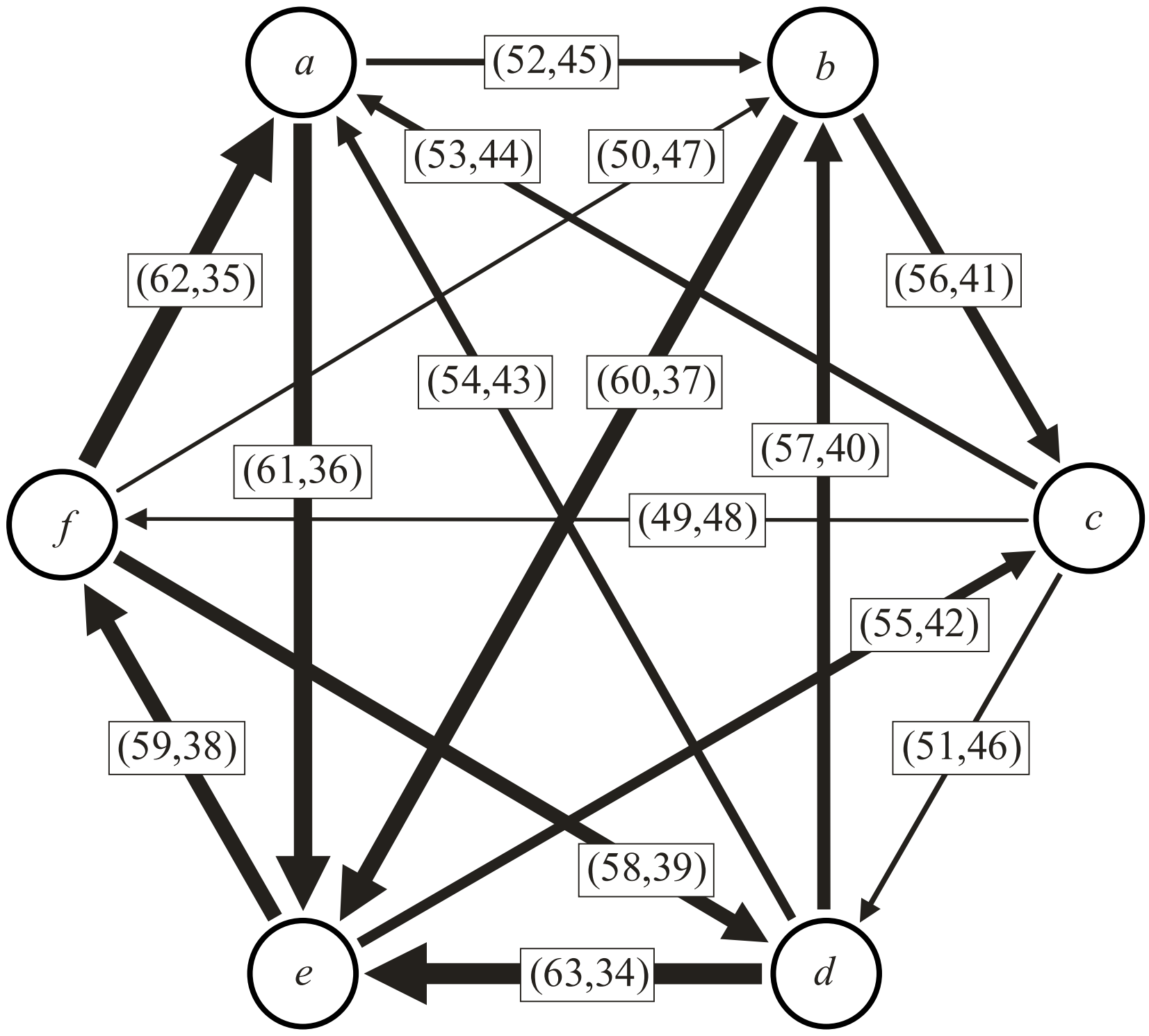

The following table lists the strongest paths, as determined by the Floyd-Warshall algorithm, as defined in section 2.3.1. The critical links of the strongest paths are underlined:

| | ... to *a* | ... to *b* | ... to *c* | ... to *d* | ... to *e* | ... to *f* |
|---|---|---|---|---|---|---|
| from *a* ... | --- | *a*, (61,36), *e*, (59,38), *f*, (58,39), *d*, (57,40), *b* | *a*, (61,36), *e*, (59,38), *f*, (58,39), *d*, (57,40), *b*, (56,41), *c* | *a*, (61,36), *e*, (59,38), *f*, (58,39), *d* | *a*, (61,36), *e* | *a*, (61,36), *e*, (59,38), *f* |
| from *b* ... | *b*, (60,37), *e*, (59,38), *f*, (62,35), *a* | --- | *b*, (56,41), *c* | *b*, (60,37), *e*, (59,38), *f*, (58,39), *d* | *b*, (60,37), *e* | *b*, (60,37), *e*, (59,38), *f* |
| from *c* ... | *c*, (53,44), *a* | *c*, (53,44), *a*, (61,36), *e*, (59,38), *f*, (58,39), *d*, (57,40), *b* | --- | *c*, (53,44), *a*, (61,36), *e*, (59,38), *f*, (58,39), *d* | *c*, (53,44), *a*, (61,36), *e* | *c*, (53,44), *a*, (61,36), *e*, (59,38), *f* |
| from *d* ... | *d*, (63,34), *e*, (59,38), *f*, (62,35), *a* | *d*, (57,40), *b* | *d*, (57,40), *b*, (56,41), *c* | --- | *d*, (63,34), *e* | *d*, (63,34), *e*, (59,38), *f* |
| from *e* ... | *e*, (59,38), *f*, (62,35), *a* | *e*, (59,38), *f*, (58,39), *d*, (57,40), *b* | *e*, (59,38), *f*, (58,39), *d*, (57,40), *b*, (56,41), *c* | *e*, (59,38), *f*, (58,39), *d* | --- | *e*, (59,38), *f* |
| from *f* ... | *f*, (62,35), *a* | *f*, (58,39), *d*, (57,40), *b* | *f*, (58,39), *d*, (57,40), *b*, (56,41), *c* | *f*, (58,39), *d* | *f*, (62,35), *a*, (61,36), *e* | --- |

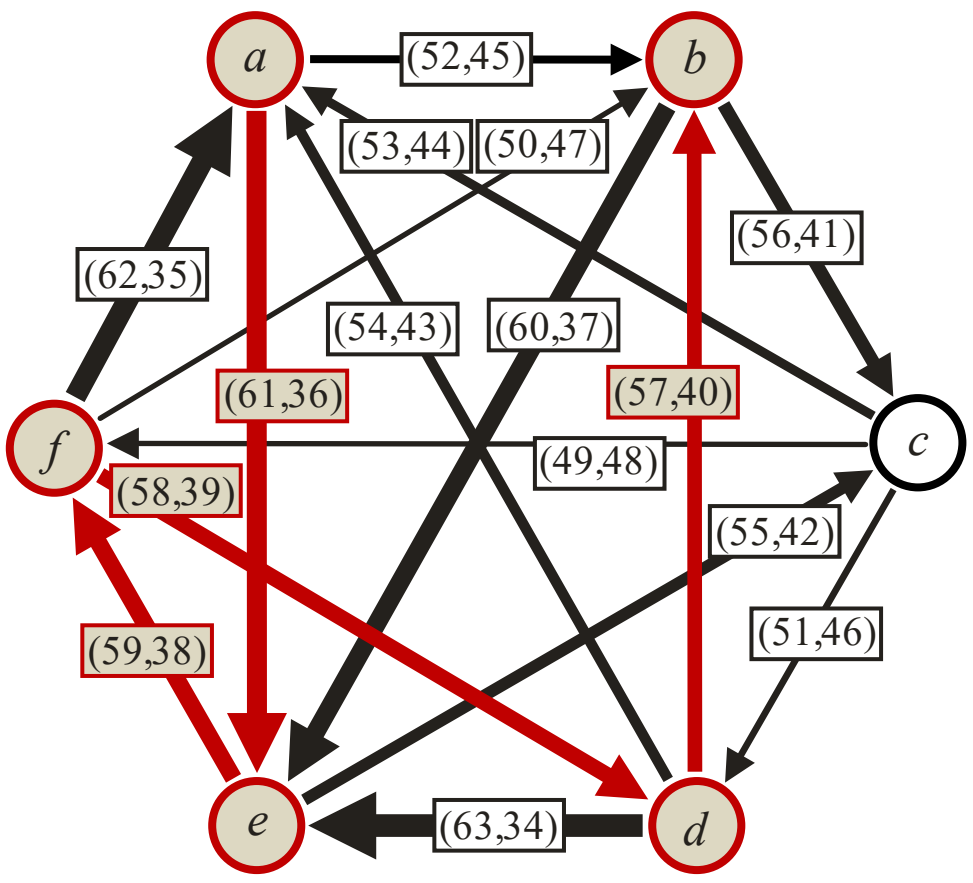

The strongest path from *a* to *b* is:
*a*, (61,36), *e*, (59,38), *f*, (58,39), *d*,
(57,40), *b*

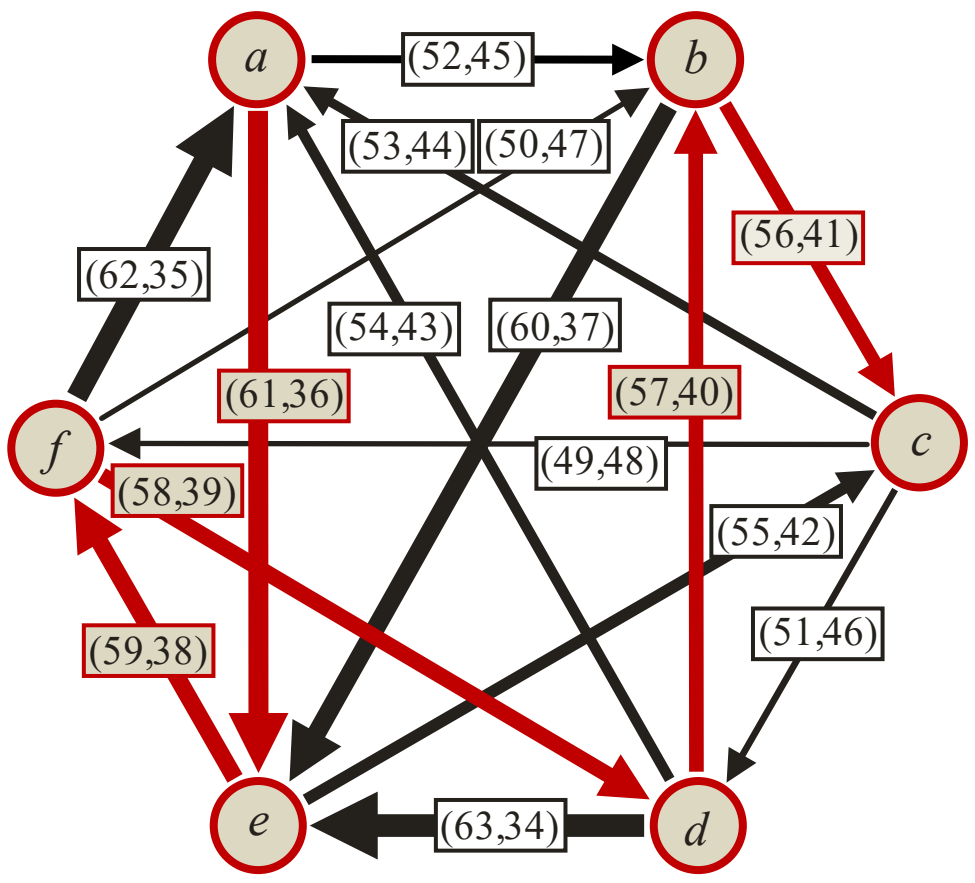

The strongest path from *a* to *c* is:
*a*, (61,36), *e*, (59,38), *f*, (58,39), *d*,
(57,40), *b*, (56,41), *c*

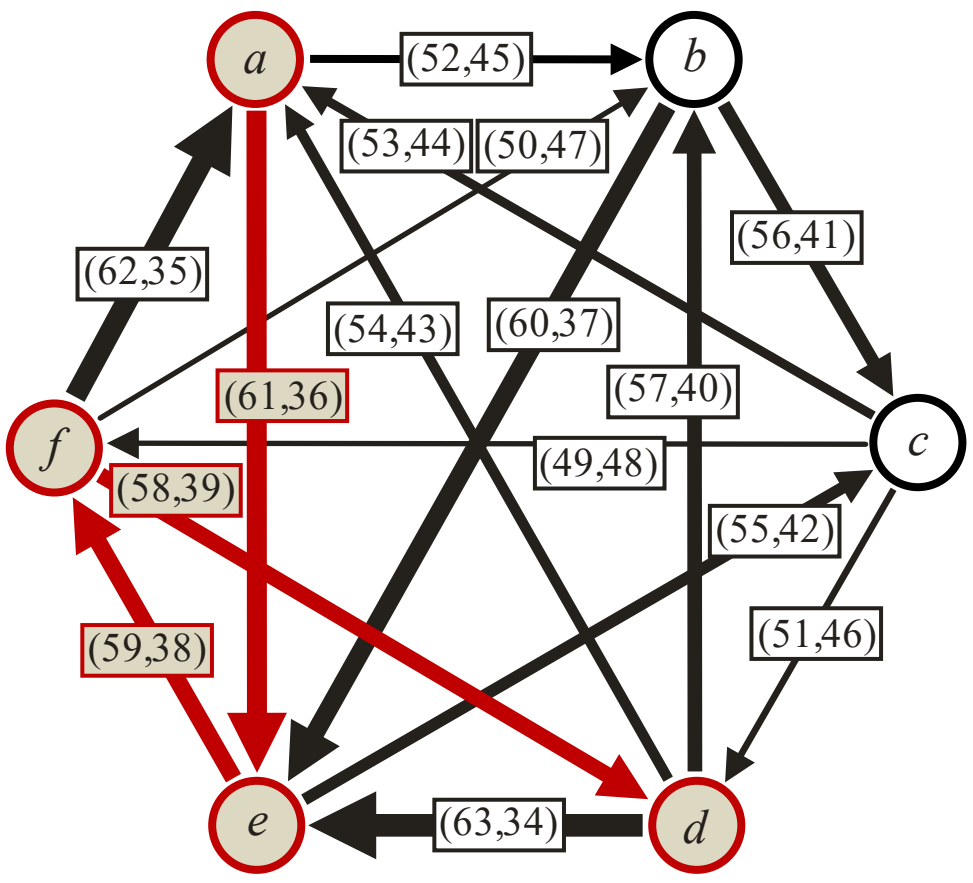

The strongest path from *a* to *d* is:
*a*, (61,36), *e*, (59,38), *f*, (58,39), *d*

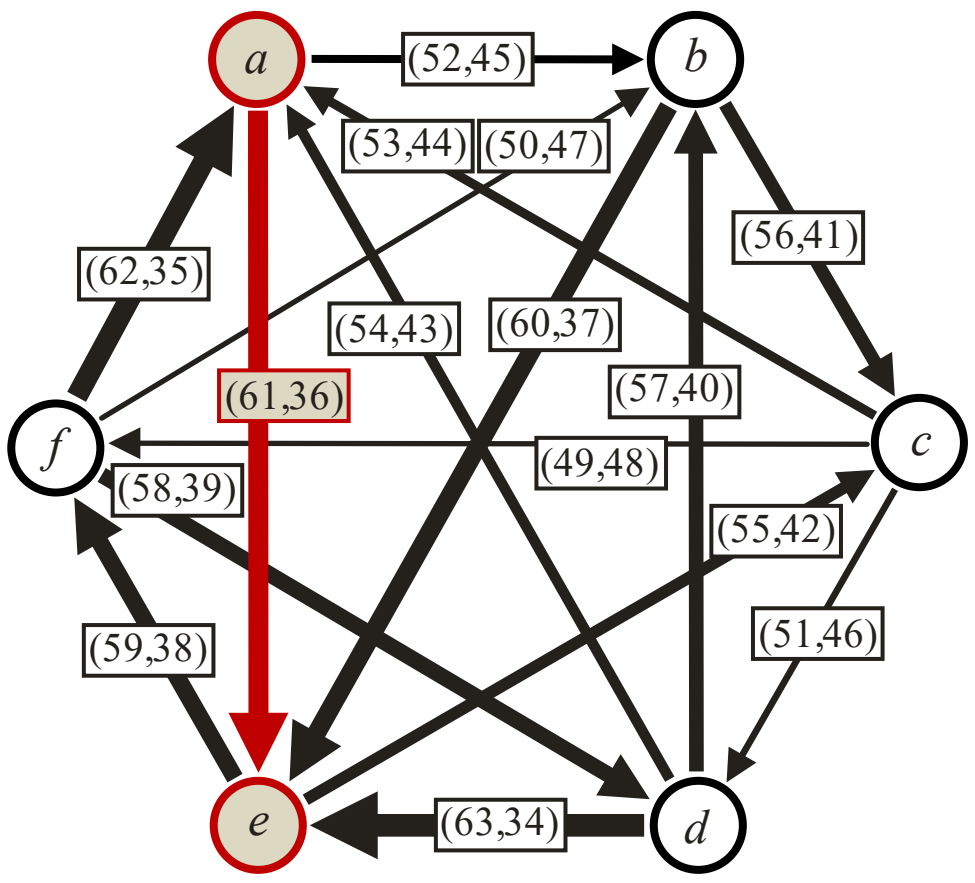

The strongest path from *a* to *e* is:
*a*, (61,36), *e*

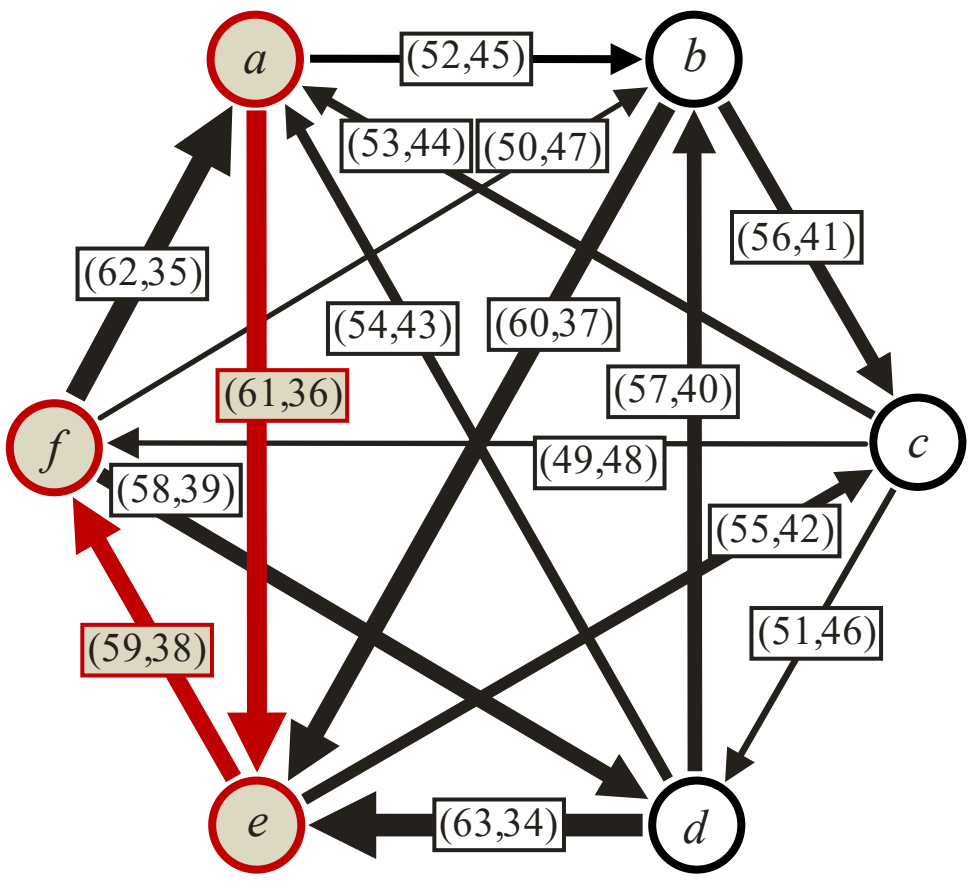

The strongest path from *a* to *f* is:
*a*, (61,36), *e*, (59,38), *f*

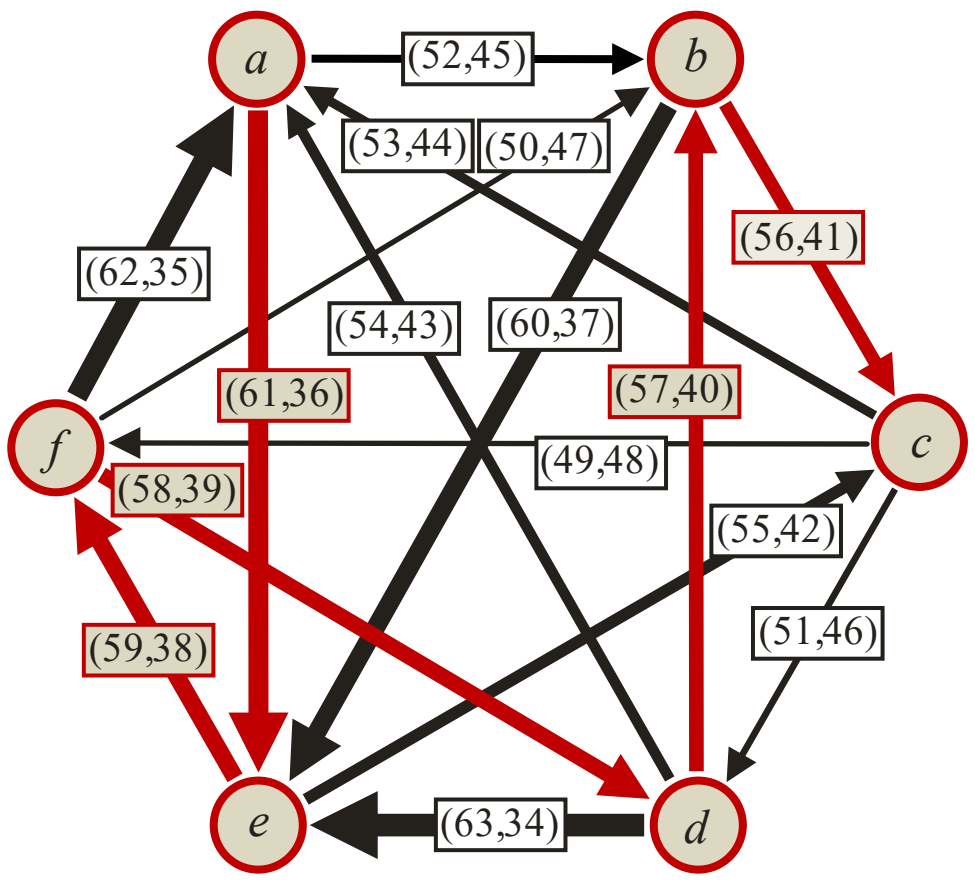

These are the strongest paths
from *a* to every other alternative.

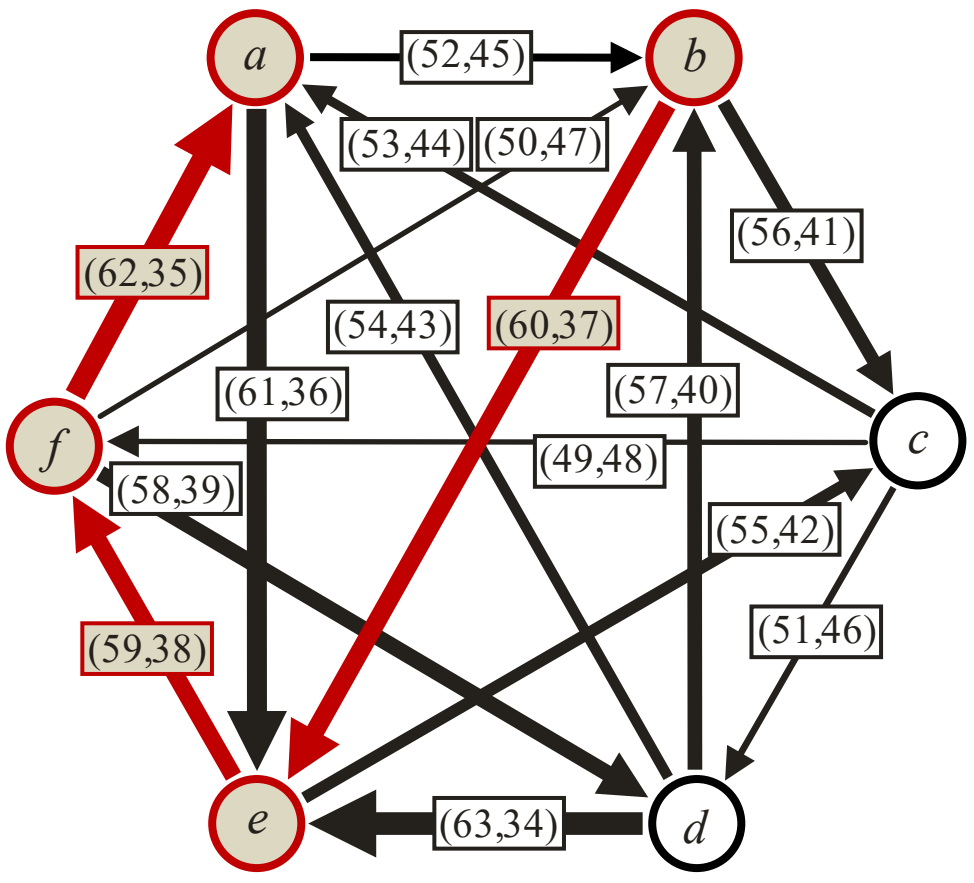

The strongest path from *b* to *a* is:
*b*, (60,37), *e*, (59,38), *f*, (62,35), *a*

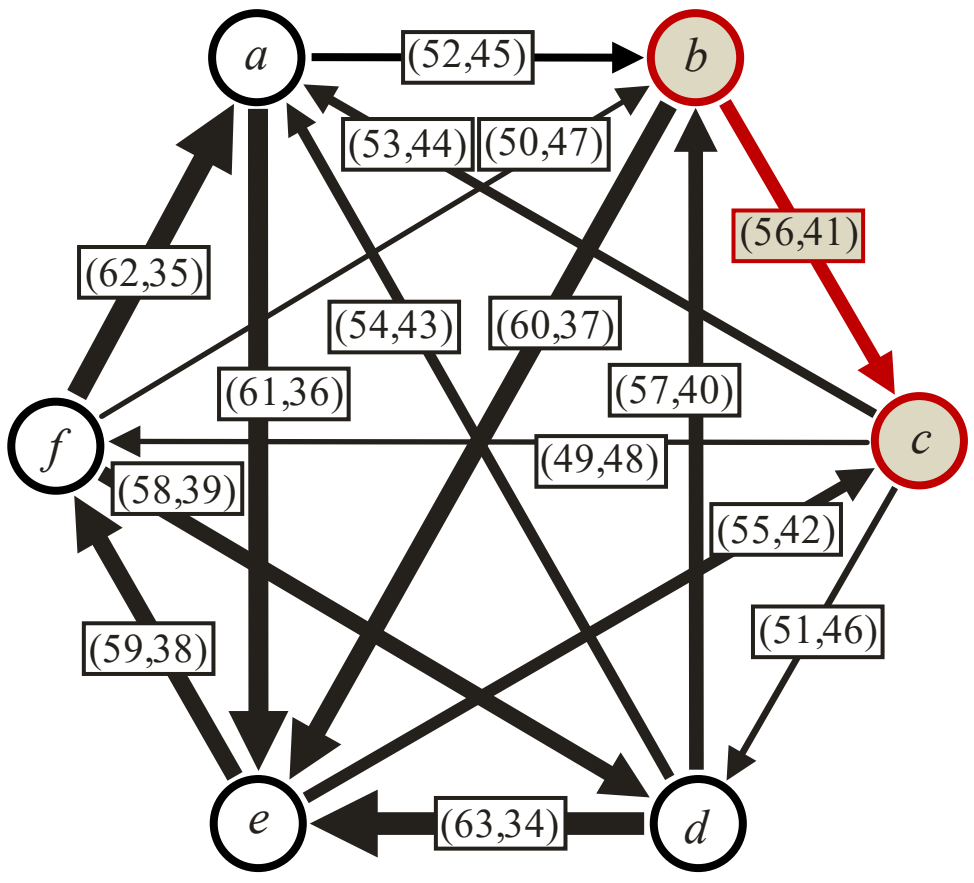

The strongest path from *b* to *c* is:
*b*, (56,41), *c*

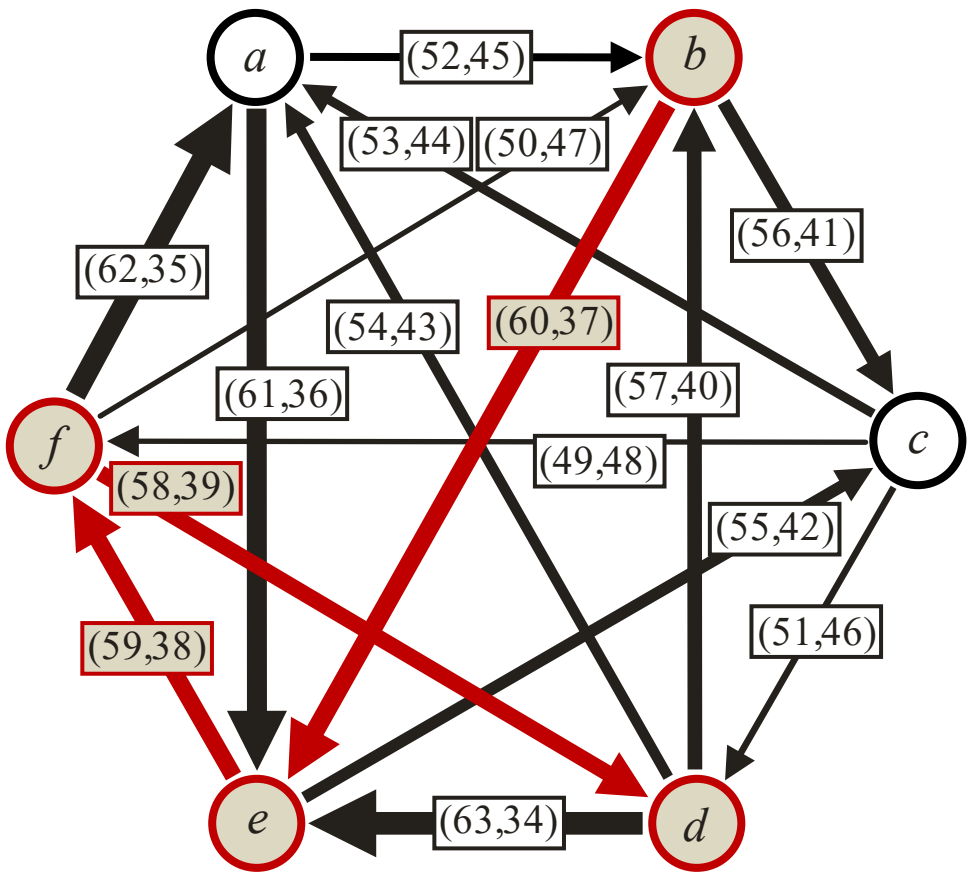

The strongest path from *b* to *d* is:
*b*, (60,37), *e*, (59,38), *f*, (58,39), *d*

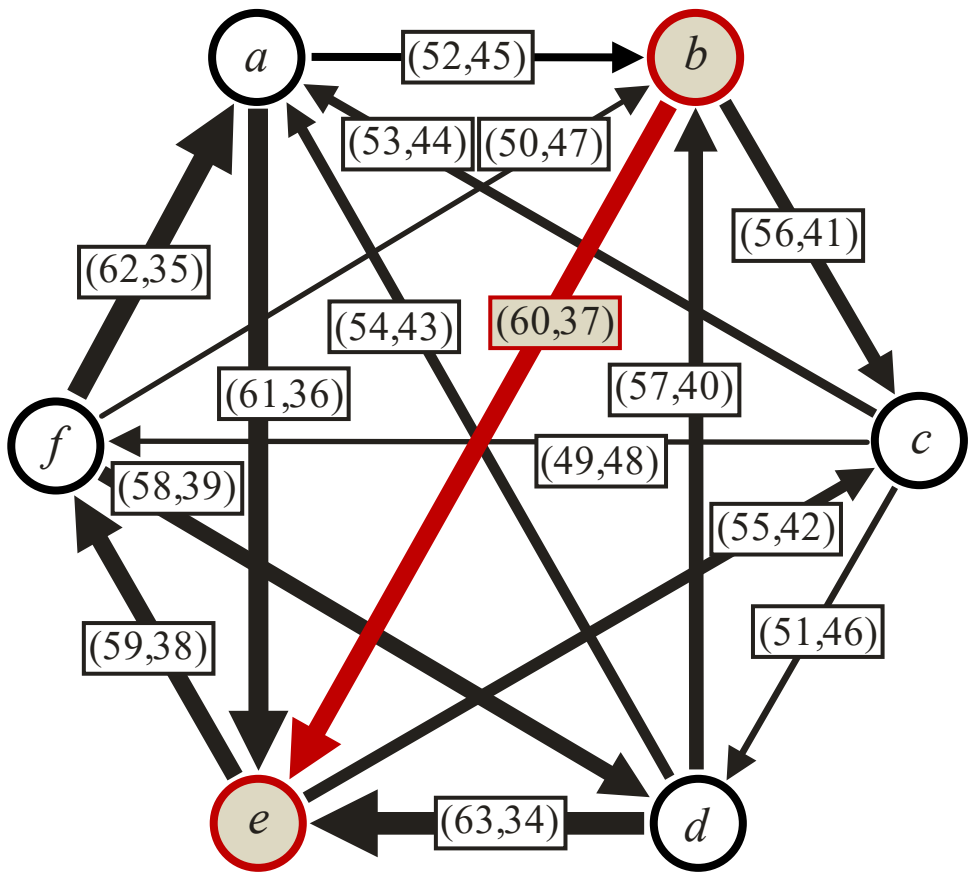

The strongest path from *b* to *e* is:
*b*, (60,37), *e*

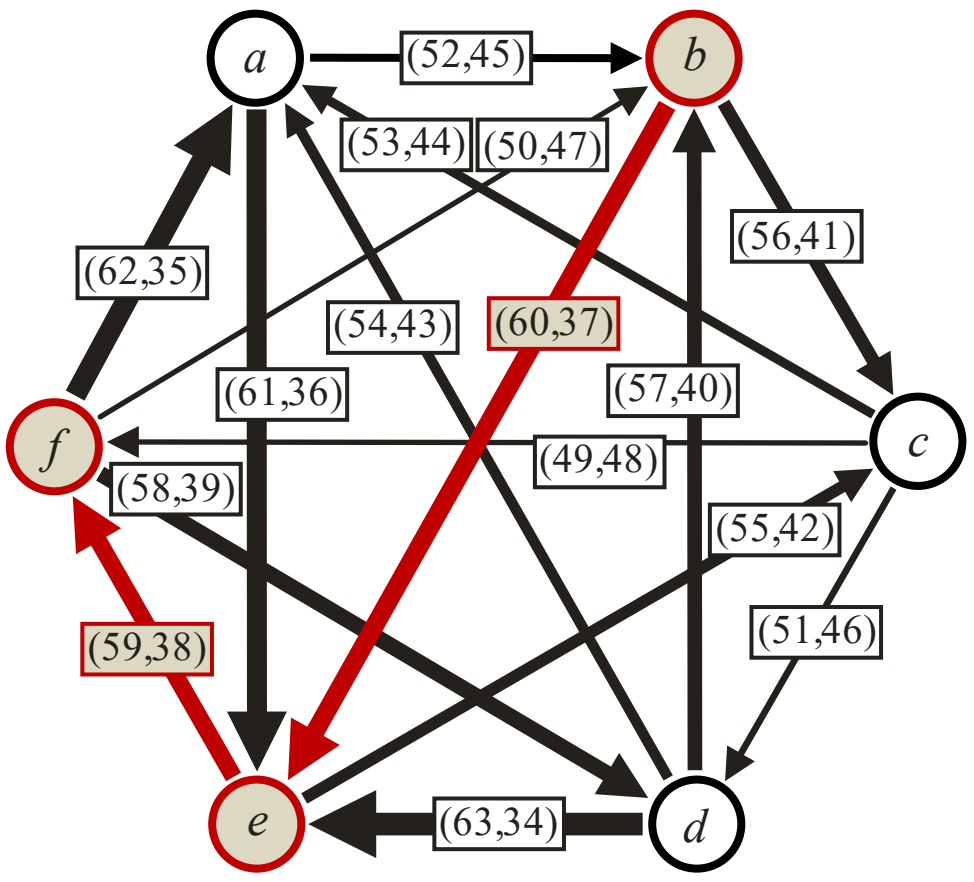

The strongest path from *b* to *f* is:
*b*, (60,37), *e*, (59,38), *f*

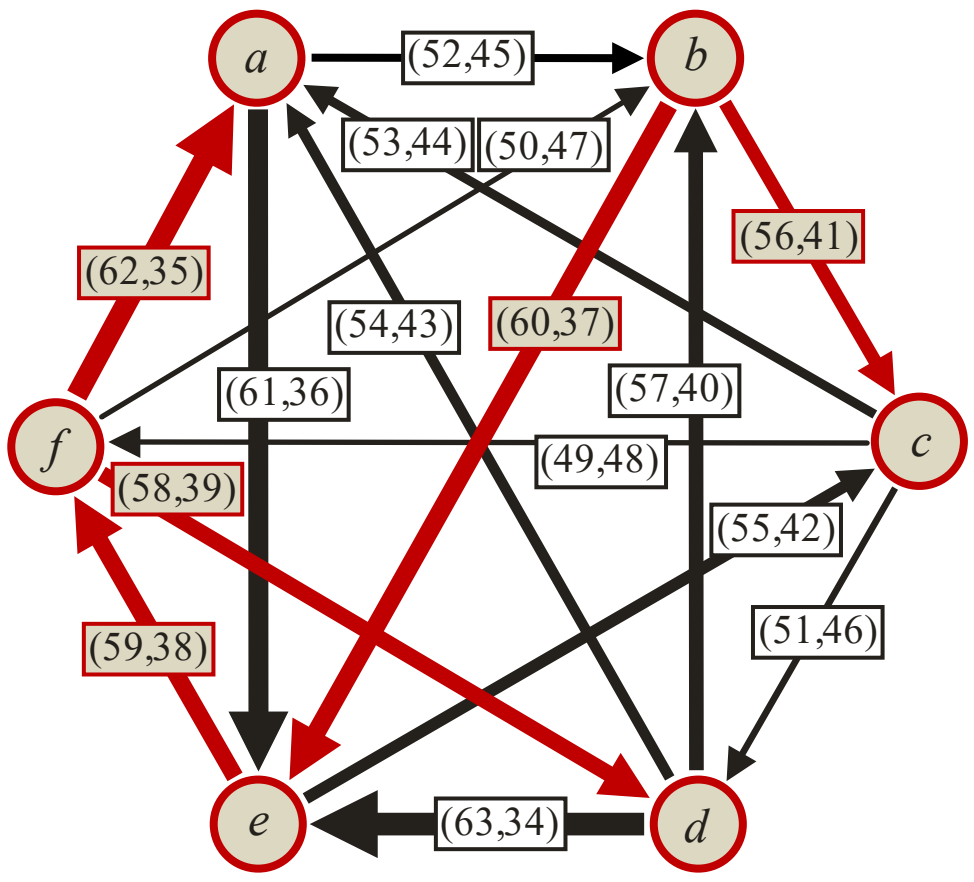

These are the strongest paths
from *b* to every other alternative.

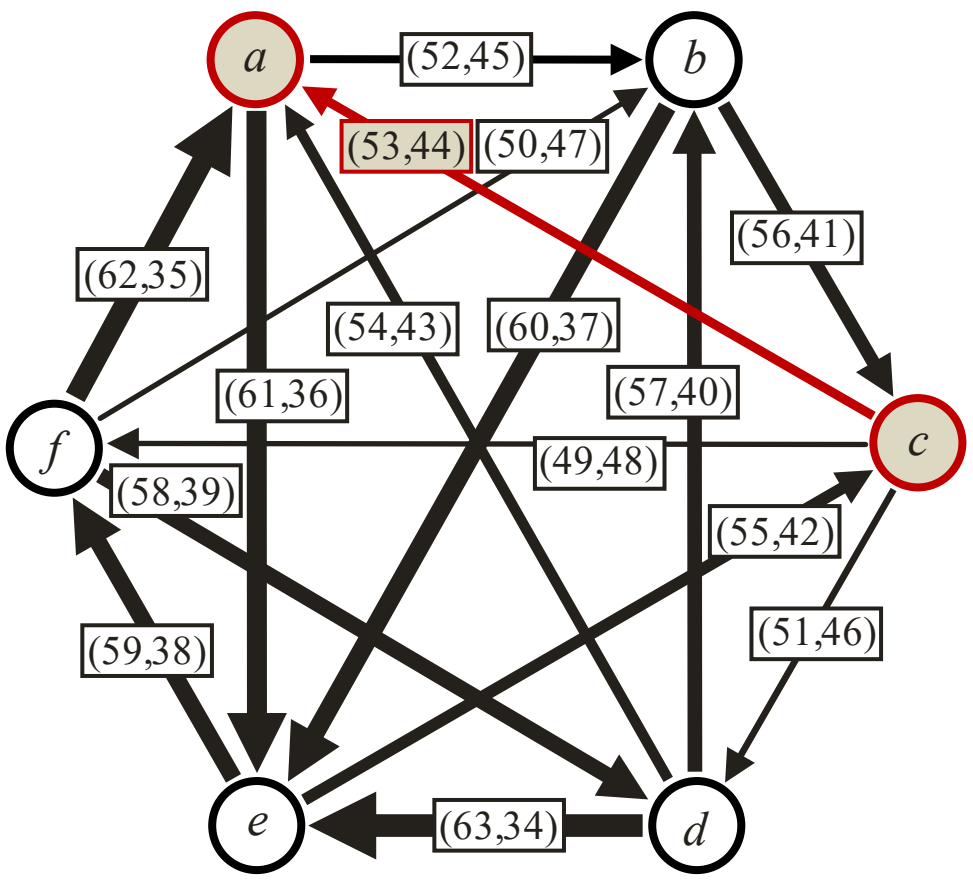

The strongest path from *c* to *a* is:
*c*, (53,44), *a*

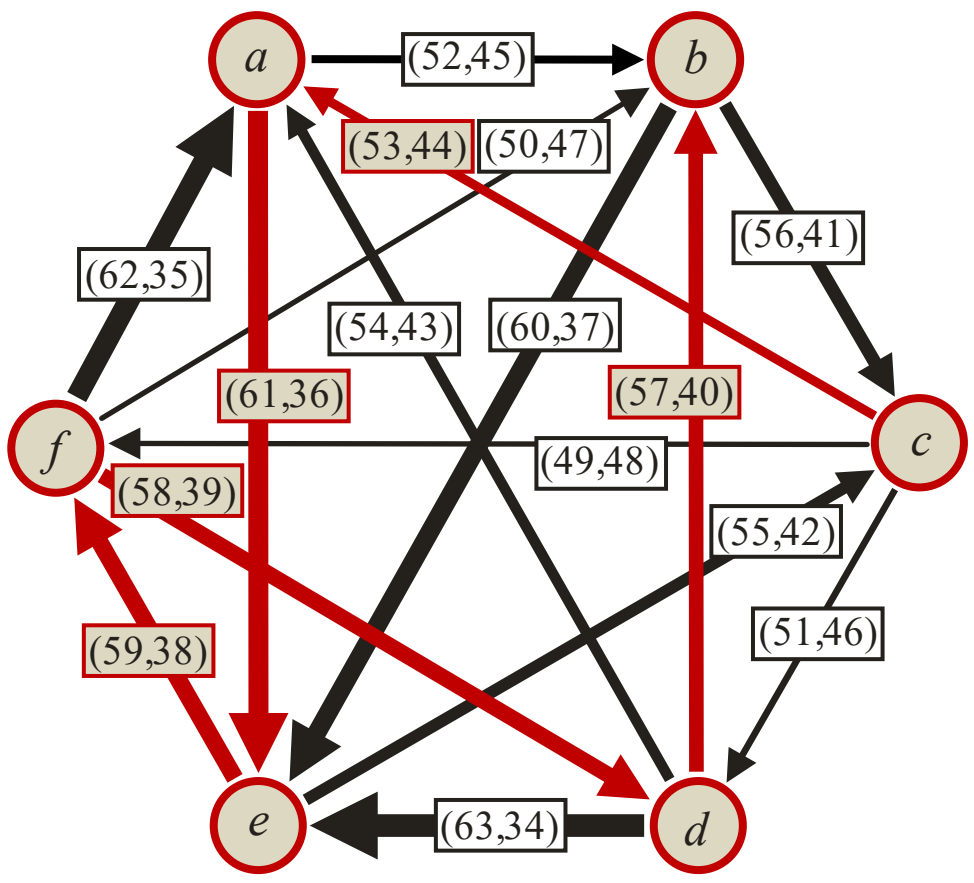

The strongest path from *c* to *b* is:
*c*, (53,44), *a*, (61,36), *e*, (59,38), *f*, (58,39), *d*, (57,40), *b*

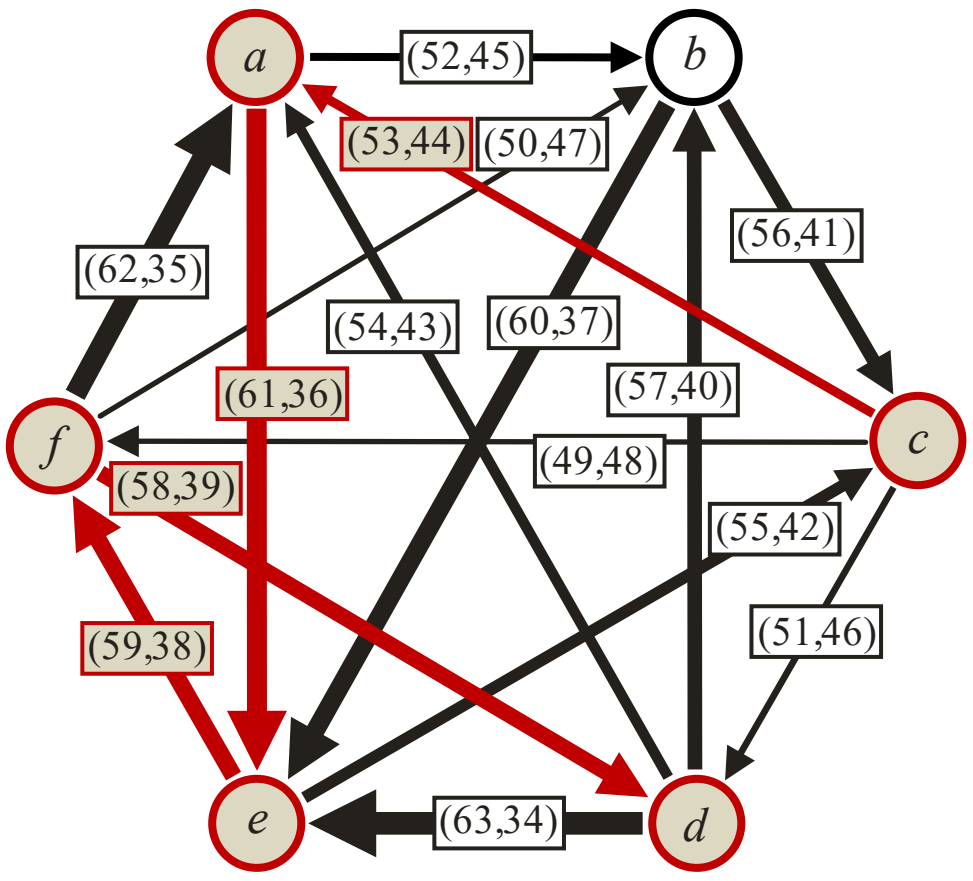

The strongest path from *c* to *d* is:
*c*, (53,44), *a*, (61,36), *e*, (59,38), *f*, (58,39), *d*

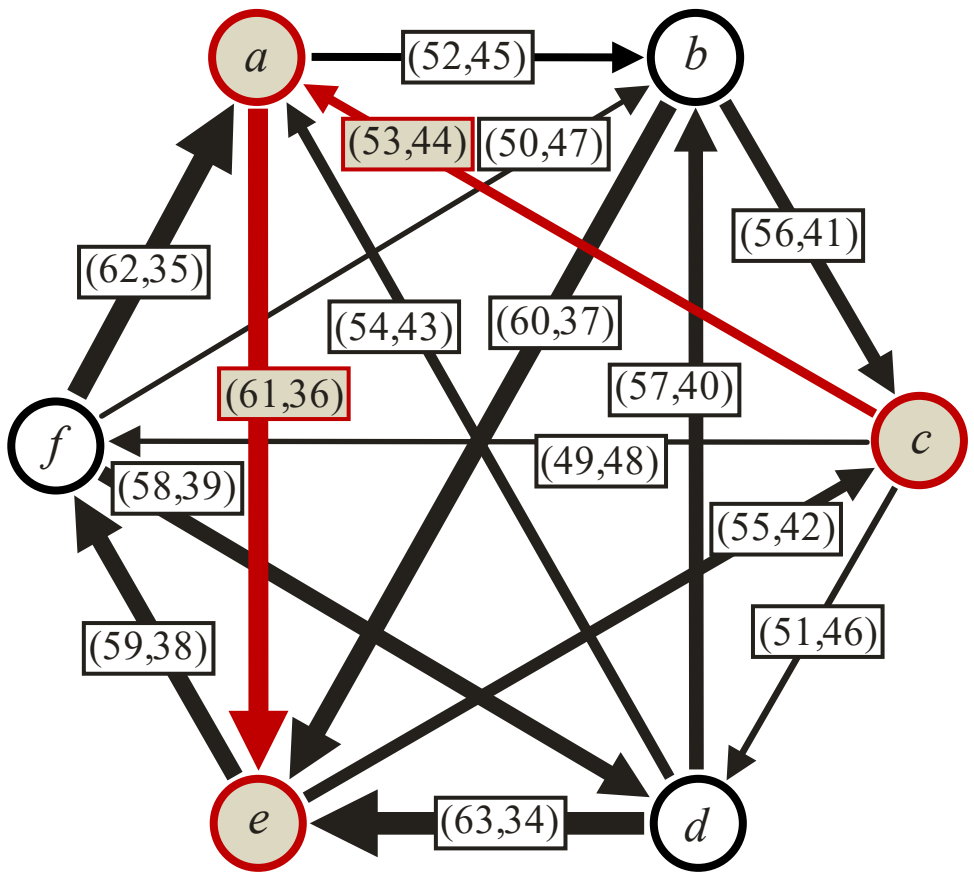

The strongest path from *c* to *e* is:
*c*, (53,44), *a*, (61,36), *e*

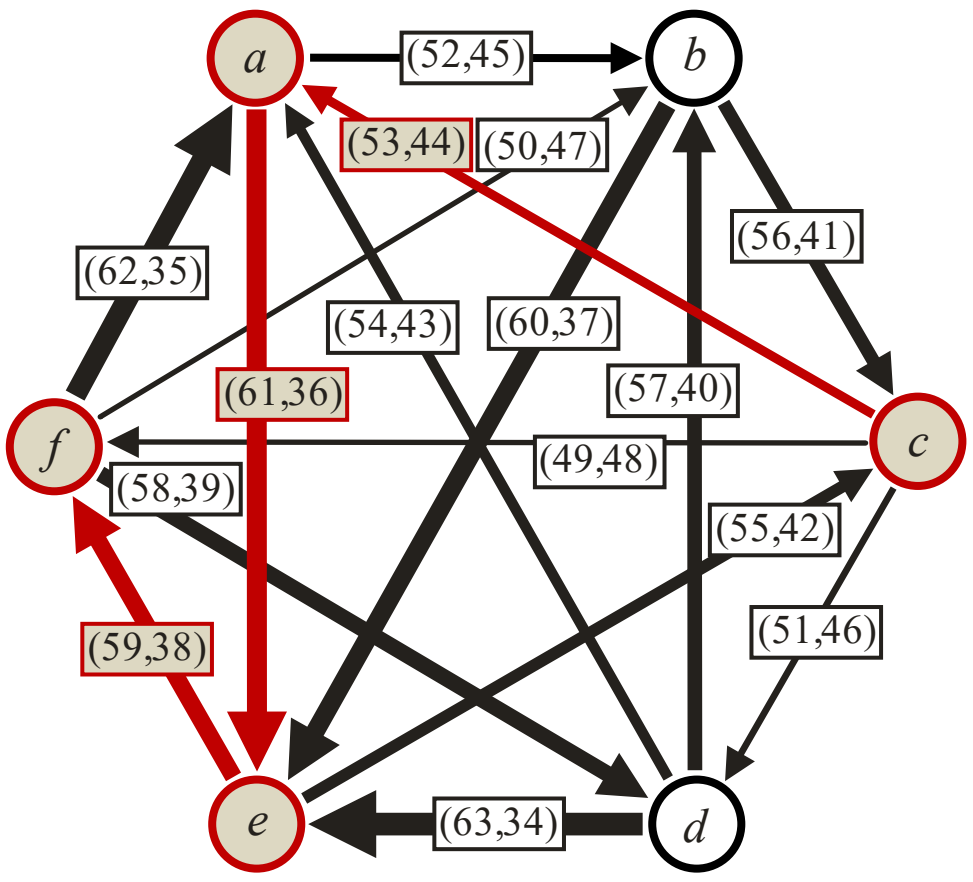

The strongest path from *c* to *f* is:
*c*, (53,44), *a*, (61,36), *e*, (59,38), *f*

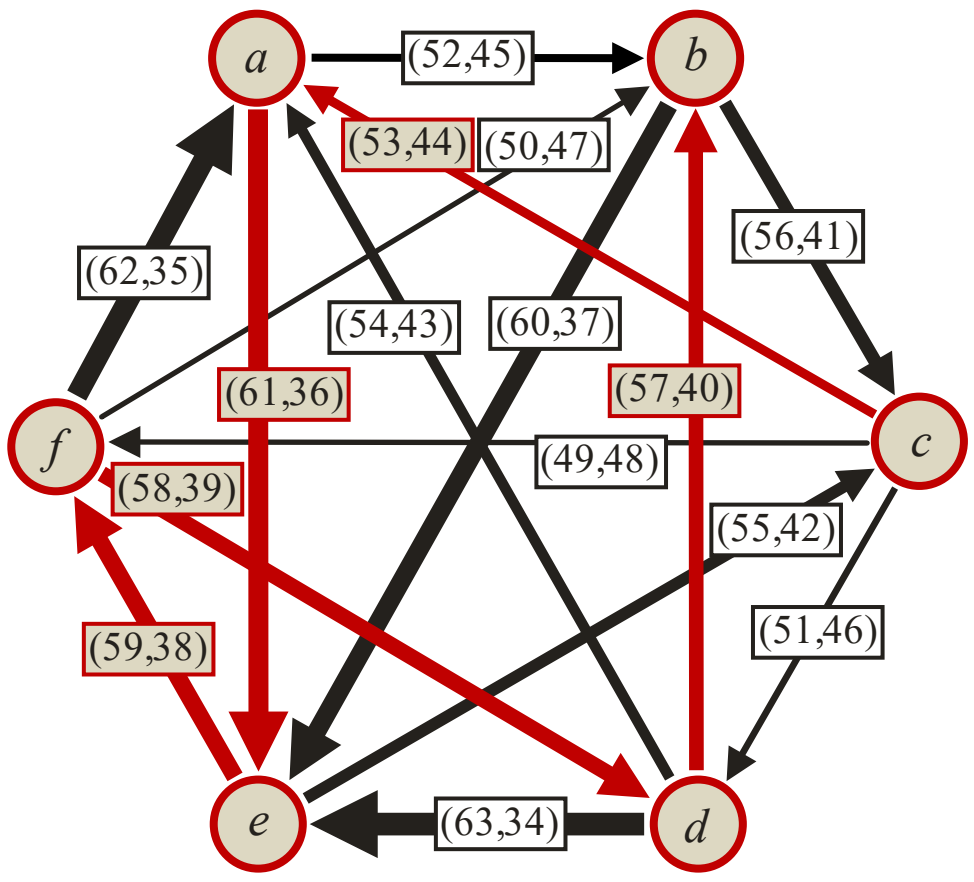

These are the strongest paths from *c* to every other alternative.

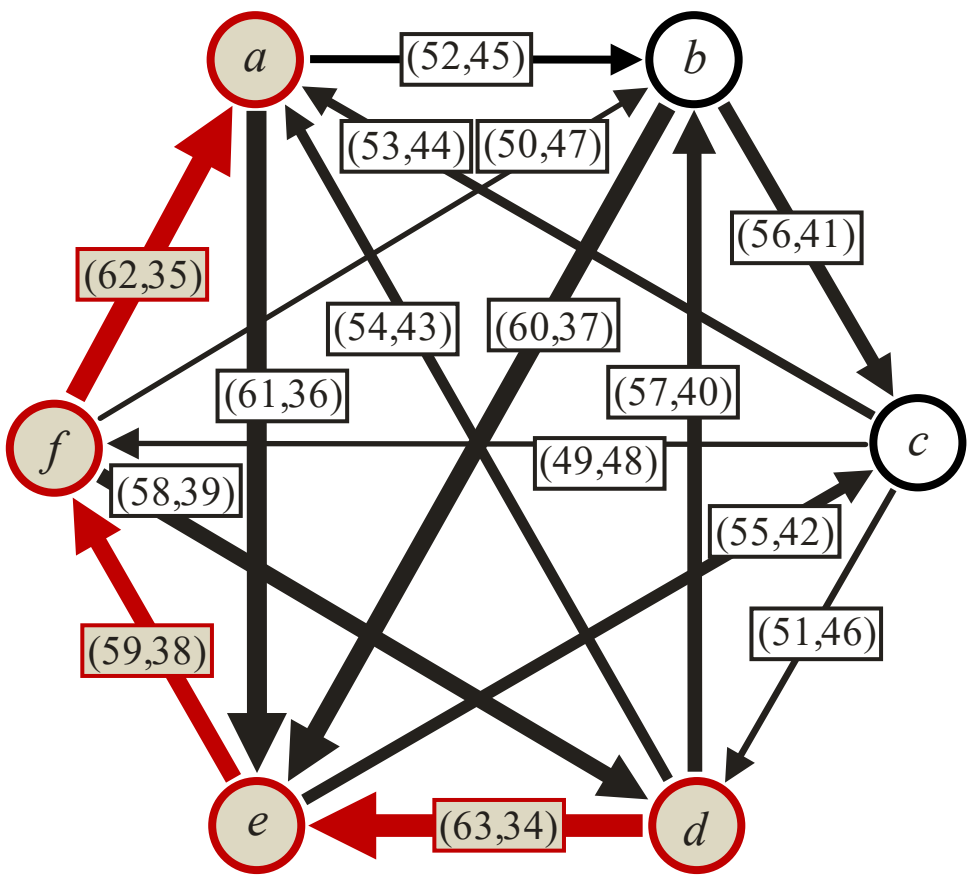

The strongest path from *d* to *a* is:
*d*, (63,34), *e*, (59,38), *f*, (62,35), *a*

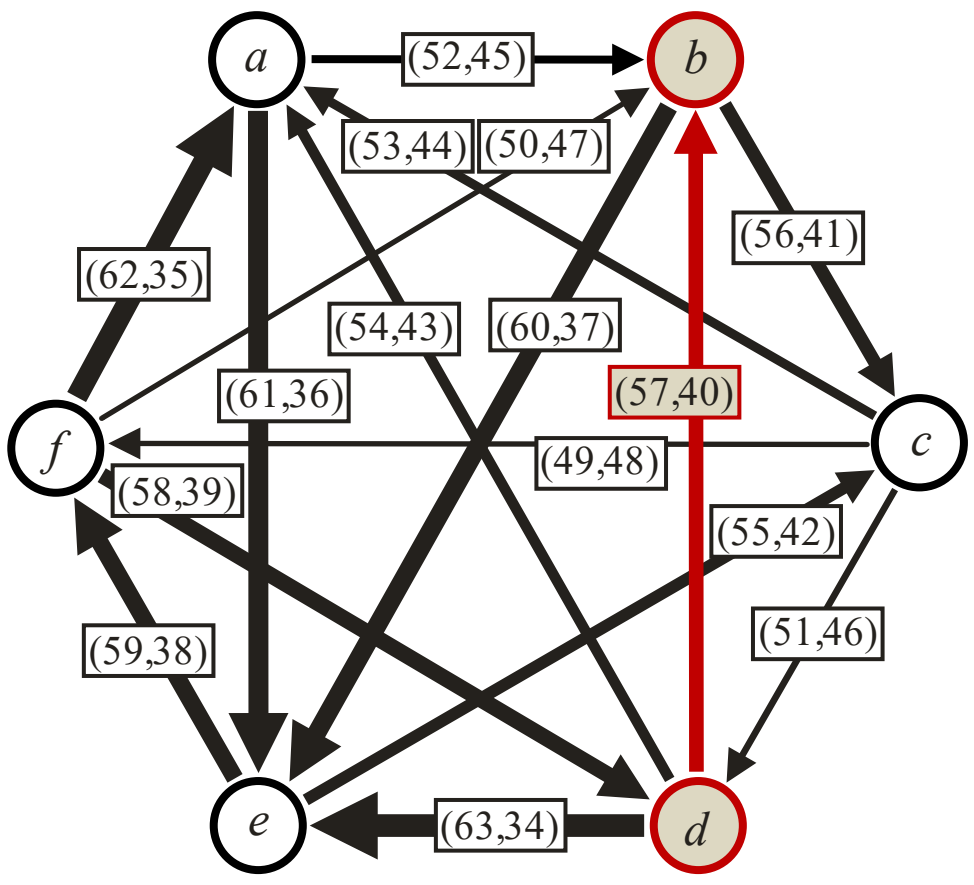

The strongest path from *d* to *b* is:
*d*, (57,40), *b*

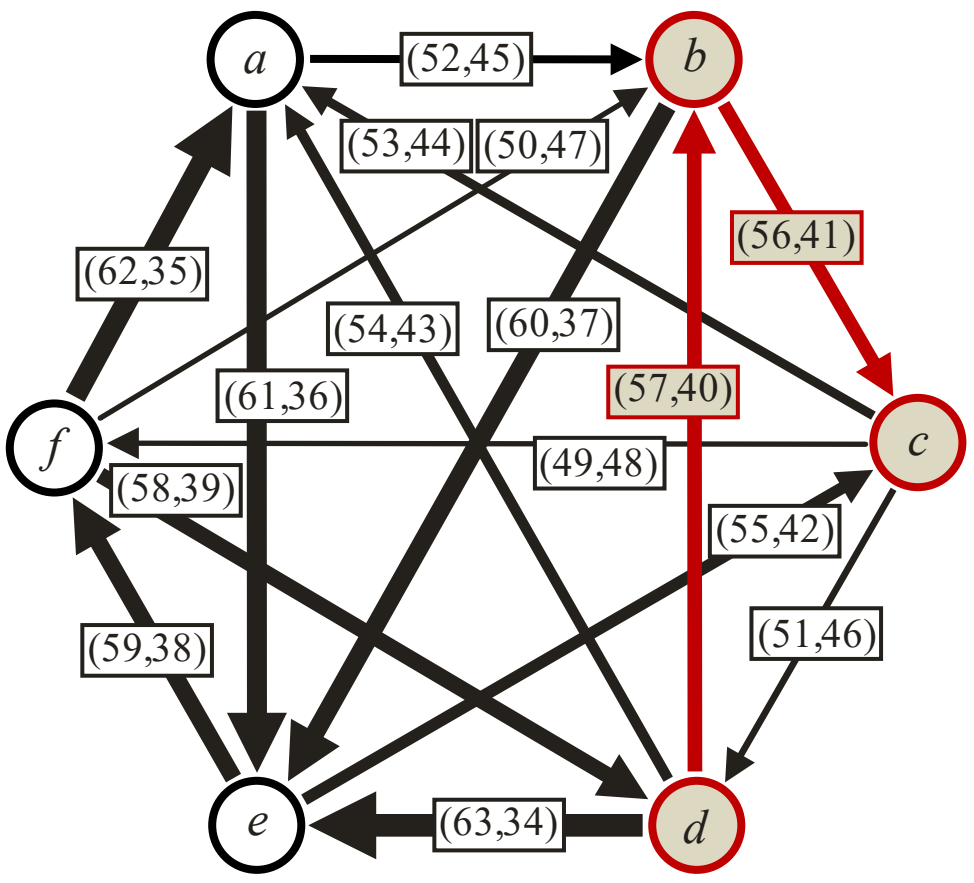

The strongest path from *d* to *c* is:
*d*, (57,40), *b*, (56,41), *c*

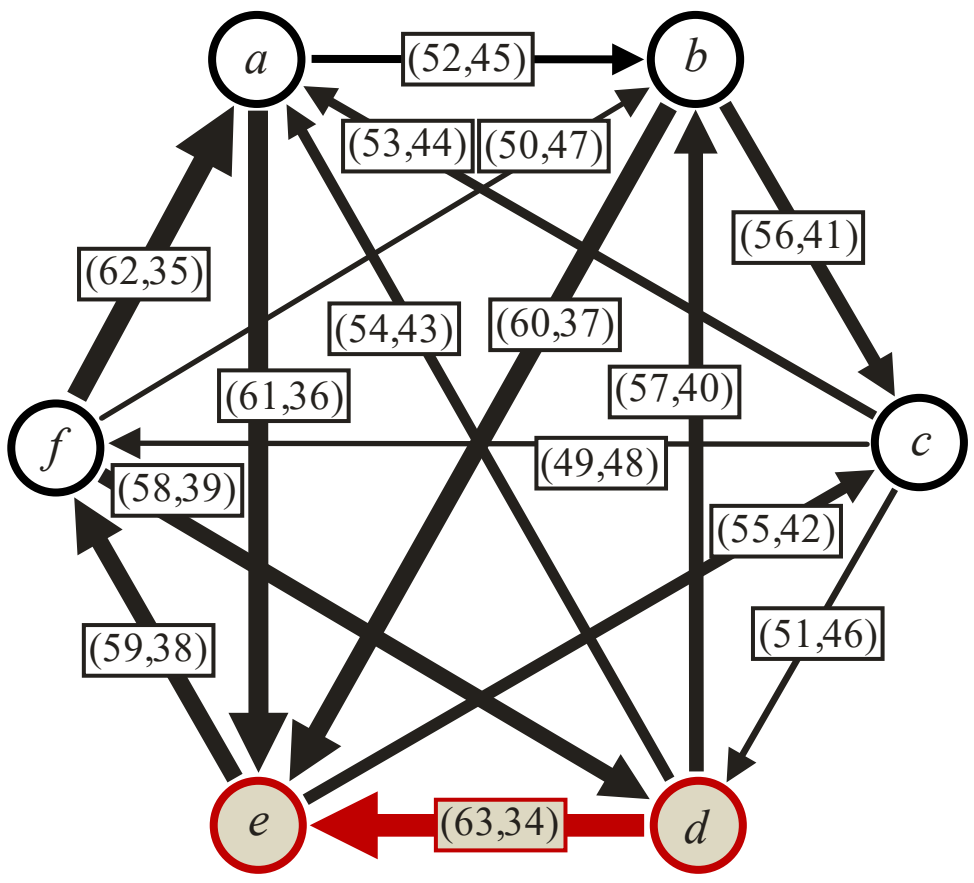

The strongest path from *d* to *e* is:
*d*, (63,34), *e*

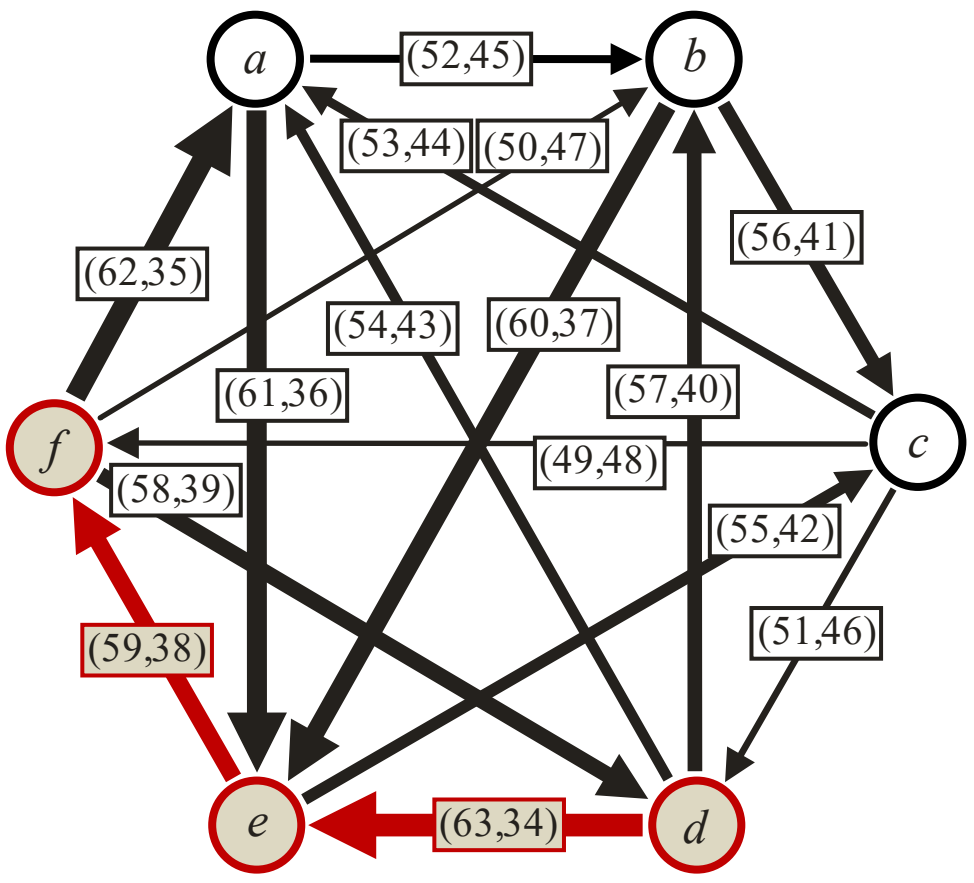

The strongest path from *d* to *f* is:
*d*, (63,34), *e*, (59,38), *f*

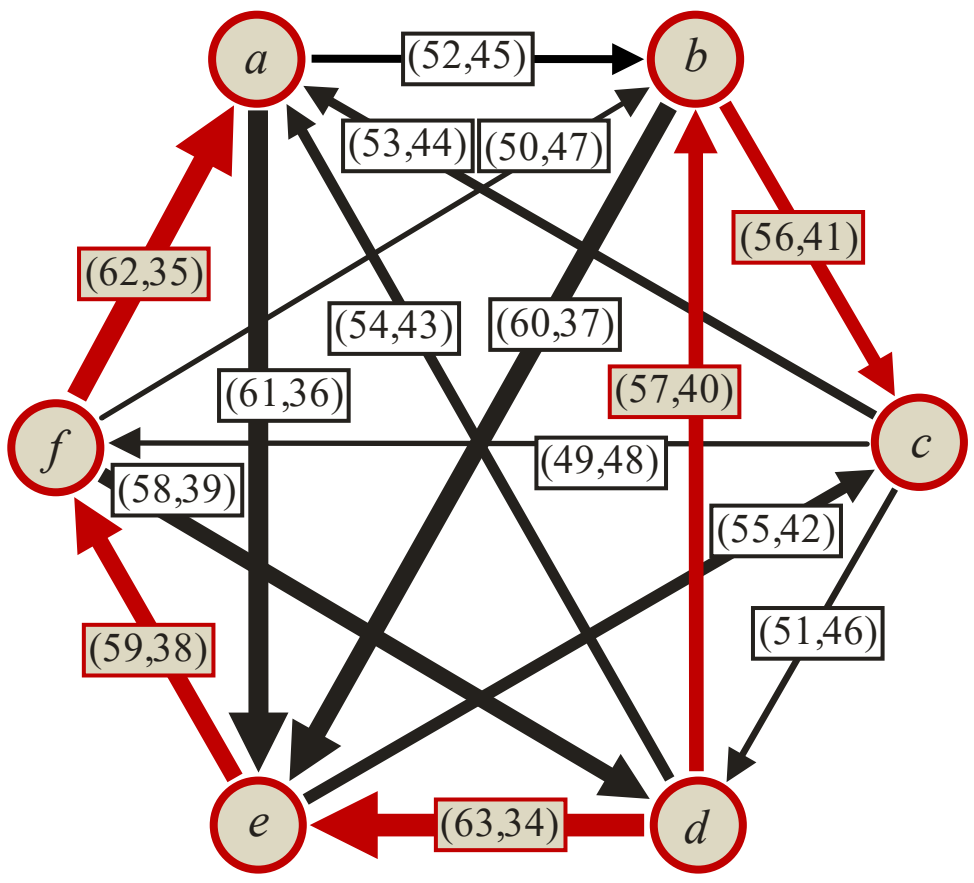

These are the strongest paths
from *d* to every other alternative.

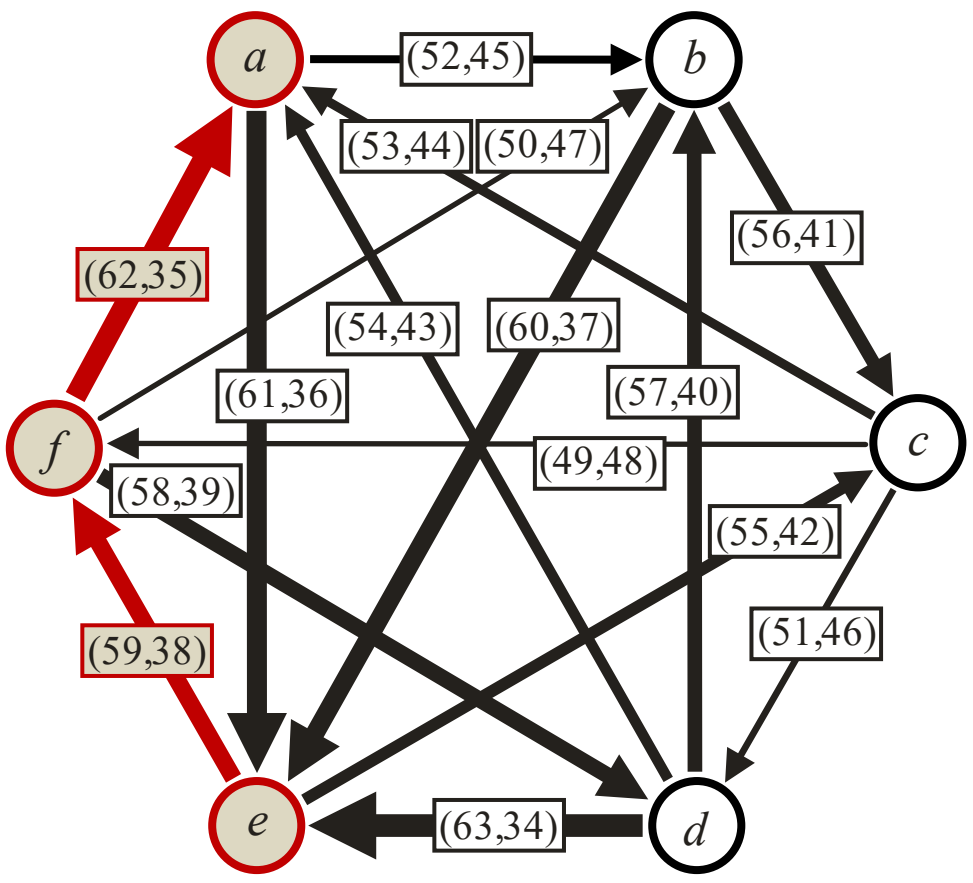

The strongest path from *e* to *a* is:
*e*, (59,38), *f*, (62,35), *a*

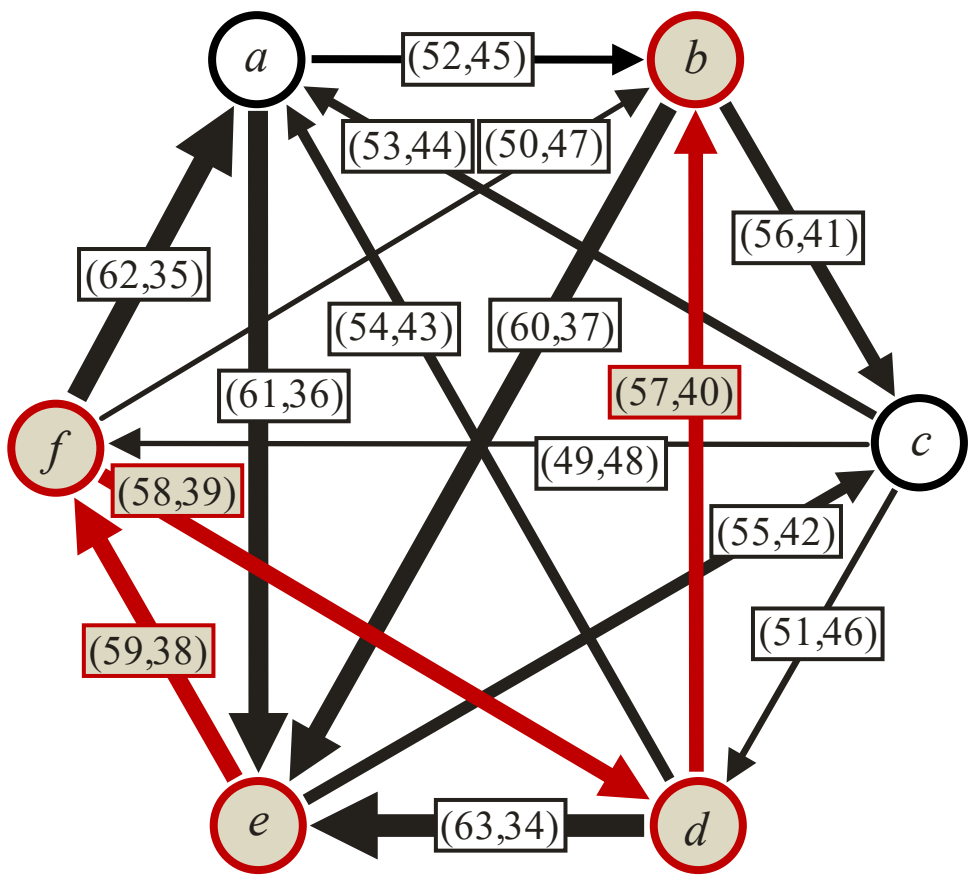

The strongest path from *e* to *b* is:
*e*, (59,38), *f*, (58,39), *d*, (57,40), *b*

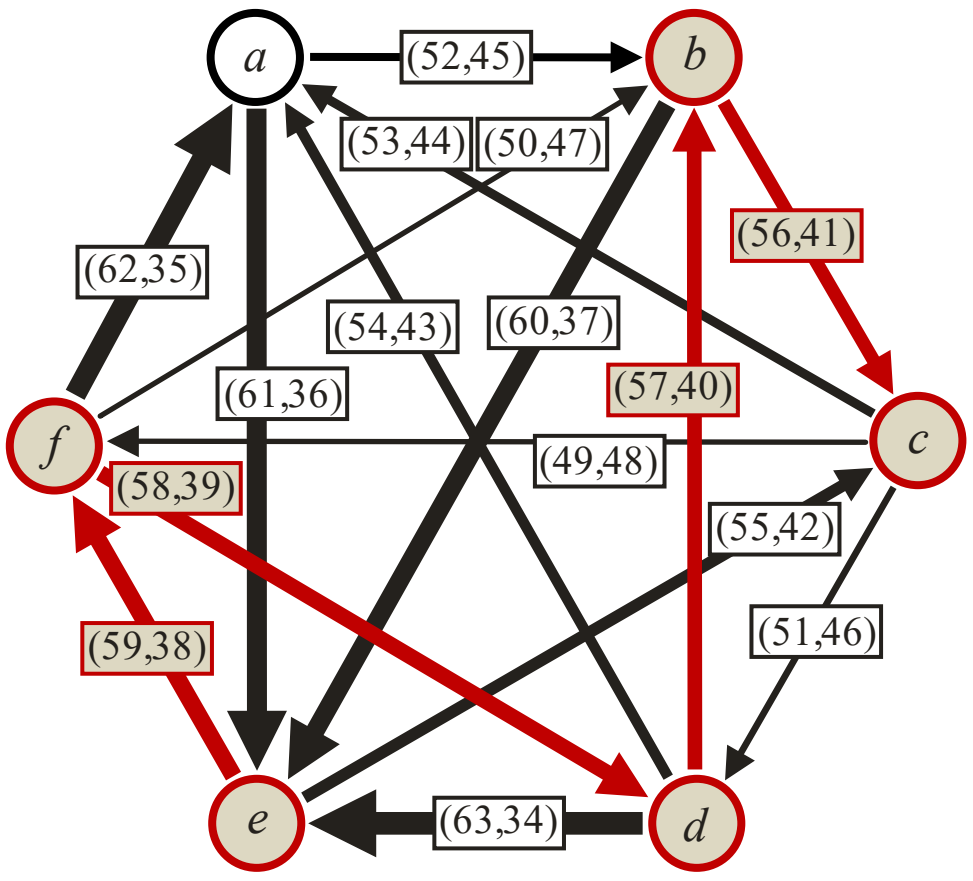

The strongest path from *e* to *c* is:
*e*, (59,38), *f*, (58,39), *d*, (57,40), *b*,
(56,41), *c*

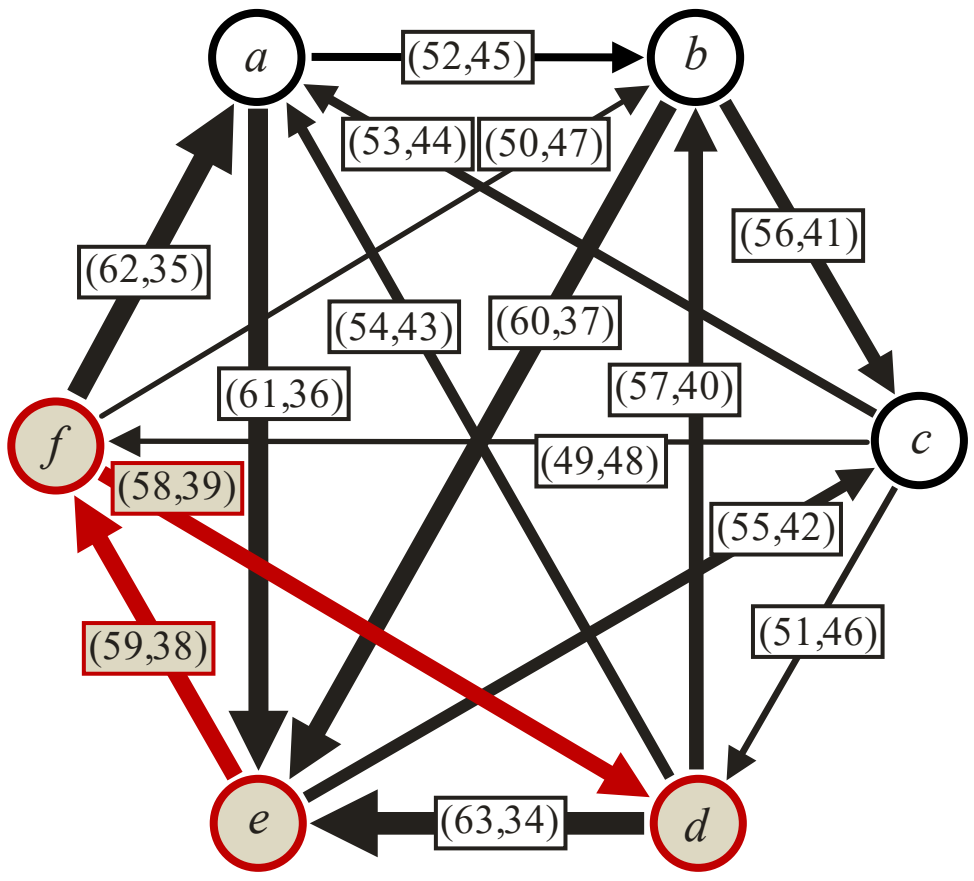

The strongest path from *e* to *d* is:
*e*, (59,38), *f*, (58,39), *d*

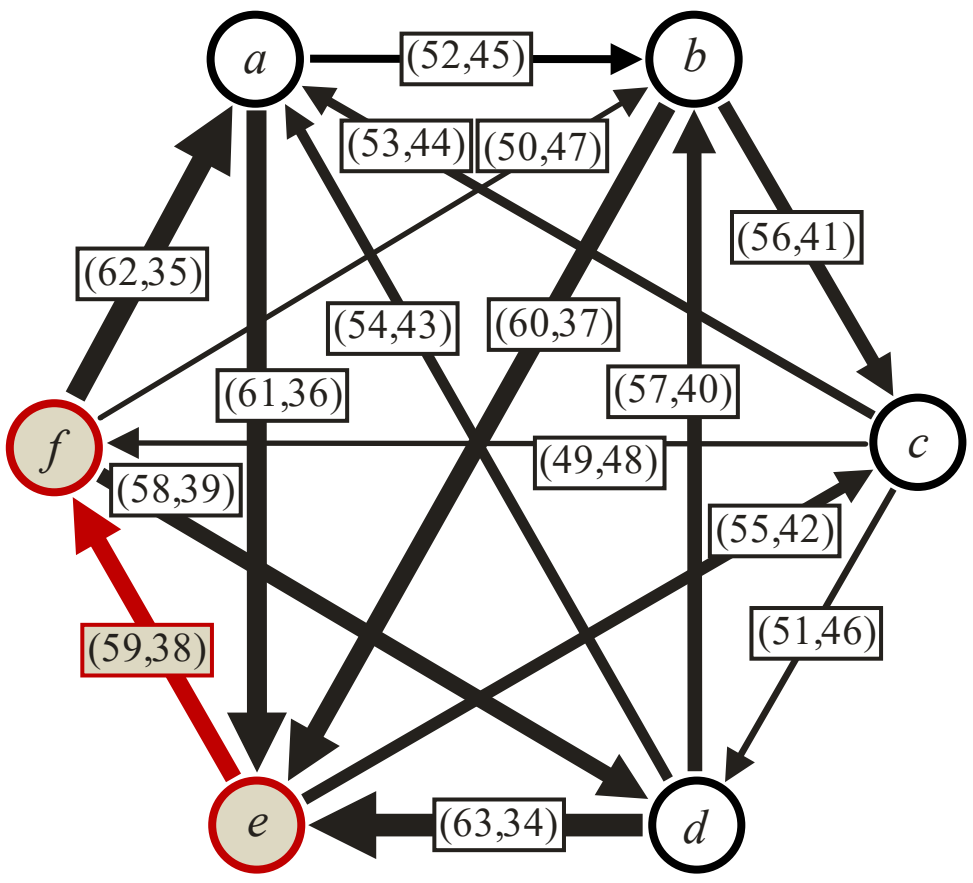

The strongest path from *e* to *f* is:
*e*, (59,38), *f*

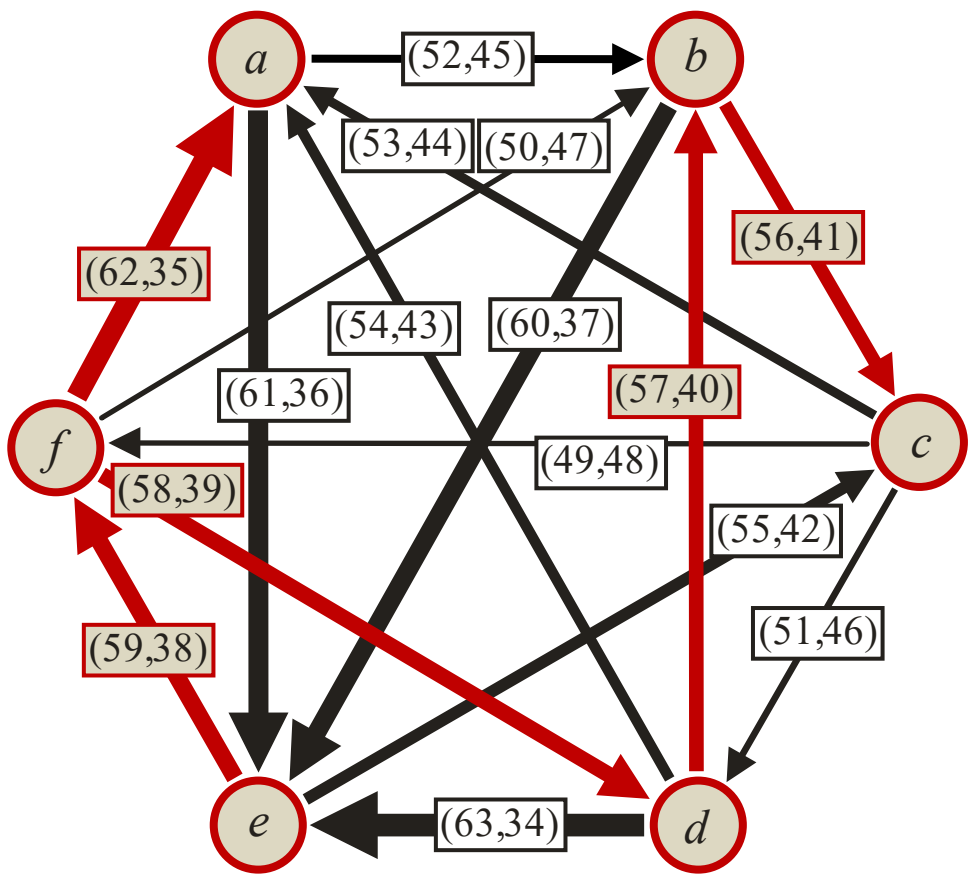

These are the strongest paths
from *e* to every other alternative.

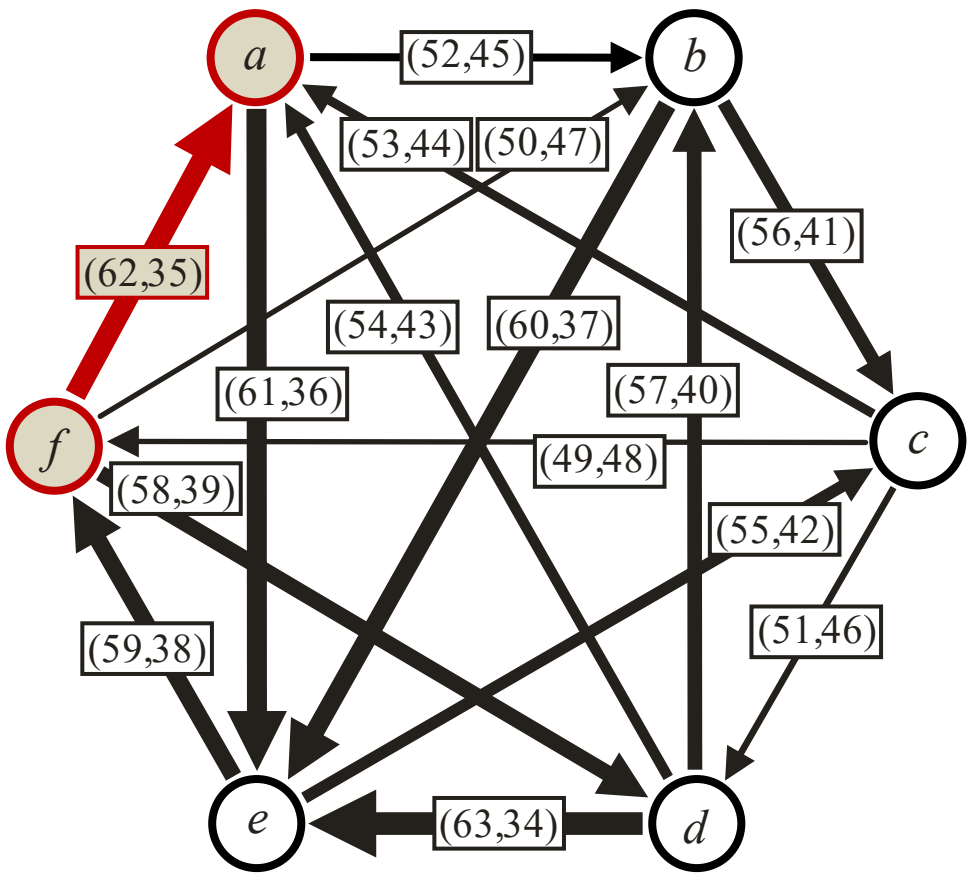

The strongest path from *f* to *a* is:
*f*, <u>(62,35)</u>, *a*

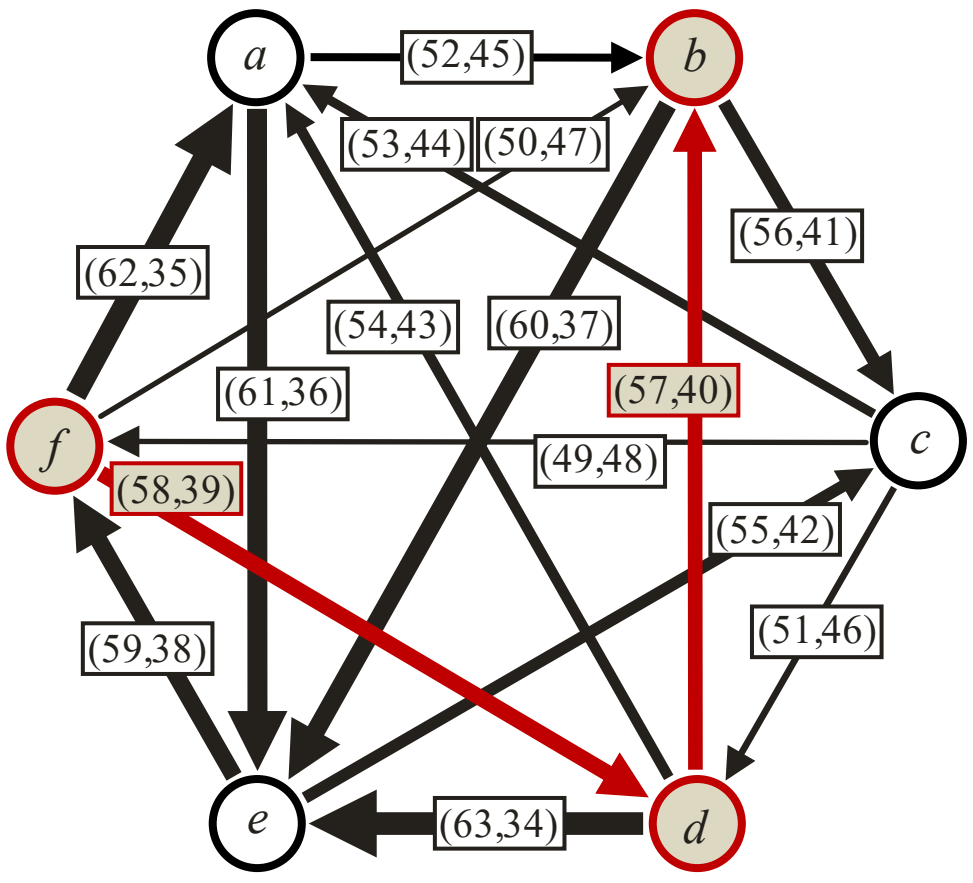

The strongest path from *f* to *b* is:
*f*, (58,39), *d*, <u>(57,40)</u>, *b*

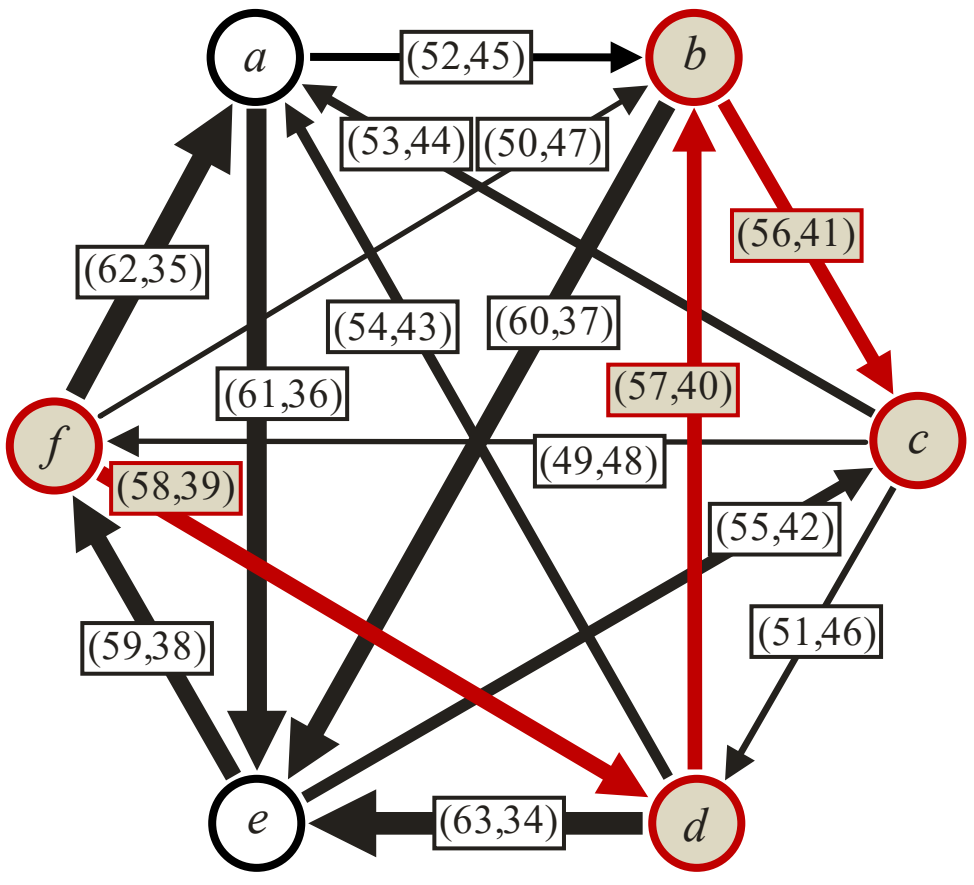

The strongest path from *f* to *c* is:
*f*, (58,39), *d*, (57,40), *b*, <u>(56,41)</u>, *c*

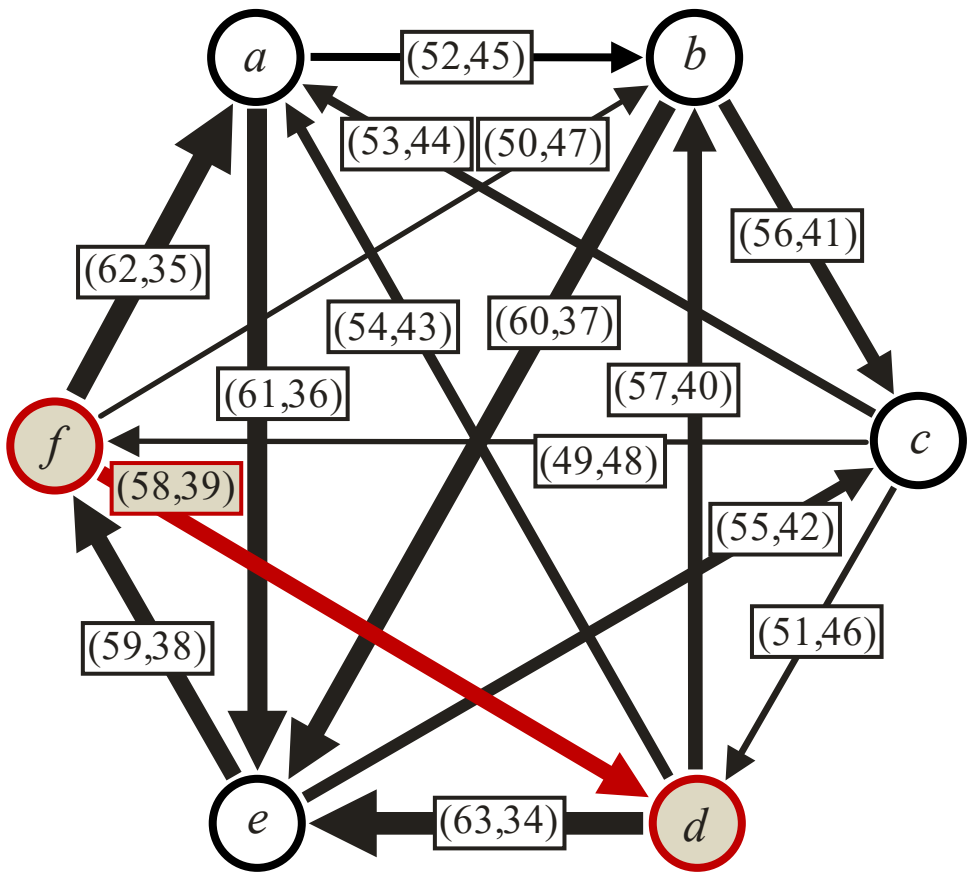

The strongest path from *f* to *d* is:
*f*, <u>(58,39)</u>, *d*

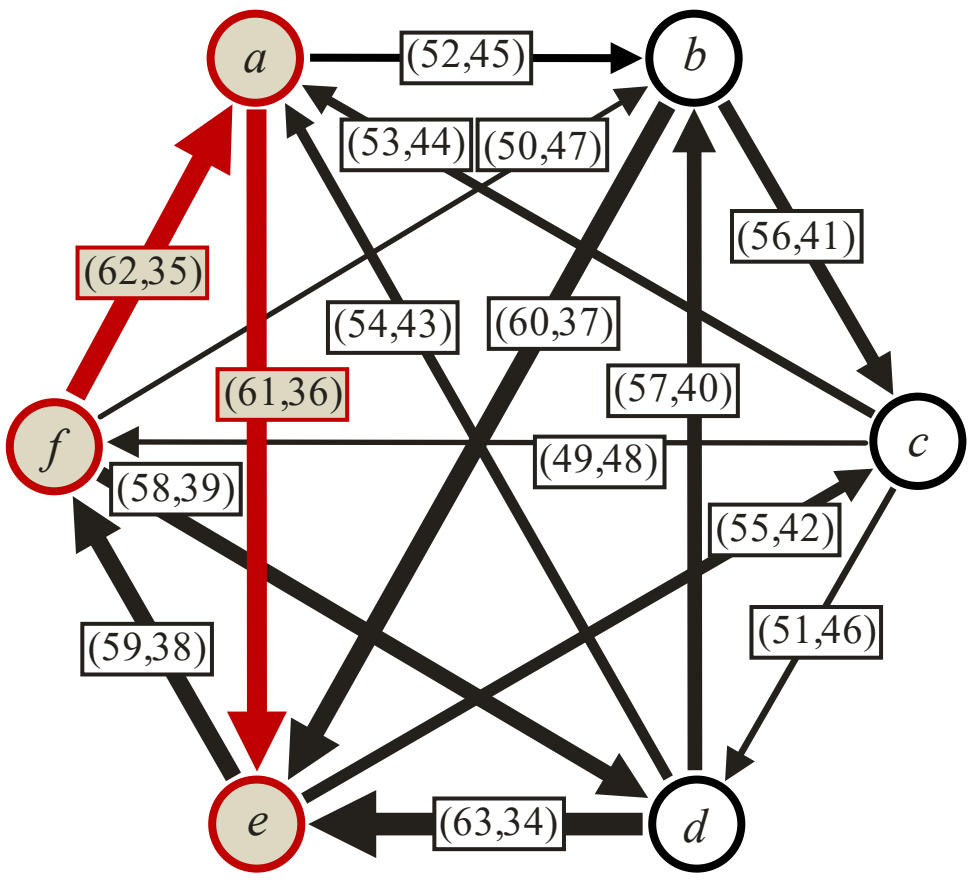

The strongest path from *f* to *e* is:
*f*, (62,35), *a*, <u>(61,36)</u>, *e*

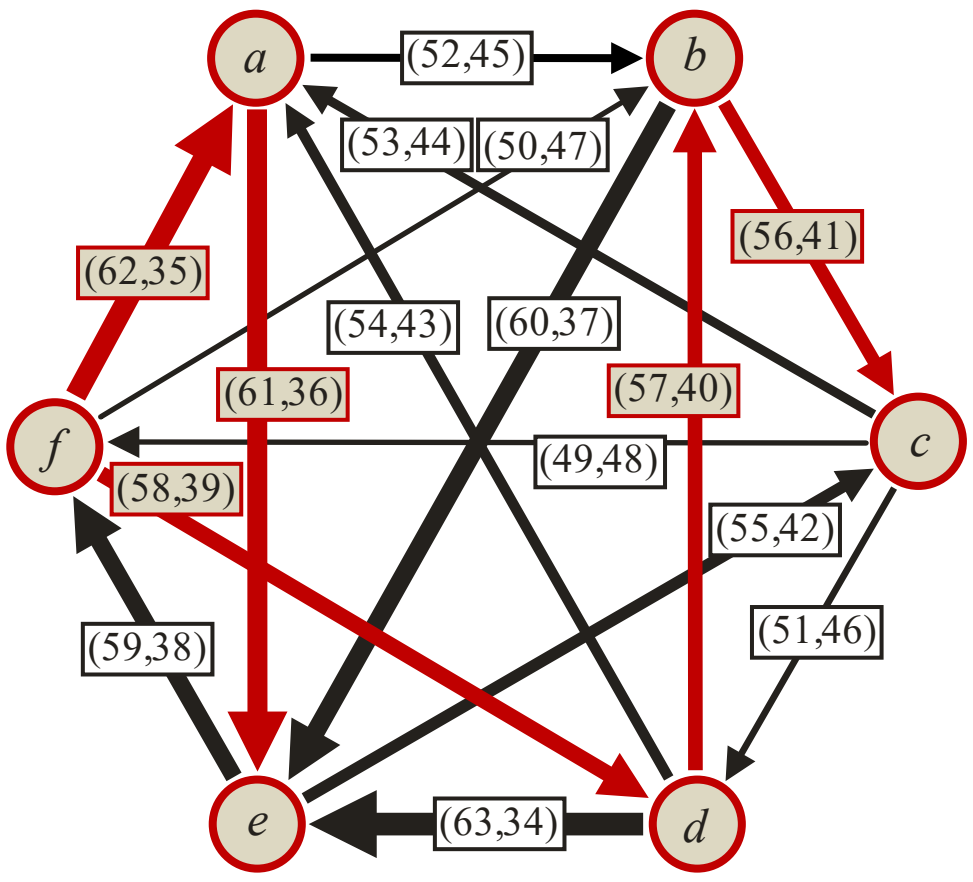

These are the strongest paths
from *f* to every other alternative.

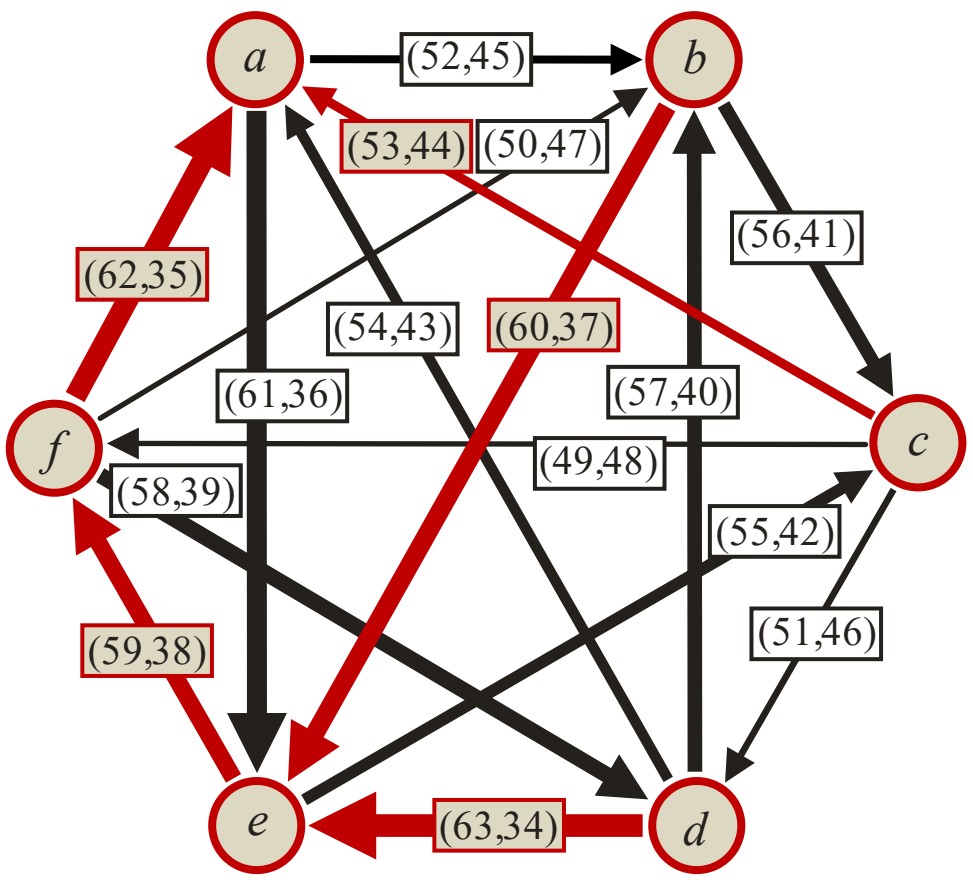

These are the strongest paths
from every other alternative to *a*.

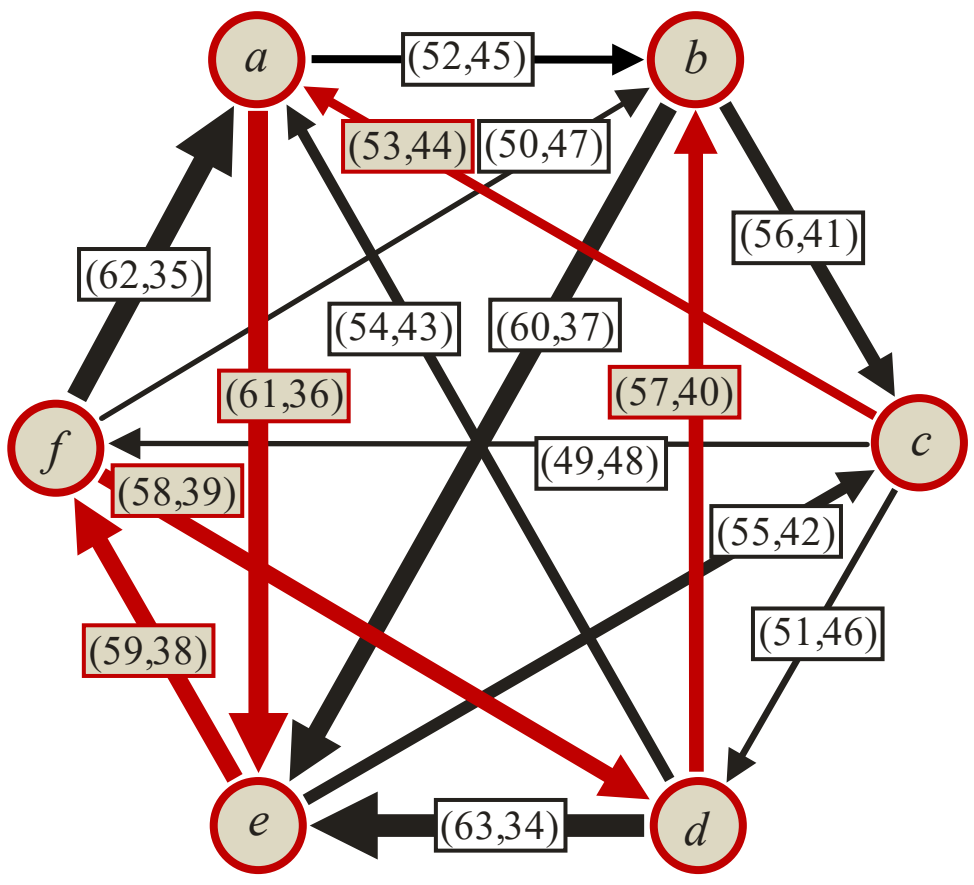

These are the strongest paths
from every other alternative to *b*.

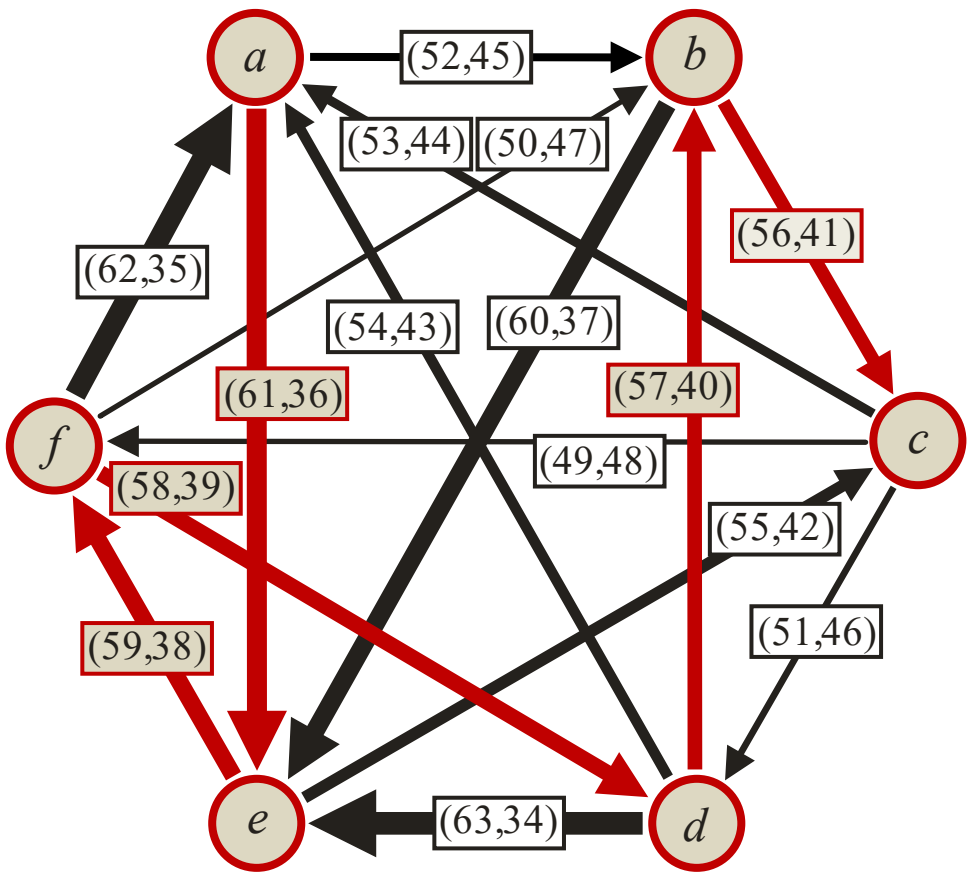

These are the strongest paths
from every other alternative to *c*.

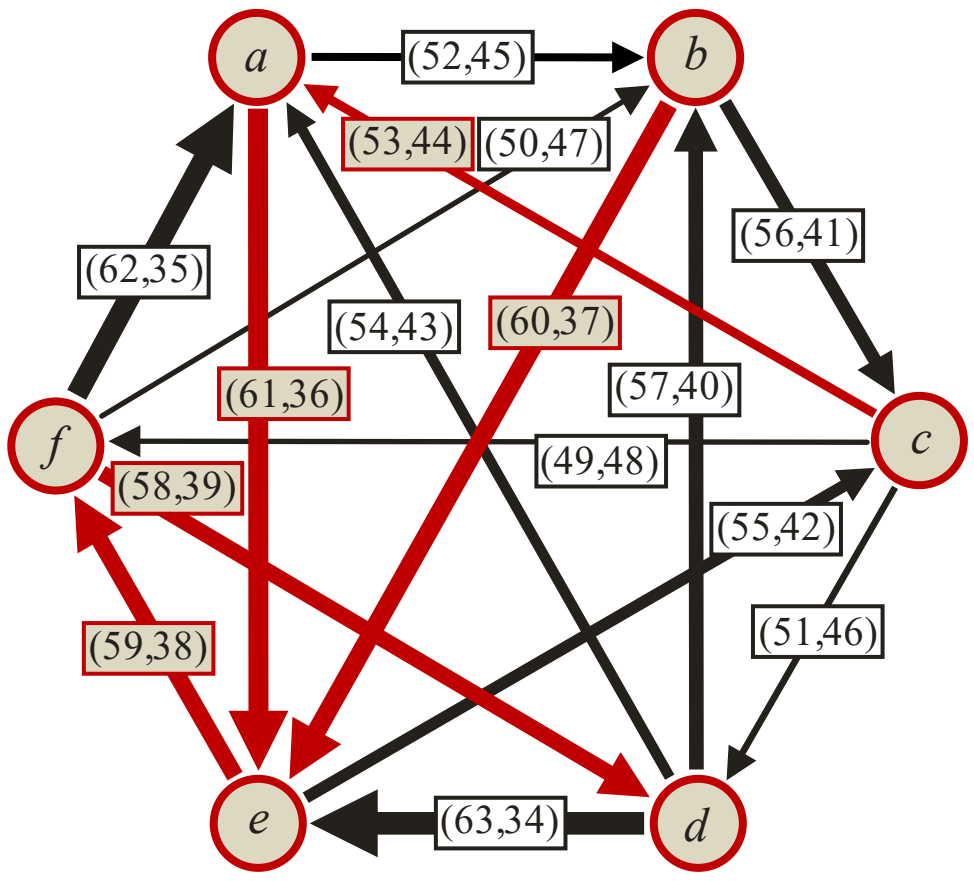

These are the strongest paths
from every other alternative to *d*.

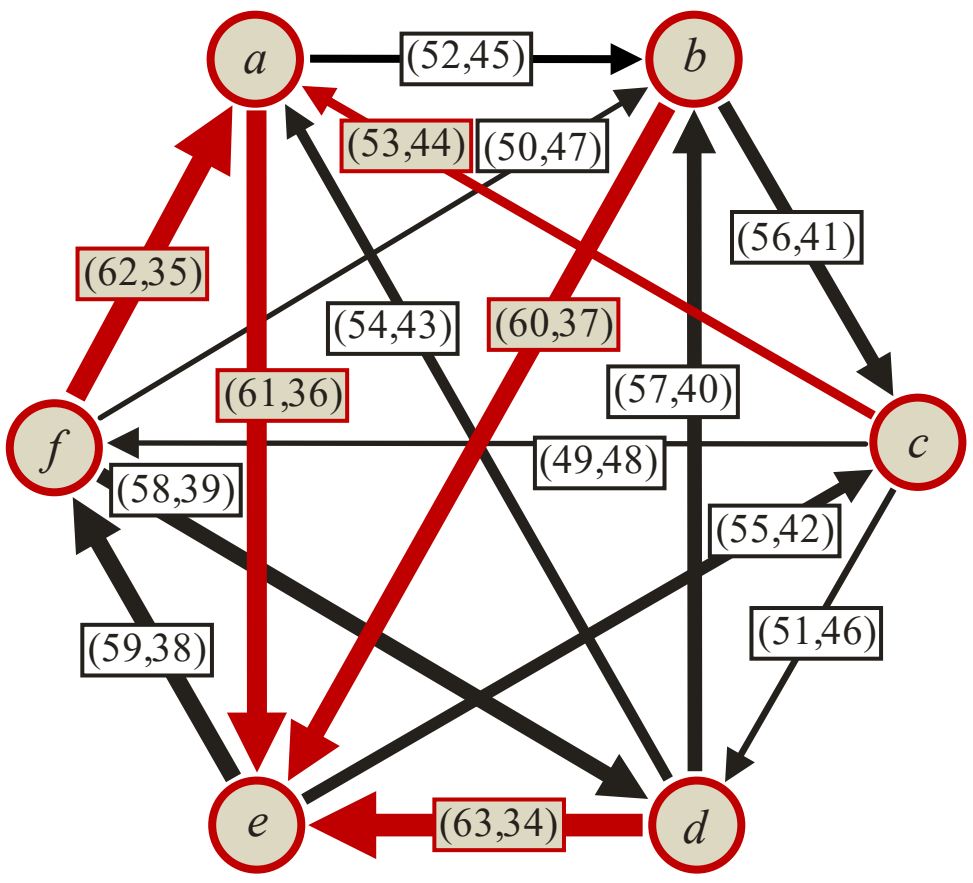

These are the strongest paths
from every other alternative to *e*.

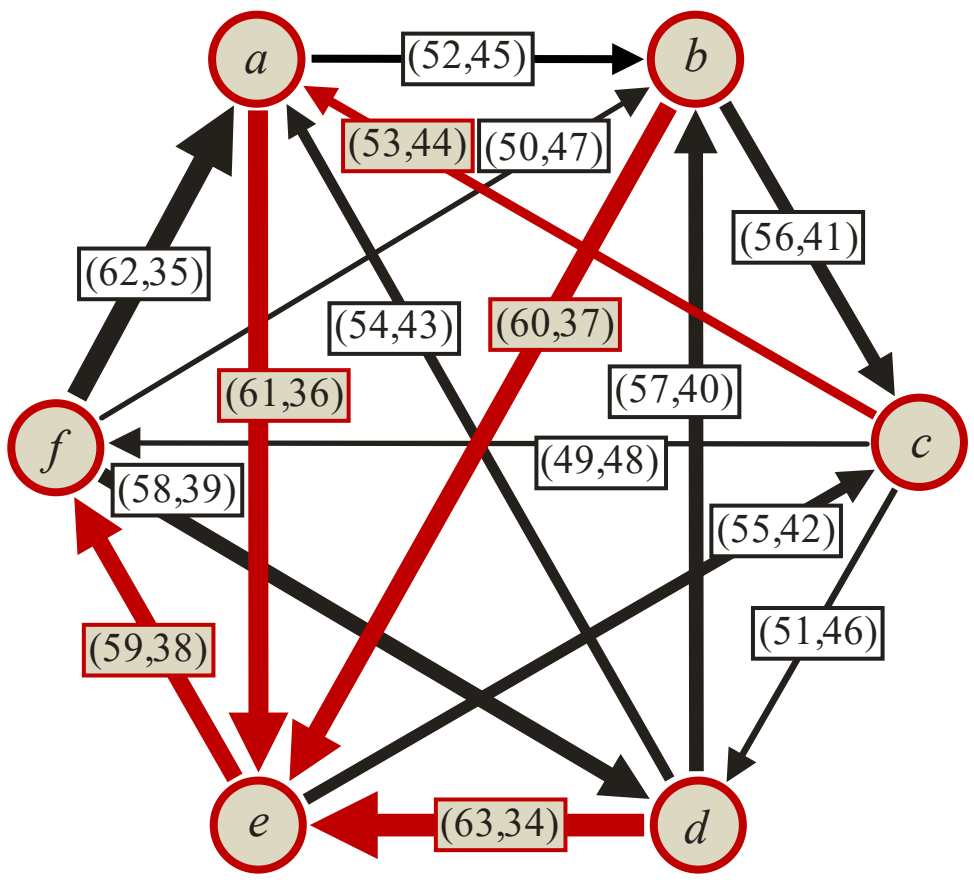

These are the strongest paths
from every other alternative to *f*.

Therefore, the strengths of the strongest paths are:

| | $P_D[*,a]$ | $P_D[*,b]$ | $P_D[*,c]$ | $P_D[*,d]$ | $P_D[*,e]$ | $P_D[*,f]$ |
|---|---|---|---|---|---|---|
| $P_D[a,*]$ | --- | (57,40) | (56,41) | (58,39) | (61,36) | (59,38) |
| $P_D[b,*]$ | (59,38) | --- | (56,41) | (58,39) | (60,37) | (59,38) |
| $P_D[c,*]$ | (53,44) | (53,44) | --- | (53,44) | (53,44) | (53,44) |
| $P_D[d,*]$ | (59,38) | (57,40) | (56,41) | --- | (63,34) | (59,38) |
| $P_D[e,*]$ | (59,38) | (57,40) | (56,41) | (58,39) | --- | (59,38) |
| $P_D[f,*]$ | (62,35) | (57,40) | (56,41) | (58,39) | (61,36) | --- |

We get $\mathcal{O} = \{ac, ae, ba, bc, bd, be, bf, da, dc, de, df, ec, fa, fc, fe\}$ and $\mathcal{S} = \{b\}$.

Suppose, the strongest paths are calculated with the Floyd-Warshall algorithm, as defined in section 2.3.1. Then the following table documents the $C \cdot (C-1) \cdot (C-2) = 120$ steps of the Floyd-Warshall algorithm.

We start with

- $P_D[i,j] := (N[i,j],N[j,i])$ for all $i \in A$ and $j \in A \setminus \{i\}$.
- $pred[i,j] := i$ for all $i \in A$ and $j \in A \setminus \{i\}$.

| | *i* | *j* | *k* | $P_D[j,k]$ | $P_D[j,i]$ | $P_D[i,k]$ | *pred*[*j*,*k*] | *pred*[*i*,*k*] | result |
|---|---|---|---|---|---|---|---|---|---|
| 1 | *a* | *b* | *c* | (56,41) | (45,52) | (44,53) | *b* | *a* | |
| 2 | *a* | *b* | *d* | (40,57) | (45,52) | (43,54) | *b* | *a* | $P_D[b,d]$ is updated from (40,57) to (43,54); *pred*[*b*,*d*] is updated from *b* to *a*. |
| 3 | *a* | *b* | *e* | (60,37) | (45,52) | (61,36) | *b* | *a* | |
| 4 | *a* | *b* | *f* | (47,50) | (45,52) | (35,62) | *b* | *a* | |
| 5 | *a* | *c* | *b* | (41,56) | (53,44) | (52,45) | *c* | *a* | $P_D[c,b]$ is updated from (41,56) to (52,45); *pred*[*c*,*b*] is updated from *c* to *a*. |
| 6 | *a* | *c* | *d* | (51,46) | (53,44) | (43,54) | *c* | *a* | |
| 7 | *a* | *c* | *e* | (42,55) | (53,44) | (61,36) | *c* | *a* | $P_D[c,e]$ is updated from (42,55) to (53,44); *pred*[*c*,*e*] is updated from *c* to *a*. |
| 8 | *a* | *c* | *f* | (49,48) | (53,44) | (35,62) | *c* | *a* | |
| 9 | *a* | *d* | *b* | (57,40) | (54,43) | (52,45) | *d* | *a* | |
| 10 | *a* | *d* | *c* | (46,51) | (54,43) | (44,53) | *d* | *a* | |
| 11 | *a* | *d* | *e* | (63,34) | (54,43) | (61,36) | *d* | *a* | |
| 12 | *a* | *d* | *f* | (39,58) | (54,43) | (35,62) | *d* | *a* | |
| 13 | *a* | *e* | *b* | (37,60) | (36,61) | (52,45) | *e* | *a* | |
| 14 | *a* | *e* | *c* | (55,42) | (36,61) | (44,53) | *e* | *a* | |
| 15 | *a* | *e* | *d* | (34,63) | (36,61) | (43,54) | *e* | *a* | $P_D[e,d]$ is updated from (34,63) to (36,61); *pred*[*e*,*d*] is updated from *e* to *a*. |
| 16 | *a* | *e* | *f* | (59,38) | (36,61) | (35,62) | *e* | *a* | |
| 17 | *a* | *f* | *b* | (50,47) | (62,35) | (52,45) | *f* | *a* | $P_D[f,b]$ is updated from (50,47) to (52,45); *pred*[*f*,*b*] is updated from *f* to *a*. |
| 18 | *a* | *f* | *c* | (48,49) | (62,35) | (44,53) | *f* | *a* | |
| 19 | *a* | *f* | *d* | (58,39) | (62,35) | (43,54) | *f* | *a* | |
| 20 | *a* | *f* | *e* | (38,59) | (62,35) | (61,36) | *f* | *a* | $P_D[f,e]$ is updated from (38,59) to (61,36); *pred*[*f*,*e*] is updated from *f* to *a*. |
| 21 | *b* | *a* | *c* | (44,53) | (52,45) | (56,41) | *a* | *b* | $P_D[a,c]$ is updated from (44,53) to (52,45); *pred*[*a*,*c*] is updated from *a* to *b*. |
| 22 | *b* | *a* | *d* | (43,54) | (52,45) | (43,54) | *a* | *a* | |
| 23 | *b* | *a* | *e* | (61,36) | (52,45) | (60,37) | *a* | *b* | |
| 24 | *b* | *a* | *f* | (35,62) | (52,45) | (47,50) | *a* | *b* | $P_D[a,f]$ is updated from (35,62) to (47,50); *pred*[*a*,*f*] is updated from *a* to *b*. |
| 25 | *b* | *c* | *a* | (53,44) | (52,45) | (45,52) | *c* | *b* | |
| 26 | *b* | *c* | *d* | (51,46) | (52,45) | (43,54) | *c* | *a* | |
| 27 | *b* | *c* | *e* | (53,44) | (52,45) | (60,37) | *a* | *b* | |
| 28 | *b* | *c* | *f* | (49,48) | (52,45) | (47,50) | *c* | *b* | |
| 29 | *b* | *d* | *a* | (54,43) | (57,40) | (45,52) | *d* | *b* | |
| 30 | *b* | *d* | *c* | (46,51) | (57,40) | (56,41) | *d* | *b* | $P_D[d,c]$ is updated from (46,51) to (56,41); *pred*[*d*,*c*] is updated from *d* to *b*. |

| | *i* | *j* | *k* | $P_D[j,k]$ | $P_D[j,i]$ | $P_D[i,k]$ | *pred*[*j*,*k*] | *pred*[*i*,*k*] | result |
|---|---|---|---|---|---|---|---|---|---|
| 31 | *b* | *d* | *e* | (63,34) | (57,40) | (60,37) | *d* | *b* | |
| 32 | *b* | *d* | *f* | (39,58) | (57,40) | (47,50) | *d* | *b* | $P_D[d,f]$ is updated from (39,58) to (47,50); *pred*[*d*,*f*] is updated from *d* to *b*. |
| 33 | *b* | *e* | *a* | (36,61) | (37,60) | (45,52) | *e* | *b* | $P_D[e,a]$ is updated from (36,61) to (37,60); *pred*[*e*,*a*] is updated from *e* to *b*. |
| 34 | *b* | *e* | *c* | (55,42) | (37,60) | (56,41) | *e* | *b* | |
| 35 | *b* | *e* | *d* | (36,61) | (37,60) | (43,54) | *a* | *a* | $P_D[e,d]$ is updated from (36,61) to (37,60). |
| 36 | *b* | *e* | *f* | (59,38) | (37,60) | (47,50) | *e* | *b* | |
| 37 | *b* | *f* | *a* | (62,35) | (52,45) | (45,52) | *f* | *b* | |
| 38 | *b* | *f* | *c* | (48,49) | (52,45) | (56,41) | *f* | *b* | $P_D[f,c]$ is updated from (48,49) to (52,45); *pred*[*f*,*c*] is updated from *f* to *b*. |
| 39 | *b* | *f* | *d* | (58,39) | (52,45) | (43,54) | *f* | *a* | |
| 40 | *b* | *f* | *e* | (61,36) | (52,45) | (60,37) | *a* | *b* | |
| 41 | *c* | *a* | *b* | (52,45) | (52,45) | (52,45) | *a* | *a* | |
| 42 | *c* | *a* | *d* | (43,54) | (52,45) | (51,46) | *a* | *c* | $P_D[a,d]$ is updated from (43,54) to (51,46); *pred*[*a*,*d*] is updated from *a* to *c*. |
| 43 | *c* | *a* | *e* | (61,36) | (52,45) | (53,44) | *a* | *a* | |
| 44 | *c* | *a* | *f* | (47,50) | (52,45) | (49,48) | *b* | *c* | $P_D[a,f]$ is updated from (47,50) to (49,48); *pred*[*a*,*f*] is updated from *b* to *c*. |
| 45 | *c* | *b* | *a* | (45,52) | (56,41) | (53,44) | *b* | *c* | $P_D[b,a]$ is updated from (45,52) to (53,44); *pred*[*b*,*a*] is updated from *b* to *c*. |
| 46 | *c* | *b* | *d* | (43,54) | (56,41) | (51,46) | *a* | *c* | $P_D[b,d]$ is updated from (43,54) to (51,46); *pred*[*b*,*d*] is updated from *a* to *c*. |
| 47 | *c* | *b* | *e* | (60,37) | (56,41) | (53,44) | *b* | *a* | |
| 48 | *c* | *b* | *f* | (47,50) | (56,41) | (49,48) | *b* | *c* | $P_D[b,f]$ is updated from (47,50) to (49,48); *pred*[*b*,*f*] is updated from *b* to *c*. |
| 49 | *c* | *d* | *a* | (54,43) | (56,41) | (53,44) | *d* | *c* | |
| 50 | *c* | *d* | *b* | (57,40) | (56,41) | (52,45) | *d* | *a* | |
| 51 | *c* | *d* | *e* | (63,34) | (56,41) | (53,44) | *d* | *a* | |
| 52 | *c* | *d* | *f* | (47,50) | (56,41) | (49,48) | *b* | *c* | $P_D[d,f]$ is updated from (47,50) to (49,48); *pred*[*d*,*f*] is updated from *b* to *c*. |
| 53 | *c* | *e* | *a* | (37,60) | (55,42) | (53,44) | *b* | *c* | $P_D[e,a]$ is updated from (37,60) to (53,44); *pred*[*e*,*a*] is updated from *b* to *c*. |
| 54 | *c* | *e* | *b* | (37,60) | (55,42) | (52,45) | *e* | *a* | $P_D[e,b]$ is updated from (37,60) to (52,45); *pred*[*e*,*b*] is updated from *e* to *a*. |
| 55 | *c* | *e* | *d* | (37,60) | (55,42) | (51,46) | *a* | *c* | $P_D[e,d]$ is updated from (37,60) to (51,46); *pred*[*e*,*d*] is updated from *a* to *c*. |
| 56 | *c* | *e* | *f* | (59,38) | (55,42) | (49,48) | *e* | *c* | |
| 57 | *c* | *f* | *a* | (62,35) | (52,45) | (53,44) | *f* | *c* | |
| 58 | *c* | *f* | *b* | (52,45) | (52,45) | (52,45) | *a* | *a* | |
| 59 | *c* | *f* | *d* | (58,39) | (52,45) | (51,46) | *f* | *c* | |
| 60 | *c* | *f* | *e* | (61,36) | (52,45) | (53,44) | *a* | *a* | |

| | *i* | *j* | *k* | $P_D[j,k]$ | $P_D[j,i]$ | $P_D[i,k]$ | *pred*[*j*,*k*] | *pred*[*i*,*k*] | result |
|---|---|---|---|---|---|---|---|---|---|
| 61 | *d* | *a* | *b* | (52,45) | (51,46) | (57,40) | *a* | *d* | |
| 62 | *d* | *a* | *c* | (52,45) | (51,46) | (56,41) | *b* | *b* | |
| 63 | *d* | *a* | *e* | (61,36) | (51,46) | (63,34) | *a* | *d* | |
| 64 | *d* | *a* | *f* | (49,48) | (51,46) | (49,48) | *c* | *c* | |
| 65 | *d* | *b* | *a* | (53,44) | (51,46) | (54,43) | *c* | *d* | |
| 66 | *d* | *b* | *c* | (56,41) | (51,46) | (56,41) | *b* | *b* | |
| 67 | *d* | *b* | *e* | (60,37) | (51,46) | (63,34) | *b* | *d* | |
| 68 | *d* | *b* | *f* | (49,48) | (51,46) | (49,48) | *c* | *c* | |
| 69 | *d* | *c* | *a* | (53,44) | (51,46) | (54,43) | *c* | *d* | |
| 70 | *d* | *c* | *b* | (52,45) | (51,46) | (57,40) | *a* | *d* | |
| 71 | *d* | *c* | *e* | (53,44) | (51,46) | (63,34) | *a* | *d* | |
| 72 | *d* | *c* | *f* | (49,48) | (51,46) | (49,48) | *c* | *c* | |
| 73 | *d* | *e* | *a* | (53,44) | (51,46) | (54,43) | *c* | *d* | |
| 74 | *d* | *e* | *b* | (52,45) | (51,46) | (57,40) | *a* | *d* | |
| 75 | *d* | *e* | *c* | (55,42) | (51,46) | (56,41) | *e* | *b* | |
| 76 | *d* | *e* | *f* | (59,38) | (51,46) | (49,48) | *e* | *c* | |
| 77 | *d* | *f* | *a* | (62,35) | (58,39) | (54,43) | *f* | *d* | |
| 78 | *d* | *f* | *b* | (52,45) | (58,39) | (57,40) | *a* | *d* | $P_D[f,b]$ is updated from (52,45) to (57,40); *pred*[*f*,*b*] is updated from *a* to *d*. |
| 79 | *d* | *f* | *c* | (52,45) | (58,39) | (56,41) | *b* | *b* | $P_D[f,c]$ is updated from (52,45) to (56,41). |
| 80 | *d* | *f* | *e* | (61,36) | (58,39) | (63,34) | *a* | *d* | |
| 81 | *e* | *a* | *b* | (52,45) | (61,36) | (52,45) | *a* | *a* | |
| 82 | *e* | *a* | *c* | (52,45) | (61,36) | (55,42) | *b* | *e* | $P_D[a,c]$ is updated from (52,45) to (55,42); *pred*[*a*,*c*] is updated from *b* to *e*. |
| 83 | *e* | *a* | *d* | (51,46) | (61,36) | (51,46) | *c* | *c* | |
| 84 | *e* | *a* | *f* | (49,48) | (61,36) | (59,38) | *c* | *e* | $P_D[a,f]$ is updated from (49,48) to (59,38); *pred*[*a*,*f*] is updated from *c* to *e*. |
| 85 | *e* | *b* | *a* | (53,44) | (60,37) | (53,44) | *c* | *c* | |
| 86 | *e* | *b* | *c* | (56,41) | (60,37) | (55,42) | *b* | *e* | |
| 87 | *e* | *b* | *d* | (51,46) | (60,37) | (51,46) | *c* | *c* | |
| 88 | *e* | *b* | *f* | (49,48) | (60,37) | (59,38) | *c* | *e* | $P_D[b,f]$ is updated from (49,48) to (59,38); *pred*[*b*,*f*] is updated from *c* to *e*. |
| 89 | *e* | *c* | *a* | (53,44) | (53,44) | (53,44) | *c* | *c* | |
| 90 | *e* | *c* | *b* | (52,45) | (53,44) | (52,45) | *a* | *a* | |

| | *i* | *j* | *k* | $P_D[j,k]$ | $P_D[j,i]$ | $P_D[i,k]$ | *pred*[*j*,*k*] | *pred*[*i*,*k*] | result |
|---|---|---|---|---|---|---|---|---|---|
| 91 | *e* | *c* | *d* | (51,46) | (53,44) | (51,46) | *c* | *c* | |
| 92 | *e* | *c* | *f* | (49,48) | (53,44) | (59,38) | *c* | *e* | $P_D[c,f]$ is updated from (49,48) to (53,44); *pred*[*c*,*f*] is updated from *c* to *e*. |
| 93 | *e* | *d* | *a* | (54,43) | (63,34) | (53,44) | *d* | *c* | |
| 94 | *e* | *d* | *b* | (57,40) | (63,34) | (52,45) | *d* | *a* | |
| 95 | *e* | *d* | *c* | (56,41) | (63,34) | (55,42) | *b* | *e* | |
| 96 | *e* | *d* | *f* | (49,48) | (63,34) | (59,38) | *c* | *e* | $P_D[d,f]$ is updated from (49,48) to (59,38); *pred*[*d*,*f*] is updated from *c* to *e*. |
| 97 | *e* | *f* | *a* | (62,35) | (61,36) | (53,44) | *f* | *c* | |
| 98 | *e* | *f* | *b* | (57,40) | (61,36) | (52,45) | *d* | *a* | |
| 99 | *e* | *f* | *c* | (56,41) | (61,36) | (55,42) | *b* | *e* | |
| 100 | *e* | *f* | *d* | (58,39) | (61,36) | (51,46) | *f* | *c* | |
| 101 | *f* | *a* | *b* | (52,45) | (59,38) | (57,40) | *a* | *d* | $P_D[a,b]$ is updated from (52,45) to (57,40); *pred*[*a*,*b*] is updated from *a* to *d*. |
| 102 | *f* | *a* | *c* | (55,42) | (59,38) | (56,41) | *e* | *b* | $P_D[a,c]$ is updated from (55,42) to (56,41); *pred*[*a*,*c*] is updated from *e* to *b*. |
| 103 | *f* | *a* | *d* | (51,46) | (59,38) | (58,39) | *c* | *f* | $P_D[a,d]$ is updated from (51,46) to (58,39); *pred*[*a*,*d*] is updated from *c* to *f*. |
| 104 | *f* | *a* | *e* | (61,36) | (59,38) | (61,36) | *a* | *a* | |
| 105 | *f* | *b* | *a* | (53,44) | (59,38) | (62,35) | *c* | *f* | $P_D[b,a]$ is updated from (53,44) to (59,38); *pred*[*b*,*a*] is updated from *c* to *f*. |
| 106 | *f* | *b* | *c* | (56,41) | (59,38) | (56,41) | *b* | *b* | |
| 107 | *f* | *b* | *d* | (51,46) | (59,38) | (58,39) | *c* | *f* | $P_D[b,d]$ is updated from (51,46) to (58,39); *pred*[*b*,*d*] is updated from *c* to *f*. |
| 108 | *f* | *b* | *e* | (60,37) | (59,38) | (61,36) | *b* | *a* | |
| 109 | *f* | *c* | *a* | (53,44) | (53,44) | (62,35) | *c* | *f* | |
| 110 | *f* | *c* | *b* | (52,45) | (53,44) | (57,40) | *a* | *d* | $P_D[c,b]$ is updated from (52,45) to (53,44); *pred*[*c*,*b*] is updated from *a* to *d*. |
| 111 | *f* | *c* | *d* | (51,46) | (53,44) | (58,39) | *c* | *f* | $P_D[c,d]$ is updated from (51,46) to (53,44); *pred*[*c*,*d*] is updated from *c* to *f*. |
| 112 | *f* | *c* | *e* | (53,44) | (53,44) | (61,36) | *a* | *a* | |
| 113 | *f* | *d* | *a* | (54,43) | (59,38) | (62,35) | *d* | *f* | $P_D[d,a]$ is updated from (54,43) to (59,38); *pred*[*d*,*a*] is updated from *d* to *f*. |
| 114 | *f* | *d* | *b* | (57,40) | (59,38) | (57,40) | *d* | *d* | |
| 115 | *f* | *d* | *c* | (56,41) | (59,38) | (56,41) | *b* | *b* | |
| 116 | *f* | *d* | *e* | (63,34) | (59,38) | (61,36) | *d* | *a* | |
| 117 | *f* | *e* | *a* | (53,44) | (59,38) | (62,35) | *c* | *f* | $P_D[e,a]$ is updated from (53,44) to (59,38); *pred*[*e*,*a*] is updated from *c* to *f*. |
| 118 | *f* | *e* | *b* | (52,45) | (59,38) | (57,40) | *a* | *d* | $P_D[e,b]$ is updated from (52,45) to (57,40); *pred*[*e*,*b*] is updated from *a* to *d*. |
| 119 | *f* | *e* | *c* | (55,42) | (59,38) | (56,41) | *e* | *b* | $P_D[e,c]$ is updated from (55,42) to (56,41); *pred*[*e*,*c*] is updated from *e* to *b*. |
| 120 | *f* | *e* | *d* | (51,46) | (59,38) | (58,39) | *c* | *f* | $P_D[e,d]$ is updated from (51,46) to (58,39); *pred*[*e*,*d*] is updated from *c* to *f*. |

# 4. Analysis of the Schulze Method

## 4.1. Transitivity

In this section, we will prove that the binary relation $\mathcal{O}$, as defined in (2.2.1), is *transitive*. This means: If $ab \in \mathcal{O}$ and $bc \in \mathcal{O}$, then $ac \in \mathcal{O}$. This guarantees that the set $\mathcal{S}$ of potential winners, as defined in (2.2.2), is non-empty. When we interpret the Schulze method as a method to find a set $\mathcal{S}$ of potential winners, rather than a method to generate a binary relation $\mathcal{O}$, then the proof of the transitivity of $\mathcal{O}$ is an essential part of the proof that the Schulze method is well defined.

**Definition:**

An election method satisfies *transitivity* if the following holds for all $a,b,c \in A$:

Suppose:

(4.1.1) $ab \in \mathcal{O}$.

(4.1.2) $bc \in \mathcal{O}$.

Then:

(4.1.3) $ac \in \mathcal{O}$.

**Claim:**

The binary relation $\mathcal{O}$, as defined in (2.2.1), is transitive.

**Proof:**

With (4.1.1), we get

(4.1.4) $P_D[a,b] \succ_D P_D[b,a]$.

With (4.1.2), we get

(4.1.5) $P_D[b,c] \succ_D P_D[c,b]$.

With (2.2.5), we get

(4.1.6) $\min_D \{ P_D[a,b], P_D[b,c] \} \precsim_D P_D[a,c]$.

(4.1.7) $\min_D \{ P_D[b,c], P_D[c,a] \} \precsim_D P_D[b,a]$.

(4.1.8) $\min_D \{ P_D[c,a], P_D[a,b] \} \precsim_D P_D[c,b]$.

Case 1: Suppose

(4.1.9a) $P_D[a,b] \succsim_D P_D[b,c]$.

Combining (4.1.5) and (4.1.9a) gives

(4.1.10a) $P_D[a,b] \succ_D P_D[c,b]$.

Combining (4.1.8) and (4.1.10a) gives

(4.1.11a) $P_D[c,a] \precsim_D P_D[c,b]$.

Combining (4.1.6) and (4.1.9a) gives

(4.1.12a) $P_D[b,c] \precsim_D P_D[a,c]$.

Combining (4.1.11a), (4.1.5), and (4.1.12a) gives

(4.1.13a) $P_D[c,a] \precsim_D P_D[c,b] \prec_D P_D[b,c] \precsim_D P_D[a,c]$.

With (4.1.13a), we get (4.1.3).

Case 2: Suppose

(4.1.9b) $P_D[a,b] \prec_D P_D[b,c]$.

Combining (4.1.4) and (4.1.9b) gives

(4.1.10b) $P_D[b,a] \prec_D P_D[b,c]$.

Combining (4.1.7) and (4.1.10b) gives

(4.1.11b) $P_D[c,a] \precsim_D P_D[b,a]$.

Combining (4.1.6) and (4.1.9b) gives

(4.1.12b) $P_D[a,b] \precsim_D P_D[a,c]$.

Combining (4.1.11b), (4.1.4), and (4.1.12b) gives

(4.1.13b) $P_D[c,a] \precsim_D P_D[b,a] \prec_D P_D[a,b] \precsim_D P_D[a,c]$.

With (4.1.13b), we get (4.1.3). □

The proof, that the Schulze method is transitive, has first been published by Schulze (1998b, 2003, 2011a). This proof can also be found in papers by Tideman (2006), Camps (2008, 2012b), Börgers (2009), and Duchin (2021, 04:26:14 – 05:05:10 hh:mm:ss).

In example 4 (section 3.4), we have $ba \notin \mathcal{O}$ and $ac \notin \mathcal{O}$ and $bc \in \mathcal{O}$. This shows that the Schulze relation, as defined in (2.2.1), is not necessarily negatively transitive.

The following corollary says that the set $\mathcal{S}$ of potential winners, as defined in (2.2.2), is non-empty.

**Corollary (4.1.14):**

For the Schulze method, as defined in section 2.2, we get

(4.1.14) $\forall\ b \notin \mathcal{S}\ \exists\ a \in \mathcal{S}: ab \in \mathcal{O}.$

**Proof of corollary (4.1.14):**

As $b \notin \mathcal{S}$, there must be a $c(1) \in A$ with $c(1),b \in \mathcal{O}$.

If $c(1) \in \mathcal{S}$, then the corollary is proven. If $c(1) \notin \mathcal{S}$, then there must be a $c(2) \in A$ with $c(2),c(1) \in \mathcal{O}$. With the asymmetry and the transitivity of $\mathcal{O}$, we get $c(2),b \in \mathcal{O}$ and $c(2) \notin \{b, c(1)\}$.

We now proceed as follows: If $c(i) \in \mathcal{S}$, then the corollary is proven. If $c(i) \notin \mathcal{S}$, then there must be a $c(i+1) \in A$ with $c(i+1),c(i) \in \mathcal{O}$. With the asymmetry and the transitivity of $\mathcal{O}$, we get $c(i+1),b \in \mathcal{O}$ and $c(i+1) \notin \{b, c(1), ..., c(i)\}$.

We proceed until $c(i) \in \mathcal{S}$ for some $i \in \mathbb{N}$. Such an $i \in \mathbb{N}$ exists because $A$ is finite. □

The following corollary says that alternative $a \in A$ is the unique winner if and only if alternative $a$ disqualifies every other alternative $b \in A \setminus \{a\}$.

**Corollary (4.1.15):**

For the Schulze method, as defined in section 2.2, we get

(4.1.15) $\mathcal{S} = \{a\} \Leftrightarrow (\ \forall\ b \in A \setminus \{a\}: ab \in \mathcal{O}\ ).$

**Proof of corollary (4.1.15):**

$\Leftarrow$ If $ab \in \mathcal{O}\ \forall\ b \in A \setminus \{a\}$, then $a \in A$ disqualifies every $b \in A \setminus \{a\}$ according to (2.2.2). Therefore, we get $\mathcal{S} = \{a\}$.

$\Rightarrow$ With (4.1.14) and $\mathcal{S} = \{a\}$, we get

(4.1.16) $\forall\ b \notin \mathcal{S}: ab \in \mathcal{O}.$

With $\mathcal{S} = \{a\}$, we get

(4.1.17) $b \notin \mathcal{S} \Leftrightarrow b \in A \setminus \{a\}.$

With (4.1.16) and (4.1.17), we get

(4.1.18) $\forall\ b \in A \setminus \{a\}: ab \in \mathcal{O}.$ □

## 4.2. Decisiveness

*Decisiveness* basically says that usually there is a unique winner $\mathcal{S} = \{a\}$. This criterion is immensely important because — although in theory, the sole purpose of a single-winner election is to fill a single seat — in real life, a single-winner election also serves other purposes. For example, one purpose of an election is to settle a dispute, at least temporarily. Another purpose is to give the winner the legitimacy he needs to run his office successfully. For example, when in a referendum the winning option is chosen by a random choice, then the supporters of the losing option will immediately request a re-vote. And a president who is elected by a random choice will be denied any legitimacy and will be quite unable to exercise his office.

**Definition:**

An election method satisfies *decisiveness* if ( for every given number of alternatives ) the proportion of profiles without a unique winner tends to zero as the number of voters in the profile tends to infinity.

**Claim:**

If $\succ_D$ satisfies (2.1.1), then the Schulze method, as defined in section 2.2, satisfies decisiveness.

**Proof (overview):**

Suppose $(x_1,x_2),(y_1,y_2) \in \mathbb{N}_0 \times \mathbb{N}_0$. Then, according to (2.1.1), there is a $v_1 \in \mathbb{N}_0$ such that for all $w_1 \in \mathbb{N}_0$:

1. $w_1 < v_1 \Rightarrow (x_1,x_2) \succ_D (w_1,y_2)$.

2. $w_1 > v_1 \Rightarrow (x_1,x_2) \prec_D (w_1,y_2)$.

When the number of voters tends to infinity ( i.e. when $x_1$, $x_2$, $y_1$, and $y_2$ become large ), then the proportion of profiles, where the condition “$y_1 = v_1$” happens to be satisfied, tends to zero. Therefore, when the number of voters tends to infinity, then the proportion of profiles, where two links *ef* and *gh* happen to have equivalent strengths $(N[e,f],N[f,e]) \approx_D (N[g,h],N[h,g])$, tends to zero.

Therefore, we will prove that, unless there are links *ef* and *gh* of equivalent strengths, there is always a unique winner. We will prove this by showing that, when we simultaneously presume (a) that there is more than one potential winner and (b) that there are no links *ef* and *gh* of equivalent strengths, then we necessarily get to a contradiction.

**Proof (details):**

Suppose that there is more than one potential winner. Suppose alternative $a \in A$ and alternative $b \in A$ are potential winners. Then

(4.2.1) $\forall\, i \in A \setminus \{a\}$: $P_D[a,i] \gtrsim_D P_D[i,a]$.

(4.2.2) $\forall\, j \in A \setminus \{b\}$: $P_D[b,j] \gtrsim_D P_D[j,b]$.

(4.2.3) $P_D[a,b] \approx_D P_D[b,a]$.

Suppose there are no links *ef* and *gh* of equivalent strengths $(N[e,f],N[f,e]) \approx_D (N[g,h],N[h,g])$. Then $P_D[a,b] \approx_D P_D[b,a]$ means that the weakest link in the strongest path from alternative *a* to alternative *b* and the weakest link in the strongest path from alternative *b* to alternative *a* must be the same link, say *cd*. Therefore, the strongest paths have the following structure:

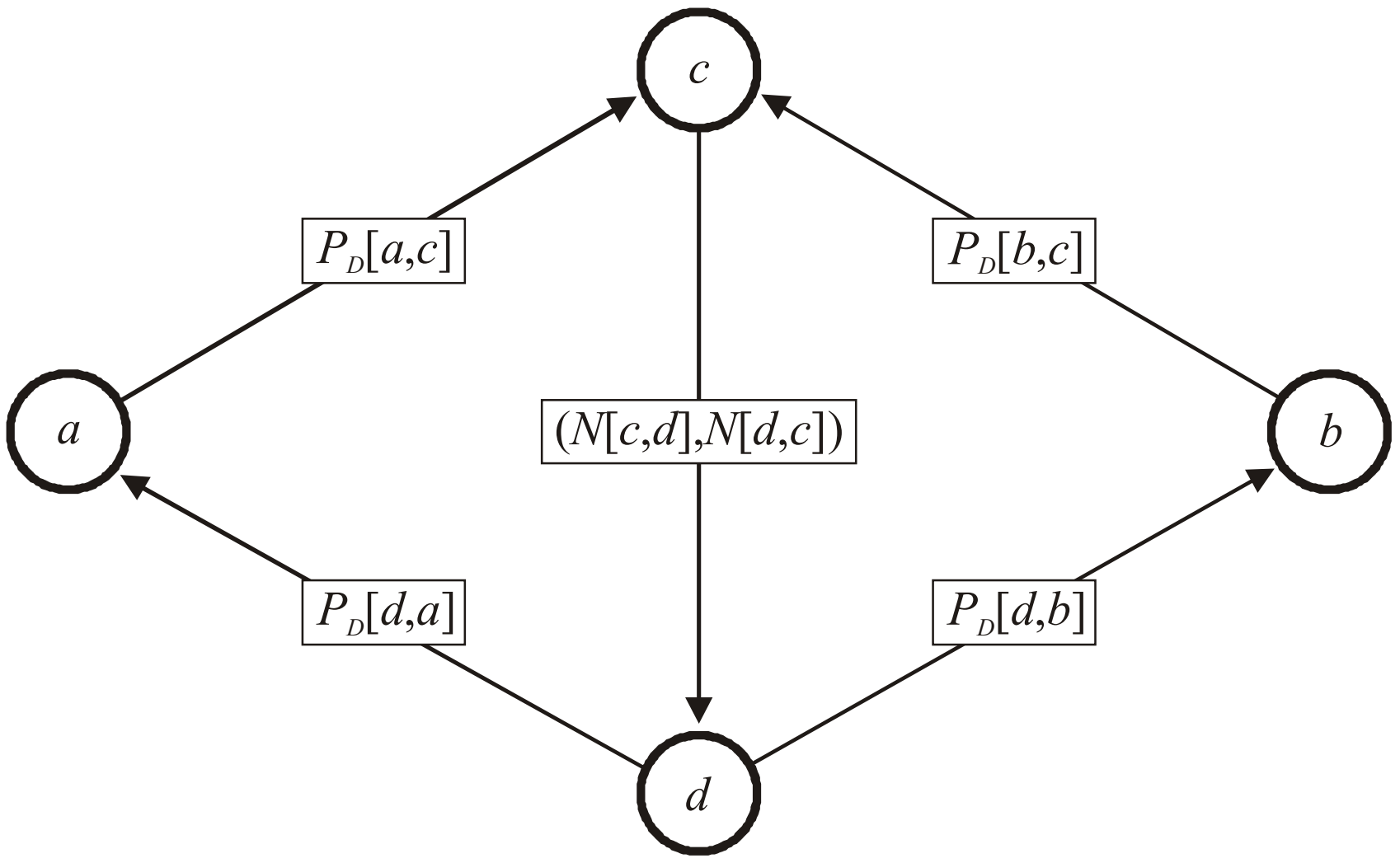

As $cd$ is the weakest link in the strongest path from alternative $a$ to alternative $b$, we get

(4.2.4) $P_D[a,d] \approx_D P_D[a,b]$.

(4.2.5) $P_D[d,b] \succ_D P_D[a,b]$.

As $cd$ is the weakest link in the strongest path from alternative $b$ to alternative $a$, we get

(4.2.6) $P_D[b,d] \approx_D P_D[b,a]$.

(4.2.7) $P_D[d,a] \succ_D P_D[b,a]$.

With (4.2.7), (4.2.3), and (4.2.4), we get

(4.2.8) $P_D[d,a] \succ_D P_D[b,a] \approx_D P_D[a,b] \approx_D P_D[a,d]$.

But (4.2.8) contradicts (4.2.1).

Similarly, with (4.2.5), (4.2.3), and (4.2.6), we get

(4.2.9) $P_D[d,b] \succ_D P_D[a,b] \approx_D P_D[b,a] \approx_D P_D[b,d]$.

But (4.2.9) contradicts (4.2.2). □

When $\succ_D$ satisfies (2.1.1), then the more detailed definition for the Schulze method, as given in (5.1.6) – (5.1.7), also satisfies the following criterion: For every given number of alternatives, the proportion of profiles where $O$ is not a unique linear order tends to zero as the number of voters in the profile tends to infinity. See section 5.3.2.

## 4.3. Resolvability

The *resolvability criterion* says that, when there is more than one potential winner, then, for every alternative $a \in \mathcal{S}$, it is sufficient to add a single ballot $w$ so that alternative $a$ becomes the unique winner.

**Definition:**

An election method satisfies *resolvability* if the following holds:

$\forall\ a \in \mathcal{S}^{\text{old}}$: It is possible to construct a strict weak order $w$ with the following two properties:

(4.3.1) $\quad \forall\ f \in A \setminus \{a\}: a \succ_w f.$

(4.3.2) $\quad \mathcal{S}^{\text{new}} = \{a\}$ for $V^{\text{new}} := V^{\text{old}} + \{w\}$.

**Claim:**

If $\succ_D$ satisfies (2.1.1), then the Schulze method, as defined in section 2.2, satisfies resolvability.

**Proof:**

Suppose $a \in \mathcal{S}^{\text{old}}$. Then we get

(4.3.3) $\quad \forall\ b \in A \setminus \{a\}: P_D^{\text{old}}[a,b] \succsim_D P_D^{\text{old}}[b,a].$

Suppose $pred^{\text{old}}[x,y]$ is the predecessor of alternative $y$ in the strongest path from alternative $x \in A$ to alternative $y \in A \setminus \{x\}$, as calculated in section 2.3.1.

Suppose the strict weak order $w$ is chosen as follows:

(4.3.4) $\quad \forall\ f \in A \setminus \{a\}: pred^{\text{old}}[a,f] \succ_w f.$

(4.3.5) $\quad \forall\ e,f \in A \setminus \{a\}: (\ P_D^{\text{old}}[e,a] \succ_D P_D^{\text{old}}[f,a] \Rightarrow e \succ_w f\ ).$

With (4.3.4), we get (4.3.1).

We will prove the following three claims:

Claim #1: It is not possible that (4.3.4) requires $e \succ_w f$ and that simultaneously (4.3.5) requires $f \succ_w e$.

Claim #2: $\forall\ g \in A \setminus \{a\}: P_D^{\text{new}}[a,g] \succ_D P_D^{\text{old}}[a,g].$

Claim #3: $\forall\ g \in A \setminus \{a\}: P_D^{\text{new}}[g,a] \prec_D P_D^{\text{old}}[a,g].$

With claim #2 and claim #3, we get

$P_D^{\text{new}}[a,g] \succ_D P_D^{\text{new}}[g,a]$ for all $g \in A \setminus \{a\}$

so that $ag \in \mathcal{O}^{\text{new}}$ for all $g \in A \setminus \{a\}$

so that $\mathcal{S}^{\text{new}} = \{a\}$.

Proof of claim #1:

Suppose $e,f \in A \setminus \{a\}$. With (2.2.3), we get

(4.3.6) $P_D^{\text{old}}[e,f] \succsim_D (N^{\text{old}}[e,f],N^{\text{old}}[f,e])$.

With (2.2.5), we get

(4.3.7) $\min_D \{ P_D^{\text{old}}[e,f], P_D^{\text{old}}[f,a] \} \precsim_D P_D^{\text{old}}[e,a]$.

With (4.3.3), we get

(4.3.8) $P_D^{\text{old}}[a,f] \succsim_D P_D^{\text{old}}[f,a]$.

Suppose (4.3.4) requires $e \succ_w f$. Then $e = pred^{\text{old}}[a,f]$. Therefore, the link $ef$ was in the strongest path from alternative $a$ to alternative $f$. Thus, we get

(4.3.9) $P_D^{\text{old}}[a,f] \precsim_D (N^{\text{old}}[e,f],N^{\text{old}}[f,e])$.

Suppose (4.3.5) requires $f \succ_w e$. Then

(4.3.10) $P_D^{\text{old}}[f,a] \succ_D P_D^{\text{old}}[e,a]$.

With (4.3.6), (4.3.9), (4.3.8), and (4.3.10), we get

(4.3.11) $P_D^{\text{old}}[e,f] \succsim_D (N^{\text{old}}[e,f],N^{\text{old}}[f,e]) \succsim_D P_D^{\text{old}}[a,f] \succsim_D P_D^{\text{old}}[f,a] \succ_D P_D^{\text{old}}[e,a]$.

But (4.3.10) and (4.3.11) together contradict (4.3.7).

Proof of claim #2:

With (2.1.1) and (4.3.4), we get: The strength of each link of the strongest paths from alternative $a$ to each other alternative $g \in A \setminus \{a\}$ is increased. Therefore

(4.3.12) $\forall\, g \in A \setminus \{a\}$: $P_D^{\text{new}}[a,g] \succ_D P_D^{\text{old}}[a,g]$.

Proof of claim #3:

Suppose $g \in A \setminus \{a\}$. Suppose

(4.3.13) $\mathfrak{T}(g) := ( \{a\} \cup \{ h \in A \setminus \{a\} \mid P_D^{\text{old}}[h,a] \succ_D P_D^{\text{old}}[a,g] \} )$.

With (4.3.3) and (4.3.13), we get

(4.3.14) $g \notin \mathfrak{T}(g)$ and $a \in \mathfrak{T}(g)$

and, therefore, $\varnothing \neq \mathfrak{T}(g) \subsetneq A$. Furthermore, we get

(4.3.15) $\forall\, i \notin \mathfrak{T}(g)\ \forall\, j \in \mathfrak{T}(g)$: $(N^{\text{old}}[i,j],N^{\text{old}}[j,i]) \precsim_D P_D^{\text{old}}[a,g]$.

Otherwise, there was a path from alternative $i$ to alternative $a$ via alternative $j$ with a strength of more than $P_D^{\text{old}}[a,g]$. But ( as $i \notin \mathfrak{T}(g)$ ) this would contradict the definition of $\mathfrak{T}(g)$.

With (4.3.5), (4.3.1), and (4.3.13), we get

(4.3.16) $\forall\, i \notin \mathfrak{T}(g)\ \forall\, j \in \mathfrak{T}(g)$: $j \succ_w i$.

With (2.1.1) and (4.3.16), we get

(4.3.17) $\forall\, i \notin \mathfrak{T}(g)\ \forall\, j \in \mathfrak{T}(g)$: $(N^{\text{new}}[i,j],N^{\text{new}}[j,i]) \prec_D (N^{\text{old}}[i,j],N^{\text{old}}[j,i])$.

With (4.3.15) and (4.3.17), we get

(4.3.18) $\forall\, i \notin \mathfrak{T}(g)\ \forall\, j \in \mathfrak{T}(g)$: $(N^{\text{new}}[i,j],N^{\text{new}}[j,i]) \prec_D P_D^{\text{old}}[a,g]$.

With (4.3.14) and (4.3.18), we get

(4.3.19) $P_D^{\text{new}}[g,a] \prec_D P_D^{\text{old}}[a,g]$. □

The proof in section 4.3 has first been published by Schulze (2011a). It immediately attracted attention, because it doesn’t only prove that there is a tie-breaking ballot *w*, it also shows how this tie-breaking ballot *w* can be calculated in a polynomial runtime. Parkes (2012) pointed to the fact that this proof can also be interpreted as saying that it is possible to calculate a voting strategy in a polynomial runtime. This observation by Parkes has been extended by Gaspers (2012), Menton (2013a, 2013b), J. Müller (2013, 2020), and Hemaspaandra (2016).

Papers on the computational manipulability of the Schulze method have also been written by Reisch (2014) and Schend (2015). Surveys on the complexity of calculating a voting strategy under the Schulze method and under other single-winner election methods have been written by Durand (2015, 2026), Baumeister (2016), Conitzer (2016), and Faliszewski (2016).

In example 6 (section 3.6), we have $\mathcal{S} = \{a, c\}$.

Suppose, we want to make alternative $a \in \mathcal{S}$ become the unique winner by adding a single ballot $w$. The strongest paths from alternative $a$ to every other alternative $y \in A \setminus \{a\}$, as calculated in section 2.3.1, form the following arborescence:

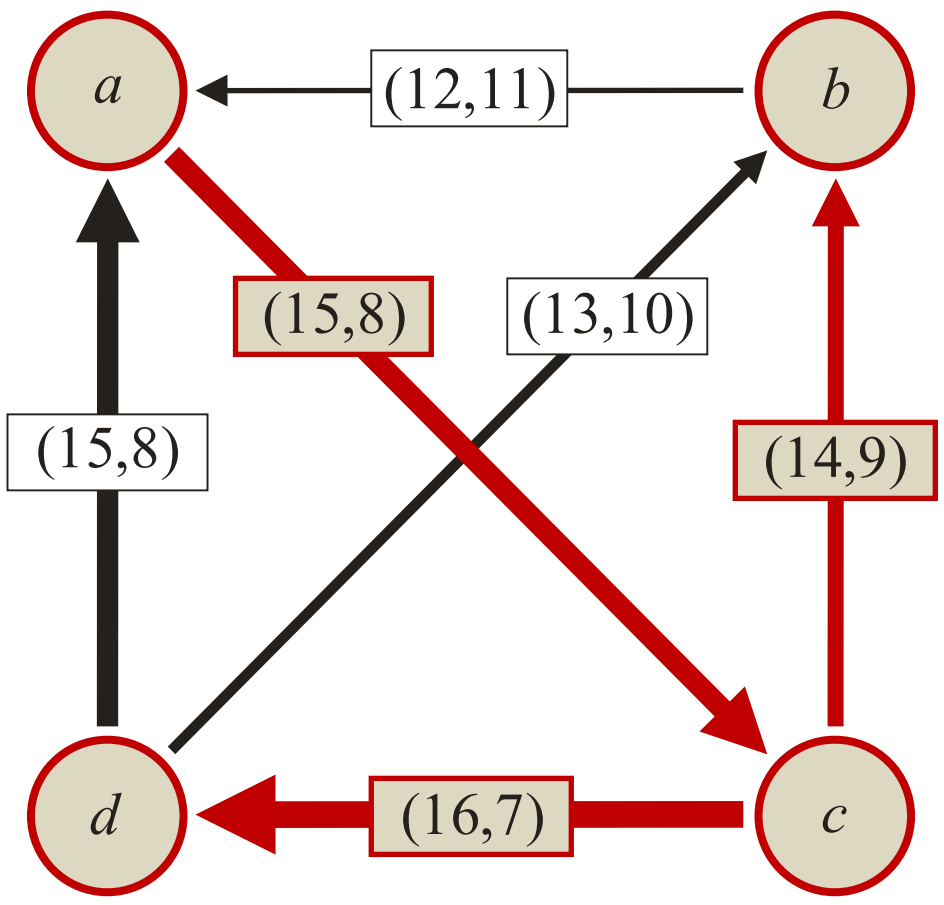

See also the row "from $a$ ..." and the column "... to every other alternative" in the table on page 99. So with (4.3.4) and the arborescence above, we get $a \succ_w c$, $c \succ_w b$ and $c \succ_w d$.

The strengths of the strongest paths in example 6 are:

| | $P_D[*,a]$ | $P_D[*,b]$ | $P_D[*,c]$ | $P_D[*,d]$ |
|---|---|---|---|---|
| $P_D[a,*]$ | --- | (14,9) | (15,8) | (15,8) |
| $P_D[b,*]$ | (12,11) | --- | (12,11) | (12,11) |
| $P_D[c,*]$ | (15,8) | (14,9) | --- | (16,7) |
| $P_D[d,*]$ | (15,8) | (14,9) | (15,8) | --- |

So (4.3.5) implies $c \succ_w b$ and $d \succ_w b$.

Therefore, we can add the ballot $acdb$ to make alternative $a$ become the unique winner.

Suppose, we want to make alternative $c \in \mathcal{S}$ become the unique winner by adding a single ballot $w$. The strongest paths from alternative $c$ to every other alternative $y \in A \setminus \{c\}$, as calculated in section 2.3.1, form the following arborescence:

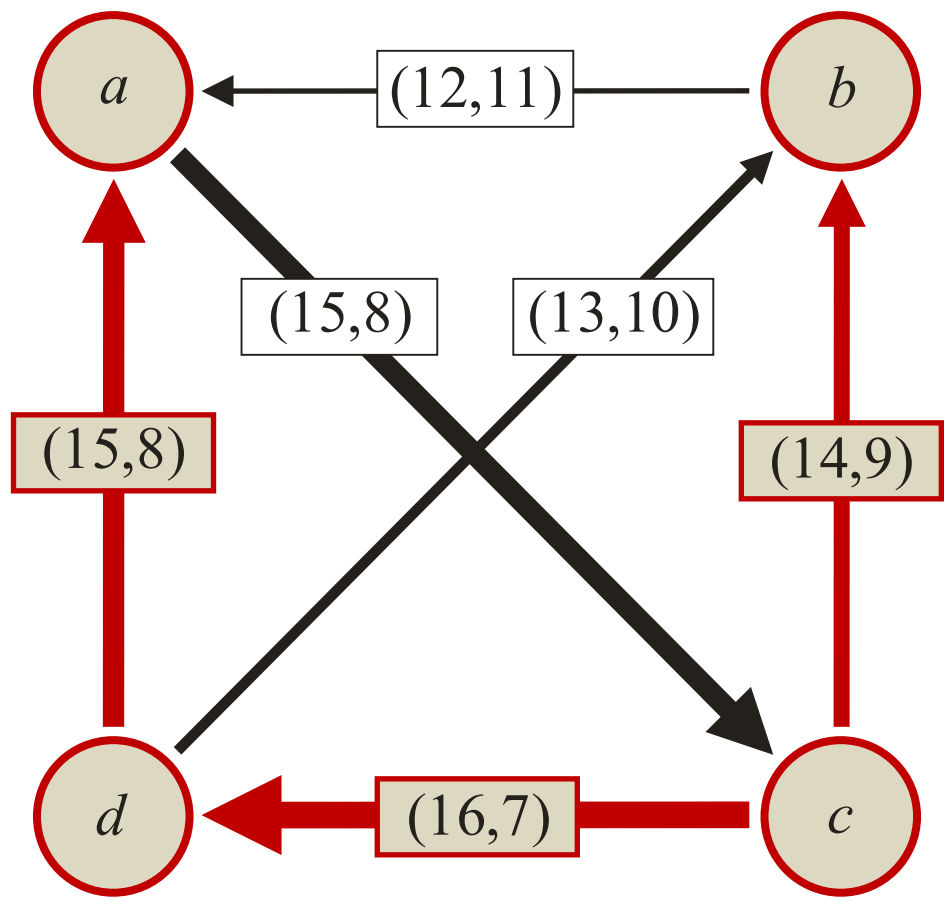

See also the row "from $c$ ..." and the column "... to every other alternative" in the table on page 99. So with (4.3.4) and the arborescence above, we get $c \succ_w b$, $c \succ_w d$ and $d \succ_w a$.

The strengths of the strongest paths in example 6 are:

| | $P_D[*,a]$ | $P_D[*,b]$ | $P_D[*,c]$ | $P_D[*,d]$ |
|---|---|---|---|---|
| $P_D[a,*]$ | --- | (14,9) | (15,8) | (15,8) |
| $P_D[b,*]$ | (12,11) | --- | (12,11) | (12,11) |
| $P_D[c,*]$ | (15,8) | (14,9) | --- | (16,7) |
| $P_D[d,*]$ | (15,8) | (14,9) | (15,8) | --- |

So (4.3.5) implies $a \succ_w b$ and $d \succ_w b$.

Therefore, we can add the ballot *cdab* to make alternative $c$ become the unique winner.

### 4.4. Pareto

The *Pareto criterion* says that the election method must respect unanimous opinions. There are two different versions of the Pareto criterion. The first version addresses situations with " $a \succ_v b$ for all $v \in V$ ", while the second version addresses situations with " $a \succsim_v b$ for all $v \in V$ " ( for some pair of alternatives $a,b \in A$ ). The first version says that, when every voter strictly prefers alternative $a$ to alternative $b$ ( i.e. $a \succ_v b$ for all $v \in V$ ), then alternative $a$ must perform better than alternative $b$. The second version says that, when no voter strictly prefers alternative $b$ to alternative $a$ ( i.e. $a \succsim_v b$ for all $v \in V$ ), then alternative $b$ must not perform better than alternative $a$. We will prove that the Schulze method, as defined in section 2.2, satisfies both versions of the Pareto criterion.

#### 4.4.1. Formulation #1

**Definition:**

An election method satisfies the first version of the *Pareto criterion* if the following holds:

Suppose:

(4.4.1.1) $\forall\, v \in V\text{: } a \succ_v b.$

Then:

(4.4.1.2) $ab \in \mathcal{O}.$

(4.4.1.3) $\forall\, f \in A \setminus \{a,b\}\text{: } bf \in \mathcal{O} \Rightarrow af \in \mathcal{O}.$

(4.4.1.4) $\forall\, f \in A \setminus \{a,b\}\text{: } fa \in \mathcal{O} \Rightarrow fb \in \mathcal{O}.$

(4.4.1.5) $b \notin \mathcal{S}.$

**Claim:**

If $\succ_D$ satisfies (2.1.1), then the Schulze method, as defined in section 2.2, satisfies the first version of the Pareto criterion.

**Proof:**

With (2.1.1) and (4.4.1.1), we get

(4.4.1.6) $\forall\, e,f \in A\text{: } (N[a,b],N[b,a]) \succsim_D (N[e,f],N[f,e]).$

(4.4.1.7) $[\ (N[a,b],N[b,a]) \approx_D (N[e,f],N[f,e])\ ] \Leftrightarrow [\ \forall\, v \in V\text{: } e \succ_v f\ ].$

With (2.2.4), we get: $ab \in \mathcal{O}$, unless the link $ab$ is in a directed cycle that consists of links of which each is at least as strong as the link $ab$.

However, as we presumed that the individual ballots $\succ_v$ are strict weak orders, it is not possible that there is a directed cycle of unanimous opinions. Therefore, it is not possible that the link $ab$ is in a directed cycle that consists of links of which each is at least as strong as the link $ab$. Therefore, with (2.2.4), (4.4.1.6), and (4.4.1.7), we get (4.4.1.2). With (4.4.1.2), we get (4.4.1.5). With (4.4.1.2) and the transitivity of $\mathcal{O}$, we get (4.4.1.3) and (4.4.1.4). □

### 4.4.2. Formulation #2

**Definition:**

An election method satisfies the second version of the *Pareto criterion* if the following holds:

Suppose:

(4.4.2.1) $\forall\, v \in V$: $a \succsim_v b$.

Then:

(4.4.2.2) $ba \notin \mathcal{O}$.

(4.4.2.3) $\forall\, f \in A \setminus \{a,b\}$: $bf \in \mathcal{O} \Rightarrow af \in \mathcal{O}$.

(4.4.2.4) $\forall\, f \in A \setminus \{a,b\}$: $fa \in \mathcal{O} \Rightarrow fb \in \mathcal{O}$.

(4.4.2.5) $b \in \mathcal{S} \Rightarrow a \in \mathcal{S}$.

**Claim:**

If $\succ_D$ satisfies (2.1.1), then the Schulze method, as defined in section 2.2, satisfies the second version of the Pareto criterion.

**Remark:**

As $\mathcal{O}$ isn't necessarily negatively transitive, (4.4.2.3) and (4.4.2.4) don't follow directly from (4.4.2.2).

**Proof:**

With (4.4.2.1), we get

(4.4.2.6) $\forall\, e \in A \setminus \{a,b\}$: $N[a,e] \geq N[b,e]$.

With (4.4.2.1), we get

(4.4.2.7) $\forall\, e \in A \setminus \{a,b\}$: $N[e,b] \geq N[e,a]$.

With (2.1.1), (4.4.2.6), and (4.4.2.7), we get

(4.4.2.8) $\forall\, e \in A \setminus \{a,b\}$: $(N[a,e],N[e,a]) \succsim_D (N[b,e],N[e,b])$.

With (2.1.1), (4.4.2.6), and (4.4.2.7), we get

(4.4.2.9) $\forall\, e \in A \setminus \{a,b\}$: $(N[e,b],N[b,e]) \succsim_D (N[e,a],N[a,e])$.

Suppose $c(1),...,c(n) \in A$ is the strongest path from alternative $b$ to alternative $a$. With (4.4.2.8) and (4.4.2.9), we get: $a,c(2),...,c(n-1),b$ is a path from alternative $a$ to alternative $b$ with at least the same strength. Therefore

(4.4.2.10) $P_D[a,b] \succsim_D P_D[b,a]$.

With (4.4.2.10), we get (4.4.2.2).

Suppose $c(1),...,c(n) \in A$ is the strongest path from alternative $b$ to alternative $f \in A \setminus \{a,b\}$. With (4.4.2.8), we get: $a,c(m+1),...,c(n)$, where $c(m)$ is the last occurrence of an alternative of the set $\{a,b\}$, is a path from alternative $a$ to alternative $f$ with at least the same strength. Therefore

(4.4.2.11) $\forall f \in A \setminus \{a,b\}: P_D[a,f] \gtrsim_D P_D[b,f]$.

Suppose $c(1),...,c(n) \in A$ is the strongest path from alternative $f \in A \setminus \{a,b\}$ to alternative $a$. With (4.4.2.9), we get: $c(1),...,c(m-1),b$, where $c(m)$ is the first occurrence of an alternative of the set $\{a,b\}$, is a path from alternative $f$ to alternative $b$ with at least the same strength. Therefore

(4.4.2.12) $\forall f \in A \setminus \{a,b\}: P_D[f,b] \gtrsim_D P_D[f,a]$.

Part 1: Suppose $f \in A \setminus \{a,b\}$. Suppose

(4.4.2.13a) $bf \in \mathcal{O}$.

With (4.4.2.13a), we get

(4.4.2.14a) $P_D[b,f] \succ_D P_D[f,b]$.

With (4.4.2.11), (4.4.2.14a), and (4.4.2.12), we get

(4.4.2.15a) $P_D[a,f] \gtrsim_D P_D[b,f] \succ_D P_D[f,b] \gtrsim_D P_D[f,a]$.

With (4.4.2.15a), we get (4.4.2.3).

Part 2: Suppose $f \in A \setminus \{a,b\}$. Suppose

(4.4.2.13b) $fa \in \mathcal{O}$.

With (4.4.2.13b), we get

(4.4.2.14b) $P_D[f,a] \succ_D P_D[a,f]$.

With (4.4.2.12), (4.4.2.14b), and (4.4.2.11), we get

(4.4.2.15b) $P_D[f,b] \gtrsim_D P_D[f,a] \succ_D P_D[a,f] \gtrsim_D P_D[b,f]$.

With (4.4.2.15b), we get (4.4.2.4).

Part 3: Suppose

(4.4.2.13c) $b \in \mathcal{S}$.

With (4.4.2.13c), we get

(4.4.2.14c) $\forall f \in A \setminus \{b\}: fb \notin \mathcal{O}$.

With (4.4.2.4) and (4.4.2.14c), we get

(4.4.2.15c) $\forall f \in A \setminus \{a,b\}: fa \notin \mathcal{O}$.

With (4.4.2.2) and (4.4.2.15c), we get

(4.4.2.16c) $\forall f \in A \setminus \{a\}: fa \notin \mathcal{O}$.

With (4.4.2.16c), we get (4.4.2.5). □

Suppose $\succ_D$ satisfies (2.1.1). Then the more detailed definition for the Schulze method, as given in section 5.1, also satisfies the following version of the Pareto criterion:

Suppose:

(4.4.2.17) $\forall\, v \in V\!: a \succsim_v b.$

(4.4.2.18) $\exists\, v \in V\!: a \succ_v b.$

Then:

(4.4.2.19) $ab \in \mathcal{O}.$

(4.4.2.20) $\forall\, f \in A \setminus \{a,b\}\!: bf \in \mathcal{O} \Rightarrow af \in \mathcal{O}.$

(4.4.2.21) $\forall\, f \in A \setminus \{a,b\}\!: fa \in \mathcal{O} \Rightarrow fb \in \mathcal{O}.$

(4.4.2.22) $b \notin \mathcal{S}.$

## 4.5. Reversal Symmetry

*Reversal symmetry* as a criterion for single-winner election methods has been proposed by Saari (1994). This criterion says that, when $\succ_v$ is reversed for all $v \in V$, then also the result of the elections must be reversed; see (4.5.2). $\mathcal{S}^{\text{old}}$ must not be a strict subset of $\mathcal{S}^{\text{new}}$; $\mathcal{S}^{\text{new}}$ must not be a strict subset of $\mathcal{S}^{\text{old}}$; see (4.5.3). It should not be possible that the same alternatives are elected in the original situation and in the reversed situation, unless all alternatives are tied; see (4.5.4).

Basic idea of this criterion is that, when there is a vote on the best alternatives and then there is a vote on the worst alternatives and when in both cases the same alternatives are chosen, then this questions the logic of the underlying heuristic of the used election method.

**<u>Definition:</u>**

An election method satisfies *reversal symmetry* if the following holds:

Suppose:

(4.5.1) $\forall\, e,f \in A\ \forall\, v \in V\!: e \succ_v^{\text{old}} f \Leftrightarrow f \succ_v^{\text{new}} e.$

Then:

(4.5.2) $\forall\, a,b \in A\!: ab \in \mathcal{O}^{\text{old}} \Leftrightarrow ba \in \mathcal{O}^{\text{new}}.$

(4.5.3) $(\, \exists\, i \in A\!: i \in \mathcal{S}^{\text{old}} \wedge i \notin \mathcal{S}^{\text{new}}\,) \Leftrightarrow$
$(\, \exists\, j \in A\!: j \notin \mathcal{S}^{\text{old}} \wedge j \in \mathcal{S}^{\text{new}}\,).$

(4.5.4) $\mathcal{S}^{\text{old}} = \mathcal{S}^{\text{new}} \Leftrightarrow \mathcal{S}^{\text{old}} = A.$

**Claim:**

The Schulze method, as defined in section 2.2, satisfies reversal symmetry.

**Proof:**

With (4.5.1), we get

(4.5.5) $\forall\ e,f \in A$: $N^{old}[e,f] = N^{new}[f,e]$.

With (4.5.5), we get

(4.5.6) $\forall\ e,f \in A$: $(N^{old}[e,f],N^{old}[f,e]) \approx_D (N^{new}[f,e],N^{new}[e,f])$.

With (4.5.6), we get: When $c(1),...,c(n) \in A$ was a path from alternative $g \in A$ to alternative $h \in A \setminus \{g\}$, then $c(n),...,c(1)$ is a path from alternative $h$ to alternative $g$ with the same strength. Therefore

(4.5.7) $\forall\ g,h \in A$: $P_D^{old}[g,h] \approx_D P_D^{new}[h,g]$.

With (4.5.7), we get (4.5.2).

Part 1:

We now prove: $(\ \exists\ i \in A$: $i \in \mathcal{S}^{old} \wedge i \notin \mathcal{S}^{new}\ ) \Rightarrow (\ \exists\ j \in A$: $j \notin \mathcal{S}^{old} \wedge j \in \mathcal{S}^{new}\ )$.

Suppose $\exists\ i \in A$: $i \in \mathcal{S}^{old}$ and $i \notin \mathcal{S}^{new}$. With $i \notin \mathcal{S}^{new}$ and (4.1.14), we get that there is a $j \in \mathcal{S}^{new}$ with $ji \in \mathcal{O}^{new}$. With (4.5.2), we get $ij \in \mathcal{O}^{old}$ and, therefore, $j \notin \mathcal{S}^{old}$. With $j \notin \mathcal{S}^{old}$ and $j \in \mathcal{S}^{new}$, we get the "$\Rightarrow$" direction of (4.5.3). The proof for the "$\Leftarrow$" direction of (4.5.3) is analogous.

Part 2:

We now prove: $\mathcal{S}^{old} = A \Rightarrow \mathcal{S}^{old} = \mathcal{S}^{new}$.

Suppose $\mathcal{S}^{old} = A$. Then we get $\mathcal{O}^{old} = \emptyset$. Otherwise, if there was an $ij \in \mathcal{O}^{old}$, we would immediately get $j \notin \mathcal{S}^{old}$ and, therefore, $\mathcal{S}^{old} \neq A$. With $\mathcal{O}^{old} = \emptyset$ and (4.5.2), we get $\mathcal{O}^{new} = \emptyset$ and, therefore, $\mathcal{S}^{new} = A$. With $\mathcal{S}^{old} = A$ and $\mathcal{S}^{new} = A$, we get $\mathcal{S}^{old} = \mathcal{S}^{new}$.

Part 3:

We now prove: $\mathcal{S}^{old} \neq A \Rightarrow \mathcal{S}^{old} \neq \mathcal{S}^{new}$.

Suppose $\mathcal{S}^{old} \neq A$. Then there was a $j \notin \mathcal{S}^{old}$. With (4.1.14), we get that there was an $i \in \mathcal{S}^{old}$ with $ij \in \mathcal{O}^{old}$. With (4.5.2), we get $ji \in \mathcal{O}^{new}$ and, therefore, $i \notin \mathcal{S}^{new}$. With $i \in \mathcal{S}^{old}$ and $i \notin \mathcal{S}^{new}$, we get $\mathcal{S}^{old} \neq \mathcal{S}^{new}$. With part 2 and part 3, we get (4.5.4). □

## 4.6. Monotonicity and Strict Monotonicity

### 4.6.1. Monotonicity

*Monotonicity* says that, when some voters rank alternative $a \in A$ higher [see (4.6.1.1) and (4.6.1.2)] without changing the order in which they rank the other alternatives relatively to each other [see (4.6.1.3)], then this must not hurt alternative *a* [see (4.6.1.4) – (4.6.1.5)]. Monotonicity is also known as *non-negative responsiveness* and *mono-raise*.

**Definition:**

An election method satisfies *monotonicity* if the following holds:

Suppose $a \in A$. Suppose the ballots are modified in such a manner that the following three statements are satisfied:

(4.6.1.1) $\forall f \in A \setminus \{a\}\ \forall v \in V: a \succ_v^{\text{old}} f \Rightarrow a \succ_v^{\text{new}} f.$

(4.6.1.2) $\forall f \in A \setminus \{a\}\ \forall v \in V: a \succsim_v^{\text{old}} f \Rightarrow a \succsim_v^{\text{new}} f.$

(4.6.1.3) $\forall e,f \in A \setminus \{a\}\ \forall v \in V: e \succ_v^{\text{old}} f \Leftrightarrow e \succ_v^{\text{new}} f.$

Then:

(4.6.1.4) $\forall b \in A \setminus \{a\}: ab \in \mathcal{O}^{\text{old}} \Rightarrow ab \in \mathcal{O}^{\text{new}}.$

(4.6.1.5) $\forall b \in A \setminus \{a\}: ba \notin \mathcal{O}^{\text{old}} \Rightarrow ba \notin \mathcal{O}^{\text{new}}.$

(4.6.1.6) $\mathcal{S}^{\text{old}} = \{a\} \Rightarrow \mathcal{S}^{\text{new}} = \{a\}.$

(4.6.1.7) $a \in \mathcal{S}^{\text{old}} \Rightarrow a \in \mathcal{S}^{\text{new}}.$

**Claim:**

If $\succ_D$ satisfies (2.1.1), then the Schulze method, as defined in section 2.2, satisfies monotonicity.

**Proof:**

With (4.6.1.1), we get

(4.6.1.8) $\forall f \in A \setminus \{a\}: N^{old}[a,f] \leq N^{new}[a,f]$.

With (4.6.1.2), we get

(4.6.1.9) $\forall f \in A \setminus \{a\}: N^{old}[f,a] \geq N^{new}[f,a]$.

With (4.6.1.3), we get

(4.6.1.10) $\forall e,f \in A \setminus \{a\}: N^{old}[e,f] = N^{new}[e,f]$.

With (2.1.1), (4.6.1.8), and (4.6.1.9), we get

(4.6.1.11) $\forall f \in A \setminus \{a\}: (N^{old}[a,f],N^{old}[f,a]) \precsim_D (N^{new}[a,f],N^{new}[f,a])$.

With (2.1.1), (4.6.1.8), and (4.6.1.9), we get

(4.6.1.12) $\forall f \in A \setminus \{a\}: (N^{old}[f,a],N^{old}[a,f]) \succsim_D (N^{new}[f,a],N^{new}[a,f])$.

With (4.6.1.10), we get

(4.6.1.13) $\forall e,f \in A \setminus \{a\}: (N^{old}[e,f],N^{old}[f,e]) \approx_D (N^{new}[e,f],N^{new}[f,e])$.

Suppose $c(1),...,c(n) \in A$ was the strongest path from alternative $a$ to alternative $b \in A \setminus \{a\}$. Then with (4.6.1.11) and (4.6.1.13), we get: $c(1),...,c(n)$ is a path from alternative $a$ to alternative $b$ with at least the same strength. Therefore

(4.6.1.14) $\forall b \in A \setminus \{a\}: P_D^{new}[a,b] \succsim_D P_D^{old}[a,b]$.

Suppose $c(1),...,c(n) \in A$ is the strongest path from alternative $b \in A \setminus \{a\}$ to alternative $a$. Then with (4.6.1.12) and (4.6.1.13), we get: $c(1),...,c(n)$ was a path from alternative $b$ to alternative $a$ with at least the same strength. Therefore

(4.6.1.15) $\forall b \in A \setminus \{a\}: P_D^{old}[b,a] \succsim_D P_D^{new}[b,a]$.

With (4.6.1.14) and (4.6.1.15), we get (4.6.1.4) and (4.6.1.5).

With (4.6.1.4), we get (4.6.1.6).

With (4.6.1.5), we get (4.6.1.7). □

### 4.6.2. Strict Monotonicity

**Definition:**

An election method satisfies *strict monotonicity* if the following holds:

Suppose $a \in A$. Suppose the ballots are modified in such a manner that (4.6.1.1) – (4.6.1.3) are satisfied.

Then:

(4.6.2.1) $\quad a \in \mathcal{S}^{\text{old}} \Rightarrow a \in \mathcal{S}^{\text{new}} \subseteq \mathcal{S}^{\text{old}}$.

**Claim:**

If $\succ_D$ satisfies (2.1.1), then the Schulze method, as defined in section 2.2, satisfies strict monotonicity.

**Proof:**

" $a \in \mathcal{S}^{\text{old}} \Rightarrow a \in \mathcal{S}^{\text{new}}$ " has already been proven in section 4.6.1.

To prove " $\mathcal{S}^{\text{new}} \subseteq \mathcal{S}^{\text{old}}$ ", we have to prove: $h \notin \mathcal{S}^{\text{old}} \Rightarrow h \notin \mathcal{S}^{\text{new}}$.

As $a \in \mathcal{S}^{\text{old}}$, we get

(4.6.2.2) $\quad \forall\, b \in A \setminus \{a\}: P_D^{\text{old}}[a,b] \succsim_D P_D^{\text{old}}[b,a]$.

Suppose $h \notin \mathcal{S}^{\text{old}}$. Then, according to (4.1.14), there must have been an alternative $g \in \mathcal{S}^{\text{old}}$ with

(4.6.2.3) $\quad P_D^{\text{old}}[g,h] \succ_D P_D^{\text{old}}[h,g]$.

With (4.6.1.11) – (4.6.1.13) and (4.6.2.3), we get: $P_D^{\text{new}}[g,h] \succ_D P_D^{\text{new}}[h,g]$, unless at least one of the following two cases occurred.

Case 1: $xa$ was a weakest link in the strongest path from alternative $g$ to alternative $h$.

Case 2: $ay$ was the unique weakest link in the strongest path from alternative $h$ to alternative $g$.

With (4.6.2.2), we get: $P_D^{\text{old}}[a,h] \succsim_D P_D^{\text{old}}[h,a]$. For $P_D^{\text{old}}[a,h] \succ_D P_D^{\text{old}}[h,a]$, we would, with (4.6.1.4), immediately get $P_D^{\text{new}}[a,h] \succ_D P_D^{\text{new}}[h,a]$, so that alternative $h$ is still not a potential winner. Therefore, without loss of generality, we can presume $g \in \mathcal{S}^{\text{old}} \setminus \{a\}$ and

(4.6.2.4) $\quad P_D^{\text{old}}[a,h] \approx_D P_D^{\text{old}}[h,a]$.

With $a \in \mathcal{S}^{\text{old}}$ and $g \in \mathcal{S}^{\text{old}} \setminus \{a\}$, we get

(4.6.2.5) $P_D^{\text{old}}[a,g] \approx_D P_D^{\text{old}}[g,a]$.

With (2.2.5), we get

(4.6.2.6) $\min_D \{ P_D^{\text{old}}[g,h], P_D^{\text{old}}[h,a] \} \precsim_D P_D^{\text{old}}[g,a]$.

(4.6.2.7) $\min_D \{ P_D^{\text{old}}[h,a], P_D^{\text{old}}[a,g] \} \precsim_D P_D^{\text{old}}[h,g]$.

Case 1: Suppose $xa$ was a weakest link in the strongest path from alternative $g$ to alternative $h$. Then

(4.6.2.8a) $P_D^{\text{old}}[g,h] \approx_D P_D^{\text{old}}[g,a]$ and

(4.6.2.9a) $P_D^{\text{old}}[a,h] \succsim_D P_D^{\text{old}}[g,h]$.

Now (4.6.2.5), (4.6.2.8a), and (4.6.2.3) give

(4.6.2.10a) $P_D^{\text{old}}[a,g] \approx_D P_D^{\text{old}}[g,a] \approx_D P_D^{\text{old}}[g,h] \succ_D P_D^{\text{old}}[h,g]$,

while (4.6.2.4), (4.6.2.9a), and (4.6.2.3) give

(4.6.2.11a) $P_D^{\text{old}}[h,a] \approx_D P_D^{\text{old}}[a,h] \succsim_D P_D^{\text{old}}[g,h] \succ_D P_D^{\text{old}}[h,g]$.

But (4.6.2.10a) and (4.6.2.11a) together contradict (4.6.2.7).

Case 2: Suppose $ay$ was the unique weakest link in the strongest path from alternative $h$ to alternative $g$. Then

(4.6.2.8b) $P_D^{\text{old}}[h,g] \approx_D P_D^{\text{old}}[a,g]$ and

(4.6.2.9b) $P_D^{\text{old}}[h,a] \succ_D P_D^{\text{old}}[h,g]$.

Now (4.6.2.9b), (4.6.2.8b), and (4.6.2.5) give

(4.6.2.10b) $P_D^{\text{old}}[h,a] \succ_D P_D^{\text{old}}[h,g] \approx_D P_D^{\text{old}}[a,g] \approx_D P_D^{\text{old}}[g,a]$,

while (4.6.2.3), (4.6.2.8b), and (4.6.2.5) give

(4.6.2.11b) $P_D^{\text{old}}[g,h] \succ_D P_D^{\text{old}}[h,g] \approx_D P_D^{\text{old}}[a,g] \approx_D P_D^{\text{old}}[g,a]$.

But (4.6.2.10b) and (4.6.2.11b) together contradict (4.6.2.6).

We have proven that neither case 1 nor case 2 is possible. Therefore

(4.6.2.12) $P_D^{\text{new}}[g,h] \succ_D P_D^{\text{new}}[h,g]$.

With (4.6.2.12), we get: $h \notin \mathcal{S}^{\text{new}}$. □

## 4.7. Independence of Clones

*Independence of clones* as a criterion for single-winner election methods has been proposed by Tideman (1987) and Zavist (1989). This criterion says that running a large number of similar alternatives, so-called *clones*, must not have any impact on the result of the elections.

The precise definition for a *set of clones* stipulates that every voter ranks all the alternatives of this set in a consecutive manner; see (4.7.1) and (4.7.2). Replacing an alternative $d \in A^{old}$ by a set of clones $K$ should not change the winner; see (4.7.7) and (4.7.8).

This criterion is very desirable especially for referendums because, while it might be difficult to find several candidates who are simultaneously sufficiently popular to campaign with them and sufficiently similar to misuse them for this strategy, it is usually very simple to formulate a large number of almost identical proposals. For example: In 1969, when the Canadian city that is now known as *Thunder Bay* was amalgamating, there was some controversy over what the name should be. In opinion polls, a majority of the voters preferred the name *The Lakehead* to the name *Thunder Bay*. But when the polls opened, there were three names on the referendum ballot: *Thunder Bay*, *Lakehead*, and *The Lakehead*. As the ballots were counted using *plurality voting*, it was not a surprise when *Thunder Bay* won. The votes were as follows: *Thunder Bay* 15870, *Lakehead* 15302, *The Lakehead* 8377 (www144).

In recent years, cloning strategies are also tried in elections. A good example was the 2019 Ukrainian presidential election. The president of Ukraine is elected by top-two runoff. To split the votes of candidate Yulia Volodymyrivna Tymoshenko (Ю́лія Володи́мирівна Тимоше́нко), her opponents nominated another candidate with almost identical name: Yuri Volodymyrovych Tymoshenko (Ю́рій Володи́мирович Тимоше́нко). The idea was to mislead the supporters of Yulia Tymoshenko so that sufficiently many of them mistakenly vote for the wrong Tymoshenko so that Yulia Tymoshenko doesn’t get into the runoff. In the end, Yulia Tymoshenko got 13.40% of the votes in the first round, while Yuri Tymoshenko got 0.62% of the votes. Volodymyr Zelenskyy (30.24%) and Petro Poroshenko (15.95%) got into the runoff. So this cloning strategy had no impact in this election.

An example for a successful cloning strategy in an election was the 2020 election to the State Senate in Florida. To split the votes of incumbent José Javier Rodríguez of State Senate District 37, a candidate with almost identical name was nominated: Alex Rodríguez. In this case, the strategy worked. José Javier Rodríguez (48.51%) lost his seat to Ileana Garcia (48.53%), while Alex Rodríguez got 2.96% of the votes.

The 2021 election to the Legislative Assembly of Saint Petersburg in Russia showed an even more extreme example of a cloning strategy in an election. To keep Boris Vishnevsky (Борис Вишневский) of District 2 from winning, his opponents searched for people who could be used as doppelgängers of Boris Vishnevsky. When they had found some, they took the two most suitable of them and adjusted their appearances so that they looked as much as possible like the original Boris Vishnevsky; these doppelgängers even thinned their hair and grew beards and mustaches similar to the one of the original Boris Vishnevsky. These doppelgängers also legally changed their names to "Boris Vishnevsky" before they submitted their candidacies. So (among other candidates), there were three identical looking candidates with the same name "Boris Vishnevsky" on the ballot. In this case, the cloning strategy had no impact on the result of the election: The original Boris Vishnevsky received 11,062 votes. The two fake candidates received 608 and 510 votes. However, Alexander Rzhanenkov (Александр Ржаненков) of the ruling "United Russia" party won the seat with 12,412 votes.

When this cloning strategy is used, then candidates like Yuri Tymoshenko or Alex Rodríguez or the new Boris Vishnevskies are called *ghost candidates* or *shadow candidates* as they don't appear in public and don't campaign.

**Definition:**

An election method is *independent of clones* if the following holds:

Suppose $d \in A^{\text{old}}$. Suppose $A^{\text{new}} := ( A^{\text{old}} \cup K ) \setminus \{d\}$.

Suppose alternative $d$ is replaced by the set of alternatives $K$ in such a manner that the following three statements are satisfied:

(4.7.1) $\forall\, e \in A^{\text{old}} \setminus \{d\}\ \forall\, g \in K\ \forall\, v \in V\!: e \succ_v^{\text{old}} d \Leftrightarrow e \succ_v^{\text{new}} g.$

(4.7.2) $\forall\, f \in A^{\text{old}} \setminus \{d\}\ \forall\, g \in K\ \forall\, v \in V\!: d \succ_v^{\text{old}} f \Leftrightarrow g \succ_v^{\text{new}} f.$

(4.7.3) $\forall\, e,f \in A^{\text{old}} \setminus \{d\}\ \forall\, v \in V\!: e \succ_v^{\text{old}} f \Leftrightarrow e \succ_v^{\text{new}} f.$

Then the following statements are satisfied:

(4.7.4) $\forall\, a \in A^{\text{old}} \setminus \{d\}\ \forall\, g \in K\!: ad \in \mathcal{O}^{\text{old}} \Leftrightarrow ag \in \mathcal{O}^{\text{new}}.$

(4.7.5) $\forall\, b \in A^{\text{old}} \setminus \{d\}\ \forall\, g \in K\!: db \in \mathcal{O}^{\text{old}} \Leftrightarrow gb \in \mathcal{O}^{\text{new}}.$

(4.7.6) $\forall\, a,b \in A^{\text{old}} \setminus \{d\}\!: ab \in \mathcal{O}^{\text{old}} \Leftrightarrow ab \in \mathcal{O}^{\text{new}}.$

(4.7.7) $d \in \mathcal{S}^{\text{old}} \Leftrightarrow ( \mathcal{S}^{\text{new}} \cap K ) \neq \emptyset.$

(4.7.8) $\forall\, a \in A^{\text{old}} \setminus \{d\}\!: a \in \mathcal{S}^{\text{old}} \Leftrightarrow a \in \mathcal{S}^{\text{new}}.$

(4.7.9) $\forall\, g,h \in K\!: gh \in \mathcal{O}|_K \Leftrightarrow gh \in \mathcal{O}^{\text{new}}.$

(4.7.10) $d \in \mathcal{S}^{\text{old}} \Leftrightarrow ( \mathcal{S}^{\text{new}} \cap \mathcal{S}|_K ) \neq \emptyset.$

(4.7.11) $\forall\, g \in K\!: ( \{d\} = \mathcal{S}^{\text{old}} \wedge g \in \mathcal{S}|_K ) \Rightarrow g \in \mathcal{S}^{\text{new}}.$

(4.7.12) $\forall\, g \in K\!: ( d \in \mathcal{S}^{\text{old}} \wedge \{g\} = \mathcal{S}|_K ) \Rightarrow g \in \mathcal{S}^{\text{new}}.$

(4.7.13) $\forall\, g \in K\!: ( d \notin \mathcal{S}^{\text{old}} \vee g \notin \mathcal{S}|_K ) \Rightarrow g \notin \mathcal{S}^{\text{new}}.$

**Claim:**

The Schulze method, as defined in section 2.2, satisfies (4.7.4) – (4.7.8). The more detailed definition for the Schulze method, as given in section 5.1, also satisfies (4.7.9) – (4.7.13).

**Proof:**

We will only prove that the Schulze method, as defined in section 2.2, satisfies (4.7.4) – (4.7.8). The proof that the Schulze method, as defined in section 5.1, also satisfies (4.7.9) – (4.7.13) will then be straight forward.

With (4.7.1), we get

(4.7.14) $\forall\, e \in A^{old} \setminus \{d\}\ \forall\, g \in K\colon N^{old}[e,d] = N^{new}[e,g].$

With (4.7.2), we get

(4.7.15) $\forall\, f \in A^{old} \setminus \{d\}\ \forall\, g \in K\colon N^{old}[d,f] = N^{new}[g,f].$

With (4.7.3), we get

(4.7.16) $\forall\, e,f \in A^{old} \setminus \{d\}\colon N^{old}[e,f] = N^{new}[e,f].$

With (4.7.14) and (4.7.15), we get

(4.7.17) $\forall\, e \in A^{old} \setminus \{d\}\ \forall\, g \in K\colon$
$(N^{old}[e,d],N^{old}[d,e]) \approx_D (N^{new}[e,g],N^{new}[g,e]).$

With (4.7.14) and (4.7.15), we get

(4.7.18) $\forall\, f \in A^{old} \setminus \{d\}\ \forall\, g \in K\colon$
$(N^{old}[d,f],N^{old}[f,d]) \approx_D (N^{new}[g,f],N^{new}[f,g]).$

With (4.7.16), we get

(4.7.19) $\forall\, e,f \in A^{old} \setminus \{d\}\colon (N^{old}[e,f],N^{old}[f,e]) \approx_D (N^{new}[e,f],N^{new}[f,e]).$

Suppose $c(1),...,c(n) \in A^{\text{old}}$ was the strongest path from alternative $a \in A^{\text{old}} \setminus \{d\}$ to alternative $d$. Then with (4.7.17) and (4.7.19), we get: $c(1),...,c(n-1),g$ is a path from alternative $a$ to alternative $g \in K$ with the same strength. Therefore

(4.7.20) $\forall\, a \in A^{\text{old}} \setminus \{d\}\ \forall\, g \in K$: $P_D^{\text{new}}[a,g] \succsim_D P_D^{\text{old}}[a,d]$.

Suppose $c(1),...,c(n) \in A^{\text{new}}$ is the strongest path from alternative $a \in A^{\text{new}} \setminus K$ to alternative $g \in K$. Then with (4.7.17) and (4.7.19), we get: $c(1),...,c(m-1),d$, where $c(m)$ is the first occurrence of an alternative of the set $K$, was a path from alternative $a$ to alternative $d$ with at least the same strength. Therefore

(4.7.21) $\forall\, a \in A^{\text{new}} \setminus K\ \forall\, g \in K$: $P_D^{\text{old}}[a,d] \succsim_D P_D^{\text{new}}[a,g]$.

Suppose $c(1),...,c(n) \in A^{\text{old}}$ was the strongest path from alternative $d$ to alternative $b \in A^{\text{old}} \setminus \{d\}$. Then with (4.7.18) and (4.7.20), we get: $g,c(2),...,c(n)$ is a path from alternative $g \in K$ to alternative $b$ with the same strength. Therefore

(4.7.22) $\forall\, b \in A^{\text{old}} \setminus \{d\}\ \forall\, g \in K$: $P_D^{\text{new}}[g,b] \succsim_D P_D^{\text{old}}[d,b]$.

Suppose $c(1),...,c(n) \in A^{\text{new}}$ is the strongest path from alternative $g \in K$ to alternative $b \in A^{\text{new}} \setminus K$. Then with (4.7.18) and (4.7.19), we get: $d,c(m+1),...,c(n)$, where $c(m)$ is the last occurrence of an alternative of the set $K$, was a path from alternative $d$ to alternative $b$ with at least the same strength. Therefore

(4.7.23) $\forall\, b \in A^{\text{new}} \setminus K\ \forall\, g \in K$: $P_D^{\text{old}}[d,b] \succsim_D P_D^{\text{new}}[g,b]$.

(α) Suppose the strongest path $c(1),...,c(n) \in A^{\text{old}}$ from alternative $a \in A^{\text{old}} \setminus \{d\}$ to alternative $b \in A^{\text{old}} \setminus \{a,d\}$ did not contain alternative $d$. Then with (4.7.19), we get: $c(1),...,c(n)$ is still a path from alternative $a$ to alternative $b$ with the same strength. Therefore: $P_D^{\text{new}}[a,b] \succsim_D P_D^{\text{old}}[a,b]$.

(β) Suppose the strongest path $c(1),...,c(n) \in A^{\text{old}}$ from alternative $a \in A^{\text{old}} \setminus \{d\}$ to alternative $b \in A^{\text{old}} \setminus \{a,d\}$ contained alternative $d$. Then with (4.7.17), (4.7.18), and (4.7.19), we get: $c(1),...,c(n)$, with alternative $d$ replaced by an arbitrarily chosen alternative $g \in K$, is still a path from alternative $a$ to alternative $b$ with the same strength. Therefore: $P_D^{\text{new}}[a,b] \succsim_D P_D^{\text{old}}[a,b]$.

With (α) and (β), we get

(4.7.24) $\forall\, a,b \in A^{\text{old}} \setminus \{d\}$: $P_D^{\text{new}}[a,b] \succsim_D P_D^{\text{old}}[a,b]$.

(γ) Suppose the strongest path $c(1),...,c(n) \in A^{\text{new}}$ from alternative $a \in A^{\text{new}} \setminus K$ to alternative $b \in A^{\text{new}} \setminus ( K \cup \{a\} )$ does not contain alternatives of the set $K$. Then with (4.7.19), we get: $c(1),...,c(n)$ was a path from alternative $a$ to alternative $b$ with the same strength. Therefore: $P_D^{\text{old}}[a,b] \gtrsim_D P_D^{\text{new}}[a,b]$.

(δ) Suppose the strongest path $c(1),...,c(n) \in A^{\text{new}}$ from alternative $a \in A^{\text{new}} \setminus K$ to alternative $b \in A^{\text{new}} \setminus ( K \cup \{a\} )$ contains some alternatives of the set $K$. Then with (4.7.17), (4.7.18), and (4.7.19), we get: $c(1),...,c(s-1),d,c(t+1),...,c(n)$, where $c(s)$ is the first occurrence of an alternative of the set $K$ and $c(t)$ is the last occurrence of an alternative of the set $K$, was a path from alternative $a$ to alternative $b$ with at least the same strength. Therefore: $P_D^{\text{old}}[a,b] \gtrsim_D P_D^{\text{new}}[a,b]$.

With (γ) and (δ), we get

(4.7.25) $\quad \forall\, a,b \in A^{\text{new}} \setminus K\colon P_D^{\text{old}}[a,b] \gtrsim_D P_D^{\text{new}}[a,b]$.

Combining (4.7.20) and (4.7.21) gives

(4.7.26) $\quad \forall\, a \in A^{\text{old}} \setminus \{d\}\ \forall\, g \in K\colon P_D^{\text{old}}[a,d] \approx_D P_D^{\text{new}}[a,g]$.

Combining (4.7.22) and (4.7.23) gives

(4.7.27) $\quad \forall\, b \in A^{\text{old}} \setminus \{d\}\ \forall\, g \in K\colon P_D^{\text{old}}[d,b] \approx_D P_D^{\text{new}}[g,b]$.

Combining (4.7.24) and (4.7.25) gives

(4.7.28) $\quad \forall\, a,b \in A^{\text{old}} \setminus \{d\}\colon P_D^{\text{old}}[a,b] \approx_D P_D^{\text{new}}[a,b]$.

With (4.7.26) – (4.7.28), we get (4.7.4) – (4.7.6).

Part 1:

Suppose $d \in \mathcal{S}^{\text{old}}$. Then

(4.7.29) $\forall\, a \in A^{\text{old}} \setminus \{d\}: ad \notin O^{\text{old}}$.

With (4.7.4) and (4.7.29), we get

(4.7.30) $\forall\, a \in A^{\text{new}} \setminus K\; \forall\, g \in K: ag \notin O^{\text{new}}$.

Since the binary relation $O^{\text{new}}$, as defined in (2.2.1), is asymmetric and transitive, there must be an alternative $k \in K$ with

(4.7.31) $\forall\, l \in K \setminus \{k\}: lk \notin O^{\text{new}}$.

With (4.7.30) and (4.7.31), we get $k \in ( \mathcal{S}^{\text{new}} \cap K )$ and, therefore, $( \mathcal{S}^{\text{new}} \cap K ) \neq \emptyset$.

Part 2:

Suppose $d \notin \mathcal{S}^{\text{old}}$. Then

(4.7.32) $\exists\, a \in A^{\text{old}} \setminus \{d\}: ad \in O^{\text{old}}$.

With (4.7.4) and (4.7.32), we get

(4.7.33) $\exists\, a \in A^{\text{new}} \setminus K\; \forall\, g \in K: ag \in O^{\text{new}}$.

With (4.7.33), we get: $( \mathcal{S}^{\text{new}} \cap K ) = \emptyset$.

With part 1 and part 2, we get (4.7.7).

Part 3:

Suppose $a \in A^{\text{old}} \setminus \{d\}$ and $a \in \mathcal{S}^{\text{old}}$. Then

(4.7.34) $da \notin O^{\text{old}}$.

(4.7.35) $\forall\, b \in A^{\text{old}} \setminus \{a,d\}: ba \notin O^{\text{old}}$.

With (4.7.26), (4.7.27) and (4.7.34), we get

(4.7.36) $\forall\, g \in K: ga \notin O^{\text{new}}$.

With (4.7.28) and (4.7.35), we get

(4.7.37) $\forall\, b \in A^{\text{new}} \setminus ( K \cup \{a\} ): ba \notin O^{\text{new}}$.

With (4.7.36) and (4.7.37), we get: $a \in \mathcal{S}^{\text{new}}$.

Part 4:

Suppose $a \in A^{\text{old}} \setminus \{d\}$ and $a \notin \mathcal{S}^{\text{old}}$. Then at least one of the following two statements must have been valid:

(4.7.38a) $da \in \mathcal{O}^{\text{old}}$.

(4.7.38b) $\exists\ b \in A^{\text{old}} \setminus \{a,d\}$: $ba \in \mathcal{O}^{\text{old}}$.

With (4.7.26), (4.7.27), (4.7.28), and (4.7.38), we get that at least one of the following two statements must be valid:

(4.7.39a) $\forall\ g \in K$: $ga \in \mathcal{O}^{\text{new}}$.

(4.7.39b) $\exists\ b \in A^{\text{new}} \setminus ( K \cup \{a\} )$: $ba \in \mathcal{O}^{\text{new}}$.

With (4.7.39), we get: $a \notin \mathcal{S}^{\text{new}}$.

With part 3 and part 4, we get (4.7.8). □

Section 4.7 is also discussed by Camps (2008, 2012b), Bouremel (2017a, 2017b), and Zedam (2018). On the other side, Felsenthal (2011, page 28, footnote 6; 2018, page 20, footnote 2; 2019, page 10, footnote 2) explicitly refuses to discuss the Schulze method because of the fact that it is independent of clones.

## 4.8. Smith Criterion, Condorcet Winners, Condorcet Losers

The *Smith set* is the smallest set $\emptyset \neq B \subseteq A$ with

(4.8.1) $\forall\, a \in B\ \forall\, b \notin B$: $N[a,b] > N[b,a]$.

The *Smith criterion* and *Smith-IIA* (where IIA means "independence of irrelevant alternatives") say that *weak* alternatives should have no impact on the result of the elections.

Suppose:

(4.8.2) $\emptyset \neq B_1 \subsetneq A$, $\emptyset \neq B_2 \subsetneq A$, $B_1 \cup B_2 = A$, $B_1 \cap B_2 = \emptyset$.

(4.8.3) $\forall\, a \in B_1\ \forall\, b \in B_2$: $N[a,b] > N[b,a]$.

Then a *weak* alternative in the Smith paradigm is an alternative $b \in B_2$. Adding or removing a weak alternative $b \in B_2$ should have no impact on the set $\mathcal{S}$ of potential winners.

**Definition:**

An election method satisfies the *Smith criterion* if the following holds:

Suppose (4.8.2) and (4.8.3). Then:

(4.8.4) $\forall\, a \in B_1\ \forall\, b \in B_2$: $ab \in \mathcal{O}$.

(4.8.5) $\emptyset \neq \mathcal{S} \subseteq B_1$.

**Remark:**

If $B_1$ consists of only one alternative $a \in A$, then this alternative is the so-called *Condorcet winner* and the Smith criterion becomes the so-called *Condorcet criterion* or *Condorcet winner criterion* (Condorcet, 1785). In short:

(4.8.6) Alternative $a \in A$ is a *Condorcet winner* :⇔
$N[a,b] > N[b,a]$ for all $b \in A \setminus \{a\}$.

(4.8.7) An election method satisfies the *Condorcet criterion* if the following holds:

(i) Alternative $a \in A$ is a Condorcet winner. $\Rightarrow \mathcal{S} = \{a\}$.

(ii) Alternative $a \in A$ is a Condorcet winner.
$\Rightarrow \forall\, b \in A \setminus \{a\}$: $ab \in \mathcal{O}$.

If $B_2$ consists of only one alternative $b \in A$, then this alternative is the so-called *Condorcet loser* and the Smith criterion becomes the so-called *Condorcet loser criterion*. In short:

(4.8.8) Alternative $b \in A$ is a *Condorcet loser* : $\Leftrightarrow$
$N[a,b] > N[b,a]$ for all $a \in A \setminus \{b\}$.

(4.8.9) An election method satisfies the *Condorcet loser criterion* if the following holds:

(i) Alternative $b \in A$ is a Condorcet loser. $\Rightarrow b \notin \mathcal{S}$.

(ii) Alternative $b \in A$ is a Condorcet loser.
$\Rightarrow \forall\ a \in A \setminus \{b\}$: $ab \in \mathcal{O}$.

**Claim:**

If $>_D$ satisfies (2.1.5), then the Schulze method, as defined in section 2.2, satisfies the Smith criterion.

**Proof:**

The proof is trivial. Presumption (2.1.5) guarantees that any pairwise victory is stronger than any pairwise defeat. If $a \in B_1$ and $b \in B_2$, then already the link $ab$ is a path from alternative $a$ to alternative $b$ that consists only of a pairwise victory. On the other side, (4.8.3) says that there cannot be a path from alternative $b$ to alternative $a$ that contains no pairwise defeat. So already the link $ab$ is stronger than any path from alternative $b$ to alternative $a$. □

**Definition:**

An election method satisfies *Smith-IIA* if the following holds:

Suppose (4.8.2) and (4.8.3). Then:

(4.8.10) If $d \in B_2$ is removed, then

(a) $\forall\ e,f \in B_1$: $ef \in \mathcal{O}^{\text{old}} \Leftrightarrow ef \in \mathcal{O}^{\text{new}}$.

(b) $\mathcal{S}^{\text{old}} = \mathcal{S}^{\text{new}}$.

(4.8.11) If $d \in B_1$ is removed, then

$\forall\ e,f \in B_2$: $ef \in \mathcal{O}^{\text{old}} \Leftrightarrow ef \in \mathcal{O}^{\text{new}}$.

**Claim:**

If $\succ_D$ satisfies (2.1.5), then the Schulze method, as defined in section 2.2, satisfies Smith-IIA.

**Proof:**

We will prove (4.8.10)(a). The proof for (4.8.11) is analogous.

(4.8.10)(b) follows directly from (4.8.5) and (4.8.10)(a).

Part 1: Suppose $e,f \in B_1$. Suppose $ef \in O^{\text{old}}$. Then

(4.8.12) $\quad P_D^{\text{old}}[e,f] \succ_D P_D^{\text{old}}[f,e]$.

With (2.2.3), we get

(4.8.13) $\quad P_D^{\text{old}}[e,f] \succsim_D (N[e,f],N[f,e])$.

With (4.8.12) and (2.2.3), we get

(4.8.14) $\quad P_D^{\text{old}}[e,f] \succ_D P_D^{\text{old}}[f,e] \succsim_D (N[f,e],N[e,f])$.

With (4.8.13) and (4.8.14), we get

(4.8.15) $\quad P_D^{\text{old}}[e,f] \succsim_D \max_D \{ (N[e,f],N[f,e]), (N[f,e],N[e,f]) \}$.

With (4.8.3), we get: Any path from alternative $e \in B_1$ to alternative $f \in B_1$ that contained alternative $d \in B_2$ necessarily contained a pairwise defeat.

As it is not possible that the link $ef$ is a pairwise defeat and that simultaneously the link $fe$ is a pairwise defeat, $\max_D \{ (N[e,f],N[f,e]), (N[f,e], N[e,f]) \}$ is stronger than any pairwise defeat [ because of (2.1.5) ]. Therefore, with (4.8.3) and (4.8.15), we get: The strongest path from alternative $e \in B_1$ to alternative $f \in B_1$ did not contain alternative $d \in B_2$. Therefore

(4.8.16) $\quad P_D^{\text{new}}[e,f] \approx_D P_D^{\text{old}}[e,f]$.

As the elimination of alternative $d \in B_2$ only removes paths, we get

(4.8.17) $\quad P_D^{\text{new}}[f,e] \precsim_D P_D^{\text{old}}[f,e]$.

With (4.8.16), (4.8.12), and (4.8.17), we get

(4.8.18) $\quad P_D^{\text{new}}[e,f] \approx_D P_D^{\text{old}}[e,f] \succ_D P_D^{\text{old}}[f,e] \succsim_D P_D^{\text{new}}[f,e]$.

With (4.8.18), we get: $ef \in O^{\text{new}}$.

Part 2: The proof for " $P_D^{\text{old}}[f,e] \succ_D P_D^{\text{old}}[e,f]$ " is analogous.

Part 3: When we have $P_D^{\text{old}}[e,f] \approx_D P_D^{\text{old}}[f,e]$ then, with the same argumentation as in Part 1, we get

(4.8.19) $$P_D^{\text{old}}[e,f] \succsim_D \max_D \{ (N[e,f],N[f,e]), (N[f,e],N[e,f]) \}.$$

(4.8.20) $$P_D^{\text{old}}[f,e] \succsim_D \max_D \{ (N[e,f],N[f,e]), (N[f,e],N[e,f]) \}.$$

So with the same argumentation as in Part 1, we can show that neither the strongest path from alternative $e \in B_1$ to alternative $f \in B_1$ nor the strongest path from alternative $f \in B_1$ to alternative $e \in B_1$ did contain alternative $d \in B_2$. □

The *majority criterion for solid coalitions* says that, when a majority of the voters strictly prefers every alternative of a given set of alternatives to every alternative outside this set of alternatives, then the winner must be chosen from this set. In short, an election method satisfies the *majority criterion for solid coalitions* if the following holds:

Suppose (4.8.2).
Suppose $\| \{ v \in V \mid \forall a \in B_1 \; \forall b \in B_2 : a >_v b \} \| > N/2$.
Then $\mathcal{S} \subseteq B_1$.

If $B_1$ consists of only one alternative $a \in A$, then this is the so-called *majority criterion* or *majority winner criterion*. If $B_2$ consists of only one alternative $b \in A$, then this is the so-called *majority loser criterion*.

The Smith criterion implies the majority criterion for solid coalitions, the Condorcet criterion, and the Condorcet loser criterion. The majority criterion for solid coalitions implies the majority criterion and the majority loser criterion. The Condorcet criterion implies the majority criterion. The Condorcet loser criterion implies the majority loser criterion.

*Participation* says that adding a list $W$ of ballots, on which every alternative of set $B_1$ ( with $\emptyset \neq B_1 \subsetneq A$ ) is strictly preferred to every alternative of set $B_2$ ( with $B_2 := A \setminus B_1$ ), must not change the winner from an alternative of set $B_1$ to an alternative of set $B_2$. In short, an election method satisfies *participation* if the following holds:

| | | |
|---|---|---|
| Suppose | (4.8.2). | |
| Suppose | (4.8.21) | $\forall\, a \in B_1\ \forall\, b \in B_2\ \forall\, w \in W$: $a >_w b$. |
| Suppose | (4.8.22) | $V^{new} := V^{old} + W$. |
| Then | (4.8.23) | $\forall\, e \in B_1\ \forall f \in B_2$: $ef \in O^{old} \Rightarrow ef \in O^{new}$. |
| | (4.8.24) | $\forall\, e \in B_1\ \forall f \in B_2$: $fe \notin O^{old} \Rightarrow fe \notin O^{new}$. |
| | (4.8.25) | $( \mathcal{S}^{old} \cap B_1 ) \neq \emptyset \Rightarrow ( \mathcal{S}^{new} \cap B_1 ) \neq \emptyset$. |
| | (4.8.26) | $( \mathcal{S}^{old} \cap B_2 ) = \emptyset \Rightarrow ( \mathcal{S}^{new} \cap B_2 ) = \emptyset$. |

Unfortunately, the Condorcet criterion is incompatible with the participation criterion (Moulin, 1988). Example 7 (section 3.7) shows a drastic violation of the participation criterion.

While the Condorcet criterion and the participation criterion are incompatible, random simulations by Xia (2021a, fig. 13(b) and 14(b)) showed that, of all Condorcet-consistent methods, the Schulze method violates the participation criterion the least frequently.

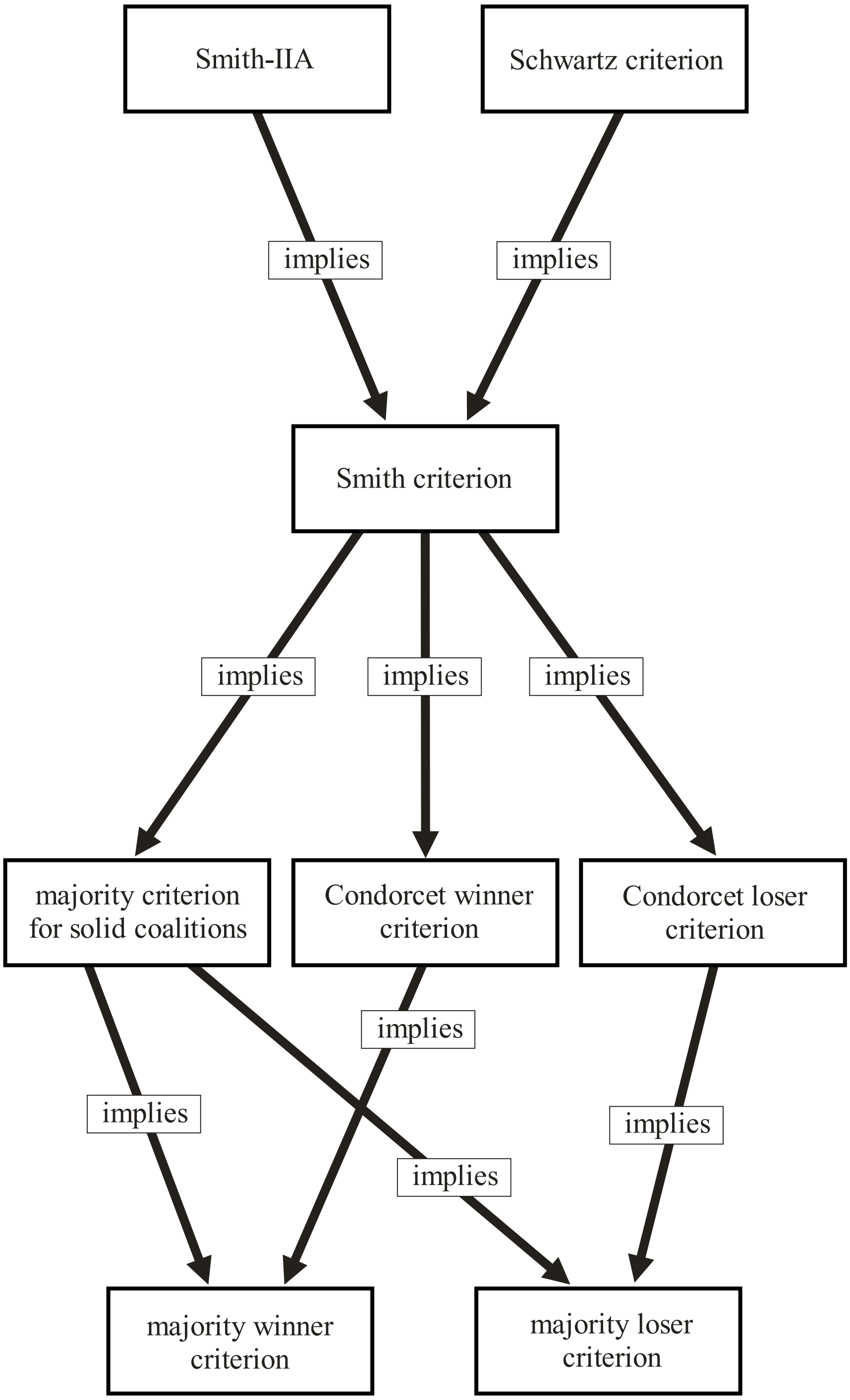

Relationships between the criteria defined in sections 4.8 and 4.11

## 4.9. MinMax Set

For all $\emptyset \neq B \subsetneq A$, we define

(4.9.1) $\Gamma_D(B) := \max_D \{ (N[x,y],N[y,x]) \mid x \notin B, y \in B \}$.

In short: $\Gamma_D(B)$ is the strongest defeat of an alternative in $B$ by an alternative from outside of $B$.

Furthermore, we define

(4.9.2) $\beta_D := \min_D \{ \Gamma_D(B) \mid \emptyset \neq B \subsetneq A \}$.

(4.9.3) $\mathfrak{B}_D := \bigcup \{ \emptyset \neq B \subsetneq A \mid \Gamma_D(B) \approx_D \beta_D \}$.

$\mathfrak{B}_D$ is the *MinMax set*. $\mathfrak{B}_D$ has the following properties:

1. $\mathfrak{B}_D \neq \emptyset$.

2. If $\mathfrak{B}_D$ consists of only one alternative $a \in A$, then alternative $a$ is the unique Simpson-Kramer MinMax winner ( i.e. that alternative $a \in A$ with minimum $\max_D \{ (N[b,a],N[a,b]) \mid b \in A \setminus \{a\} \}$ ).

3. If $d \in \mathfrak{B}_D$ is replaced by a set of alternatives $K$ as described in (4.7.1) – (4.7.3), then $\mathfrak{B}_D^{new} = ( \mathfrak{B}_D \cup K ) \setminus \{d\}$.

4. If $d \notin \mathfrak{B}_D$ is replaced by a set of alternatives $K$ as described in (4.7.1) – (4.7.3), then $\mathfrak{B}_D^{new} = \mathfrak{B}_D$.

So, in some sense, the MinMax set $\mathfrak{B}_D$ is a cloneproof generalization of the Simpson-Kramer MinMax winner.

When we want primarily that the used election method is independent of clones and secondarily that the strongest link *ef*, that is overruled when determining the winner, is minimized, then we have to demand that the winner is always chosen from the MinMax set $\mathfrak{B}_D$.

**Claim:**

The Schulze method, as defined in section 2.2, has the following properties:

(4.9.4) $\forall\ a \in \mathfrak{B}_D\ \forall\ b \notin \mathfrak{B}_D$: $ab \in \mathcal{O}$.

(4.9.5) $\mathcal{S} \subseteq \mathfrak{B}_D$.

**Proof:**

Suppose $a \in \mathfrak{B}_D$. Then, with (4.9.3), we get

(4.9.6) $\exists\ \emptyset \neq B \subsetneq A$: $\Gamma_D(B) \approx_D \beta_D$ and $a \in B$.

Suppose $b \notin \mathfrak{B}_D$. Then, with (4.9.3), we get

(4.9.7) $\gamma_D := \min_D \{\ \Gamma_D(B) \mid \emptyset \neq B \subsetneq A$ and $b \in B\ \} \succ_D \beta_D$.

We will prove the following claims:

Claim #1: $P_D[b,a] \precsim_D \beta_D$.
Claim #2: $P_D[a,b] \succsim_D \gamma_D$.

With claim #1, claim #2 and (4.9.7), we get

(4.9.8) $P_D[a,b] \succsim_D \gamma_D \succ_D \beta_D \succsim_D P_D[b,a]$.

With (4.9.8), we get (4.9.4). With (4.9.4), we get (4.9.5).

Proof of claim #1:

With (4.9.6) and (4.9.7), we get

(4.9.9) $\exists\ \emptyset \neq B \subsetneq A$: $\Gamma_D(B) \approx_D \beta_D$ and $a \in B$ and $b \notin B$.

Suppose $c(1),...,c(n) \in A$ is the strongest path from alternative $b$ to alternative $a$. Suppose $c(i)$ is the last alternative with $c(i) \notin B$. Then we get $(N[c(i),c(i+1)],N[c(i+1),c(i)]) \precsim_D \beta_D$. Therefore, we get

(4.9.10) $P_D[b,a] \precsim_D \beta_D$.

Proof of claim #2:

We can construct a path from alternative $a$ to alternative $b$ with a strength of at least $\gamma_D$ as follows:

(1) We start with $E_1 := \{a\}$ and $i := 1$. Trivially, we get $b \notin E_1$ and $P_D[a,h] \succsim_D \gamma_D$ for all $h \in E_1 \setminus \{a\}$.

(2) At each stage, we consider the set $B_i := A \setminus E_i$.

With $b \in B_i$ and with (4.9.7), we get

(4.9.11) $\Gamma_D(B_i) \approx_D \max_D \{ (N[x,y],N[y,x]) \mid x \notin B_i, y \in B_i \} \succsim_D \gamma_D$.

We choose $f \in E_i$ and $g \in B_i$ with

(4.9.12) $(N[f,g],N[g,f]) \approx_D \max_D \{ (N[x,y],N[y,x]) \mid x \notin B_i, y \in B_i \} \succsim_D \gamma_D$.

We define $E_{i+1} := E_i \cup \{g\}$.

With $f \in E_i$, with $P_D[a,h] \succsim_D \gamma_D$ for all $h \in E_i \setminus \{a\}$, with $(N[f,g], N[g,f]) \succsim_D \gamma_D$, and with $E_{i+1} := E_i \cup \{g\}$, we get

(4.9.13) $P_D[a,h] \succsim_D \gamma_D$ for all $h \in E_{i+1} \setminus \{a\}$.

(3) We repeat stage 2 with $i \to i+1$, until $g \equiv b$.

Therefore, we get

(4.9.14) $P_D[a,b] \succsim_D \gamma_D$. □

Example 8 (section 3.8) shows that IPDA and the desideratum, that the winner is always chosen from the MinMax set $\mathfrak{B}_D$, are incompatible. In example 8(old), we get $\mathfrak{B}_D^{old} = \{a, c, d\}$. In example 8(new), we get $\mathfrak{B}_D^{new} = \{b\}$. Therefore, $(\mathfrak{B}_D^{old} \cap \mathfrak{B}_D^{new}) = \emptyset$. Thus, the desideratum, that the winner is always chosen from the MinMax set $\mathfrak{B}_D$, implies that the winner must change.

Actually, the Schulze method can be described completely with the desideratum to find a binary relation $O$ on $A$ that, primarily, is independent of clones (as defined in section 4.7) and that, secondarily, tries to rank the alternatives according to their worst defeats.

For all $a,b \in A$, we define

(4.9.15) $\gamma_D[a,b] := \min_D \{ \Gamma_D(B) \mid \emptyset \neq B \subsetneq A \text{ and } a \notin B \text{ and } b \in B \}$.

(4.9.16) $ab \in O :\Leftrightarrow \gamma_D[a,b] \succ_D \gamma_D[b,a]$.

To prove that (4.9.16) is identical to (2.2.1), we have to prove $\gamma_D[a,b] = P_D[a,b]$. This proof is identical to the proof for (4.9.4).

## Example 1

With $\Gamma_D(B) := \max_D \{ (N[x,y],N[y,x]) \mid x \notin B, y \in B \}$,
we get in example 1 (section 3.1):

$\Gamma_D(\{a\}) = (N[b,a],N[a,b]) = (13,8).$
$\Gamma_D(\{b\}) = (N[d,b],N[b,d]) = (19,2).$
$\Gamma_D(\{c\}) = (N[a,c],N[c,a]) = (14,7).$
$\Gamma_D(\{d\}) = (N[c,d],N[d,c]) = (12,9).$
$\Gamma_D(\{a,b\}) = (N[d,b],N[b,d]) = (19,2).$
$\Gamma_D(\{a,c\}) = (N[b,a],N[a,b]) = (13,8).$
$\Gamma_D(\{a,d\}) = (N[b,a],N[a,b]) = (13,8).$
$\Gamma_D(\{b,c\}) = (N[d,b],N[b,d]) = (19,2).$
$\Gamma_D(\{b,d\}) = (N[c,b],N[b,c]) = (15,6).$
$\Gamma_D(\{c,d\}) = (N[a,c],N[c,a]) = (14,7).$
$\Gamma_D(\{a,b,c\}) = (N[d,b],N[b,d]) = (19,2).$
$\Gamma_D(\{a,b,d\}) = (N[c,b],N[b,c]) = (15,6).$
$\Gamma_D(\{a,c,d\}) = (N[b,a],N[a,b]) = (13,8).$
$\Gamma_D(\{b,c,d\}) = (N[a,c],N[c,a]) = (14,7).$

With $\beta_D := \min_D \{ \Gamma_D(B) \mid \emptyset \neq B \subsetneq A \}$, we get: $\beta_D = (12,9)$.

With $\mathfrak{B}_D := \bigcup \{ \emptyset \neq B \subsetneq A \mid \Gamma_D(B) \approx_D \beta_D \}$, we get $\mathfrak{B}_D = \{d\}$.

So with (4.9.5), we get $\mathcal{S} = \{d\}$.

With $\gamma_D[x,y] := \min_D \{ \Gamma_D(B) \mid \emptyset \neq B \subsetneq A \text{ and } x \notin B \text{ and } y \in B \}$, we get:

$\gamma_D[a,b] = \Gamma_D(\{b,c,d\}) = (14,7).$
$\gamma_D[a,c] = \Gamma_D(\{c\}) = \Gamma_D(\{c,d\}) = \Gamma_D(\{b,c,d\}) = (14,7).$
$\gamma_D[a,d] = \Gamma_D(\{d\}) = (12,9).$
$\gamma_D[b,a] = \Gamma_D(\{a\}) = \Gamma_D(\{a,c\}) = \Gamma_D(\{a,d\}) = \Gamma_D(\{a,c,d\}) = (13,8).$
$\gamma_D[b,c] = \Gamma_D(\{a,c\}) = \Gamma_D(\{a,c,d\}) = (13,8).$
$\gamma_D[b,d] = \Gamma_D(\{d\}) = (12,9).$
$\gamma_D[c,a] = \Gamma_D(\{a\}) = \Gamma_D(\{a,d\}) = (13,8).$
$\gamma_D[c,b] = \Gamma_D(\{b,d\}) = \Gamma_D(\{a,b,d\}) = (15,6).$
$\gamma_D[c,d] = \Gamma_D(\{d\}) = (12,9).$
$\gamma_D[d,a] = \Gamma_D(\{a\}) = \Gamma_D(\{a,c\}) = (13,8).$
$\gamma_D[d,b] = \Gamma_D(\{b\}) = \Gamma_D(\{a,b\}) = \Gamma_D(\{b,c\}) = \Gamma_D(\{a,b,c\}) = (19,2).$
$\gamma_D[d,c] = \Gamma_D(\{a,c\}) = (13,8).$

## Example 2

With $\Gamma_D(B) := \max_D \{ (N[x,y],N[y,x]) \mid x \notin B, y \in B \}$,
we get in example 2 (section 3.2):

$\Gamma_D(\{a\}) = (N[d,a],N[a,d]) = (18,12)$.
$\Gamma_D(\{b\}) = (N[d,b],N[b,d]) = (21,9)$.
$\Gamma_D(\{c\}) = (N[b,c],N[c,b]) = (19,11)$.
$\Gamma_D(\{d\}) = (N[c,d],N[d,c]) = (20,10)$.
$\Gamma_D(\{a,b\}) = (N[d,b],N[b,d]) = (21,9)$.
$\Gamma_D(\{a,c\}) = (N[b,c],N[c,b]) = (19,11)$.
$\Gamma_D(\{a,d\}) = (N[c,d],N[d,c]) = (20,10)$.
$\Gamma_D(\{b,c\}) = (N[d,b],N[b,d]) = (21,9)$.
$\Gamma_D(\{b,d\}) = (N[c,d],N[d,c]) = (20,10)$.
$\Gamma_D(\{c,d\}) = (N[b,c],N[c,b]) = (19,11)$.
$\Gamma_D(\{a,b,c\}) = (N[d,b],N[b,d]) = (21,9)$.
$\Gamma_D(\{a,b,d\}) = (N[c,d],N[d,c]) = (20,10)$.
$\Gamma_D(\{a,c,d\}) = (N[b,c],N[c,b]) = (19,11)$.
$\Gamma_D(\{b,c,d\}) = (N[a,c],N[c,a]) = (17,13)$.

With $\beta_D := \min_D \{ \Gamma_D(B) \mid \emptyset \neq B \subsetneq A \}$, we get: $\beta_D = (17,13)$.

With $\mathfrak{B}_D := \bigcup \{ \emptyset \neq B \subsetneq A \mid \Gamma_D(B) \approx_D \beta_D \}$, we get $\mathfrak{B}_D = \{b,c,d\}$.

So with (4.9.5), we get $\mathcal{S} \subseteq \{b,c,d\}$.

With $\gamma_D[x,y] := \min_D \{ \Gamma_D(B) \mid \emptyset \neq B \subsetneq A$ and $x \notin B$ and $y \in B \}$, we get:

$\gamma_D[a,b] = \Gamma_D(\{b,c,d\}) = (17,13)$.
$\gamma_D[a,c] = \Gamma_D(\{b,c,d\}) = (17,13)$.
$\gamma_D[a,d] = \Gamma_D(\{b,c,d\}) = (17,13)$.
$\gamma_D[b,a] = \Gamma_D(\{a\}) = (18,12)$.
$\gamma_D[b,c] = \Gamma_D(\{c\}) = \Gamma_D(\{a,c\}) = \Gamma_D(\{c,d\}) = \Gamma_D(\{a,c,d\}) = (19,11)$.
$\gamma_D[b,d] = \Gamma_D(\{c,d\}) = \Gamma_D(\{a,c,d\}) = (19,11)$.
$\gamma_D[c,a] = \Gamma_D(\{a\}) = (18,12)$.
$\gamma_D[c,b] = \Gamma_D(\{b,d\}) = \Gamma_D(\{a,b,d\}) = (20,10)$.
$\gamma_D[c,d] = \Gamma_D(\{d\}) = \Gamma_D(\{a,d\}) = \Gamma_D(\{b,d\}) = \Gamma_D(\{a,b,d\}) = (20,10)$.
$\gamma_D[d,a] = \Gamma_D(\{a\}) = (18,12)$.
$\gamma_D[d,b] = \Gamma_D(\{b\}) = \Gamma_D(\{a,b\}) = \Gamma_D(\{b,c\}) = \Gamma_D(\{a,b,c\}) = (21,9)$.
$\gamma_D[d,c] = \Gamma_D(\{c\}) = \Gamma_D(\{a,c\}) = (19,11)$.

## Example 3

With $\Gamma_D(B) := \max_D \{ (N[x,y],N[y,x]) \mid x \notin B, y \in B \}$,
we get in example 3 (section 3.3):

$\Gamma_D(\{a\}) = (N[b,a],N[a,b]) = (25,20)$.
$\Gamma_D(\{b\}) = (N[c,b],N[b,c]) = (29,16)$.
$\Gamma_D(\{c\}) = (N[d,c],N[c,d]) = (28,17)$.
$\Gamma_D(\{d\}) = (N[b,d],N[d,b]) = (33,12)$.
$\Gamma_D(\{e\}) = (N[c,e],N[e,c]) = (24,21)$.
$\Gamma_D(\{a,b\}) = (N[c,b],N[b,c]) = (29,16)$.
$\Gamma_D(\{a,c\}) = (N[d,c],N[c,d]) = (28,17)$.
$\Gamma_D(\{a,d\}) = (N[b,d],N[d,b]) = (33,12)$.
$\Gamma_D(\{a,e\}) = (N[b,a],N[a,b]) = (25,20)$.
$\Gamma_D(\{b,c\}) = (N[d,c],N[c,d]) = (28,17)$.
$\Gamma_D(\{b,d\}) = (N[e,d],N[d,e]) = (31,14)$.
$\Gamma_D(\{b,e\}) = (N[c,b],N[b,c]) = (29,16)$.
$\Gamma_D(\{c,d\}) = (N[b,d],N[d,b]) = (33,12)$.
$\Gamma_D(\{c,e\}) = (N[d,c],N[c,d]) = (28,17)$.
$\Gamma_D(\{d,e\}) = (N[b,d],N[d,b]) = (33,12)$.
$\Gamma_D(\{a,b,c\}) = (N[d,c],N[c,d]) = (28,17)$.
$\Gamma_D(\{a,b,d\}) = (N[e,d],N[d,e]) = (31,14)$.
$\Gamma_D(\{a,b,e\}) = (N[c,b],N[b,c]) = (29,16)$.
$\Gamma_D(\{a,c,d\}) = (N[b,d],N[d,b]) = (33,12)$.
$\Gamma_D(\{a,c,e\}) = (N[d,c],N[c,d]) = (28,17)$.
$\Gamma_D(\{a,d,e\}) = (N[b,d],N[d,b]) = (33,12)$.
$\Gamma_D(\{b,c,d\}) = (N[e,d],N[d,e]) = (31,14)$.
$\Gamma_D(\{b,c,e\}) = (N[d,c],N[c,d]) = (28,17)$.
$\Gamma_D(\{b,d,e\}) = (N[a,d],N[d,a]) = (30,15)$.
$\Gamma_D(\{c,d,e\}) = (N[b,d],N[d,b]) = (33,12)$.
$\Gamma_D(\{a,b,c,d\}) = (N[e,d],N[d,e]) = (31,14)$.
$\Gamma_D(\{a,b,c,e\}) = (N[d,c],N[c,d]) = (28,17)$.
$\Gamma_D(\{a,b,d,e\}) = (N[c,b],N[b,c]) = (29,16)$.
$\Gamma_D(\{a,c,d,e\}) = (N[b,d],N[d,b]) = (33,12)$.
$\Gamma_D(\{b,c,d,e\}) = (N[a,d],N[d,a]) = (30,15)$.

With $\beta_D := \min_D \{ \Gamma_D(B) \mid \varnothing \neq B \subsetneq A \}$, we get: $\beta_D = (24,21)$.

With $\mathfrak{B}_D := \cup \{ \varnothing \neq B \subsetneq A \mid \Gamma_D(B) \approx_D \beta_D \}$, we get $\mathfrak{B}_D = \{e\}$.

So with (4.9.5), we get $\mathcal{S} = \{e\}$.

With $\gamma_D[x,y] := \min_D \{ \Gamma_D(B) \mid \varnothing \neq B \subsetneq A \text{ and } x \notin B \text{ and } y \in B \}$, we get:

$\gamma_D[a,b] = \Gamma_D(\{b,c\}) = \Gamma_D(\{b,c,e\}) = (28,17)$.

$\gamma_D[a,c] = \Gamma_D(\{c\}) = \Gamma_D(\{b,c\}) = \Gamma_D(\{c,e\}) = \Gamma_D(\{b,c,e\}) = (28,17)$.

$\gamma_D[a,d] = \Gamma_D(\{b,d,e\}) = \Gamma_D(\{b,c,d,e\}) = (30,15)$.

$\gamma_D[a,e] = \Gamma_D(\{e\}) = (24,21)$.

$\gamma_D[b,a] = \Gamma_D(\{a\}) = \Gamma_D(\{a,e\}) = (25,20)$.

$\gamma_D[b,c] = \Gamma_D(\{c\}) = \Gamma_D(\{a,c\}) = \Gamma_D(\{c,e\}) = \Gamma_D(\{a,c,e\}) = (28,17)$.

$\gamma_D[b,d] = \Gamma_D(\{d\}) = \Gamma_D(\{a,d\}) = \Gamma_D(\{c,d\}) = \Gamma_D(\{d,e\}) = \Gamma_D(\{a,c,d\}) =$
$\Gamma_D(\{a,d,e\}) = \Gamma_D(\{c,d,e\}) = \Gamma_D(\{a,c,d,e\}) = (33,12)$.

$\gamma_D[b,e] = \Gamma_D(\{e\}) = (24,21)$.

$\gamma_D[c,a] = \Gamma_D(\{a\}) = \Gamma_D(\{a,e\}) = (25,20)$.

$\gamma_D[c,b] = \Gamma_D(\{b\}) = \Gamma_D(\{a,b\}) = \Gamma_D(\{b,e\}) = \Gamma_D(\{a,b,e\}) =$
$\Gamma_D(\{a,b,d,e\}) = (29,16)$.

$\gamma_D[c,d] = \Gamma_D(\{a,b,d,e\}) = (29,16)$.

$\gamma_D[c,e] = \Gamma_D(\{e\}) = (24,21)$.

$\gamma_D[d,a] = \Gamma_D(\{a\}) = \Gamma_D(\{a,e\}) = (25,20)$.

$\gamma_D[d,b] = \Gamma_D(\{b,c\}) = \Gamma_D(\{a,b,c\}) = \Gamma_D(\{b,c,e\}) = \Gamma_D(\{a,b,c,e\}) = (28,17)$.

$\gamma_D[d,c] = \Gamma_D(\{c\}) = \Gamma_D(\{a,c\}) = \Gamma_D(\{b,c\}) = \Gamma_D(\{c,e\}) = \Gamma_D(\{a,b,c\}) =$
$\Gamma_D(\{a,c,e\}) = \Gamma_D(\{b,c,e\}) = \Gamma_D(\{a,b,c,e\}) = (28,17)$.

$\gamma_D[d,e] = \Gamma_D(\{e\}) = (24,21)$.

$\gamma_D[e,a] = \Gamma_D(\{a\}) = (25,20)$.

$\gamma_D[e,b] = \Gamma_D(\{b,c\}) = \Gamma_D(\{a,b,c\}) = (28,17)$.

$\gamma_D[e,c] = \Gamma_D(\{c\}) = \Gamma_D(\{a,c\}) = \Gamma_D(\{b,c\}) = \Gamma_D(\{a,b,c\}) = (28,17)$.

$\gamma_D[e,d] = \Gamma_D(\{b,d\}) = \Gamma_D(\{a,b,d\}) = \Gamma_D(\{b,c,d\}) = \Gamma_D(\{a,b,c,d\}) = (31,14)$.

## 4.10. Prudence

*Prudence* as a criterion for single-winner election methods has been proposed by Köhler (1978) and generalized by Arrow and Raynaud (1986). *Prudence* says (1) that the collective ranking $\mathcal{O}$ should be a strict partial order on $A$ and (2) that the strength $\lambda_D$ of the strongest link $ef$, that is not respected by $\mathcal{O}$, should be as weak as possible. So

$$\lambda_D := \max_D \{ (N[e,f],N[f,e]) \mid ef \notin \mathcal{O} \}$$

should be minimized.

A *directed cycle* is a sequence of alternatives $c(1),\dots,c(n) \in A$ with the following properties:

1. $c(1) \equiv c(n)$.
2. $n \in \mathbb{N}$ with $2 < n \leq (C+1)$.
3. For all $i,j \in \{1,\dots,n\}$: $( i < j \wedge (i,j) \neq (1,n) ) \Rightarrow c(i) \in A \setminus \{c(j)\}$.

A *majority cycle* is a directed cycle with the following additional property:

4. For all $i = 1,\dots,(n-1)$: $N[c(i),c(i+1)] > N[c(i+1),c(i)]$.

It is obvious that, when there is a directed cycle $c(1),\dots,c(n)$, then the strongest link, that is not respected by the binary relation $\mathcal{O}$, is at least as strong as the weakest link $c(i),c(i+1)$ of this directed cycle. Therefore, we get:

(4.10.1) $\lambda_D \succsim_D \min_D \{ (N[c(i),c(i+1)],N[c(i+1),c(i)]) \mid i = 1,\dots,(n-1) \}$.

As we have to make this consideration for all directed cycles, the maximum, that we can ask for, is the following criterion.

**Definition:**

Suppose $\lambda_D \in \mathbb{N}_0 \times \mathbb{N}_0$ is the strength of the strongest directed cycle.

(4.10.2) $\lambda_D := \max_D \{ \min_D \{ (N[c(i),c(i+1)],N[c(i+1),c(i)]) \mid i = 1,\dots,(n-1) \} \mid c(1),\dots,c(n) \text{ is a directed cycle} \}$.

Then an election method is *prudent* if the following holds:

(4.10.3) $\forall\, a,b \in A$: $(N[a,b],N[b,a]) \succ_D \lambda_D \Rightarrow ab \in \mathcal{O}$.

(4.10.4) $\forall\, a,b \in A$: $(N[a,b],N[b,a]) \succ_D \lambda_D \Rightarrow b \notin \mathcal{S}$.

**Claim:**

The Schulze method, as defined in section 2.2, is prudent.

**Proof:**

The proof is trivial. With (2.2.4), we get: $ab \in \mathcal{O}$, unless the link $ab$ is in a directed cycle that consists of links of which each is at least as strong as the link $ab$. □

## Example 1

In example 1 (section 3.1), the strongest directed cycle (measured by the strength of its weakest link) is $a$,(14,7),$c$,(15,6),$b$,(13,8),$a$ with a strength of $\lambda_D \approx_D (13,8)$. So prudence says that the collective ranking $\mathcal{O}$ must respect all links that are stronger than (13,8).

$(N[d,b],N[b,d]) = (19,2) \succ_D (13,8) \approx_D \lambda_D \Rightarrow db \in \mathcal{O}$.

$(N[c,b],N[b,c]) = (15,6) \succ_D (13,8) \approx_D \lambda_D \Rightarrow cb \in \mathcal{O}$.

$(N[a,c],N[c,a]) = (14,7) \succ_D (13,8) \approx_D \lambda_D \Rightarrow ac \in \mathcal{O}$.

With $db \in \mathcal{O}$, $cb \in \mathcal{O}$, and $ac \in \mathcal{O}$, we get $b \notin \mathcal{S}$ and $c \notin \mathcal{S}$.

With $ac \in \mathcal{O}$ and $cb \in \mathcal{O}$ and the transitivity of $\mathcal{O}$, we get $ab \in \mathcal{O}$.

## Example 2

In example 2 (section 3.2), the strongest directed cycle (measured by the strength of its weakest link) is $b$,(19,11),$c$,(20,10),$d$,(21,9),$b$ with a strength of $\lambda_D \approx_D (19,11)$. So prudence says that the collective ranking $\mathcal{O}$ must respect all links that are stronger than (19,11).

$(N[d,b],N[b,d]) = (21,9) \succ_D (19,11) \approx_D \lambda_D \Rightarrow db \in \mathcal{O}$.

$(N[c,d],N[d,c]) = (20,10) \succ_D (19,11) \approx_D \lambda_D \Rightarrow cd \in \mathcal{O}$.

With $db \in \mathcal{O}$ and $cd \in \mathcal{O}$, we get $b \notin \mathcal{S}$ and $d \notin \mathcal{S}$.

With $cd \in \mathcal{O}$ and $db \in \mathcal{O}$ and the transitivity of $\mathcal{O}$, we get $cb \in \mathcal{O}$.

## Example 3

In example 3 (section 3.3), the strongest directed cycle (measured by the strength of its weakest link) is $b$,(33,12),$d$,(28,17),$c$,(29,16),$b$ with a strength of $\lambda_D \approx_D (28,17)$. So prudence says that the collective ranking $\mathcal{O}$ must respect all links that are stronger than (28,17).

$(N[b,d],N[d,b]) = (33,12) \succ_D (28,17) \approx_D \lambda_D \Rightarrow bd \in \mathcal{O}$.

$(N[e,d],N[d,e]) = (31,14) \succ_D (28,17) \approx_D \lambda_D \Rightarrow ed \in \mathcal{O}$.

$(N[a,d],N[d,a]) = (30,15) \succ_D (28,17) \approx_D \lambda_D \Rightarrow ad \in \mathcal{O}$.

$(N[c,b],N[b,c]) = (29,16) \succ_D (28,17) \approx_D \lambda_D \Rightarrow cb \in \mathcal{O}$.

With $bd \in \mathcal{O}$, $ed \in \mathcal{O}$, $ad \in \mathcal{O}$, and $cb \in \mathcal{O}$, we get $b \notin \mathcal{S}$ and $d \notin \mathcal{S}$.

With $cb \in \mathcal{O}$ and $bd \in \mathcal{O}$ and the transitivity of $\mathcal{O}$, we get $cd \in \mathcal{O}$.

## 4.11. Schwartz

The Schwartz criterion as a criterion for single-winner election methods has been proposed by Schwartz (1986). The Schwartz criterion implies the Smith criterion.

A *chain* from alternative $x \in A$ to alternative $y \in A \setminus \{x\}$ is a sequence of alternatives $c(1),\dots,c(n) \in A$ with the following properties:

1. $x \equiv c(1)$.
2. $y \equiv c(n)$.
3. $n \in \mathbb{N}$ with $2 \leq n \leq C$.
4. For all $i,j \in \{1,\dots,n\}$: $i \neq j \Rightarrow c(i) \in A \setminus \{c(j)\}$.
5. For all $i = 1,\dots,(n-1)$: $N[c(i),c(i+1)] > N[c(i+1),c(i)]$.

**Definition:**

An election method satisfies the *Schwartz criterion* if the following holds:

Suppose there is a chain from alternative $a \in A$ to alternative $b \in A$ and no chain from alternative $b$ to alternative $a$. Then:

(4.11.1) $ab \in \mathcal{O}$.

(4.11.2) $b \notin \mathcal{S}$.

**Claim:**

If $\succ_D$ satisfies (2.1.5), then the Schulze method, as defined in section 2.2, satisfies the Schwartz criterion.

**Proof:**

The proof is trivial. □

**Definition:**

The *Schwartz set* $\emptyset \neq B \subseteq A$ is defined as follows:

(4.11.3) $a \in B :\Leftrightarrow (\ \forall\ b \in A \setminus \{a\}$:
( There is a chain from alternative $b$ to alternative $a$.
$\Rightarrow$ There is a chain from alternative $a$ to alternative $b$. ) )

A real-life example that illustrates the difference between the Smith set, as defined in (4.8.1), and the Schwartz set, as defined in (4.11.3), was the nomination (on 25/26 February 2011) of the top candidate for the party list of the Pirate Party for the elections (on 18 September 2011) to the Berlin House of Representatives. The Schulze method was used for this nomination. The following digraph shows the result of the nomination (www145):

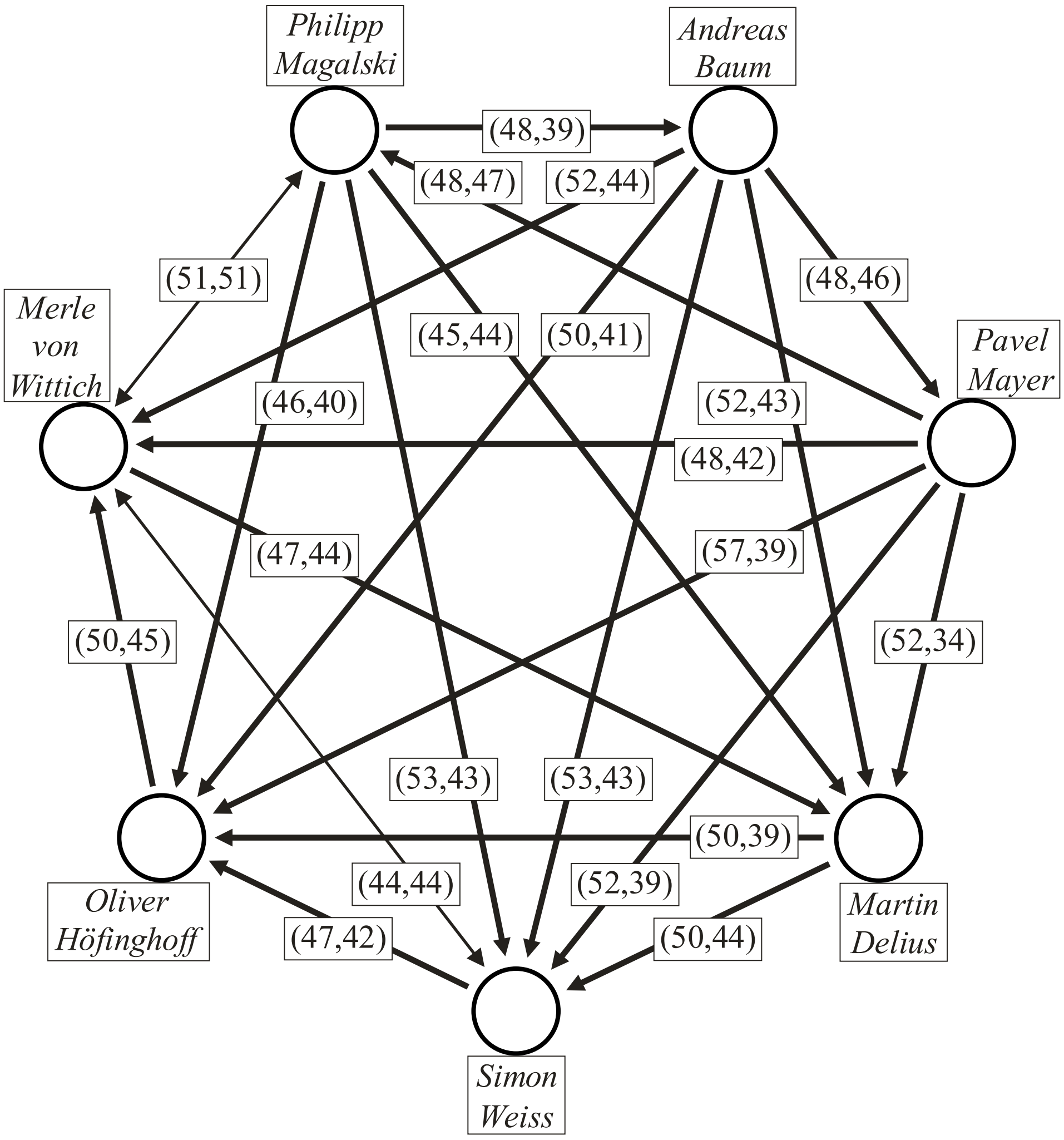

There were 106 valid ballots in total. There were three majority cycles (as defined in section 4.10): Magalski → Baum → Mayer → Magalski, Delius → Weiss → Höfinghoff → Wittich → Delius, and Delius → Höfinghoff → Wittich → Delius. There were two pairwise ties: Magalski ↔ Wittich and Weiss ↔ Wittich. The Smith set consists of all 7 candidates, while the Schwartz set contains only 3 candidates (Magalski, Baum, Mayer). The Schulze ranking is Magalski $\succ_S$ Baum $\succ_S$ Mayer $\succ_S$ Delius $\succ_S$ Weiss $\succ_S$ Höfinghoff $\succ_S$ Wittich for $\succ_{margin}$, $\succ_{ratio}$, and $\succ_{win}$ and Magalski $\succ_S$ Baum $\succ_S$ Mayer $\succ_S$ Wittich $\succ_S$ Delius $\succ_S$ Weiss $\succ_S$ Höfinghoff for $\succ_{los}$.

## 4.12. Weak Condorcet Winners and Weak Condorcet Losers

### 4.12.1. Weak Condorcet Winners

A *Condorcet winner* is an alternative $a \in A$ that wins every head-to-head contest with some other alternative $b \in A \setminus \{a\}$. See (4.8.6). In other words:

(4.12.1.1) Alternative $a \in A$ is a *Condorcet winner* $:\Leftrightarrow$
$N[a,b] > N[b,a]$ for all $b \in A \setminus \{a\}$.

A *weak Condorcet winner* is an alternative $a \in A$ that doesn't lose any head-to-head contest with some other alternative $b \in A \setminus \{a\}$. In other words:

(4.12.1.2) Alternative $a \in A$ is a *weak Condorcet winner* $:\Leftrightarrow$
$N[a,b] \geq N[b,a]$ for all $b \in A \setminus \{a\}$.

Suppose $\mathcal{E}$ is the set of weak Condorcet winners. Then we get:

(4.12.1.3) $a \in \mathcal{E} :\Leftrightarrow N[a,b] \geq N[b,a]$ for all $b \in A \setminus \{a\}$.

A frequently stated desideratum says that, when there is a weak Condorcet winner, then it should win.

When there happens to be exactly one potential winner $x \in A$ and exactly one weak Condorcet winner $y \in A$, it is obvious what the above desideratum means: Alternative $x$ and alternative $y$ must be the same alternative.

In other words:

(4.12.1.4) $( \| \mathcal{E} \| = 1 \text{ and } \| \mathcal{S} \| = 1 ) \Rightarrow \mathcal{E} = \mathcal{S}.$

However, when there happens to be more than one potential winner or more than one weak Condorcet winner, the proper formulation for the above desideratum isn't obvious. The most intuitive formulation is:

(4.12.1.5) $\mathcal{E} \neq \varnothing \Rightarrow \mathcal{S} \subseteq \mathcal{E}.$

Formulation (4.12.1.5) says that, when there is at least one weak Condorcet winner, then every potential winner should be a weak Condorcet winner. Unfortunately, the following example demonstrates that (4.12.1.5) is incompatible with reversal symmetry:

> Suppose there are four alternatives $A = \{a,b,c,d\}$. Suppose $N^{old}[a,b] = N^{old}[b,a]$, $N^{old}[a,c] = N^{old}[c,a]$, $N^{old}[a,d] = N^{old}[d,a]$, $N^{old}[b,c] > N^{old}[c,b]$, $N^{old}[c,d] > N^{old}[d,c]$, and $N^{old}[d,b] > N^{old}[b,d]$. Then we get $\mathcal{E}^{old} = \{a\}$. With (4.12.1.5) and the requirement that $\mathcal{S}^{old}$ must not be empty, we get $\mathcal{S}^{old} = \{a\}$.
>
> When the individual preferences are reversed, as defined in (4.5.1), we get $N^{new}[a,b] = N^{new}[b,a]$, $N^{new}[a,c] = N^{new}[c,a]$, $N^{new}[a,d] = N^{new}[d,a]$, $N^{new}[b,c] < N^{new}[c,b]$, $N^{new}[c,d] < N^{new}[d,c]$, and $N^{new}[d,b] < N^{new}[b,d]$. Therefore, we get $\mathcal{E}^{new} = \{a\}$. With (4.12.1.5) and the requirement that $\mathcal{S}^{new}$ must not be empty, we get $\mathcal{S}^{new} = \{a\}$.
>
> But $\mathcal{S}^{old} = \{a\}$ and $\mathcal{S}^{new} = \{a\}$ together contradict (4.5.4).

In short: It can happen that the same alternative is the unique weak Condorcet winner in the original situation and, simultaneously, the unique weak Condorcet winner in the reversed situation. Therefore, (4.12.1.5) cannot be compatible with reversal symmetry.

Furthermore, the following example demonstrates that (4.12.1.5) is incompatible with independence of clones:

> Suppose there are only two alternatives $A^{old} = \{a,b\}$. Suppose $N[a,b] = N[b,a]$. Then we get $\mathcal{E}^{old} = \{a,b\}$. With (4.12.1.5), we get $\mathcal{S}^{old} \subseteq \{a,b\}$.
>
> Case I: Suppose $a \in \mathcal{S}^{old}$. When alternative $a$ is replaced by alternatives $a_1,a_2,a_3$ such that $N[a_1,a_2] > N[a_2,a_1]$, $N[a_2,a_3] > N[a_3,a_2]$, and $N[a_3,a_1] > N[a_1,a_3]$ and such that (4.7.1) – (4.7.3) are satisfied, we get $\mathcal{E}^{new} = \{b\}$. With (4.12.1.5) and the requirement that $\mathcal{S}^{new}$ must not be empty, we get $\mathcal{S}^{new} = \{b\}$. But with (4.7.7) and $a \in \mathcal{S}^{old}$, we get $(\mathcal{S}^{new} \cap \{a_1,a_2,a_3\}) \neq \emptyset$. As $\mathcal{S}^{new} = \{b\}$ and $(\mathcal{S}^{new} \cap \{a_1,a_2,a_3\}) \neq \emptyset$ are incompatible, we get $a \notin \mathcal{S}^{old}$.
>
> Case II: Suppose $b \in \mathcal{S}^{old}$. When alternative $b$ is replaced by alternatives $b_1,b_2,b_3$ such that $N[b_1,b_2] > N[b_2,b_1]$, $N[b_2,b_3] > N[b_3,b_2]$, and $N[b_3,b_1] > N[b_1,b_3]$ and such that (4.7.1) – (4.7.3) are satisfied, we get $\mathcal{E}^{new} = \{a\}$. With (4.12.1.5) and the requirement that $\mathcal{S}^{new}$ must not be empty, we get $\mathcal{S}^{new} = \{a\}$. But with (4.7.7) and $b \in \mathcal{S}^{old}$, we get $(\mathcal{S}^{new} \cap \{b_1,b_2,b_3\}) \neq \emptyset$. As $\mathcal{S}^{new} = \{a\}$ and $(\mathcal{S}^{new} \cap \{b_1,b_2,b_3\}) \neq \emptyset$ are incompatible, we get $b \notin \mathcal{S}^{old}$.
>
> However, $a \notin \mathcal{S}^{old}$ and $b \notin \mathcal{S}^{old}$ together are incompatible with the requirement that $\mathcal{S}^{old}$ must not be empty.

In short: When a weak Condorcet winner is replaced by a set of clones, as defined in (4.7.1) – (4.7.3), it is not guaranteed that at least one of these clones is a weak Condorcet winner. Therefore, (4.12.1.5) cannot be compatible with independence of clones.

The above examples demonstrate that, to satisfy reversal symmetry and independence of clones, we have, in some situations, to allow alternatives, which are not weak Condorcet winners, to be among the potential winners.

So the maximum, that we could ask for, is:

(4.12.1.6) $\mathcal{E} \subseteq \mathcal{S}$.

Formulation (4.12.1.6) says that every weak Condorcet winner should be a potential winner, but it makes no stipulations about those alternatives which are not weak Condorcet winners. In (4.12.1.6), the presumption " $\mathcal{E} \neq \emptyset$ " is not needed. We don't have to write " $\mathcal{E} \neq \emptyset \Rightarrow \mathcal{E} \subseteq \mathcal{S}$ " because the empty set is, by definition, subset of every set.

The following proof demonstrates that the Schulze method satisfies (4.12.1.6) and that, therefore, (4.12.1.6) is compatible with reversal symmetry and independence of clones.

**Claim:**

If $\succ_D$ satisfies (2.1.4) and (2.1.5), then the Schulze method, as defined in section 2.2, satisfies (4.12.1.6).

**Proof:**

Step 1:

(2.1.4) says that all ties have equivalent strengths. So without loss of generality, we can set

(4.12.1.7) $\forall\, x \in \mathbb{N}_0$: $(x,x) \approx_D (1,1)$.

Step 2:

With (2.2.3), we get

(4.12.1.8) $\forall\, a \in \mathcal{E}\ \forall\, b \in A \setminus \{a\}$: $P_D[a,b] \succsim_D (N[a,b],N[b,a])$.

With (4.12.1.3), we get

(4.12.1.9) $\forall\, a \in \mathcal{E}\ \forall\, b \in A \setminus \{a\}$: $N[a,b] \geq N[b,a]$.

With (2.1.5), (4.12.1.7), and (4.12.1.9), we get

(4.12.1.10) $\forall\, a \in \mathcal{E}\ \forall\, b \in A \setminus \{a\}$: $(N[a,b],N[b,a]) \succsim_D (1,1)$.

Step 3:

Suppose $a \in \mathcal{E}$. Suppose $b \in A \setminus \{a\}$. Suppose the link $ca$ is the last link in the strongest path from alternative $b$ to alternative $a$. Then we get

(4.12.1.11) $P_D[b,a] \precsim_D (N[c,a],N[a,c])$.

With (4.12.1.9), we get

(4.12.1.12) $N[a,c] \geq N[c,a]$.

With (2.1.5), (4.12.1.7), and (4.12.1.12), we get

(4.12.1.13) $(N[c,a],N[a,c]) \precsim_D (1,1)$.

With (4.12.1.8), (4.12.1.10), (4.12.1.13), and (4.12.1.11), we get

(4.12.1.14) $P_D[a,b] \succsim_D (N[a,b],N[b,a]) \succsim_D (1,1) \succsim_D (N[c,a],N[a,c]) \succsim_D P_D[b,a]$.

The considerations in (4.12.1.11) – (4.12.1.14) can be repeated for every $a \in \mathcal{E}$ and every $b \in A \setminus \{a\}$. Therefore, with (4.12.1.14), we get

(4.12.1.15) $\forall\, a \in \mathcal{E}\ \forall\, b \in A \setminus \{a\}$: $P_D[a,b] \succsim_D P_D[b,a]$.

With (4.12.1.15), we get

(4.12.1.16) $a \in \mathcal{E} \Rightarrow a \in \mathcal{S}$.

With (4.12.1.16), we get (4.12.1.6). □

The following desideratum further reduces the scenarios where some alternative, that is not a weak Condorcet winner, can be a potential winner:

(4.12.1.17) $\forall\ a \in \mathcal{E}\ \forall\ b \in (\ \mathcal{S} \setminus \mathcal{E}\ ): N[a,b] = N[b,a]$.

Desideratum (4.12.1.17) says that an alternative, that is not a weak Condorcet winner, can be a potential winner only when it pairwise ties all weak Condorcet winners.

**Claim:**

If $\succ_D$ satisfies (2.1.5), then the Schulze method, as defined in section 2.2, satisfies (4.12.1.17).

**Proof:**

Suppose $a \in \mathcal{E}$ and $b \in (\ \mathcal{S} \setminus \mathcal{E}\ )$.

Step 1:

$N[a,b] < N[b,a]$ is a contradiction to the presumption that alternative $a$ is a weak Condorcet winner.

Step 2:

It remains to be proven that $N[a,b] > N[b,a]$ is not possible.

So suppose

(4.12.1.18) $N[a,b] > N[b,a]$.

With (2.2.3), we get

(4.12.1.19) $P_D[a,b] \succsim_D (N[a,b],N[b,a])$.

Suppose the link $ca$ is the last link in the strongest path from alternative $b$ to alternative $a$. Then we get

(4.12.1.20) $P_D[b,a] \precsim_D (N[c,a],N[a,c])$.

With $a \in \mathcal{E}$ and (4.12.1.3), we get

(4.12.1.21) $N[a,c] \geq N[c,a]$.

With (2.1.5), (4.12.1.18), and (4.12.1.21), we get

(4.12.1.22) $(N[a,b],N[b,a]) \succ_D (N[c,a],N[a,c])$.

With (4.12.1.19), (4.12.1.22), and (4.12.1.20), we get

(4.12.1.23) $P_D[a,b] \succsim_D (N[a,b],N[b,a]) \succ_D (N[c,a],N[a,c]) \succsim_D P_D[b,a]$.

So alternative $a$ disqualifies alternative $b$. But this is a contradiction to the presumption that alternative $b$ is a potential winner. □

## 4.12.2. Weak Condorcet Losers

A *Condorcet loser* is an alternative $b \in A$ that loses every head-to-head contest with some other alternative $a \in A \setminus \{b\}$. See (4.8.8). In other words:

(4.12.2.1) Alternative $b \in A$ is a *Condorcet loser* $:\Leftrightarrow$
$N[a,b] > N[b,a]$ for all $a \in A \setminus \{b\}$.

A *weak Condorcet loser* is an alternative $b \in A$ that doesn't win any head-to-head contest with some other alternative $a \in A \setminus \{b\}$. In other words:

(4.12.2.2) Alternative $b \in A$ is a *weak Condorcet loser* $:\Leftrightarrow$
$N[a,b] \geq N[b,a]$ for all $a \in A \setminus \{b\}$.

Suppose $\mathcal{F}$ is the set of weak Condorcet losers. Then we get:

(4.12.2.3) $b \in \mathcal{F} :\Leftrightarrow N[a,b] \geq N[b,a]$ for all $a \in A \setminus \{b\}$.

A frequently stated desideratum says that a weak Condorcet loser should not be a potential winner. So with (4.12.2.3), we get

(4.12.2.4) $\forall\ b \in A$: ( $b \in \mathcal{F} \Rightarrow b \notin \mathcal{S}$ ).

However, a problem with desideratum (4.12.2.4) is that it can happen that a weak Condorcet loser is, simultaneously, a weak Condorcet winner. In this case, (4.12.2.4) is incompatible with (4.12.1.6).

Example: Suppose there are only $C = 2$ alternatives $a,b \in A$. Suppose there is a pairwise tie, $N[a,b] = N[b,a]$. Then both alternatives are weak Condorcet losers and, simultaneously, weak Condorcet winners. (4.12.1.6) says: $a \in \mathcal{S}$ and $b \in \mathcal{S}$. (4.12.2.4) says: $a \notin \mathcal{S}$ and $b \notin \mathcal{S}$.

So the maximum, that we could ask for, is:

(4.12.2.5) $\forall\ b \in A$: ( ( $b \in \mathcal{F}$ and $b \notin \mathcal{E}$ ) $\Rightarrow b \notin \mathcal{S}$ ).

Desideratum (4.12.2.5) says that a weak Condorcet loser should not win, unless it is also a weak Condorcet winner. The following proof demonstrates that the Schulze method satisfies (4.12.2.5) and that, therefore, there is no need to weaken (4.12.2.5) any further.

**Claim:**

If $\succ_D$ satisfies (2.1.5), then the Schulze method, as defined in section 2.2, satisfies (4.12.2.5).

**Proof:**

With (2.2.3), we get

(4.12.2.6) $\forall\ a,b \in A$: $P_D[a,b] \succsim_D (N[a,b],N[b,a])$.

Suppose $b \in \mathcal{F}$ and $b \notin \mathcal{E}$.

With $b \in \mathcal{F}$ and (4.12.2.3), we get

(4.12.2.7) $\forall\ a \in A \setminus \{b\}$: $N[a,b] \geq N[b,a]$.

With $b \notin \mathcal{E}$ and (4.12.1.3), we get

(4.12.2.8) $\exists\ a \in A \setminus \{b\}$: $N[a,b] > N[b,a]$.

For the rest of this proof, we choose an alternative $a \in A \setminus \{b\}$ with (4.12.2.8). Suppose the link $bc$ is the first link in the strongest path from alternative $b$ to alternative $a$. Then we get

(4.12.2.9) $P_D[b,a] \precsim_D (N[b,c],N[c,b])$.

With (4.12.2.7), we get

(4.12.2.10) $N[c,b] \geq N[b,c]$.

With (2.1.5), (4.12.2.8), and (4.12.2.10), we get

(4.12.2.11) $(N[b,c],N[c,b]) \prec_D (N[a,b],N[b,a])$.

With (4.12.2.6), (4.12.2.11), and (4.12.2.9), we get

(4.12.2.12) $P_D[a,b] \succsim_D (N[a,b],N[b,a]) \succ_D (N[b,c],N[c,b]) \succsim_D P_D[b,a]$.

So alternative $a$ disqualifies alternative $b$. So $b \notin \mathcal{S}$. □

The following desideratum says that a weak Condorcet loser cannot be a unique winner:

(4.12.2.13) $\forall\ b \in A$: ( $b \in \mathcal{F} \Rightarrow \mathcal{S} \neq \{b\}$ ).

**Claim:**

If $\succ_D$ satisfies (2.1.4) and (2.1.5), then the Schulze method, as defined in section 2.2, satisfies (4.12.2.13).

**Proof:**

Step 1:

(2.1.4) says that all ties have equivalent strengths. So without loss of generality, we can set

(4.12.2.14) $\forall\ x \in \mathbb{N}_0$: $(x,x) \approx_D (1,1)$.

Step 2:

With (2.2.3), we get

(4.12.2.15) $\forall\ a,b \in A$: $P_D[a,b] \succsim_D (N[a,b],N[b,a])$.

Suppose $b \in \mathcal{F}$. With (4.12.2.3), we get

(4.12.2.16) $\forall\ a \in A \setminus \{b\}$: $N[a,b] \geq N[b,a]$.

With (2.1.5), (4.12.2.14), and (4.12.2.16), we get

(4.12.2.17) $\forall\ a \in A \setminus \{b\}$: $(N[a,b],N[b,a]) \succsim_D (1,1)$.

For the rest of this proof, we choose a concrete alternative $a \in A \setminus \{b\}$. Suppose the link *bc* is the first link in the strongest path from alternative *b* to alternative *a*. Then we get

(4.12.2.18) $P_D[b,a] \precsim_D (N[b,c],N[c,b])$.

With (4.12.2.16), we get

(4.12.2.19) $N[c,b] \geq N[b,c]$.

With (2.1.5), (4.12.2.14), and (4.12.2.19), we get

(4.12.2.20) $(N[b,c],N[c,b]) \precsim_D (1,1)$.

With (4.12.2.15), (4.12.2.17), (4.12.2.20), and (4.12.2.18), we get

(4.12.2.21) $P_D[a,b] \succsim_D (N[a,b],N[b,a]) \succsim_D (1,1) \succsim_D (N[b,c],N[c,b]) \succsim_D P_D[b,a]$.

The considerations in (4.12.2.18) – (4.12.2.21) can be repeated for every $a \in A \setminus \{b\}$. Therefore, with (4.12.2.21), we get

(4.12.2.22) $\forall\ a \in A \setminus \{b\}: P_D[a,b] \succsim_D P_D[b,a]$.

<u>Step 3:</u>

As $\mathcal{O}$ is transitive, there is an alternative *d* in $A \setminus \{b\}$ that is not disqualified by any other alternative in $A \setminus \{b\}$. We get

(4.12.2.23) $\exists\ d \in A \setminus \{b\}\ \forall\ e \in A \setminus \{b,d\}: ed \notin \mathcal{O}$.

With (4.12.2.22), we get that alternative *b* doesn't disqualify alternative *d*. With (4.12.2.23), we get that no other alternative $e \in A \setminus \{b,d\}$ disqualifies alternative *d*. Therefore, alternative *d* is a potential winner. Therefore, we get $d \in \mathcal{S}$. Therefore, we get $\mathcal{S} \neq \{b\}$. Therefore, we get (4.12.2.13). □

The proofs of section 4.12 have recently been reproduced by Chavrimootoo (2026).

## 4.13. Increasing Sequential Independence

*Increasing sequential independence* says that, when alternative $a \in A$ is a winner, then there must be an alternative $d \in A \setminus \{a\}$ such that, when the used election method is applied to $A \setminus \{d\}$, then alternative $a$ is still a winner.

The name for this criterion comes from the fact that — when the used election method satisfies this criterion and when alternative $a \in A$ is a winner and alternative $d(1) \in A \setminus \{a\}$ is an alternative such that, when the used election method is applied to $A \setminus \{d(1)\}$, then alternative $a$ is still a winner — the same criterion can then be applied to $A \setminus \{d(1)\}$ to identify an alternative $d(2) \in A \setminus \{a,d(1)\}$ such that, when the used election method is applied to $A \setminus \{d(1),d(2)\}$, then alternative $a$ is still a winner. When we continue applying this criterion, we get a linear order $d(1),\ldots,d(C-1)$ of the alternatives in $A \setminus \{a\}$ such that, for every $i \in \{1,\ldots,(C-1)\}$, alternative $a$ is still a winner when the used election method is applied to $A \setminus \{d(1),\ldots,d(i)\}$.

The motivation for this criterion is that an alternative $a \in A$ should be able to win only by disqualifying all the other alternatives directly or indirectly in some manner. It should not be possible that some alternatives $\emptyset \neq \{d(1),\ldots,d(i)\} \subsetneq A$ disqualify each other in such a manner that the final winner comes from outside of $\{d(1),\ldots,d(i)\}$. When increasing sequential independence is satisfied, then one alternative after the other is disqualified, so that the final winner $a \in A$ can come from outside of $\{d(1),\ldots,d(i)\}$ only when the last remaining alternative $d(j) \in \{d(1),\ldots,d(i)\}$ is disqualified by some alternatives outside of $\{d(1),\ldots,d(i)\}$.

Increasing sequential independence and decreasing sequential independence (section 4.15) as criteria for single-winner election methods have been proposed by Arrow and Raynaud (1986).

**Definition #1:**

An election method satisfies the first version of *increasing sequential independence* if the following holds:

Suppose alternative $a \in A$ is a unique winner when this election method is applied to $A$. Then there must be a (not necessarily unique) alternative $d \in A \setminus \{a\}$ such that, when this election method is applied to $A \setminus \{d\}$, then alternative $a$ is still a unique winner.

In short:

(4.13.1) $\forall\, a \in A$: $( ( \mathcal{S} = \{a\} ) \Rightarrow ( \exists\, d \in A \setminus \{a\}: \mathcal{S}|_{(A \setminus \{d\})} = \{a\} ) )$.

**Claim #1:**

The Schulze method, as defined in section 2.2, satisfies the first version of increasing sequential independence.

**Proof of claim #1:**

Suppose alternative $a \in A$ is a unique winner when this election method is applied to $A$. Then, according to (4.1.15), alternative $a$ disqualifies every other alternative $b \in A \setminus \{a\}$. Therefore, we get

(4.13.2) $\forall\, b \in A \setminus \{a\}$: $P_D^{\text{old}}[a,b] \succ_D P_D^{\text{old}}[b,a]$.

Suppose $pred^{\text{old}}[a,x]$ is the predecessor of alternative $x \in A \setminus \{a\}$ in the strongest path from alternative $a$ to alternative $x$, as calculated in section 2.3.1. Then a *leaf* is an alternative $y \in A \setminus \{a\}$ such that there is no alternative $x \in A \setminus \{a\}$ with $pred^{\text{old}}[a,x] = y$. As the strongest paths from alternative $a$ to every other alternative $x \in A \setminus \{a\}$, as calculated by the Floyd-Warshall algorithm, form an arborescence, there must be at least one leaf. Alternative $d$ is chosen arbitrarily from these leaves.

Suppose alternative $d$ is removed. As alternative $d$ is a leaf, alternative $d$ is not in the strongest path from alternative $a$ to any other alternative $b \in A \setminus \{a,d\}$. Therefore, we get

(4.13.3) $\forall\, b \in A \setminus \{a,d\}$: $P_D^{\text{new}}[a,b] \approx_D P_D^{\text{old}}[a,b]$.

On the other side, when an alternative is removed, then the strengths of the strongest paths can only decrease. Therefore, we get

(4.13.4) $\forall\, b \in A \setminus \{a,d\}$: $P_D^{\text{new}}[b,a] \precsim_D P_D^{\text{old}}[b,a]$.

With (4.13.3), (4.13.2), and (4.13.4), we get

(4.13.5) $\forall\, b \in A \setminus \{a,d\}$:

$$P_D^{\text{new}}[a,b] \approx_D P_D^{\text{old}}[a,b] \succ_D P_D^{\text{old}}[b,a] \succsim_D P_D^{\text{new}}[b,a]$$

so that alternative $a$ is still a unique winner when alternative $d$ is removed. □

**Definition #2:**

An election method satisfies the second version of *increasing sequential independence* if the following holds:

Suppose alternative $a \in A$ is a potential winner when this election method is applied to $A$. Then there must be a (not necessarily unique) alternative $d \in A \setminus \{a\}$ such that, when this election method is applied to $A \setminus \{d\}$, then alternative $a$ is still a potential winner.

In short:

(4.13.6) $\forall\, a \in A: ( ( a \in \mathcal{S} ) \Rightarrow ( \exists\, d \in A \setminus \{a\} : a \in \mathcal{S}|_{(A \setminus \{d\})} ) ).$

**Claim #2:**

The Schulze method, as defined in section 2.2, satisfies the second version of increasing sequential independence.

**Proof of claim #2:**

Suppose alternative $a \in A$ is a potential winner when this election method is applied to $A$. Then, we get

(4.13.7) $\forall\, b \in A \setminus \{a\}: P_D^{\text{old}}[a,b] \succsim_D P_D^{\text{old}}[b,a].$

The rest of this proof is identical to the proof of claim #1. □

## 4.14. *k*-Consistency

The Condorcet criterion says that, when some candidate $a \in A$ wins every head-to-head contest, then this candidate $a$ should also be the overall winner (Condorcet, 1785).

However, many countries have a strong 3-party, 4-party or 5-party system where no single party can win a majority and where every party is willing to coalesce with every other party. In such a scenario, it seems to be rather uninteresting which candidate might win in a head-to-head contest. It is more interesting to ask whether there is some candidate who wins regardless of which candidates are nominated by the other parties.

So for example in the 3-party case with party α, party β, and party γ, it might be more interesting to ask whether there is a candidate from party α who wins every 3-way contest between himself and a candidate from party β and a candidate from party γ. If there is such a candidate, then this candidate should also be the overall winner.

More generally, if there is a $k \in \mathbb{N}$ with $k \geq 2$ such that there is an alternative $a \in A$ such that alternative $a$ wins every $k$-way contest, then alternative $a$ should also be the overall winner. This criterion is called *k-set-consistency* (Heitzig, 2004b) or *k-consistency* (Simmons, 2004).

*k*-consistency as a criterion for single-winner election methods has been proposed by Heitzig (2004b) and Simmons (2004). However, a similar idea had already been formulated by Saari (Saari, 2001, pages 154–156; Lagerspetz, 2015, page 207, footnote 20). To question the relevance of the Condorcet criterion, Saari argued that it could happen that some alternative $a \in A$ wins every 2-way contest, some other alternative $b \in A \setminus \{a\}$ wins every 3-way contest, some other alternative $c \in A \setminus \{a,b\}$ wins every 4-way contest, etc., so that, with the same justification, every alternative could claim to be the overall winner. However, the fact that the Schulze method satisfies $k$-consistency for every $k \in \mathbb{N}$ with $k \geq 2$ means that there are election methods where it is impossible to create examples such that there are $m,n \in \mathbb{N}$ with $2 \leq m < n \leq C$ such that some alternative $a \in A$ wins every $m$-way contest and some other alternative $b \in A \setminus \{a\}$ wins every $n$-way contest. So for these election methods, Saari’s scenario is not possible, so that his criticism of the Condorcet criterion doesn’t work.

There are five different versions for $k$-consistency.

The first version addresses unique winners. This version says that, when alternative $a \in A$ is a unique winner in every $k$-way contest, then alternative $a$ should also be a unique winner overall. For $k = 2$, the first version of $k$-consistency is identical to the Condorcet criterion; equation (4.8.7)(i).

The second version addresses potential winners. This version says that, when alternative $a \in A$ is a potential winner in every $k$-way contest, then alternative $a$ should also be a potential winner overall. For $k = 2$, the second version of $k$-consistency is identical to the desideratum that weak Condorcet winners should always be potential winners; equation (4.12.1.6).

The third version addresses the set of potential winners. This version says that, when in every $k$-way contest (that contains at least one alternative of the set $\emptyset \neq B \subsetneq A$) the winner comes from the set $B$, then the winner must also come from the set $B$ when the method is applied to $A$. For $k = 2$, the third version of $k$-consistency is identical to the Smith criterion; equation (4.8.5).

The fourth version says that, when alternative $a \in A$ is not a unique winner in any $k$-way contest, then alternative $a$ should also be not a unique winner overall. For $k = 2$, the fourth version of $k$-consistency is identical to the desideratum that a weak Condorcet loser should not be a unique winner; equation (4.12.2.13).

The fifth version says that, when alternative $a \in A$ is not a potential winner in any $k$-way contest, then alternative $a$ should also be not a potential winner overall. For $k = 2$, the fifth version of $k$-consistency is identical to the Condorcet loser criterion; equation (4.8.9)(i).

## 4.14.1. Formulation #1

**Definition:**

Suppose $k \in \mathbb{N}$ with $k \geq 2$. An election method satisfies the first version of *k-consistency* if the following holds:

Suppose $C$ with $C > k$ is the number of alternatives in $A$. Suppose alternative $a \in A$ is a unique winner whenever this election method is applied to some subset $\tilde{A} \subset A$ with $\| \tilde{A} \| = k$ and $a \in \tilde{A}$. Then alternative $a$ is also a unique winner when this election method is applied to $A$.

In short:

(4.14.1.1) $\forall\, a \in A$: ( ( $\forall\, \tilde{A} \subset A$ with $\| \tilde{A} \| = k$ and $a \in \tilde{A}$: $\mathcal{S}|_{\tilde{A}} = \{a\}$ ) $\Rightarrow$ ( $\mathcal{S} = \{a\}$ ) ).

**Claim:**

If $\succ_D$ satisfies (2.1.5), then the Schulze method, as defined in section 2.2, satisfies the first version of $k$-consistency for every $k \in \mathbb{N}$ with $k \geq 2$.

**Proof (overview):**

We will show how, when alternative $a \in A$ is not a unique winner (when this election method is applied to $A$), we can create, for every $k \in \mathbb{N}$ with $2 \leq k < C$, a subset $\tilde{A} \subset A$ with $\| \tilde{A} \| = k$ and $a \in \tilde{A}$ such that, when the Schulze method is applied to $\tilde{A}$, alternative $a$ is not a unique winner.

**Proof (details):**

Suppose alternative $a \in A$ is not a unique winner when the Schulze method is applied to $A$. Then, with (4.1.15), we get that there must be an alternative $b \in A \setminus \{a\}$ with

(4.14.1.2) $P_D[b,a] \succsim_D P_D[a,b]$.

We set

(4.14.1.3) $(z_1,z_2) := P_D[b,a]$

to stress that this value is constant for the rest of this proof.

Suppose $c(1),\dots,c(n)$ is the shortest path (in terms of its number of links) from alternative $b \equiv c(1)$ to alternative $a \equiv c(n)$ of strength $(z_1,z_2)$. Then we get

(4.14.1.4) $\forall\, i = 1,\dots,(n-1)$: $(N[c(i),c(i+1)],N[c(i+1),c(i)]) \succsim_D (z_1,z_2)$.

Especially, we get

(4.14.1.5) $(N[c(n-1),c(n)],N[c(n),c(n-1)]) \succsim_D (z_1,z_2)$.

Furthermore, we get

(4.14.1.6) $\forall\ i,j \in \{1,...,n\}$ with $j - i \geq 2$: $(N[c(i),c(j)],N[c(j),c(i)]) \prec_D (z_1,z_2)$.

Otherwise, if there was a link $c(i),c(j)$ with $(N[c(i),c(j)],N[c(j),c(i)]) \succsim_D (z_1,z_2)$ and $j - i \geq 2$, then we could find a shorter path of strength $(z_1,z_2)$ by omitting the alternatives $c(i+1),...,c(j-1)$. This would be a contradiction to the presumption that $c(1),...,c(n)$ is the shortest path of strength $(z_1,z_2)$.

With (2.1.5), we get that every path that contains no defeat is always stronger than every path that contains a defeat.

It is easy to prove that the path $c(1),...,c(n)$ contains no defeat. Therefore, we get

(4.14.1.7) $\forall\ i = 1,...,(n-1)$: $N[c(i),c(i+1)] \geq N[c(i+1),c(i)]$.

> Proof for (4.14.1.7):
>
> To prove that the path $c(1),...,c(n)$ contains no defeat, we presume that (2.1.5) and (4.14.1.2) are satisfied and that the path $c(1),...,c(n)$ contains a defeat and then we will show that this leads to a contradiction.
>
> To get to a contradiction, it is sufficient to consider the link *ab*.
>
> Case #A: If the link *ab* is a victory ( i.e. $N[a,b] > N[b,a]$ ), then this link is already a path from alternative *a* to alternative *b* that contains no defeat. Therefore, with (2.1.5) and the fact that the path $c(1),...,c(n)$ contains a defeat, we get $P_D[a,b] \succsim_D (N[a,b],N[b,a]) \succ_D P_D[b,a]$. But this is a contradiction to (4.14.1.2).
>
> Case #B: If the link *ab* is a defeat ( i.e. $N[a,b] < N[b,a]$ ) or a tie ( i.e. $N[a,b] = N[b,a]$ ), then the link *ba* is a path from alternative *b* to alternative *a* that contains no defeat. But then, according to (2.1.5), the link *ba* is stronger than the path $c(1),...,c(n)$ that contains a defeat. But this is a contradiction to the presumption that the path $c(1),...,c(n)$ is a strongest path from alternative *b* to alternative *a*.

With (4.14.1.7), we get

(4.14.1.8) $N[c(n-1),c(n)] \geq N[c(n),c(n-1)]$.

With (2.1.5) and (4.14.1.8), we get

(4.14.1.9) $(N[c(n-1),c(n)],N[c(n),c(n-1)])$
$\succsim_D (N[c(n),c(n-1)],N[c(n-1),c(n)])$.

With the above considerations, we can now show how the subset $\tilde{A} \subset A$ can be chosen.

Case #1: $k = 2$.

The first version of 2-consistency is identical to the Condorcet criterion, as formulated in (4.8.7)(i). However, it has already been proven in section 4.8 that, when $\succ_D$ satisfies (2.1.5), then the Schulze method satisfies (4.8.7)(i).

Case #2: $3 \leq k < n$.

Here, we choose $\tilde{A} := \{c(1),...,c(k-2),c(n-1),c(n)\}$.

When the Schulze method is applied to $\tilde{A}$, then there is a path from $c(n-1)$ to $a \equiv c(n)$ of at least $(N[c(n-1),c(n)],N[c(n),c(n-1)]) \succsim_D (z_1,z_2)$ because, according to (4.14.1.5), already the link $c(n-1),c(n)$ is a path from $c(n-1)$ to $c(n)$ of this strength.

On the other side, there cannot be a path in $\tilde{A}$ from $a \equiv c(n)$ to $c(n-1)$ of more than $(N[c(n-1),c(n)],N[c(n),c(n-1)])$ because, according to (4.14.1.6), every link from $c(1)$, ..., $c(k-2)$ to $c(n-1)$ is weaker than $(z_1,z_2)$ and, according to (4.14.1.9), the link $c(n),c(n-1)$ is not stronger than $(N[c(n-1),c(n)],N[c(n),c(n-1)])$.

Therefore, alternative $a \equiv c(n)$ cannot disqualify alternative $c(n-1)$. So either alternative $c(n-1)$ is also a potential winner or, according to (4.1.14), alternative $c(n-1)$ must be disqualified by some other potential winner. In both cases, alternative $a \equiv c(n)$ is not a unique winner.

Case #3: $k \geq n$.

Here, $\tilde{A}$ consists of the alternatives $c(1),...,c(n)$ and $k-n$ additional alternatives from $A$.

As $\{c(1),...,c(n)\} \subseteq \tilde{A}$, there is a path in $\tilde{A}$ from alternative $b \equiv c(1)$ to alternative $a \equiv c(n)$ of strength $(z_1,z_2)$. On the other side, we get, with (4.14.1.2), that there cannot be a path in $\tilde{A}$ from alternative $a \equiv c(n)$ to alternative $b \equiv c(1)$ of more than $(z_1,z_2)$ because, when alternatives are removed from $A$, then the strength of the strongest path from alternative $a \equiv c(n)$ to alternative $b \equiv c(1)$ can only decrease.

Therefore, alternative $a \equiv c(n)$ cannot disqualify alternative $b \equiv c(1)$ in $\tilde{A}$. So either alternative $b \equiv c(1)$ is also a potential winner or, according to (4.1.14), alternative $b \equiv c(1)$ must be disqualified by some other potential winner. In both cases, alternative $a \equiv c(n)$ is not a unique winner. □

### 4.14.2. Formulation #2

**Definition:**

Suppose $k \in \mathbb{N}$ with $k \geq 2$. An election method satisfies the second version of *k-consistency* if the following holds:

Suppose $C$ with $C > k$ is the number of alternatives in $A$. Suppose alternative $a \in A$ is a potential winner whenever this election method is applied to some subset $\tilde{A} \subset A$ with $\| \tilde{A} \| = k$ and $a \in \tilde{A}$. Then alternative $a$ is also a potential winner when this election method is applied to $A$.

In short:

(4.14.2.1) $\forall\, a \in A$: ( ( $\forall\, \tilde{A} \subset A$ with $\| \tilde{A} \| = k$ and $a \in \tilde{A}$: $a \in \mathcal{S}|_{\tilde{A}}$ ) $\Rightarrow$ ( $a \in \mathcal{S}$ ) ).

**Claim:**

If $\succ_D$ satisfies (2.1.4) and (2.1.5), then the Schulze method, as defined in section 2.2, satisfies the second version of $k$-consistency for every $k \in \mathbb{N}$ with $k \geq 2$.

**Proof (overview):**

We will show how, when alternative $a \in A$ is not a potential winner (when this election method is applied to $A$), we can create, for every $k \in \mathbb{N}$ with $2 \leq k < C$, a subset $\tilde{A} \subset A$ with $\| \tilde{A} \| = k$ and $a \in \tilde{A}$ such that, when the Schulze method is applied to $\tilde{A}$, alternative $a$ is not a potential winner.

**Proof (details):**

Suppose alternative $a \in A$ is not a potential winner when the Schulze method is applied to $A$. Then, with (2.2.2), we get that there must be an alternative $b \in A \setminus \{a\}$ with

(4.14.2.2) $P_D[b,a] \succ_D P_D[a,b]$.

We set

(4.14.2.3) $(z_1,z_2) := P_D[b,a]$

to stress that this value is constant for the rest of this proof.

Suppose $c(1),\dots,c(n)$ is the shortest path (in terms of its number of links) from alternative $b \equiv c(1)$ to alternative $a \equiv c(n)$ of strength $(z_1,z_2)$. Then we get

(4.14.2.4) $\forall\, i = 1,\dots,(n-1)$: $(N[c(i),c(i+1)],N[c(i+1),c(i)]) \succsim_D (z_1,z_2)$.

Especially, we get

(4.14.2.5) $(N[c(n-1),c(n)],N[c(n),c(n-1)]) \succsim_D (z_1,z_2)$.

Furthermore, we get

(4.14.2.6) $\forall$ $i,j \in \{1,...,n\}$ with $j - i \geq 2$: $(N[c(i),c(j)],N[c(j),c(i)]) \prec_D (z_1,z_2)$.

Otherwise, if there was a link $c(i),c(j)$ with $(N[c(i),c(j)],N[c(j),c(i)]) \succsim_D (z_1,z_2)$ and $j - i \geq 2$, then we could find a shorter path of strength $(z_1,z_2)$ by omitting the alternatives $c(i+1),...,c(j-1)$. This would be a contradiction to the presumption that $c(1),...,c(n)$ is the shortest path of strength $(z_1,z_2)$.

With (2.1.4) and (2.1.5), we get that every path that contains no defeat or tie is always stronger than every path that contains a defeat or tie.

It is easy to prove that the path $c(1),...,c(n)$ contains no defeat or tie. Therefore, we get

(4.14.2.7) $\forall$ $i = 1,...,(n-1)$: $N[c(i),c(i+1)] > N[c(i+1),c(i)]$.

> Proof for (4.14.2.7):
>
> To prove that the path $c(1),...,c(n)$ contains no defeat or tie, we presume that (2.1.4), (2.1.5), and (4.14.2.2) are satisfied and that the path $c(1),...,c(n)$ contains a defeat or tie and then we will show that this leads to a contradiction.
>
> (2.1.4) says that all ties have equivalent strengths. (2.1.5) says that every victory is stronger than every tie and that every tie is stronger than every defeat. So when the path $c(1),...,c(n)$ contains a defeat or tie then, without loss of generality, we can set
>
> (4.14.2.8) $P_D[b,a] \precsim_D (1,1)$.
>
> To get to a contradiction, it is sufficient to consider the link *ab*.
>
> Case #A: If the link *ab* is a victory ( i.e. $N[a,b] > N[b,a]$ ) or a tie ( i.e. $N[a,b] = N[b,a]$ ), then this link is already a path from alternative *a* to alternative *b* that contains no defeat. Therefore, with (2.1.4), (2.1.5), and (4.14.2.8), we get $P_D[a,b] \succsim_D (N[a,b],N[b,a]) \succsim_D (1,1) \succsim_D P_D[b,a]$. But this is a contradiction to (4.14.2.2).
>
> Case #B: If the link *ab* is a defeat ( i.e. $N[a,b] < N[b,a]$ ), then the link *ba* is a path from alternative *b* to alternative *a* that contains no defeat or tie. But then, according to (2.1.5), the link *ba* is stronger than the path $c(1),...,c(n)$ that contains a defeat or tie. But this is a contradiction to the presumption that the path $c(1),...,c(n)$ is a strongest path from alternative *b* to alternative *a*.

With (4.14.2.7), we get

(4.14.2.9) $N[c(n-1),c(n)] > N[c(n),c(n-1)]$.

With (2.1.5) and (4.14.2.9), we get

(4.14.2.10) $(N[c(n-1),c(n)],N[c(n),c(n-1)])$
$\succ_D (N[c(n),c(n-1)],N[c(n-1),c(n)])$.

With the above considerations, we can now show how the subset $\tilde{A} \subset A$ can be chosen.

Case #1: $k = 2$.

The second version of 2-consistency says that a weak Condorcet winner should always be a potential winner, as formulated in (4.12.1.6). However, it has already been proven in section 4.12.1 that, when $\succ_D$ satisfies (2.1.4) and (2.1.5), then the Schulze method satisfies (4.12.1.6).

Case #2: $3 \leq k < n$.

Here, we choose $\tilde{A} := \{c(1),...,c(k-2),c(n-1),c(n)\}$.

When the Schulze method is applied to $\tilde{A}$, then there is a path from $c(n-1)$ to $a \equiv c(n)$ of at least $(N[c(n-1),c(n)],N[c(n),c(n-1)]) \gtrsim_D (z_1,z_2)$ because, according to (4.14.2.5), already the link $c(n-1),c(n)$ is a path from $c(n-1)$ to $c(n)$ of this strength.

On the other side, there cannot be a path in $\tilde{A}$ from $a \equiv c(n)$ to $c(n-1)$ of at least $(N[c(n-1),c(n)],N[c(n),c(n-1)])$ because, according to (4.14.2.6), every link from $c(1)$, ..., $c(k-2)$ to $c(n-1)$ is weaker than $(z_1,z_2)$ and, according to (4.14.2.10), the link $c(n),c(n-1)$ is weaker than $(N[c(n-1),c(n)],N[c(n),c(n-1)])$.

Therefore, alternative $c(n-1)$ disqualifies alternative $a \equiv c(n)$, so that alternative $a \equiv c(n)$ is not a potential winner.

Case #3: $k \geq n$.

Here, $\tilde{A}$ consists of the alternatives $c(1),...,c(n)$ and $k-n$ additional alternatives from $A$.

As $\{c(1),...,c(n)\} \subseteq \tilde{A}$, there is a path in $\tilde{A}$ from alternative $b \equiv c(1)$ to alternative $a \equiv c(n)$ of strength $(z_1,z_2)$. On the other side, we get, with (4.14.2.2), that there cannot be a path in $\tilde{A}$ from alternative $a \equiv c(n)$ to alternative $b \equiv c(1)$ of at least $(z_1,z_2)$ because, when alternatives are removed from $A$, then the strength of the strongest path from alternative $a \equiv c(n)$ to alternative $b \equiv c(1)$ can only decrease.

Therefore, alternative $b \equiv c(1)$ disqualifies alternative $a \equiv c(n)$ in $\tilde{A}$, so that alternative $a \equiv c(n)$ is not a potential winner. □

### 4.14.3. Formulation #3

**Definition:**

Suppose $k \in \mathbb{N}$ with $k \geq 2$. An election method satisfies the third version of *k-consistency* if the following holds:

Suppose $C$ with $C > k$ is the number of alternatives in $A$. Suppose $\mathcal{S}|_{\tilde{A}}$ is the set of potential winners when this election method is applied to $\emptyset \neq \tilde{A} \subseteq A$. Suppose $\emptyset \neq B \subsetneq A$. Suppose $\mathcal{S}|_{\tilde{A}} \subseteq B$ whenever this election method is applied to some subset $\tilde{A} \subset A$ with $\| \tilde{A} \| = k$ and $( B \cap \tilde{A} ) \neq \emptyset$. Then we must also get $\mathcal{S} \subseteq B$.

In short:

$$\text{(4.14.3.1)} \qquad \forall \emptyset \neq B \subsetneq A \text{: } ( ( \forall \tilde{A} \subset A \text{ with } \| \tilde{A} \| = k \text{ and } ( B \cap \tilde{A} ) \neq \emptyset \text{: } \mathcal{S}|_{\tilde{A}} \subseteq B ) \Rightarrow ( \mathcal{S} \subseteq B ) ).$$

**Claim:**

If $>_D$ satisfies (2.1.5), then the Schulze method, as defined in section 2.2, satisfies the third version of $k$-consistency for every $k \in \mathbb{N}$ with $k \geq 2$.

**Proof (overview):**

We will show how, when $\mathcal{S} \not\subseteq B$, we can create, for every $k \in \mathbb{N}$ with $2 \leq k < C$, a subset $\tilde{A} \subset A$ with $\| \tilde{A} \| = k$ and $( B \cap \tilde{A} ) \neq \emptyset$ such that, when the Schulze method is applied to $\tilde{A}$, we get $\mathcal{S}|_{\tilde{A}} \not\subseteq B$.

**Proof (details):**

Suppose $r := \| B \|$ is the number of alternatives in $B$. With $\emptyset \neq B \subsetneq A$, we get: $0 < r < C$.

Case #1: $k = 2$.

The third version of 2-consistency is identical to the Smith criterion, as formulated in (4.8.5). However, it has already been proven in section 4.8 that, when $>_D$ satisfies (2.1.5), then the Schulze method satisfies the Smith criterion.

Case #2: $k > C - r$.

As $k > C - r$, every set $\tilde{A} \subset A$ with $\| \tilde{A} \| = k$ contains at least $k + r - C \geq 1$ alternatives of $B$. Therefore, we get $( B \cap \tilde{A} ) \neq \emptyset$ for every set $\tilde{A} \subset A$ with $\| \tilde{A} \| = k$.

Suppose $\mathcal{S} \not\subseteq B$. Then there is an alternative $a \in A$ with $a \in \mathcal{S}$ and $a \notin B$. In section 4.13, we have proven that, when alternative $a \in A$ is a potential winner, then there is a linear order $d(1),\ldots,d(C-1)$ of the alternatives in $A \setminus \{a\}$ such that, when the Schulze method is applied to $A \setminus \{d(1),\ldots,d(C-k)\}$, then alternative $a$ is still a potential winner. Therefore, when we choose $\tilde{A} := A \setminus \{d(1),\ldots,d(C-k)\}$, we get $a \in \mathcal{S}|_{\tilde{A}}$. With $a \notin B$, we get $\mathcal{S}|_{\tilde{A}} \not\subseteq B$.

Case #3: $3 \leq k \leq C - r$.

Suppose $\mathcal{S} \not\subseteq B$. Then there must be an alternative $b \in A$ with $b \in \mathcal{S}$ and $b \notin B$. With $b \in \mathcal{S}$ we get

(4.14.3.2) $\forall\ a \in A \setminus \{b\}: P_D[b,a] \succsim_D P_D[a,b]$.

Without loss of generality, we can assume that the Floyd-Warshall algorithm is used to calculate the strongest paths (as defined in section 2.3.1). We then sort the alternatives $\{a(1),...,a(C\text{–}1)\}$ in $A \setminus \{b\}$ such that

(4.14.3.3) $pred[b,a(j)] = a(i) \Rightarrow i < j$.

Suppose $y \in \mathbb{N}$ with $1 \leq y < C$ is the smallest number with $a(y) \in B$. Then we get $a(x) \notin B$ for all $x \in \mathbb{N}$ with $1 \leq x < y$. Furthermore, when $g(1),...,g(m)$ is the strongest path from alternative $b \equiv g(1)$ to alternative $a(y) \equiv g(m)$ then, with the definition for $pred[i,j]$ and with the definition for the order of $\{a(1),...,a(C\text{–}1)\}$, we get $\{g(1),...,g(m\text{–}1)\} \subseteq \{b,a(1),...,a(y\text{–}1)\} \subseteq A \setminus B$.

We set

(4.14.3.4) $(z_1,z_2) := P_D[b,a(y)]$

to stress that this value is constant for the rest of this proof.

We now shorten the path $g(1),...,g(m)$ by removing possible short cuts. So when there is a link $g(i),g(j)$ with $(N[g(i),g(j)],N[g(j),g(i)]) \succsim_D (z_1,z_2)$ and $j - i \geq 2$, we remove the alternatives $g(i+1),...,g(j\text{–}1)$ from this path. We continue removing possible short cuts, until the resulting path contains no short cuts anymore. The resulting path will be called $c(1),...,c(n)$.

We get

(4.14.3.5) $c(i) \notin B$ for all $i \in \mathbb{N}$ with $1 \leq i < n$,

because we have already established $g(i) \notin B$ for all $i \in \mathbb{N}$ with $1 \leq i < m$ and because, when we shortened the path $g(1),...,g(m)$, we only removed and didn’t add alternatives.

With the same arguments as for (4.14.1.4) – (4.14.1.9), we get (4.14.3.6) – (4.14.3.11):

(4.14.3.6) $\forall\ i = 1,...,(n\text{–}1): (N[c(i),c(i+1)],N[c(i+1),c(i)]) \succsim_D (z_1,z_2)$.

(4.14.3.7) $(N[c(n\text{–}1),c(n)],N[c(n),c(n\text{–}1)]) \succsim_D (z_1,z_2)$.

(4.14.3.8) $\forall\ i,j \in \{1,...,n\}$ with $j - i \geq 2$: $(N[c(i),c(j)],N[c(j),c(i)]) \prec_D (z_1,z_2)$.

(4.14.3.9) $\forall\ i = 1,...,(n\text{–}1): N[c(i),c(i+1)] \geq N[c(i+1),c(i)]$.

(4.14.3.10) $N[c(n\text{–}1),c(n)] \geq N[c(n),c(n\text{–}1)]$.

(4.14.3.11) $(N[c(n\text{–}1),c(n)],N[c(n),c(n\text{–}1)]) \succsim_D (N[c(n),c(n\text{–}1)],N[c(n\text{–}1),c(n)])$.

With the above considerations, we can now show how the subset $\tilde{A} \subset A$ can be chosen.

Case #3a: $3 \leq k \leq C - r$ with $k < n$.

Here, we choose $\tilde{A} := \{c(1),...,c(k-2),c(n-1),c(n)\}$.

With $c(n) \equiv a(y) \in B$, we get $( B \cap \tilde{A} ) \neq \emptyset$.

When the Schulze method is applied to $\tilde{A}$, then there is a path from $c(n-1)$ to $c(n)$ of at least $(N[c(n-1),c(n)],N[c(n),c(n-1)]) \approx_D (z_1,z_2)$ because, according to (4.14.3.7), already the link $c(n-1),c(n)$ is a path from $c(n-1)$ to $c(n)$ of this strength.

On the other side, there cannot be a path in $\tilde{A}$ from $c(n)$ to $c(n-1)$ of more than $(N[c(n-1),c(n)],N[c(n),c(n-1)])$ because, according to (4.14.3.8), every link from $c(1)$, ..., $c(k-2)$ to $c(n-1)$ is weaker than $(z_1,z_2)$ and, according to (4.14.3.11), the link $c(n),c(n-1)$ is not stronger than $(N[c(n-1),c(n)],N[c(n),c(n-1)])$.

Therefore, alternative $c(n)$ cannot disqualify alternative $c(n-1)$. So either alternative $c(n-1)$ is also a potential winner or, according to (4.1.14), alternative $c(n-1)$ must be disqualified by some other potential winner in $\tilde{A}$. With (4.14.3.5) we get that this potential winner comes from outside $B$. So we get $\mathcal{S}|_{\tilde{A}} \not\subseteq B$.

Case #3b: $3 \leq k \leq C - r$ with $n \leq k$.

Here, $\tilde{A}$ consists of the alternatives $c(1),...,c(n)$ and $k-n$ additional alternatives from $A \setminus B$.

With $c(n) \equiv a(y) \in B$, we get $( B \cap \tilde{A} ) \neq \emptyset$.

As $\{c(1),...,c(n)\} \subseteq \tilde{A}$, there is a path in $\tilde{A}$ from alternative $b \equiv c(1)$ to alternative $a(y) \equiv c(n)$ of strength $(z_1,z_2)$. On the other side, we get, with (4.14.3.2), that there cannot be a path in $\tilde{A}$ from alternative $a(y) \equiv c(n)$ to alternative $b \equiv c(1)$ of more than $(z_1,z_2)$ because, when alternatives are removed from $A$, then the strength of the strongest path from alternative $a(y) \equiv c(n)$ to alternative $b \equiv c(1)$ can only decrease.

Therefore, alternative $a(y) \equiv c(n)$ cannot disqualify alternative $b \equiv c(1)$. So either alternative $b \equiv c(1)$ is also a potential winner or, according to (4.1.14), alternative $b \equiv c(1)$ must be disqualified by some other potential winner in $\tilde{A}$. With (4.14.3.5) we get that this potential winner comes from outside $B$. So we get $\mathcal{S}|_{\tilde{A}} \not\subseteq B$. □

## 4.14.4. Formulation #4

**Definition:**

Suppose $k \in \mathbb{N}$ with $k \geq 2$. An election method satisfies the fourth version of *k-consistency* if the following holds:

Suppose $C$ with $C > k$ is the number of alternatives in $A$. Suppose alternative $a \in A$ is not a unique winner whenever this election method is applied to some subset $\tilde{A} \subset A$ with $\| \tilde{A} \| = k$ and $a \in \tilde{A}$. Then alternative $a$ is also not a unique winner when this election method is applied to $A$.

In short:

(4.14.4.1) $\forall\, a \in A$: ( ( $\forall\, \tilde{A} \subset A$ with $\| \tilde{A} \| = k$ and $a \in \tilde{A}$: $\mathcal{S}|_{\tilde{A}} \neq \{a\}$ ) $\Rightarrow$ ( $\mathcal{S} \neq \{a\}$ ) ).

**Claim:**

The Schulze method, as defined in section 2.2, satisfies the fourth version of $k$-consistency for every $k \in \mathbb{N}$ with $k \geq 2$.

**Remark:**

Presumptions (2.1.4) and (2.1.5) are not needed in the following proof. However, only when $>_D$ satisfies (2.1.4) and (2.1.5), the fourth version of $k$-consistency with $k = 2$ is identical to the desideratum that a weak Condorcet loser should not be a unique winner; equation (4.12.2.13).

**Proof (overview):**

We will show how, when alternative $a \in A$ is a unique winner (when this election method is applied to $A$), we can create, for every $k \in \mathbb{N}$ with $2 \leq k < C$, a subset $\tilde{A} \subset A$ with $\| \tilde{A} \| = k$ and $a \in \tilde{A}$ such that, when the Schulze method is applied to $\tilde{A}$, alternative $a$ is a unique winner.

**Proof (details):**

In section 4.13 (definition #1), we have proven that, when alternative $a \in A$ is a unique winner, then there is a linear order $d(1),\dots,d(C-1)$ of the alternatives in $A \setminus \{a\}$ such that, for every $i \in \{1,\dots,(C-1)\}$, alternative $a$ is still a unique winner when the Schulze method is applied to $A \setminus \{d(1),\dots,d(i)\}$.

Therefore, for $k \in \mathbb{N}$ with $2 \leq k < C$, we can simply choose $\tilde{A} := A \setminus \{d(1),\dots,d(C-k)\}$. □

### 4.14.5. Formulation #5

**Definition:**

Suppose $k \in \mathbb{N}$ with $k \geq 2$. An election method satisfies the fifth version of *k-consistency* if the following holds:

Suppose $C$ with $C > k$ is the number of alternatives in $A$. Suppose alternative $a \in A$ is not a potential winner whenever this election method is applied to some subset $\tilde{A} \subset A$ with $\| \tilde{A} \| = k$ and $a \in \tilde{A}$. Then alternative $a$ is also not a potential winner when this election method is applied to $A$.

In short:

(4.14.5.1) $$\forall\, a \in A: ( ( \forall\, \tilde{A} \subset A \text{ with } \| \tilde{A} \| = k \text{ and } a \in \tilde{A}: a \notin \mathcal{S}|_{\tilde{A}} ) \Rightarrow ( a \notin \mathcal{S} ) ).$$

**Claim:**

The Schulze method, as defined in section 2.2, satisfies the fifth version of $k$-consistency for every $k \in \mathbb{N}$ with $k \geq 2$.

**Remark:**

Presumption (2.1.5) is not needed in the following proof. However, only when $>_D$ satisfies (2.1.5), the fifth version of $k$-consistency with $k = 2$ is identical to the Condorcet loser criterion; equation (4.8.9)(i).

**Proof (overview):**

We will show how, when alternative $a \in A$ is a potential winner (when this election method is applied to $A$), we can create, for every $k \in \mathbb{N}$ with $2 \leq k < C$, a subset $\tilde{A} \subset A$ with $\| \tilde{A} \| = k$ and $a \in \tilde{A}$ such that, when the Schulze method is applied to $\tilde{A}$, alternative $a$ is a potential winner.

**Proof (details):**

In section 4.13 (definition #2), we have proven that, when alternative $a \in A$ is a potential winner, then there is a linear order $d(1),\ldots,d(C-1)$ of the alternatives in $A \setminus \{a\}$ such that, for every $i \in \{1,\ldots,(C-1)\}$, alternative $a$ is still a potential winner when the Schulze method is applied to $A \setminus \{d(1),\ldots,d(i)\}$.

Therefore, for $k \in \mathbb{N}$ with $2 \leq k < C$, we can simply choose $\tilde{A} := A \setminus \{d(1),\ldots,d(C-k)\}$. □

## 4.15. Decreasing Sequential Independence

*Decreasing sequential independence* says that, when alternative $a \in A$ is not a winner, then there must be an alternative $e \in A \setminus \{a\}$ such that, when the used election method is applied to $A \setminus \{e\}$, then alternative $a$ is still not a winner.

The name for this criterion comes from the fact that — when the used election method satisfies this criterion and when alternative $a \in A$ is not a winner and alternative $e(1) \in A \setminus \{a\}$ is an alternative such that, when the used election method is applied to $A \setminus \{e(1)\}$, then alternative $a$ is still not a winner — the same criterion can then be applied to $A \setminus \{e(1)\}$ to identify an alternative $e(2) \in A \setminus \{a,e(1)\}$ such that, when the used election method is applied to $A \setminus \{e(1),e(2)\}$, then alternative $a$ is still not a winner. When we continue applying this criterion, we get a linear order $e(1),...,e(C\text{–}2)$ of the alternatives in $A \setminus \{a,e(C\text{–}1)\}$ ( where $e(C\text{–}1) \in A \setminus \{a,e(1),...,e(C\text{–}2)\}$ is the last remaining alternative to disqualify alternative $a$ ) such that, for every $i \in \{1,...,(C\text{–}2)\}$, alternative $a$ is still not a winner when the used election method is applied to $A \setminus \{e(1),...,e(i)\}$.

Increasing sequential independence (section 4.13) and decreasing sequential independence address opposite problems. On the one side, *increasing sequential independence* says that it should not be possible that alternatives $\varnothing \neq \{d(1),...,d(i)\} \subsetneq A$ *harm* each other in such a manner that the final winner comes from outside of $\{d(1),...,d(i)\}$. On the other side, *decreasing sequential independence* says that, when no proper subset of $\{e(1),...,e(i)\}$ can disqualify alternative $a \in A \setminus \{e(1),...,e(i)\}$, then the alternatives $\{e(1),...,e(i)\}$ should not *help* each other in such a manner that $\{e(1),...,e(i)\}$ together disqualify this alternative.

The fact that the Schulze method satisfies decreasing sequential independence follows directly from the fact that the Schulze method satisfies the first and the second version of $k$-consistency for every $k \in \mathbb{N}$ with $2 \leq k < C$ (sections 4.14.1 and 4.14.2).

**Definition #1:**

An election method satisfies the first version of *decreasing sequential independence* if the following holds:

Suppose there are at least $C \geq 3$ alternatives. Suppose alternative $a \in A$ is not a unique winner when this election method is applied to $A$. Then there must be a (not necessarily unique) alternative $e \in A \setminus \{a\}$ such that, when this election method is applied to $A \setminus \{e\}$, then alternative $a$ is still not a unique winner.

In short:

(4.15.1) $\forall\, C \geq 3\ \forall\, a \in A$:
$$( ( \mathcal{S} \neq \{a\} ) \Rightarrow ( \exists\, e \in A \setminus \{a\} : \mathcal{S}|_{(A \setminus \{e\})} \neq \{a\} ) ).$$

**Claim #1:**

If $\succ_D$ satisfies (2.1.5), then the Schulze method, as defined in section 2.2, satisfies the first version of decreasing sequential independence.

**Proof of claim #1:**

Suppose that the Schulze method violates (4.15.1). Then there must be a $C \geq 3$ such that there is an $a \in A$ with $\mathcal{S} \neq \{a\}$ and

(4.15.2) $\quad \forall\ e \in A \setminus \{a\}: \mathcal{S}|_{(A \setminus \{e\})} = \{a\}.$

With (4.14.1.1), (4.15.2) and $k := C - 1 \geq 2$, we get: $\mathcal{S} = \{a\}$. But this is a contradiction to the presumption $\mathcal{S} \neq \{a\}$. □

**Definition #2:**

An election method satisfies the second version of *decreasing sequential independence* if the following holds:

Suppose there are at least $C \geq 3$ alternatives. Suppose alternative $a \in A$ is <u>not a potential winner</u> when this election method is applied to $A$. Then there must be a (not necessarily unique) alternative $e \in A \setminus \{a\}$ such that, when this election method is applied to $A \setminus \{e\}$, then alternative $a$ is still <u>not a potential winner</u>.

In short:

(4.15.3) $\quad \forall\ C \geq 3\ \forall\ a \in A:$
$$( ( a \notin \mathcal{S} ) \Rightarrow ( \exists\ e \in A \setminus \{a\}: a \notin \mathcal{S}|_{(A \setminus \{e\})} ) ).$$

**Claim #2:**

If $\succ_D$ satisfies (2.1.4) and (2.1.5), then the Schulze method, as defined in section 2.2, satisfies the second version of decreasing sequential independence.

**Proof of claim #2:**

Suppose that the Schulze method violates (4.15.3). Then there must be a $C \geq 3$ such that there is an $a \in A$ with $a \notin \mathcal{S}$ and

(4.15.4) $\quad \forall\ e \in A \setminus \{a\}: a \in \mathcal{S}|_{(A \setminus \{e\})}.$

With (4.14.2.1), (4.15.4) and $k := C - 1 \geq 2$, we get: $a \in \mathcal{S}$. But this is a contradiction to the presumption $a \notin \mathcal{S}$. □

## 4.16. Weak Independence from Pareto-Dominated Alternatives

Suppose an alternative $j$ is added such that:

(3.8.1) $\exists\, i \in A^{\text{old}}\ \forall\, v \in V{:}\ i \succsim_v^{\text{new}} j.$

(3.8.2) $\forall\, g,h \in A^{\text{old}}\ \forall\, v \in V{:}\ g \succ_v^{\text{old}} h \Leftrightarrow g \succ_v^{\text{new}} h.$

Then *independence from Pareto-dominated alternatives* (IPDA) says that we must get:

(3.8.3) $\forall\, g,h \in A^{\text{old}}{:}\ gh \in \mathcal{O}^{\text{old}} \Leftrightarrow gh \in \mathcal{O}^{\text{new}}.$

(3.8.4) $\forall\, g \in A^{\text{old}}{:}\ g \in \mathcal{S}^{\text{old}} \Leftrightarrow g \in \mathcal{S}^{\text{new}}.$

In example 8 (section 3.8) and example 9 (section 3.9), we have seen that the Schulze method violates IPDA. In example 8, the winner is changed from alternative $a \in A^{\text{old}}$ to alternative $b \in A^{\text{old}} \setminus \{a\}$ by adding an alternative $e$ with

(4.16.1) $\exists\, d \in A^{\text{old}} \setminus \{a,b\}\ \forall\, v \in V{:}\ d \succsim_v^{\text{new}} e.$

In example 9, the winner is changed from alternative $a \in A^{\text{old}}$ to alternative $b \in A^{\text{old}} \setminus \{a\}$ by adding an alternative $e$ with

(4.16.2) $\forall\, v \in V{:}\ a \succsim_v^{\text{new}} e.$

It has already been mentioned in section 4.9 that IPDA and (4.9.5) are incompatible. In example 8(old), we have $\mathfrak{B}_D^{\text{old}} = \{a, c, d\}$. In example 8(new), we have $\mathfrak{B}_D^{\text{new}} = \{b\}$. Therefore, $(\mathfrak{B}_D^{\text{old}} \cap \mathfrak{B}_D^{\text{new}}) = \emptyset$. So (4.9.5) says that the winner must change. In example 9(old), we have $\mathfrak{B}_D^{\text{old}} = \{a, c, d\}$. In example 9(new), we have $\mathfrak{B}_D^{\text{new}} = \{b\}$ so that, again, (4.9.5) says that the winner must change.

So we cannot exclude that the winner is changed from alternative $a \in A^{\text{old}}$ to alternative $b \in A^{\text{old}} \setminus \{a\}$ by adding an alternative $e$ with (4.16.1) or (4.16.2). But we will prove that the winner cannot be changed from alternative $a \in A^{\text{old}}$ to alternative $b \in A^{\text{old}} \setminus \{a\}$ by adding an alternative $e$ with

(4.16.3) $\forall\, v \in V{:}\ b \succsim_v^{\text{new}} e.$

**Definition:**

> An election method satisfies *weak independence from Pareto-dominated alternatives* (wIPDA) if the following holds:
>
> Suppose $b \notin \mathcal{S}^{\text{old}}$.
>
> Suppose an alternative $e$ is added with (3.8.2) and
>
> (4.16.4) $\forall\, v \in V{:}\ b \succsim_v^{\text{new}} e.$
>
> Then we get: $b \notin \mathcal{S}^{\text{new}}$.

**Claim:**

If $\succ_D$ satisfies (2.1.1), then the Schulze method, as defined in section 2.2, satisfies *weak independence from Pareto-dominated alternatives*.

**Proof:**

Suppose $b \notin \mathcal{S}^{old}$. Then, there was an alternative $a \in A^{old} \setminus \{b\}$ with $ab \in \mathcal{O}^{old}$. With $ab \in \mathcal{O}^{old}$, we get

(4.16.5) $P_D^{old}[a,b] \succ_D P_D^{old}[b,a]$.

Suppose an alternative $e$ is added with (3.8.2) and (4.16.4).

Suppose $c(1),...,c(n)$ was the strongest path from alternative $a$ to alternative $b$ in $A^{old}$. Then $c(1),...,c(n)$ is still a path from alternative $a$ to alternative $b$ in $A^{new}$ of the same strength. Therefore, we get

(4.16.6) $P_D^{new}[a,b] \succsim_D P_D^{old}[a,b]$.

Suppose $d(1),...,d(m)$ is the strongest path from alternative $b$ to alternative $a$ in $A^{new}$.

Case I: Suppose $d(1),...,d(m)$ does not contain alternative $e$. Then $d(1),...,d(m)$ was a path from alternative $b$ to alternative $a$ in $A^{old}$ with the same strength. Therefore, we get: $P_D^{old}[b,a] \succsim_D P_D^{new}[b,a]$.

Case II: Suppose $d(1),...,d(m)$ contains alternative $e$. Suppose $d(s)$ is the last occurrence of alternative $e$ in the path $d(1),...,d(m)$. With (2.1.1), (4.16.4), and $d(s) \equiv e$, we get: $(N[b,d(s+1)],N[d(s+1),b]) \succsim_D (N[d(s),d(s+1)],N[d(s+1),d(s)])$. So $b,d(s+1),...,d(m)$ was a path from alternative $b$ to alternative $a$ in $A^{old}$ of at least the same strength as $d(1),...,d(m)$. Therefore, we get: $P_D^{old}[b,a] \succsim_D P_D^{new}[b,a]$.

So, with Case I and Case II, we get

(4.16.7) $P_D^{old}[b,a] \succsim_D P_D^{new}[b,a]$.

With (4.16.6), (4.16.5), and (4.16.7), we get

(4.16.8) $P_D^{new}[a,b] \succsim_D P_D^{old}[a,b] \succ_D P_D^{old}[b,a] \succsim_D P_D^{new}[b,a]$.

With (4.16.8), we get $ab \in \mathcal{O}^{new}$ and, therefore, $b \notin \mathcal{S}^{new}$. □

## 4.17. Interpolatability

Suppose $\mathcal{LO}_A$ is the set of linear orders on $A$.

Suppose $\emptyset \neq \mathcal{R} \subseteq \mathcal{LO}_A$ is the set of potential winning rankings for a given election method and a given profile.

An election method is *interpolatable* if, whenever $O_1,O_2 \in \mathcal{LO}_A$ are potential winning rankings of this election method, then also every $O_3 \in \mathcal{LO}_A$ in between must be a potential winning ranking.

**Definition:**

An election method satisfies *interpolatability* if the following holds for every profile:

(4.17.1) $\forall\, O_1,O_2,O_3 \in \mathcal{LO}_A$:
$(\, O_1,O_2 \in \mathcal{R} \wedge (\, O_1 \cap O_2 \,) \subseteq O_3 \,) \Rightarrow (\, O_3 \in \mathcal{R} \,)$.

**Claim:**

The Schulze method, as defined in section 2.2, satisfies interpolatability.

**Proof:**

Suppose $O_0$ is the binary relation defined in (2.2.1). Suppose $O \in \mathcal{LO}_A$. Then, with (2.2.6), we get

(4.17.2) $O \in \mathcal{R} :\Leftrightarrow O_0 \subseteq O$.

In section 4.1, it has been proven that the binary relation $O_0$, as defined in (2.2.1), is a strict partial order. Therefore, it is guaranteed that $\mathcal{R}$ is not empty. Suppose $O_1,O_2 \in \mathcal{R}$. Then, with (4.17.2), we get

(4.17.3) $O_0 \subseteq O_1$.

(4.17.4) $O_0 \subseteq O_2$.

With (4.17.3) and (4.17.4), we get

(4.17.5) $O_0 \subseteq (\, O_1 \cap O_2 \,)$.

So for $O_3 \in \mathcal{LO}_A$ with $(\, O_1 \cap O_2 \,) \subseteq O_3$, we get

(4.17.6) $O_0 \subseteq (\, O_1 \cap O_2 \,) \subseteq O_3$.

With (4.17.6), we get

(4.17.7) $O_0 \subseteq O_3$.

With (4.17.2) and (4.17.7), we get that also $O_3$ is a potential winning ranking of the Schulze method. □

Interpolatability is important when we want an election method to respond in a continuous manner to a change of the ballots. Not all election methods are interpolatable. An example for an election method that is not interpolatable is the Kemeny-Young method. This method calculates for every $O \in \mathcal{LO}_A$ the Kemeny score $K(O)$:

(4.17.8) $K(O) := \sum ( N[a,b] - N[b,a] \mid ab \in O )$.

The potential winning rankings of the Kemeny-Young method are the linear orders $O \in \mathcal{LO}_A$ with maximum Kemeny score $K(O)$. The potential winners of the Kemeny-Young method are the top-ranked alternatives of the linear orders $O \in \mathcal{LO}_A$ with maximum Kemeny score $K(O)$.

Example 4.17.9:

| | |
|---|---|
| 3 voters | $a \succ_v c \succ_v b \succ_v d$ |
| 8 voters | $a \succ_v c \succ_v d \succ_v b$ |
| 3 voters | $b \succ_v a \succ_v d \succ_v c$ |
| 6 voters | $c \succ_v b \succ_v d \succ_v a$ |
| 6 voters | $d \succ_v b \succ_v a \succ_v c$ |
| 4 voters | $d \succ_v c \succ_v b \succ_v a$ |

The pairwise matrix $N$ looks as follows:

| | $N$[*,$a$] | $N$[*,$b$] | $N$[*,$c$] | $N$[*,$d$] |
|---|---|---|---|---|
| $N$[$a$,*] | --- | 11 | 20 | 14 |
| $N$[$b$,*] | 19 | --- | 9 | 12 |
| $N$[$c$,*] | 10 | 21 | --- | 17 |
| $N$[$d$,*] | 16 | 18 | 13 | --- |

The following table lists the linear orders $O \in \mathcal{LO}_A$ and their Kemeny scores $K(O)$:

| linear order $O$ | Kemeny score $K(O)$ |
|---|---|
| *abcd* | −14 |
| *abdc* | −22 |
| *acbd* | 10 |
| *acdb* | 22 |
| *adbc* | −10 |
| *adcb* | 14 |
| *bacd* | 2 |
| *badc* | −6 |
| *bcad* | −18 |
| *bcda* | −14 |
| *bdac* | −2 |
| *bdca* | −22 |
| *cabd* | −10 |
| *cadb* | 2 |
| *cbad* | 6 |
| *cbda* | 10 |
| *cdab* | 6 |
| *cdba* | 22 |
| *dabc* | −6 |
| *dacb* | 18 |
| *dbac* | 10 |
| *dbca* | −10 |
| *dcab* | −2 |
| *dcba* | 14 |

In example 4.17.9, the linear orders with maximum Kemeny scores are $O_1 = acdb = \{ab, ac, ad, cb, cd, db\}$ and $O_2 = cdba = \{ba, ca, cb, cd, da, db\}$ each with a Kemeny score of 22. We get $( O_1 \cap O_2 ) = \{cb, cd, db\}$. However, $O_3 = cadb = \{ab, ad, ca, cb, cd, db\}$ with a Kemeny score of 2 and $O_4 = cdab = \{ab, ca, cb, cd, da, db\}$ with a Kemeny score of 6 are not potential winning rankings although $( O_1 \cap O_2 ) \subseteq O_3$ and $( O_1 \cap O_2 ) \subseteq O_4$. So interpolatability is violated. When example 4.17.9 is modified slightly, the result goes abruptly from *acdb* to *cdba* without going through *cadb* and *cdab*.

Tideman's ranked pairs method works from the strongest to the weakest link. The link *xy* is locked if and only if it doesn't create a directed cycle with already locked links. Otherwise, this link is locked in its opposite direction (Tideman, 1987).

Example 4.17.10:

| | |
|---|---|
| 1 voter | $a \succ_v c \succ_v d \succ_v b$ |
| 5 voters | $a \succ_v d \succ_v b \succ_v c$ |
| 2 voters | $a \succ_v d \succ_v c \succ_v b$ |
| 4 voters | $b \succ_v a \succ_v c \succ_v d$ |
| 3 voters | $c \succ_v b \succ_v a \succ_v d$ |
| 5 voters | $c \succ_v d \succ_v b \succ_v a$ |
| 3 voters | $d \succ_v c \succ_v b \succ_v a$ |

The pairwise matrix *N* looks as follows:

| | *N*[*,*a*] | *N*[*,*b*] | *N*[*,*c*] | *N*[*,*d*] |
|---|---|---|---|---|
| *N*[*a*,*] | --- | 8 | 12 | 15 |
| *N*[*b*,*] | 15 | --- | 9 | 7 |
| *N*[*c*,*] | 11 | 14 | --- | 13 |
| *N*[*d*,*] | 8 | 16 | 10 | --- |

In example 4.17.10, two links have the same strength: *ad* and *ba*.

When the ranked pairs method chooses the link *ad* over the link *ba*, then it proceeds as follows: It locks *db* (with a strength of 16:7). Then, it locks *ad* (with a strength of 15:8). Then it locks *ab*, since locking *ba* in its original direction would create a directed cycle with the already locked links *ad* and *db*. Then it locks *cb* (with a strength of 14:9). Then it locks *cd* (with a strength of 13:10). Then it locks *ac* (with a strength of 12:11). So the winning ranking is $O_1$ = *acdb*.

When the ranked pairs method chooses the link *ba* over the link *ad*, then it proceeds as follows: It locks *db* (with a strength of 16:7). Then, it locks *ba* (with a strength of 15:8). Then it locks *da*, since locking *ad* in its original direction would create a directed cycle with the already locked links *db* and *ba*. Then it locks *cb* (with a strength of 14:9). Then it locks *cd* (with a strength of 13:10). Then it locks *ca*, since locking *ac* in its original direction would create a directed cycle with the already locked links *cb* and *ba*. So the winning ranking is $O_2$ = *cdba*.

So the potential winning rankings of the ranked pairs method in example 4.17.10 are $O_1 = acdb = \{ab, ac, ad, cb, cd, db\}$ and $O_2 = cdba = \{ba, ca, cb, cd, da, db\}$. We get $( O_1 \cap O_2 ) = \{cb, cd, db\}$. However, $O_3 = cadb = \{ab, ad, ca, cb, cd, db\}$ and $O_4 = cdab = \{ab, ca, cb, cd, da, db\}$ are not potential winning rankings although $( O_1 \cap O_2 ) \subseteq O_3$ and $( O_1 \cap O_2 ) \subseteq O_4$. So interpolatability is violated. When example 4.17.10 is modified slightly, the result goes abruptly from *acdb* to *cdba* without going through *cadb* and *cdab*.

## 4.18. Reversal Cancellation

*Reversal cancellation* says that adding a ballot and its reverse should not change the result of the elections. In other words, a ballot and its reverse should always cancel each other out.

**Definition:**

Suppose $w_1$ and $w_2$ are strict weak orders with

(4.18.1) $\forall\ a,b \in A\text{: } a \succ_{w_1} b \Leftrightarrow b \succ_{w_2} a.$

Suppose $V^{\text{new}} := V^{\text{old}} + \{w_1\} + \{w_2\}$.

Then, an election method satisfies *reversal cancellation* if the following holds:

(4.18.2) $\mathcal{O}^{\text{new}} = \mathcal{O}^{\text{old}}$.

(4.18.3) $\mathcal{S}^{\text{new}} = \mathcal{S}^{\text{old}}$.

**Claim:**

If $\succ_{margin}$ is being used, then the Schulze method, as defined in section 2.2, satisfies reversal cancellation.

**Proof:**

The proof is trivial. When $w_1$ and $w_2$, as defined in (4.18.1), are added, then $N^{\text{new}}[a,b] - N^{\text{new}}[b,a] = N^{\text{old}}[a,b] - N^{\text{old}}[b,a]$ for all $a,b \in A$. Therefore

(4.18.4) $\forall\ (e,f),(g,h) \in A \times A$:

$$( ( N^{\text{new}}[e,f] - N^{\text{new}}[f,e] > N^{\text{new}}[g,h] - N^{\text{new}}[h,g] )$$

$$\Leftrightarrow ( N^{\text{old}}[e,f] - N^{\text{old}}[f,e] > N^{\text{old}}[g,h] - N^{\text{old}}[h,g] ) ).$$

Therefore

(4.18.5) $\forall\ (e,f),(g,h) \in A \times A$:

$$(N^{\text{new}}[e,f],N^{\text{new}}[f,e]) \succ_{margin} (N^{\text{new}}[g,h],N^{\text{new}}[h,g])$$

$$\Leftrightarrow (N^{\text{old}}[e,f],N^{\text{old}}[f,e]) \succ_{margin} (N^{\text{old}}[g,h],N^{\text{old}}[h,g]).$$

With (2.2.1), (2.2.2), and (4.18.5), we get (4.18.2) and (4.18.3). □

## 4.19. Woodall’s Plurality Criterion

*Woodall’s plurality criterion* says: “If some candidate $b$ has strictly fewer votes in total than some other candidate $a$ has first-preference votes, then candidate $b$ should not be elected” (Woodall, 1994, 1996, 1997).

**<u>Definition:</u>**

Suppose

(4.19.1) $R_{first}[a] := \| \{ v \in V \mid \forall c \in A \setminus \{a\}: a >_v c \} \|$

is the number of voters who strictly prefer alternative $a \in A$ to every other alternative (“first-preference votes”).

Suppose

(4.19.2) $R_{total}[b] := \| \{ v \in V \mid \exists c \in A \setminus \{b\}: b >_v c \} \|$

is the number of voters who strictly prefer alternative $b \in A$ to at least one other alternative (“votes in total”).

Suppose

(4.19.3) $R_{first}[a] > R_{total}[b]$.

Then, an election method satisfies *Woodall’s plurality criterion* if the following holds:

(4.19.4) $ab \in \mathcal{O}$.

(4.19.5) $b \notin \mathcal{S}$.

**Claim:**

If $\succ_{win}$ or $\succ_{los}$ is being used, then the Schulze method satisfies Woodall’s plurality criterion.

**Proof:**

Suppose $a,b \in A$ with (4.19.1) – (4.19.3).

With (4.19.1), we get

(4.19.6) $\forall\ c \in A \setminus \{a\}: N[a,c] \geq R_{first}[a]$.

Especially, we get

(4.19.7) $N[a,b] \geq R_{first}[a]$.

With (4.19.2), we get

(4.19.8) $\forall\ c \in A \setminus \{b\}: N[b,c] \leq R_{total}[b]$.

Especially, we get

(4.19.9) $N[b,a] \leq R_{total}[b]$.

$\succ_{win}$ and $\succ_{los}$ each satisfy (2.1.1).

With (2.1.1), (4.19.7), and (4.19.9), we get

(4.19.10) $(N[a,b],N[b,a]) \succsim_D (R_{first}[a],R_{total}[b])$.

With (2.2.3) and (4.19.10), we get

(4.19.11) $P_D[a,b] \succsim_D (N[a,b],N[b,a]) \succsim_D (R_{first}[a],R_{total}[b])$.

Case I: Suppose $\succ_{win}$ is used.

With (4.19.3) and with the definition for $\succ_{win}$, we get

(4.19.12a) $(R_{first}[a],R_{total}[b]) \succ_{win} (R_{total}[b],0)$.

With (4.19.8) and with the definition for $\succ_{win}$, we get

(4.19.13a) $\forall\ c \in A \setminus \{b\}: (N[b,c],N[c,b]) \precsim_{win} (R_{total}[b],0)$.

Suppose, $bd$ is the first link in the strongest path from alternative $b$ to alternative $a$. Then, with (4.19.13a), we get

(4.19.14a) $P_{win}[b,a] \precsim_{win} (N[b,d],N[d,b]) \precsim_{win} (R_{total}[b],0)$.

With (4.19.11), (4.19.12a), and (4.19.14a), we get

(4.19.15a) $P_{win}[a,b] \succsim_{win} (R_{first}[a],R_{total}[b]) \succ_{win} (R_{total}[b],0) \succsim_{win} P_{win}[b,a]$.

With (4.19.15a), we get (4.19.4) and (4.19.5).

Case II: Suppose $\succ_{los}$ is used.

With (4.19.3) and with the definition for $\succ_{los}$, we get

(4.19.12b) $(R_{first}[a],R_{total}[b]) \succ_{los} (N,R_{first}[a])$.

With (4.19.6) and with the definition for $\succ_{los}$, we get

(4.19.13b) $\forall\ c \in A \setminus \{a\}: (N[c,a],N[a,c]) \precsim_{los} (N,R_{first}[a])$.

Suppose, $da$ is the last link in the strongest path from alternative $b$ to alternative $a$. Then, with (4.19.13b), we get

(4.19.14b) $P_{los}[b,a] \precsim_{los} (N[d,a],N[a,d]) \precsim_{los} (N,R_{first}[a])$.

With (4.19.11), (4.19.12b), and (4.19.14b), we get

(4.19.15b) $P_{los}[a,b] \succsim_{los} (R_{first}[a],R_{total}[b]) \succ_{los} (N,R_{first}[a]) \succsim_{los} P_{los}[b,a]$.

With (4.19.15b), we get (4.19.4) and (4.19.5). □

## 4.20. Perturbation

*Perturbation* says that alternative $a \in A$ is a potential winner if and only if it is sufficient to add a voter with infinitesimally small weight $\varepsilon \rightarrow 0$ to turn alternative $a$ into a unique winner. In other words: The potential winners are all those alternatives that are unique winners in a perturbed scenario of the original situation.

**Claim:**

If $\succ_{margin}$ is being used, then the Schulze method satisfies perturbation.

**Proof:**

Claim #1: If alternative $a \in A$ is a potential winner, then it is sufficient to add a voter with infinitesimally small weight $\varepsilon \rightarrow 0$ to turn alternative $a$ into a unique winner.

Proof: $\succ_{margin}$ satisfies (2.1.3). Therefore, the Schulze method with $\succ_{margin}$ satisfies homogeneity.

$\succ_{margin}$ satisfies (2.1.1). Therefore, the Schulze method with $\succ_{margin}$ satisfies resolvability, as defined in section 4.3.

The fact that Schulze method with $\succ_{margin}$ satisfies claim #1 follows directly from the fact that it satisfies homogeneity and resolvability.

Claim #2: If alternative $a \in A$ it not a potential winner, then it is not sufficient to add a voter with infinitesimally small weight $\varepsilon \rightarrow 0$ to turn alternative $a$ into a unique winner.

Proof: The proof makes use of the fact that, for all $a,b = 1,...,C$, $P_{margin}[a,b]$ is continuous in $N[i,j]$ for all $i,j = 1,...,C$. When a voter with weight $\varepsilon$ is added, then $N[i,j]$ increases by $\varepsilon$ at most, $N[i,j] - N[j,i]$ increases or decreases by $\varepsilon$ at most, and $P_{margin}[i,j]$ increases or decreases by $\varepsilon$ at most for all $i,j = 1,...,C$.

Suppose $a \in A$ it not a potential winner. Then there is an alternative $b \in A \setminus \{a\}$ with $P^{old}_{margin}[a,b] \prec_{margin} P^{old}_{margin}[b,a]$. $P^{old}_{margin}[a,b]$ and $P^{old}_{margin}[b,a]$ differ by at least 1 vote. When a voter with infinitesimally small weight $\varepsilon \rightarrow 0$ is added, then $P^{old}_{margin}[a,b]$ increases by $\varepsilon \rightarrow 0$ at best and $P^{old}_{margin}[b,a]$ decreases by $\varepsilon \rightarrow 0$ at best. Therefore, we still have: $P^{new}_{margin}[a,b] \prec_{margin} P^{new}_{margin}[b,a]$ for $\varepsilon < 0.5$. □

## 4.21. Munger’s Scenario

In Munger’s scenario, alternative $a$ is identified as best alternative and alternative $b$ is identified as second best alternative. But when alternative $a$ is removed, then alternative $b$ is identified as worst alternative. Munger (2023c) writes: “If the order of candidates is to have something to do with merit, then if the candidate who placed first had not run, then the candidate who placed second should have won. Kemeny-Young and Ranked Pairs have that property, but Beatpath fails it in the worst possible way; the candidate who placed second can become ranked *last* of the remaining candidates.” However, we will prove that Munger’s scenario is not possible under the Schulze method, so that Munger’s criticism doesn’t work. The full correspondence is here: Munger (2023a) → Munger (2023b) → Munger (2023c) → Schulze (2024b) → Munger (2024a) → Munger (2024b) → Munger (2024c) → Munger (2025).

As the Schulze method satisfies decisiveness (section 4.2), there is usually a unique winner. As the Schulze method satisfies decisiveness (section 4.2) and reversal symmetry (section 4.5), there is usually a unique worst alternative. However, the Schulze method usually doesn’t find a unique linear order of all alternatives (section 3.5). Therefore, the term “second best” is open to interpretation. I presume that, when alternative $a \in A$ is identified as best alternative, then the term “second best” refers to every alternative $b \in A \setminus \{a\}$ that is disqualified only by alternative $a$.

**Claim:**

Suppose there were at least 3 alternatives in the original situation. Suppose $\succ_D$ satisfies (2.1.5). Then Munger’s scenario is not possible under the Schulze method.

**Proof:**

We will presume that alternative $a \in A$ was a unique winner, that alternative $b \in A \setminus \{a\}$ was a second best alternative and that, when alternative $a$ is removed, then alternative $b$ becomes a unique worst alternative. We will show that these presumptions ultimately lead to contradictions (step 4).

Step 1:

Suppose alternative $a$ was the unique winner in the original situation. Then, with (4.1.15), we get:

(4.21.1) $$\forall\, x \in A \setminus \{a\}: P_D^{\text{old}}[a,x] \succ_D P_D^{\text{old}}[x,a].$$

Suppose alternative $b$ was a second best alternative in the original situation. Then we get:

(4.21.2) $$\forall\, x \in A \setminus \{a,b\}: P_D^{\text{old}}[b,x] \succsim_D P_D^{\text{old}}[x,b].$$

Suppose alternative $b$ is the unique worst alternative when alternative $a$ is removed. Then we get:

(4.21.3) $$\forall\, x \in A \setminus \{a,b\}: P_D^{\text{new}}[x,b] \succ_D P_D^{\text{new}}[b,x].$$

When alternative $a$ is removed, then the strength of the strongest path from alternative $x \in A \setminus \{a\}$ to alternative $y \in A \setminus \{a,x\}$ either decreases or stays the same. But $P_D[x,y]$ cannot increase, since the set of possible paths is only reduced. Therefore, we get:

(4.21.4) $\forall\, x \in A \setminus \{a,b\}$: $P^{\text{old}}_D[x,b] \gtrsim_D P^{\text{new}}_D[x,b]$.

With (4.21.2), (4.21.4) and (4.21.3), we get:

(4.21.5) $\forall\, x \in A \setminus \{a,b\}$: $P^{\text{old}}_D[b,x] \succ_D P^{\text{new}}_D[b,x]$.

But (4.21.5) is possible only when, for every $x \in A \setminus \{a,b\}$, alternative $a$ was in every strongest path from alternative $b$ to alternative $x$.

Step 2:

Without loss of generality, we can assume that the Floyd-Warshall algorithm is used to calculate the strongest paths (as defined in section 2.3.1). The strongest paths then have the property that every sub-sequence is also a strongest path. So when the sequence $c(1),\ldots,c(n)$ is a strongest path from alternative $c(1)$ to alternative $c(n)$ then, for every $1 \leq i < j \leq n$, the sub-sequence $c(i),\ldots,c(j)$ is a strongest path from alternative $c(i)$ to alternative $c(j)$.

Suppose there were alternatives $x_1,x_2 \in A \setminus \{a,b\}$ such that the link $bx_1$ was the first link in the strongest path from alternative $b$ to alternative $x_2$. Then already the link $bx_1$ was a strongest path from alternative $b$ to alternative $x_1$. But this is a contradiction to the result of step 1 that, for every $x_1 \in A \setminus \{a,b\}$, alternative $a$ was in every strongest path from alternative $b$ to alternative $x_1$. Therefore, without loss of generality, we can assume (1) that, for every $x \in A \setminus \{a,b\}$, the link $ba$ was the first link in every strongest path from alternative $b$ to alternative $x$ and (2) that the unique strongest path from alternative $b$ to alternative $a$ was the link $ba$.

Step3:

Suppose $ax_3$ with $x_3 \in A \setminus \{a,b\}$ was the first link in the strongest path from alternative $a$ to alternative $b$. Then, with the result of step 2, we get that the strongest paths must have had the following structure:

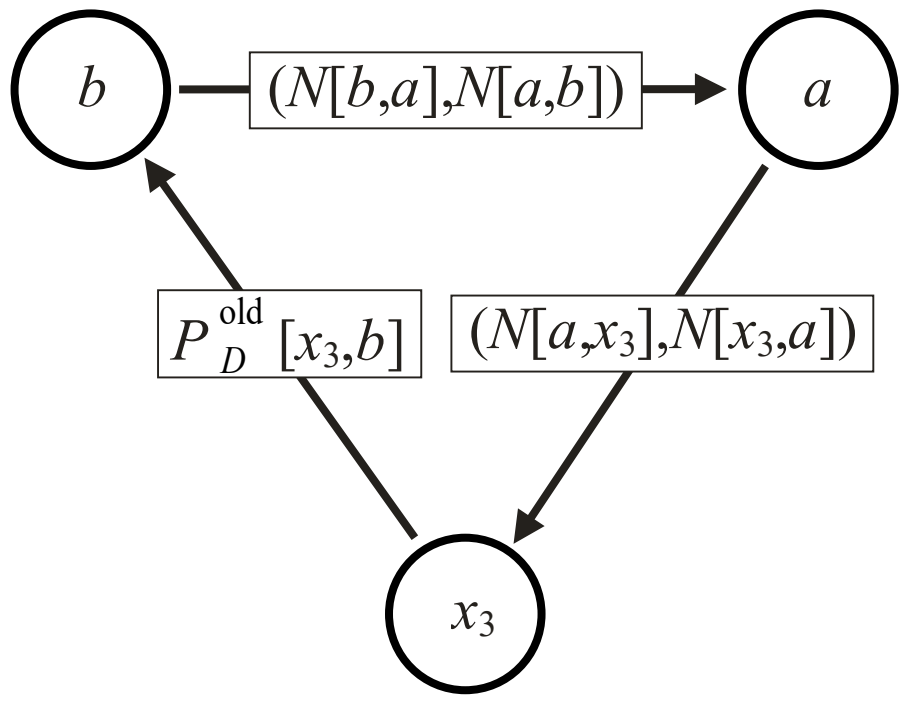

With (4.21.1), we get $P^{\text{old}}_D[a,b] \succ_D P^{\text{old}}_D[b,a]$ and therefore:

(4.21.6) $\min_D\{\ (N[a,x_3],N[x_3,a]),\ P^{\text{old}}_D[x_3,b]\ \} \succ_D (N[b,a],N[a,b])$

and therefore:

(4.21.7) $P^{\text{old}}_D[x_3,b] \succ_D (N[b,a],N[a,b])$.

With (4.21.2), we get $P^{\text{old}}_D[b,x_3] \succsim_D P^{\text{old}}_D[x_3,b]$ and therefore:

(4.21.8) $\min_D\{\ (N[b,a],N[a,b]),\ (N[a,x_3],N[x_3,a])\ \} \succsim_D P^{\text{old}}_D[x_3,b]$

and therefore:

(4.21.9) $(N[b,a],N[a,b]) \succsim_D P^{\text{old}}_D[x_3,b]$.

But (4.21.7) and (4.21.9) are incompatible. Therefore, it is not possible that the link $ax_3$ with some alternative $x_3 \in A \setminus \{a,b\}$ was the first link in the strongest path from alternative $a$ to alternative $b$. Therefore, the strongest path from alternative $a$ to alternative $b$ must have been the link $ab$.

Step 4:

So in step 2, we have seen that the strongest path from alternative $b$ to alternative $a$ was the link $ba$. And in step 3, we have seen that the strongest path from alternative $a$ to alternative $b$ was the link $ab$.

Therefore, we get:

(4.21.10) $P^{\text{old}}_D[b,a] \approx_D (N[b,a],N[a,b])$.

(4.21.11) $P^{\text{old}}_D[a,b] \approx_D (N[a,b],N[b,a])$.

Case I: Suppose

(4.21.12) $N[a,b] \leq N[b,a]$.

With (2.1.5) and (4.21.12), we get:

(4.21.13) $(N[a,b],N[b,a]) \precsim_D (N[b,a],N[a,b])$.

With (4.21.11), (4.21.13) and (4.21.10), we get:

(4.21.14) $P^{\text{old}}_D[a,b] \precsim_D P^{\text{old}}_D[b,a]$.

But (4.21.14) contradicts (4.21.1).

Case II: Suppose

(4.21.15) $N[a,b] > N[b,a]$.

Suppose $c \in A \setminus \{a,b\}$. Such an alternative exists, as we had presumed that there were at least 3 alternatives in the original situation.

Case IIa: Suppose

(4.21.16) $N[b,c] \geq N[c,b]$.

Then, with (2.1.5), (4.21.15) and (4.21.16), we get:

(4.21.17) $(N[b,c],N[c,b]) \succ_D (N[b,a],N[a,b])$.

So already the link *bc* was a stronger path from alternative *b* to alternative *c* than every path that starts with the link *ba*. But this contradicts the observation in step 2 that, for every $x \in A \setminus \{a,b\}$, the link *ba* was the first link in every strongest path from alternative *b* to alternative *x*.

Case IIb: Suppose

(4.21.18) $N[b,c] < N[c,b]$.

Then, with (2.1.5), (4.21.15) and (4.21.18), we get:

(4.21.19) $(N[c,b],N[b,c]) \succ_D (N[b,a],N[a,b])$.

With (4.21.19) and the observation in step 2 that, for every $x \in A \setminus \{a,b\}$, the link *ba* was the first link in every strongest path from alternative *b* to alternative *x*, we get that the link *cb* was already a stronger path from alternative *c* to alternative *b* than every path from alternative *b* to alternative *c*. But this contradicts (4.21.2). □

The fact, that Munger's scenario isn't possible under the Schulze method, also means that it isn't possible that the collective ranking of the remaining alternatives is reversed when the best alternative is removed (since reversing the collective ranking would imply that the second best alternative of the original situation becomes the worst in the new situation).

The proof that, when the worst alternative is removed, the second worst alternative of the original situation cannot become the best alternative in the new situation is equivalent to the proof that Munger's scenario isn't possible under the Schulze method.

So the fact that Munger's scenario isn't possible under the Schulze method basically means that removing the best or the worst alternative cannot reverse the collective ranking of the remaining alternatives.

## 5. Tie-Breaking

In this section, we will propose a more elaborate definition for the Schulze method. This more elaborate version addresses 3 problems of the simpler definition of section 2.2.

**Problem #1:** It can happen that the weakest link in the strongest path from alternative *a* to alternative *b* and the weakest link in the strongest path from alternative *b* to alternative *a* are the same link, say *cd*. In this case, the Schulze method is indifferent between alternative *a* and alternative *b*, i.e. $ab \notin O$ and $ba \notin O$. See sections 3.5, 3.11, 3.12, and 4.2.

We recommend that, to resolve this indifference, the link *cd* should be declared *forbidden* and the strongest paths from alternative *a* to alternative *b* and from alternative *b* to alternative *a*, that don’t contain *forbidden* links, should be calculated. Either this indifference is now resolved or, again, the weakest link in the strongest path from alternative *a* to alternative *b* and the weakest link in the strongest path from alternative *b* to alternative *a* are the same link, say *ef*. In the latter case, the link *ef* is declared *forbidden* and the strongest paths that don’t contain *forbidden* links are calculated. This procedure is repeated until this indifference is resolved. The resulting Schulze relation will be called $O_{final}$.

In example 5 (section 3.5), the link *cd* is the weakest link in the strongest path from alternative *a* to alternative *b* and the weakest link in the strongest path from alternative *b* to alternative *a*. Therefore, the link *cd* is declared *forbidden*. The strongest path from alternative *a* to alternative *b*, that doesn’t contain *forbidden* links, is *a*,(33,30),*b*. The strongest path from alternative *b* to alternative *a*, that doesn’t contain *forbidden* links, is *b*,(30,33),*a*. Therefore, we get $ab \in O_{final}$.

**Problem #2:** We want the Schulze method to satisfy *composition consistency*. That means: When $\emptyset \neq K \subsetneq A$ is a set of clones, as defined in (4.7.1) – (4.7.3), then, when we apply the election method only to the alternatives in *K*, the result should be the same as if we had applied the election method to *A* and then looked only at the alternatives in *K*. See (4.7.9) – (4.7.13) and (5.3.6.8) – (5.3.6.10). See also Öztürk (2020) and Berker (2025).

The simpler definition for the Schulze method in section 2.2 violates composition consistency because it can happen that the strongest path from $g \in K$ to $h \in K \setminus \{g\}$ passes through alternatives $a,b \in A \setminus K$.

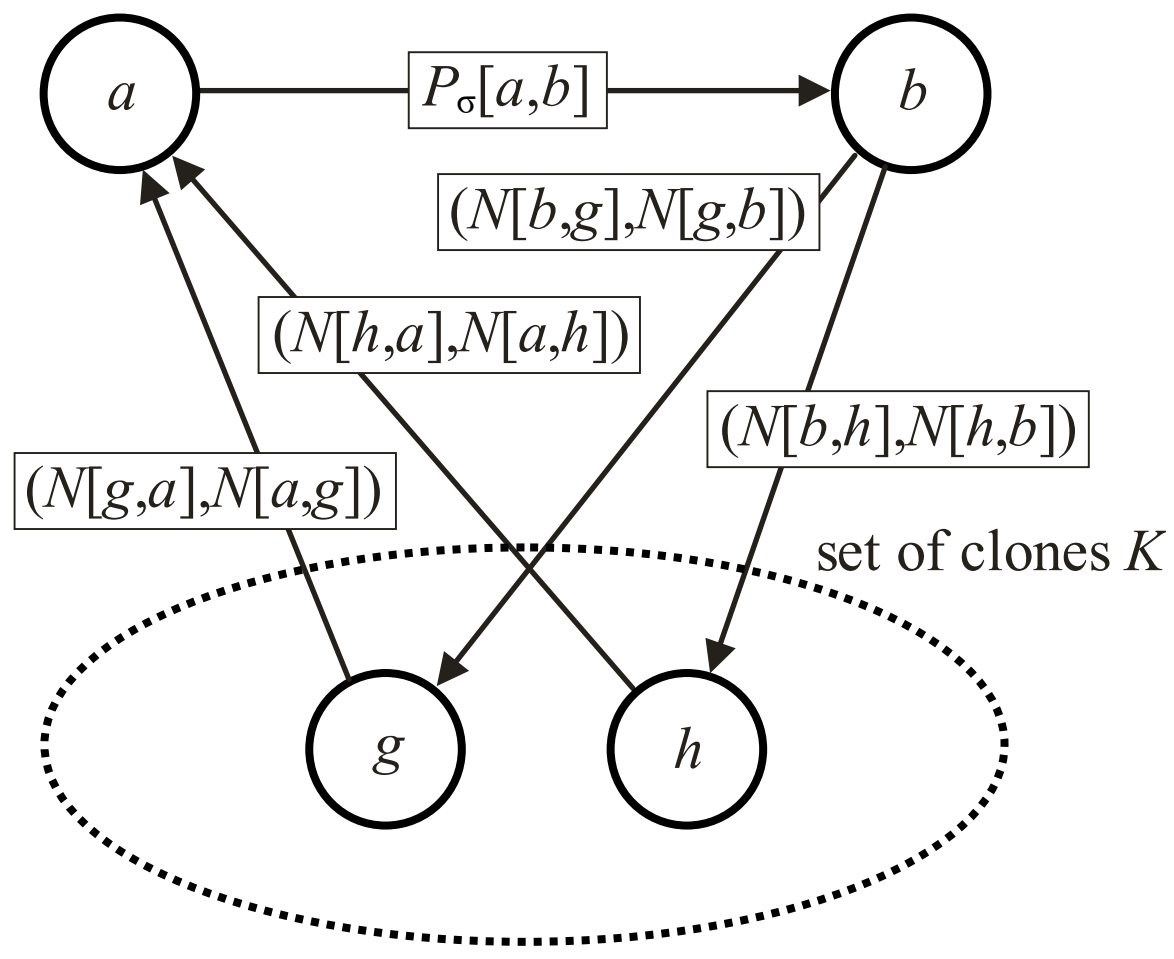

As $\emptyset \neq K \subsetneq A$ is a set of clones, we get $(N[g,a],N[a,g]) \approx_D (N[h,a],N[a,h])$ and $(N[b,g],N[g,b]) \approx_D (N[b,h],N[h,b])$ for all $g,h \in K$ and for all $a,b \in A \setminus K$. Therefore, we get:

> When $g,(N[g,a],N[a,g]),a,P_\sigma[a,b],b,(N[b,h],N[h,b]),h$ is the strongest path from alternative $g$ to alternative $h$, then $h,(N[h,a],N[a,h]),a,P_\sigma[a,b],b,(N[b,g],N[g,b]),g$ is the strongest path from alternative $h$ to alternative $g$.

When the strongest path from $g \in K$ to $h \in K \setminus \{g\}$ passes through alternatives $a,b \in A \setminus K$, there are three possible cases:

(i) The weakest link in the strongest path from $g$ to $h$ could be the weakest link in the strongest path from $a$ to $b$. As the sequence $a,P_\sigma[a,b],b$ is in the strongest path from $g$ to $h$ and simultaneously in the strongest path from $h$ to $g$, this means that the weakest link in the strongest path from $g$ to $h$ is the same link as the weakest link in the strongest path from $h$ to $g$. However, this case has already been addressed in problem #1.

(ii) The weakest link in the strongest path from $g$ to $h$ could be the link $ga$. In this case, $ha$ with $(N[g,a],N[a,g]) \approx_D (N[h,a],N[a,h])$ is the weakest link in the strongest path from $h$ to $g$. In this case, we recommend that $O|_K$ should be calculated and that, when $gh \in O|_K$, we arbitrarily set $(N[g,a],N[a,g]) \succ_D (N[h,a],N[a,h])$, so that the weakest link in the strongest path from $g$ to $h$ arbitrarily becomes stronger than the weakest link in the strongest path from $h$ to $g$ so that $gh \in O$. See section 5.1 step 2 (b) (1).

(iii) The weakest link in the strongest path from $g$ to $h$ could be the link $bh$. In this case, $bg$ with $(N[b,g],N[g,b]) \approx_D (N[b,h],N[h,b])$ is the weakest link in the strongest path from $h$ to $g$. In this case, we recommend that $O|_K$ should be calculated and that, when $gh \in O|_K$, we arbitrarily set $(N[b,g],N[g,b]) \prec_D (N[b,h],N[h,b])$, so that the weakest link in the strongest path from $g$ to $h$ arbitrarily becomes stronger than the weakest link in the strongest path from $h$ to $g$ so that $gh \in O$. See section 5.1 step 2 (b) (2).

**Problem #3:** We want to know how to choose the final winner from among the potential winners when the simpler definition for the Schulze method finds more than one potential winner. We want to solve these situations without having to sacrifice any of the desirable properties of the Schulze method.

To solve these situations, we recommend that a Tie-Breaking Ranking of the Voters (TBRV) and a Tie-Breaking Ranking of the Candidates (TBRC) should be calculated and that indecisive situations should be perturbed in favour of those voters who are ranked higher in this TBRV resp. in favour of those alternatives that are ranked higher in this TBRC.

## 5.1. Calculating a Complete Ranking Using a Tie-Breaking Ranking of the Links

The Schulze relation $\mathcal{O}$, as defined in (2.2.1), is only a strict partial order. However, sometimes, a linear order is needed. In this section, we will show how the Schulze relation $\mathcal{O}$ can be completed to a linear order $\mathcal{O}_{final}(\sigma)$ without having to sacrifice any of the desired criteria. The following four steps describe how the linear order $\mathcal{O}_{final}(\sigma)$ is calculated.

**Step 1 (calculating a TBRV $\succ_{\xi}$ and a first TBRC $\succ_{\omega}$):**

A *Tie-Breaking Ranking of the Voters* (TBRV), a linear order $\succ_{\xi}$ on $V$, is calculated by ranking the voters in $V$ randomly.

A first *Tie-Breaking Ranking of the Candidates* (TBRC), a linear order $\succ_{\omega}$ on $A$, is calculated by ranking the alternatives in $A$ randomly.

When the bylaws require that the chairperson decides in the case of a tie, then, for the calculation of the TBRV, the chairperson has to be chosen first.

**Step 2 (calculating a TBRL $\succ_{\sigma}$ and a second TBRC $\succ_{\mu}$):**

We now use the TBRV $\succ_{\xi}$ and the first TBRC $\succ_{\omega}$ to calculate a *Tie-Breaking Ranking of the Links* (TBRL), a linear order $\succ_{\sigma}$ on $A \times A$, and a second TBRC $\succ_{\mu}$. The second TBRC $\succ_{\mu}$ is only needed to complete the TBRL $\succ_{\sigma}$. So the final result $O_{final}(\sigma)$ depends only on $\succ_{\sigma}$.

(a) We start with:

- $\forall\ (i,j),(m,n) \in A \times A$: $(N[i,j],N[j,i]) \succ_D (N[m,n],N[n,m]) \Rightarrow ij \succ_{\sigma} mn$.
- $\forall\ (i,j),(m,n) \in A \times A$: $(N[i,j],N[j,i]) \approx_D (N[m,n],N[n,m]) \Rightarrow ij \approx_{\sigma} mn$.
- $\forall\ i,j \in A$: $i \approx_{\mu} j$.

(b) (1) It can happen (e.g. when there are clones) that there are links with equivalent strengths and the same ending alternative: $ij \approx_D mj$.

In these cases, we determine $K(i,j) := \{ k \in A \mid kj \approx_D ij \}$. If $\| K(i,j) \| > 1$, then (before we calculate $O_{final}(\sigma)$ for $A$ as defined in stages 3 and 4) we first calculate $\psi := \sigma|_{K(i,j)}$ and $O_{final}(\psi)|_{K(i,j)}$ and then we set: If $mn \in O_{final}(\psi)|_{K(i,j)}$ then $mj \succ_{\sigma} nj$.

(2) It can happen (e.g. when there are clones) that there are links with equivalent strengths and the same starting alternative: $ij \approx_D im$.

In these cases, we determine $L(i,j) := \{ k \in A \mid ik \approx_D ij \}$. If $\| L(i,j) \| > 1$, then (before we calculate $O_{final}(\sigma)$ for $A$ as defined in stages 3 and 4) we first calculate $\psi := \sigma|_{L(i,j)}$ and $O_{final}(\psi)|_{L(i,j)}$ and then we set: If $mn \in O_{final}(\psi)|_{L(i,j)}$, then $in \succ_{\sigma} im$.

(3) It can happen that $K(i,j)$ or $L(i,j)$ contains links with equivalent strengths and the same starting alternative or the same ending alternative. In these cases, we first calculate $\psi := \sigma|_{\mathcal{K}(m,n)}$ resp. $\psi := \sigma|_{\mathcal{L}(m,n)}$ and $O_{final}(\psi)|_{\mathcal{K}(m,n)}$ resp. $O_{final}(\psi)|_{\mathcal{L}(m,n)}$ for $\mathcal{K}(m,n) := \{ k \in K(i,j) \mid kn \approx_D mn \}$ resp. $\mathcal{K}(m,n) := \{ k \in L(i,j) \mid kn \approx_D mn \}$ and $\mathcal{L}(m,n) := \{ k \in K(i,j) \mid mk \approx_D mn \}$ resp. $\mathcal{L}(m,n) := \{ k \in L(i,j) \mid mk \approx_D mn \}$. If $st \in O_{final}(\psi)|_{\mathcal{K}(m,n)}$ then we set: $sn \succ_{\sigma|_{K(i,j)}} tn$ resp. $sn \succ_{\sigma|_{L(i,j)}} tn$. If $st \in O_{final}(\psi)|_{\mathcal{L}(m,n)}$, then we set: $mt \succ_{\sigma|_{K(i,j)}} ms$ resp. $mt \succ_{\sigma|_{L(i,j)}} ms$. We proceed until there are no links anymore with equivalent strengths and the same starting alternative or the same ending alternative.

(c) From the first voter to the last voter, according to $\succ_\xi$, we use his ranking to complete $\succ_\sigma$. Suppose $v \in V$ is the current voter. Then we proceed:

Step 1: $\forall\ i,j \in A$: If $i \approx_\mu j$ and $i \succ_v j$, then replace “ $i \approx_\mu j$ ” by “ $i \succ_\mu j$ ”.

Step 2: $\forall\ (i,j),(m,n) \in A \times A$:

If $ij \approx_\sigma mn$ and

(5.1.1) $( ( i \succsim_v j ) \wedge ( m \prec_v n ) ) \vee ( ( i \succ_v j ) \wedge ( m \precsim_v n ) )$,

then replace “ $ij \approx_\sigma mn$ ” by “ $ij \succ_\sigma mn$ ”.

Step 3: It can happen, that $\succ_\sigma$ contains a directed cycle. In this case, we replace all $\succ_\sigma$ of this directed cycle (except for those that had already been locked before we used the ranking of voter $v \in V$) by $\approx_\sigma$.

Step 4: $\succ_\sigma$ is updated to its transitive closure.

(d) If you go through all ballots and there are still alternatives $i,j \in A$ with $i \approx_\mu j$, then proceed as follows: If $i \approx_\mu j$ and $i \succ_\omega j$, then replace “ $i \approx_\mu j$ ” by “ $i \succ_\mu j$ ”.

(e) Suppose there are still $(i,j),(m,n) \in A \times A$ with $ij \approx_\sigma mn$, then proceed as follows:

**<u>Variant 1:</u>** When at least one of the following conditions is satisfied, then replace “ $ij \approx_\sigma mn$ ” by “ $ij \succ_\sigma mn$ ”:

(5.1.2a) $i \succ_\mu j$ and $n \succ_\mu m$.
(5.1.3a) $i \succ_\mu j$ and $m \succ_\mu n$ and $i \succ_\mu m$.
(5.1.4a) $j \succ_\mu i$ and $n \succ_\mu m$ and $n \succ_\mu j$.
(5.1.5a) $i \equiv m$ and $n \succ_\mu j$.
(5.1.6a) $j \equiv n$ and $i \succ_\mu m$.

**<u>Variant 2:</u>** When at least one of the following conditions is satisfied, then replace “ $ij \approx_\sigma mn$ ” by “ $ij \succ_\sigma mn$ ”:

(5.1.2b) $i \succ_\mu j$ and $n \succ_\mu m$.
(5.1.3b) $i \succ_\mu j$ and $m \succ_\mu n$ and $n \succ_\mu j$.
(5.1.4b) $j \succ_\mu i$ and $n \succ_\mu m$ and $i \succ_\mu m$.
(5.1.5b) $i \equiv m$ and $n \succ_\mu j$.
(5.1.6b) $j \equiv n$ and $i \succ_\mu m$.

(5.1.2a) – (5.1.6a) and (5.1.2b) – (5.1.6b) are chosen in such a manner that e.g. when the TBRC $\succ_\mu$ is *abcdefgh* then links of otherwise equivalent strengths are sorted *ah*, *ag*, *af*, *ae*, *ad*, *ac*, *ab*, *bh*, *bg*, *bf*, *be*, *bd*, *bc*, *ch*, *cg*, *cf*, *ce*, *cd*, *dh*, *dg*, *df*, *de*, *eh*, *eg*, *ef*, *fh*, *fg*, *gh*, *hg*, *gf*, *hf*, *fe*, *ge*, *he*, *ed*, *fd*, *gd*, *hd*, *dc*, *ec*, *fc*, *gc*, *hc*, *cb*, *db*, *eb*, *fb*, *gb*, *hb*, *ba*, *ca*, *da*, *ea*, *fa*, *ga*, *ha* in variant 1 resp. *ah*, *bh*, *ch*, *dh*, *eh*, *fh*, *gh*, *ag*, *bg*, *cg*, *dg*, *eg*, *fg*, *af*, *bf*, *cf*, *df*, *ef*, *ae*, *be*, *ce*, *de*, *ad*, *bd*, *cd*, *ac*, *bc*, *ab*, *ba*, *cb*, *ca*, *dc*, *db*, *da*, *ed*, *ec*, *eb*, *ea*, *fe*, *fd*, *fc*, *fb*, *fa*, *gf*, *ge*, *gd*, *gc*, *gb*, *ga*, *hg*, *hf*, *he*, *hd*, *hc*, *hb*, *ha* in variant 2.

**Stage 3 (initializing the strongest paths):**

```
 1  for i : = 1 to C
 2  begin
 3      for j : = 1 to C
 4      begin
 5          if ( i ≠ j ) then
 6          begin
 7              P_σ[i,j] : = ij
 8          end
 9      end
10  end

11  for i : = 1 to C
12  begin
13      for j : = 1 to C
14      begin
15          if ( i ≠ j ) then
16          begin
17              for k : = 1 to C
18              begin
19                  if ( i ≠ k ) then
20                  begin
21                      if ( j ≠ k ) then
22                      begin
23                          if ( P_σ[j,k] ≺_σ min_σ { P_σ[j,i], P_σ[i,k] } ) then
24                          begin
25                              P_σ[j,k] : = min_σ { P_σ[j,i], P_σ[i,k] }
26                          end
27                      end
28                  end
29              end
30          end
31      end
32  end

33  O_final(σ) : = ∅
34  S_final(σ) : = A
35  for i : = 1 to C
36  begin
37      for j : = 1 to C
38      begin
39          if ( i ≠ j ) then
40          begin
41              if ( P_σ[j,i] ≻_σ P_σ[i,j] ) then
42              begin
43                  O_final(σ) : = O_final(σ) + {ji}
44                  S_final(σ) : = S_final(σ) \ {i}
45              end
46          end
47      end
48  end
```

**Stage 4 (applying the TBRL $\succ_{\sigma}$):**

```
xy : = min_σ { ij | i,j ∈ {1,...,C}, i ≠ j }
for m : = 1 to C–1
begin
    for n : = m+1 to C
    begin
        if ( P_σ[m,n] ≈_σ P_σ[n,m] ) then
        begin
            Q_σ[m,n] : = P_σ[m,n]
            for i : = 1 to C
            begin
                for j : = 1 to C
                begin
                    if ( i ≠ j ) then
                    begin
                        forbidden[i,j] : = false
                    end
                end
            end
            bool1 : = false
            while ( bool1 = false )
            begin
                for i : = 1 to C
                begin
                    for j : = 1 to C
                    begin
                        if ( i ≠ j ) then
                        begin
                            if ( Q_σ[m,n] ≈_σ ij ) then
                            begin
                                forbidden[i,j] : = true
                            end
                        end
                    end
                end
                for i : = 1 to C
                begin
                    for j : = 1 to C
                    begin
                        if ( i ≠ j ) then
                        begin
                            if ( forbidden[i,j] = true ) then
                            begin
                                Q_σ[i,j] : = xy
                            end
                            else
                            begin
                                Q_σ[i,j] : = ij
                            end
                        end
                    end
                end
```

```
100             for i : = 1 to C
101             begin
102                for j : = 1 to C
103                begin
104                   if ( i ≠ j ) then
105                   begin
106                      for k : = 1 to C
107                      begin
108                         if ( i ≠ k ) then
109                         begin
110                            if ( j ≠ k ) then
111                            begin
112                               if ( Qσ[j,k] ≺σ minσ { Qσ[j,i], Qσ[i,k] } ) then
113                               begin
114                                  Qσ[j,k] : = minσ { Qσ[j,i], Qσ[i,k] }
115                               end
116                            end
117                         end
118                      end
119                   end
120                end
121             end
122             if ( Qσ[m,n] ≻σ Qσ[n,m] ) then
123             begin
124                Ofinal(σ) : = Ofinal(σ) + {mn}
125                Sfinal(σ) : = Sfinal(σ) \ {n}
126                bool1 : = true
127             end
128             else
129             begin
130                if ( Qσ[m,n] ≺σ Qσ[n,m] ) then
131                begin
132                   Ofinal(σ) : = Ofinal(σ) + {nm}
133                   Sfinal(σ) : = Sfinal(σ) \ {m}
134                   bool1 : = true
135                end
136                else
137                if ( Qσ[m,n] = xy and Qσ[n,m] = xy ) then
138                begin
139                   bool1 : = true
140                end
141             end
142          end
143       end
144    end
145 end
```

For each pair of alternatives $m,n \in A$, we check whether $P_\sigma[m,n] \approx_\sigma P_\sigma[n,m]$ (lines 50–55). In this case, the link $ij$ with $P_\sigma[m,n] \approx_\sigma ij$ is declared *forbidden* (lines 70–82) and the strongest paths, that don't contain *forbidden* links, are calculated (lines 83–121). This procedure is repeated (lines 67–68) until this indifference is resolved (lines 122–135) or all links have been declared *forbidden* (lines 136–140).

### Example 5.1.1

Example 5.1.1:

| | |
|---|---|
| 6 voters | $a \succ_v b \succ_v c \succ_v d$ |
| 2 voters | $b \succ_v a \succ_v c \succ_v d$ |
| 6 voters | $c \succ_v d \succ_v b \succ_v a$ |
| 4 voters | $d \succ_v a \succ_v b \succ_v c$ |
| 1 voter | $d \succ_v b \succ_v c \succ_v a$ |

The pairwise matrix $N$ looks as follows:

| | $N[*,a]$ | $N[*,b]$ | $N[*,c]$ | $N[*,d]$ |
|---|---|---|---|---|
| $N[a,*]$ | --- | 10 | 12 | 8 |
| $N[b,*]$ | 9 | --- | 13 | 8 |
| $N[c,*]$ | 7 | 6 | --- | 14 |
| $N[d,*]$ | 11 | 11 | 5 | --- |

The corresponding digraph looks as follows:

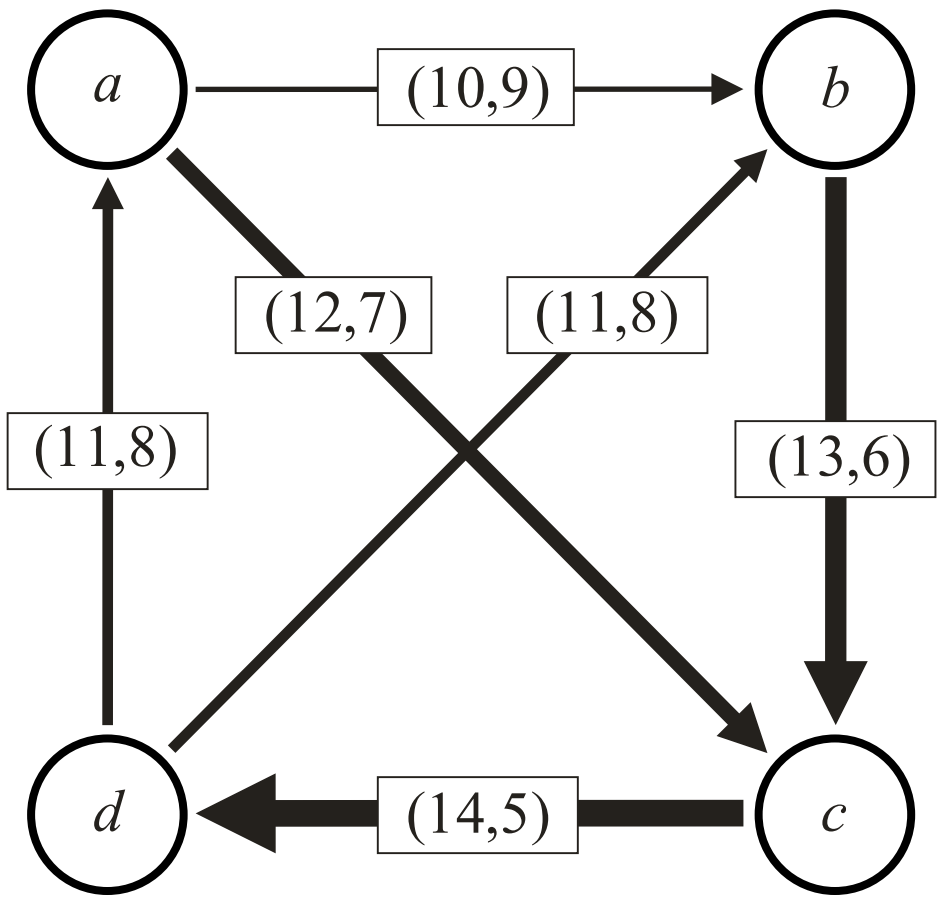

In section 5.1 step 2 (a), we get:

$$(N[c,d],N[d,c]) \succ_\sigma (N[b,c],N[c,b]) \succ_\sigma (N[a,c],N[c,a]) \succ_\sigma (N[d,a],N[a,d]) \approx_\sigma (N[d,b],N[b,d]) \succ_\sigma (N[a,b],N[b,a]).$$

The links $da$ and $db$ have equivalent strengths and the same start $d \in A$. Therefore, we apply section 5.1 step 2 (b) (2). When we apply the Schulze method only to $\{a,b\}$, we get $a \succ_S b$. Thus, we set $(N[d,b],N[b,d]) \succ_\sigma (N[d,a],N[a,d])$.

When we apply section 5.1 step 3, we finally get $O_{final}(\sigma) = \{ab, ac, ad, bc, bd, cd\}$ and $S_{final}(\sigma) = \{a\}$.

## 5.2. Transitivity

In section 4.1, we have proven that the binary relation $O$, as defined in (2.2.1), is transitive. Nevertheless, it isn't intuitively clear whether also the binary relation $O_{final}(\sigma)$, as defined in section 5.1, is transitive. It seems to be possible that ties $P_\sigma[x,y] \approx_\sigma P_\sigma[y,x]$ are resolved based on different sets of *non-forbidden* links, so that the transitivity of $O_{final}(\sigma)$ doesn't follow directly from the transitivity of $O$.

However, in the following proof, we will see that also the binary relation $O_{final}(\sigma)$, as defined in section 5.1, is transitive. We will prove that ties $P_\sigma[x,y] \approx_\sigma P_\sigma[y,x]$ are either resolved based on the same set of *non-forbidden* links (sections 5.2.1, 5.2.4, and 5.2.5) or — in those cases, where these ties happen to be resolved based on different sets of *non-forbidden* links — they cannot violate transitivity (sections 5.2.2 and 5.2.3).

### 5.2.1. Part 1

Suppose, before we start declaring links *forbidden*, we have:

(5.2.1.1) $P_\sigma[a,b] \succ_\sigma P_\sigma[b,a]$.

(5.2.1.2) $P_\sigma[b,c] \succ_\sigma P_\sigma[c,b]$.

(5.2.1.3) $P_\sigma[c,a] \approx_\sigma P_\sigma[a,c]$.

With (5.2.1.1), we get $ab \in O$ and, therefore, $ab \in O_{final}(\sigma)$.

With (5.2.1.2), we get $bc \in O$ and, therefore, $bc \in O_{final}(\sigma)$.

This situation is not possible because, when no link has been declared *forbidden*, then all paths are calculated based on the same set of *non-forbidden* links. But in section 4.1, we have proven that, when all paths are calculated based on the same set of links, then the binary relation $O$, as defined by $P_\sigma[x,y] \succ_\sigma P_\sigma[y,x]$, is transitive. So, with $P_\sigma[a,b] \succ_\sigma P_\sigma[b,a]$ and $P_\sigma[b,c] \succ_\sigma P_\sigma[c,b]$, we immediately get $P_\sigma[a,c] \succ_\sigma P_\sigma[c,a]$.

### 5.2.2. Part 2

Suppose, before we start declaring links *forbidden*, we have:

(5.2.2.1) $P_\sigma[a,b] \prec_\sigma P_\sigma[b,a]$.

(5.2.2.2) $P_\sigma[b,c] \succ_\sigma P_\sigma[c,b]$.

(5.2.2.3) $P_\sigma[c,a] \approx_\sigma P_\sigma[a,c]$.

With (5.2.2.1), we get $ba \in O$ and, therefore, $ba \in O_{final}(\sigma)$.

With (5.2.2.2), we get $bc \in O$ and, therefore, $bc \in O_{final}(\sigma)$.

Suppose there are no pairwise links of equivalent strengths. Suppose (5.2.2.1) – (5.2.2.3). With (5.2.2.3), we get that the weakest link in the strongest path from alternative $a$ to alternative $c$ and the weakest link in the strongest path from alternative $c$ to alternative $a$ must be the same link, say $de$.

Therefore, the strongest paths have the following structure:

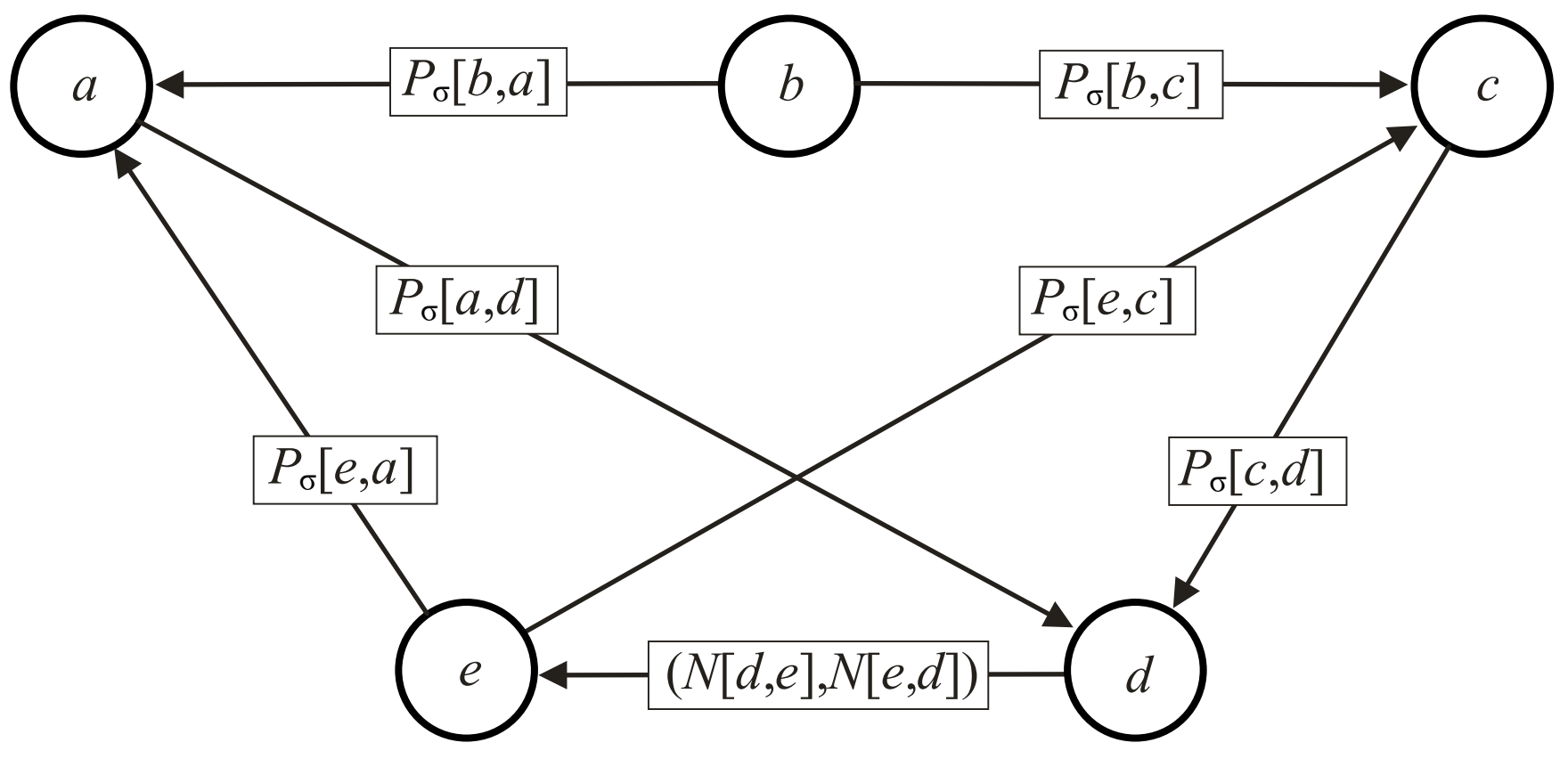

In this case, it can actually happen that the paths are based on different sets of *non-forbidden* links. In example 11 (section 3.11), we have a situation with $P_\sigma[a,b] \prec_\sigma P_\sigma[b,a]$, $P_\sigma[b,c] \succ_\sigma P_\sigma[c,b]$, and $P_\sigma[c,a] \approx_\sigma P_\sigma[a,c]$ and where the link *de* is the weakest link in the strongest path from alternative *a* to alternative *c* and simultaneously the weakest link in the strongest path from alternative *c* to alternative *a*. So when we resolve $P_\sigma[c,a] \approx_\sigma P_\sigma[a,c]$, the link *de* has to be declared *forbidden*. The strongest path from alternative *a* to alternative *c*, that doesn't contain the link *de*, is *a*,(24,21),*c*. The strongest path from alternative *c* to alternative *a*, that doesn't contain the link *de*, is *c*,(25,20),*b*,(22,23),*e*,(30,15),*a*. So $P_\sigma[c,a] \approx_\sigma P_\sigma[a,c]$ is resolved to $ac \in O_{final}(\sigma)$.

Now the interesting observation is that the link *de* is also in the strongest path from alternative *b* to alternative *a*. And the strongest path *b*,(22,23),*e*, (30,15),*a* from alternative *b* to alternative *a*, that doesn't contain the link *de*, is weaker than the strongest path *a*,(26,19),*b* from alternative *a* to alternative *b*, that doesn't contain the link *de*. Therefore, if we had to recalculate the strengths of the strongest paths from alternative *a* to alternative *b* and from alternative *b* to alternative *a* based on the fact that the link *de* has been declared *forbidden* { what we don't have to do, because each of (5.2.2.1) – (5.2.2.3) is resolved separately, based on its own set of *non-forbidden* links }, we would get $P_\sigma[a,b] \succ_\sigma P_\sigma[b,a]$.

Furthermore, the link *de* is in the strongest path from alternative *b* to alternative *c*. And the strongest path *b*,(22,23),*e*,(32,13),*c* from alternative *b* to alternative *c*, that doesn't contain the link *de*, is weaker than the strongest path *c*,(25,20),*b* from alternative *c* to alternative *b*, that doesn't contain the link *de*. Therefore, if we had to recalculate the strengths of the strongest paths from alternative *b* to alternative *c* and from alternative *c* to alternative *b* based on the fact that the link *de* has been declared *forbidden*, we would get $P_\sigma[b,c] \prec_\sigma P_\sigma[c,b]$.

So example 11 (section 3.11) demonstrates that it can happen that (5.2.2.1) – (5.2.2.3) are resolved based on different sets of *non-forbidden* links. However, this is not a problem because — it doesn't matter whether $P_\sigma[c,a] \approx_\sigma P_\sigma[a,c]$ is resolved to $P_\sigma[c,a] \succ_\sigma P_\sigma[a,c]$ or to $P_\sigma[c,a] \prec_\sigma P_\sigma[a,c]$ — transitivity will never be violated.

### 5.2.3. Part 3

Suppose, before we start declaring links *forbidden*, we have:

(5.2.3.1) $P_\sigma[a,b] \succ_\sigma P_\sigma[b,a]$.

(5.2.3.2) $P_\sigma[b,c] \prec_\sigma P_\sigma[c,b]$.

(5.2.3.3) $P_\sigma[c,a] \approx_\sigma P_\sigma[a,c]$.

With (5.2.3.1), we get $ab \in O$ and, therefore, $ab \in O_{final}(\sigma)$.

With (5.2.3.2), we get $cb \in O$ and, therefore, $cb \in O_{final}(\sigma)$.

Suppose there are no pairwise links of equivalent strengths. Suppose (5.2.3.1) – (5.2.3.3). With (5.2.3.3), we get that the weakest link in the strongest path from alternative *a* to alternative *c* and the weakest link in the strongest path from alternative *c* to alternative *a* must be the same link, say *de*.

Therefore, the strongest paths have the following structure:

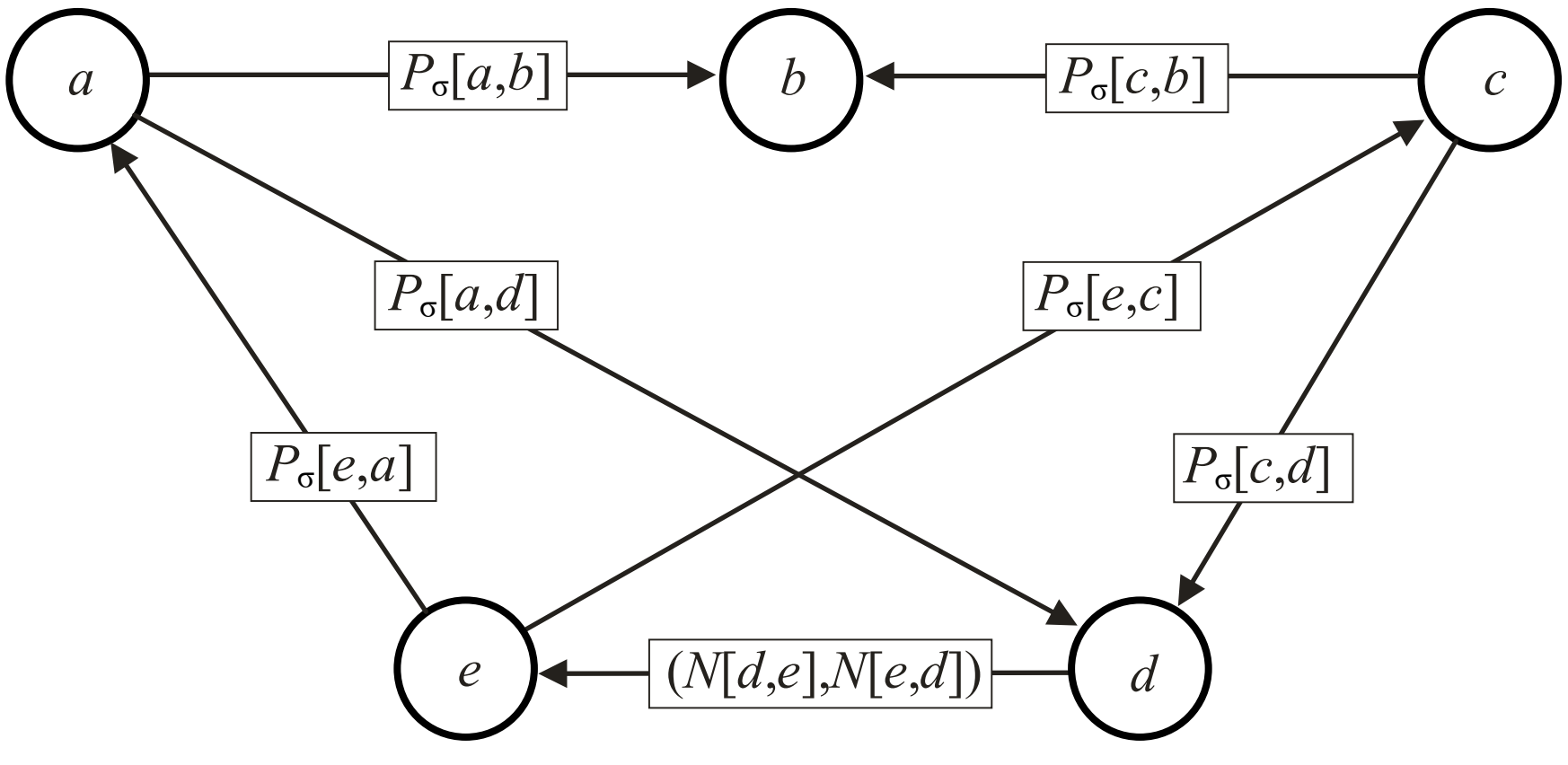

In this case, it can actually happen that the paths are based on different sets of *non-forbidden* links. In example 12 (section 3.12), we have a situation with $P_\sigma[a,b] \succ_\sigma P_\sigma[b,a]$, $P_\sigma[b,c] \prec_\sigma P_\sigma[c,b]$, and $P_\sigma[c,a] \approx_\sigma P_\sigma[a,c]$ and where the link *de* is the weakest link in the strongest path from alternative *a* to alternative *c* and simultaneously the weakest link in the strongest path from alternative *c* to alternative *a*. So when we resolve $P_\sigma[c,a] \approx_\sigma P_\sigma[a,c]$, the link *de* has to be declared *forbidden*. The strongest path from alternative *a* to alternative *c*, that doesn't contain the link *de*, is *a*,(24,21),*c*. The strongest path from alternative *c* to alternative *a*, that doesn't contain the link *de*, is *c*,(30,15),*d*,(22,23),*b*,(25,20),*a*. So $P_\sigma[c,a] \approx_\sigma P_\sigma[a,c]$ is resolved to $ac \in O_{final}(\sigma)$.

Now the interesting observation is that the link *de* is also in the strongest path from alternative *a* to alternative *b*. And the strongest path *a*,(32,13),*d*,(22,23),*b* from alternative *a* to alternative *b*, that doesn't contain the link *de*, is weaker than the strongest path *b*,(25,20),*a* from alternative *b* to alternative *a*, that doesn't contain the link *de*. Therefore, if we had to recalculate the strengths of the strongest paths from alternative *a* to alternative *b* and from alternative *b* to alternative *a* based on the fact that the link *de* has been declared *forbidden* { what we don't have to do, because each of (5.2.3.1) – (5.2.3.3) is resolved separately, based on its own set of *non-forbidden* links }, we would get $P_\sigma[a,b] \prec_\sigma P_\sigma[b,a]$.

Furthermore, the link *de* is in the strongest path from alternative *c* to alternative *b*. And the strongest path *c*,(30,15),*d*,(22,23),*b* from alternative *c* to alternative *b*, that doesn't contain the link *de*, is weaker than the strongest path *b*,(26,19),*c* from alternative *b* to alternative *c*, that doesn't contain the link *de*. Therefore, if we had to recalculate the strengths of the strongest paths from alternative *b* to alternative *c* and from alternative *c* to alternative *b* based on the fact that the link *de* has been declared *forbidden*, we would get $P_\sigma[b,c] \succ_\sigma P_\sigma[c,b]$.

So example 12 (section 3.12) demonstrates that it can happen that (5.2.3.1) – (5.2.3.3) are resolved based on different sets of *non-forbidden* links. However, this is not a problem because — it doesn't matter whether $P_\sigma[c,a] \approx_\sigma P_\sigma[a,c]$ is resolved to $P_\sigma[c,a] \succ_\sigma P_\sigma[a,c]$ or to $P_\sigma[c,a] \prec_\sigma P_\sigma[a,c]$ — transitivity will never be violated.

### 5.2.4. Part 4

Suppose, before we start declaring links *forbidden*, we have:

(5.2.4.1) $P_\sigma[a,b] \approx_\sigma P_\sigma[b,a]$.

(5.2.4.2) $P_\sigma[b,c] \approx_\sigma P_\sigma[c,b]$.

(5.2.4.3) $P_\sigma[c,a] \succ_\sigma P_\sigma[a,c]$.

With (5.2.4.3), we get $ca \in O$ and, therefore, $ca \in O_{final}(\sigma)$.

As the tie (5.2.4.1) and the tie (5.2.4.2) are resolved separately, it seems to be possible that they are resolved based on different sets of *non-forbidden* links, so that the transitivity of $O_{final}(\sigma)$ doesn't follow directly from the transitivity of $O$. It seems to be possible that the tie (5.2.4.1) is resolved to $P_\sigma[a,b] \succ_\sigma P_\sigma[b,a]$ and that simultaneously — as other links are declared *forbidden* during the process of resolving the tie (5.2.4.2), so that the strengths of the strongest paths are determined based on different sets of *non-forbidden* links — the tie (5.2.4.2) is resolved to $P_\sigma[b,c] \succ_\sigma P_\sigma[c,b]$, so that the transitivity of $O_{final}(\sigma)$ is violated. However, the following proof shows that transitivity will never be violated.

**Claim:**

Suppose (5.2.4.1) – (5.2.4.3) are resolved as prescribed in section 5.1. Then transitivity will never be violated.

**Proof:**

Suppose there are no pairwise links of equivalent strengths. Suppose (5.2.4.1) – (5.2.4.3). With (5.2.4.1), we get that the weakest link in the strongest path from alternative $a$ to alternative $b$ and the weakest link in the strongest path from alternative $b$ to alternative $a$ must be the same link, say $de$. With (5.2.4.2), we get that the weakest link in the strongest path from alternative $b$ to alternative $c$ and the weakest link in the strongest path from alternative $c$ to alternative $b$ must be the same link, say $fg$.

Therefore, the strongest paths have the following structure:

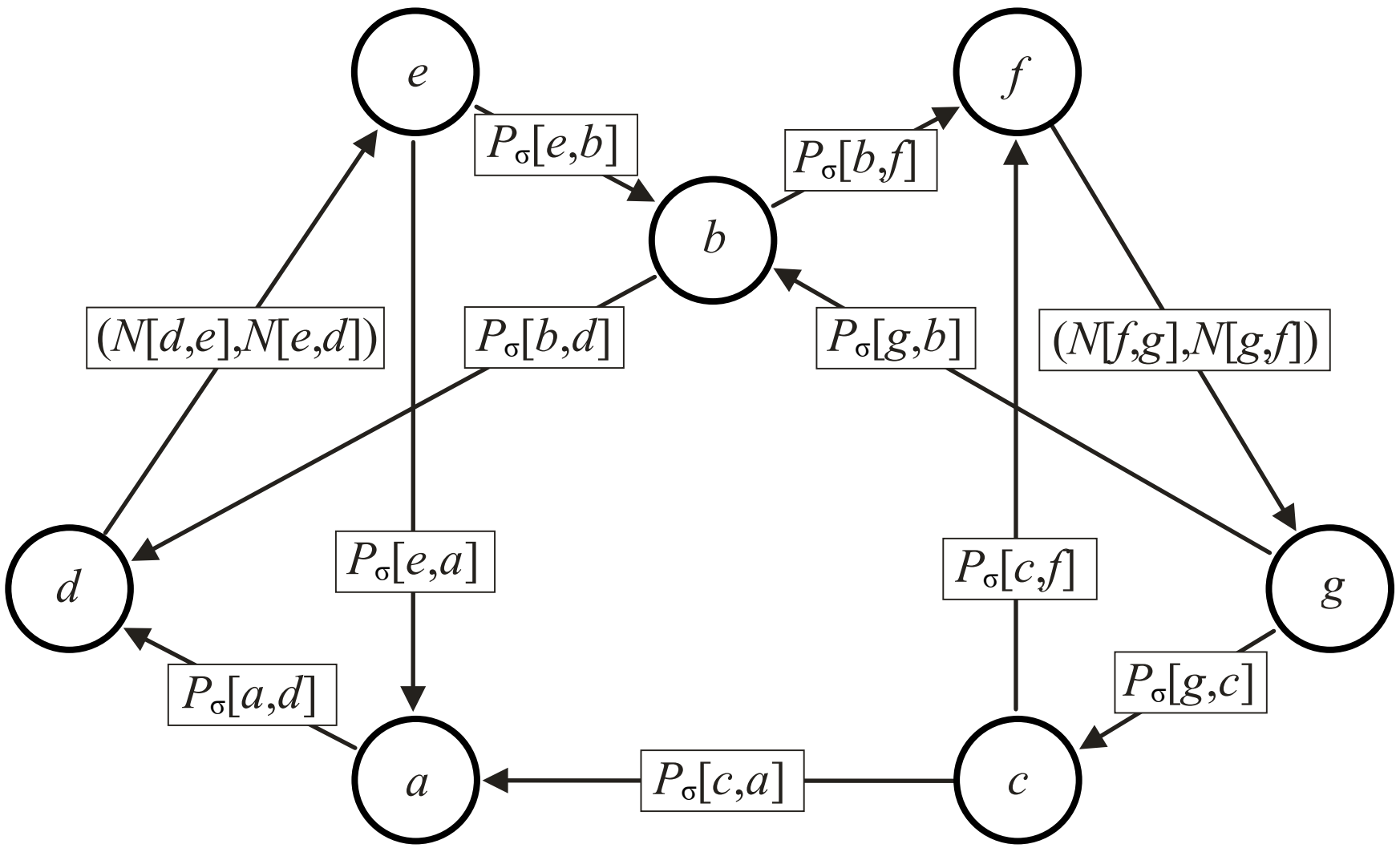

As $de$ is the weakest link in the strongest path from alternative $a$ to alternative $b$, we get

(5.2.4.4) $P_\sigma[a,d] \succ_\sigma (N[d,e],N[e,d])$.

(5.2.4.5) $P_\sigma[e,b] \succ_\sigma (N[d,e],N[e,d])$.

As $de$ is the weakest link in the strongest path from alternative $b$ to alternative $a$, we get

(5.2.4.6) $P_\sigma[b,d] \succ_\sigma (N[d,e],N[e,d])$.

(5.2.4.7) $P_\sigma[e,a] \succ_\sigma (N[d,e],N[e,d])$.

As *fg* is the weakest link in the strongest path from alternative *b* to alternative *c*, we get

(5.2.4.8) $P_\sigma[b,f] >_\sigma (N[f,g],N[g,f])$.

(5.2.4.9) $P_\sigma[g,c] >_\sigma (N[f,g],N[g,f])$.

As *fg* is the weakest link in the strongest path from alternative *c* to alternative *b*, we get

(5.2.4.10) $P_\sigma[c,f] >_\sigma (N[f,g],N[g,f])$.

(5.2.4.11) $P_\sigma[g,b] >_\sigma (N[f,g],N[g,f])$.

With (5.2.4.4), (5.2.4.5), (5.2.4.8), and (5.2.4.9), we get: $a \to d \to e \to b \to f \to g \to c$ is a path from alternative *a* to alternative *c* with a strength of $\min_\sigma \{ (N[d,e],N[e,d]), (N[f,g],N[g,f]) \}$. Therefore, with (5.2.4.3), we get

(5.2.4.12) $P_\sigma[c,a] >_\sigma \min_\sigma \{ (N[d,e],N[e,d]), (N[f,g],N[g,f]) \}$.

<u>Case 1:</u> Suppose

(5.2.4.13a) $(N[d,e],N[e,d]) >_\sigma (N[f,g],N[g,f])$.

Then, with (5.2.4.12), (5.2.4.4), (5.2.4.13a), and (5.2.4.5), we get: $c \to a \to d \to e \to b$ is a path from alternative *c* to alternative *b* with a strength of more than $(N[f,g],N[g,f])$. But this is a contradiction to the presumption that *fg* is the weakest link in the strongest path from alternative *c* to alternative *b*.

<u>Case 2:</u> Suppose

(5.2.4.13b) $(N[d,e],N[e,d]) <_\sigma (N[f,g],N[g,f])$.

Then, with (5.2.4.8), (5.2.4.13b), (5.2.4.9), and (5.2.4.12), we get: $b \to f \to g \to c \to a$ is a path from alternative *b* to alternative *a* with a strength of more than $(N[d,e],N[e,d])$. But this is a contradiction to the presumption that *de* is the weakest link in the strongest path from alternative *b* to alternative *a*.

As (5.2.4.13a) and (5.2.4.13b) are not possible, we get

(5.2.4.13c) $(N[d,e],N[e,d]) \approx_\sigma (N[f,g],N[g,f])$.

As there are no links of equivalent strengths, (5.2.4.13c) means that *de* and *fg* are the same link. So to resolve (5.2.4.1) and (5.2.4.2), the same link is declared *forbidden*.

Therefore, the strongest paths have the following structure:

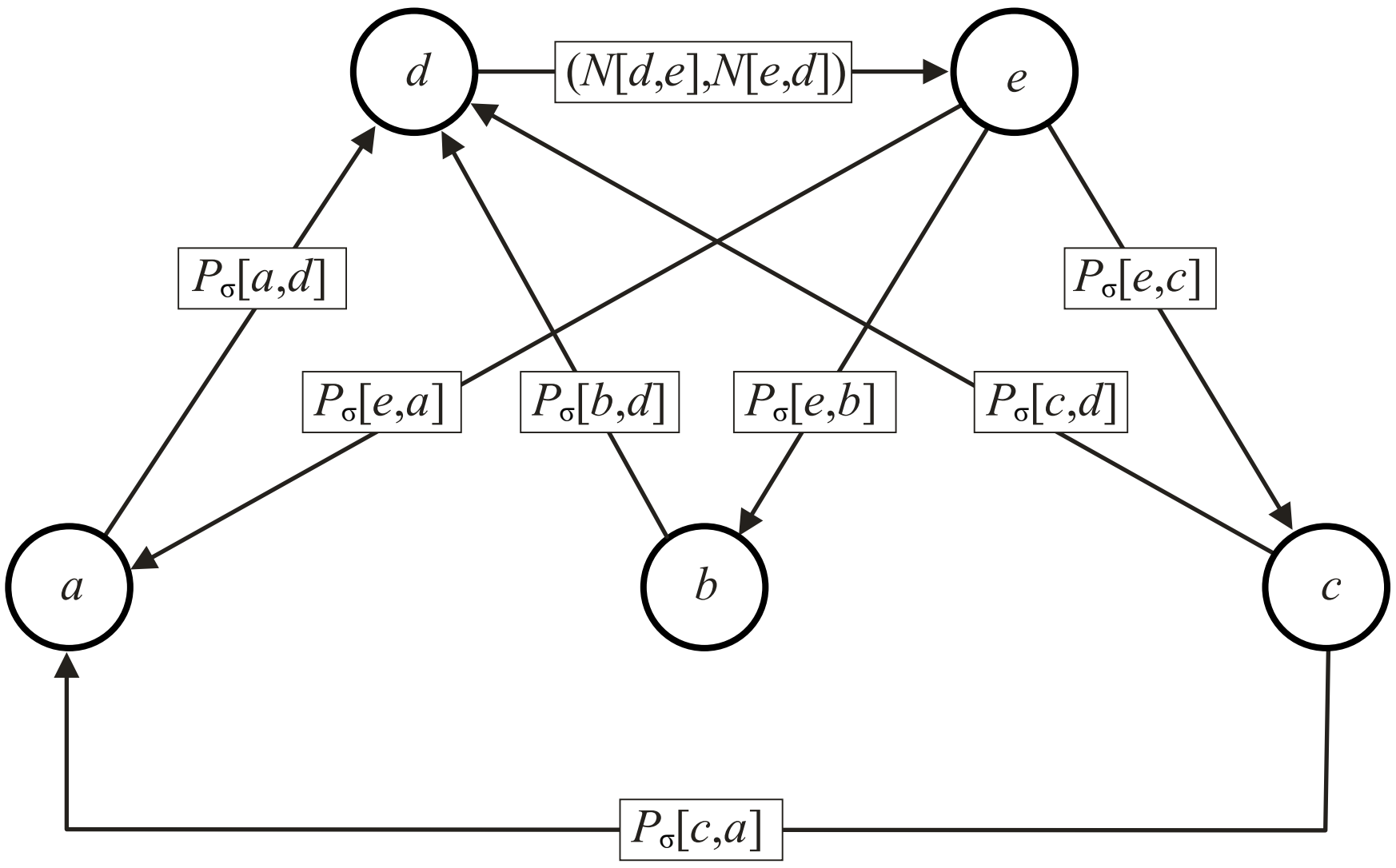

Without loss of generality, we can also say that the same link is declared *forbidden* in the process of resolving (5.2.4.3). The reason: With (5.2.4.12), we get that the link *de* cannot be in the strongest path from alternative *c* to alternative *a*. Therefore, the strongest path from alternative *c* to alternative *a* cannot be weakened by declaring the link *de forbidden*. The strongest path from alternative *a* to alternative *c* can be weakened by declaring the link *de forbidden*. But as we already know from (5.2.4.3) that the strongest path from alternative *c* to alternative *a* is stronger than the strongest path from alternative *a* to alternative *c*, declaring the link *de forbidden* cannot have an impact on the resolution of (5.2.4.3).

When the link *de* is declared *forbidden*, we get one of the following cases:

Case A: We still get $P_\sigma[a,b] \approx_\sigma P_\sigma[b,a]$ and $P_\sigma[b,c] \approx_\sigma P_\sigma[c,b]$. In this case, with the same argumentation as in cases 1–2 we get that the same link, say *de*, is the weakest link in the strongest path from alternative *a* to alternative *b*, the weakest link in the strongest path from alternative *b* to alternative *a*, the weakest link in the strongest path from alternative *b* to alternative *c*, and the weakest link in the strongest path from alternative *c* to alternative *b*. So we can proceed with declaring the link *de forbidden* until we get one of the cases B–G.

Case B: We get ( $P_\sigma[a,b] \prec_\sigma P_\sigma[b,a]$ and $P_\sigma[b,c] \prec_\sigma P_\sigma[c,b]$ ) or ( $P_\sigma[a,b] \prec_\sigma P_\sigma[b,a]$ and $P_\sigma[b,c] \succ_\sigma P_\sigma[c,b]$ ) or ( $P_\sigma[a,b] \succ_\sigma P_\sigma[b,a]$ and $P_\sigma[b,c] \prec_\sigma P_\sigma[c,b]$ ). In this case, we succeeded in resolving (5.2.4.1) – (5.2.4.3) without violating transitivity.

Case C: We get $P_\sigma[a,b] \succ_\sigma P_\sigma[b,a]$ and $P_\sigma[b,c] \approx_\sigma P_\sigma[c,b]$. This case is not possible because, after the link *de* has been declared *forbidden*, (5.2.4.1) – (5.2.4.3) are still calculated based on the same set of *non-forbidden* links. So with $P_\sigma[c,a] \succ_\sigma P_\sigma[a,c]$ and $P_\sigma[a,b] \succ_\sigma P_\sigma[b,a]$ and the transitivity, as proven in section 4.1 for cases where all paths are based on the same set of *non-forbidden* links, we would immediately get $P_\sigma[b,c] \prec_\sigma P_\sigma[c,b]$.

Case D: We get $P_\sigma[a,b] \approx_\sigma P_\sigma[b,a]$ and $P_\sigma[b,c] \succ_\sigma P_\sigma[c,b]$. This case is not possible because, after the link *de* has been declared *forbidden*, (5.2.4.1) – (5.2.4.3) are still calculated based on the same set of *non-forbidden* links. So with $P_\sigma[c,a] \succ_\sigma P_\sigma[a,c]$ and $P_\sigma[b,c] \succ_\sigma P_\sigma[c,b]$ and the transitivity, as proven in section 4.1 for cases where all paths are based on the same set of *non-forbidden* links, we would immediately get $P_\sigma[a,b] \prec_\sigma P_\sigma[b,a]$.

Case E: We get $P_\sigma[a,b] \succ_\sigma P_\sigma[b,a]$ and $P_\sigma[b,c] \succ_\sigma P_\sigma[c,b]$. This case is not possible because, after the link *de* has been declared *forbidden*, (5.2.4.1) – (5.2.4.3) are still calculated based on the same set of *non-forbidden* links. So $P_\sigma[a,b] \succ_\sigma P_\sigma[b,a]$, $P_\sigma[b,c] \succ_\sigma P_\sigma[c,b]$, and $P_\sigma[c,a] \succ_\sigma P_\sigma[a,c]$ together violate transitivity, as proven in section 4.1 for cases where all paths are based on the same set of *non-forbidden* links.

Case F: We get $P_\sigma[a,b] \approx_\sigma P_\sigma[b,a]$ and $P_\sigma[b,c] \prec_\sigma P_\sigma[c,b]$. This case is identical to the situation in section 5.2.2. It is possible that $P_\sigma[a,b] \approx_\sigma P_\sigma[b,a]$ is resolved based on a different set of *non-forbidden* links. However, this is not a problem because — it doesn't matter whether $P_\sigma[a,b] \approx_\sigma P_\sigma[b,a]$ is resolved to $P_\sigma[a,b] \succ_\sigma P_\sigma[b,a]$ or to $P_\sigma[a,b] \prec_\sigma P_\sigma[b,a]$ — transitivity will never be violated.

Case G: We get $P_\sigma[a,b] \prec_\sigma P_\sigma[b,a]$ and $P_\sigma[b,c] \approx_\sigma P_\sigma[c,b]$. This case is identical to the situation in section 5.2.3. It is possible that $P_\sigma[b,c] \approx_\sigma P_\sigma[c,b]$ is resolved based on a different set of *non-forbidden* links. However, this is not a problem because — it doesn't matter whether $P_\sigma[b,c] \approx_\sigma P_\sigma[c,b]$ is resolved to $P_\sigma[b,c] \succ_\sigma P_\sigma[c,b]$ or to $P_\sigma[b,c] \prec_\sigma P_\sigma[c,b]$ — transitivity will never be violated.

The following table shows that cases A–G cover all possible combinations. Therefore, it has been proven for every possible situation that, when we resolve (5.2.4.1) – (5.2.4.3) as prescribed in section 5.1, then transitivity will never be violated.

| | |
|---|---|
| $P_\sigma[a,b] \approx_\sigma P_\sigma[b,a]$ and $P_\sigma[b,c] \approx_\sigma P_\sigma[c,b]$ | → case A |
| $P_\sigma[a,b] \approx_\sigma P_\sigma[b,a]$ and $P_\sigma[b,c] \succ_\sigma P_\sigma[c,b]$ | → case D |
| $P_\sigma[a,b] \approx_\sigma P_\sigma[b,a]$ and $P_\sigma[b,c] \prec_\sigma P_\sigma[c,b]$ | → case F |
| $P_\sigma[a,b] \succ_\sigma P_\sigma[b,a]$ and $P_\sigma[b,c] \approx_\sigma P_\sigma[c,b]$ | → case C |
| $P_\sigma[a,b] \succ_\sigma P_\sigma[b,a]$ and $P_\sigma[b,c] \succ_\sigma P_\sigma[c,b]$ | → case E |
| $P_\sigma[a,b] \succ_\sigma P_\sigma[b,a]$ and $P_\sigma[b,c] \prec_\sigma P_\sigma[c,b]$ | → case B |
| $P_\sigma[a,b] \prec_\sigma P_\sigma[b,a]$ and $P_\sigma[b,c] \approx_\sigma P_\sigma[c,b]$ | → case G |
| $P_\sigma[a,b] \prec_\sigma P_\sigma[b,a]$ and $P_\sigma[b,c] \succ_\sigma P_\sigma[c,b]$ | → case B |
| $P_\sigma[a,b] \prec_\sigma P_\sigma[b,a]$ and $P_\sigma[b,c] \prec_\sigma P_\sigma[c,b]$ | → case B |

□

### 5.2.5. Part 5

Suppose, before we start declaring links *forbidden*, we have:

(5.2.5.1) $P_{\sigma}[a,b] \approx_{\sigma} P_{\sigma}[b,a]$.

(5.2.5.2) $P_{\sigma}[b,c] \approx_{\sigma} P_{\sigma}[c,b]$.

(5.2.5.3) $P_{\sigma}[c,a] \approx_{\sigma} P_{\sigma}[a,c]$.

**Claim:**

Suppose (5.2.5.1) – (5.2.5.3) are resolved as prescribed in section 5.1. Then transitivity will never be violated.

**Proof:**

Suppose there are no pairwise links of equivalent strengths. Suppose (5.2.5.1) – (5.2.5.3). With (5.2.5.1), we get that the weakest link in the strongest path from alternative *a* to alternative *b* and the weakest link in the strongest path from alternative *b* to alternative *a* must be the same link, say *de*. With (5.2.5.2), we get that the weakest link in the strongest path from alternative *b* to alternative *c* and the weakest link in the strongest path from alternative *c* to alternative *b* must be the same link, say *fg*. With (5.2.5.3), we get that the weakest link in the strongest path from alternative *c* to alternative *a* and the weakest link in the strongest path from alternative *a* to alternative *c* must be the same link, say *hi*.

Therefore, the strongest paths have the following structure:

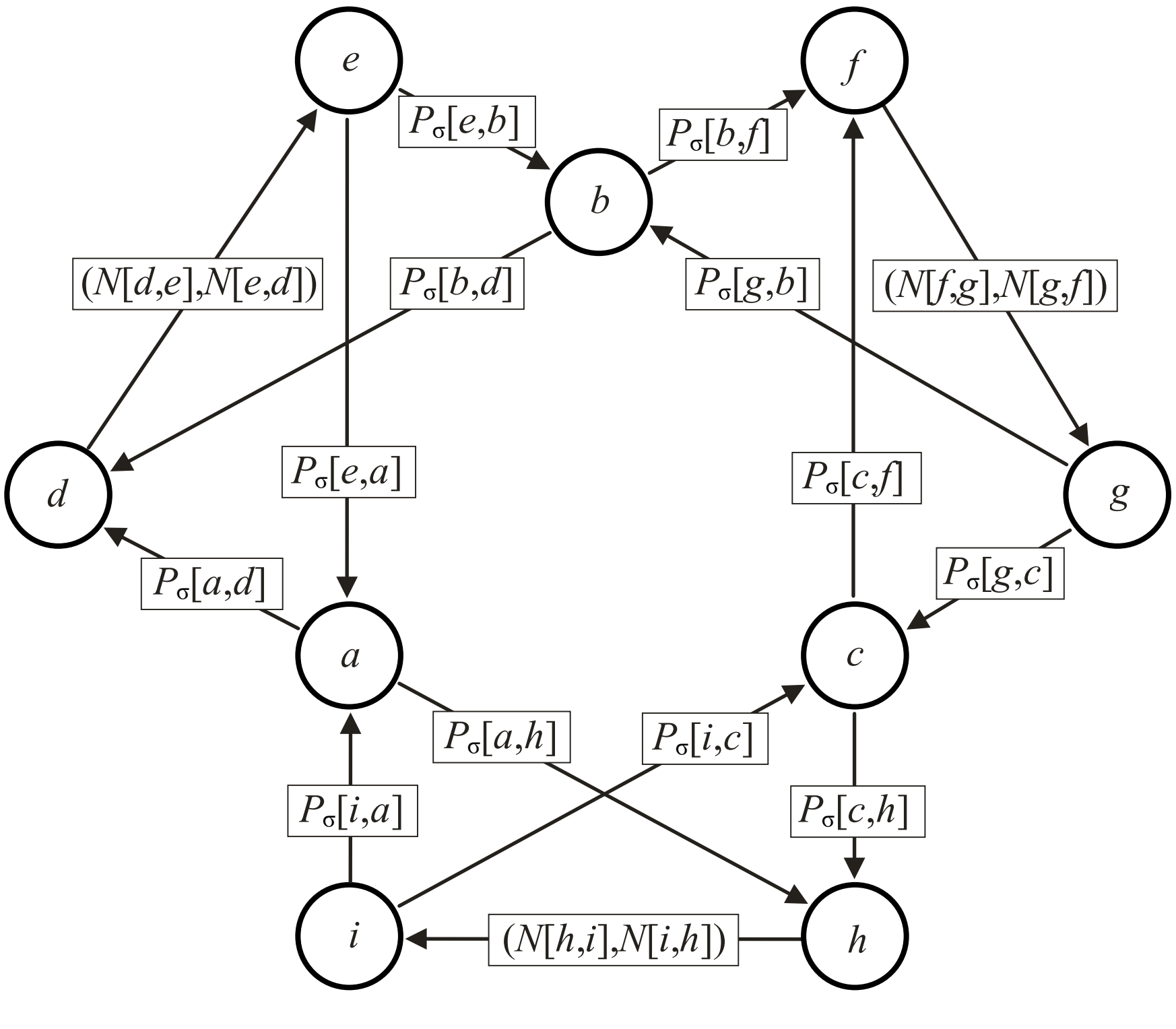

As *de* is the weakest link in the strongest path from alternative *a* to alternative *b*, we get

(5.2.5.4) $P_{\sigma}[a,d] >_{\sigma} (N[d,e],N[e,d])$.

(5.2.5.5) $P_{\sigma}[e,b] >_{\sigma} (N[d,e],N[e,d])$.

As *de* is the weakest link in the strongest path from alternative *b* to alternative *a*, we get

(5.2.5.6) $P_{\sigma}[b,d] >_{\sigma} (N[d,e],N[e,d])$.

(5.2.5.7) $P_{\sigma}[e,a] >_{\sigma} (N[d,e],N[e,d])$.

As *fg* is the weakest link in the strongest path from alternative *b* to alternative *c*, we get

(5.2.5.8) $P_{\sigma}[b,f] >_{\sigma} (N[f,g],N[g,f])$.

(5.2.5.9) $P_{\sigma}[g,c] >_{\sigma} (N[f,g],N[g,f])$.

As *fg* is the weakest link in the strongest path from alternative *c* to alternative *b*, we get

(5.2.5.10) $P_{\sigma}[c,f] >_{\sigma} (N[f,g],N[g,f])$.

(5.2.5.11) $P_{\sigma}[g,b] >_{\sigma} (N[f,g],N[g,f])$.

As *hi* is the weakest link in the strongest path from alternative *c* to alternative *a*, we get

(5.2.5.12) $P_{\sigma}[c,h] >_{\sigma} (N[h,i],N[i,h])$.

(5.2.5.13) $P_{\sigma}[i,a] >_{\sigma} (N[h,i],N[i,h])$.

As *hi* is the weakest link in the strongest path from alternative *a* to alternative *c*, we get

(5.2.5.14) $P_{\sigma}[a,h] >_{\sigma} (N[h,i],N[i,h])$.

(5.2.5.15) $P_{\sigma}[i,c] >_{\sigma} (N[h,i],N[i,h])$.

Case 1: Suppose

(5.2.5.16a) $(N[d,e],N[e,d]) <_{\sigma} (N[f,g],N[g,f])$.

(5.2.5.17a) $(N[d,e],N[e,d]) <_{\sigma} (N[h,i],N[i,h])$.

Then, with (5.2.5.14), (5.2.5.17a), (5.2.5.15), (5.2.5.10), (5.2.5.16a), and (5.2.5.11), we get: $a \rightarrow h \rightarrow i \rightarrow c \rightarrow f \rightarrow g \rightarrow b$ is a path from alternative *a* to alternative *b* with a strength of more than $(N[d,e],N[e,d])$. But this is a contradiction to the presumption that *de* is the weakest link in the strongest path from alternative *a* to alternative *b*.

Similarly, with (5.2.5.8), (5.2.5.16a), (5.2.5.9), (5.2.5.12), (5.2.5.17a), and (5.2.5.13), we get: $b \rightarrow f \rightarrow g \rightarrow c \rightarrow h \rightarrow i \rightarrow a$ is a path from alternative *b* to alternative *a* with a strength of more than $(N[d,e],N[e,d])$. But this is a contradiction to the presumption that *de* is the weakest link in the strongest path from alternative *b* to alternative *a*.

Case 2: Suppose

(5.2.5.16b) $(N[f,g],N[g,f]) <_{\sigma} (N[d,e],N[e,d])$.

(5.2.5.17b) $(N[f,g],N[g,f]) <_{\sigma} (N[h,i],N[i,h])$.

Then, with (5.2.5.6), (5.2.5.16b), (5.2.5.7), (5.2.5.14), (5.2.5.17b), and (5.2.5.15), we get: $b \rightarrow d \rightarrow e \rightarrow a \rightarrow h \rightarrow i \rightarrow c$ is a path from alternative *b* to alternative *c* with a strength of more than $(N[f,g],N[g,f])$. But this is a contradiction to the presumption that *fg* is the weakest link in the strongest path from alternative *b* to alternative *c*.

Similarly, with (5.2.5.12), (5.2.5.17b), (5.2.5.13), (5.2.5.4), (5.2.5.16b), and (5.2.5.5), we get: $c \rightarrow h \rightarrow i \rightarrow a \rightarrow d \rightarrow e \rightarrow b$ is a path from alternative *c* to alternative *b* with a strength of more than $(N[f,g],N[g,f])$. But this is a contradiction to the presumption that *fg* is the weakest link in the strongest path from alternative *c* to alternative *b*.

Case 3: Suppose

(5.2.5.16c) $(N[h,i],N[i,h]) <_{\sigma} (N[d,e],N[e,d])$.

(5.2.5.17c) $(N[h,i],N[i,h]) <_{\sigma} (N[f,g],N[g,f])$.

Then, with (5.2.5.10), (5.2.5.17c), (5.2.5.11), (5.2.5.6), (5.2.5.16c), and (5.2.5.7), we get: $c \rightarrow f \rightarrow g \rightarrow b \rightarrow d \rightarrow e \rightarrow a$ is a path from alternative *c* to alternative *a* with a strength of more than $(N[h,i],N[i,h])$. But this is a contradiction to the presumption that *hi* is the weakest link in the strongest path from alternative *c* to alternative *a*.

Similarly, with (5.2.5.4), (5.2.5.16c), (5.2.5.5), (5.2.5.8), (5.2.5.17c), and (5.2.5.9), we get: $a \rightarrow d \rightarrow e \rightarrow b \rightarrow f \rightarrow g \rightarrow c$ is a path from alternative *a* to alternative *c* with a strength of more than $(N[h,i],N[i,h])$. But this is a contradiction to the presumption that *hi* is the weakest link in the strongest path from alternative *a* to alternative *c*.

With cases 1–3, we get that none of the links *de*, *fg*, *hi* can be weaker than each of the other two links. Without loss of generality, we can presume that the link *hi* is the strongest one of the links *de*, *fg*, *hi*. So we get

(5.2.5.18) $(N[d,e],N[e,d]) \approx_\sigma (N[f,g],N[g,f]) \precsim_\sigma (N[h,i],N[i,h])$.

We can ignore the case $(N[d,e],N[e,d]) \approx_\sigma (N[f,g],N[g,f]) \approx_\sigma (N[h,i], N[i,h])$ because in this case the links *de*, *fg*, *hi* are the same link so that for each of (5.2.5.1) – (5.2.5.3) the same link is declared *forbidden* first so that, afterwards, each of (5.2.5.1) – (5.2.5.3) is still resolved based on the same set of *non-forbidden* links.

So without loss of generality, we get

(5.2.5.19) $(N[d,e],N[e,d]) \approx_\sigma (N[f,g],N[g,f]) \prec_\sigma (N[h,i],N[i,h])$.

As there are no links of equivalent strengths, (5.2.5.19) means that the link *de* and the link *fg* must be the same link. Therefore, the strongest paths have the following structure:

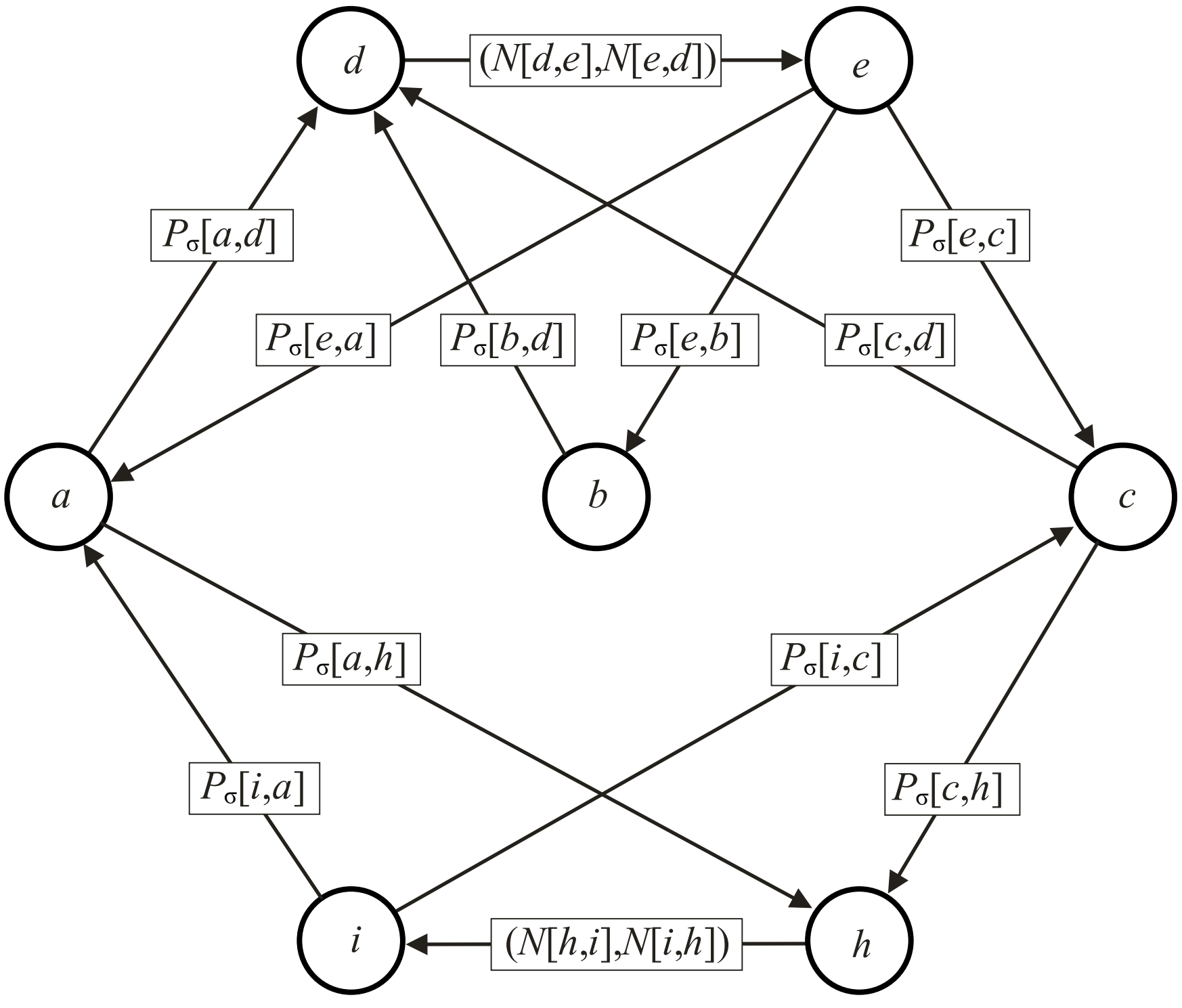

Without loss of generality, we can also say that, when we resolve (5.2.5.1) – (5.2.5.3), then, at each stage, the weakest of the weakest links of the current strongest paths is declared *forbidden*. So in our situation, the link *de* is declared *forbidden* next.

Since $(N[d,e],N[e,d]) \prec_\sigma (N[h,i],N[i,h]) \approx_\sigma P_\sigma[c,a] \approx_\sigma P_\sigma[a,c]$, the link *de* cannot be in the strongest path from alternative *c* to alternative *a* or in the strongest path from alternative *a* to alternative *c*. Therefore, declaring the link *de forbidden* cannot have an impact on the strongest path from alternative *c* to alternative *a* or on the strongest path from alternative *a* to alternative *c*.

When the link *de* is declared *forbidden*, we get one of the following cases:

Case A: We still get $P_\sigma[a,b] \approx_\sigma P_\sigma[b,a]$ and $P_\sigma[b,c] \approx_\sigma P_\sigma[c,b]$. In this case, with the same argumentation as in cases 1–2 we get that the same link, say *de*, is the weakest link in the strongest path from alternative *a* to alternative *b*, the weakest link in the strongest path from alternative *b* to alternative *a*, the weakest link in the strongest path from alternative *b* to alternative *c*, and the weakest link in the strongest path from alternative *c* to alternative *b*. So we can proceed with declaring the link *de forbidden* until we get one of the cases B–F.

Case B: We get $P_\sigma[a,b] \succ_\sigma P_\sigma[b,a]$ and $P_\sigma[b,c] \succ_\sigma P_\sigma[c,b]$. This case is not possible because, after the link *de* has been declared *forbidden*, (5.2.5.1) – (5.2.5.3) are still calculated based on the same set of *non-forbidden* links. With $P_\sigma[a,b] \succ_\sigma P_\sigma[b,a]$ and $P_\sigma[b,c] \succ_\sigma P_\sigma[c,b]$ and the transitivity, as proven in section 4.1 for cases where all paths are based on the same set of *non-forbidden* links, we would immediately get $P_\sigma[c,a] \prec_\sigma P_\sigma[a,c]$. But this is a contradiction to the fact that the link *de* cannot have been in the strongest path from alternative *c* to alternative *a* or in the strongest path from alternative *a* to alternative *c*, so that declaring the link *de forbidden* cannot have an impact on $P_\sigma[c,a] \approx_\sigma P_\sigma[a,c]$.

Case C: We get $P_\sigma[a,b] \prec_\sigma P_\sigma[b,a]$ and $P_\sigma[b,c] \prec_\sigma P_\sigma[c,b]$. This case is not possible because, after the link *de* has been declared *forbidden*, (5.2.5.1) – (5.2.5.3) are still calculated based on the same set of *non-forbidden* links. With $P_\sigma[a,b] \prec_\sigma P_\sigma[b,a]$ and $P_\sigma[b,c] \prec_\sigma P_\sigma[c,b]$ and the transitivity, as proven in section 4.1 for cases where all paths are based on the same set of *non-forbidden* links, we would immediately get $P_\sigma[c,a] \succ_\sigma P_\sigma[a,c]$. But this is a contradiction to the fact that the link *de* cannot have been in the strongest path from alternative *c* to alternative *a* or in the strongest path from alternative *a* to alternative *c*, so that declaring the link *de forbidden* cannot have an impact on $P_\sigma[c,a] \approx_\sigma P_\sigma[a,c]$.

Case D: We get ( $P_\sigma[a,b] \succ_\sigma P_\sigma[b,a]$ and $P_\sigma[b,c] \approx_\sigma P_\sigma[c,b]$ ) or ( $P_\sigma[a,b] \prec_\sigma P_\sigma[b,a]$ and $P_\sigma[b,c] \approx_\sigma P_\sigma[c,b]$ ) or ( $P_\sigma[a,b] \approx_\sigma P_\sigma[b,a]$ and $P_\sigma[b,c] \succ_\sigma P_\sigma[c,b]$ ) or ( $P_\sigma[a,b] \approx_\sigma P_\sigma[b,a]$ and $P_\sigma[b,c] \prec_\sigma P_\sigma[c,b]$ ). This case is not possible because we have seen in (5.2.4.13a) – (5.2.4.13c) that, when we have a situation with $P_\sigma[x,y] \approx_\sigma P_\sigma[y,x]$, $P_\sigma[y,z] \approx_\sigma P_\sigma[z,y]$, and $P_\sigma[z,x] \succ_\sigma P_\sigma[x,z]$, then the weakest link in the strongest path from alternative $x$ to alternative $y$, the weakest link in the strongest path from alternative $y$ to alternative $x$, the weakest link in the strongest path from alternative $y$ to alternative $z$, and the weakest link in the strongest path from alternative $z$ to alternative $y$ must be the same link. But this is not possible because (5.2.5.19) says that the link *hi* is stronger than the link *de*.

Case E: We get $P_\sigma[a,b] \prec_\sigma P_\sigma[b,a]$ and $P_\sigma[b,c] \succ_\sigma P_\sigma[c,b]$. This case is identical to the situation in section 5.2.2. It is possible that $P_\sigma[c,a] \approx_\sigma P_\sigma[a,c]$ is resolved based on a different set of *non-forbidden* links. However, this is not a problem because — it doesn't matter whether $P_\sigma[a,c] \approx_\sigma P_\sigma[c,a]$ is resolved to $P_\sigma[a,c] \succ_\sigma P_\sigma[c,a]$ or to $P_\sigma[a,c] \prec_\sigma P_\sigma[c,a]$ — transitivity will never be violated.

Case F: We get $P_\sigma[a,b] \succ_\sigma P_\sigma[b,a]$ and $P_\sigma[b,c] \prec_\sigma P_\sigma[c,b]$. This case is identical to the situation in section 5.2.3. It is possible that $P_\sigma[c,a] \approx_\sigma P_\sigma[a,c]$ is resolved based on a different set of *non-forbidden* links. However, this is not a problem because — it doesn't matter whether $P_\sigma[a,c] \approx_\sigma P_\sigma[c,a]$ is resolved to $P_\sigma[a,c] \succ_\sigma P_\sigma[c,a]$ or to $P_\sigma[a,c] \prec_\sigma P_\sigma[c,a]$ — transitivity will never be violated.

The following table shows that cases A–F cover all possible combinations. Therefore, it has been proven for every possible situation that, when we resolve (5.2.5.1) – (5.2.5.3) as prescribed in section 5.1, then transitivity will never be violated.

| | |
|---|---|
| $P_\sigma[a,b] \approx_\sigma P_\sigma[b,a]$ and $P_\sigma[b,c] \approx_\sigma P_\sigma[c,b]$ | → case A |
| $P_\sigma[a,b] \approx_\sigma P_\sigma[b,a]$ and $P_\sigma[b,c] \succ_\sigma P_\sigma[c,b]$ | → case D |
| $P_\sigma[a,b] \approx_\sigma P_\sigma[b,a]$ and $P_\sigma[b,c] \prec_\sigma P_\sigma[c,b]$ | → case D |
| $P_\sigma[a,b] \succ_\sigma P_\sigma[b,a]$ and $P_\sigma[b,c] \approx_\sigma P_\sigma[c,b]$ | → case D |
| $P_\sigma[a,b] \succ_\sigma P_\sigma[b,a]$ and $P_\sigma[b,c] \succ_\sigma P_\sigma[c,b]$ | → case B |
| $P_\sigma[a,b] \succ_\sigma P_\sigma[b,a]$ and $P_\sigma[b,c] \prec_\sigma P_\sigma[c,b]$ | → case F |
| $P_\sigma[a,b] \prec_\sigma P_\sigma[b,a]$ and $P_\sigma[b,c] \approx_\sigma P_\sigma[c,b]$ | → case D |
| $P_\sigma[a,b] \prec_\sigma P_\sigma[b,a]$ and $P_\sigma[b,c] \succ_\sigma P_\sigma[c,b]$ | → case E |
| $P_\sigma[a,b] \prec_\sigma P_\sigma[b,a]$ and $P_\sigma[b,c] \prec_\sigma P_\sigma[c,b]$ | → case C |

□

## 5.3. Analysis

### 5.3.1. The Probabilistic Framework

An election method is simply a mapping from some input to some output. In section 2.1, we presumed that the output is (1) a strict partial order $\mathcal{O}$ on $A$ and (2) a set $\emptyset \neq \mathcal{S} \subseteq A$ of potential winners. In the probabilistic framework, the output of an election method is a probability distribution $p[\mathcal{O}] \in \mathbb{R}$ on $\mathcal{LO}_A$, where $\mathcal{LO}_A$ is the set of linear orders on $A$.

We get

(5.3.1.1) $\forall\ \mathcal{O} \in \mathcal{LO}_A$: $p[\mathcal{O}] \geq 0$.

(5.3.1.2) $\sum$ ( $p[\mathcal{O}]$ | $\mathcal{O} \in \mathcal{LO}_A$ ) = 1.

Suppose $q[a,b] \in \mathbb{R}$ is the probability for $ab \in \mathcal{O}$ ( i.e. the probability that alternative $a \in A$ is ranked ahead of alternative $b \in A \setminus \{a\}$ in the collective ranking $\mathcal{O}$ ).

Then, we get

(5.3.1.3) $q[a,b]$ : = $\sum$ ( $p[\mathcal{O}]$ | $\mathcal{O} \in \mathcal{LO}_A$ with $ab \in \mathcal{O}$ ).

(5.3.1.4) $\forall\ a,b \in A$: $q[a,b] \geq 0$.

(5.3.1.5) $\forall\ a,b \in A$: $q[a,b] + q[b,a] = 1$.

Suppose $r[a] \in \mathbb{R}$ is the probability that alternative $a \in A$ is elected.

Then, we get

(5.3.1.6) $r[a]$ : = $\sum$ ( $p[\mathcal{O}]$ | $\mathcal{O} \in \mathcal{LO}_A$ with $ab \in \mathcal{O}$ for all $b \in A \setminus \{a\}$ ).

(5.3.1.7) $\forall\ a \in A$: $r[a] \geq 0$.

(5.3.1.8) $\sum$ ( $r[a]$ | $a \in A$ ) = 1.

### 5.3.2. Decisiveness

**Definition:**

An election method satisfies *decisiveness* if ( for every given number of alternatives ) the proportion of profiles without a linear order $O \in \mathcal{LO}_A$ with $p[O] = 1$ tends to zero as the number of voters in the profile tends to infinity.

**Claim:**

If $\succ_D$ satisfies (2.1.1), then the Schulze method $O_{final} := \cap\, O_{final}(\sigma)$, as defined in section 5.1, satisfies decisiveness.

**Proof (overview):**

1. Suppose the number of alternatives is fixed. We prove that, when the number of voters in the profile tends to infinity, the probability, that there are links with equivalent strengths, goes to zero. We then get: The probability, that there are links $ef$ and $gh$ with $ef \approx_\sigma gh$, goes to zero.

2. We prove that (1) the link $ij$ cannot be in the strongest path from alternative $j$ to alternative $i$ and (2) the link $ji$ cannot be in the strongest path from alternative $i$ to alternative $j$. Therefore, when we resolve the tie $P_\sigma[i,j] \approx_\sigma P_\sigma[j,i]$, it can neither happen that the link $ij$ is declared *forbidden* nor that the link $ji$ is declared *forbidden*. Therefore, in worst case, when there are no other paths of *non-forbidden* links anymore, $P_\sigma[i,j] \approx_\sigma P_\sigma[j,i]$ is resolved to $ij \in O$ when $ij \succ_\sigma ji$ and to $ji \in O$ when $ij \prec_\sigma ji$. So the algorithm in section 5.1 step 4 always terminates before all links have been declared *forbidden*.

**Remark:**

When there is a unique linear order ( i.e. a linear order $O \in \mathcal{LO}_A$ with $p[O] = 1$ ) then, with (5.3.1.6), we get that there is also a unique winner ( i.e. an alternative $a \in A$ with $r[a] = 1$ ):

$$( \exists\, O \in \mathcal{LO}_A\colon p[O] = 1 ) \Rightarrow ( \exists\, a \in A\colon r[a] = 1 ).$$

### 5.3.3. Pareto

In the probabilistic framework, the *Pareto criterion* says that, when no voter strictly prefers alternative $b \in A$ to alternative $a \in A$ [see (5.3.3.1)] and at least one voter strictly prefers alternative $a$ to alternative $b$ [see (5.3.3.2)], then $r[b] = 0$.

**Definition:**

An election method satisfies the *Pareto criterion* if the following holds:

Suppose:

(5.3.3.1) $\forall\, v \in V: a \succsim_v b.$

(5.3.3.2) $\exists\, v \in V: a \succ_v b.$

Then:

(5.3.3.3) $q[a,b] = 1.$

(5.3.3.4) $\forall\, f \in A \setminus \{a,b\}: q[a,f] \geq q[b,f].$

(5.3.3.5) $\forall\, f \in A \setminus \{a,b\}: q[f,a] \leq q[f,b].$

(5.3.3.6) $r[b] = 0.$

**Claim:**

If $\succ_D$ satisfies (2.1.1), then the Schulze method $O_{final}(\sigma)$, as defined in section 5.1, satisfies the Pareto criterion.

**Proof (overview):**

With (5.1.1) and (5.3.3.1), we prove

(5.3.3.7) $\forall\, e \in A \setminus \{a,b\}: ae \succsim_\sigma be$ with certainty.

With (5.1.1) and (5.3.3.1), we prove

(5.3.3.8) $\forall\, e \in A \setminus \{a,b\}: eb \succsim_\sigma ea$ with certainty.

With (5.1.1), (5.3.3.1), and (5.3.3.2), we prove

(5.3.3.9) $ab \succ_\sigma ba$ with certainty.

With (5.3.3.7), (5.3.3.8), and (5.3.3.9), we prove

(5.3.3.10) $ab \in O_{final}(\sigma)$ with certainty.

The rest of the proof is identical to the proof in section 4.4.2.

### 5.3.4. Reversal Symmetry

In the probabilistic framework, *reversal symmetry* says that, when $\succ_v$ is reversed for all $v \in V$, then $r^{old}[a] + r^{new}[a] \leq 1$ for all $a \in A$. Otherwise, if $r^{old}[a] + r^{new}[a]$ was larger than 1 for some alternative $a \in A$, then this would mean that, with a probability of at least $r^{old}[a] + r^{new}[a] - 1 > 0$, alternative $a$ is identified as best alternative and, simultaneously, identified as worst alternative.

Suppose $O^{reverse} \in \mathcal{LO}_A$ is the reversal of $O \in \mathcal{LO}_A$.

That means:

(5.3.4.1) $\quad \forall\ a,b \in A: ab \in O \Leftrightarrow ba \in O^{reverse}.$

**<u>Definition:</u>**

An election method satisfies *reversal symmetry* if the following holds:

Suppose:

(5.3.4.2) $\quad \forall\ e,f \in A\ \forall\ v \in V: e \succ_v^{old} f \Leftrightarrow f \succ_v^{new} e.$

Then:

(5.3.4.3) $\quad \forall\ O \in \mathcal{LO}_A: p^{old}[O] = p^{new}[O^{reverse}].$

(5.3.4.4) $\quad \forall\ a,b \in A: q^{old}[a,b] = q^{new}[b,a].$

(5.3.4.5) $\quad \forall\ a \in A: r^{old}[a] + r^{new}[a] \leq 1.$

**<u>Claim:</u>**

Suppose $\succ_D$ satisfies (2.1.2). Then the Schulze method $O_{final}(\sigma)$, as defined in sections 5.1, satisfies reversal symmetry.

**<u>Proof (overview):</u>**

(2.1.2) guarantees that, when $\succ_v$ is reversed for all $v \in V$, then also the TBRL $\succ_\sigma$, as defined in section 5.1, is reversed. So the probability that $O$ is chosen in the original situation is identical to the probability that $O^{reverse}$ is chosen in the reversed situation. As we have presumed in section 2.1 that there are at least 2 alternatives in $A$, $a \in A$ cannot be the maximum element of $O$ and simultaneously the maximum element of $O^{reverse}$. Therefore, we get (5.3.4.5).

**Example 13:**

(α) When we apply the proposed method to example 13 (section 3.13), we first calculate the TBRL $\succ_\sigma$.

We have:

$(N[b,c],N[c,b]) \approx_D (4,1)$.
$(N[a,b],N[b,a]) \approx_D (3,2)$.
$(N[c,a],N[a,c]) \approx_D (3,2)$.
$(N[a,c],N[c,a]) \approx_D (2,3)$.
$(N[b,a],N[a,b]) \approx_D (2,3)$.
$(N[c,b],N[b,c]) \approx_D (1,4)$.

So we start with $bc \succ_\sigma ab \approx_\sigma ca \succ_\sigma ac \approx_\sigma ba \succ_\sigma cb$.

Case I: With a probability of 2/5, one of the $a \succ_v b \succ_v c$ voters is chosen first. $ab \approx_\sigma ca$ is then completed to $ab \succ_\sigma ca$ because this voter supports the link *ab* and opposes the link *ca*. $ac \approx_\sigma ba$ is completed to $ac \succ_\sigma ba$ because this voter supports the link *ac* and opposes the link *ba*. So the TBRL $\succ_\sigma$ is completed to $bc \succ_\sigma ab \succ_\sigma ca \succ_\sigma ac \succ_\sigma ba \succ_\sigma cb$.

Case II: With a probability of 2/5, one of the $b \succ_v c \succ_v a$ voters is chosen first. $ab \approx_\sigma ca$ is then completed to $ca \succ_\sigma ab$ because this voter supports the link *ca* and opposes the link *ab*. $ac \approx_\sigma ba$ is completed to $ba \succ_\sigma ac$ because this voter supports the link *ba* and opposes the link *ac*. So the TBRL $\succ_\sigma$ is completed to $bc \succ_\sigma ca \succ_\sigma ab \succ_\sigma ba \succ_\sigma ac \succ_\sigma cb$.

Case III: With a probability of 1/5, the $c \succ_v a \succ_v b$ voter is chosen first. As this voter supports both links *ab* and *ca*, this voter cannot be used to complete $ab \approx_\sigma ca$. As this voter opposes both links *ac* and *ba*, this voter cannot be used to complete $ac \approx_\sigma ba$. With a probability of 1/2, one of the $a \succ_v b \succ_v c$ voters is chosen second; the TBRL $\succ_\sigma$ is then completed to $bc \succ_\sigma ab \succ_\sigma ca \succ_\sigma ac \succ_\sigma ba \succ_\sigma cb$ as described in Case I. With a probability of 1/2, one of the $b \succ_v c \succ_v a$ voters is chosen second; the TBRL $\succ_\sigma$ is then completed to $bc \succ_\sigma ca \succ_\sigma ab \succ_\sigma ba \succ_\sigma ac \succ_\sigma cb$ as described in Case II.

So with a probability of 1/2, the TBRL $\succ_\sigma$ is completed to $bc \succ_\sigma ab \succ_\sigma ca \succ_\sigma ac \succ_\sigma ba \succ_\sigma cb$ and, with a probability of 1/2, the TBRL $\succ_\sigma$ is completed to $bc \succ_\sigma ca \succ_\sigma ab \succ_\sigma ba \succ_\sigma ac \succ_\sigma cb$.

(β) Suppose the TBRL $bc \succ_\sigma ab \succ_\sigma ca \succ_\sigma ac \succ_\sigma ba \succ_\sigma cb$ is used. The weakest link in the strongest path from alternative *a* to alternative *b* is *ab*. The weakest link in the strongest path from alternative *b* to alternative *a* is *ca*. As $ab \succ_\sigma ca$, we get $P_D[a,b] \succ_D P_D[b,a]$. Alternative *a* is the final winner.

Suppose the TBRL $bc \succ_\sigma ca \succ_\sigma ab \succ_\sigma ba \succ_\sigma ac \succ_\sigma cb$ is used. The weakest link in the strongest path from alternative *a* to alternative *b* is *ab*. The weakest link in the strongest path from alternative *b* to alternative *a* is *ca*. As $ca \succ_\sigma ab$, we get $P_D[b,a] \succ_D P_D[a,b]$. Alternative *b* is the final winner.

So in example 13, we get: $r^{old}[a] = 0.5$ and $r^{old}[b] = 0.5$.

(γ) When the individual ballots are reversed, we get:

Example 13 (new):

| | | |
|---|---|---|
| 2 | voters | $c \succ_v b \succ_v a$ |
| 2 | voters | $a \succ_v c \succ_v b$ |
| 1 | voter | $b \succ_v a \succ_v c$ |

When we rename the alternatives *b* and *c* and reorder the voters, we see that example 13 (new) is identical to example 13. So with anonymity and neutrality, we get $r^{new}[a] = r^{old}[a]$, $r^{new}[b] = r^{old}[c]$, and $r^{new}[c] = r^{old}[b]$. So we get: $r^{new}[a] = 0.5$ and $r^{new}[c] = 0.5$.

(δ) The interesting conclusion is that anonymity, neutrality, and reversal symmetry together imply $r^{old}[a] \leq 0.5$ in example 13, because anonymity and neutrality together imply $r^{new}[a] = r^{old}[a]$ and reversal symmetry implies $r^{old}[a] + r^{new}[a] \leq 1$.

### 5.3.5. Monotonicity

In the probabilistic framework, *monotonicity* says that, when some voters rank alternative $a \in A$ higher [see (4.6.1.1) and (4.6.1.2)] without changing the order in which they rank the other alternatives relatively to each other [see (4.6.1.3)], then $r[a]$ must not decrease.

**Definition:**

An election method satisfies *monotonicity* if the following holds:

Suppose $a \in A$. Suppose the ballots are modified as described in (4.6.1.1) – (4.6.1.3). Then

(5.3.5.1) $\forall \emptyset \neq B \subseteq A \setminus \{a\}$:

$\sum ( p^{old}[O] \mid O \in \mathcal{LO}_A$ with $ab \in O$ for all $b \in B )$

$\leq \sum ( p^{new}[O] \mid O \in \mathcal{LO}_A$ with $ab \in O$ for all $b \in B )$.

(5.3.5.2) $\forall b \in A \setminus \{a\}$: $q^{old}[a,b] \leq q^{new}[a,b]$.

(5.3.5.3) $r^{old}[a] \leq r^{new}[a]$.

**Claim:**

If $>_D$ satisfies (2.1.1), then the Schulze method $O_{final}(\sigma)$, as defined in sections 5.1, satisfies monotonicity.

**Proof (overview):**

We prove that, when the ballots are modified as described in (4.6.1.1) – (4.6.1.3), then we get

(5.3.5.4) $\forall c,d,e \in A \setminus \{a\}$: $ac \succ_{\sigma}^{old} de \Rightarrow ac \succ_{\sigma}^{new} de$.

(5.3.5.5) $\forall c,d,e \in A \setminus \{a\}$: $ac \succsim_{\sigma}^{old} de \Rightarrow ac \succsim_{\sigma}^{new} de$.

(5.3.5.6) $\forall c,d,e \in A \setminus \{a\}$: $ca \prec_{\sigma}^{old} de \Rightarrow ca \prec_{\sigma}^{new} de$.

(5.3.5.7) $\forall c,d,e \in A \setminus \{a\}$: $ca \precsim_{\sigma}^{old} de \Rightarrow ca \precsim_{\sigma}^{new} de$.

(5.3.5.8) $\forall c,d \in A \setminus \{a\}$: $ac \succ_{\sigma}^{old} da \Rightarrow ac \succ_{\sigma}^{new} da$.

(5.3.5.9) $\forall c,d \in A \setminus \{a\}$: $ac \succsim_{\sigma}^{old} da \Rightarrow ac \succsim_{\sigma}^{new} da$.

(5.3.5.10) $\forall c \in A \setminus \{a\}$: $ac \succ_{\sigma}^{old} ca \Rightarrow ac \succ_{\sigma}^{new} ca$.

(5.3.5.11) $\forall c \in A \setminus \{a\}$: $ac \succsim_{\sigma}^{old} ca \Rightarrow ac \succsim_{\sigma}^{new} ca$.

(5.3.5.12) $\forall b,c,d,e \in A \setminus \{a\}$: $bc \succ_{\sigma}^{old} de \Leftrightarrow bc \succ_{\sigma}^{new} de$.

The rest of the proof is identical to the proof in section 4.6.1.

### 5.3.6. Independence of Clones

*Independence of clones* says that running a large number of similar alternatives, so-called *clones*, must not have any impact on the result of the elections (section 4.7).

When we define independence of clones for the probabilistic framework, we have to explicitly address the case where there are alternatives $a \in A$ such that every voter is indifferent between alternative $a$ and alternative $d$. According to neutrality, both alternatives must then be elected with the same probability ( $r^{old}[a] = r^{old}[d]$ ). When alternative $d$ is replaced by a set of clones $K$ ( with $\| K \| > 1$ ), as defined in (4.7.1) – (4.7.3), and, again, every voter is indifferent between alternative $a$ and every alternative $g \in K$ then, again, every alternative in $K \cup \{a\}$ must be elected with the same probability ( $\forall\ g \in K$: $r^{new}[a] = r^{new}[g]$ ) so that the probability, that alternative $a$ is chosen, necessarily drops ( i.e.: $r^{old}[a] > r^{new}[a]$ ).

In an earlier version of this paper (Schulze, 2003), I addressed this case by explicitly presuming (1) $r^{old}[a] = 0$ or (2) that, for every alternative $a \in A \setminus \{d\}$, there is at least one voter $v \in V$ with $a \not\approx_v d$.

In this version, I address this case by explicitly specifying how $r^{old}[a]$ changes. Suppose $B \subseteq A^{old} \setminus \{d\}$ is the set of alternatives such that every voter is indifferent between alternative $d$ and every alternative $a \in B$. See (5.3.6.1). Then, when alternative $d$ is replaced by a set of clones $K$, $r[a]$ changes by the factor ( $1 + \| B \|$ ) / ( $\| K \| + \| B \|$ ). See (5.3.6.6)(ii) and (5.3.6.7)(ii).

The conditions (4.7.4) – (4.7.13) and (5.3.6.2) – (5.3.6.10) are not independent conditions. Rather, I have arranged them according to the addressed aspects. For example, it is easy to see that (4.7.10) implies (4.7.7) or that (5.3.6.2) implies (5.3.6.3).

The conditions (4.7.4) – (4.7.13) and (5.3.6.2) – (5.3.6.10) are not very intuitive. The best way to understand them is by asking: “What would *random dictatorship* (with *random alternative* as a tie-breaker for those situations where there is a set of alternatives such that every voter is indifferent between every alternative of this set) do?”

*Random dictatorship* means that we randomly choose a voter $v \in V$ and the collective ranking is this voter’s individual ranking. When this voter is indifferent between some alternatives, then we randomly choose another voter and use his ranking to complete the collective ranking. We continue to randomly choose voters until either (1) the collective ranking is a complete order or (2) all voters have been used to complete the collective ranking.

*Random alternative* means that the alternatives are ranked randomly, regardless of how the voters vote.

**Definition:**

An election method is *independent of clones* if the following holds:

Suppose $d \in A^{old}$. Suppose $A^{new} := ( A^{old} \cup K ) \setminus \{d\}$.

Suppose $B \subseteq A^{old} \setminus \{d\}$ is defined as follows:

(5.3.6.1) $a \in B :\Leftrightarrow ( \forall v \in V: a \approx_v d )$.

Suppose alternative $d$ is replaced by the set of alternatives $K$ in such a manner that (4.7.1) – (4.7.3) are satisfied.

Then:

(5.3.6.2) $\forall O_1 \in \mathcal{LO}_{A^{old}} \; \forall g \in K$:

$p^{old}[O_1] = \sum ( p^{new}[O] \mid O \in \mathcal{LO}_{A^{new}}$ with
(1) $\forall a,b \in A^{new} \setminus K: ab \in O_1 \Leftrightarrow ab \in O$.
(2) $\forall a \in A^{new} \setminus K: ad \in O_1 \Leftrightarrow ag \in O$.
(3) $\forall b \in A^{new} \setminus K: db \in O_1 \Leftrightarrow gb \in O$. )

(5.3.6.3) $\forall a,b \in A^{new} \setminus K: q^{old}[a,b] = q^{new}[a,b]$.

(5.3.6.4) $\forall a \in A^{new} \setminus K \; \forall g \in K: q^{old}[a,d] = q^{new}[a,g]$.

(5.3.6.5) $\forall b \in A^{new} \setminus K \; \forall g \in K: q^{old}[d,b] = q^{new}[g,b]$.

(5.3.6.6) $\forall a \in A^{new} \setminus K$:

(i) $( a \notin B ) \Rightarrow ( r^{old}[a] = r^{new}[a] )$.

(ii) $( a \in B ) \Rightarrow ( r^{old}[a] \cdot ( 1 + \| B \| ) = r^{new}[a] \cdot ( \| K \| + \| B \| ) )$.

(5.3.6.7) (i) $( B = \emptyset ) \Rightarrow ( r^{old}[d] = \sum ( r^{new}[g] \mid g \in K ) )$.

(ii) $( B \neq \emptyset ) \Rightarrow \forall g \in K: ( r^{old}[d] \cdot ( 1 + \| B \| ) = r^{new}[g] \cdot ( \| K \| + \| B \| ) )$.

(5.3.6.8) $\forall O_1 \in \mathcal{LO}_K$:

$p[O_1]|_K = \sum ( p^{new}[O] \mid O \in \mathcal{LO}_{A^{new}}$ with $O_1 \subseteq O$ ).

(5.3.6.9) $\forall g,h \in K: q[g,h]|_K = q^{new}[g,h]$.

(5.3.6.10) $\forall \emptyset \neq K_1 \subseteq K$:

Suppose $X := \sum ( r[g]|_K \mid g \in K_1 )$.

Suppose $Y := \sum ( r^{new}[g] \mid g \in K_1 )$.

Suppose $B = \emptyset$.

Then we get:

$\max \{ 0, r^{old}[d] + X - 1 \} \leq Y \leq \min \{ r^{old}[d], X \}$.

**Claim:**

The Schulze method $O_{final}(\sigma)$, as defined in sections 5.1, is independent of clones.

**Proof (overview):**

The proof for (5.3.6.2) – (5.3.6.7) is identical to the proof for (4.7.4) – (4.7.8) in section 4.7. (5.3.6.8) – (5.3.6.10) follows directly from section 5.1 Step 2(b).

### 5.3.7. Smith

**Definition:**

An election method satisfies *Smith* if the following holds:

Suppose (4.8.2) and (4.8.3).

Then we get:

(5.3.7.1) $\forall\, a \in B_1\ \forall\, b \in B_2$: $q[a,b] = 1$.

(5.3.7.2) $\sum\, (\, r[a] \mid a \in B_1\, ) = 1$.

An election method satisfies *Smith-IIA* if the following holds:

Suppose (4.8.2) and (4.8.3).

Suppose $d \in B_2$ is removed. Then we get:

(5.3.7.3) $\forall\, O_1 \in \mathcal{LO}_{B_1}$:
$\sum\, (\, p^{old}[O] \mid O \in \mathcal{LO}_A$ with $O_1 \subset O\, ) =$
$\sum\, (\, p^{new}[O] \mid O \in \mathcal{LO}_{(A \setminus \{d\})}$ with $O_1 \subset O\, )$.

(5.3.7.4) $\forall\, a,b \in B_1$: $q^{old}[a,b] = q^{new}[a,b]$.

(5.3.7.5) $\forall\, a \in B_1$: $r^{old}[a] = r^{new}[a]$.

Suppose $d \in B_1$ is removed. Then we get:

(5.3.7.6) $\forall\, O_1 \in \mathcal{LO}_{B_2}$:
$\sum\, (\, p^{old}[O] \mid O \in \mathcal{LO}_A$ with $O_1 \subset O\, ) =$
$\sum\, (\, p^{new}[O] \mid O \in \mathcal{LO}_{(A \setminus \{d\})}$ with $O_1 \subset O\, )$.

(5.3.7.7) $\forall\, a,b \in B_2$: $q^{old}[a,b] = q^{new}[a,b]$.

**Claim:**

If $\succ_D$ satisfies (2.1.5), then the Schulze method $O_{final}(\sigma)$, as defined in sections 5.1, satisfies Smith and Smith-IIA.

**Proof (overview):**

The proof is identical to the proofs in section 4.8.

In the probabilic framework, *participation* says that, when we have (4.8.2), (4.8.21), and (4.8.22), then we get:

(5.3.7.8) $\forall\ e \in B_1\ \forall f \in B_2$: $q^{\text{old}}[e,f] \leq q^{\text{new}}[e,f]$.

(5.3.7.9) $\sum\ (\ r^{\text{old}}[e] \mid e \in B_1\ ) \leq \sum\ (\ r^{\text{new}}[e] \mid e \in B_1\ )$.

Example 7 (section 3.7) can be used to show a violation of the probabilistic version of the participation criterion.

## 5.3.8. Runtime

The runtime to calculate the pairwise matrix is O($N$·($C$^2)).

The runtime to calculate the TBRL $>_\sigma$, as defined in section 5.1 step 2, is O($N$·($C$^4)) because, in worst case, O($N$) ballots have to be picked and, each time, O($C$^2) links are compared with O($C$^2) other links.

On closer examination, to sort the O($C$^2) links according to their strengths, it is not necessary to compare each of the O($C$^2) links with each other of the O($C$^2) links directly. As the fastest algorithms to sort $X$ items according to their strengths have a runtime of O($X$·log($X$)), the runtime of the fastest algorithms to sort the O($C$^2) links according to their strengths is O(($C$^2)·log($C$)).

Therefore, the runtime to calculate the TBRL $>_\sigma$, as defined in section 5.1 step 2, reduces to O($N$·($C$^2)·log($C$)).

The runtime to calculate the binary relation $O_{final}(\sigma)$, as defined in section 5.1 step 4, is O($C$^7) because, in worst case, there are O($C$^2) pairwise ties "$P_\sigma[m,n] \approx_\sigma P_\sigma[n,m]$" (line 54). In worst case, O($C$^2) links have to be declared *forbidden* to resolve a pairwise tie. Each time, the runtime of the Floyd-Warshall algorithm to calculate the strength of the strongest path from every alternative to every other alternative is O($C$^3).

On closer examination, to resolve the pairwise tie "$P_\sigma[m,n] \approx_\sigma P_\sigma[n,m]$", it is not necessary to calculate the strength of the strongest path from every alternative to every other alternative. It is sufficient to calculate the strength of the strongest path from alternative $m$ to alternative $n$ and the strength of the strongest path from alternative $n$ to alternative $m$. This can be done with the Dijkstra algorithm in a runtime O($C$^2).

Therefore, the runtime to calculate the binary relation $O_{final}(\sigma)$, as defined in section 5.1 step 4, reduces to O($C$^6).

Thus, the total runtime to calculate the binary relation $O_{final}(\sigma)$, as defined in section 5.1, is O(($N$·($C$^2)·log($C$)) + ($C$^6)).

## 6. Supermajority Requirements

When preferential ballots are being used in referendums, then sometimes alternatives have to fulfill some supermajority requirements to qualify. Typical supermajority requirements define some $M_1 \in \mathbb{N}$ or some $1 \leq M_2 \in \mathbb{R}$ and say that $N[a,b]$ must be strictly larger than max $\{ N[b,a], M_1 \}$ or that $N[a,b]$ must be strictly larger than $M_2 \cdot N[b,a]$ to replace alternative $b \in A$ by alternative $a \in A$. Or they say that $N[a,b]$ must be strictly larger than $N[b,a]$ not only in the electorate as a whole, but also in a majority of its geographic parts or even in each of its geographic parts. It is also possible that in the same referendum the voters have to choose between alternatives that have to fulfill different supermajority requirements to qualify. In this section, we discuss a possible way to combine the Schulze method with supermajority requirements. Suppose $s \in A$ is the *status quo*.

These are the two tasks of supermajority requirements:

Task #1 (*protecting the status quo*):

> Supermajority requirements protect the status quo from accidental majorities. They make it more difficult to replace the status quo $s$ by alternative $a \in A \setminus \{s\}$. Therefore, an important property of all supermajority requirements is that, when $s$ had won in the absence of these requirements, then it also wins in the presence of these requirements.

Task #2 (*preventing the status quo from cycling*):

> Supermajority requirements prevent the status quo from cycling. Suppose $s(0)$ is the starting status quo. Suppose $s(k+1)$ is the new status quo when the method is applied to the same set of alternatives $A$, to the same set of ballots $V$, and to the status quo $s(k)$. Then we would expect that ( for every possible set of alternatives $A$, for every possible set of ballots $V$, and for every possible starting status quo $s(0) \in A$ ) there is an $m < C$ such that $s(k) \equiv s(m)$ for all $k \geq m$.

We recommend the following method:

> The Schulze relation $\mathcal{O}$, as defined in section 2.2, and the Schulze relation $\mathcal{O}_{final}(\sigma)$, as defined in section 5.1, are calculated.
>
> Alternative $a \in A \setminus \{s\}$ is *attainable* if and only if $N[a,s] > N[s,a]$ and (a) there is no supermajority requirement to replace the status quo $s$ by alternative $a$ or (b) alternative $a$ has the supermajority required to replace the status quo $s$ by alternative $a$.
>
> Alternative $a \in A$ is *eligible* if and only if ( $a \equiv s$ ) or ( ( $a$ is attainable ) and ( $as \in \mathcal{O}$ ) ).
>
> A winner is an alternative $a \in A$ with (1) alternative $a$ is eligible and (2) $ab \in \mathcal{O}_{final}(\sigma)$ for every other eligible alternative $b$.

The condition “$as \in O$” in the definition of eligibility implies that alternative $a$ can win only if it had disqualified the status quo $s$ in the absence of supermajority requirements. This guarantees that, if $s$ had won in the absence of supermajority requirements, then $s$ also wins in the presence of these supermajority requirements.

In the above suggestion, the status quo $s$ can only be replaced by an alternative $a$ with $as \in O$. As $O$ is transitive, it is guaranteed that the status quo cannot be changed in a cyclic manner.

Section 6 is also discussed by Behrens (2014b).

## 7. Electoral College

There has been some debate about how to combine the Schulze method with the Electoral College for the elections of the President of the USA. In my opinion, the Electoral College serves two important purposes:

> Purpose #1: The Electoral College gives more power to the smaller states.
>
> > The Senate, where each state has the same voting power regardless of its population, is more powerful than the House of Representatives, where each state has a voting power in proportion of its population. This is true especially for decisions that are close to the executive. For example, the President needs the consent of the Senate for treaties and for the appointment of officers and judges. Because of this reason, it is more important that the President has a reliable support in the Senate than that he has a reliable support in the House of Representatives.
>
> Purpose #2: The Electoral College makes it possible to count the ballots on the state levels and then to add up the electoral votes.
>
> > The Electoral College makes it possible that, to guarantee that all voters are treated in an equal manner, it is only necessary to guarantee that all voters *in the same state* are treated in an equal manner. However, if the ballots were added up on the national level, it would be necessary to guarantee that *all voters all over the USA* are treated in an equal manner. In the latter case, many provisions (e.g. the rules to gain suffrage or to be excluded from suffrage, the ballot access rules, the rules for postal voting, the opening hours of the polling places) would have to be harmonized all over the USA, leading to a very powerful central election authority.
> >
> > This property is desirable especially for the elections to the National Conventions for the nominations of the presidential candidates. Here, the election rules and the set of candidates differ significantly from state to state.

To combine the Schulze method with the Electoral College without losing any of its purposes, we recommend that, for each pair of candidates $a$ and $b$ separately, we should determine how many electoral votes $N_{electors}[a,b]$ candidate $a$ would get and how many electoral votes $N_{electors}[b,a]$ candidate $b$ would get when only these two candidates were running. We then apply the Schulze method to the matrix $N_{electors}$.

So we recommend the following method:

**Stage 1:**

Suppose $A$ is the set of candidates who are running in at least one state.

Suppose $A_X \subseteq A$ is the set of candidates who are running in state $X$.

For $a,b \in A_X$: $N_X[a,b] \in \mathbb{N}_0$ is the number of voters in state $X$ who strictly prefer candidate $a$ to candidate $b$.

**Stage 2:**

Suppose $y \in \mathbb{R}$ with $y > 0$. Then "smaller_or_equal($y$)" is the largest integer that is smaller than or equal to $y$. In other words: "smaller_or_equal($y$)" is that integer $z \in \mathbb{N}_0$ with $z \leq y < ( z + 1 )$.

Suppose $y \in \mathbb{R}$ with $y > 0$. Then "strictly_smaller($y$)" is the largest integer that is strictly smaller than $y$. In other words: "strictly_smaller($y$)" is that integer $z \in \mathbb{N}_0$ with $z < y \leq ( z + 1 )$.

Suppose $E_X \in \mathbb{N}$ is the number of electors of state $X$.

Suppose:

(a) $F_X[a,b] := E_X$,

if $\{ a \in A_X$ and $b \notin A_X \}$ or $\{ a,b \in A_X$ and $N_X[a,b] > N_X[b,a] = 0 \}$.

(b) $F_X[a,b] := 0$,

if $\{ a \notin A_X$ and $b \in A_X \}$ or $\{ a,b \in A_X$ and $N_X[b,a] > N_X[a,b] = 0 \}$.

(c) $F_X[a,b] := E_X / 2$,

if $\{ a,b \notin A_X \}$ or $\{ a,b \in A_X$ and $N_X[a,b] = N_X[b,a] \}$.

(d) $F_X[a,b] := 0.01 \cdot \text{smaller_or_equal} \left( \dfrac{N_X[a,b] \cdot (1 + 100 \cdot E_X)}{N_X[a,b] + N_X[b,a]} \right)$,

if $a,b \in A_X$ and $N_X[a,b] > N_X[b,a] > 0$.

(e) $F_X[a,b] := 0.01 \cdot \text{strictly_smaller} \left( \dfrac{N_X[a,b] \cdot (1 + 100 \cdot E_X)}{N_X[a,b] + N_X[b,a]} \right)$,

if $a,b \in A_X$ and $N_X[b,a] > N_X[a,b] > 0$.

$N_{electors}[a,b] := \sum_X F_X[a,b]$.

**Stage 3:**

The Schulze method, as defined in section 2.2, is applied to $N_{electors}$.

Suppose the Schulze method is used for presidential primaries. Suppose some candidate $g$ withdraws and doesn't take part in the remaining primaries. Then candidate $g$ is not removed from the pairwise matrix. Rather he is treated as described at stage 2 (a) – (c). This regulation is necessary because removing a loser can still change the winner.

Stage 2 (a) – (b) guarantees that it can never be advantageous for a candidate not to run in a state. Stage 2 (c) guarantees that, when neither candidate $a$ nor candidate $b$ is running in a state, then $N_{electors}[a,b] - N_{electors}[b,a]$ doesn't change.

Stage 2 (a) – (e) guarantees that, for every state $X$, we get

(7.1) $\forall\ a,b \in A$: $0 \leq F_X[a,b] \leq E_X$.

(7.2) $\forall\ a,b \in A$: $F_X[a,b] + F_X[b,a] = E_X$.

So the weight of each state $X$ in each pairwise contest is given by $E_X$.

## 8. Proportional Representation by the Single Transferable Vote

The term "Proportional Representation by the Single Transferable Vote" (STV) refers to preferential multi-winner election methods where the winning alternatives represent the electorate in a proportional manner. What exactly "in a proportional manner" means in this context is debatable and will be discussed in section 8.4.

$A$ is a finite and non-empty set of alternatives. $M \in \mathbb{N}$ with $0 < M < \infty$ is the number of seats. $C \in \mathbb{N}$ with $M < C < \infty$ is the number of alternatives. $N \in \mathbb{N}$ with $0 < N < \infty$ is the number of voters.

$A_M$ is the set of the $(C!)/((M!)\cdot((C-M)!))$ possible ways to choose $M$ different alternatives from the set $A$. The elements of $A_M$ are indicated with *wedding* letters $\mathfrak{A}$, $\mathfrak{B}$, $\mathfrak{C}$, ...

Input of an STV method is a profile, as defined in section 2.1. Output of an STV method is a subset $\emptyset \neq \mathcal{S}_M \subseteq A_M$ of potential winning sets.

### 8.1. Schulze STV

In Schulze STV, we only compare every set of $M$ alternatives with every other set of $M$ alternatives that differs in exactly one alternative.

There are $(C!)/((M!)\cdot(C-M)!)$ sets of exactly $M$ alternatives.

There are $(C!)/(((M+1)!)\cdot((C-M-1)!))$ possible $(M+1)$-way contests. Each $(M+1)$-way contest leads to $M\cdot(M+1)$ links in that digraph where each node represents a set of $M$ alternatives. See e.g. page 416.

So we have a digraph with $(C!)/((M!)\cdot(C-M)!)$ nodes and $M\cdot(M+1)\cdot(C!)/(((M+1)!)\cdot((C-M-1)!)) = (C!)/(((M-1)!)\cdot((C-M-1)!))$ links. This digraph is strongly connected. (A digraph is *strongly connected* : $\Leftrightarrow$ For every pair of two different nodes $\mathfrak{A}$ and $\mathfrak{B}$, there is a directed path from node $\mathfrak{A}$ to node $\mathfrak{B}$ and a directed path from node $\mathfrak{B}$ to node $\mathfrak{A}$.) We then apply the Schulze method, as defined in section 2.2, to this digraph. This works because, for the proof in section 4.1, it is sufficient that the digraph, that the Schulze method is applied to, is strongly connected. It is not necessary that this digraph is complete.

Schulze STV is motivated by the fact that we want a generalization of the Condorcet criterion from single-winner elections to multi-winner elections that is as strong as possible (section 8.3), so that the possibility, that an additional alternative changes the result of the election without being elected, is minimized. In section 8.4, we will see that the Condorcet criterion, that we get by this manner, is so strong that we almost always have $M$ Condorcet winners or, at least, $(M-1)$ Condorcet winners.

Like in section 2, two degrees of freedom have to be addressed: (1) How is the strength $(a_1;...;a_{(M-1)};b;c)$ of the link $(a_1;...;a_{(M-1)};b) \rightarrow (a_1;...;a_{(M-1)};c)$ measured? (2) How do we solve situations without a unique winning set?

Suppose $\mathcal{U} := \{u_1,...,u_N\}$ is a list of $N$ strict weak orders each on the same set of $(M+1)$ alternatives $(a_1;...;a_{(M-1)};b;c)$. Suppose $\mathcal{V} := \{v_1,...,v_N\}$ is a list of $N$ strict weak orders each on the same set of $(M+1)$ alternatives $(d_1;...;d_{(M-1)};e;f)$. Then, we presume that $\succ_D$ is a strict weak order that compares $\mathcal{U}$ and $\mathcal{V}$. We presume that $\succ_D$ satisfies at least the following presumptions:

(8.1.1) (*independence of permutating* $a_1,...,a_{(M-1)}$)

Suppose the alternatives in $a_1,...,a_{(M-1)}$ are permutated. Then we get: $\mathcal{U}^{\text{old}} \approx_D \mathcal{U}^{\text{new}}$.

Presumption (8.1.1) allows us to write $(\{a_1,...,a_{(M-1)}\};b;c)$, instead of $(a_1;...;a_{(M-1)};b;c)$, for the strength of the link $(\{a_1,...,a_{(M-1)}\};b) \rightarrow (\{a_1,...,a_{(M-1)}\};c)$.

(8.1.2) (*anonymity*)

Suppose $\{\tau(1),...,\tau(N)\}$ is a permutation of $\{1,...,N\}$. Then for every list $\mathcal{U} := \{u_1,...,u_N\}$ of $N$ elements, we get

$$\{u_{\tau(1)},...,u_{\tau(N)}\} \approx_D \{u_1,...,u_N\}.$$

(8.1.3) (*independence of empty ballots*)

Suppose an empty ballot (i.e. a ballot that is indifferent between all alternatives $a_1,...,a_{(M-1)},b,c$ resp. between all alternatives $d_1,...,d_{(M-1)},e,f$) is added at the end of $\mathcal{U}$ and $\mathcal{V}$. Then we get

$$\mathcal{U}^{\text{old}} \succ_D \mathcal{V}^{\text{old}} \Leftrightarrow \mathcal{U}^{\text{new}} \succ_D \mathcal{V}^{\text{new}}.$$

(8.1.4) (*non-negative responsiveness*)

Suppose some ballots in $\mathcal{U} := \{u_1,...,u_N\}$ are replaced by ballots where alternative $b$ is ranked higher and/or where alternative $c$ is ranked lower without changing the order in which the other alternatives $a_1,...,a_{(M-1)}$ are ranked relatively to each other. Then the strength of $\mathcal{U}$ must not decrease.

Suppose

(8.1.4a) $\forall\, v \in \mathcal{U}\ \forall\, a_i,a_j \in a_1,...,a_{(M-1)}$: $a_i \succ_v^{\text{old}} a_j \Leftrightarrow a_i \succ_v^{\text{new}} a_j$.

(8.1.4b) $\forall\, v \in \mathcal{U}\ \forall\, a_i \in a_1,...,a_{(M-1)}$: $b \succ_v^{\text{old}} a_i \Rightarrow b \succ_v^{\text{new}} a_i$.

(8.1.4c) $\forall\, v \in \mathcal{U}\ \forall\, a_i \in a_1,...,a_{(M-1)}$: $b \succsim_v^{\text{old}} a_i \Rightarrow b \succsim_v^{\text{new}} a_i$.

(8.1.4d) $\forall\, v \in \mathcal{U}\ \forall\, a_i \in a_1,...,a_{(M-1)}$: $a_i \succ_v^{\text{old}} c \Rightarrow a_i \succ_v^{\text{new}} c$.

(8.1.4e) $\forall\, v \in \mathcal{U}\ \forall\, a_i \in a_1,...,a_{(M-1)}$: $a_i \succsim_v^{\text{old}} c \Rightarrow a_i \succsim_v^{\text{new}} c$.

(8.1.4f) $\forall\, v \in \mathcal{U}$: $b \succ_v^{\text{old}} c \Rightarrow b \succ_v^{\text{new}} c$.

(8.1.4g) $\forall\, v \in \mathcal{U}$: $b \succsim_v^{\text{old}} c \Rightarrow b \succsim_v^{\text{new}} c$.

Then: $\mathcal{U}^{\text{old}} \precsim_D \mathcal{U}^{\text{new}}$.

(8.1.5) (*positive responsiveness*)

Suppose (8.1.4a) – (8.1.4g).

Suppose

(8.1.5a) $\exists\, v \in \mathcal{U}$: ( $b \prec_v^{\text{old}} c \wedge b \succ_v^{\text{new}} c$ ).

Suppose

(8.1.5b) $\forall\, a_i \in a_1,\dots,a_{(M-1)}\ \exists\, v \in \mathcal{U}$: ( $b \prec_v^{\text{old}} a_i \wedge b \succ_v^{\text{new}} a_i$ ).

or

(8.1.5c) $\forall\, a_i \in a_1,\dots,a_{(M-1)}\ \exists\, v \in \mathcal{U}$: ( $a_i \prec_v^{\text{old}} c \wedge a_i \succ_v^{\text{new}} c$ ).

Then: $\mathcal{U}^{\text{old}} \prec_D \mathcal{U}^{\text{new}}$.

(8.1.6) (*reversal symmetry*)

Suppose the strength $(\{a_1,\dots,a_{(M-1)}\};b;c)$ of the link $(\{a_1,\dots,a_{(M-1)}\};b) \to (\{a_1,\dots,a_{(M-1)}\};c)$ is stronger than the strength $(\{d_1,\dots,d_{(M-1)}\};e;f)$ of the link $(\{d_1,\dots,d_{(M-1)}\};e) \to (\{d_1,\dots,d_{(M-1)}\};f)$. Then the strength $(\{d_1,\dots,d_{(M-1)}\};f;e)$ of the link $(\{d_1,\dots,d_{(M-1)}\};f) \to (\{d_1,\dots,d_{(M-1)}\};e)$ is stronger than the strength $(\{a_1,\dots,a_{(M-1)}\};c;b)$ of the link $(\{a_1,\dots,a_{(M-1)}\};c) \to (\{a_1,\dots,a_{(M-1)}\};b)$.

(8.1.7) (*homogeneity*)

For every $x_1,x_2 \in \mathbb{N}$, we get

$$\underbrace{\mathcal{U}+\dots+\mathcal{U}}_{x_1\text{ times}} \succ_D \underbrace{\mathcal{V}+\dots+\mathcal{V}}_{x_1\text{ times}} \Rightarrow \underbrace{\mathcal{U}+\dots+\mathcal{U}}_{x_2\text{ times}} \succ_D \underbrace{\mathcal{V}+\dots+\mathcal{V}}_{x_2\text{ times}}.$$

(8.1.8) (*independence of strong winners*)

Suppose that some voters change the order in which they rank the alternatives $a_1,...,a_{(M-1)}$ relatively to each other. Suppose we have

(8.1.8a) $\forall\ v \in \mathcal{U}\ \forall\ a_i \in a_1,...,a_{(M-1)}$: $a_i \succ_v^{\text{old}} b \Leftrightarrow a_i \succ_v^{\text{new}} b$.

(8.1.8b) $\forall\ v \in \mathcal{U}\ \forall\ a_i \in a_1,...,a_{(M-1)}$: $b \succ_v^{\text{old}} a_i \Leftrightarrow b \succ_v^{\text{new}} a_i$.

(8.1.8c) $\forall\ v \in \mathcal{U}\ \forall\ a_i \in a_1,...,a_{(M-1)}$: $a_i \succ_v^{\text{old}} c \Leftrightarrow a_i \succ_v^{\text{new}} c$.

(8.1.8d) $\forall\ v \in \mathcal{U}\ \forall\ a_i \in a_1,...,a_{(M-1)}$: $c \succ_v^{\text{old}} a_i \Leftrightarrow c \succ_v^{\text{new}} a_i$.

(8.1.8e) $\forall\ v \in \mathcal{U}$: $b \succ_v^{\text{old}} c \Leftrightarrow b \succ_v^{\text{new}} c$.

(8.1.8f) $\forall\ v \in \mathcal{U}$: $c \succ_v^{\text{old}} b \Leftrightarrow c \succ_v^{\text{new}} b$.

Then we get: $\mathcal{U}^{\text{old}} \approx_D \mathcal{U}^{\text{new}}$.

(8.1.9) (*transitivity and negative transitivity in the M-seat* (*M*+1)*-alternative case*)

(a) Suppose $(\{a_1,...,a_{(M-2)},z\};x;y) \succ_D (\{a_1,...,a_{(M-2)},z\};y;x)$ and $(\{a_1,...,a_{(M-2)},x\};y;z) \succ_D (\{a_1,...,a_{(M-2)},x\};z;y)$. Then we get: $(\{a_1,...,a_{(M-2)},y\};x;z) \succ_D (\{a_1,...,a_{(M-2)},y\};z;x)$.

(b) Suppose $(\{a_1,...,a_{(M-2)},z\};x;y) \succsim_D (\{a_1,...,a_{(M-2)},z\};y;x)$ and $(\{a_1,...,a_{(M-2)},x\};y;z) \succsim_D (\{a_1,...,a_{(M-2)},x\};z;y)$. Then we get: $(\{a_1,...,a_{(M-2)},y\};x;z) \succsim_D (\{a_1,...,a_{(M-2)},y\};z;x)$.

(c) Suppose $(\{a_1,...,a_{(M-2)},z\};x;y) \succsim_D (\{a_1,...,a_{(M-2)},z\};y;x)$ and $(\{a_1,...,a_{(M-2)},x\};y;z) \succsim_D (\{a_1,...,a_{(M-2)},x\};z;y)$. Then we get: $(\{a_1,...,a_{(M-2)},y\};x;z) \succsim_D \max_D \{(\{a_1,...,a_{(M-2)},z\};x;y), (\{a_1,...,a_{(M-2)},x\};y;z)\}$.

(d) Suppose $(\{a_1,...,a_{(M-2)},z\};x;y) \precsim_D (\{a_1,...,a_{(M-2)},z\};y;x)$ and $(\{a_1,...,a_{(M-2)},x\};y;z) \precsim_D (\{a_1,...,a_{(M-2)},x\};z;y)$. Then we get: $(\{a_1,...,a_{(M-2)},y\};x;z) \precsim_D \min_D \{(\{a_1,...,a_{(M-2)},z\};x;y), (\{a_1,...,a_{(M-2)},x\};y;z)\}$.

The importance of the presumptions (8.1.1) – (8.1.7) follows directly from the considerations in sections 2 – 5.

Presumption (8.1.8) is motivated by the considerations in Schulze (2004, 2011b): In multi-winner elections, but not in single-winner elections, it is a useful strategy for a voter not to give a good preference to an alternative that wins with certainty even without this voter's vote (*Hylland free riding*). By using this strategy, this voter increases his impact on which the other winners are. When the voters have understood this strategic loophole well, the order in which the individual voter ranks the strong alternatives relatively to each other doesn't say anything anymore about the sincere opinion of this voter, but only about his strategic skills and his information about the opinions of the other voters. So the order, in which the individual voter ranks the strong alternatives relatively to each other, doesn't contain any information and should, therefore, have no impact on the result of the election.

Presumption (8.1.9) says that, at least in the $M$-seat ($M$+1)-alternative case, pairwise comparisons must be transitive [presumption (8.1.9)(a)] and negatively transitive [presumption (8.1.9)(b)] so that, at least in this case, the result cannot be cyclic.

In sections 8.1.1 and 8.1.2, we will introduce a concrete definition for the strength of links that satisfies (8.1.1) – (8.1.9). In section 8.1.3, we will describe how Schulze STV will look like with this definition for the strength of links.

### 8.1.1. Proportional Completion

*Proportional completion* means that non-linear individual orders are completed to linear orders in such a manner that, for each set of alternatives, the proportions of the individual orders, restricted to these alternatives, are not changed.

Example: Suppose a voter is indifferent between alternative $a$ and alternative $b$. Suppose of the other voters $X_1 = 56$ strictly prefer alternative $a$ to alternative $b$ and $X_2 = 44$ strictly prefer alternative $b$ to alternative $a$, then this voter is replaced by $X_1/(X_1+X_2) = 0.56$ voters who rank these alternatives $a \succ_v b$ and by $X_2/(X_1+X_2) = 0.44$ voters who rank these alternatives $b \succ_v a$ and who rank the other alternatives in the same manner as the original voter did.

Basic idea behind proportional completion is that, on the one side, adding a voter who is indifferent between all alternatives, that have chances to win, should not change the result of the election as this additional voter doesn't add new information. On the other side, the definition for the strengths of the links between sets of alternatives (section 8.1.2) requires that each voter casts a linear order.

The following 3 stages give a precise definition for proportional completion.

**Stage 1:**

$W$ shall be the proportional completion of $V$. $\rho(w) \in \mathbb{R}$ shall be the weight of voter $w \in W$. Then we start with

(8.1.1.1) $W := V$.

(8.1.1.2) $\forall\, w \in W$: $\rho(w) := 1$.

**Stage 2:**

Suppose there is a voter $w \in W$ and a set of alternatives $f_1,...,f_n \in A$ with

(8.1.1.3) $n > 1$.

(8.1.1.4) $\forall\, f_i,f_j \in \{f_1,...,f_n\} : f_i \approx_w f_j$.

(8.1.1.5) $\forall\, f_i \in \{f_1,...,f_n\}\ \forall\, e \in A \setminus \{f_1,...,f_n\} : f_i \not\approx_w e$.

Suppose X $\in \mathbb{N}_0$ is the number of voters $v \in V$ with

(8.1.1.6) $\exists\, f_i,f_j \in \{f_1,...,f_n\} : f_i \not\approx_v f_j$.

Case 1: X > 0.

For each voter $v \in V$ with (8.1.1.6), a voter $u$ is added to $W$ with

(8.1.1.7) $\forall\, g,h \in A \setminus \{f_1,...,f_n\} : g \succ_w h \Leftrightarrow g \succ_u h$.

(8.1.1.8) $\forall\, f_i \in \{f_1,...,f_n\}\ \forall\, g \in A \setminus \{f_1,...,f_n\} : g \succ_w f_i \Leftrightarrow g \succ_u f_i$.

(8.1.1.9) $\forall\, f_i \in \{f_1,...,f_n\}\ \forall\, h \in A \setminus \{f_1,...,f_n\} : f_i \succ_w h \Leftrightarrow f_i \succ_u h$.

(8.1.1.10) $\forall\, f_i,f_j \in \{f_1,...,f_n\} : f_i \succ_v f_j \Leftrightarrow f_i \succ_u f_j$.

(8.1.1.11) $\rho(u) := \rho(w) / \mathrm{X}$.

Case 2: X = 0.

For each of the $n!$ possible permutations $\{\tau(1),...,\tau(n)\}$ of $\{1,...,n\}$, a voter $u$ is added to $W$ with (8.1.1.7) – (8.1.1.9) and

(8.1.1.12) $\forall\, f_i,f_j \in \{f_1,...,f_n\}$: $\tau(i) > \tau(j) \Leftrightarrow f_i \succ_u f_j$.

(8.1.1.13) $\rho(u) := \rho(w) / (n!)$.

After all these voters $u$ have been added to $W$, the original voter $w$ is removed from $W$.

**Stage 3:**

Stage 2 is repeated until $a \not\approx_w b\ \forall\, a \in A\ \forall\, b \in A \setminus \{a\}\ \forall\, w \in W$.

So in each iteration of proportional completion, we look whether there is still a voter who casts a non-linear order. When there is still such a voter, then we take a voter $w \in W$ and a set of alternatives $\emptyset \neq \{f_1,...,f_n\} \subseteq A$ ( with $n > 1$ ) where voter $w$ is indifferent between all the alternatives in $\{f_1,...,f_n\}$ [ see (8.1.1.4) ] and different between any alternative in $\{f_1,...,f_n\}$ and any alternative in $A \setminus \{f_1,...,f_n\}$ [ see (8.1.1.5) ]. We then look how those voters, who are not indifferent between all the alternatives in $\{f_1,...,f_n\}$ [ see (8.1.1.6) ], rank the alternatives in $\{f_1,...,f_n\}$. Voter $w$ is then replaced, in a proportional manner [ see (8.1.1.11) ], by voters who rank the alternatives in $A \setminus \{f_1,...,f_n\}$ in the same order as voter $w$ did [ see (8.1.1.7) – (8.1.1.9) ] and who rank the alternatives in $\{f_1,...,f_n\}$ in the same order as the other voters do [ see (8.1.1.10) ].

## 8.1.2. Links between Sets of Alternatives

In this section, we will propose a concrete definition for $\succ_D$ that satisfies (8.1.1) – (8.1.9).

Suppose $\{a_1,...,a_M,g\} \subset A$. We will define $N[\{a_1,...,a_M\};g] \in \mathbb{R}_{\geq 0}$. We then get the *support* of the link $(\{a_1,...,a_{(M-1)}\};b) \to (\{a_1,...,a_{(M-1)}\};c)$ by replacing $a_M$ by $b$ and by replacing $g$ by $c$ in the definition of $N[\{a_1,...,a_M\};g]$. We get the *opposition* of the link $(\{a_1,...,a_{(M-1)}\};b) \to (\{a_1,...,a_{(M-1)}\};c)$ by replacing $a_M$ by $c$ and by replacing $g$ by $b$ in the definition of $N[\{a_1,...,a_M\};g]$.

The link $(\{a_1,...,a_{(M-1)}\};b) \to (\{a_1,...,a_{(M-1)}\};c)$ is then stronger than the link $(\{d_1,...,d_{(M-1)}\};e) \to (\{d_1,...,d_{(M-1)}\};f)$ if and only if

$$(N[\{a_1,...,a_{(M-1)},b\};c], N[\{a_1,...,a_{(M-1)},c\};b]) \succ_{D2} (N[\{d_1,...,d_{(M-1)},e\};f], N[\{d_1,...,d_{(M-1)},f\};e])$$

where $\succ_{D2}$ is a strict weak order on $\mathbb{R}_{\geq 0} \times \mathbb{R}_{\geq 0}$ that satisfies (2.1.1) – (2.1.3).

So we get

(8.1.2.1) $$(\{a_1,...,a_{(M-1)}\};b;c) \succ_D (\{d_1,...,d_{(M-1)}\};e;f) :\Leftrightarrow$$

$$(N[\{a_1,...,a_{(M-1)},b\};c], N[\{a_1,...,a_{(M-1)},c\};b]) \succ_{D2} (N[\{d_1,...,d_{(M-1)},e\};f], N[\{d_1,...,d_{(M-1)},f\};e]).$$

Basic idea for the definition for the support $N[\{a_1,...,a_M\};g]$ of the link $\{a_1,...,a_M\} \rightarrow (\{a_1,...,a_{(M-1)}\};g)$ is that a defeat of alternative $a$ against alternative $g$ of strength $N[a,g]$ in a single-winner election corresponds to a situation where each of the alternatives $\{a_1,...,a_M\}$ has a "separate quota" against alternative $g$ of strength $N[\{a_1,...,a_M\};g]$ in an $M$-seat election. See (8.1.2.6) – (8.1.2.7).

$W$ is the proportional completion of $V$. $\rho(w) \in \mathbb{R}$ is the weight of voter $w \in W$. $N_W$ is the number of voters in $W$. Then the support $N[\{a_1,...,a_M\};g] \in \mathbb{R}_{\geq 0}$ of the link from $\{a_1,...,a_M\}$ to $(\{a_1,...,a_{(M-1)}\};g)$ is defined as follows:

> $N[\{a_1,...,a_M\};g] \in \mathbb{R}_{\geq 0}$ is the largest value such that there is a $t \in \mathbb{R}^{(N_W \times M)}$ such that

(8.1.2.2) $\forall\, i \in \{1,...,N_W\}\ \forall\, j \in \{1,...,M\}$: $t_{ij} \geq 0$.

(8.1.2.3) $\forall\, i \in \{1,...,N_W\}$: $\sum_{j=1}^{M} t_{ij} \leq \rho(i)$.

(8.1.2.4) $\forall\, i \in \{1,...,N_W\}\ \forall\, j \in \{1,...,M\}$: $g \succ_i a_j \Rightarrow t_{ij} = 0$.

(8.1.2.5) $\forall\, j \in \{1,...,M\}$: $\sum_{i=1}^{N_W} t_{ij} \geq N[\{a_1,...,a_M\};g]$.

So the support $N[\{a_1,...,a_M\};g] \in \mathbb{R}_{\geq 0}$ of the link $\{a_1,...,a_M\} \rightarrow (\{a_1,...,a_{(M-1)}\};g)$ is the largest number such that the electorate can be divided into $M$+1 disjoint sets $T_1,...,T_{(M+1)}$ such that:

(8.1.2.6) $\forall\, j \in \{1,...,M\}$: Every voter in $T_j$ prefers alternative $a_j$ to alternative $g$.

(8.1.2.7) $\forall\, j \in \{1,...,M\}$: The total weight of the voters in $T_j$ is at least $N[\{a_1,...,a_M\};g]$.

### 8.1.3. Definition of Schulze STV

Proportional completion is defined in section 8.1.1. $N[\{a_1,...,a_M\};g]$ is defined in section 8.1.2. $\succ_{D1}$ and $\succ_{D2}$ are two binary relations that each satisfy (2.1.1) – (2.1.3).

**Stage 1:**

We calculate a Schulze single-winner ranking $O_{final}(\sigma)$ on $A$, as defined in section 5.1, with $\succ_{D1}$.

**Stage 2:**

Proportional completion is used to complete $V$ to $W$.

**Stage 3:**

A *path* from set $\mathfrak{X} \in A_M$ to set $\mathfrak{Y} \in A_M \setminus \{\mathfrak{X}\}$ is a sequence of sets $\mathfrak{C}(1),...,\mathfrak{C}(n) \in A_M$ with the following properties:

1. $\mathfrak{X} \equiv \mathfrak{C}(1)$.
2. $\mathfrak{Y} \equiv \mathfrak{C}(n)$.
3. $n \in \mathbb{N}$ with $2 \leq n \leq (C!)/((M!)\cdot((C-M)!))$.
4. For all $i,j \in \{1,...,n\}$: $i \neq j \Rightarrow \mathfrak{C}(i) \in A_M \setminus \{\mathfrak{C}(j)\}$.
5. For all $i = 1,...,(n-1)$: $\mathfrak{C}(i)$ and $\mathfrak{C}(i+1)$ differ in exactly one alternative. That means: $\| \mathfrak{C}(i) \cap \mathfrak{C}(i+1) \| = M - 1$ and $\| \mathfrak{C}(i) \cup \mathfrak{C}(i+1) \| = M + 1$.

The *strength* of the path $\mathfrak{C}(1),...,\mathfrak{C}(n)$ is

$$\min_{D2} \{ (N[\{a_1,...,a_{(M-1)},b\};c], N[\{a_1,...,a_{(M-1)},c\};b]) \text{ with } \{a_1,...,a_{(M-1)}\} := \mathfrak{C}(i) \cap \mathfrak{C}(i+1), b := \mathfrak{C}(i) \setminus \mathfrak{C}(i+1), \text{ and } c := \mathfrak{C}(i+1) \setminus \mathfrak{C}(i) \mid i = 1,...,(n-1) \}.$$

In other words: The strength of a path is the strength of its weakest link.

$$P_{D2}[\mathfrak{A},\mathfrak{B}] := \max_{D2} \{ \min_{D2} \{ (N[\{a_1,...,a_{(M-1)},b\};c], N[\{a_1,...,a_{(M-1)},c\};b]) \text{ with } \{a_1,...,a_{(M-1)}\} := \mathfrak{C}(i) \cap \mathfrak{C}(i+1), b := \mathfrak{C}(i) \setminus \mathfrak{C}(i+1), \text{ and } c := \mathfrak{C}(i+1) \setminus \mathfrak{C}(i) \mid i = 1,...,(n-1) \} \mid \mathfrak{C}(1),...,\mathfrak{C}(n) \text{ is a path from set } \mathfrak{A} \text{ to set } \mathfrak{B} \}.$$

In other words: $P_{D2}[\mathfrak{A},\mathfrak{B}] \in \mathbb{N}_0 \times \mathbb{N}_0$ is the strength of the strongest path from set $\mathfrak{A} \in A_M$ to set $\mathfrak{B} \in A_M \setminus \{\mathfrak{A}\}$.

(8.1.3.1) The binary relation $O_M$ on $A_M$ is defined as follows:

$$\mathfrak{A}\mathfrak{B} \in O_M :\Leftrightarrow P_{D2}[\mathfrak{A},\mathfrak{B}] \succ_{D2} P_{D2}[\mathfrak{B},\mathfrak{A}].$$

(8.1.3.2) $\mathcal{S}_M := \{ \mathfrak{A} \in A_M \mid \forall \mathfrak{B} \in A_M \setminus \{\mathfrak{A}\}: \mathfrak{B}\mathfrak{A} \notin O_M \}$ is the *set of potential winning sets*.

**Stage 4:**

For all $\mathfrak{A},\mathfrak{B} \in \mathcal{S}_M$: Suppose there is an alternative $a \in \mathfrak{A} \setminus \mathfrak{B}$ with $ab \in O_{final}(\sigma)$ for every alternative $b \in \mathfrak{B} \setminus \mathfrak{A}$, then the set $\mathfrak{A}$ *disqualifies* the set $\mathfrak{B}$.

The winning set of Schulze STV is that set $\mathfrak{A} \in \mathcal{S}_M$ that is not disqualified by some other set $\mathfrak{B} \in \mathcal{S}_M$.

## 8.2. Example A53

To illustrate Schulze STV, we will use a rather large example because smaller examples usually don't address all aspects of an STV election. We will use example A53 from Tideman's database. This example is analysed in great detail by Tideman (2000). Example A53 consists of $V = 460$ voters and $C = 10$ alternatives running for $M = 4$ seats.

Example A53 is interesting because the Newland-Britton (1997) method, the Meek (1969, 1970; I.D. Hill, 1987) method, and the Warren (1994) method each find a different set of winners. The Newland-Britton method chooses *a*, *b*, *g*, and *j*. The Meek method chooses *a*, *d*, *g*, and *j*. The Warren method chooses *a*, *f*, *g*, and *j*.

Example A53 is decribed in the following table 8.2.1. For example, row 233 says that voter 233 gives a "1" to alternative *b*, a "2" to alternative *c*, a "3" to alternative *d*, a "4" to alternative *a*, and a "5" to alternative *j*. Voter 233 doesn't rank any of the other alternatives.

| | *a* | *b* | *c* | *d* | *e* | *f* | *g* | *h* | *i* | *j* |
|---|---|---|---|---|---|---|---|---|---|---|
| 1 | 1 | - | - | - | - | 4 | 3 | 2 | - | - |
| 2 | 1 | 2 | 4 | 5 | 3 | 9 | 6 | 10 | 8 | 7 |
| 3 | 2 | 6 | 10 | 7 | 3 | 8 | 5 | 9 | 1 | 4 |
| 4 | - | - | - | - | - | - | - | - | - | 1 |
| 5 | - | - | - | - | - | - | - | - | - | 1 |
| 6 | 3 | - | - | - | 5 | 4 | 6 | 7 | 2 | 1 |
| 7 | 4 | - | 3 | - | - | - | - | - | 2 | 1 |
| 8 | 3 | - | 1 | - | - | - | - | - | 4 | 2 |
| 9 | 2 | - | 1 | - | - | - | - | - | - | - |
| 10 | 3 | - | - | - | - | - | - | 2 | - | 1 |
| 11 | - | 5 | - | - | 1 | 4 | 2 | - | 3 | 6 |
| 12 | - | 4 | 5 | - | 1 | - | 2 | 3 | - | - |
| 13 | 7 | 9 | 6 | 10 | 1 | 5 | 3 | 4 | 8 | 2 |
| 14 | 4 | 5 | 3 | 9 | 1 | 10 | 2 | 6 | 8 | 7 |
| 15 | - | - | - | - | 2 | - | 3 | - | 1 | 4 |
| 16 | - | - | 4 | - | 2 | - | 3 | - | 1 | - |
| 17 | 2 | - | 5 | - | 1 | - | 4 | - | 6 | 3 |
| 18 | - | - | - | - | 1 | 4 | 2 | - | - | 3 |
| 19 | 3 | - | - | 6 | - | 4 | 5 | 1 | - | 2 |
| 20 | 4 | - | - | 5 | - | 2 | 3 | - | - | 1 |
| 21 | 4 | 9 | 7 | 5 | 8 | 2 | 10 | 6 | 3 | 1 |
| 22 | 4 | 7 | 3 | 6 | 8 | 2 | 10 | 5 | 9 | 1 |
| 23 | 4 | - | - | 6 | 3 | 2 | - | 5 | - | 1 |
| 24 | - | 5 | - | 4 | - | 3 | - | - | 2 | 1 |
| 25 | - | - | - | 4 | - | 3 | - | - | 2 | 1 |
| 26 | 4 | 10 | 9 | 8 | 7 | 3 | 5 | 6 | 2 | 1 |
| 27 | 4 | - | - | 6 | - | 3 | - | 5 | 2 | 1 |
| 28 | 3 | 4 | - | 2 | - | - | - | 5 | - | 1 |
| 29 | 3 | - | - | 2 | - | - | - | - | - | 1 |
| 30 | 8 | 9 | 7 | 2 | 3 | 4 | 5 | 6 | 10 | 1 |
| 31 | 5 | 7 | 6 | 2 | 3 | 4 | 10 | 9 | 8 | 1 |
| 32 | 10 | 8 | 4 | 2 | 3 | 9 | 7 | 5 | 6 | 1 |
| 33 | 4 | 6 | - | 2 | 5 | 3 | - | - | - | 1 |
| 34 | - | - | - | 2 | - | - | 3 | - | - | 1 |
| 35 | 3 | 9 | 7 | 2 | 4 | 8 | 6 | 10 | 5 | 1 |
| 36 | 1 | 5 | 2 | - | - | - | 3 | - | - | 4 |
| 37 | 1 | 7 | 5 | 3 | 9 | 6 | 8 | 4 | 10 | 2 |
| 38 | 1 | 7 | 5 | 6 | 3 | 2 | 8 | 9 | 10 | 4 |
| 39 | 1 | 2 | 4 | 3 | 10 | 9 | 5 | 8 | 7 | 6 |
| 40 | 1 | - | - | 3 | 6 | 2 | - | 4 | - | 5 |
| 41 | 1 | - | - | - | - | - | 2 | - | - | - |
| 42 | 1 | 8 | 5 | 2 | 4 | 3 | 10 | 9 | 7 | 6 |
| 43 | 1 | - | - | 2 | 3 | 5 | - | - | 4 | - |
| 44 | 1 | 2 | 3 | 5 | 7 | 4 | 8 | 9 | 10 | 6 |
| 45 | 1 | 2 | 7 | 4 | 6 | 5 | 9 | 10 | 8 | 3 |
| 46 | 1 | 2 | 3 | 5 | 4 | 6 | 10 | 11 | 12 | 7 |
| 47 | 1 | 3 | 8 | 7 | 6 | 5 | 4 | 2 | 9 | 10 |
| 48 | 1 | 7 | 6 | 4 | 5 | 2 | 8 | 10 | 9 | 3 |
| 49 | 1 | 3 | 7 | 6 | 10 | 4 | 9 | 5 | 8 | 2 |
| 50 | 1 | 8 | 10 | 4 | 7 | 2 | 3 | 9 | 6 | 5 |
| 51 | 1 | - | 6 | - | 2 | 3 | - | 7 | 5 | - |
| 52 | 1 | 3 | 6 | 10 | 7 | 9 | 5 | 2 | 8 | 4 |
| 53 | 1 | 10 | 4 | 9 | 7 | 2 | 5 | 8 | 3 | 6 |
| 54 | 1 | - | - | - | - | 2 | - | 3 | - | 4 |
| 55 | 1 | 2 | - | - | - | - | 3 | - | - | 4 |
| 56 | 4 | 5 | - | - | 6 | - | - | 3 | 2 | 1 |
| 57 | 4 | 5 | 6 | 10 | 8 | 9 | 7 | 3 | 2 | 1 |
| 58 | 4 | 5 | 9 | 6 | 7 | 8 | 10 | 2 | 3 | 1 |
| 59 | 6 | 3 | 5 | 7 | 10 | 4 | 8 | 2 | 9 | 1 |
| 60 | 10 | 3 | 4 | 9 | 5 | 6 | 8 | 2 | 7 | 1 |
| 61 | 3 | 4 | 2 | 7 | 8 | 9 | 10 | 5 | 6 | 1 |
| 62 | 7 | 3 | 6 | 9 | 2 | 8 | 5 | 10 | 4 | 1 |
| 63 | 4 | 3 | 7 | 9 | 2 | 6 | 10 | 5 | 8 | 1 |
| 64 | 5 | 3 | 10 | 7 | 2 | 9 | 8 | 6 | 4 | 1 |
| 65 | - | 1 | 2 | - | - | - | - | - | - | 3 |
| 66 | - | - | 1 | 6 | 2 | - | 3 | - | 4 | 5 |
| 67 | - | - | 4 | 3 | 1 | - | 2 | - | - | - |
| 68 | 2 | 9 | 8 | 5 | 1 | 6 | 3 | 7 | 10 | 4 |
| 69 | 6 | 3 | 7 | 9 | 2 | 8 | 1 | 10 | 4 | 5 |
| 70 | 4 | 9 | 6 | 10 | 3 | 7 | 1 | 8 | 5 | 2 |
| 71 | 3 | 9 | 5 | 6 | 4 | 8 | 1 | 7 | 10 | 2 |
| 72 | 2 | - | - | - | - | - | 1 | 4 | - | 3 |
| 73 | 7 | 4 | 8 | 5 | 9 | 6 | 1 | 2 | 10 | 3 |
| 74 | 9 | 10 | 8 | 5 | 7 | 6 | 1 | 2 | 3 | 4 |
| 75 | - | - | - | 2 | 3 | - | 1 | - | - | - |
| 76 | 7 | 8 | 10 | 3 | 2 | 6 | 1 | 9 | 4 | 5 |
| 77 | 4 | 3 | 2 | - | - | - | 1 | - | - | - |
| 78 | 3 | 7 | 4 | 5 | 6 | 8 | 1 | 9 | 10 | 2 |
| 79 | 5 | 10 | 6 | 7 | 2 | 8 | 1 | 9 | 3 | 4 |
| 80 | 2 | 5 | 4 | 6 | 7 | 10 | 1 | 8 | 9 | 3 |
| 81 | - | - | - | - | - | - | 1 | - | - | 2 |
| 82 | - | - | - | 3 | - | - | 1 | 2 | - | - |
| 83 | 2 | 4 | 3 | 9 | 10 | 8 | 1 | 5 | 7 | 6 |
| 84 | 4 | 7 | 2 | 6 | 5 | 8 | 1 | 9 | 10 | 3 |
| 85 | - | 2 | - | - | 3 | - | 1 | - | - | - |
| 86 | 2 | 5 | - | - | 4 | - | 1 | - | - | 3 |
| 87 | 5 | - | - | 2 | 4 | - | 1 | 3 | - | 6 |
| 88 | - | - | - | - | - | - | 1 | - | 2 | - |
| 89 | 2 | - | 4 | - | - | - | 1 | - | - | 3 |
| 90 | 7 | 3 | 4 | 2 | 9 | 6 | 1 | 8 | 10 | 5 |
| 91 | 2 | 8 | 5 | 7 | 6 | 3 | 1 | 10 | 9 | 4 |
| 92 | 7 | 10 | 6 | 9 | 5 | 4 | 1 | 8 | 3 | 2 |
| 93 | - | 2 | - | - | - | - | 1 | 3 | - | - |
| 94 | 3 | - | - | 2 | - | - | 1 | - | - | - |
| 95 | - | - | - | - | - | 3 | 1 | - | 2 | 4 |
| 96 | - | - | - | 2 | - | 3 | 1 | 4 | - | - |
| 97 | - | - | - | - | - | 2 | 1 | - | 3 | 4 |
| 98 | 6 | 10 | 8 | 9 | 7 | 5 | 1 | 4 | 2 | 3 |
| 99 | 3 | - | - | 2 | - | - | 1 | - | - | - |
| 100 | 5 | - | - | 4 | - | - | 1 | 3 | - | 2 |
| 101 | 4 | - | - | 3 | 5 | - | 1 | - | - | 2 |
| 102 | 3 | - | - | - | - | 2 | 1 | - | 5 | 4 |
| 103 | 4 | 3 | - | 2 | - | - | 1 | - | 5 | 6 |
| 104 | 3 | - | - | 4 | - | 2 | 1 | - | 6 | 5 |
| 105 | - | 4 | - | 2 | - | 3 | 1 | - | - | - |
| 106 | 3 | - | 4 | - | 5 | - | 1 | 6 | - | 2 |
| 107 | 10 | 5 | 3 | 2 | 4 | 9 | 1 | 7 | 8 | 6 |
| 108 | 3 | - | - | 1 | - | - | - | - | - | 2 |
| 109 | - | 3 | 2 | 1 | - | 4 | - | - | - | - |
| 110 | 2 | - | - | 1 | - | - | - | - | 4 | 3 |
| 111 | - | - | - | 1 | - | - | - | - | - | 2 |
| 112 | 5 | 4 | - | 1 | - | - | - | 3 | - | 2 |
| 113 | - | - | - | 1 | - | - | 2 | - | - | - |
| 114 | 6 | 5 | 10 | 1 | 3 | 7 | 2 | 8 | 9 | 4 |
| 115 | - | - | - | 1 | 4 | - | - | - | 3 | 2 |
| 116 | 2 | - | - | 1 | - | - | - | - | - | 3 |
| 117 | - | 4 | - | 1 | - | 3 | - | - | - | 2 |
| 118 | 4 | 3 | - | 1 | - | - | 2 | - | - | - |
| 119 | 2 | 5 | 6 | 1 | 3 | 4 | 7 | 10 | 9 | 8 |
| 120 | - | - | - | 1 | - | - | 2 | - | - | - |

Table 8.2.1 (part 1 of 4): Example A53

| | *a* | *b* | *c* | *d* | *e* | *f* | *g* | *h* | *i* | *j* |
|---|---|---|---|---|---|---|---|---|---|---|
| 121 | 2 | - | - | 1 | - | 3 | - | - | - | 4 |
| 122 | - | - | - | 1 | - | - | 2 | - | - | 3 |
| 123 | 3 | 4 | - | 1 | - | - | 2 | - | - | - |
| 124 | - | - | - | 1 | - | - | - | - | - | - |
| 125 | - | - | - | 1 | - | 2 | 4 | - | 3 | - |
| 126 | - | - | - | 1 | - | - | - | - | - | - |
| 127 | 4 | 5 | 7 | 1 | 2 | 9 | 6 | 8 | 3 | 10 |
| 128 | 3 | - | - | 1 | - | - | - | 2 | - | 4 |
| 129 | 2 | 3 | - | 1 | - | - | 5 | - | - | 4 |
| 130 | 3 | 10 | 6 | 1 | 4 | 8 | 7 | 9 | 5 | 2 |
| 131 | 2 | - | 4 | 1 | - | 5 | - | - | - | 3 |
| 132 | - | - | - | 1 | - | - | - | 3 | - | 2 |
| 133 | - | - | - | 1 | 2 | 3 | - | - | - | 4 |
| 134 | 3 | 2 | 7 | 1 | 6 | 9 | 10 | 5 | 8 | 4 |
| 135 | 2 | 5 | 6 | 1 | - | 3 | 4 | - | - | - |
| 136 | 5 | 4 | 8 | 1 | 6 | 9 | 7 | 3 | 2 | 10 |
| 137 | 2 | 9 | 5 | 1 | 3 | 10 | 8 | 6 | 4 | 7 |
| 138 | - | - | - | 1 | - | 2 | - | 3 | - | 4 |
| 139 | - | - | - | 1 | 2 | 3 | - | - | - | - |
| 140 | 9 | 7 | 8 | 1 | 2 | 6 | 3 | 10 | 5 | 4 |
| 141 | 3 | 4 | 6 | 1 | 7 | 9 | 2 | 10 | 8 | 5 |
| 142 | 5 | 6 | 7 | 1 | 2 | 9 | 3 | 10 | 4 | 8 |
| 143 | 3 | 9 | 6 | 1 | 10 | 4 | 2 | 7 | 8 | 5 |
| 144 | - | 4 | - | 1 | 5 | 3 | - | - | 2 | - |
| 145 | - | - | - | 1 | - | - | - | - | - | - |
| 146 | 2 | 6 | 9 | 1 | 8 | 5 | 10 | 3 | 7 | 4 |
| 147 | 2 | - | 3 | 1 | - | - | - | - | 4 | 5 |
| 148 | - | - | - | 1 | - | 2 | - | - | - | 3 |
| 149 | 2 | - | - | 1 | 4 | - | - | - | - | 3 |
| 150 | 3 | 6 | 5 | 2 | 7 | 10 | 8 | 9 | 1 | 4 |
| 151 | 8 | 6 | 7 | 4 | 5 | 10 | 9 | 3 | 1 | 2 |
| 152 | - | - | - | 3 | - | - | 4 | 1 | - | 2 |
| 153 | 7 | - | 1 | 3 | 6 | 5 | 4 | - | - | 2 |
| 154 | - | - | 1 | 2 | - | - | - | - | - | - |
| 155 | - | 5 | 2 | 3 | - | 4 | - | - | 1 | 6 |
| 156 | - | - | 2 | 3 | 5 | - | 4 | - | 1 | 6 |
| 157 | 5 | 4 | 9 | 2 | 1 | 6 | 7 | 8 | 10 | 3 |
| 158 | 4 | 10 | 5 | 2 | 1 | 6 | 3 | 9 | 8 | 7 |
| 159 | - | - | - | 2 | 1 | - | - | - | - | - |
| 160 | 2 | - | - | 3 | 1 | - | 5 | - | - | 4 |
| 161 | 2 | 9 | 7 | 5 | 1 | 8 | 6 | 4 | 10 | 3 |
| 162 | - | - | 1 | 3 | 2 | 4 | - | - | - | - |
| 163 | 3 | 8 | 9 | 4 | 10 | 1 | 5 | 2 | 6 | 7 |
| 164 | - | - | 6 | 4 | 3 | 1 | - | 5 | - | 2 |
| 165 | 2 | 8 | 7 | 3 | 4 | 1 | 5 | 9 | 10 | 6 |
| 166 | 8 | 6 | 5 | 3 | 10 | 1 | 4 | 7 | 9 | 2 |
| 167 | 5 | - | 6 | 3 | - | 1 | - | 4 | - | 2 |
| 168 | 6 | 8 | 5 | 7 | 4 | 1 | 9 | 3 | 10 | 2 |
| 169 | 4 | - | - | 3 | - | 1 | - | - | 5 | 2 |
| 170 | 6 | 8 | 7 | 2 | 9 | 1 | 10 | 5 | 4 | 3 |
| 171 | 2 | - | - | 3 | - | 1 | - | - | 4 | - |
| 172 | 8 | 9 | 6 | 2 | 4 | 1 | 3 | 10 | 5 | 7 |
| 173 | 2 | - | - | 5 | 3 | 1 | 4 | - | - | - |
| 174 | - | 5 | - | 3 | 4 | 1 | - | - | - | 2 |
| 175 | 2 | 6 | 7 | 3 | 5 | 1 | 8 | 10 | 9 | 4 |
| 176 | 9 | 10 | 3 | 8 | 2 | 1 | 4 | 7 | 5 | 6 |
| 177 | 4 | - | - | 2 | - | 1 | 3 | - | - | - |
| 178 | 9 | 6 | 4 | 5 | 2 | 3 | 10 | 8 | 1 | 7 |
| 179 | 5 | - | 4 | - | - | 3 | - | 2 | - | 1 |
| 180 | - | - | 4 | - | 6 | 3 | 5 | 2 | - | 1 |

| | *a* | *b* | *c* | *d* | *e* | *f* | *g* | *h* | *i* | *j* |
|---|---|---|---|---|---|---|---|---|---|---|
| 181 | 3 | - | - | - | - | 5 | 6 | 2 | 4 | 1 |
| 182 | 3 | - | - | - | - | 4 | - | 2 | - | 1 |
| 183 | 4 | - | - | - | - | 2 | - | 3 | - | 1 |
| 184 | - | - | - | - | - | 2 | 3 | - | - | 1 |
| 185 | 4 | - | - | - | - | 2 | 3 | - | - | 1 |
| 186 | 3 | - | - | - | - | 2 | 4 | - | - | 1 |
| 187 | 3 | 4 | 5 | 8 | 7 | 2 | 6 | 9 | 10 | 1 |
| 188 | 7 | 8 | 9 | 10 | 3 | 6 | 4 | 2 | 5 | 1 |
| 189 | - | 4 | - | - | 2 | 3 | - | - | - | 1 |
| 190 | 4 | 9 | 7 | 10 | 2 | 5 | 8 | 3 | 6 | 1 |
| 191 | 3 | 5 | 8 | 6 | 10 | 2 | 7 | 9 | 4 | 1 |
| 192 | 10 | 5 | 6 | 9 | 4 | 3 | 7 | 8 | 2 | 1 |
| 193 | 6 | 4 | - | 5 | - | 3 | - | 7 | 2 | 1 |
| 194 | 5 | 8 | 6 | 10 | 4 | 3 | 9 | 7 | 2 | 1 |
| 195 | - | 7 | - | - | 5 | 3 | 4 | 6 | 2 | 1 |
| 196 | 5 | 10 | 9 | 4 | 6 | 7 | 3 | 8 | 1 | 2 |
| 197 | - | - | - | - | 3 | - | 2 | - | 1 | - |
| 198 | - | - | - | - | - | 3 | 2 | - | 1 | 4 |
| 199 | - | - | - | - | - | - | 2 | 1 | - | - |
| 200 | 3 | 9 | 1 | 5 | 10 | 6 | 2 | 4 | 8 | 7 |
| 201 | 2 | 7 | 1 | 5 | 6 | 4 | 3 | 8 | 9 | 10 |
| 202 | 2 | 6 | 1 | 7 | 9 | 10 | 3 | 8 | 5 | 4 |
| 203 | - | 5 | 1 | - | - | - | 2 | 3 | 4 | - |
| 204 | - | - | 1 | - | 4 | - | 3 | - | 5 | 2 |
| 205 | 9 | 10 | 1 | 8 | 7 | 3 | 2 | 5 | 4 | 6 |
| 206 | 2 | - | 3 | 5 | - | - | 4 | 1 | - | - |
| 207 | 4 | - | - | - | 3 | - | - | 2 | - | 1 |
| 208 | 4 | - | - | - | 3 | - | - | 2 | - | 1 |
| 209 | - | - | - | - | 2 | - | - | - | - | 1 |
| 210 | - | - | - | - | 2 | - | - | 3 | 4 | 1 |
| 211 | 2 | 6 | 5 | 8 | 3 | 9 | 7 | 1 | 10 | 4 |
| 212 | 3 | - | 4 | - | - | 2 | 1 | - | - | - |
| 213 | - | - | - | - | - | - | 1 | 2 | 3 | - |
| 214 | - | 4 | - | - | - | 2 | 1 | 3 | - | - |
| 215 | - | - | 5 | - | 6 | 4 | 1 | - | 3 | 2 |
| 216 | 5 | 4 | 6 | - | 8 | 2 | 1 | 7 | - | 3 |
| 217 | 8 | 10 | 3 | 6 | 7 | 2 | 1 | 4 | 9 | 5 |
| 218 | - | - | - | 3 | - | 2 | 1 | - | 4 | 5 |
| 219 | 3 | 4 | 5 | 10 | 6 | 9 | 1 | 8 | 7 | 2 |
| 220 | 2 | 6 | 10 | 8 | 7 | 4 | 1 | 5 | 9 | 3 |
| 221 | - | 3 | - | 2 | - | 4 | 1 | - | - | - |
| 222 | - | - | - | 3 | 4 | - | - | - | 2 | 1 |
| 223 | 3 | - | 5 | 4 | - | - | - | 2 | - | 1 |
| 224 | 4 | - | - | 3 | - | - | - | 2 | - | 1 |
| 225 | 3 | - | 2 | 4 | - | - | - | - | - | 1 |
| 226 | 4 | 9 | 2 | 6 | 5 | 7 | 8 | 3 | 10 | 1 |
| 227 | 3 | 7 | 4 | 6 | 5 | 9 | 8 | 2 | 10 | 1 |
| 228 | 5 | 4 | - | 3 | 2 | - | - | - | - | 1 |
| 229 | 3 | 1 | - | - | - | 5 | 4 | - | 6 | 2 |
| 230 | 2 | 1 | - | - | - | - | 4 | - | 3 | - |
| 231 | 2 | 1 | - | - | - | 3 | - | - | 4 | - |
| 232 | - | 1 | - | - | - | - | 3 | - | - | 2 |
| 233 | 4 | 1 | 2 | 3 | - | - | - | - | - | 5 |
| 234 | 6 | 1 | 5 | 7 | 4 | 8 | 2 | 9 | 10 | 3 |
| 235 | 9 | 1 | 5 | 10 | 4 | 2 | 6 | 8 | 7 | 3 |
| 236 | 4 | 1 | 5 | 9 | 3 | 6 | 8 | 7 | 10 | 2 |
| 237 | - | 1 | - | - | 3 | - | 2 | - | - | - |
| 238 | 2 | 1 | 3 | - | - | 4 | 5 | - | - | - |
| 239 | - | 1 | - | - | 2 | - | - | 3 | - | 4 |
| 240 | - | 1 | - | - | - | - | - | - | - | - |

Table 8.2.1 (part 2 of 4): Example A53

| | *a* | *b* | *c* | *d* | *e* | *f* | *g* | *h* | *i* | *j* |
|---|---|---|---|---|---|---|---|---|---|---|
| 241 | 3 | 1 | 7 | 2 | 8 | 9 | 4 | 10 | 6 | 5 |
| 242 | - | 1 | - | - | 2 | - | 3 | 4 | 5 | 6 |
| 243 | 2 | 1 | 9 | 6 | 5 | 7 | 10 | 3 | 8 | 4 |
| 244 | 7 | 1 | 8 | 3 | 6 | 4 | 5 | 9 | 10 | 2 |
| 245 | 5 | 1 | 6 | 7 | 2 | 10 | 3 | 9 | 8 | 4 |
| 246 | 2 | 1 | 8 | 5 | 9 | 6 | 10 | 7 | 3 | 4 |
| 247 | 2 | 1 | 3 | 10 | 4 | 5 | 7 | 6 | 8 | 9 |
| 248 | - | 1 | - | - | 2 | 4 | - | 3 | - | - |
| 249 | 6 | 1 | 4 | 8 | 7 | 2 | 5 | 10 | 9 | 3 |
| 250 | 2 | 1 | 9 | 10 | 7 | 6 | 4 | 5 | 8 | 3 |
| 251 | 4 | 1 | 5 | 6 | - | - | 2 | 7 | - | 3 |
| 252 | - | 1 | - | - | - | 2 | - | - | 3 | 4 |
| 253 | - | 1 | 5 | - | 2 | - | 3 | - | - | 4 |
| 254 | 3 | 1 | 5 | 2 | 8 | 4 | 10 | 7 | 9 | 6 |
| 255 | 6 | 1 | 5 | 7 | 4 | 8 | 3 | 9 | 2 | 10 |
| 256 | 5 | 1 | 6 | 3 | 2 | 9 | 4 | 7 | 8 | 10 |
| 257 | 9 | 1 | 6 | 5 | 3 | 10 | 2 | 4 | 7 | 8 |
| 258 | 6 | 1 | 3 | 5 | 4 | 10 | 2 | 8 | 9 | 7 |
| 259 | - | 1 | - | - | - | 2 | - | - | 3 | 4 |
| 260 | - | 1 | - | - | 4 | 3 | 5 | - | - | 2 |
| 261 | - | 1 | 4 | - | - | 2 | - | - | - | 3 |
| 262 | - | 1 | - | - | 2 | - | - | 3 | 4 | - |
| 263 | - | 1 | - | - | - | 4 | 2 | - | - | 3 |
| 264 | 5 | 1 | 9 | 4 | 2 | 6 | 10 | 7 | 8 | 3 |
| 265 | 2 | 1 | 8 | 7 | 6 | 5 | 9 | 10 | 3 | 4 |
| 266 | 3 | 1 | 7 | 6 | 9 | 5 | 4 | 8 | 10 | 2 |
| 267 | 9 | 2 | 8 | 3 | 7 | 10 | 4 | 6 | 1 | 5 |
| 268 | - | 3 | - | 4 | - | 5 | - | 1 | - | 2 |
| 269 | - | 3 | 2 | - | 6 | - | 5 | 1 | - | 4 |
| 270 | 7 | 4 | 2 | 10 | 5 | 9 | 6 | 1 | 3 | 8 |
| 271 | 6 | 4 | 3 | 10 | 5 | 7 | 9 | 1 | 8 | 2 |
| 272 | 3 | 2 | 1 | - | - | - | - | - | 4 | - |
| 273 | - | 2 | 1 | - | - | 3 | - | - | - | 4 |
| 274 | 9 | 5 | 8 | 7 | 2 | 6 | 10 | 1 | 3 | 4 |
| 275 | 4 | 2 | 8 | 7 | 1 | 10 | 9 | 5 | 3 | 6 |
| 276 | - | 2 | - | - | 1 | - | - | - | - | - |
| 277 | 4 | 5 | 6 | 8 | 3 | 9 | 10 | 1 | 7 | 2 |
| 278 | 6 | 3 | 5 | 10 | 1 | 4 | 9 | 7 | 8 | 2 |
| 279 | 2 | 6 | 9 | 10 | 1 | 7 | 8 | 3 | 4 | 5 |
| 280 | 6 | 4 | - | - | 3 | 5 | - | 1 | - | 2 |
| 281 | 7 | 2 | 8 | 6 | 4 | 9 | 3 | 10 | 5 | 1 |
| 282 | 3 | 2 | - | - | - | - | 4 | - | - | 1 |
| 283 | 5 | 2 | 8 | 4 | 3 | 6 | 10 | 7 | 9 | 1 |
| 284 | 7 | 2 | 9 | 8 | 3 | 10 | 4 | 5 | 6 | 1 |
| 285 | - | 2 | - | 3 | - | - | - | 4 | - | 1 |
| 286 | 5 | 2 | 10 | 7 | 4 | 3 | 8 | 6 | 9 | 1 |
| 287 | 4 | 2 | 6 | 5 | 3 | 8 | 7 | 10 | 9 | 1 |
| 288 | 2 | 3 | 9 | 5 | 10 | 6 | 7 | 4 | 8 | 1 |
| 289 | 2 | 4 | - | 6 | 3 | - | - | 5 | - | 1 |
| 290 | 2 | 4 | - | - | 5 | 6 | 3 | - | - | 1 |
| 291 | 2 | 3 | - | - | - | - | 5 | 4 | - | 1 |
| 292 | 2 | 3 | 10 | 4 | 9 | 5 | 6 | 7 | 8 | 1 |
| 293 | 2 | - | - | - | 3 | - | 4 | 5 | - | 1 |
| 294 | 2 | - | - | - | - | 4 | - | 3 | - | 1 |
| 295 | 2 | 3 | 4 | 8 | 7 | 10 | 5 | 6 | 9 | 1 |
| 296 | 2 | - | - | - | - | 3 | - | 4 | - | 1 |
| 297 | 2 | - | - | - | - | 3 | 4 | - | - | 1 |
| 298 | 2 | - | 3 | 4 | - | 5 | - | - | - | 1 |
| 299 | 2 | - | - | - | 3 | - | - | - | 4 | 1 |
| 300 | 2 | 5 | 4 | 3 | 6 | - | - | - | - | 1 |

| | *a* | *b* | *c* | *d* | *e* | *f* | *g* | *h* | *i* | *j* |
|---|---|---|---|---|---|---|---|---|---|---|
| 301 | 2 | - | - | - | - | - | 4 | - | 3 | 1 |
| 302 | 2 | 10 | 3 | 4 | 6 | 9 | 5 | 7 | 8 | 1 |
| 303 | 2 | - | 5 | - | - | 3 | - | - | 4 | 1 |
| 304 | 2 | 3 | 6 | 7 | 8 | 4 | 9 | 5 | 10 | 1 |
| 305 | 2 | 5 | - | - | - | - | - | 4 | 3 | 1 |
| 306 | 2 | 3 | - | - | - | - | - | 4 | - | 1 |
| 307 | 2 | 3 | 10 | 8 | 7 | 4 | 9 | 5 | 6 | 1 |
| 308 | 2 | - | - | - | - | 3 | - | 2 | 3 | 1 |
| 309 | 2 | 3 | - | - | - | - | 4 | - | - | 1 |
| 310 | 2 | 9 | 4 | 5 | 6 | 7 | 3 | 10 | 8 | 1 |
| 311 | 2 | 3 | 6 | 4 | - | - | - | - | 5 | 1 |
| 312 | 2 | - | - | 4 | - | - | - | - | 3 | 1 |
| 313 | 2 | 10 | 3 | 9 | 6 | 5 | 8 | 7 | 4 | 1 |
| 314 | 2 | - | - | - | - | 3 | - | 4 | - | 1 |
| 315 | 2 | 3 | 5 | 8 | 7 | 6 | 9 | 4 | 10 | 1 |
| 316 | 2 | - | - | 3 | - | - | 4 | - | - | 1 |
| 317 | 2 | 3 | 7 | 6 | 8 | 4 | 9 | 5 | 10 | 1 |
| 318 | 2 | 4 | - | 3 | - | - | 5 | - | - | 1 |
| 319 | 2 | - | 5 | 3 | 4 | - | 6 | - | - | 1 |
| 320 | 2 | 3 | 6 | 10 | 5 | 4 | 7 | 8 | 9 | 1 |
| 321 | 2 | 4 | 7 | 8 | 5 | 9 | 10 | 3 | 6 | 1 |
| 322 | 2 | - | - | 5 | 3 | - | 4 | - | - | 1 |
| 323 | - | - | 4 | - | 1 | - | - | - | 3 | 2 |
| 324 | - | - | - | - | 1 | - | - | 3 | 4 | 2 |
| 325 | - | - | - | - | 1 | - | - | - | - | - |
| 326 | - | - | - | - | 1 | - | - | - | - | 2 |
| 327 | 2 | - | 3 | - | 1 | - | - | - | - | - |
| 328 | 4 | - | 5 | 3 | - | - | 1 | - | - | 2 |
| 329 | 5 | - | 2 | 4 | - | 3 | 1 | - | - | - |
| 330 | - | - | - | 2 | - | - | 1 | - | - | 3 |
| 331 | - | - | - | 2 | - | - | 1 | 4 | - | 3 |
| 332 | - | - | 3 | 2 | 4 | 5 | 1 | - | - | - |
| 333 | 3 | 5 | - | - | - | 4 | 1 | 6 | - | 2 |
| 334 | 5 | - | 6 | 4 | 2 | - | 1 | - | 3 | - |
| 335 | 2 | 3 | - | - | - | - | 1 | - | - | - |
| 336 | - | 3 | 2 | - | - | - | 1 | - | - | - |
| 337 | 6 | 3 | 7 | 2 | 4 | 10 | 1 | 8 | 9 | 5 |
| 338 | - | 4 | 5 | - | - | 2 | 1 | - | 3 | - |
| 339 | - | 3 | - | 4 | - | - | 1 | 2 | - | - |
| 340 | - | - | - | - | - | - | 1 | - | - | - |
| 341 | 8 | 9 | 7 | 2 | 3 | 4 | 1 | 10 | 5 | 6 |
| 342 | 2 | 3 | 5 | 6 | 9 | 10 | 1 | 4 | 8 | 7 |
| 343 | 2 | 7 | 3 | 4 | 9 | 8 | 1 | 10 | 6 | 5 |
| 344 | 5 | 4 | 9 | 3 | 8 | 10 | 1 | 7 | 2 | 6 |
| 345 | - | - | - | - | - | - | 1 | - | - | - |
| 346 | 2 | - | - | 4 | 3 | - | 1 | 6 | 7 | 5 |
| 347 | 4 | 9 | 7 | 10 | 3 | 2 | 1 | 8 | 6 | 5 |
| 348 | 9 | 8 | 7 | 2 | 10 | 6 | 1 | 5 | 4 | 3 |
| 349 | 3 | - | - | - | - | - | 2 | - | - | 1 |
| 350 | - | 4 | - | - | - | - | 2 | - | 3 | 1 |
| 351 | - | - | - | - | - | - | 2 | 3 | 4 | 1 |
| 352 | - | - | - | - | - | - | 2 | - | - | 1 |
| 353 | 3 | - | - | - | 4 | - | 2 | - | - | 1 |
| 354 | 3 | 4 | 8 | 7 | 10 | 9 | 2 | 6 | 5 | 1 |
| 355 | 3 | - | - | - | - | - | 2 | - | 4 | 1 |
| 356 | - | 4 | - | 3 | - | - | 2 | - | - | 1 |
| 357 | 5 | 8 | 7 | 3 | 10 | 4 | 2 | 9 | 6 | 1 |
| 358 | - | - | - | 3 | 4 | - | 2 | - | - | 1 |
| 359 | - | 4 | - | 3 | - | - | 2 | - | - | 1 |
| 360 | - | 4 | - | - | - | 3 | 2 | - | - | 1 |

Table 8.2.1 (part 3 of 4): Example A53

| | *a* | *b* | *c* | *d* | *e* | *f* | *g* | *h* | *i* | *j* |
|---|---|---|---|---|---|---|---|---|---|---|
| 361 | 4 | 3 | - | - | - | - | 2 | - | - | 1 |
| 362 | - | 3 | - | 4 | - | - | 2 | - | - | 1 |
| 363 | - | - | - | - | - | - | 3 | - | 2 | 1 |
| 364 | - | - | - | - | 4 | - | 2 | 3 | 1 | - |
| 365 | 3 | 7 | 8 | 10 | 9 | 6 | 4 | 2 | 5 | 1 |
| 366 | 5 | 9 | 10 | 6 | 8 | 4 | 3 | 2 | 7 | 1 |
| 367 | 7 | 4 | 2 | 6 | 8 | 5 | 3 | 10 | 9 | 1 |
| 368 | - | 4 | - | - | 3 | 1 | 2 | - | - | - |
| 369 | 3 | 7 | 6 | 10 | 9 | 1 | 2 | 4 | 8 | 5 |
| 370 | 3 | 4 | - | - | - | 1 | 2 | - | - | 5 |
| 371 | 4 | 2 | - | - | - | 1 | - | 3 | - | - |
| 372 | 5 | 6 | 7 | 10 | 8 | 1 | 2 | 3 | 9 | 4 |
| 373 | 4 | 3 | 5 | 9 | 8 | 1 | 6 | 10 | 7 | 2 |
| 374 | 4 | 8 | 5 | 9 | 7 | 1 | 3 | - | 2 | 6 |
| 375 | 2 | 4 | 3 | 5 | 6 | 1 | 10 | 8 | 9 | 7 |
| 376 | 5 | 9 | 2 | 10 | 3 | 1 | 6 | 7 | 4 | 8 |
| 377 | - | 2 | - | - | - | 1 | 3 | - | 4 | 5 |
| 378 | 6 | 7 | 10 | 9 | 5 | 1 | 2 | 9 | 4 | 3 |
| 379 | 3 | 2 | - | 4 | - | 1 | - | - | - | - |
| 380 | 5 | 6 | 4 | 10 | 3 | 1 | 2 | 7 | 9 | 8 |
| 381 | 5 | 4 | 7 | 8 | 1 | 3 | 6 | 9 | 10 | 2 |
| 382 | 2 | 5 | - | - | 1 | 4 | - | - | 3 | - |
| 383 | 3 | 6 | 7 | 9 | 1 | 2 | 8 | 5 | 4 | 10 |
| 384 | - | 2 | 5 | - | 3 | 1 | 4 | - | - | - |
| 385 | 4 | - | 3 | - | - | 1 | - | - | - | 2 |
| 386 | 4 | - | - | - | 2 | 1 | - | - | - | 3 |
| 387 | - | - | 3 | - | - | 1 | 2 | - | - | - |
| 388 | - | - | 3 | - | - | 1 | 4 | - | - | 2 |
| 389 | - | - | - | - | - | 1 | 3 | 4 | - | 2 |
| 390 | - | - | 5 | - | - | 1 | - | 3 | 4 | 2 |
| 391 | 5 | - | 3 | - | - | 1 | 4 | - | 6 | 2 |
| 392 | - | - | - | - | 2 | 1 | 5 | - | 3 | 4 |
| 393 | 4 | - | 5 | - | - | 2 | - | - | 1 | 3 |
| 394 | - | - | - | - | 1 | 2 | 3 | - | 4 | - |
| 395 | - | - | 1 | - | 2 | 3 | - | 4 | - | - |
| 396 | - | - | 2 | - | 1 | 3 | - | 4 | - | 5 |
| 397 | - | - | 4 | - | 3 | 1 | 2 | - | - | - |
| 398 | 3 | - | - | - | - | 1 | - | - | - | 2 |
| 399 | - | - | - | - | - | 1 | - | - | - | 2 |
| 400 | - | - | 2 | - | - | 1 | - | - | - | - |
| 401 | 1 | 7 | 5 | 2 | 10 | 8 | 6 | 9 | 3 | 4 |
| 402 | 1 | - | 3 | - | - | - | 2 | 4 | - | - |
| 403 | 1 | 6 | 8 | 2 | 9 | 5 | 7 | 4 | 10 | 3 |
| 404 | 1 | - | - | 2 | - | - | - | - | - | 3 |
| 405 | 1 | - | - | - | - | - | 2 | - | - | 3 |
| 406 | 1 | 4 | - | - | - | - | - | 2 | - | 3 |
| 407 | 1 | - | - | 5 | 4 | - | - | - | 3 | 2 |
| 408 | 1 | - | 4 | 5 | - | 6 | 3 | 7 | 8 | 2 |
| 409 | 1 | - | 5 | 3 | 4 | - | 2 | - | - | - |
| 410 | 1 | 10 | 9 | 8 | 6 | 4 | 7 | 2 | 5 | 3 |

| | *a* | *b* | *c* | *d* | *e* | *f* | *g* | *h* | *i* | *j* |
|---|---|---|---|---|---|---|---|---|---|---|
| 411 | 1 | - | 3 | 2 | - | - | 4 | - | - | - |
| 412 | 1 | 3 | - | - | 2 | 4 | - | - | - | - |
| 413 | 1 | 10 | 6 | 4 | 9 | 8 | 3 | 7 | 2 | 5 |
| 414 | 1 | - | - | - | - | 2 | - | - | - | - |
| 415 | 1 | - | 2 | - | - | - | - | 3 | - | 4 |
| 416 | 1 | 4 | - | - | - | 3 | 2 | - | - | 5 |
| 417 | 1 | 3 | - | - | - | 2 | - | - | 4 | - |
| 418 | 1 | - | - | 4 | - | 5 | 3 | - | - | 2 |
| 419 | 1 | 7 | 4 | 7 | 8 | 9 | 3 | 5 | 6 | 2 |
| 420 | 1 | - | - | - | - | 2 | 3 | - | - | 4 |
| 421 | 1 | 6 | 3 | 7 | 2 | 9 | 5 | 4 | 10 | 8 |
| 422 | 1 | - | 2 | 7 | 8 | 9 | 3 | 4 | 6 | 5 |
| 423 | 1 | - | - | 2 | - | - | 3 | - | - | - |
| 424 | 1 | 10 | 9 | 2 | 5 | 3 | 7 | 6 | 8 | 4 |
| 425 | 1 | 3 | - | - | - | - | - | 4 | - | 2 |
| 426 | 1 | - | - | - | - | 2 | - | - | - | 3 |
| 427 | 1 | 6 | - | - | 3 | - | - | 4 | 5 | 2 |
| 428 | 1 | 8 | 9 | 4 | 7 | 2 | 3 | 10 | 6 | 5 |
| 429 | 1 | 6 | 10 | 3 | 9 | 7 | 8 | 2 | 4 | 5 |
| 430 | 1 | - | - | - | - | 4 | 3 | - | - | 2 |
| 431 | 1 | 9 | 2 | 8 | 3 | 10 | 4 | 6 | 7 | 5 |
| 432 | 1 | 4 | - | - | - | - | - | 3 | - | 2 |
| 433 | 1 | - | 3 | - | - | - | 4 | - | - | 2 |
| 434 | 1 | - | - | - | - | - | - | - | - | 2 |
| 435 | 1 | - | 3 | - | - | 2 | - | - | 5 | 4 |
| 436 | 1 | - | 2 | 3 | 5 | - | 4 | - | - | 6 |
| 437 | 1 | - | - | - | - | 2 | 3 | - | - | 4 |
| 438 | 1 | 4 | 10 | 6 | 5 | 8 | 2 | 9 | 3 | 7 |
| 439 | 1 | - | - | 3 | - | - | 4 | - | - | 2 |
| 440 | 1 | - | - | - | - | - | 2 | 4 | 3 | - |
| 441 | 1 | 6 | 2 | 5 | 3 | 9 | 10 | 7 | 4 | 8 |
| 442 | 1 | 8 | 9 | 2 | 4 | 7 | 10 | 5 | 6 | 3 |
| 443 | 1 | - | - | - | - | - | 2 | - | - | - |
| 444 | 1 | 2 | - | - | - | - | 4 | - | 5 | 3 |
| 445 | 1 | 7 | 8 | 9 | 6 | 4 | 10 | 3 | 5 | 2 |
| 446 | 1 | 3 | - | - | - | 5 | - | 4 | - | 2 |
| 447 | 1 | - | 2 | 4 | 3 | - | - | - | - | - |
| 448 | - | 1 | - | - | - | - | - | - | - | - |
| 449 | 1 | 4 | - | 5 | - | 3 | - | - | - | 2 |
| 450 | 1 | 2 | 6 | 9 | 5 | 7 | 8 | 3 | 10 | 4 |
| 451 | 1 | 2 | 3 | - | - | - | 4 | - | - | - |
| 452 | 1 | 3 | - | 4 | - | - | - | - | - | 2 |
| 453 | 1 | - | - | 2 | 3 | - | - | - | - | 4 |
| 454 | 1 | 3 | 2 | 8 | 7 | 10 | 9 | 6 | 4 | 5 |
| 455 | 1 | - | - | 3 | - | 4 | - | - | - | 2 |
| 456 | 1 | - | - | - | - | 2 | - | 4 | - | 3 |
| 457 | 1 | 6 | 10 | 2 | 5 | 8 | 3 | 9 | 4 | 7 |
| 458 | 1 | 4 | 10 | 5 | 9 | 8 | 6 | 2 | 7 | 3 |
| 459 | 1 | - | 3 | 2 | - | - | - | - | - | - |
| 460 | 1 | 3 | 7 | 2 | 10 | 8 | 6 | 9 | 4 | 5 |

Table 8.2.1 (part 4 of 4): Example A53

### 8.2.1. Proportional Completion

We apply proportional completion separately for the calculation of each link. The strength of the links $\{b,c,e,j\} \rightarrow \{a,c,e,j\}$, $\{b,c,e,j\} \rightarrow \{a,b,e,j\}$, $\{b,c,e,j\} \rightarrow \{a,b,c,j\}$, and $\{b,c,e,j\} \rightarrow \{a,b,c,e\}$ depends only on whether the individual voter strictly prefers the different candidates of the set $\{b,c,e,j\}$ to candidate $a$ or strictly prefers candidate $a$ to the different candidates of the set $\{b,c,e,j\}$ or is indifferent between the different candidates of the set $\{b,c,e,j\}$ and candidate $a$. Therefore, the fact, that we apply proportional completion for every link separately, means that only 3^$C$ = 81 possible voting patterns need to be considered. Table 8.2.1.1 lists these 81 possible voting patterns, where "1" means that a voter with this voting pattern strictly prefers this candidate to candidate $a$, a "2" means that this voter is indifferent between this candidate and candidate $a$, and a "3" means that this voter strictly prefers candidate $a$ to this candidate.

Throughout section 8.2.1, $w_j^i$ is the number of voters at stage $j$ who are using voting pattern $i$.

| voting pattern | *b* | *c* | *e* | *j* |
|---|---|---|---|---|
| #1 | 1 | 1 | 1 | 1 |
| #2 | 1 | 1 | 1 | 2 |
| #3 | 1 | 1 | 1 | 3 |
| #4 | 1 | 1 | 2 | 1 |
| #5 | 1 | 1 | 2 | 2 |
| #6 | 1 | 1 | 2 | 3 |
| #7 | 1 | 1 | 3 | 1 |
| #8 | 1 | 1 | 3 | 2 |
| #9 | 1 | 1 | 3 | 3 |
| #10 | 1 | 2 | 1 | 1 |
| #11 | 1 | 2 | 1 | 2 |
| #12 | 1 | 2 | 1 | 3 |
| #13 | 1 | 2 | 2 | 1 |
| #14 | 1 | 2 | 2 | 2 |
| #15 | 1 | 2 | 2 | 3 |
| #16 | 1 | 2 | 3 | 1 |
| #17 | 1 | 2 | 3 | 2 |
| #18 | 1 | 2 | 3 | 3 |
| #19 | 1 | 3 | 1 | 1 |
| #20 | 1 | 3 | 1 | 2 |
| #21 | 1 | 3 | 1 | 3 |
| #22 | 1 | 3 | 2 | 1 |
| #23 | 1 | 3 | 2 | 2 |
| #24 | 1 | 3 | 2 | 3 |
| #25 | 1 | 3 | 3 | 1 |
| #26 | 1 | 3 | 3 | 2 |
| #27 | 1 | 3 | 3 | 3 |
| #28 | 2 | 1 | 1 | 1 |
| #29 | 2 | 1 | 1 | 2 |
| #30 | 2 | 1 | 1 | 3 |
| #31 | 2 | 1 | 2 | 1 |
| #32 | 2 | 1 | 2 | 2 |
| #33 | 2 | 1 | 2 | 3 |
| #34 | 2 | 1 | 3 | 1 |
| #35 | 2 | 1 | 3 | 2 |
| #36 | 2 | 1 | 3 | 3 |
| #37 | 2 | 2 | 1 | 1 |
| #38 | 2 | 2 | 1 | 2 |
| #39 | 2 | 2 | 1 | 3 |
| #40 | 2 | 2 | 2 | 1 |

| voting pattern | *b* | *c* | *e* | *j* |
|---|---|---|---|---|
| #41 | 2 | 2 | 2 | 2 |
| #42 | 2 | 2 | 2 | 3 |
| #43 | 2 | 2 | 3 | 1 |
| #44 | 2 | 2 | 3 | 2 |
| #45 | 2 | 2 | 3 | 3 |
| #46 | 2 | 3 | 1 | 1 |
| #47 | 2 | 3 | 1 | 2 |
| #48 | 2 | 3 | 1 | 3 |
| #49 | 2 | 3 | 2 | 1 |
| #50 | 2 | 3 | 2 | 2 |
| #51 | 2 | 3 | 2 | 3 |
| #52 | 2 | 3 | 3 | 1 |
| #53 | 2 | 3 | 3 | 2 |
| #54 | 2 | 3 | 3 | 3 |
| #55 | 3 | 1 | 1 | 1 |
| #56 | 3 | 1 | 1 | 2 |
| #57 | 3 | 1 | 1 | 3 |
| #58 | 3 | 1 | 2 | 1 |
| #59 | 3 | 1 | 2 | 2 |
| #60 | 3 | 1 | 2 | 3 |
| #61 | 3 | 1 | 3 | 1 |
| #62 | 3 | 1 | 3 | 2 |
| #63 | 3 | 1 | 3 | 3 |
| #64 | 3 | 2 | 1 | 1 |
| #65 | 3 | 2 | 1 | 2 |
| #66 | 3 | 2 | 1 | 3 |
| #67 | 3 | 2 | 2 | 1 |
| #68 | 3 | 2 | 2 | 2 |
| #69 | 3 | 2 | 2 | 3 |
| #70 | 3 | 2 | 3 | 1 |
| #71 | 3 | 2 | 3 | 2 |
| #72 | 3 | 2 | 3 | 3 |
| #73 | 3 | 3 | 1 | 1 |
| #74 | 3 | 3 | 1 | 2 |
| #75 | 3 | 3 | 1 | 3 |
| #76 | 3 | 3 | 2 | 1 |
| #77 | 3 | 3 | 2 | 2 |
| #78 | 3 | 3 | 2 | 3 |
| #79 | 3 | 3 | 3 | 1 |
| #80 | 3 | 3 | 3 | 2 |
| #81 | 3 | 3 | 3 | 3 |

Table 8.2.1.1: The 81 possible voting patterns

## Step 1

At first, we determine which profile is used by how many voters. Table 8.2.1.2 lists, for every voting pattern, how many voters (column "number of voters") and which voters (column "voters") are using this voting pattern.

| voting pattern | number of voters | *b* | *c* | *e* | *j* | voters |
|---|---|---|---|---|---|---|
| #1 | $w_1^1 = 17$ | 1 | 1 | 1 | 1 | 32, 60, 62, 107, 140, 151, 178, 192, 234, 235, 253, 257, 267, 269, 271, 274, 278 |
| #2 | $w_1^2 = 2$ | 1 | 1 | 1 | 2 | 12, 384 |
| #3 | $w_1^3 = 3$ | 1 | 1 | 1 | 3 | 255, 258, 270 |
| #4 | $w_1^4 = 4$ | 1 | 1 | 2 | 1 | 65, 155, 261, 273 |
| #5 | $w_1^5 = 4$ | 1 | 1 | 2 | 2 | 109, 203, 336, 338 |
| #7 | $w_1^7 = 6$ | 1 | 1 | 3 | 1 | 59, 90, 166, 249, 348, 367 |
| #9 | $w_1^9 = 3$ | 1 | 1 | 3 | 3 | 77, 233, 272 |
| #10 | $w_1^{10} = 7$ | 1 | 2 | 1 | 1 | 11, 174, 189, 195, 239, 242, 260 |
| #11 | $w_1^{11} = 7$ | 1 | 2 | 1 | 2 | 85, 144, 237, 248, 262, 276, 368 |
| #13 | $w_1^{13} = 14$ | 1 | 2 | 2 | 1 | 24, 117, 232, 252, 259, 263, 268, 285, 350, 356, 359, 360, 362, 377 |
| #14 | $w_1^{14} = 7$ | 1 | 2 | 2 | 2 | 93, 105, 214, 221, 240, 339, 448 |
| #19 | $w_1^{19} = 18$ | 1 | 3 | 1 | 1 | 63, 64, 69, 114, 157, 228, 236, 244, 245, 264, 280, 281, 283, 284, 286, 287, 337, 381 |
| #21 | $w_1^{21} = 2$ | 1 | 3 | 1 | 3 | 256, 275 |
| #25 | $w_1^{25} = 10$ | 1 | 3 | 3 | 1 | 73, 112, 193, 216, 229, 251, 266, 282, 361, 373 |
| #27 | $w_1^{27} = 17$ | 1 | 3 | 3 | 3 | 103, 118, 134, 136, 230, 231, 238, 241, 243, 246, 247, 250, 254, 265, 344, 371, 379 |
| #28 | $w_1^{28} = 8$ | 2 | 1 | 1 | 1 | 66, 156, 164, 180, 204, 215, 323, 396 |
| #29 | $w_1^{29} = 6$ | 2 | 1 | 1 | 2 | 16, 67, 162, 332, 395, 397 |
| #31 | $w_1^{31} = 2$ | 2 | 1 | 2 | 1 | 388, 390 |
| #32 | $w_1^{32} = 3$ | 2 | 1 | 2 | 2 | 154, 387, 400 |
| #37 | $w_1^{37} = 11$ | 2 | 2 | 1 | 1 | 15, 18, 115, 133, 209, 210, 222, 324, 326, 358, 392 |
| #38 | $w_1^{38} = 7$ | 2 | 2 | 1 | 2 | 75, 139, 159, 197, 325, 364, 394 |
| #40 | $w_1^{40} = 23$ | 2 | 2 | 2 | 1 | 4, 5, 25, 34, 81, 95, 97, 111, 122, 132, 138, 148, 152, 184, 198, 218, 330, 331, 351, 352, 363, 389, 399 |

Table 8.2.1.2 (1 of 2): voting patterns is example A53

| voting pattern | number of voters | *b* | *c* | *e* | *j* | voters |
|---|---|---|---|---|---|---|
| #41 | $w_1^{41} = 13$ | 2 | 2 | 2 | 2 | 82, 88, 96, 113, 120, 124, 125, 126, 145, 199, 213, 340, 345 |
| #55 | $w_1^{55} = 11$ | 3 | 1 | 1 | 1 | 13, 30, 74, 92, 153, 168, 172, 176, 205, 217, 341 |
| #57 | $w_1^{57} = 3$ | 3 | 1 | 1 | 3 | 14, 376, 380 |
| #61 | $w_1^{61} = 10$ | 3 | 1 | 3 | 1 | 7, 8, 22, 61, 84, 179, 225, 226, 385, 391 |
| #63 | $w_1^{63} = 5$ | 3 | 1 | 3 | 3 | 9, 200, 201, 202, 329 |
| #73 | $w_1^{73} = 13$ | 3 | 3 | 1 | 1 | 23, 31, 70, 76, 79, 188, 190, 194, 207, 208, 277, 378, 386 |
| #75 | $w_1^{75} = 14$ | 3 | 3 | 1 | 3 | 17, 68, 87, 127, 142, 158, 160, 161, 279, 327, 334, 347, 382, 383 |
| #79 | $w_1^{79} = 84$ | 3 | 3 | 3 | 1 | 6, 10, 19, 20, 21, 26, 27, 28, 29, 33, 35, 56, 57, 58, 71, 78, 98, 100, 101, 106, 108, 130, 167, 169, 170, 181, 182, 183, 185, 186, 187, 191, 196, 219, 223, 224, 227, 288, 289, 290, 291, 292, 293, 294, 295, 296, 297, 298, 299, 300, 301, 302, 303, 304, 305, 306, 307, 308, 309, 310, 311, 312, 313, 314, 315, 316, 317, 318, 319, 320, 321, 322, 328, 333, 349, 353, 354, 355, 357, 365, 366, 372, 393, 398 |
| #81 | $w_1^{81} = 126$ | 3 | 3 | 3 | 3 | 1, 2, 3, 36, 37, 38, 39, 40, 41, 42, 43, 44, 45, 46, 47, 48, 49, 50, 51, 52, 53, 54, 55, 72, 80, 83, 86, 89, 91, 94, 99, 102, 104, 110, 116, 119, 121, 123, 128, 129, 131, 135, 137, 141, 143, 146, 147, 149, 150, 163, 165, 171, 173, 175, 177, 206, 211, 212, 220, 335, 342, 343, 346, 369, 370, 374, 375, 401, 402, 403, 404, 405, 406, 407, 408, 409, 410, 411, 412, 413, 414, 415, 416, 417, 418, 419, 420, 421, 422, 423, 424, 425, 426, 427, 428, 429, 430, 431, 432, 433, 434, 435, 436, 437, 438, 439, 440, 441, 442, 443, 444, 445, 446, 447, 449, 450, 451, 452, 453, 454, 455, 456, 457, 458, 459, 460 |

Table 8.2.1.2 (2 of 2): voting patterns is example A53

## Step 2

Each time, when we apply proportional completion to a voting pattern, we apply it to a voting pattern, where the number of alternatives with a “2” is the maximum. As, in each stage, a voting pattern is replaced by voting patterns with smaller numbers of alternatives with a “2”, it is guaranteed that those voting patterns, to which proportional completion has already been applied at earlier stages of the proportional completion procedure, cannot reappear at later stages.

So first, we apply proportional completion to voting pattern #41. Applying proportional completion to a voting pattern where voters are indifferent between all candidates simply means that the weight of every other voting pattern is multiplicated by the same factor.

Therefore, we get:

| voting pattern | number of voters | $b$ | $c$ | $e$ | $j$ |
|---|---|---|---|---|---|
| #1 | $w_2^1 = (1 + w_1^{41} / (N - w_1^{41})) \cdot w_1^1 = 17.494407$ | 1 | 1 | 1 | 1 |
| #2 | $w_2^2 = (1 + w_1^{41} / (N - w_1^{41})) \cdot w_1^2 = 2.058166$ | 1 | 1 | 1 | 2 |
| #3 | $w_2^3 = (1 + w_1^{41} / (N - w_1^{41})) \cdot w_1^3 = 3.087248$ | 1 | 1 | 1 | 3 |
| #4 | $w_2^4 = (1 + w_1^{41} / (N - w_1^{41})) \cdot w_1^4 = 4.116331$ | 1 | 1 | 2 | 1 |
| #5 | $w_2^5 = (1 + w_1^{41} / (N - w_1^{41})) \cdot w_1^5 = 4.116331$ | 1 | 1 | 2 | 2 |
| #7 | $w_2^7 = (1 + w_1^{41} / (N - w_1^{41})) \cdot w_1^7 = 6.174497$ | 1 | 1 | 3 | 1 |
| #9 | $w_2^9 = (1 + w_1^{41} / (N - w_1^{41})) \cdot w_1^9 = 3.087248$ | 1 | 1 | 3 | 3 |
| #10 | $w_2^{10} = (1 + w_1^{41} / (N - w_1^{41})) \cdot w_1^{10} = 7.203579$ | 1 | 2 | 1 | 1 |
| #11 | $w_2^{11} = (1 + w_1^{41} / (N - w_1^{41})) \cdot w_1^{11} = 7.203579$ | 1 | 2 | 1 | 2 |
| #13 | $w_2^{13} = (1 + w_1^{41} / (N - w_1^{41})) \cdot w_1^{13} = 14.407159$ | 1 | 2 | 2 | 1 |
| #14 | $w_2^{14} = (1 + w_1^{41} / (N - w_1^{41})) \cdot w_1^{14} = 7.203579$ | 1 | 2 | 2 | 2 |
| #19 | $w_2^{19} = (1 + w_1^{41} / (N - w_1^{41})) \cdot w_1^{19} = 18.523490$ | 1 | 3 | 1 | 1 |
| #21 | $w_2^{21} = (1 + w_1^{41} / (N - w_1^{41})) \cdot w_1^{21} = 2.058166$ | 1 | 3 | 1 | 3 |
| #25 | $w_2^{25} = (1 + w_1^{41} / (N - w_1^{41})) \cdot w_1^{25} = 10.290828$ | 1 | 3 | 3 | 1 |
| #27 | $w_2^{27} = (1 + w_1^{41} / (N - w_1^{41})) \cdot w_1^{27} = 17.494407$ | 1 | 3 | 3 | 3 |
| #28 | $w_2^{28} = (1 + w_1^{41} / (N - w_1^{41})) \cdot w_1^{28} = 8.232662$ | 2 | 1 | 1 | 1 |
| #29 | $w_2^{29} = (1 + w_1^{41} / (N - w_1^{41})) \cdot w_1^{29} = 6.174497$ | 2 | 1 | 1 | 2 |
| #31 | $w_2^{31} = (1 + w_1^{41} / (N - w_1^{41})) \cdot w_1^{31} = 2.058166$ | 2 | 1 | 2 | 1 |
| #32 | $w_2^{32} = (1 + w_1^{41} / (N - w_1^{41})) \cdot w_1^{32} = 3.087248$ | 2 | 1 | 2 | 2 |
| #37 | $w_2^{37} = (1 + w_1^{41} / (N - w_1^{41})) \cdot w_1^{37} = 11.319911$ | 2 | 2 | 1 | 1 |
| #38 | $w_2^{38} = (1 + w_1^{41} / (N - w_1^{41})) \cdot w_1^{38} = 7.203579$ | 2 | 2 | 1 | 2 |
| #40 | $w_2^{40} = (1 + w_1^{41} / (N - w_1^{41})) \cdot w_1^{40} = 23.668904$ | 2 | 2 | 2 | 1 |
| #55 | $w_2^{55} = (1 + w_1^{41} / (N - w_1^{41})) \cdot w_1^{55} = 11.319911$ | 3 | 1 | 1 | 1 |
| #57 | $w_2^{57} = (1 + w_1^{41} / (N - w_1^{41})) \cdot w_1^{57} = 3.087248$ | 3 | 1 | 1 | 3 |
| #61 | $w_2^{61} = (1 + w_1^{41} / (N - w_1^{41})) \cdot w_1^{61} = 10.290828$ | 3 | 1 | 3 | 1 |
| #63 | $w_2^{63} = (1 + w_1^{41} / (N - w_1^{41})) \cdot w_1^{63} = 5.145414$ | 3 | 1 | 3 | 3 |
| #73 | $w_2^{73} = (1 + w_1^{41} / (N - w_1^{41})) \cdot w_1^{73} = 13.378076$ | 3 | 3 | 1 | 1 |
| #75 | $w_2^{75} = (1 + w_1^{41} / (N - w_1^{41})) \cdot w_1^{75} = 14.407159$ | 3 | 3 | 1 | 3 |
| #79 | $w_2^{79} = (1 + w_1^{41} / (N - w_1^{41})) \cdot w_1^{79} = 86.442953$ | 3 | 3 | 3 | 1 |
| #81 | $w_2^{81} = (1 + w_1^{41} / (N - w_1^{41})) \cdot w_1^{81} = 129.664430$ | 3 | 3 | 3 | 3 |
| | 460.000000 | | | | |

## Step 3

We now apply proportional completion to voting pattern #14. In voting pattern #14, the voters are indifferent between the alternatives in $\{a, c, e, j\}$. At stage 1, $Y := w_1^{14} + w_1^{41} = 20$ voters were indifferent between all the alternatives in $\{a, c, e, j\}$. The following $N - Y = 440$ voters were not indifferent between all the alternatives in $\{a, c, e, j\}$:

| number of voters | $c$ | $e$ | $j$ |
|---|---|---|---|
| $w_1^{1} + w_1^{28} + w_1^{55} = 36$ | 1 | 1 | 1 |
| $w_1^{2} + w_1^{29} = 8$ | 1 | 1 | 2 |
| $w_1^{3} + w_1^{57} = 6$ | 1 | 1 | 3 |
| $w_1^{4} + w_1^{31} = 6$ | 1 | 2 | 1 |
| $w_1^{5} + w_1^{32} = 7$ | 1 | 2 | 2 |
| $w_1^{7} + w_1^{61} = 16$ | 1 | 3 | 1 |
| $w_1^{9} + w_1^{63} = 8$ | 1 | 3 | 3 |
| $w_1^{10} + w_1^{37} = 18$ | 2 | 1 | 1 |
| $w_1^{11} + w_1^{38} = 14$ | 2 | 1 | 2 |
| $w_1^{13} + w_1^{40} = 37$ | 2 | 2 | 1 |
| $w_1^{19} + w_1^{73} = 31$ | 3 | 1 | 1 |
| $w_1^{21} + w_1^{75} = 16$ | 3 | 1 | 3 |
| $w_1^{25} + w_1^{79} = 94$ | 3 | 3 | 1 |
| $w_1^{27} + w_1^{81} = 143$ | 3 | 3 | 3 |
| $N - Y = 440$ | | | |

Therefore, the $w_2^{14}$ = 7.203579 voters with voting pattern #14 are replaced by the following voters:

| voting pattern | number of voters | $b$ | $c$ | $e$ | $j$ |
|---|---|---|---|---|---|
| #1 | $(w_1^{1} + w_1^{28} + w_1^{55}) \cdot w_2^{14} / (N - Y) = 0.589384$ | 1 | 1 | 1 | 1 |
| #2 | $(w_1^{2} + w_1^{29}) \cdot w_2^{14} / (N - Y) = 0.130974$ | 1 | 1 | 1 | 2 |
| #3 | $(w_1^{3} + w_1^{57}) \cdot w_2^{14} / (N - Y) = 0.098231$ | 1 | 1 | 1 | 3 |
| #4 | $(w_1^{4} + w_1^{31}) \cdot w_2^{14} / (N - Y) = 0.098231$ | 1 | 1 | 2 | 1 |
| #5 | $(w_1^{5} + w_1^{32}) \cdot w_2^{14} / (N - Y) = 0.114602$ | 1 | 1 | 2 | 2 |
| #7 | $(w_1^{7} + w_1^{61}) \cdot w_2^{14} / (N - Y) = 0.2619483$ | 1 | 1 | 3 | 1 |
| #9 | $(w_1^{9} + w_1^{63}) \cdot w_2^{14} / (N - Y) = 0.130974$ | 1 | 1 | 3 | 3 |
| #10 | $(w_1^{10} + w_1^{37}) \cdot w_2^{14} / (N - Y) = 0.294692$ | 1 | 2 | 1 | 1 |
| #11 | $(w_1^{11} + w_1^{38}) \cdot w_2^{14} / (N - Y) = 0.229205$ | 1 | 2 | 1 | 2 |
| #13 | $(w_1^{13} + w_1^{40}) \cdot w_2^{14} / (N - Y) = 0.605756$ | 1 | 2 | 2 | 1 |
| #19 | $(w_1^{19} + w_1^{73}) \cdot w_2^{14} / (N - Y) = 0.507525$ | 1 | 3 | 1 | 1 |
| #21 | $(w_1^{21} + w_1^{75}) \cdot w_2^{14} / (N - Y) = 0.261948$ | 1 | 3 | 1 | 3 |
| #25 | $(w_1^{25} + w_1^{79}) \cdot w_2^{14} / (N - Y) = 1.538947$ | 1 | 3 | 3 | 1 |
| #27 | $(w_1^{27} + w_1^{81}) \cdot w_2^{14} / (N - Y) = 2.341163$ | 1 | 3 | 3 | 3 |
| | $w_2^{14}$ = 7.203579 | | | | |

Therefore, we get:

| voting pattern | number of voters | *b* | *c* | *e* | *j* |
|---|---|---|---|---|---|
| #1 | $w_3^1 = w_2^1 + 0.589384 = 18.083791$ | 1 | 1 | 1 | 1 |
| #2 | $w_3^2 = w_2^2 + 0.130974 = 2.189140$ | 1 | 1 | 1 | 2 |
| #3 | $w_3^3 = w_2^3 + 0.098231 = 3.185479$ | 1 | 1 | 1 | 3 |
| #4 | $w_3^4 = w_2^4 + 0.098231 = 4.214562$ | 1 | 1 | 2 | 1 |
| #5 | $w_3^5 = w_2^5 + 0.114602 = 4.230933$ | 1 | 1 | 2 | 2 |
| #7 | $w_3^7 = w_2^7 + 0.261948 = 6.436445$ | 1 | 1 | 3 | 1 |
| #9 | $w_3^9 = w_2^9 + 0.130974 = 3.218222$ | 1 | 1 | 3 | 3 |
| #10 | $w_3^{10} = w_2^{10} + 0.294692 = 7.498271$ | 1 | 2 | 1 | 1 |
| #11 | $w_3^{11} = w_2^{11} + 0.229205 = 7.432784$ | 1 | 2 | 1 | 2 |
| #13 | $w_3^{13} = w_2^{13} + 0.605756 = 15.012914$ | 1 | 2 | 2 | 1 |
| #19 | $w_3^{19} = w_2^{19} + 0.507525 = 19.031015$ | 1 | 3 | 1 | 1 |
| #21 | $w_3^{21} = w_2^{21} + 0.261948 = 2.320114$ | 1 | 3 | 1 | 3 |
| #25 | $w_3^{25} = w_2^{25} + 1.538947 = 11.829774$ | 1 | 3 | 3 | 1 |
| #27 | $w_3^{27} = w_2^{27} + 2.341163 = 19.835570$ | 1 | 3 | 3 | 3 |
| #28 | $w_3^{28} = w_2^{28} = 8.232662$ | 2 | 1 | 1 | 1 |
| #29 | $w_3^{29} = w_2^{29} = 6.174497$ | 2 | 1 | 1 | 2 |
| #31 | $w_3^{31} = w_2^{31} = 2.058166$ | 2 | 1 | 2 | 1 |
| #32 | $w_3^{32} = w_2^{32} = 3.087248$ | 2 | 1 | 2 | 2 |
| #37 | $w_3^{37} = w_2^{37} = 11.319911$ | 2 | 2 | 1 | 1 |
| #38 | $w_3^{38} = w_2^{38} = 7.203579$ | 2 | 2 | 1 | 2 |
| #40 | $w_3^{40} = w_2^{40} = 23.668904$ | 2 | 2 | 2 | 1 |
| #55 | $w_3^{55} = w_2^{55} = 11.319911$ | 3 | 1 | 1 | 1 |
| #57 | $w_3^{57} = w_2^{57} = 3.087248$ | 3 | 1 | 1 | 3 |
| #61 | $w_3^{61} = w_2^{61} = 10.290828$ | 3 | 1 | 3 | 1 |
| #63 | $w_3^{63} = w_2^{63} = 5.145414$ | 3 | 1 | 3 | 3 |
| #73 | $w_3^{73} = w_2^{73} = 13.378076$ | 3 | 3 | 1 | 1 |
| #75 | $w_3^{75} = w_2^{75} = 14.407159$ | 3 | 3 | 1 | 3 |
| #79 | $w_3^{79} = w_2^{79} = 86.442953$ | 3 | 3 | 3 | 1 |
| #81 | $w_3^{81} = w_2^{81} = 129.664430$ | 3 | 3 | 3 | 3 |
| | 460.000000 | | | | |

## Step 4

We now apply proportional completion to voting pattern #32. In voting pattern #32, the voters are indifferent between the alternatives in $\{a, b, e, j\}$. At stage 1, $Y := w_1^{32} + w_1^{41} = 16$ voters were indifferent between all the alternatives in $\{a, b, e, j\}$. The following $N - Y = 444$ voters were not indifferent between all the alternatives in $\{a, b, e, j\}$:

| number of voters | $b$ | $e$ | $j$ |
|---|---|---|---|
| $w_1^{1} + w_1^{10} + w_1^{19} = 42$ | 1 | 1 | 1 |
| $w_1^{2} + w_1^{11} = 9$ | 1 | 1 | 2 |
| $w_1^{3} + w_1^{21} = 5$ | 1 | 1 | 3 |
| $w_1^{4} + w_1^{13} = 18$ | 1 | 2 | 1 |
| $w_1^{5} + w_1^{14} = 11$ | 1 | 2 | 2 |
| $w_1^{7} + w_1^{25} = 16$ | 1 | 3 | 1 |
| $w_1^{9} + w_1^{27} = 20$ | 1 | 3 | 3 |
| $w_1^{28} + w_1^{37} = 19$ | 2 | 1 | 1 |
| $w_1^{29} + w_1^{38} = 13$ | 2 | 1 | 2 |
| $w_1^{31} + w_1^{40} = 25$ | 2 | 2 | 1 |
| $w_1^{55} + w_1^{73} = 24$ | 3 | 1 | 1 |
| $w_1^{57} + w_1^{75} = 17$ | 3 | 1 | 3 |
| $w_1^{61} + w_1^{79} = 94$ | 3 | 3 | 1 |
| $w_1^{63} + w_1^{81} = 131$ | 3 | 3 | 3 |
| $N - Y = 444$ | | | |

Therefore, the $w_3^{32}$ = 3.087248 voters with voting pattern #32 are replaced by the following voters:

| voting pattern | number of voters | *b* | *c* | *e* | *j* |
|---|---|---|---|---|---|
| #1 | $(w_1^1 + w_1^{10} + w_1^{19}) \cdot w_3^{32} / (N - Y) = 0.292037$ | 1 | 1 | 1 | 1 |
| #2 | $(w_1^2 + w_1^{11}) \cdot w_3^{32} / (N - Y) = 0.062579$ | 1 | 1 | 1 | 2 |
| #3 | $(w_1^3 + w_1^{21}) \cdot w_3^{32} / (N - Y) = 0.034766$ | 1 | 1 | 1 | 3 |
| #4 | $(w_1^4 + w_1^{13}) \cdot w_3^{32} / (N - Y) = 0.125159$ | 1 | 1 | 2 | 1 |
| #5 | $(w_1^5 + w_1^{14}) \cdot w_3^{32} / (N - Y) = 0.076486$ | 1 | 1 | 2 | 2 |
| #7 | $(w_1^7 + w_1^{25}) \cdot w_3^{32} / (N - Y) = 0.111252$ | 1 | 1 | 3 | 1 |
| #9 | $(w_1^9 + w_1^{27}) \cdot w_3^{32} / (N - Y) = 0.139065$ | 1 | 1 | 3 | 3 |
| #28 | $(w_1^{28} + w_1^{37}) \cdot w_3^{32} / (N - Y) = 0.132112$ | 2 | 1 | 1 | 1 |
| #29 | $(w_1^{29} + w_1^{38}) \cdot w_3^{32} / (N - Y) = 0.090392$ | 2 | 1 | 1 | 2 |
| #31 | $(w_1^{31} + w_1^{40}) \cdot w_3^{32} / (N - Y) = 0.173832$ | 2 | 1 | 2 | 1 |
| #55 | $(w_1^{55} + w_1^{73}) \cdot w_3^{32} / (N - Y) = 0.166878$ | 3 | 1 | 1 | 1 |
| #57 | $(w_1^{57} + w_1^{75}) \cdot w_3^{32} / (N - Y) = 0.118205$ | 3 | 1 | 1 | 3 |
| #61 | $(w_1^{61} + w_1^{79}) \cdot w_3^{32} / (N - Y) = 0.653607$ | 3 | 1 | 3 | 1 |
| #63 | $(w_1^{63} + w_1^{81}) \cdot w_3^{32} / (N - Y) = 0.910877$ | 3 | 1 | 3 | 3 |
| | $w_3^{32}$ = 3.087248 | | | | |

Therefore, we get:

| voting pattern | number of voters | *b* | *c* | *e* | *j* |
|---|---|---|---|---|---|
| #1 | $w_4^1 = w_3^1 + 0.292037 = 18.375828$ | 1 | 1 | 1 | 1 |
| #2 | $w_4^2 = w_3^2 + 0.062579 = 2.251719$ | 1 | 1 | 1 | 2 |
| #3 | $w_4^3 = w_3^3 + 0.034766 = 3.220245$ | 1 | 1 | 1 | 3 |
| #4 | $w_4^4 = w_3^4 + 0.125159 = 4.339720$ | 1 | 1 | 2 | 1 |
| #5 | $w_4^5 = w_3^5 + 0.076486 = 4.307419$ | 1 | 1 | 2 | 2 |
| #7 | $w_4^7 = w_3^7 + 0.111252 = 6.547697$ | 1 | 1 | 3 | 1 |
| #9 | $w_4^9 = w_3^9 + 0.139065 = 3.357288$ | 1 | 1 | 3 | 3 |
| #10 | $w_4^{10} = w_3^{10} = 7.498271$ | 1 | 2 | 1 | 1 |
| #11 | $w_4^{11} = w_3^{11} = 7.432784$ | 1 | 2 | 1 | 2 |
| #13 | $w_4^{13} = w_3^{13} = 15.012914$ | 1 | 2 | 2 | 1 |
| #19 | $w_4^{19} = w_3^{19} = 19.031015$ | 1 | 3 | 1 | 1 |
| #21 | $w_4^{21} = w_3^{21} = 2.320114$ | 1 | 3 | 1 | 3 |
| #25 | $w_4^{25} = w_3^{25} = 11.829774$ | 1 | 3 | 3 | 1 |
| #27 | $w_4^{27} = w_3^{27} = 19.835570$ | 1 | 3 | 3 | 3 |
| #28 | $w_4^{28} = w_3^{28} + 0.132112 = 8.364774$ | 2 | 1 | 1 | 1 |
| #29 | $w_4^{29} = w_3^{29} + 0.090392 = 6.264889$ | 2 | 1 | 1 | 2 |
| #31 | $w_4^{31} = w_3^{31} + 0.173832 = 2.231997$ | 2 | 1 | 2 | 1 |
| #37 | $w_4^{37} = w_3^{37} = 11.319911$ | 2 | 2 | 1 | 1 |
| #38 | $w_4^{38} = w_3^{38} = 7.203579$ | 2 | 2 | 1 | 2 |
| #40 | $w_4^{40} = w_3^{40} = 23.668904$ | 2 | 2 | 2 | 1 |
| #55 | $w_4^{55} = w_3^{55} + 0.166878 = 11.486789$ | 3 | 1 | 1 | 1 |
| #57 | $w_4^{57} = w_3^{57} + 0.118205 = 3.205454$ | 3 | 1 | 1 | 3 |
| #61 | $w_4^{61} = w_3^{61} + 0.653607 = 10.944434$ | 3 | 1 | 3 | 1 |
| #63 | $w_4^{63} = w_3^{63} + 0.910877 = 6.056291$ | 3 | 1 | 3 | 3 |
| #73 | $w_4^{73} = w_3^{73} = 13.378076$ | 3 | 3 | 1 | 1 |
| #75 | $w_4^{75} = w_3^{75} = 14.407159$ | 3 | 3 | 1 | 3 |
| #79 | $w_4^{79} = w_3^{79} = 86.442953$ | 3 | 3 | 3 | 1 |
| #81 | $w_4^{81} = w_3^{81} = 129.664430$ | 3 | 3 | 3 | 3 |
| | 460.000000 | | | | |

## Step 5

We now apply proportional completion to voting pattern #38. In voting pattern #38, the voters are indifferent between the alternatives in $\{a, b, c, j\}$. At stage 1, $Y := w_1^{38} + w_1^{41} = 20$ voters were indifferent between all the alternatives in $\{a, b, c, j\}$. The following $N - Y = 440$ voters were not indifferent between all the alternatives in $\{a, b, c, j\}$:

| number of voters | $b$ | $c$ | $j$ |
|---|---|---|---|
| $w_1^1 + w_1^4 + w_1^7 = 27$ | 1 | 1 | 1 |
| $w_1^2 + w_1^5 = 6$ | 1 | 1 | 2 |
| $w_1^3 + w_1^9 = 6$ | 1 | 1 | 3 |
| $w_1^{10} + w_1^{13} = 21$ | 1 | 2 | 1 |
| $w_1^{11} + w_1^{14} = 14$ | 1 | 2 | 2 |
| $w_1^{19} + w_1^{25} = 28$ | 1 | 3 | 1 |
| $w_1^{21} + w_1^{27} = 19$ | 1 | 3 | 3 |
| $w_1^{28} + w_1^{31} = 10$ | 2 | 1 | 1 |
| $w_1^{29} + w_1^{32} = 9$ | 2 | 1 | 2 |
| $w_1^{37} + w_1^{40} = 34$ | 2 | 2 | 1 |
| $w_1^{55} + w_1^{61} = 21$ | 3 | 1 | 1 |
| $w_1^{57} + w_1^{63} = 8$ | 3 | 1 | 3 |
| $w_1^{73} + w_1^{79} = 97$ | 3 | 3 | 1 |
| $w_1^{75} + w_1^{81} = 140$ | 3 | 3 | 3 |
| $N - Y = 440$ | | | |

Therefore, the $w_4^{38}$ = 7.203579 voters with voting pattern #38 are replaced by the following voters:

| voting pattern | number of voters | $b$ | $c$ | $e$ | $j$ |
|---|---|---|---|---|---|
| #1 | $(w_1^1 + w_1^4 + w_1^7) \cdot w_4^{38} / (N - Y) = 0.442038$ | 1 | 1 | 1 | 1 |
| #2 | $(w_1^2 + w_1^5) \cdot w_4^{38} / (N - Y) = 0.098231$ | 1 | 1 | 1 | 2 |
| #3 | $(w_1^3 + w_1^9) \cdot w_4^{38} / (N - Y) = 0.098231$ | 1 | 1 | 1 | 3 |
| #10 | $(w_1^{10} + w_1^{13}) \cdot w_4^{38} / (N - Y) = 0.343807$ | 1 | 2 | 1 | 1 |
| #11 | $(w_1^{11} + w_1^{14}) \cdot w_4^{38} / (N - Y) = 0.229205$ | 1 | 2 | 1 | 2 |
| #19 | $(w_1^{19} + w_1^{25}) \cdot w_4^{38} / (N - Y) = 0.458410$ | 1 | 3 | 1 | 1 |
| #21 | $(w_1^{21} + w_1^{27}) \cdot w_4^{38} / (N - Y) = 0.311064$ | 1 | 3 | 1 | 3 |
| #28 | $(w_1^{28} + w_1^{31}) \cdot w_4^{38} / (N - Y) = 0.163718$ | 2 | 1 | 1 | 1 |
| #29 | $(w_1^{29} + w_1^{32}) \cdot w_4^{38} / (N - Y) = 0.147346$ | 2 | 1 | 1 | 2 |
| #37 | $(w_1^{37} + w_1^{40}) \cdot w_4^{38} / (N - Y) = 0.556640$ | 2 | 2 | 1 | 1 |
| #55 | $(w_1^{55} + w_1^{61}) \cdot w_4^{38} / (N - Y) = 0.343807$ | 3 | 1 | 1 | 1 |
| #57 | $(w_1^{57} + w_1^{63}) \cdot w_4^{38} / (N - Y) = 0.130974$ | 3 | 1 | 1 | 3 |
| #73 | $(w_1^{73} + w_1^{79}) \cdot w_4^{38} / (N - Y) = 1.588062$ | 3 | 3 | 1 | 1 |
| #75 | $(w_1^{75} + w_1^{81}) \cdot w_4^{38} / (N - Y) = 2.292048$ | 3 | 3 | 1 | 3 |
| | $w_4^{38}$ = 7.203579 | | | | |

Therefore, we get:

| voting pattern | number of voters | *b* | *c* | *e* | *j* |
|---|---|---|---|---|---|
| #1 | $w_5^1 = w_4^1 + 0.442038 = 18.817866$ | 1 | 1 | 1 | 1 |
| #2 | $w_5^2 = w_4^2 + 0.098231 = 2.349950$ | 1 | 1 | 1 | 2 |
| #3 | $w_5^3 = w_4^3 + 0.098231 = 3.318476$ | 1 | 1 | 1 | 3 |
| #4 | $w_5^4 = w_4^4 = 4.339720$ | 1 | 1 | 2 | 1 |
| #5 | $w_5^5 = w_4^5 = 4.307419$ | 1 | 1 | 2 | 2 |
| #7 | $w_5^7 = w_4^7 = 6.547697$ | 1 | 1 | 3 | 1 |
| #9 | $w_5^9 = w_4^9 = 3.357288$ | 1 | 1 | 3 | 3 |
| #10 | $w_5^{10} = w_4^{10} + 0.343807 = 7.842079$ | 1 | 2 | 1 | 1 |
| #11 | $w_5^{11} = w_4^{11} + 0.229205 = 7.661989$ | 1 | 2 | 1 | 2 |
| #13 | $w_5^{13} = w_4^{13} = 15.012914$ | 1 | 2 | 2 | 1 |
| #19 | $w_5^{19} = w_4^{19} + 0.458410 = 19.489424$ | 1 | 3 | 1 | 1 |
| #21 | $w_5^{21} = w_4^{21} + 0.311064 = 2.631178$ | 1 | 3 | 1 | 3 |
| #25 | $w_5^{25} = w_4^{25} = 11.829774$ | 1 | 3 | 3 | 1 |
| #27 | $w_5^{27} = w_4^{27} = 19.835570$ | 1 | 3 | 3 | 3 |
| #28 | $w_5^{28} = w_4^{28} + 0.163718 = 8.528492$ | 2 | 1 | 1 | 1 |
| #29 | $w_5^{29} = w_4^{29} + 0.147346 = 6.412235$ | 2 | 1 | 1 | 2 |
| #31 | $w_5^{31} = w_4^{31} = 2.231997$ | 2 | 1 | 2 | 1 |
| #37 | $w_5^{37} = w_4^{37} + 0.556640 = 11.876551$ | 2 | 2 | 1 | 1 |
| #40 | $w_5^{40} = w_4^{40} = 23.668904$ | 2 | 2 | 2 | 1 |
| #55 | $w_5^{55} = w_4^{55} + 0.343807 = 11.830596$ | 3 | 1 | 1 | 1 |
| #57 | $w_5^{57} = w_4^{57} + 0.130974 = 3.336428$ | 3 | 1 | 1 | 3 |
| #61 | $w_5^{61} = w_4^{61} = 10.944434$ | 3 | 1 | 3 | 1 |
| #63 | $w_5^{63} = w_4^{63} = 6.056291$ | 3 | 1 | 3 | 3 |
| #73 | $w_5^{73} = w_4^{73} + 1.588062 = 14.966138$ | 3 | 3 | 1 | 1 |
| #75 | $w_5^{75} = w_4^{75} + 2.292048 = 16.699207$ | 3 | 3 | 1 | 3 |
| #79 | $w_5^{79} = w_4^{79} = 86.442953$ | 3 | 3 | 3 | 1 |
| #81 | $w_5^{81} = w_4^{81} = 129.664430$ | 3 | 3 | 3 | 3 |
| | 460.000000 | | | | |

## Step 6

We now apply proportional completion to voting pattern #40. In voting pattern #40, the voters are indifferent between the alternatives in $\{a, b, c, e\}$. At stage 1, $Y := w_1^{40} + w_1^{41} = 36$ voters were indifferent between all the alternatives in $\{a, b, c, e\}$. The following $N - Y = 424$ voters were not indifferent between all the alternatives in $\{a, b, c, e\}$:

| number of voters | $b$ | $c$ | $e$ |
|---|---|---|---|
| $w_1^1 + w_1^2 + w_1^3 = 22$ | 1 | 1 | 1 |
| $w_1^4 + w_1^5 = 8$ | 1 | 1 | 2 |
| $w_1^7 + w_1^9 = 9$ | 1 | 1 | 3 |
| $w_1^{10} + w_1^{11} = 14$ | 1 | 2 | 1 |
| $w_1^{13} + w_1^{14} = 21$ | 1 | 2 | 2 |
| $w_1^{19} + w_1^{21} = 20$ | 1 | 3 | 1 |
| $w_1^{25} + w_1^{27} = 27$ | 1 | 3 | 3 |
| $w_1^{28} + w_1^{29} = 14$ | 2 | 1 | 1 |
| $w_1^{31} + w_1^{32} = 5$ | 2 | 1 | 2 |
| $w_1^{37} + w_1^{38} = 18$ | 2 | 2 | 1 |
| $w_1^{55} + w_1^{57} = 14$ | 3 | 1 | 1 |
| $w_1^{61} + w_1^{63} = 15$ | 3 | 1 | 3 |
| $w_1^{73} + w_1^{75} = 27$ | 3 | 3 | 1 |
| $w_1^{79} + w_1^{81} = 210$ | 3 | 3 | 3 |
| $N - Y = 424$ | | | |

Therefore, the $w_5^{40}$ = 23.668904 voters with voting pattern #40 are replaced by the following voters:

| voting pattern | number of voters | *b* | *c* | *e* | *j* |
|---|---|---|---|---|---|
| #1 | $(w_1^1 + w_1^2 + w_1^3) \cdot w_5^{40} / (N - Y) = 1.228103$ | 1 | 1 | 1 | 1 |
| #4 | $(w_1^4 + w_1^5) \cdot w_5^{40} / (N - Y) = 0.446583$ | 1 | 1 | 2 | 1 |
| #7 | $(w_1^7 + w_1^9) \cdot w_5^{40} / (N - Y) = 0.502406$ | 1 | 1 | 3 | 1 |
| #10 | $(w_1^{10} + w_1^{11}) \cdot w_5^{40} / (N - Y) = 0.781520$ | 1 | 2 | 1 | 1 |
| #13 | $(w_1^{13} + w_1^{14}) \cdot w_5^{40} / (N - Y) = 1.172281$ | 1 | 2 | 2 | 1 |
| #19 | $(w_1^{19} + w_1^{21}) \cdot w_5^{40} / (N - Y) = 1.116458$ | 1 | 3 | 1 | 1 |
| #25 | $(w_1^{25} + w_1^{27}) \cdot w_5^{40} / (N - Y) = 1.507218$ | 1 | 3 | 3 | 1 |
| #28 | $(w_1^{28} + w_1^{29}) \cdot w_5^{40} / (N - Y) = 0.781520$ | 2 | 1 | 1 | 1 |
| #31 | $(w_1^{31} + w_1^{32}) \cdot w_5^{40} / (N - Y) = 0.279114$ | 2 | 1 | 2 | 1 |
| #37 | $(w_1^{37} + w_1^{38}) \cdot w_5^{40} / (N - Y) = 1.004812$ | 2 | 2 | 1 | 1 |
| #55 | $(w_1^{55} + w_1^{57}) \cdot w_5^{40} / (N - Y) = 0.781520$ | 3 | 1 | 1 | 1 |
| #61 | $(w_1^{61} + w_1^{63}) \cdot w_5^{40} / (N - Y) = 0.837343$ | 3 | 1 | 3 | 1 |
| #73 | $(w_1^{73} + w_1^{75}) \cdot w_5^{40} / (N - Y) = 1.507218$ | 3 | 3 | 1 | 1 |
| #79 | $(w_1^{79} + w_1^{81}) \cdot w_5^{40} / (N - Y) = 11.722806$ | 3 | 3 | 3 | 1 |
| | $w_5^{40} = 23.668904$ | | | | |

Therefore, we get:

| voting pattern | number of voters | *b* | *c* | *e* | *j* |
|---|---|---|---|---|---|
| #1 | $w_6^1 = w_5^1 + 1.228103 = 20.045969$ | 1 | 1 | 1 | 1 |
| #2 | $w_6^2 = w_5^2 = 2.349950$ | 1 | 1 | 1 | 2 |
| #3 | $w_6^3 = w_5^3 = 3.318476$ | 1 | 1 | 1 | 3 |
| #4 | $w_6^4 = w_5^4 + 0.446583 = 4.786304$ | 1 | 1 | 2 | 1 |
| #5 | $w_6^5 = w_5^5 = 4.307419$ | 1 | 1 | 2 | 2 |
| #7 | $w_6^7 = w_5^7 + 0.502406 = 7.050103$ | 1 | 1 | 3 | 1 |
| #9 | $w_6^9 = w_5^9 = 3.357288$ | 1 | 1 | 3 | 3 |
| #10 | $w_6^{10} = w_5^{10} + 0.781520 = 8.623599$ | 1 | 2 | 1 | 1 |
| #11 | $w_6^{11} = w_5^{11} = 7.661989$ | 1 | 2 | 1 | 2 |
| #13 | $w_6^{13} = w_5^{13} + 1.172281 = 16.185195$ | 1 | 2 | 2 | 1 |
| #19 | $w_6^{19} = w_5^{19} + 1.116458 = 20.605882$ | 1 | 3 | 1 | 1 |
| #21 | $w_6^{21} = w_5^{21} = 2.631178$ | 1 | 3 | 1 | 3 |
| #25 | $w_6^{25} = w_5^{25} + 1.507218 = 13.336992$ | 1 | 3 | 3 | 1 |
| #27 | $w_6^{27} = w_5^{27} = 19.835570$ | 1 | 3 | 3 | 3 |
| #28 | $w_6^{28} = w_5^{28} + 0.781520 = 9.310012$ | 2 | 1 | 1 | 1 |
| #29 | $w_6^{29} = w_5^{29} = 6.412235$ | 2 | 1 | 1 | 2 |
| #31 | $w_6^{31} = w_5^{31} + 0.279114 = 2.511112$ | 2 | 1 | 2 | 1 |
| #37 | $w_6^{37} = w_5^{37} + 1.004812 = 12.881363$ | 2 | 2 | 1 | 1 |
| #55 | $w_6^{55} = w_5^{55} + 0.781520 = 12.612116$ | 3 | 1 | 1 | 1 |
| #57 | $w_6^{57} = w_5^{57} = 3.336428$ | 3 | 1 | 1 | 3 |
| #61 | $w_6^{61} = w_5^{61} + 0.837343 = 11.781778$ | 3 | 1 | 3 | 1 |
| #63 | $w_6^{63} = w_5^{63} = 6.056291$ | 3 | 1 | 3 | 3 |
| #73 | $w_6^{73} = w_5^{73} + 1.507218 = 16.473356$ | 3 | 3 | 1 | 1 |
| #75 | $w_6^{75} = w_5^{75} = 16.699207$ | 3 | 3 | 1 | 3 |
| #79 | $w_6^{79} = w_5^{79} + 11.722806 = 98.165759$ | 3 | 3 | 3 | 1 |
| #81 | $w_6^{81} = w_5^{81} = 129.664430$ | 3 | 3 | 3 | 3 |
| | 460.000000 | | | | |

## Step 7

We now apply proportional completion to voting pattern #5. In voting pattern #5, the voters are indifferent between the alternatives in $\{a, e, j\}$. At stage 1, $Y := w_1^5 + w_1^{14} + w_1^{32} + w_1^{41} = 27$ voters were indifferent between all the alternatives in $\{a, e, j\}$. The following $N - Y = 433$ voters were not indifferent between all the alternatives in $\{a, e, j\}$:

| number of voters | $e$ | $j$ |
|---|---|---|
| $w_1^1 + w_1^{10} + w_1^{19} + w_1^{28} + w_1^{37} + w_1^{55} + w_1^{73} = 85$ | 1 | 1 |
| $w_1^2 + w_1^{11} + w_1^{29} + w_1^{38} = 22$ | 1 | 2 |
| $w_1^3 + w_1^{21} + w_1^{57} + w_1^{75} = 22$ | 1 | 3 |
| $w_1^4 + w_1^{13} + w_1^{31} + w_1^{40} = 43$ | 2 | 1 |
| $w_1^7 + w_1^{25} + w_1^{61} + w_1^{79} = 110$ | 3 | 1 |
| $w_1^9 + w_1^{27} + w_1^{63} + w_1^{81} = 151$ | 3 | 3 |
| $N - Y = 433$ | | |

Therefore, the $w_6^5 = 4.307419$ voters with voting pattern #5 are replaced by the following voters:

| voting pattern | number of voters | $b$ | $c$ | $e$ | $j$ |
|---|---|---|---|---|---|
| #1 | $(w_1^1 + w_1^{10} + w_1^{19} + w_1^{28} + w_1^{37} + w_1^{55} + w_1^{73}) \cdot w_6^5 / (N - Y) = 0.845567$ | 1 | 1 | 1 | 1 |
| #2 | $(w_1^2 + w_1^{11} + w_1^{29} + w_1^{38}) \cdot w_6^5 / (N - Y) = 0.218853$ | 1 | 1 | 1 | 2 |
| #3 | $(w_1^3 + w_1^{21} + w_1^{57} + w_1^{75}) \cdot w_6^5 / (N - Y) = 0.218853$ | 1 | 1 | 1 | 3 |
| #4 | $(w_1^4 + w_1^{13} + w_1^{31} + w_1^{40}) \cdot w_6^5 / (N - Y) = 0.427758$ | 1 | 1 | 2 | 1 |
| #7 | $(w_1^7 + w_1^{25} + w_1^{61} + w_1^{79}) \cdot w_6^5 / (N - Y) = 1.094264$ | 1 | 1 | 3 | 1 |
| #9 | $(w_1^9 + w_1^{27} + w_1^{63} + w_1^{81}) \cdot w_6^5 / (N - Y) = 1.502125$ | 1 | 1 | 3 | 3 |
| | $w_6^5 = 4.307419$ | | | | |

Therefore, we get:

| voting pattern | number of voters | *b* | *c* | *e* | *j* |
|---|---|---|---|---|---|
| #1 | $w_7^1 = w_6^1 + 0.845567 = 20.891537$ | 1 | 1 | 1 | 1 |
| #2 | $w_7^2 = w_6^2 + 0.218853 = 2.568802$ | 1 | 1 | 1 | 2 |
| #3 | $w_7^3 = w_6^3 + 0.218853 = 3.537329$ | 1 | 1 | 1 | 3 |
| #4 | $w_7^4 = w_6^4 + 0.427758 = 5.214061$ | 1 | 1 | 2 | 1 |
| #7 | $w_7^7 = w_6^7 + 1.094264 = 8.144367$ | 1 | 1 | 3 | 1 |
| #9 | $w_7^9 = w_6^9 + 1.502125 = 4.859413$ | 1 | 1 | 3 | 3 |
| #10 | $w_7^{10} = w_6^{10} = 8.623599$ | 1 | 2 | 1 | 1 |
| #11 | $w_7^{11} = w_6^{11} = 7.661989$ | 1 | 2 | 1 | 2 |
| #13 | $w_7^{13} = w_6^{13} = 16.185195$ | 1 | 2 | 2 | 1 |
| #19 | $w_7^{19} = w_6^{19} = 20.605882$ | 1 | 3 | 1 | 1 |
| #21 | $w_7^{21} = w_6^{21} = 2.631178$ | 1 | 3 | 1 | 3 |
| #25 | $w_7^{25} = w_6^{25} = 13.336992$ | 1 | 3 | 3 | 1 |
| #27 | $w_7^{27} = w_6^{27} = 19.835570$ | 1 | 3 | 3 | 3 |
| #28 | $w_7^{28} = w_6^{28} = 9.310012$ | 2 | 1 | 1 | 1 |
| #29 | $w_7^{29} = w_6^{29} = 6.412235$ | 2 | 1 | 1 | 2 |
| #31 | $w_7^{31} = w_6^{31} = 2.511112$ | 2 | 1 | 2 | 1 |
| #37 | $w_7^{37} = w_6^{37} = 12.881363$ | 2 | 2 | 1 | 1 |
| #55 | $w_7^{55} = w_6^{55} = 12.612116$ | 3 | 1 | 1 | 1 |
| #57 | $w_7^{57} = w_6^{57} = 3.336428$ | 3 | 1 | 1 | 3 |
| #61 | $w_7^{61} = w_6^{61} = 11.781778$ | 3 | 1 | 3 | 1 |
| #63 | $w_7^{63} = w_6^{63} = 6.056291$ | 3 | 1 | 3 | 3 |
| #73 | $w_7^{73} = w_6^{73} = 16.473356$ | 3 | 3 | 1 | 1 |
| #75 | $w_7^{75} = w_6^{75} = 16.699207$ | 3 | 3 | 1 | 3 |
| #79 | $w_7^{79} = w_6^{79} = 98.165759$ | 3 | 3 | 3 | 1 |
| #81 | $w_7^{81} = w_6^{81} = 129.664430$ | 3 | 3 | 3 | 3 |
| | 460.000000 | | | | |

## Step 8

We now apply proportional completion to voting pattern #11. In voting pattern #11, the voters are indifferent between the alternatives in $\{a, c, j\}$. At stage 1, $Y := w_1^{11} + w_1^{14} + w_1^{38} + w_1^{41} = 34$ voters were indifferent between all the alternatives in $\{a, c, j\}$. The following $N - Y = 426$ voters were not indifferent between all the alternatives in $\{a, c, j\}$:

| number of voters | $c$ | $j$ |
|---|---|---|
| $w_1^{1} + w_1^{4} + w_1^{7} + w_1^{28} + w_1^{31} + w_1^{55} + w_1^{61} = 58$ | 1 | 1 |
| $w_1^{2} + w_1^{5} + w_1^{29} + w_1^{32} = 15$ | 1 | 2 |
| $w_1^{3} + w_1^{9} + w_1^{57} + w_1^{63} = 14$ | 1 | 3 |
| $w_1^{10} + w_1^{13} + w_1^{37} + w_1^{40} = 55$ | 2 | 1 |
| $w_1^{19} + w_1^{25} + w_1^{73} + w_1^{79} = 125$ | 3 | 1 |
| $w_1^{21} + w_1^{27} + w_1^{75} + w_1^{81} = 159$ | 3 | 3 |
| $N - Y = 426$ | | |

Therefore, the $w_7^{11} = 7.661989$ voters with voting pattern #11 are replaced by the following voters:

| voting pattern | number of voters | $b$ | $c$ | $e$ | $j$ |
|---|---|---|---|---|---|
| #1 | $(w_1^{1} + w_1^{4} + w_1^{7} + w_1^{28} + w_1^{31} + w_1^{55} + w_1^{61}) \cdot w_7^{11} / (N - Y) = 1.043182$ | 1 | 1 | 1 | 1 |
| #2 | $(w_1^{2} + w_1^{5} + w_1^{29} + w_1^{32}) \cdot w_7^{11} / (N - Y) = 0.269788$ | 1 | 1 | 1 | 2 |
| #3 | $(w_1^{3} + w_1^{9} + w_1^{57} + w_1^{63}) \cdot w_7^{11} / (N - Y) = 0.251802$ | 1 | 1 | 1 | 3 |
| #10 | $(w_1^{10} + w_1^{13} + w_1^{37} + w_1^{40}) \cdot w_7^{11} / (N - Y) = 0.989224$ | 1 | 2 | 1 | 1 |
| #19 | $(w_1^{19} + w_1^{25} + w_1^{73} + w_1^{79}) \cdot w_7^{11} / (N - Y) = 2.248236$ | 1 | 3 | 1 | 1 |
| #21 | $(w_1^{21} + w_1^{27} + w_1^{75} + w_1^{81}) \cdot w_7^{11} / (N - Y) = 2.859756$ | 1 | 3 | 1 | 3 |
| | $w_7^{11} = 7.661989$ | | | | |

Therefore, we get:

| voting pattern | number of voters | *b* | *c* | *e* | *j* |
|---|---|---|---|---|---|
| #1 | $w_8^1 = w_7^1 + 1.043182 = 21.934718$ | 1 | 1 | 1 | 1 |
| #2 | $w_8^2 = w_7^2 + 0.269788 = 2.838591$ | 1 | 1 | 1 | 2 |
| #3 | $w_8^3 = w_7^3 + 0.251802 = 3.789131$ | 1 | 1 | 1 | 3 |
| #4 | $w_8^4 = w_7^4 = 5.214061$ | 1 | 1 | 2 | 1 |
| #7 | $w_8^7 = w_7^7 = 8.144367$ | 1 | 1 | 3 | 1 |
| #9 | $w_8^9 = w_7^9 = 4.859413$ | 1 | 1 | 3 | 3 |
| #10 | $w_8^{10} = w_7^{10} + 0.989224 = 9.612823$ | 1 | 2 | 1 | 1 |
| #13 | $w_8^{13} = w_7^{13} = 16.185199$ | 1 | 2 | 2 | 1 |
| #19 | $w_8^{19} = w_7^{19} + 2.248236 = 22.854118$ | 1 | 3 | 1 | 1 |
| #21 | $w_8^{21} = w_7^{21} + 2.859756 = 5.490934$ | 1 | 3 | 1 | 3 |
| #25 | $w_8^{25} = w_7^{25} = 13.336992$ | 1 | 3 | 3 | 1 |
| #27 | $w_8^{27} = w_7^{27} = 19.835570$ | 1 | 3 | 3 | 3 |
| #28 | $w_8^{28} = w_7^{28} = 9.310012$ | 2 | 1 | 1 | 1 |
| #29 | $w_8^{29} = w_7^{29} = 6.412235$ | 2 | 1 | 1 | 2 |
| #31 | $w_8^{31} = w_7^{31} = 2.511112$ | 2 | 1 | 2 | 1 |
| #37 | $w_8^{37} = w_7^{37} = 12.881363$ | 2 | 2 | 1 | 1 |
| #55 | $w_8^{55} = w_7^{55} = 12.612116$ | 3 | 1 | 1 | 1 |
| #57 | $w_8^{57} = w_7^{57} = 3.336428$ | 3 | 1 | 1 | 3 |
| #61 | $w_8^{61} = w_7^{61} = 11.781778$ | 3 | 1 | 3 | 1 |
| #63 | $w_8^{63} = w_7^{63} = 6.056291$ | 3 | 1 | 3 | 3 |
| #73 | $w_8^{73} = w_7^{73} = 16.473356$ | 3 | 3 | 1 | 1 |
| #75 | $w_8^{75} = w_7^{75} = 16.699207$ | 3 | 3 | 1 | 3 |
| #79 | $w_8^{79} = w_7^{79} = 98.165759$ | 3 | 3 | 3 | 1 |
| #81 | $w_8^{81} = w_7^{81} = 129.664430$ | 3 | 3 | 3 | 3 |
| | 460.000000 | | | | |

## Step 9

We now apply proportional completion to voting pattern #13. In voting pattern #13, the voters are indifferent between the alternatives in {$a$, $c$, $e$}. At stage 1, $Y := w_1^{13} + w_1^{14} + w_1^{40} + w_1^{41} = 57$ voters were indifferent between all the alternatives in {$a$, $c$, $e$}. The following $N - Y = 403$ voters were not indifferent between all the alternatives in {$a$, $c$, $e$}:

| number of voters | $c$ | $e$ |
|---|---|---|
| $w_1^{1} + w_1^{2} + w_1^{3} + w_1^{28} + w_1^{29} + w_1^{55} + w_1^{57} = 50$ | 1 | 1 |
| $w_1^{4} + w_1^{5} + w_1^{31} + w_1^{32} = 13$ | 1 | 2 |
| $w_1^{7} + w_1^{9} + w_1^{61} + w_1^{63} = 24$ | 1 | 3 |
| $w_1^{10} + w_1^{11} + w_1^{37} + w_1^{38} = 32$ | 2 | 1 |
| $w_1^{19} + w_1^{21} + w_1^{73} + w_1^{75} = 47$ | 3 | 1 |
| $w_1^{25} + w_1^{27} + w_1^{79} + w_1^{81} = 237$ | 3 | 3 |
| $N - Y = 403$ | | |

Therefore, the $w_8^{13} = 16.185195$ voters with voting pattern #13 are replaced by the following voters:

| voting pattern | number of voters | $b$ | $c$ | $e$ | $j$ |
|---|---|---|---|---|---|
| #1 | $(w_1^{1} + w_1^{2} + w_1^{3} + w_1^{28} + w_1^{29} + w_1^{55} + w_1^{57}) \cdot w_8^{13} / (N - Y) = 2.008089$ | 1 | 1 | 1 | 1 |
| #4 | $(w_1^{4} + w_1^{5} + w_1^{31} + w_1^{32}) \cdot w_8^{13} / (N - Y) = 0.522103$ | 1 | 1 | 2 | 1 |
| #7 | $(w_1^{7} + w_1^{9} + w_1^{61} + w_1^{63}) \cdot w_8^{13} / (N - Y) = 0.963883$ | 1 | 1 | 3 | 1 |
| #10 | $(w_1^{10} + w_1^{11} + w_1^{37} + w_1^{38}) \cdot w_8^{13} / (N - Y) = 1.285177$ | 1 | 2 | 1 | 1 |
| #19 | $(w_1^{19} + w_1^{21} + w_1^{73} + w_1^{75}) \cdot w_8^{13} / (N - Y) = 1.887603$ | 1 | 3 | 1 | 1 |
| #25 | $(w_1^{25} + w_1^{27} + w_1^{79} + w_1^{81}) \cdot w_8^{13} / (N - Y) = 9.518340$ | 1 | 3 | 3 | 1 |
| | $w_8^{13} = 16.185195$ | | | | |

Therefore, we get:

| voting pattern | number of voters | *b* | *c* | *e* | *j* |
|---|---|---|---|---|---|
| #1 | $w_9^1 = w_8^1 + 2.008089 = 23.942807$ | 1 | 1 | 1 | 1 |
| #2 | $w_9^2 = w_8^2 = 2.838591$ | 1 | 1 | 1 | 2 |
| #3 | $w_9^3 = w_8^3 = 3.789131$ | 1 | 1 | 1 | 3 |
| #4 | $w_9^4 = w_8^4 + 0.522103 = 5.736164$ | 1 | 1 | 2 | 1 |
| #7 | $w_9^7 = w_8^7 + 0.963883 = 9.108249$ | 1 | 1 | 3 | 1 |
| #9 | $w_9^9 = w_8^9 = 4.859413$ | 1 | 1 | 3 | 3 |
| #10 | $w_9^{10} = w_8^{10} + 1.285177 = 10.898000$ | 1 | 2 | 1 | 1 |
| #19 | $w_9^{19} = w_8^{19} + 1.887603 = 24.741722$ | 1 | 3 | 1 | 1 |
| #21 | $w_9^{21} = w_8^{21} = 5.490934$ | 1 | 3 | 1 | 3 |
| #25 | $w_9^{25} = w_8^{25} + 9.518340 = 22.855333$ | 1 | 3 | 3 | 1 |
| #27 | $w_9^{27} = w_8^{27} = 19.835570$ | 1 | 3 | 3 | 3 |
| #28 | $w_9^{28} = w_8^{28} = 9.310012$ | 2 | 1 | 1 | 1 |
| #29 | $w_9^{29} = w_8^{29} = 6.412235$ | 2 | 1 | 1 | 2 |
| #31 | $w_9^{31} = w_8^{31} = 2.511112$ | 2 | 1 | 2 | 1 |
| #37 | $w_9^{37} = w_8^{37} = 12.881363$ | 2 | 2 | 1 | 1 |
| #55 | $w_9^{55} = w_8^{55} = 12.612116$ | 3 | 1 | 1 | 1 |
| #57 | $w_9^{57} = w_8^{57} = 3.336428$ | 3 | 1 | 1 | 3 |
| #61 | $w_9^{61} = w_8^{61} = 11.781778$ | 3 | 1 | 3 | 1 |
| #63 | $w_9^{63} = w_8^{63} = 6.056291$ | 3 | 1 | 3 | 3 |
| #73 | $w_9^{73} = w_8^{73} = 16.473356$ | 3 | 3 | 1 | 1 |
| #75 | $w_9^{75} = w_8^{75} = 16.699207$ | 3 | 3 | 1 | 3 |
| #79 | $w_9^{79} = w_8^{79} = 98.165759$ | 3 | 3 | 3 | 1 |
| #81 | $w_9^{81} = w_8^{81} = 129.664430$ | 3 | 3 | 3 | 3 |
| | 460.000000 | | | | |

## Step 10

We now apply proportional completion to voting pattern #29. In voting pattern #29, the voters are indifferent between the alternatives in $\{a, b, j\}$. At stage 1, $Y := w_1^{29} + w_1^{32} + w_1^{38} = 29$ voters were indifferent between all the alternatives in $\{a, b, j\}$. The following $N - Y = 431$ voters were not indifferent between all the alternatives in $\{a, b, j\}$:

| number of voters | $b$ | $j$ |
|---|---|---|
| $w_1^{1} + w_1^{4} + w_1^{7} + w_1^{10} + w_1^{13} + w_1^{19} + w_1^{25} = 76$ | 1 | 1 |
| $w_1^{2} + w_1^{5} + w_1^{11} + w_1^{14} = 20$ | 1 | 2 |
| $w_1^{3} + w_1^{9} + w_1^{21} + w_1^{27} = 25$ | 1 | 3 |
| $w_1^{28} + w_1^{31} + w_1^{37} + w_1^{40} = 44$ | 2 | 1 |
| $w_1^{55} + w_1^{61} + w_1^{73} + w_1^{79} = 118$ | 3 | 1 |
| $w_1^{57} + w_1^{63} + w_1^{75} + w_1^{81} = 148$ | 3 | 3 |
| $N - Y = 431$ | | |

Therefore, the $w_9^{29} = 6.412235$ voters with voting pattern #29 are replaced by the following voters:

| voting pattern | number of voters | $b$ | $c$ | $e$ | $j$ |
|---|---|---|---|---|---|
| #1 | $(w_1^{1} + w_1^{4} + w_1^{7} + w_1^{10} + w_1^{13} + w_1^{19} + w_1^{25}) \cdot w_9^{29} / (N - Y) = 1.130696$ | 1 | 1 | 1 | 1 |
| #2 | $(w_1^{2} + w_1^{5} + w_1^{11} + w_1^{14}) \cdot w_9^{29} / (N - Y) = 0.297552$ | 1 | 1 | 1 | 2 |
| #3 | $(w_1^{3} + w_1^{9} + w_1^{21} + w_1^{27}) \cdot w_9^{29} / (N - Y) = 0.371939$ | 1 | 1 | 1 | 3 |
| #28 | $(w_1^{28} + w_1^{31} + w_1^{37} + w_1^{40}) \cdot w_9^{29} / (N - Y) = 0.654613$ | 2 | 1 | 1 | 1 |
| #55 | $(w_1^{55} + w_1^{61} + w_1^{73} + w_1^{79}) \cdot w_9^{29} / (N - Y) = 1.755554$ | 3 | 1 | 1 | 1 |
| #57 | $(w_1^{57} + w_1^{63} + w_1^{75} + w_1^{81}) \cdot w_9^{29} / (N - Y) = 2.201881$ | 3 | 1 | 1 | 3 |
| | $w_9^{29} = 6.412235$ | | | | |

Therefore, we get:

| voting pattern | number of voters | *b* | *c* | *e* | *j* |
|---|---|---|---|---|---|
| #1 | $w_{10}^{1} = w_{9}^{1} + 1.130696 = 25.073503$ | 1 | 1 | 1 | 1 |
| #2 | $w_{10}^{2} = w_{9}^{2} + 0.297552 = 3.136142$ | 1 | 1 | 1 | 2 |
| #3 | $w_{10}^{3} = w_{9}^{3} + 0.371939 = 4.161070$ | 1 | 1 | 1 | 3 |
| #4 | $w_{10}^{4} = w_{9}^{4} = 5.736164$ | 1 | 1 | 2 | 1 |
| #7 | $w_{10}^{7} = w_{9}^{7} = 9.108249$ | 1 | 1 | 3 | 1 |
| #9 | $w_{10}^{9} = w_{9}^{9} = 4.859413$ | 1 | 1 | 3 | 3 |
| #10 | $w_{10}^{10} = w_{9}^{10} = 10.898000$ | 1 | 2 | 1 | 1 |
| #19 | $w_{10}^{19} = w_{9}^{19} = 24.741722$ | 1 | 3 | 1 | 1 |
| #21 | $w_{10}^{21} = w_{9}^{21} = 5.490934$ | 1 | 3 | 1 | 3 |
| #25 | $w_{10}^{25} = w_{9}^{25} = 22.855333$ | 1 | 3 | 3 | 1 |
| #27 | $w_{10}^{27} = w_{9}^{27} = 19.835570$ | 1 | 3 | 3 | 3 |
| #28 | $w_{10}^{28} = w_{9}^{28} + 0.654613 = 9.964626$ | 2 | 1 | 1 | 1 |
| #31 | $w_{10}^{31} = w_{9}^{31} = 2.511112$ | 2 | 1 | 2 | 1 |
| #37 | $w_{10}^{37} = w_{9}^{37} = 12.881363$ | 2 | 2 | 1 | 1 |
| #55 | $w_{10}^{55} = w_{9}^{55} + 1.755554 = 14.367670$ | 3 | 1 | 1 | 1 |
| #57 | $w_{10}^{57} = w_{9}^{57} + 2.201881 = 5.538309$ | 3 | 1 | 1 | 3 |
| #61 | $w_{10}^{61} = w_{9}^{61} = 11.781778$ | 3 | 1 | 3 | 1 |
| #63 | $w_{10}^{63} = w_{9}^{63} = 6.056291$ | 3 | 1 | 3 | 3 |
| #73 | $w_{10}^{73} = w_{9}^{73} = 16.473356$ | 3 | 3 | 1 | 1 |
| #75 | $w_{10}^{75} = w_{9}^{75} = 16.699207$ | 3 | 3 | 1 | 3 |
| #79 | $w_{10}^{79} = w_{9}^{79} = 98.165759$ | 3 | 3 | 3 | 1 |
| #81 | $w_{10}^{81} = w_{9}^{81} = 129.664430$ | 3 | 3 | 3 | 3 |
| | 460.000000 | | | | |

## Step 11

We now apply proportional completion to voting pattern #31. In voting pattern #31, the voters are indifferent between the alternatives in {*a*, *b*, *e*}. At stage 1, $Y := w_1^{31} + w_1^{32} + w_1^{40} + w_1^{41} = 41$ voters were indifferent between all the alternatives in {*a*, *b*, *e*}. The following $N - Y = 419$ voters were not indifferent between all the alternatives in {*a*, *b*, *e*}:

| number of voters | *b* | *e* |
|---|---|---|
| $w_1^1 + w_1^2 + w_1^3 + w_1^{10} + w_1^{11} + w_1^{19} + w_1^{21} = 56$ | 1 | 1 |
| $w_1^4 + w_1^5 + w_1^{13} + w_1^{14} = 29$ | 1 | 2 |
| $w_1^7 + w_1^9 + w_1^{25} + w_1^{27} = 36$ | 1 | 3 |
| $w_1^{28} + w_1^{29} + w_1^{37} + w_1^{38} = 32$ | 2 | 1 |
| $w_1^{55} + w_1^{57} + w_1^{73} + w_1^{75} = 41$ | 3 | 1 |
| $w_1^{61} + w_1^{63} + w_1^{79} + w_1^{81} = 225$ | 3 | 3 |
| $N - Y = 419$ | | |

Therefore, the $w_{10}^{31} = 2.511112$ voters with voting pattern #31 are replaced by the following voters:

| voting pattern | number of voters | *b* | *c* | *e* | *j* |
|---|---|---|---|---|---|
| #1 | $(w_1^1 + w_1^2 + w_1^3 + w_1^{10} + w_1^{11} + w_1^{19} + w_1^{21}) \cdot w_{10}^{31} / (N - Y) = 0.335614$ | 1 | 1 | 1 | 1 |
| #4 | $(w_1^4 + w_1^5 + w_1^{13} + w_1^{14}) \cdot w_{10}^{31} / (N - Y) = 0.173800$ | 1 | 1 | 2 | 1 |
| #7 | $(w_1^7 + w_1^9 + w_1^{25} + w_1^{27}) \cdot w_{10}^{31} / (N - Y) = 0.215752$ | 1 | 1 | 3 | 1 |
| #28 | $(w_1^{28} + w_1^{29} + w_1^{37} + w_1^{38}) \cdot w_{10}^{31} / (N - Y) = 0.191779$ | 2 | 1 | 1 | 1 |
| #55 | $(w_1^{55} + w_1^{57} + w_1^{73} + w_1^{75}) \cdot w_{10}^{31} / (N - Y) = 0.245717$ | 3 | 1 | 1 | 1 |
| #61 | $(w_1^{61} + w_1^{63} + w_1^{79} + w_1^{81}) \cdot w_{10}^{31} / (N - Y) = 1.348449$ | 3 | 1 | 3 | 1 |
| | $w_{10}^{31} = 2.511112$ | | | | |

Therefore, we get:

| voting pattern | number of voters | *b* | *c* | *e* | *j* |
|---|---|---|---|---|---|
| #1 | $w_{11}^{1} = w_{10}^{1} + 0.335614 = 25.409117$ | 1 | 1 | 1 | 1 |
| #2 | $w_{11}^{2} = w_{10}^{2} = 3.136142$ | 1 | 1 | 1 | 2 |
| #3 | $w_{11}^{3} = w_{10}^{3} = 4.161070$ | 1 | 1 | 1 | 3 |
| #4 | $w_{11}^{4} = w_{10}^{4} + 0.173800 = 5.909964$ | 1 | 1 | 2 | 1 |
| #7 | $w_{11}^{7} = w_{10}^{7} + 0.215752 = 9.324001$ | 1 | 1 | 3 | 1 |
| #9 | $w_{11}^{9} = w_{10}^{9} = 4.859413$ | 1 | 1 | 3 | 3 |
| #10 | $w_{11}^{10} = w_{10}^{10} = 10.898000$ | 1 | 2 | 1 | 1 |
| #19 | $w_{11}^{19} = w_{10}^{19} = 24.741722$ | 1 | 3 | 1 | 1 |
| #21 | $w_{11}^{21} = w_{10}^{21} = 5.490934$ | 1 | 3 | 1 | 3 |
| #25 | $w_{11}^{25} = w_{10}^{25} = 22.855333$ | 1 | 3 | 3 | 1 |
| #27 | $w_{11}^{27} = w_{10}^{27} = 19.835570$ | 1 | 3 | 3 | 3 |
| #28 | $w_{11}^{28} = w_{10}^{28} + 0.191779 = 10.156405$ | 2 | 1 | 1 | 1 |
| #37 | $w_{11}^{37} = w_{10}^{37} = 12.881363$ | 2 | 2 | 1 | 1 |
| #55 | $w_{11}^{55} = w_{10}^{55} + 0.245717 = 14.613388$ | 3 | 1 | 1 | 1 |
| #57 | $w_{11}^{57} = w_{10}^{57} = 5.538309$ | 3 | 1 | 1 | 3 |
| #61 | $w_{11}^{61} = w_{10}^{61} + 1.348449 = 13.130227$ | 3 | 1 | 3 | 1 |
| #63 | $w_{11}^{63} = w_{10}^{63} = 6.056291$ | 3 | 1 | 3 | 3 |
| #73 | $w_{11}^{73} = w_{10}^{73} = 16.473356$ | 3 | 3 | 1 | 1 |
| #75 | $w_{11}^{75} = w_{10}^{75} = 16.699207$ | 3 | 3 | 1 | 3 |
| #79 | $w_{11}^{79} = w_{10}^{79} = 98.165759$ | 3 | 3 | 3 | 1 |
| #81 | $w_{11}^{81} = w_{10}^{81} = 129.664430$ | 3 | 3 | 3 | 3 |
| | 460.000000 | | | | |

## Step 12

We now apply proportional completion to voting pattern #37. In voting pattern #37, the voters are indifferent between the alternatives in $\{a, b, c\}$. At stage 1, $Y := w_1^{37} + w_1^{38} + w_1^{40} + w_1^{41} = 54$ voters were indifferent between all the alternatives in $\{a, b, c\}$. The following $N - Y = 406$ voters were not indifferent between all the alternatives in $\{a, b, c\}$:

| number of voters | $b$ | $c$ |
|---|---|---|
| $w_1^{1} + w_1^{2} + w_1^{3} + w_1^{4} + w_1^{5} + w_1^{7} + w_1^{9} = 39$ | 1 | 1 |
| $w_1^{10} + w_1^{11} + w_1^{13} + w_1^{14} = 35$ | 1 | 2 |
| $w_1^{19} + w_1^{21} + w_1^{25} + w_1^{27} = 47$ | 1 | 3 |
| $w_1^{28} + w_1^{29} + w_1^{31} + w_1^{32} = 19$ | 2 | 1 |
| $w_1^{55} + w_1^{57} + w_1^{61} + w_1^{63} = 29$ | 3 | 1 |
| $w_1^{73} + w_1^{75} + w_1^{79} + w_1^{81} = 237$ | 3 | 3 |
| $N - Y = 406$ | | |

Therefore, the $w_{11}^{37} = 12.881363$ voters with voting pattern #37 are replaced by the following voters:

| voting pattern | number of voters | $b$ | $c$ | $e$ | $j$ |
|---|---|---|---|---|---|
| #1 | $(w_1^{1} + w_1^{2} + w_1^{3} + w_1^{4} + w_1^{5} + w_1^{7} + w_1^{9}) \cdot w_{11}^{37} / (N - Y) = 1.237372$ | 1 | 1 | 1 | 1 |
| #10 | $(w_1^{10} + w_1^{11} + w_1^{13} + w_1^{14}) \cdot w_{11}^{37} / (N - Y) = 1.110462$ | 1 | 2 | 1 | 1 |
| #19 | $(w_1^{19} + w_1^{21} + w_1^{25} + w_1^{27}) \cdot w_{11}^{37} / (N - Y) = 1.491192$ | 1 | 3 | 1 | 1 |
| #28 | $(w_1^{28} + w_1^{29} + w_1^{31} + w_1^{32}) \cdot w_{11}^{37} / (N - Y) = 0.602822$ | 2 | 1 | 1 | 1 |
| #55 | $(w_1^{55} + w_1^{57} + w_1^{61} + w_1^{63}) \cdot w_{11}^{37} / (N - Y) = 0.920097$ | 3 | 1 | 1 | 1 |
| #73 | $(w_1^{73} + w_1^{75} + w_1^{79} + w_1^{81}) \cdot w_{11}^{37} / (N - Y) = 7.519416$ | 3 | 3 | 1 | 1 |
| | $w_{11}^{37} = 12.881363$ | | | | |

Therefore, we get:

| voting pattern | number of voters | *b* | *c* | *e* | *j* |
|---|---|---|---|---|---|
| #1 | $w_{12}^{1} = w_{11}^{1} + 1.237372 = 26.646489$ | 1 | 1 | 1 | 1 |
| #2 | $w_{12}^{2} = w_{11}^{2} = 3.136142$ | 1 | 1 | 1 | 2 |
| #3 | $w_{12}^{3} = w_{11}^{3} = 4.161070$ | 1 | 1 | 1 | 3 |
| #4 | $w_{12}^{4} = w_{11}^{4} = 5.909964$ | 1 | 1 | 2 | 1 |
| #7 | $w_{12}^{7} = w_{11}^{7} = 9.324001$ | 1 | 1 | 3 | 1 |
| #9 | $w_{12}^{9} = w_{11}^{9} = 4.859413$ | 1 | 1 | 3 | 3 |
| #10 | $w_{12}^{10} = w_{11}^{10} + 1.110462 = 12.008462$ | 1 | 2 | 1 | 1 |
| #19 | $w_{12}^{19} = w_{11}^{19} + 1.491192 = 26.232914$ | 1 | 3 | 1 | 1 |
| #21 | $w_{12}^{21} = w_{11}^{21} = 5.490934$ | 1 | 3 | 1 | 3 |
| #25 | $w_{12}^{25} = w_{11}^{25} = 22.855333$ | 1 | 3 | 3 | 1 |
| #27 | $w_{12}^{27} = w_{11}^{27} = 19.835570$ | 1 | 3 | 3 | 3 |
| #28 | $w_{12}^{28} = w_{11}^{28} + 0.602822 = 10.759227$ | 2 | 1 | 1 | 1 |
| #55 | $w_{12}^{55} = w_{11}^{55} + 0.920097 = 15.533485$ | 3 | 1 | 1 | 1 |
| #57 | $w_{12}^{57} = w_{11}^{57} = 5.538309$ | 3 | 1 | 1 | 3 |
| #61 | $w_{12}^{61} = w_{11}^{61} = 13.130227$ | 3 | 1 | 3 | 1 |
| #63 | $w_{12}^{63} = w_{11}^{63} = 6.056291$ | 3 | 1 | 3 | 3 |
| #73 | $w_{12}^{73} = w_{11}^{73} + 7.519416 = 23.992772$ | 3 | 3 | 1 | 1 |
| #75 | $w_{12}^{75} = w_{11}^{75} = 16.699207$ | 3 | 3 | 1 | 3 |
| #79 | $w_{12}^{79} = w_{11}^{79} = 98.165759$ | 3 | 3 | 3 | 1 |
| #81 | $w_{12}^{81} = w_{11}^{81} = 129.664430$ | 3 | 3 | 3 | 3 |
| | 460.000000 | | | | |

## Step 13

We now apply proportional completion to voting pattern #2. In voting pattern #2, the voters are indifferent between the alternatives in $\{a, j\}$. At stage 1, $Y := w_1^2 + w_1^5 + w_1^{11} + w_1^{14} + w_1^{29} + w_1^{32} + w_1^{38} + w_1^{41} = 49$ voters were indifferent between all the alternatives in $\{a, j\}$. The following $N - Y = 411$ voters were not indifferent between all the alternatives in $\{a, j\}$:

| number of voters | $j$ |
|---|---|
| $w_1^1 + w_1^4 + w_1^7 + w_1^{10} + w_1^{13} + w_1^{19} + w_1^{25} + w_1^{28} + w_1^{31} + w_1^{37} + w_1^{40} + w_1^{55} + w_1^{61} + w_1^{73} + w_1^{79} = 238$ | 1 |
| $w_1^3 + w_1^9 + w_1^{21} + w_1^{27} + w_1^{57} + w_1^{63} + w_1^{75} + w_1^{81} = 173$ | 3 |
| $N - Y = 411$ | |

Therefore, the $w_{12}^2 = 3.136142$ voters with voting pattern #2 are replaced by the following voters:

| voting pattern | number of voters | $b$ | $c$ | $e$ | $j$ |
|---|---|---|---|---|---|
| #1 | $(w_1^1 + w_1^4 + w_1^7 + w_1^{10} + w_1^{13} + w_1^{19} + w_1^{25} + w_1^{28} + w_1^{31} + w_1^{37} + w_1^{40} + w_1^{55} + w_1^{61} + w_1^{73} + w_1^{79}) \cdot w_{12}^2 / (N - Y) = 1.816063$ | 1 | 1 | 1 | 1 |
| #3 | $(w_1^3 + w_1^9 + w_1^{21} + w_1^{27} + w_1^{57} + w_1^{63} + w_1^{75} + w_1^{81}) \cdot w_{12}^2 / (N - Y) = 1.320079$ | 1 | 1 | 1 | 3 |
| | $w_{12}^2 = 3.136142$ | | | | |

Therefore, we get:

| voting pattern | number of voters | *b* | *c* | *e* | *j* |
|---|---|---|---|---|---|
| #1 | $w_{13}^{1} = w_{12}^{1} + 1.816063 = 28.462552$ | 1 | 1 | 1 | 1 |
| #3 | $w_{13}^{3} = w_{12}^{3} + 1.320079 = 5.481150$ | 1 | 1 | 1 | 3 |
| #4 | $w_{13}^{4} = w_{12}^{4} = 5.909964$ | 1 | 1 | 2 | 1 |
| #7 | $w_{13}^{7} = w_{12}^{7} = 9.324001$ | 1 | 1 | 3 | 1 |
| #9 | $w_{13}^{9} = w_{12}^{9} = 4.859413$ | 1 | 1 | 3 | 3 |
| #10 | $w_{13}^{10} = w_{12}^{10} = 12.008462$ | 1 | 2 | 1 | 1 |
| #19 | $w_{13}^{19} = w_{12}^{19} = 26.232914$ | 1 | 3 | 1 | 1 |
| #21 | $w_{13}^{21} = w_{12}^{21} = 5.490934$ | 1 | 3 | 1 | 3 |
| #25 | $w_{13}^{25} = w_{12}^{25} = 22.855333$ | 1 | 3 | 3 | 1 |
| #27 | $w_{13}^{27} = w_{12}^{27} = 19.835570$ | 1 | 3 | 3 | 3 |
| #28 | $w_{13}^{28} = w_{12}^{28} = 10.759227$ | 2 | 1 | 1 | 1 |
| #55 | $w_{13}^{55} = w_{12}^{55} = 15.533485$ | 3 | 1 | 1 | 1 |
| #57 | $w_{13}^{57} = w_{12}^{57} = 5.538309$ | 3 | 1 | 1 | 3 |
| #61 | $w_{13}^{61} = w_{12}^{61} = 13.130227$ | 3 | 1 | 3 | 1 |
| #63 | $w_{13}^{63} = w_{12}^{63} = 6.056291$ | 3 | 1 | 3 | 3 |
| #73 | $w_{13}^{73} = w_{12}^{73} = 23.992772$ | 3 | 3 | 1 | 1 |
| #75 | $w_{13}^{75} = w_{12}^{75} = 16.699207$ | 3 | 3 | 1 | 3 |
| #79 | $w_{13}^{79} = w_{12}^{79} = 98.165759$ | 3 | 3 | 3 | 1 |
| #81 | $w_{13}^{81} = w_{12}^{81} = 129.664430$ | 3 | 3 | 3 | 3 |
| | 460.000000 | | | | |

## Step 14

We now apply proportional completion to voting pattern #4. In voting pattern #4, the voters are indifferent between the alternatives in $\{a, e\}$. At stage 1, $Y := w_1^4 + w_1^5 + w_1^{13} + w_1^{14} + w_1^{31} + w_1^{32} + w_1^{40} + w_1^{41} = 70$ voters were indifferent between all the alternatives in $\{a, e\}$. The following $N - Y =$ 390 voters were not indifferent between all the alternatives in $\{a, e\}$:

| number of voters | $e$ |
|---|---|
| $w_1^1 + w_1^2 + w_1^3 + w_1^{10} + w_1^{11} + w_1^{19} + w_1^{21} + w_1^{28} + w_1^{29} + w_1^{37} + w_1^{38} + w_1^{55} + w_1^{57} + w_1^{73} + w_1^{75} = 129$ | 1 |
| $w_1^7 + w_1^9 + w_1^{25} + w_1^{27} + w_1^{61} + w_1^{63} + w_1^{79} + w_1^{81} = 261$ | 3 |
| $N - Y = 390$ | |

Therefore, the $w_{13}^4 = 5.909964$ voters with voting pattern #4 are replaced by the following voters:

| voting pattern | number of voters | $b$ | $c$ | $e$ | $j$ |
|---|---|---|---|---|---|
| #1 | $(w_1^1 + w_1^2 + w_1^3 + w_1^{10} + w_1^{11} + w_1^{19} + w_1^{21} + w_1^{28} + w_1^{29} + w_1^{37} + w_1^{38} + w_1^{55} + w_1^{57} + w_1^{73} + w_1^{75}) \cdot w_{13}^4 / (N - Y) = 1.954834$ | 1 | 1 | 1 | 1 |
| #7 | $(w_1^7 + w_1^9 + w_1^{25} + w_1^{27} + w_1^{61} + w_1^{63} + w_1^{79} + w_1^{81}) \cdot w_{13}^4 / (N - Y) = 3.955130$ | 1 | 1 | 3 | 1 |
| | $w_{13}^4 = 5.909964$ | | | | |

Therefore, we get:

| voting pattern | number of voters | *b* | *c* | *e* | *j* |
|---|---|---|---|---|---|
| #1 | $w^{1}_{14} = w^{1}_{13} + 1.954834 = 30.417386$ | 1 | 1 | 1 | 1 |
| #3 | $w^{3}_{14} = w^{3}_{13} = 5.481150$ | 1 | 1 | 1 | 3 |
| #7 | $w^{7}_{14} = w^{7}_{13} + 3.955130 = 13.279131$ | 1 | 1 | 3 | 1 |
| #9 | $w^{9}_{14} = w^{9}_{13} = 4.859413$ | 1 | 1 | 3 | 3 |
| #10 | $w^{10}_{14} = w^{10}_{13} = 12.008462$ | 1 | 2 | 1 | 1 |
| #19 | $w^{19}_{14} = w^{19}_{13} = 26.232914$ | 1 | 3 | 1 | 1 |
| #21 | $w^{21}_{14} = w^{21}_{13} = 5.490934$ | 1 | 3 | 1 | 3 |
| #25 | $w^{25}_{14} = w^{25}_{13} = 22.855333$ | 1 | 3 | 3 | 1 |
| #27 | $w^{27}_{14} = w^{27}_{13} = 19.835570$ | 1 | 3 | 3 | 3 |
| #28 | $w^{28}_{14} = w^{28}_{13} = 10.759227$ | 2 | 1 | 1 | 1 |
| #55 | $w^{55}_{14} = w^{55}_{13} = 15.533485$ | 3 | 1 | 1 | 1 |
| #57 | $w^{57}_{14} = w^{57}_{13} = 5.538309$ | 3 | 1 | 1 | 3 |
| #61 | $w^{61}_{14} = w^{61}_{13} = 13.130227$ | 3 | 1 | 3 | 1 |
| #63 | $w^{63}_{14} = w^{63}_{13} = 6.056291$ | 3 | 1 | 3 | 3 |
| #73 | $w^{73}_{14} = w^{73}_{13} = 23.992772$ | 3 | 3 | 1 | 1 |
| #75 | $w^{75}_{14} = w^{75}_{13} = 16.699207$ | 3 | 3 | 1 | 3 |
| #79 | $w^{79}_{14} = w^{79}_{13} = 98.165759$ | 3 | 3 | 3 | 1 |
| #81 | $w^{81}_{14} = w^{81}_{13} = 129.664430$ | 3 | 3 | 3 | 3 |
| | 460.000000 | | | | |

## Step 15

We now apply proportional completion to voting pattern #10. In voting pattern #10, the voters are indifferent between the alternatives in $\{a, c\}$. At stage 1, $Y := w_1^{10} + w_1^{11} + w_1^{13} + w_1^{14} + w_1^{37} + w_1^{38} + w_1^{40} + w_1^{41}$ = 89 voters were indifferent between all the alternatives in $\{a, c\}$. The following $N - Y$ = 371 voters were not indifferent between all the alternatives in $\{a, c\}$:

| number of voters | $c$ |
|---|---|
| $w_1^{1} + w_1^{2} + w_1^{3} + w_1^{4} + w_1^{5} + w_1^{7} + w_1^{9} + w_1^{28} + w_1^{29} + w_1^{31} + w_1^{32} + w_1^{55} + w_1^{57} + w_1^{61} + w_1^{63} = 87$ | 1 |
| $w_1^{19} + w_1^{21} + w_1^{25} + w_1^{27} + w_1^{73} + w_1^{75} + w_1^{79} + w_1^{81} = 284$ | 3 |
| $N - Y = 371$ | |

Therefore, the $w_{14}^{10}$ = 12.008462 voters with voting pattern #10 are replaced by the following voters:

| voting pattern | number of voters | $b$ | $c$ | $e$ | $j$ |
|---|---|---|---|---|---|
| #1 | $(w_1^{1} + w_1^{2} + w_1^{3} + w_1^{4} + w_1^{5} + w_1^{7} + w_1^{9} + w_1^{28} + w_1^{29} + w_1^{31} + w_1^{32} + w_1^{55} + w_1^{57} + w_1^{61} + w_1^{63}) \cdot w_{14}^{10} / (N - Y) = 2.816001$ | 1 | 1 | 1 | 1 |
| #19 | $(w_1^{19} + w_1^{21} + w_1^{25} + w_1^{27} + w_1^{73} + w_1^{75} + w_1^{79} + w_1^{81}) \cdot w_{14}^{10} / (N - Y) = 9.192461$ | 1 | 3 | 1 | 1 |
| | $w_{14}^{10}$ = 12.008462 | | | | |

Therefore, we get:

| voting pattern | number of voters | *b* | *c* | *e* | *j* |
|---|---|---|---|---|---|
| #1 | $w_{15}^{1} = w_{14}^{1} + 2.816001 = 33.233387$ | 1 | 1 | 1 | 1 |
| #3 | $w_{15}^{3} = w_{14}^{3} = 5.481150$ | 1 | 1 | 1 | 3 |
| #7 | $w_{15}^{7} = w_{14}^{7} = 13.279131$ | 1 | 1 | 3 | 1 |
| #9 | $w_{15}^{9} = w_{14}^{9} = 4.859413$ | 1 | 1 | 3 | 3 |
| #19 | $w_{15}^{19} = w_{14}^{19} + 9.192461 = 35.425375$ | 1 | 3 | 1 | 1 |
| #21 | $w_{15}^{21} = w_{14}^{21} = 5.490934$ | 1 | 3 | 1 | 3 |
| #25 | $w_{15}^{25} = w_{14}^{25} = 22.855333$ | 1 | 3 | 3 | 1 |
| #27 | $w_{15}^{27} = w_{14}^{27} = 19.835570$ | 1 | 3 | 3 | 3 |
| #28 | $w_{15}^{28} = w_{14}^{28} = 10.759227$ | 2 | 1 | 1 | 1 |
| #55 | $w_{15}^{55} = w_{14}^{55} = 15.533485$ | 3 | 1 | 1 | 1 |
| #57 | $w_{15}^{57} = w_{14}^{57} = 5.538309$ | 3 | 1 | 1 | 3 |
| #61 | $w_{15}^{61} = w_{14}^{61} = 13.130227$ | 3 | 1 | 3 | 1 |
| #63 | $w_{15}^{63} = w_{14}^{63} = 6.056291$ | 3 | 1 | 3 | 3 |
| #73 | $w_{15}^{73} = w_{14}^{73} = 23.992772$ | 3 | 3 | 1 | 1 |
| #75 | $w_{15}^{75} = w_{14}^{75} = 16.699207$ | 3 | 3 | 1 | 3 |
| #79 | $w_{15}^{79} = w_{14}^{79} = 98.165759$ | 3 | 3 | 3 | 1 |
| #81 | $w_{15}^{81} = w_{14}^{81} = 129.664430$ | 3 | 3 | 3 | 3 |
| | 460.000000 | | | | |

### Step 16

We now apply proportional completion to voting pattern #28. In voting pattern #28, the voters are indifferent between the alternatives in $\{a, b\}$. At stage 1, $Y := w_1^{28} + w_1^{29} + w_1^{31} + w_1^{32} + w_1^{37} + w_1^{38} + w_1^{40} + w_1^{41}$ = 73 voters were indifferent between all the alternatives in $\{a, b\}$. The following $N - Y$ = 387 voters were not indifferent between all the alternatives in $\{a, b\}$:

| number of voters | $b$ |
|---|---|
| $w_1^{1} + w_1^{2} + w_1^{3} + w_1^{4} + w_1^{5} + w_1^{7} + w_1^{9} + w_1^{10} + w_1^{11} + w_1^{13} + w_1^{14} + w_1^{19} + w_1^{21} + w_1^{25} + w_1^{27} = 121$ | 1 |
| $w_1^{55} + w_1^{57} + w_1^{61} + w_1^{63} + w_1^{73} + w_1^{75} + w_1^{79} + w_1^{81} = 266$ | 3 |
| $N - Y = 387$ | |

Therefore, the $w_{15}^{28}$ = 10.759227 voters with voting pattern #28 are replaced by the following voters:

| voting pattern | number of voters | $b$ | $c$ | $e$ | $j$ |
|---|---|---|---|---|---|
| #1 | $(w_1^{1} + w_1^{2} + w_1^{3} + w_1^{4} + w_1^{5} + w_1^{7} + w_1^{9} + w_1^{10} + w_1^{11} + w_1^{13} + w_1^{14} + w_1^{19} + w_1^{21} + w_1^{25} + w_1^{27}) \cdot w_{15}^{28} / (N - Y) = 3.363996$ | 1 | 1 | 1 | 1 |
| #55 | $(w_1^{55} + w_1^{57} + w_1^{61} + w_1^{63} + w_1^{73} + w_1^{75} + w_1^{79} + w_1^{81}) \cdot w_{15}^{28} / (N - Y) = 7.395231$ | 3 | 1 | 1 | 1 |
| | $w_{15}^{28}$ = 10.759227 | | | | |

Therefore, we get:

| voting pattern | number of voters | *b* | *c* | *e* | *j* |
|---|---|---|---|---|---|
| #1 | $w_{16}^{1} = w_{15}^{1} + 3.363996 = 36.597383$ | 1 | 1 | 1 | 1 |
| #3 | $w_{16}^{3} = w_{15}^{3} = 5.481150$ | 1 | 1 | 1 | 3 |
| #7 | $w_{16}^{7} = w_{15}^{7} = 13.279131$ | 1 | 1 | 3 | 1 |
| #9 | $w_{16}^{9} = w_{15}^{9} = 4.859413$ | 1 | 1 | 3 | 3 |
| #19 | $w_{16}^{19} = w_{15}^{19} = 35.425375$ | 1 | 3 | 1 | 1 |
| #21 | $w_{16}^{21} = w_{15}^{21} = 5.490934$ | 1 | 3 | 1 | 3 |
| #25 | $w_{16}^{25} = w_{15}^{25} = 22.855333$ | 1 | 3 | 3 | 1 |
| #27 | $w_{16}^{27} = w_{15}^{27} = 19.835570$ | 1 | 3 | 3 | 3 |
| #55 | $w_{16}^{55} = w_{15}^{55} + 7.395231 = 22.928716$ | 3 | 1 | 1 | 1 |
| #57 | $w_{16}^{57} = w_{15}^{57} = 5.538309$ | 3 | 1 | 1 | 3 |
| #61 | $w_{16}^{61} = w_{15}^{61} = 13.130227$ | 3 | 1 | 3 | 1 |
| #63 | $w_{16}^{63} = w_{15}^{63} = 6.056291$ | 3 | 1 | 3 | 3 |
| #73 | $w_{16}^{73} = w_{15}^{73} = 23.992772$ | 3 | 3 | 1 | 1 |
| #75 | $w_{16}^{75} = w_{15}^{75} = 16.699207$ | 3 | 3 | 1 | 3 |
| #79 | $w_{16}^{79} = w_{15}^{79} = 98.165759$ | 3 | 3 | 3 | 1 |
| #81 | $w_{16}^{81} = w_{15}^{81} = 129.664430$ | 3 | 3 | 3 | 3 |
| | 460.000000 | | | | |

## 8.2.2. Links between Sets of Alternatives

In section 8.2.2, we will show how the strengths of the links are calculated.

According to (8.1.2.1), the strength of the link $(\{a_1,...,a_{(M-1)}\};b) \rightarrow (\{a_1,...,a_{(M-1)}\};c)$ is given by $(N[\{a_1,...,a_{(M-1)},b\};c], N[\{a_1,...,a_{(M-1)},c\};b])$, where $N[\{a_1,...,a_{(M-1)},b\};c]$ is the support and $N[\{a_1,...,a_{(M-1)},c\};b])$ is the opposition of this link.

$N[\{a_1,...,a_M\};g]$ is defined as follows:

$N[\{a_1,...,a_M\};g] \in \mathbb{R}_{\geq 0}$ is the largest value such that there is a $t \in \mathbb{R}^{(N_W \times M)}$ such that:

(8.1.2.2) $\forall\, i \in \{1,...,N_W\}\ \forall\, j \in \{1,...,M\}: t_{ij} \geq 0.$

(8.1.2.3) $\forall\, i \in \{1,...,N_W\}: \sum_{j=1}^{M} t_{ij} \leq \rho(i).$

(8.1.2.4) $\forall\, i \in \{1,...,N_W\}\ \forall\, j \in \{1,...,M\}: g \succ_i a_j \Rightarrow t_{ij} = 0.$

(8.1.2.5) $\forall\, j \in \{1,...,M\}: \sum_{i=1}^{N_W} t_{ij} \geq N[\{a_1,...,a_M\};g].$

Suppose $N^*[\{a_1,...,a_M\};g] \in \mathbb{R}_{\geq 0}$ is the largest value such that there is a $t^* \in \mathbb{R}^{(N_W \times M)}$ such that:

(8.2.2.1) $\forall\, i \in \{1,...,N_W\}\ \forall\, j \in \{1,...,M\}: t^*_{ij} \geq 0.$

(8.2.2.2) $\forall\, i \in \{1,...,N_W\}: \sum_{j=1}^{M} t^*_{ij} \leq \rho(i).$

(8.2.2.3) $\forall\, i \in \{1,...,N_W\}\ \forall\, j \in \{1,...,M\}: g \succ_i a_j \Rightarrow t^*_{ij} = 0.$

(8.2.2.4) $\sum_{i=1}^{N_W} \sum_{j=1}^{M} t^*_{ij} \geq M \cdot N^*[\{a_1,...,a_M\};g].$

As (8.2.2.4) is weaker than (8.1.2.5), we get:

$$N[\{a_1,...,a_M\};g] \leq N^*[\{a_1,...,a_M\};g].$$

Suppose $t^* \in \mathbb{R}^{(N_W \times M)}$ is a solution of (8.2.2.1) – (8.2.2.4). Then we define:

$$N\hat{}[\{a_1,...,a_M\};g] := \min \{ \sum_{i=1}^{N_W} t^*_{ij} \mid 1 \leq j \leq M \}.$$

So we get:

$$N\hat{}[\{a_1,...,a_M\};g] \leq N[\{a_1,...,a_M\};g] \leq N^*[\{a_1,...,a_M\};g].$$

Compared to (8.1.2.2) – (8.1.2.5), (8.2.2.1) – (8.2.2.4) has the advantage that it describes a trivial max-flow problem. A max-flow problem can be solved significantly faster than a general linear program. Therefore, we solve (8.1.2.2) – (8.1.2.5) by solving a series of max-flow problems as follows:

Suppose $\mathbf{W}$ is the number of voters who strictly prefer candidate $g$ to every candidate of the set $\{a_1,...,a_M\}$. Then we know that $N[\{a_1,...,a_M\};g]$ cannot be larger than $( N - \mathbf{W} ) / M$.

Therefore, we start with

$$r^{(0)} := ( N - \mathbf{W} ) / M.$$

$$s^{(0)} := 0.$$

For $z = 1, 2, 3, ...$, we solve the following linear programs $\mathsf{LP}^{(z)}$:

Find the maximum $r^{(z)} \in \mathbb{R}$ such that there is a $t^{(z)} \in \mathbb{R}^{(N_W \times M)}$ such that

(8.2.2.5) $$\forall\, i \in \{1,...,N_W\}\ \forall\, j \in \{1,...,M\}: t_{ij}^{(z)} \geq 0.$$

(8.2.2.6) $$\forall\, i \in \{1,...,N_W\}: \sum_{j=1}^{M} t_{ij}^{(z)} \leq \rho(i).$$

(8.2.2.7) $$\forall\, i \in \{1,...,N_W\}\ \forall\, j \in \{1,...,M\}: g \succ_i a_j \Rightarrow t_{ij}^{(z)} = 0.$$

(8.2.2.8) $$\sum_{i=1}^{N_W} \sum_{j=1}^{M} t_{ij}^{(z)} \geq M \cdot r^{(z)}.$$

(8.2.2.9) $$\forall\, j \in \{1,...,M\}: \sum_{i=1}^{N_W} t_{ij}^{(z)} \leq r^{(z-1)}.$$

Furthermore, we define for $z = 1, 2, 3, ...$ :

(8.2.2.10) $$s^{(z)} := \max \{ s^{(z-1)}, \min \{ \sum_{i=1}^{N_W} t_{ij}^{(z)} \mid 1 \leq j \leq M \} \}.$$

When we solve (8.2.2.5) – (8.2.2.9), then we get a decreasing sequence $r^{(0)}$, $r^{(1)}$, $r^{(2)}$, $r^{(3)}$, ... and an increasing sequence $s^{(0)}$, $s^{(1)}$, $s^{(2)}$, $s^{(3)}$, ... These two sequences converge to the same limit. This limit is the solution of (8.1.2.2) – (8.1.2.5).

Now, we use this algorithm to calculate the support of link $\{b,c,e,j\} \rightarrow (\{b,c,e\};a)$ which is identical to the support of links $\{b,c,e,j\} \rightarrow (\{b,c,j\};a)$, $\{b,c,e,j\} \rightarrow (\{b,e,j\};a)$ and $\{b,c,e,j\} \rightarrow (\{c,e,j\};a)$. After proportional completion, the voter profile looks as follows:

| | | *b* | *c* | *e* | *j* |
|---|---|---|---|---|---|
| voter01 | 36.597383 | 1 | 1 | 1 | 1 |
| voter02 | 5.481150 | 1 | 1 | 1 | 3 |
| voter03 | 13.279131 | 1 | 1 | 3 | 1 |
| voter04 | 4.859413 | 1 | 1 | 3 | 3 |
| voter05 | 35.425375 | 1 | 3 | 1 | 1 |
| voter06 | 5.490934 | 1 | 3 | 1 | 3 |
| voter07 | 22.855333 | 1 | 3 | 3 | 1 |
| voter08 | 19.835570 | 1 | 3 | 3 | 3 |
| voter09 | 22.928716 | 3 | 1 | 1 | 1 |
| voter10 | 5.538309 | 3 | 1 | 1 | 3 |
| voter11 | 13.130227 | 3 | 1 | 3 | 1 |
| voter12 | 6.056291 | 3 | 1 | 3 | 3 |
| voter13 | 23.992772 | 3 | 3 | 1 | 1 |
| voter14 | 16.699207 | 3 | 3 | 1 | 3 |
| voter15 | 98.165759 | 3 | 3 | 3 | 1 |
| voter16 | 129.664430 | 3 | 3 | 3 | 3 |
| | 460.000000 | | | | |

The corresponding max-flow problem has the following form:

> Each voting pattern, where voters strictly prefer at least one alternative of the set $\{b,c,e,j\}$ to alternative $a$, is represented by a vertex. Each alternative of the set $\{b,c,e,j\}$ is represented by a vertex. Furthermore, there is a vertex “source” and a vertex “drain”.
>
> From the vertex “source” we draw a link to each vertex that represents a voting pattern. The maximum capacity of this link is the number of voters with this voting pattern.
>
> From each vertex, that represents a voting pattern, we draw a link to each vertex that represents an alternative that is strictly preferred to alternative $a$ by voters with this voting pattern. The maximum capacity of this link is the number of voters with this voting pattern.
>
> From each vertex, that represents an alternative, we draw a link to the vertex “drain”. The maximum capacity of this link is $r^{(z-1)}$.
>
> The task is: Maximize the total flow from the vertex “source” to the vertex “drain”.

In our case, we get a digraph with 21 vertices and 51 links.

Furthermore, we get:

$$r^{(0)} := ( N - W ) / M = ( 460 - 129.664430 ) / 4 = 82.583893$$

Our digraph has the following form:

| link | start | end | capacity |
|---|---|---|---|
| 1 | source | voter01 | 36.597383 |
| 2 | source | voter02 | 5.481150 |
| 3 | source | voter03 | 13.279131 |
| 4 | source | voter04 | 4.859413 |
| 5 | source | voter05 | 35.425375 |
| 6 | source | voter06 | 5.490934 |
| 7 | source | voter07 | 22.855333 |
| 8 | source | voter08 | 19.835570 |
| 9 | source | voter09 | 22.928716 |
| 10 | source | voter10 | 5.538309 |
| 11 | source | voter11 | 13.130227 |
| 12 | source | voter12 | 6.056291 |
| 13 | source | voter13 | 23.992772 |
| 14 | source | voter14 | 16.699207 |
| 15 | source | voter15 | 98.165759 |
| 16 | voter01 | alternative *b* | 36.597383 |
| 17 | voter01 | alternative *c* | 36.597383 |
| 18 | voter01 | alternative *e* | 36.597383 |
| 19 | voter01 | alternative *j* | 36.597383 |
| 20 | voter02 | alternative *b* | 5.481150 |
| 21 | voter02 | alternative *c* | 5.481150 |
| 22 | voter02 | alternative *e* | 5.481150 |
| 23 | voter03 | alternative *b* | 13.279131 |
| 24 | voter03 | alternative *c* | 13.279131 |
| 25 | voter03 | alternative *j* | 13.279131 |
| 26 | voter04 | alternative *b* | 4.859413 |
| 27 | voter04 | alternative *c* | 4.859413 |
| 28 | voter05 | alternative *b* | 35.425375 |
| 29 | voter05 | alternative *e* | 35.425375 |
| 30 | voter05 | alternative *j* | 35.425375 |
| 31 | voter06 | alternative *b* | 5.490934 |
| 32 | voter06 | alternative *e* | 5.490934 |
| 33 | voter07 | alternative *b* | 22.855333 |
| 34 | voter07 | alternative *j* | 22.855333 |
| 35 | voter08 | alternative *b* | 19.835570 |
| 36 | voter09 | alternative *c* | 22.928716 |
| 37 | voter09 | alternative *e* | 22.928716 |
| 38 | voter09 | alternative *j* | 22.928716 |
| 39 | voter10 | alternative *c* | 5.538309 |
| 40 | voter10 | alternative *e* | 5.538309 |
| 41 | voter11 | alternative *c* | 13.130227 |
| 42 | voter11 | alternative *j* | 13.130227 |
| 43 | voter12 | alternative *c* | 6.056291 |
| 44 | voter13 | alternative *e* | 23.992772 |
| 45 | voter13 | alternative *j* | 23.992772 |
| 46 | voter14 | alternative *e* | 16.699207 |
| 47 | voter15 | alternative *j* | 98.165759 |
| 48 | alternative *b* | drain | $r^{(z-1)}$ |
| 49 | alternative *c* | drain | $r^{(z-1)}$ |
| 50 | alternative *e* | drain | $r^{(z-1)}$ |
| 51 | alternative *j* | drain | $r^{(z-1)}$ |

The following 13 pages document the solutions for (8.2.2.5) – (8.2.2.10). Those entries that change from one table to the next table are **<u>fat and underlined</u>**.

We get:

$r^{(0)} = 82.583893; \quad s^{(0)} = 0.000000$
$r^{(1)} = 78.688426; \quad s^{(1)} = 71.469640$
$r^{(2)} = 77.714559; \quad s^{(2)} = 75.365107$
$r^{(3)} = 77.471093; \quad s^{(3)} = 76.740693$
$r^{(4)} = 77.410226; \quad s^{(4)} = 77.227626$
$r^{(5)} = 77.395009; \quad s^{(5)} = 77.349359$
$r^{(6)} = 77.391205; \quad s^{(6)} = 77.379793$
$r^{(7)} = 77.390254; \quad s^{(7)} = 77.387401$
$r^{(8)} = 77.390016; \quad s^{(8)} = 77.389303$
$r^{(9)} = 77.389957; \quad s^{(9)} = 77.389779$
$r^{(10)} = 77.389942; \quad s^{(10)} = 77.389897$
$r^{(11)} = 77.389938; \quad s^{(11)} = 77.389927$
$r^{(12)} = 77.389937; \quad s^{(12)} = 77.389935$

We get:

$$r = \lim_{z \to \infty} r^{(z)} = \lim_{z \to \infty} s^{(z)} = 77.389937$$

Stage $z = 1$:

| link | start | end | capacity | flow |
|---|---|---|---|---|
| 1 | source | voter01 | 36.597383 | 36.597383 |
| 2 | source | voter02 | 5.481150 | 5.481150 |
| 3 | source | voter03 | 13.279131 | 13.279131 |
| 4 | source | voter04 | 4.859413 | 4.859413 |
| 5 | source | voter05 | 35.425375 | 35.425375 |
| 6 | source | voter06 | 5.490934 | 5.490934 |
| 7 | source | voter07 | 22.855333 | 22.855333 |
| 8 | source | voter08 | 19.835570 | 19.835570 |
| 9 | source | voter09 | 22.928716 | 22.928716 |
| 10 | source | voter10 | 5.538309 | 5.538309 |
| 11 | source | voter11 | 13.130227 | 13.130227 |
| 12 | source | voter12 | 6.056291 | 6.056291 |
| 13 | source | voter13 | 23.992772 | 23.992772 |
| 14 | source | voter14 | 16.699207 | 16.699207 |
| 15 | source | voter15 | 98.165759 | 82.583893 |
| 16 | voter01 | alternative *b* | 36.597383 | 0.000000 |
| 17 | voter01 | alternative *c* | 36.597383 | 34.239372 |
| 18 | voter01 | alternative *e* | 36.597383 | 2.358011 |
| 19 | voter01 | alternative *j* | 36.597383 | 0.000000 |
| 20 | voter02 | alternative *b* | 5.481150 | 0.000000 |
| 21 | voter02 | alternative *c* | 5.481150 | 5.481150 |
| 22 | voter02 | alternative *e* | 5.481150 | 0.000000 |
| 23 | voter03 | alternative *b* | 13.279131 | 0.000000 |
| 24 | voter03 | alternative *c* | 13.279131 | 13.279131 |
| 25 | voter03 | alternative *j* | 13.279131 | 0.000000 |
| 26 | voter04 | alternative *b* | 4.859413 | 0.000000 |
| 27 | voter04 | alternative *c* | 4.859413 | 4.859413 |
| 28 | voter05 | alternative *b* | 35.425375 | 35.425375 |
| 29 | voter05 | alternative *e* | 35.425375 | 0.000000 |
| 30 | voter05 | alternative *j* | 35.425375 | 0.000000 |
| 31 | voter06 | alternative *b* | 5.490934 | 0.000000 |
| 32 | voter06 | alternative *e* | 5.490934 | 5.490934 |
| 33 | voter07 | alternative *b* | 22.855333 | 22.855333 |
| 34 | voter07 | alternative *j* | 22.855333 | 0.000000 |
| 35 | voter08 | alternative *b* | 19.835570 | 19.835570 |
| 36 | voter09 | alternative *c* | 22.928716 | 0.000000 |
| 37 | voter09 | alternative *e* | 22.928716 | 22.928716 |
| 38 | voter09 | alternative *j* | 22.928716 | 0.000000 |
| 39 | voter10 | alternative *c* | 5.538309 | 5.538309 |
| 40 | voter10 | alternative *e* | 5.538309 | 0.000000 |
| 41 | voter11 | alternative *c* | 13.130227 | 13.130227 |
| 42 | voter11 | alternative *j* | 13.130227 | 0.000000 |
| 43 | voter12 | alternative *c* | 6.056291 | 6.056291 |
| 44 | voter13 | alternative *e* | 23.992772 | 23.992772 |
| 45 | voter13 | alternative *j* | 23.992772 | 0.000000 |
| 46 | voter14 | alternative *e* | 16.699207 | 16.699207 |
| 47 | voter15 | alternative *j* | 98.165759 | 82.583893 |
| 48 | alternative *b* | drain | $r^{(0)} = 82.583893$ | 78.116279 |
| 49 | alternative *c* | drain | $r^{(0)} = 82.583893$ | 82.583893 |
| 50 | alternative *e* | drain | $r^{(0)} = 82.583893$ | 71.469640 |
| 51 | alternative *j* | drain | $r^{(0)} = 82.583893$ | 82.583893 |

$r^{(1)} = ( 78.116279 + 82.583893 + 71.469640 + 82.583893 ) / 4 = 78.688426$
$s^{(1)} = \max \{ 0.000000; \min \{ 78.116279; 82.583893; 71.469640; 82.583893 \} \} = 71.469640$

Stage $z = 2$:

| link | start | end | capacity | flow |
|---|---|---|---|---|
| 1 | source | voter01 | 36.597383 | 36.597383 |
| 2 | source | voter02 | 5.481150 | 5.481150 |
| 3 | source | voter03 | 13.279131 | 13.279131 |
| 4 | source | voter04 | 4.859413 | 4.859413 |
| 5 | source | voter05 | 35.425375 | 35.425375 |
| 6 | source | voter06 | 5.490934 | 5.490934 |
| 7 | source | voter07 | 22.855333 | 22.855333 |
| 8 | source | voter08 | 19.835570 | 19.835570 |
| 9 | source | voter09 | 22.928716 | 22.928716 |
| 10 | source | voter10 | 5.538309 | 5.538309 |
| 11 | source | voter11 | 13.130227 | 13.130227 |
| 12 | source | voter12 | 6.056291 | 6.056291 |
| 13 | source | voter13 | 23.992772 | 23.992772 |
| 14 | source | voter14 | 16.699207 | 16.699207 |
| 15 | source | voter15 | 98.165759 | **78.688426** |
| 16 | voter01 | alternative *b* | 36.597383 | 0.000000 |
| 17 | voter01 | alternative *c* | 36.597383 | **30.343905** |
| 18 | voter01 | alternative *e* | 36.597383 | **6.253478** |
| 19 | voter01 | alternative *j* | 36.597383 | 0.000000 |
| 20 | voter02 | alternative *b* | 5.481150 | 0.000000 |
| 21 | voter02 | alternative *c* | 5.481150 | 5.481150 |
| 22 | voter02 | alternative *e* | 5.481150 | 0.000000 |
| 23 | voter03 | alternative *b* | 13.279131 | 0.000000 |
| 24 | voter03 | alternative *c* | 13.279131 | 13.279131 |
| 25 | voter03 | alternative *j* | 13.279131 | 0.000000 |
| 26 | voter04 | alternative *b* | 4.859413 | 0.000000 |
| 27 | voter04 | alternative *c* | 4.859413 | 4.859413 |
| 28 | voter05 | alternative *b* | 35.425375 | 35.425375 |
| 29 | voter05 | alternative *e* | 35.425375 | 0.000000 |
| 30 | voter05 | alternative *j* | 35.425375 | 0.000000 |
| 31 | voter06 | alternative *b* | 5.490934 | 0.000000 |
| 32 | voter06 | alternative *e* | 5.490934 | 5.490934 |
| 33 | voter07 | alternative *b* | 22.855333 | 22.855333 |
| 34 | voter07 | alternative *j* | 22.855333 | 0.000000 |
| 35 | voter08 | alternative *b* | 19.835570 | 19.835570 |
| 36 | voter09 | alternative *c* | 22.928716 | 0.000000 |
| 37 | voter09 | alternative *e* | 22.928716 | 22.928716 |
| 38 | voter09 | alternative *j* | 22.928716 | 0.000000 |
| 39 | voter10 | alternative *c* | 5.538309 | 5.538309 |
| 40 | voter10 | alternative *e* | 5.538309 | 0.000000 |
| 41 | voter11 | alternative *c* | 13.130227 | 13.130227 |
| 42 | voter11 | alternative *j* | 13.130227 | 0.000000 |
| 43 | voter12 | alternative *c* | 6.056291 | 6.056291 |
| 44 | voter13 | alternative *e* | 23.992772 | 23.992772 |
| 45 | voter13 | alternative *j* | 23.992772 | 0.000000 |
| 46 | voter14 | alternative *e* | 16.699207 | 16.699207 |
| 47 | voter15 | alternative *j* | 98.165759 | **78.688426** |
| 48 | alternative *b* | drain | **$r^{(1)}$ = 78.688426** | 78.116279 |
| 49 | alternative *c* | drain | **$r^{(1)}$ = 78.688426** | **78.688426** |
| 50 | alternative *e* | drain | **$r^{(1)}$ = 78.688426** | **75.365107** |
| 51 | alternative *j* | drain | **$r^{(1)}$ = 78.688426** | **78.688426** |

$r^{(2)}$ = ( 78.116279 + 78.688426 + 75.365107 + 78.688426 ) / 4 = 77.714559
$s^{(2)}$ = max { 71.469640; min { 78.116279; 78.688426; 75.365107 ; 78.688426 } } = 75.365107

Stage $z = 3$:

| link | start | end | capacity | flow |
|---|---|---|---|---|
| 1 | source | voter01 | 36.597383 | 36.597383 |
| 2 | source | voter02 | 5.481150 | 5.481150 |
| 3 | source | voter03 | 13.279131 | 13.279131 |
| 4 | source | voter04 | 4.859413 | 4.859413 |
| 5 | source | voter05 | 35.425375 | 35.425375 |
| 6 | source | voter06 | 5.490934 | 5.490934 |
| 7 | source | voter07 | 22.855333 | 22.855333 |
| 8 | source | voter08 | 19.835570 | 19.835570 |
| 9 | source | voter09 | 22.928716 | 22.928716 |
| 10 | source | voter10 | 5.538309 | 5.538309 |
| 11 | source | voter11 | 13.130227 | 13.130227 |
| 12 | source | voter12 | 6.056291 | 6.056291 |
| 13 | source | voter13 | 23.992772 | 23.992772 |
| 14 | source | voter14 | 16.699207 | 16.699207 |
| 15 | source | voter15 | 98.165759 | **77.714559** |
| 16 | voter01 | alternative *b* | 36.597383 | 0.000000 |
| 17 | voter01 | alternative *c* | 36.597383 | **29.370038** |
| 18 | voter01 | alternative *e* | 36.597383 | **7.227344** |
| 19 | voter01 | alternative *j* | 36.597383 | 0.000000 |
| 20 | voter02 | alternative *b* | 5.481150 | 0.000000 |
| 21 | voter02 | alternative *c* | 5.481150 | 5.481150 |
| 22 | voter02 | alternative *e* | 5.481150 | 0.000000 |
| 23 | voter03 | alternative *b* | 13.279131 | 0.000000 |
| 24 | voter03 | alternative *c* | 13.279131 | 13.279131 |
| 25 | voter03 | alternative *j* | 13.279131 | 0.000000 |
| 26 | voter04 | alternative *b* | 4.859413 | 0.000000 |
| 27 | voter04 | alternative *c* | 4.859413 | 4.859413 |
| 28 | voter05 | alternative *b* | 35.425375 | **35.023656** |
| 29 | voter05 | alternative *e* | 35.425375 | **0.401719** |
| 30 | voter05 | alternative *j* | 35.425375 | 0.000000 |
| 31 | voter06 | alternative *b* | 5.490934 | 0.000000 |
| 32 | voter06 | alternative *e* | 5.490934 | 5.490934 |
| 33 | voter07 | alternative *b* | 22.855333 | 22.855333 |
| 34 | voter07 | alternative *j* | 22.855333 | 0.000000 |
| 35 | voter08 | alternative *b* | 19.835570 | 19.835570 |
| 36 | voter09 | alternative *c* | 22.928716 | 0.000000 |
| 37 | voter09 | alternative *e* | 22.928716 | 22.928716 |
| 38 | voter09 | alternative *j* | 22.928716 | 0.000000 |
| 39 | voter10 | alternative *c* | 5.538309 | 5.538309 |
| 40 | voter10 | alternative *e* | 5.538309 | 0.000000 |
| 41 | voter11 | alternative *c* | 13.130227 | 13.130227 |
| 42 | voter11 | alternative *j* | 13.130227 | 0.000000 |
| 43 | voter12 | alternative *c* | 6.056291 | 6.056291 |
| 44 | voter13 | alternative *e* | 23.992772 | 23.992772 |
| 45 | voter13 | alternative *j* | 23.992772 | 0.000000 |
| 46 | voter14 | alternative *e* | 16.699207 | 16.699207 |
| 47 | voter15 | alternative *j* | 98.165759 | **77.714559** |
| 48 | alternative *b* | drain | $r^{(2)}$ **= 77.714559** | **77.714559** |
| 49 | alternative *c* | drain | $r^{(2)}$ **= 77.714559** | **77.714559** |
| 50 | alternative *e* | drain | $r^{(2)}$ **= 77.714559** | **76.740693** |
| 51 | alternative *j* | drain | $r^{(2)}$ **= 77.714559** | **77.714559** |

$r^{(3)}$ = ( 77.714559 + 77.714559 + 76.740693 + 77.714559 ) / 4 = 77.471093
$s^{(3)}$ = max { 75.365107; min { 77.714559; 77.714559; 76.740693 ; 77.714559 } } = 76.740693

Stage $z = 4$:

| link | start | end | capacity | flow |
|---|---|---|---|---|
| 1 | source | voter01 | 36.597383 | 36.597383 |
| 2 | source | voter02 | 5.481150 | 5.481150 |
| 3 | source | voter03 | 13.279131 | 13.279131 |
| 4 | source | voter04 | 4.859413 | 4.859413 |
| 5 | source | voter05 | 35.425375 | 35.425375 |
| 6 | source | voter06 | 5.490934 | 5.490934 |
| 7 | source | voter07 | 22.855333 | 22.855333 |
| 8 | source | voter08 | 19.835570 | 19.835570 |
| 9 | source | voter09 | 22.928716 | 22.928716 |
| 10 | source | voter10 | 5.538309 | 5.538309 |
| 11 | source | voter11 | 13.130227 | 13.130227 |
| 12 | source | voter12 | 6.056291 | 6.056291 |
| 13 | source | voter13 | 23.992772 | 23.992772 |
| 14 | source | voter14 | 16.699207 | 16.699207 |
| 15 | source | voter15 | 98.165759 | **77.471093** |
| 16 | voter01 | alternative *b* | 36.597383 | 0.000000 |
| 17 | voter01 | alternative *c* | 36.597383 | **29.126572** |
| 18 | voter01 | alternative *e* | 36.597383 | **7.470811** |
| 19 | voter01 | alternative *j* | 36.597383 | 0.000000 |
| 20 | voter02 | alternative *b* | 5.481150 | 0.000000 |
| 21 | voter02 | alternative *c* | 5.481150 | 5.481150 |
| 22 | voter02 | alternative *e* | 5.481150 | 0.000000 |
| 23 | voter03 | alternative *b* | 13.279131 | 0.000000 |
| 24 | voter03 | alternative *c* | 13.279131 | 13.279131 |
| 25 | voter03 | alternative *j* | 13.279131 | 0.000000 |
| 26 | voter04 | alternative *b* | 4.859413 | 0.000000 |
| 27 | voter04 | alternative *c* | 4.859413 | 4.859413 |
| 28 | voter05 | alternative *b* | 35.425375 | **34.780190** |
| 29 | voter05 | alternative *e* | 35.425375 | **0.645186** |
| 30 | voter05 | alternative *j* | 35.425375 | 0.000000 |
| 31 | voter06 | alternative *b* | 5.490934 | 0.000000 |
| 32 | voter06 | alternative *e* | 5.490934 | 5.490934 |
| 33 | voter07 | alternative *b* | 22.855333 | 22.855333 |
| 34 | voter07 | alternative *j* | 22.855333 | 0.000000 |
| 35 | voter08 | alternative *b* | 19.835570 | 19.835570 |
| 36 | voter09 | alternative *c* | 22.928716 | 0.000000 |
| 37 | voter09 | alternative *e* | 22.928716 | 22.928716 |
| 38 | voter09 | alternative *j* | 22.928716 | 0.000000 |
| 39 | voter10 | alternative *c* | 5.538309 | 5.538309 |
| 40 | voter10 | alternative *e* | 5.538309 | 0.000000 |
| 41 | voter11 | alternative *c* | 13.130227 | 13.130227 |
| 42 | voter11 | alternative *j* | 13.130227 | 0.000000 |
| 43 | voter12 | alternative *c* | 6.056291 | 6.056291 |
| 44 | voter13 | alternative *e* | 23.992772 | 23.992772 |
| 45 | voter13 | alternative *j* | 23.992772 | 0.000000 |
| 46 | voter14 | alternative *e* | 16.699207 | 16.699207 |
| 47 | voter15 | alternative *j* | 98.165759 | **77.471093** |
| 48 | alternative *b* | drain | $r^{(3)}$ **= 77.471093** | **77.471093** |
| 49 | alternative *c* | drain | $r^{(3)}$ **= 77.471093** | **77.471093** |
| 50 | alternative *e* | drain | $r^{(3)}$ **= 77.471093** | **77.227626** |
| 51 | alternative *j* | drain | $r^{(3)}$ **= 77.471093** | **77.471093** |

$r^{(4)}$ = ( 77.471093 + 77.471093 + 77.227626 + 77.471093 ) / 4 = 77.410226
$s^{(4)}$ = max { 76.740693; min { 77.471093; 77.471093; 77.227626 ; 77.471093 } } = 77.227626

Stage $z = 5$:

| link | start | end | capacity | flow |
|---|---|---|---|---|
| 1 | source | voter01 | 36.597383 | 36.597383 |
| 2 | source | voter02 | 5.481150 | 5.481150 |
| 3 | source | voter03 | 13.279131 | 13.279131 |
| 4 | source | voter04 | 4.859413 | 4.859413 |
| 5 | source | voter05 | 35.425375 | 35.425375 |
| 6 | source | voter06 | 5.490934 | 5.490934 |
| 7 | source | voter07 | 22.855333 | 22.855333 |
| 8 | source | voter08 | 19.835570 | 19.835570 |
| 9 | source | voter09 | 22.928716 | 22.928716 |
| 10 | source | voter10 | 5.538309 | 5.538309 |
| 11 | source | voter11 | 13.130227 | 13.130227 |
| 12 | source | voter12 | 6.056291 | 6.056291 |
| 13 | source | voter13 | 23.992772 | 23.992772 |
| 14 | source | voter14 | 16.699207 | 16.699207 |
| 15 | source | voter15 | 98.165759 | **77.410226** |
| 16 | voter01 | alternative *b* | 36.597383 | 0.000000 |
| 17 | voter01 | alternative *c* | 36.597383 | **29.065705** |
| 18 | voter01 | alternative *e* | 36.597383 | **7.531678** |
| 19 | voter01 | alternative *j* | 36.597383 | 0.000000 |
| 20 | voter02 | alternative *b* | 5.481150 | 0.000000 |
| 21 | voter02 | alternative *c* | 5.481150 | 5.481150 |
| 22 | voter02 | alternative *e* | 5.481150 | 0.000000 |
| 23 | voter03 | alternative *b* | 13.279131 | 0.000000 |
| 24 | voter03 | alternative *c* | 13.279131 | 13.279131 |
| 25 | voter03 | alternative *j* | 13.279131 | 0.000000 |
| 26 | voter04 | alternative *b* | 4.859413 | 0.000000 |
| 27 | voter04 | alternative *c* | 4.859413 | 4.859413 |
| 28 | voter05 | alternative *b* | 35.425375 | **34.719323** |
| 29 | voter05 | alternative *e* | 35.425375 | **0.706053** |
| 30 | voter05 | alternative *j* | 35.425375 | 0.000000 |
| 31 | voter06 | alternative *b* | 5.490934 | 0.000000 |
| 32 | voter06 | alternative *e* | 5.490934 | 5.490934 |
| 33 | voter07 | alternative *b* | 22.855333 | 22.855333 |
| 34 | voter07 | alternative *j* | 22.855333 | 0.000000 |
| 35 | voter08 | alternative *b* | 19.835570 | 19.835570 |
| 36 | voter09 | alternative *c* | 22.928716 | 0.000000 |
| 37 | voter09 | alternative *e* | 22.928716 | 22.928716 |
| 38 | voter09 | alternative *j* | 22.928716 | 0.000000 |
| 39 | voter10 | alternative *c* | 5.538309 | 5.538309 |
| 40 | voter10 | alternative *e* | 5.538309 | 0.000000 |
| 41 | voter11 | alternative *c* | 13.130227 | 13.130227 |
| 42 | voter11 | alternative *j* | 13.130227 | 0.000000 |
| 43 | voter12 | alternative *c* | 6.056291 | 6.056291 |
| 44 | voter13 | alternative *e* | 23.992772 | 23.992772 |
| 45 | voter13 | alternative *j* | 23.992772 | 0.000000 |
| 46 | voter14 | alternative *e* | 16.699207 | 16.699207 |
| 47 | voter15 | alternative *j* | 98.165759 | **77.410226** |
| 48 | alternative *b* | drain | **$r^{(4)}$ = 77.410226** | **77.410226** |
| 49 | alternative *c* | drain | **$r^{(4)}$ = 77.410226** | **77.410226** |
| 50 | alternative *e* | drain | **$r^{(4)}$ = 77.410226** | **77.349359** |
| 51 | alternative *j* | drain | **$r^{(4)}$ = 77.410226** | **77.410226** |

$r^{(5)}$ = ( 77.410226 + 77.410226 + 77.349359 + 77.410226 ) / 4 = 77.395009
$s^{(5)}$ = max { 77.227626; min { 77.410226; 77.410226; 77.349359; 77.410226 } } = 77.349359

Stage $z = 6$:

| link | start | end | capacity | flow |
|---|---|---|---|---|
| 1 | source | voter01 | 36.597383 | 36.597383 |
| 2 | source | voter02 | 5.481150 | 5.481150 |
| 3 | source | voter03 | 13.279131 | 13.279131 |
| 4 | source | voter04 | 4.859413 | 4.859413 |
| 5 | source | voter05 | 35.425375 | 35.425375 |
| 6 | source | voter06 | 5.490934 | 5.490934 |
| 7 | source | voter07 | 22.855333 | 22.855333 |
| 8 | source | voter08 | 19.835570 | 19.835570 |
| 9 | source | voter09 | 22.928716 | 22.928716 |
| 10 | source | voter10 | 5.538309 | 5.538309 |
| 11 | source | voter11 | 13.130227 | 13.130227 |
| 12 | source | voter12 | 6.056291 | 6.056291 |
| 13 | source | voter13 | 23.992772 | 23.992772 |
| 14 | source | voter14 | 16.699207 | 16.699207 |
| 15 | source | voter15 | 98.165759 | **77.395009** |
| 16 | voter01 | alternative *b* | 36.597383 | 0.000000 |
| 17 | voter01 | alternative *c* | 36.597383 | **29.050488** |
| 18 | voter01 | alternative *e* | 36.597383 | **7.546894** |
| 19 | voter01 | alternative *j* | 36.597383 | 0.000000 |
| 20 | voter02 | alternative *b* | 5.481150 | 0.000000 |
| 21 | voter02 | alternative *c* | 5.481150 | 5.481150 |
| 22 | voter02 | alternative *e* | 5.481150 | 0.000000 |
| 23 | voter03 | alternative *b* | 13.279131 | 0.000000 |
| 24 | voter03 | alternative *c* | 13.279131 | 13.279131 |
| 25 | voter03 | alternative *j* | 13.279131 | 0.000000 |
| 26 | voter04 | alternative *b* | 4.859413 | 0.000000 |
| 27 | voter04 | alternative *c* | 4.859413 | 4.859413 |
| 28 | voter05 | alternative *b* | 35.425375 | **34.704106** |
| 29 | voter05 | alternative *e* | 35.425375 | **0.721269** |
| 30 | voter05 | alternative *j* | 35.425375 | 0.000000 |
| 31 | voter06 | alternative *b* | 5.490934 | 0.000000 |
| 32 | voter06 | alternative *e* | 5.490934 | 5.490934 |
| 33 | voter07 | alternative *b* | 22.855333 | 22.855333 |
| 34 | voter07 | alternative *j* | 22.855333 | 0.000000 |
| 35 | voter08 | alternative *b* | 19.835570 | 19.835570 |
| 36 | voter09 | alternative *c* | 22.928716 | 0.000000 |
| 37 | voter09 | alternative *e* | 22.928716 | 22.928716 |
| 38 | voter09 | alternative *j* | 22.928716 | 0.000000 |
| 39 | voter10 | alternative *c* | 5.538309 | 5.538309 |
| 40 | voter10 | alternative *e* | 5.538309 | 0.000000 |
| 41 | voter11 | alternative *c* | 13.130227 | 13.130227 |
| 42 | voter11 | alternative *j* | 13.130227 | 0.000000 |
| 43 | voter12 | alternative *c* | 6.056291 | 6.056291 |
| 44 | voter13 | alternative *e* | 23.992772 | 23.992772 |
| 45 | voter13 | alternative *j* | 23.992772 | 0.000000 |
| 46 | voter14 | alternative *e* | 16.699207 | 16.699207 |
| 47 | voter15 | alternative *j* | 98.165759 | **77.395009** |
| 48 | alternative *b* | drain | **$r^{(5)}$ = 77.395009** | **77.395009** |
| 49 | alternative *c* | drain | **$r^{(5)}$ = 77.395009** | **77.395009** |
| 50 | alternative *e* | drain | **$r^{(5)}$ = 77.395009** | **77.379793** |
| 51 | alternative *j* | drain | **$r^{(5)}$ = 77.395009** | **77.395009** |

$r^{(6)}$ = ( 77.395009 + 77.395009 + 77.379793 + 77.395009 ) / 4 = 77.391205
$s^{(6)}$ = max { 77.349359; min { 77.395009; 77.395009; 77.379793; 77.395009 } } = 77.379793

Stage $z = 7$:

| link | start | end | capacity | flow |
|---|---|---|---|---|
| 1 | source | voter01 | 36.597383 | 36.597383 |
| 2 | source | voter02 | 5.481150 | 5.481150 |
| 3 | source | voter03 | 13.279131 | 13.279131 |
| 4 | source | voter04 | 4.859413 | 4.859413 |
| 5 | source | voter05 | 35.425375 | 35.425375 |
| 6 | source | voter06 | 5.490934 | 5.490934 |
| 7 | source | voter07 | 22.855333 | 22.855333 |
| 8 | source | voter08 | 19.835570 | 19.835570 |
| 9 | source | voter09 | 22.928716 | 22.928716 |
| 10 | source | voter10 | 5.538309 | 5.538309 |
| 11 | source | voter11 | 13.130227 | 13.130227 |
| 12 | source | voter12 | 6.056291 | 6.056291 |
| 13 | source | voter13 | 23.992772 | 23.992772 |
| 14 | source | voter14 | 16.699207 | 16.699207 |
| 15 | source | voter15 | 98.165759 | **77.391205** |
| 16 | voter01 | alternative *b* | 36.597383 | 0.000000 |
| 17 | voter01 | alternative *c* | 36.597383 | **29.046684** |
| 18 | voter01 | alternative *e* | 36.597383 | **7.550699** |
| 19 | voter01 | alternative *j* | 36.597383 | 0.000000 |
| 20 | voter02 | alternative *b* | 5.481150 | 0.000000 |
| 21 | voter02 | alternative *c* | 5.481150 | 5.481150 |
| 22 | voter02 | alternative *e* | 5.481150 | 0.000000 |
| 23 | voter03 | alternative *b* | 13.279131 | 0.000000 |
| 24 | voter03 | alternative *c* | 13.279131 | 13.279131 |
| 25 | voter03 | alternative *j* | 13.279131 | 0.000000 |
| 26 | voter04 | alternative *b* | 4.859413 | 0.000000 |
| 27 | voter04 | alternative *c* | 4.859413 | 4.859413 |
| 28 | voter05 | alternative *b* | 35.425375 | **34.700302** |
| 29 | voter05 | alternative *e* | 35.425375 | **0.725073** |
| 30 | voter05 | alternative *j* | 35.425375 | 0.000000 |
| 31 | voter06 | alternative *b* | 5.490934 | 0.000000 |
| 32 | voter06 | alternative *e* | 5.490934 | 5.490934 |
| 33 | voter07 | alternative *b* | 22.855333 | 22.855333 |
| 34 | voter07 | alternative *j* | 22.855333 | 0.000000 |
| 35 | voter08 | alternative *b* | 19.835570 | 19.835570 |
| 36 | voter09 | alternative *c* | 22.928716 | 0.000000 |
| 37 | voter09 | alternative *e* | 22.928716 | 22.928716 |
| 38 | voter09 | alternative *j* | 22.928716 | 0.000000 |
| 39 | voter10 | alternative *c* | 5.538309 | 5.538309 |
| 40 | voter10 | alternative *e* | 5.538309 | 0.000000 |
| 41 | voter11 | alternative *c* | 13.130227 | 13.130227 |
| 42 | voter11 | alternative *j* | 13.130227 | 0.000000 |
| 43 | voter12 | alternative *c* | 6.056291 | 6.056291 |
| 44 | voter13 | alternative *e* | 23.992772 | 23.992772 |
| 45 | voter13 | alternative *j* | 23.992772 | 0.000000 |
| 46 | voter14 | alternative *e* | 16.699207 | 16.699207 |
| 47 | voter15 | alternative *j* | 98.165759 | **77.391205** |
| 48 | alternative *b* | drain | **$r^{(6)}$ = 77.391205** | **77.391205** |
| 49 | alternative *c* | drain | **$r^{(6)}$ = 77.391205** | **77.391205** |
| 50 | alternative *e* | drain | **$r^{(6)}$ = 77.391205** | **77.387401** |
| 51 | alternative *j* | drain | **$r^{(6)}$ = 77.391205** | **77.391205** |

$r^{(7)}$ = ( 77.391205 + 77.391205 + 77.387401 + 77.391205 ) / 4 = 77.390254
$s^{(7)}$ = max { 77.379793; min { 77.391205; 77.391205; 77.387401; 77.391205 } } = 77.387401

Stage $z = 8$:

| link | start | end | capacity | flow |
|---|---|---|---|---|
| 1 | source | voter01 | 36.597383 | 36.597383 |
| 2 | source | voter02 | 5.481150 | 5.481150 |
| 3 | source | voter03 | 13.279131 | 13.279131 |
| 4 | source | voter04 | 4.859413 | 4.859413 |
| 5 | source | voter05 | 35.425375 | 35.425375 |
| 6 | source | voter06 | 5.490934 | 5.490934 |
| 7 | source | voter07 | 22.855333 | 22.855333 |
| 8 | source | voter08 | 19.835570 | 19.835570 |
| 9 | source | voter09 | 22.928716 | 22.928716 |
| 10 | source | voter10 | 5.538309 | 5.538309 |
| 11 | source | voter11 | 13.130227 | 13.130227 |
| 12 | source | voter12 | 6.056291 | 6.056291 |
| 13 | source | voter13 | 23.992772 | 23.992772 |
| 14 | source | voter14 | 16.699207 | 16.699207 |
| 15 | source | voter15 | 98.165759 | **77.390254** |
| 16 | voter01 | alternative *b* | 36.597383 | 0.000000 |
| 17 | voter01 | alternative *c* | 36.597383 | **29.045733** |
| 18 | voter01 | alternative *e* | 36.597383 | **7.551650** |
| 19 | voter01 | alternative *j* | 36.597383 | 0.000000 |
| 20 | voter02 | alternative *b* | 5.481150 | 0.000000 |
| 21 | voter02 | alternative *c* | 5.481150 | 5.481150 |
| 22 | voter02 | alternative *e* | 5.481150 | 0.000000 |
| 23 | voter03 | alternative *b* | 13.279131 | 0.000000 |
| 24 | voter03 | alternative *c* | 13.279131 | 13.279131 |
| 25 | voter03 | alternative *j* | 13.279131 | 0.000000 |
| 26 | voter04 | alternative *b* | 4.859413 | 0.000000 |
| 27 | voter04 | alternative *c* | 4.859413 | 4.859413 |
| 28 | voter05 | alternative *b* | 35.425375 | **34.699351** |
| 29 | voter05 | alternative *e* | 35.425375 | **0.726024** |
| 30 | voter05 | alternative *j* | 35.425375 | 0.000000 |
| 31 | voter06 | alternative *b* | 5.490934 | 0.000000 |
| 32 | voter06 | alternative *e* | 5.490934 | 5.490934 |
| 33 | voter07 | alternative *b* | 22.855333 | 22.855333 |
| 34 | voter07 | alternative *j* | 22.855333 | 0.000000 |
| 35 | voter08 | alternative *b* | 19.835570 | 19.835570 |
| 36 | voter09 | alternative *c* | 22.928716 | 0.000000 |
| 37 | voter09 | alternative *e* | 22.928716 | 22.928716 |
| 38 | voter09 | alternative *j* | 22.928716 | 0.000000 |
| 39 | voter10 | alternative *c* | 5.538309 | 5.538309 |
| 40 | voter10 | alternative *e* | 5.538309 | 0.000000 |
| 41 | voter11 | alternative *c* | 13.130227 | 13.130227 |
| 42 | voter11 | alternative *j* | 13.130227 | 0.000000 |
| 43 | voter12 | alternative *c* | 6.056291 | 6.056291 |
| 44 | voter13 | alternative *e* | 23.992772 | 23.992772 |
| 45 | voter13 | alternative *j* | 23.992772 | 0.000000 |
| 46 | voter14 | alternative *e* | 16.699207 | 16.699207 |
| 47 | voter15 | alternative *j* | 98.165759 | **77.390254** |
| 48 | alternative *b* | drain | **$r^{(7)}$ = 77.390254** | **77.390254** |
| 49 | alternative *c* | drain | **$r^{(7)}$ = 77.390254** | **77.390254** |
| 50 | alternative *e* | drain | **$r^{(7)}$ = 77.390254** | **77.389303** |
| 51 | alternative *j* | drain | **$r^{(7)}$ = 77.390254** | **77.390254** |

$r^{(8)}$ = ( 77.390254 + 77.390254 + 77.389303 + 77.390254 ) / 4 = 77.390016
$s^{(8)}$ = max { 77.387401; min { 77.390254; 77.390254; 77.389303; 77.390254 } } = 77.389303

Stage $z = 9$:

| link | start | end | capacity | flow |
|---|---|---|---|---|
| 1 | source | voter01 | 36.597383 | 36.597383 |
| 2 | source | voter02 | 5.481150 | 5.481150 |
| 3 | source | voter03 | 13.279131 | 13.279131 |
| 4 | source | voter04 | 4.859413 | 4.859413 |
| 5 | source | voter05 | 35.425375 | 35.425375 |
| 6 | source | voter06 | 5.490934 | 5.490934 |
| 7 | source | voter07 | 22.855333 | 22.855333 |
| 8 | source | voter08 | 19.835570 | 19.835570 |
| 9 | source | voter09 | 22.928716 | 22.928716 |
| 10 | source | voter10 | 5.538309 | 5.538309 |
| 11 | source | voter11 | 13.130227 | 13.130227 |
| 12 | source | voter12 | 6.056291 | 6.056291 |
| 13 | source | voter13 | 23.992772 | 23.992772 |
| 14 | source | voter14 | 16.699207 | 16.699207 |
| 15 | source | voter15 | 98.165759 | **77.390016** |
| 16 | voter01 | alternative *b* | 36.597383 | 0.000000 |
| 17 | voter01 | alternative *c* | 36.597383 | **29.045495** |
| 18 | voter01 | alternative *e* | 36.597383 | **7.551887** |
| 19 | voter01 | alternative *j* | 36.597383 | 0.000000 |
| 20 | voter02 | alternative *b* | 5.481150 | 0.000000 |
| 21 | voter02 | alternative *c* | 5.481150 | 5.481150 |
| 22 | voter02 | alternative *e* | 5.481150 | 0.000000 |
| 23 | voter03 | alternative *b* | 13.279131 | 0.000000 |
| 24 | voter03 | alternative *c* | 13.279131 | 13.279131 |
| 25 | voter03 | alternative *j* | 13.279131 | 0.000000 |
| 26 | voter04 | alternative *b* | 4.859413 | 0.000000 |
| 27 | voter04 | alternative *c* | 4.859413 | 4.859413 |
| 28 | voter05 | alternative *b* | 35.425375 | **34.699113** |
| 29 | voter05 | alternative *e* | 35.425375 | **0.726262** |
| 30 | voter05 | alternative *j* | 35.425375 | 0.000000 |
| 31 | voter06 | alternative *b* | 5.490934 | 0.000000 |
| 32 | voter06 | alternative *e* | 5.490934 | 5.490934 |
| 33 | voter07 | alternative *b* | 22.855333 | 22.855333 |
| 34 | voter07 | alternative *j* | 22.855333 | 0.000000 |
| 35 | voter08 | alternative *b* | 19.835570 | 19.835570 |
| 36 | voter09 | alternative *c* | 22.928716 | 0.000000 |
| 37 | voter09 | alternative *e* | 22.928716 | 22.928716 |
| 38 | voter09 | alternative *j* | 22.928716 | 0.000000 |
| 39 | voter10 | alternative *c* | 5.538309 | 5.538309 |
| 40 | voter10 | alternative *e* | 5.538309 | 0.000000 |
| 41 | voter11 | alternative *c* | 13.130227 | 13.130227 |
| 42 | voter11 | alternative *j* | 13.130227 | 0.000000 |
| 43 | voter12 | alternative *c* | 6.056291 | 6.056291 |
| 44 | voter13 | alternative *e* | 23.992772 | 23.992772 |
| 45 | voter13 | alternative *j* | 23.992772 | 0.000000 |
| 46 | voter14 | alternative *e* | 16.699207 | 16.699207 |
| 47 | voter15 | alternative *j* | 98.165759 | **77.390016** |
| 48 | alternative *b* | drain | **$r^{(8)}$ = 77.390016** | **77.390016** |
| 49 | alternative *c* | drain | **$r^{(8)}$ = 77.390016** | **77.390016** |
| 50 | alternative *e* | drain | **$r^{(8)}$ = 77.390016** | **77.389779** |
| 51 | alternative *j* | drain | **$r^{(8)}$ = 77.390016** | **77.390016** |

$r^{(9)}$ = ( 77.390016 + 77.390016 + 77.389779 + 77.390016 ) / 4 = 77.389957
$s^{(9)}$ = max { 77.389303; min { 77.390016; 77.390016; 77.389779; 77.390016 } } = 77.389779

Stage $z = 10$:

| link | start | end | capacity | flow |
|---|---|---|---|---|
| 1 | source | voter01 | 36.597383 | 36.597383 |
| 2 | source | voter02 | 5.481150 | 5.481150 |
| 3 | source | voter03 | 13.279131 | 13.279131 |
| 4 | source | voter04 | 4.859413 | 4.859413 |
| 5 | source | voter05 | 35.425375 | 35.425375 |
| 6 | source | voter06 | 5.490934 | 5.490934 |
| 7 | source | voter07 | 22.855333 | 22.855333 |
| 8 | source | voter08 | 19.835570 | 19.835570 |
| 9 | source | voter09 | 22.928716 | 22.928716 |
| 10 | source | voter10 | 5.538309 | 5.538309 |
| 11 | source | voter11 | 13.130227 | 13.130227 |
| 12 | source | voter12 | 6.056291 | 6.056291 |
| 13 | source | voter13 | 23.992772 | 23.992772 |
| 14 | source | voter14 | 16.699207 | 16.699207 |
| 15 | source | voter15 | 98.165759 | **77.389957** |
| 16 | voter01 | alternative *b* | 36.597383 | 0.000000 |
| 17 | voter01 | alternative *c* | 36.597383 | **29.045436** |
| 18 | voter01 | alternative *e* | 36.597383 | **7.551947** |
| 19 | voter01 | alternative *j* | 36.597383 | 0.000000 |
| 20 | voter02 | alternative *b* | 5.481150 | 0.000000 |
| 21 | voter02 | alternative *c* | 5.481150 | 5.481150 |
| 22 | voter02 | alternative *e* | 5.481150 | 0.000000 |
| 23 | voter03 | alternative *b* | 13.279131 | 0.000000 |
| 24 | voter03 | alternative *c* | 13.279131 | 13.279131 |
| 25 | voter03 | alternative *j* | 13.279131 | 0.000000 |
| 26 | voter04 | alternative *b* | 4.859413 | 0.000000 |
| 27 | voter04 | alternative *c* | 4.859413 | 4.859413 |
| 28 | voter05 | alternative *b* | 35.425375 | **34.699054** |
| 29 | voter05 | alternative *e* | 35.425375 | **0.726322** |
| 30 | voter05 | alternative *j* | 35.425375 | 0.000000 |
| 31 | voter06 | alternative *b* | 5.490934 | 0.000000 |
| 32 | voter06 | alternative *e* | 5.490934 | 5.490934 |
| 33 | voter07 | alternative *b* | 22.855333 | 22.855333 |
| 34 | voter07 | alternative *j* | 22.855333 | 0.000000 |
| 35 | voter08 | alternative *b* | 19.835570 | 19.835570 |
| 36 | voter09 | alternative *c* | 22.928716 | 0.000000 |
| 37 | voter09 | alternative *e* | 22.928716 | 22.928716 |
| 38 | voter09 | alternative *j* | 22.928716 | 0.000000 |
| 39 | voter10 | alternative *c* | 5.538309 | 5.538309 |
| 40 | voter10 | alternative *e* | 5.538309 | 0.000000 |
| 41 | voter11 | alternative *c* | 13.130227 | 13.130227 |
| 42 | voter11 | alternative *j* | 13.130227 | 0.000000 |
| 43 | voter12 | alternative *c* | 6.056291 | 6.056291 |
| 44 | voter13 | alternative *e* | 23.992772 | 23.992772 |
| 45 | voter13 | alternative *j* | 23.992772 | 0.000000 |
| 46 | voter14 | alternative *e* | 16.699207 | 16.699207 |
| 47 | voter15 | alternative *j* | 98.165759 | **77.389957** |
| 48 | alternative *b* | drain | **$r^{(9)}$ = 77.389957** | **77.389957** |
| 49 | alternative *c* | drain | **$r^{(9)}$ = 77.389957** | **77.389957** |
| 50 | alternative *e* | drain | **$r^{(9)}$ = 77.389957** | **77.389897** |
| 51 | alternative *j* | drain | **$r^{(9)}$ = 77.389957** | **77.389957** |

$r^{(10)}$ = ( 77.389957 + 77.389957 + 77.389897 + 77.389957 ) / 4 = 77.389942
$s^{(10)}$ = max { 77.389779; min { 77.389957; 77.389957; 77.389897; 77.389957 } } = 77.389897

Stage $z = 11$:

| link | start | end | capacity | flow |
|---|---|---|---|---|
| 1 | source | voter01 | 36.597383 | 36.597383 |
| 2 | source | voter02 | 5.481150 | 5.481150 |
| 3 | source | voter03 | 13.279131 | 13.279131 |
| 4 | source | voter04 | 4.859413 | 4.859413 |
| 5 | source | voter05 | 35.425375 | 35.425375 |
| 6 | source | voter06 | 5.490934 | 5.490934 |
| 7 | source | voter07 | 22.855333 | 22.855333 |
| 8 | source | voter08 | 19.835570 | 19.835570 |
| 9 | source | voter09 | 22.928716 | 22.928716 |
| 10 | source | voter10 | 5.538309 | 5.538309 |
| 11 | source | voter11 | 13.130227 | 13.130227 |
| 12 | source | voter12 | 6.056291 | 6.056291 |
| 13 | source | voter13 | 23.992772 | 23.992772 |
| 14 | source | voter14 | 16.699207 | 16.699207 |
| 15 | source | voter15 | 98.165759 | **77.389942** |
| 16 | voter01 | alternative *b* | 36.597383 | 0.000000 |
| 17 | voter01 | alternative *c* | 36.597383 | **29.045421** |
| 18 | voter01 | alternative *e* | 36.597383 | **7.551962** |
| 19 | voter01 | alternative *j* | 36.597383 | 0.000000 |
| 20 | voter02 | alternative *b* | 5.481150 | 0.000000 |
| 21 | voter02 | alternative *c* | 5.481150 | 5.481150 |
| 22 | voter02 | alternative *e* | 5.481150 | 0.000000 |
| 23 | voter03 | alternative *b* | 13.279131 | 0.000000 |
| 24 | voter03 | alternative *c* | 13.279131 | 13.279131 |
| 25 | voter03 | alternative *j* | 13.279131 | 0.000000 |
| 26 | voter04 | alternative *b* | 4.859413 | 0.000000 |
| 27 | voter04 | alternative *c* | 4.859413 | 4.859413 |
| 28 | voter05 | alternative *b* | 35.425375 | **34.699039** |
| 29 | voter05 | alternative *e* | 35.425375 | **0.726336** |
| 30 | voter05 | alternative *j* | 35.425375 | 0.000000 |
| 31 | voter06 | alternative *b* | 5.490934 | 0.000000 |
| 32 | voter06 | alternative *e* | 5.490934 | 5.490934 |
| 33 | voter07 | alternative *b* | 22.855333 | 22.855333 |
| 34 | voter07 | alternative *j* | 22.855333 | 0.000000 |
| 35 | voter08 | alternative *b* | 19.835570 | 19.835570 |
| 36 | voter09 | alternative *c* | 22.928716 | 0.000000 |
| 37 | voter09 | alternative *e* | 22.928716 | 22.928716 |
| 38 | voter09 | alternative *j* | 22.928716 | 0.000000 |
| 39 | voter10 | alternative *c* | 5.538309 | 5.538309 |
| 40 | voter10 | alternative *e* | 5.538309 | 0.000000 |
| 41 | voter11 | alternative *c* | 13.130227 | 13.130227 |
| 42 | voter11 | alternative *j* | 13.130227 | 0.000000 |
| 43 | voter12 | alternative *c* | 6.056291 | 6.056291 |
| 44 | voter13 | alternative *e* | 23.992772 | 23.992772 |
| 45 | voter13 | alternative *j* | 23.992772 | 0.000000 |
| 46 | voter14 | alternative *e* | 16.699207 | 16.699207 |
| 47 | voter15 | alternative *j* | 98.165759 | **77.389942** |
| 48 | alternative *b* | drain | **$r^{(10)}$ = 77.389942** | **77.389942** |
| 49 | alternative *c* | drain | **$r^{(10)}$ = 77.389942** | **77.389942** |
| 50 | alternative *e* | drain | **$r^{(10)}$ = 77.389942** | **77.389927** |
| 51 | alternative *j* | drain | **$r^{(10)}$ = 77.389942** | **77.389942** |

$r^{(11)}$ = ( 77.389942 + 77.389942 + 77.389927 + 77.389942 ) / 4 = 77.389938
$s^{(11)}$ = max { 77.389897; min { 77.389942; 77.389942; 77.389927; 77.389942 } } = 77.389927

Stage $z = 12$:

| link | start | end | capacity | flow |
|---|---|---|---|---|
| 1 | source | voter01 | 36.597383 | 36.597383 |
| 2 | source | voter02 | 5.481150 | 5.481150 |
| 3 | source | voter03 | 13.279131 | 13.279131 |
| 4 | source | voter04 | 4.859413 | 4.859413 |
| 5 | source | voter05 | 35.425375 | 35.425375 |
| 6 | source | voter06 | 5.490934 | 5.490934 |
| 7 | source | voter07 | 22.855333 | 22.855333 |
| 8 | source | voter08 | 19.835570 | 19.835570 |
| 9 | source | voter09 | 22.928716 | 22.928716 |
| 10 | source | voter10 | 5.538309 | 5.538309 |
| 11 | source | voter11 | 13.130227 | 13.130227 |
| 12 | source | voter12 | 6.056291 | 6.056291 |
| 13 | source | voter13 | 23.992772 | 23.992772 |
| 14 | source | voter14 | 16.699207 | 16.699207 |
| 15 | source | voter15 | 98.165759 | **77.389938** |
| 16 | voter01 | alternative *b* | 36.597383 | 0.000000 |
| 17 | voter01 | alternative *c* | 36.597383 | **29.045417** |
| 18 | voter01 | alternative *e* | 36.597383 | **7.551965** |
| 19 | voter01 | alternative *j* | 36.597383 | 0.000000 |
| 20 | voter02 | alternative *b* | 5.481150 | 0.000000 |
| 21 | voter02 | alternative *c* | 5.481150 | 5.481150 |
| 22 | voter02 | alternative *e* | 5.481150 | 0.000000 |
| 23 | voter03 | alternative *b* | 13.279131 | 0.000000 |
| 24 | voter03 | alternative *c* | 13.279131 | 13.279131 |
| 25 | voter03 | alternative *j* | 13.279131 | 0.000000 |
| 26 | voter04 | alternative *b* | 4.859413 | 0.000000 |
| 27 | voter04 | alternative *c* | 4.859413 | 4.859413 |
| 28 | voter05 | alternative *b* | 35.425375 | **34.699035** |
| 29 | voter05 | alternative *e* | 35.425375 | **0.726340** |
| 30 | voter05 | alternative *j* | 35.425375 | 0.000000 |
| 31 | voter06 | alternative *b* | 5.490934 | 0.000000 |
| 32 | voter06 | alternative *e* | 5.490934 | 5.490934 |
| 33 | voter07 | alternative *b* | 22.855333 | 22.855333 |
| 34 | voter07 | alternative *j* | 22.855333 | 0.000000 |
| 35 | voter08 | alternative *b* | 19.835570 | 19.835570 |
| 36 | voter09 | alternative *c* | 22.928716 | 0.000000 |
| 37 | voter09 | alternative *e* | 22.928716 | 22.928716 |
| 38 | voter09 | alternative *j* | 22.928716 | 0.000000 |
| 39 | voter10 | alternative *c* | 5.538309 | 5.538309 |
| 40 | voter10 | alternative *e* | 5.538309 | 0.000000 |
| 41 | voter11 | alternative *c* | 13.130227 | 13.130227 |
| 42 | voter11 | alternative *j* | 13.130227 | 0.000000 |
| 43 | voter12 | alternative *c* | 6.056291 | 6.056291 |
| 44 | voter13 | alternative *e* | 23.992772 | 23.992772 |
| 45 | voter13 | alternative *j* | 23.992772 | 0.000000 |
| 46 | voter14 | alternative *e* | 16.699207 | 16.699207 |
| 47 | voter15 | alternative *j* | 98.165759 | **77.389938** |
| 48 | alternative *b* | drain | $\mathbf{r^{(11)}}$ **= 77.389938** | **77.389938** |
| 49 | alternative *c* | drain | $\mathbf{r^{(11)}}$ **= 77.389938** | **77.389938** |
| 50 | alternative *e* | drain | $\mathbf{r^{(11)}}$ **= 77.389938** | **77.389935** |
| 51 | alternative *j* | drain | $\mathbf{r^{(11)}}$ **= 77.389938** | **77.389938** |

$r^{(12)}$ = ( 77.389938 + 77.389938 + 77.389935 + 77.389938 ) / 4 = 77.389937
$s^{(12)}$ = max { 77.389927; min { 77.389938; 77.389938; 77.389935; 77.389938 } } = 77.389935

Stage $z = 13$:

| link | start | end | capacity | flow |
|---|---|---|---|---|
| 1 | source | voter01 | 36.597383 | 36.597383 |
| 2 | source | voter02 | 5.481150 | 5.481150 |
| 3 | source | voter03 | 13.279131 | 13.279131 |
| 4 | source | voter04 | 4.859413 | 4.859413 |
| 5 | source | voter05 | 35.425375 | 35.425375 |
| 6 | source | voter06 | 5.490934 | 5.490934 |
| 7 | source | voter07 | 22.855333 | 22.855333 |
| 8 | source | voter08 | 19.835570 | 19.835570 |
| 9 | source | voter09 | 22.928716 | 22.928716 |
| 10 | source | voter10 | 5.538309 | 5.538309 |
| 11 | source | voter11 | 13.130227 | 13.130227 |
| 12 | source | voter12 | 6.056291 | 6.056291 |
| 13 | source | voter13 | 23.992772 | 23.992772 |
| 14 | source | voter14 | 16.699207 | 16.699207 |
| 15 | source | voter15 | 98.165759 | **77.389937** |
| 16 | voter01 | alternative *b* | 36.597383 | 0.000000 |
| 17 | voter01 | alternative *c* | 36.597383 | **29.045416** |
| 18 | voter01 | alternative *e* | 36.597383 | **7.551966** |
| 19 | voter01 | alternative *j* | 36.597383 | 0.000000 |
| 20 | voter02 | alternative *b* | 5.481150 | 0.000000 |
| 21 | voter02 | alternative *c* | 5.481150 | 5.481150 |
| 22 | voter02 | alternative *e* | 5.481150 | 0.000000 |
| 23 | voter03 | alternative *b* | 13.279131 | 0.000000 |
| 24 | voter03 | alternative *c* | 13.279131 | 13.279131 |
| 25 | voter03 | alternative *j* | 13.279131 | 0.000000 |
| 26 | voter04 | alternative *b* | 4.859413 | 0.000000 |
| 27 | voter04 | alternative *c* | 4.859413 | 4.859413 |
| 28 | voter05 | alternative *b* | 35.425375 | **34.699034** |
| 29 | voter05 | alternative *e* | 35.425375 | **0.726341** |
| 30 | voter05 | alternative *j* | 35.425375 | 0.000000 |
| 31 | voter06 | alternative *b* | 5.490934 | 0.000000 |
| 32 | voter06 | alternative *e* | 5.490934 | 5.490934 |
| 33 | voter07 | alternative *b* | 22.855333 | 22.855333 |
| 34 | voter07 | alternative *j* | 22.855333 | 0.000000 |
| 35 | voter08 | alternative *b* | 19.835570 | 19.835570 |
| 36 | voter09 | alternative *c* | 22.928716 | 0.000000 |
| 37 | voter09 | alternative *e* | 22.928716 | 22.928716 |
| 38 | voter09 | alternative *j* | 22.928716 | 0.000000 |
| 39 | voter10 | alternative *c* | 5.538309 | 5.538309 |
| 40 | voter10 | alternative *e* | 5.538309 | 0.000000 |
| 41 | voter11 | alternative *c* | 13.130227 | 13.130227 |
| 42 | voter11 | alternative *j* | 13.130227 | 0.000000 |
| 43 | voter12 | alternative *c* | 6.056291 | 6.056291 |
| 44 | voter13 | alternative *e* | 23.992772 | 23.992772 |
| 45 | voter13 | alternative *j* | 23.992772 | 0.000000 |
| 46 | voter14 | alternative *e* | 16.699207 | 16.699207 |
| 47 | voter15 | alternative *j* | 98.165759 | **77.389937** |
| 48 | alternative *b* | drain | **$r^{(12)}$ = 77.389937** | **77.389937** |
| 49 | alternative *c* | drain | **$r^{(12)}$ = 77.389937** | **77.389937** |
| 50 | alternative *e* | drain | **$r^{(12)}$ = 77.389937** | **77.389936** |
| 51 | alternative *j* | drain | **$r^{(12)}$ = 77.389937** | **77.389937** |

$r^{(13)}$ = ( 77.389937 + 77.389937 + 77.389936 + 77.389937 ) / 4 = 77.389937
$s^{(13)}$ = max { 77.389935; min { 77.389937; 77.389937; 77.389936; 77.389937 } } = 77.389936

The following table 8.2.2.1 summarizes the other tables of sections 8.2.1 and 8.2.2. The column *X* is the voting profile (before proportional completion) according to table 8.2.1.2. The column *Y* is the number of voters with this profile.

<table>
<tr><th colspan="6">before proportional completion</th><th colspan="13">after proportional completion</th></tr>
<tr><th>X</th><th>Y</th><th>b</th><th>c</th><th>e</th><th>j</th><th>Y</th><th>b</th><th>c</th><th>e</th><th>j</th><th>b</th><th>c</th><th>e</th><th>j</th><th>b</th><th>c</th><th>e</th><th>j</th></tr>
<tr><td rowspan="4">#11</td><td rowspan="4">7</td><td rowspan="4">1</td><td rowspan="4">2</td><td rowspan="4">1</td><td rowspan="4">2</td><td>1.307713</td><td>1</td><td>1</td><td>1</td><td>1</td><td></td><td>1.037863</td><td>0.269850</td><td></td><td rowspan="4">2.689516</td><td rowspan="4">1.371659</td><td rowspan="4">2.938825</td><td rowspan="4"></td></tr>
<tr><td>0.333796</td><td>1</td><td>1</td><td>1</td><td>3</td><td></td><td>0.333796</td><td></td><td></td></tr>
<tr><td>2.745815</td><td>1</td><td>3</td><td>1</td><td>1</td><td>2.689516</td><td></td><td>0.056299</td><td></td></tr>
<tr><td>2.612676</td><td>1</td><td>3</td><td>1</td><td>3</td><td></td><td></td><td>2.612676</td><td></td></tr>
</table>

For example, the above row says that, before proportional completion, there are 7 voters who strictly prefer alternative *b* (“1” in column *b*) and alternative *e* (“1” in column *e*) to alternative *a* and who are indifferent between alternative *a*, alternative *c* (“2” in column *c*), and alternative *j* (“2” in column *j*).

Proportional completion replaces these 7 voters by

- 1.307713 voters who strictly prefer alternatives *b*, *c*, *e*, and *j* to alternative *a* (voting profile “1111”),
- 0.333796 voters who strictly prefer alternatives *b*, *c*, and *e* to alternative *a* and who strictly prefer alternative *a* to alternative *j* (voting profile “1113”),
- 2.745815 voters who strictly prefer alternatives *b*, *e*, and *j* to alternative *a* and who strictly prefer alternative *a* to alternative *c* (voting profile “1311”),
- 2.612676 voters who strictly prefer alternatives *b* and *e* to alternative *a* and who strictly prefer alternative *a* to alternatives *c* and *j* (voting profile “1313”).

When we solve (8.1.2.2) – (8.1.2.5) to calculate $N[\{b,c,e,j\};a]$, then

- 1.037863 of the 1.307713 voters with profile “1111” are allocated to alternative *c* and 0.269850 are allocated to alternative *e*,
- 0.333796 of the 0.333796 voters with profile “1113” are allocated to alternative *c*,
- 2.689516 of the 2.745815 voters with profile “1311” are allocated to alternative *b* and 0.056299 are allocated to alternative *e*,
- 2.612676 of the 2.612676 voters with profile “1313” are allocated to alternative *e*.

In total, 2.689516 of the 7 voters with voting profile #11 are allocated to alternative *b*, 1.371659 are allocated to alternative *c*, and 2.938825 of these voters are allocated to alternative *e*.

| before proportional completion | | | | | | after proportional completion | | | | | | | | | | | | |
|---|---|---|---|---|---|---|---|---|---|---|---|---|---|---|---|---|---|---|
| *X* | *Y* | *b* | *c* | *e* | *j* | *Y* | *b* | *c* | *e* | *j* | *b* | *c* | *e* | *j* | *b* | *c* | *e* | *j* |
| #1 | 17 | 1 | 1 | 1 | 1 | 17.000000 | 1 | 1 | 1 | 1 | | 13.492005 | 3.507995 | | | 13.492005 | 3.507995 | |
| #2 | 2 | 1 | 1 | 1 | 2 | 1.158151 | 1 | 1 | 1 | 1 | | 0.919163 | 0.238987 | | | 1.761013 | 0.238987 | |
| | | | | | | 0.841849 | 1 | 1 | 1 | 3 | | 0.841849 | | | | | | |
| #3 | 3 | 1 | 1 | 1 | 3 | 3.000000 | 1 | 1 | 1 | 3 | | 3.000000 | | | | 3.000000 | | |
| #4 | 4 | 1 | 1 | 2 | 1 | 1.323077 | 1 | 1 | 1 | 1 | | 1.050057 | 0.273020 | | | 3.726980 | 0.273020 | |
| | | | | | | 2.676923 | 1 | 1 | 3 | 1 | | 2.676923 | | | | | | |
| #5 | 4 | 1 | 1 | 2 | 2 | 1.034298 | 1 | 1 | 1 | 1 | | 0.820868 | 0.213430 | | | 3.786570 | 0.213430 | |
| | | | | | | 0.288779 | 1 | 1 | 1 | 3 | | 0.288779 | | | | | | |
| | | | | | | 1.282004 | 1 | 1 | 3 | 1 | | 1.282004 | | | | | | |
| | | | | | | 1.394919 | 1 | 1 | 3 | 3 | | 1.394919 | | | | | | |
| #7 | 6 | 1 | 1 | 3 | 1 | 6.000000 | 1 | 1 | 3 | 1 | | 6.000000 | | | | 6.000000 | | |
| #9 | 3 | 1 | 1 | 3 | 3 | 3.000000 | 1 | 1 | 3 | 3 | | 3.000000 | | | | 3.000000 | | |
| #10 | 7 | 1 | 2 | 1 | 1 | 1.641509 | 1 | 1 | 1 | 1 | | 1.302780 | 0.338730 | | 5.248623 | 1.302780 | 0.448597 | |
| | | | | | | 5.358491 | 1 | 3 | 1 | 1 | 5.248623 | | 0.109867 | | | | | |
| #11 | 7 | 1 | 2 | 1 | 2 | 1.307713 | 1 | 1 | 1 | 1 | | 1.037863 | 0.269850 | | 2.689516 | 1.371659 | 2.938825 | |
| | | | | | | 0.333796 | 1 | 1 | 1 | 3 | | 0.333796 | | | | | | |
| | | | | | | 2.745815 | 1 | 3 | 1 | 1 | 2.689516 | | 0.056299 | | | | | |
| | | | | | | 2.612676 | 1 | 3 | 1 | 3 | | | 2.612676 | | | | | |
| #13 | 14 | 1 | 2 | 2 | 1 | 2.147039 | 1 | 1 | 1 | 1 | | 1.703992 | 0.443047 | | 10.666056 | 2.839972 | 0.493972 | |
| | | | | | | 1.135980 | 1 | 1 | 3 | 1 | | 1.135980 | | | | | | |
| | | | | | | 2.483731 | 1 | 3 | 1 | 1 | 2.432806 | | 0.050925 | | | | | |
| | | | | | | 8.233251 | 1 | 3 | 3 | 1 | 8.233251 | | | | | | | |
| #14 | 7 | 1 | 2 | 2 | 2 | 0.905832 | 1 | 1 | 1 | 1 | | 0.718911 | 0.186921 | | 5.002276 | 1.454589 | 0.543136 | |
| | | | | | | 0.167687 | 1 | 1 | 1 | 3 | | 0.167687 | | | | | | |
| | | | | | | 0.401882 | 1 | 1 | 3 | 1 | | 0.401882 | | | | | | |
| | | | | | | 0.166109 | 1 | 1 | 3 | 3 | | 0.166109 | | | | | | |
| | | | | | | 0.904189 | 1 | 3 | 1 | 1 | 0.885650 | | 0.018539 | | | | | |
| | | | | | | 0.337676 | 1 | 3 | 1 | 3 | | | 0.337676 | | | | | |
| | | | | | | 1.841625 | 1 | 3 | 3 | 1 | 1.841625 | | | | | | | |
| | | | | | | 2.275000 | 1 | 3 | 3 | 3 | 2.275000 | | | | | | | |
| #19 | 18 | 1 | 3 | 1 | 1 | 18.000000 | 1 | 3 | 1 | 1 | 17.630938 | | 0.369062 | | 17.630938 | | 0.369062 | |
| #21 | 2 | 1 | 3 | 1 | 3 | 2.000000 | 1 | 3 | 1 | 3 | | | 2.000000 | | | | 2.000000 | |
| #25 | 10 | 1 | 3 | 3 | 1 | 10.000000 | 1 | 3 | 3 | 1 | 10.000000 | | | | 10.000000 | | | |
| #27 | 17 | 1 | 3 | 3 | 3 | 17.000000 | 1 | 3 | 3 | 3 | 17.000000 | | | | 17.000000 | | | |
| #28 | 8 | 2 | 1 | 1 | 1 | 2.501292 | 1 | 1 | 1 | 1 | | 1.985144 | 0.516148 | | | 1.985144 | 6.014856 | |
| | | | | | | 5.498708 | 3 | 1 | 1 | 1 | | | 5.498708 | | | | | |
| #29 | 6 | 2 | 1 | 1 | 2 | 1.410746 | 1 | 1 | 1 | 1 | | 1.119635 | 0.291111 | | | 3.645183 | 2.354817 | |
| | | | | | | 0.465223 | 1 | 1 | 1 | 3 | | 0.465223 | | | | | | |
| | | | | | | 2.063706 | 3 | 1 | 1 | 1 | | | 2.063706 | | | | | |
| | | | | | | 2.060325 | 3 | 1 | 1 | 3 | | 2.060325 | | | | | | |
| #31 | 2 | 2 | 1 | 2 | 1 | 0.360847 | 1 | 1 | 1 | 1 | | 0.286385 | 0.074462 | | | 1.624847 | 0.375153 | |
| | | | | | | 0.264476 | 1 | 1 | 3 | 1 | | 0.264476 | | | | | | |
| | | | | | | 0.300691 | 3 | 1 | 1 | 1 | | | 0.300691 | | | | | |
| | | | | | | 1.073986 | 3 | 1 | 3 | 1 | | 1.073986 | | | | | | |
| #32 | 3 | 2 | 1 | 2 | 2 | 0.469714 | 1 | 1 | 1 | 1 | | 0.372787 | 0.096927 | | | 2.597064 | 0.402936 | |
| | | | | | | 0.071557 | 1 | 1 | 1 | 3 | | 0.071557 | | | | | | |
| | | | | | | 0.235660 | 1 | 1 | 3 | 1 | | 0.235660 | | | | | | |
| | | | | | | 0.161054 | 1 | 1 | 3 | 3 | | 0.161054 | | | | | | |
| | | | | | | 0.306010 | 3 | 1 | 1 | 1 | | | 0.306010 | | | | | |
| | | | | | | 0.145027 | 3 | 1 | 1 | 3 | | 0.145027 | | | | | | |
| | | | | | | 0.725843 | 3 | 1 | 3 | 1 | | 0.725843 | | | | | | |
| | | | | | | 0.885135 | 3 | 1 | 3 | 3 | | 0.885135 | | | | | | |

Table 8.2.2.1 (part 1 of 2): Voting patterns and allocation of votes in example A53

| before proportional completion | | | | | | after proportional completion | | | | | | | | | | | | |
|---|---|---|---|---|---|---|---|---|---|---|---|---|---|---|---|---|---|---|
| *X* | *Y* | *b* | *c* | *e* | *j* | *Y* | *b* | *c* | *e* | *j* | *b* | *c* | *e* | *j* | *b* | *c* | *e* | *j* |
| #37 | 11 | 2 | 2 | 1 | 1 | 1.439974 | 1 | 1 | 1 | 1 | | 1.142831 | 0.297142 | | 1.958310 | 1.142831 | 7.898858 | |
| | | | | | | 1.999303 | 1 | 3 | 1 | 1 | 1.958310 | | 0.040993 | | | | | |
| | | | | | | 1.139541 | 3 | 1 | 1 | 1 | | | 1.139541 | | | | | |
| | | | | | | 6.421182 | 3 | 3 | 1 | 1 | | | 6.421182 | | | | | |
| #38 | 7 | 2 | 2 | 1 | 2 | 0.758990 | 1 | 1 | 1 | 1 | | 0.602371 | 0.156620 | | 0.868696 | 0.936167 | 5.195137 | |
| | | | | | | 0.157356 | 1 | 1 | 1 | 3 | | 0.157356 | | | | | | |
| | | | | | | 0.886880 | 1 | 3 | 1 | 1 | 0.868696 | | 0.018184 | | | | | |
| | | | | | | 0.385403 | 1 | 3 | 1 | 3 | | | 0.385403 | | | | | |
| | | | | | | 0.548723 | 3 | 1 | 1 | 1 | | | 0.548723 | | | | | |
| | | | | | | 0.176440 | 3 | 1 | 1 | 3 | | 0.176440 | | | | | | |
| | | | | | | 1.858934 | 3 | 3 | 1 | 1 | | | 1.858934 | | | | | |
| | | | | | | 2.227273 | 3 | 3 | 1 | 3 | | | 2.227273 | | | | | |
| #40 | 23 | 2 | 2 | 2 | 1 | 2.103927 | 1 | 1 | 1 | 1 | | 1.669776 | 0.434151 | | 4.138414 | 3.536029 | 3.934047 | 8.980608 |
| | | | | | | 0.906927 | 1 | 1 | 3 | 1 | | 0.906927 | | | | | | |
| | | | | | | 2.045815 | 1 | 3 | 1 | 1 | 2.003869 | | 0.041946 | | | | | |
| | | | | | | 2.134545 | 1 | 3 | 3 | 1 | 2.134545 | | | | | | | |
| | | | | | | 1.423351 | 3 | 1 | 1 | 1 | | | 1.423351 | | | | | |
| | | | | | | 0.959326 | 3 | 1 | 3 | 1 | | 0.959326 | | | | | | |
| | | | | | | 2.034599 | 3 | 3 | 1 | 1 | | | 2.034599 | | | | | |
| | | | | | | 11.391509 | 3 | 3 | 3 | 1 | | | | 8.980608 | | | | |
| #41 | 13 | 2 | 2 | 2 | 2 | 1.034274 | 1 | 1 | 1 | 1 | | 0.820849 | 0.213425 | | 2.187107 | 2.187107 | 2.187107 | 2.187107 |
| | | | | | | 0.154902 | 1 | 1 | 1 | 3 | | 0.154902 | | | | | | |
| | | | | | | 0.375280 | 1 | 1 | 3 | 1 | | 0.375280 | | | | | | |
| | | | | | | 0.137331 | 1 | 1 | 3 | 3 | | 0.137331 | | | | | | |
| | | | | | | 1.001152 | 1 | 3 | 1 | 1 | 0.980625 | | 0.020527 | | | | | |
| | | | | | | 0.155179 | 1 | 3 | 1 | 3 | | | 0.155179 | | | | | |
| | | | | | | 0.645912 | 1 | 3 | 3 | 1 | 0.645912 | | | | | | | |
| | | | | | | 0.560570 | 1 | 3 | 3 | 3 | 0.560570 | | | | | | | |
| | | | | | | 0.647985 | 3 | 1 | 1 | 1 | | | 0.647985 | | | | | |
| | | | | | | 0.156517 | 3 | 1 | 1 | 3 | | 0.156517 | | | | | | |
| | | | | | | 0.371072 | 3 | 1 | 3 | 1 | | 0.371072 | | | | | | |
| | | | | | | 0.171156 | 3 | 1 | 3 | 3 | | 0.171156 | | | | | | |
| | | | | | | 0.678057 | 3 | 3 | 1 | 1 | | | 0.678057 | | | | | |
| | | | | | | 0.471934 | 3 | 3 | 1 | 3 | | | 0.471934 | | | | | |
| | | | | | | 2.774250 | 3 | 3 | 3 | 1 | | | | 2.187107 | | | | |
| | | | | | | 3.664430 | 3 | 3 | 3 | 3 | | | | | | | | |
| #55 | 11 | 3 | 1 | 1 | 1 | 11.000000 | 3 | 1 | 1 | 1 | | | 11.000000 | | | | 11.000000 | |
| #57 | 3 | 3 | 1 | 1 | 3 | 3.000000 | 3 | 1 | 1 | 3 | | 3.000000 | | | | 3.000000 | | |
| #61 | 10 | 3 | 1 | 3 | 1 | 10.000000 | 3 | 1 | 3 | 1 | | 10.000000 | | | | 10.000000 | | |
| #63 | 5 | 3 | 1 | 3 | 3 | 5.000000 | 3 | 1 | 3 | 3 | | 5.000000 | | | | 5.000000 | | |
| #73 | 13 | 3 | 3 | 1 | 1 | 13.000000 | 3 | 3 | 1 | 1 | | | 13.000000 | | | | 13.000000 | |
| #75 | 14 | 3 | 3 | 1 | 3 | 14.000000 | 3 | 3 | 1 | 3 | | | 14.000000 | | | | 14.000000 | |
| #79 | 84 | 3 | 3 | 3 | 1 | 84.000000 | 3 | 3 | 3 | 1 | | | | 66.222222 | | | | 66.222222 |
| #81 | 126 | 3 | 3 | 3 | 3 | 126.000000 | 3 | 3 | 3 | 3 | | | | | | | | |
| | 460 | | | | | 460.000000 | | | | | 77.389937 | 77.389937 | 77.389937 | 77.389937 | 77.389937 | 77.389937 | 77.389937 | 77.389937 |

Table 8.2.2.1 (part 2 of 2): Voting patterns and allocation of votes in example A53

### 8.2.3. Applying the Schulze Tie-Breaking Method

Table 8.2.3.1 lists the links in example A53.

| | | | | | | | | | | |
|---|---|---|---|---|---|---|---|---|---|---|
| 11 | *a* | *b* | *c* | *e* | *j* | 77.389937 | 99.563758 | **107.281879** | 101.107383 | 69.463087 |

For example, row 11 of table 8.2.3.1 contains the following information:

- $N[\{b,c,e,j\};a] = 77.389937$
- $N[\{a,c,e,j\};b] = 99.563758$
- $N[\{a,b,e,j\};c] = 107.281879$
- $N[\{a,b,c,j\};e] = 101.107383$
- $N[\{a,b,c,e\};j] = 69.463087$

- The link $\{b, c, e, j\} \rightarrow \{a, c, e, j\}$ has a strength of $(N[\{b,c,e,j\};a], N[\{a,c,e,j\};b])$.
- The link $\{b, c, e, j\} \rightarrow \{a, b, e, j\}$ has a strength of $(N[\{b,c,e,j\};a], N[\{a,b,e,j\};c])$.
- The link $\{b, c, e, j\} \rightarrow \{a, b, c, j\}$ has a strength of $(N[\{b,c,e,j\};a], N[\{a,b,c,j\};e])$.
- The link $\{b, c, e, j\} \rightarrow \{a, b, c, e\}$ has a strength of $(N[\{b,c,e,j\};a], N[\{a,b,c,e\};j])$.

- The link $\{a, c, e, j\} \rightarrow \{b, c, e, j\}$ has a strength of $(N[\{a,c,e,j\};b], N[\{b,c,e,j\};a])$.
- The link $\{a, c, e, j\} \rightarrow \{a, b, e, j\}$ has a strength of $(N[\{a,c,e,j\};b], N[\{a,b,e,j\};c])$.
- The link $\{a, c, e, j\} \rightarrow \{a, b, c, j\}$ has a strength of $(N[\{a,c,e,j\};b], N[\{a,b,c,j\};e])$.
- The link $\{a, c, e, j\} \rightarrow \{a, b, c, e\}$ has a strength of $(N[\{a,c,e,j\};b], N[\{a,b,c,e\};j])$.

- The link $\{a, b, e, j\} \rightarrow \{b, c, e, j\}$ has a strength of $(N[\{a,b,e,j\};c], N[\{b,c,e,j\};a])$.
- The link $\{a, b, e, j\} \rightarrow \{a, c, e, j\}$ has a strength of $(N[\{a,b,e,j\};c], N[\{a,c,e,j\};b])$.
- The link $\{a, b, e, j\} \rightarrow \{a, b, c, j\}$ has a strength of $(N[\{a,b,e,j\};c], N[\{a,b,c,j\};e])$.
- The link $\{a, b, e, j\} \rightarrow \{a, b, c, e\}$ has a strength of $(N[\{a,b,e,j\};c], N[\{a,b,c,e\};j])$.

- The link $\{a, b, c, j\} \rightarrow \{b, c, e, j\}$ has a strength of $(N[\{a,b,c,j\};e], N[\{b,c,e,j\};a])$.
- The link $\{a, b, c, j\} \rightarrow \{a, c, e, j\}$ has a strength of $(N[\{a,b,c,j\};e], N[\{a,c,e,j\};b])$.
- The link $\{a, b, c, j\} \rightarrow \{a, b, e, j\}$ has a strength of $(N[\{a,b,c,j\};e], N[\{a,b,e,j\};c])$.
- The link $\{a, b, c, j\} \rightarrow \{a, b, c, e\}$ has a strength of $(N[\{a,b,c,j\};e], N[\{a,b,c,e\};j])$.

- The link $\{a, b, c, e\} \rightarrow \{b, c, e, j\}$ has a strength of $(N[\{a,b,c,e\};j], N[\{b,c,e,j\};a])$.
- The link $\{a, b, c, e\} \rightarrow \{a, c, e, j\}$ has a strength of $(N[\{a,b,c,e\};j], N[\{a,c,e,j\};b])$.
- The link $\{a, b, c, e\} \rightarrow \{a, b, e, j\}$ has a strength of $(N[\{a,b,c,e\};j], N[\{a,b,e,j\};c])$.
- The link $\{a, b, c, e\} \rightarrow \{a, b, c, j\}$ has a strength of $(N[\{a,b,c,e\};j], N[\{a,b,c,j\};e])$.

So, row 11 of table 8.2.3.1 represents the following links:

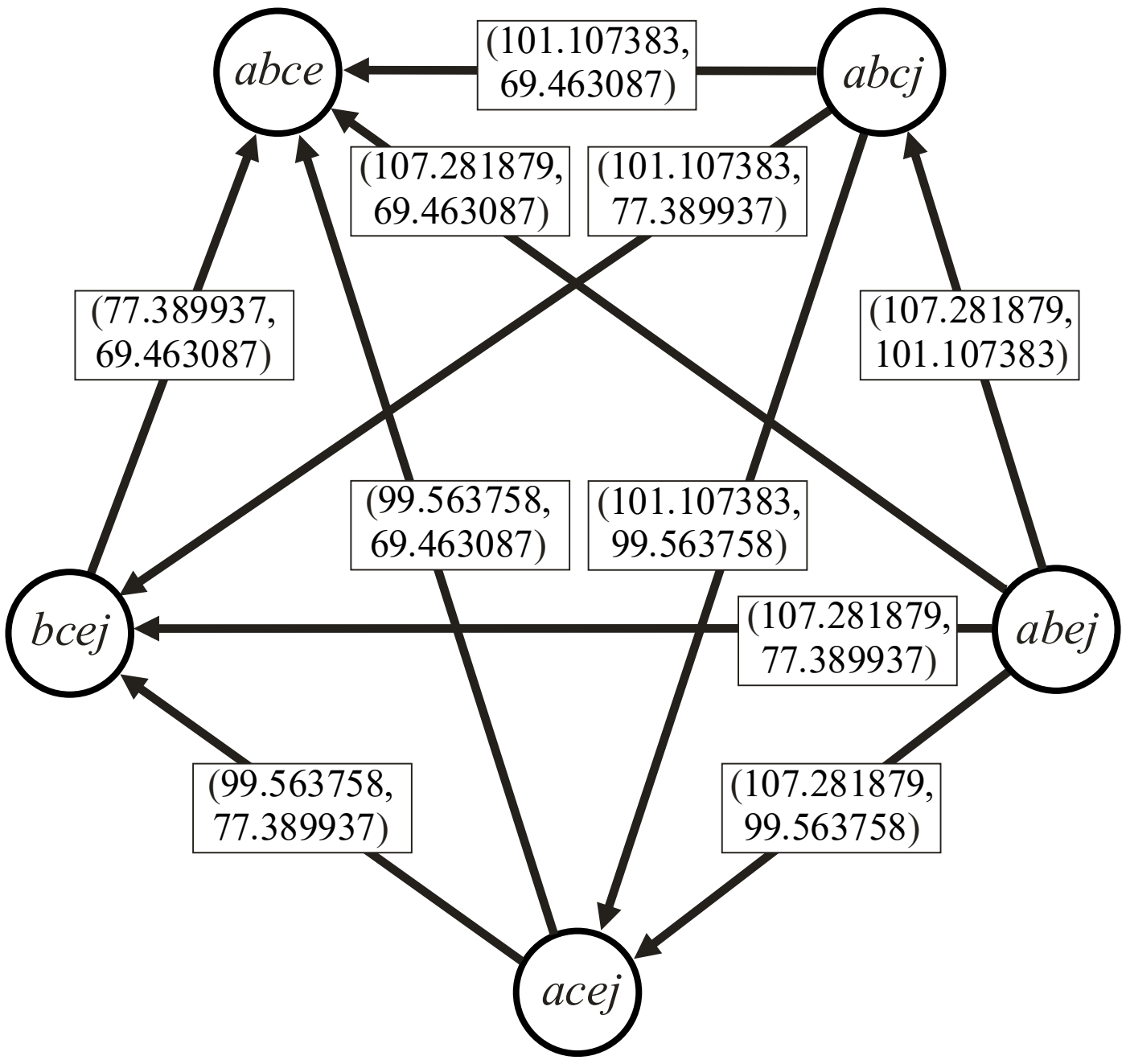

When we apply the Schulze tie-breaker, as defined at stage 3 of section 8.1.3, to the links of table 8.2.3.1 with $\succ_{margin}$ for $\succ_{D2}$, we get $\{a, d, g, j\}$ as winning set.

For example, we have:

- Line 33: The link $\{a, b, g, j\} \rightarrow \{a, d, g, j\}$ has a strength of $(N[\{a,b,g,j\};d], N[\{a,d,g,j\};b])$ = ( 101.411379, 102.166302 ).
- Line 33: The link $\{a, d, g, j\} \rightarrow \{a, b, g, j\}$ has a strength of $(N[\{a,d,g,j\};b], N[\{a,b,g,j\};d])$ = ( 102.166302, 101.411379 ).
- Line 49: The link $\{a, b, g, j\} \rightarrow \{a, f, g, j\}$ has a strength of $(N[\{a,b,g,j\};f], N[\{a,f,g,j\};b])$ = ( 101.068282, 102.334802 ).
- Line 49: The link $\{a, f, g, j\} \rightarrow \{a, b, g, j\}$ has a strength of $(N[\{a,f,g,j\};b], N[\{a,b,g,j\};f])$ = ( 102.334802, 101.068282 ).
- Line 104: The link $\{a, d, g, j\} \rightarrow \{a, f, g, j\}$ has a strength of $(N[\{a,d,g,j\};f], N[\{a,f,g,j\};d])$ = ( 101.351648, 101.098901 ).
- Line 104: The link $\{a, f, g, j\} \rightarrow \{a, d, g, j\}$ has a strength of $(N[\{a,f,g,j\};d], N[\{a,d,g,j\};f])$ = ( 101.098901, 101.351648 ).

So $\{a, d, g, j\}$ beats $\{a, b, g, j\}$ in the direct comparison, $\{a, f, g, j\}$ beats $\{a, b, g, j\}$ in the direct comparison, and $\{a, d, g, j\}$ beats $\{a, f, g, j\}$ in the direct comparison.

When there are $C$ alternatives, then there are $(C!)/(((M+1)!)\cdot((C-M-1)!))$ possible $(M+1)$-way contests. For $C = 10$ and $M = 4$, we get 252 possible 5-way contests. Table 8.2.3.1 lists these 252 possible 5-way contests for example A53.

When Schulze STV is used to choose $M$ from $(M+1)$ alternatives $\{a_1,...,a_{(M+1)}\}$, then that alternative $k \in \{1,...,(M+1)\}$ is eliminated for which $N[(\{a_1,...,a_{(M+1)}\}\backslash\{a_k\});a_k]$ is the maximum, while the other $M$ alternatives are elected. In table 8.2.3.1, the maximum $N[(\{a_1,...,a_{(M+1)}\}\backslash\{a_k\});a_k]$ of each 5-way contest is **fat and underlined**.

Suppose the maximum $N[(\{a_1,...,a_{(M+1)}\}\backslash\{a_k\});a_k]$ of a $(M+1)$-way contest is not unique. Suppose $1 < m \leq (M+1)$ entries are tied for maximum $N[(\{a_1,...,a_{(M+1)}\}\backslash\{a_k\});a_k]$, then the $m$ alternatives with maximum $N[(\{a_1,...,a_{(M+1)}\}\backslash\{a_k\});a_k]$ are tied for winning one of the remaining $(m-1)$ seats, while the other $(M+1-m)$ alternatives are elected. In table 8.2.3.1 for those 5-way contests, where the maximum $N[(\{a_1,...,a_{(M+1)}\}\backslash\{a_k\});a_k]$ is not unique, those $N[(\{a_1,...,a_{(M+1)}\}\backslash\{a_k\});a_k]$, that are tied for maximum $N[(\{a_1,...,a_{(M+1)}\}\backslash\{a_k\});a_k]$, are *italic and underlined* (only lines 27, 149, and 155).

In table 8.2.3.1, we see:

- Alternatives *a*, *g*, and *j* each win in every 5-way contest.
- Alternative *d* is tied for winning in one 5-way contest (line 27) and wins in every other 5-way contest.
- Alternative *f* loses in one 5-way contest (line 104) and wins in every other 5-way contest.
- Alternative *b* wins in 121 5-way contests, is tied for winning in one 5-way contest (line 27), and loses in four 5-way contests (lines 30, 33, 49, and 174).
- Alternative *e* wins 111 times and loses 15 times.
- Alternative *h* wins 59 times and loses 67 times.
- Alternative *c* wins 45 times, is tied twice (lines 149 and 155), and loses 79 times.
- Alternative *i* wins 41 times, is tied twice (lines 149 and 155), and loses 83 times.

| | *k* | *l* | *m* | *n* | *o* | *N*[{*l*,*m*,*n*,*o*};*k*] | *N*[{*k*,*m*,*n*,*o*};*l*] | *N*[{*k*,*l*,*n*,*o*};*m*] | *N*[{*k*,*l*,*m*,*o*};*n*] | *N*[{*k*,*l*,*m*,*n*};*o*] |
|---|---|---|---|---|---|---|---|---|---|---|
| 1 | *a* | *b* | *c* | *d* | *e* | 69.311512 | 97.347630 | **104.356659** | 91.117381 | 97.866817 |
| 2 | *a* | *b* | *c* | *d* | *f* | 72.494331 | 97.267574 | **106.394558** | 91.791383 | 92.052154 |
| 3 | *a* | *b* | *c* | *d* | *g* | 74.292035 | 97.699115 | **105.077434** | 97.444690 | 85.486726 |
| 4 | *a* | *b* | *c* | *d* | *h* | 69.482146 | 95.615034 | **103.473804** | 90.375854 | 101.053161 |
| 5 | *a* | *b* | *c* | *d* | *i* | 68.329596 | 95.403587 | **103.912556** | 91.535874 | 100.818386 |
| 6 | *a* | *b* | *c* | *d* | *j* | 83.765432 | 100.720621 | **106.330377** | 96.895787 | 70.631929 |
| 7 | *a* | *b* | *c* | *e* | *f* | 68.050459 | 96.800459 | **106.559633** | 97.327982 | 91.261468 |
| 8 | *a* | *b* | *c* | *e* | *g* | 71.971047 | 98.864143 | **106.035635** | 98.608018 | 84.521158 |
| 9 | *a* | *b* | *c* | *e* | *h* | 65.248069 | 95.126728 | **104.665899** | 95.391705 | 99.567599 |
| 10 | *a* | *b* | *c* | *e* | *i* | 63.064516 | 95.126728 | **104.400922** | 96.186636 | 101.221198 |
| 11 | *a* | *b* | *c* | *e* | *j* | 77.389937 | 99.563758 | **107.281879** | 101.107383 | 69.463087 |
| 12 | *a* | *b* | *c* | *f* | *g* | 73.393258 | 98.202247 | **107.505618** | 95.101124 | 85.797753 |
| 13 | *a* | *b* | *c* | *f* | *h* | 68.320236 | 95.877598 | **105.704388** | 88.972286 | 101.125492 |
| 14 | *a* | *b* | *c* | *f* | *i* | 65.979263 | 94.596774 | **106.255760** | 92.741935 | 100.426267 |
| 15 | *a* | *b* | *c* | *f* | *j* | 82.285264 | 100.495495 | **107.229730** | 97.646396 | 72.004505 |
| 16 | *a* | *b* | *c* | *g* | *h* | 72.748673 | 96.828442 | **106.173815** | 81.252822 | 102.996248 |
| 17 | *a* | *b* | *c* | *g* | *i* | 70.450450 | 96.869369 | **105.675676** | 83.141892 | 103.862613 |
| 18 | *a* | *b* | *c* | *g* | *j* | 86.629956 | 102.334802 | **108.667401** | 88.403084 | 73.964758 |
| 19 | *a* | *b* | *c* | *h* | *i* | 63.805224 | 93.221709 | **103.845266** | 99.797547 | 99.330254 |
| 20 | *a* | *b* | *c* | *h* | *j* | 76.937668 | 98.977528 | 105.438202 | **108.022472** | 67.449438 |
| 21 | *a* | *b* | *c* | *i* | *j* | 75.764706 | 99.529148 | **106.233184** | 105.201794 | 67.719298 |
| 22 | *a* | *b* | *d* | *e* | *f* | 74.020045 | 97.839644 | 92.973274 | **100.913140** | 94.253898 |
| 23 | *a* | *b* | *d* | *e* | *g* | 75.571429 | 99.329670 | 97.813187 | **100.846154** | 86.439560 |
| 24 | *a* | *b* | *d* | *e* | *h* | 70.771762 | 97.646396 | 91.430180 | 98.423423 | **101.728238** |
| 25 | *a* | *b* | *d* | *e* | *i* | 69.205817 | 96.733781 | 92.360179 | 99.049217 | **102.651007** |
| 26 | *a* | *b* | *d* | *e* | *j* | 86.821192 | 100.529801 | 97.483444 | **102.814570** | 72.350993 |
| 27 | *a* | *b* | *d* | *f* | *g* | 77.090708 | *98.716814* | *98.716814* | 97.444690 | 88.030973 |
| 28 | *a* | *b* | *d* | *f* | *h* | 74.397888 | 98.164414 | 91.948198 | 92.725225 | **102.764274** |
| 29 | *a* | *b* | *d* | *f* | *i* | 72.322222 | 96.600000 | 93.277778 | 95.833333 | **101.966667** |
| 30 | *a* | *b* | *d* | *f* | *j* | 87.716186 | **100.975610** | 96.895787 | 99.190687 | 75.221729 |
| 31 | *a* | *b* | *d* | *g* | *h* | 76.388633 | 98.462389 | 96.681416 | 83.960177 | **104.507385** |
| 32 | *a* | *b* | *d* | *g* | *i* | 73.946785 | 97.660754 | 97.915743 | 85.421286 | **105.055432** |
| 33 | *a* | *b* | *d* | *g* | *j* | 89.332604 | **102.166302** | 101.411379 | 90.842451 | 76.247265 |
| 34 | *a* | *b* | *d* | *h* | *i* | 69.217708 | 96.092342 | 91.430180 | 101.469229 | **101.790541** |
| 35 | *a* | *b* | *d* | *h* | *j* | 84.333333 | 100.433333 | 95.577778 | **108.100000** | 71.555556 |
| 36 | *a* | *b* | *d* | *i* | *j* | 84.176158 | 100.243363 | 96.935841 | **106.095133** | 72.256637 |
| 37 | *a* | *b* | *e* | *f* | *g* | 75.055310 | 99.734513 | **100.243363** | 97.444690 | 87.522124 |
| 38 | *a* | *b* | *e* | *f* | *h* | 70.311453 | 97.307692 | 98.088235 | 92.104072 | **102.188547** |
| 39 | *a* | *b* | *e* | *f* | *i* | 67.847380 | 95.876993 | 97.972665 | 95.091116 | **103.211845** |
| 40 | *a* | *b* | *e* | *f* | *j* | 84.966518 | 99.598214 | **101.651786** | 98.828125 | 74.955357 |
| 41 | *a* | *b* | *e* | *g* | *h* | 74.337778 | 99.120267 | 99.632517 | 82.984410 | **103.925029** |
| 42 | *a* | *b* | *e* | *g* | *i* | 72.131696 | 98.828125 | 99.598214 | 84.196429 | **105.245536** |
| 43 | *a* | *b* | *e* | *g* | *j* | 88.208791 | 101.351648 | **104.131868** | 90.230769 | 76.076923 |
| 44 | *a* | *b* | *e* | *h* | *i* | 64.914754 | 95.308219 | 96.358447 | 101.021319 | **102.397260** |
| 45 | *a* | *b* | *e* | *h* | *j* | 81.744689 | 98.828125 | 100.111607 | **107.555804** | 71.104911 |
| 46 | *a* | *b* | *e* | *i* | *j* | 78.449612 | 99.306488 | 100.850112 | **107.281879** | 70.492170 |
| 47 | *a* | *b* | *f* | *g* | *h* | 76.384893 | 99.529148 | 94.630045 | 84.831839 | **104.624076** |
| 48 | *a* | *b* | *f* | *g* | *i* | 73.671875 | 98.058036 | 97.287946 | 85.993304 | **104.988839** |
| 49 | *a* | *b* | *f* | *g* | *j* | 87.643172 | **102.334802** | 101.068282 | 90.429515 | 78.524229 |
| 50 | *a* | *b* | *f* | *h* | *i* | 68.484353 | 95.000000 | 92.105263 | **102.305121** | 102.105263 |

Table 8.2.3.1 (part 1 of 5): links in example A53

| | *k* | *l* | *m* | *n* | *o* | *N*[{*l*,*m*,*n*,*o*};*k*] | *N*[{*k*,*m*,*n*,*o*};*l*] | *N*[{*k*,*l*,*n*,*o*};*m*] | *N*[{*k*,*l*,*m*,*o*};*n*] | *N*[{*k*,*l*,*m*,*n*};*o*] |
|---|---|---|---|---|---|---|---|---|---|---|
| 51 | *a* | *b* | *f* | *h* | *j* | 82.438202 | 99.752809 | 95.876404 | **109.056180** | 72.876404 |
| 52 | *a* | *b* | *f* | *i* | *j* | 82.769058 | 99.529148 | 97.724215 | **106.748879** | 73.228700 |
| 53 | *a* | *b* | *g* | *h* | *i* | 73.008267 | 97.347630 | 81.512415 | 103.255842 | **104.875847** |
| 54 | *a* | *b* | *g* | *h* | *j* | 86.441242 | 101.230599 | 86.951220 | **110.410200** | 74.966741 |
| 55 | *a* | *b* | *g* | *i* | *j* | 86.377778 | 101.966667 | 88.677778 | **109.122222** | 73.855556 |
| 56 | *a* | *b* | *h* | *i* | *j* | 78.376979 | 98.608597 | **107.454751** | 104.852941 | 68.028169 |
| 57 | *a* | *c* | *d* | *e* | *f* | 72.781532 | **106.452703** | 91.430180 | 97.128378 | 92.207207 |
| 58 | *a* | *c* | *d* | *e* | *g* | 74.635762 | **104.845475** | 96.975717 | 97.737307 | 85.805740 |
| 59 | *a* | *c* | *d* | *e* | *h* | 69.115667 | **104.235160** | 90.582192 | 93.995434 | 102.071548 |
| 60 | *a* | *c* | *d* | *e* | *i* | 67.753950 | **103.837472** | 91.117381 | 95.011287 | 102.279910 |
| 61 | *a* | *c* | *d* | *e* | *j* | 84.830247 | **107.095344** | 97.405765 | 100.465632 | 69.866962 |
| 62 | *a* | *c* | *d* | *f* | *g* | 74.698661 | **107.299107** | 97.544643 | 94.720982 | 85.736607 |
| 63 | *a* | *c* | *d* | *f* | *h* | 71.415141 | **105.329545** | 90.693182 | 91.215909 | 101.346222 |
| 64 | *a* | *c* | *d* | *f* | *i* | 69.728507 | **105.893665** | 90.542986 | 93.404977 | 100.429864 |
| 65 | *a* | *c* | *d* | *f* | *j* | 86.314607 | **106.988764** | 96.134831 | 98.460674 | 72.101124 |
| 66 | *a* | *c* | *d* | *g* | *h* | 73.476924 | **104.988839** | 95.747768 | 81.629464 | 104.157004 |
| 67 | *a* | *c* | *d* | *g* | *i* | 71.361607 | **104.475446** | 96.517857 | 83.939732 | 103.705357 |
| 68 | *a* | *c* | *d* | *g* | *j* | 87.389868 | **108.414097** | 101.574890 | 88.909692 | 73.711454 |
| 69 | *a* | *c* | *d* | *h* | *i* | 66.115932 | **102.894737** | 90.000000 | 100.463016 | 100.526316 |
| 70 | *a* | *c* | *d* | *h* | *j* | 81.284987 | 105.245536 | 96.004464 | **107.555804** | 67.767857 |
| 71 | *a* | *c* | *d* | *i* | *j* | 80.402166 | **105.995526** | 96.219239 | 105.480984 | 69.205817 |
| 72 | *a* | *c* | *e* | *f* | *g* | 72.833333 | **107.588889** | 97.877778 | 95.066667 | 86.633333 |
| 73 | *a* | *c* | *e* | *f* | *h* | 67.058858 | **106.295872** | 94.690367 | 90.206422 | 101.748481 |
| 74 | *a* | *c* | *e* | *f* | *i* | 64.303944 | **105.928074** | 94.454756 | 92.053364 | 103.259861 |
| 75 | *a* | *c* | *e* | *f* | *j* | 81.705790 | **107.247191** | 99.752809 | 97.943820 | 73.134831 |
| 76 | *a* | *c* | *e* | *g* | *h* | 71.904859 | **106.471910** | 96.134831 | 81.921348 | 103.567051 |
| 77 | *a* | *c* | *e* | *g* | *i* | 69.775281 | **105.438202** | 96.393258 | 83.730337 | 104.662921 |
| 78 | *a* | *c* | *e* | *g* | *j* | 85.428571 | **108.934066** | 102.615385 | 88.967033 | 74.054945 |
| 79 | *a* | *c* | *e* | *h* | *i* | 61.699912 | **104.060325** | 92.053364 | 100.794287 | 101.392111 |
| 80 | *a* | *c* | *e* | *h* | *j* | 76.021251 | 106.510067 | 98.020134 | **107.796421** | 66.018519 |
| 81 | *a* | *c* | *e* | *i* | *j* | 72.631579 | 106.693002 | 98.905192 | **106.952596** | 65.612403 |
| 82 | *a* | *c* | *f* | *g* | *h* | 72.163286 | **107.764045** | 93.292135 | 82.696629 | 104.083905 |
| 83 | *a* | *c* | *f* | *g* | *i* | 69.414414 | **107.747748** | 94.538288 | 84.177928 | 104.121622 |
| 84 | *a* | *c* | *f* | *g* | *j* | 84.911308 | **109.135255** | 100.465632 | 88.481153 | 77.006652 |
| 85 | *a* | *c* | *f* | *h* | *i* | 62.378284 | **105.372093** | 90.662791 | 100.761251 | 100.825581 |
| 86 | *a* | *c* | *f* | *h* | *j* | 77.297595 | 106.433409 | 96.049661 | **107.731377** | 70.349887 |
| 87 | *a* | *c* | *f* | *i* | *j* | 75.027712 | **106.394558** | 97.267574 | 106.133787 | 70.147392 |
| 88 | *a* | *c* | *g* | *h* | *i* | 68.278778 | **105.852273** | 79.454545 | 102.914404 | 103.500000 |
| 89 | *a* | *c* | *g* | *h* | *j* | 84.077778 | 108.611111 | 84.844444 | **109.888889** | 72.577778 |
| 90 | *a* | *c* | *g* | *i* | *j* | 82.672811 | **108.866667** | 86.888889 | 108.355556 | 72.066667 |
| 91 | *a* | *c* | *h* | *i* | *j* | 69.181244 | 105.351474 | **106.655329** | 104.308390 | 63.411215 |
| 92 | *a* | *d* | *e* | *f* | *g* | 77.087912 | 98.318681 | **99.329670** | 97.054945 | 88.208791 |
| 93 | *a* | *d* | *e* | *f* | *h* | 73.290722 | 92.567265 | 97.724215 | 92.567265 | **103.850533** |
| 94 | *a* | *d* | *e* | *f* | *i* | 71.521253 | 92.617450 | 96.733781 | 94.932886 | **104.194631** |
| 95 | *a* | *d* | *e* | *f* | *j* | 87.144444 | 97.366667 | **101.200000** | 99.155556 | 75.133333 |
| 96 | *a* | *d* | *e* | *g* | *h* | 75.618401 | 96.559020 | 98.351893 | 84.008909 | **105.461777** |
| 97 | *a* | *d* | *e* | *g* | *i* | 73.691796 | 97.915743 | 97.405765 | 85.166297 | **105.820399** |
| 98 | *a* | *d* | *e* | *g* | *j* | 88.829322 | 101.663020 | **102.921225** | 91.345733 | 75.240700 |
| 99 | *a* | *d* | *e* | *h* | *i* | 67.494687 | 91.477273 | 95.136364 | 102.653041 | **103.238636** |
| 100 | *a* | *d* | *e* | *h* | *j* | 84.656319 | 96.640798 | 99.700665 | **108.370288** | 70.631929 |

Table 8.2.3.1 (part 2 of 5): links in example A53

| | *k* | *l* | *m* | *n* | *o* | *N*[{*l*,*m*,*n*,*o*};*k*] | *N*[{*k*,*m*,*n*,*o*};*l*] | *N*[{*k*,*l*,*n*,*o*};*m*] | *N*[{*k*,*l*,*m*,*o*};*n*] | *N*[{*k*,*l*,*m*,*n*};*o*] |
|---|---|---|---|---|---|---|---|---|---|---|
| 101 | *a* | *d* | *e* | *i* | *j* | 83.348624 | 97.366667 | 100.688889 | **107.333333** | 70.533333 |
| 102 | *a* | *d* | *f* | *g* | *h* | 76.983694 | 97.111111 | 95.577778 | 85.100000 | **105.227417** |
| 103 | *a* | *d* | *f* | *g* | *i* | 74.366667 | 97.622222 | 96.600000 | 86.377778 | **105.033333** |
| 104 | *a* | *d* | *f* | *g* | *j* | 88.714286 | 101.098901 | **101.351648** | 90.736264 | 78.098901 |
| 105 | *a* | *d* | *f* | *h* | *i* | 70.051272 | 91.323529 | 93.665158 | 102.188547 | **102.771493** |
| 106 | *a* | *d* | *f* | *h* | *j* | 84.966518 | 95.491071 | 97.544643 | **108.325893** | 73.671875 |
| 107 | *a* | *d* | *f* | *i* | *j* | 85.223214 | 96.261161 | 98.571429 | **106.785714** | 73.158482 |
| 108 | *a* | *d* | *g* | *h* | *i* | 73.032875 | 96.177130 | 81.737668 | 104.108381 | **104.943946** |
| 109 | *a* | *d* | *g* | *h* | *j* | 87.197802 | 100.087912 | 87.956044 | **110.197802** | 74.560440 |
| 110 | *a* | *d* | *g* | *i* | *j* | 86.821192 | 101.291391 | 89.359823 | **108.907285** | 73.620309 |
| 111 | *a* | *d* | *h* | *i* | *j* | 80.164441 | 95.661435 | **106.748879** | 105.717489 | 69.876682 |
| 112 | *a* | *e* | *f* | *g* | *h* | 74.683694 | 98.900000 | 95.322222 | 85.611111 | **105.482973** |
| 113 | *a* | *e* | *f* | *g* | *i* | 72.131696 | 98.058036 | 97.031250 | 86.506696 | **106.272321** |
| 114 | *a* | *e* | *f* | *g* | *j* | 86.123348 | **102.841410** | 101.321586 | 91.189427 | 78.524229 |
| 115 | *a* | *e* | *f* | *h* | *i* | 65.476289 | 94.690367 | 92.052752 | 103.331050 | **104.449541** |
| 116 | *a* | *e* | *f* | *h* | *j* | 81.372768 | 98.828125 | 96.774554 | **109.095982** | 73.928571 |
| 117 | *a* | *e* | *f* | *i* | *j* | 80.965732 | 99.752809 | 97.943820 | **108.539326** | 72.617978 |
| 118 | *a* | *e* | *g* | *h* | *i* | 71.191111 | 97.088036 | 82.031603 | 104.553810 | **105.135440** |
| 119 | *a* | *e* | *g* | *h* | *j* | 85.486726 | 101.769912 | 88.030973 | **110.674779** | 74.037611 |
| 120 | *a* | *e* | *g* | *i* | *j* | 84.723451 | 102.533186 | 89.811947 | **109.402655** | 73.528761 |
| 121 | *a* | *e* | *h* | *i* | *j* | 74.210623 | 97.646396 | **107.747748** | 107.229730 | 66.179245 |
| 122 | *a* | *f* | *g* | *h* | *i* | 71.646432 | 94.842697 | 83.471910 | 104.600759 | **105.438202** |
| 123 | *a* | *f* | *g* | *h* | *j* | 84.656319 | 99.190687 | 87.461197 | **110.665188** | 78.026608 |
| 124 | *a* | *f* | *g* | *i* | *j* | 84.911308 | 99.955654 | 89.246120 | **109.390244** | 76.496674 |
| 125 | *a* | *f* | *h* | *i* | *j* | 77.294626 | 95.486425 | **107.975113** | 107.194570 | 70.509050 |
| 126 | *a* | *g* | *h* | *i* | *j* | 83.612975 | 85.671141 | **109.597315** | 108.568233 | 72.550336 |
| 127 | *b* | *c* | *d* | *e* | *f* | 89.066059 | **101.116173** | 86.708428 | 97.710706 | 85.398633 |
| 128 | *b* | *c* | *d* | *e* | *g* | 90.135135 | **101.272523** | 90.394144 | 97.387387 | 80.810811 |
| 129 | *b* | *c* | *d* | *e* | *h* | 88.255814 | **97.616279** | 82.906977 | 94.941860 | 96.279070 |
| 130 | *b* | *c* | *d* | *e* | *i* | 85.845070 | 97.453052 | 82.605634 | 95.833333 | **98.262911** |
| 131 | *b* | *c* | *d* | *e* | *j* | 97.877778 | **103.755556** | 92.255556 | 101.711111 | 64.400000 |
| 132 | *b* | *c* | *d* | *f* | *g* | 89.587054 | **103.705357** | 93.180804 | 90.100446 | 83.426339 |
| 133 | *b* | *c* | *d* | *f* | *h* | 90.057078 | **99.771689** | 85.593607 | 86.381279 | 98.196347 |
| 134 | *b* | *c* | *d* | *f* | *i* | 87.494305 | **100.068337** | 85.922551 | 87.861509 | 98.653298 |
| 135 | *b* | *c* | *d* | *f* | *j* | 98.864143 | **104.755011** | 93.229399 | 95.278396 | 67.873051 |
| 136 | *b* | *c* | *d* | *g* | *h* | 90.613839 | **101.138393** | 90.613839 | 78.292411 | 99.341518 |
| 137 | *b* | *c* | *d* | *g* | *i* | 88.542141 | **101.116173** | 90.113895 | 79.897494 | 100.330296 |
| 138 | *b* | *c* | *d* | *g* | *j* | 99.432314 | **105.207424** | 97.674672 | 88.133188 | 69.552402 |
| 139 | *b* | *c* | *d* | *h* | *i* | 86.516204 | 97.164352 | 82.256944 | 96.099537 | **97.962963** |
| 140 | *b* | *c* | *d* | *h* | *j* | 97.150776 | 101.995565 | 90.776053 | **106.330377** | 63.747228 |
| 141 | *b* | *c* | *d* | *i* | *j* | 97.190265 | 103.042035 | 91.084071 | **105.586283** | 63.097345 |
| 142 | *b* | *c* | *e* | *f* | *g* | 90.598194 | **102.279910** | 95.530474 | 88.521445 | 83.069977 |
| 143 | *b* | *c* | *e* | *f* | *h* | 88.211765 | **98.764706** | 92.000000 | 83.611765 | 97.411765 |
| 144 | *b* | *c* | *e* | *f* | *i* | 85.771971 | 98.337292 | 92.327791 | 84.242761 | **99.320184** |
| 145 | *b* | *c* | *e* | *f* | *j* | 96.828442 | **104.356659** | 99.943567 | 93.713318 | 65.158014 |
| 146 | *b* | *c* | *e* | *g* | *h* | 91.052632 | **101.052632** | 92.894737 | 77.105263 | 97.894737 |
| 147 | *b* | *c* | *e* | *g* | *i* | 88.790698 | 100.290698 | 92.802326 | 77.558140 | **100.558140** |
| 148 | *b* | *c* | *e* | *g* | *j* | 99.329670 | **105.648352** | 101.857143 | 86.692308 | 66.472527 |
| 149 | *b* | *c* | *e* | *h* | *i* | 84.166667 | *96.111111* | 88.888889 | 94.722222 | *96.111111* |
| 150 | *b* | *c* | *e* | *h* | *j* | 95.011287 | 102.799097 | 97.866817 | **105.135440** | 59.187359 |

Table 8.2.3.1 (part 3 of 5): links in example A53

| | *k* | *l* | *m* | *n* | *o* | *N*[{*l*,*m*,*n*,*o*};*k*] | *N*[{*k*,*m*,*n*,*o*};*l*] | *N*[{*k*,*l*,*n*,*o*};*m*] | *N*[{*k*,*l*,*m*,*o*};*n*] | *N*[{*k*,*l*,*m*,*n*};*o*] |
|---|---|---|---|---|---|---|---|---|---|---|
| 151 | *b* | *c* | *e* | *i* | *j* | 95.226244 | 103.031674 | 98.608597 | **105.633484** | 57.500000 |
| 152 | *b* | *c* | *f* | *g* | *h* | 91.578947 | **101.842105** | 87.631579 | 79.210526 | 99.736842 |
| 153 | *b* | *c* | *f* | *g* | *i* | 88.947368 | **101.842105** | 88.526779 | 80.263158 | 100.420590 |
| 154 | *b* | *c* | *f* | *g* | *j* | 99.835165 | **105.901099** | 96.549451 | 88.461538 | 69.252747 |
| 155 | *b* | *c* | *f* | *h* | *i* | 85.357995 | *97.159905* | 83.711217 | 96.610979 | *97.159905* |
| 156 | *b* | *c* | *f* | *h* | *j* | 96.049661 | 102.539503 | 91.117381 | **106.693002** | 63.600451 |
| 157 | *b* | *c* | *f* | *i* | *j* | 96.568849 | 103.058691 | 92.934537 | **104.875847** | 62.562077 |
| 158 | *b* | *c* | *g* | *h* | *i* | 87.909931 | **100.392610** | 74.896074 | 97.205543 | 99.595843 |
| 159 | *b* | *c* | *g* | *h* | *j* | 98.425721 | 104.290466 | 83.636364 | **107.860310** | 65.787140 |
| 160 | *b* | *c* | *g* | *i* | *j* | 99.006623 | 105.099338 | 85.044150 | **106.876380** | 63.973510 |
| 161 | *b* | *c* | *h* | *i* | *j* | 94.659864 | 100.918367 | **105.351474** | 102.743764 | 56.326531 |
| 162 | *b* | *d* | *e* | *f* | *g* | 91.338496 | 93.119469 | **98.971239** | 92.101770 | 84.469027 |
| 163 | *b* | *d* | *e* | *f* | *h* | 90.170455 | 85.727273 | 97.488636 | 87.818182 | **98.795455** |
| 164 | *b* | *d* | *e* | *f* | *i* | 87.879819 | 85.793651 | 96.746032 | 89.288441 | **100.292058** |
| 165 | *b* | *d* | *e* | *f* | *j* | 97.877778 | 93.533333 | **102.477778** | 96.088889 | 70.022222 |
| 166 | *b* | *d* | *e* | *g* | *h* | 92.258427 | 90.191011 | 97.943820 | 79.078652 | **100.528090** |
| 167 | *b* | *d* | *e* | *g* | *i* | 89.909091 | 90.431818 | 97.488636 | 79.715909 | **102.454545** |
| 168 | *b* | *d* | *e* | *g* | *j* | 99.146608 | 97.636761 | **102.921225** | 88.326039 | 71.969365 |
| 169 | *b* | *d* | *e* | *h* | *i* | 86.918605 | 82.104651 | 95.209302 | 96.279070 | **99.488372** |
| 170 | *b* | *d* | *e* | *h* | *j* | 97.111111 | 91.233333 | 100.688889 | **105.288889** | 65.677778 |
| 171 | *b* | *d* | *e* | *i* | *j* | 96.517857 | 91.127232 | 101.395089 | **107.299107** | 63.660714 |
| 172 | *b* | *d* | *f* | *g* | *h* | 93.180804 | 91.897321 | 91.897321 | 81.629464 | **101.395089** |
| 173 | *b* | *d* | *f* | *g* | *i* | 89.843750 | 91.897321 | 93.027237 | 82.399554 | **102.832138** |
| 174 | *b* | *d* | *f* | *g* | *j* | **99.901532** | 97.888403 | 99.146608 | 89.080963 | 73.982495 |
| 175 | *b* | *d* | *f* | *h* | *i* | 88.401361 | 84.229025 | 89.705215 | 97.528345 | **100.136054** |
| 176 | *b* | *d* | *f* | *h* | *j* | 98.133333 | 91.744444 | 94.300000 | **106.311111** | 69.511111 |
| 177 | *b* | *d* | *f* | *i* | *j* | 97.150776 | 92.560976 | 95.620843 | **106.585366** | 68.082040 |
| 178 | *b* | *d* | *g* | *h* | *i* | 90.449438 | 89.932584 | 78.044944 | 99.235955 | **102.337079** |
| 179 | *b* | *d* | *g* | *h* | *j* | 99.076923 | 96.043956 | 85.428571 | **107.923077** | 71.527473 |
| 180 | *b* | *d* | *g* | *i* | *j* | 98.788546 | 96.762115 | 86.629956 | **108.414097** | 69.405286 |
| 181 | *b* | *d* | *h* | *i* | *j* | 96.004464 | 89.843750 | 104.732143 | **105.245536** | 64.174107 |
| 182 | *b* | *e* | *f* | *g* | *h* | 92.784091 | 95.920455 | 90.170455 | 79.977273 | **101.147727** |
| 183 | *b* | *e* | *f* | *g* | *i* | 90.113895 | 95.353075 | 90.743058 | 80.683371 | **103.106601** |
| 184 | *b* | *e* | *f* | *g* | *j* | 99.295154 | **102.081498** | 98.535242 | 88.403084 | 71.685022 |
| 185 | *b* | *e* | *f* | *h* | *i* | 85.910165 | 90.531915 | 85.638298 | 97.872340 | **100.047281** |
| 186 | *b* | *e* | *f* | *h* | *j* | 96.049661 | 98.386005 | 92.934537 | **105.914221** | 66.715576 |
| 187 | *b* | *e* | *f* | *i* | *j* | 95.659091 | 99.318182 | 93.568182 | **106.897727** | 64.556818 |
| 188 | *b* | *e* | *g* | *h* | *i* | 90.357143 | 93.536866 | 75.518433 | 99.101382 | **101.486175** |
| 189 | *b* | *e* | *g* | *h* | *j* | 98.425721 | 100.465632 | 84.401330 | **108.115299** | 68.592018 |
| 190 | *b* | *e* | *g* | *i* | *j* | 98.462389 | 101.006637 | 85.486726 | **108.639381** | 66.404867 |
| 191 | *b* | *e* | *h* | *i* | *j* | 93.995434 | 96.358447 | 104.235160 | **106.073059** | 59.337900 |
| 192 | *b* | *f* | *g* | *h* | *i* | 90.526316 | 90.000000 | 77.631579 | 100.000000 | **101.842105** |
| 193 | *b* | *f* | *g* | *h* | *j* | 99.155556 | 95.322222 | 85.611111 | **108.355556** | 71.555556 |
| 194 | *b* | *f* | *g* | *i* | *j* | 99.155556 | 96.855556 | 86.377778 | **107.844444** | 69.766667 |
| 195 | *b* | *f* | *h* | *i* | *j* | 95.000000 | 90.789474 | **105.789474** | 105.263158 | 63.157895 |
| 196 | *b* | *g* | *h* | *i* | *j* | 97.982063 | 82.253363 | **107.264574** | 106.748879 | 65.751121 |
| 197 | *c* | *d* | *e* | *f* | *g* | **102.193764** | 91.180401 | 96.559020 | 88.106904 | 81.959911 |
| 198 | *c* | *d* | *e* | *f* | *h* | **99.366359** | 85.322581 | 94.066820 | 83.997696 | 97.246544 |
| 199 | *c* | *d* | *e* | *f* | *i* | 97.696759 | 84.386574 | 93.437500 | 84.227320 | **100.251846** |
| 200 | *c* | *d* | *e* | *f* | *j* | **103.680089** | 93.903803 | 100.335570 | 94.932886 | 67.147651 |

Table 8.2.3.1 (part 4 of 5): links in example A53

| | *k* | *l* | *m* | *n* | *o* | *N*[{*l*,*m*,*n*,*o*};*k*] | *N*[{*k*,*m*,*n*,*o*};*l*] | *N*[{*k*,*l*,*n*,*o*};*m*] | *N*[{*k*,*l*,*m*,*o*};*n*] | *N*[{*k*,*l*,*m*,*n*};*o*] |
|---|---|---|---|---|---|---|---|---|---|---|
| 201 | *c* | *d* | *e* | *g* | *h* | **100.335570** | 88.501119 | 95.190157 | 77.953020 | 98.020134 |
| 202 | *c* | *d* | *e* | *g* | *i* | 98.684211 | 87.894737 | 94.210526 | 78.157895 | **101.052632** |
| 203 | *c* | *d* | *e* | *g* | *j* | **104.867841** | 97.268722 | 102.081498 | 87.643172 | 68.138767 |
| 204 | *c* | *d* | *e* | *h* | *i* | 95.293427 | 81.525822 | 91.514085 | 93.943662 | **97.723005** |
| 205 | *c* | *d* | *e* | *h* | *j* | 102.165179 | 91.640625 | 98.571429 | **105.758929** | 61.863839 |
| 206 | *c* | *d* | *e* | *i* | *j* | 102.393736 | 91.588367 | 99.563758 | **105.738255** | 60.715884 |
| 207 | *c* | *d* | *f* | *g* | *h* | **102.733333** | 90.977778 | 88.677778 | 79.222222 | 98.388889 |
| 208 | *c* | *d* | *f* | *g* | *i* | 100.981941 | 89.559819 | 88.106552 | 79.954853 | **101.396834** |
| 209 | *c* | *d* | *f* | *g* | *j* | **105.142857** | 98.318681 | 97.560440 | 88.461538 | 70.516484 |
| 210 | *c* | *d* | *f* | *h* | *i* | 97.205543 | 83.660508 | 85.519630 | 94.815242 | **98.799076** |
| 211 | *c* | *d* | *f* | *h* | *j* | 102.881166 | 92.051570 | 93.598655 | **105.201794** | 66.266816 |
| 212 | *c* | *d* | *f* | *i* | *j* | 102.651007 | 92.617450 | 94.418345 | **105.480984** | 64.832215 |
| 213 | *c* | *d* | *g* | *h* | *i* | 99.200450 | 88.322072 | 75.889640 | 95.833333 | **100.754505** |
| 214 | *c* | *d* | *g* | *h* | *j* | 103.780488 | 96.385809 | 84.656319 | **107.350333** | 67.827051 |
| 215 | *c* | *d* | *g* | *i* | *j* | 104.314159 | 96.426991 | 86.250000 | **106.603982** | 66.404867 |
| 216 | *c* | *d* | *h* | *i* | *j* | 101.076233 | 90.246637 | **104.428251** | 103.654709 | 60.594170 |
| 217 | *c* | *e* | *f* | *g* | *h* | **101.531532** | 94.279279 | 86.509009 | 78.738739 | 98.941441 |
| 218 | *c* | *e* | *f* | *g* | *i* | 100.459770 | 92.793103 | 85.761385 | 78.517241 | **102.468500** |
| 219 | *c* | *e* | *f* | *g* | *j* | **104.890110** | 101.098901 | 97.054945 | 88.208791 | 68.747253 |
| 220 | *c* | *e* | *f* | *h* | *i* | 96.156627 | 88.674699 | 81.192771 | 95.602410 | **98.373494** |
| 221 | *c* | *e* | *f* | *h* | *j* | 102.482993 | 97.006803 | 91.791383 | **106.133787** | 62.585034 |
| 222 | *c* | *e* | *f* | *i* | *j* | 101.902050 | 98.234624 | 92.209567 | **106.093394** | 61.560364 |
| 223 | *c* | *e* | *g* | *h* | *i* | 99.438073 | 90.470183 | 74.380734 | 95.745413 | **99.965596** |
| 224 | *c* | *e* | *g* | *h* | *j* | 104.011111 | 99.411111 | 84.077778 | **107.588889** | 64.911111 |
| 225 | *c* | *e* | *g* | *i* | *j* | 104.059735 | 100.243363 | 85.486726 | **106.858407** | 63.351770 |
| 226 | *c* | *e* | *h* | *i* | *j* | 100.559361 | 95.570776 | **104.497717** | 103.710046 | 55.662100 |
| 227 | *c* | *f* | *g* | *h* | *i* | **100.821918** | 85.593607 | 76.141553 | 96.883562 | 100.559361 |
| 228 | *c* | *f* | *g* | *h* | *j* | 104.314159 | 94.391593 | 84.723451 | **107.621681** | 68.949115 |
| 229 | *c* | *f* | *g* | *i* | *j* | 104.314159 | 95.409292 | 86.504425 | **106.858407** | 66.913717 |
| 230 | *c* | *f* | *h* | *i* | *j* | 100.526316 | 90.526316 | **105.000000** | 103.947368 | 60.000000 |
| 231 | *c* | *g* | *h* | *i* | *j* | 102.935268 | 81.886161 | **106.785714** | 104.988839 | 63.404018 |
| 232 | *d* | *e* | *f* | *g* | *h* | 91.233333 | 97.366667 | 89.955556 | 80.755556 | **100.688889** |
| 233 | *d* | *e* | *f* | *g* | *i* | 91.073826 | 94.932886 | 90.148104 | 80.525727 | **103.319458** |
| 234 | *d* | *e* | *f* | *g* | *j* | 98.355263 | **101.381579** | 98.607456 | 89.024123 | 72.631579 |
| 235 | *d* | *e* | *f* | *h* | *i* | 83.926097 | 92.690531 | 86.050808 | 96.674365 | **100.658199** |
| 236 | *d* | *e* | *f* | *h* | *j* | 93.082960 | 98.755605 | 93.856502 | **105.717489** | 68.587444 |
| 237 | *d* | *e* | *f* | *i* | *j* | 93.082960 | 100.044843 | 94.114350 | **107.006726** | 65.751121 |
| 238 | *d* | *e* | *g* | *h* | *i* | 88.640449 | 95.101124 | 76.235955 | 98.202247 | **101.820225** |
| 239 | *d* | *e* | *g* | *h* | *j* | 95.918142 | 100.243363 | 85.741150 | **108.639381** | 69.457965 |
| 240 | *d* | *e* | *g* | *i* | *j* | 96.508811 | 100.814978 | 86.883260 | **107.907489** | 67.885463 |
| 241 | *d* | *e* | *h* | *i* | *j* | 90.394144 | 97.646396 | 104.380631 | **105.416667** | 62.162162 |
| 242 | *d* | *f* | *g* | *h* | *i* | 89.587054 | 90.100446 | 78.805804 | 98.828125 | **102.678571** |
| 243 | *d* | *f* | *g* | *h* | *j* | 95.960265 | 96.214128 | 86.567329 | **107.891832** | 73.366446 |
| 244 | *d* | *f* | *g* | *i* | *j* | 97.015419 | 96.762115 | 87.389868 | **107.907489** | 70.925110 |
| 245 | *d* | *f* | *h* | *i* | *j* | 91.171171 | 92.725225 | 103.862613 | **105.934685** | 66.306306 |
| 246 | *d* | *g* | *h* | *i* | *j* | 94.600887 | 83.891353 | **106.585366** | 106.330377 | 68.592018 |
| 247 | *e* | *f* | *g* | *h* | *i* | 93.568182 | 88.079545 | 76.840909 | 99.318182 | **102.193182** |
| 248 | *e* | *f* | *g* | *h* | *j* | 99.700665 | 95.620843 | 85.676275 | **108.625277** | 70.376940 |
| 249 | *e* | *f* | *g* | *i* | *j* | 100.243363 | 96.426991 | 86.758850 | **107.876106** | 68.694690 |
| 250 | *e* | *f* | *h* | *i* | *j* | 95.745413 | 90.733945 | 105.240826 | **106.032110** | 62.247706 |
| 251 | *e* | *g* | *h* | *i* | *j* | 98.058036 | 83.169643 | **107.555804** | 106.272321 | 64.944196 |
| 252 | *f* | *g* | *h* | *i* | *j* | 93.131991 | 83.355705 | **107.539150** | 106.767338 | 69.205817 |

Table 8.2.3.1 (part 5 of 5): links in example A53

## 8.3. Condorcet Criterion for Multi-Winner Elections

In this section, we will propose a generalization of the Condorcet criterion to multi-winner elections. The Condorcet criterion for single-winner elections (section 4.8) is important because, when there is a Condorcet winner $b \in A$, then it is still a Condorcet winner when alternatives $a_1,...,a_n \in A \setminus \{b\}$ are removed. So an alternative $b \in A$ doesn't owe his property of being a Condorcet winner to the presence of some other alternatives. Therefore, when we declare a Condorcet winner $b \in A$ elected whenever a Condorcet winner exists, we know that no other alternatives $a_1,...,a_n \in A \setminus \{b\}$ have changed the result of the election without being elected.

Therefore, a generalization of the Condorcet criterion to multi-winner elections should have the following properties:

- It should not be possible that there are more than $M$ Condorcet winners (where $M$ is the number of seats). This property is important because the Condorcet winners will later be declared the winners.

- Suppose $b \in A$ is a Condorcet winner. Then it should still be a Condorcet winner when alternatives $a_1,...,a_n \in A \setminus \{b\}$ are removed.

- The requirement for being a Condorcet winner should be as weak as possible, so that there are as many Condorcet winners as possible, so that the resulting Condorcet criterion becomes as strong as possible.

We propose the following generalization:

(8.3.1) In multi-winner elections, a *Condorcet winner* is an alternative $b \in A$ that wins in every $(M+1)$-way contest. Suppose $\mathcal{S}_M|_B$ ( with $\varnothing \neq \mathcal{S}_M|_B \subseteq A_M$ ) is the set of potential winning sets when the used method to fill $M$ seats is applied to the set $B$ ( with $\varnothing \neq B \subseteq A$ and $\| B \| > M$ ). Then we get:

(8.3.1a) $b \in A$ is a *Condorcet winner* : $\Leftrightarrow$

$\forall \varnothing \neq B \subseteq A$ with $b \in B$ and $\| B \| = (M+1)$ $\forall \mathbb{A} \in \mathcal{S}_M|_B$: $b \in \mathbb{A}$.

The *Condorcet criterion* says that, when there is a Condorcet winner, then it should also be a winner overall. In short:

(8.3.1b) $b \in A$ is a Condorcet winner. $\Rightarrow$ ( $\forall \mathbb{A} \in \mathcal{S}_M$: $b \in \mathbb{A}$. )

When $\succ_D$ satisfies (2.1.5) then for $M = 1$:

- (8.3.1a) is identical to (4.8.6) and (4.12.1.1).

- (8.3.1b) is identical to (4.8.7)(i).

(8.3.2) In multi-winner elections, a *weak Condorcet winner* is an alternative $b \in A$ that wins or is tied for winning/losing in every $(M+1)$-way contest. In short:

(8.3.2a) $b \in A$ is a *weak Condorcet winner* : $\Leftrightarrow$

$\forall\ \emptyset \neq B \subseteq A$ with $b \in B$ and $\| B \| = (M+1)\ \exists\ \mathbb{A} \in \mathcal{S}_M|_B$: $b \in \mathbb{A}$.

A weak Condorcet winner should win or be tied for winning/ losing overall. In short:

(8.3.2b) $b \in A$ is a weak Condorcet winner. $\Rightarrow$ ( $\exists\ \mathbb{A} \in \mathcal{S}_M$: $b \in \mathbb{A}$. )

When $\succ_D$ satisfies (2.1.4) and (2.1.5) then for $M = 1$:

- (8.3.2a) is identical to (4.12.1.2).
- (8.3.2b) is identical to (4.12.1.6).

(8.3.3) In multi-winner elections, A *Condorcet loser* is an alternative $b \in A$ that loses in every $(M+1)$-way contest. In short:

(8.3.3a) $b \in A$ is a *Condorcet loser* : $\Leftrightarrow$

$\forall\ \emptyset \neq B \subseteq A$ with $b \in B$ and $\| B \| = (M+1)\ \forall\ \mathbb{A} \in \mathcal{S}_M|_B$: $b \notin \mathbb{A}$.

A Condorcet loser should be a loser overall. In short:

(8.3.3b) $b \in A$ is a Condorcet loser. $\Rightarrow$ ( $\forall\ \mathbb{A} \in \mathcal{S}_M$: $b \notin \mathbb{A}$. )

When $\succ_D$ satisfies (2.1.5) then for $M = 1$:

- (8.3.3a) is identical to (4.8.8) and (4.12.2.1).
- (8.3.3b) is identical to (4.8.9)(i).

(8.3.4) In multi-winner elections, a *weak Condorcet loser* is an alternative $b \in A$ that loses or is tied for winning/losing in every $(M+1)$-way contest. In short:

(8.3.4a) $b \in A$ is a *weak Condorcet loser* : $\Leftrightarrow$

$\forall\ \emptyset \neq B \subseteq A$ with $b \in B$ and $\| B \| = (M+1)\ \exists\ \mathbb{A} \in \mathcal{S}_M|_B$: $b \notin \mathbb{A}$.

A weak Condorcet loser should lose or be tied for winning/ losing overall. In short:

(8.3.4b) $b \in A$ is a weak Condorcet loser. $\Rightarrow$ ( $\exists\ \mathbb{A} \in \mathcal{S}_M$: $b \notin \mathbb{A}$. )

When $\succ_D$ satisfies (2.1.4) and (2.1.5) then for $M = 1$:

- (8.3.4a) is identical to (4.12.2.2).
- (8.3.4b) is identical to (4.12.2.13).

It is important to keep in mind that, in multi-winner elections, the terms “Condorcet winner”, “weak Condorcet winner”, “Condorcet loser”, and “weak Condorcet loser” always refer to the specific election method. For example, plurality-at-large will lead to different Condorcet winners than an STV method. So in multi-winner elections, the Condorcet criterion rather refers to the inner logic of the specific election method than to alternatives that must be elected regardless of the election method used.

If $\succ_{D2}$ satisfies (2.1.4) and (2.1.5), then we get for Schulze STV:

(8.3.5) $b \in A$ is a Condorcet winner $\Leftrightarrow$
$$\forall \{a_1,...,a_M\} \subseteq A \setminus \{b\} \; \exists \, a_i \in \{a_1,...,a_M\}: N[\{a_1,...,a_M\};b] < N[(\{a_1,...,a_M,b\} \setminus \{a_i\});a_i].$$

(8.3.6) $b \in A$ is a weak Condorcet winner $\Leftrightarrow$
$$\forall \{a_1,...,a_M\} \subseteq A \setminus \{b\} \; \exists \, a_i \in \{a_1,...,a_M\}: N[\{a_1,...,a_M\};b] \leq N[(\{a_1,...,a_M,b\} \setminus \{a_i\});a_i].$$

(8.3.7) $b \in A$ is a Condorcet loser $\Leftrightarrow$
$$\forall \{a_1,...,a_M\} \subseteq A \setminus \{b\} \; \forall \, a_i \in \{a_1,...,a_M\}: N[\{a_1,...,a_M\};b] > N[(\{a_1,...,a_M,b\} \setminus \{a_i\});a_i].$$

(8.3.8) $b \in A$ is a weak Condorcet loser $\Leftrightarrow$
$$\forall \{a_1,...,a_M\} \subseteq A \setminus \{b\} \; \forall \, a_i \in \{a_1,...,a_M\}: N[\{a_1,...,a_M\};b] \geq N[(\{a_1,...,a_M,b\} \setminus \{a_i\});a_i].$$

In example A53, the alternatives $a$, $g$, and $j$ win in every 5-way contest; therefore, these alternatives should also win overall. The alternative $d$ is tied for winning in one case (line 27) and wins in every other case; therefore, this alternative should win or be tied for winning/losing overall.

While there can be up to $M$ Condorcet winners, there cannot be more than one Condorcet loser.

**<u>Claim:</u>**

If $\succ_{D2}$ satisfies (2.1.5), then Schulze STV, as defined in section 8.1, satisfies the Condorcet criterion for multi-winner electons, as defined in (8.3.1).

**<u>Proof:</u>**

Suppose alternative $b \in A$ is a Condorcet winner. Suppose $\{a_1,...,a_M\} \subseteq A \setminus \{b\}$.

We apply Schulze STV, as defined in section 8.1, on $\{a_1,...,a_M,b\}$. Suppose $c \in \{a_1,...,a_M,b\}$ is an alternative with maximum $N[(\{a_1,...,a_M,b\} \setminus \{c\});c]$. Then $(N[(\{a_1,...,a_M,b\} \setminus \{c\});c], N[\{a_1,...,a_M\};b])$ is a win. With (8.3.5), we get that alternative $c$ cannot be identical to alternative $b$. Therefore, the link $(\{a_1,...,a_M,b\} \setminus \{c\}) \rightarrow \{a_1,...,a_M\}$ is a path from $(\{a_1,...,a_M,b\} \setminus \{c\})$ to $\{a_1,...,a_M\}$ that contains only wins.

On the other side, there cannot be a path from $\{a_1,...,a_M\}$ to $(\{a_1,...,a_M,b\} \setminus \{c\})$ that contains only wins because any path from $\{a_1,...,a_M\}$ to $(\{a_1,...,a_M,b\} \setminus \{c\})$ must contain a link from a set $\mathfrak{C}(i)$ with $b \notin \mathfrak{C}(i)$ to a set $\mathfrak{C}(i+1)$ with $b \in \mathfrak{C}(i+1)$. But because of the definition of Condorcet winners, the link $\mathfrak{C}(i) \rightarrow \mathfrak{C}(i+1)$ must be a tie or a defeat.

With (2.1.5), we get that every path that contains only wins is stronger than every path that contains a tie or a defeat.

Therefore, every set $\{a_1,...,a_M\}$, that does not contain alternative $b$, is disqualified by some set that contains alternative $b$. □

The proofs that Schulze STV satisfies (8.3.2b), (8.3.2c), and (8.3.2d) are analogue to the proofs for (4.12.1.6), (4.12.2.13), and (8.3.2a).

In a similar manner, we can generalize the Smith criterion (section 4.8) to multi-winner elections.

**Definition:**

A multi-winner election method, where $M$ is the number of seats, satisfies the *Smith criterion for multi-winner elections*, if the following holds:

Suppose $\emptyset \neq B \subsetneq A$. Suppose $x \in \mathbb{N}$ with $1 \leq x \leq \| B \|$ and $x \leq M$.

(8.3.9) Suppose, for every $y \in \mathbb{N}$ with $1 \leq y \leq x$, we have: In every ($M$+1)-contest between $y$ alternatives of the set $B$ and $M$+1–$y$ alternatives of $A \setminus B$ each of the alternatives of the set $B$ is in every potential winning set.

Then every potential winning set contains at least $x$ alternatives of the set $B$.

In short, a multi-winner election method, where $M$ is the number of seats, satisfies the *Smith criterion for multi-winner elections*, if the following holds:

(8.3.10) $\forall \emptyset \neq B \subsetneq A \; \forall x \in \mathbb{N}$ with $1 \leq x \leq \| B \|$ and $x \leq M$:

$$( ( \forall y \in \mathbb{N} \text{ with } 1 \leq y \leq x$$

$$\forall \emptyset \neq \tilde{A} \subseteq A \text{ with } \| \tilde{A} \| = (M+1) \text{ and } \| \tilde{A} \cap B \| = y$$

$$\forall \mathbb{A} \in \mathcal{S}_M|_{\tilde{A}}: \| \mathbb{A} \cap B \| = y. )$$

$$\Rightarrow ( \forall \mathbb{A} \in \mathcal{S}_M: \| \mathbb{A} \cap B \| \geq x. ) )$$

So (8.3.10) basically says that, when all alternatives of the set $B$ are elected whenever exactly $y$ alternatives of the set $B$ and exactly $M$+1–$y$ alternatives of $A \setminus B$ are running and when this is true for all $y \in \mathbb{N}$ with $1 \leq y \leq x$ ( for some $x \in \mathbb{N}$ with $1 \leq x \leq \| B \|$ and $x \leq M$ ), then at least $x$ alternatives of the set $B$ must be elected. This means: Every potential winning set $\mathbb{A} \in \mathcal{S}_M$ must contain at least $x$ alternatives of set $B$.

For $M$ = 1, (8.3.10) is identical to (4.8.5), the Smith criterion for single-winner elections.

**Claim:**

If $\succ_{D2}$ satisfies (2.1.5), then Schulze STV, as defined in section 8.1, satisfies the Smith criterion for multi-winner electons.

**Proof (overview):**

The proof that Schulze STV satisfies the Smith criterion for multi-winner elections is analogous to the proof that Schulze STV satisfies the Condorcet criterion for multi-winner elections.

Part 1: Suppose $z \in \mathbb{N}_0$ with $0 \leq z < x$. Suppose $\{a_1,...,a_{(M-z)},b_1,...,b_z\}$ is a set of $M-z$ alternatives $a_1,...,a_{(M-z)} \in A \setminus B$ and $z$ alternatives $b_1,...,b_z \in B$. Suppose $b_{(z+1)} \in B \setminus \{b_1,...,b_z\}$ is an arbitrarily chosen alternative. Suppose $c \in \{a_1,...,a_{(M-z)},b_1,...,b_z,b_{(z+1)}\}$ is an alternative with maximum $N[(\{a_1,..., a_{(M-z)},b_1,...,b_z,b_{(z+1)}\}\setminus\{c\});c]$. Then $(N[(\{a_1,...,a_{(M-z)},b_1,...,b_z,b_{(z+1)}\}\setminus\{c\});c], N[\{a_1,...,a_{(M-z)},b_1,...,b_z,b_{(z+1)}\};b_{(z+1)}])$ is a win. With (8.3.9), we get $c \notin \{b_1,...,b_z,b_{(z+1)}\}$. Therefore, the link $(\{a_1,...,a_{(M-z)},b_1,...,b_z,b_{(z+1)}\}\setminus\{c\}) \rightarrow \{a_1,...,a_{(M-z)},b_1,...,b_z\}$ is a path from $(\{a_1,...,a_{(M-z)},b_1,...,b_z,b_{(z+1)}\}\setminus\{c\})$ to $\{a_1,...,a_{(M-z)},b_1,...,b_z\}$ that contains only wins.

On the other side, there cannot be a path from $\{a_1,...,a_{(M-z)},b_1,...,b_z\}$ to $(\{a_1,...,a_{(M-z)},b_1,...,b_z,b_{(z+1)}\}\setminus\{c\})$ that contains only wins because any path from $\{a_1,...,a_{(M-z)},b_1,...,b_z\}$ to $(\{a_1,...,a_{(M-z)},b_1,...,b_z,b_{(z+1)}\}\setminus\{c\})$ must contain a link from a set $\mathfrak{C}(i)$ with $z$ alternatives from the set $B$ to a set $\mathfrak{C}(i+1)$ with $z+1$ alternatives from the set $B$. But with (8.3.9), we get that the link $\mathfrak{C}(i) \rightarrow \mathfrak{C}(i+1)$ must be a tie or a defeat.

With (2.1.5), we get that every path that contains only wins is stronger than every path that contains a tie or a defeat.

Therefore, every set $\{a_1,...,a_{(M-z)},b_1,...,b_z\}$, that contains only $z$ alternatives from the set $B$ is disqualified by some set that contains $z+1$ alternatives from the set $B$.

Part 2: Part 1 is applied to $z := 0,...,(x-1)$. As indirect defeats are transitive (section 4.1), we get that every set with less than $x$ alternatives from the set $B$ is disqualified by some set with $x$ alternatives from the set $B$.□

The Smith criterion for multi-winner elections implies the Condorcet criterion for multi-winner elections. We get the Condorcet criterion for multi-winner elections when we restrict the Smith criterion for multi-winner elections to sets with exactly one alternative.

In example A53, the Smith criterion for multi-winner elections implies that at least one winner must come from the set $\{d,f\}$ because, whenever exactly one alternative from the set $\{d,f\}$ and exactly four alternatives from $A \setminus \{d,f\}$ are running, the alternative from $\{d,f\}$ is a winner of Schulze STV.

### 8.3.1: Example A

Suppose $M = 10$ and $x = 6$.

Then the Smith criterion for multi-winner elections has the following form:

> Suppose all of the following conditions are satisfied:
>
> > $\varnothing \neq B \subsetneq A$ consists of at least $x = 6$ candidates.
> >
> > ( $y = 1$ ): Whenever exactly $y = 1$ candidate of the set $B$ and exactly $M + 1 - y = 10$ candidates of $A \setminus B$ are running, the candidate of the set $B$ is in every potential winning set.
> >
> > ( $y = 2$ ): Whenever exactly $y = 2$ candidate of the set $B$ and exactly $M + 1 - y = 9$ candidates of $A \setminus B$ are running, the candidates of the set $B$ are in every potential winning set.
> >
> > ( $y = 3$ ): Whenever exactly $y = 3$ candidate of the set $B$ and exactly $M + 1 - y = 8$ candidates of $A \setminus B$ are running, the candidates of the set $B$ are in every potential winning set.
> >
> > ( $y = 4$ ): Whenever exactly $y = 4$ candidate of the set $B$ and exactly $M + 1 - y = 7$ candidates of $A \setminus B$ are running, the candidates of the set $B$ are in every potential winning set.
> >
> > ( $y = 5$ ): Whenever exactly $y = 5$ candidate of the set $B$ and exactly $M + 1 - y = 6$ candidates of $A \setminus B$ are running, the candidates of the set $B$ are in every potential winning set.
> >
> > ( $y = 6$ ): Whenever exactly $y = 6$ candidate of the set $B$ and exactly $M + 1 - y = 5$ candidates of $A \setminus B$ are running, the candidates of the set $B$ are in every potential winning set.
>
> Then when the used election method is applied on $A$, every potential winning set must contain at least $x = 6$ candidates of the set $B$.

Conditions ( $y = 1$ ) to ( $y = 5$ ) seem to be superfluous. Condition ( $y = 6$ ) seems to be sufficient to guarantee that the set $B$ should get at least $x = 6$ seats. However, example B (section 8.3.2) shows that, when we drop the conditions $y = 1, \ldots, ( x - 1 )$, then the resulting criterion is not satisfiable anymore.

## 8.3.2: Example B

Suppose $M = 2$.

Then the Smith criterion for multi-winner elections has the following form:

Suppose

(a1) $\emptyset \neq B \subsetneq A$ consists of at least $x = 1$ candidate.

(a2) Whenever exactly one candidate of the set $B$ and exactly two candidates of $A \setminus B$ are running, the candidate of the set $B$ is in every potential winning set.

Then, when the used election method is applied on $A$, every potential winning set must contain at least $x = 1$ candidate of the set $B$.

Suppose

(b1) $\emptyset \neq B \subsetneq A$ consists of at least $x = 2$ candidates.

(b2) Whenever exactly one candidate of the set $B$ and exactly two candidates of $A \setminus B$ are running, the candidate of the set $B$ is in every potential winning set.

(b3) Whenever exactly two candidates of the set $B$ and exactly one candidates of $A \setminus B$ are running, both candidates of the set $B$ are in every potential winning set.

Then, when the used election method is applied on $A$, every potential winning set must contain at least $x = 2$ candidates of the set $B$.

Condition b2 seems to be superfluous. Conditions b1 and b3 seem to be sufficient to guarantee that the set $B$ should get at least 2 seats. However, the following example shows that, when we drop condition b2, then there can be more than one set such that both winners must come from set $B_1$ and, simultaneously, both winners must come from set $B_2$ with $( B_1 \cap B_2 ) = \emptyset$. In the following example, we have $B_1 = \{a,b\}$ and $B_2 = \{c,d\}$.

The following example has been proposed by I.D. Hill (1995).

There are $N = 54$ voters and $C = 4$ alternatives for $M = 2$ seats:

| | *a* | *b* | *c* | *d* |
|---|---|---|---|---|
| 1 | 1 | - | - | - |
| 2 | 1 | - | - | - |
| 3 | 1 | - | - | - |
| 4 | 1 | - | - | - |
| 5 | 1 | - | - | - |
| 6 | 1 | - | - | - |
| 7 | - | 1 | - | - |
| 8 | - | 1 | - | - |
| 9 | - | 1 | - | - |
| 10 | - | 1 | - | - |
| 11 | - | 1 | - | - |
| 12 | - | 1 | - | - |
| 13 | - | - | 1 | - |
| 14 | - | - | 1 | - |
| 15 | - | - | 1 | - |
| 16 | - | - | 1 | - |
| 17 | - | - | 1 | - |
| 18 | - | - | - | 1 |
| 19 | - | - | - | 1 |
| 20 | - | - | - | 1 |
| 21 | - | - | - | 1 |
| 22 | - | - | - | 1 |
| 23 | 2 | - | - | 1 |
| 24 | 2 | - | - | 1 |
| 25 | 2 | - | - | 1 |
| 26 | 2 | - | - | 1 |
| 27 | - | 2 | - | 1 |

| | *a* | *b* | *c* | *d* |
|---|---|---|---|---|
| 28 | - | 2 | - | 1 |
| 29 | - | 2 | - | 1 |
| 30 | - | 2 | - | 1 |
| 31 | 2 | - | 1 | - |
| 32 | 2 | - | 1 | - |
| 33 | 2 | - | 1 | - |
| 34 | 2 | - | 1 | - |
| 35 | - | 2 | 1 | - |
| 36 | - | 2 | 1 | - |
| 37 | - | 2 | 1 | - |
| 38 | - | 2 | 1 | - |
| 39 | - | 1 | 2 | - |
| 40 | - | 1 | 2 | - |
| 41 | - | 1 | 2 | - |
| 42 | - | 1 | 2 | - |
| 43 | - | 1 | - | 2 |
| 44 | - | 1 | - | 2 |
| 45 | - | 1 | - | 2 |
| 46 | - | 1 | - | 2 |
| 47 | 1 | - | 2 | - |
| 48 | 1 | - | 2 | - |
| 49 | 1 | - | 2 | - |
| 50 | 1 | - | 2 | - |
| 51 | 1 | - | - | 2 |
| 52 | 1 | - | - | 2 |
| 53 | 1 | - | - | 2 |
| 54 | 1 | - | - | 2 |

The links are:

| | *k* | *l* | *m* | $N[\{l,m\};k]$ | $N[\{k,m\};l]$ | $N[\{k,l\};m]$ |
|---|---|---|---|---|---|---|
| 1 | *a* | *b* | *c* | 17.081633 | 17.081633 | **19.836735** |
| 2 | *a* | *b* | *d* | 17.081633 | 17.081633 | **19.836735** |
| 3 | *a* | *c* | *d* | **19.125000** | 17.437500 | 17.437500 |
| 4 | *b* | *c* | *d* | **19.125000** | 17.437500 | 17.437500 |

The corresponding digraph looks as follows:

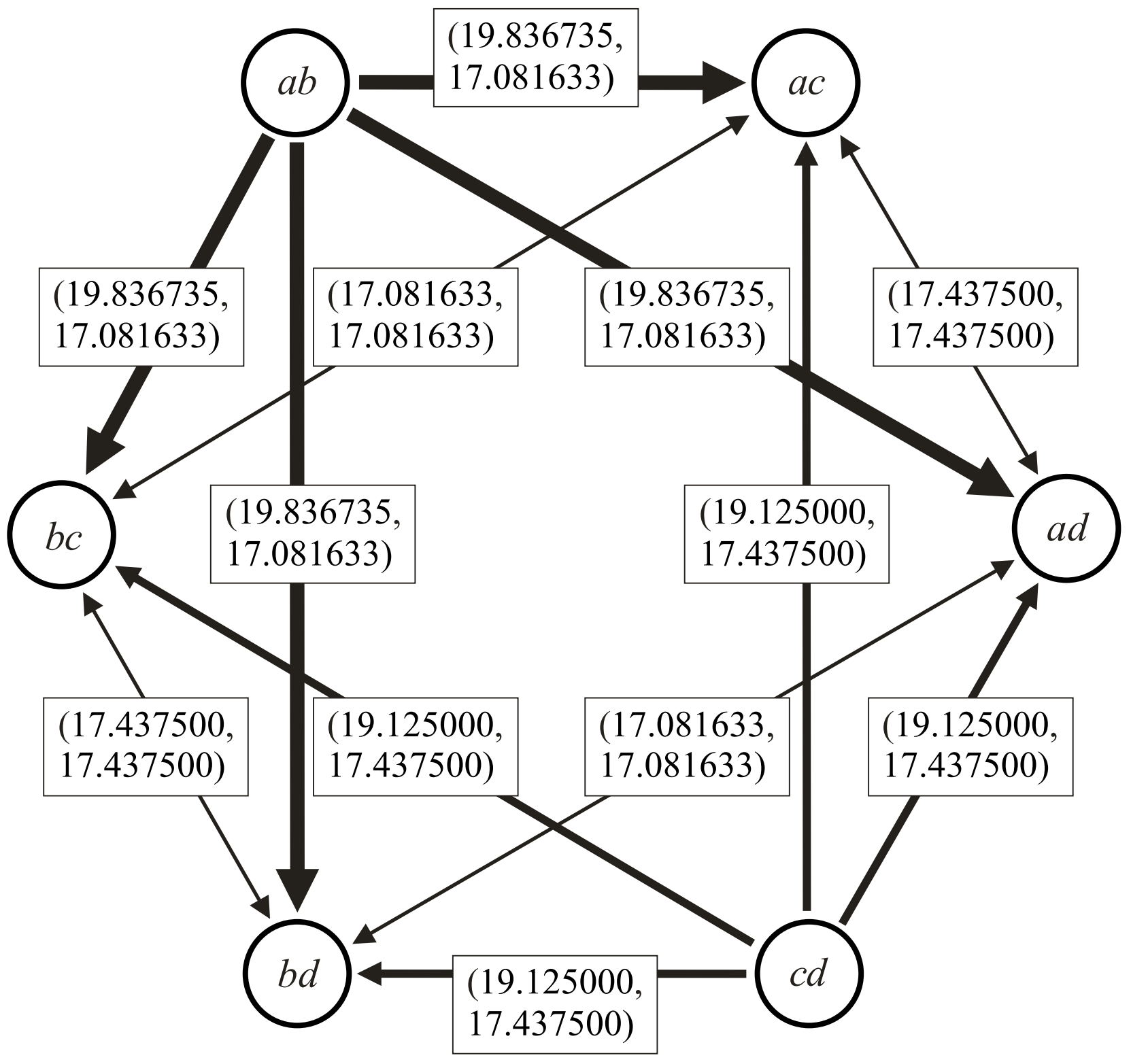

When $\{a,b,c\}$ are running, the unique winning set is $\{a,b\}$.
When $\{a,b,d\}$ are running, the unique winning set is $\{a,b\}$.
When $\{a,c,d\}$ are running, the unique winning set is $\{c,d\}$.
When $\{b,c,d\}$ are running, the unique winning set is $\{c,d\}$.

In the example above, alternatives *a* and *b* are winners whenever they and exactly one other alternative are running. Furthermore, alternatives *c* and *d* are winners whenever they and exactly one other alternative are running. The Smith criterion for multi-winner elections says that at least one alternative of the set $\{a,c\}$ must be elected, at least one alternative of the set $\{b,c\}$ must be elected, at least one alternative of the set $\{a,d\}$ must be elected, and at least one alternative of the set $\{b,d\}$ must be elected. So the Smith criterion for multi-winner elections says that either the set $\{a,b\}$ or the set $\{c,d\}$ must be the winner.

So this example shows that, for a set of *M* alternatives to be the unique winning set, it is not sufficient to win every contest between itself and some other set of *M* alternatives that differs in exactly one alternative.

In example B, Schulze STV chooses the set $\{a,b\}$.

## 8.4. Proportionality

**Definition (Dummett-Droop Proportionality):**

A preferential multi-winner election method satisfies *Dummett-Droop proportionality* (DDP) if the following holds for every $\emptyset \neq B \subsetneq A$ and for every $x \in \mathbb{N}$ with $x \leq \| B \|$:

Suppose that strictly more than $x \cdot N / (M+1)$ voters strictly prefer every alternative in $B$ to every other alternative. In other words:

$$\| \{ v \in V \mid \forall a \in B \; \forall b \notin B \colon a \succ_v b \} \| > x \cdot N / (M+1).$$

Then at least $x$ alternatives of set $B$ must be elected.

It has been proposed by Droop (1881) that an alternative should be elected as soon as it has received more than $N / (M+1)$ votes. This idea has been generalized by Dummett to sets of alternatives (Dummett, 1984; Schulze, 2002). Today, DDP is considered a necessary and sufficient criterion for every preferential multi-winner election method to qualify as an STV method.

**Claim:**

Schulze STV, as defined in section 8.1, satisfies Dummett-Droop proportionality.

**Proof (overview):**

The proof is very similar to the proof that Schulze STV satisfies the Smith criterion for multi-winner elections (section 8.3). Therefore, we give only an overview.

Step 1: We prove that Schulze STV satisfies DDP when there are $M$ seats and $C = M + 1$ alternatives. This step is trivial because, in the $C = M + 1$ case, we simply calculate $N[(\{a_1,...,a_{(M+1)}\} \backslash \{a_i\});a_i]$ for every $i \in \{1,...,(M+1)\}$ and eliminate the alternative $i$ with maximum $N[(\{a_1,...,a_{(M+1)}\} \backslash \{a_i\});a_i]$.

Step 2: DDP leads to constraints of the form "*abcdef*(*k*)", saying that at least $k$ alternatives of the set $\{a,b,c,d,e,f\}$ must be elected according to DDP. See the column "Dummett" in table 9.4.1.

Suppose $\mathfrak{A} \in A_M$ violates DDP. Then there is a DDP constraint *abcdef*(*k*), such that $\mathfrak{A}$ contains only $j$ alternatives of the set $\{a,b,c,d,e,f\}$ with $j < k$.

We then prove that there is a $\mathfrak{B} \in A_M \backslash \{\mathfrak{A}\}$ such that there is a path from $\mathfrak{B}$ to $\mathfrak{A}$ that contains only wins. We choose $\mathfrak{B}$ as follows: We take an arbitrarily chosen alternative $g$ from the set $\{a,b,c,d,e,f\}$ that is not in $\mathfrak{A}$; we then apply Schulze STV to ( $\mathfrak{A} \cup \{g\}$ ); the resulting winning set is then the set $\mathfrak{B}$; the link $\mathfrak{B} \rightarrow \mathfrak{A}$ is then a path from $\mathfrak{B}$ to $\mathfrak{A}$ that contains only wins.

Step 3: We prove that there cannot be a path from $\mathfrak{A}$ to $\mathfrak{B}$ that contains only wins because such a path would necessarily contain a link $\mathfrak{C}(i) \rightarrow \mathfrak{C}(i+1)$ where $\mathfrak{C}(i)$ contains only $j$ alternatives of the set $\{a,b,c,d,e,f\}$ and $\mathfrak{C}(i+1)$ contains $j + 1$ alternatives of the set $\{a,b,c,d,e,f\}$. But as Schulze STV satisfies DDP in the $C = M + 1$ case, the link $\mathfrak{C}(i) \rightarrow \mathfrak{C}(i+1)$ must be a tie or a defeat.

As path defeats are transitive, there must be a $\mathfrak{X} \in A_M$ that is not disqualified by some other $\mathfrak{Y} \in A_M \setminus \{\mathfrak{X}\}$. As every $\mathfrak{A} \in A_M$ that violates DDP is disqualified by some other $\mathfrak{B} \in A_M \setminus \{\mathfrak{A}\}$, this $\mathfrak{X}$ must satisfy DDP.□

## 9. Proportional Ranking

In party-list proportional representation, each party must submit a linear order of its candidates beforehand, without knowing how many seats it will win. Parties are often interested in ensuring that — no matter how many candidates are elected — the candidates elected should reflect the strengths of the different wings of the party as proportionally as possible (Otten, 1998, 2000; Rosenstiel, 1998; Warren, 1999; Skowron, 2017). A linear order with this property is called a *proportional ranking*. The two most important approaches to create a proportional ranking are the *bottom-up* approach (Rosenstiel, 1998) and the *top-down* approach (Otten, 1998, 2000).

The *bottom-up* approach says that we start with the situation where all $C$ candidates are elected. Then, for $k = C$ to 2, we ask which candidate can be eliminated (without changing who is already eliminated) so that the distortion of the proportionality of the remaining candidates is as small as possible; the newly eliminated candidate then gets the $k$-th place of this party list.

The *top-down* approach says that we use a single-winner election method to fill the first place of this party list. Then, for $k = 2$ to $C$, we ask which candidate can be added to the already elected candidates (without changing who is already elected) so that the distortion of the proportionality is as small as possible; the newly added candidate then gets the $k$-th place of this party list.

I prefer the top-down approach to the bottom-up approach, because the bottom-up approach starts with the lowest and, therefore, (as the number of candidates is usually significantly larger than the number of seats this party can realistically hope to win) least important places so that slight fluctuations in the filling of the lowest places can have an enormous impact on the filling of the best places. Therefore, in this paper we presume that the top-down approach is being used.

In section 9.1, we will propose a new proportional ranking method. In sections 9.3 and 9.4, we will apply this method to the examples of Tideman's database. The proposed proportional ranking method is based on the following idea:

- Suppose $a_1,\ldots,a_{(k-1)} \in A$ are already elected.
- Suppose there are candidates $\varnothing \neq \{b_1,\ldots,b_z\} \subseteq A \setminus \{a_1,\ldots,a_{(k-1)}\}$ such that, whenever some candidate $b_j \in \{b_1,\ldots,b_z\}$ is added to $\{a_1,\ldots,a_{(k-1)}\}$, then choosing the set $\{a_1,\ldots,a_{(k-1)},b_j\}$ is compatible to the Smith criterion for $k$-winner elections (section 8.3).
- Then the $k$-th seat should go to one of the candidates in $\{b_1,\ldots,b_z\}$.

## 9.1. Schulze Proportional Ranking

Proportional completion is defined in section 8.1.1. $N[\{a_1,...,a_k\};g]$ is defined in section 8.1.2. $\succ_{D1}$ and $\succ_{D2}$ are two binary relations that each satisfy (2.1.1) – (2.1.3).

**Stage 1:**

We calculate a Schulze single-winner ranking $O_{final}(\sigma)$ on $A$, as defined in section 5.1, with $\succ_{D1}$.

**Stage 2:**

Proportional completion is used to complete $V$ to $W$.

**Stage 3:**

For $k := 1$ to $(C-1)$ do

{

Suppose $a_1,...,a_{(k-1)}$ are already elected.

For each pair of alternatives $b,c \notin \{a_1,...,a_{(k-1)}\}$, we define:

$$H[b,c] := N[\{a_1,...,a_{(k-1)},b\};c].$$

We apply the Schulze single-winner election method, as defined in section 2.3.1 stage 2, on $H[i,j]$, instead of $N[i,j]$, and with $\succ_{D2}$. If there is only one potential winner, then it gets the $k$-th place. If there is more than one potential winner, then the $k$-th place goes to that potential winner $b$ with $bc \in O_{final}(\sigma)$ for every other potential winner $c$.

}

## 9.2. Independence of Clones

In general, independence of clones means that, when some candidate $d \in A$ is replaced by a set of clones $K$ as defined in (4.7.1) – (4.7.3), then this must neither help nor harm candidate $d$ or any other candidate. In context of proportional ranking methods, this means that, when candidate $d$ had the $z$-th place in the proportional ranking, then the highest ranked clone $g \in K$ must get the $z$-th place of the proportional ranking and there must be no change in the places 1, ..., $(z-1)$ of the proportional ranking.

The Schulze proportional ranking method satisfies independence of clones. To prove this, the proof in section 4.7 has to be applied to places 1, ..., $z$ of the proportional ranking and to $H[x,y]$ instead of $N[x,y]$.

## 9.3. Example A53

The following series of tables illustrates the Schulze proportional ranking method when applied to example A53 of Tideman's database with $\succ_{ratio}$ for $\succ_{D1}$ and $\succ_{margin}$ for $\succ_{D2}$. Pairwise wins are **fat and underlined**. Pairwise ties are *italic and underlined*.

| | *N*[*;*a*] | *N*[*;*b*] | *N*[*;*c*] | *N*[*;*d*] | *N*[*;*e*] | *N*[*;*f*] | *N*[*;*g*] | *N*[*;*h*] | *N*[*;*i*] | *N*[*;*j*] |
|---|---|---|---|---|---|---|---|---|---|---|
| *N*[*a*;*] | -- | **316.175711** | **352.129380** | **303.100775** | **307.846154** | **298.883249** | **266.374696** | **349.351351** | **348.337731** | 193.625304 |
| *N*[*b*;*] | 143.824289 | -- | **262.462462** | 221.153846 | **240.176991** | 222.913165 | 199.414894 | **265.438066** | **263.253012** | 128.845209 |
| *N*[*c*;*] | 107.870620 | 197.537538 | -- | 197.906977 | 201.703470 | 183.197674 | 171.397260 | **240.747664** | **241.022364** | 105.891089 |
| *N*[*d*;*] | 156.899225 | **238.846154** | **262.093023** | -- | **248.295455** | **242.234043** | 197.142857 | **264.532578** | **276.260623** | 146.975610 |
| *N*[*e*;*] | 152.153846 | 219.823009 | **258.296530** | 211.704545 | -- | 214.494382 | 191.152815 | **259.814815** | **274.842767** | 120.992556 |
| *N*[*f*;*] | 161.116751 | **237.086835** | **276.802326** | 217.765957 | **245.505618** | -- | 207.817259 | **280.229885** | **275.190616** | 139.803922 |
| *N*[*g*;*] | 193.625304 | **260.585106** | **288.602740** | **262.857143** | **268.847185** | **252.182741** | -- | **306.259947** | **314.604905** | 183.785047 |
| *N*[*h*;*] | 110.648649 | 194.561934 | 219.252336 | 195.467422 | 200.185185 | 179.770115 | 153.740053 | -- | **250.125000** | 87.058824 |
| *N*[*i*;*] | 111.662269 | 196.746988 | 218.977636 | 183.739377 | 185.157233 | 184.809384 | 145.395095 | 209.875000 | -- | 97.150127 |
| *N*[*j*;*] | **266.374696** | **331.154791** | **354.108911** | **313.024390** | **339.007444** | **320.196078** | **276.214953** | **372.941176** | **362.849873** | -- |

The 1. place goes to alternative *j*.

| | *N*[{*j*,*};*a*] | *N*[{*j*,*};*b*] | *N*[{*j*,*};*c*] | *N*[{*j*,*};*d*] | *N*[{*j*,*};*e*] | *N*[{*j*,*};*f*] | *N*[{*j*,*};*g*] | *N*[{*j*,*};*h*] | *N*[{*j*,*};*i*] |
|---|---|---|---|---|---|---|---|---|---|
| *N*[{*a*,*j*};*] | -- | **188.909513** | **204.084507** | **184.640371** | **188.568129** | **185.604651** | **164.582393** | **208.844340** | **201.113744** |
| *N*[{*b*,*j*};*] | 143.824289 | -- | **193.995327** | 171.444954 | **185.831382** | 174.389671 | 155.588235 | **200.287081** | **196.515513** |
| *N*[{*c*,*j*};*] | 107.870620 | 178.948598 | -- | 172.097902 | 181.492891 | 173.849765 | 153.507973 | **201.318945** | **192.673031** |
| *N*[{*d*,*j*};*] | 156.357309 | **183.050459** | **193.006993** | -- | **187.645688** | **179.186047** | 156.628959 | **196.988235** | **197.605634** |
| *N*[{*e*,*j*};*] | 149.792148 | 179.367681 | **195.118483** | 172.097902 | -- | 175.754717 | 157.313770 | **201.802885** | **200.555556** |
| *N*[{*f*,*j*};*] | 148.162791 | **180.328638** | **192.206573** | 171.162791 | **182.264151** | -- | 157.123596 | **200.428571** | **198.162291** |
| *N*[{*g*,*j*};*] | 160.948081 | **188.891403** | **200.136674** | **184.208145** | **191.060948** | **182.449438** | -- | **207.159353** | **204.210526** |
| *N*[{*h*,*j*};*] | 110.648649 | 173.325359 | 183.669065 | 165.058824 | 175.264423 | 165.928571 | 144.480370 | -- | 191.477833 |
| *N*[{*i*,*j*};*] | 111.662269 | 176.754177 | 187.732697 | 165.751174 | 178.333333 | 169.069212 | 145.395095 | **194.876847** | -- |

The 2. place goes to alternative *a*.

| | *N*[{*a*,*j*,*};*b*] | *N*[{*a*,*j*,*};*c*] | *N*[{*a*,*j*,*};*d*] | *N*[{*a*,*j*,*};*e*] | *N*[{*a*,*j*,*};*f*] | *N*[{*a*,*j*,*};*g*] | *N*[{*a*,*j*,*};*h*] | *N*[{*a*,*j*,*};*i*] |
|---|---|---|---|---|---|---|---|---|
| *N*[{*a*,*b*,*j*};*] | -- | **139.742424** | 126.860987 | **132.267267** | 127.603930 | 115.511111 | **142.105263** | **137.929985** |
| *N*[{*a*,*c*,*j*};*] | 130.333333 | -- | 126.560847 | 129.287879 | 127.308869 | 112.604167 | **141.376147** | 137.081413 |
| *N*[{*a*,*d*,*j*};*] | **131.674141** | **139.077853** | -- | **131.674141** | **129.342404** | 116.356932 | **141.511716** | **138.348485** |
| *N*[{*a*,*e*,*j*};*] | 129.504505 | **141.136364** | 127.892377 | -- | 128.702866 | 117.035398 | **142.554800** | **141.430746** |
| *N*[{*a*,*f*,*j*};*] | **130.733182** | **140.321101** | 125.865457 | **130.090498** | -- | 115.851852 | **143.204252** | **140.000000** |
| *N*[{*a*,*g*,*j*};*] | **133.911111** | **144.092262** | **133.318584** | **135.014749** | **131.866667** | -- | **145.769806** | **143.685393** |
| *N*[{*a*,*h*,*j*};*] | 128.771930 | 137.859327 | 124.822373 | 127.603930 | 124.692483 | 110.702541 | -- | 137.324053 |
| *N*[{*a*,*i*,*j*};*] | 129.878234 | **138.847926** | 126.151515 | 129.178082 | 126.666667 | 113.707865 | **140.525909** | -- |

The 3. place goes to alternative *g*.

| | *N*[{*a*,*g*,*j*,*};*b*] | *N*[{*a*,*g*,*j*,*};*c*] | *N*[{*a*,*g*,*j*,*};*d*] | *N*[{*a*,*g*,*j*,*};*e*] | *N*[{*a*,*g*,*j*,*};*f*] | *N*[{*a*,*g*,*j*,*};*h*] | *N*[{*a*,*g*,*j*,*};*i*] |
|---|---|---|---|---|---|---|---|
| *N*[{*a*,*b*,*g*,*j*};*] | -- | **108.667401** | 101.411379 | **104.131868** | 101.068282 | **110.410200** | **109.122222** |
| *N*[{*a*,*c*,*g*,*j*};*] | 102.334802 | -- | 101.574890 | 102.615385 | 100.465632 | **109.888889** | 108.355556 |
| *N*[{*a*,*d*,*g*,*j*};*] | **102.166302** | **108.414097** | -- | **102.921225** | **101.351648** | **110.197802** | **108.907285** |
| *N*[{*a*,*e*,*g*,*j*};*] | 101.351648 | **108.934066** | 101.663020 | -- | 101.321586 | **110.674779** | **109.402655** |
| *N*[{*a*,*f*,*g*,*j*};*] | **102.334802** | **109.135255** | 101.098901 | **102.841410** | -- | **110.665188** | **109.390244** |
| *N*[{*a*,*g*,*h*,*j*};*] | 101.230599 | 108.611111 | 100.087912 | 101.769912 | 99.190687 | -- | 108.568233 |
| *N*[{*a*,*g*,*i*,*j*};*] | 101.966667 | **108.866667** | 101.291391 | 102.533186 | 99.955654 | **109.597315** | -- |

The 4. place goes to alternative *d*.

| | N[{a,d,g,j,*};b] | N[{a,d,g,j,*};c] | N[{a,d,g,j,*};e] | N[{a,d,g,j,*};f] | N[{a,d,g,j,*};h] | N[{a,d,g,j,*};i] |
|---|---|---|---|---|---|---|
| N[{a,b,d,g,j};*] | -- | **87.189542** | **84.383442** | *82.579521* | **88.986900** | **88.175055** |
| N[{a,c,d,g,j};*] | 82.579521 | -- | 83.362445 | 81.687912 | **88.570175** | **87.551648** |
| N[{a,d,e,g,j};*] | 82.178649 | **87.379913** | -- | 82.358079 | **89.181619** | **88.175055** |
| N[{a,d,f,g,j};*] | *82.579521* | **87.551648** | **83.362445** | -- | **88.973684** | **88.166667** |
| N[{a,d,g,h,j};*] | 82.157205 | 87.157895 | 82.739606 | 81.307018 | -- | 87.753846 |
| N[{a,d,g,i,j};*] | 82.135667 | 87.349451 | 83.142232 | 81.508772 | **88.158242** | -- |

The 5. place goes to alternative *f* ( because alternative *f* is ranked above alternative *b* in the single-winner ranking; i.e. $fb \in O_{final}(\sigma)$ ).

| | N[{a,d,f,g,j,*};b] | N[{a,d,f,g,j,*};c] | N[{a,d,f,g,j,*};e] | N[{a,d,f,g,j,*};h] | N[{a,d,f,g,j,*};i] |
|---|---|---|---|---|---|
| N[{a,b,d,f,g,j};*] | -- | **73.326071** | **71.000000** | **74.662309** | **73.994190** |
| N[{a,c,d,f,g,j};*] | 69.484386 | -- | 70.138282 | **74.144737** | **73.640351** |
| N[{a,d,e,f,g,j};*] | 69.166667 | **73.151383** | -- | **74.657933** | **73.988355** |
| N[{a,d,f,g,h,j};*] | 69.150327 | 73.304094 | 69.803493 | -- | 73.976608 |
| N[{a,d,f,g,i,j};*] | 69.150327 | 73.304094 | 70.138282 | **74.144737** | -- |

The 6. place goes to alternative *b*.

| | N[{a,b,d,f,g,j,*};c] | N[{a,b,d,f,g,j,*};e] | N[{a,b,d,f,g,j,*};h] | N[{a,b,d,f,g,j,*};i] |
|---|---|---|---|---|
| N[{a,b,c,d,f,g,j};*] | -- | 61.285714 | **63.996265** | **63.566760** |
| N[{a,b,d,e,f,g,j};*] | **63.000000** | -- | **64.000000** | **63.571429** |
| N[{a,b,d,f,g,h,j};*] | 63.137255 | 61.000000 | -- | 63.709928 |
| N[{a,b,d,f,g,i,j};*] | 63.137255 | 61.285714 | **63.996265** | -- |

The 7. place goes to alternative *e*.

| | N[{a,b,d,e,f,g,j,*};c] | N[{a,b,d,e,f,g,j,*};h] | N[{a,b,d,e,f,g,j,*};i] |
|---|---|---|---|
| N[{a,b,c,d,e,f,g,j};*] | -- | **56.000000** | **55.750000** |
| N[{a,b,d,e,f,g,h,j};*] | 55.375000 | -- | 55.750000 |
| N[{a,b,d,e,f,g,i,j};*] | 55.375000 | **56.000000** | -- |

The 8. place goes to alternative *c*.

| | N[{a,b,c,d,e,f,g,j,*};h] | N[{a,b,c,d,e,f,g,j,*};i] |
|---|---|---|
| N[{a,b,c,d,e,f,g,h,j};*] | -- | 49.555556 |
| N[{a,b,c,d,e,f,g,i,j};*] | **49.777778** | -- |

The 9. place goes to alternative *i*.

The 10. place goes to alternative *h*.

So, the Schulze proportional ranking is *j a g d f b e c i h*.

## 9.4. Tideman’s Database

In table 9.4.1, Schulze STV and Schulze proportional ranking are applied to the instances of Tideman’s (2000) database. We use $\succ_{ratio}$ for $\succ_{D1}$ and $\succ_{margin}$ for $\succ_{D2}$ because $\succ_{ratio}$ corresponds to proportional completion; the fact that we use $\succ_{ratio}$ for $\succ_{D1}$ means that it makes no difference whether we first calculate a Schulze single-winner ranking $O_{final}(\sigma)$ and then apply proportional completion or first apply proportional completion and then calculate a Schulze single-winner ranking $O_{final}(\sigma)$.

The column “name #1” contains the name of the instance. The column “name #2” contains the name of the same instance in Wichmann’s (1994) database. $N$ is the number of voters. $C$ is the number of alternatives. $M$ is the number of seats.

Column “Dummett” contains the constraints given by “Dummett-Droop proportionality” (DDP), as defined in section 8.4. The constraints are separated by spaces. If this constraint consists of a single alternative, then this means that this alternative must be elected according to DDP. If this constraint has the form “*abcdef*(*k*)” then this means that at least *k* alternatives of the set $\{a,b,c,d,e,f\}$ must be elected according to DDP. For example, in instance A35 the constraints are “*f eijkq*(1)” so that (1) alternative *f* must be elected and (2) at least one alternative of the set $\{e,i,j,k,q\}$ must be elected according to DDP. In 3 instances (A64, A72, A83), there is only one set of $M$ alternatives that can be elected according to DDP.

The column “Condorcet winners” contains the Condorcet winners in Schulze STV [according to (8.3.5)]; alternatives, that are only weak Condorcet winners [according to (8.3.6)], are listed in brackets ( ). The column “Condorcet losers” contains the Condorcet losers in Schulze STV [according to (8.3.7)]; alternatives, that are only weak Condorcet losers [according to (8.3.8)], are listed in brackets ( ). It is important to keep in mind that, as long as $\succ_{D2}$ satisfies (2.1.5), the Condorcet winners in Schulze STV, the Condorcet losers in Schulze STV, and the possible winning sets according to the Smith criterion (for multi-winner elections) in Schulze STV do not depend on the specific choice for $\succ_{D2}$. As long as $\succ_{D2}$ satisfies (2.1.4) and (2.1.5), the weak Condorcet winners in Schulze STV and the weak Condorcet losers in Schulze STV do not depend on the specific choice for $\succ_{D2}$.

The column "Schulze STV" contains the winning set of Schulze STV, as defined in section 8.1.3. When several sets are tied for winning, then (rather than listing all potential winning sets) the winning set chosen by the tie-breaker, as defined in section 8.1.3 stage 4, is listed. In 3 instances (A34, A88, A97), an alternative, that is a weak Condorcet winner, is not elected. In instances A34 and A97, this is due to the fact that the number of alternatives, that are weak Condorcet winners or non-weak Condorcet winners, is larger than the number of seats. In instance A34, the sets $\{a,b,c,d,e,f,g,h,j,k,m,n\}$, $\{a,b,c,d,e,f,h,j,k,l,m,n\}$, $\{a,b,c,d,e,g,h,j,k,l,m,n\}$, $\{a,b,c,e,f,g,h,j,k,l,m,n\}$, and $\{b,c,d,e,f,g,h,j,k,l,m,n\}$ are tied for winning; the tie-breaker chooses $\{a,b,c,d,e,f,h,j,k,l,m,n\}$, because (1) $da \in O_{final}(\sigma)$, $dl \in O_{final}(\sigma)$, $df \in O_{final}(\sigma)$, and $dg \in O_{final}(\sigma)$ so that the set $\{a,b,c,e,f,g,h,j,k,l,m,n\}$ is disqualified at the first stage for not containing alternative $d$, (2) $al \in O_{final}(\sigma)$, $af \in O_{final}(\sigma)$, and $ag \in O_{final}(\sigma)$ so that the set $\{b,c,d,e,f,g,h,j,k,l,m,n\}$ is disqualified at the second stage for not containing alternative $a$, (3) $lf \in O_{final}(\sigma)$ and $lg \in O_{final}(\sigma)$ so that the set $\{a,b,c,d,e,f,g,h,j,k,m,n\}$ is disqualified at the third stage for not containing alternative $l$, and (4) $fg \in O_{final}(\sigma)$ so that the set $\{a,b,c,d,e,g,h,j,k,l,m,n\}$ is disqualified at the fourth stage for not containing alternative $f$. In instance A88, the sets $\{a,c,e,f,g,h\}$, $\{b,c,e,f,g,h\}$, and $\{c,d,e,f,g,h\}$ are tied for winning; while only alternative $d$ is a weak Condorcet winner, the tie-breaker chooses $\{a,c,e,f,g,h\}$, because $ab \in O_{final}(\sigma)$ and $ad \in O_{final}(\sigma)$. In instance A97, the sets $\{a,b\}$ and $\{a,c\}$ are tied for winning; the tie-breaker chooses $\{a,b\}$, because $bc \in O_{final}(\sigma)$.

The column "Schulze proportional ranking" contains the Schulze proportional ranking, as defined in section 9.1. When several rankings are tied for winning, then (rather than listing all potential rankings) the ranking chosen by the tie-breaker, as defined in section 9.1 stage 3 last sentence, is listed. In 5 instances (A04, A10, A12, A33, A67), the Schulze proportional ranking is not unique even with the proposed tie-breaker. This is due to the fact that, in these instances, even the Schulze single-winner ranking $O_{final}(\sigma)$ is not unique. Only in 6 of the 66 instances of Tideman's database (A10, A11, A13, A33, A34, A59), the winning set of Schulze STV differs from the first $M$ alternatives of Schulze proportional ranking.

The column "runtime" contains the runtime to calculate the Schulze STV winners. A Fujitsu RX 350S8 with two 6-core "E5-2630v2 @ 2.60 GHz" processors was used for the calculations. Hyper-threading was disabled. The programs to calculate the STV winners and the Schulze proportional ranking were written in Microsoft Visual C++ 2010.

| | name #1 | name #2 | *N* | *C* | *M* | Dummett | Condorcet winners | Condorcet losers | Schulze STV | Schulze proportional ranking | runtime |
|---|---|---|---|---|---|---|---|---|---|---|---|
| 1 | A01 | R006 | 380 | 10 | 3 | *a* | *a h i* | --- | *a h i* | *a i h d b c g j f e* | < 0.1 s |
| 2 | A02 | R007 | 371 | 9 | 2 | --- | *c d* | *g* | *c d* | *c d e b f a h i g* | < 0.1 s |
| 3 | A03 | R008 | 989 | 15 | 7 | *d f h* | *b d e f h k* | --- | *b d e f h k n* | *f h d k b e n g*<br>*a l c i j o m* | 5.3 s |
| 4 | A04 | R009 | 43 | 14 | 2 | --- | *i* | *d* | *f i* | *i f*<br>( ( *a e* ) or ( *e a* ) )<br>*k c b g d h m j l n* | < 0.1 s |
| 5 | A05 | R010 | 762 | 16 | 7 | *a* | *a c d e g l m* | --- | *a c d e g l m* | *a c m e d g l k*<br>*f o p h i j b n* | 7.4 s |
| 6 | A06 | R011 | 280 | 9 | 5 | *i* | *c e h i* | --- | *b c e h i* | *i h e c b f g a d* | < 0.1 s |
| 7 | A07 | R012 | 79 | 17 | 2 | --- | (*d*) *i* | *f* | *d i* | *i d c o m p h a*<br>*k g e j l n f b q* | < 0.1 s |
| 8 | A08 | R013 | 78 | 7 | 2 | *d* | *d g* | (*a*) | *d g* | *d g c b f e a* | < 0.1 s |
| 9 | A10 | R015 | 83 | 19 | 3 | --- | *m n p* | --- | *m n p* | *n* ( ( *a p m q* ) or<br>( *m p q a* ) )<br>*g f s r l i b*<br>*d j k e h o c* | < 0.1 s |
| 10 | A11 | R016 | 963 | 10 | 6 | *a c* | *a c* (*e*) *h* | --- | *a c d e g h* | *a c e h j g d i b f* | < 0.1 s |
| 11 | A12 | R017 | 76 | 20 | 2 | --- | *i r* | --- | *i r* | *r i l e g s a m*<br>*p b h t n o k d*<br>( ( *f j* ) or ( *j f* ) )<br>*c q* | < 0.1 s |
| 12 | A13 | R018 | 104 | 26 | 2 | --- | *t* | --- | *k t* | *i t k m s j c f y*<br>*z l u n a g e b p*<br>*r d h v x o q w* | < 0.1 s |
| 13 | A14 | R019 | 73 | 17 | 2 | --- | *b j* | --- | *b j* | *j b c n h q o i a*<br>*l e d g k p m f* | < 0.1 s |
| 14 | A15 | R020 | 77 | 21 | 2 | --- | (*g*) *l* | --- | *g l* | *l g t r m i c h p k j*<br>*q s a b o d u n f e* | < 0.1 s |
| 15 | A17 | R022 | 867 | 13 | 8 | *a b j* | *a b d e f j l* | --- | *a b d e f i j l* | *j b a e l f d*<br>*i m h k c g* | 0.5 s |
| 16 | A18 | R023 | 976 | 6 | 4 | *b c* | *a b c f* | *e* | *a b c f* | *b c f a d e* | < 0.1 s |
| 17 | A19 | R024 | 860 | 7 | 3 | --- | *a e g* | *f* | *a e g* | *e a g c d b f* | < 0.1 s |
| 18 | A20 | R025 | 2785 | 5 | 4 | *a d* | *a c d e* | *b* | *a c d e* | *a d c e b* | < 0.1 s |
| 19 | A22 | R027 | 44 | 11 | 2 | *ck*(1) | (*c*) *k* | *f* | *c k* | *k c a g b d*<br>*i j h e f* | < 0.1 s |
| 20 | A23 | R028 | 91 | 29 | 2 | --- | 3 5 | --- | 3 5 | 3-5-21-7-27-<br>26-22-9-17-14-<br>15-24-4-16-19-<br>20-6-11-18-28-<br>2-23-29-1-13-<br>8-10-12-25 | < 0.1 s |

Table 9.4.1 (part 1 of 3): Schulze STV applied to instances of Tideman’s database

| | name #1 | name #2 | *N* | *C* | *M* | Dummett | Condorcet winners | Condorcet losers | Schulze STV | Schulze proportional ranking | runtime |
|---|---|---|---|---|---|---|---|---|---|---|---|
| 21 | A33 | R038 | 9 | 18 | 3 | --- | (*o*) | (*j*) | *e o q* | *o a e i h c l n q f r d g* ( ( *b m p* ) or ( *b p m* ) or ( *m b p* ) ) *k j* | < 0.1 s |
| 22 | A34 | R039 | 63 | 14 | 12 | *b e h j n* | (*a*) *b c* (*d*) *e* (*f*) (*g*) *h j k* (*l*) *m n* | (*i*) | *a b c d e f h j k l m n* | *j b h e k n l g m c d a f i* | < 0.1 s |
| 23 | A35 | R040 | 176 | 17 | 5 | *f eijkq*(1) | *a* (*d*) *e f* | --- | *a d e f q* | *f e a q d k b i m n c h j p o g l* | 2.9 s |
| 24 | A48 | R041 | 923 | 10 | 9 | *b c d e f* | *a b c d e f g h j* | *i* | *a b c d e f g h j* | *d f b e c h j g a i* | < 0.1 s |
| 25 | A49 | R042 | 575 | 13 | 3 | *h* | *a c h* | *k* | *a c h* | *h c a j l d m g b i e f k* | < 0.1 s |
| 26 | A51 | R044 | 42 | 6 | 3 | *d* | *a d e* | *b* | *a d e* | *d a e f c b* | < 0.1 s |
| 27 | A52 | R045 | 667 | 10 | 6 | *d e* | *a b c d e g* | *h* | *a b c d e g* | *e d b g a c j f i h* | < 0.1 s |
| 28 | A53 | R046 | 460 | 10 | 4 | *j* | *a* (*d*) *g j* | --- | *a d g j* | *j a g d f b e c i h* | < 0.1 s |
| 29 | A54 | R047 | 924 | 11 | 9 | *a d e f k* | *a b d e f g h j k* | --- | *a b d e f g h j k* | *e d f a k g h j b i c* | < 0.1 s |
| 30 | A55 | R048 | 302 | 10 | 5 | *i* | *a* (*d*) *f i j* | *b* | *a d f i j* | *i a j f d e h c g b* | < 0.1 s |
| 31 | A56 | R049 | 685 | 13 | 2 | --- | *j k* | --- | *j k* | *j k f h m g d a e c b l i* | < 0.1 s |
| 32 | A57 | R050 | 310 | 9 | 2 | *de*(1) | *d e* | --- | *d e* | *d e i b h c g f a* | < 0.1 s |
| 33 | A59 | R052 | 694 | 7 | 4 | *d f* | *d f g* | --- | *b d f g* | *f d e g b c a* | < 0.1 s |
| 34 | A63 | R056 | 156 | 7 | 2 | --- | *c f* | --- | *c f* | *c f e d b a g* | < 0.1 s |
| 35 | A64 | R057 | 196 | 3 | 2 | *b c* | *b c* | *a* | *b c* | *b c a* | < 0.1 s |
| 36 | A65 | R058 | 198 | 10 | 6 | *b g* | *b e f g j* | --- | *a b e f g j* | *g b f e j a d c h i* | < 0.1 s |
| 37 | A66 | R059 | 193 | 6 | 4 | *f* | *b d e f* | *a* | *b d e f* | *f d e b c a* | < 0.1 s |
| 38 | A67 | R060 | 183 | 14 | 10 | *b f g k* | *b c e f g i j k l* | --- | *b c e f g h i j k l* | ( ( *f g* ) or ( *g f* ) ) *k b i e j l c h n m d a* | 4.0 s |
| 39 | A68 | R061 | 50 | 4 | 3 | *a c* | *a c d* | *b* | *a c d* | *a c d b* | < 0.1 s |
| 40 | A69 | R062 | 86 | 9 | 3 | --- | *a c e* | --- | *a c e* | *e c a f i d b h g* | < 0.1 s |

Table 9.4.1 (part 2 of 3): Schulze STV applied to instances of Tideman’s database

| | name #1 | name #2 | *N* | *C* | *M* | Dummett | Condorcet winners | Condorcet losers | Schulze STV | Schulze proportional ranking | runtime |
|---|---|---|---|---|---|---|---|---|---|---|---|
| 41 | A70 | R063 | 529 | 9 | 3 | *e* | *e h i* | --- | *e h i* | *e i h c d b a g f* | < 0.1 s |
| 42 | A71 | R064 | 500 | 8 | 7 | *d g* | *a b c d e f g* | *h* | *a b c d e f g* | *d c g e a b f h* | < 0.1 s |
| 43 | A72 | R065 | 272 | 3 | 2 | *a c* | *a c* | *b* | *a c* | *a c b* | < 0.1 s |
| 44 | A73 | R066 | 525 | 5 | 2 | --- | *c d* | --- | *c d* | *d c b a e* | < 0.1 s |
| 45 | A74 | R067 | 253 | 3 | 2 | *a* | *a c* | *b* | *a c* | *a c b* | < 0.1 s |
| 46 | A76 | R069 | 403 | 5 | 2 | *c* | *a c* | --- | *a c* | *c a d b e* | < 0.1 s |
| 47 | A78 | R071 | 486 | 4 | 3 | *c d* | *b c d* | *a* | *b c d* | *c d b a* | < 0.1 s |
| 48 | A79 | R072 | 362 | 8 | 4 | *g* | *a c e g* | --- | *a c e g* | *g a e c f d b h* | < 0.1 s |
| 49 | A80 | R073 | 269 | 7 | 5 | *a* | *a b c e g* | --- | *a b c e g* | *a e c g b f d* | < 0.1 s |
| 50 | A81 | R074 | 902 | 11 | 9 | *b c e h j* | *a b c e g h i j k* | *f* | *a b c e g h i j k* | *h e c b j g a i k d f* | < 0.1 s |
| 51 | A83 | R076 | 1123 | 4 | 3 | *a b c* | *a b c* | *d* | *a b c* | *c a b d* | < 0.1 s |
| 52 | A84 | R077 | 277 | 7 | 6 | *b c e* | *a b c d e g* | *f* | *a b c d e g* | *e b c d g a f* | < 0.1 s |
| 53 | A85 | R078 | 158 | 4 | 3 | *a d* | *a b d* | *c* | *a b d* | *d a b c* | < 0.1 s |
| 54 | A86 | R079 | 157 | 5 | 4 | *c* | *a c d e* | *b* | *a c d e* | *c a d e b* | < 0.1 s |
| 55 | A87 | R080 | 120 | 4 | 3 | *b d* | *a b d* | *c* | *a b d* | *d b a c* | < 0.1 s |
| 56 | A88 | R081 | 135 | 9 | 6 | *e h* | *c* (*d*) *e f g h* | --- | *a c e f g h* | *h e g c f a d b i* | < 0.1 s |
| 57 | A89 | R082 | 256 | 5 | 3 | *e acd*(1) | *a d e* | *c* | *a d e* | *e d a b c* | < 0.1 s |
| 58 | A90 | R083 | 366 | 20 | 12 | --- | *a* (*b*) *c d e f i l* (*n*) (*o*) *t* | --- | *a b c d e f i l n o s t* | *a i t l e c s d f n o b p j m g k r h q* | 49.0 s |
| 59 | A92 | R085 | 540 | 13 | 3 | *d* | *d f i* | --- | *d f i* | *d f i e b h a m c j g k l* | < 0.1 s |
| 60 | A93 | R086 | 561 | 4 | 2 | --- | *b d* | *a* | *b d* | *b d c a* | < 0.1 s |
| 61 | A94 | R087 | 579 | 4 | 2 | --- | *a d* | *b* | *a d* | *a d b c* | < 0.1 s |
| 62 | A95 | R088 | 587 | 7 | 2 | --- | *a* (*b*) | *c* | *a b* | *a b f g d e c* | < 0.1 s |
| 63 | A96 | R089 | 564 | 6 | 2 | --- | *a b* | *c* | *a b* | *a b e f d c* | < 0.1 s |
| 64 | A97 | R090 | 284 | 4 | 2 | --- | *a* (*b*) (*c*) | *d* | *a b* | *a b c d* | < 0.1 s |
| 65 | A98 | R091 | 279 | 4 | 2 | --- | *a c* | *d* | *a c* | *a c b d* | < 0.1 s |
| 66 | A99 | R092 | 275 | 4 | 2 | --- | *a b* | --- | *a b* | *b a c d* | < 0.1 s |

Table 9.4.1 (part 3 of 3): Schulze STV applied to instances of Tideman's database

In 45 instances (A01, A02, A05, A08, A10, A12, A14, A18, A19, A20, A23, A48, A49, A51, A52, A54, A56, A57, A63, A64, A66, A68, A69, A70, A71, A72, A73, A74, A76, A78, A79, A80, A81, A83, A84, A85, A86, A87, A89, A92, A93, A94, A96, A98, A99), there are $M$ Condorcet winners [according to (8.3.5)].

In 16 instances, there are $M$–1 Condorcet winners. For the remaining seat, there is a set $\emptyset \neq B \subsetneq A$ such that the Smith criterion for multi-winner elections (section 8.3) says that every winning set must contain at least one alternative from the set $B$. Table 9.4.2 lists these instances and the set $B$.

| | name #1 | $N$ | $C$ | $M$ | Condorcet winners | set $B$ |
|---|---|---|---|---|---|---|
| 1 | A03 | 989 | 15 | 7 | *b d e f h k* | *g n* |
| 2 | A04 | 43 | 14 | 2 | *i* | *a f k* |
| 3 | A06 | 280 | 9 | 5 | *c e h i* | *b f g* |
| 4 | A07 | 79 | 17 | 2 | *i* | *c d* |
| 5 | A13 | 104 | 26 | 2 | *t* | *i k* |
| 6 | A15 | 77 | 21 | 2 | *l* | *a b c d g h i j k m n p q r s t* |
| 7 | A17 | 867 | 13 | 8 | *a b d e f j l* | *i m* |
| 8 | A22 | 44 | 11 | 2 | *k* | *a b c d e g h i j* |
| 9 | A53 | 460 | 10 | 4 | *a g j* | *d f* |
| 10 | A55 | 302 | 10 | 5 | *a f i j* | *d e h* |
| 11 | A59 | 694 | 7 | 4 | *d f g* | *b e* |
| 12 | A65 | 198 | 10 | 6 | *b e f g j* | *a d* |
| 13 | A67 | 183 | 14 | 10 | *b c e f g i j k l* | *h m n* |
| 14 | A88 | 135 | 9 | 6 | *c e f g h* | *a b d* |
| 15 | A95 | 587 | 7 | 2 | *a* | *b f* |
| 16 | A97 | 284 | 4 | 2 | *a* | *b c* |

Table 9.4.2: Instances of Tideman's database with $M$–1 Condorcet winners

The 45 instances with $M$ Condorcet winners are interesting because, in these instances, we know that we can remove all other alternatives or any subset of the other alternatives and the result will not change. This observation follows directly from the definition of Condorcet winners.

The 16 instances with $M$–1 Condorcet winners and a set $B$ such that the last remaining winner must come from the set $B$ are interesting because, again, we know that we can remove all other alternatives or any subset of the other alternatives and the result will not change. The proof for this is identical to the proof that the Schulze method satisfies Smith-IIA (section 4.8).

Only in 5 instances (A11, A33, A34, A35, A90), there are less than $M$–1 Condorcet winners. This shows that we succeeded in defining the Condorcet criterion for multi-winner elections in such a manner that there are always many Condorcet winners (section 8.3).

## 10. Mixed Member Proportional Representation

Today, many countries use *proportional representation by party lists* (PRPL). This election method has two serious problems. First, it gives too much power to the party machines that control the nomination processes. Second, it allows proportionality only according to one criterion: party affiliations.

A possible way to circumvent the shortcomings of PRPL is through *proportional representation by the single transferable vote* (STV). To cope with the large number of candidates who are typically running for a parliament, the electorate is divided into districts of e.g. 7 seats each and the same candidate must not run in more than one district. Droop proportionality is then satisfied only on the district level, so that it can happen that a party with about 10% of the votes does not win a single seat.

Therefore, people who promote STV in countries that use PRPL are frequently confronted with the defamation that they were dishonest and that their real aim was to increase the threshold for small parties to gain parliamentary representation from a *legal* threshold of typically about 5% in countries that are using PRPL to a *natural* threshold of typically about 10% in countries that are using STV. Because of this reason, it might be a useful strategy to include into the STV proposal a compensation of party proportionality on the national level.

One way to achieve this compensation of party proportionality is through *mixed member proportional representation* (MMP). Here, each voter gets two ballots: a *district ballot* and a *party ballot*. With the district ballots, the voters elect their district representatives. With the party ballots, the voters indicate their party support. When, on the national level, the elected district representatives do not reflect party proportionality (as indicated by the party ballots) in an appropriate manner, then *additional representatives* are added to the parliament to compensate party proportionality (so that the total size of the parliament increases). Usually, these additional representatives are chosen from candidate lists that have been submitted by the parties in advance of the elections. The main problem of this way to choose these additional representatives is that, again, it gives too much power to the party machines that control the nomination processes.

One way to avoid this problem of the common way to choose these additional representatives is to use MMP with the “*best loser*” *method*. Here, when a party gets additional representatives, then these additional representatives are those candidates who performed best in their respective districts without being elected. Advantage of the “best loser” method is that, with their district votes, the voters do not only elect their district representatives, they also decide (in those cases where a party gets additional representatives) who these additional representatives are.

Where concrete numbers are needed, we use the elections to the *Berlin House of Representatives* (“*Abgeordnetenhaus von Berlin*”) to illustrate the proposed method. The constitution of Berlin says that its House has a standard size of 130 seats; this is usually interpreted as saying that the House must have 130 seats plus eventual overhang seats plus eventual compensation seats. We recommend that 80% of these seats should be filled by STV on the district level.

## 10.1. The Districts

Berlin is currently divided into 12 boroughs (sing.: *Bezirk*, plur.: *Bezirke*).

| | borough | eligible voters |
|---|---|---|
| 1 | Mitte | 197,148 |
| 2 | Friedrichshain-Kreuzberg | 171,249 |
| 3 | Pankow | 283,368 |
| 4 | Charlottenburg-Wilmersdorf | 216,762 |
| 5 | Spandau | 162,922 |
| 6 | Steglitz-Zehlendorf | 217,191 |
| 7 | Tempelhof-Schöneberg | 232,529 |
| 8 | Neukölln | 200,578 |
| 9 | Treptow-Köpenick | 199,830 |
| 10 | Marzahn-Hellersdorf | 202,868 |
| 11 | Lichtenberg | 203,709 |
| 12 | Reinickendorf | 181,562 |
| | total: | 2,425,480 |

Table 10.1.1: The 12 Berlin boroughs

(Source: Amt für Statistik Berlin-Brandenburg, “Statistischer Bericht B VII 2-1 – 5j / 16: Wahlen zum Abgeordnetenhaus von Berlin und zu den Bezirksverordnetenversammlungen 2016 — Vorwahldaten, Strukturdaten”, Potsdam, Germany, 2016, page 7 https://www.wahlen-berlin.de/Wahlen/Be2016/SB_B07-02-01_2016j05_BE.pdf)

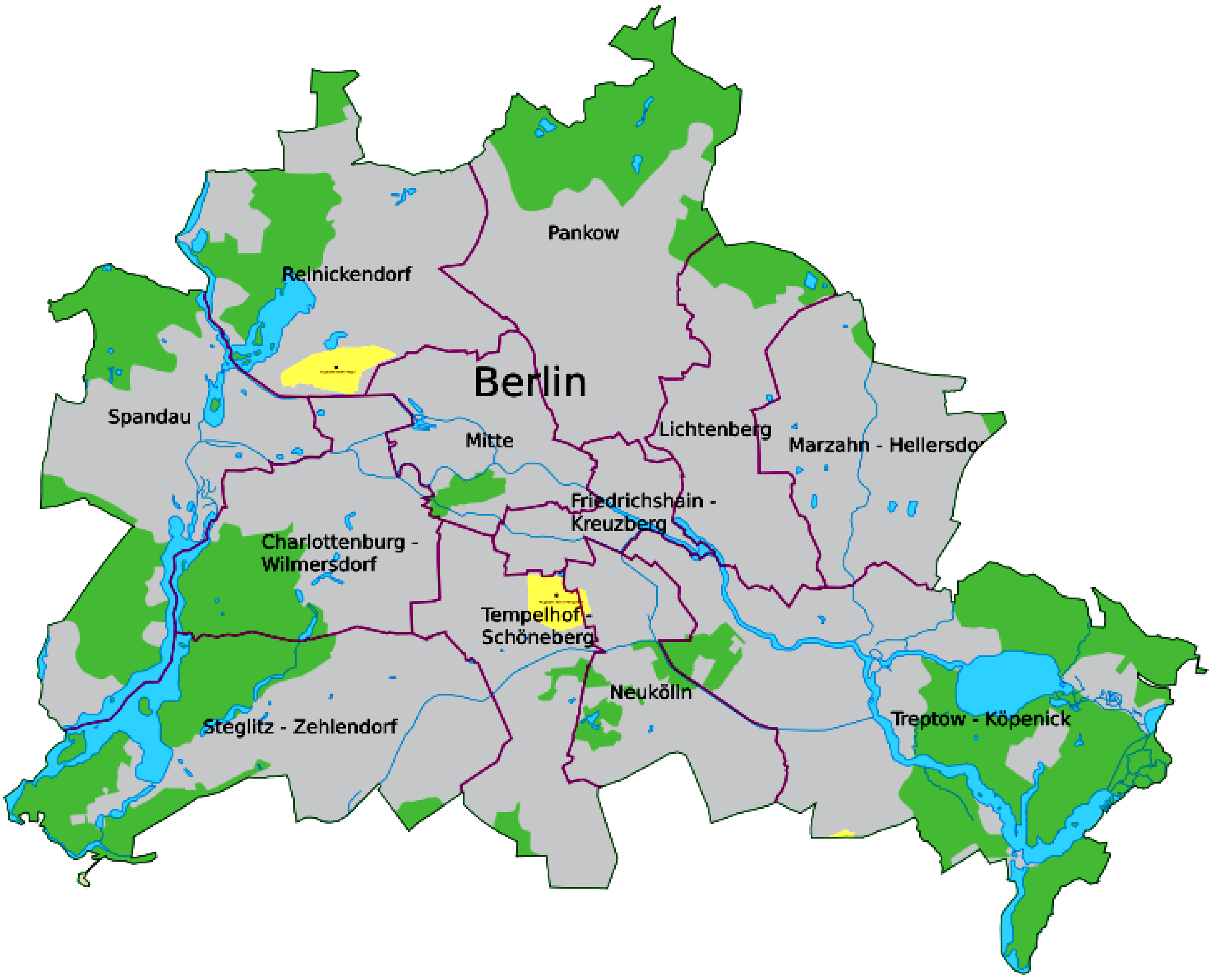

Graphic 10.1.2: The 12 Berlin boroughs

(Source: Von Nodder - Based on Water map of Berlin, CC BY-SA 3.0, https://commons.wikimedia.org/w/index.php?curid=4332728)

We recommend that the districts for the elections to the House should be the 12 Berlin boroughs. In the Hill-Huntington method, each districts first gets one seat; then the numbers of eligible voters for each district are divided by $\sqrt{(1 \cdot 2)}$, $\sqrt{(2 \cdot 3)}$, $\sqrt{(3 \cdot 4)}$, ... and the remaining seats go to the largest quotients. When the Hill-Huntington method is being used to allocate the 104 district seats to the 12 districts, then we get two 7-seat districts (Friedrichshain-Kreuzberg, Spandau), four 8-seat districts (Mitte, Neukölln, Treptow-Köpenick, Reinickendorf), four 9-seat districts (Charlottenburg-Wilmersdorf, Steglitz-Zehlendorf, Marzahn-Hellersdorf, Lichtenberg), one 10-seat district (Tempelhof-Schöneberg), and one 12-seat district (Pankow) [seats 1–104 in table 10.1.3]. The compensation seats go to Mitte (2 seats), Friedrichshain-Kreuzberg (2 seats), Pankow (3 seats), Charlottenburg-Wilmersdorf (2 seats), Spandau (2 seats), Steglitz-Zehlendorf (2 seats), Tempelhof-Schöneberg (2 seats), Neukölln (3 seats), Treptow-Köpenick (2 seats), Marzahn-Hellersdorf (2 seats), Lichtenberg (2 seats), and Reinickendorf (2 seats) [seats 105–130 in table 10.1.3]. If additional 15 compensation seats should be needed to achieve proportionality, then these seats go to Mitte (one seat), Friedrichshain-Kreuzberg (one seat), Pankow (one seat), Charlottenburg-Wilmersdorf (2 seats), Steglitz-Zehlendorf (2 seats), Tempelhof-Schöneberg (2 seats), Neukölln (one seat), Treptow-Köpenick (2 seats), Marzahn-Hellersdorf (one seat), Lichtenberg (one seat), and Reinickendorf (one seat) [seats 131–145 in table 10.1.3]. This guarantees that, regardless of the final number of seats, each district gets its proportional share of seats.

| district | eligible voters | number of eligible voters divided by ... | | | | | | | | | | | | | | | |
|---|---|---|---|---|---|---|---|---|---|---|---|---|---|---|---|---|---|
| | | ... √(1·2) | ... √(2·3) | ... √(3·4) | ... √(4·5) | ... √(5·6) | ... √(6·7) | ... √(7·8) | ... √(8·9) | ... √(9·10) | ... √(10·11) | ... √(11·12) | ... √(12·13) | ... √(13·14) | ... √(14·15) | ... √(15·16) | ...√(16·17) |
| Mitte | 197,148 | 139,405<br>(21. seat) | 80,485<br>(34. seat) | 56,912<br>(46. seat) | 44,084<br>(58. seat) | 35,994<br>(71. seat) | 30,421<br>(83. seat) | 26,345<br>(96. seat) | 23,234<br>(107. seat) | 20,781<br>(120. seat) | 18,797<br>(134. seat) | 17,160 | 15,784 | 14,614 | 13,605 | 12,726 | 11,954 |
| Friedrichshain-Kreuzberg | 171,249 | 121,091<br>(23. seat) | 69,912<br>(36. seat) | 49,435<br>(50. seat) | 38,292<br>(64. seat) | 31,266<br>(79. seat) | 26,424<br>(95. seat) | 22,884<br>(109. seat) | 20,182<br>(124. seat) | 18,051<br>(137. seat) | 16,328 | 14,905 | 13,711 | 12,694 | 11,817 | 11,054 | 10,383 |
| Pankow | 283,368 | 200,371<br>(13. seat) | 115,685<br>(24. seat) | 81,801<br>(32. seat) | 63,363<br>(39. seat) | 51,736<br>(49. seat) | 43,725<br>(59. seat) | 37,867<br>(65. seat) | 33,395<br>(75. seat) | 29,870<br>(84. seat) | 27,018<br>(92. seat) | 24,664<br>(100. seat) | 22,688<br>(111. seat) | 21,005<br>(119. seat) | 19,554<br>(125. seat) | 18,291<br>(136. seat) | 17,182 |
| Charlottenburg-Wilmersdorf | 216,762 | 153,274<br>(16. seat) | 88,493<br>(28. seat) | 62,574<br>(41. seat) | 48,469<br>(52. seat) | 39,575<br>(63. seat) | 33,447<br>(74. seat) | 28,966<br>(87. seat) | 25,546<br>(98. seat) | 22,849<br>(110. seat) | 20,667<br>(122. seat) | 18,867<br>(133. seat) | 17,355<br>(143. seat) | 16,067 | 14,958 | 13,992 | 13,143 |
| Spandau | 162,922 | 115,203<br>(25. seat) | 66,513<br>(38. seat) | 47,032<br>(53. seat) | 36,430<br>(70. seat) | 29,745<br>(85. seat) | 25,139<br>(99. seat) | 21,771<br>(113. seat) | 19,201<br>(128. seat) | 17,173 | 15,534 | 14,181 | 13,044 | 12,077 | 11,243 | 10,517 | 9,879 |
| Steglitz-Zehlendorf | 217,191 | 153,577<br>(15. seat) | 88,668<br>(27. seat) | 62,698<br>(40. seat) | 48,565<br>(51. seat) | 39,653<br>(62. seat) | 33,513<br>(73. seat) | 29,023<br>(86. seat) | 25,596<br>(97. seat) | 22,894<br>(108. seat) | 20,708<br>(121. seat) | 18,904<br>(132. seat) | 17,389<br>(142. seat) | 16,099 | 14,988 | 14,020 | 13,169 |
| Tempelhof-Schöneberg | 232,529 | 164,423<br>(14. seat) | 94,930<br>(26. seat) | 67,125<br>(37. seat) | 51,995<br>(48. seat) | 42,454<br>(60. seat) | 35,880<br>(72. seat) | 31,073<br>(80. seat) | 27,404<br>(89. seat) | 24,511<br>(101. seat) | 22,171<br>(112. seat) | 20,239<br>(123. seat) | 18,617<br>(135. seat) | 17,236<br>(145. seat) | 16,046 | 15,010 | 14,099 |
| Neukölln | 200,578 | 141,830<br>(19. seat) | 81,886<br>(31. seat) | 57,902<br>(44. seat) | 44,851<br>(56. seat) | 36,620<br>(68. seat) | 30,950<br>(81. seat) | 26,803<br>(93. seat) | 23,638<br>(105. seat) | 21,143<br>(117. seat) | 19,124<br>(130. seat) | 17,458<br>(140. seat) | 16,059 | 14,868 | 13,841 | 12,947 | 12,162 |
| Treptow-Köpenick | 199,830 | 141,301<br>(20. seat) | 81,580<br>(33. seat) | 57,686<br>(45. seat) | 44,683<br>(57. seat) | 36,484<br>(69. seat) | 30,834<br>(82. seat) | 26,703<br>(94. seat) | 23,550<br>(106. seat) | 21,064<br>(118. seat) | 19,053<br>(131. seat) | 17,393<br>(141. seat) | 15,999 | 14,812 | 13,790 | 12,899 | 12,116 |
| Marzahn-Hellersdorf | 202,868 | 143,449<br>(18. seat) | 82,821<br>(30. seat) | 58,563<br>(43. seat) | 45,363<br>(55. seat) | 37,038<br>(67. seat) | 31,303<br>(78. seat) | 27,109<br>(91. seat) | 23,908<br>(104. seat) | 21,384<br>(116. seat) | 19,343<br>(127. seat) | 17,657<br>(139. seat) | 16,242 | 15,038 | 13,999 | 13,095 | 12,301 |
| Lichtenberg | 203,709 | 144,044<br>(17. seat) | 83,164<br>(29. seat) | 58,806<br>(42. seat) | 45,551<br>(54. seat) | 37,192<br>(66. seat) | 31,433<br>(77. seat) | 27,222<br>(90. seat) | 24,007<br>(103. seat) | 21,473<br>(114. seat) | 19,423<br>(126. seat) | 17,731<br>(138. seat) | 16,310 | 15,100 | 14,057 | 13,149 | 12,352 |
| Reinickendorf | 181,562 | 128,384<br>(22. seat) | 74,122<br>(35. seat) | 52,412<br>(47. seat) | 40,598<br>(61. seat) | 33,149<br>(76. seat) | 28,016<br>(88. seat) | 24,262<br>(102. seat) | 21,397<br>(115. seat) | 19,138<br>(129. seat) | 17,311<br>(144. seat) | 15,803 | 14,537 | 13,458 | 12,529 | 11,720 | 11,009 |

Table 10.1.3: Allocation of the seats to the 12 districts according to the Hill-Huntington method

## 10.2. The District Ballot

The same candidate must not run in more than one district. The same candidate must not run simultaneously as an independent candidate and as a party candidate. The same candidate must not run for more than one party simultaneously.

On the district ballot, the candidates are sorted according to their party affiliations. Candidates with the same party affiliation are sorted in an order determined by this party. Each party can nominate as many candidates as it wants to.

The individual voter ranks the candidates in order of preference. The individual voter may ...

... give the same preference to more than one candidate.

... keep candidates unranked. When a given voter does not rank all candidates, then this means (1) that this voter strictly prefers all ranked candidates to all not ranked candidates and (2) that this voter is indifferent between all not ranked candidates.

... skip preferences. However, skipping some preferences does not have any impact on the result of the elections, since the result of the elections depends only on the order in which the individual voter ranks the candidates and not on the absolute preferences of the individual voters.

... give preferences to party labels. When a voter gives a preference to a party label, then this means that each candidate of this party gets this preference unless this voter explicitly gives a different preference to this candidate.

Graphic 10.2.1 shows how a cast district ballot could look like. Graphic 10.2.2 illustrates how this district ballot would be interpreted.

The winners of the district seats are calculated by Schulze STV (section 8).

Elections to the **Berlin House of Representatives**
on 17. September 2006

# District Ballot

for district Friedrichshain-Kreuzberg

*please rank the candidates*
*in order of preference* ⇩

| | |
|---|---|
| 01: **Social Democratic Party of Germany (SPD)** | *14* |
| 01.001: **Junge-Reyer**, Ingeborg | *23* |
| 01.002: **Zackenfels**, Stefan | *7* |
| 01.003: **Kitschun**, Susanne | |
| 01.004: **Eggert**, Björn | |
| 01.005: **Bayram**, Canan | |
| 01.006: **Fischer**, Silke | |
| 01.007: **Heinemann**, Sven | *23* |
| 01.008: **Miethke**, Petra | *23* |
| 01.009: **Kayhan**, Sevgi | *6* |
| 01.010: **Erdem**, Hediye | *8* |
| 01.011: **Klebba**, Sigrid | *13* |
| 01.012: **Postler**, Lorenz | |
| 01.013: **Hehmke**, Andy | |
| 01.014: **Dr. Beckers**, Peter | *19* |
| 01.015: **Lorenz**, Dorit | *24* |
| 01.016: **Borchard**, Andreas | |
| 02: **Christian Democratic Union of Germany (CDU)** | |
| 02.001: **Wansner**, Kurt | |
| 02.002: **Bleiler**, Rainer | |
| 02.003: **Ruhland**, Thomas | |
| 02.004: **Samuray**, Sedat | *16* |
| 02.005: **Stry**, Ernst-Uwe | |
| 02.006: **Rösner**, Helga | *14* |
| 02.007: **Glatzel**, Edgar | *20* |
| 02.008: **Schill**, Michael | |
| 02.009: **Müller**, Götz | |
| 02.010: **Freitag**, Jens-Matthias | |
| 02.011: **Husein**, Timur | |
| 02.012: **Wöhrn**, Marina | |
| 02.013: **Taşkıran**, Ertan | |
| 02.014: **Przewieslik**, Wolfgang | |
| 02.015: **Konschak**, Benjamin | |
| 02.016: **Bohl**, Daniel-Stephan | |
| 03: **Left Party** | *15* |
| 03.001: **Michels**, Martina | *11* |
| 03.002: **Wolf**, Udo | *14* |
| 03.003: **Matuschek**, Jutta | *23* |
| 03.004: **Zillich**, Steffen | *16* |
| 03.005: **İzgin**, Figen | *5* |
| 03.006: **Günther**, Andreas | |
| 03.007: **Vordenbäumen**, Vera | *13* |
| 03.008: **Krüger**, Wolfgang | |
| 03.009: **Reinauer**, Cornelia | |
| 03.010: **Bauer**, Kerstin | |
| 03.011: **Mildner-Spindler**, Knut | |
| 03.012: **Richter**, Claudia | *6* |
| 03.013: **Schüssler**, Lothar | *16* |
| 03.014: **Thimm**, Helga | *10* |
| 03.015: **Pempel**, Joachim | *21* |
| 03.016: **Sommer-Wetter**, Regine | |
| 04: **Alliance ‘90 / The Greens (B‘90G)** | *15* |
| 04.001: **Ratzmann**, Volker | *16* |
| 04.002: **Mutlu**, Özcan | *2* |
| 04.003: **Dr. Klotz**, Sibyll-Anka | *12* |
| 04.004: **Lux**, Benedikt | |
| 04.005: **Herrmann**, Clara | *17* |
| 04.006: **Stephan**, André | *9* |
| 04.007: **Pohner**, Wolfgang | |
| 04.008: **Dr. Altug**, Mehmet | |
| 04.009: **Burkert-Eulitz**, Marianne | *13* |
| 04.010: **Kosche**, Heidi | *23* |
| 04.011: **Behrendt**, Dirk | *1* |
| 04.012: **Hauser-Jabs**, Christine | *12* |
| 04.013: **Schulz**, Franz | *4* |
| 04.014: **Kapek**, Antje | *3* |
| 04.015: **Wesener**, Daniel | *22* |
| 04.016: **Çetinkaya**, İstikbal | *13* |
| 05: **Free Democratic Party of Germany (FDP)** | *20* |
| 05.001: **Peters**, Frank | *21* |
| 05.002: **Dr. Hansen**, Nikoline | |
| 05.003: **Eydner**, John | |
| 05.004: **Hohl**, Heinrich | |
| 05.005: **Salonek**, Gumbert-Olaf | *23* |
| 05.006: **Diener**, Thomas | |
| 05.007: **Schaefer**, Martina | |
| 05.008: **Wolf**, Tobias | |
| 05.009: **Dr. Stolz**, Peter | |
| 05.010: **Lauf**, Sebastian | |
| 05.011: **Paun**, Christopher | *21* |
| 05.012: **Joecken**, Ilka | |
| 06: **The Republicans (REP)** | |
| 06.001: **Dr. Clemens**, Björn | |
| 06.002: **Kuhn**, Daniel | |
| 06.003: **Hinze**, Harald Björn Gunnar | |
| 06.004: **Nestmann**, Günther | |
| 07: **Ecological Democratic Party (ödp)**<br>07.001: **Machel-Ebeling**, Johannes | *18* |
| 08: **Civil Rights Movement Solidarity (BüSo)**<br>08.001: **Hinz**, Björn | *22* |
| 09: **Humane Economy Party**<br>09.001: **Dr. Heinrichs**, Johannes | *18* |
| 10.001: **Eisner**, Udo (independent) | *21* |
| 11.001: **Stiewe**, Hauke (independent) | *20* |

Graphic 10.2.1: Example of a cast district ballot

Elections to the **Berlin House of Representatives**
on 17. September 2006

# District Ballot

for district Friedrichshain-Kreuzberg

*please rank the candidates in order of preference* ⇩

| | |
|---|---|
| 01: **Social Democratic Party of Germany (SPD)** | |
| 01.001: **Junge-Reyer**, Ingeborg | *23* |
| 01.002: **Zackenfels**, Stefan | *7* |
| 01.003: **Kitschun**, Susanne | *14* |
| 01.004: **Eggert**, Björn | *14* |
| 01.005: **Bayram**, Canan | *14* |
| 01.006: **Fischer**, Silke | *14* |
| 01.007: **Heinemann**, Sven | *23* |
| 01.008: **Miethke**, Petra | *23* |
| 01.009: **Kayhan**, Sevgi | *6* |
| 01.010: **Erdem**, Hediye | *8* |
| 01.011: **Klebba**, Sigrid | *13* |
| 01.012: **Postler**, Lorenz | *14* |
| 01.013: **Hehmke**, Andy | *14* |
| 01.014: **Dr. Beckers**, Peter | *19* |
| 01.015: **Lorenz**, Dorit | *24* |
| 01.016: **Borchard**, Andreas | *14* |
| 02: **Christian Democratic Union of Germany (CDU)** | |
| 02.001: **Wansner**, Kurt | *25* |
| 02.002: **Bleiler**, Rainer | *25* |
| 02.003: **Ruhland**, Thomas | *25* |
| 02.004: **Samuray**, Sedat | *16* |
| 02.005: **Stry**, Ernst-Uwe | *25* |
| 02.006: **Rösner**, Helga | *14* |
| 02.007: **Glatzel**, Edgar | *20* |
| 02.008: **Schill**, Michael | *25* |
| 02.009: **Müller**, Götz | *25* |
| 02.010: **Freitag**, Jens-Matthias | *25* |
| 02.011: **Husein**, Timur | *25* |
| 02.012: **Wöhrn**, Marina | *25* |
| 02.013: **Taşkıran**, Ertan | *25* |
| 02.014: **Przewieslik**, Wolfgang | *25* |
| 02.015: **Konschak**, Benjamin | *25* |
| 02.016: **Bohl**, Daniel-Stephan | *25* |
| 03: **Left Party** | |
| 03.001: **Michels**, Martina | *11* |
| 03.002: **Wolf**, Udo | *14* |
| 03.003: **Matuschek**, Jutta | *23* |
| 03.004: **Zillich**, Steffen | *16* |
| 03.005: **İzgin**, Figen | *5* |
| 03.006: **Günther**, Andreas | *15* |
| 03.007: **Vordenbäumen**, Vera | *13* |
| 03.008: **Krüger**, Wolfgang | *15* |
| 03.009: **Reinauer**, Cornelia | *15* |
| 03.010: **Bauer**, Kerstin | *15* |
| 03.011: **Mildner-Spindler**, Knut | *15* |
| 03.012: **Richter**, Claudia | *6* |
| 03.013: **Schüssler**, Lothar | 16 |
| 03.014: **Thimm**, Helga | 10 |
| 03.015: **Pempel**, Joachim | 21 |
| 03.016: **Sommer-Wetter**, Regine | 15 |
| 04: **Alliance ‘90 / The Greens (B‘90G)** | |
| 04.001: **Ratzmann**, Volker | 16 |
| 04.002: **Mutlu**, Özcan | 2 |
| 04.003: **Dr. Klotz**, Sibyll-Anka | 12 |
| 04.004: **Lux**, Benedikt | 15 |
| 04.005: **Herrmann**, Clara | 17 |
| 04.006: **Stephan**, André | 9 |
| 04.007: **Pohner**, Wolfgang | 15 |
| 04.008: **Dr. Altug**, Mehmet | 15 |
| 04.009: **Burkert-Eulitz**, Marianne | 13 |
| 04.010: **Kosche**, Heidi | 23 |
| 04.011: **Behrendt**, Dirk | 1 |
| 04.012: **Hauser-Jabs**, Christine | 12 |
| 04.013: **Schulz**, Franz | 4 |
| 04.014: **Kapek**, Antje | 3 |
| 04.015: **Wesener**, Daniel | 22 |
| 04.016: **Çetinkaya**, İstikbal | 13 |
| 05: **Free Democratic Party of Germany (FDP)** | |
| 05.001: **Peters**, Frank | 21 |
| 05.002: **Dr. Hansen**, Nikoline | 20 |
| 05.003: **Eydner**, John | 20 |
| 05.004: **Hohl**, Heinrich | 20 |
| 05.005: **Salonek**, Gumbert-Olaf | 23 |
| 05.006: **Diener**, Thomas | 20 |
| 05.007: **Schaefer**, Martina | 20 |
| 05.008: **Wolf**, Tobias | 20 |
| 05.009: **Dr. Stolz**, Peter | 20 |
| 05.010: **Lauf**, Sebastian | 20 |
| 05.011: **Paun**, Christopher | 21 |
| 05.012: **Joecken**, Ilka | 20 |
| 06: **The Republicans (REP)** | |
| 06.001: **Dr. Clemens**, Björn | 25 |
| 06.002: **Kuhn**, Daniel | 25 |
| 06.003: **Hinze**, Harald Björn Gunnar | 25 |
| 06.004: **Nestmann**, Günther | 25 |
| 07: **Ecological Democratic Party (ödp)**<br>07.001: **Machel-Ebeling**, Johannes | 18 |
| 08: **Civil Rights Movement Solidarity (BüSo)**<br>08.001: **Hinz**, Björn | 22 |
| 09: **Humane Economy Party**<br>09.001: **Dr. Heinrichs**, Johannes | 18 |
| 10.001: **Eisner**, Udo (independent) | 21 |
| 11.001: **Stiewe**, Hauke (independent) | 20 |

Graphic 10.2.2: Interpretation of the cast district ballot

## 10.3. The Party Ballot

On the party ballot of a given district, all those parties are listed that have nominated district candidates. The individual voter can vote for one and only one party. Graphic 10.3.1 shows how a party ballot for district Friedrichshain-Kreuzberg could look like.

The district ballot and the party ballot are on the same sheet of paper. This is necessary because the weight of a voter’s party vote can depend on this voter’s district vote.

Elections to the **Berlin House of Representatives**
on 17. September 2006

**Party Ballot**

for district Friedrichshain-Kreuzberg

*please vote for one and only one party* ⇩

| | |
|---|---|
| 01: Social Democratic Party of Germany (SPD) | |
| 02: Christian Democratic Union of Germany (CDU) | |
| 03: Left Party | |
| 04: Alliance ‘90 / The Greens (B‘90G) | |
| 05: Free Democratic Party of Germany (FDP) | |
| 06: The Republicans (REP) | |
| 07: Ecological Democratic Party (ödp) | |
| 08: Civil Rights Movement Solidarity (BüSo) | |
| 09: Humane Economy Party | |

Graphic 10.3.1: Party ballot for the district Friedrichshain-Kreuzberg

## 10.4. Allocation of District Voters to District Winners

For each district, we calculate an allocation of the voters of this district to the district winners of this district. The purpose of this allocation will be explained in section 10.5.

Table 10.4.1 shows the voting patterns in example A53 before and after proportional completion (restricted to the district winners).

| before proportional completion | | | | | | voters | after proportional completion | | | | | |
|---|---|---|---|---|---|---|---|---|---|---|---|---|
| voting pattern | number of voters | *a* | *d* | *g* | *j* | | number of voters | *a* | *d* | *g* | *j* | voting pattern |
| 1 | 8 | 1 | 3 | 2 | 3 | 1, 41, 230, 238, 402, 440, 443, 451 | 2.556098 | 1 | 3 | 2 | 4 | 1.001 |
| | | | | | | | 5.443902 | 1 | 4 | 2 | 3 | 1.002 |
| 2 | 8 | 1 | 2 | 3 | 4 | 2, 39, 163, 165, 411, 423, 436, 457 | 8.000000 | 1 | 2 | 3 | 4 | 2.001 |
| 3 | 14 | 1 | 4 | 3 | 2 | 3, 17, 52, 211, 250, 279, 408, 410, 418, 419, 430, 433, 444, 450 | 14.000000 | 1 | 4 | 3 | 2 | 3.001 |
| 4 | 17 | 2 | 2 | 2 | 1 | 4, 5, 65, 189, 209, 210, 239, 252, 259, 261, 273, 323, 324, 326, 390, 396, 399 | 3.885714 | 2 | 3 | 4 | 1 | 4.001 |
| | | | | | | | 3.891297 | 2 | 4 | 3 | 1 | 4.002 |
| | | | | | | | 2.067271 | 3 | 2 | 4 | 1 | 4.003 |
| | | | | | | | 3.424539 | 3 | 4 | 2 | 1 | 4.004 |
| | | | | | | | 1.332729 | 4 | 2 | 3 | 1 | 4.005 |
| | | | | | | | 2.398450 | 4 | 3 | 2 | 1 | 4.006 |
| 5 | 29 | 2 | 4 | 3 | 1 | 6, 19, 26, 57, 181, 186, 187, 190, 194, 229, 236, 266, 271, 278, 282, 290, 291, 293, 295, 297, 301, 309, 310, 313, 320, 322, 365, 373, 381 | 29.000000 | 2 | 4 | 3 | 1 | 5.001 |
| 6 | 22 | 2 | 3 | 3 | 1 | 7, 8, 10, 56, 179, 182, 183, 207, 208, 280, 294, 296, 299, 303, 305, 306, 308, 314, 385, 386, 393, 398 | 9.428571 | 2 | 3 | 4 | 1 | 6.001 |
| | | | | | | | 12.571429 | 2 | 4 | 3 | 1 | 6.002 |
| 7 | 10 | 1 | 2 | 2 | 2 | 9, 51, 231, 272, 327, 371, 382, 412, 414, 417 | 0.805282 | 1 | 2 | 3 | 4 | 7.001 |
| | | | | | | | 1.185668 | 1 | 2 | 4 | 3 | 7.002 |
| | | | | | | | 1.204172 | 1 | 3 | 2 | 4 | 7.003 |
| | | | | | | | 2.294764 | 1 | 3 | 4 | 2 | 7.004 |
| | | | | | | | 1.985874 | 1 | 4 | 2 | 3 | 7.005 |
| | | | | | | | 2.524240 | 1 | 4 | 3 | 2 | 7.006 |
| 8 | 12 | 3 | 3 | 1 | 2 | 11, 15, 18, 81, 95, 97, 198, 215, 242, 253, 263, 377 | 7.906977 | 3 | 4 | 1 | 2 | 8.001 |
| | | | | | | | 4.093023 | 4 | 3 | 1 | 2 | 8.002 |
| 9 | 21 | 2 | 2 | 1 | 2 | 12, 16, 85, 88, 93, 197, 199, 203, 213, 214, 237, 336, 338, 340, 345, 364, 368, 384, 387, 394, 397 | 2.568225 | 2 | 3 | 1 | 4 | 9.001 |
| | | | | | | | 4.155673 | 2 | 4 | 1 | 3 | 9.002 |
| | | | | | | | 2.115518 | 3 | 2 | 1 | 4 | 9.003 |
| | | | | | | | 7.113311 | 3 | 4 | 1 | 2 | 9.004 |
| | | | | | | | 2.026013 | 4 | 2 | 1 | 3 | 9.005 |
| | | | | | | | 3.021259 | 4 | 3 | 1 | 2 | 9.006 |
| 10 | 15 | 3 | 4 | 2 | 1 | 13, 20, 62, 185, 188, 235, 249, 284, 349, 353, 354, 355, 361, 366, 391 | 15.000000 | 3 | 4 | 2 | 1 | 10.001 |
| 11 | 15 | 2 | 4 | 1 | 3 | 14, 72, 80, 83, 86, 89, 91, 102, 220, 270, 347, 369, 370, 374, 380 | 15.000000 | 2 | 4 | 1 | 3 | 11.001 |
| 12 | 34 | 2 | 3 | 4 | 1 | 21, 22, 23, 27, 58, 59, 61, 63, 64, 168, 191, 223, 225, 226, 227, 277, 286, 287, 288, 289, 292, 298, 300, 302, 304, 307, 311, 312, 315, 316, 317, 318, 319, 321 | 34.000000 | 2 | 3 | 4 | 1 | 12.001 |
| 13 | 7 | 3 | 2 | 3 | 1 | 24, 25, 164, 174, 222, 268, 285 | 4.053528 | 3 | 2 | 4 | 1 | 13.001 |
| | | | | | | | 2.946472 | 4 | 2 | 3 | 1 | 13.002 |

Table 10.4.1 (part 1 of 3): the voting patterns in example A53

| before proportional completion | | | | | | voters | after proportional completion | | | | | |
|---|---|---|---|---|---|---|---|---|---|---|---|---|
| voting pattern | number of voters | *a* | *d* | *g* | *j* | | number of voters | *a* | *d* | *g* | *j* | voting pattern |
| 14 | 14 | 3 | 2 | 4 | 1 | 28, 29, 31, 33, 35, 151, 167, 169, 193, 224, 228, 264, 274, 283 | 14.000000 | 3 | 2 | 4 | 1 | 14.001 |
| 15 | 7 | 4 | 2 | 3 | 1 | 30, 32, 34, 152, 153, 166, 244 | 7.000000 | 4 | 2 | 3 | 1 | 15.001 |
| 16 | 13 | 1 | 4 | 2 | 3 | 36, 53, 55, 68, 202, 247, 376, 405, 416, 420, 422, 431, 437 | 13.000000 | 1 | 4 | 2 | 3 | 16.001 |
| 17 | 18 | 1 | 3 | 4 | 2 | 37, 38, 45, 48, 49, 161, 243, 246, 265, 275, 407, 439, 445, 449, 452, 454, 455, 458 | 18.000000 | 1 | 3 | 4 | 2 | 17.001 |
| 18 | 16 | 1 | 2 | 4 | 3 | 40, 42, 44, 46, 160, 175, 375, 401, 403, 404, 424, 429, 441, 442, 453, 460 | 16.000000 | 1 | 2 | 4 | 3 | 18.001 |
| 19 | 5 | 1 | 2 | 3 | 3 | 43, 171, 379, 447, 459 | 1.997664 | 1 | 2 | 3 | 4 | 19.001 |
| | | | | | | | 3.002336 | 1 | 2 | 4 | 3 | 19.002 |
| 20 | 11 | 1 | 3 | 2 | 4 | 47, 50, 173, 201, 206, 383, 409, 413, 421, 428, 438 | 11.000000 | 1 | 3 | 2 | 4 | 20.001 |
| 21 | 11 | 1 | 3 | 3 | 2 | 54, 406, 415, 425, 426, 427, 432, 434, 435, 446, 456 | 4.714286 | 1 | 3 | 4 | 2 | 21.001 |
| | | | | | | | 6.285714 | 1 | 4 | 3 | 2 | 21.002 |
| 22 | 10 | 4 | 3 | 2 | 1 | 60, 192, 196, 281, 356, 357, 358, 359, 362, 367 | 10.000000 | 4 | 3 | 2 | 1 | 22.001 |
| 23 | 9 | 4 | 3 | 1 | 2 | 66, 73, 74, 100, 101, 176, 205, 217, 328 | 9.000000 | 4 | 3 | 1 | 2 | 23.001 |
| 24 | 8 | 3 | 2 | 1 | 3 | 67, 75, 82, 96, 105, 221, 332, 339 | 3.367397 | 3 | 2 | 1 | 4 | 24.001 |
| | | | | | | | 4.632603 | 4 | 2 | 1 | 3 | 24.002 |
| 25 | 17 | 3 | 4 | 1 | 2 | 69, 70, 71, 78, 79, 84, 92, 98, 106, 216, 219, 234, 245, 251, 333, 372, 378 | 17.000000 | 3 | 4 | 1 | 2 | 25.001 |
| 26 | 10 | 4 | 2 | 1 | 3 | 76, 90, 107, 218, 257, 330, 331, 337, 341, 348 | 10.000000 | 4 | 2 | 1 | 3 | 26.001 |
| 27 | 3 | 2 | 3 | 1 | 3 | 77, 212, 335 | 0.958537 | 2 | 3 | 1 | 4 | 27.001 |
| | | | | | | | 2.041463 | 2 | 4 | 1 | 3 | 27.002 |
| 28 | 8 | 3 | 2 | 1 | 4 | 87, 94, 99, 103, 258, 329, 334, 344 | 8.000000 | 3 | 2 | 1 | 4 | 28.001 |
| 29 | 6 | 2 | 3 | 1 | 4 | 104, 200, 255, 342, 343, 346 | 6.000000 | 2 | 3 | 1 | 4 | 29.001 |
| 30 | 6 | 3 | 1 | 4 | 2 | 108, 112, 130, 157, 170, 178 | 6.000000 | 3 | 1 | 4 | 2 | 30.001 |
| 31 | 9 | 2 | 1 | 2 | 2 | 109, 124, 126, 139, 144, 145, 154, 159, 162 | 1.035975 | 2 | 1 | 3 | 4 | 31.001 |
| | | | | | | | 1.666057 | 2 | 1 | 4 | 3 | 31.002 |
| | | | | | | | 1.086290 | 3 | 1 | 2 | 4 | 31.003 |
| | | | | | | | 2.509647 | 3 | 1 | 4 | 2 | 31.004 |
| | | | | | | | 1.473530 | 4 | 1 | 2 | 3 | 31.005 |
| | | | | | | | 1.228502 | 4 | 1 | 3 | 2 | 31.006 |
| 32 | 14 | 2 | 1 | 4 | 3 | 110, 116, 121, 128, 129, 131, 134, 137, 146, 147, 149, 150, 233, 254 | 14.000000 | 2 | 1 | 4 | 3 | 32.001 |
| 33 | 8 | 3 | 1 | 3 | 2 | 111, 115, 117, 132, 133, 138, 148, 155 | 4.632603 | 3 | 1 | 4 | 2 | 33.001 |
| | | | | | | | 3.367397 | 4 | 1 | 3 | 2 | 33.002 |

Table 10.4.1 (part 2 of 3): the voting patterns in example A53

| before proportional completion | | | | | | voters | after proportional completion | | | | | |
|---|---|---|---|---|---|---|---|---|---|---|---|---|
| voting pattern | number of voters | *a* | *d* | *g* | *j* | | number of voters | *a* | *d* | *g* | *j* | voting pattern |
| 34 | 3 | 3 | 1 | 2 | 3 | 113, 120, 125 | 1.262774 | 3 | 1 | 2 | 4 | 34.001 |
| | | | | | | | 1.737226 | 4 | 1 | 2 | 3 | 34.002 |
| 35 | 6 | 4 | 1 | 2 | 3 | 114, 122, 140, 156, 172, 267 | 6.000000 | 4 | 1 | 2 | 3 | 35.001 |
| 36 | 8 | 3 | 1 | 2 | 4 | 118, 123, 141, 142, 143, 158, 177, 256 | 8.000000 | 3 | 1 | 2 | 4 | 36.001 |
| 37 | 5 | 2 | 1 | 3 | 4 | 119, 127, 135, 136, 241 | 5.000000 | 2 | 1 | 3 | 4 | 37.001 |
| 38 | 15 | 3 | 3 | 2 | 1 | 180, 184, 195, 204, 232, 260, 269, 350, 351, 352, 360, 363, 388, 389, 392 | 9.883721 | 3 | 4 | 2 | 1 | 38.001 |
| | | | | | | | 5.116279 | 4 | 3 | 2 | 1 | 38.002 |
| 39 | 8 | 1 | 1 | 1 | 1 | 240, 248, 262, 276, 325, 395, 400, 448 | 0.191203 | 1 | 2 | 3 | 4 | 39.001 |
| | | | | | | | 0.357310 | 1 | 2 | 4 | 3 | 39.002 |
| | | | | | | | 0.261244 | 1 | 3 | 2 | 4 | 39.003 |
| | | | | | | | 0.442638 | 1 | 3 | 4 | 2 | 39.004 |
| | | | | | | | 0.361589 | 1 | 4 | 2 | 3 | 39.005 |
| | | | | | | | 0.403716 | 1 | 4 | 3 | 2 | 39.006 |
| | | | | | | | 0.106831 | 2 | 1 | 3 | 4 | 39.007 |
| | | | | | | | 0.277275 | 2 | 1 | 4 | 3 | 39.008 |
| | | | | | | | 0.168615 | 2 | 3 | 1 | 4 | 39.009 |
| | | | | | | | 0.837421 | 2 | 3 | 4 | 1 | 39.010 |
| | | | | | | | 0.375171 | 2 | 4 | 1 | 3 | 39.011 |
| | | | | | | | 0.804650 | 2 | 4 | 3 | 1 | 39.012 |
| | | | | | | | 0.183169 | 3 | 1 | 2 | 4 | 39.013 |
| | | | | | | | 0.232606 | 3 | 1 | 4 | 2 | 39.014 |
| | | | | | | | 0.238636 | 3 | 2 | 1 | 4 | 39.015 |
| | | | | | | | 0.356120 | 3 | 2 | 4 | 1 | 39.016 |
| | | | | | | | 0.566731 | 3 | 4 | 1 | 2 | 39.017 |
| | | | | | | | 0.501031 | 3 | 4 | 2 | 1 | 39.018 |
| | | | | | | | 0.163022 | 4 | 1 | 2 | 3 | 39.019 |
| | | | | | | | 0.081343 | 4 | 1 | 3 | 2 | 39.020 |
| | | | | | | | 0.294843 | 4 | 2 | 1 | 3 | 39.021 |
| | | | | | | | 0.199632 | 4 | 2 | 3 | 1 | 39.022 |
| | | | | | | | 0.285209 | 4 | 3 | 1 | 2 | 39.023 |
| | | | | | | | 0.309995 | 4 | 3 | 2 | 1 | 39.024 |
| | 460 | | | | | | 460.000000 | | | | | |

Table 10.4.1 (part 3 of 3): the voting patterns in example A53

To calculate the allocation of the voters to the winners, we proceed as follows:

Step #1: Suppose $a_m$ is the number of voters with voting profile $m$. For example, line 16 of table 10.4.2 says that, after proportional completion, $a_m$ = 20.476919 voters have the voting pattern $j \succ d \succ a \succ g$.

Suppose $x_{mn}$ is the share of voter $m$ that is allocated to candidate $n$. Then we must have

(10.4.1) $$\sum_{m=1}^{24} (a_m \cdot x_{mn}) = 115$$ for every candidate $n$.

(10.4.2) $$\sum_{n=1}^{4} x_{mn} = 1$$ for every voter $m$.

(10.4.3) $$x_{mn} \geq 0$$ for every voter $m$ and every candidate $n$.

We minimize [ subject to (10.4.1) – (10.4.3) ] max { $x_{mr}$ } where candidate $r$ is the fourth preference of voter $m$. We then fix the $x_{mr}$ with maximum $x_{mr}$. We then minimize [ subject to (10.4.1) – (10.4.3) and subject to those $x_{mr}$ that have already been fixed ] max { $x_{mr}$ } (again where candidate $r$ is the fourth preference of voter $m$) for the remaining voters. We proceed until $x_{mr}$ is fixed wherever candidate $r$ is the fourth preference of voter $m$.

For example A53, we get $x_{mr}$ = 0 wherever candidate $r$ is the fourth preference of voter $m$. So a voter is never allocated to his fourth preference.

Step #2: We minimize [ subject to (10.4.1) – (10.4.3) and subject to those $x_{mr}$ that have been fixed in Step #1 ] max { $x_{mr} + x_{ms}$ } where candidate $r$ is the fourth preference and candidate $s$ is the third preference of voter $m$. We fix the $x_{ms}$ with maximum $x_{mr} + x_{ms}$. We then minimize [ subject to (10.4.1) – (10.4.3) and subject to those $x_{mr}$ and $x_{ms}$ that have already been fixed ] max { $x_{mr} + x_{ms}$ } for the remaining voters. We proceed until $x_{ms}$ is fixed wherever candidate $s$ is the third preference of voter $m$.

For example A53, we get $x_{ms}$ = 0 wherever candidate $s$ is the third preference of voter $m$. So a voter is never allocated to his third preference.

Step #3: We minimize [ subject to (10.4.1) – (10.4.3) and subject to those $x_{mr}$ and $x_{ms}$ that have been fixed in Step #1 and Step #2 ] max { $x_{mr} + x_{ms} + x_{mt}$ } where candidate $r$ is the fourth preference, candidate $s$ is the third preference, and candidate $t$ is the second preference of voter $m$. We fix the $x_{mt}$ with maximum $x_{mr} + x_{ms} + x_{mt}$. We then minimize [ subject to (10.4.1) – (10.4.3) and subject to those $x_{mr}$, $x_{ms}$, and $x_{mt}$ that have already been fixed ] max { $x_{mr} + x_{ms} + x_{mt}$ } for the remaining voters. We proceed until $x_{mt}$ is fixed wherever candidate $t$ is the second preference of voter $m$.

For example A53, we get max { $x_{mr}$ + $x_{ms}$ + $x_{mt}$ } = 0.583579 where candidate $r$ is the fourth preference, candidate $s$ is the third preference, and candidate $t$ is the second preference of voter $m$. So 58% of these votes are allocated to this voter’s fourth, third or second preference. See the profiles #1, #2, #15, #16, #21, and #22 in table 10.4.2. We fix $x_{mr} + x_{ms} + x_{mt}$ for these voters.

Next, we minimize [ subject to (10.4.1) – (10.4.3) and subject to those $x_{mr}$, $x_{ms}$ and $x_{mt}$ that have already been fixed ] max { $x_{mr} + x_{ms} + x_{mt}$ } for the remaining voters. We get max { $x_{mr} + x_{ms} + x_{mt}$ } = 0.279045 where candidate $r$ is the fourth preference, candidate $s$ is the third preference, and candidate $t$ is the second preference of voter $m$. So 28% of these votes are allocated to this voter’s fourth, third or second preference. See the profiles #10, #12, #18, and #24 in table 10.4.2. We now fix $x_{mr} + x_{ms} + x_{mt}$ also for these voters.

Next, we minimize [ subject to (10.4.1) – (10.4.3) and subject to those $x_{mr}$, $x_{ms}$ and $x_{mt}$ that have already been fixed ] max { $x_{mr} + x_{ms} + x_{mt}$ } for the remaining voters. We get max { $x_{mr} + x_{ms} + x_{mt}$ } = 0.250164 where candidate $r$ is the fourth preference, candidate $s$ is the third preference, and candidate $t$ is the second preference of voter $m$. So 25% of these votes are allocated to this voter’s fourth, third or second preference. See the profiles #3 and #5 in table 10.4.2. We now fix $x_{mr} + x_{ms} + x_{mt}$ also for these voters.

Next, we minimize [ subject to (10.4.1) – (10.4.3) and subject to those $x_{mr}$, $x_{ms}$ and $x_{mt}$ that have already been fixed ] max { $x_{mr} + x_{ms} + x_{mt}$ } for the remaining voters. We get max { $x_{mr} + x_{ms} + x_{mt}$ } = 0.000000 where candidate $r$ is the fourth preference, candidate $s$ is the third preference, and candidate $t$ is the second preference of voter $m$. So 0% of these votes are allocated to this voter’s fourth, third or second preference. See the profiles #4, #6, #7, #8, #9, #11, #13, #14, #17, #19, #20, and #23 in table 10.4.2.

The above allocation procedure is motivated by the fact that the winners of an STV election are spaced roughly at equal distance.

| | number of voters | | | | | share | | | | votes | | | |
|---|---|---|---|---|---|---|---|---|---|---|---|---|---|
| | | *a* | *d* | *g* | *j* | *a* | *d* | *g* | *j* | *a* | *d* | *g* | *j* |
| 1 | 10.994148 | 1 | 2 | 3 | 4 | 0.416421 | 0.583579 | | | 4.578195 | 6.415952 | | |
| 2 | 20.545315 | 1 | 2 | 4 | 3 | 0.416421 | 0.583579 | | | 8.555500 | 11.989811 | | |
| 3 | 15.021513 | 1 | 3 | 2 | 4 | 0.749836 | | 0.250164 | | 11.263672 | | 3.757842 | |
| 4 | 25.451689 | 1 | 3 | 4 | 2 | 1.000000 | | | | 25.451688 | | | |
| 5 | 20.791365 | 1 | 4 | 2 | 3 | 0.749836 | | 0.250164 | | 15.590114 | | 5.201251 | |
| 6 | 23.213671 | 1 | 4 | 3 | 2 | 1.000000 | | | | 23.213670 | | | |
| 7 | 6.142806 | 2 | 1 | 3 | 4 | | 1.000000 | | | | 6.142806 | | |
| 8 | 15.943332 | 2 | 1 | 4 | 3 | | 1.000000 | | | | 15.943332 | | |
| 9 | 9.695377 | 2 | 3 | 1 | 4 | | | 1.000000 | | | | 9.695377 | |
| 10 | 48.151707 | 2 | 3 | 4 | 1 | 0.279045 | | | 0.720955 | 13.436493 | | | 34.715213 |
| 11 | 21.572307 | 2 | 4 | 1 | 3 | | | 1.000000 | | | | 21.572307 | |
| 12 | 46.267376 | 2 | 4 | 3 | 1 | 0.279045 | | | 0.720955 | 12.910680 | | | 33.356696 |
| 13 | 10.532233 | 3 | 1 | 2 | 4 | | 1.000000 | | | | 10.532233 | | |
| 14 | 13.374857 | 3 | 1 | 4 | 2 | | 1.000000 | | | | 13.374856 | | |
| 15 | 13.721550 | 3 | 2 | 1 | 4 | | 0.583579 | 0.416421 | | | 8.007609 | 5.713942 | |
| 16 | 20.476919 | 3 | 2 | 4 | 1 | | 0.583579 | | 0.416421 | | 11.949900 | | 8.527019 |
| 17 | 32.587019 | 3 | 4 | 1 | 2 | | | 1.000000 | | | | 32.587019 | |
| 18 | 28.809291 | 3 | 4 | 2 | 1 | | | 0.279045 | 0.720955 | | | 8.039089 | 20.770202 |
| 19 | 9.373778 | 4 | 1 | 2 | 3 | | 1.000000 | | | | 9.373778 | | |
| 20 | 4.677242 | 4 | 1 | 3 | 2 | | 1.000000 | | | | 4.677242 | | |
| 21 | 16.953459 | 4 | 2 | 1 | 3 | | 0.583579 | 0.416421 | | | 9.893683 | 7.059776 | |
| 22 | 11.478833 | 4 | 2 | 3 | 1 | | 0.583579 | | 0.416421 | | 6.698806 | | 4.780027 |
| 23 | 16.399491 | 4 | 3 | 1 | 2 | | | 1.000000 | | | | 16.399491 | |
| 24 | 17.824724 | 4 | 3 | 2 | 1 | | | 0.279045 | 0.720955 | | | 4.973900 | 12.850824 |
| | 460.000000 | | | | | | | | | 115.000000 | 115.000000 | 115.000000 | 115.000000 |

Table 10.4.2: Allocation of the voters to the district winners in example A53

In the general case with $M$ seats, $N$ voters, and $W$ different voting profiles after proportional completion, we have:

(10.4.4) $\sum_{m=1}^{W} (a_m \cdot x_{mn}) = N / M$ for every candidate $n$.

(10.4.5) $\sum_{n=1}^{M} x_{mn} = 1$ for every voting profile $m$.

(10.4.6) $x_{mn} \geq 0$ for every voting profile $m$ and every candidate $n$.

$x_{mn}$ is the share of voting profile $m$ that is allocated to candidate $n$.
$a_m$ is the number of voters in voting profile $m$.

For $i := M$ to 2, we proceed as follows:

> Subject to (10.4.4) – (10.4.6) and subject to those $x_{mn}$ that have already been fixed, we minimize
>
> $$\max \{ \sum_{r=i}^{M} x_{m,R_m(r)} \mid m \in \{ 1, ..., W \} \text{ with } x_{m,R_m(i)} \text{ has not yet been fixed} \}$$
>
> where $R_m(r)$ is the candidate with the $r$-th preference in voting profile $m$. We fix $x_{m,R_m(i)}$ with maximum $\sum_{r=i}^{M} x_{m,R_m(r)}$. We proceed until $x_{m,R_m(i)}$ is fixed for all $m \in \{ 1, ..., W \}$.

For example, row 7 of table 10.4.1 says that, before proportional completion, there are 10 voters (voters #9, #51, #231, #272, #327, #371, #382, #412, #414, #417) who ranked the candidates $a \succ d \approx g \approx j$. These voters are replaced by 0.805282 voters who rank the candidates $a \succ d \succ g \succ j$ (row 7.001), 1.185668 voters who rank the candidates $a \succ d \succ j \succ g$ (row 7.002), 1.204172 voters who rank the candidates $a \succ g \succ d \succ j$ (row 7.003), 2.294764 voters who rank the candidates $a \succ j \succ d \succ g$ (row 7.004), 1.985874 voters who rank the candidates $a \succ g \succ j \succ d$ (row 7.005), and 2.524240 voters who rank the candidates $a \succ j \succ g \succ d$ (row 7.006) after proportional completion.

| before proportional completion | | | | | | voters | after proportional completion | | | | | |
|---|---|---|---|---|---|---|---|---|---|---|---|---|
| voting pattern | number of voters | *a* | *d* | *g* | *j* | | number of voters | *a* | *d* | *g* | *j* | voting pattern |
| 7 | 10 | 1 | 2 | 2 | 2 | 9, 51, 231, 272, 327, 371, 382, 412, 414, 417 | 0.805282 | 1 | 2 | 3 | 4 | 7.001 |
| | | | | | | | 1.185668 | 1 | 2 | 4 | 3 | 7.002 |
| | | | | | | | 1.204172 | 1 | 3 | 2 | 4 | 7.003 |
| | | | | | | | 2.294764 | 1 | 3 | 4 | 2 | 7.004 |
| | | | | | | | 1.985874 | 1 | 4 | 2 | 3 | 7.005 |
| | | | | | | | 2.524240 | 1 | 4 | 3 | 2 | 7.006 |

Row 7 of table 10.4.1

Row 1 of table 10.4.2 says that, from every voter who votes $a \succ d \succ g \succ j$, 0.416421 is allocated to candidate *a* and 0.583579 is allocated to candidate *d*. Row 2 of table 10.4.2 says that, from every voter who votes $a \succ d \succ j \succ g$, 0.416421 is allocated to candidate *a* and 0.583579 is allocated to candidate *d*. Row 3 of table 10.4.2 says that, from every voter who votes $a \succ g \succ d \succ j$, 0.749836 is allocated to candidate *a* and 0.250164 is allocated to candidate *g*. Row 4 of table 10.4.2 says that, from every voter who votes $a \succ j \succ d \succ g$, 1.000000 is allocated to candidate *a*. Row 5 of table 10.4.2 says that, from every voter who votes $a \succ g \succ j \succ d$, 0.749836 is allocated to candidate *a* and 0.250164 is allocated to candidate *g*. Row 6 of table 10.4.2 says that, from every voter who votes $a \succ j \succ g \succ d$, 1.000000 is allocated to candidate *a*.

| | number of voters | | | | | share | | | | votes | | | |
|---|---|---|---|---|---|---|---|---|---|---|---|---|---|
| | | *a* | *d* | *g* | *j* | *a* | *d* | *g* | *j* | *a* | *d* | *g* | *j* |
| 1 | 10.994148 | 1 | 2 | 3 | 4 | 0.416421 | 0.583579 | | | 4.578195 | 6.415952 | | |
| 2 | 20.545315 | 1 | 2 | 4 | 3 | 0.416421 | 0.583579 | | | 8.555500 | 11.989811 | | |
| 3 | 15.021513 | 1 | 3 | 2 | 4 | 0.749836 | | 0.250164 | | 11.263672 | | 3.757842 | |
| 4 | 25.451689 | 1 | 3 | 4 | 2 | 1.000000 | | | | 25.451688 | | | |
| 5 | 20.791365 | 1 | 4 | 2 | 3 | 0.749836 | | 0.250164 | | 15.590114 | | 5.201251 | |
| 6 | 23.213671 | 1 | 4 | 3 | 2 | 1.000000 | | | | 23.213670 | | | |

Rows 1–6 of table 10.4.2

When we combine table 10.4.1 and table 10.4.2, we get that, from every voter who votes $a \succ d \approx g \approx j$, a share of ( 0.805282 / 10 ) · 0.416421 + ( 1.185668 / 10 ) · 0.416421 + ( 1.204172 / 10 ) · 0.749836 + ( 2.294764 / 10 ) · 1.000000 + ( 1.985874 / 10 ) · 0.749836 + ( 2.524240 / 10 ) · 1.000000 = 0.804009 is allocated to candidate *a*, a share of ( 0.805282 / 10 ) · 0.583579 + ( 1.185668 / 10 ) · 0.583579 = 0.116188 is allocated to candidate *d*, and a share of ( 1.204172 / 10 ) · 0.250164 + ( 1.985874 / 10 ) · 0.250164 = 0.079804 is allocated to candidate *g*. In other words: A vote $a \succ d \approx g \approx j$ contributes with 80% to the election of candidate *a*, with 12% to the election of candidate *d*, and with 8% to the election of candidate *g*.

| | number of voters | | | | | share | | | | votes | | | |
|---|---|---|---|---|---|---|---|---|---|---|---|---|---|
| | | *a* | *d* | *g* | *j* | *a* | *d* | *g* | *j* | *a* | *d* | *g* | *j* |
| 1 | 8 | 1 | 3 | 2 | 3 | 0.749836 | | 0.250164 | | 5.998687 | | 2.001313 | |
| 2 | 8 | 1 | 2 | 3 | 4 | 0.416421 | 0.583579 | | | 3.331369 | 4.668631 | | |
| 3 | 14 | 1 | 4 | 3 | 2 | 1.000000 | | | | 14.000000 | | | |
| 4 | 17 | 2 | 2 | 2 | 1 | 0.127655 | 0.116716 | 0.095581 | 0.660048 | 2.170135 | 1.984168 | 1.624875 | 11.220821 |
| 5 | 29 | 2 | 4 | 3 | 1 | 0.279045 | | | 0.720955 | 8.092302 | | | 20.907698 |
| 6 | 22 | 2 | 3 | 3 | 1 | 0.279045 | | | 0.720955 | 6.138988 | | | 15.861012 |
| 7 | 10 | 1 | 2 | 2 | 2 | 0.804009 | 0.116188 | 0.079804 | | 8.040088 | 1.161876 | 0.798035 | |
| 8 | 12 | 3 | 3 | 1 | 2 | | | 1.000000 | | | | 12.000000 | |
| 9 | 21 | 2 | 2 | 1 | 2 | | 0.115091 | 0.884909 | | | 2.416910 | 18.583090 | |
| 10 | 15 | 3 | 4 | 2 | 1 | | | 0.279045 | 0.720955 | | | 4.185674 | 10.814326 |
| 11 | 15 | 2 | 4 | 1 | 3 | | | 1.000000 | | | | 15.000000 | |
| 12 | 34 | 2 | 3 | 4 | 1 | 0.279045 | | | 0.720955 | 9.487527 | | | 24.512473 |
| 13 | 7 | 3 | 2 | 3 | 1 | | 0.583579 | | 0.416421 | | 4.085052 | | 2.914948 |
| 14 | 14 | 3 | 2 | 4 | 1 | | 0.583579 | | 0.416421 | | 8.170104 | | 5.829896 |
| 15 | 7 | 4 | 2 | 3 | 1 | | 0.583579 | | 0.416421 | | 4.085052 | | 2.914948 |
| 16 | 13 | 1 | 4 | 2 | 3 | 0.749836 | | 0.250164 | | 9.747866 | | 3.252134 | |
| 17 | 18 | 1 | 3 | 4 | 2 | 1.000000 | | | | 18.000000 | | | |
| 18 | 16 | 1 | 2 | 4 | 3 | 0.416421 | 0.583579 | | | 6.662738 | 9.337262 | | |
| 19 | 5 | 1 | 2 | 3 | 3 | 0.416421 | 0.583579 | | | 2.082106 | 2.917894 | | |
| 20 | 11 | 1 | 3 | 2 | 4 | 0.749836 | | 0.250164 | | 8.248194 | | 2.751806 | |
| 21 | 11 | 1 | 3 | 3 | 2 | 1.000000 | | | | 11.000000 | | | |
| 22 | 10 | 4 | 3 | 2 | 1 | | | 0.279045 | 0.720955 | | | 2.790449 | 7.209551 |
| 23 | 9 | 4 | 3 | 1 | 2 | | | 1.000000 | | | | 9.000000 | |
| 24 | 8 | 3 | 2 | 1 | 3 | | 0.583579 | 0.416421 | | | 4.668631 | 3.331369 | |
| 25 | 17 | 3 | 4 | 1 | 2 | | | 1.000000 | | | | 17.000000 | |
| 26 | 10 | 4 | 2 | 1 | 3 | | 0.583579 | 0.416421 | | | 5.835789 | 4.164211 | |
| 27 | 3 | 2 | 3 | 1 | 3 | | | 1.000000 | | | | 3.000000 | |
| 28 | 8 | 3 | 2 | 1 | 4 | | 0.583579 | 0.416421 | | | 4.668631 | 3.331369 | |
| 29 | 6 | 2 | 3 | 1 | 4 | | | 1.000000 | | | | 6.000000 | |
| 30 | 6 | 3 | 1 | 4 | 2 | | 1.000000 | | | | 6.000000 | | |
| 31 | 9 | 2 | 1 | 2 | 2 | | 1.000000 | | | | 9.000000 | | |
| 32 | 14 | 2 | 1 | 4 | 3 | | 1.000000 | | | | 14.000000 | | |
| 33 | 8 | 3 | 1 | 3 | 2 | | 1.000000 | | | | 8.000000 | | |
| 34 | 3 | 3 | 1 | 2 | 3 | | 1.000000 | | | | 3.000000 | | |
| 35 | 6 | 4 | 1 | 2 | 3 | | 1.000000 | | | | 6.000000 | | |
| 36 | 8 | 3 | 1 | 2 | 4 | | 1.000000 | | | | 8.000000 | | |
| 37 | 5 | 2 | 1 | 3 | 4 | | 1.000000 | | | | 5.000000 | | |
| 38 | 15 | 3 | 3 | 2 | 1 | | | 0.279045 | 0.720955 | | | 4.185674 | 10.814326 |
| 39 | 8 | 1 | 1 | 1 | 1 | 0.250000 | 0.250000 | 0.250000 | 0.250000 | 2.000000 | 2.000000 | 2.000000 | 2.000000 |
| | | | | | | | | | | 115.000000 | 115.000000 | 115.000000 | 115.000000 |

Table 10.4.3: Allocation of the voters to the district winners in example A53

Table 10.4.3 summarizes tables 10.4.1 and 10.4.2. Table 10.4.3 lists, for every voter of table 10.4.1, how he is allocated to the winners in example A53.

## 10.5. Allocation of Seats to Parties

We use the Sainte-Laguë method to determine how many seats each party gets. We use 0.8 as first divisor because, as 80% of the seats are district seats, an independent candidate needs 80% of a quota to get elected. So we take the party votes of each party and divide them by 0.8, 1.5, 2.5, 3.5, ... and the 130 seats of the House go to the 130 largest quotients.

Now it can happen that a party $a$ has already won more district seats than it deserves according to its proportional share of party votes (*overhang seats*). In this case, we increase the number of party votes of party $a$ by adding some of those voters, whose district vote has been allocated in section 10.4 to a candidate of party $a$, to party $a$.

To use a more formal language: Suppose $0 \leq T(x) \leq 1$ is the share of each voter whose ballot vote has been allocated to a candidate of party $x$ and whose party vote will be allocated to party $x$ (regardless to which party this voter has actually given his party vote). Then we start with $T(x) = 0$ for each party $x$. When some party $a$ has won overhang seats, then we increase $T(a)$ until party $a$ has enough party votes so that it has no overhang anymore; increasing $T(a)$ means that the share $T(a)$ of each voter, whose district vote has been allocated to a candidate of party $a$, is automatically interpreted as a party vote for party $a$ and $1-T(a)$ of each voter, whose district vote has been allocated to a candidate of party $a$, is interpreted as a party vote for that party for which this voter has actually voted with his party ballot (section 10.3), which might also be party $a$. If e.g. a voter contributed with 0.804009 to the election of district winners of party $a$ with his district vote (according to section 10.4) and gave his party vote to party $b$, then $0.804009 \cdot T(a)$ of this voter's party vote goes to party $a$. When we increase $T(a)$, it can happen that we create an overhang for some other party $b$; in this case, we also have to increase $T(b)$. Increasing $T(b)$ however can, again, lead to an overhang for party $a$. So we have to apply this procedure several times until either (i) it converges to a distribution of party votes without an overhang for any party or (ii) $T(x) = 1$ for some party $x$. In the latter case, the total number of seats is increased by 1 and the calculation of $T(x)$ is restarted.

Independent candidates are treated as parties with no initial party votes.

Basic idea behind this procedure is the idea that, when party $a$ has an uncompensated overhang, then those voters, who contributed to the election of some candidate of party $a$ with their district vote and who voted for some other party $b$ with their party vote, would have a double voting power. Therefore, a share $0 \leq T(a) \leq 1$ of these votes should be counted for party $a$ and not for party $b$. This share $T(a)$ should be chosen just large enough that this double voting power disappears.

## 10.6. Allocation of the Party Seats to its District Organizations

Suppose $a_i$ is the number of party votes (including those voters who have been allocated to party $a$ by raising $T(a)$ and excluding those voters who have been allocated to some other party $b$ by raising $T(b)$ in section 10.5) for party $a$ in district $i$. Suppose $s_i(a)$ is the number of district seats that the party $a$ has won in district $i$ (according to section 10.2).

To allocate the seats of party $a$ to its district organizations, first each district organization gets $s_i(a)$ seats. We then divide the numbers of party votes for party $a$ in each district $i$ by $( s_i(a) + 0.5 )$, $( s_i(a) + 1.5 )$, $( s_i(a) + 2.5 )$, $( s_i(a) + 3.5 )$, ... The remaining seats go to the largest quotients.

Now it can happen that the total number of seats for a district differs from the number of seats this district must get according to table 10.1.3. In this case, we apply Pukelsheim's biproportional method (Pukelsheim, 2004).

To use a more formal language: For each district $i$, we calculate a factor $F_i$. This factor depends only on the district $i$ and is the same for all parties. The factor $F_i$ is chosen in such a manner that, when each party $a$ applies the Sainte-Laguë method to $a_i \cdot F_i$ (instead of $a_i$), then each district gets exactly as many seats as it should get according to table 10.1.3.

## 10.7. Best Losers

When a district organization of party $a$ has won more seats (according to section 10.6) than it has won district seats (according to section 10.2), then these additional seats go to the "best losers" of this party. To determine these best losers, we apply Schulze STV, but we restrict it to those voters who voted for party $a$ with their party vote (including those voters who have been allocated to party $a$ by raising $T(a)$ and excluding those voters who have been allocated to some other party $b$ by raising $T(b)$ in section 10.5) and those candidates who have been nominated by party $a$ (according to section 10.2).

For example, if the candidates $c_1$ and $c_2$ are those candidates who have run for candidate $a$ on the district ballot and have won district seats and if the district organization has won additional 2 seats according to section 10.6, then the final winners for party $a$ are the candidates $(c_1,c_2,c_m,c_n)$ with $P_{D2}[(c_1,c_2,c_m,c_n),(c_1,c_2,c_r,c_s)] \gtrsim_{D2} P_{D2}[(c_1,c_2,c_r,c_s),(c_1,c_2,c_m,c_n)]$ for every other set of candidates $(c_1,c_2,c_r,c_s)$ according to (8.1.3.1).

When the list of candidates of a district organization gets exhausted, then the remaining seats stay vacant.

## 10.8. Vacant Seats

(1) Suppose a seat, that has been won by the district vote (according to section 10.2), gets vacant. Then this seat is filled by recounting the district ballots. Those candidates who have died or who have won a compensation seat (according to section 10.7) are ignored. So when $(c_1,...,c_n)$ are the current winners. Then the vacant seat goes to candidate $c_i \notin \{c_1,...,c_n\}$ with $P_{D2}[(c_1,...,c_n,c_i),(c_1,...,c_n,c_j)] \gtrsim_{D2} P_{D2}[(c_1,...,c_n,c_j),(c_1,...,c_n,c_i)]$ for every other candidate $c_j \notin \{c_1,...,c_n\}$ according to (8.1.3.1).

(2) Suppose a compensation seat (according to section 10.7) gets vacant. Then this seat is filled by recounting the district ballots. Suppose this seat went to party $a$. Then the recount is restricted to those voters who voted for party $a$ with their party vote (including those voters who have been allocated to party $a$ by raising $T(a)$ and excluding those voters who have been allocated to some other party $b$ by raising $T(b)$ in section 10.5) and those candidates who have been nominated by party $a$ (according to section 10.2). Those candidates who have died are ignored. So when $(c_1,...,c_n)$ are those candidates of party $a$ who are currently holding a district seat (according to section 10.2) or a compensation seat (according to section 10.7). Then the vacant seat goes to candidate $c_i \notin \{c_1,...,c_n\}$ with $P_{D2}[(c_1,...,c_n,c_i),(c_1,...,c_n,c_j)] \gtrsim_{D2} P_{D2}[(c_1,...,c_n,c_j),(c_1,...,c_n,c_i)]$ for every other candidate $c_j \notin \{c_1,...,c_n\}$ according to (8.1.3.1).

## 11. Comparison with other Methods

Table 11.2 compares the Schulze method (with margins) with its main contenders. Extensive descriptions of the different methods can be found in publications by Fishburn (1977), Nurmi (1987), Kopfermann (1991), Levin and Nalebuff (1995), and Tideman (2006). As most of these methods only generate a set $\mathcal{S}$ of potential winners and don’t generate a binary relation $\mathcal{O}$, only that part of the different criteria is considered that refers to the set $\mathcal{S}$ of potential winners.

In terms of satisfied and violated criteria, that election method, that comes closest to the Schulze method, is Tideman’s ranked pairs method (Tideman, 1987). The only difference is that the ranked pairs method doesn’t choose from the MinMax set $\mathcal{B}_D$.

The ranked pairs method works from the strongest to the weakest link. The link *xy* is locked if and only if it doesn’t create a directed cycle with already locked links. Otherwise, this link is locked in its opposite direction.

In example 1 (section 3.1), the ranked pairs method locks *db*. Then it locks *cb*. Then it locks *ac*. Then it locks *ab*, since locking *ba* in its original direction would create a directed cycle with the already locked links *ac* and *cb*. Then it locks *cd*. Then it locks *ad*, since locking *da* in its original direction would create a directed cycle with the already locked links *ac* and *cd*.

The winner of the ranked pairs method is alternative $a \notin \mathcal{B}_D = \{d\}$, because there is no locked link that ends in alternative *a*.

Although Tideman’s ranked pairs method is that election method that comes closest to the Schulze method in terms of satisfied and violated criteria, random simulations by Wright (2009) showed that that election method, that agrees the most frequently with the Schulze method, is the Simpson-Kramer MinMax method (table 11.1). This observation has been confirmed by Petry (2000), Heitzig (2004a), Darlington (2018, 2023), and Pacuit (2021). See also table 11.3.

| A | B | C | D |
|---|---|---|---|
| 3 | 100.0 % | 100.0 % | 100.0 % |
| 4 | 99.7 % | 98.5 % | 98.2 % |
| 5 | 99.2 % | 96.0 % | 95.3 % |
| 6 | 99.1 % | 93.0 % | 92.3 % |
| 7 | 98.9 % | 90.0 % | 89.1 % |

Table 11.1: Simulations by Wright (2009)

A: number of alternatives

B: Probability that the Schulze method conforms with the Simpson-Kramer MinMax method

C: Probability that the Schulze method conforms with the ranked pairs method

D: Probability that the ranked pairs method conforms with the Simpson-Kramer MinMax method

| | decisiveness | resolvability | Pareto | reversal symmetry | monotonicity | independence of clones | Smith | Smith-IIA | Condorcet winner | Condorcet loser | majority for solid coalitions | majority winner | majority loser | participation | MinMax set | prudence | reversal cancellation | polynomial runtime |
|---|---|---|---|---|---|---|---|---|---|---|---|---|---|---|---|---|---|---|
| Baldwin | Y | Y | Y | N | N | N | Y | N | Y | Y | Y | Y | Y | N | N | N | Y | Y |
| Black | Y | Y | Y | Y | Y | N | N | N | Y | Y | N | Y | Y | N | N | N | Y | Y |
| Borda | Y | Y | Y | Y | Y | N | N | N | N | Y | N | N | Y | Y | N | N | Y | Y |
| Bucklin | Y | Y | Y | N | Y | N | N | N | N | N | Y | Y | Y | N | N | N | N | Y |
| Copeland | N | N | Y | Y | Y | N | Y | Y | Y | Y | Y | Y | Y | N | N | N | Y | Y |
| Dodgson | Y | Y | Y | N | N | N | N | N | Y | N | N | Y | N | N | N | N | N | N |
| instant runoff | Y | Y | Y | N | N | Y | N | N | N | Y | Y | Y | Y | N | N | N | N | Y |
| Kemeny-Young | Y | Y | Y | Y | Y | N | Y | Y | Y | Y | Y | Y | Y | N | N | N | Y | N |
| Nanson | Y | Y | Y | Y | N | N | Y | N | Y | Y | Y | Y | Y | N | N | N | Y | Y |
| plurality | Y | Y | Y | N | Y | N | N | N | N | N | N | Y | N | Y | N | N | N | Y |
| ranked pairs | Y | Y | Y | Y | Y | Y | Y | Y | Y | Y | Y | Y | Y | N | N | Y | Y | Y |
| river | Y | Y | Y | Y | Y | Y | Y | Y | Y | Y | Y | Y | Y | N | N | Y | Y | Y |
| Schulze | Y | Y | Y | Y | Y | Y | Y | Y | Y | Y | Y | Y | Y | N | Y | Y | Y | Y |
| Simpson-Kramer MinMax | Y | Y | Y | N | Y | N | N | N | Y | N | N | Y | N | N | N | Y | Y | Y |
| Slater | N | N | Y | Y | Y | N | Y | Y | Y | Y | Y | Y | Y | N | N | N | Y | N |
| Young | Y | Y | Y | N | Y | N | N | N | Y | N | N | Y | N | N | N | N | N | N |

Table 11.2: Comparison of Election Methods

“Y” = compliance<br>
“N” = violation

In the introduction of this paper, I mentioned that the purpose of the Schulze method is to create an election method that satisfies the Smith criterion, reversal symmetry and independence of clones and that agrees with the Simpson-Kramer MinMax method whenever possible. To check whether the Schulze method delivers, we will compare the Schulze method with two other methods that satisfy the Smith criterion, reversal symmetry and independence of clones: Tideman’s ranked pairs method (Tideman, 1987) and Heitzig’s river method (Döring, 2026).

Example 14 (section 3.14) demonstrates the difference between the Simpson-Kramer MinMax method, the Schulze method, Tideman’s ranked pairs method and Heitzig’s river method. The winner of the Schulze method in example 14 is alternative *b*.

**Simpson-Kramer MinMax method:** This method calculates, for every alternative $x \in A$, the strongest link from some other alternative $y \in A \setminus \{x\}$ to alternative $x$. The winner is that alternative $x \in A$ for which

$$\max_D \{ (N[y,x],N[x,y]) \mid y \in A \setminus \{x\} \}$$

is the weakest. In example 14, we have :

$$\max_D \{ (N[y,a],N[a,y]) \mid y \in A \setminus \{a\} \} \approx_D (62,35).$$
$$\max_D \{ (N[y,b],N[b,y]) \mid y \in A \setminus \{b\} \} \approx_D (57,40).$$
$$\max_D \{ (N[y,c],N[c,y]) \mid y \in A \setminus \{c\} \} \approx_D (56,41).$$
$$\max_D \{ (N[y,d],N[d,y]) \mid y \in A \setminus \{d\} \} \approx_D (58,39).$$
$$\max_D \{ (N[y,e],N[e,y]) \mid y \in A \setminus \{e\} \} \approx_D (63,34).$$
$$\max_D \{ (N[y,f],N[f,y]) \mid y \in A \setminus \{f\} \} \approx_D (59,38).$$

So the winner of the Simpson-Kramer MinMax method in example 14 is alternative *c*.

**Tideman’s ranked pairs method:** From the strongest to the weakest link, we proceed as follows: When locking this link doesn’t create a directed cycle with already locked links, then we lock this link in its original direction. When locking this link creates a directed cycle with already locked links, then we lock this link in its opposite direction.

In example 14, we lock $d \rightarrow e$ with a strength of (63,34). We then lock $f \rightarrow a$ with a strength of (62,35). We then lock $a \rightarrow e$ with a strength of (61,36). We then lock $b \rightarrow e$ with a strength of (60,37). We then lock $f \rightarrow e$ with a strength of (59,38), since locking *ef* in its original direction would create a directed cycle with the already locked links $f \rightarrow a$ and $a \rightarrow e$. We then lock $f \rightarrow d$ with a strength of (58,39). We then lock $d \rightarrow b$ with a strength of (57,40). We then lock $b \rightarrow c$ with a strength of (56,41). We then lock $e \rightarrow c$ with a strength of (55,42). We then lock $d \rightarrow a$ with a strength of (54,43). We then lock $a \rightarrow c$ with a strength of (53,44), since locking *ca* in its original direction would create a directed cycle with the already locked links $a \rightarrow e$ and $e \rightarrow c$. We then lock $a \rightarrow b$ with a strength of (52,45). We then lock $d \rightarrow c$ with a strength of (51,46), since locking *cd* in its original direction would create a directed cycle with the already locked links $d \rightarrow b$ and $b \rightarrow c$. We then lock $f \rightarrow b$ with a strength of (50,47). We then lock $f \rightarrow c$ with a strength of (49,48), since locking *cf* in its original direction would create a directed cycle with the already locked links $f \rightarrow a$, $a \rightarrow e$ and $e \rightarrow c$.

The winner of Tideman’s ranked pairs method is alternative *f* because $f \rightarrow y$ for every other alternative $y \in A \setminus \{f\}$.

**Heitzig’s river method:** From the strongest to the weakest link, we proceed as follows: When (1) locking the link *xy* creates a directed cycle with already locked links or when (2) some link *zy* has already been locked, then we skip *xy*. Otherwise we lock *xy*.

In example 14, we lock $d \rightarrow e$ with a strength of (63,34). We then lock $f \rightarrow a$ with a strength of (62,35). We then skip *ae* with a strength of (61,36), since there is already a locked link *ze*. We then skip *be* with a strength of (60,37), since there is already a locked link *ze*. We then lock $e \rightarrow f$ with a strength of (59,38). We then skip *fd* with a strength of (58,39), since locking *fd* would create a directed cycle with the already locked links $d \rightarrow e$ and $e \rightarrow f$. We then lock $d \rightarrow b$ with a strength of (57,40). We then lock $b \rightarrow c$ with a strength of (56,41). We then skip *ec* with a strength of (55,42), since there is already a locked link *zc*. We then skip *da* with a strength of (54,43), since there is already a locked link *za*. We then skip *ca* with a strength of (53,44), since there is already a locked link *za*. We then skip *ab* with a strength of (52,45), since there is already a locked link *zb*. We then skip *cd* with a strength of (51,46), since locking *cd* would create a directed cycle with the already locked links $d \rightarrow b$ and $b \rightarrow c$. We then skip *fb* with a strength of (50,47), since there is already a locked link *zb*. We then skip *cf* with a strength of (49,48), since there is already a locked link *zf*.

The winner of Heitzig’s river method is alternative *d* because there is no alternative $y \in A \setminus \{d\}$ with $y \rightarrow d$.

Tables 11.3 and 11.4 show how frequently the Simpson-Kramer MinMax method, the Schulze method, Tideman’s ranked pairs method and Heitzig’s river method agree or disagree with each other in a random simulation. The tests for tables 11.3 and 11.4 are created as follows:

1. At first, for each pair of alternatives $ij \in A$, we decide whether alternative *i* pairwise beats alternative *j* or alternative *j* pairwise beats alternative *i*.

   To make sure that all alternatives are in the Smith set, we demand that alternative *i* pairwise beats alternative *i*+1 for every $i = 1,...,(C–1)$. Furthermore, we demand that alternative *C* pairwise beats alternative 1.

   For the remaining pairs of alternatives $ij \in A$, we decide randomly whether alternative *i* pairwise beats alternative *j* or alternative *j* pairwise beats alternative *i*.

2. Then, the strengths of the $(C \cdot (C–1))/2$ pairwise wins are ordered randomly. This is sufficient because the winners of the Schulze method, the ranked pairs method and the Simpson-Kramer MinMax method depend only on the order of the strengths of the pairwise wins, not on their absolute strengths. This is equivalent to saying that the number of voters goes to infinity.

   As there are no wins with equivalent strengths, the Simpson-Kramer MinMax method, the Schulze method, Tideman’s ranked pairs method and Heitzig’s river method always have unique winners. Therefore, there is no need to specify how to handle indecisive cases.

The columns in table 11.3 are:

Column "A": number of alternatives. For each number in column "A", 1,000,000 tests were run.

Column "B": number of tests where ...

- the Simpson-Kramer MinMax method chooses alternative $a_1 \in A$,
- the Schulze method chooses alternative $a_1$,
- Tideman's ranked pairs method chooses alternative $a_1$,
- Heitzig's river method chooses alternative $a_1$.

Column "C": number of tests where ...

- the Simpson-Kramer MinMax method chooses alternative $a_1 \in A$,
- the Schulze method chooses alternative $a_1$,
- Tideman's ranked pairs method chooses alternative $a_1$,
- Heitzig's river method chooses alternative $a_2 \in A \setminus \{a_1\}$.

Column "D": number of tests where ...

- the Simpson-Kramer MinMax method chooses alternative $a_1 \in A$,
- the Schulze method chooses alternative $a_1$,
- Tideman's ranked pairs method chooses alternative $a_2 \in A \setminus \{a_1\}$,
- Heitzig's river method chooses alternative $a_1$.

Column "E": number of tests where ...

- the Simpson-Kramer MinMax method chooses alternative $a_1 \in A$,
- the Schulze method chooses alternative $a_1$,
- Tideman's ranked pairs method chooses alternative $a_2 \in A \setminus \{a_1\}$,
- Heitzig's river method chooses alternative $a_2$.

Column "F": number of tests where ...

- the Simpson-Kramer MinMax method chooses alternative $a_1 \in A$,
- the Schulze method chooses alternative $a_1$,
- Tideman's ranked pairs method chooses alternative $a_2 \in A \setminus \{a_1\}$,
- Heitzig's river method chooses alternative $a_3 \in A \setminus \{a_1,a_2\}$.

Column "G": number of tests where ...

- the Simpson-Kramer MinMax method chooses alternative $a_1 \in A$,
- the Schulze method chooses alternative $a_2 \in A \setminus \{a_1\}$,
- Tideman's ranked pairs method chooses alternative $a_1$,
- Heitzig's river method chooses alternative $a_1$.

Column "H": number of tests where ...

- the Simpson-Kramer MinMax method chooses alternative $a_1 \in A$,
- the Schulze method chooses alternative $a_2 \in A \setminus \{a_1\}$,
- Tideman's ranked pairs method chooses alternative $a_1$,
- Heitzig's river method chooses alternative $a_2$.

Column "I": number of tests where ...

- the Simpson-Kramer MinMax method chooses alternative $a_1 \in A$,
- the Schulze method chooses alternative $a_2 \in A \setminus \{a_1\}$,
- Tideman's ranked pairs method chooses alternative $a_1$,
- Heitzig's river method chooses alternative $a_3 \in A \setminus \{a_1,a_2\}$.

Column “J”: number of tests where ...

- the Simpson-Kramer MinMax method chooses alternative $a_1 \in A$,
- the Schulze method chooses alternative $a_2 \in A \setminus \{a_1\}$,
- Tideman’s ranked pairs method chooses alternative $a_2$,
- Heitzig’s river method chooses alternative $a_1$.

Column “K”: number of tests where ...

- the Simpson-Kramer MinMax method chooses alternative $a_1 \in A$,
- the Schulze method chooses alternative $a_2 \in A \setminus \{a_1\}$,
- Tideman’s ranked pairs method chooses alternative $a_2$,
- Heitzig’s river method chooses alternative $a_2$.

Column “L”: number of tests where ...

- the Simpson-Kramer MinMax method chooses alternative $a_1 \in A$,
- the Schulze method chooses alternative $a_2 \in A \setminus \{a_1\}$,
- Tideman’s ranked pairs method chooses alternative $a_2$,
- Heitzig’s river method chooses alternative $a_3 \in A \setminus \{a_1,a_2\}$.

Column “M”: number of tests where ...

- the Simpson-Kramer MinMax method chooses alternative $a_1 \in A$,
- the Schulze method chooses alternative $a_2 \in A \setminus \{a_1\}$,
- Tideman’s ranked pairs method chooses alternative $a_3 \in A \setminus \{a_1,a_2\}$,
- Heitzig’s river method chooses alternative $a_1$.

Column “N”: number of tests where ...

- the Simpson-Kramer MinMax method chooses alternative $a_1 \in A$,
- the Schulze method chooses alternative $a_2 \in A \setminus \{a_1\}$,
- Tideman’s ranked pairs method chooses alternative $a_3 \in A \setminus \{a_1,a_2\}$,
- Heitzig’s river method chooses alternative $a_2$.

Column “O”: number of tests where ...

- the Simpson-Kramer MinMax method chooses alternative $a_1 \in A$,
- the Schulze method chooses alternative $a_2 \in A \setminus \{a_1\}$,
- Tideman’s ranked pairs method chooses alternative $a_3 \in A \setminus \{a_1,a_2\}$,
- Heitzig’s river method chooses alternative $a_3$.

Column “P”: number of tests where ...

- the Simpson-Kramer MinMax method chooses alternative $a_1 \in A$,
- the Schulze method chooses alternative $a_2 \in A \setminus \{a_1\}$,
- Tideman’s ranked pairs method chooses alternative $a_3 \in A \setminus \{a_1,a_2\}$,
- Heitzig’s river method chooses alternative $a_4 \in A \setminus \{a_1,a_2,a_3\}$.

The standard deviation is in brackets. When there are *pos* positive tests and *neg* negative tests, then the standard deviation is:

$$\sqrt{((\,pos \cdot neg\,) \,/\, (\,pos + neg\,))}.$$

| A | B | C | D | E | F | G | H |
|---|---|---|---|---|---|---|---|
| 4 | 809,269 (±393) | 6,900 (±83) | 62,845 (±243) | 71,195 (±257) | 0 | 0 | 0 |
| 5 | 728,775 (±445) | 22,037 (±147) | 80,887 (±273) | 98,699 (±298) | 2,067 (±45) | 0 | 0 |
| 6 | 651,628 (±476) | 35,636 (±185) | 100,983 (±301) | 127,493 (±334) | 6,862 (±83) | 210 (±14) | 458 (±21) |
| 7 | 587,815 (±492) | 48,364 (±215) | 118,622 (±323) | 150,776 (±358) | 13,651 (±116) | 330 (±18) | 736 (±27) |
| 8 | 534,736 (±499) | 59,434 (±236) | 133,592 (±340) | 167,967 (±374) | 22,194 (±147) | 416 (±20) | 961 (±31) |
| 9 | 488,145 (±500) | 69,406 (±254) | 145,450 (±353) | 183,900 (±387) | 32,764 (±178) | 426 (±21) | 1,022 (±32) |
| 10 | 449,753 (±497) | 77,693 (±268) | 155,929 (±363) | 195,503 (±397) | 44,410 (±206) | 422 (±21) | 1,087 (±33) |
| 11 | 415,491 (±493) | 84,375 (±278) | 165,400 (±372) | 204,842 (±404) | 56,348 (±231) | 400 (±20) | 1,107 (±33) |
| 12 | 384,530 (±486) | 90,783 (±287) | 172,280 (±378) | 212,946 (±409) | 69,738 (±255) | 406 (±20) | 1,064 (±33) |
| 13 | 359,385 (±480) | 95,770 (±294) | 177,920 (±382) | 217,758 (±413) | 83,467 (±277) | 355 (±19) | 995 (±32) |
| 14 | 335,102 (±472) | 100,436 (±301) | 184,252 (±388) | 222,431 (±416) | 96,609 (±295) | 290 (±17) | 857 (±29) |
| 15 | 313,982 (±464) | 104,032 (±305) | 188,820 (±391) | 224,898 (±418) | 109,916 (±313) | 288 (±17) | 763 (±28) |
| 16 | 296,543 (±457) | 106,951 (±309) | 191,811 (±394) | 225,957 (±418) | 123,433 (±329) | 259 (±16) | 757 (±28) |
| 17 | 278,882 (±448) | 109,786 (±313) | 195,093 (±396) | 227,881 (±419) | 136,391 (±343) | 208 (±14) | 629 (±25) |
| 18 | 264,661 (±441) | 111,502 (±315) | 197,821 (±398) | 227,894 (±419) | 149,090 (±356) | 184 (±14) | 593 (±24) |
| 19 | 250,834 (±433) | 112,796 (±316) | 201,332 (±401) | 227,044 (±419) | 162,048 (±368) | 166 (±13) | 553 (±24) |
| 20 | 237,845 (±426) | 114,709 (±319) | 202,763 (±402) | 225,973 (±418) | 174,903 (±380) | 167 (±13) | 443 (±21) |
| 21 | 225,895 (±418) | 115,802 (±320) | 204,527 (±403) | 225,456 (±418) | 187,214 (±390) | 141 (±12) | 391 (±20) |
| 22 | 215,844 (±411) | 116,358 (±321) | 205,988 (±404) | 224,314 (±417) | 198,274 (±399) | 115 (±11) | 343 (±19) |
| 23 | 207,349 (±405) | 117,192 (±322) | 207,402 (±405) | 221,392 (±415) | 209,766 (±407) | 116 (±11) | 303 (±17) |
| 24 | 198,033 (±399) | 117,319 (±322) | 208,080 (±406) | 219,416 (±414) | 221,516 (±415) | 95 (±10) | 305 (±17) |
| 25 | 189,553 (±392) | 117,894 (±322) | 207,973 (±406) | 217,585 (±413) | 232,728 (±423) | 73 (±9) | 280 (±17) |
| 26 | 181,496 (±385) | 118,204 (±323) | 209,252 (±407) | 215,753 (±411) | 242,751 (±429) | 88 (±9) | 226 (±15) |
| 27 | 173,880 (±379) | 117,896 (±322) | 209,661 (±407) | 213,771 (±410) | 253,467 (±435) | 64 (±8) | 209 (±14) |
| 28 | 167,718 (±374) | 118,031 (±323) | 210,028 (±407) | 210,406 (±408) | 263,450 (±441) | 55 (±7) | 205 (±14) |
| 29 | 162,040 (±368) | 117,667 (±322) | 210,243 (±407) | 208,176 (±406) | 272,457 (±445) | 63 (±8) | 175 (±13) |
| 30 | 155,968 (±363) | 117,525 (±322) | 210,445 (±408) | 206,066 (±404) | 281,976 (±450) | 63 (±8) | 139 (±12) |
| 31 | 150,230 (±357) | 117,725 (±322) | 210,823 (±408) | 203,238 (±402) | 290,991 (±454) | 52 (±7) | 107 (±10) |
| 32 | 145,854 (±353) | 116,623 (±321) | 210,239 (±407) | 201,491 (±401) | 299,250 (±458) | 43 (±7) | 131 (±11) |
| 33 | 140,487 (±347) | 117,096 (±322) | 210,440 (±408) | 198,501 (±399) | 308,113 (±462) | 42 (±6) | 98 (±10) |
| 34 | 136,189 (±343) | 116,084 (±320) | 210,205 (±407) | 196,709 (±398) | 316,127 (±465) | 42 (±6) | 124 (±11) |
| 35 | 131,975 (±338) | 115,222 (±319) | 210,401 (±408) | 194,737 (±396) | 324,034 (±468) | 32 (±6) | 77 (±9) |
| 36 | 128,052 (±334) | 115,261 (±319) | 209,734 (±407) | 191,662 (±394) | 332,278 (±471) | 37 (±6) | 86 (±9) |
| 37 | 123,717 (±329) | 115,299 (±319) | 209,291 (±407) | 190,441 (±393) | 338,999 (±473) | 23 (±5) | 87 (±9) |
| 38 | 120,245 (±325) | 114,398 (±318) | 208,423 (±406) | 187,563 (±390) | 347,626 (±476) | 40 (±6) | 94 (±10) |
| 39 | 117,601 (±322) | 113,823 (±318) | 208,069 (±406) | 184,916 (±388) | 354,362 (±478) | 21 (±5) | 61 (±8) |
| 40 | 113,811 (±318) | 112,883 (±316) | 207,824 (±406) | 182,750 (±386) | 362,148 (±481) | 25 (±5) | 66 (±8) |
| 41 | 110,774 (±314) | 112,120 (±316) | 207,099 (±405) | 181,210 (±385) | 368,647 (±482) | 13 (±4) | 55 (±7) |
| 42 | 108,316 (±311) | 111,171 (±314) | 206,811 (±405) | 179,454 (±384) | 374,627 (±484) | 15 (±4) | 63 (±8) |
| 43 | 105,239 (±307) | 110,710 (±314) | 206,780 (±405) | 176,411 (±381) | 381,699 (±486) | 24 (±5) | 44 (±7) |
| 44 | 102,670 (±304) | 110,332 (±313) | 206,626 (±405) | 175,262 (±380) | 386,483 (±487) | 10 (±3) | 58 (±8) |
| 45 | 100,149 (±300) | 110,349 (±313) | 206,258 (±405) | 172,298 (±378) | 392,569 (±488) | 13 (±4) | 52 (±7) |
| 46 | 97,223 (±296) | 108,938 (±312) | 205,911 (±404) | 170,670 (±376) | 399,657 (±490) | 10 (±3) | 33 (±6) |
| 47 | 94,915 (±293) | 108,678 (±311) | 204,737 (±404) | 168,751 (±375) | 405,342 (±491) | 14 (±4) | 43 (±7) |
| 48 | 93,095 (±291) | 108,121 (±311) | 204,848 (±404) | 166,401 (±372) | 410,662 (±492) | 25 (±5) | 42 (±6) |
| 49 | 90,691 (±287) | 107,127 (±309) | 204,122 (±403) | 164,858 (±371) | 416,440 (±493) | 9 (±3) | 31 (±6) |
| 50 | 88,398 (±284) | 107,137 (±309) | 203,875 (±403) | 162,705 (±369) | 421,475 (±494) | 10 (±3) | 40 (±6) |

Table 11.3 (part 1 of 8): comparison of the Simpson-Kramer MinMax method, the Schulze method, Tideman’s ranked pairs method and Heitzig’s river method (source: own calculation)

| I | J | K | L | M | N | O | P | A |
|---|---|---|---|---|---|---|---|---|
| 0 | 0 | 49,791 (±218) | 0 | 0 | 0 | 0 | 0 | 4 |
| 0 | 0 | 63,921 (±245) | 269 (±16) | 0 | 1,316 (±36) | 2,029 (±45) | 0 | 5 |
| 0 | 75 (±9) | 68,351 (±252) | 856 (±29) | 0 | 2,985 (±55) | 4,402 (±66) | 61 (±8) | 6 |
| 14 (±4) | 150 (±12) | 66,647 (±249) | 1,670 (±41) | 8 (±3) | 4,414 (±66) | 6,595 (±81) | 208 (±14) | 7 |
| 32 (±6) | 222 (±15) | 62,922 (±243) | 2,375 (±49) | 20 (±4) | 5,681 (±75) | 8,901 (±94) | 547 (±23) | 8 |
| 52 (±7) | 255 (±16) | 57,202 (±232) | 3,176 (±56) | 34 (±6) | 6,868 (±83) | 10,432 (±102) | 868 (±29) | 9 |
| 71 (±8) | 259 (±16) | 50,853 (±220) | 3,655 (±60) | 44 (±7) | 7,497 (±86) | 11,504 (±107) | 1,320 (±36) | 10 |
| 89 (±9) | 261 (±16) | 44,994 (±207) | 4,097 (±64) | 44 (±7) | 8,266 (±91) | 12,467 (±111) | 1,819 (±43) | 11 |
| 93 (±10) | 295 (±17) | 39,760 (±195) | 4,348 (±66) | 66 (±8) | 8,389 (±91) | 12,822 (±113) | 2,480 (±50) | 12 |
| 97 (±10) | 290 (±17) | 34,828 (±183) | 4,484 (±67) | 78 (±9) | 8,632 (±93) | 12,981 (±113) | 2,960 (±54) | 13 |
| 104 (±10) | 290 (±17) | 29,647 (±170) | 4,637 (±68) | 66 (±8) | 8,808 (±93) | 12,975 (±113) | 3,496 (±59) | 14 |
| 124 (±11) | 223 (±15) | 26,352 (±160) | 4,680 (±68) | 49 (±7) | 8,746 (±93) | 12,952 (±113) | 4,175 (±64) | 15 |
| 87 (±9) | 224 (±15) | 23,272 (±151) | 4,674 (±68) | 60 (±8) | 8,675 (±93) | 12,660 (±112) | 4,637 (±68) | 16 |
| 103 (±10) | 194 (±14) | 20,412 (±141) | 4,614 (±68) | 89 (±9) | 8,334 (±91) | 12,339 (±110) | 5,045 (±71) | 17 |
| 95 (±10) | 188 (±14) | 17,803 (±132) | 4,530 (±67) | 62 (±8) | 8,193 (±90) | 11,880 (±108) | 5,504 (±74) | 18 |
| 90 (±9) | 161 (±13) | 15,409 (±123) | 4,450 (±67) | 70 (±8) | 7,730 (±88) | 11,442 (±106) | 5,875 (±76) | 19 |
| 94 (±10) | 173 (±13) | 13,911 (±117) | 4,299 (±65) | 75 (±9) | 7,419 (±86) | 11,093 (±105) | 6,133 (±78) | 20 |
| 82 (±9) | 153 (±12) | 12,078 (±109) | 4,199 (±65) | 63 (±8) | 7,201 (±85) | 10,462 (±102) | 6,336 (±79) | 21 |
| 85 (±9) | 122 (±11) | 10,816 (±103) | 3,988 (±63) | 59 (±8) | 6,956 (±83) | 10,062 (±100) | 6,676 (±81) | 22 |
| 64 (±8) | 134 (±12) | 9,414 (±97) | 3,848 (±62) | 52 (±7) | 6,534 (±81) | 9,515 (±97) | 6,919 (±83) | 23 |
| 83 (±9) | 93 (±10) | 8,684 (±93) | 3,768 (±61) | 66 (±8) | 6,313 (±79) | 9,086 (±95) | 7,143 (±84) | 24 |
| 60 (±8) | 95 (±10) | 7,652 (±87) | 3,495 (±59) | 49 (±7) | 6,301 (±79) | 8,843 (±94) | 7,419 (±86) | 25 |
| 58 (±8) | 82 (±9) | 6,938 (±83) | 3,355 (±58) | 57 (±8) | 5,977 (±77) | 8,371 (±91) | 7,392 (±86) | 26 |
| 39 (±6) | 91 (±10) | 6,220 (±79) | 3,319 (±58) | 46 (±7) | 5,623 (±75) | 8,050 (±89) | 7,664 (±87) | 27 |
| 29 (±5) | 79 (±9) | 5,642 (±75) | 3,143 (±56) | 56 (±7) | 5,551 (±74) | 7,823 (±88) | 7,784 (±88) | 28 |
| 34 (±6) | 72 (±8) | 5,348 (±73) | 3,094 (±56) | 53 (±7) | 5,337 (±73) | 7,372 (±86) | 7,869 (±88) | 29 |
| 39 (±6) | 60 (±8) | 4,795 (±69) | 2,971 (±54) | 44 (±7) | 5,152 (±72) | 6,906 (±83) | 7,851 (±88) | 30 |
| 34 (±6) | 69 (±8) | 4,411 (±66) | 2,836 (±53) | 58 (±8) | 4,924 (±70) | 6,626 (±81) | 7,876 (±88) | 31 |
| 40 (±6) | 67 (±8) | 4,034 (±63) | 2,822 (±53) | 41 (±6) | 4,839 (±69) | 6,422 (±80) | 8,104 (±90) | 32 |
| 34 (±6) | 47 (±7) | 3,723 (±61) | 2,598 (±51) | 42 (±6) | 4,621 (±68) | 6,160 (±78) | 7,998 (±89) | 33 |
| 31 (±6) | 46 (±7) | 3,361 (±58) | 2,555 (±50) | 44 (±7) | 4,420 (±66) | 5,832 (±76) | 8,231 (±90) | 34 |
| 32 (±6) | 34 (±6) | 3,064 (±55) | 2,489 (±50) | 38 (±6) | 4,311 (±66) | 5,730 (±75) | 7,824 (±88) | 35 |
| 36 (±6) | 23 (±5) | 2,927 (±54) | 2,321 (±48) | 35 (±6) | 4,201 (±65) | 5,451 (±74) | 7,896 (±89) | 36 |
| 23 (±5) | 45 (±7) | 2,742 (±52) | 2,321 (±48) | 43 (±7) | 4,024 (±63) | 5,091 (±71) | 7,854 (±88) | 37 |
| 30 (±5) | 32 (±6) | 2,522 (±50) | 2,183 (±47) | 26 (±5) | 3,871 (±62) | 5,002 (±71) | 7,945 (±89) | 38 |
| 25 (±5) | 28 (±5) | 2,359 (±49) | 2,141 (±46) | 40 (±6) | 3,851 (±62) | 4,727 (±69) | 7,976 (±89) | 39 |
| 24 (±5) | 36 (±6) | 2,252 (±47) | 2,071 (±45) | 28 (±5) | 3,580 (±60) | 4,587 (±68) | 7,915 (±89) | 40 |
| 17 (±4) | 39 (±6) | 2,090 (±46) | 2,084 (±46) | 23 (±5) | 3,627 (±60) | 4,459 (±67) | 7,743 (±88) | 41 |
| 21 (±5) | 28 (±5) | 1,943 (±44) | 1,966 (±44) | 36 (±6) | 3,401 (±58) | 4,260 (±65) | 7,888 (±88) | 42 |
| 26 (±5) | 27 (±5) | 1,857 (±43) | 1,874 (±43) | 29 (±5) | 3,339 (±58) | 4,131 (±64) | 7,810 (±88) | 43 |
| 19 (±4) | 22 (±5) | 1,774 (±42) | 1,845 (±43) | 30 (±5) | 3,230 (±57) | 3,876 (±62) | 7,763 (±88) | 44 |
| 9 (±3) | 19 (±4) | 1,723 (±41) | 1,819 (±43) | 25 (±5) | 3,214 (±57) | 3,783 (±61) | 7,720 (±88) | 45 |
| 18 (±4) | 24 (±5) | 1,551 (±39) | 1,755 (±42) | 25 (±5) | 3,082 (±55) | 3,559 (±60) | 7,544 (±87) | 46 |
| 21 (±5) | 23 (±5) | 1,419 (±38) | 1,703 (±41) | 30 (±5) | 3,090 (±56) | 3,598 (±60) | 7,636 (±87) | 47 |
| 19 (±4) | 17 (±4) | 1,378 (±37) | 1,625 (±40) | 20 (±4) | 2,944 (±54) | 3,337 (±58) | 7,466 (±86) | 48 |
| 9 (±3) | 12 (±3) | 1,319 (±36) | 1,571 (±40) | 24 (±5) | 2,948 (±54) | 3,271 (±57) | 7,568 (±87) | 49 |
| 14 (±4) | 24 (±5) | 1,270 (±36) | 1,544 (±39) | 15 (±4) | 2,743 (±52) | 3,180 (±56) | 7,570 (±87) | 50 |

Table 11.3 (part 2 of 8): comparison of the Simpson-Kramer MinMax method, the Schulze method, Tideman’s ranked pairs method and Heitzig’s river method (source: own calculation)

| A | B | C | D | E | F | G | H |
|---|---|---|---|---|---|---|---|
| 51 | 86,302 (±281) | 106,744 (±309) | 202,719 (±402) | 161,178 (±368) | 426,845 (±495) | 9 (±3) | 31 (±6) |
| 52 | 85,916 (±280) | 105,720 (±307) | 201,958 (±401) | 159,290 (±366) | 431,469 (±495) | 20 (±4) | 24 (±5) |
| 53 | 83,135 (±276) | 104,810 (±306) | 201,601 (±401) | 156,845 (±364) | 438,263 (±496) | 13 (±4) | 29 (±5) |
| 54 | 81,199 (±273) | 104,546 (±306) | 200,867 (±401) | 156,050 (±363) | 442,101 (±497) | 11 (±3) | 27 (±5) |
| 55 | 79,388 (±270) | 103,805 (±305) | 200,369 (±400) | 154,225 (±361) | 447,015 (±497) | 10 (±3) | 28 (±5) |
| 56 | 78,600 (±269) | 103,108 (±304) | 199,570 (±400) | 151,677 (±359) | 452,299 (±498) | 11 (±3) | 15 (±4) |
| 57 | 75,926 (±265) | 102,645 (±303) | 200,161 (±400) | 151,416 (±358) | 455,363 (±498) | 7 (±3) | 30 (±5) |
| 58 | 74,594 (±263) | 102,027 (±303) | 199,058 (±399) | 149,906 (±357) | 460,145 (±498) | 4 (±2) | 19 (±4) |
| 59 | 73,283 (±261) | 101,386 (±302) | 198,417 (±399) | 148,430 (±356) | 464,655 (±499) | 11 (±3) | 16 (±4) |
| 60 | 72,175 (±259) | 100,828 (±301) | 197,491 (±398) | 146,255 (±353) | 469,464 (±499) | 9 (±3) | 16 (±4) |
| 61 | 71,247 (±257) | 99,509 (±299) | 197,102 (±398) | 144,942 (±352) | 473,792 (±499) | 3 (±2) | 18 (±4) |
| 62 | 70,093 (±255) | 99,398 (±299) | 196,566 (±397) | 143,467 (±351) | 477,049 (±499) | 8 (±3) | 23 (±5) |
| 63 | 68,360 (±252) | 98,741 (±298) | 196,147 (±397) | 142,316 (±349) | 481,382 (±500) | 5 (±2) | 17 (±4) |
| 64 | 67,433 (±251) | 97,900 (±297) | 195,242 (±396) | 140,984 (±348) | 485,500 (±500) | 9 (±3) | 18 (±4) |
| 65 | 65,618 (±248) | 97,981 (±297) | 194,913 (±396) | 139,342 (±346) | 489,376 (±500) | 4 (±2) | 13 (±4) |
| 66 | 64,870 (±246) | 96,745 (±296) | 194,170 (±396) | 138,357 (±345) | 493,118 (±500) | 6 (±2) | 14 (±4) |
| 67 | 63,328 (±244) | 96,876 (±296) | 193,633 (±395) | 137,236 (±344) | 496,571 (±500) | 5 (±2) | 10 (±3) |
| 68 | 62,502 (±242) | 96,193 (±295) | 193,530 (±395) | 135,333 (±342) | 500,321 (±500) | 5 (±2) | 15 (±4) |
| 69 | 61,432 (±240) | 95,239 (±294) | 192,656 (±394) | 133,883 (±341) | 504,674 (±500) | 6 (±2) | 9 (±3) |
| 70 | 60,109 (±238) | 95,028 (±293) | 192,412 (±394) | 133,246 (±340) | 507,411 (±500) | 5 (±2) | 15 (±4) |
| 71 | 59,090 (±236) | 94,519 (±293) | 191,759 (±394) | 131,404 (±338) | 511,614 (±500) | 5 (±2) | 9 (±3) |
| 72 | 58,108 (±234) | 94,348 (±292) | 191,708 (±394) | 130,485 (±337) | 513,822 (±500) | 3 (±2) | 14 (±4) |
| 73 | 57,511 (±233) | 93,288 (±291) | 190,405 (±393) | 130,109 (±336) | 517,360 (±500) | 4 (±2) | 14 (±4) |
| 74 | 56,477 (±231) | 92,848 (±290) | 190,403 (±393) | 128,096 (±334) | 520,772 (±500) | 1 (±1) | 9 (±3) |
| 75 | 55,528 (±229) | 92,180 (±289) | 189,752 (±392) | 127,173 (±333) | 524,351 (±499) | 5 (±2) | 6 (±2) |
| 76 | 54,563 (±227) | 91,674 (±289) | 188,995 (±392) | 126,067 (±332) | 527,637 (±499) | 5 (±2) | 7 (±3) |
| 77 | 54,567 (±227) | 90,832 (±287) | 188,710 (±391) | 124,663 (±330) | 530,464 (±499) | 4 (±2) | 5 (±2) |
| 78 | 53,597 (±225) | 91,093 (±288) | 187,459 (±390) | 123,524 (±329) | 533,730 (±499) | 3 (±2) | 10 (±3) |
| 79 | 52,556 (±223) | 90,606 (±287) | 187,371 (±390) | 122,856 (±328) | 535,998 (±499) | 4 (±2) | 5 (±2) |
| 80 | 51,498 (±221) | 89,995 (±286) | 187,230 (±390) | 121,979 (±327) | 538,876 (±498) | 3 (±2) | 7 (±3) |
| 81 | 50,623 (±219) | 89,886 (±286) | 186,698 (±390) | 120,998 (±326) | 541,460 (±498) | 7 (±3) | 9 (±3) |
| 82 | 50,460 (±219) | 88,963 (±285) | 185,707 (±389) | 119,950 (±325) | 544,912 (±498) | 2 (±1) | 7 (±3) |
| 83 | 49,941 (±218) | 88,207 (±284) | 184,859 (±388) | 118,758 (±324) | 548,127 (±498) | 0 | 5 (±2) |
| 84 | 48,693 (±215) | 87,649 (±283) | 184,512 (±388) | 117,725 (±322) | 551,363 (±497) | 0 | 9 (±3) |
| 85 | 47,948 (±214) | 87,283 (±282) | 184,302 (±388) | 117,114 (±322) | 553,501 (±497) | 3 (±2) | 6 (±2) |
| 86 | 47,262 (±212) | 86,863 (±282) | 183,751 (±387) | 116,554 (±321) | 555,909 (±497) | 2 (±1) | 8 (±3) |
| 87 | 46,801 (±211) | 86,448 (±281) | 183,299 (±387) | 115,620 (±320) | 558,377 (±497) | 1 (±1) | 6 (±2) |
| 88 | 46,092 (±210) | 85,795 (±280) | 182,719 (±386) | 113,714 (±317) | 562,251 (±496) | 2 (±1) | 5 (±2) |
| 89 | 45,580 (±209) | 85,826 (±280) | 182,433 (±386) | 113,215 (±317) | 563,581 (±496) | 1 (±1) | 8 (±3) |
| 90 | 44,921 (±207) | 85,150 (±279) | 182,193 (±386) | 112,648 (±316) | 565,748 (±496) | 3 (±2) | 4 (±2) |
| 91 | 44,656 (±207) | 84,833 (±279) | 180,662 (±385) | 111,242 (±314) | 569,290 (±495) | 4 (±2) | 7 (±3) |
| 92 | 43,531 (±204) | 84,220 (±278) | 181,384 (±385) | 110,642 (±314) | 571,048 (±495) | 1 (±1) | 6 (±2) |
| 93 | 42,988 (±203) | 83,819 (±277) | 180,707 (±385) | 109,824 (±313) | 573,476 (±495) | 1 (±1) | 5 (±2) |
| 94 | 42,671 (±202) | 83,870 (±277) | 179,300 (±384) | 109,578 (±312) | 575,608 (±494) | 4 (±2) | 3 (±2) |
| 95 | 41,961 (±201) | 83,289 (±276) | 179,363 (±384) | 108,515 (±311) | 577,803 (±494) | 3 (±2) | 4 (±2) |
| 96 | 41,707 (±200) | 82,859 (±276) | 178,992 (±383) | 107,633 (±310) | 579,889 (±494) | 1 (±1) | 5 (±2) |
| 97 | 40,873 (±198) | 82,483 (±275) | 178,671 (±383) | 105,966 (±308) | 583,261 (±493) | 1 (±1) | 3 (±2) |
| 98 | 40,773 (±198) | 81,464 (±274) | 178,264 (±383) | 106,392 (±308) | 584,486 (±493) | 0 | 3 (±2) |
| 99 | 40,178 (±196) | 82,041 (±274) | 177,496 (±382) | 105,424 (±307) | 586,288 (±492) | 1 (±1) | 4 (±2) |
| 100 | 39,552 (±195) | 81,427 (±273) | 176,721 (±381) | 104,260 (±306) | 589,604 (±492) | 4 (±2) | 2 (±1) |

Table 11.3 (part 3 of 8): comparison of the Simpson-Kramer MinMax method, the Schulze method, Tideman's ranked pairs method and Heitzig's river method (source: own calculation)

| I | J | K | L | M | N | O | P | A |
|---|---|---|---|---|---|---|---|---|
| 8 (±3) | 20 (±4) | 1,225 (±35) | 1,559 (±39) | 29 (±5) | 2,780 (±53) | 3,112 (±56) | 7,439 (±86) | 51 |
| 13 (±4) | 22 (±5) | 1,076 (±33) | 1,452 (±38) | 19 (±4) | 2,721 (±52) | 2,908 (±54) | 7,392 (±86) | 52 |
| 11 (±3) | 14 (±4) | 1,100 (±33) | 1,424 (±38) | 17 (±4) | 2,508 (±50) | 2,891 (±54) | 7,339 (±85) | 53 |
| 15 (±4) | 11 (±3) | 1,055 (±32) | 1,384 (±37) | 30 (±5) | 2,633 (±51) | 2,773 (±53) | 7,298 (±85) | 54 |
| 17 (±4) | 10 (±3) | 1,025 (±32) | 1,327 (±36) | 17 (±4) | 2,581 (±51) | 2,771 (±53) | 7,412 (±86) | 55 |
| 11 (±3) | 11 (±3) | 994 (±32) | 1,389 (±37) | 25 (±5) | 2,428 (±49) | 2,640 (±51) | 7,222 (±85) | 56 |
| 6 (±2) | 15 (±4) | 889 (±30) | 1,291 (±36) | 16 (±4) | 2,495 (±50) | 2,569 (±51) | 7,171 (±84) | 57 |
| 10 (±3) | 12 (±3) | 877 (±30) | 1,322 (±36) | 12 (±3) | 2,305 (±48) | 2,605 (±51) | 7,104 (±84) | 58 |
| 14 (±4) | 10 (±3) | 822 (±29) | 1,224 (±35) | 17 (±4) | 2,333 (±48) | 2,366 (±49) | 7,016 (±83) | 59 |
| 10 (±3) | 13 (±4) | 833 (±29) | 1,190 (±34) | 13 (±4) | 2,354 (±48) | 2,392 (±49) | 6,957 (±83) | 60 |
| 9 (±3) | 13 (±4) | 757 (±28) | 1,188 (±34) | 10 (±3) | 2,196 (±47) | 2,261 (±47) | 6,953 (±83) | 61 |
| 12 (±3) | 8 (±3) | 740 (±27) | 1,190 (±34) | 19 (±4) | 2,155 (±46) | 2,295 (±48) | 6,977 (±83) | 62 |
| 8 (±3) | 6 (±2) | 712 (±27) | 1,121 (±33) | 20 (±4) | 2,120 (±46) | 2,197 (±47) | 6,848 (±82) | 63 |
| 6 (±2) | 11 (±3) | 722 (±27) | 1,111 (±33) | 20 (±4) | 2,119 (±46) | 2,148 (±46) | 6,777 (±82) | 64 |
| 10 (±3) | 8 (±3) | 686 (±26) | 1,134 (±34) | 11 (±3) | 2,100 (±46) | 2,063 (±45) | 6,741 (±82) | 65 |
| 8 (±3) | 8 (±3) | 625 (±25) | 1,050 (±32) | 20 (±4) | 2,089 (±46) | 2,091 (±46) | 6,829 (±82) | 66 |
| 6 (±2) | 11 (±3) | 565 (±24) | 1,070 (±33) | 15 (±4) | 2,014 (±45) | 2,011 (±45) | 6,649 (±81) | 67 |
| 6 (±2) | 6 (±2) | 590 (±24) | 1,001 (±32) | 11 (±3) | 1,967 (±44) | 1,865 (±43) | 6,655 (±81) | 68 |
| 10 (±3) | 6 (±2) | 619 (±25) | 955 (±31) | 16 (±4) | 2,014 (±45) | 1,809 (±42) | 6,672 (±81) | 69 |
| 8 (±3) | 12 (±3) | 553 (±24) | 941 (±31) | 17 (±4) | 1,923 (±44) | 1,868 (±43) | 6,452 (±80) | 70 |
| 7 (±3) | 6 (±2) | 558 (±24) | 940 (±31) | 17 (±4) | 1,806 (±42) | 1,764 (±42) | 6,502 (±80) | 71 |
| 3 (±2) | 7 (±3) | 545 (±23) | 911 (±30) | 8 (±3) | 1,878 (±43) | 1,762 (±42) | 6,398 (±80) | 72 |
| 6 (±2) | 3 (±2) | 520 (±23) | 926 (±30) | 9 (±3) | 1,820 (±43) | 1,657 (±41) | 6,368 (±80) | 73 |
| 6 (±2) | 2 (±1) | 496 (±22) | 874 (±30) | 12 (±3) | 1,741 (±42) | 1,684 (±41) | 6,579 (±81) | 74 |
| 3 (±2) | 7 (±3) | 467 (±22) | 839 (±29) | 12 (±3) | 1,791 (±42) | 1,641 (±40) | 6,245 (±79) | 75 |
| 8 (±3) | 5 (±2) | 483 (±22) | 887 (±30) | 7 (±3) | 1,734 (±42) | 1,603 (±40) | 6,325 (±79) | 76 |
| 7 (±3) | 4 (±2) | 429 (±21) | 814 (±29) | 5 (±2) | 1,754 (±42) | 1,486 (±39) | 6,256 (±79) | 77 |
| 6 (±2) | 2 (±1) | 438 (±21) | 785 (±28) | 7 (±3) | 1,690 (±41) | 1,555 (±39) | 6,101 (±78) | 78 |
| 3 (±2) | 4 (±2) | 411 (±20) | 849 (±29) | 8 (±3) | 1,675 (±41) | 1,475 (±38) | 6,179 (±78) | 79 |
| 7 (±3) | 8 (±3) | 384 (±20) | 804 (±28) | 6 (±2) | 1,602 (±40) | 1,415 (±38) | 6,186 (±78) | 80 |
| 5 (±2) | 3 (±2) | 386 (±20) | 776 (±28) | 8 (±3) | 1,530 (±39) | 1,492 (±39) | 6,119 (±78) | 81 |
| 4 (±2) | 2 (±1) | 441 (±21) | 733 (±27) | 15 (±4) | 1,534 (±39) | 1,332 (±36) | 5,938 (±77) | 82 |
| 9 (±3) | 6 (±2) | 422 (±21) | 762 (±28) | 12 (±3) | 1,573 (±40) | 1,372 (±37) | 5,947 (±77) | 83 |
| 4 (±2) | 3 (±2) | 376 (±19) | 765 (±28) | 6 (±2) | 1,569 (±40) | 1,299 (±36) | 6,027 (±77) | 84 |
| 1 (±1) | 0 | 355 (±19) | 732 (±27) | 8 (±3) | 1,479 (±38) | 1,283 (±36) | 5,985 (±77) | 85 |
| 3 (±2) | 2 (±1) | 359 (±19) | 716 (±27) | 8 (±3) | 1,464 (±38) | 1,253 (±35) | 5,846 (±76) | 86 |
| 2 (±1) | 4 (±2) | 361 (±19) | 729 (±27) | 8 (±3) | 1,429 (±38) | 1,261 (±35) | 5,654 (±75) | 87 |
| 2 (±1) | 5 (±2) | 359 (±19) | 696 (±26) | 10 (±3) | 1,435 (±38) | 1,182 (±34) | 5,733 (±75) | 88 |
| 3 (±2) | 4 (±2) | 326 (±18) | 671 (±26) | 17 (±4) | 1,448 (±38) | 1,158 (±34) | 5,729 (±75) | 89 |
| 5 (±2) | 8 (±3) | 280 (±17) | 685 (±26) | 8 (±3) | 1,390 (±37) | 1,147 (±34) | 5,810 (±76) | 90 |
| 3 (±2) | 4 (±2) | 300 (±17) | 667 (±26) | 7 (±3) | 1,359 (±37) | 1,176 (±34) | 5,790 (±76) | 91 |
| 2 (±1) | 3 (±2) | 320 (±18) | 667 (±26) | 7 (±3) | 1,339 (±37) | 1,147 (±34) | 5,683 (±75) | 92 |
| 2 (±1) | 4 (±2) | 274 (±17) | 637 (±25) | 9 (±3) | 1,328 (±36) | 1,139 (±34) | 5,787 (±76) | 93 |
| 2 (±1) | 4 (±2) | 285 (±17) | 573 (±24) | 7 (±3) | 1,336 (±37) | 1,100 (±33) | 5,659 (±75) | 94 |
| 4 (±2) | 5 (±2) | 272 (±16) | 632 (±25) | 9 (±3) | 1,384 (±37) | 1,081 (±33) | 5,675 (±75) | 95 |
| 4 (±2) | 1 (±1) | 266 (±16) | 623 (±25) | 4 (±2) | 1,321 (±36) | 1,076 (±33) | 5,619 (±75) | 96 |
| 5 (±2) | 1 (±1) | 247 (±16) | 630 (±25) | 6 (±2) | 1,295 (±36) | 1,039 (±32) | 5,519 (±74) | 97 |
| 1 (±1) | 3 (±2) | 275 (±17) | 596 (±24) | 7 (±3) | 1,268 (±36) | 968 (±31) | 5,500 (±74) | 98 |
| 5 (±2) | 1 (±1) | 270 (±16) | 571 (±24) | 7 (±3) | 1,222 (±35) | 978 (±31) | 5,514 (±74) | 99 |
| 1 (±1) | 0 | 239 (±15) | 576 (±24) | 5 (±2) | 1,244 (±35) | 962 (±31) | 5,403 (±73) | 100 |

Table 11.3 (part 4 of 8): comparison of the Simpson-Kramer MinMax method, the Schulze method, Tideman’s ranked pairs method and Heitzig’s river method (source: own calculation)

| A | B | C | D | E | F | G | H |
|---|---|---|---|---|---|---|---|
| 101 | 38,828 (±193) | 81,130 (±273) | 176,795 (±381) | 103,526 (±305) | 591,319 (±492) | 1 (±1) | 2 (±1) |
| 102 | 38,825 (±193) | 80,728 (±272) | 175,638 (±381) | 103,649 (±305) | 592,822 (±491) | 2 (±1) | 3 (±2) |
| 103 | 38,219 (±192) | 79,545 (±271) | 175,311 (±380) | 102,945 (±304) | 595,800 (±491) | 3 (±2) | 5 (±2) |
| 104 | 37,864 (±191) | 80,098 (±271) | 175,522 (±380) | 101,319 (±302) | 596,885 (±491) | 1 (±1) | 3 (±2) |
| 105 | 37,715 (±191) | 79,162 (±270) | 175,203 (±380) | 100,477 (±301) | 599,223 (±490) | 1 (±1) | 2 (±1) |
| 106 | 37,276 (±189) | 78,827 (±269) | 174,452 (±379) | 100,193 (±300) | 601,225 (±490) | 2 (±1) | 4 (±2) |
| 107 | 36,918 (±189) | 78,782 (±269) | 174,336 (±379) | 100,002 (±300) | 602,238 (±489) | 2 (±1) | 3 (±2) |
| 108 | 35,978 (±186) | 77,847 (±268) | 173,819 (±379) | 99,118 (±299) | 605,193 (±489) | 1 (±1) | 5 (±2) |
| 109 | 35,940 (±186) | 77,690 (±268) | 173,184 (±378) | 98,317 (±298) | 607,110 (±488) | 1 (±1) | 2 (±1) |
| 110 | 35,454 (±185) | 77,523 (±267) | 172,897 (±378) | 97,410 (±297) | 608,961 (±488) | 3 (±2) | 3 (±2) |
| 111 | 35,167 (±184) | 77,310 (±267) | 172,674 (±378) | 97,277 (±296) | 610,014 (±488) | 0 | 0 |
| 112 | 34,819 (±183) | 76,510 (±266) | 171,918 (±377) | 96,055 (±295) | 613,155 (±487) | 1 (±1) | 5 (±2) |
| 113 | 34,235 (±182) | 76,861 (±266) | 171,724 (±377) | 95,583 (±294) | 614,086 (±487) | 0 | 2 (±1) |
| 114 | 33,973 (±181) | 76,153 (±265) | 171,240 (±377) | 94,496 (±293) | 616,569 (±486) | 0 | 3 (±2) |
| 115 | 33,420 (±180) | 75,630 (±264) | 170,998 (±377) | 94,664 (±293) | 617,763 (±486) | 0 | 3 (±2) |
| 116 | 33,182 (±179) | 75,187 (±264) | 170,127 (±376) | 94,219 (±292) | 619,929 (±485) | 0 | 3 (±2) |
| 117 | 32,977 (±179) | 75,219 (±264) | 169,806 (±375) | 93,195 (±291) | 621,378 (±485) | 2 (±1) | 2 (±1) |
| 118 | 32,378 (±177) | 74,799 (±263) | 169,863 (±376) | 92,470 (±290) | 623,252 (±485) | 0 | 2 (±1) |
| 119 | 32,253 (±177) | 74,310 (±262) | 168,836 (±375) | 91,769 (±289) | 625,571 (±484) | 0 | 2 (±1) |
| 120 | 31,826 (±176) | 74,476 (±263) | 168,949 (±375) | 91,770 (±289) | 625,743 (±484) | 1 (±1) | 0 |
| 121 | 31,328 (±174) | 74,018 (±262) | 169,068 (±375) | 91,260 (±288) | 627,199 (±484) | 0 | 1 (±1) |
| 122 | 31,247 (±174) | 73,273 (±261) | 167,697 (±374) | 91,114 (±288) | 629,607 (±483) | 1 (±1) | 4 (±2) |
| 123 | 31,286 (±174) | 72,559 (±259) | 168,033 (±374) | 89,805 (±286) | 631,313 (±482) | 0 | 3 (±2) |
| 124 | 30,738 (±173) | 72,424 (±259) | 167,404 (±373) | 89,166 (±285) | 633,340 (±482) | 0 | 1 (±1) |
| 125 | 30,359 (±172) | 72,614 (±260) | 166,444 (±372) | 89,150 (±285) | 634,575 (±482) | 3 (±2) | 1 (±1) |
| 126 | 30,140 (±171) | 72,451 (±259) | 166,435 (±372) | 88,481 (±284) | 635,792 (±481) | 0 | 1 (±1) |
| 127 | 29,946 (±170) | 71,900 (±258) | 164,987 (±371) | 88,429 (±284) | 638,059 (±481) | 0 | 3 (±2) |
| 128 | 29,314 (±169) | 71,747 (±258) | 166,296 (±372) | 87,297 (±282) | 638,577 (±480) | 0 | 0 |
| 129 | 29,472 (±169) | 71,025 (±257) | 165,775 (±372) | 86,816 (±282) | 640,271 (±480) | 2 (±1) | 0 |
| 130 | 28,918 (±168) | 70,999 (±257) | 164,958 (±371) | 86,399 (±281) | 642,163 (±479) | 0 | 1 (±1) |
| 131 | 28,944 (±168) | 70,880 (±257) | 164,512 (±371) | 85,958 (±280) | 643,186 (±479) | 0 | 2 (±1) |
| 132 | 28,540 (±167) | 70,161 (±255) | 164,521 (±371) | 84,792 (±279) | 645,383 (±478) | 1 (±1) | 4 (±2) |
| 133 | 28,510 (±166) | 69,992 (±255) | 163,835 (±370) | 84,790 (±279) | 646,547 (±478) | 0 | 1 (±1) |
| 134 | 27,996 (±165) | 69,805 (±255) | 163,051 (±369) | 84,549 (±278) | 648,196 (±478) | 1 (±1) | 1 (±1) |
| 135 | 27,786 (±164) | 69,658 (±255) | 163,406 (±370) | 84,380 (±278) | 648,296 (±478) | 1 (±1) | 1 (±1) |
| 136 | 27,454 (±163) | 69,374 (±254) | 163,106 (±369) | 83,890 (±277) | 649,918 (±477) | 0 | 1 (±1) |
| 137 | 27,360 (±163) | 68,788 (±253) | 162,737 (±369) | 82,787 (±276) | 652,096 (±476) | 1 (±1) | 2 (±1) |
| 138 | 26,817 (±162) | 68,744 (±253) | 162,357 (±369) | 82,855 (±276) | 652,933 (±476) | 0 | 3 (±2) |
| 139 | 26,828 (±162) | 68,404 (±252) | 162,113 (±369) | 82,725 (±275) | 653,745 (±476) | 1 (±1) | 1 (±1) |
| 140 | 26,538 (±161) | 68,056 (±252) | 161,340 (±368) | 81,536 (±274) | 656,439 (±475) | 1 (±1) | 0 |
| 141 | 26,319 (±160) | 67,735 (±251) | 161,395 (±368) | 81,207 (±273) | 657,276 (±475) | 0 | 3 (±2) |
| 142 | 26,001 (±159) | 67,571 (±251) | 161,218 (±368) | 80,929 (±273) | 658,200 (±474) | 0 | 1 (±1) |
| 143 | 25,841 (±159) | 66,957 (±250) | 161,146 (±368) | 80,351 (±272) | 659,614 (±474) | 0 | 0 |
| 144 | 25,723 (±158) | 67,269 (±250) | 160,658 (±367) | 79,913 (±271) | 660,478 (±474) | 0 | 1 (±1) |
| 145 | 25,334 (±157) | 66,937 (±250) | 159,599 (±366) | 80,030 (±271) | 662,135 (±473) | 1 (±1) | 1 (±1) |
| 146 | 25,142 (±157) | 66,821 (±250) | 158,957 (±366) | 79,100 (±270) | 664,038 (±472) | 0 | 0 |
| 147 | 25,015 (±156) | 66,542 (±249) | 159,018 (±366) | 78,398 (±269) | 665,132 (±472) | 0 | 2 (±1) |
| 148 | 24,985 (±156) | 66,228 (±249) | 159,036 (±366) | 78,062 (±268) | 665,806 (±472) | 0 | 1 (±1) |
| 149 | 24,905 (±156) | 65,616 (±248) | 159,206 (±366) | 78,003 (±268) | 666,517 (±471) | 0 | 0 |
| 150 | 24,478 (±155) | 65,462 (±247) | 158,704 (±365) | 77,714 (±268) | 667,877 (±471) | 0 | 1 (±1) |

Table 11.3 (part 5 of 8): comparison of the Simpson-Kramer MinMax method, the Schulze method, Tideman’s ranked pairs method and Heitzig’s river method (source: own calculation)

| I | J | K | L | M | N | O | P | A |
|---|---|---|---|---|---|---|---|---|
| 3 (±2) | 2 (±1) | 241 (±16) | 545 (±23) | 6 (±2) | 1,238 (±35) | 1,008 (±32) | 5,356 (±73) | 101 |
| 1 (±1) | 2 (±1) | 221 (±15) | 546 (±23) | 7 (±3) | 1,238 (±35) | 973 (±31) | 5,345 (±73) | 102 |
| 2 (±1) | 2 (±1) | 203 (±14) | 574 (±24) | 2 (±1) | 1,208 (±35) | 943 (±31) | 5,238 (±72) | 103 |
| 6 (±2) | 5 (±2) | 236 (±15) | 565 (±24) | 3 (±2) | 1,136 (±34) | 935 (±31) | 5,422 (±73) | 104 |
| 1 (±1) | 2 (±1) | 218 (±15) | 560 (±24) | 4 (±2) | 1,173 (±34) | 893 (±30) | 5,366 (±73) | 105 |
| 1 (±1) | 0 | 205 (±14) | 520 (±23) | 6 (±2) | 1,147 (±34) | 949 (±31) | 5,193 (±72) | 106 |
| 7 (±3) | 4 (±2) | 195 (±14) | 489 (±22) | 5 (±2) | 1,105 (±33) | 851 (±29) | 5,063 (±71) | 107 |
| 1 (±1) | 1 (±1) | 226 (±15) | 524 (±23) | 10 (±3) | 1,144 (±34) | 879 (±30) | 5,254 (±72) | 108 |
| 1 (±1) | 0 | 210 (±14) | 526 (±23) | 2 (±1) | 1,063 (±33) | 855 (±29) | 5,099 (±71) | 109 |
| 2 (±1) | 2 (±1) | 184 (±14) | 485 (±22) | 7 (±3) | 1,090 (±33) | 793 (±28) | 5,186 (±72) | 110 |
| 4 (±2) | 0 | 186 (±14) | 519 (±23) | 1 (±1) | 990 (±31) | 839 (±29) | 5,019 (±71) | 111 |
| 2 (±1) | 2 (±1) | 179 (±13) | 469 (±22) | 1 (±1) | 1,059 (±33) | 843 (±29) | 4,982 (±70) | 112 |
| 0 | 1 (±1) | 182 (±13) | 451 (±21) | 3 (±2) | 1,065 (±33) | 772 (±28) | 5,035 (±71) | 113 |
| 4 (±2) | 1 (±1) | 168 (±13) | 468 (±22) | 10 (±3) | 1,031 (±32) | 790 (±28) | 5,094 (±71) | 114 |
| 2 (±1) | 1 (±1) | 174 (±13) | 461 (±21) | 3 (±2) | 1,089 (±33) | 769 (±28) | 5,023 (±71) | 115 |
| 3 (±2) | 2 (±1) | 170 (±13) | 463 (±22) | 1 (±1) | 1,072 (±33) | 769 (±28) | 4,873 (±70) | 116 |
| 1 (±1) | 0 | 187 (±14) | 497 (±22) | 9 (±3) | 1,015 (±32) | 743 (±27) | 4,969 (±70) | 117 |
| 1 (±1) | 0 | 178 (±13) | 439 (±21) | 6 (±2) | 987 (±31) | 731 (±27) | 4,894 (±70) | 118 |
| 4 (±2) | 3 (±2) | 178 (±13) | 438 (±21) | 5 (±2) | 994 (±32) | 776 (±28) | 4,861 (±70) | 119 |
| 1 (±1) | 2 (±1) | 175 (±13) | 404 (±20) | 2 (±1) | 998 (±32) | 721 (±27) | 4,932 (±70) | 120 |
| 0 | 0 | 166 (±13) | 450 (±21) | 3 (±2) | 993 (±31) | 679 (±26) | 4,835 (±69) | 121 |
| 1 (±1) | 0 | 167 (±13) | 417 (±20) | 5 (±2) | 1,002 (±32) | 702 (±26) | 4,763 (±69) | 122 |
| 1 (±1) | 1 (±1) | 169 (±13) | 428 (±21) | 1 (±1) | 933 (±31) | 650 (±25) | 4,818 (±69) | 123 |
| 1 (±1) | 2 (±1) | 125 (±11) | 403 (±20) | 2 (±1) | 932 (±31) | 673 (±26) | 4,789 (±69) | 124 |
| 1 (±1) | 1 (±1) | 141 (±12) | 398 (±20) | 2 (±1) | 919 (±30) | 666 (±26) | 4,726 (±69) | 125 |
| 0 | 2 (±1) | 152 (±12) | 400 (±20) | 4 (±2) | 937 (±31) | 627 (±25) | 4,578 (±68) | 126 |
| 4 (±2) | 0 | 143 (±12) | 383 (±20) | 2 (±1) | 912 (±30) | 631 (±25) | 4,601 (±68) | 127 |
| 3 (±2) | 2 (±1) | 141 (±12) | 404 (±20) | 6 (±2) | 892 (±30) | 697 (±26) | 4,624 (±68) | 128 |
| 3 (±2) | 0 | 152 (±12) | 349 (±19) | 0 | 915 (±30) | 608 (±25) | 4,612 (±68) | 129 |
| 3 (±2) | 2 (±1) | 127 (±11) | 390 (±20) | 4 (±2) | 865 (±29) | 640 (±25) | 4,531 (±67) | 130 |
| 2 (±1) | 2 (±1) | 134 (±12) | 404 (±20) | 3 (±2) | 870 (±29) | 612 (±25) | 4,491 (±67) | 131 |
| 3 (±2) | 0 | 135 (±12) | 366 (±19) | 8 (±3) | 891 (±30) | 600 (±24) | 4,595 (±68) | 132 |
| 3 (±2) | 2 (±1) | 111 (±11) | 365 (±19) | 4 (±2) | 818 (±29) | 570 (±24) | 4,452 (±67) | 133 |
| 1 (±1) | 0 | 125 (±11) | 399 (±20) | 6 (±2) | 857 (±29) | 610 (±25) | 4,403 (±66) | 134 |
| 1 (±1) | 4 (±2) | 108 (±10) | 374 (±19) | 2 (±1) | 892 (±30) | 562 (±24) | 4,529 (±67) | 135 |
| 3 (±2) | 3 (±2) | 122 (±11) | 355 (±19) | 3 (±2) | 799 (±28) | 557 (±24) | 4,415 (±66) | 136 |
| 1 (±1) | 2 (±1) | 129 (±11) | 381 (±20) | 3 (±2) | 805 (±28) | 586 (±24) | 4,322 (±66) | 137 |
| 1 (±1) | 2 (±1) | 124 (±11) | 348 (±19) | 2 (±1) | 822 (±29) | 550 (±23) | 4,442 (±67) | 138 |
| 0 | 1 (±1) | 100 (±10) | 327 (±18) | 4 (±2) | 816 (±29) | 548 (±23) | 4,387 (±66) | 139 |
| 1 (±1) | 0 | 115 (±11) | 348 (±19) | 4 (±2) | 784 (±28) | 552 (±23) | 4,286 (±65) | 140 |
| 2 (±1) | 1 (±1) | 110 (±10) | 317 (±18) | 6 (±2) | 843 (±29) | 508 (±23) | 4,278 (±65) | 141 |
| 0 | 0 | 116 (±11) | 304 (±17) | 6 (±2) | 850 (±29) | 504 (±22) | 4,300 (±65) | 142 |
| 0 | 1 (±1) | 113 (±11) | 367 (±19) | 2 (±1) | 785 (±28) | 544 (±23) | 4,279 (±65) | 143 |
| 0 | 0 | 101 (±10) | 322 (±18) | 1 (±1) | 780 (±28) | 515 (±23) | 4,239 (±65) | 144 |
| 1 (±1) | 2 (±1) | 109 (±10) | 337 (±18) | 3 (±2) | 804 (±28) | 511 (±23) | 4,196 (±65) | 145 |
| 2 (±1) | 3 (±2) | 104 (±10) | 325 (±18) | 2 (±1) | 770 (±28) | 499 (±22) | 4,237 (±65) | 146 |
| 1 (±1) | 1 (±1) | 108 (±10) | 331 (±18) | 6 (±2) | 796 (±28) | 493 (±22) | 4,157 (±64) | 147 |
| 1 (±1) | 0 | 101 (±10) | 314 (±18) | 3 (±2) | 776 (±28) | 455 (±21) | 4,232 (±65) | 148 |
| 2 (±1) | 2 (±1) | 93 (±10) | 318 (±18) | 2 (±1) | 690 (±26) | 514 (±23) | 4,132 (±64) | 149 |
| 2 (±1) | 0 | 95 (±10) | 296 (±17) | 3 (±2) | 759 (±28) | 467 (±22) | 4,142 (±64) | 150 |

Table 11.3 (part 6 of 8): comparison of the Simpson-Kramer MinMax method, the Schulze method, Tideman’s ranked pairs method and Heitzig’s river method (source: own calculation)

| A | B | C | D | E | F | G | H |
|---|---|---|---|---|---|---|---|
| 151 | 24,360 (±154) | 65,548 (±247) | 158,149 (±365) | 76,652 (±266) | 669,484 (±470) | 2 (±1) | 1 (±1) |
| 152 | 24,006 (±153) | 65,141 (±247) | 157,362 (±364) | 76,738 (±266) | 671,099 (±470) | 0 | 0 |
| 153 | 23,814 (±152) | 64,757 (±246) | 157,108 (±364) | 76,358 (±266) | 672,226 (±469) | 0 | 1 (±1) |
| 154 | 23,671 (±152) | 64,738 (±246) | 157,449 (±364) | 76,091 (±265) | 672,218 (±469) | 0 | 0 |
| 155 | 23,514 (±152) | 63,975 (±245) | 156,857 (±364) | 74,923 (±263) | 675,024 (±468) | 0 | 0 |
| 156 | 23,241 (±151) | 63,865 (±245) | 156,885 (±364) | 75,302 (±264) | 675,319 (±468) | 0 | 1 (±1) |
| 157 | 22,835 (±149) | 63,884 (±245) | 156,976 (±364) | 74,902 (±263) | 675,900 (±468) | 1 (±1) | 0 |
| 158 | 22,663 (±149) | 63,980 (±245) | 155,495 (±362) | 74,563 (±263) | 677,857 (±467) | 0 | 0 |
| 159 | 22,839 (±149) | 63,222 (±243) | 155,466 (±362) | 73,874 (±262) | 679,001 (±467) | 1 (±1) | 1 (±1) |
| 160 | 22,542 (±148) | 63,426 (±244) | 156,024 (±363) | 73,569 (±261) | 678,910 (±467) | 1 (±1) | 1 (±1) |
| 161 | 22,364 (±148) | 62,802 (±243) | 154,881 (±362) | 73,129 (±260) | 681,444 (±466) | 2 (±1) | 1 (±1) |
| 162 | 22,261 (±148) | 62,365 (±242) | 154,485 (±361) | 73,534 (±261) | 681,971 (±466) | 0 | 0 |
| 163 | 21,921 (±146) | 62,713 (±242) | 154,657 (±362) | 72,977 (±260) | 682,372 (±466) | 0 | 0 |
| 164 | 21,866 (±146) | 62,483 (±242) | 154,138 (±361) | 72,538 (±259) | 683,658 (±465) | 0 | 1 (±1) |
| 165 | 21,676 (±146) | 62,091 (±241) | 154,084 (±361) | 72,301 (±259) | 684,562 (±465) | 0 | 0 |
| 166 | 21,626 (±145) | 61,801 (±241) | 153,600 (±361) | 71,646 (±258) | 686,029 (±464) | 0 | 1 (±1) |
| 167 | 21,417 (±145) | 61,798 (±241) | 153,795 (±361) | 71,564 (±258) | 686,159 (±464) | 0 | 0 |
| 168 | 21,261 (±144) | 61,389 (±240) | 152,730 (±360) | 71,262 (±257) | 688,199 (±463) | 0 | 0 |
| 169 | 21,248 (±144) | 60,804 (±239) | 153,243 (±360) | 70,689 (±256) | 688,764 (±463) | 0 | 1 (±1) |
| 170 | 21,055 (±144) | 61,488 (±240) | 152,550 (±360) | 70,456 (±256) | 689,257 (±463) | 1 (±1) | 0 |
| 171 | 20,815 (±143) | 61,333 (±240) | 152,408 (±359) | 69,884 (±255) | 690,351 (±462) | 0 | 2 (±1) |
| 172 | 20,606 (±142) | 60,661 (±239) | 152,129 (±359) | 69,730 (±255) | 691,878 (±462) | 0 | 0 |
| 173 | 20,471 (±142) | 60,091 (±238) | 151,931 (±359) | 69,432 (±254) | 692,976 (±461) | 0 | 1 (±1) |
| 174 | 20,318 (±141) | 59,969 (±237) | 151,173 (±358) | 69,347 (±254) | 694,242 (±461) | 0 | 0 |
| 175 | 20,169 (±141) | 59,673 (±237) | 150,831 (±358) | 68,913 (±253) | 695,489 (±460) | 0 | 1 (±1) |
| 176 | 20,169 (±141) | 59,730 (±237) | 150,682 (±358) | 68,513 (±253) | 695,963 (±460) | 0 | 0 |
| 177 | 19,894 (±140) | 59,964 (±237) | 150,813 (±358) | 68,328 (±252) | 695,991 (±460) | 0 | 1 (±1) |
| 178 | 20,133 (±140) | 59,730 (±237) | 149,608 (±357) | 68,004 (±252) | 697,592 (±459) | 0 | 0 |
| 179 | 19,449 (±138) | 59,591 (±237) | 149,836 (±357) | 67,459 (±251) | 698,799 (±459) | 0 | 1 (±1) |
| 180 | 19,813 (±139) | 59,121 (±236) | 149,742 (±357) | 66,928 (±250) | 699,353 (±459) | 0 | 0 |
| 181 | 19,585 (±139) | 58,920 (±235) | 149,391 (±356) | 66,693 (±249) | 700,658 (±458) | 0 | 0 |
| 182 | 19,248 (±137) | 58,888 (±235) | 149,083 (±356) | 66,786 (±250) | 701,240 (±458) | 1 (±1) | 2 (±1) |
| 183 | 19,186 (±137) | 58,215 (±234) | 148,911 (±356) | 66,371 (±249) | 702,483 (±457) | 0 | 0 |
| 184 | 19,183 (±137) | 58,154 (±234) | 148,550 (±356) | 66,029 (±248) | 703,455 (±457) | 0 | 0 |
| 185 | 18,736 (±136) | 58,165 (±234) | 148,281 (±355) | 65,629 (±248) | 704,497 (±456) | 0 | 0 |
| 186 | 18,735 (±136) | 58,126 (±234) | 148,638 (±356) | 65,928 (±248) | 703,960 (±457) | 1 (±1) | 1 (±1) |
| 187 | 18,851 (±136) | 57,787 (±233) | 147,076 (±354) | 65,303 (±247) | 706,276 (±455) | 0 | 0 |
| 188 | 18,474 (±135) | 57,502 (±233) | 147,693 (±355) | 64,764 (±246) | 706,794 (±455) | 0 | 0 |
| 189 | 18,585 (±135) | 57,454 (±233) | 147,320 (±354) | 65,059 (±247) | 706,945 (±455) | 0 | 1 (±1) |
| 190 | 18,465 (±135) | 56,982 (±232) | 147,233 (±354) | 65,134 (±247) | 707,649 (±455) | 0 | 1 (±1) |
| 191 | 17,999 (±133) | 57,101 (±232) | 147,530 (±355) | 64,315 (±245) | 708,367 (±455) | 0 | 0 |
| 192 | 18,138 (±133) | 57,079 (±232) | 146,192 (±353) | 64,407 (±245) | 709,573 (±454) | 0 | 2 (±1) |
| 193 | 17,929 (±133) | 56,946 (±232) | 145,662 (±353) | 64,125 (±245) | 710,818 (±453) | 0 | 1 (±1) |
| 194 | 17,791 (±132) | 56,859 (±232) | 146,296 (±353) | 63,732 (±244) | 710,645 (±453) | 0 | 1 (±1) |
| 195 | 17,471 (±131) | 56,665 (±231) | 145,453 (±353) | 63,333 (±244) | 712,524 (±453) | 1 (±1) | 0 |
| 196 | 17,962 (±133) | 56,157 (±230) | 144,911 (±352) | 63,124 (±243) | 713,316 (±452) | 0 | 0 |
| 197 | 17,431 (±131) | 55,650 (±229) | 146,049 (±353) | 62,783 (±243) | 713,496 (±452) | 0 | 0 |
| 198 | 17,533 (±131) | 55,579 (±229) | 145,122 (±352) | 62,694 (±242) | 714,495 (±452) | 0 | 0 |
| 199 | 17,314 (±130) | 55,510 (±229) | 144,611 (±352) | 62,539 (±242) | 715,514 (±451) | 0 | 0 |
| 200 | 17,362 (±131) | 55,589 (±229) | 144,222 (±351) | 61,930 (±241) | 716,419 (±451) | 0 | 0 |

Table 11.3 (part 7 of 8): comparison of the Simpson-Kramer MinMax method, the Schulze method, Tideman’s ranked pairs method and Heitzig’s river method (source: own calculation)

| I | J | K | L | M | N | O | P | A |
|---|---|---|---|---|---|---|---|---|
| 1 (±1) | 1 (±1) | 98 (±10) | 293 (±17) | 4 (±2) | 739 (±27) | 440 (±21) | 4,228 (±65) | 151 |
| 2 (±1) | 0 | 84 (±9) | 306 (±17) | 1 (±1) | 729 (±27) | 500 (±22) | 4,032 (±63) | 152 |
| 0 | 0 | 109 (±10) | 312 (±18) | 4 (±2) | 794 (±28) | 481 (±22) | 4,036 (±63) | 153 |
| 0 | 0 | 111 (±11) | 341 (±18) | 3 (±2) | 729 (±27) | 477 (±22) | 4,172 (±64) | 154 |
| 1 (±1) | 0 | 86 (±9) | 316 (±18) | 4 (±2) | 721 (±27) | 446 (±21) | 4,133 (±64) | 155 |
| 1 (±1) | 0 | 74 (±9) | 287 (±17) | 2 (±1) | 723 (±27) | 432 (±21) | 3,868 (±62) | 156 |
| 0 | 0 | 103 (±10) | 297 (±17) | 1 (±1) | 697 (±26) | 455 (±21) | 3,949 (±63) | 157 |
| 1 (±1) | 0 | 78 (±9) | 298 (±17) | 3 (±2) | 700 (±26) | 423 (±21) | 3,939 (±63) | 158 |
| 2 (±1) | 0 | 78 (±9) | 299 (±17) | 1 (±1) | 698 (±26) | 461 (±21) | 4,057 (±64) | 159 |
| 0 | 1 (±1) | 89 (±9) | 289 (±17) | 5 (±2) | 709 (±27) | 425 (±21) | 4,009 (±63) | 160 |
| 2 (±1) | 0 | 73 (±9) | 288 (±17) | 1 (±1) | 673 (±26) | 444 (±21) | 3,896 (±62) | 161 |
| 1 (±1) | 0 | 100 (±10) | 280 (±17) | 1 (±1) | 665 (±26) | 394 (±20) | 3,943 (±63) | 162 |
| 0 | 0 | 103 (±10) | 279 (±17) | 1 (±1) | 671 (±26) | 386 (±20) | 3,920 (±62) | 163 |
| 2 (±1) | 2 (±1) | 82 (±9) | 247 (±16) | 1 (±1) | 690 (±26) | 390 (±20) | 3,902 (±62) | 164 |
| 1 (±1) | 0 | 68 (±8) | 262 (±16) | 2 (±1) | 686 (±26) | 422 (±21) | 3,845 (±62) | 165 |
| 0 | 1 (±1) | 67 (±8) | 285 (±17) | 1 (±1) | 608 (±25) | 382 (±20) | 3,953 (±63) | 166 |
| 1 (±1) | 1 (±1) | 76 (±9) | 273 (±17) | 2 (±1) | 644 (±25) | 382 (±20) | 3,888 (±62) | 167 |
| 1 (±1) | 0 | 80 (±9) | 245 (±16) | 0 | 636 (±25) | 414 (±20) | 3,783 (±61) | 168 |
| 2 (±1) | 0 | 92 (±10) | 255 (±16) | 2 (±1) | 663 (±26) | 398 (±20) | 3,839 (±62) | 169 |
| 0 | 1 (±1) | 81 (±9) | 260 (±16) | 0 | 647 (±25) | 409 (±20) | 3,795 (±61) | 170 |
| 0 | 1 (±1) | 91 (±10) | 248 (±16) | 1 (±1) | 649 (±25) | 387 (±20) | 3,830 (±62) | 171 |
| 2 (±1) | 0 | 70 (±8) | 241 (±16) | 2 (±1) | 624 (±25) | 374 (±19) | 3,683 (±61) | 172 |
| 0 | 0 | 78 (±9) | 253 (±16) | 4 (±2) | 633 (±25) | 368 (±19) | 3,762 (±61) | 173 |
| 1 (±1) | 0 | 65 (±8) | 234 (±15) | 0 | 612 (±25) | 392 (±20) | 3,647 (±60) | 174 |
| 2 (±1) | 0 | 73 (±9) | 232 (±15) | 1 (±1) | 603 (±25) | 355 (±19) | 3,658 (±60) | 175 |
| 1 (±1) | 0 | 63 (±8) | 251 (±16) | 1 (±1) | 598 (±24) | 341 (±18) | 3,688 (±61) | 176 |
| 0 | 0 | 69 (±8) | 219 (±15) | 3 (±2) | 620 (±25) | 374 (±19) | 3,724 (±61) | 177 |
| 0 | 1 (±1) | 69 (±8) | 238 (±15) | 0 | 584 (±24) | 369 (±19) | 3,672 (±60) | 178 |
| 0 | 1 (±1) | 55 (±7) | 257 (±16) | 1 (±1) | 600 (±24) | 349 (±19) | 3,602 (±60) | 179 |
| 1 (±1) | 0 | 68 (±8) | 233 (±15) | 2 (±1) | 596 (±24) | 384 (±20) | 3,759 (±61) | 180 |
| 0 | 0 | 70 (±8) | 229 (±15) | 3 (±2) | 587 (±24) | 338 (±18) | 3,526 (±59) | 181 |
| 1 (±1) | 0 | 84 (±9) | 233 (±15) | 2 (±1) | 543 (±23) | 344 (±19) | 3,545 (±59) | 182 |
| 0 | 0 | 64 (±8) | 230 (±15) | 0 | 587 (±24) | 358 (±19) | 3,595 (±60) | 183 |
| 0 | 1 (±1) | 50 (±7) | 215 (±15) | 3 (±2) | 559 (±24) | 328 (±18) | 3,473 (±59) | 184 |
| 0 | 0 | 53 (±7) | 232 (±15) | 1 (±1) | 582 (±24) | 361 (±19) | 3,463 (±59) | 185 |
| 0 | 0 | 45 (±7) | 204 (±14) | 4 (±2) | 554 (±24) | 302 (±17) | 3,502 (±59) | 186 |
| 1 (±1) | 0 | 39 (±6) | 204 (±14) | 0 | 581 (±24) | 297 (±17) | 3,585 (±60) | 187 |
| 1 (±1) | 0 | 70 (±8) | 215 (±15) | 0 | 547 (±23) | 356 (±19) | 3,584 (±60) | 188 |
| 1 (±1) | 1 (±1) | 63 (±8) | 212 (±15) | 3 (±2) | 542 (±23) | 317 (±18) | 3,497 (±59) | 189 |
| 3 (±2) | 0 | 53 (±7) | 243 (±16) | 1 (±1) | 549 (±23) | 313 (±18) | 3,374 (±58) | 190 |
| 1 (±1) | 0 | 59 (±8) | 215 (±15) | 0 | 576 (±24) | 313 (±18) | 3,524 (±59) | 191 |
| 2 (±1) | 1 (±1) | 53 (±7) | 241 (±16) | 1 (±1) | 566 (±24) | 304 (±17) | 3,441 (±59) | 192 |
| 1 (±1) | 1 (±1) | 49 (±7) | 208 (±14) | 1 (±1) | 571 (±24) | 304 (±17) | 3,384 (±58) | 193 |
| 1 (±1) | 0 | 59 (±8) | 200 (±14) | 4 (±2) | 531 (±23) | 306 (±17) | 3,575 (±60) | 194 |
| 5 (±2) | 0 | 58 (±8) | 183 (±14) | 0 | 530 (±23) | 330 (±18) | 3,447 (±59) | 195 |
| 1 (±1) | 0 | 42 (±6) | 195 (±14) | 1 (±1) | 521 (±23) | 285 (±17) | 3,485 (±59) | 196 |
| 0 | 0 | 53 (±7) | 226 (±15) | 1 (±1) | 503 (±22) | 321 (±18) | 3,487 (±59) | 197 |
| 0 | 0 | 43 (±7) | 215 (±15) | 2 (±1) | 525 (±23) | 318 (±18) | 3,474 (±59) | 198 |
| 0 | 0 | 46 (±7) | 220 (±15) | 0 | 500 (±22) | 302 (±17) | 3,444 (±59) | 199 |
| 2 (±1) | 0 | 53 (±7) | 202 (±14) | 1 (±1) | 519 (±23) | 279 (±17) | 3,422 (±58) | 200 |

Table 11.3 (part 8 of 8): comparison of the Simpson-Kramer MinMax method, the Schulze method, Tideman’s ranked pairs method and Heitzig’s river method (source: own calculation)

With table 11.3, it is possible to calculate how frequently the different election methods agree with each other. The columns in table 11.4 are:

Column “Q” (= “B” + “C” + “D” + “E” + “F”): number of tests, where the Simpson-Kramer MinMax method and the Schulze method choose the same alternative.

Column “R” (= “B” + “C” + “G” + “H” + “I”): number of tests, where the Simpson-Kramer MinMax method and Tideman’s ranked pairs method choose the same alternative.

Column “S” (= “B” + “D” + “G” + “J” + “M”): number of tests, where the Simpson-Kramer MinMax method and Heitzig’s river method choose the same alternative.

Column “T” (= “B” + “C” + “J” + “K” + “L”): number of tests, where the Schulze method and Tideman’s ranked pairs method choose the same alternative.

Column “U” (= “B” + “D” + “H” + “K” + “N”): number of tests, where the Schulze method and Heitzig’s river method choose the same alternative.

Column “V” (= “B” + “E” + “G” + “K” + “O”): number of tests, where Tideman’s ranked pairs method and Heitzig’s river method choose the same alternative.

| A | Q | R | S | T | U | V |
|---|---|---|---|---|---|---|
| 4 | 950,209 (±218) | 816,169 (±387) | 872,114 (±334) | 865,960 (±341) | 921,905 (±268) | 930,255 (±255) |
| 5 | 932,465 (±251) | 750,812 (±433) | 809,662 (±393) | 815,002 (±388) | 874,899 (±331) | 893,424 (±309) |
| 6 | 922,602 (±267) | 687,932 (±463) | 752,896 (±431) | 756,546 (±429) | 824,405 (±380) | 852,084 (±355) |
| 7 | 919,228 (±272) | 637,259 (±481) | 706,925 (±455) | 704,646 (±456) | 778,234 (±415) | 812,163 (±391) |
| 8 | 917,923 (±274) | 595,579 (±491) | 668,986 (±471) | 659,689 (±474) | 737,892 (±440) | 774,942 (±418) |
| 9 | 919,665 (±272) | 559,051 (±497) | 634,310 (±482) | 618,184 (±486) | 698,687 (±459) | 740,105 (±439) |
| 10 | 923,288 (±266) | 529,026 (±499) | 606,407 (±489) | 582,213 (±493) | 665,119 (±472) | 708,035 (±455) |
| 11 | 926,456 (±261) | 501,462 (±500) | 581,596 (±493) | 549,218 (±498) | 635,258 (±481) | 678,194 (±467) |
| 12 | 930,277 (±255) | 476,876 (±499) | 557,577 (±497) | 519,716 (±500) | 606,023 (±489) | 650,464 (±477) |
| 13 | 934,300 (±248) | 456,602 (±498) | 538,028 (±499) | 494,757 (±500) | 581,760 (±493) | 625,307 (±484) |
| 14 | 938,830 (±240) | 436,789 (±496) | 520,000 (±500) | 470,112 (±499) | 558,666 (±497) | 600,445 (±490) |
| 15 | 941,648 (±234) | 419,189 (±493) | 503,362 (±500) | 449,269 (±497) | 538,663 (±499) | 578,472 (±494) |
| 16 | 944,695 (±229) | 404,597 (±491) | 488,897 (±500) | 431,664 (±495) | 521,058 (±500) | 558,691 (±497) |
| 17 | 948,033 (±222) | 389,608 (±488) | 474,466 (±499) | 413,888 (±493) | 503,350 (±500) | 539,722 (±498) |
| 18 | 950,968 (±216) | 377,035 (±485) | 462,916 (±499) | 398,684 (±490) | 489,071 (±500) | 522,422 (±499) |
| 19 | 954,054 (±209) | 364,439 (±481) | 452,563 (±498) | 383,650 (±486) | 475,858 (±499) | 504,895 (±500) |
| 20 | 956,193 (±205) | 353,258 (±478) | 441,023 (±497) | 370,937 (±483) | 462,381 (±499) | 488,989 (±500) |
| 21 | 958,894 (±199) | 342,311 (±474) | 430,779 (±495) | 358,127 (±479) | 450,092 (±498) | 474,032 (±499) |
| 22 | 960,778 (±194) | 332,745 (±471) | 422,128 (±494) | 347,128 (±476) | 439,947 (±496) | 461,151 (±498) |
| 23 | 963,101 (±189) | 325,024 (±468) | 415,053 (±493) | 337,937 (±473) | 431,002 (±495) | 447,786 (±497) |
| 24 | 964,364 (±185) | 315,835 (±465) | 406,367 (±491) | 327,897 (±469) | 421,415 (±494) | 435,314 (±496) |
| 25 | 965,733 (±182) | 307,860 (±462) | 397,743 (±489) | 318,689 (±466) | 411,759 (±492) | 423,706 (±494) |
| 26 | 967,456 (±177) | 300,072 (±458) | 390,975 (±488) | 310,075 (±463) | 403,889 (±491) | 412,646 (±492) |
| 27 | 968,675 (±174) | 292,088 (±455) | 383,742 (±486) | 301,406 (±459) | 395,593 (±489) | 401,985 (±490) |
| 28 | 969,633 (±172) | 286,038 (±452) | 377,936 (±485) | 294,613 (±456) | 389,144 (±488) | 391,644 (±488) |
| 29 | 970,583 (±169) | 279,979 (±449) | 372,471 (±483) | 288,221 (±453) | 383,143 (±486) | 382,999 (±486) |
| 30 | 971,980 (±165) | 273,734 (±446) | 366,580 (±482) | 281,319 (±450) | 376,499 (±485) | 373,798 (±484) |
| 31 | 973,007 (±162) | 268,148 (±443) | 361,232 (±480) | 275,271 (±447) | 370,495 (±483) | 364,557 (±481) |
| 32 | 973,457 (±161) | 262,691 (±440) | 356,244 (±479) | 269,400 (±444) | 365,097 (±481) | 357,844 (±479) |
| 33 | 974,637 (±157) | 257,757 (±437) | 351,058 (±477) | 263,951 (±441) | 359,369 (±480) | 348,913 (±477) |
| 34 | 975,314 (±155) | 252,470 (±434) | 346,526 (±476) | 258,235 (±438) | 354,299 (±478) | 342,133 (±474) |
| 35 | 976,369 (±152) | 247,338 (±431) | 342,480 (±475) | 252,784 (±435) | 349,828 (±477) | 335,538 (±472) |
| 36 | 976,987 (±150) | 243,472 (±429) | 337,881 (±473) | 248,584 (±432) | 345,000 (±475) | 328,129 (±470) |
| 37 | 977,747 (±148) | 239,149 (±427) | 333,119 (±471) | 244,124 (±430) | 339,861 (±474) | 322,014 (±467) |
| 38 | 978,255 (±146) | 234,807 (±424) | 328,766 (±470) | 239,380 (±427) | 335,155 (±472) | 315,372 (±465) |
| 39 | 978,771 (±144) | 231,531 (±422) | 325,759 (±469) | 235,952 (±425) | 331,941 (±471) | 309,624 (±462) |
| 40 | 979,416 (±142) | 226,809 (±419) | 321,724 (±467) | 231,053 (±422) | 327,533 (±469) | 303,425 (±460) |
| 41 | 979,850 (±141) | 222,979 (±416) | 317,948 (±466) | 227,107 (±419) | 323,645 (±468) | 298,546 (±458) |
| 42 | 980,379 (±139) | 219,586 (±414) | 315,206 (±465) | 223,424 (±417) | 320,534 (±467) | 293,988 (±456) |
| 43 | 980,839 (±137) | 216,043 (±412) | 312,099 (±463) | 219,707 (±414) | 317,259 (±465) | 287,662 (±453) |
| 44 | 981,373 (±135) | 213,089 (±409) | 309,358 (±462) | 216,643 (±412) | 314,358 (±464) | 283,592 (±451) |
| 45 | 981,623 (±134) | 210,572 (±408) | 306,464 (±461) | 214,059 (±410) | 311,396 (±463) | 277,966 (±448) |
| 46 | 982,399 (±131) | 206,222 (±405) | 303,193 (±460) | 209,491 (±407) | 307,800 (±462) | 273,013 (±446) |
| 47 | 982,423 (±131) | 203,671 (±403) | 299,719 (±458) | 206,738 (±405) | 304,204 (±460) | 268,697 (±443) |
| 48 | 983,127 (±129) | 201,302 (±401) | 298,005 (±457) | 204,236 (±403) | 302,307 (±459) | 264,236 (±441) |
| 49 | 983,238 (±128) | 197,867 (±398) | 294,858 (±456) | 200,720 (±401) | 299,111 (±458) | 260,148 (±439) |
| 50 | 983,590 (±127) | 195,599 (±397) | 292,322 (±455) | 198,373 (±399) | 296,326 (±457) | 255,563 (±436) |

Table 11.4 (part 1 of 4): comparison of the Simpson-Kramer MinMax method, the Schulze method, Tideman's ranked pairs method and Heitzig's river method (source: own calculation)

| A | Q | R | S | T | U | V |
|---|---|---|---|---|---|---|
| 51 | 983,788 (±126) | 193,094 (±395) | 289,079 (±453) | 195,850 (±397) | 293,057 (±455) | 251,826 (±434) |
| 52 | 984,353 (±124) | 191,693 (±394) | 287,935 (±453) | 194,186 (±396) | 291,695 (±455) | 249,210 (±433) |
| 53 | 984,654 (±123) | 187,998 (±391) | 284,780 (±451) | 190,483 (±393) | 288,373 (±453) | 243,984 (±429) |
| 54 | 984,763 (±122) | 185,798 (±389) | 282,118 (±450) | 188,195 (±391) | 285,781 (±452) | 241,088 (±428) |
| 55 | 984,802 (±122) | 183,248 (±387) | 279,794 (±449) | 185,555 (±389) | 283,391 (±451) | 237,419 (±426) |
| 56 | 985,254 (±121) | 181,745 (±386) | 278,217 (±448) | 184,102 (±388) | 281,607 (±450) | 233,922 (±423) |
| 57 | 985,511 (±119) | 178,614 (±383) | 276,125 (±447) | 180,766 (±385) | 279,501 (±449) | 230,807 (±421) |
| 58 | 985,730 (±119) | 176,654 (±381) | 273,680 (±446) | 178,832 (±383) | 276,853 (±447) | 227,986 (±420) |
| 59 | 986,171 (±117) | 174,710 (±380) | 271,738 (±445) | 176,725 (±381) | 274,871 (±446) | 224,912 (±418) |
| 60 | 986,213 (±117) | 173,038 (±378) | 269,701 (±444) | 175,039 (±380) | 272,869 (±445) | 221,664 (±415) |
| 61 | 986,592 (±115) | 170,786 (±376) | 268,375 (±443) | 172,714 (±378) | 271,320 (±445) | 219,210 (±414) |
| 62 | 986,573 (±115) | 169,534 (±375) | 266,694 (±442) | 171,429 (±377) | 269,577 (±444) | 216,603 (±412) |
| 63 | 986,946 (±114) | 167,131 (±373) | 264,538 (±441) | 168,940 (±375) | 267,356 (±443) | 213,590 (±410) |
| 64 | 987,059 (±113) | 165,366 (±372) | 262,715 (±440) | 167,177 (±373) | 265,534 (±442) | 211,296 (±408) |
| 65 | 987,230 (±112) | 163,626 (±370) | 260,554 (±439) | 165,427 (±372) | 263,330 (±440) | 207,713 (±406) |
| 66 | 987,260 (±112) | 161,643 (±368) | 259,074 (±438) | 163,298 (±370) | 261,768 (±440) | 205,949 (±404) |
| 67 | 987,644 (±110) | 160,225 (±367) | 256,992 (±437) | 161,850 (±368) | 259,550 (±438) | 203,145 (±402) |
| 68 | 987,879 (±109) | 158,721 (±365) | 256,054 (±436) | 160,292 (±367) | 258,604 (±438) | 200,295 (±400) |
| 69 | 987,884 (±109) | 156,696 (±364) | 254,116 (±435) | 158,251 (±365) | 256,730 (±437) | 197,749 (±398) |
| 70 | 988,206 (±108) | 155,165 (±362) | 252,555 (±434) | 156,643 (±363) | 255,012 (±436) | 195,781 (±397) |
| 71 | 988,386 (±107) | 153,630 (±361) | 250,877 (±434) | 155,113 (±362) | 253,222 (±435) | 192,821 (±395) |
| 72 | 988,471 (±107) | 152,476 (±359) | 249,834 (±433) | 153,919 (±361) | 252,253 (±434) | 190,903 (±393) |
| 73 | 988,673 (±106) | 150,823 (±358) | 247,932 (±432) | 152,248 (±359) | 250,270 (±433) | 189,801 (±392) |
| 74 | 988,596 (±106) | 149,341 (±356) | 246,895 (±431) | 150,697 (±358) | 249,126 (±433) | 186,754 (±390) |
| 75 | 988,984 (±104) | 147,722 (±355) | 245,304 (±430) | 149,021 (±356) | 247,544 (±432) | 184,814 (±388) |
| 76 | 988,936 (±105) | 146,257 (±353) | 243,575 (±429) | 147,612 (±355) | 245,782 (±431) | 182,721 (±386) |
| 77 | 989,236 (±103) | 145,415 (±353) | 243,290 (±429) | 146,646 (±354) | 245,465 (±430) | 181,149 (±385) |
| 78 | 989,403 (±102) | 144,709 (±352) | 241,068 (±428) | 145,915 (±353) | 243,194 (±429) | 179,117 (±383) |
| 79 | 989,387 (±102) | 143,174 (±350) | 239,943 (±427) | 144,426 (±352) | 242,018 (±428) | 177,302 (±382) |
| 80 | 989,578 (±102) | 141,510 (±349) | 238,745 (±426) | 142,689 (±350) | 240,721 (±428) | 175,279 (±380) |
| 81 | 989,665 (±101) | 140,530 (±348) | 237,339 (±425) | 141,674 (±349) | 239,246 (±427) | 173,506 (±379) |
| 82 | 989,992 (±100) | 139,436 (±346) | 236,186 (±425) | 140,599 (±348) | 238,149 (±426) | 172,185 (±378) |
| 83 | 989,892 (±100) | 138,162 (±345) | 234,818 (±424) | 139,338 (±346) | 236,800 (±425) | 170,493 (±376) |
| 84 | 989,942 (±100) | 136,355 (±343) | 233,214 (±423) | 137,486 (±344) | 235,159 (±424) | 168,093 (±374) |
| 85 | 990,148 (±99) | 135,241 (±342) | 232,261 (±422) | 136,318 (±343) | 234,090 (±423) | 166,703 (±373) |
| 86 | 990,339 (±98) | 134,138 (±341) | 231,025 (±421) | 135,202 (±342) | 232,844 (±423) | 165,430 (±372) |
| 87 | 990,545 (±97) | 133,258 (±340) | 230,113 (±421) | 134,343 (±341) | 231,896 (±422) | 164,044 (±370) |
| 88 | 990,571 (±97) | 131,896 (±338) | 228,828 (±420) | 132,947 (±340) | 230,610 (±421) | 161,349 (±368) |
| 89 | 990,635 (±96) | 131,418 (±338) | 228,035 (±420) | 132,407 (±339) | 229,795 (±421) | 160,280 (±367) |
| 90 | 990,660 (±96) | 130,083 (±336) | 227,133 (±419) | 131,044 (±337) | 228,788 (±420) | 158,999 (±366) |
| 91 | 990,683 (±96) | 129,503 (±336) | 225,333 (±418) | 130,460 (±337) | 226,984 (±419) | 157,378 (±364) |
| 92 | 990,825 (±95) | 127,760 (±334) | 224,926 (±418) | 128,741 (±335) | 226,580 (±419) | 155,641 (±363) |
| 93 | 990,814 (±95) | 126,815 (±333) | 223,709 (±417) | 127,722 (±334) | 225,302 (±418) | 154,226 (±361) |
| 94 | 991,027 (±94) | 126,550 (±332) | 221,986 (±416) | 127,403 (±333) | 223,595 (±417) | 153,638 (±361) |
| 95 | 990,931 (±95) | 125,261 (±331) | 221,341 (±415) | 126,159 (±332) | 222,984 (±416) | 151,832 (±359) |
| 96 | 991,080 (±94) | 124,576 (±330) | 220,705 (±415) | 125,456 (±331) | 222,291 (±416) | 150,683 (±358) |
| 97 | 991,254 (±93) | 123,365 (±329) | 219,552 (±414) | 124,234 (±330) | 221,089 (±415) | 148,126 (±355) |
| 98 | 991,379 (±92) | 122,241 (±328) | 219,047 (±414) | 123,111 (±329) | 220,583 (±415) | 148,408 (±356) |
| 99 | 991,427 (±92) | 122,229 (±328) | 217,683 (±413) | 123,061 (±329) | 219,170 (±414) | 146,851 (±354) |
| 100 | 991,564 (±91) | 120,986 (±326) | 216,282 (±412) | 121,794 (±327) | 217,758 (±413) | 145,017 (±352) |

Table 11.4 (part 2 of 4): comparison of the Simpson-Kramer MinMax method, the Schulze method, Tideman’s ranked pairs method and Heitzig’s river method (source: own calculation)

| A | Q | R | S | T | U | V |
|---|---|---|---|---|---|---|
| 101 | 991,598 (±91) | 119,964 (±325) | 215,632 (±411) | 120,746 (±326) | 217,104 (±412) | 143,604 (±351) |
| 102 | 991,662 (±91) | 119,559 (±324) | 214,474 (±410) | 120,322 (±325) | 215,925 (±411) | 143,670 (±351) |
| 103 | 991,820 (±90) | 117,774 (±322) | 213,537 (±410) | 118,543 (±323) | 214,946 (±411) | 142,313 (±349) |
| 104 | 991,688 (±91) | 117,972 (±323) | 213,395 (±410) | 118,768 (±324) | 214,761 (±411) | 140,355 (±347) |
| 105 | 991,780 (±90) | 116,881 (±321) | 212,925 (±409) | 117,657 (±322) | 214,311 (±410) | 139,304 (±346) |
| 106 | 991,973 (±89) | 116,110 (±320) | 211,736 (±409) | 116,828 (±321) | 213,084 (±409) | 138,625 (±346) |
| 107 | 992,276 (±88) | 115,712 (±320) | 211,265 (±408) | 116,388 (±321) | 212,557 (±409) | 137,968 (±345) |
| 108 | 991,955 (±89) | 113,832 (±318) | 209,809 (±407) | 114,576 (±319) | 211,172 (±408) | 136,202 (±343) |
| 109 | 992,241 (±88) | 113,634 (±317) | 209,127 (±407) | 114,366 (±318) | 210,399 (±408) | 135,323 (±342) |
| 110 | 992,245 (±88) | 112,985 (±317) | 208,363 (±406) | 113,648 (±317) | 209,628 (±407) | 133,844 (±340) |
| 111 | 992,442 (±87) | 112,481 (±316) | 207,842 (±406) | 113,182 (±317) | 209,017 (±407) | 133,469 (±340) |
| 112 | 992,457 (±87) | 111,337 (±315) | 206,741 (±405) | 111,979 (±315) | 207,980 (±406) | 131,897 (±338) |
| 113 | 992,489 (±86) | 111,098 (±314) | 205,963 (±404) | 111,730 (±315) | 207,208 (±405) | 130,772 (±337) |
| 114 | 992,431 (±87) | 110,133 (±313) | 205,224 (±404) | 110,763 (±314) | 206,415 (±405) | 129,427 (±336) |
| 115 | 992,475 (±86) | 109,055 (±312) | 204,422 (±403) | 109,686 (±312) | 205,684 (±404) | 129,027 (±335) |
| 116 | 992,644 (±85) | 108,375 (±311) | 203,312 (±402) | 109,004 (±312) | 204,554 (±403) | 128,340 (±334) |
| 117 | 992,575 (±86) | 108,201 (±311) | 202,794 (±402) | 108,880 (±311) | 203,987 (±403) | 127,104 (±333) |
| 118 | 992,762 (±85) | 107,180 (±309) | 202,247 (±402) | 107,794 (±310) | 203,408 (±403) | 125,757 (±332) |
| 119 | 992,739 (±85) | 106,569 (±309) | 201,097 (±401) | 107,182 (±309) | 202,263 (±402) | 124,976 (±331) |
| 120 | 992,764 (±85) | 106,304 (±308) | 200,780 (±401) | 106,883 (±309) | 201,948 (±401) | 124,493 (±330) |
| 121 | 992,873 (±84) | 105,347 (±307) | 200,399 (±400) | 105,962 (±308) | 201,556 (±401) | 123,433 (±329) |
| 122 | 992,938 (±84) | 104,526 (±306) | 198,950 (±399) | 105,104 (±307) | 200,117 (±400) | 123,231 (±329) |
| 123 | 992,996 (±83) | 103,849 (±305) | 199,321 (±399) | 104,443 (±306) | 200,424 (±400) | 121,910 (±327) |
| 124 | 993,072 (±83) | 103,164 (±304) | 198,146 (±399) | 103,692 (±305) | 199,200 (±399) | 120,702 (±326) |
| 125 | 993,142 (±83) | 102,978 (±304) | 196,809 (±398) | 103,513 (±305) | 197,864 (±398) | 120,319 (±325) |
| 126 | 993,299 (±82) | 102,592 (±303) | 196,581 (±397) | 103,145 (±304) | 197,665 (±398) | 119,400 (±324) |
| 127 | 993,321 (±81) | 101,853 (±302) | 194,935 (±396) | 102,372 (±303) | 195,991 (±397) | 119,149 (±324) |
| 128 | 993,231 (±82) | 101,064 (±301) | 195,618 (±397) | 101,608 (±302) | 196,643 (±397) | 117,449 (±322) |
| 129 | 993,359 (±81) | 100,502 (±301) | 195,249 (±396) | 100,998 (±301) | 196,314 (±397) | 117,050 (±321) |
| 130 | 993,437 (±81) | 99,921 (±300) | 193,882 (±395) | 100,436 (±301) | 194,869 (±396) | 116,084 (±320) |
| 131 | 993,480 (±80) | 99,828 (±300) | 193,461 (±395) | 100,364 (±300) | 194,462 (±396) | 115,648 (±320) |
| 132 | 993,397 (±81) | 98,709 (±298) | 193,070 (±395) | 99,202 (±299) | 194,091 (±395) | 114,068 (±318) |
| 133 | 993,674 (±79) | 98,506 (±298) | 192,351 (±394) | 98,980 (±299) | 193,275 (±395) | 113,981 (±318) |
| 134 | 993,597 (±80) | 97,804 (±297) | 191,054 (±393) | 98,325 (±298) | 192,030 (±394) | 113,281 (±317) |
| 135 | 993,526 (±80) | 97,447 (±297) | 191,199 (±393) | 97,930 (±297) | 192,193 (±394) | 112,837 (±316) |
| 136 | 993,742 (±79) | 96,832 (±296) | 190,566 (±393) | 97,308 (±296) | 191,482 (±393) | 112,023 (±315) |
| 137 | 993,768 (±79) | 96,152 (±295) | 190,103 (±392) | 96,660 (±295) | 191,033 (±393) | 110,863 (±314) |
| 138 | 993,706 (±79) | 95,565 (±294) | 189,178 (±392) | 96,035 (±295) | 190,123 (±392) | 110,346 (±313) |
| 139 | 993,815 (±78) | 95,234 (±294) | 188,947 (±391) | 95,660 (±294) | 189,858 (±392) | 110,202 (±313) |
| 140 | 993,909 (±78) | 94,596 (±293) | 187,883 (±391) | 95,057 (±293) | 188,777 (±391) | 108,742 (±311) |
| 141 | 993,932 (±78) | 94,059 (±292) | 187,721 (±390) | 94,482 (±292) | 188,670 (±391) | 108,144 (±311) |
| 142 | 993,919 (±78) | 93,573 (±291) | 187,225 (±390) | 93,992 (±292) | 188,186 (±391) | 107,550 (±310) |
| 143 | 993,909 (±78) | 92,798 (±290) | 186,990 (±390) | 93,279 (±291) | 187,885 (±391) | 106,849 (±309) |
| 144 | 994,041 (±77) | 92,993 (±290) | 186,382 (±389) | 93,415 (±291) | 187,263 (±390) | 106,252 (±308) |
| 145 | 994,035 (±77) | 92,274 (±289) | 184,939 (±388) | 92,719 (±290) | 185,847 (±389) | 105,985 (±308) |
| 146 | 994,058 (±77) | 91,965 (±289) | 184,104 (±388) | 92,395 (±290) | 184,973 (±388) | 104,845 (±306) |
| 147 | 994,105 (±77) | 91,560 (±288) | 184,040 (±388) | 91,997 (±289) | 184,939 (±388) | 104,014 (±305) |
| 148 | 994,117 (±76) | 91,215 (±288) | 184,024 (±388) | 91,628 (±289) | 184,899 (±388) | 103,603 (±305) |
| 149 | 994,247 (±76) | 90,523 (±287) | 184,115 (±388) | 90,934 (±288) | 184,894 (±388) | 103,515 (±305) |
| 150 | 994,235 (±76) | 89,943 (±286) | 183,185 (±387) | 90,331 (±287) | 184,037 (±388) | 102,754 (±304) |

Table 11.4 (part 3 of 4): comparison of the Simpson-Kramer MinMax method, the Schulze method, Tideman’s ranked pairs method and Heitzig’s river method (source: own calculation)

| A | Q | R | S | T | U | V |
|---|---|---|---|---|---|---|
| 151 | 994,193 (±76) | 89,912 (±286) | 182,516 (±386) | 90,300 (±287) | 183,347 (±387) | 101,552 (±302) |
| 152 | 994,346 (±75) | 89,149 (±285) | 181,369 (±385) | 89,537 (±286) | 182,181 (±386) | 101,328 (±302) |
| 153 | 994,263 (±76) | 88,572 (±284) | 180,926 (±385) | 88,992 (±285) | 181,826 (±386) | 100,762 (±301) |
| 154 | 994,167 (±76) | 88,409 (±284) | 181,123 (±385) | 88,861 (±285) | 181,960 (±386) | 100,350 (±300) |
| 155 | 994,293 (±75) | 87,490 (±283) | 180,375 (±384) | 87,891 (±283) | 181,178 (±385) | 98,969 (±299) |
| 156 | 994,612 (±73) | 87,108 (±282) | 180,128 (±384) | 87,467 (±283) | 180,924 (±385) | 99,049 (±299) |
| 157 | 994,497 (±74) | 86,720 (±281) | 179,813 (±384) | 87,119 (±282) | 180,611 (±385) | 98,296 (±298) |
| 158 | 994,558 (±74) | 86,644 (±281) | 178,161 (±383) | 87,019 (±282) | 178,936 (±383) | 97,727 (±297) |
| 159 | 994,402 (±75) | 86,065 (±280) | 178,307 (±383) | 86,438 (±281) | 179,082 (±383) | 97,253 (±296) |
| 160 | 994,471 (±74) | 85,970 (±280) | 178,573 (±383) | 86,347 (±281) | 179,365 (±384) | 96,626 (±295) |
| 161 | 994,620 (±73) | 85,171 (±279) | 177,248 (±382) | 85,527 (±280) | 177,992 (±383) | 96,012 (±295) |
| 162 | 994,616 (±73) | 84,627 (±278) | 176,747 (±381) | 85,006 (±279) | 177,511 (±382) | 96,289 (±295) |
| 163 | 994,640 (±73) | 84,634 (±278) | 176,579 (±381) | 85,016 (±279) | 177,352 (±382) | 95,387 (±294) |
| 164 | 994,683 (±73) | 84,352 (±278) | 176,007 (±381) | 84,680 (±278) | 176,777 (±381) | 94,876 (±293) |
| 165 | 994,714 (±73) | 83,768 (±277) | 175,762 (±381) | 84,097 (±278) | 176,514 (±381) | 94,467 (±292) |
| 166 | 994,702 (±73) | 83,428 (±277) | 175,228 (±380) | 83,780 (±277) | 175,902 (±381) | 93,721 (±291) |
| 167 | 994,733 (±72) | 83,216 (±276) | 175,215 (±380) | 83,565 (±277) | 175,932 (±381) | 93,439 (±291) |
| 168 | 994,841 (±72) | 82,651 (±275) | 173,991 (±379) | 82,975 (±276) | 174,707 (±380) | 93,017 (±290) |
| 169 | 994,748 (±72) | 82,055 (±274) | 174,493 (±380) | 82,399 (±275) | 175,247 (±380) | 92,427 (±290) |
| 170 | 994,806 (±72) | 82,544 (±275) | 173,607 (±379) | 82,885 (±276) | 174,333 (±379) | 92,002 (±289) |
| 171 | 994,791 (±72) | 82,150 (±275) | 173,225 (±378) | 82,488 (±275) | 173,965 (±379) | 91,177 (±288) |
| 172 | 995,004 (±71) | 81,269 (±273) | 172,737 (±378) | 81,578 (±274) | 173,429 (±379) | 90,780 (±287) |
| 173 | 994,901 (±71) | 80,563 (±272) | 172,406 (±378) | 80,893 (±273) | 173,114 (±378) | 90,349 (±287) |
| 174 | 995,049 (±70) | 80,288 (±272) | 171,491 (±377) | 80,586 (±272) | 172,168 (±378) | 90,122 (±286) |
| 175 | 995,075 (±70) | 79,845 (±271) | 171,001 (±377) | 80,147 (±272) | 171,677 (±377) | 89,510 (±285) |
| 176 | 995,057 (±70) | 79,900 (±271) | 170,852 (±376) | 80,213 (±272) | 171,512 (±377) | 89,086 (±285) |
| 177 | 994,990 (±71) | 79,859 (±271) | 170,710 (±376) | 80,146 (±272) | 171,397 (±377) | 88,665 (±284) |
| 178 | 995,067 (±70) | 79,863 (±271) | 169,742 (±375) | 80,171 (±272) | 170,394 (±376) | 88,575 (±284) |
| 179 | 995,134 (±70) | 79,041 (±270) | 169,287 (±375) | 79,353 (±270) | 169,941 (±376) | 87,312 (±282) |
| 180 | 994,957 (±71) | 78,935 (±270) | 169,557 (±375) | 79,235 (±270) | 170,219 (±376) | 87,193 (±282) |
| 181 | 995,247 (±69) | 78,505 (±269) | 168,979 (±375) | 78,804 (±269) | 169,633 (±375) | 86,686 (±281) |
| 182 | 995,245 (±69) | 78,140 (±268) | 168,334 (±374) | 78,453 (±269) | 168,960 (±375) | 86,463 (±281) |
| 183 | 995,166 (±69) | 77,401 (±267) | 168,097 (±374) | 77,695 (±268) | 168,748 (±375) | 85,979 (±280) |
| 184 | 995,371 (±68) | 77,337 (±267) | 167,737 (±374) | 77,603 (±268) | 168,342 (±374) | 85,590 (±280) |
| 185 | 995,308 (±68) | 76,901 (±266) | 167,018 (±373) | 77,186 (±267) | 167,652 (±374) | 84,779 (±279) |
| 186 | 995,387 (±68) | 76,863 (±266) | 167,378 (±373) | 77,110 (±267) | 167,973 (±374) | 85,011 (±279) |
| 187 | 995,293 (±68) | 76,639 (±266) | 165,927 (±372) | 76,881 (±266) | 166,547 (±373) | 84,490 (±278) |
| 188 | 995,227 (±69) | 75,977 (±265) | 166,167 (±372) | 76,261 (±265) | 166,784 (±373) | 83,664 (±277) |
| 189 | 995,363 (±68) | 76,041 (±265) | 165,909 (±372) | 76,315 (±266) | 166,511 (±373) | 84,024 (±277) |
| 190 | 995,463 (±67) | 75,451 (±264) | 165,699 (±372) | 75,743 (±265) | 166,301 (±372) | 83,965 (±277) |
| 191 | 995,312 (±68) | 75,101 (±264) | 165,529 (±372) | 75,374 (±264) | 166,164 (±372) | 82,686 (±275) |
| 192 | 995,389 (±68) | 75,221 (±264) | 164,332 (±371) | 75,512 (±264) | 164,951 (±371) | 82,902 (±276) |
| 193 | 995,480 (±67) | 74,877 (±263) | 163,593 (±370) | 75,133 (±264) | 164,212 (±370) | 82,407 (±275) |
| 194 | 995,323 (±68) | 74,652 (±263) | 164,091 (±370) | 74,909 (±263) | 164,678 (±371) | 81,888 (±274) |
| 195 | 995,446 (±67) | 74,142 (±262) | 162,925 (±369) | 74,377 (±262) | 163,512 (±370) | 81,193 (±273) |
| 196 | 995,470 (±67) | 74,120 (±262) | 162,874 (±369) | 74,356 (±262) | 163,436 (±370) | 81,413 (±273) |
| 197 | 995,409 (±68) | 73,081 (±260) | 163,481 (±370) | 73,360 (±261) | 164,036 (±370) | 80,588 (±272) |
| 198 | 995,423 (±67) | 73,112 (±260) | 162,657 (±369) | 73,370 (±261) | 163,223 (±370) | 80,588 (±272) |
| 199 | 995,488 (±67) | 72,824 (±260) | 161,925 (±368) | 73,090 (±260) | 162,471 (±369) | 80,201 (±272) |
| 200 | 995,522 (±67) | 72,953 (±260) | 161,585 (±368) | 73,206 (±260) | 162,156 (±369) | 79,624 (±271) |

Table 11.4 (part 4 of 4): comparison of the Simpson-Kramer MinMax method, the Schulze method, Tideman’s ranked pairs method and Heitzig’s river method (source: own calculation)

Table 11.4 shows that, when the number of alternatives increases (column “A”), the probability that the Schulze method confirms with the Simpson-Kramer MinMax method goes to certainty (column “Q”) while, for every other pair of election methods, the probability that they agree with each other goes to zero. For every number of alternatives, the probability that the Schulze method confirms with the Simpson-Kramer MinMax method is above 90%. This demonstrates that (unlike Tideman’s ranked pairs method and Heitzig’s river method) the Schulze method doesn’t only satisfy the Smith criterion, reversal symmetry and independence of clones; it also gives a justification why the Schulze winner is a desirable alternative.

Table 11.5 lists the runtimes to calculate the winners of the different methods. The columns in table 11.5 are:

Column “W”: runtime to calculate the winners of the Simpson-Kramer MinMax method in the 1,000,000 test cases.

Column “X”: runtime to calculate the winners of the Schulze method in the 1,000,000 test cases.

Column “Y”: runtime to calculate the winners of Tideman’s ranked pairs method in the 1,000,000 test cases.

Column “Z”: runtime to calculate the winners of Heitzig’s river method in the 1,000,000 test cases.

A LENOVO ThinkBook 16p G5 IRX with an Intel Core i9-14900HX processor was used for these calculations. The program was written in g++ 15.2.0. The program was single-threading.

| A | W | X | Y | Z |
|---|---|---|---|---|
| 4 | < 0.1 s | < 0.1 s | 0.2 s | 0.2 s |
| 5 | < 0.1 s | 0.2 s | 0.4 s | 0.4 s |
| 6 | < 0.1 s | 0.3 s | 0.8 s | 0.7 s |
| 7 | < 0.1 s | 0.5 s | 1.3 s | 1.0 s |
| 8 | < 0.1 s | 0.7 s | 1.8 s | 1.4 s |
| 9 | 0.2 s | 0.9 s | 2.4 s | 2.0 s |
| 10 | 0.2 s | 0.9 s | 3.0 s | 2.5 s |
| 11 | 0.2 s | 1.2 s | 3.8 s | 3.1 s |
| 12 | 0.3 s | 1.5 s | 4.8 s | 3.9 s |
| 13 | 0.3 s | 1.7 s | 5.7 s | 4.7 s |
| 14 | 0.4 s | 1.8 s | 6.9 s | 5.5 s |
| 15 | 0.4 s | 2.1 s | 8.4 s | 6.8 s |
| 16 | 0.5 s | 2.4 s | 10.6 s | 9.5 s |
| 17 | 0.5 s | 2.9 s | 11.0 s | 8.8 s |
| 18 | 0.5 s | 2.8 s | 12.6 s | 10.4 s |
| 19 | 0.5 s | 3.0 s | 14.1 s | 11.5 s |
| 20 | 0.7 s | 3.5 s | 17.1 s | 14.0 s |
| 21 | 0.6 s | 3.4 s | 17.3 s | 14.5 s |
| 22 | 0.8 s | 4.0 s | 20.5 s | 17.2 s |
| 23 | 0.8 s | 3.9 s | 21.7 s | 18.1 s |
| 24 | 0.8 s | 4.2 s | 24.0 s | 20.2 s |
| 25 | 0.9 s | 4.6 s | 26.2 s | 22.2 s |
| 26 | 1.0 s | 5.2 s | 30.4 s | 25.7 s |
| 27 | 1.0 s | 5.5 s | 33.4 s | 27.9 s |
| 28 | 1.2 s | 6.0 s | 36.0 s | 29.9 s |
| 29 | 1.1 s | 5.8 s | 37.7 s | 30.7 s |
| 30 | 1.1 s | 6.0 s | 40.1 s | 33.0 s |
| 31 | 1.2 s | 6.4 s | 43.6 s | 35.6 s |
| 32 | 1.2 s | 6.7 s | 46.1 s | 38.4 s |
| 33 | 1.3 s | 7.1 s | 49.3 s | 41.1 s |
| 34 | 1.5 s | 7.6 s | 53.9 s | 44.6 s |
| 35 | 1.4 s | 7.9 s | 58.6 s | 47.4 s |
| 36 | 1.7 s | 9.0 s | 66.4 s | 54.9 s |
| 37 | 1.8 s | 9.3 s | 69.2 s | 57.9 s |
| 38 | 1.8 s | 9.7 s | 73.7 s | 62.4 s |
| 39 | 1.8 s | 10.1 s | 77.3 s | 65.8 s |
| 40 | 1.9 s | 10.4 s | 85.2 s | 71.3 s |
| 41 | 2.1 s | 10.9 s | 88.0 s | 74.5 s |
| 42 | 1.9 s | 10.4 s | 89.2 s | 74.3 s |
| 43 | 2.0 s | 10.9 s | 93.8 s | 76.9 s |
| 44 | 2.2 s | 12.6 s | 107.6 s | 87.8 s |
| 45 | 2.1 s | 11.7 s | 105.1 s | 85.8 s |
| 46 | 2.2 s | 12.2 s | 110.3 s | 90.0 s |
| 47 | 2.3 s | 12.6 s | 116.2 s | 95.1 s |
| 48 | 2.4 s | 13.1 s | 122.6 s | 100.0 s |
| 49 | 2.6 s | 13.6 s | 128.5 s | 104.6 s |
| 50 | 2.6 s | 14.0 s | 134.4 s | 109.6 s |

Table 11.5 (part 1 of 4): runtimes to calculate the winners of the Simpson-Kramer MinMax method, the Schulze method, Tideman’s ranked pairs method and Heitzig’s river method in the 1,000,000 test cases (source: own calculation)

| A | W | X | Y | Z |
|---|---|---|---|---|
| 51 | 3.0 s | 15.5 s | 148.3 s | 124.2 s |
| 52 | 3.0 s | 15.9 s | 156.6 s | 130.9 s |
| 53 | 2.8 s | 15.2 s | 151.6 s | 126.4 s |
| 54 | 2.8 s | 15.5 s | 160.1 s | 133.8 s |
| 55 | 3.1 s | 17.2 s | 178.5 s | 148.3 s |
| 56 | 3.1 s | 16.6 s | 172.8 s | 144.9 s |
| 57 | 2.9 s | 16.9 s | 181.9 s | 152.9 s |
| 58 | 3.3 s | 18.6 s | 200.1 s | 167.1 s |
| 59 | 3.1 s | 18.1 s | 201.9 s | 170.1 s |
| 60 | 3.3 s | 18.4 s | 207.0 s | 173.9 s |
| 61 | 3.7 s | 20.3 s | 231.7 s | 191.1 s |
| 62 | 3.4 s | 19.2 s | 230.0 s | 188.3 s |
| 63 | 3.5 s | 20.2 s | 237.1 s | 194.5 s |
| 64 | 3.9 s | 21.5 s | 247.6 s | 202.9 s |
| 65 | 4.1 s | 22.5 s | 266.2 s | 222.8 s |
| 66 | 4.2 s | 22.9 s | 274.8 s | 228.5 s |
| 67 | 4.1 s | 22.0 s | 274.6 s | 224.2 s |
| 68 | 4.1 s | 22.5 s | 284.5 s | 234.1 s |
| 69 | 4.5 s | 23.1 s | 285.8 s | 240.5 s |
| 70 | 4.9 s | 25.5 s | 312.9 s | 262.0 s |
| 71 | 5.0 s | 26.8 s | 327.7 s | 273.4 s |
| 72 | 4.8 s | 25.0 s | 322.8 s | 269.4 s |
| 73 | 5.2 s | 27.3 s | 346.4 s | 278.0 s |
| 74 | 5.3 s | 27.8 s | 357.8 s | 299.1 s |
| 75 | 5.3 s | 28.5 s | 381.6 s | 317.4 s |
| 76 | 5.6 s | 29.3 s | 381.7 s | 317.9 s |
| 77 | 5.6 s | 30.0 s | 414.6 s | 338.8 s |
| 78 | 5.1 s | 28.4 s | 396.0 s | 329.2 s |
| 79 | 5.4 s | 29.1 s | 411.0 s | 341.7 s |
| 80 | 6.3 s | 32.8 s | 440.9 s | 350.2 s |
| 81 | 5.6 s | 30.2 s | 419.3 s | 351.1 s |
| 82 | 5.5 s | 31.2 s | 464.4 s | 393.7 s |
| 83 | 5.9 s | 34.0 s | 453.8 s | 376.7 s |
| 84 | 5.8 s | 32.5 s | 464.1 s | 384.7 s |
| 85 | 6.1 s | 33.1 s | 496.0 s | 411.8 s |
| 86 | 6.5 s | 33.9 s | 499.1 s | 413.0 s |
| 87 | 6.5 s | 36.4 s | 522.1 s | 419.0 s |
| 88 | 7.0 s | 38.3 s | 580.4 s | 473.2 s |
| 89 | 6.6 s | 38.2 s | 573.6 s | 462.4 s |
| 90 | 6.9 s | 39.1 s | 593.7 s | 469.5 s |
| 91 | 6.6 s | 37.2 s | 582.0 s | 481.3 s |
| 92 | 7.3 s | 40.8 s | 620.9 s | 512.1 s |
| 93 | 6.9 s | 38.4 s | 610.4 s | 508.0 s |
| 94 | 7.0 s | 39.4 s | 612.9 s | 508.0 s |
| 95 | 7.3 s | 40.6 s | 626.7 s | 520.0 s |
| 96 | 8.5 s | 47.4 s | 678.0 s | 556.4 s |
| 97 | 7.6 s | 44.2 s | 679.9 s | 542.2 s |
| 98 | 7.3 s | 42.1 s | 678.8 s | 561.7 s |
| 99 | 7.7 s | 43.1 s | 716.4 s | 593.2 s |
| 100 | 8.5 s | 47.4 s | 760.3 s | 624.7 s |

Table 11.5 (part 2 of 4): runtimes to calculate the winners of the Simpson-Kramer MinMax method, the Schulze method, Tideman’s ranked pairs method and Heitzig’s river method in the 1,000,000 test cases (source: own calculation)

| A | W | X | Y | Z |
|---|---|---|---|---|
| 101 | 7.9 s | 44.8 s | 733.0 s | 608.5 s |
| 102 | 8.1 s | 45.7 s | 754.6 s | 625.9 s |
| 103 | 8.3 s | 46.5 s | 775.8 s | 631.0 s |
| 104 | 8.6 s | 47.9 s | 791.1 s | 654.8 s |
| 105 | 8.8 s | 48.4 s | 829.8 s | 689.0 s |
| 106 | 8.9 s | 49.2 s | 827.1 s | 682.4 s |
| 107 | 9.2 s | 50.0 s | 834.3 s | 688.8 s |
| 108 | 9.5 s | 51.3 s | 870.9 s | 719.7 s |
| 109 | 10.2 s | 48.4 s | 881.1 s | 725.5 s |
| 110 | 9.7 s | 52.8 s | 924.8 s | 765.8 s |
| 111 | 9.7 s | 52.4 s | 946.6 s | 786.4 s |
| 112 | 10.9 s | 56.4 s | 955.8 s | 791.0 s |
| 113 | 8.3 s | 53.5 s | 1,000.1 s | 831.3 s |
| 114 | 11.0 s | 59.9 s | 1,044.0 s | 857.2 s |
| 115 | 10.8 s | 56.9 s | 1,012.1 s | 832.3 s |
| 116 | 10.8 s | 57.9 s | 1,047.0 s | 866.7 s |
| 117 | 11.1 s | 59.1 s | 1,049.6 s | 889.8 s |
| 118 | 11.1 s | 59.7 s | 1,095.4 s | 908.9 s |
| 119 | 11.9 s | 61.1 s | 1,117.6 s | 926.6 s |
| 120 | 12.0 s | 62.5 s | 1,124.1 s | 930.9 s |
| 121 | 11.5 s | 62.3 s | 1,164.5 s | 957.3 s |
| 122 | 12.8 s | 68.0 s | 1,241.6 s | 1,010.1 s |
| 123 | 13.7 s | 68.8 s | 1,245.9 s | 1,011.6 s |
| 124 | 13.3 s | 70.0 s | 1,271.2 s | 1,001.9 s |
| 125 | 12.6 s | 65.9 s | 1,244.8 s | 1,020.7 s |
| 126 | 14.3 s | 70.3 s | 1,279.3 s | 1,050.5 s |
| 127 | 14.1 s | 71.2 s | 1,324.1 s | 1,081.1 s |
| 128 | 15.2 s | 74.3 s | 1,365.4 s | 1,099.7 s |
| 129 | 13.1 s | 69.9 s | 1,376.6 s | 1,127.9 s |
| 130 | 15.5 s | 72.3 s | 1,408.4 s | 1,163.5 s |
| 131 | 15.0 s | 73.6 s | 1,469.2 s | 1,183.9 s |
| 132 | 15.2 s | 74.2 s | 1,445.4 s | 1,179.1 s |
| 133 | 15.2 s | 73.6 s | 1,480.3 s | 1,200.3 s |
| 134 | 15.1 s | 74.5 s | 1,495.9 s | 1,232.7 s |
| 135 | 15.9 s | 76.6 s | 1,582.1 s | 1,273.6 s |
| 136 | 15.8 s | 77.9 s | 1,549.6 s | 1,261.2 s |
| 137 | 16.0 s | 82.9 s | 1,607.7 s | 1,301.9 s |
| 138 | 15.1 s | 78.7 s | 1,608.7 s | 1,295.5 s |
| 139 | 17.5 s | 86.8 s | 1,758.4 s | 1,398.8 s |
| 140 | 16.6 s | 81.3 s | 1,694.8 s | 1,382.9 s |
| 141 | 15.9 s | 80.9 s | 1,711.1 s | 1,391.6 s |
| 142 | 16.4 s | 83.0 s | 1,712.3 s | 1,375.2 s |
| 143 | 17.8 s | 85.5 s | 1,790.3 s | 1,441.1 s |
| 144 | 19.4 s | 88.7 s | 1,827.8 s | 1,490.0 s |
| 145 | 18.9 s | 91.9 s | 1,938.5 s | 1,551.4 s |
| 146 | 17.4 s | 93.9 s | 1,952.4 s | 1,576.7 s |
| 147 | 19.0 s | 95.0 s | 2,009.6 s | 1,597.2 s |
| 148 | 18.8 s | 90.0 s | 1,963.0 s | 1,597.2 s |
| 149 | 17.3 s | 87.6 s | 1,981.1 s | 1,607.4 s |
| 150 | 17.3 s | 91.0 s | 2,002.5 s | 1,627.8 s |

Table 11.5 (part 3 of 4): runtimes to calculate the winners of the Simpson-Kramer MinMax method, the Schulze method, Tideman’s ranked pairs method and Heitzig’s river method in the 1,000,000 test cases (source: own calculation)

| A | W | X | Y | Z |
|---|---|---|---|---|
| 151 | 19.7 s | 93.8 s | 2,044.8 s | 1,632.7 s |
| 152 | 19.4 s | 94.9 s | 2,079.8 s | 1,688.2 s |
| 153 | 15.0 s | 88.6 s | 2,086.5 s | 1,674.9 s |
| 154 | 20.0 s | 104.7 s | 2,129.3 s | 1,687.5 s |
| 155 | 20.0 s | 99.2 s | 2,210.2 s | 1,746.0 s |
| 156 | 20.0 s | 99.2 s | 2,207.5 s | 1,787.9 s |
| 157 | 20.5 s | 99.5 s | 2,257.3 s | 1,818.4 s |
| 158 | 21.4 s | 103.6 s | 2,293.8 s | 1,840.1 s |
| 159 | 23.0 s | 112.8 s | 2,501.1 s | 1,903.0 s |
| 160 | 23.4 s | 108.6 s | 2,366.6 s | 1,904.7 s |
| 161 | 20.5 s | 102.2 s | 2,385.2 s | 1,910.8 s |
| 162 | 20.0 s | 104.7 s | 2,428.6 s | 1,942.0 s |
| 163 | 22.4 s | 107.3 s | 2,496.3 s | 1,986.1 s |
| 164 | 23.8 s | 117.0 s | 2,659.3 s | 2,017.3 s |
| 165 | 21.3 s | 107.4 s | 2,681.5 s | 2,151.7 s |
| 166 | 24.4 s | 118.0 s | 2,764.9 s | 2,105.7 s |
| 167 | 23.2 s | 111.0 s | 2,683.8 s | 2,118.8 s |
| 168 | 24.8 s | 123.2 s | 2,707.3 s | 2,146.9 s |
| 169 | 22.6 s | 114.2 s | 2,724.3 s | 2,144.8 s |
| 170 | 15.4 s | 105.7 s | 2,783.4 s | 2,231.6 s |
| 171 | 23.8 s | 117.3 s | 2,813.3 s | 2,197.4 s |
| 172 | 24.6 s | 118.6 s | 2,980.2 s | 2,464.5 s |
| 173 | 23.9 s | 117.7 s | 2,875.5 s | 2,277.2 s |
| 174 | 22.7 s | 110.9 s | 2,967.0 s | 2,383.2 s |
| 175 | 25.9 s | 121.7 s | 2,969.7 s | 2,345.4 s |
| 176 | 27.6 s | 129.6 s | 3,007.9 s | 2,398.6 s |
| 177 | 24.0 s | 122.2 s | 3,054.5 s | 2,432.4 s |
| 178 | 29.3 s | 135.2 s | 3,225.2 s | 2,516.7 s |
| 179 | 28.1 s | 137.4 s | 3,334.5 s | 2,553.5 s |
| 180 | 27.9 s | 128.9 s | 3,206.3 s | 2,546.4 s |
| 181 | 25.4 s | 127.7 s | 3,215.7 s | 2,594.6 s |
| 182 | 25.8 s | 129.1 s | 3,351.2 s | 2,660.5 s |
| 183 | 28.2 s | 132.3 s | 3,404.0 s | 2,654.5 s |
| 184 | 30.9 s | 139.6 s | 3,438.3 s | 2,748.2 s |
| 185 | 34.0 s | 146.7 s | 3,506.2 s | 2,818.5 s |
| 186 | 28.6 s | 145.3 s | 3,747.1 s | 2,840.7 s |
| 187 | 30.2 s | 127.5 s | 3,571.8 s | 2,825.9 s |
| 188 | 33.1 s | 151.8 s | 3,661.4 s | 2,940.0 s |
| 189 | 28.9 s | 138.1 s | 3,744.9 s | 2,975.6 s |
| 190 | 35.4 s | 142.2 s | 3,747.3 s | 3,019.6 s |
| 191 | 37.0 s | 163.3 s | 3,971.8 s | 2,954.6 s |
| 192 | 39.1 s | 158.5 s | 3,782.8 s | 3,022.4 s |
| 193 | 31.0 s | 146.0 s | 3,938.1 s | 3,124.9 s |
| 194 | 32.1 s | 148.4 s | 4,003.9 s | 3,162.1 s |
| 195 | 32.8 s | 151.3 s | 4,055.2 s | 3,222.5 s |
| 196 | 37.6 s | 158.0 s | 4,087.9 s | 3,184.1 s |
| 197 | 32.4 s | 153.4 s | 4,137.9 s | 3,273.9 s |
| 198 | 34.2 s | 155.4 s | 4,361.7 s | 3,369.5 s |
| 199 | 36.5 s | 159.5 s | 4,353.1 s | 3,388.6 s |
| 200 | 44.9 s | 182.6 s | 4,363.3 s | 3,572.6 s |

Table 11.5 (part 4 of 4): runtimes to calculate the winners of the Simpson-Kramer MinMax method, the Schulze method, Tideman's ranked pairs method and Heitzig's river method in the 1,000,000 test cases (source: own calculation)

## Acknowledgments

I want to thank Lowell Bruce Anderson, Ratip Emin Berker, Blake Cretney, James Green-Armytage, Jobst Heitzig, Ross Hyman, Aleksei Kondratev, Rob Lanphier, Rob LeGrand, Andrew Myers, Norman Petry, Nic Tideman, Kevin Venzke, Douglas R. Woodall, and Thomas Zavist for fruitful discussions.

## References

- Amir Babak Aazami, Hubert Lewis Bray (2023), “Classification of preferential ballot voting methods”, *Constitutional Political Economy*, volume 34, issue 4, pages 510–523, DOI: 10.1007/s10602-022-09384-8 (link)

- Adiya Abisheva (2012), “Crowdsourced order: getting top N values from the crowd”, master thesis, Eidgenössische Technische Hochschule Zürich, Switzerland, DOI: 10.3929/ethz-a-007568951 (link) {In this paper, the Schulze method is called “MinMax decision function”.}

- Сергей Геннадьевич Абрамов / Sergey Gennadievich Abramov (2019), “Выбор и Оценка Экономических Целей Деятельности, их Юридическое Формулирование Субъектами Предпринимательской Деятельности (к Вопросу о Саморегулировании в Сфере Предпринимательской Деятельности)” / “Selection and Assessment of the Economic Objectives of the Activity, Their Legal Formulation by Business Entities (on the Issue of Self-Regulation in the Field of Entrepreneurial Activity)”, *Проблемы экономики и юридической практики / Problems of Economics and Legal Practice*, volume XV, issue 4, pages 106–110 (link, link)

- Antal Ádám (2016), “Választási játékok”, bachelor thesis, University of Budapest, Hungary (link, link)

- Avraham Adler (2014), “Condorcet Voting with Rcpp” (link, link, link, link)

- Nico Agon (2021), “Other Voting Methods (Instant Run Off, Majority Judgement, Schulze Voting and Quadratic Voting)” (video: 00:08:18 – 00:13:51 hh:mm:ss)

- Maha Akbib, Ouafae Baida, Abdelouahid Lyhyaoui, Abdellatif Ghacham Amrani, Abdelfettah Sedqui (2014a), “Workshop Layout by the Method of Vote and Comparison to the Average Ranks Method”, *Hamburg International Conference of Logistics* (HICL), Hamburg, Germany, 18–19 September 2014; *Innovative Methods in Logistics and Supply Chain Management: Current Issues and Emerging Practices*, eds. Thorsten Blecker, Wolfgang Kersten, Christian M. Ringle, volume 19, pages 555–575, DOI: 10419/209248, DOI: 10.15480/882.1189 (link, link, link, link, link, link, link, link)

- Maha Akbib (2014b), “La Technologie de Groupe Appliquée à la Chaîne Logistique”, doctoral dissertation, Abdelmalek Essaâdi University, National School of Applied Sciences Tangier, Morocco (link, link)

- Şahin Albayrak, Stefan Fricke (2012), "Agentenorientierte Technologien in der Forschung — Soziale Entscheidungen: Wahlen" (link)

- Николай Сергеевич Алексеев / Nikolay Sergeevich Alekseev, В.Л. Садов / V.L. Sadov (2019), "Принцип Согласования, Основанный на Информации о Сравнительной Важности Мнений Экспертов (Правило ИВМ), Метод Шульца" / "Principle of Agreement Based on Information about the Comparative Importance of Expert Opinions (ITU Rule), the Schulze Method", *Colloquium Journal*, volume 10(34), issue 2, pages 224–226 (link)

- Caroline Alves da Silveira, Marceli Adriane Schvartz, Letícia Oestreich, Carmen Brum Rosa, Bárbara Giaccom, Alejandro Ruiz-Padillo (2021), "Avaliação ponderada sobre a percepção da infraestrutura de calçadas por meio da técnica Delphi-Fuzzy e análises geoespaciais", *Revista de Morfologia Urbana*, volume 9, number 2, article e00186, DOI: 10.47235/rmu.v9i2.186 (link, link, link, link, link, link)

- Anshul Anand (2019), "Private Computation of Schulze Voting Method", master thesis, Indian Institute of Information Technology, Allahabad, India (link, link)

- Anandita, Mahek Goyal, Vaishali, Vijay Kumar Yadav (2024), "Secure Schulze Voting Method with Ring Signature using Elliptic Curve Cryptography" (link, link)

- Елизавета Д. Анурова / Elizaveta D. Anurova (2026a), "Исследование эффективности модифицированного метода Шульце для взвешенного голосования" / "A Study of the Efficiency of a Modified Schulze Method for Weighted Voting", *Студенческая Научная Весна / Student Scientific Spring Conference*, Moscow, 1–30 April 2026; proceedings, pages 784–785 (link, link)

- Елизавета Д. Анурова / Elizaveta D. Anurova (2026b), "Исследование устойчивости к воздействию со стороны организатора модифицированного метода Шульце для взвешенного голосования" / "A study of the robustness of the modified Schulze method for weighted voting to the influence of the organizer", 17 June 2026 (video, video)

- Jaime Arguello, Fernando Diaz, Jamie Callan, Ben Carterette (2011a), "A Methodology for Evaluating Aggregated Search Results", *33rd European Conference on Information Retrieval* (ECIR 2011), Dublin, Ireland, 18–21 April 2011; eds. Paul Clough, Colum Foley, Cathal Gurrin, Gareth J.F. Jones, Wessel Kraaij, Hyowon Lee, Vanessa Mudoch, *Advances in Information Retrieval*, volume 6611 of the series *Lecture Notes in Computer Science*, pages 141–152, Springer, DOI: 10.1007/978-3-642-20161-5_15 (link, link, link, link, link, link, link, link, link)

- Jaime Arguello, Fernando Diaz, Jamie Callan (2011b), "Learning to Aggregate Vertical Results into Web Search Results", *20th ACM International Conference on Information and Knowledge Management* (CIKM '11), Glasgow, United Kingdom, 24–28 October 2011, proceedings, pages 201–210, DOI: 10.1145/2063576.2063611 (link, link, link, link, link, link, link, link, link, link)

- Jaime Arguello (2011c), "Federated Search for Heterogeneous Environments", doctoral dissertation, Carnegie Mellon University, Pittsburgh, Pennsylvania, USA (link, link, link, link)

- Jaime Arguello, Fernando Diaz (2013), "Vertical Selection and Aggregation" (link, link, link)

- Jaime Arguello (2017), "Aggregated Search", *Foundations and Trends in Information Retrieval*, volume 10, number 5, pages 365–502, DOI: 10.1561/1500000052 (link, link, link)

- Arushi Arora, David Eppstein, Randy Le Huynh (2026), "Fast Schulze Voting Using Quickselect", *Journal of Graph Algorithms and Applications*, volume 30, number 1, pages 237–252, DOI: 10.7155/jgaa.v30i1.3030 (arXiv:2411.18790) (link, link, link, link, link)

- Kenneth J. Arrow, Hervé Raynaud (1986), "Social Choice and Multicriterion Decision-Making", MIT Press, Cambridge, Massachusetts, USA

- Shashaank M. Aswatha, Jayanta Mukherjee, Partha Bhowmick (2016), "An Integrated Repainting System for Digital Restoration of Vijayanagara Murals", *International Journal of Image and Graphics*, volume 16, number 1, article 1650005, DOI: 10.1142/S0219467816500054 (link) {In this paper, I am called "S. Markus".}

- Kartik Audhkhasi, Shrikanth S. Narayanan (2011), "Emotion Classification from Speech Using Evaluator Reliability-Weighted Combination of Ranked Lists", *2011 IEEE International Conference on Acoustics, Speech and Signal Processing* (ICASSP), Prague, Czech Republic, 22–27 May 2011, proceedings, pages 4956–4959, DOI: 10.1109/ICASSP.2011.5947468 (link, link, link)

- Kartik Audhkhasi, Shrikanth S. Narayanan (2013), "A Globally-Variant Locally-Constant Model for Fusion of Labels from Multiple Diverse Experts without Using Reference Labels", *IEEE Transactions on Pattern Analysis and Machine Intelligence*, volume 35, issue 4, pages 769–783, DOI: 10.1109/TPAMI.2012.139 (link, link, link, link, link)

- О.О. Авраменко / O.O. Avramenko, В.А. Голембо / V.A. Golembo (2017), "Комп'ютерна система підтримки та прийняття колективних рішень" / "Computer Group Decision Support Systems" (link, link)

- Raisa Axenie (2026), "Social Choice Theory in Protein Folding" (link, link)

- Haris Aziz, Barton E. Lee (2020), "The expanding approvals rule: improving proportional representation and monotonicity", *Social Choice and Welfare*, volume 54, issue 1, pages 1–45, DOI: 10.1007/s00355-019-01208-3 (arXiv:1708.07580) (link, link, link, link)

- Alexandra Baer (2015a), "Perception-Guided Evaluation of 3D Medical Visualizations", doctoral dissertation, Otto von Guericke University Magdeburg, Germany, DOI: 10.25673/4325 (link, link, link, link)

- Alexandra Baer, Kai Lawonn, Patrick Saalfeld, Bernhard Preim (2015b), "Statistical Analysis of a Qualitative Evaluation on Feature Lines", *Bildverarbeitung für die Medizin*, eds. Heinz Handels, Thomas Martin Deserno, Hans-Peter Meinzer, Thomas Tolxdorff, pages 71–76, Springer Vieweg, DOI: 10.1007/978-3-662-46224-9_14 (link, link, link, link)

- Babak Bagheri, Mohammad Rezaalipour, Mojtaba Vahidi-Asl (2019), "An Approach to Generate Effective Fault Localization Methods for Programs", *8th International Conference on Fundamentals of Software Engineering* (FSEN 2019), Tehran, Iran, 1–3 May 2019; eds. Hossein Hojjat, Mieke Massink, *Fundamentals of Software Engineering*, volume 11761 of the series *Lecture Notes in Computer Science*, pages 244–259, Springer, DOI: 10.1007/978-3-030-31517-7_17 (link, link, link)

- Jason Barbour, Stephany Rajeh, Sara Najem, Hocine Cherifi (2023), "Evaluating Network Embeddings Through the Lens of Community Structure", *12th International Conference on Complex Networks and their Applications* (COMPLEX NETWORKS 2023), Menton, France, 28–30 November 2023; eds. Hocine Cherifi, Luis M. Rocha, Chantal Cherifi, Murat Donduran, volume 1 of *Complex Networks & Their Applications XII*, volume 1141 of the series *Studies in Computational Intelligence*, pages 440–451, Springer, DOI: 10.1007/978-3-031-53468-3_37 (link, link)

- Jason Barbour, Stephany Rajeh, Sara Najem, Hocine Cherifi (2024), "Evaluating Community Structure Preservation of Network Embedding Algorithms", *French Regional Conference on Complex Systems*, Montpellier, France, 29–31 May 2024 (link, link)

- Didier Barradas-Bautista, Iain H. Moal, Juan Fernández-Recio (2016), "Docking through Democracy: Re-ranking protein-protein decoys with a voting system", *BSC International Doctoral Symposium — 3rd BSC International Doctoral Symposium*, Barcelona, Spain, 4–6 May 2016, DOI: 2117/91048 (link, link, link)

- Didier Barradas-Bautista (2017), "Protein-protein docking for interactomic studies and its application to personalized medicine", doctoral dissertation, University of Barcelona, Spain, DOI: 10261/175026 (link, link, link, link, link)

- Matthew Basilico (2013), "Microeconomics Review Sheet: Green" (link, link)

- Dorothea Baumeister, Jörg Rothe (2016), "Preference Aggregation by Voting", ed. Jörg Rothe, *Economics and Computation: An Introduction to Algorithmic Game Theory, Computational Social Choice, and Fair Division*, Springer Texts in Business and Economics, pages 197–325, DOI: 10.1007/978-3-662-47904-9_4; pages 233–367, DOI: 10.1007/978-3-031-60099-9_4 (link, link, link)

- BBGLab (2019), "IntOGen" (link, link, link, link)

- Ema Becirovic (2017), "On Social Choice in Social Networks", doctoral dissertation, Linköping University, Sweden (link, link)

- Jan Behrens, Axel Kistner, Andreas Nitsche, Björn Swierczek (2014a), “The Principles of LiquidFeedback”, Berlin (link, link, link, link)

- Jan Behrens (2014b), “How Chaos Protected the Status Quo (more than we intended to)”, *Liquid Democracy Journal*, issue 2, pages 27–30 (link, link, link, link)

- Raouia Belayadi (2019), “Contribution à l’étude des axiomes du choix social: la symétrie inverse et l’homogénéité des procédures de vote”, doctoral dissertation, University of Caen Normandy, France (link, link)

- Raouia Belayadi, Boniface Mbih (2021), “Violations of Reversal Symmetry Under Simple and Runoff Scoring Rules”, *Evaluating Voting Systems with Probability Models*, eds. Mostapha Diss, Vincent Merlin, pages 137–160, Springer, DOI: 10.1007/978-3-030-48598-6_7 (link)

- Barış Mert Bener, Berke Turanlıoğlu, Enes Eryarsoy (2021), “Digital Reputation Ranking” (link)

- Ratip Emin Berker, Sílvia Casacuberta, Isaac Robinson, Christopher Ong, Vincent Conitzer, Edith Elkind (2025), “From Independence of Clones to Composition Consistency: A Hierarchy of Barriers to Strategic Nomination”, *26th ACM Conference on Economics and Computation* (EC ‘25), Stanford, California, USA, 7–10 July 2025, proceedings, page 1109; *10th International Workshop on Computational Social Choice* (COMSOC 2025), Vienna, Austria, 17–19 September 2025, DOI: 10.1145/3736252.3742673 (arXiv:2502.16973) (link, link, link, video)

- Bilal Telman Bilalov, Zakir Cümşüd Zabidov (2016), “Şuls Metodu ilə Müxtəlif Istiqamətlər üçün Şəhər Havasinin Optik Vəziyyətinin Informativliyinin Qiymətləndirilməsi” / “Estimation of Informative Significance with a Schulze Method Optic State of Urban Air in Different Directions”, *Scientific News of Sumqayit State University*, volume 16, number 4, pages 22–26 (link, link)

- Sourav Biswas, Aman Bansal, Pabitra Mitra, Rajib Mall (2023), “Fault-Based Regression Test Case Prioritization”, *IEEE Transactions on Reliability*, volume 72, issue 3, pages 1176–1190, DOI: 10.1109/TR.2022.3205483 (link)

- Анна Ивановна Блохина / Anna Ivanovna Blokhina (2024a), “Применение Метода Шульце при Решении Задач Голосования в Малых Группах” / “Use of the Schulze Method in Small Committees”, *Студенческая Научная Весна / Student Scientific Spring Conference*, Moscow, 1–30 April 2024; proceedings, pages 573–574 (link, link)

- Анна Ивановна Блохина / Anna Ivanovna Blokhina, Ирина Александровна Самойлова / Irina Aleksandrovna Samoilova (2024b), “Контроль Голосования Путем Разделения Избирателей при Использовании Метода Шульце” / “Electoral Control by Partition Voters Using the Schulze Method”, *Управление Развитием Крупномасштабных Систем / 2024 17th International Conference on Management of Large-Scale System Development* (MLSD 2024), 24–26 September 2024, Moscow, Russia; proceedings, pages 575–580, DOI: 10.1109/MLSD61779.2024.10739529 (link, link)

- Alexander Bochkov, Nikolay Zhigirev, Alla Kuzminova (2025), “Expert Method of Rank Penalties”, *Reliability: Theory & Applications*, volume 20, number 3 (86), pages 1014–1021 (link, link, link, link)

- Thomas Bohne, Uwe M. Borghoff (2013), “Data Fusion: Boosting Performance in Keyword Extraction”, *20th Annual IEEE International Conference and Workshops on the Engineering of Computer Based Systems* (ECBS 2013), Phoenix/Scottsdale, Arizona, USA, 22–24 April 2013, proceedings, pages 166–173, DOI: 10.1109/ECBS.2013.12 (link, link)

- Thomas Bohne (2015), “Heuristic Strategies for Single Document Analysis”, doctoral dissertation, Bundeswehr University Munich, Germany (link, link, link)

- Christoph Börgers (2009), “Mathematics of Social Choice: Voting, Compensation, and Division”, SIAM, DOI: 10.1137/1.9780898717624 (link, link, video: 00:23:14 – 00:30:04 hh:mm:ss)

- Panagiotis Bountris, Charalampos Tsirmpas, Maria Haritou, Abraham Pouliakis, Petros Karakitsos, Dimitrios Koutsouris (2015a), “Evaluation of Cervical Cancer Detection Tests using Feature Ranking Aggregation”, *37th Annual International Conference of the IEEE Engineering in Medicine and Biology Society* (EMBC 2015), Milan, Italy, 25–29 August 2015, presentation FrFPoT6.25 (link)

- Panagiotis Bountris, Charalampos Tsirmpas, Maria Haritou, Abraham Pouliakis, Ioannis Kouris, Petros Karakitsos, Dimitrios Koutsouris (2015b), “An Ensemble Feature Ranking Framework for the Assessment of the Efficacy of Cervical Cancer Detection Tests and Human Papillomavirus Genotypes in the Detection of High-grade Cervical Intraepithelial Neoplasia and Cervical Carcinoma”, *2015 IEEE 15th International Conference on Bioinformatics and Bioengineering* (BIBE), Belgrade, Serbia, 2–4 November 2015, proceedings, pages 1–5, DOI: 10.1109/BIBE.2015.7367657 (link)

- Panagiotis Bountris (2017), “Development of artificial intelligence algorithms and intelligent risk assessment and clinical decision support systems for the early diagnosis and the personalised management of women with cervical intraepithelial lesions”, doctoral dissertation, National Technical University of Athens, Greece, DOI: 10.26240/heal.ntua.2724 (link, link)

- Hassane Bouremel (2017a), “Clonal sets of a binary relation: Theory and Applications”, doctoral dissertation, Msila University, Algeria (link, link)

- Hassane Bouremel, Raúl Pérez-Fernández, Lemnaouar Zedam, Bernard DeBaets (2017b), “The clone relation of a binary relation”, *Information Sciences*, volumes 382–383, pages 308–325, DOI: 10.1016/j.ins.2016.12.008 (link, link)

- Hubert Lewis Bray (2018), “The Schulze Method” (video, video)

- Colin James Brearley (1999), “Properties of Single-Seat Preferential Election Rules”, doctoral dissertation, Nottingham University, United Kingdom {In this paper, the Schulze method is called “descending minimum gross score” (DminGS), “descending minimum augmented gross score” (DminAGS), and “descending minimum doubly augmented gross score” (DminDAGS).}

- Daniela Bubboloni, Michele Gori (2018), “The flow network method”, *Social Choice and Welfare*, volume 51, issue 4, pages 621–656, DOI: 10.1007/s00355-018-1131-7 (arXiv:1606.03898) (link, link, link, link, link)

- Daniela Bubboloni, Stefano Moretti (2026), “Comparison games and ranking of players” (arXiv:2606.31955) (link, link)

- Sam Bucovetsky (2012), “Many Social Choice Rules” (link, link)

- Chris Budd (2019), “Maths and Voting”, 12 November 2019 (link, link, link, link, link, video: 00:41:04 – 00:44:18 hh:mm:ss)

- Lukáš Bureš, Luděk Müller (2021), “Semantic Segmentation in the Task of Long-Term Visual Localization”, *6th International Conference on Interactive Collaborative Robotics* (ICR 2021), Saint Petersburg, Russia, 27–30 September 2021; eds. Andrey Ronzhin, Gerhard Rigoll, Roman Meshcheryakov, *Interactive Collaborative Robotics*, volume 12998 of the series *Lecture Notes in Computer Science*, pages 27–39, Springer, DOI: 10.1007/978-3-030-87725-5_3 (link)

- Lukáš Bureš (2022), “Semantic Segmentation in Long-term Visual Localization”, doctoral dissertation, University of West Bohemia, Pilsen, Czech Republic, DOI: 11025/50759 (link, link)

- Kathleen Cachel (2021), “MANI-Rank: Multiple Attribute and Intersectional Group Fairness for Consensus Ranking”, master thesis, Worcester Polytechnic Institute, Worcester, Massachusetts, USA (link, link)

- Kathleen Cachel, Elke Rundensteiner, Lane Harrison (2022), “MANI-Rank: Multiple Attribute and Intersectional Group Fairness for Consensus Ranking”, *2022 IEEE 38th International Conference on Data Engineering* (ICDE 2022), Kuala Lumpur, Malaysia, 9–12 May 2022, proceedings, pages 1124–1137, DOI: 10.1109/ICDE53745.2022.00089 (arXiv:2207.10020) (link, link, link, link, link, link, video)

- Wei Cai, Xiaodong Fu, Li Liu, Lijun Liu, Qingsong Huang (2016), “A Method of Competition Result Evaluation Based on Social Choice Theory”, *2016 Chinese Control and Decision Conference* (CCDC), Yinchuan, China, 28–30 May 2016, proceedings, pages 2126–2131, DOI: 10.1109/CCDC.2016.7531336 (link)

- Chris Callison-Burch (2009), “Fast, Cheap, and Creative: Evaluating Translation Quality Using Amazon’s Mechanical Turk”, *2009 Conference on Empirical Methods in Natural Language Processing* (EMNLP ‘09), Singapore, 6–7 August 2009, proceedings, pages 286–295 (link, link, link, link, link, link, link, link, link)

- Betania Silva Carneiro Campello, Guilherme Dean Pelegrina, Renata Pelissari, Ricardo Suyama, Leonardo Tomazeli Duarte (2024), "Mitigating subjectivity and bias in AI development indices: A robust approach to redefining country rankings", *Expert Systems with Applications*, volume 255, part D, article 124803, DOI: 10.1016/j.eswa.2024.124803 (arXiv:2402.10122) (link)

- Rosa Camps, Xavier Mora, Laia Saumell (2008), "A continuous rating method for preferential voting" (arXiv:0810.2263) (link, link, link, link)

- Rosa Camps, Xavier Mora, Laia Saumell (2012a), "A general method for deciding about logically constrained issues", *Annals of Mathematics and Artificial Intelligence*, volume 64, issue 1, pages 39–72, DOI: 10.1007/s10472-012-9292-z (arXiv:1007.2534) (link, link, link, link)

- Rosa Camps, Xavier Mora, Laia Saumell (2012b), "A continuous rating method for preferential voting: the complete case", *Social Choice and Welfare*, volume 39, issue 1, pages 141–170, DOI: 10.1007/s00355-011-0548-z (arXiv:0912.2190) (link, link, link, link, link)

- Rosa Camps, Xavier Mora, Laia Saumell (2013), "A continuous rating method for preferential voting: the incomplete case", *Social Choice and Welfare*, volume 40, issue 4, pages 1111–1142, DOI: 10.1007/s00355-012-0663-5 (arXiv:0912.2195) (link, link, link)

- Rosa Camps, Xavier Mora, Laia Saumell (2014a), "Social choice rules driven by propositional logic", *Annals of Mathematics and Artificial Intelligence*, volume 70, issue 3, pages 279–312, DOI: 10.1007/s10472-013-9395-1 (arXiv:1109.4335) (link, link, link)

- Rosa Camps, Xavier Mora, Laia Saumell (2014b), "Choosing by means of approval-preferential voting. The path-revised approval choice" (arXiv:1411.1367) (link, link, link, link, link)

- Rosa Camps, Xavier Mora, Laia Saumell (2015), "Fraction-Like Ratings from Preferential Voting", *Publicacions Matemàtiques*, volume 59, issue 1, pages 99–136, DOI: 10.5565/PUBLMAT_59115_06 (arXiv:1001.3931) (link, link, link, link)

- Ioannis Caragiannis, Ariel D. Procaccia, Nisarg Shah (2016), "When Do Noisy Votes Reveal the Truth?", *14th ACM Conference on Electronic Commerce* (EC '13), Philadelphia, Pennsylvania, USA, 16–20 June 2013, proceedings, pages 143–160, DOI: 10.1145/2482540.2482570; *ACM Transactions on Economics and Computation*, volume 4, number 3, article 15, DOI: 10.1145/2892565 (link, link, link, link, link, link, link, link)

- Pierre Casadebaig, Ronan Trépos, Victor Picheny, Nicolas B. Langlade, Patrick Vincourt, Philippe Debaeke (2014), "Increased genetic diversity improves crop yield stability under climate variability: a computational study on sunflower" (arXiv:1403.2825) (link, link, link, link)

- Pieter Cawood, Terence L. van Zyl (2022), "Evaluating State of the Art, Forecasting Ensembles- and Meta-learning Strategies for Model Fusion", *Forecasting*, volume 4, issue 3, pages 732–751, DOI: 10.3390/forecast4030040 (arXiv:2203.03279) (link, link, link, link)

- Katarína Cechlárová, Jörg Rothe, Šimon Schierreich, Ildikó Schlotter (2026), "Navigating the Complexity Landscape of Nominee Selection in Schulze Voting" (arXiv:2604.11933) (link, link)

- Jacques Cellier (2020), "La face cachée des urnes", Presses Universitaires de Rennes (link, link, link)

- Erin Wolf Chambers, Xavier Goaoc, Michael Kerber, Suresh Venkatasubramanian (2021), "Implementing double blind reviews in SoCG" (link, link)

- David Chandler (2008), "Voting for more than just either-or", *MIT Tech Talk*, volume 52, number 19, page 2, 12 March 2008 (link, link, link, link)

- Silvia D. Chang (2018), "Canadian Radiologists' Perception of the PGY-1 Basic Clinical Year: Results of a National Survey", *2018 Annual Scientific Meeting of the Canadian Association of Radiologists* (CAR 2018), Montréal, Québec, Canada, 26–29 April 2018, proceedings, page 36 (link, link, link)

- Silvia D. Chang, Daniel J.A. Margolis, Baris Turkbey, Abigail A. Arnold, Sadhna Verma (2021), "Practice Patterns and Challenges of Performing and Interpreting Prostate MRI: A Survey by the Society of Abdominal Radiology Prostate Disease-Focused Panel", *American Journal of Roentgenology*, volume 216, number 4, pages 952–959, DOI: 10.2214/AJR.20.23256 (link, link, link)

- Cyril Charignon (2021), "Graphes" (link, link)

- Michael C. Chavrimootoo, Ian Clingerman, Ethan Ferland, Erin Gibson, Lane A. Hemaspaandra, Quan Luu, David E. Narváez, Yanfei Wang (2026), "Axiomatic Tools for Separating Electoral Control Types, with Applications to Concrete Systems" (arXiv:2606.12039) (link)

- Zbigniew Chechliński (2019), "The effects of the hypothetical implementation of preferential voting methods in Poland on the Polish political stage and national integrity", *European Journal of Geopolitics*, volume 7, pages 34–66 (link, link, link, link, link)

- Jiehua Chen, Joanna Kaczmarek, Paul Nüsken, Jörg Rothe, Ildikó Schlotter, Tessa Seeger (2025), "Control in Computational Social Choice", *34th International Joint Conference on Artificial Intelligence* (IJCAI 2025), Montréal, Québec, Canada, 16–22 August 2025; proceedings, pages 10391–10399, DOI: 10.24963/ijcai.2025/1154 (link, link, link, link, link, link)

- Ying Chen, Xiaodong Fu, Kun Yue, Li Liu, Li-jun Liu (2016), “Ranking Online Services by Aggregating Ordinal Preferences”, *17th International Conference on Web-Age Information Management* (WAIM 2016), Nanchang, China, 3–5 June 2016; eds. Shaoxu Song, Yongxin Tong, *Web-Age Information Management*, volume 9998 of the series *Lecture Notes in Computer Science*, pages 41–53, Springer, DOI: 10.1007/978-3-319-47121-1_4 (link)

- Huey Eng Chua (2015), “Trecento: *in silico* network-driven identification of target combinations for combination therapy”, doctoral dissertation, Nanyang Technological University, Singapore, DOI: 10.32657/10356/64995 (link, link, link)

- Maxime Chupin (2022), “Retour sur le choix du nouveau nom de domaine”, *Lettre GUTenberg*, number 45, pages 12–14 (link, link, link)

- Fiorella De Cindio, Andrea Trentini (2024), “Dal Tecnocivismo alla Cittadinanza Digitale” (link, link)

- Abdoulaye Cisse, Alain de Janvry, Joel Ferguson, Marco Gonzalez-Navarro, Samba Mbaye, Elisabeth Sadoulet, Mame Mor Anta Syll (2026), “Irrigation infrastructure and satellite-measured land cultivation impacts: Evidence from the Senegal river valley”, *Journal of Development Economics*, volume 179, article 103615, DOI: 10.1016/j.jdeveco.2025.103615 (link, link, link, link, link)

- Alessandra Comparetto (2022), “Piattaforma innovativa per i processi di democrazia diretta e partecipativa degli Enti Comunali”, doctoral dissertation, Polytechnic University of Turin, Italy (link, link)

- Marie Jean Antoine Nicolas de Caritat de Condorcet (1785), “Essai sur l’application de l’analyse à la probabilité des décisions rendues à la pluralité des voix”, Imprimerie Royale, Paris, France (link)

- Vincent Conitzer, Toby Walsh (2016), “Barriers to Manipulation in Voting”, eds. Felix Brandt, Vincent Conitzer, Ulle Endriss, Jérôme Lang, Ariel D. Procaccia, *Handbook of Computational Social Choice*, pages 127–145, Cambridge University Press, DOI: 10.1017/CBO9781107446984.007 (link, link, link, link, link, link)

- Pierluigi Contucci, Emanuele Panizzi, Federico Ricci-Tersenghi, Alina Sîrbu (2016), “Egalitarianism in the rank aggregation problem: a new dimension for democracy”, *Quality & Quantity*, volume 50, issue 3, pages 1185–1200, DOI: 10.1007/s11135-015-0197-x (arXiv:1406.7642) (link, link, link, link, link, link, link, link)

- Pierluigi Contucci, Alina Sîrbu (2019), “Egalitarianism vs. Utilitarianism in Preferential Voting”, eds. Pierluigi Contucci, Andrea Omicini, Danilo Pianini, Alina Sîrbu, *The Future of Digital Democracy*, volume 11300 of the series *Lecture Notes in Computer Science*, pages 24–37, Springer, DOI: 10.1007/978-3-030-05333-8_3; DOI: 11568/942362 (link, link, link, link, link)

- Jacopo Corina (2024), “EleKtion: una Libreria Kotlin Multiplatform per la Democrazia Digitale in Nuovi Contesti”, doctoral dissertation, University of Bologna, Italy (link, link)

- Alex J. Cornish, Andreas J. Gruber, Ben Kinnersley, Daniel Chubb, Anna Frangou, Giulio Caravagna, Boris Noyvert, Eszter Lakatos, Henry M. Wood, Steve Thorn, Richard Culliford, Claudia Arnedo-Pac, Jacob Househam, William Cross, Amit Sud, Philip Law, Maire Ni Leathlobhair, Aliah Hawari, Connor Woolley, Kitty Sherwood, Nathalie Feeley, Güler Gül, Juan Fernandez-Tajes, Luis Zapata, Ludmil B. Alexandrov, Nirupa Murugaesu, Alona Sosinsky, Jonathan Mitchell, Nuria Lopez-Bigas, Philip Quirke, David N. Church, Ian P.M. Tomlinson, Andrea Sottoriva, Trevor A. Graham, David C. Wedge, Richard S. Houlston (2024), “The genomic landscape of 2,023 colorectal cancers”, *nature*, volume 633, pages 127–136, DOI: 10.1038/s41586-024-07747-9 (link, link, link, link, link, link, link, link, link, link, link, link, link)

- Véronique Cortier, Pierrick Gaudry, Quentin Yang (2022), “A Toolbox for Verifiable Tally-Hiding E-Voting Systems”, *27th European Symposium on Research in Computer Security* (ESORICS 2022), Copenhagen, Denmark, 26–30 September 2022; eds. Vijayalakshmi Atluri, Roberto Di Pietro, Christian D. Jensen, Weizhi Meng, proceedings, part II, volume 13555 of the series *Lecture Notes in Computer Science*, pages 631–652, Springer, DOI: 10.1007/978-3-031-17146-8_31 (link, link, link, link, link, link)

- Lucas Gabriel Lage Costa (2020), “ECT: An Offline Metric to Evaluate Case-Based Explanations for Recommender Systems”, doctoral dissertation, Federal Center for Technological Education of Minas Gerais, Belo Horizonte, Brazil (link, link)

- Markos Cramer (2014), “Voĉdonsistemoj, grupaj decidoj kaj iliaj problemoj” (link, link)

- Blake Cretney (1998), personal communication, 10 November 1998

- Julia Theresa Csar (2018a), “Cloud Computing Techniques for Winner Determination in Computational Social Choice”, doctoral dissertation, Vienna University of Technology, Austria, DOI: 10.34726/hss.2018.58041 (link, link)

- Julia Theresa Csar, Martin Lackner, Reinhard Pichler (2018b), “Computing the Schulze Method for Large-Scale Preference Data Sets”, *27th International Joint Conference on Artificial Intelligence* (IJCAI 2018), Stockholm, Sweden, 13–19 July 2018, proceedings, pages 180–187, DOI: 10.24963/ijcai.2018/25 (arXiv:2505.12976) (link, link, link, link, link, link, link)

- Richard Culliford, Samuel E.D. Lawrence, Charlie Mills, Zayd Tippu, Daniel Chubb, Alex J. Cornish, Lisa Browning, Ben Kinnersley, Robert Bentham, Amit Sud, Husayn Pallikonda, Anna Frangou, Andreas J. Gruber, Kevin Litchfield, David C. Wedge, James Larkin, Samra Turajlic, Richard S. Houlston (2024), “Whole genome sequencing refines stratification and therapy of patients with clear cell renal cell carcinoma”, *nature communications*, volume 15, article 5935, DOI: 10.1038/s41467-024-49692-1 {In this paper, the Schulze method is mentioned only in the supplementary material (link).}

- Jonathan Cummings (2020), “The Rating Game”, Post Hill Press (link)

- Richard B. Darlington (2016), “Minimax is the best electoral system after all” (arXiv:1606.04371) (link)

- Richard B. Darlington (2018), “Are Condorcet and Minimax Voting Systems the Best?” (arXiv:1807.01366) (link, link, link)

- Richard B. Darlington (2023), “The case for minimax-TD”, *Constitutional Political Economy*, volume 34, issue 3, pages 410–420, DOI: 10.1007/s10602-022-09390-w (link, link, link, link)

- Kathryn E. Darras, Rebecca Spouge, Abigail A. Arnold, Anique B.H. de Bruin, Savvas Nicolaou, Claudia Krebs, Rose Hatala, Bruce B. Forster (2018a), “Virtual Dissection Adds Educational Value to a Traditional Medical Undergraduate Cadaveric Anatomy Course”, *Experimental Biology*, San Diego, California, USA, 21–25 April 2018; *FASEB Journal*, volume 32, issue S1, number 635.2 (link)

- Kathryn E. Darras, Rebecca Spouge, Abigail A. Arnold, Anique B.H. de Bruin, Claudia Krebs, Rose Hatala, Savvas Nicolaou, Bruce B. Forster (2018b), “Radiology Adds Value to Medical Undergraduate Anatomy Education Through Virtual Dissection”, *2018 Annual Scientific Meeting of the Canadian Association of Radiologists* (CAR 2018), Montréal, Québec, Canada, 26–29 April 2018, presentation SRP037, proceedings, page 115 (link, link)

- Kathryn E. Darras, Abigail A. Arnold, Colin Mar, Bruce B. Forster, Linda Probyn, Silvia D. Chang (2018c), “Rethinking the PGY-1 Basic Clinical Year: A Canadian National Survey of Its Educational Value for Diagnostic Radiology Residents”, *2017 International Conference on Residency Education* (ICRE 2017), Québec City, Québec, Canada, 17–22 October 2017, abstract 152; *2018 Annual Scientific Meeting of the Canadian Association of Radiologists* (CAR 2018), Montréal, Québec, Canada, 26–29 April 2018, presentation SRP038, proceedings, page 116; *Academic Radiology*, volume 25, issue 9, pages 1213–1218, DOI: 10.1016/j.acra.2018.03.012 (link, link, link)

- Kathryn E. Darras, Rebecca Spouge, Anique B.H. de Bruin, Jeroen J.G. van Merriënboer, Savvas Nicolaou, Rose Hatala, Bruce B. Forster (2018d), “Beyond Cadaveric Dissection: Through Virtual Dissection, Radiology Adds Educational Value to Medical Undergraduate Anatomy Education”, *2018 Scientific Assembly and Annual Meeting of the Radiological Society of North America*, Chicago, Illinois, USA, 25–30 November 2018, presentation SSC08-06 (link, link)

- Kathryn E. Darras, Bruce B. Forster, Rebecca Spouge, Anique B.H. de Bruin, Abigail A. Arnold, Savvas Nicolaou, Jeff Hu, Rose Hatala, Jeroen J.G. van Merriënboer (2020), “Virtual Dissection with Clinical Radiology Cases Provides Educational Value to First Year Medical Students”, *Academic Radiology*, volume 27, issue 11, pages 1633–1640, DOI: 10.1016/j.acra.2019.09.031 (link, link)

- Alexandra Degeest, Michel Verleysen, Benoît Frénay (2015), “Feature Ranking in Changing Environments where New Features are Introduced”, *2015 International Joint Conference on Neural Networks* (IJCNN), Killarney, Ireland, 12–17 July 2015, proceedings, pages 1–8, DOI: 10.1109/IJCNN.2015.7280533 (link, link)

- Alexandra Degeest (2021), “Relevance Criteria for Feature Selection using Filters and Dynamic Environments”, doctoral dissertation, Catholic University of Leuven, Belgium, DOI: 2078.1/252868 (link, link, link)

- Holger Dell, Thore Husfeldt, Bart M.P. Jansen, Petteri Kaski, Christian Komusiewicz, Frances A. Rosamond (2016), “The First Parameterized Algorithms and Computational Experiments Challenge”, *11th International Symposium on Parameterized and Exact Computation* (IPEC 2016), *Leibniz International Proceedings in Informatics* (LIPIcs), volume 63, article 30, DOI: 10.4230/LIPIcs.IPEC.2016.30 (link, link, link, link, link, link, link)

- Holger Dell, Christian Komusiewicz, Nimrod Talmon, Mathias Weller (2017), “The PACE 2017 Parameterized Algorithms and Computational Experiments Challenge: The Second Iteration”, *12th International Symposium on Parameterized and Exact Computation* (IPEC 2017), *Leibniz International Proceedings in Informatics* (LIPIcs), volume 89, article 30, DOI: 10.4230/LIPIcs.IPEC.2017.30 (link, link, link, link, link, link, link)

- Christopher R. Devlin (2019), “Voting Systems: From Method to Algorithm”, doctoral dissertation, California State University Channel Islands, USA, DOI: 10211.3/214761 (link, link, link)

- Kai Diethelm (2016), “Gemeinschaftliches Entscheiden: Untersuchung von Entscheidungsverfahren mit mathematischen Hilfsmitteln”, Springer, DOI: 10.1007/978-3-662-48780-8 (link)

- Edsger W. Dijkstra (1959), “A Note on Two Problems in Connexion with Graphs”, *Numerische Mathematik*, volume 1, issue 1, pages 269–271, DOI: 10.1007/BF01386390 (link, link, link, link, link, link)

- Michelle Döring, Markus Brill, Jobst Heitzig (2026), “The River Voting Method”, *40th Annual AAAI Conference on Artificial Intelligence* (AAAI-26), Singapore, 20–27 January 2026; proceedings, pages 16846–16854, DOI: 10.1609/aaai.v40i20.38729 (arXiv:2504.14195) (link, link, link, link)

- Henry Richmond Droop (1881), “On Methods of Electing Representatives”, *Journal of the Statistical Society of London*, volume 44, number 2, pages 141–202, DOI: 10.2307/2339223 (link, link)

- Moon Duchin (2021), “Math of Social Choice” (video: 04:13:22 – 05:15:54 hh:mm:ss, 10:26:23 – 10:37:17 hh:mm:ss, 11:20:31 – 11:27:00 hh:mm:ss) {In this lecture, the Schulze method is called “beatpath”.}

- Michael A.E. Dummett (1984), “Voting Procedures”, Clarendon Press, Oxford

- François Durand (2015), “Vers des modes de scrutin moins manipulables”, doctoral dissertation, Pierre and Marie Curie University, Paris, France (link, link, link, link, link, link, link)

- François Durand (2026), “Which Voting Rules Are More Resilient to Coalitional Manipulation?” (arXiv:2607.00758) (link, link, link)

- Stefan Eng, Jennifer Tan, Markus Iseli (2018), “Adjective Intensity Ordering by Representing Word Definitions as a System of Linear Equations”, *2018 IEEE 12th International Conference on Semantic Computing* (ICSC), Laguna Hills, California, USA, 31 January – 2 February 2018; proceedings, pages 265–268, DOI: 10.1109/ICSC.2018.00047 (link)

- Steve Eppley (2003), “Independence from Pareto-dominated alternatives: A criterion for voting rules” (link) {In this paper, the Schulze method is called “PathWinner”.}

- David Eppstein (2021), “Widest Paths and Voting Systems” (video1: 00:50:07 – 00:57:12 hh:mm:ss) (video2: 00:49:54 – 00:54:53 hh:mm:ss) {In this lecture, the Schulze method is also called “widest path method”.}

- Lorne Erhardt, David Waller (2017), “Radiological Nuclear Defence R&D Priorities: Proposal for a Long Range Plan for DRDC” (link)

- Behrooz Etesamipour, Robert J. Hammell II (2020), “Investigation of Ranking Methods Within the Military Value of Information (VoI) Problem Domain”, *18th International Conference on Information Processing and Management of Uncertainty in Knowledge-Based Systems* (IPMU 2020), Lisbon, Portugal, 15–19 June 2020; proceedings, part I, eds. Marie-Jeanne Lesot, Susana Vieira, Marek Z. Reformat, João Paulo Carvalho, Anna Wilbik, Bernadette Bouchon-Meunier, Ronald R. Yager, pages 129–142, Springer, DOI: 10.1007/978-3-030-50146-4_11 (link, link, link)

- Maria Evita, Mitra Djamal, Bernd Zimanowski, Klaus Schilling (2015), “Mobile Monitoring System for Indonesian Volcano”, *2015 4th International Conference on Instrumentation, Communications, Information Technology, and Biomedical Engineering* (ICICI-BME), Bandung, Indonesia, 2–3 November 2015, proceedings, pages 278–281, DOI: 10.1109/ICICI-BME.2015.7401378 (link)

- Piotr Faliszewski, Jörg Rothe (2016), “Control and Bribery in Voting”, eds. Felix Brandt, Vincent Conitzer, Ulle Endriss, Jérôme Lang, Ariel D. Procaccia, *Handbook of Computational Social Choice*, pages 146–168, Cambridge University Press, DOI: 10.1017/CBO9781107446984.008 (link, link, link, link, link)

- Marios A. Fanourakis, Guillaume Chanel, Rayan Elalamy, Phil Lopes (2021), “Towards Recognizing Emotion in the Latent Space”, *2021 IEEE International Conference on Pervasive Computing and Communications* (PERCOM), Kassel, Germany, 22–26 March 2021, proceedings, pages 364–367, DOI: 10.1109/PerComWorkshops51409.2021.9430963 (link, link, link)

- Dan S. Felsenthal (2011), “Review of Paradoxes Afflicting Procedures for Electing a Single Candidate”, eds. Dan S. Felsenthal, Moshé Machover, *Electoral Systems: Paradoxes, Assumptions, and Procedures*, pages 19–91, Springer, DOI: 10.1007/978-3-642-20441-8_3 (link)

- Dan S. Felsenthal, T. Nicolaus Tideman (2014), “Weak Condorcet Winner(s) Revisited”, *Public Choice*, volume 160, issues 3–4, pages 313–326, DOI: 10.1007/s11127-014-0180-4 (link, link)

- Dan S. Felsenthal, Hannu Nurmi (2018), “Voting Procedures for Electing a Single Candidate”, Springer, DOI: 10.1007/978-3-319-74033-1 (link)

- Dan S. Felsenthal, Hannu Nurmi (2019), “Voting Procedures Under a Restricted Domain”, Springer, DOI: 10.1007/978-3-030-12627-8 (link)

- Silvia Fernandez (2018), “Schulze voting method: choosing a hike” (video)

- Felix Fischer, Olivier Hudry, Rolf Niedermeier (2016), “Weighted Tournament Solutions”, eds. Felix Brandt, Vincent Conitzer, Ulle Endriss, Jérôme Lang, Ariel D. Procaccia, *Handbook of Computational Social Choice*, pages 85–102, Cambridge University Press, DOI: 10.1017/CBO9781107446984.005 (link, link, link, link)

- Peter C. Fishburn (1973), “The Theory of Social Choice”, Princeton University Press

- Peter C. Fishburn (1977), “Condorcet Social Choice Functions”, *SIAM Journal on Applied Mathematics*, volume 33, issue 3, pages 469–489, DOI: 10.1137/0133030

- Robert W. Floyd (1962), “Algorithm 97: Shortest Path”, *Communications of the ACM*, volume 5, number 6, page 345, DOI: 10.1145/367766.368168 (link)

- Cameron Foale (2010), “The Directional Propagation Cache: Real-time Acoustic Simulation for Immersive Computer Games”, doctoral dissertation, University of Ballarat, Victoria, Australia (link, link)

- Bruce B. Forster, Anique B.H. de Bruin, Abigail A. Arnold, Rose Hatala, Jeroen J.G. van Merriënboer, Kathryn E. Darras, Rebecca Spouge, Savvas Nicolaou (2020), “Medical Student Perceptions of a Case-Based Virtual Dissection Anatomy Curriculum”, *Canadian Medical Education Journal*, volume 11, number 2, page e155, article DP 1-1, DOI: 10.36834/cmej.70106 (link, link, link, link, link, link)

- Dominique Fortin (2020), “Clustering Analysis of a Dissimilarity: a Review of Algebraic and Geometric Representation”, *Journal of Classification*, volume 37, issue 1, pages 180–202, DOI: 10.1007/s00357-019-09315-7 (link)

- Cléber Vinícius de Freitas (2021), “Sistema de mensuração de desempenho para estruturação do processo de aprovação de conexões de sistemas de geração distribuída em distribuidoras de energia elétrica”, doctoral dissertation, Federal University of Santa Maria, Brazil (link, link)

- Cléber Vinícius de Freitas, Carmen Brum Rosa, Júlia Possebon Spellmeier, Julio Cezar Mairesse Siluk, Paula Donaduzzi Rigo, Rogério Moreira Alves Júnior (2022), “Definição de Ações Estratégicas orientadas ao desempenho de concessionárias de energia elétrica na ótica da Micro e Minigeração Distribuída”, *International Conference on Production Research — Americas* (ICPR Americas 2022), Curitiba, Brazil, 23–25 November 2022, DOI: 10.29327/foreigners_subscription_icpr_americas22.664561 (link, link)

- Елена Александровна Фролова / Elena Alexandrovna Frolova, Владимир Александрович Тушавин / Vladimir Alexandrovich Tushavin, Альбина Сергеевна Тур / Albina Sergeevna Tur (2022), “Проблемы Ранжирования при Управлении Качеством Процесса Закупок Комплектующих” / “Problems of Ranking when Managing the Quality of the Process Purchasing of Components”, *Математические Методы и Модели в Высокотехнологичном Производстве / 1st International Forum on Mathematical Methods and Models in High-Tech Production*, Saint Petersburg, Russia, 9 November 2022, proceedings, pages 431–433 (link)

- Satoru Fujishige, Leo Goto, Satoshi Nakada (2026), “A New Method for Finding the Schulze Winner Set” (arXiv:2606.02213) (link, link)

- Xiaozhuan Gao, Lipeng Pan (2025), “An information fusion model of mutual influence between focal elements: A perspective on interference effects in Dempster-Shafer evidence theory”, *Information Fusion*, volume 124, issue C, article 103286, DOI: 10.1016/j.inffus.2025.103286 (link, link)

- Gazal Garg, Prasenjit Mondal, Shashaank M. Aswatha, Jit Mukherjee, Tapas Maji, Jayanta Mukherjee (2014), “VIMARSHAK — A Web Based Subjective Image Evaluation System”, *Fifth International Conference on Signal and Image Processing* (ICSIP), Bangalore, India, 8–10 January 2014, proceedings, pages 73–76, DOI: 10.1109/ICSIP.2014.16 (link, link, link)

- Serge Gaspers, Thomas Kalinowski, Nina Narodytska, Toby Walsh (2012), “Coalitional Manipulation for Schulze’s Rule”, *36th Australasian Conference on Combinatorial Mathematics and Combinatorial Computing* (36ACCMCC), Sydney, Australia, 10–14 December 2012; *12th International Conference on Autonomous Agents and Multiagent Systems* (AAMAS 2013), Saint Paul, Minnesota, USA, 6–10 May 2013, proceedings, pages 431–438 (arXiv:1304.0844) (link, link, link, link, link, link, link, link, link)

- Allen van Gelder (2011), “Careful Ranking of Multiple Solvers with Timeouts and Ties”, *14th International Conference on Theory and Application of Satisfiability Testing* (SAT 2011), Ann Arbor, Michigan, 19–22 June 2011; eds. Karem A. Sakallah, Laurent Simon, volume 6695 of the series *Lecture Notes in Computer Science*, pages 317–328, Springer, DOI: 10.1007/978-3-642-21581-0_25 (link, link, link, link)

- Felix Gervits, Gordon Briggs, Matthias Scheutz (2017), “The Pragmatic Parliament: A Framework for Socially-Appropriate Utterance Selection in Artificial Agents”, *Annual Conference of the Cognitive Science Society* (CogSci 2017), London, 26–29 July 2017, proceedings, pages 2085–2090 (link, link, link, link)

- Fatemeh Ghovanlooy Ghajar, Axel Sikora, Dominik Welte (2022), “Schloss: Blockchain-Based System Architecture for Secure Industrial IoT”, Electronics, volume 11, issue 10, article 1629, DOI: 10.3390/electronics11101629 (link, link, link, link, link)

- Milad K. Ghale (2019), “Verified and Verifiable Computation with STV Algorithms”, doctoral dissertation, Australian National University, DOI: 10.25911/5f58affacfc58 (link, link, link)

- Milad K. Ghale, Rajeev Goré, Dirk Pattinson (2021), “Formally Verified and Publicly Verifiable E-counting For Complex Voting Schemes”, 15 December 2021 (video, video)

- Cristiano Ghersi, Luca Pulina, Armando Tacchella (2007), “Which system should I buy? A case study about the QBF solvers competition”, *International Workshop on Scheduling a Scheduling Competition at the 17th International Conference on Automated Planning and Scheduling* (SSC@ICAPS 2007), Providence, Rhode Island, USA, 22–26 September 2007 (link, link, link)

- Xavier Goaoc (2022), “Algorithmes et Complexité” (link, link, link, link, link, link)

- Ashish Goel, Anilesh Kollagunta Krishnaswamy, Kamesh Munagala (2017), “Metric Distortion of Social Choice Rules: Lower Bounds and Fairness Properties”, *18th ACM Conference on Economics and Computation* (EC ‘17), Cambridge, Massachusetts, USA, 26–30 June 2017, proceedings, pages 287–304, DOI: 10.1145/3033274.3085138 (arXiv:1612.02912) (link, link, link, link, link)

- Илья Алексеевич Голышков / Ilya Alekseevich Golyshkov, Михаил Витальевич Фуфаев / Mikhail Vitalievich Fufaev (2017), “Теорема Эрроу, и как с ней бороться?” / “Arrow’s Theorem, and how to deal with it?”, *VIII Международного Научного Студенческого Конгресса / VIII International Scientific Students Congress*, *Секция 5: Лучшие Работы Математической Секции / Chapter 5: Best Papers of the Mathematics Section*, Moscow, Russia, 15 April 2017, proceedings, pages 315–318 (link)

- Víctor Gómez Álvarez (2018), “Método de Schulze: Teoría y práctica”, bachelor thesis, University of Valladolid, Spain, DOI: 10324/34325 (link, link)

- Juozas Gordevičius (2010a), “Summarizing Temporal Data with Approximate Temporal Aggregation and Ranking”, doctoral dissertation, Free University of Bozen, Italy (link, link)

- Juozas Gordevičius, Francisco J. Estrada, Hyun Chul Lee, Periklis Andritsos, Johann Gamper (2010b), “Ranking of Evolving Stories Through Meta-Aggregation”, *19th ACM International Conference on Information and Knowledge Management* (CIKM ‘10), Toronto, Ontario, Canada, 26–30 October 2010, proceedings, pages 1909–1912, DOI: 10.1145/1871437.1871761 (link, link, link, link, link)

- Michele Gori (2023), “Families of abstract decision problems whose admissible sets intersect in a singleton”, *Social Choice and Welfare*, volume 61, issue 1, pages 131–154, DOI: 10.1007/s00355-022-01443-1 (link)

- Michele Gori (2025), “Some solution concepts for finite strategic games based on Kalai and Schmeidler’s admissible set” (link, link)

- Michele Gori (2026), “The maximum-weight arborescence network solution” (link, link)

- Johannes Grabmeier (2009), “Condorcet Computations”, *Permutation Patterns* (PP 2009), Florence, Italy, 13–17 July 2009 (link)

- Alfonso Gracia-Saz (2015), “Voting Theory”, Canada/USA Mathcamp 2015 (link, link)

- James Green-Armytage (2004), “Cardinal-weighted pairwise comparison”, *Voting Matters*, issue 19, pages 6–13 (link, link, link, link)

- Anja Grünheid, Donald Kossmann, Sukriti Ramesh, Florian Widmer (2012), “Crowdsourcing Entity Resolution: When is A=B?”, Technical Report No. 785, Systems Group, Department of Computer Science, Eidgenössische Technische Hochschule Zürich, Switzerland, DOI: 10.3929/ethz-a-009761323 (link)

- Anja Grünheid, Besmira Nushi, Donald Kossmann (2014), “Cost-Efficient Querying Strategies for the Crowd”, *First International Workshop on Big Uncertain Data* (BUDA 2014) in conjunction with the *2014 International Conference on Management of Data* (SIGMOD/PODS 2014), Snowbird, Utah, USA, 22–27 June 2014 (link, link)

- Anja Grünheid, Besmira Nushi, Donald Kossmann, Wolfgang Gatterbauer, Tim Kraska (2015), “Fault-Tolerant Entity Resolution with the Crowd” (arXiv:1512.00537) (link, link)

- Anja Grünheid (2016), “Data Integration with Dynamic Data Sources”, doctoral dissertation, Eidgenössische Technische Hochschule Zürich, Switzerland, DOI: 10.3929/ethz-a-010861625 (link) {In the papers by Grünheid, the Schulze method is called “MinMax decision function”.}

- Daniel Gruss (2012), “Schulze-Methode” (link, link)

- Thomas Edmund Haines, Dirk Pattinson, Mukesh Tiwari (2019), “Verifiable Homomorphic Tallying for the Schulze Vote Counting Scheme”, *11th International Conference on Verified Software: Theories, Tools, and Experiments* (VSTTE 2019), New York City, USA, 13–14 July 2019; volume 12031 of the series *Lecture Notes in Computer Science*, pages 36–53, Springer, DOI: 10.1007/978-3-030-41600-3_4, DOI: 11250/2726543 (link, link, link, link, slides, slides, poster, poster)

- Sheng Han, Wei Gao, Zhanyi Hu (2023), “Why You Cannot Rank First: Modifications for Benchmarking Six-Degree-of-Freedom Visual Localization Algorithms”, *Sensors*, volume 23, issue 23, paper 9580, DOI: 10.3390/s23239580 (link, link, link, link)

- Steve Hardt, Lia C.R. Lopes (2015), “Google Votes: A Liquid Democracy Experiment on a Corporate Social Network”, 5 June 2015 (link, link, link, video: 00:19:54 – 00:22:28 hh:mm:ss)

- Luke Harrison, Samiran Bag, Hang Luo, Feng Hao (2022), “VERICONDOR: End-to-End Verifiable Condorcet Voting without Tallying Authorities”, *17th ACM Asia Conference on Computer and Communications Security* (ACM ASIA CCS 2022), Nagasaki, Japan, 30 May – 3 June 2022, proceedings, pages 1113–1125, DOI: 10.1145/3488932.3497758 (link, link, link, link, link, link)

- Luke Harrison, Samiran Bag, Hang Luo, Feng Hao (2024a), “VERICONDOR: End-to-End Verifiable Condorcet Voting with support for Strict Preference and Indifference”, *Digital Threats: Research and Practice*, volume 5, issue 4, article 35, DOI: 10.1145/3676267 (link, link)

- Luke Harrison (2024b), “Self-Enforcing E-Voting and its Applications to Ranked Voting”, doctoral dissertation, University of Warwick, United Kingdom (link, link)

- Thomas Hausberger, Thomas François, Patrice Marie-Jeanne, Dominique Guin (2018), “Le vote et la démocratie” (link, link)

- Tanmoy Hazra, C.R.S. Kumar, Manisha J. Nene (2017), “Strategies for Searching Targets Using Mobile Sensors in Defense Scenarios”, *International Journal of Information Technology and Computer Science*, volume 9, number 5, pages 61–70, DOI: 10.5815/ijitcs.2017.05.08 (link, link, link)

- Sophie Heilmaier (2020), “Optimal Voting Rules for Few Candidates”, master thesis, Technical University of Munich, Germany (link, link)

- Jobst Heitzig (2002), “Social Choice Under Incomplete, Cyclic Preferences” (arXiv:math/0201285) (link, link, link, link) {In this paper, the Schulze method is called “strong immunity from binary arguments” ($sIm_A$).}

- Jobst Heitzig (2004a), “Examples with 4 options for immune methods”, *Election-Methods mailing list*, 1 May 2004 (link, link) {In this paper, the Schulze method is called “beatpath”.}

- Jobst Heitzig (2004b), “attempt of a grand compromise”, *Election-Methods mailing list*, 11 October 2004 (link) {In this paper, the Schulze method is called “beatpath”.}

- Lane A. Hemaspaandra, Rahman Lavaee, Curtis Menton (2016), “Schulze and ranked-pairs voting are fixed-parameter tractable to bribe, manipulate, and control”, *12th International Conference on Autonomous Agents and Multiagent Systems* (AAMAS 2013), Saint Paul, Minnesota, USA, 6–10 May 2013, proceedings, pages 1345–1346; *Annals of Mathematics and Artificial Intelligence*, volume 77, issue 3, pages 191–223, 2016, DOI: 10.1007/s10472-015-9479-1 (arXiv:1210.6963) (link, link, link, link, link, link, link, link, poster, poster)

- Fabian Hennecke (2013), “Effekte und Potenziale eines gebogenen interaktiven Displays”, doctoral dissertation, Ludwig Maximilian University of Munich, Germany, DOI: 10.5282/edoc.16782 (link, link, link)

- Fabian Hertel (2020), “Erweiterung des E-Voting-Systems Ordinos um weitere Wahlverfahren”, bachelor thesis, University of Stuttgart, Germany, DOI: 10.18419/opus-11417 (link, link, link)

- Fabian Hertel, Nicolas Huber, Jonas Kittelberger, Ralf Küsters, Julian Liedtke, Daniel Rausch (2021), “Extending the Tally-Hiding Ordinos System: Implementations for Borda, Hare-Niemeyer, Condorcet, and Instant-Runoff Voting”, *Sixth International Joint Conference on Electronic Voting* (E-Vote-ID 2021), Bregenz, Austria, 5–8 October 2021, proceedings, pages 269–284, DOI: 10062/74533 (link, link, link, link, link, link, link, link, link, link)

- Benjamin Mako Hill (2008), “Voting Machinery for the Masses”, *O’Reilly Open Source Convention*, Portland, Oregon, USA, 21–25 July 2008 (link)

- I.D. Hill, Brian A. Wichmann, Douglas R. Woodall (1987), “Algorithm 123 — The Single Transferable Vote by Meek’s Method”, *Computer Journal*, volume 30, issue 3, pages 277–281, DOI: 10.1093/comjnl/30.3.277 (link)

- I.D. Hill (1995), “Trying to find a winning set of candidates”, *Voting Matters*, issue 4, page 3 (link, link, link)

- Clarence G. Hoag, George H. Hallett (1926), “Proportional Representation”, Macmillan (link)

- Lê Nguyên Hoang (2017), “Strategy-proofness of the randomized Condorcet voting system”, *Social Choice and Welfare*, volume 48, issue 3, pages 679–701, DOI: 10.1007/s00355-017-1031-2 (link, link, link, link, link, link, link)

- Wesley H. Holliday, Eric Pacuit (2021a), “Axioms for defeat in democratic elections”, *Journal of Theoretical Politics*, volume 33, issue 4, pages 475–524, DOI: 10.1177/09516298211043236 (arXiv:2008.08451) (link, link, link, link, link)

- Wesley H. Holliday, Eric Pacuit (2021b), “Measuring Violations of Positive Involvement in Voting”, *Theoretical Aspects of Rationality and Knowledge* (TARK 2021), Beijing, China, 25–27 June 2021; *Electronic Proceedings in Theoretical Computer Science*, volume 335, pages 189–209, DOI: 10.4204/EPTCS.335.17 (arXiv:2106.11502) (link, link, link, link, link, video)

- Wesley H. Holliday, Chase Norman, Eric Pacuit (2021c), “Voting Theory in the Lean Theorem Prover”, *Eighth International Conference on Logic, Rationality and Interaction* (LORI 2017), Xi’an, China, 16–18 October 2021; eds. Sujata Ghosh, Thomas Icard, *Logic, Rationality and Interaction*, volume 13039 of the series *Lecture Notes in Computer Science*, pages 111–127, Springer, DOI: 10.1007/978-3-030-88708-7_9 (arXiv:2110.08453) (link, link, link, link)

- Wesley H. Holliday, Eric Pacuit (2023a), "Split Cycle: a new Condorcet-consistent voting method independent of clones and immune to spoilers", *Public Choice*, volume 197, issues 1–2, pages 1–62, DOI: 10.1007/s11127-023-01042-3 (arXiv:2004.02350) (link, link, link, link, link, link)

- Wesley H. Holliday, Eric Pacuit (2023b), "Stable Voting", *Constitutional Political Economy*, volume 34, issue 3, pages 421–433, DOI: 10.1007/s10602-022-09383-9 (arXiv:2108.00542) (link, link, link, link)

- Wesley H. Holliday, Mikayla Kelley (2025), "Escaping Arrow's theorem: the Advantage-Standard model", *Theory and Decision*, volume 98, issue 2, pages 165–204, DOI: 10.1007/s11238-024-10004-0 (arXiv:2108.01134) (link, link, link, link, link, link) {In the papers by Holliday, the Schulze method is called "Beat Path".}

- Steffen Holtkamp (2018), "Schulze-Methode: Algorithmus zum Finden eines eindeutigen Siegers", Westfälische Hochschule, Germany (link, link)

- Hadi Hosseini, Debmalya Mandal, Amrit Puhan (2024), "The Surprising Effectiveness of SP Voting with Partial Preferences", *38th International Conference on Neural Information Processing Systems* (NIPS '24), proceedings, article 92, pages 2787–2829 (arXiv:2406.00870) (link, link, link, link, link, link, link, link)

- Erkan Işıklı, Nezir Aydın, Erkan Çelik, Alev Taşkın Gümüş (2017), "Identifying Key Factors of Rail Transit Service Quality: An Empirical Analysis for Istanbul", *Journal of Public Transportation*, volume 20, number 1, pages 63–90, DOI: 10.5038/2375-0901.20.1.4 (link, link, link, link, link, link)

- Erkan Işıklı, Şeyda Serdar Asan (2023), "The Power of Combination Models in Energy Demand Forecasting", *Decision Making Using AI in Energy and Sustainability*, eds. Gülgün Kayakutlu, Mehmet Özgür Kayalıca, pages 153–167, Springer, DOI: 10.1007/978-3-031-38387-8_9 (link)

- Javier Izetta, Pablo F. Verdes, Pablo M. Granitto (2017), "Improved multiclass feature selection via list combination", *Expert Systems with Applications*, volume 88, pages 205–216, DOI: 10.1016/j.eswa.2017.06.043, DOI: 11336/50349 (link, link, link)

- Nemanja D. Jelić (2022), "Algoritmi za Rešavanje Grafovskog Problema Najšireg Puta", master thesis, University of Belgrade, Serbia (link, link, link)

- Andrew Jennings (2010), "Monotonicity and Manipulability of Ordinal and Cardinal Social Choice Functions", doctoral dissertation, Arizona State University, USA (link, link, link, link)

- Junpeng Jing, Jiankun Li, Pengfei Xiong, Jiangyu Liu, Shuaicheng Liu, Yichen Guo, Xin Deng, Mai Xu, Lai Jiang, Leonid Sigal (2023), “Uncertainty Guided Adaptive Warping for Robust and Efficient Stereo Matching”, *IEEE/CVF International Conference on Computer Vision* (ICCV 2023), Paris, France, 1–6 October 2023; proceedings, pages 3318–3327, DOI: 10.1109/ICCV51070.2023.00307 (arXiv:2307.14071) (link, link, link, supplementary material (mirror))

- Inwon Kang, Qishen Han, Lirong Xia (2023), “Learning to Explain Voting Rules”, *2023 22nd International Conference on Autonomous Agents and Multiagent Systems* (AAMAS ‘23), London, United Kingdom, 29 May – 2 June 2023, proceedings, pages 2883–2885 (link, link, link, link, link, link)

- Hooman Katirai (2006), “A Theory and Toolkit for the Mathematics of Privacy: Methods for Anonymizing Data while Minimizing Information Loss”, master thesis, Massachusetts Institute of Technology, Cambridge, Massachusetts, USA (link, link, link)

- David Kempe (2020), “An Analysis Framework for Metric Voting based on LP Duality”, *34th AAAI Conference on Artificial Intelligence* (AAAI-20), New York City, USA, 7–12 February 2020; proceedings, volume 34, number 2, pages 2079–2086, DOI: 10.1609/aaai.v34i02.5581 (arXiv:1911.07162) (link, link, link, link)

- Дмитрий Михайлович Худолей / Dmitry Mikhailovich Khudoley (2017), “Парадоксы Кондорсе и их Решение” / “The Condorcet Paradoxes and their Solution”, *Вестник Пермского Университета, Юридические Науки / Bulletin of the Perm State University, Juridical Sciences*, volume 37, pages 288–302, DOI: 10.17072/1995-4190-2017-37-288-302 (link, link, link, link, link)

- Дмитрий Михайлович Худолей / Dmitry Mikhailovich Khudoley, Константин Михайлович Худолей / Konstantin Mikhailovich Khudoley (2021), “Предложение по Реформированию Избирательной Системы Российской Федерации” / “A Proposal on Reformation of the Election System of the Russian Federation”, *Вестник Пермского Университета, Юридические Науки / Bulletin of the Perm State University, Juridical Sciences*, volume 53, pages 562–601, DOI: 10.17072/1995-4190-2021-53-562-601 (link, link, link, link, link, link)

- Ryul Kim, Jonghan Yu, Joonoh Lim, Brian Baek-Lok Oh, Seok Jin Nam, Seok Won Kim, Jeong Eon Lee, Byung Joo Chae, Ji-Yeon Kim, Ga Eun Park, Bong Joo Kang, Pill Sun Paik, Soo Yeon Bae, Chang Ik Yoon, Young Joo Lee, Dooreh Kim, Kabsoo Shin, Ji Eun Lee, Jun Kang, Ahwon Lee, Erin Connolly-Strong, Sangmoon Lee, Bo Rahm Lee, Yuna Lee, Ki Jong Yi, Young Oh Kwon, In Hwan Chun, Junggil Park, Jihye Kim, Chahyun Choi, Jong Yeon Shin, Hyungjung Lee, Minji Kim, Hansol Park, Ilecheon Jeong, Boram Yi, Won-Chul Lee, Jeong Seok Lee, Woo Chan Park, Sung Hun Kim, Yoon-La Choi, Jeongmin Lee, Young Seok Ju, Yeon Hee Park (2026), “Whole-genome landscapes of 1,364 breast cancers”, *nature*, volume 649, pages 1282–1291, DOI: 10.1038/s41586-025-09812-3 (link, link, link)

- Ben Kinnersley, Amit Sud, Andrew Everall, Alex J. Cornish, Daniel Chubb, Richard Culliford, Andreas J. Gruber, Adrian Lärkeryd, Costas Mitsopoulos, David C. Wedge, Richard S. Houlston (2024), “Analysis of 10,478 cancer genomes identifies candidate driver genes and opportunities for precision oncology”, *nature genetics*, volume 56, pages 1868–1877, DOI: 10.1038/s41588-024-01785-9 {In this paper, the Schulze method is mentioned only in the supplementary material (link, link).}

- Gert Köhler (1978), “Choix multicritère et analyse algébrique de données ordinales”, doctoral dissertation, Joseph Fourier University, Grenoble, France (link, link)

- Klaus Kopfermann (1991), “Mathematische Aspekte der Wahlverfahren: Mandatsverteilung bei Abstimmungen”, BI-Wissenschaftsverlag, Mannheim

- Ярослав Олександрович Костенко / Iarislav Oleksandrovich Kostenko, Микола Володимир-Станіславович Сидоров / Mykola Volodymyr-Stanislavovych Sydorov (2017), “Аналіз Рангових Шкал у Масових Опитуваннях” / “Analysis of Rank Scales in Mass Surveys”, *Вісник Харківського Національного Університету Імені В.Н. Каразіна / Bulletin of the V.N. Karazin Kharkiv National University*, *Соціологічні Дослідження Сучасного Суспільства: Методологія, Теорія, Методи / Sociological Studies of Contemporary Society: Methodology, Theory, Methods*, volume 39, pages 110–117 (link, link)

- Robert Kowalski, Sebastian Loehmann, Doris Hausen (2013), “cubble: A Multi-Device Hybrid Approach Supporting Communication in Long-Distance Relationships”, *Seventh International Conference on Tangible, Embedded and Embodied Interaction* (TEI 2013), Barcelona, Spain, 10–13 February 2013, proceedings, pages 201–204, DOI: 10.1145/2460625.2460656 (link, link, link, link, link, link)

- Gerald H. Kramer (1977), “A Dynamical Model of Political Equilibrium”, *Journal of Economic Theory*, volume 16, issue 2, pages 310–334, DOI: 10.1016/0022-0531(77)90011-4 (link)

- Benjamin Krenn (2019), “Algorithms for Implicit Delegation to Predict Preferences”, master thesis, Vienna University of Technology, Austria, DOI: 20.500.12708/3256 (link, link)

- Anilesh Kollagunta Krishnaswamy (2019), “Scaling up Collective Decision Making”, doctoral dissertation, Stanford University, California, USA (link, link)

- Sascha Kurz, Alexander Mayer, Stefan Napel (2019), “Influence in Weighted Committees” (arXiv:1912.10208) (link, link)

- Eerik Lagerspetz (2015), “Social Choice and Democratic Values”, Springer, DOI: 10.1007/978-3-319-23261-4 (link, link)

- John Lambert, Zhuang Liu, Ozan Sener, James Hays, Vladlen Koltun (2023), "MSeg: A Composite Dataset for Multi-Domain Semantic Segmentation", *2020 IEEE/CVF Conference on Computer Vision and Pattern Recognition* (CVPR), Seattle, Washington, USA, 13–19 June 2020, DOI: 10.1109/CVPR42600.2020.00295; *IEEE Transactions on Pattern Analysis and Machine Intelligence*, volume 45, issue 1, pages 796–810, DOI: 10.1109/TPAMI.2022.3151200 (arXiv:2112.13762) (link, link, link, link)

- Marc Lanctot, Kate Larson, Yoram Bachrach, Luke Marris, Zun Li, Avishkar Bhoopchand, Thomas Anthony, Brian Tanner, Anna Koop (2023), "Evaluating Agents using Social Choice Theory" (arXiv:2312.03121) (link, link, link, video)

- Jean-François Laslier (2019), "Voter Autrement" (link, link)

- Kai Lawonn, Alexandra Baer, Patrick Saalfeld, Bernhard Preim (2014), "Comparative Evaluation of Feature Line Techniques for Shape Depiction", eds. Jan Bender, Arjan Kuijper, Tatiana von Landesberger, Holger Theisel, Philipp Urban, *19th International Workshop on Vision, Modeling and Visualization* (VMV 2014), Darmstadt, Germany, 8–10 October 2014, proceedings, pages 31–38, DOI: 10.2312/vmv.20141273 (link, link, link, link, link)

- Christopher N. Lawrence (2013), "Group Decision-Making in Open Source Development: Putting Condorcet's Method in Practice", *84th Annual Conference of the Southern Political Science Association*, Orlando, Florida, USA, 2–5 January 2013 (link, link, link, link, link)

- Clément Lesaege, William George, Federico Ast (2021), "Kleros" (link, link)

- Jonathan Levin, Barry Nalebuff (1995), "An Introduction to Vote-Counting Schemes", *Journal of Economic Perspectives*, volume 9, number 1, pages 3–26, DOI: 10.1257/jep.9.1.3 (link, link, link)

- Minyi Li, Quoc Bao Vo, Ryszard Kowalczyk (2014), "A distributed social choice protocol for combinatorial domains", *Journal of Heuristics*, volume 20, issue 4, pages 453–481, DOI: 10.1007/s10732-014-9246-1 (link)

- Julian Liedtke (2023), "Verifiable Tally-Hiding Remote Electronic Voting", doctoral dissertation, University of Stuttgart, Germany, DOI: 10.18419/opus-14080 (link, link)

- Andrei Lihu, Jincheng Du, Igor Barjaktarević, Patrick Gerzanics, Mark Harvilla (2020), "A Proof of Useful Work for Artificial Intelligence on the Blockchain" (arXiv:2001.09244) (link, link, link)

- Andreas Lommatzsch (2009), "Eine offene Architektur für die agentenbasierte, adaptive, personalisierte Informationsfilterung", doctoral dissertation, Berlin University of Technology, Germany, DOI: 10.14279/depositonce-2162 (link, link, link)

- Mohammad Lotfi, Behzad Tokhmechi (2019), "Paradigm Shift in Studying Joint Micro-Roughness Coefficients using Graph Theory", *International Journal of Mining and Geo-Engineering*, volume 53, issue 2, pages 165–174, DOI: 10.22059/ijmge.2019.267783.594763 (link, link, link, link, link) {In this paper, the Schulze method is called "clone-proof Schwartz sequential dropping".}

- Panos Louridas (2017), "Real-World Algorithms: A Beginner's Guide", MIT Press (link, link, link)

- Joseph Low, Oscar Duys, Claude Formanek, Michiel A. Bakker, Lewis Hammond (2026), "HABERMOLT: Delegating Deliberation to AI Representatives", *Trustworthy AI for Good Workshop* (AI4GOOD) (arXiv:2605.24413) (link, link)

- A. Uma Maheswari, P. Kumari (2012), "A Fuzzy Mathematical Model for Multi Criteria Group Decision Making — An Application in Supply Chain Management", *International Journal of Computer Applications*, volume 54, number 7, pages 5–10, DOI: 10.5120/8576-2314 (link, link, link, link) {In this paper, the Schulze method is called "beat path method" and "Schwartz sequential dropping".}

- A. Uma Maheswari, P. Kumari (2013), "A Performance Appraisal Model using Fuzzy Multi Criteria Group Decision Making", *International Journal of Scientific & Engineering Research*, volume 4, issue 5, pages 442–449 (link) {In this paper, the Schulze method is called "Schwartz beat path method" and "Schwartz sequential dropping".}

- Hardikkumar Harishbhai Maheta, Chauhan Pareshbhai Mansangbhai, Chintan Makwana (2024), "An Improved Feature Removal Approach for Classification of High Dimensional Feature Dataset", *Journal of Electrical Systems*, volume 20, number 3, pages 1567–1576, DOI: 10.52783/jes.3653 (link, link, link, link)

- Saeed Mahmoudpour, Peter Schelkens (2021), "On the performance of objective quality metrics for lightfields", *Signal Processing: Image Communication*, volume 93, article 116179, DOI: 10.1016/j.image.2021.116179 (link)

- Francesco Di Maio, Alessandro Bandini, Enrico Zio, Sofia Carlos Alberola, Francisco Sanchez-Saez, Sebastián Martorell (2016), "Bootstrapped-ensemble-based Sensitivity Analysis of a trace thermal-hydraulic model based on a limited number of PWR large break loca simulations", *Reliability Engineering & System Safety*, volume 153, pages 122–134, DOI: 10.1016/j.ress.2016.04.013, DOI: 11311/1020669, DOI: 10251/112944 (link, link, link, link)

- Amirabbas Majd, Mojtaba Vahidi-Asl, Alireza Khalilian, Babak Bagheri (2022), "ConsilientSFL: using preferential voting system to generate combinatorial ranking metrics for spectrum-based fault localization", *Applied Intelligence*, volume 52, issue 10, pages 11068–11088, DOI: 10.1007/s10489-021-02954-7 (link)

- Eduardo Freitas Mangeli de Brito (2016), "An Aesthetic Metric for Multiplayer Turn-based Games", doctoral dissertation, Federal University of Rio de Janeiro, Brazil (link, link)

- Shiwen Mao, Tommaso Melodia, Prasun Sinha, Danda B. Rawat (2018), "Message from TPC Chairs", *IEEE INFOCOM 2018 — IEEE Conference on Computer Communications*, 16–19 April 2018, Honolulu, Hawaii, USA, DOI: 10.1109/INFOCOM.2018.8486347 (link)

- Victor Marian, Knud Jahnke, Mira Mechtley, Seth Cohen, Bernd Husemann, Victoria Jones, Anton M. Koekemoer, Andreas Schulze, Arjen van der Wel, Carolin Villforth, Rogier A. Windhorst (2019), "Major mergers are not the dominant trigger for high-accretion AGNs at $z \sim 2$", *Astrophysical Journal*, volume 882, number 2, article 141, DOI: 10.3847/1538-4357/ab385b (arXiv:1904.00037) (link)

- Victor Marian, Knud Jahnke, Irham Andika, Eduardo Bañados, Vardha N. Bennert, Seth Cohen, Bernd Husemann, Melanie Kaasinen, Anton M. Koekemoer, Mira Mechtley, Masafusa Onoue, Jan-Torge Schindler, Malte Schramm, Andreas Schulze, John D. Silverman, Irina Smirnova-Pinchukova, Arjen van der Wel, Carolin Villforth, Rogier A. Windhorst (2020), "A significant excess in major merger rate for AGNs with the highest Eddington ratios at $z < 0.2$", *Astrophysical Journal*, volume 904, number 1, article 79, DOI: 10.3847/1538-4357/abbd3e (arXiv:2010.00022) (link, link, link)

- Victor Marian (2021), "The Intricate Connection Between Major Mergers and AGN with the Highest Eddington Ratios", doctoral dissertation, Heidelberg, Germany, DOI: 10.11588/heidok.00029888 (link, link, link)

- Francisco Martínez-Jiménez, Ferran Muiños, Inés Sentís, Jordi Deu-Pons, Iker Reyes-Salazar, Claudia Arnedo-Pac, Loris Mularoni, Oriol Pich, José Bonet, Hanna Kranas, Abel Gonzalez-Perez, Núria López-Bigas (2020), "A compendium of mutational cancer driver genes", *Nature Reviews Cancer*, volume 20, issue 10, pages 555–572, DOI: 10.1038/s41568-020-0290-x, DOI: 2445/191783 (link, link) {In this paper, the Schulze method is mentioned only in the supplementary material (link).}

- Anurita Mathur, Arnab Bhattacharyya (2017), "On the Gap between Outcomes of Voting Rules", *16th International Conference on Autonomous Agents and Multiagent Systems* (AAMAS 2017), São Paulo, Brazil, 8–12 May 2017, proceedings, pages 1631–1633 (link, link, link)

- Leonardo Matone, Ben Abramowitz, Ben Armstrong, Avinash Balakrishnan, Nicholas Mattei (2024), "DeepVoting: Learning Voting Rules with Tailored Embeddings" (arXiv:2408.13630) (link, link)

- Nicholas Mattei, Nina Narodytska, Toby Walsh (2014), "How Hard Is It to Control an Election by Breaking Ties?", *21st biennial European Conference on Artificial Intelligence* (ECAI 2014), Prague, Czech Republic, 18–22 August 2014; *Frontiers in Artificial Intelligence and Applications*, volume 263, pages 1067–1068, DOI: 10.3233/978-1-61499-419-0-1067 (arXiv:1304.6174) (link, link, link, link)

- Cynthia Maushagen, Jörg Rothe (2020), "The Last Voting Rule Is Home: Complexity of Control by Partition of Candidates or Voters in Maximin Elections", *24th European Conference on Artificial Intelligence* (ECAI 2020), Santiago de Compostela, Spain, 29 August – 5 September 2020; *Frontiers in Artificial Intelligence and Applications*, volume 325, pages 163–170, DOI: 10.3233/FAIA200089 (link, link, link, video)

- Cynthia Maushagen (2021), "Wahlbetrug mit festen und variablen Präferenzen: Eine Komplexitätsanalyse von Kontroll- und Bestechungsproblemen", doctoral dissertation, Heinrich Heine University Düsseldorf, Germany (link, link, link, link)

- Cynthia Maushagen, David Niclaus, Paul Nüsken, Jörg Rothe, Tessa Seeger (2024), "Toward Completing the Picture of Control in Schulze and Ranked Pairs Elections", *International Symposium on Artificial Intelligence and Mathematics* (ISAIM 2024), Fort Lauderdale, Florida, USA, 8–10 January 2024; *33rd International Joint Conference on Artificial Intelligence* (IJCAI 2024), Jeju, South Korea, 3–9 August 2024, proceedings, pages 2940–2948, DOI: 10.24963/ijcai.2024/326 (arXix:2405.08956) (link, link, link, link, link, link)

- Alexander Mayer (2018), "Essays on Voting Power", doctoral dissertation, University of Bayreuth, Germany (link, link, link)

- James D. McCaffrey (2008a), "Collaborative Techniques for the Determination of a Best Alternative in a Software Quality Environment", *Twenty-Sixth Annual Pacific Northwest Software Quality Conference* (PNSQC 2008), Portland, Oregon, USA, 13–15 October 2008, proceedings, pages 281–287 (link)

- James D. McCaffrey (2008b), "Test Run: Group Determination in Software Testing", *MSDN Magazine*, volume 23, number 12, pages 91–97, November 2008 (link, link, link, link, link)

- David C. McGarvey (1953), "A Theorem on the Construction of Voting Paradoxes", *Econometrica*, volume 21, number 4, pages 608–610, DOI: 10.2307/1907926 (link, link)

- Aonghus McGovern (2019), "Investigating the application of structured representations of unstructured content in personalisation tasks", doctoral dissertation, University of Dublin, Ireland, DOI: 2262/86069 (link, link)

- Ewan McInnes (2019), "Theoretical Underpinnings of a Pure Digital Democracy Model of Government: A Hypothetically Superior Way of Structuring the Legislative and Executive Branches of Government" (link)

- Aaron McKenna, Jay Shendure (2018), "FlashFry: a fast and flexible tool for large-scale CRISPR target design", *BMC Biology*, volume 16, article 74, DOI: 10.1186/s12915-018-0545-0 (link, link, link, link, link, link)

- Brian Lawrence Meek (1969), “Une nouvelle approche du scrutin transférable”, *Mathématiques et Sciences Humaines*, volume 25, pages 13–23 (link, link) [Reprinted as: “A New Approach to the Single Transferable Vote — Part I”, *Voting Matters*, issue 1, pages 1–7, 1994 (link, link, link)]

- Brian Lawrence Meek (1970), “Une nouvelle approche du scrutin transférable (fin)”, *Mathématiques et Sciences Humaines*, volume 29, pages 33–39 (link, link) [Reprinted as: “A New Approach to the Single Transferable Vote — Part II”, *Voting Matters*, issue 1, pages 7–11, 1994 (link, link, link)]

- Iaroslav Melekhov, Gabriel J. Brostow, Juho Kannala, Daniyar Turmukhambetov (2020), “Image Stylization for Robust Features” (arXiv:2008.06959) (link)

- Curtis Menton (2013a), “Attacking and Defending Popular Election Systems”, doctoral dissertation, University of Rochester, New York, USA (link, link)

- Curtis Menton, Preetjot Singh (2013b), “Control Complexity of Schulze Voting”, *23rd International Joint Conference on Artificial Intelligence* (IJCAI 2013), Beijing, China, 3–9 August 2013, proceedings, pages 286–292 (arXiv:1206.2111) (link, link, link, poster, poster)

- Andrés Aguado Merino (2025), “Matemática del Voto”, doctoral dissertation, Technical University of Madrid, Spain (link, link)

- Katharine Merow (2016), “Three Mathematicians Walk into a Carriage House ... / Moon Duchin: Math and the Vote: A Geometer Examines Elections”, *MAA Focus*, volume 36, number 6, pages 28–29, December 2016/January 2017 (link, link, link, link) {In this paper, the Schulze method is called “beatpath”.}

- Tommi Meskanen, Hannu Nurmi (2006a), “Analyzing Political Disagreement”, *20th IPSA World Congress*, Fukuoka, Japan, 9–13 July 2006 (link, link, link)

- Tommi Meskanen, Hannu Nurmi (2006b), “Distance from Consensus: A Theme and Variations”, eds. Bruno Simeone, Friedrich Pukelsheim, *Mathematics and Democracy: Recent Advances in Voting Systems and Collective Choice*, pages 117–132, Springer, DOI: 10.1007/3-540-35605-3_9 (link, link, link, link, link)

- Tommi Meskanen, Hannu Nurmi (2008), “Closeness Counts in Social Choice”, eds. Matthew Braham, Frank Steffen, *Power, Freedom, and Voting*, pages 289–306, Springer, DOI: 10.1007/978-3-540-73382-9_15 (link)

- Sebastian Meyer, Ben Kinnersley, Katarzyna Kedzierska, Ignacio Soriano, Eszter Lakatos, Claudia Arnedo-Pac, Richard Culliford, Avraam Tapinos, Laura Knight, Yannick Comoglio, Anna Frangou, Alex J. Cornish, Aliah Hawari, Daniel Chubb, Amit Sud, Boris Noyvert, Steve Thorn, Helen White, Alona Sosinsky, Ahmed Ahmed, James D. Brenton, Nuria Lopez-Bigas, Andrea Sottoriva, Tjalling Bosse, Emma J. Davidson, Richard Edmondson, Trevor A. Graham, Ian P.M. Tomlinson, Richard S. Houlston, Andreas J. Gruber, David C. Wedge, David N. Church (2026), “Whole genome sequencing of endometrial cancer identifies novel subgroups, drivers, and actionable alterations”, DOI: 10.64898/2026.06.15.730391 (link, link)

- Charlie Mills, Amit Sud, Andrew Everall, Daniel Chubb, Samuel E.D. Lawrence, Ben Kinnersley, Alex J. Cornish, Robert Bentham, Richard S. Houlston (2024), “Genetic landscape of interval and screen detected breast cancer”, *npj Precision Oncology*, volume 8, article 122, DOI: 10.1038/s41698-024-00618-6 {In this paper, the Schulze method is mentioned only in the supplementary material (link).}

- Robin Milosz (2018), “Étude algorithmique et combinatoire de la méthode de Kemeny-Young et du consensus de classements”, doctoral dissertation, University of Montréal, Canada, DOI: 1866/21741 (link, link)

- Iain H. Moal, Didier Barradas-Bautista, Brian Jiménez-García, Mieczyslaw Torchala, Arjan van der Velde, Thom Vreven, Zhiping Weng, Paul A. Bates, Juan Fernández-Recio (2017), “IRaPPA: information retrieval based integration of biophysical models for protein assembly selection”, *Bioinformatics*, volume 33, issue 12, pages 1806–1813, DOI: 10.1093/bioinformatics/btx068, DOI: 2117/106834 (link, link, link, link, link, link, link, link, link, link)

- Rohit Mohan, Abhinav Valada (2020), “Robust Vision Challenge 2020 — 1st Place Report for Panoptic Segmentation” (arXiv:2008.10112) (link, link)

- Xavier Mora (2010a), “Votar: no tan fàcil com sembla, però podríem fer-ho millor!”, *Materials mathemàtics*, volume 2010, pages 1–35 (link, link, link, link)

- Xavier Mora (2010b), “Votar: No tan fácil como parece, ¡pero podríamos hacerlo mejor!”, *La Gaceta de la RSME*, volume 13, number 3, pages 471–498 (link, link)

- José F. Morales, Manuel Carro, Manuel Hermenegildo (2008), “Comparing Tag Scheme Variations Using an Abstract Machine Generator”, *10th International ACM SIGPLAN Conference on Principles and Practice of Declarative Programming* (PPDP 2008), Valencia, Spain, 15–17 July 2008, proceedings, pages 32–43, DOI: 10.1145/1389449.1389455 (link, link, link, link, link, link)

- José F. Morales (2010), “Advanced Compilation Techniques for Logic Programming”, doctoral dissertation, University of Madrid, Spain, DOI: 10.20868/UPM.thesis.7778 (link, link, link, link)

- Jairo Alexander Moreno Calderón (2021), “Modelo para la evaluación multicriterio de tecnologías en salud”, doctoral dissertation, National University of Columbia, Bogotá, Colombia (link, link)

- John Mori (2020), “Asymmetric Critical Level Theories” (link, link)

- Nathan Moroney, Ingeborg Tastl, Melanie Gottwals, Michael Ludwig, Gary W. Meyer (2018), “Single Anchor Sorting of Visual Appearance as an Oriented Graph”, *26th Color and Imaging Conference: Color Science and Engineering Systems, Technologies, and Applications* (CIC 2018), Vancouver, British Columbia, Canada, 12–16 November 2018, proceedings, pages 365–370, Society for Imaging Science and Technology, DOI: 10.2352/ISSN.2169-2629.2018.26.365 (link, link)

- Lyria Bennett Moses, Rajeev Goré, Ron Levy, Dirk Pattinson, Mukesh Tiwari (2017), “No More Excuses: Automated Synthesis of Practical and Verifiable Vote-Counting Programs for Complex Voting Schemes”, *Second International Joint Conference on Electronic Voting* (E-Vote-ID 2017), Bregenz, Austria, 24–27 October 2017, proceedings, pages 256–273; eds. Robert Krimmer, Melanie Volkamer, Nadja Braun Binder, Norbert Kersting, Olivier Pereira, Carsten Schürmann, *Electronic Voting*, volume 10615 of the series *Lecture Notes in Computer Science*, pages 66–83, Springer, DOI: 10.1007/978-3-319-68687-5_5 (link, link, link, link, link, link, link, slides)

- Hervé Moulin (1988), “Condorcet’s principle implies the no show paradox”, *Journal of Economic Theory*, volume 45, number 1, pages 53–64, DOI: 10.1016/0022-0531(88)90253-0

- Ferran Muiños, Francisco Martínez-Jiménez (2018), “Voting-based Ranking Combination using Python”, *PyCon Nove*, PyCon Italia 2018, Florence, Italy, 19–22 April 2018 (video)

- Conor Muldoon, Gregory M.P. O’Hare (2012), “Social Choice in Sensor Networks”, *ECAI 2012 Workshop on Artificial Intelligence in Telecommunications and Sensor Networks* (WAITS 2012), Montpellier, France, 28 August 2012, DOI: 10197/4407 (link, link, link, link, link, link)

- David Müller (2014), “Investitionscontrolling”, Springer Gabler, DOI: 10.1007/978-3-642-41990-4 (link)

- David Müller (2015), “Abstimmungen gerechter gestalten”, *Controlling & Management Review*, volume 59, issue 5, pages 50–56, DOI: 10.1007/s12176-015-0609-8 (link)

- David Müller (2019), “Investitionsrechnung und Investitionscontrolling”, Springer Gabler, DOI: 10.1007/978-3-662-57609-0 (link)

- David Müller (2022), “Investitionscontrolling: Entscheidungsfindung bei Investitionen II”, Springer Gabler, DOI: 10.1007/978-3-658-36597-4 (link)

- Julian Müller (2013), “The Complexity of Manipulating Schulze Voting”, bachelor thesis, University of Constance, Germany (link)

- Julian Müller, Sven Kosub (2020), “A note on the complexity of manipulating weighted Schulze voting”, *Information Processing Letters*, volume 162, article 105989, DOI: 10.1016/j.ipl.2020.105989 (arXiv:1808.09226) (link, link)

- Kamesh Munagala, Kangning Wang (2019), “Improved Metric Distortion for Deterministic Social Choice Rules”, *20th ACM Conference on Economics and Computation* (EC ‘19), Phoenix, Arizona, USA, 24–28 June 2019, proceedings, pages 245–262, DOI: 10.1145/3328526.3329550 (arXiv:1905.01401) (link, link, link)

- Charles T. Munger Jr. (2023a), “The right way to read Ranked-choice ballots: not Instant Runoff, but Ranked Pairs” (part A (mirror), part B (mirror), part C (mirror), part D (mirror)) {The Schulze method is mainly discussed in part C.}

- Charles T. Munger Jr. (2023b), “‘Spoilers’ spoil nothing” (link, link)

- Charles T. Munger Jr. (2023c), “The best Condorcet-compatible election method: Ranked Pairs”, *Constitutional Political Economy*, volume 34, issue 3, pages 434–444, DOI: 10.1007/s10602-022-09382-w (link, link, link, link, link, link, link)

- Charles T. Munger Jr. (2024a), “Correction to: The best Condorcet-compatible election method: Ranked Pairs”, *Constitutional Political Economy*, volume 35, issue 3, pages 457–459, DOI: 10.1007/s10602-024-09431-6 (link, link, link, link, link)

- Charles T. Munger Jr. (2024b), “Dropping one candidate under Beatpath and Ranked Pairs” (link, link)

- Charles T. Munger Jr. (2024c), “Tiebreaking in Beatpath, Split Cycle, and Ranked Pairs that is free of influence by clones” (link, link)

- Charles T. Munger Jr. (2025), “Tactical voting in three-candidate polar elections under Instant Runoff and under Condorcet methods” (link, link) {In the papers by Munger, the Schulze method is called “Beatpath”.}

- Suranga Chandima Nanayakkara (2009), “Enhancing Musical Experience for the Hearing-impaired using Visual and Haptic Displays”, doctoral dissertation, National University of Singapore (link)

- Massimo Narizzano, Luca Pulina, Armando Tacchella (2006a), “Scoring Methods for the Evaluation of Automated Reasoning Systems”, *Analisi sperimentale e benchmark di algoritmi per l’intelligenza artificiale*, Workshop organized by *Gruppo AI*IA sulla Rappresentazione della Conoscenza e Ragionamento Automatico*, 23 June 2006, Udine, Italy (link, link)

- Massimo Narizzano, Luca Pulina, Armando Tacchella (2006b), “Competitive evaluation of automated reasoning tools: Empirical scoring and statistical testing”, *Workshop on Empirical Methods for the Analysis of Algorithms* (EMAA 2006), Reykjavik, Iceland, 9 September 2006, proceedings, pages 21–26 (link, link, link, link)

- Massimo Narizzano, Luca Pulina, Armando Tacchella (2006c), “Voting Systems and Automated Reasoning: the QBFEVAL Case Study”, *1st International Workshop on Computational Social Choice* (COMSOC 2006), Amsterdam, Netherlands, 6–8 December 2006, proceedings, pages 366–379 (link, link, link, link, link, link, link, link, link)

- Massimo Narizzano, Luca Pulina, Armando Tacchella (2006d), “Competitive evaluation of QBF solvers: noisy data and scoring methods” (link)

- Massimo Narizzano, Luca Pulina, Armando Tacchella (2007), “Ranking and Reputation Systems in the QBF Competition”, *10th Congress of the Italian Association for Artificial Intelligence: Artificial Intelligence and Human-Oriented Computing* (AI*IA 2007), Rome, Italy, 10–13 September 2007; eds. Roberto Basili, Maria Teresa Pazienza, volume 4733 of the series *Lecture Notes in Computer Science*, pages 97–108, Springer, DOI: 10.1007/978-3-540-74782-6_10 (link, link)

- Anghel Negriu, Cyrille Piatecki (2012), “On the performance of voting systems in spatial voting simulations”, *2010 Workshop of the Society for Economic Science with Heterogeneous Interacting Agents* (ESHIA/WEHIA 2010), Alessandria, Italy, 23–25 June 2010; *Journal of Economic Interaction and Coordination*, volume 7, number 1, pages 63–77, DOI: 10.1007/s11403-011-0082-1 (link)

- Robert A. Newland, Frank S. Britton (1997), “How to conduct an election by the Single Transferable Vote”, Electoral Reform Society of Great Britain and Ireland (ERS), 3rd Edition

- Ruffin-Benoît M. Ngoie, Selain K. Kasereka, Jean-Aimé B. Sakulu, Kyandoghere Kyamakya (2024), “Mean-Median Compromise Method: A Novel Deepest Voting Function Balancing Range Voting and Majority Judgment”, *Mathematics*, volume 12, issue 22, article 3631, DOI: 10.3390/math12223631 (link, link, link)

- Phuong-Mai Nguyen, Cédric Lyathaud, Olivier Vitrac (2015), “A Two-Scale Pursuit Method for the Tailored Identification and Quantification of Unknown Polymer Additives and Contaminants by $^{1}$H NMR”, *Industrial & Engineering Chemistry Research*, volume 54, issue 10, pages 2667–2681, DOI: 10.1021/ie503592z (link)

- Hannu J. Nurmi (1987), “Comparing Voting Systems”, Springer, DOI: 10.1007/978-94-009-3985-1

- Victor Oei, Jenny Schmalfuss, Lukas Mehl, Madlen Bartsch, Shashank Agnihotri, Margret Keuper, Andreas Bulling, Andrés Bruhn (2026), “RobustSpring: Benchmarking Robustness to Image Corruptions for Optical Flow, Scene Flow and Stereo”, *14th International Conference on Learning Representations* (ICLR 2026), Rio de Janeiro, Brazil, 23–27 April 2026 (arXiv:2505.09368) (link, link, link, link, link, link, link)

- Anders Ohrn (2017), “Electoral Reform Models” (link)

- Yasunori Okumura (2026), “Respecting linear orders for supermajority rules”, *Social Choice and Welfare*, DOI: 10.1007/s00355-026-01666-6 (arXiv:2304.09419) (link, link, link)

- Zuzanna Opała, Mateusz Sieniawski (2020), “Schulze method” (link)

- Grzegorz Oryńczak, Zbigniew Kotulski (2012), “Notary-based self-healing mechanism for centralized peer-to-peer infrastructures”, *Annales UMCS*, *Informatica*, Sectio AI, volume XII, number 4, pages 97–112, DOI: 10.2478/v10065-012-0023-1 (link, link, link, link, link, link, link)

- Grzegorz Oryńczak (2014), “System agentowy dla wspomagania bezpiecznych usług czasu rzeczywistego”, doctoral dissertation, Jagiellonian University, Warsaw, Poland (link, link)

- Mike Ossipoff (1998), “Party List P.S.”, *Election-Methods mailing list*, 27 July 1998 (link, link)

- Joseph Otten (1998), “Ordered List Selection”, *Voting Matters*, issue 9, pages 8–9 (link, link, link)

- Joseph Otten (2000), “Ordered List Selection — revisited”, *Voting Matters*, issue 12, pages 2–5 (link, link, link)

- Z. Emel Öztürk (2020), “Consistency of scoring rules: a reinvestigation of composition-consistency”, *International Journal of Game Theory*, volume 49, issue 3, pages 801–831, DOI: 10.1007/s00182-020-00711-7 (link, link, link, link)

- Eric Pacuit (2021), “Probabilistic Methods in Social Choice”, 10 August 2021 (link) {In this paper, the Schulze method is called “Beat Path”.}

- Pamela Pallett, Aleix Martinez (2014), “Beyond the basics: Facial expressions of compound emotions”, *Journal of Vision*, volume 14, issue 10, page 1401, DOI: 10.1167/14.10.1401 (link)

- Dongbo Pan, Feng Liu, Yong Deng (2016), “An improved AHP method in preferential voting system”, *2016 11th International Conference on Computer Science & Education* (ICCSE 2016), Nagoya, Japan, 23–25 August 2016, proceedings, pages 82–85, DOI: 10.1109/ICCSE.2016.7581559 (link)

- David C. Parkes, Lirong Xia (2012), “A Complexity-of-Strategic-Behavior Comparison between Schulze’s Rule and Ranked Pairs”, *26th AAAI Conference on Artificial Intelligence* (AAAI-12), Toronto, Ontario, Canada, 22–26 July 2012, proceedings, pages 1429–1435, DOI: 10.1609/aaai.v26i1.8258 (link, link, link, link, link, link, link)

- David C. Parkes, Sven Seuken (2013), “Economics and Computation”, Cambridge University Press (link, link)

- Elias A. Parzen, Christian D. Johnson (2021), "Harnessing Condorcet Methods to Improve Decision-making Based on Ranked Data", Pacific Northwest National Laboratory, Richland, Washington, USA, DOI: 10.2172/1875637 (link, link)

- Joanna Paszkowska (2022), "Analysis of Voting Rules in the 1D-Euclidean Model with Strategic Candidates", master thesis, University of Warsaw, Poland (link, link)

- Dirk Pattinson, Mukesh Tiwari (2017), "Schulze Voting as Evidence Carrying Computation", *8th International Conference on Interactive Theorem Proving* (ITP 2017), Brasília, Brazil, 26–29 September 2017; eds. Mauricio Ayala-Rincón, César A. Muñoz, *Interactive Theorem Proving*, volume 10499 of the series *Lecture Notes in Computer Science*, pages 410–426, Springer, DOI: 10.1007/978-3-319-66107-0_26 (link, link, link)

- Alois Paulin (2019), "Controlling Citizens or Controlling the State?", *Smart City Governance*, pages 61–79, Elsevier, DOI: 10.1016/B978-0-12-816224-8.00003-0 (link)

- Rui Pereira, Marco Couto, Francisco Ribeiro, Rui Rua, Jácome Cunha, João Paulo Fernandes, João Saraiva (2021), "Ranking Programming Languages by Energy Efficiency", *Science of Computer Programming*, volume 205, article 102609, DOI: 10.1016/j.scico.2021.102609, DOI: 1822/69044 (link, link, link, link, link, link, link, link, link, link)

- Raúl Pérez-Fernández (2017a), "Monotonicity-based consensus states for the monometric rationalisation of ranking rules with application in decision making", doctoral dissertation, University of Oviedo and Ghent University, DOI: 10651/44390 (link, link, link)

- Raúl Pérez-Fernández, Bernard DeBaets (2017b), "Representations of votes based on pairwise information: Monotonicity versus consistency", *Information Sciences*, volumes 412–413, pages 87–100, DOI: 10.1016/j.ins.2017.05.039 (link)

- Raúl Pérez-Fernández, Bernard DeBaets (2019), "The acclamation consensus state and an associated ranking rule", *International Journal of Intelligent Systems*, volume 34, issue 6, pages 1223–1247, DOI: 10.1002/int.22093 (link)

- Norman Petry (1998), "Tideman vs. Schulze", *Election-Methods mailing list*, 31 July 1998 (link, link)

- Norman Petry (2000), "PC, Tideman & Schulze — Winner Comparison", *Election-Methods mailing list*, 3 November 2000 (link, link, link)

- Rémi Peyre (2012), "Et le vainqueur du second tour est ...", 10 May 2012 (link, link)

- Grzegorz Pierczyński (2019a), "Approval-Based Elections and Distortion of Voting Rules", master thesis, University of Warsaw, Poland (link, link)

- Grzegorz Pierczyński, Piotr Skowron (2019b), “Approval-Based Elections and Distortion of Voting Rules”, *28th International Joint Conference on Artificial Intelligence* (IJCAI 2019), Macao, China, 10–16 August 2019, proceedings, pages 543–549, DOI: 10.24963/ijcai.2019/77 (arXiv:1901.06709) (link, link, link, link)

- Grzegorz Pierczyński, Stanisław Szufa (2024), “Single-Winner Voting with Alliances: Avoiding the Spoiler Effect”, *23rd International Conference on Autonomous Agents and Multiagent Systems* (AAMAS ‘24), Auckland, New Zealand, 6–10 May 2024; proceedings, pages 1567–1575 (arXiv:2401.16399) (link, link, link, link, link)

- Sabina Pintar (2022), “O paradoksima društvenih izbora s preferencijalnim glasanjem”, master thesis, University of Zagreb, Croatia (link, link)

- Reinhold Plösch, Severin Schürz, Christian Körner (2015), “On the Validity of the IT-CISQ Quality Model for Automatic Measurement of Maintainability”, *2015 IEEE 39th Annual International Computer Software and Applications Conference* (COMPSAC), Taichung, Taiwan, 1–5 July 2015, proceedings, volume 2, pages 326–334, DOI: 10.1109/COMPSAC.2015.47 (link)

- Вячеслав Анатольевич Подвербный / Vyacheslav Anatolyevich Podverbnyy, Никита Михайлович Протасов / Nikita Mikhailovich Protasov, Анастасия Анатольевна Перелыгина / Anastasia Anatolyevna Perelygina (2021), “Эколого-Экономическая Оценка Вариантов Городских Пассажирских Линий Легкорельсового Транспорта в Процедуре Групповых Решений на Примере Стамбула” / “Ecological and economic assessment of options for urban passenger light rail lines in the procedure of group decisions on the example of Istanbul”, *Бюллетень Транспортной Информации / Bulletin of Transportation Information*, number 2021.5 (= number 311), pages 29–43 (link)

- Christian Potschka, Christian Heise (2013), “E-Partizipation in Unternehmen, Politik und Verwaltung — Möglichkeiten und Grenzen”, *Wirtschaftsmediation*, issue 1, pages 52–56 (link, link)

- William Poundstone (2008), “Gaming the Vote: Why Elections Aren’t Fair (and What We Can Do About It)”, Hill and Wang (link)

- Ronaldo C. Prati (2010), “Improving Biomarker Identification through Ensemble Feature Rank Aggregation”, *Brazilian Symposium on Bioinformatics* (BSB 2010), Búzios, Brazil, 31 August – 3 September 2010, proceedings, pages 52–55 (link, link) {In this paper, the Schulze method is called “Schwartz sequential dropping”.}

- Ronaldo C. Prati (2012), “Combining feature ranking algorithms through rank aggregation”, *International Joint Conference on Neural Networks* (IJCNN 2012) in conjunction with the *IEEE World Congress on Computational Intelligence* (WCCI 2012), Brisbane, Australia, 10–15 June 2012, proceedings, pages 754–761, DOI: 10.1109/IJCNN.2012.6252467 (link, link, link, link)

- Satya-Lekh Proag (2015), “Evaluation d’une politique publique et mise en œuvre d’une méthodologie supplémentant le calcul du ‘Net Social Benefit’: Cas de l’expérimentation par la Ville de Paris d’une Zone d’Actions Prioritaires pour l’Air (ZAPA)”, doctoral dissertation, Panthéon-Sorbonne University, Paris, France (link, link)

- Amrit Puhan (2024), “Recovering Ground Truth Rankings When The Majority Is Misinformed”, master thesis, Pennsylvania State University, USA (link, link)

- Friedrich Pukelsheim, Christian Schuhmacher (2004), “Das neue Zürcher Zuteilungsverfahren für Parlamentswahlen”, *Aktuelle Juristische Praxis/ Pratique Juridique Actuelle*, volume 13, issue 5, pages 505–522 (link, link)

- Stephany Rajeh, Marinette Savonnet, Eric Leclercq, Hocine Cherifi (2020a), “Interplay Between Hierarchy and Centrality in Complex Networks”, *IEEE Access*, volume 8, pages 129717–129742, DOI: 10.1109/ACCESS.2020.3009525 (arXiv:2103.01376) (link, link, link, link)

- Stephany Rajeh, Marinette Savonnet, Eric Leclercq, Hocine Cherifi (2020b), “Investigating the Relationship Between Hierarchy and Centrality Measures in Complex Networks”, *9th Belgian Network Research Meeting* (BENet 2020), Ghent, Belgium, 12 November 2020 (link, link, video)

- Stephany Rajeh, Marinette Savonnet, Eric Leclercq, Hocine Cherifi (2020c), “Assessing the Relationship Between Centrality and Hierarchy in Complex Networks”, *9th International Conference on Complex Networks and their Applications*, Madrid, Spain, 1–3 December 2020; Book of Abstracts, pages 153–155, DOI: 10.5281/zenodo.4744934 (link, link, link, link)

- Stephany Rajeh, Marinette Savonnet, Eric Leclercq, Hocine Cherifi (2020d), “Hierarchy and Centrality: Two Sides of The Same Coin?”, *Conference on Complex Systems* (CCS 2020), 7–11 December 2020; Book of Abstracts, page 249, DOI: 10.5281/zenodo.4419177 (link, link, link, link, link, link)

- Stephany Rajeh, Marinette Savonnet, Eric Leclercq, Hocine Cherifi (2021a), “Analyse des mesures de hiérarchie et de centralité dans les grands graphes de terrain”, *Revue des Nouvelles Technologies de l’Information* (RNTI-E-37), *Extraction et Gestion des Connaissances* (EGC 2021), Montpellier, France, 25–29 January 2021, proceedings, pages 397–404 (link, link, video: 00:40:40 – 00:55:49 hh:mm:ss)

- Stephany Rajeh, Ali Yassin, Ali Jaber, Hocine Cherifi (2021b), "Analyzing Community-Aware Centrality Measures Using the Linear Threshold Model", *10th International Conference on Complex Networks and their Applications* (COMPLEX NETWORKS 2021), Madrid, Spain, 30 November – 2 December 2021; eds. Rosa Maria Benito, Chantal Cherifi, Hocine Cherifi, Esteban Moro, Luis M. Rocha, Marta Sales-Pardo, volume 1 of *Complex Networks & Their Applications X*, volume 1015 of the series *Studies in Computational Intelligence*, pages 342–353, Springer, DOI: 10.1007/978-3-030-93409-5_29 (arXiv:2202.00514) (link, link, link)

- Torben Rapp, Philipp Tröster (2024), "Achieving Accurate Printing Conditions Through Machine Learning", *Technical Association of the Graphic Arts*, *2024 TAGA Conference*; proceedings, pages 39–45 (link, link)

- Yannick Reisch, Jörg Rothe, Lena Schend (2014), "The Margin of Victory in Schulze, Cup, and Copeland Elections: Complexity of the Regular and Exact Variants", *7th European Starting AI Researcher Symposium* (STAIRS 2014), Prague, Czech Republic, 18–22 August 2014, *Frontiers in Artificial Intelligence and Applications*, eds. Ulle Endriss, João Leite, volume 264, pages 250–259, DOI: 10.3233/978-1-61499-421-3-250 (link, link)

- Manon Revel, Théophile Pénigaud (2026), "AI-Enhanced Deliberative Democracy and the Future of the Collective Will", *Philosophy & Technology*, volume 39, article 81, DOI: 10.1007/s13347-026-01044-1 (arXiv:2503.05830) (link, link, link, link, link)

- Veronica G. Reynolds, Devon H. Callan, Kumar Saurabh, Elizabeth A. Murphy, Kaitlin R. Albanese, Yan-Qiao Chen, Claire Wu, Eliot Gann, Craig J. Hawker, Baskar Ganapathysubramanian, Christopher M. Bates, Michael L. Chabinyc (2022), "Simulation-guided analysis of resonant soft X-ray scattering for determining the microstructure of triblock copolymers", *Molecular Systems Design & Engineering*, volume 7, issue 11, pages 1449–1458, DOI: 10.1039/D2ME00096B (link, link, link, link, link, link)

- Jeffrey T. Richelson (1978), "A Characterization Result for the Plurality Rule", *Journal of Economic Theory*, volume 19, issue 2, pages 548–550, DOI: 10.1016/0022-0531(78)90108-4 (link)

- Wyko Rijnsburger (2015), "Personalized presentation annotations using optical HMDs", doctoral dissertation, University of Amsterdam, Netherlands (link)

- Wyko Rijnsburger, Sven Kratz (2017), "Personalized presentation annotations using optical HMDs", *Multimedia Tools and Applications*, volume 76, issue 4, pages 5607–5629, DOI: 10.1007/s11042-016-4064-0 (link)

- Ronald L. Rivest, Emily Shen (2010), "An Optimal Single-Winner Preferential Voting System Based on Game Theory", *3rd International Workshop on Computational Social Choice* (COMSOC 2010), 13–16 September 2010, proceedings, pages 399–410, Düsseldorf University Press (link, link, link, link, link, link, link, link, link, link, link, link, link)

- Colin Rosenstiel (1998), "Producing a Party List using STV", *Voting Matters*, issue 9, pages 7–8 (link, link, link)

- Riyad Al-Rousan, Mohd Shahrizal Sunar, Hoshang Kolivand (2018), "Geometry-based shading for shape depiction enhancement", *Multimedia Tools and Applications*, volume 77, issue 5, pages 5737–5766, DOI: 10.1007/s11042-017-4486-3 (link, link, link, link)

- Alejandro Ruiz-Padillo, Thiago B.F. de Oliveira, Matheus Alves, Ana L.C. Bazzan, Diego P. Ruiz (2016), "Social choice functions: A tool for ranking variables involved in action plans against road noise", *Journal of Environmental Management*, volume 178, pages 1–10, DOI: 10.1016/j.jenvman.2016.04.038 (link)

- Alejandro Ruiz-Padillo, Francisco Minella Pasqual, Ana Margarita Larranaga Uriarte, Helena Beatriz Bettella Cybis (2018), "Application of multi-criteria decision analysis methods for assessing walkability: A case study in Porto Alegre, Brazil", *Transportation Research Part D: Transport and Environment*, volume 63, pages 855–871, DOI: 10.1016/j.trd.2018.07.016 (link, link)

- Donald G. Saari (1994), "Geometry of Voting", Springer, DOI: 10.1007/978-3-642-48644-9

- Donald G. Saari (2001), "Decisions and Elections: Explaining the Unexpected", Cambridge University Press, DOI: 10.1017/CBO9780511606076

- Saria Safdar, Shoab Ahmed Khan, Arslan Shaukat, M. Usman Akram (2021), "Genetic Algorithm Based Automatic Out-Patient Experience Management System (GAPEM) Using RFIDs and Sensors", *IEEE Access*, volume 9, pages 8961–8976, DOI: 10.1109/ACCESS.2020.3046839 (link, link, link)

- Julia Sageder, Ariane Demleitner, Oliver Irlbacher, Raphael Wimmer (2019), "Applying Voting Methods in User Research", *Mensch und Computer 2019* (MuC '19), Hamburg, Germany, 8–11 September 2019, proceedings, pages 571–575, DOI: 10.1145/3340764.3344461, DOI: 10.5283/epub.40682 (link, link, link, poster, poster)

- Adrian Salamon (2017), "Software implementations and analysis of three voting systems", senior thesis, Kungsholmens Gymnasium, Stockholm, Sweden (link, link)

- Fabio Marcos de Abreu Santos, Bianca Trinkenreich, João Felipe Nicolaci Pimentel, Igor Scaliante Wiese, Igor Steinmacher, Anita Sarma, Marco Aurelio Gerosa (2022), "How to Choose a Task? Mismatches in Perspectives of Newcomers and Existing Contributors", *16th ACM/IEEE International Symposium on Empirical Software Engineering and Measurement* (ESEM 2022), Helsinki, Finland, 18–23 September 2022, proceedings, pages 114–124, DOI: 10.1145/3544902.3546236 (link, link, link, link, link, link, link, link)

- Fabio Marcos de Abreu Santos (2023a), “Supporting the Task-driven Skill Identification in Open Source Project Issue Tracking Systems”, doctoral dissertation, Northern Arizona University, Flagstaff, Arizona, USA (link, link, link)

- Fabio Marcos de Abreu Santos (2023b), “Supporting the Task-driven Skill Identification in Open Source Project Issue Tracking Systems”, *International Doctoral Symposium on Empirical Software Engineering* (IDoESE 2022) at the *Empirical Software Engineering International Week* (ESEIW 2022) and the *16th ACM/IEEE International Symposium on Empirical Software Engineering and Measurement* (ESEM 2022), Helsinki, Finland, 18–23 September 2022; *ACM SIGSOFT Software Engineering Notes*, volume 48, issue 1, pages 54–58, DOI: 10.1145/3573074.3573088 (arXiv:2211.08143) (link, link)

- Алексей Владимирович Савватеев / Alexey Vladimirovich Savvateev, Александр Юрьевич Филатов / Alexander Yuryevich Filatov, Дмитрий Александрович Шварц / Dmitry Alexandrovich Schwartz (2018), “Проблема коллективного выбора, теорема Эрроу и ее короткое доказательство” / “The Problem of Collective Choice, Arrow’s Impossibility Theorem, and it’s Short Proof”, *Известия ДВФУ: Экономика и управление / Bulletin of the Far Eastern Federal University: Economics and Management*, volume 4, pages 5–22, DOI: 10.24866/2311-2271/2018-4/5-22 (link, link, link, link)

- Lena Schend (2015), “From Election Fraud to Finding the Dream Team: A Study of the Computational Complexity in Voting Problems and Stability in Hedonic Games”, doctoral dissertation, Heinrich Heine University Düsseldorf, Germany (link, link, link, link, link)

- Markus Schulze (1997), “Condorect sub-cycle rule”, *Election-Methods mailing list*, 3 October 1997 (link, link)

- Markus Schulze (1998a), “Tiebreakers, Subcycle Rules”, *Election-Methods mailing list*, 10 August 1998 (link, link)

- Markus Schulze (1998b), “Maybe Schulze is decisive”, *Election-Methods mailing list*, 31 August 1998 (link, link, link)

- Markus Schulze (1998c), “Schulze Method”, *Election-Methods mailing list*, 14 November 1998 (link, link)

- Markus Schulze (2002), “On Dummett’s ‘Quota Borda System’”, *Voting Matters*, issue 15, pages 10–13 (link, link, link)

- Markus Schulze (2003), “A new monotonic and clone-independent single-winner election method”, *Voting Matters*, issue 17, pages 9–19 (link, link, link)

- Markus Schulze (2004), “Free Riding”, *Voting Matters*, issue 18, pages 2–8 (link, link, link)

- Markus Schulze (2008a), “A New MMP Method” (link, link, link, link)

- Markus Schulze (2008b), “A New MMP Method (Part 2)” (link, link, link, link)

- Markus Schulze (2009), “Proposed Statutory Rules for the Schulze Single-Winner Election Method” (link, link)

- Markus Schulze (2011a), “A new monotonic, clone-independent, reversal symmetric, and Condorcet-consistent single-winner election method”, *Social Choice and Welfare*, volume 36, issue 2, pages 267–303, DOI: 10.1007/s00355-010-0475-4 (link, link)

- Markus Schulze (2011b), “Free Riding and Vote Management under Proportional Representation by the Single Transferable Vote” (link, link, link, link)

- Markus Schulze (2011c), “Implementing the Schulze STV Method” (link, link, link, link)

- Markus Schulze (2016a), “Condorcet/Schulze Voting in Single-Winner Districts”, *Submission to the Canadian House of Commons Special Committee on Electoral Reform* (link, link)

- Markus Schulze (2016b), “Mode de scrutin Condorcet/Schulze dans les circonscriptions uninominales”, *Submission to the Canadian House of Commons Special Committee on Electoral Reform* (link, link)

- Markus Schulze (2019), “Condorcet/Schulze Voting for Burlington” (link, part1, part2, part3)

- Markus Schulze (2024a), “Comment on ‘The case for minimax-TD’”, *Constitutional Political Economy*, volume 35, issue 3, pages 437–438, DOI: 10.1007/s10602-023-09416-x (link, link)

- Markus Schulze (2024b), “Comment on ‘The best Condorcet-compatible election method: Ranked Pairs’”, *Constitutional Political Economy*, volume 35, issue 3, pages 439–442, DOI: 10.1007/s10602-023-09415-y (link, link)

- Thomas Schwartz (1986), “The Logic of Collective Choice”, Columbia University Press, New York

- Shreyas Sekar, Sujoy Sikdar, Lirong Xia (2017), “Condorcet Consistent Bundling with Social Choice”, *16th International Conference on Autonomous Agents and Multiagent Systems* (AAMAS 2017), São Paulo, Brazil, 8–12 May 2017, proceedings, pages 33–41 (link, link, link, link, link, link)

- Paolo Serafini (2020), “Mathematics to the Rescue of Democracy”, Springer, DOI: 10.1007/978-3-030-38368-8 (link, link)

- Nisarg Shah (2016), “Optimal Social Decision Making”, doctoral dissertation, Carnegie Mellon University, Pittsburgh, USA (link, link, link, link, link)

- Zhelun Shen, Yuchao Dai, Zhibo Rao (2021), “CFNet: Cascade and Fused Cost Volume for Robust Stereo Matching”, *2021 IEEE/CVF Conference on Computer Vision and Pattern Recognition* (CVPR), Nashville, Tennessee, USA, 20–25 June 2021, proceedings, pages 13906–13915, DOI: 10.1109/CVPR46437.2021.01369 (arXiv:2104.04314) (link, link, link, link, link)

- Zhelun Shen, Xibin Song, Yuchao Dai, Dingfu Zhou, Zhibo Rao, Liangjun Zhang (2023), “Digging Into Uncertainty-Based Pseudo-Label for Robust Stereo Matching”, *IEEE Transactions on Pattern Analysis and Machine Intelligence*, volume 45, issue 12, pages 14301–14320, DOI: 10.1109/TPAMI.2023.3300976 (arXiv:2307.16509) (link)

- Андрей Николаевич Шешко / Andrey Nikolaevich Sheshko (2023), “Разработка структуры ПО с помощью метода Шульце” / “Software Structure Development using the Schulze Method” (link, link, link, link)

- Raffael Boymian Siahaan (2023), “Analisis Kompleksitas Algoritma dalam implementasi *Schulze Methods* berbasis teori graf pada *Preferential Single-Winner Election system*”, Bandung Institute of Technology, Indonesia (link, link)

- Forest W. Simmons (2004), “Consistency in PR methods”, *Election-Methods mailing list*, 11 December 2004 (link)

- Paul B. Simpson (1969), “On Defining Areas of Voter Choice: Professor Tullock on Stable Voting”, *Quarterly Journal of Economics*, volume 83, number 3, pages 478–490, DOI: 10.2307/1880533 (link)

- Dominic J. Skinner, Patrick Lemaire, Madhav Mani (2025), “Physical Modeling of Embryonic Transcriptomes Identifies Collective Modes of Gene Expression”, *PRX LIFE*, volume 3, article 033028, DOI: 10.1103/7gs5-w48q; DOI: 10.1101/2024.07.26.605398 (link, link, link, link, link) {See also the supplementary material (link, link, link).}

- Piotr Skowron, Martin Lackner, Markus Brill, Dominik Peters, Edith Elkind (2017), “Proportional Rankings”, *26th International Joint Conference on Artificial Intelligence* (IJCAI 2017), Melbourne, Australia, 19–25 August 2017, proceedings, pages 409–415, DOI: 10.24963/ijcai.2017/58 (arXiv:1612.01434) (link, link, link, link, link)

- Piotr Skowron (2024), “Computational Social Choice” (link, link)

- John H. Smith (1973), “Aggregation of Preferences with Variable Electorate”, *Econometrica*, volume 41, issue 6, pages 1027–1041, DOI: 10.2307/1914033

- Warren D. Smith (2006), “Descriptions of single-winner voting systems” (link, link)

- Yonatan Sompolinsky (2015), “Breaking free from the chains of blockchain protocols”, *Scaling Bitcoin*, Hong Kong, China, 6–7 December 2015 (link, link, link, video: 00:08:59 – 00:34:46 hh:mm:ss)

- Krzysztof Sornat, Virginia Vassilevska Williams, Yinzhan Xu (2021), "Fine-Grained Complexity and Algorithms for the Schulze Voting Method", *22nd ACM Conference on Economics and Computation* (EC '21), Budapest, Hungary, 18–23 July 2021, proceedings, pages 841–859, DOI: 10.1145/3465456.3467630; *9th International Workshop on Computational Social Choice* (COMSOC 2023), Beersheba, Israel, 3–5 July 2023 (arXiv:2103.03959) (link, link, link, link, link, link, link, link, link, video)

- Stephen H. Sosnick (2016), "Which Election System is Best: A Comprehensive Comparison of World Wide Election Systems", CreateSpace Independent Publishing Platform (link)

- Paulo Cesar Ferreira de Souza (2009), "Seleção de construtora como parceira para empreendimento de energia elétrica: utilização dos métodos ordinais do apoio multicritério à decisão", doctoral dissertation, Rio de Janeiro, Brazil (link)

- Michele Staffieri (2024), "Studio e sviluppo di strumenti per la partecipazione civica basati su blockchain per la piattaforma FirstLife", doctoral dissertation, University of Turin, Italy (link, link)

- Saul Stahl, Paul E. Johnson (2006), "Understanding Modern Mathematics", Jones & Bartlett Publishing (link, link, link, link)

- Saul Stahl, Paul E. Johnson (2017), "Mathematics Old and New", Courier Dover Publications (link, link)

- Dennis Stanke, Peer Schroth, Michael Rohs (2022), "TrackballWatch: Trackball and Rotary Knob as a Non-Occluding Input Method for Smartwatches in Map Navigation Scenarios", *Proceedings of the ACM on Human-Computer Interaction*, volume 6, number MHCI (MobileHCI '22), article 199, DOI: 10.1145/3546734 (link, link, link)

- Dennis Stanke, Tim Duente, Kerem Can Demir, Michael Rohs (2023), "Can You Ear Me? — A Comparison of Different Private and Public Notification Channels for the Earlobe", *Proceedings of the ACM on Interactive, Mobile, Wearable and Ubiquitous Technologies*, volume 7, issue 3 (IMWUT '23), article 123, DOI: 10.1145/3610925 (link, link, link)

- Dennis Stanke, Benjamin Simon, Sergej Löwen, Michael Rohs (2024), "CaseTouch: Occlusion-Free Touch Input by adding a Thin Sensor Stripe to the Smartwatch Case", *International Conference on Mobile and Ubiquitous Multimedia* (MUM '24), Stockholm, Sweden, 1–4 December 2024; proceedings, pages 172–183, DOI: 10.1145/3701571.3701583 (link, link, link)

- Anna Sterzik, Nils Lichtenberg, Michael Krone, Daniel Baum, Douglas W. Cunningham, Kai Lawonn (2023), "Enhancing molecular visualization: Perceptual evaluation of line variables with application to uncertainty visualization", *Computers & Graphics*, volume 114, pages 401–413, DOI: 10.1016/j.cag.2023.06.006 (link, link, link)

- Fatemeh Stodt, Christoph Reich (2023), "Introducing a Fair Tax Method to Harden Industrial Blockchain Applications against Network Attacks: A Game Theory Approach", *Computers*, volume 12, issue 3, article 64, DOI: 10.3390/computers12030064 (link, link, link, link, link)

- Ben Strasser (2017), "Computing Tree Decompositions with FlowCutter: PACE 2017 Submission" (arXiv:1709.08949) (link, link)

- Radek Švec (2011), "Bordovo hlasování jako alternativa k ostatním typům ordinálních většinových volebních systémů", master thesis, University of Prague, Czech Republic, DOI: 20.500.11956/34151 (link, link)

- Oleg A. Sychev, Mikhail E. Denisov, Anton V. Anikin (2020), "Ways To Write Algorithms And Their Execution Traces For Teaching Programming", *National Interest, National Identity and National Security* (NININS 2020), Nizhny Novgorod, Russia, 10–12 March 2020; *European Proceedings of Social and Behavioural Sciences* (EpSBS), volume 102, article 127, pages 1029–1039, DOI: 10.15405/epsbs.2021.02.02.127 (link, link, link)

- Alan G. Szepieniec (2013), "Systemen voor Online Stemmingen", master thesis, Catholic University of Leuven, Belgium (link, link)

- Andrew Taylor (2012), "Electoral reforms and non-transitive dice", 14 April 2012 (link, link)

- Peter A. Taylor (2004), "Election Systems 101", *SWUUSI 2004* (link, link, link, link, link)

- Michael Henry Tessler, Michiel A. Bakker, Daniel Jarrett, Hannah Sheahan, Martin J. Chadwick, Raphael Koster, Georgina Evans, Lucy Campbell-Gillingham, Tantum Collins, David C. Parkes, Matthew Botvinick, Christopher Summerfield (2024), "AI can help humans find common ground in democratic deliberation", *Science*, volume 386, issue 6719, DOI: 10.1126/science.adq2852 (link, link, video) {The Schulze method is mainly discussed in the supplementary material (link).}

- Eliott Thomas, Mickael Coustaty, Aurélie Joseph, Gaspar Deloin, Vincent Poulain d'Andecy, Jean-Marc Ogier (2026), "Active Learning for Cascaded Object Detection: Balancing Coverage and Uncertainty in Table Extraction Pipelines", *20th International Conference on Document Analysis and Recognition* (ICDAR 2026), Vienna, Austria, 31 August – 2 September 2026; proceedings, part IV, volume 16975 of the series *Lecture Notes in Computer Science*, pages 21–37, DOI: 10.1007/978-3-032-36042-7_2 (arXiv:2607.00747) (link)

- Teruji Thomas (2019), "The Asymmetry, Uncertainty, and the Long Term" (link, link, link)

- T. Nicolaus Tideman (1987), "Independence of Clones as a Criterion for Voting Rules", *Social Choice and Welfare*, volume 4, issue 3, pages 185–206, DOI: 10.1007/BF00433944 (link, link)

- T. Nicolaus Tideman, Daniel Richardson (2000), “Better voting methods through technology: The refinement-manageability trade-off in the single transferable vote”, *Public Choice*, volume 103, issue 1, pages 13–34, DOI: 10.1023/A:1005082925477

- T. Nicolaus Tideman (2006), “Collective Decisions and Voting: The Potential for Public Choice”, Ashgate Publishing (link, link)

- T. Nicolaus Tideman (2019), “How Should Votes Be Cast and Counted?”, *Oxford Handbook of Public Choice*, eds. Roger D. Congleton, Bernard N. Grofman, Stefan Voigt, volume 2, pages 5–23, Oxford University Press, DOI: 10.1093/oxfordhb/9780190469771.013.1 (link, link, link, link, link, link)

- Mukesh Tiwari (2021a), “Formally Verified Verifiable Electronic Voting Scheme”, doctoral dissertation, Australian National University, DOI: 1885/227674, DOI: 10.25911/X41N-PM15 (link, link, link, slides, slides)

- Mukesh Tiwari, Dirk Pattinson (2021b), “Machine Checked Properties of the Schulze Method”, *7th Workshop on Hot Issues in Security Principles and Trust* (HotSpot 2021), affiliated with the *6th IEEE European Symposium on Security and Privacy* (EuroS&P 2021), Vienna, Austria, 6–10 September 2021 (link, link, slides, slides)

- Mukesh Tiwari (2022), “Theorem Provers to Protect Democracy: Formally Verified Verifiable Vote Counting Software”, University of Cambridge, United Kingdom, 21 January 2022 (link, link, slides, slides)

- С.С. Токар / S.S. Tokar, Олег С. Вєтров / Oleh S. Vetrov (2025), “Реалізація методу Шульце мовою програмування JavaScript” / “Implementation of the Schulze Method in JavaScript”, *Прикладні Інформаційні Технології / Applied Information Technologies*, pages 106–107 (link, link)

- Fabio Tosi, Luca Bartolomei, Matteo Poggi (2025), “A Survey on Deep Stereo Matching in the Twenties”, *International Journal of Computer Vision*, volume 133, issue 7, pages 4245–4276, DOI: 10.1007/s11263-024-02331-0 (arXiv:2407.07816) (link, link, link, link, link)

- Bentz P. Tozer III (2017), “Many Objective Sequential Decision Making”, doctoral dissertation, George Washington University, Washington, D.C., USA (link, link)

- Alasdair Tran, Cheng Soon Ong, Christian Wolf (2018), “Combining active learning suggestions”, *PeerJ Computer Science*, volume 4, article e157, DOI: 10.7717/peerj-cs.157 (link, link, link, link, link, link, link, link, link, link, link)

- Thomas Trapanese (2018), “Modelli e piattaforme per la democrazia digitale: analisi e confronto”, master thesis, University of Bologna, Italy (link, link, link)

- Владимир Александрович Тушавин / Vladimir Alexandrovich Tushavin (2015), "Ранжирование Показателей Качества с Использованием Методов Кемени-Янга и Шульце" / "Ranking of Quality Indicators Using the Kemeny-Young Method and the Schulze Method", *Экономика и Менеджмент Систем Управления / Economics and Management of Control Systems*, issue 18, number 4, pages 497–503 (link)

- Владимир Александрович Тушавин / Vladimir Alexandrovich Tushavin (2016), "К Вопросу о Сравнении Эффективности Алгоритмов Ранжирования" / "To the Problem of Efficiency Comparison of Ranking Algorithms", *Современная Наука: Актуальные Проблемы Теории и Практики / Modern Science: Current Problems in Theory and Practice*, *Естественные и Технические Науки / Natural and Technical Sciences*, number 1, pages 67–70 (link, link)

- Владимир Александрович Тушавин / Vladimir Alexandrovich Tushavin, Марина Сергеевна Смирнова / Maria Sergeevna Smirnova (2018), "Сравнение Методов Ранжирования Объектов при Решении Квалиметрических Задач" / "Comparison of Methods for Ranking Objects in Solving Qualimetric Tasks", *Моделирование и Ситуационное Управление Качеством Сложных Систем / Modeling and Situation Quality Control of Complex Systems*, Saint Petersburg, Russia, 9–13 April 2018, proceedings, pages 163–165 (link)

- Vladimir Alexandrovich Tushavin (2020), "Complex Quality Indicators and Ranking in Uncertainty Conditions", *Components of Scientific and Technological Progress*, volume 47, issue 5, pages 15–18 (link, link)

- Ян Владимирович Тушавин / Yan Vladimirovich Tushavin, Александра Алексеевна Каплиева / Alexandra Alekseevna Kaplieva (2023), "Методика Ранжирования Показателей Качества с Использованием Методов Борда, Шульце и Кемени-Янга" / "Method for Ranking Quality Indicators Using Borda, Schulze and Kemeny-Young Methods", *Взгляд Молодых Исследователей: Экономика, Управление, Инновации / View of Young Researchers: Economics, Management, Innovations*, *Общероссийская Научно-Практическая Конференция / All-Russian Scientific and Practical Conference*, Mytishchi, Russia, 27–28 April 2023, proceedings, pages 274–277 (link, link)

- Adrienne Tuynman, Timothée Mathieu (2026), "Optimal Posterior E-values with Non-Convex Parameter Sets with Applications to Voting Systems" (arXiv:2606.29998) (link, link)

- Brent Vansprengel (2025), "A short study of pairwise comparison methods applied to the ranking of sport teams", master thesis, Catholic University of Leuven, DOI: 2078.1/thesis:49872 (link, link)

- Ash Vardanian (2024), "GPU-accelerated Schulze voting method in Python, Numba, and CUDA, using ideas from Algebraic Graph Theory" (link, link, link, link, link)

- Luis G. Vargas (2016), “Voting with Intensity of Preferences”, *International Journal of Information Technology & Decision Making*, volume 15, number 4, pages 839–859, DOI: 10.1142/S0219622016400058 (link)

- Jim Vaughan, Sven Kratz, Don Kimber (2016), “Look where you’re going: Visual interfaces for robot teleoperation”, *2016 25th IEEE International Symposium on Robot and Human Interactive Communication* (RO-MAN 2016), 26–31 August 2016, New York, USA, proceedings, pages 273–280, DOI: 10.1109/ROMAN.2016.7745142 (link, link)

- C. Venkatesan, S. Jeevanantham, B. Rebekka (2026), “Credibility-Aware Hierarchical Blockchain Consensus Protocol for Bandwidth-Limited IoT Edge Networks”, *IEEE Transactions on Network and Service Management*, volume 23, pages 4067–4081, DOI: 10.1109/TNSM.2026.3683555

- Ilse Verdiesen, Martijn Cligge, Jan Timmermans, Lennard Segers, Virginia Dignum, Jeroen van den Hoven (2016), “MOOD: Massive Open Online Deliberation Platform — A practical application”, *1st Workshop on Ethics in the Design of Intelligent Agents* (EDIA 2016) in conjunction with the *22th European Conference on Artificial Intelligence* (ECAI 2016), The Hague, Netherlands, 29 August – 2 September 2016, proceedings, eds. Grégory Bonnet, Maaike Harbers, Koen Hindriks, Mike Katell, Catherine Tessier, pages 6–11 (link, link, link, link, link, link, link)

- Ilse Verdiesen, Virginia Dignum, Jeroen van den Hoven (2018), “Measuring Moral Acceptability in E-deliberation: A Practical Application of Ethics by Participation”, *ACM Transactions on Internet Technology*, volume 18, number 4, article 43, DOI: 10.1145/3183324 (link)

- KaiYu Wan, Vangalur Alagar (2017), “Analyzing healthcare big data for patient satisfaction”, *2017 13th International Conference on Natural Computation, Fuzzy Systems and Knowledge Discovery* (ICNC-FSKD 2017), Guilin, China, 29–31 July 2017, proceedings, pages 2084–2091, DOI: 10.1109/FSKD.2017.8393093 (link)

- Shuaiqiang Wang, Jiankai Sun, Byron J. Gao, Jun Ma (2014), “VSRank: A Novel Framework for Ranking-Based Collaborative Filtering”, *ACM Transactions on Intelligent Systems and Technology* (TIST 2019), volume 5, issue 3, article 51, DOI: 10.1145/2542048 (link, link, link)

- Yujie Wang (2022), “Game Theory Survey: Voting Methods of Preferential Ballot”, *Journal of Physics: Conference Series*, volume 2386, article 012001, DOI: 10.1088/1742-6596/2386/1/012001 (link, link)

- Yun Wang, Longguang Wang, Chenghao Zhang, Yongjian Zhang, Zhanjie Zhang, Ao Ma, Chenyou Fan, Tin Lun Lam, Junjie Hu (2025), “Learning Robust Stereo Matching in the Wild with Selective Mixture-of-Experts”, *International Conference on Computer Vision* (ICCV 2025), Honolulu, Hawaii, USA, 19–23 October 2025; proceedings, pages 21276–21287, DOI: 10.1109/ICCV51701.2025.01976 (arXiv:2507.04631) (link, link, link)

- C. Hugh E. Warren (1994), “Counting in STV Elections”, *Voting Matters*, issue 1, pages 12–13 (link, link, link)

- C. Hugh E. Warren (1999), “An example of ordering elected candidates”, *Voting Matters*, issue 10, page 3 (link, link, link)

- Stephen Warshall (1962), “A Theorem on Boolean Matrices”, *Journal of the ACM*, volume 9, issue 1, pages 11–12, DOI: 10.1145/321105.321107 (link)

- Thomas F. Wenisch (2018), “Top Picks from the 2017 Computer Architecture Conferences”, *IEEE Micro*, volume 38, issue 3, pages 5–9, DOI: 10.1109/MM.2018.032271056 (link, link, link)

- Brian A. Wichmann (1994), “An STV Database”, *Voting Matters*, issue 2, page 9 (link, link, link)

- Martin Wilke (2007), “Personalisierung der Verhältniswahl durch Varianten der Single Transferable Vote”, master thesis, Otto Suhr Institute for Political Science, Free University of Berlin, Germany (link, link)

- Kyle S. Wilkinson (2018), “Analysis of a Voting Method for Ranking Network Centrality Measures on a Node-Aligned Multiplex Network”, master thesis, Air Force Institute of Technology, Wright-Patterson Air Force Base, Ohio, USA (link, link, link, link)

- Raphael Wimmer, Florian Schulz, Fabian Hennecke, Sebastian Boring, Heinrich Hussmann (2009), “Curve: Blending Horizontal and Vertical Interactive Surfaces”, *Adjunct Proceedings of the 4th IEEE Workshop on Tabletops and Interactive Surfaces* (IEEE Tabletop 2009), Banff, Alberta, Canada, 23–25 November 2009 (link, link, link, link, link)

- Raphael Wimmer, Fabian Hennecke, Florian Schulz, Sebastian Boring, Andreas Butz, Heinrich Hussmann (2010), “Curve: Revisiting the Digital Desk”, *6th Nordic Conference on Human-Computer Interaction* (NordiCHI 2010), Reykjavik, Iceland, 16–20 October 2010, proceedings, pages 561–570, DOI: 10.1145/1868914.1868977 (link, link, link, link)

- Jana Woitaschik (2026), “Complexity of the Possible President Problem in Schulze Voting”, *International Symposium on Artificial Intelligence and Mathematics* (ISAIM 2026), Fort Lauderdale, Florida, USA, 7–9 January 2026 (link, link)

- Douglas R. Woodall (1994), “Properties of Preferential Election Rules”, *Voting Matters*, issue 3, pages 8–15 (link, link, link)

- Douglas R. Woodall (1996), “Monotonicity and Single-Seat Election Rules”, *Voting Matters*, issue 6, pages 9–14 (link, link, link)

- Douglas R. Woodall (1997), “Monotonicity of single-seat preferential election rules”, *Discrete Applied Mathematics*, volume 77, issue 1, pages 81–98, DOI: 10.1016/S0166-218X(96)00100-X (link)

- Barry Wright (2009), “Objective Measures of Preferential Ballot Voting Systems”, doctoral dissertation, Duke University, Durham, North Carolina, USA (link, link, link, link)

- Geraldo Xexéo, Eduardo Freitas Mangeli de Brito, Luís Fernando Oliveira (2016), “Drama Measures Applied to a Large Scale Business Game”, *Developments in Business Simulation and Experiential Learning*, volume 43, number 1, pages 325–333 (link, link, link)

- Lirong Xia (2021a), “The Smoothed Satisfaction of Voting Axioms”, *8th International Workshop on Computational Social Choice* (COMSOC 2021), Haifa, Israel, 7–10 June 2021 (arXiv:2106.01947) (link, link, link, link, link, link)

- Lirong Xia (2021b), “How Likely Are Large Elections Tied?”, *22nd ACM Conference on Economics and Computation* (EC ‘21), Budapest, Hungary, 18–23 July 2021, proceedings, pages 884–885, DOI: 10.1145/3465456.3467592 (arXiv:2011.03791) (link, video)

- Lirong Xia (2021c), “The Semi-Random Satisfaction of Voting Axioms”, *35th Conference on Neural Information Processing Systems* (NIPS ‘21) (link, link, link); supplementary material: (link, link, link)

- Lirong Xia (2022), “How Likely A Coalition of Voters Can Influence A Large Election?”, *1st Workshop on Learning with Strategic Agents* (LSA 2022) at the *21st International Conference on Autonomous Agents and Multiagent Systems* (AAMAS 2022), Auckland, New Zealand, 9–10 May 2022 (arXiv:2202.06411) (link, link, link)

- Lirong Xia (2023), “Most Equitable Voting Rules”, *19th Conference On Web And InterNet Economics* (WINE 2023), Shanghai, China, 4–8 December 2023; proceedings, pages 711–712 (arXiv:2205.14838) (link)

- Fan Xue, Geoffrey Qiping Shen (2017), “Design of an efficient hyper-heuristic algorithm CMA-VNS for combinatorial black-box optimization problems”, *Genetic and Evolutionary Computation Conference* (GECCO 2017), Berlin, Germany, 15–19 July 2017, proceedings, pages 1157–1162, DOI: 10.1145/3067695.3082054 (link, link, link, link, link)

- Vijay Kumar Yadav, Anshul Anand, Shekhar Verma, Sharannya Venkatesan (2020), “Private computation of the Schulze voting method over the cloud”, *Cluster Computing*, volume 23, issue 4, pages 2517–2531, DOI: 10.1007/s10586-019-03025-w (link)

- Sen Yan, Catherine Soladié, Jean-Julien Aucouturier, Renaud Seguier (2023a), “Combining GAN with reverse correlation to construct personalized facial expressions”, *PLoS ONE*, volume 18, issue 8, article e0290612, DOI: 10.1371/journal.pone.0290612 (link, link, link, link, link, supplementary material (mirror))

- Sen Yan (2023b), “Personalizing facial expressions by exploring emotional mental prototypes”, doctoral dissertation, Paris-Saclay University, France (link, link)

- Quentin Yang (2023), “Coercion-resistance in electronic voting: design and analysis”, doctoral dissertation, University of Lorraine, France (link, link, link)

- Vadym Serhiyovych Yaremenko, Oleksandr Vasylovych Syrotiuk (2020), “Development of a Multi-Agent System for Solving Domain Dictionary Construction Problem”, *Technology Audit and Production Reserves*, volume 2020.4 (= volume 54), number 2, pages 27–30, DOI: 10.15587/2706-5448.2020.208400 (link, link, link, link, link)

- Вадим Сергійович Яременко / Vadym Serhiyovych Yaremenko (2025), “Модель мультиагентної системи для автоматизованої побудови словника предметної області при обробці потокових даних” / “A Multi-Agent System Model for Automated Domain Dictionary Construction in Stream Data Processing”, doctoral dissertation, National Technical University of Ukraine, Kyiv (link, link)

- Anbu Yue, Weiru Liu, Anthony Hunter (2007), “Approaches to Constructing a Stratified Merged Knowledge Base”, *9th European Conference on Symbolic and Quantitative Approaches to Reasoning with Uncertainty* (ECSQARU 2007), Hammamet, Tunisia, 31 October – 2 November 2007; ed. Khaled Mellouli, volume 4724 of the series *Lecture Notes in Computer Science*, pages 54–65, Springer, DOI: 10.1007/978-3-540-75256-1_8 (link, link, link, link, link)

- Yohan Yun, Jisu Yim, Geehyuk Lee (2025), “AutoPark: Adaptive Friction Mouse Facilitating Stable Multi-touch Gesture Input”, *Proceedings of the ACM on Human-Computer Interaction*, volume 9, issue 8, pages 145–169, DOI: 10.1145/3773064 (link)

- Zakir Cümşüd Zabidov, B.C. Maharramov, Khayala M. Gadirova (2022), “Assessment of informativity data of the Water Flow in observation points by Schulz method”, *International Conference dedicated to the 110th anniversary of Academician Ibrahim Ibrahimov*, Baku, Azerbaijan, 29 June – 1 July 2022 (link, link)

- Shahrooz Zarbafian (2018), “Optimization and machine learning methods for Computational Protein Docking”, doctoral dissertation, Boston University, Massachusetts, USA, DOI: 2144/32673 (link, link)

- Thomas M. Zavist, T. Nicolaus Tideman (1989), “Complete Independence of Clones in the Ranked Pairs Rule”, *Social Choice and Welfare*, volume 6, issue 2, pages 167–173, DOI: 10.1007/BF00303170 (link, link)

- Lemnaouar Zedam, Raúl Pérez-Fernández, Hassane Bouremel, Bernard DeBaets (2018), "Clonal sets of a binary relation", *International Journal of General Systems*, volume 47, issue 4, pages 329–349, DOI: 10.1080/03081079.2018.1445738 (link, link)

- Oleg Zendel (2024), "New Perspectives on Query Performance Prediction", doctoral dissertation, Royal Melbourne Institute of Technology, Australia, DOI: 10.25439/rmt.26104630 (link, link, link)

- Oliver Zendel (2023), "Test Data for Dependable Computer Vision Applications", doctoral dissertation, Vienna University of Technology, Austria, DOI: 10.34726/hss.2023.116751 (link, link)

- Wanwan Zheng, Mingzhe Jin (2018), "Comparing Feature Selection Methods by Using Rank Aggregation", *2018 Sixteenth International Conference on ICT and Knowledge Engineering* (ICT&KE), Bangkok, Thailand, 21–23 November 2018, DOI: 10.1109/ICTKE.2018.8612429 (link)

- Wanwan Zheng, Mingzhe Jin (2020), "Comparing multiple categories of feature selection methods for text classification", *Digital Scholarship in the Humanities*, volume 35, issue 1, pages 208–224, DOI: 10.1093/llc/fqz003 (link)

- Ke Zhou, Ronan Cummins, Mounia Lalmas, Joemon M. Jose (2013), "Which Vertical Search Engines are Relevant?", *22nd International Conference on World Wide Web* (WWW 2013), Rio de Janeiro, Brazil, 13–17 May 2013, proceedings, pages 1557–1568, DOI: 10.1145/2488388.2488524 (link, link, link, link, link, link, link)

- Ke Zhou (2014), "On the Evaluation of Aggregated Web Search", doctoral dissertation, University of Glasgow, United Kingdom (link, link, link)

- Adam Zizien, Karel Fliegel (2022), "Regarding the quality of disparity estimation from distorted light fields", *Signal Processing: Image Communication*, volume 109, issue C, article 116867, DOI: 10.1016/j.image.2022.116867 (link, link)

- Aviv Zohar (2015), "Breaking Free from Chains", 12 December 2015 (video: 00:10:06 – 00:12:55 hh:mm:ss)

- Marko Zovko (2021), "Funkcije društvenog izbora s preferencijalnim glasanjem", master thesis, University of Zagreb, Croatia (link, link, link)

- Terence L. van Zyl (2023), "Late Meta-learning Fusion Using Representation Learning for Time Series Forecasting", *2023 26th International Conference on Information Fusion* (FUSION), Charleston, South Carolina, USA, 27–30 June 2023, DOI: 10.23919/FUSION52260.2023.10224217 (arXiv:2303.11000) (link, link, link)

- Department of Computer Science and Engineering at the Indian Institute of Technology Kanpur (2017), "Social Choice" (link, link)

www001: https://www.debian.org/vote/2003/vote_0002.en.html (mirror)

www002: https://wiki.gentoo.org/index.php?oldid=955146#Condorcet_method_of_voting (mirror)

www003: https://meta.wikimedia.org/wiki/Wikimedia_Foundation_elections/Board_elections/2008/en (mirror)

https://meta.wikimedia.org/wiki/Wikimedia_Foundation_elections/Board_elections/2009/en (mirror)

https://meta.wikimedia.org/wiki/Wikimedia_Foundation_elections/Board_elections/2011/en (mirror)

www004: https://lists.ubuntu.com/archives/ubuntu-irc/2012-May/001538.html (mirror)

www005: Section 3.4.1 of the Rules of Procedures for Online Voting
https://ev.kde.org/rules/online_voting/#341-multiple-options (mirror)

www006: https://web.archive.org/web/20201111215950/http://ledning.piratpartiet.se/medlemsomrostning-om-riksdagslista-dags-att-foresla-kandidater/

https://web.archive.org/web/20181106171911/https://forum.piratpartiet.se/showthread.php?p=190840

www007: https://eudec.org/about-us/guidance-document/ (mirror)

www008: https://governance.openstack.org/election/#election-system (mirror)

https://wiki.openstack.org/w/images/4/41/Condorcet-briefing_10_16_2013.pdf (mirror)

www009: Constitution
https://wiki.piratenpartei.at/wiki/Satzung (mirror)

Election Bylaws
https://wiki.piratenpartei.at/wiki/Bundeswahlordnung (mirror)

www010: Kooperationsvereinbarung KPÖ/Piraten/Wandel, 20 January 2013

https://wiki.piratenpartei.at/w/images/5/5b/Kooperationsvereinbarung_KPOE_Pirat_Wandel.pdf (mirror)

www011: https://web.archive.org/web/20120227033121/
https://pirateparty.org.au/2011/11/18/national-congress-2011-results/

Article 10.1 Section 9 of the Constitution
https://pirateparty.org.au/constitution/#part-iii-10.1(9) (mirror)

www012: https://www.partito-pirata.it/statute/

www013: Article IV Section 3 of the Bylaws of the Pirate Party of the United States of America
https://wiki.uspirates.org/w/index.php?oldid=129907#Section_3:_Election (mirror)

Article V Section 3.a.ii.i of the Bylaws of the Pirate Party of Pennsylvania
https://web.archive.org/web/20240823020421/https://papirates.org/bylaws/

www014: https://log.piratar.is/modurfelag/log-pirata.html (mirror)

https://web.archive.org/web/20231211195901/
https://piratar.is/piratafraedarinn/schulze-adferdin/

https://piratar.is/primaries-rvk-2026 (mirror)

https://log.piratar.is/adildarfelog/svaedisfelog/log-pirata-i-kopavogi.html (mirror)

https://log.piratar.is/adildarfelog/svaedisfelog/log-pirata-a-sudurlandi.html (mirror)

https://log.piratar.is/adildarfelog/svaedisfelog/log-pirata-i-reykjavik.html (mirror)

https://log.piratar.is/adildarfelog/svaedisfelog/slitin/log-pirata-i-hafnarfirdi.html (mirror)

https://log.piratar.is/adildarfelog/svaedisfelog/slitin/log-pirata-a-vestfjordum.html (mirror)

https://log.piratar.is/adildarfelog/svaedisfelog/slitin/log-pirata-a-vesturlandi.html (mirror)

https://log.piratar.is/adildarfelog/kjordaemisfelog/log-pirata-i-nordvesturkjordaemi.html
(mirror)

https://log.piratar.is/adildarfelog/kjordaemisfelog/log-pirata-i-sudvesturkjordaemi.html
(mirror)

www015: https://web.archive.org/web/20160601183414/
http://www.northbynorthwestern.com/story/ajith-van-atta-win-asg-election/

https://web.archive.org/web/20140205223106/
https://asg.northwestern.edu/news/2013/04/
announcing-2013-asg-executive-elections-results

www016: https://www.h24notizie.com/2015/03/03/fondi-punto-sui-candidati-sindaco-certezze-novita-colpi-scena/ (mirror)

https://www.nextquotidiano.it/crescimanno-daniele-diaco/ (mirror)

https://campobasso5stelle.it/liberiamocampobasso/ (mirror)

https://campobasso5stelle.it/candidato-sindaco-movimento-5-stelle-campobasso/ (mirror)

https://www.osservatoreitalia.eu/vetralla-m5s-lorena-ciucci-e-il-candidato-sindaco/ (mirror)

https://www.termolionline.it/news/politica/596508/restitution-day-decidi-come-spendere-in-citta-le-indennita-del-movimento-5-stelle (mirror)

https://www.viterbonews24.it/news/il-movimento-5-stelle-cerca-candidati-credibili_57946.htm (mirror)

https://www.viterbonews24.it/news/lorena-ciucci-%C3%A8-il-candidato-sindaco-del-m5s_61112.htm (mirror)

https://www.molise5stelle.it/eventi/gazebo-informativo-sul-reddito-di-cittadinanza-24/ (mirror)

https://www.blitzquotidiano.it/politica-italiana/cose-il-metodo-schulze-liquid-feedback-grillo-1351235/ (mirror)

https://www.informamolise.com/prima-pagina/scelta-del-candidato-sindaco-nel-movimento-5-stelle-massima-trasparenza-e-partecipazione/ (mirror)

https://www.iltirreno.it/prato/cronaca/2013/11/06/news/comunarie-di-montemurlo-ecco-i-tre-candidati-a-sindaco-1.8064802 (mirror)

https://www.iltirreno.it/grosseto/cronaca/2015/12/23/news/il-sondaggio-alternativo-dice-della-negra-1.12670115 (mirror)

https://2017.gonews.it/2013/11/06/comunarie-parte-seconda-ecco-i-tre-candidati-del-m5s/ (mirror)

https://www.ilgiornaledelmolise.it/2014/02/27/comunali-campobasso-il-movimento-5-stelle-presenta-il-metodo-di-scelta-dei-proprio-candidati/ (mirror)

https://www.ilcentro.it/pescara/e-i-grillini-presentano-le-loro-primarie-1.261638 (mirror)

https://mynews.it/elezioni-campobasso-m5stelle-sceglie-con-voto-palese-il-candidato-ecco-come/ (mirror)

https://www.primonumero.it/2014/03/zero-esperienza-e-lontano-dai-partiti-i-grillini-trovano-il-candidato-perfetto/16084/ (mirror)

https://web.archive.org/web/20140221225515/http://www.5stellemontecompatri.it/wp-content/uploads/2013/11/REGOLAMENTO-M5S-Montecompatri1.doc

https://web.archive.org/web/20150701030529/https://www.meetup.com/MoVimento-5-Stelle-San-Cesareo/events/166108572/

https://web.archive.org/web/20150408072712/http://movimento5stellepescara.it/wp-content/uploads/2015/02/Regolamento-m5S-pescara_rev.03_del_30_01_-2015.pdf

www017: https://web.archive.org/web/20220309103049/
https://www.cde-ev.de/wir/satzung/mb_ws_2013-12-09.pdf

https://www.cde-ev.de/vereinsleben/mgv/how-to/#abstimmungen-und-wahlen
(mirror)

https://db.cde-ev.de/doc/Realm_Assembly_Voting-Details.html#id2 (mirror)

https://github.com/cde-ev/schulze-condorcet

www018: Article III.3.4 of the Statutory Rules

https://wiki.pirateparty.be/images/9/9f/Ppbe-statutes-fr_be.pdf (mirror)

https://wiki.pirateparty.be/images/e/e7/Ppbe-statutes-nl_be.pdf (mirror)

www019: https://www.stampmedia.be/artikel/
lijst-paars-voor-gemeenteraadsverkiezingen-antwerpen-bekendgemaakt (mirror)

www020: https://www.stura.uni-freiburg.de/wp-content/uploads/
2024/11/Stura-GO_Stand_06_06_2019.pdf (mirror)

https://www.stura.uni-freiburg.de/wp-content/uploads/
2024/11/How-to-StuRa_final_final.pdf (mirror)

https://web.archive.org/web/20250708150351/
https://www.stura.uni-freiburg.de/gremien/studierendenrat/abstimmungserklaerung

www021: https://web.archive.org/web/20160526162450/
https://noord-holland.piratenpartij.nl/711/

https://piratenpartij.nl/tweede-kamerverkiezingen-2017/
vacatures-kieslijst/stemprocedure/ (mirror)

www022: https://sillaparticipa.com/

https://web.archive.org/web/20170322161112/
http://www.sillaendemocracia.es/un-ano-profundizacion-democratica-silla/

https://web.archive.org/web/20160223122017/
http://www.silla.es/noticies/i/82830/1342/
22-26-de-febrer-la-primera-enquesta-popular-de-la-historia-de-silla

https://www.hortanoticias.com/al-voltant-de-2-000-participants-en-dos-dies-en-la-primera-enquesta-popular-de-silla-que-decidira-sobre-espectacles-taurins/ (mirror)

https://www.hortanoticias.com/silla-firma-un-hito-en-la-participacion-ciudadana-en-espana-al-usar-el-voto-preferencial-en-su-primera-consulta-popular/ (mirror)

https://sistemaencrisis.es/2016/05/06/silla-innova-en-participacion-ciudadana-en-espana-usando-el-voto-preferencial/ (mirror)

https://www.hortanoticias.com/els-veins-de-silla-opinen-a-la-segona-enquesta-popular-sobre-com-ajudar-a-refugiats-o-la-frequencia-de-canvi-de-banda-destacionament/ (mirror)

https://www.hortanoticias.com/el-lunes-se-abren-las-votaciones-de-la-segunda-encuesta-popular-de-silla/ (mirror)

https://www.hortanoticias.com/mas-de-600-personas-participan-en-la-segunda-encuesta-popular-de-silla/ (mirror)

https://www.hortanoticias.com/silla-tir-i-arrossegament-enquesta/ (mirror)

https://www.hortanoticias.com/els-veins-de-silla-opinaran-a-una-enquesta-municipal-sobre-la-practica-del-tir-i-arrossegament/ (mirror)

www023: https://www.zeus.aegee.org/magazine/2017/09/24/the-schulze-method-agora-101/ (mirror)

www024: https://equalitybylot.files.wordpress.com/2017/06/organizativo-umac.pdf (mirror)

https://web.archive.org/web/20220709110633/
https://regiondemurcia.podemos.info/wp-content/uploads/
2017/09/2017.08.31.-Documento-organizativo-1.pdf

https://web.archive.org/web/20190328065847/
https://regiondemurcia.podemos.info/wp-content/uploads/
2018/05/11.05.2018.-REGLAMENTO-CCA.pdf

https://web.archive.org/web/20220709095907/
https://profundizaciondemocratica.org/sites/profundizaciondemocratica.org/
files/documents/profundizacion_democratica_comunidad.pdf

https://web.archive.org/web/20210227205020/
http://profundizaciondemocratica.org/sites/profundizaciondemocratica.org/
files/documents/profundizacion_democratica_aragon.pdf

https://archive.org/details/dende-abaxo-documento-organizativo

https://archive.org/details/organizativo-canarias-310517

https://archive.org/details/documento-organizativo-podemos-4.03

https://archive.org/details/2017.05.31.-mpmpd-organizativodef

https://archive.org/details/profu_ib_organizativo

www025: Section 9.4.7.3 of the Operating Procedures of the
Address Council of the Address Supporting Organization

https://web.archive.org/web/20230606191239/
https://aso.icann.org/documents/operational-documents/
operating-procedures-of-the-address-council-of-the-address-supporting-organization/
#A_9.4.7.3._Election_Counting

https://aso.icann.org/documents/operational-documents/
aso-ac-icann-board-selection-procedures/description-schulze-method/ (mirror)

www026: Article 9.4.5.h of the Charter

https://web.archive.org/web/20220419114023/
https://associatedstudents.minerva.community/
wp-content/uploads/2017/11/ASM-Charter-November-2017.pdf

https://github.com/neila/Minerva-ASM-elections

www027: https://archive.ph/20240820145337/
https://web.archive.org/web/20211113112113/
https://medium.com/volt-espa%C3%B1a/algunas-consideraciones-sobre-en-qu%C3%A9-grupo-estar%C3%A1-volt-europa-en-el-parlamento-europeo-b10b7431d947

https://www.reddit.com/r/europe/comments/bymmk5/comment/eqjdxp0/

https://www.reddit.com/r/de/comments/bymnlk/comment/eqk00je/

www028: Article 31.2 of the Bylaws
https://drive.google.com/file/d/1y4FJ7zgnTV2T4KQzd6m3vMkgyr2bVYxs/view
(mirror)

www029: https://github.com/knative/community/blob/
64bf6dcaa83b8df98114db63d7c30e9f755dfd15/
mechanics/SC.md#election-process (mirror)

www030: https://github.com/kubernetes/community/blob/
3d0b06028939524ad0d8c06ae460a1d56d759d0b/
committee-code-of-conduct/election.md#election-process (mirror)

www031: https://web.archive.org/web/20260529145614/
https://www.kumoricon.org/mascot-contest

www032: https://github.com/graphql/graphql-wg/blob/
1a3a2eb71bde2e586ee698739ce2339c1381ee78/
GraphQL-TSC.md#voting-process (mirror)

www033: https://github.com/dapr/community/blob/
67ad4233b0ccbca491d092c5d4e3b9efa9029019/
steering-and-technical-committee-charter.md#voting-procedure (mirror)

www034: https://helm.sh/de/community/governance/governance/#helm-org-maintainers
(mirror)

https://hackmd.io/@mattfarina/B1BksVYW7?type=view#Helm-Org-Maintainers
(mirror)

www035: Article VI Section 3(8) of the Bylaws
https://wiki.hacksburg.org/index.php?oldid=1229#Section_3:_Election_Procedure
(mirror)

www036: Article 6 Section 2 of the Constitution

https://github.com/Homebrew/brew/blob/
683bb6d73acb409d7ce4d04e54566443fecf2b50/
docs/Homebrew-Governance.md#6-project-leader (mirror)

https://docs.brew.sh/Homebrew-Governance#6-project-leader (mirror)

www037: https://docs.starlingx.io/election/#election-system (mirror)

www038: https://airshipit.readthedocs.io/_/downloads/airship-election/en/latest/pdf/ (mirror)

www039: https://github.com/brigadecore/community/blob/
535af0ff9f1e6f8658566acf139884eab0737ddc/
governance.md#brigade-org-maintainers (mirror)

www040: Section 7.4 of the Bylaws
https://www.netbsd.org/foundation/bylaws.html (mirror)

www041: Section 7.2.3 of the Charter
https://fosdem.org/2026/about/charter/ (mirror)

www042: https://petsc.org/release/community/governance/#petsc-council (mirror)

www043: https://bepo.fr/wiki/index.php?oldid=28656#Assembl.C3.A9e_G.C3.A9n.C3.A9rale_.28AG.29 (mirror)

www044: https://evriichi.fr/statuts (mirror)

www045: https://wiki.squeak.org/squeak/6162 (mirror)

www046: https://github.com/tektoncd/community/blob/4a4b7c0d4f909a895cea7e393aa1ff83c6afd9f8/governance.md#election-process (mirror)

www047: https://lists.sugarlabs.org/archives/list/iaep@lists.sugarlabs.org/message/7QJVU66EXFJMSVI72TWVY4SMNQC2YCCP/ (mirror)

www048: https://web.archive.org/web/20120607064836/http://www.rllmukforum.com/index.php?/topic/260622-committee-elections-2012/

www049: https://www.mail-archive.com/ffmpeg-devel@ffmpeg.org/msg104455.html (mirror)

www050: https://foopgp.org/about/rules-of-procedures/ (mirror)

https://foopgp.org/solutions/theme-vote/ (mirror)

www051: OpenSwitch Project Charter
https://web.archive.org/web/20211111192934/https://lists.linuxfoundation.org/pipermail/ops-partners/2016-May/000030.html

www052: https://www.openembedded.org/w/index.php?title=Online_Voting_Policy&oldid=10518#Multiple_options (mirror)

www053: https://github.com/ocaml/RFCs/blob/7bde352116ad23eb9d0d0f98d7a12c7ee3b26217/Committee.md#term-limits-and-committee-selection (mirror)

www054: Section 9.6 of the Constitution
https://wiki.datashield.org/constitution_v20260912.pdf (mirror)

www055: https://dcentproject.eu/wp-content/uploads/2014/11/D4.3-final.pdf (mirror)

www056: https://web.archive.org/web/20120131134802/http://testpacpleaseignore.org/about/official-test-pac-bylaws/

www057: https://web.archive.org/web/20200809234302/https://ithacagenerator.org/alternate-membership-terms/bylaws/

www058: Article XI Section 2 of the Bylaws

https://web.archive.org/web/20120823170044/http://fairtradenorthwest.org/FTNW%20Bylaws.pdf

www059: §14 of the Statutes
https://wiki.gbghackerspace.se/general/forfattning/ (mirror)

www060: https://olten.jetzt/oltner-innen-wunschen-sich-eine-schwimmstadt/ (mirror)

www061: http://wiki.freegeek.org/images/7/7a/Voters_guide.pdf (mirror)

http://wiki.freegeek.org/images/5/53/Town-Hall-Meeting.pdf (mirror)

www062: https://boardgamegeek.com/thread/3696572/article/47578664

www063: https://2025.vidyagaemawards.com/results/pairwise (mirror)

www064: https://www.fsfla.org/pipermail/anti-drm/2006-September/000739.html (mirror)

www065: https://github.com/onnx/onnx/blob/850a81b0b77786bf99ea90580242b084f86a6235/community/sc-election-guidelines.md#voting-mechanics-and-algorithm (mirror)

www066: https://discuss.python.org/t/should-we-consider-ranked-choice-voting-for-sc-elections/61880/5 (mirror)

www067: Article 7(e)(iii)(2) of the Charter
https://www.cloudfoundry.org/wp-content/uploads/CFF-Charter-September-2023.pdf (mirror)

https://github.com/cloudfoundry/community/blob/2a5b2a2fd102a8d98720dab2f9434d1257c07f56/governing-board/charter.md#7technical-oversight-committee-toc (mirror)

https://www.cloudfoundry.org/blog/results-of-the-toc-elections-2022 (mirror)

www068: CNCF Code of Conduct Committee Charter
https://github.com/cncf/foundation/blob/35cbcc6886915e8bf2b2a662f516ddd578787b74/code-of-conduct/coc-committee-charter.md#nomination-of-community-member-representatives (mirror)

CNCF Code of Conduct Elections for 2023
https://www.cncf.io/blog/2023/09/08/cncf-code-of-conduct-elections-for-2023/ (mirror)

https://www.cncf.io/wp-content/uploads/2024/07/CNCF-Governing-Board-Approved-Resolution-July-2023.pdf (mirror)

www069: Section 3.2 of the Bylaws
https://computational-geometry.org/documents/Bylaws%20of%20the%20CG%20Society%2018.12.2024.pdf (mirror)

www070: Section 8.5.1.3 of the Bylaws

https://web.archive.org/web/20260908132941/https://dcc.ligo-wa.caltech.edu/public/0162/M1900139/012/LSC_Policies_and_Procedures_V12_September_04_2026.pdf

www071: https://github.com/finopsfoundation/foundation/blob/636534374a41146a32cf481889bceb8a7c132359/01-charter.md#election-process (mirror)

www072: Article IV Section 3 of the Bylaws

https://docs.google.com/document/d/
1FwzWJpL8kOr1wWvdA3XLD5NxxDlm1uWGFlPSxqAftZo/view
(mirror)

https://web.archive.org/web/20240125013632/
https://www.boisedsa.org/statements/yjmtibx8okazm7i7042c10tuld0gm5

www073: Article 21 of the Statutory Rules
https://www.partidopiratapt.eu/o-movimento/estatutos/ (mirror)

www074: Article 10.1.9 of the Constitution

https://web.archive.org/web/20220708221706/
https://www.aec.gov.au/parties_and_representatives/party_registration/
applications/files/2018/australian-better-familes-constitution.pdf

www075: https://web.archive.org/web/20230309022748/
http://www.burgerlijst.be/assets/docs/Burgerlijst_whitepaper_v1.3.pdf

www076: §6(9) of the bylaws
https://dboe.at/wp-content/uploads/2020/09/Satzung-dboe-Stand-1.6.2020.pdf
(mirror)

www077: https://www.demokratiezentrum.org/wp-content/uploads/2021/04/
WP1_DirekteDemokratie_final_download.pdf (mirror)

www078: Julie Flaherty, “The Voting Puzzle”, *TuftsNow*, 2 May 2016
https://now.tufts.edu/articles/voting-puzzle (mirror)

https://web.archive.org/web/20181116052628/
https://ase.tufts.edu/europeanCenter/programs/talloires/syllabi/MATH19.pdf

www079: Section 7.2.2 of the Bylaws

https://development-ys-ypa-yale-edu.pantheonsite.io/
about/mission-statement-and-bylaws (mirror)

www080: https://web.archive.org/web/20220708223943/
https://lists.ufl.edu/cgi-bin/wa?A2=ind16&L=IUSSI-NAS-L&P=R78019

www081: Article 1.5.6 of the Policies

https://web.archive.org/web/20210103130846/
https://mathsoc.uwaterloo.ca/wp-content/uploads/2020/06/March-9-2020.pdf

www082: Article 3.5 of the Constitution
https://www.math.uwo.ca/macaw/misc/constitution.pdf (mirror)

www083: Article IV of the Constitution
https://www.um-fencing.com/constitution.pdf (mirror)

www084: Article 5.3.1 of the Constitution

https://www.ocf.berkeley.edu/docs/docs/operatingrules/constitution/#amended-2020-09-09-5-3-1 (mirror)

https://github.com/ocf/ocfweb/blob/d02c7b9ccc78c9a3465d520ca537148aaee8e89d/ocfweb/docs/docs/docs/operatingrules/constitution.md#531-election-of-officers-amended-2020-09-09-5-3-1 (mirror)

www085: Article 40 Section 1 and Article 62 Section 2
https://drive.google.com/file/d/1HjTXSJHu9wtrzHuE7OnA2v17sW_Ywpfd/view (mirror)

Article 50 Section 1
https://vtk.be/page/file/380585475db3a1f6028ead784be8609f5d0b9b7f/ (mirror)

www086: §6(6) of the Statutory Rules
https://www.fachschaft.informatik.uni-kl.de/assets/documents/go.pdf (mirror)

www087: https://www.gazzettamatin.com/gazzettamatin/2021/06/09/impresa-simulata-bit-company-non-solo-banca-del-tempo/ (mirror)

www088: https://www.dg.ens.fr/media/archives/depouillement.pdf (mirror)

https://www.dg.ens.fr/documents/148/note_explicative.pdf (mirror)

www089: page 72–73
https://www.yumpu.com/en/document/read/6769890/kelowna-and-to-ubcs-okanagan-campus-cumc-canadian- (mirror)

page 50
https://www.yumpu.com/en/document/read/44261916/20e-anniversaire-20th-anniversary-cumc (mirror)

www090: https://ioinformatics.org/files/regulations24.pdf (mirror)

https://ioinformatics.org/files/regulations24-web.pdf (mirror)

https://www.linkedin.com/posts/sk-sscs_michal-fori%C5%A1ek-a-member-of-the-slovak-society-activity-7362135358957899776-KTfz (mirror)

www091: https://sfs.se/wp-content/uploads/2019/04/fum-handboken_2019.pdf (mirror)

Article 6.3 of the Rules of Order
https://sfs.se/wp-content/uploads/2019/01/arbetsordning_sfsfum_2019.pdf (mirror)

www092: §4.4 of the Statutory Rules
https://styrdokument.datasektionen.se/reglemente#4-4-urnval (mirror)

www093: Article 2.3.5 of the Bylaws
https://storage.googleapis.com/medieteknik-static/documents/en-GB/2025/06/Stadgar%20%28reviderat%20av%20SM%232%2024_25%29%20.pdf (mirror)

www094: https://web.archive.org/web/20251204143607/
https://f.kth.se/en/about-the-chapter/the-election-committee/

§4.4.1 of the Bylaws
https://drive.google.com/file/d/1fB4_xz2JReqvYFqEtp_-WwAsqDfNOg5_/view
(mirror)

www095: §2.1 and §2.2 of the Bylaws
https://drive.google.com/drive/folders/1QroQ4mXSu0-HkPVnO1-6fktnefzhK9FH
(mirror)

www096: §3.8.1 of the Statutory Rules
https://flygsektionen.se/wp-content/uploads/2025/05/
Flygsektionens_Reglemente_SM1_2025.pdf (mirror)

§4.10.3 and 4.10.5 of the Meeting Rules
https://flygsektionen.se/wp-content/uploads/2025/01/
Flygsektionens_Motespraxis-2_2024_sm4.pdf (mirror)

www097: https://www.svenskaturistforeningen.se/app/uploads/2020/03/
7-faststallande-av-motesordning-for-riksstamman.pdf (mirror)

https://www.svenskaturistforeningen.se/app/uploads/2020/03/
handlingar-riksstamma-2020-1.pdf (mirror)

https://www.svenskaturistforeningen.se/app/uploads/2022/03/
6-motesordning-for-riksstamman.pdf (mirror)

https://www.svenskaturistforeningen.se/app/uploads/2022/03/
utskriftsversion-handlingar_ny.pdf (mirror)

www098: https://lommabk.com/protokoll/VoteIT.pdf (mirror)

www099: https://web.archive.org/web/20240602084001/
https://docplayer.se/208071148-Innehallsforteckning-praktisk-information-skapa-anvandarkonto-rattigheter-i-motet-dagordning-hitta-tidsschema-forslag-och-diskussion.html

www100: https://web.archive.org/web/20220709053918/
https://vision.se/globalassets/om-vision/
forbundsmoten/2019/arbetsordning_digitalt_arsmote.pdf

www101: https://web.archive.org/web/20220812224750/
https://www.akavia.se/rad-och-stod/
fortroendevald/utbildning-i-motesverktyget-voteit/

www102: https://www.vardforbundet.se/SysSiteAssets/lokala-avdelningar/
_gemensamma-filer/kort-intro-till-voteit.pdf (mirror)

www103: https://web.archive.org/web/20240602084326/
https://www.seko.se/siteassets/voteit-handlingar/
sekovoteit_deltagarmanual_2021.pdf

https://web.archive.org/web/20240602083930/
https://www.seko.se/siteassets/voteit-handlingar/
klubbrr/seko-voteit-deltagarmanual.pdf

www104: https://gnesta.scout.se/scoutkar/demokratijamboree-2020/

https://web.cdn.scouterna.net/uploads/sites/79/2020/04/handlingar-distriktsstmma-vt2020.pdf (mirror)

https://web.cdn.scouterna.net/uploads/sites/698/2020/09/handlingar-demokratijamboree-2020.pdf (mirror)

https://web.cdn.scouterna.net/uploads/sites/74/2021/03/forhandlingsordning-for-sodra-skane-scoutdistrikts-varstamma-2021.pdf (mirror)

https://web.cdn.scouterna.net/uploads/sites/79/2021/03/handlingar-distriktsstmma-ht2020_full.pdf (mirror)

https://web.cdn.scouterna.net/uploads/sites/79/2021/03/handlingar_vdst_2021_v1_0.pdf (mirror)

https://web.cdn.scouterna.net/uploads/sites/79/2021/04/handlingar_vt21_komplett.pdf (mirror)

https://web.cdn.scouterna.net/uploads/sites/82/2021/07/handlingar-till-scouternas-forvaltningsmote-2021.pdf (mirror)

https://web.cdn.scouterna.net/uploads/sites/79/2021/09/bilaga-5-voteit-instruktion.pdf (mirror)

https://web.cdn.scouterna.net/uploads/sites/79/2022/03/handlingar-ht2021.pdf (mirror)

https://media.scoutcontent.se/uploads/2022/08/Stammoarkiv-2020-comp.pdf (mirror)

https://media.scoutcontent.se/uploads/2022/09/Forhandlingsordning-DJ22.pdf (mirror)

https://media.scoutcontent.se/uploads/2023/09/Kallelse-till-Scouternas-forvaltningsmote-2023.pdf (mirror)

https://scoutkansliet.se/wp-content/uploads/2023/10/Handlingar-Stockholms-distriktsstamma-2023-10.pdf (mirror)

https://media.scoutcontent.se/uploads/2024/09/6-Forhandlingsordning-2024.pdf (mirror)

https://media.scoutcontent.se/uploads/2024/09/Handlingar-till-Scouternas-stamma-2024-2.pdf (mirror)

www105: https://funktionsrattstockholmslan.se/wp-content/uploads/2020/09/2-Funktionsratt-Stockholms-lans-manual-for-VoteIT.pdf (mirror)

www106: https://equmeniakyrkan.se/wp-content/uploads/2022/01/protokoll-kyrkokonferensen-2021-med-underskrifter.pdf (mirror)

www107: https://infobank.sverok.se/wp-content/uploads/2019/12/Protokoll-ordinarie-Riksm%C3%B6te-2019.pdf (mirror)

https://infobank.sverok.se/wp-content/uploads/2021/01/Protokoll-forbundsstyrelsemote-2020-5-3-4-oktober-receipt-gikgx.pdf (mirror)

www108: https://www.djurensratt.se/vanliga-fragor/vad-ar-voteit

www109: https://www.iogt.se/wp-content/uploads/2021/03/remiss-arbetsordning-20210322.pdf (mirror)

www110: https://library.fes.de/pdf-files/id/ipa/15887.pdf (mirror)

www111: https://socialdemokraternaistockholm.se/wp-content/uploads/2021/09/Utbildningsdokument-VoteIt.pdf

www112: https://assets.ctfassets.net/vkygthqfrquy/2aPlm2BP1g95TzLUh7A5tG/6a005e39473d86127db6f57800e98b3b/Kongresshandlingar-2021.pdf (mirror)

www113: https://web.archive.org/web/20230126212419/https://www.sv.se/globalassets/nationella-sv_se/forbundsstamman-2021/filer/svs-forbundsstamma-anvandarguide-voteit.pdf

www114: https://aik.se/wp-content/uploads/2017/10/Information-VoteIT-extra-medlemsm%c3%b6te-171106.pdf (mirror)

www115: https://www.svenskakyrkan.se/filer/698270ed-e543-4c8a-93b5-822904785ad3.pdf (mirror)

www116: https://norden.se/wp-content/uploads/2020/09/Hur-VoteIT-fungerar.pdf (mirror)

www117: https://slf.se/sylf/app/uploads/2021/03/voteit.pdf (mirror)

https://slf.se/sylf/app/uploads/2024/05/protokoll-sylf-fullmaktige-2024.pdf (mirror)

https://slf.se/sylf/app/uploads/2025/04/arbetsordning-sylf-fum-2025.pdf (mirror)

https://slf.se/sylf/app/uploads/2026/04/arbetsordning-fullmaktige-2026.pdf (mirror)

www118: https://www.unf.se/wp-content/uploads/2021/03/Handlingar-till-m%C3%B6te-med-styrelsen-f%C3%B6r-Ungdomens-nykterhetsf%C3%B6rbund-mars-2021-uppdaterad.pdf (mirror)

https://www.unf.se/wp-content/uploads/2021/05/Utbildningsdokument-VoteIt.pdf (mirror)

https://www.unf.se/wp-content/uploads/2021/09/FS04-20210815-offentlig-version.pdf (mirror)

https://www.unf.se/wp-content/uploads/2022/01/FS07-211014.pdf (mirror)

https://www.unf.se/wp-content/uploads/2022/01/FS08-211102.pdf (mirror)

https://www.unf.se/wp-content/uploads/2022/03/FS10-220129-220130-offentlig-version.pdf (mirror)

https://www.unf.se/wp-content/uploads/2023/04/Handlingar-till-m%C3%B6te-med-styrelsen-f%C3%B6r-Ungdomens-Nykterhetsf%C3%B6rbund-27-april-1-1.pdf (mirror)

https://www.unf.se/wp-content/uploads/2023/06/Kongresshandlingar_UNFs-30e-kongress-2023-Uppsala.pdf (mirror)

https://www.unf.se/wp-content/uploads/2025/08/Handlingar-FS16.pdf (mirror)

https://www.unf.se/wp-content/uploads/2025/09/Arbetsordning-Kongress-2025-1.pdf (mirror)

www119: https://www.rfsl.se/wp-content/uploads/2025/01/kongress-protokoll2023-slutligt-receipt-nm5w7.pdf (mirror)

www120: https://www.biblioteksforeningen.se/nyheter/bekanta-dig-med-voteit/ (mirror)

www121: https://klimatriksdagen.se/wp-content/uploads/2018/06/VB-2017-20180531-KR.pdf (mirror)

www122: https://www.rfsu.se/globalassets/pdf/kongressen/arbetsordning-kongress-2021-forslag.pdf (mirror)

www123: https://www.4h.se/kloverbladet/kallelse-till-riksforbundsstamma-2021/ (mirror)

www124: https://www.centerpartiet.se/download/18.215500fb1781f2997271e07/1616599568094/Kort%20guide%20till%20VoteIT%20inf%C3%B6r%20distriktsst%C3%A4mman%202021.pdf (mirror)

www125: https://ieeevis.org/year/2021/info/elections (mirror)

https://ieeevis.org/year/2022/info/elections (mirror)

https://ieeevis.org/year/2023/info/elections (mirror)

https://ieeevis.org/year/2024/info/elections (mirror)

https://ieeevis.org/year/2025/info/elections (mirror)

https://ieeevis.org/year/2026/info/elections (mirror)

www126: Ka-Ping Yee, “Kingman adopts Condorcet voting”, 14 April 2005
https://zestyping.livejournal.com/111588.html (mirror)

www127: https://archive.org/details/thconst-720

www128: Section 5 of the Hillegass-Parker House Bylaws

https://sites.google.com/a/bsc.coop/hip-house/
bylaws-and-policies/bylaws-bylaw-appendix

https://sites.google.com/a/bsc.coop/hip-house/
bylaws-and-policies/parliamentary-procedure

https://www.hip.coop/gov:bylaws (mirror)

https://www.hip.coop/gov:parliament (mirror)

www129: Melina Lang, “Der Traum vom selbstorganisierten Wohnen”, *Stimme:ab*, Frankreich-Zentrum der Albert-Ludwigs-Universität Freiburg, pages 4–5, April 2018
https://www.fz.uni-freiburg.de/contenu/dateien-studium/dateien-dfj/stimme_ab (mirror)

www130: https://ru.wikipedia.org/w/index.php?oldid=34284315 (mirror)

https://web.archive.org/web/20150222225428/http://y.kalan.cc/schulze

www131: https://fa.wikipedia.org/w/index.php?oldid=33798284 (mirror)

www132: https://diff.wikimedia.org/2022/12/06/vote-for-the-sound-of-all-human-knowledge/ (mirror)

https://commons.wikimedia.org/wiki/Commons:Sound_Logo_Vote/statistics (mirror)

www133: https://de.wikipedia.org/wiki/Wikipedia:Meinungsbilder/
Genealogische_Zeichen (mirror)

www134: https://www.heise.de/hintergrund/Synaxon-bleibt-experimentierfreudig-1721030.html (mirror)

www135: https://github.com/PartiPirate/congressus/blob/05944a13ca0da6c3f25374243f8821b78faa6629/application/assets/js/perpage/meeting.js#L1119 (mirror)

https://archive.org/details/manualzilla-id-6448421

https://archive.org/details/manualzilla-id-6448422

https://web.archive.org/web/20130323015009/http://partipirate.org/ri.pdf

www136: https://ec.europa.eu/futurium/sites/futurium/files/wegovnow_draft_d5-4_oprp.pdf (mirror)

www137: https://web.archive.org/web/20220709025641/https://www.projectco3.eu/wp-content/uploads/2021/01/CO3_D2.3_with_annexes_compressed.pdf

www138: https://egov.formez.it/sites/all/files/intervista_a_marieva_favoino.pdf (mirror)

https://atti.desio.org/bundles/Delibere%20di%20Giunta/2015/64_URP%20-%20COMUNICAZIONE/0__64.1-Bilancio%20Partecipativo%20Desio%20-%20regole%20e%20funzionamento.pdf (mirror)

https://atti.desio.org/bundles/Delibere%20di%20Giunta/2015/64_URP%20-%20COMUNICAZIONE/0__64.2-Bilancio%20Partecipativo%20JUNIOR%20Desio%20-%20regole%20e%20funzionamento.pdf (mirror)

www139: https://x.com/habermolt/status/2034639377427558425

https://www.reddit.com/r/EffectiveAltruism/comments/1ryerfp/we_built_an_opensource_ai_deliberation_platform/ (mirror)

https://www.linkedin.com/posts/oscar-duys-448792192_habermolt-a-deliberation-platform-for-ai-activity-7440406273390678016-OWOc (mirror)

https://www.habermolt.com/blog/what-is-habermolt (mirror)

www140: https://web.archive.org/web/20211015033650/http://www.linux-france.org/~dmentre/demexp/latest-src/demexp-book-0.8.2.pdf

www141: https://www.dokuwiki.org/plugin:schulzevote (mirror)

www142: https://www.forcerank.it/blog/counting-votes-is-hard (mirror)

www143: https://help.hotcrp.com/help/ranking (mirror)

www144: page B-2

https://web.archive.org/web/20070809121412/http://www.thunderbay.ca/docs/business/1012.pdf

www145: https://wiki.piratenpartei.de/BE:Parteitag/2011.1/Rohdaten/Auswertung (mirror)

During the *Great National Debate* (“Grand Débat National”) in France 2019, there had been 6 submissions in favour of Schulze voting (link, link) (#1/3395, #1/37876, #1/48197, #1/66277, #1/72501, #1/118243). During the debate on *Abstention and Electoral Participation* (“Abstention et Participation Electorale”) in France 2021, there had been 6 submissions in favour of Schulze voting (link, link).

Tutorials:

Bray: part1 (mirror), part2 (mirror), part3, part4, part5, part6, part7

Duchin: part1 (mirror), part2 (mirror), part3 (mirror), part4 (mirror), part5 (mirror), part6 (mirror), part7 (mirror), part8 (mirror), part9 (mirror), part10 (mirror), part11 (mirror), part12 (mirror), part13, part14

Eppstein: part1, part2

MetricGate: video

Nebel: part1 (mirror), part2 (mirror), part3 (mirror), part4 (mirror), part5 (mirror), part6 (mirror), part7 (mirror), part8 (mirror), part9 (mirror), part10 (mirror), part11 (mirror), part12 (mirror), part13 (mirror), part14 (mirror), part15 (mirror), part16 (mirror), part17 (mirror), part18 (mirror)

Pritchard: part1 (mirror), part2 (mirror)

Procaccia: part1 (mirror), part2 (mirror), part3 (mirror)

Seuken: part1 (mirror), part2 (mirror)

Singh: part1

Vaish: part1 (mirror), part2 (mirror), part3 (mirror), part4 (mirror), part5 (mirror), part6, part7, video1 (mirror: 01:02:00 – 01:19:05 hh:mm:ss), video2 (mirror: 00:51:38 – 01:07:36 hh:mm:ss), video3 (mirror: 00:57:09 – 01:17:45 hh:mm:ss)

California State University, Northridge: part1 (mirror), part2 (mirror)

Concours Centrale-Supélec: part1, part2, part3, part4, part5, part6, part7, part8, part9, part10

University of Stuttgart: part1 (mirror)

# Appendix: Proposed Statutory Rules for the Schulze Single-Winner Election Method with Margins

## Article 1
## [Preferential Ballot]

Each ballot shall contain a complete list of all qualified candidates. Furthermore, each voter may write in {*number of write-in options*} additional candidates. Each voter ranks these candidates in order of preference. The individual voter may give the same preference to more than one candidate, he may keep candidates unranked, and he may skip numbers. When a given voter does not rank all candidates, then it is presumed that this voter strictly prefers all ranked candidates to all not ranked candidates and that this voter is indifferent between all not ranked candidates.

## Article 2
## [Schulze Method]

§1: Suppose $N[x,y]$ with $x \neq y$ is the number of valid ballots on which candidate $x$ is preferred to candidate $y$. For all pairs of candidates $x$ and $y$ with $x \neq y$, $N[x,y]$ shall be calculated.

§2: A “path from candidate $x$ to candidate $y$” with $x \neq y$ is a sequence of candidates $c(1),\ldots,c(n)$ that starts with $x = c(1)$ and ends with $y = c(n)$ and that visits the same candidate not more than once. That means: If $i \neq j$, then $c(i) \neq c(j)$.

§3: The “strength” of the path $c(1),\ldots,c(n)$ is the minimum of

$$N[c(i),c(i+1)] - N[c(i+1),c(i)] \text{ for all } i = 1,\ldots,(n-1).$$

§4: $P[a,b]$ with $a \neq b$ is the maximum value such that there is a path from candidate $a$ to candidate $b$ with this strength.

§5: For all pairs of candidates $d$ and $e$ with $d \neq e$, $P[d,e]$ shall be calculated. Candidate $f$ is a “potential winner” if and only if $P[f,g] \geq P[g,f]$ for every other candidate $g$.

## Article 3
## [Tie-Breaking]

§1: When there is only one potential winner, then he is the final winner.

§2: When there is more than one potential winner, then we pick a random ballot. All those potential winners *b* for whom there is another potential winner *a* such that candidate *a* is preferred to candidate *b* on this randomly chosen ballot lose their status of being a potential winner.

§3: We continue picking ballots randomly from those that have not yet been picked and use them to reduce the set of potential winners, as described in §2.

§4: When all ballots have been picked and there is still more than one potential winner, then the final winner is chosen randomly from these potential winners.

## Article 4
## [Example]

§1: There are 4 candidates and 21 voters.

8 voters prefer *a* to *c* to *d* to *b*.
2 voters prefer *b* to *a* to *d* to *c*.
4 voters prefer *c* to *d* to *b* to *a*.
4 voters prefer *d* to *b* to *a* to *c*.
3 voters prefer *d* to *c* to *b* to *a*.

§2: The pairwise matrix *N* looks as follows:

| | *N*[*,*a*] | *N*[*,*b*] | *N*[*,*c*] | *N*[*,*d*] |
|---|---|---|---|---|
| *N*[*a*,*] | --- | 8 | 14 | 10 |
| *N*[*b*,*] | 13 | --- | 6 | 2 |
| *N*[*c*,*] | 7 | 15 | --- | 12 |
| *N*[*d*,*] | 11 | 19 | 9 | --- |

§3: The pairwise matrix can also be written as a graph. The strength of the link from candidate $i$ to candidate $j$ is $N[i,j] - N[j,i]$.

§4: The corresponding graph looks as follows:

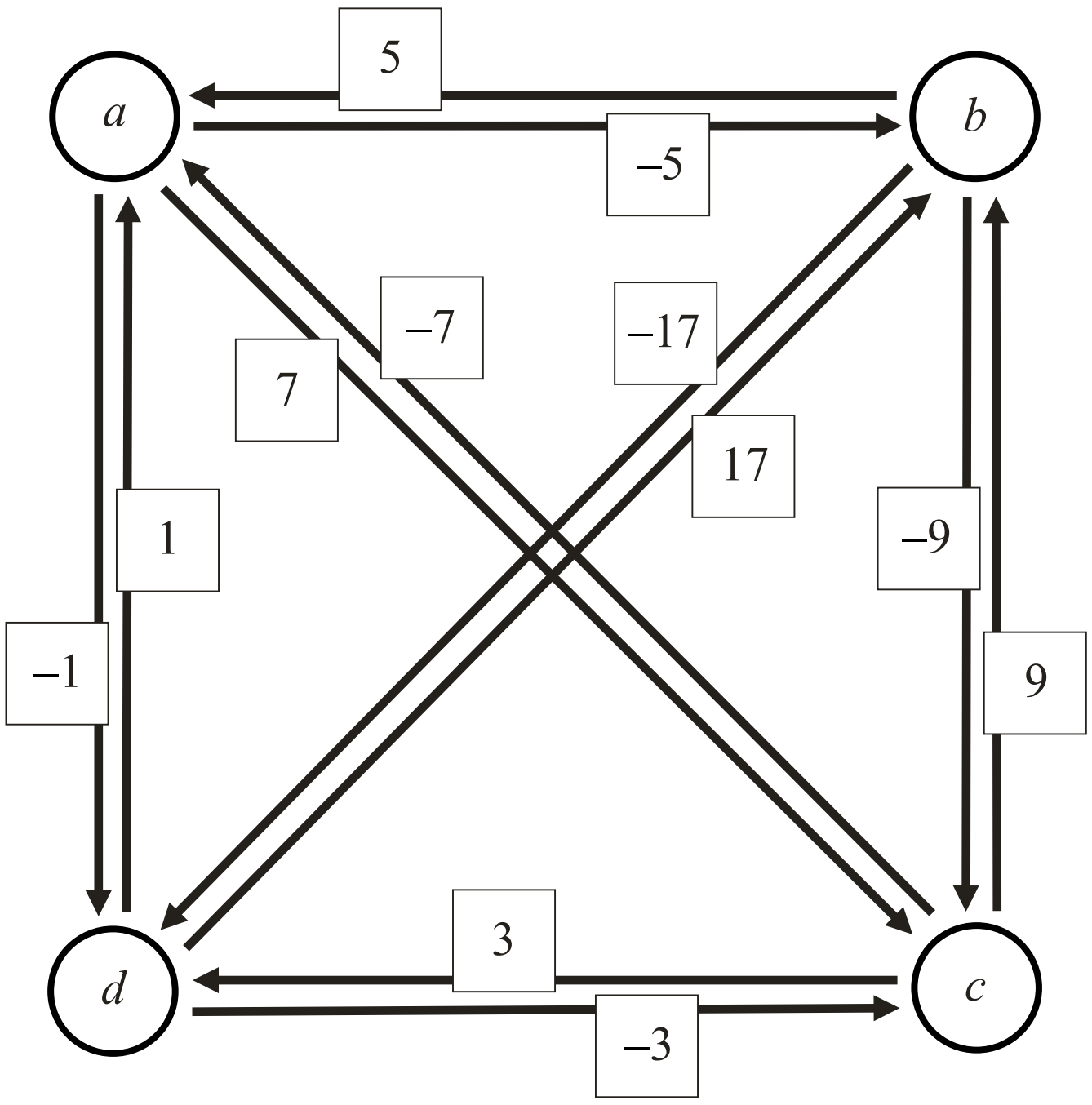

§5: The strongest path ...
... from candidate $a$ to candidate $b$ is $a \xrightarrow{\textcircled{7}} c \xrightarrow{9} b$ with a strength of $P[a,b] = 7$.
... from candidate $a$ to candidate $c$ is $a \xrightarrow{\textcircled{7}} c$ with a strength of $P[a,c] = 7$.
... from candidate $a$ to candidate $d$ is $a \xrightarrow{7} c \xrightarrow{\textcircled{3}} d$ with a strength of $P[a,d] = 3$.
... from candidate $b$ to candidate $a$ is $b \xrightarrow{\textcircled{5}} a$ with a strength of $P[b,a] = 5$.
... from candidate $b$ to candidate $c$ is $b \xrightarrow{\textcircled{5}} a \xrightarrow{7} c$ with a strength of $P[b,c] = 5$.
... from candidate $b$ to candidate $d$ is $b \xrightarrow{5} a \xrightarrow{7} c \xrightarrow{\textcircled{3}} d$ with a strength of $P[b,d] = 3$.
... from candidate $c$ to candidate $a$ is $c \xrightarrow{9} b \xrightarrow{\textcircled{5}} a$ with a strength of $P[c,a] = 5$.
... from candidate $c$ to candidate $b$ is $c \xrightarrow{\textcircled{9}} b$ with a strength of $P[c,b] = 9$.
... from candidate $c$ to candidate $d$ is $c \xrightarrow{\textcircled{3}} d$ with a strength of $P[c,d] = 3$.
... from candidate $d$ to candidate $a$ is $d \xrightarrow{17} b \xrightarrow{\textcircled{5}} a$ with a strength of $P[d,a] = 5$.
... from candidate $d$ to candidate $b$ is $d \xrightarrow{\textcircled{17}} b$ with a strength of $P[d,b] = 17$.
... from candidate $d$ to candidate $c$ is $d \xrightarrow{17} b \xrightarrow{\textcircled{5}} a \xrightarrow{7} c$ with a strength of $P[d,c] = 5$.

§6: The unique potential winner is candidate $d$ because, for every other candidate $x$, we have $P[d,x] \geq P[x,d]$.